\pdfoutput=1
\documentclass[fleqn,reqno,a4paper,parskip=half]{scrbook}

\usepackage{charter} %Charter fuer englische Texte
\usepackage[utf8]{inputenc}
\usepackage{pmboxdraw}

\input{binhex}
\makeatletter
\def\uc@dclc#1#2#3{%
  \ifnum\pdfstrcmp{#2}{mathletters}=\z@
    \begingroup\edef\x{\endgroup
      \noexpand\DeclareUnicodeCharacter{\hex{#1}}}\x{#3}%
  \fi
}
\input{uni-3.def}
\def\uc@dclc#1#2#3{%
  \ifnum\pdfstrcmp{#2}{default}=\z@
    \begingroup\edef\x{\endgroup
      \noexpand\DeclareUnicodeCharacter{\hex{#1}}}\x{#3}%
  \fi
}
\input{uni-34.def}
\makeatother

\usepackage[tbtags,sumlimits,intlimits,namelimits]{amsmath}

\usepackage{amsfonts}
\usepackage{amssymb}
\usepackage{bbm}
\usepackage{ulem}
\usepackage{tikz}
\usepackage{pgf}
\usepackage{ifpdf}
\usepackage{color}
\usepackage{esint}
\usepackage{framed}
\usepackage{wasysym}  % \ocircle
\usepackage{makecell}  % \makecell{...\\...} to break in table cell
\usepackage{textcomp}
\usepackage{fancyvrb}  % improved verbatim 

\usepackage[top=2.3cm, bottom=3.45cm, left=2.3cm, right=2.3cm]{geometry}
\usepackage{chngcntr}
\counterwithin*{section}{part}
\usepackage{graphicx}       % Bilder in Tabellen
\usepackage{float}          % eigene Float-Umgebungen, H-Option, um Bilder an der aktuellen Stelle anzuzeigen
\usepackage{caption}
\usepackage{subcaption,array}

\restylefloat{figure}       % Bilder an der Stelle, wo sie eingebunden werden
\usepackage{multirow}
\usepackage{listings}       % Darstellung von Source-Code
\usepackage{framed}         % Rahmen um Text
\usepackage{arydshln}       % gestrichelte Linie in Tabelle mit \hdashline
\usepackage{longtable}     % Tabellen mit automatischen Seitenumbruch
\usepackage{layouts}        % textwidth in cm: \printinunitsof{cm}\prntlen{\textwidth}

\usepackage{url}          % Links
\definecolor{darkblue}{rgb}{0,0,.5}
\definecolor{black}{rgb}{0,0,0}

\usepackage[framemethod=TikZ]{mdframed}       % Rahmen um Text und Gleichungen

\usepackage{dirtytalk}          % \say{...} erzeugt (deutsche) Anführungszeichen

\usepackage{tipa}
\usepackage{transparent}    % needed for inkscape generated pdf_tex files
\usepackage{multicol}       % multiple columns
\usepackage{moreverb}       % verbatimwrite
\usepackage{verbatimbox}    % \begin{verbbox}
\usepackage{booktabs}
\usepackage{morefloats}     % Increase the number of simultaneous LaTeX floats
\usepackage{enumitem}       % more control oover itemize

\newsavebox\lstbox
\mdfdefinestyle{MyFrame}{%
    innertopmargin=0pt,
    innerbottommargin=10pt,
    innerrightmargin=20pt,
    innerleftmargin=20pt}

\definecolor{darkgreen}{HTML}{009900}
    
\usepackage[ngerman,main=american]{babel}

\usepackage{csquotes}

\usepackage[markcase=upper]{scrlayer-scrpage}

\usepackage[math]{blindtext}

\usepackage{mathtools}

\usepackage{thmtools}

\usepackage{geometry}

\usepackage[
  type1,
  scale=0.88,
]{GoMono}

\usepackage{tgheros}

\usepackage[T1]{fontenc}

\usepackage[bitstream-charter]{mathdesign}

\DeclareSymbolFont{usualmathcal}{OMS}{cmsy}{m}{n}
\DeclareSymbolFontAlphabet{\mathcal}{usualmathcal}

\usepackage[
  final,
  tracking=smallcaps,
  stretch=10,
  shrink=10,
]{microtype}

\usepackage{lastpage}

\usepackage{refcount}

\usepackage{graphicx}

\usepackage{contour}

\usepackage{cellspace}

\usepackage{subcaption}

\usepackage{sidecap}

\usepackage{array}

\usepackage{tabularx}

\usepackage{booktabs}

\usepackage{multirow}

\usepackage{enumitem}

\usepackage{expl3}
\ExplSyntaxOn
\msg_redirect_name:nnn{siunitx}{moved-cellspace-column}{none}
\ExplSyntaxOff

\usepackage[binary-units=true]{siunitx}

\usepackage{everypage}

\usepackage{tikz}

\usepackage{tocbasic}

\usepackage{etoc}

\usepackage{wrapfig}

\usepackage{lettrine}

\usepackage{pdfpages}

\usepackage{nameref}

\usepackage[onehalfspacing]{setspace}

\usepackage{regexpatch}

\usepackage{hyperref}

\usepackage[all]{hypcap}

\usepackage[
  capitalise,
  nameinlink,
  ngerman,english,
]{cleveref}

\crefmultiformat{equation}{#2Eqs.~(#1)#3}{ and~#2(#1)#3}{, #2(#1)#3}{ and~#2(#1)#3}
\usepackage{algorithm}
\usepackage{algorithmicx}

\usepackage[noend]{algpseudocode}

\usepackage{bookmark}

\newcommand*\cleartoleftpage{%
  \clearpage
  \ifodd\value{page}\hbox{}\newpage\fi
}

\usepackage[
  giveninits=true,
  date=year,
  style=alphabetic,
  maxalphanames=1,
]{biblatex}

\newcommand*{\thetitle}{%
  Scalable Biophysical Simulations\texorpdfstring{\\}{}
  of the Neuromuscular System%
}
\newcommand*{\theapproval}{%
  Vom Stuttgarter Zentrum für Simulationswissenschaften (SC SimTech) und\\
  der Fakultät für Informatik, Elektrotechnik und Informationstechnik\\
  der Universität Stuttgart zur Erlangung der Würde eines Doktors\\
  der Naturwissenschaften (Dr.\ rer.\ nat.) genehmigte Abhandlung%
}
\newcommand*{\theauthor}{Benjamin Maier}
\newcommand*{\thebirthplace}{Waiblingen}
\newcommand*{\thedefensedate}{22. Juni 2021}

\newcommand*{\theyear}{2021}
\newcommand*{\theuniversity}{Universität Stuttgart}
\newcommand*{\theinstitute}{Institut für Parallele und Verteilte Systeme}
\newcommand*{\theadvisor}{Prof.\ Dr.\ Miriam Schulte}
\newcommand*{\theexamineri}{Prof.\ Dr.\ Hans-Joachim Bungartz}
\newcommand*{\theexaminerii}{}
\KOMAoptions{
  fontsize=12pt,
  DIV=12,
  numbers=noendperiod,
  onpsinit=\onehalfspacing,
  listof=leveldown,
}

\newcommand*{\bindingoffset}{10mm}

\graphicspath{{../gfx/}}

\graphicspath{
{images/summer_school_study/png/}
{images/summer_school_study/}
{images/summer_school_study/plots/}
{images/summer_school_study/2018/}
{images/fiber_creation/}
{images/parallel_fiber_estimation/}
{images/motor_unit_assignment/}
{images/implementation/}
{images/theory/}
{images/results/basic}
{images/results/application}
{images/results/studies}
{images/logos}
}

\usetikzlibrary{
  arrows.meta,
  calc,
  decorations.pathmorphing,
  decorations.pathreplacing,
  decorations.text,
  fadings,
  shadows,
}

\tikzset{every picture/.style={line width=1pt}}

\tikzset{>={Stealth[length=5.5pt,width=5.5pt]}}

\def\centerarc[#1](#2)(#3:#4:#5){
  \draw[#1] ($(#2)+({(#5)*cos(#3)},{(#5)*sin(#3)})$) arc (#3:#4:#5);
}

\contourlength{1.5pt}
\crefname{algorithm}{Alg.}{Algorithms}
\Crefname{algorithm}{Algorithm}{Algorithms}

\crefname{chapter}{Chap.}{Chapters}
\Crefname{chapter}{Chapter}{Chapters}

\crefname{corollary}{Cor.}{Corollaries}
\Crefname{corollary}{Corollary}{Corollaries}

\crefname{definition}{Def.}{Definitions}
\Crefname{definition}{Definition}{Definitions}

\crefname{equation}{Eq.}{Equations}
\Crefname{equation}{Equation}{Equations}

\crefname{figure}{Fig.}{Figures}
\Crefname{figure}{Figure}{Figures}

\crefname{lemma}{Lemma}{Lemmas}
\Crefname{lemma}{Lemma}{Lemmas}

\crefname{line}{line}{lines}
\Crefname{line}{Line}{Lines}

\crefname{proposition}{Prop.}{Propositions}
\Crefname{proposition}{Proposition}{Propositions}

\crefname{section}{Sec.}{Sections}
\Crefname{section}{Section}{Sections}

\crefname{table}{Tab.}{Tables}
\Crefname{table}{Table}{Tables}

\crefname{theorem}{Thm.}{Theorems}
\Crefname{theorem}{Theorem}{Theorems}

\DeclareTOCStyleEntry[indent=0em]{tocline}{figure}
\DeclareTOCStyleEntry[indent=0em]{tocline}{table}

\DeclareNewTOC[
  type=mytheorem,
  name=Theorem,
  listname={List of Theorems},
  tocentrynumwidth=2.3em,
  tocentryindent=0em,
]{lop}

\addtotheorempostheadhook{%
  \addxcontentsline{lop}{mytheorem}{\csname ll@\thmt@envname\endcsname}%
}

\AfterTOCHead{\begin{spacing}{1}}
\AfterStartingTOC{\end{spacing}}

\addtokomafont{dictum}{\rmfamily}

\setkomafont{pageheadfoot}{%
  \normalfont\normalcolor\sffamily%
  \fontsize{9.5}{12}\selectfont\lsstyle%
}
\setkomafont{pagenumber}{\usekomafont{pageheadfoot}\bfseries}

\clearscrheadfoot
\lehead[\pagemark]{\pagemark}
\rehead[]{\headmark}
\lohead[]{\headmark}
\rohead[\pagemark]{\pagemark}
\lefoot[]{}
\rofoot[]{}

\newcolumntype{L}[1]{%
  >{\raggedright\let\newline\\\arraybackslash\hspace{0pt}}p{#1}%
}
\newcolumntype{D}[1]{%
  >{\centering\let\newline\\\arraybackslash\hspace{0pt}}p{#1}%
}
\newcolumntype{R}[1]{%
  >{\raggedleft\let\newline\\\arraybackslash\hspace{0pt}}p{#1}%
}

\newcommand*{\headerrowcolor}{mittelblau!40}

\newcommand*{\oddrowcolor}{mittelblau!10}
\newcommand*{\evenrowcolor}{mittelblau!20}

\newcommand*{\autorowcolor}{%
  \ifnum\rownum<1%
    \headerrowcolor%
  \else%
    \ifodd\rownum\oddrowcolor\else\evenrowcolor\fi%
  \fi%
}

\newcommand*{\prevautorowcolor}{%
  \ifnum\rownum<2%
    \headerrowcolor%
  \else%
    \ifodd\rownum\evenrowcolor\else\oddrowcolor\fi%
  \fi%
}

\DisableLigatures{encoding=*,family=tt*}

\algrenewcommand{\alglinenumber}[1]{\footnotesize\color{anthrazit}{\texttt{#1}}}
\algrenewcommand{\algorithmicrequire}{\qquad\textbf{Input:}}
\algrenewcommand{\algorithmicensure}{\qquad\textbf{Output:}}

\algrenewcommand{\textproc}{}

\algnewcommand{\Break}{\textbf{break}}
\algnewcommand{\Continue}{\textbf{continue}}
\algnewcommand{\True}{\textbf{true}}
\algnewcommand{\False}{\textbf{false}}
\algnewcommand{\Null}{\textbf{null}}
\algrenewcommand{\algorithmicend}{\textbf{end}}
\algrenewcommand{\algorithmicdo}{\textbf{do}}
\algrenewcommand{\algorithmicwhile}{\textbf{while}}
\algrenewcommand{\algorithmicfor}{\textbf{for}}
\algrenewcommand{\algorithmicforall}{\textbf{for all}}
\algrenewcommand{\algorithmicloop}{\textbf{loop}}
\algrenewcommand{\algorithmicrepeat}{\textbf{repeat}}
\algrenewcommand{\algorithmicuntil}{\textbf{until}}
\algrenewcommand{\algorithmicprocedure}{\textbf{procedure}}
\algrenewcommand{\algorithmicfunction}{\textbf{function}}
\algrenewcommand{\algorithmicif}{\textbf{if}}
\algrenewcommand{\algorithmicthen}{\textbf{then}}
\algrenewcommand{\algorithmicelse}{\textbf{else}}
\algrenewcommand{\algorithmicreturn}{\textbf{return}}
\algnewcommand{\algorithmicforever}{\textbf{for ever}}
\algdef{S}[FOR]{ForEver}{\algorithmicforever\ \algorithmicdo}
\algnewcommand{\algorithmicgoto}{\textbf{go to}}
\algnewcommand{\Goto}[1]{\algorithmicgoto\ \cref*{#1}}
\algnewcommand{\ForOneLine}[2]{%
  \State\algorithmicfor\ #1\ \algorithmicdo\ #2%
}
\algnewcommand{\IfOneLine}[2]{%
  \State\algorithmicif\ #1\ \algorithmicthen\ #2%
}
\algnewcommand{\ElseOneLine}[1]{%
  \State\algorithmicelse\ #1%
}

\algrenewcomment[1]{\hfill$\rightsquigarrow$\;{\normalfont\emph{#1}}}

\algrenewcommand{\algorithmicindent}{\widthof{AB}}

\newcounter{algorithmicH}
\let\oldalgorithmic\algorithmic
\renewcommand*{\algorithmic}{\stepcounter{algorithmicH}\oldalgorithmic}
\DeclareSIUnit{\permille}{\text{\textperthousand}}

\ExplSyntaxOn
\NewDocumentCommand\hms{o>{\SplitArgument{2}{;}}m}{
  \group_begin:
  \IfNoValueF{#1}{\keys_set:nn{siunitx}{#1}}
  \siunitx_hms_output:nnn #2
  \group_end:
}
\cs_new_protected:Npn \siunitx_hms_output:nnn #1#2#3
{
  \IfNoValueF{#1}{
    \tl_if_blank:nF{#1}{
      \SI{#1}{\hour}
      \IfNoValueF{#2}{~}
    }
  }
  \IfNoValueF{#2}{
    \tl_if_blank:nF{#2}{
      \IfNoValueF{#1}{\tl_if_blank:nF{#1}{\sisetup{minimum-integer-digits=2}}}
      \SI{#2}{\minute}
      \IfNoValueF{#3}{~}
    }
  }
  \IfNoValueF{#3}{
    \tl_if_blank:nF{#3}{
      \IfNoValueF{#1}{\tl_if_blank:nF{#1}{\sisetup{minimum-integer-digits=2}}}
      \IfNoValueF{#2}{\tl_if_blank:nF{#2}{\sisetup{minimum-integer-digits=2}}}
      \SI{#3}{\second}
    }
  }
}
\ExplSyntaxOff

\edef\partial{\mathchar\number\partial\noexpand\mkern-2mu}
\newcommand*{\gitCommitText}[1][]{%
  \texttt{#1\gitCommitHash}%
  \ifx\gitTag\empty\else{} (\texttt{#1\gitTag})\fi%
}

\setlist{
  leftmargin=\parindent,
  itemsep=0pt,
  topsep=0.5em,
}

\newlength{\halfhphantomlength}
\newcommand*{\halfhphantom}[1]{%
  \ifmmode%
    \mathsettowidth{\halfhphantomlength}{#1}%
  \else%
    \settowidth{\halfhphantomlength}{#1}%
  \fi%
  \setlength{\halfhphantomlength}{\halfhphantomlength/2}%
  \hspace{\halfhphantomlength}%
}

\newcommand*{\lefthphantom}[2]{%
  \ifmmode\mathrlap{#1}\else\rlap{#1}\fi%
  \hphantom{#2}%
}

\newcommand*{\centerhphantom}[2]{%
  \halfhphantom{#2}%
  \ifmmode\mathclap{#1}\else\clap{#1}\fi%
  \halfhphantom{#2}%
}

\newcommand*{\righthphantom}[2]{%
  \hphantom{#2}%
  \ifmmode\mathllap{#1}\else\llap{#1}\fi%
}

\definecolor{C0}{rgb}{0.000,0.447,0.741}
\definecolor{C1}{rgb}{0.850,0.325,0.098}
\definecolor{C2}{rgb}{0.749,0.561,0.102}
\definecolor{C3}{rgb}{0.494,0.184,0.556}
\definecolor{C4}{rgb}{0.466,0.674,0.188}
\definecolor{C5}{rgb}{0.301,0.745,0.933}
\definecolor{C6}{rgb}{0.635,0.078,0.184}
\definecolor{C7}{rgb}{0.887,0.465,0.758}
\definecolor{C8}{rgb}{0.496,0.496,0.496}

\definecolor{anthrazit}{RGB}{62,68,76}
\definecolor{mittelblau}{RGB}{0,81,158}
\definecolor{hellblau}{RGB}{0,190,255}
\hypersetup{
  pdftitle={\thetitle},
  pdfauthor={\theauthor},
  pdfcreator={LaTeX, KOMA-Script, hyperref},
  pdfborderstyle={/S/U/W 1},
  citebordercolor=C1,
  filebordercolor=C1,
  linkbordercolor=C1,
  menubordercolor=C1,
  runbordercolor=C1,
  urlbordercolor=C0,
  bookmarksnumbered,
  bookmarksopen,
}

\DeclareCiteCommand{\cite}[\mkbibbrackets]{\usebibmacro{prenote}}{%
  \usebibmacro{citeindex}\usebibmacro{cite}%
}{\multicitedelim}{\usebibmacro{postnote}}

\defbibheading{myheading}[\bibname]{\addchap{#1}}

\defbibnote{mypostnote}{%
  \renewcommand*{\texttt}[1]{\oldtexttt{##1}}%
}

\DeclareNameAlias{sortname}{last-first}
\DeclareNameAlias{default}{last-first}

\DeclareFieldFormat[article]{title}{\mkbibemph{#1}}
\DeclareFieldFormat[article]{journaltitle}{#1}
\DeclareFieldFormat[article]{issuetitle}{#1}
\DeclareFieldFormat[article]{issuesubtitle}{#1}
\DeclareFieldFormat[inbook]{title}{\mkbibemph{#1}}
\DeclareFieldFormat[inbook]{booktitle}{#1}
\DeclareFieldFormat[inbook]{maintitle}{#1}
\DeclareFieldFormat[incollection]{title}{\mkbibemph{#1}}
\DeclareFieldFormat[incollection]{booktitle}{#1}
\DeclareFieldFormat[incollection]{maintitle}{#1}
\DeclareFieldFormat[inproceedings]{title}{\mkbibemph{#1}}
\DeclareFieldFormat[inproceedings]{booktitle}{#1}
\DeclareFieldFormat[inproceedings]{maintitle}{#1}
\DeclareFieldFormat[inreference]{booktitle}{#1}
\DeclareFieldFormat[inreference]{maintitle}{#1}
\DeclareFieldFormat[thesis]{title}{\mkbibemph{#1}}
\DeclareFieldFormat[unpublished]{title}{\mkbibemph{#1}}

\DeclareFieldFormat{titlecase}{\MakeTitleCase{#1}}

\newrobustcmd{\MakeTitleCase}[1]{%
  \ifthenelse{%
    \ifcurrentfield{booktitle}\OR\ifcurrentfield{booksubtitle}%
    \OR\ifcurrentfield{maintitle}\OR\ifcurrentfield{mainsubtitle}%
    \OR\ifcurrentfield{journaltitle}\OR\ifcurrentfield{journalsubtitle}%
    \OR\ifcurrentfield{issuetitle}\OR\ifcurrentfield{issuesubtitle}%
    \OR\ifentrytype{book}\OR\ifentrytype{mvbook}\OR\ifentrytype{bookinbook}%
    \OR\ifentrytype{booklet}\OR\ifentrytype{suppbook}%
    \OR\ifentrytype{collection}\OR\ifentrytype{mvcollection}%
    \OR\ifentrytype{suppcollection}\OR\ifentrytype{manual}%
    \OR\ifentrytype{periodical}\OR\ifentrytype{suppperiodical}%
    \OR\ifentrytype{proceedings}\OR\ifentrytype{mvproceedings}%
    \OR\ifentrytype{reference}\OR\ifentrytype{mvreference}%
    \OR\ifentrytype{thesis}\OR\ifentrytype{online}%
  }{%
    #1%
  }{%
    \MakeSentenceCase*{#1}%
  }%
}

\renewbibmacro*{in:}{}

\renewbibmacro*{issue+date}{%
  \setunit{\addcomma\space}%
  \iffieldundef{issue}{%
    \usebibmacro{date}%
  }{%
    \printfield{issue}\setunit*{\addspace}\usebibmacro{date}%
  }%
  \newunit%
}

\DeclareFieldFormat{url}{\scriptsize\url{#1}}
\DeclareFieldFormat{doi}{%
  \scriptsize%
  \ifhyperref{%
    \href{https://doi.org/#1}{\nolinkurl{doi:#1}}%
  }{\nolinkurl{doi:#1}}%
}
\DeclareFieldFormat{isbn}{%
  \scriptsize%
  \ifhyperref{%
    \href{https://www.amazon.com/s/?field-keywords=#1}{\nolinkurl{isbn:#1}}%
  }{\nolinkurl{isbn:#1}}%
}

\AtEveryBibitem{\clearfield{urlyear}}

\DeclareBibliographyCategory{own}
\DeclareBibliographyCategory{own_other}
\DeclareBibliographyCategory{own_forthcoming}
\newcommand{\opendihu}{OpenDiHu}
\newcommand{\Opendihu}{OpenDiHu}

\def\beqno{\begin{equation}}
\def\eeqno{\end{equation}}
\def\beq{\begin{equation*}}
\def\eeq{\end{equation*}}
\def\ba#1{\begin{array}{#1}}
\def\ea{\end{array}}
\def\mat#1{\left(\begin{matrix}#1\end{matrix}\right)}
\def\matt#1{\left[\begin{matrix}#1\end{matrix}\right]}
\newcommand{\bal}{\begin{align}}
\def\ealll{ \end{align} }

\renewcommand{\emph}[1]{\textit{#1}\/}
\def\clap#1{\hbox  to  0pt{\hss#1\hss}}                 % für underbrace
\def\mathclap{\mathpalette\mathclapinternal}
\def\mathclapinternal#1#2{\clap{$\mathsurround=0pt#1{#2}$}}
    
\newcommand{\ds}{\displaystyle}                         % displaystyle
\renewcommand{\dfrac}[2]{\ds\frac{\ds{#1}}{\ds{#2}}\,}  % nach Bruch Abstand

\usepackage{setspace}
\newcommand{\code}[1]{{\small\lstinline[basicstyle=\small\ttfamily\color{Maroon},breaklines=true,upquote=true]!#1!}}

\newcommand{\emphcode}[1]{{\small\lstinline[basicstyle=\small\itshape\ttfamily\color{Maroon},breaklines=true]!#1!}}

\definecolor{Maroon}{HTML}{801010}
\newcounter{reproduce}[section]
\newenvironment{reproduce}{%
  \stepcounter{reproduce}%
  \mdfsetup{%
    frametitle={%
      \tikz[baseline=(current bounding box.east),outer sep=0pt]
      \node[anchor=east,rectangle,fill=red!20]
      {How To Reproduce};},nobreak=false
  }
  \mdfsetup{innertopmargin=10pt,linecolor=Maroon!20,nobreak=false,backgroundcolor=Maroon!05,%
            linewidth=2pt,topline=true,roundcorner=5pt,frametitleaboveskip=\dimexpr-\ht\strutbox\relax,%
  }
  \vfill
  \begin{mdframed}\relax%
}
{
  \end{mdframed}%
}

\newenvironment{reproduce_no_break}{%
  \stepcounter{reproduce}%
  \mdfsetup{%
    frametitle={%
      \tikz[baseline=(current bounding box.east),outer sep=0pt]
      \node[anchor=east,rectangle,fill=red!20]
      {How To Reproduce};},nobreak=true
  }
  \mdfsetup{innertopmargin=10pt,linecolor=Maroon!20,nobreak=true,backgroundcolor=Maroon!05,%
            linewidth=2pt,topline=true,roundcorner=5pt,frametitleaboveskip=\dimexpr-\ht\strutbox\relax,%
  }
  \vfill
  \begin{mdframed}\relax%
}
{
  \end{mdframed}%
}

\def\det{\hbox{det} \,}

\def\div{\hbox{div} \,}
\def\grad{\hbox{grad} \,}

\def\cof{\hbox{cof} \,}
\def\tr{\hbox{tr} \,}
\def\sym{\hbox{sym} \,}

\def\dt{d\noexpand\mkern-3mu t}
\DeclareSIUnit[number-unit-product = {}]{\flop}{FLOP}   % flops unit for unitx
\DeclareSIUnit[number-unit-product = {}]{\flops}{FLOPS}   % flops unit for unitx
\makeatletter
\def\d{\futurelet\next\start@i}\def\start@i{\ifx\next\bgroup\expandafter\abl@\else\expandafter\abl@d\fi}\def\abl@#1{\def\tempa{#1}\futurelet\next\abl@i}\def\abl@i{\ifx\next\bgroup\expandafter\abl@ii\else\expandafter\abl@a\fi}\def\abl@ii#1{\def\tempb{#1}\futurelet\next\abl@iii}\def\abl@iii{\ifx\next\bgroup\expandafter\abl@c\else\expandafter\abl@b\fi}
\def\abl@d{\mathrm{d}}                                          % keine Argumente
\def\abl@a{\ds\frac{\mathrm{d}}{\mathrm{d}\tempa}\,}            % 1 Argument \d{x} -> d/dx
\def\abl@b{\ds\frac{\mathrm{d}\tempa}{\mathrm{d}\tempb}\,}  % 2 Argumente: \d{f}{x} -> df/dx
\def\abl@c#1{\ds\frac{\mathrm{d}^{#1} {\tempa}}{\mathrm{d} {\tempb}^{#1}}\,}        % 3 Argumente: \d{f}{x}{2} -> d^2f/dx^2

\def\p{\futurelet\next\startp@i}\def\startp@i{\ifx\next\bgroup\expandafter\pabl@\else\expandafter\pabl@d\fi}\def\pabl@#1{\def\tempa{#1}\futurelet\next\pabl@i}\def\pabl@i{\ifx\next\bgroup\expandafter\pabl@ii\else\expandafter\pabl@a\fi}\def\pabl@ii#1{\def\tempb{#1}\futurelet\next\pabl@iii}\def\pabl@iii{\ifx\next\bgroup\expandafter\pabl@c\else\expandafter\pabl@b\fi}
\def\pabl@d{\partial}                                           % keine Argumente
\def\pabl@a{\ds\frac{\partial}{\partial\tempa}\,}           % 1 Argument \d{x} -> d/dx
\def\pabl@b{\ds\frac{\partial\tempa}{\partial\tempb}\,} % 2 Argumente: \d{f}{x} -> df/dx
\def\pabl@c#1{\ds\frac{\partial^{#1} {\tempa}}{\partial {\tempb}^{#1}}\,}       % 3 Argumente: \d{f}{x}{2} -> d^2f/dx^2
\makeatother

\def\eps{\varepsilon}
\def\N{\mathbb{N}}  %nat. Zahlen
\def\R{\mathbb{R}}  %reelle Zahlen
\def\C{\mathbb{C}}  %komplexe Zahlen
\renewcommand{\i}[2]{\ds\int\limits_{#1}^{#2}} %Integral, %TODO_Lorin:das überschreibt "interpolierende" \I, %FIX_Benni: zweimal kleiner Buchstabe (Großbuchstaben sind eher für Räume)
\renewcommand{\s}[2]{\ds\sum\limits_{#1}^{#2}} %Summe %EDIT_Georg: mit renewcommand hat's nicht compiliert, deshalb jetzt newcommand

\renewcommand{\O}{\mathcal{O}}      %O-Notation

\newcommand{\CC}{\mathcal{C}}       %Raum der stetig diff.baren Fkt
\newcommand{\cm}{\,\mathrm{cm}}

\newcommand{\bfa}{\textbf{a}}
\newcommand{\bfb}{\textbf{b}}
\newcommand{\bfc}{\textbf{c}}

\newcommand{\bfe}{\textbf{e}}
\newcommand{\bff}{\textbf{f}}
\newcommand{\bfg}{\textbf{g}}
\newcommand{\bfh}{\textbf{h}}

\newcommand{\bfk}{\textbf{k}}
\newcommand{\bfl}{\textbf{l}}
\newcommand{\bfm}{\textbf{m}}
\newcommand{\bfn}{\textbf{n}}

\newcommand{\bfp}{\textbf{p}}
\newcommand{\bfq}{\textbf{q}}
\newcommand{\bfr}{\textbf{r}}
\newcommand{\bfs}{\textbf{s}}
\newcommand{\bft}{\textbf{t}}
\newcommand{\bfu}{\textbf{u}}
\newcommand{\bfv}{\textbf{v}}

\newcommand{\bfx}{\textbf{x}}
\newcommand{\bfy}{\textbf{y}}
\newcommand{\bfz}{\textbf{z}}
\newcommand{\bfA}{\textbf{A}}
\newcommand{\bfB}{\textbf{B}}
\newcommand{\bfC}{\textbf{C}}
\newcommand{\bfD}{\textbf{D}}
\newcommand{\bfE}{\textbf{E}}
\newcommand{\bfF}{\textbf{F}}

\newcommand{\bfH}{\textbf{H}}
\newcommand{\bfI}{\textbf{I}}
\newcommand{\bfJ}{\textbf{J}}
\newcommand{\bfK}{\textbf{K}}

\newcommand{\bfM}{\textbf{M}}

\newcommand{\bfP}{\textbf{P}}

\newcommand{\bfR}{\textbf{R}}
\newcommand{\bfS}{\textbf{S}}
\newcommand{\bfT}{\textbf{T}}
\newcommand{\bfU}{\textbf{U}}
\newcommand{\bfV}{\textbf{V}}
\newcommand{\bfW}{\textbf{W}}
\newcommand{\bfX}{\textbf{X}}

\newcommand{\bfzero}{\textbf{0}}

\newcommand{\bfbeta}{\boldsymbol{\beta}}

\newcommand{\bfeps}{\boldsymbol{\eps}}
\newcommand{\bftheta}{\boldsymbol{\theta}}

\newcommand{\bfphi}{\boldsymbol{\phi}}
\newcommand{\bfvarphi}{\boldsymbol{\varphi}}
\newcommand{\bfpsi}{\boldsymbol{\psi}}
\newcommand{\bfsigma}{\boldsymbol{\sigma}}

\newcommand{\bfxi}{\boldsymbol{\xi}}

\newcommand{\bflambda}{\boldsymbol{\lambda}}
\newcommand{\bfomega}{\boldsymbol{\omega}}

\begin{document}
\frontmatter

% title page
\begin{titlepage}
  \begin{spacing}{1}
    \begin{center}
      \begin{otherlanguage}{ngerman}
        % no indentation of paragraphs
        \setlength{\parindent}{0pt}
        
        {\bfseries\huge\thetitle\par}
        
        \vfill
        
        \theapproval
        
        \vfill
        
        Vorgelegt von
        
        {\bfseries\Large\theauthor\par}
        
        aus \thebirthplace
        
        \vfill
        
        \begin{tabular}{ll}
          Hauptberichterin:&
          \theadvisor\\[0.5em]
          Mitberichter:&
          \theexamineri\\
          &\theexaminerii\\[1em]
          \multicolumn{2}{l}{%
            Tag der mündlichen Prüfung:\quad%
            \thedefensedate%
          }
        \end{tabular}
        
        \vfill
        
        \includegraphics[width=60mm]{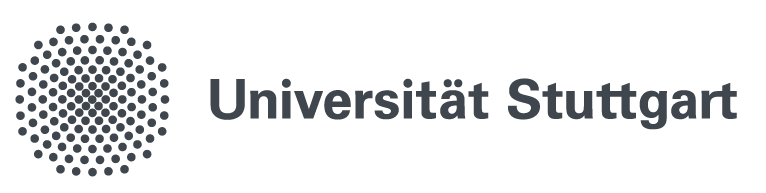}
        
        \vspace{2em}
        
        \theinstitute{} der \theuniversity
        
        \vspace{1em}
        
        \theyear
      \end{otherlanguage}
    \end{center}
  \end{spacing}
\end{titlepage}

\pdfbookmark[section]{\contentsname}{toc}
\tableofcontents

%\cleardoublepage

% abstract

% this alters "before" spacing (the second length argument) to 0
%\titlespacing*{\chapter}{0pt}{0pt}{40pt}

\RedeclareSectionCommand[
  %runin=false,
  afterindent=false,
  beforeskip=0pt,
  afterskip=40pt]{chapter}
  
\addchap{Abstract/\foreignlanguage{ngerman}{Kurzzusammenfassung}}
  
%\disableornamentsfornextheadingtrue
\section*{Abstract}

% howto write the abstract
% Context – provides background for less specialized readers and, so doing, establishes or recalls the importance of the problem.
The human neuromuscular system consisting of skeletal muscles and neural circuits is a complex system that is not yet fully understood. 
Surface electromyography (EMG) can be used to study muscle behavior from the outside. 
Computer simulations with detailed biophysical models provide a non-invasive tool to interpret EMG signals and gain new insights into the system. 

The numerical solution of such multi-scale models imposes high computational work loads, which restricts their application to short simulation time spans or coarse resolutions.
We tackled this challenge by providing scalable software employing instruction-level and task-level parallelism, suitable numerical methods and efficient data handling. We implemented a comprehensive, state-of-the-art, multi-scale multi-physics model framework that can simulate surface EMG signals and muscle contraction as a result of neuromuscular stimulation.

This work describes the model framework and its numerical discretization, develops new algorithms for mesh generation and parallelization, covers the use and implementation of our software OpenDiHu, and evaluates its computational performance in numerous use cases.

We obtain a speedup of several hundred compared to a baseline solver from the literature and demonstrate, that our distributed-memory parallelization and the use of High Performance Computing resources enables us to simulate muscular surface EMG of the biceps brachii muscle with realistic muscle fiber counts of several hundred thousands. We find that certain model effects are only visible with such high resolution.

In conclusion, our software contributes to more realistic simulations of the neuromuscular system and provides a tool for applied researchers to complement in vivo experiments with in-silico studies. It can serve as a building block to set up comprehensive models for more organs in the musculoskeletal system.

% Need – motivates the audience by stating the difference between the desired and actual situations.
% Task – states what the authors undertook to address the need, in the first person (we), past tense.
% Object – clarifies what the paper covers without repeating the task, in the active voice, present tense.
% Findings – state the main results in a way that both less and more specialized readers find helpful.
% Conclusion – interprets findings (states the so what) – in this case, all the way to a recommendation.
% Perspectives – broaden the view with any further needs and tasks.

\begin{otherlanguage}{ngerman}
  %\disableornamentsfornextheadingtrue
  \section*{Kurzzusammenfassung}

Beim neuromuskulären System, bestehend aus Skelettmuskeln und Nervenbahnen, handelt es sich um ein komplexes System, welches noch nicht komplett verstanden ist. Das Muskelverhalten kann von außen durch Oberflächen-Elektromyografie (EMG) untersucht werden. Computersimulationen mit detaillierten, biophysikalischen Modellen stellen eine nichtinvasive Methode dar, um EMG-Signale zu interpretieren und neue Erkenntnisse über das System zu erlangen.

Die numerische Lösung solcher Mehrskalenmodelle erfordert eine große Rechenleistung, sodass die Modelle nur für kurze Simulationszeitspannen oder grobe Auflösungen geeignet sind. 
Wir lösen dieses Problem durch das Bereitstellen skalierbarer Software, welche Parallelität auf Instruktions- und Taskebene ausnutzt, geeignete numerische Methoden einsetzt und eine  effiziente Datenverarbeitung sicherstellt. Wir setzen ein umfassendes, dem Stand der Wissenschaft entsprechendes Mehrskalen- und Mehrphysik-Modell um, welches Oberflächen-EMG-Signale simuliert und die Kontraktion eines Muskels als Folge neuromuskulärer Stimulation berechnet.
  
Diese Arbeit beschreibt das Modell und seine numerische Diskretisierung, entwickelt neue Algorithmen zur Gittererzeugung und Parallelisierung, behandelt die Anwendung und Umsetzung unserer Software OpenDiHu und wertet ihre Berechnungseffizienz in vielen Anwendungsbeispielen aus.

Wir erreichen eine um einen Faktor von mehreren Hundert schnellere Berechnung verglichen mit einem Referenzlöser aus der Literatur. Unsere Parallelisierung für Parallelrechner mit verteiltem Speicher und die Verwendung von Hochleistungsrechnern erlauben es uns, Oberflächen-EMG des Biceps Brachii mit einer realistischen Anzahl an Muskelfasern von mehreren Hunderttausend zu simulieren. Wir stellen fest, dass bestimmte Modelleffekte nur mit solch hoher Auflösung sichtbar werden.

Unsere Software trägt zu realistischeren Simulationen des neuromuskulären Systems bei und stellt ein Werkzeug für die angewandte Wissenschaft zur Verfügung, um In-vivo-Experimente mit In-silico-Studien zu verknüpfen. Sie kann als Baustein zur Erstellung umfassender Modelle für weitere Organe im muskuloskelettalen System dienen.
  
\end{otherlanguage}

\cleardoublepage

% publication list
% prepare publications for bibtex

\addtocategory{own}{Bradley:2018:EDB}
\addtocategory{own}{Maier2019}
\addtocategory{own}{Maier2021}
\addtocategory{own}{summerschool2019}

\addtocategory{own_forthcoming}{hlrs2020}
\addtocategory{own_forthcoming}{MaierSubmitted2}
\addtocategory{own_forthcoming}{MaierSubmitted}
\addtocategory{own_forthcoming}{hlrs2021}

\addtocategory{own_other}{opendihu}
\addtocategory{own_other}{maier_benjamin_2021_4705945}

% cite all of these categories
\nocite{Bradley:2018:EDB}
\nocite{Maier2019}
\nocite{Maier2021}
\nocite{summerschool2019}
\nocite{hlrs2020}
\nocite{MaierSubmitted2}
\nocite{MaierSubmitted}
\nocite{hlrs2021}
\nocite{maier_benjamin_2021_4705945}
\nocite{opendihu}
% ------------------------------------------------

%\thispagestyle{empty} % remove number
%\cleartoleftpage

\RedeclareSectionCommand[
  %runin=false,
  afterindent=false,
  beforeskip=-\baselineskip,
  afterskip=0pt]{chapter}
  
%\RedeclareSectionCommand[
  %%runin=false,
  %afterindent=false,
  %beforeskip=0pt, %\baselineskip,
  %afterskip=1.5\baselineskip]{section}
  
\addchap{Publications}
The following publications discuss various topics that are covered in this thesis.

% ---------------------------------------
% peer reviewed
\section*{Peer-Reviewed Publications}
%\vspace{-5mm}
\printbibliography[category=own,heading=none,title=]

% ---------------------------------------
% submitted
\section*{Forthcoming Publications}
%\vspace{-5mm}
\printbibliography[category=own_forthcoming,heading=none,title=]

% ---------------------------------------
% other publications
\section*{Source Code and Data}
%\vspace{-5mm}
The most recent version of the OpenDiHu software is made available in the GitHub repository at \url{https://github.com/maierbn/opendihu}. 
The released version and the packaged data that are needed to reproduce the results in this work, containing, e.g., mesh files and CellML models, are given below.

\printbibliography[category=own_other,heading=none,title=]

%\thispagestyle{empty} % remove number

% this changes "before" spacing back to its default of 50pt
\RedeclareSectionCommand[
  %runin=false,
  afterindent=false,
  beforeskip=50pt,
  afterskip=40pt]{chapter}

\cleardoublepage

\mainmatter
% chapters
\chapter{Introduction}\label{chap:introduction}

%\section{Motivation}

% Introductory paragraph:
%  Give a general introduction to the topic for broad audience
%  Narrow the focus to your particular topic
%  State your research problem and aims

% hinführung
Tying the shoestrings, running to catch the train, quickly slipping through the closing door, and then lifting a heavy suitcase to the luggage rack over the seat\textemdash all actions that are only possible because of the versatility of the musculoskeletal system.
Voluntary contractions of skeletal muscles enable humans to perform a variety of tasks: finely controlled and coordinated actions, endurance tasks, fast and vigorous actions, and exercises requiring high forces.

% learn from healthy system and treat diseased system
Moreover, skeletal muscles are able to be trained and adapt to requirements, can self-repair, and usually keep their capabilities for an entire lifetime. Understanding this remarkable system that has evolved over millions of years can advance both engineering and healthcare.

From an engineering view, derived biomimetic systems such as powered exoskeletons or robot arms with muscle-like actuators exhibit promising properties such as being lightweight, inexpensive, resilient, damage tolerant, noiseless, and agile and, thus, are potentially emerging field in robotics and medicine \cite{BarCohen2003,BarCohen2004Electroactivepolymer,Mirvakili2018}.

In the fields of healthcare and medicine, research is interested in obtaining a better understanding of muscular diseases such as muscular dystrophies \cite{Emery2002}. Studies show that disabling inherited neuromuscular diseases are prevalent in 1 out of 3500 of the population \cite{Emery1991}. However, for most of the neuromuscular disorders no cure is known and treatment focuses on reducing symptoms \cite{Emery2002,Heidlauf2015Diss}. Developing treatments to neuromuscular disorders is only possible with an extensive understanding of the neuromuscular system. Similarly, for diagnosing the type of disorder from symptoms and clinically available examination tools such as electromyographic recordings, a comprehensive understanding of muscle physiology is needed.

Surface electromyography (sEMG) measures the temporally changing electric potentials on the skin surface that are induced by activation of the muscle fibers \cite{Merletti2004}. It is one of the few non-invasive diagnostic tools to gain insights into the functioning of the neuromuscular system.
High-density surface EMG (HD-sEMG) involves the signal acquisition by an array of electrodes on the skin surface and, thus, enriches the traditional, monopolar EMG by spatial information about the muscular activity. 

Another application, where insights into the neuromuscular system advance technology, bridges the two fields of engineering and healthcare: Exoskeletons for rehabilitation, e.g., of stroke patients, can be controlled by sEMG or HD-sEMG signals from the patients to accurately support the intended movements (e.g., \cite{Leonardis2015,Mulas2005,Andreasen2005}).

% need for simulation, in-vivo - in-silico
Despite the need to gain comprehensive insights, experimental in vivo investigations of the neuromuscular system have severe limitations: Boundary conditions, such as contraction velocities, often cannot be accurately controlled, studies are not repeatable because of fatigue effects, material parameters of individual subjects are not known and cannot be measured precisely, and the quantities of interest, such as activation values and active stresses cannot be measured easily. Moreover, experiments are strongly limited to the ethical bounds of natural movements. 

A controlled environment for such investigations can be provided by in silico experiments, i.e., using computer simulations. The main advantages of using simulations are unlimited access to all computed quantities, reproducibility, and freedom in the experimental protocol.
With appropriate models, predictions can be made even for pathological conditions.

Employing in silico experiments demands a careful formulation and composition of mathematical models, using experimentally found evidence about the functioning of various aspects in the muscular system. Once a model is set up, its execution requires suitable numerical methods and efficient implementation to utilize the available compute hardware in the best way.

%Ich glaube, hier wäre ein Überleitungssatz zum nächsten Absatz gut: Diese Arbeit beschäftigt sich genau mit solchen num. Methoden und deren effizienter Implementierung auf paralleler Hardware. In diesem Kapitel werden Grundlagen eingeführt: .... (Auflistung wie im nächsten Absatz, was in welchem Abschnitt kommt)

This work discusses these numerical methods and their efficient implementation on parallel hardware. The present chapter introduces the fundamentals:
\Cref{sec:anatomy_physiology} presents the basic anatomy and physiology of a skeletal muscle.
\Cref{sec:challenges_in_silico} takes a closer look at the application of the in silico laboratory and derives requirements on the simulation technology. 
\Cref{sec:intro_related_works} outlines the current state of the art in skeletal muscle simulation. One of the most promising model frameworks that we use is described in more detail in \cref{sec:the_multi_scale_model_of} before contributions of this work are summarized in \cref{sec:intro_contributions}.

\section{Anatomy and Physiology of the Human Skeletal Muscle}\label{sec:anatomy_physiology}

\begin{figure}
  \centering%
  \includegraphics[width=0.5\textwidth]{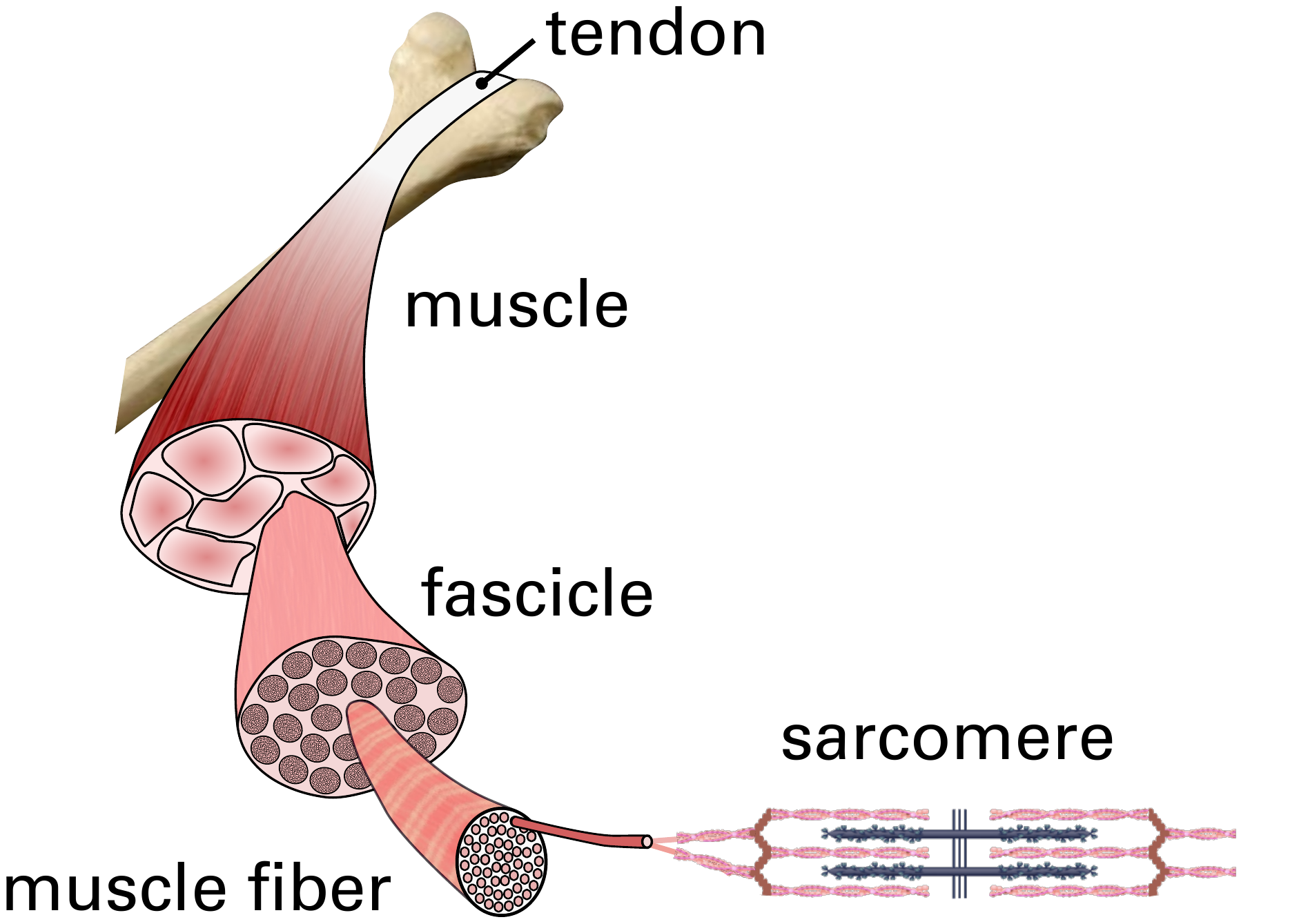}%
  \caption{Hierarchical structure of a skeletal muscle consisting of fascicles, muscle fibers and sarcomeres.}%
  \label{fig:hierarchical_structure}%
\end{figure}

Skeletal muscles have a hierarchical structure as shown in \cref{fig:hierarchical_structure}. On the macroscopic level, fibro-elastic tendons connect the muscle to the skeletal system. The muscle is composed of tens of fascicles with the exact number strongly depending on the muscle (numbers according to \cite{MacIntosh2006}). Each fascicle contains between ten and 10,000 muscle fibers, yielding a total of up to one million fibers in a muscle. Each muscle fiber usually runs through the whole length of the muscle. A muscle fiber consists of numerous parallel myofibrils, which each consists of series of sarcomeres, the smallest contractile unit of a muscle. A muscle fiber contains approximately 50,000 sarcomeres and, thus, there are from millions up to billions of sarcomeres in a whole muscle.

The contraction of the muscle is controlled by motor neurons in the spinal cord. The axons of each alpha motor neuron innervate multiple fibers in the muscle. In consequence, all connected fibers are always activated simultaneously. The set of fibers together with their motor neuron form a motor unit (MU).

The neuromuscular junctions where the axons innervate the muscle fibers are mostly located in a band within the mid-belly of the muscle \cite{Childers2004}. Upon activation of a muscle fiber, an electric stimulus, the action potential, travels from the neuromuscular junction towards both ends of the muscle. The action potential triggers subcellular processes and leads to force generation in the sarcomeres. The fibers are electrically isolated to each other but mechanically coupled through the fascicles and the extracellular matrix.

\begin{figure}
  \centering%
  \includegraphics[width=0.6\textwidth]{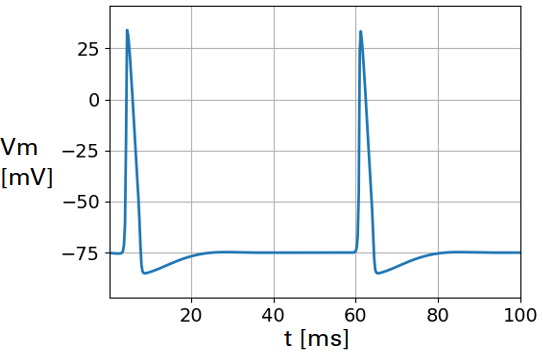}%
  \caption{Action potentials over time at a fixed point on a muscle fiber, calculated by the monodomain equation with a subcellular model of Hodgkin and Huxley using the software OpenDiHu (More details are given in later sections.).}%
  \label{fig:action_potentials}%
\end{figure}

The propagation of action potentials is governed by ionic currents through ion channels in the fiber membranes and is driven by ion pumps and the activation and deactivation of the ion channels. \Cref{fig:action_potentials} shows the shapes of two subsequent action potentials over time at a fixed point on a muscle fiber. The transmembrane potential $V_m$ initially equals its resting state of $\SI{-75}{\milli\volt}$. After stimulation occurs, the potential rapidly depolarizes to a maximum value of approximately $\SI{30}{\milli\volt}$, followed by the repolarization and a small overshoot, before returning to the resting potential. After approximately $\SI{30}{\milli\second}$, the system is again in equilibrium, and the action potential induced by the next stimulus has the exact same shape.
 
% action potential over time

%The muscle cells that form a skeletal muscle are the muscle fibres. They are bundled to fascicles whereas the whole muscle
%consists of multiple fascicles, as depicted in \cref{fig:muscle}. Tendons connect the ends of a muscle to the bones.
%Inside each muscle fibre multiple long myofibrils are responsible for the force generation.
%They consist of chains of sarcomeres which are the contractile elements of the cell.
%Multiple layers of parallel sarcomere chains areare arranged side by side, triangular in radial direction.

%The sarcomeres consist of thin filaments of the molecule actin and interleaved thick filaments out of myosin, as depicted in \cref{fig:sarcomere}.
%The thick filament contains myosin heads that can bind to the thin filament.
%In a multistep process called \say{crossbridge cycling} this binding occurs, subsequently the angle in which the head is attached changes and leads to a relative motion of the thin and thick filaments after which the head detaches again.
%Involved are, among others, calcium ions and adenosine triphosphate (ATP) which supplies the needed chemical energy.

The MUs are activated according to the size principle, starting with the smallest ones that connect to the least fibers and successively adding larger MUs \cite{Milner-Brown1973b}. The amount of muscle activation is controlled by the number of MUs and the rate-encoded stimulation signals for every MU. This finely controlled level is further modulated by the feedback loops of the neuromuscular system. Sensory organs are located within the muscle sense stretch, contraction velocity, and forces and influence the motor neuron pool.
A more elaborate description of the anatomy and physiology of the neuromuscular system can be found in the book of MacIntosh et al. \cite{MacIntosh2006}.

Considering the origin of EMG signals, all action potentials on the muscle fibers contribute to the electric potential in the muscle volume. While electric conduction is directed inside the muscle fibers, anisotropic conduction occurs in the volume of extracellular space. In addition, electric conduction in adipose tissue above the muscle belly influences the electric potential on the skin surface, which can be measured by EMG.

\section{ Use-Cases and Requirements for Simulations Used in in-Silico Experiments}\label{sec:challenges_in_silico}

With a basic understanding of the physiology of the neuromuscular system, we can now define use-cases for in-silico experiments and derive the requirements for models and simulation software.

A simulation should be able to accurately predict the response of the neuromuscular system to different recruitment strategies of the MUs. 
Also, different organizations of muscle fibers in MUs could be investigated. Similarly, the sensory feedback loop within a single muscle is by far not yet fully understood. Various assumptions could be tested in simulations, and the resulting force and EMG outputs could be compared to experiments. By complementing in vivo and in silico experiments, more comprehensive insights can be generated.

A second use case for simulations of the neuromuscular system lies in the decomposition of EMG recordings. Traditionally, signal processing techniques are used to draw conclusions from EMG data \cite{Merletti2004,Farina2010}. Decomposition algorithms exist that identify discharge patterns of individual motor units in HD-sEMG and additively decompose the recording \cite{DeLuca2006,Nawab2010,Holobar2007}. 
Novel, data-based techniques exist that employ deep learning methods \cite{Clarke2021}.

However, these techniques have limitations. The recorded signals are typically weak and noisy because of the layer of body fat between the muscle and the EMG electrodes. Cross-talk from adjacent muscles and destructive interference between signals of spatially close muscle fibers make the decomposition more difficult. Often, only isometric contractions can be considered in experiments since large movements of the muscles with respect to the electrodes would add additional uncertainties to the recorded signal.

Simulations can provide a controlled testing environment for such EMG decomposition algorithms. For data-based methods, simulations are unavoidable to generate training and validation data.

The listed use-cases for in-silico experiments demand detailed, biophysically informed models. Phenomenological descriptions cannot predict unseen scenarios or pathological conditions. The hypotheses to test related to MU organizations, recruitment, or sensory feedback have to reflect in the choice of the model description.
A suitable model usually needs to take into account the multi-scale nature of the neuromuscular system. The geometric structure of muscle fibers embedded in the muscle belly and the layer of adipose tissue have to be part of accurate models.

Multi-scale multi-physics simulations with high resolutions involve high computational loads. The simulated processes on a molecular scale, e.g., in the sarcomere require small timestep widths in the range of microseconds. At the same time, macroscopic quantities such as EMG signals and muscle contraction should be computed, leading to desired overall simulation time spans in the range of seconds.
Fine three-dimensional (3D) meshes are needed to achieve high spatial resolutions. To resolve individual muscle fibers, additional one-dimensional (1D) fiber meshes are considered.

To account for fine resolutions and a high number of timesteps in an acceptable runtime, the potential of today's and tomorrow's computer hardware has to be fully exploited. This requires task-level and instruction-level parallelism. For example, the latest processor of the Intel Core X series (the Intel Core i9-10980XE Extreme Edition Processor), which is listed at a customer price below \SI{1000}[\$]{} contains 18 hardware cores, allowing to run 16 tasks in parallel. It supports Intel AVX-512, a technology with which eight double precision floating point operations can be executed per instruction. 
In a higher price segment, it is, e.g., possible to combine two AMD EPYC 7742 server processors into a shared memory compute node with \num{128} cores. Distributed memory clusters allow the combination of almost any number of compute nodes to achieve higher total core counts. The supercomputer Hawk at the High Performance Computing Center Stuttgart combines \num{5632} of the mentioned AMD nodes into an overall cluster of \num{720896} cores.

Thus, a requirement to the simulation software is to be able to run on distributed memory computer systems. This requires efficient data management and a domain decomposition approach where the computational domain is partitioned into one subdomain for each process.
Highly parallel domain decomposition requires appropriately structured meshes and efficient parallel linear solvers. At the same time, muscle geometries obtained from medical imaging should be used to obtain a realistic setting. In a preprocessing step, the required highly resolved meshes have to be generated from imaging datasets.

A highly resolved simulation model can be used to estimate the accuracy of reduced models that do not include all biophysical processes or have reduced spatial resolutions. The advantage of such reduced models is that they can be solved with lower resources or in shorter runtimes. To assess the error of the reduced resolution, comparisons with results of the full model can be carried out. In this sense, the full model should be able to be used with a realistic number in the order of several \num{100000} muscle fibers and hundreds of MUs to allow for a comparison with simulations of smaller numbers of fibers.

% numerics
Highly resolved simulations are known to exhibit numerical instabilities, poor conditioning, or other causes for divergence in the numerical solvers. Therefore, numerical schemes have to be chosen carefully. At the same time, timestep widths can be increased and runtimes reduced by choosing, e.g., second order timestepping schemes instead of first order schemes.

% flexibility
Another important requirement of the simulation software can be formulated from the user's perspective. 
Configuring a simulation and exchanging material and numerical parameters should be possible in a convenient way. Simulation results should be accessible in various established file formats, to be examined in dedicated visualization software or used in further post-processing. On the modeling side, comprehensive state-of-the-art models should be implemented while maintaining the possibility to extend given multi-scale models later on as research advances. 
Standards in the biochemical modeling community should be respected and incorporated, such as the description language CellML \cite{Cellml2003,Lloyd2004} for subcellular models.
To find the most suited numerical solvers, the software should be flexible enough to, e.g., easily exchange timestepping schemes or employ different linear system solvers.

%  State your research problem and aims
% what we do
Our contribution is to implement and employ software that fulfills all these requirements. We aim at
simulating EMG and muscle contraction with detailed, biophysically informed multi-scale models. The software runs efficiently on the previously described hardware, ranging from workstation computers to supercomputers.

\section{Related Work and Software}\label{sec:intro_related_works}

%Literature review (usually several paragraphs):
%  Summarize the relevant literature on your topic
%  Describe the current state of the art
%  Note any gaps in the literature that your study will address
In the following, we give an overview of existing approaches for modeling the neuromuscular system. The overview involves literature and software frameworks and focuses on the multi-scale model that is the basis for the present work. For a recent, comprehensive review on all aspects of neuromuscular modeling, we refer to \cite{Rohrle2019Review}.

\subsection{Related Work}
The lowest computational effort is required when analytically solvable models are used to simulate skeletal muscle forces. The twitch force of a single motor unit can be described by the impulse response of a critically damped, second-order system, for which an analytical solution exists. For the given superposition of all motor unit action potentials, the transient output force of the muscle is computed  \cite{Cisi2008,Dideriksen2010}.
% muscle control

On the next level of detail, phenomenological Hill-type muscle models, which have to be solved numerically are used to describe muscle forces along a one-dimensional line of action. They are often used for systemic simulations of larger parts of the musculoskeletal system \cite{Zajac1989,OpenSim2007,Hilltype2014,Bayer2017}. We use Hill-type models in our case study on predicting forces of the upper arm. However, this type of model is not suited for simulations of EMG and neglects structural properties of the muscle tissue.

While phenomenological models describe a whole muscle by only a few parameters, continuum-based models exist that also take into account structural features and spatial heterogeneity \cite{Johansson2000,Blemker2005a,Roehrle2007,Boel2008}.

A commonly used approach to model muscle contraction in continuum-mechanics is to additively compose the stress tensor of a passive and an active stress term \cite{blemker2005three,Johansson2000,Roehrle2008}. The passive muscle behavior can be parametrized using experimentally found relations, though this is challenging in practice \cite{Boel2012,Takaza2013,VanLoocke2008,VanLoocke2006}.

Multi-scale models exist that combine formulations of continuum-mechanics with a description of electrophysiology \cite{Roehrle2008,Roehrle2012,Heidlauf2013,HernandezGascon2013}.
These models couple various physical phenomena that occur on different temporal and spatial scales on cell, tissue and organ levels, such as subcellular ion dynamics in scales of milliseconds and micrometers and mechanical stresses and electric potentials in scales of seconds and multiple centimeters.% on the whole muscle with a length  scale of tens of centimeters.

Model order reduction techniques and surrogate modeling have been applied to these complex, full models to speed up the computations \cite{Mordhorst2017,Valentin2018}.

EMG signals of activated muscles can be computed by volume conductor models \cite{Mesin2013}. Both 
 analytic \cite{Dimitrov1998, Farina2001, Mesin2006} and numerical methods exist \cite{Lowery2002, Mordhorst2015, Mordhorst2017, Klotz2020}.

We combine existing multi-scale models of electrophysiology, muscle contraction and generation of EMG with different subcellular models and electrophysiology formulations as well as motor neuron and afferent feedback models to form a novel, comprehensive multi-scale modeling framework for the neuromuscular system. We solve these models using numerically efficient schemes that are implemented in our unified simulation software environment named OpenDiHu.
 
\subsection{Related Software}\label{sec:intro_related_software}
Few software packages exist in the open source world that can be used for comprehensive multi-scale modeling of the neuromuscular system. In the following, CellML, OpenCMISS, Chaste, FEBio, and the generic frameworks OpenFOAM and FEniCS will be briefly evaluated.

A useful and widespread technology for biochemical models is CellML \cite{Cellml2003,Lloyd2004}. The open standard CellML language allows defining differential-algebraic equations with physical units. It can be used to develop mathematical descriptions of biophysical processes such as subcellular or neuron models. The description language has also been used for broader applications, e.g., for constitutive material laws. 

CellML provides an online repository where mathematical models and metadata such as figures and related publications are collected. The models can be downloaded as code in various formats and programming languages. Existing models can be combined into new models in a hierarchical manner. Dedicated modeling environments for CellML models exist. As an example, OpenCOR \cite{OpenCOR2015} can be used to edit, simulate and visualize CellML models. Moreover, application programming interfaces (APIs) exist, which provide low-level access to models in CellML format and allow software frameworks to integrate CellML functionality.
Among the software frameworks with CellML support are OpenCMISS and Chaste.

OpenCMISS (Continuum Mechanics, Imaging, Signal processing, and System identification) \cite{Bradley2011} provides a set of open source libraries and applications for modeling and visualization of bioengineering problems. The frameworks allow using CellML models \cite{Nickerson2014}.
OpenCMISS Iron, the computational engine, can solve finite element models, discretized also with higher order elements and using Cartesian or curvilinear coordinates. For example, the ventricles of the heart were modeled with a low number of cubic Hermite elements in a prolate spheroidal coordinate frame \cite{smith2004multiscale}.
Various nested timestepping loops and solvers can be configured to create multi-scale models. 

The library is programmed largely in the Fortran-90 standard, wrappers for the Python programming language can be automatically generated. It supports parallel execution on distributed memory systems.

The development of OpenCMISS Iron started in 2005 as a rewrite of the computational modeling tool CMISS, whose history dates back to 1980. It is part of the Physiome Project, an international collaborative open-source effort to provide a public domain framework for computational physiology \cite{Hunter2004}.
OpenCMISS has been used for multi-scale modeling of the lungs and heart \cite{smith2004multiscale}, vascular and thermoregulatory system \cite{ladd2016open,ghadam2020modeling} and skeletal muscle \cite{Heidlauf2013}. 

% Chaste
The \say{Cancer, Heart, and Soft Tissue Environment} (Chaste) is an open source C++ library targeted at simulations of physiology and biology in general \cite{Chaste2013}. The code development is driven by cardiac electrophysiology and cancer growth simulations, but the framework is also capable of solving ordinary and partial differential equations from other fields. This involves solvers for CellML models, which have been used to simulate cellular cardiac electrophysiology \cite{ChasteCellML2015}.

Chaste \cite{ChasteCellML2015} also uses the approach of first converting a CellML description into C++ code using the tool \emph{PyCml} \cite{Cooper2006}. Chaste features adaptive timestepping solvers such as the \emph{CVODE} solver from the \emph{SUNDIALS} package \cite{cohen1996cvode} and infers analytic Jacobians from the model equations. The CellML support of Chaste targets \say{automated use} by automatically inferring standard variable names, e.g., for membrane voltage and stimulation current.

While cardiac and skeletal muscle tissue is similar with respect to its multi-scale structure, significant differences exist regarding electrophysiology and recruitment of MUs. On cardiac tissue, propagation of action potentials occurs uniformly on a three-dimensional domain, whereas in the skeletal muscle, a multitude of electrically isolated one-dimensional muscle fibers are recruited independently.
Thus, significant development efforts are needed to transform a cardiac simulation into a simulation of skeletal muscles.

In contrast to OpenCMISS Iron, Chaste advertises its test-driven development process to ensure code quality, correctness and reusability \cite{Chaste2009}. Similar to Iron, the Chaste code runs in parallel on distributed memory systems and uses external numeric libraries for linear system solvers.
It also implements a solver for 3D incompressible nonlinear elasticity, which is needed for simulating muscle contraction. However, this solver is not yet parallelized.

A simulation tool specialized in the field of biomechanics is the FEBio project \cite{Maas2012,maas2017febio}. It provides an advanced finite element solver for continuum mechanics of muscle tissue and implements a well-documented library of material models, from basic to advanced and state-of-the-art models. The most recent version includes graphical pre-processing and post-processing tools. Whereas OpenCMISS Iron and Chaste require some knowledge of programming and command line usage, FEBio can be used right away also by application scientists. 
Unlike OpenCMISS and Chaste, FEBio only runs in parallel on shared memory computers, which makes it unsuited for High Performance Computing. FEBio contains no electrophysiology models. Prescribing different levels of activation at different locations in a muscle currently is only possible by a workaround of defining separate materials for every finite element. However, FEBio is extensible by user-defined plugins, and multi-scale models would have to be implemented in this way.

More generic simulation frameworks exist that can also be considered for simulations of the musculoskeletal system.

OpenFOAM \cite{jasak2007openfoam} is a well-known C++ software framework that provides methods for \say{Field Operation And Manipulation}. It is mainly designed for continuum mechanics problems in the field of computational fluid dynamics and uses the finite volume method. This method can also be used to solve nonlinear solid mechanics problems \cite{cardiff2014nonlinear}.

Another established general framework for solving partial differential equations is $\text{FEniCS}$ \cite{alnaes2015fenics}. It provides a high-level Python interface to directly describe the model in variational form using predefined operators. Then, it derives finite element discretizations, which it is also able to solve in parallel.

Advantages of such generic frameworks are their mature and efficient solvers and infrastructure such as output file formats and their comprehensive documentation and support. Disadvantages are the missing domain-specific functionality. For example, no solvers for CellML models exist in OpenFOAM and FEniCS. For FEniCS, all existing parts of the desired multi-scale model would have to be formulated in the unified form language. This task needs a more in-depth understanding of the framework for the special requirements of the complex model, e.g., the dynamic, incompressible, nonlinear solid mechanics muscle contraction model with active stress contribution, for which mixed finite element formulations and possibly special numerical preconditioners and solvers are needed. This relativizes the advantages of the high-level interface.
Furthermore, in generic frameworks, it is more difficult to bring own problem-specific contributions to the core code trunk to make them publicly available and reusable.

Therefore, we select OpenCMISS Iron as the starting point for implementing multi-scale models. We use the multi-scale chemo-electro-mechanical model that was introduced in \cite{Roehrle2012} and initially implemented in the software OpenCMISS for the tibialis anterior muscle \cite{Heidlauf2013}. However, this implementation did not fully exploit the parallel capabilities of Iron as it was hard-coded for four processes. We remove this restriction and further improve runtimes of the existing electrophysiology model by implementing second order timestepping schemes.

We evaluate the performance regarding parallel scaling and memory consumption on a supercomputer. While the performance is good for small degrees of parallelism, we see an unavoidable barrier for larger parallelism and High Performance Computing (HPC). This barrier arises due to fundamental design decisions in the memory layout and is difficult to overcome in the existing OpenCMISS Iron code.

Furthermore, some of the functionality we require for our multi-scale framework is not available: Arbitrary data mapping between meshes is needed for a 3D muscle domain with embedded 1D fibers. Output files use a text-based format that is not suited for HPC, established parallel file formats such as defined by VTK or ADIOS are not available. The nonlinear mechanics solver can only solve static problems. At the same time, the implementation assumes generic coordinate frames and element shapes, which are not necessarily needed in our models. Due to the lack of modern programming language features such as object-orientation and polymorphism, it is very difficult to extend the solid mechanics solver to a fully dynamic formulation.

Therefore, we move the existing models to the new code base \emph{OpenDiHu} and expand the multi-scale framework by new model components. This gives us the flexibility to implement all requirements listed in \cref{sec:challenges_in_silico} and target towards HPC from the beginning. For compatibility, OpenDiHu can write the same output file format as OpenCMISS. Furthermore, an adapter in OpenDiHu allows to integrate the FEBio mechanics solver with the electrophysiology solver of OpenDiHu.

% our multi-scale approach:
\section{The Multi-scale Model of the Neuromuscular System}\label{sec:the_multi_scale_model_of}

Next, we briefly describe the existing multi-scale model framework that we base our work on.
The chemo-electro-mechanical model was introduced by \cite{Roehrle2012} and described in more detail in \cite{Heidlauf2013} and \cite{Heidlauf2015Diss}. It has been used to investigate different muscle fibers lengths \cite{Heidlauf2014} and later was enhanced by also modeling actin-titin interactions \cite{Heidlauf2016}. The work of \cite{Mordhorst2015} extended the framework by a description of EMG signals.

This model reflects the structural and functional aspects of skeletal muscle tissue and describes its mechanical and electrophysiological properties. Different biophysical processes are realized by sub-models that are linked together to form the overall multi-scale and multi-physics model.

\begin{figure}
  \centering%
  \includegraphics[width=\textwidth]{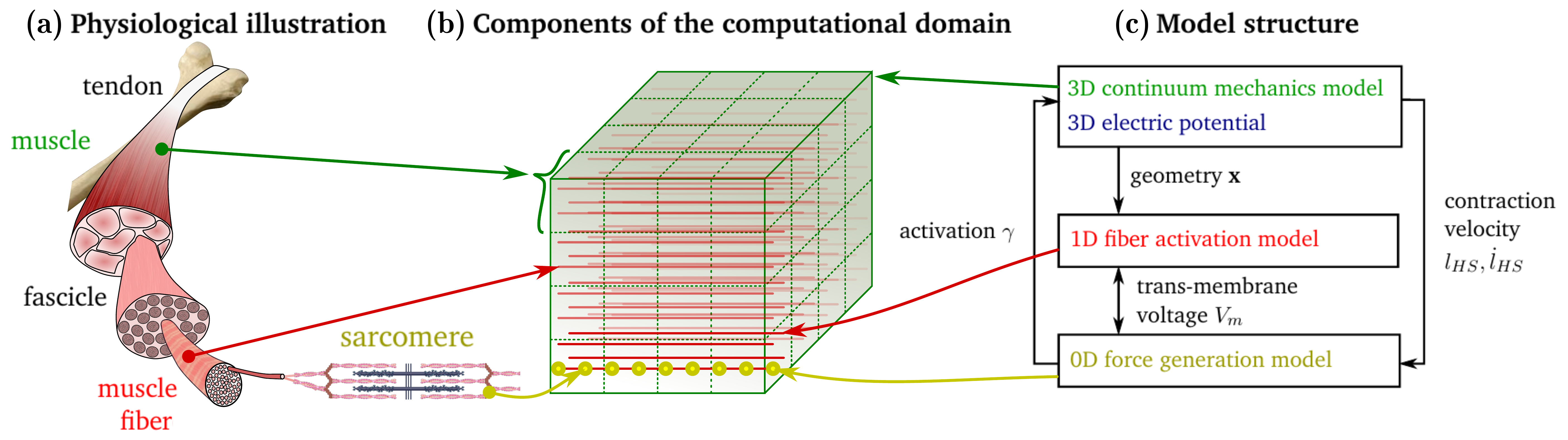}%
  \caption{Modeling skeletal muscle physiology: From the anatomy (a) over a multi-scale discretization (b) to the multi-physics model (c) of \cite{Roehrle2012}.}%
  \label{fig:model_scheme_overview_full}%
\end{figure}%

\Cref{fig:model_scheme_overview_full} visualizes the model structure. \Cref{fig:model_scheme_overview_full} (a) depicts the hierarchical skeletal muscle anatomy consisting of muscle, muscle fibers and sarcomeres. \Cref{fig:model_scheme_overview_full} (b) shows the finite element discretization of these three scales. The muscle is represented by a 3D mesh of hexahedral elements (green). Muscle fibers are modeled as 1D fiber meshes (red) that are embedded in the muscle domain. The nodes of the fiber meshes are locations of 0D sarcomere models (yellow). \Cref{fig:model_scheme_overview_full} (b) depicts them only for one fiber, however, the nodes of all fibers feature instances of this model. The cube-shaped 3D domain was chosen for the sake of a clear visualization, our simulations use real muscle geometries instead.

\Cref{fig:model_scheme_overview_full} (c) shows details of the model parts and their exchanged physical quantities. In summary, the model consists of 3D, 1D, and 0D components, which are given in different colors.
The green, blue, red, and yellow colors are used throughout this work to indicate these sub-models or the three different spatial scales.

The continuum mechanics model describes muscle contraction and is defined on the 3D mesh. The same mesh is also used for computing the 3D electric potential fields within the volume. The deforming 3D muscle domain defines the geometry, i.e., node positions $\bfx$ of the embedded 1D fibers meshes. The 1D fiber activation model computes the propagation of action potentials on every fiber. It is strongly coupled via the transmembrane voltage $V_m$ to the 0D force generation model on the sarcomeres, which is also called the subcellular model. It depends on the length $l_\text{HS}$ of the half-sarcomere and the contraction velocity $\dot{l}_\text{HS}$. These quantities are computed in the 3D model and mapped to the 0D points. The result of the subcellular model is the activation parameter $\gamma$ that is homogenized and used as input for the active stress term in the 3D continuum mechanics model.

\section{Contributions and Scope of This Work}\label{sec:intro_contributions}

In the following, we give a summarizing preview of the main contributions of this work to the world of existing in-silico models and tools. The contributions include:
\begin{enumerate}[label=(\roman*)]
  \item \textbf{Model extensions.} In addition to the previously existing model components depicted in \cref{fig:model_scheme_overview_full} (c), we add a mesh for adipose tissue to simulate EMG signals on the skin surface. The corresponding model was formulated by \cite{Mordhorst2015}, however, it has not been implemented together with the other components in a simulation program prior to our work.

Furthermore, we add the multidomain model \cite{Klotz2020}, an alternative homogenized 3D description of electrophysiology that can replace the 1D fiber activation model and the 3D electric potential in \cref{fig:model_scheme_overview_full} (c). 

Recruitment of the muscle fibers was previously done in a preprocessing step by simulating motor neuron models such as \cite{Cisi2008,Negro2011}. In OpenDiHu, we explicitly couple models of the motor neuron pool as well as models of sensory organs such as muscle spindles and Golgi tendon organs. This allows us to close the loop of afferent neural feedback.

Another extension is the consideration of tendons together with the contracting muscle. We add separate models and meshes for the tendons that are mechanically coupled to the muscle belly.

The previous quasi-static mechanics formulation in OpenCMISS Iron is also implemented in OpenDiHu and extended to a fully dynamic formulation. Instead of a numerical approximation of the Jacobian matrix in the nonlinear system in Iron, we automatically derive an analytic description in OpenDiHu. This significantly reduces the runtime and allows simulating finer meshes than is possible with OpenCMISS.

Current limitations among the implemented models are convergence difficulties for solid mechanics problems with more than approximately 1000 elements. These are not a result of the implementation but a numerical problem and could be addressed by different nonlinear solver schemes in the future.

Whereas the computation using the fiber based description of electrophysiology is close to its optimum performance and scales near-optimally for any degree of parallelism and number of fibers, the corresponding multidomain implementation exhibits high memory consumption, is less robust with respect to numerical errors and more difficult to parallelize. This restricts its application to approximately 20 motor units, 128 processes, and timespans of below a second. If scenarios above these limits are required, the fiber based models should be used.

\item \textbf{Preprocessing Algorithms.} 
A serial and a parallel algorithm are developed to generate the high-quality 1D and 3D meshes that are required for the simulation from imaging data. The algorithms are applied on the biceps and triceps brachii muscles.

Furthermore, a method is derived to assign fibers to motor units according to physiological properties. Both implementations are made publicly available together with the open-source software OpenDiHu.
\item \textbf{Improvements in OpenCMISS Iron.} Improvements include the parallelization to an arbitrary number of cores of the existing implementation of the chemo-electro-mechanical model, an algorithmic improvement from quadratic to linear time complexity in the homogenization functionality of the activation parameter and the introduction of configuration files such that different parametrizations can be simulated without recompiling. Further, numerical experiments concerning employed solvers and timestep widths are conducted. The revised choices, e.g., a conjugate gradient scheme instead of the GMRES solver lead to faster computation times.
\item \textbf{Development of OpenDiHu.} Implementation of an efficient, flexible framework for simulating the full models of surface EMG, muscle contraction as well as subsets of the mathematical multi-scale modeling framework. The software can employ CPUs and GPUs and run on small workstation computers, compute clusters, and supercomputers.
As this is the most comprehensive contribution, we refer to the implementation and results chapters, \cref{sec:implementation,sec:results} for details.
Highlights are the simulation of EMG signals with \num{270000} muscle fibers on \num{27000} cores of the supercomputer Hazel Hen and an overall speedup factor of 200 with respect to the community standard software OpenCMISS Iron.
\end{enumerate}

\begin{figure}[H]
  \centering%
  \begin{subfigure}[t]{0.485\textwidth}%
    \centering%
    \includegraphics[width=\textwidth]{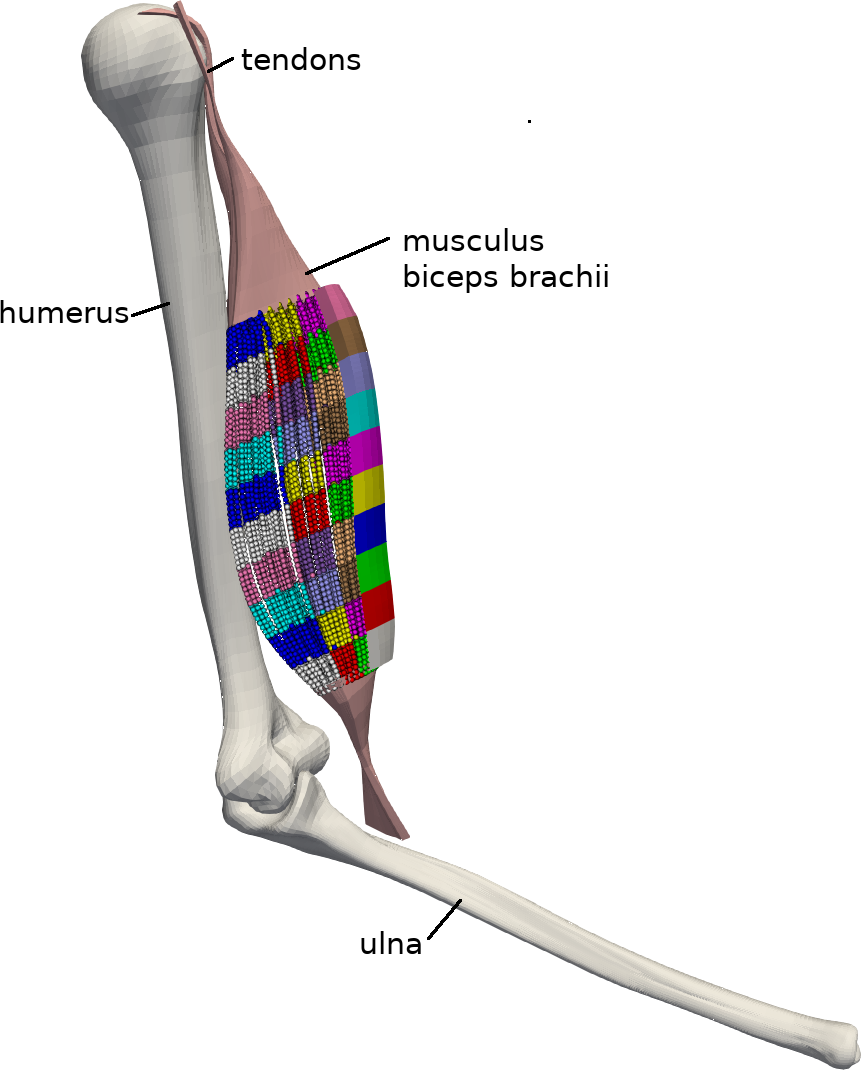}%
  \caption{The bones of the upper arm with tendons and muscle tissue of the biceps brachii muscle. The colored patches show the domain decomposition of the muscle and of the body fat layer domains.}%
    \label{fig:partitioning_names}%
  \end{subfigure}
  \,
  \begin{subfigure}[t]{0.495\textwidth}%
    \centering%
    \includegraphics[width=\textwidth]{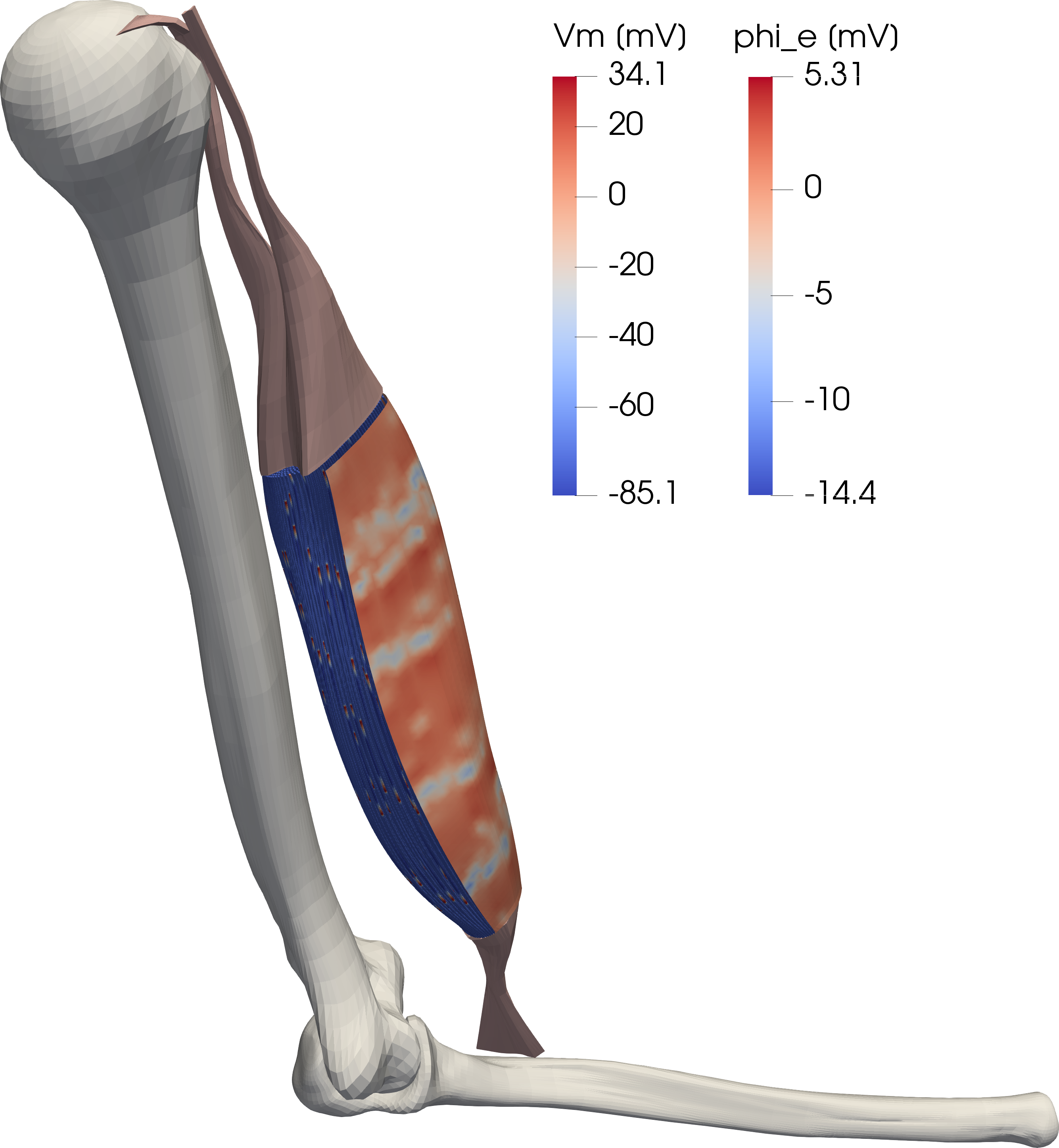}%
  \caption{Simulation of action potential propagation on the muscle fibers (mainly blue, value $V_m$ according to legend) and EMG on the skin surface (value $\phi_e$).}%
    \label{fig:full_muscle_emg}%
  \end{subfigure}   
  \caption{Preview on setting and simulations of a biceps muscle in this work.}%
  \label{fig:partitioning_and_full_muscle_emg}%
\end{figure}

\Cref{fig:partitioning_and_full_muscle_emg} shows two snapshots of a simulation that are characteristic for this work: \Cref{fig:partitioning_names} depicts the biceps brachii muscle with a body fat layer. The muscle belly and the fat layer are discretized by nodes and partitioned to multiple processes, indicated by different colors.
\Cref{fig:full_muscle_emg} shows a simulation of EMG signals on the skin surface. Solutions of the electrophysiology models can be seen on the fibers and on the surface above the muscle.

% Partitioning and Muscle emg

The scope of this work is to efficiently compute the described models and provide an environment to carry out processing and investigations. The models themselves, as well as their parameters, are taken from the literature. It is known that the properties of human organs vary greatly between individuals. For example, the number of muscle fibers in a biceps muscle varies between \num{172000} and \num{419000} \cite{MacDougall1984}.
By parameter fitting, the simulations could be adjusted to represent a particular individual. Within this work, this was done in the initial case study about the upper arm movement for a data-based and a Hill-type based model. However, parameter fitting for the multi-scale model and validation experiments or even preclinical studies on patients with musculoskeletal diseases are beyond the scope of this work. Similarly, our software could be used to design studies that foster the understanding of the neuromuscular system. However, such investigations are also beyond the scope of this thesis.

The developed methods in this work were applied to simulations of the biceps brachii muscle, as this muscle allows straightforward EMG recordings and is well-studied in literature. Nevertheless, most of the methods and results are also applicable to other muscles. The anatomical match of the used simulation models could be improved by additionally considering the aponeurosis in the biceps muscle or by differentiating between the two muscle heads during motor unit recruitment. However, these model extensions are also not within the scope of the present work.
Some notes for future work can be found in \cref{sec:future_work}.

% explain "we"
The presented findings and conclusions were partly shaped by discussions with various researchers with expertise from different disciplines. Yet, this doctoral thesis lists essentially own contributions, marks collaborative work in the text, and indicates others' work by citations.
Using the pronoun \say{we}, the author refers to the group of the originator, potentially the supervisors, and certainly the interested reader.

% overview of remainder
The remainder of this work contains the following chapters:
\Cref{chap:comparative_study} compares two model approaches to simulate upper arm movement. \Cref{sec:generation_of_meshes_for_multiscale} develops algorithms to generate the meshes that are required in the solution of the multi-scale model. \Cref{sec:muscle_fibers_and_motor_units} addresses the assignment of motor units to muscle fibers. \Cref{chap:models_and_discretization} describes all used model equations and their discretization. \Cref{chap:usage} introduces the software OpenDiHu and describes its usage. \Cref{sec:implementation} gives details on the implementation of OpenDiHu. \Cref{sec:results} presents and discusses numerical results. \Cref{sec:performance_analysis} studies the computational performance of the solvers. \Cref{sec:conclusion_and_future_work} concludes the work and gives an outlook to future work.

%\cref{chap:introduction} % #1 introduction
%\cref{chap:comparative_study} % #2 comparative study
%\cref{sec:generation_of_meshes_for_multiscale} % #3 generation of  meshes
%\cref{sec:muscle_fibers_and_motor_units} % #4 motor units
%\cref{chap:models_and_discretization} % #5 motor units
%\cref{chap:usage} % #6 usage
%\cref{sec:implementation}  % #7 implementation
%\Cref{sec:results} % #8 numerical results and discussion
%\cref{sec:performance_analysis}  % #9 performance analysis
%\cref{sec:conclusion_and_future_work} % #10 conclusion

% ---- end ----

%Research targets (usually one paragraph):
%  State your hypothesis or research question
%  Briefly describe how you will accomplish your aims
%  Give a preview of your main results and state the contribution of the work (optional)

% Contributions
% Limitations (Future Work), Grenzen, Abgrenzung (nicht Gehirn, nicht im Fokus)

%Paper overview (optional; one paragraph):
%  Give a section-by-section overview of the paper's contents
% struktur über weitere Kapitel
% ca. 4-6 Seiten

\chapter{Comparative Study: Modeling Upper Arm Movement}\label{chap:comparative_study}

Moving one's upper arms and forearms is an action that is performed unnoticed every day. 
What seems like a trivial task involves a sophisticated interplay of muscles, tendons, bones and joints. Macroscopic behavior such as mechanical properties of fibers and tissues as well as microscopic mechanisms such as molecular-scale processes inside biological cells and changes in electric potential across muscle fiber membranes contribute to the overall, versatile human musculoskeletal system.

Understanding this system well allows to use observations to make predictions.
Using observations and predictions of the musculoskeletal system, we can for example design safe assistive robotic devices.
Such robotic devices, in the form of exoskeletons, can potentially support humans in strenuous, unhealthy tasks that pose high loads on the human skeleton. An example is the precise handling of heavy objects that can only be done by humans, which is required in various industries. Moreover, exoskeletons can help to restore muscle function in a rehabilitation therapy.

In order to develop models for such predictions, various choices have to be made. 
Relevant properties of the muscular system have to be identified. Based on a physiological understanding, essential relationships have to be selected. Phenomenological relations can be incorporated. Mathematical formulations and numerical algorithms have to be found.

Model formulations can be differentiated by how much they are based on biophysical insights compared to raw experimental observations. The following sections present two different approaches that use a relatively high proportion of experimental observations combined with some biophysically justified relations. The two approaches are compared in an experimental study of forearm movement. 

The first of the two presented models completely relies on experimental data. The second model adds physiological knowledge at a high level. In the remainder of this thesis after the current chapter, this trend continues: More details of the functioning of the musculoskeletal system get included. More advanced models are introduced that have finer model resolutions.

\section{Introduction}
% definition of the task: movement prediction EMG->torque
Actuated orthoses and prosthesis help rehabilitation patients to regain their ability to move arms and legs when muscles have lost their full function. The dysfunction can be a result of muscular or nervous diseases, e.g., after a stroke, or originate from amputation of parts of the limb \cite{Krebs2002, Zhang2018}.

Rehabilitation or prosthetic devices are firmly attached to the body. Powered actuators at the joints support or replicate the natural movements of the limb.
Exoskeletons are similar devices that typically extend to a larger part of the human body. Apart from rehabilitation, they are used as assistive, haptic or teleoperation device. Details can be found in \cite{Perry2007}.

For the actuation to be supportive and helpful, the device has to determine the intended movement of the limb. If muscles are functioning at least residually, EMG signals can be captured from the skin surface. They can be interpreted to determine finely graduated levels of force.

For this purpose, mathematical models are required that, given EMG measurements, predict joint torques for the system of limb segments and muscles. 
Because of the variety of muscle characteristics among humans, such models have to be patient-specific in order to be safe and effective for the particular individual.

In the following study, two different approaches for formulating such models, A and B, are developed. Instances of these two models are parametrized for a particular healthy subject. 

The specific task in this study is to predict movements of a human upper arm. The arm is flexed and extended under varying loads and with varying velocities. Measured EMG signals on the agonist and antagonist muscles are used to predict the torque in the elbow joint. Considering the application of a supportive orthosis or exoskeleton, the predicted torque value is the control input to the actuator at the joint.

Prior to online application of the models, an offline training-phase is carried out, where all required parameter values get identified for the subject. After the two computational models have been trained, we perform validation experiments and compare the output of the models with measurements from the real system.
 
The first model approach, A, is a non-parametric, data-driven model. It uses the captured information from the training phase to construct a map between input and output values. It is based on Gaussian Process Regression.

The second approach, B, uses biophysically informed models of individual muscles together with the kinematics of the overall system. This approach requires a set of subject-specific parameters which is determined in the training phase. The model is based on the commonly used Hill-type muscle model.

\subsection{Related Works}

%The authors of \cite{Perry2007} develop a cable-actuated exoskeleton for the arm with 7 degrees of freedom. It supports the full natural movements of shoulder, elbow and wrist.
%
Numerous experimental studies of flexion and extension of the upper arm with the aim to predict elbow torques can be found in the literature. The studies presented in the following all include a Hill-type muscle model; such a model is also present in approach B of the present study.

In \cite{Rosen1999}, an exoskeleton across the elbow joint on the forearm is used as a passive measurement device. Experiments with lifting weights are performed and EMG is captured.
Two different models are compared with respect to their performance in predicting moments and, thus, their suitability for exoskeleton control.
The first model is Hill-based, similar to model B in the study of this work.
The second model is data-driven, as is model A in our study. However, the method is different, the authors use a neural network.

The study reveals that the neural network is easier to set up but only works for the space defined by the learning data set. The advantage of the Hill-based model is that it is universal and not task dependent. 
Further studies using neural networks to estimate muscle activations and elbow torques are presented by \cite{Wang2002} and \cite{Song2005}.

The paper of \cite{Rosen2001} focuses on an exoskeleton that supports the forearm in lifting heavy weights. A generic Hill-type model is the base for the model predictions. Different control strategies are investigated. 
A naturally feeling human machine interface is achieved when
control input is taken from processed EMG measurements and moment feedback of the external load. 
This result is promising as it shows that neural control of exoskeletons is possible, even using non-customized models. 

The goal and setup of all these studies is similar to the work presented in the following sections. Differences are, apart from different setups, that they use state-less Hill-type models instead of the Hill-based models in our study that are more advanced. Furthermore, they do not use subject specific parametrizations. Both improvements can lead to better predictions of the moments in the elbow.

However, several studies predicting joint torques using Hill-type models exist in the literature that optimize model parameters to fit a specific subject.
The authors of \cite{Cavallaro2005, Cavallaro2006} study a scenario where a weight is lifted by the forearm. They use a genetic algorithm to find subject specific parameters to the models. Similar studies are given in \cite{Lloyd2003,Venture2005,Pontonnier2009,Sartori2012}.

%\cite{Chadwick2014} present a real-time simulation of movements of the upper limb with eleven degrees of freedom. The input to the model consists of the muscle activation levels.
A further study is performed by \cite{Heine2003}. The authors include models of activation dynamics, Hill-type muscle contraction and musculoskeletal geometry and restrict the scenario to isometric tasks. They optimize parameters for different subjects and determine the importance of parameters for good model predictions. It is found that the predictive quality of the model decreases with its complexity, but a model with seven parameters still has reasonable validity. 
In contrast to this study that only predicts static cases, our study also includes muscle dynamics and has more parameters to describe all required muscle properties.

\cite{Falisse2016} estimate muscle model parameters of the knee joint actuators involving 23 degrees of freedom considering eight flexors and four extensors. Just like our study with model approach B, EMG signals and motion capture data are used to solve an optimization problem to fit the model. A difference is that three-element Hill-type models are used, whereas our study is based on more detailed, four-element Hill-type models but includes a smaller number of muscles.

Hill-based models of the muscle-tendon complex can also be parametrized without using EMG data.
\cite{Garner2003} estimate characteristic parameters of 26 major muscles around shoulder, elbow and  wrist in a two-phase optimization procedure. This approach uses individual experiments to identify different parameters. A method that requires fewer experiments is the ISOFIT method presented by \cite{Wagner2005}. They use non-linear regression to fit Hill-type model parameters for various muscles from only 6-8 isovelocity contractions.

The authors of \cite{Campen2014} develop a new method for estimating a subject-specific model of muscles around the knee which achieves higher accuracy than \cite{Garner2003} and is robust with respect to noisy data. Two improvements are that they use physiological constraints in the parameter optimization process and a heuristic for the initial guess of the parameters. The former is also done in our study. Concerning the latter, we base our initial guess on literature values which, too, is an improvement compared to using random initial values.

\subsection{Contribution Statement}

The work presented in this section was performed in collaboration with several members of the IRTG \say{Soft Tissue Robotics}. A summary can be found in the publication \cite{summerschool2019}, where also the list of contributors is given.
The experiments were conducted during the IRTG's Summer School 2019 at the University of Auckland, New Zealand, where all involved researchers were present. Derivation and programming of the models as well as evaluation and discussion of the results was done in smaller expert groups as well as plenary video calls before and after this event at both the Universities of Stuttgart and Auckland.

\begin{figure}%
  \centering%
  \includegraphics[width=0.8\textwidth]{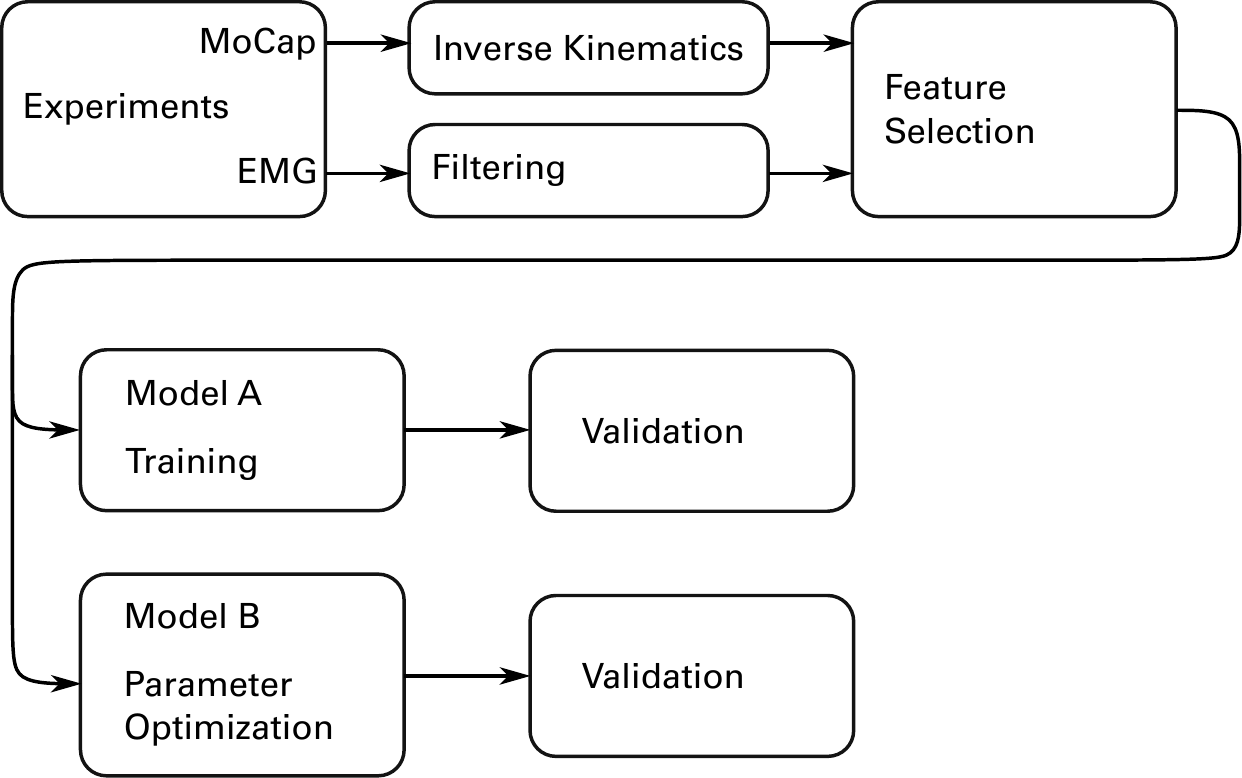}%
  \caption{Upper Arm Movement Modeling:\\ Schematic workflow of data processing. In experiments, Motion Capture (MoCap) data and EMG signals were recorded. They were processed using inverse kinematics and filtering techniques, respectively. The size of the large datasets was reduced by selecting specific features. Those were used as training inputs for the two models A and B. The trained models were validated using a validation dataset that also originated from experiments.}%
  \label{fig:schematitc}%
\end{figure}

An overview of the workflow from experiments to the models A and B and their validation is visualized in \cref{fig:schematitc}.
I contributed mainly to the fields of data processing, especially filtering of EMG signals and feature selection, derivation and training of both models, A and B, their validation and the overall programming and visualization of the results. The respective fields are presented more detailed in the following, corresponding sections.

\subsection{Structure of this Chapter}
\Cref{sec:exp_study} gives an overview of the experiments, data processing and feature selection, which resulted in the required datasets. In \cref{sec:study_models}, the two models, A and B, are described. Results including the validation of the models and a discussion are given in \cref{sec:evaluation}. Conclusions follow in \cref{sec:study_conclusion}.

\section{Experimental Study}\label{sec:exp_study}

\begin{figure}%
    \centering%
    \def\svgwidth{5cm}%
    %% Creator: Inkscape inkscape 0.92.3, www.inkscape.org
%% PDF/EPS/PS + LaTeX output extension by Johan Engelen, 2010
%% Accompanies image file 'summer_school_study.pdf' (pdf, eps, ps)
%%
%% To include the image in your LaTeX document, write
%%   \input{<filename>.pdf_tex}
%%  instead of
%%   \includegraphics{<filename>.pdf}
%% To scale the image, write
%%   \def\svgwidth{<desired width>}
%%   \input{<filename>.pdf_tex}
%%  instead of
%%   \includegraphics[width=<desired width>]{<filename>.pdf}
%%
%% Images with a different path to the parent latex file can
%% be accessed with the `import' package (which may need to be
%% installed) using
%%   \usepackage{import}
%% in the preamble, and then including the image with
%%   \import{<path to file>}{<filename>.pdf_tex}
%% Alternatively, one can specify
%%   \graphicspath{{<path to file>/}}
%% 
%% For more information, please see info/svg-inkscape on CTAN:
%%   http://tug.ctan.org/tex-archive/info/svg-inkscape
%%
\begingroup%
  \makeatletter%
  \providecommand\color[2][]{%
    \errmessage{(Inkscape) Color is used for the text in Inkscape, but the package 'color.sty' is not loaded}%
    \renewcommand\color[2][]{}%
  }%
  \providecommand\transparent[1]{%
    \errmessage{(Inkscape) Transparency is used (non-zero) for the text in Inkscape, but the package 'transparent.sty' is not loaded}%
    \renewcommand\transparent[1]{}%
  }%
  \providecommand\rotatebox[2]{#2}%
  \newcommand*\fsize{\dimexpr\f@size pt\relax}%
  \newcommand*\lineheight[1]{\fontsize{\fsize}{#1\fsize}\selectfont}%
  \ifx\svgwidth\undefined%
    \setlength{\unitlength}{141.45847951bp}%
    \ifx\svgscale\undefined%
      \relax%
    \else%
      \setlength{\unitlength}{\unitlength * \real{\svgscale}}%
    \fi%
  \else%
    \setlength{\unitlength}{\svgwidth}%
  \fi%
  \global\let\svgwidth\undefined%
  \global\let\svgscale\undefined%
  \makeatother%
  \begin{picture}(1,0.85612663)%
    \lineheight{1}%
    \setlength\tabcolsep{0pt}%
    \put(0,0){\includegraphics[width=\unitlength,page=1]{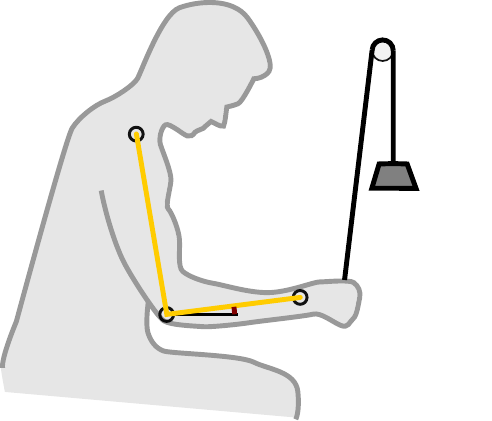}}%
    \put(0.74081946,0.3253194){\color[rgb]{0.50196078,0,0}\makebox(0,0)[lt]{\lineheight{1.25}\smash{\begin{tabular}[t]{l}$\phi_w$\end{tabular}}}}%
    \put(0,0){\includegraphics[width=\unitlength,page=2]{summer_school_study.pdf}}%
    \put(0.494186,0.20649992){\color[rgb]{0.50196078,0,0}\makebox(0,0)[lt]{\lineheight{1.25}\smash{\begin{tabular}[t]{l}$\phi_a$\end{tabular}}}}%
    \put(0.36400524,0.15111369){\color[rgb]{0.50196078,0,0}\makebox(0,0)[lt]{\lineheight{1.25}\smash{\begin{tabular}[t]{l}$\phi_e$\end{tabular}}}}%
    \put(0,0){\includegraphics[width=\unitlength,page=3]{summer_school_study.pdf}}%
    \put(0.85731572,0.48402634){\color[rgb]{0,0,0}\makebox(0,0)[lt]{\lineheight{1.25}\smash{\begin{tabular}[t]{l}$m_w$\end{tabular}}}}%
    \put(0,0){\includegraphics[width=\unitlength,page=4]{summer_school_study.pdf}}%
    \put(0.42748236,0.33727716){\color[rgb]{0,0,0}\makebox(0,0)[lt]{\lineheight{1.25}\smash{\begin{tabular}[t]{l}$\ell_u$\end{tabular}}}}%
  \end{picture}%
\endgroup%
    \caption{Upper Arm Movement Modeling: \\Experimental setup for the triceps trials. The subject pulls down a rope over a pulley which is connected to a weight with mass $m_w$.
    Angles required for the kinematic formulation are the elbow angle, $\phi_e$, the forearm angle, $\phi_a$, and the angle of the weight, $\phi_w$. The length of the ulna bone is denoted by $\ell_u$.}%
    \label{fig:summer_school_study}%
\end{figure}%

Experiments are required to identify the model parameters for the particular subject. First, the experimental setup is described, then, details on the processing of the measured values are given. Then, the selection of feature points from the experimental data is described.

\subsection{Experimental Trials}

In a series of experiments, eight different actions of flexing and extending the elbow were performed by the subject. Weights of 3 kg and 5 kg were held in the hand during the elbow flexion trials. For the elbow extension trials, a pulley system was installed that redirected the force of the weight such that the downward movement of the forearm acted against the direction of the force. This is shown in \cref{fig:summer_school_study}. A detailed description of the experimental trials can be found in \cite{summerschool2019}.

Time series of position and velocity of the upper arm and the forearm were recorded using a Motion Capture system. It consisted of eight cameras that tracked three markers placed on shoulder, elbow and wrist of the subject. 

The elbow torque $\tau$ was computed as
\begin{equation*}
  \begin{array}{lll}
    \tau = m_w\,g\,\ell_u\,\sin(\phi_w) - m_{a}\,g\,\dfrac{\ell_u}{2}\,\sin(\phi_a),
  \end{array}
\end{equation*}
where $(m_w\,g)$ is the force of the weight, $m_{a}$ is the mass of forearm and hand, $\ell_u$ is the length of the ulna bone and $\phi_a$ and $\phi_w$ are the angles of the forearm and rope, as visualized in \cref{fig:summer_school_study}.

\subsection{Data Processing}
From the captured data, derived quantities of biceps ($B$) and triceps ($T$) muscles were estimated using a geometric model of the upper arm. 
The geometric model is available in the software OpenSim \cite{OpenSim2007} and was customized for the particular subject. The inverse kinematics module of OpenSim was used to estimate the muscle tendon unit lengths, $\ell_{\text{MTU},M}$, contraction velocities, $v_M=\dot{\ell}_M$, and moment arms, $r_{M}$, of the two muscles, $M\in\{B,T\}$.

EMG signals were captured by two electrodes on the skin at the biceps and triceps muscles. For both signals, several preprocessing steps were applied to obtain the inputs for the two models, A and B.

The raw signal was filtered with the same procedure as in \cite{Falisse2016}. First, a fourth-order Butterworth high pass filter with cutoff frequency 30 Hz was applied to reduce non-zero average voltages. Second, the resulting signal was full-wave rectified by taking the absolute value of every measured data point. Third, application of a fourth-order Butterworth low pass filter with 10 Hz cutoff frequency yielded a smoothed signal.
Forth, the resulting filtered EMG signals were normalized to the interval $[0,1]$, such that the value of 1 corresponds to the experimentally determined value of maximum voluntary contraction.

The measured EMG signals on the skin directly correspond to the electric excitation level $u$ in the muscle. Excitation leads to the release of free calcium ions within the sarcomere. Binding of calcium ions to myosin increases the concentration of cross-bridges. This concentration is commonly known as the muscular activation $\alpha$. The muscular activation directly corresponds to the produced force of the muscle \cite{Bayer2017}.

\begin{figure}%
  \centering%
  \includegraphics[width=0.8\textwidth]{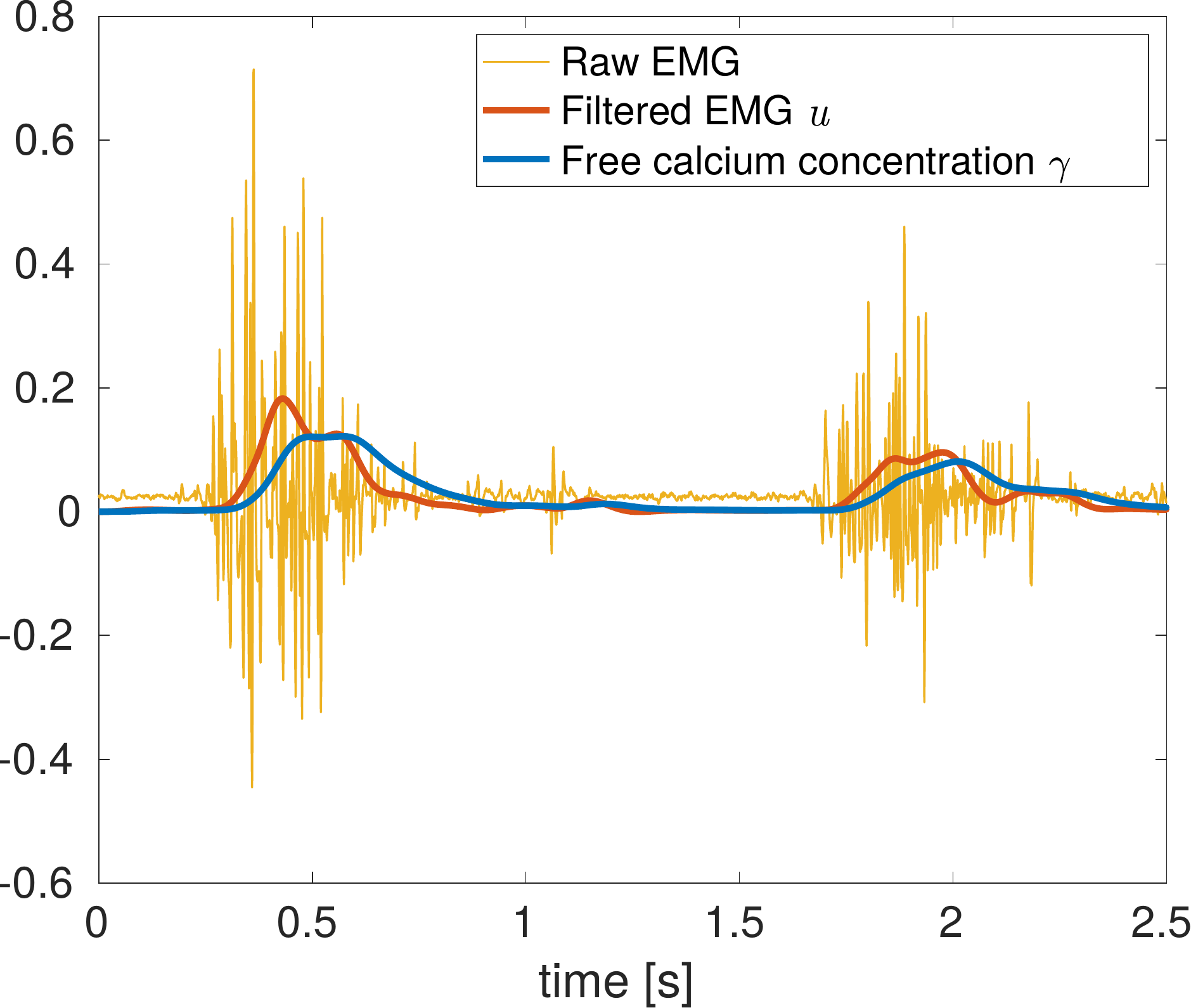}%
  \caption{From raw EMG data of the biceps (yellow) to the filtered signal $u$ (red) and the free calcium concentration $\gamma$ (blue). The data are taken from the beginning of the first elbow flexion experiment. It can be seen that the filtering smooths out the initial signal and removes the constant offset. The free calcium ion concentration follows the filtered EMG with a short delay.}%
  \label{fig:emg_filtering}%
\end{figure}

The concentration of free calcium ions is denoted as $\gamma$ and can be computed from the excitation $u$ by the following first order differential equation \cite{Hatze1977}
\begin{equation*}
  \begin{array}{lll}
    \dot{\gamma} = m\,(u - \gamma).
  \end{array}
\end{equation*}
We used the filtered EMG signal $u$ to obtain values for $\gamma$. \Cref{fig:emg_filtering} shows the raw and filtered EMG signals and the resulting free calcium concentration for a sample of the experimental data.

The activation of the muscle $\alpha$ does not only depend on the free ion concentration $\gamma$ but also on the current state of muscle contraction. This excitation-contraction coupling has to be described by a dynamic system of ODEs and is included in model B.
Therefore, preprocessing is completed with computing the free calcium concentration $\gamma$ and not the activation $\alpha$.

\subsection{Feature Selection} \label{sec:study_feature_selection}

The eight experimental trials were split into $n_\text{trials}=7$ experiments to be used for model identification and one for validation.
The total number $N$ of captured values in the training experiments was large, such that not all points could be used for training of models A and B.
To reduce the amount of data and, thus, speed up the computation, we selected $n \ll N$ featured values with the assumption that they are representative for the whole data set.
For every experimental trial, we choose the same fixed number $n_\text{per\_trial}$ of data points.
Our selection algorithm identifies $n_\text{per\_trial}$ timesteps, $t_i, i=1, \dots, n$ such that the summed values of the free calcium concentrations $\gamma_B(t_i) + \gamma_T(t_i)$ for biceps and triceps are evenly distributed along the value range.
This leads to $n = n_\text{trials} \cdot n_\text{per\_trial}$ selected data points.

The set of training data $\mathcal{D} = \mathcal{X} \times \mathcal{Y}$ consists of the $n$ selected vectors of the experimental values that are the input to the system of muscles, $\bfx_i \in \mathcal{X}, i=1,\dots, n$, together with the observed output values, $y_i \in \mathcal{Y}, i=1,\dots,n$.
The input vectors contain values for muscle tendon unit lengths, contraction velocities, moment arms and free calcium ion concentrations for biceps and triceps each, $\bfx_i = (\ell_{\text{MTU},B}, \ell_{\text{MTU},T}, v_B, v_T, r_{B}, r_{T}, \gamma_B, \gamma_T)^\top(t_i)$. The output values consist of the elbow torques, $y_i = \tau(t_i)$. This data set, $\mathcal{D}$, serves as training input for both models, A and B.

\section{Models}\label{sec:study_models}

The current section describes the two model approaches that can predict elbow torques from experimental input data. \Cref{sec:data_driven_model} introduces the non-parametric, data driven model A. \Cref{sec:biophysical_model} presents the biophysically based model B. It requires a parameter optimization, which is described in \cref{sec:parameter_optimization}.

\subsection{Data-driven Model A}\label{sec:data_driven_model}

The first modeling approach uses a non-parametric model. Such a model approximates the function $f$ that maps from input to output data points. The function $f$ is learned from the training data set. Regression is used to obtain predictions for new data points. In our case, we use a stochastic model that considers the probability distribution of the model function.

We use the method of \emph{Gaussian Process Regression}. A Gaussian process is a collection of random variables such that the joint distribution of every finite subset of these random variables is multivariate normal (Gaussian).
In our example, each input data point in the space of measured values, $\bfx \in \mathcal{X}$ has an associated random variable $f(\bfx)$ that describes the output of the model for this point.

A Gaussian process, $\mathcal{GP}$, is characterized by a mean function $m(\bfx)$ and a kernel function $k(\bfx,\bfx')$ that models the covariance between any pair $(\bfx, \bfx') \in \mathcal{X} \times \mathcal{X}$ of points. Different choices of kernel functions are possible and can depend on hyperparameters $\bfpsi$. Describing observed values $y$ by a Gaussian process distribution can be expressed as
\begin{equation*}
  \begin{array}{lll}
    f(x) \approx y \sim \mathcal{GP}\big(m(\bfx), k(\bfx,\bfx',\bfpsi)\big).
  \end{array}
\end{equation*}
This representation is non-parametric in the sense that no particular parametric form of the function $y=f(\bfx)$ is assumed whose (biophysical) parameters would be determined. Instead, a generic probabilistic model is constructed using the observed function values at measured inputs $\bfx_i \in \mathcal{X}$.
%More specifically, the conditional distribution $p(y|\bfx)$ considered.

Gaussian Process Regression is based on \emph{Bayesian Inference} to update a prior belief of the model to a posterior model using information contained in observations of the process.
The observed data are the set of measurements $\mathcal{D}$. 

The \emph{prior} distribution $p(\bff\mid\mathcal{X},\bfpsi)$ for the vector of function values $\bff$ is described by the Gaussian process,
\begin{equation*}
  \begin{array}{lll}
    p(\bff\mid\mathcal{X},\bfpsi) = \mathcal{N}(\bff\mid\bfm,\bfK),
  \end{array}
\end{equation*}
with mean values $\bfm = (m(\bfx_i))^\top_{i=1,\dots,n}$ and covariance matrix $\bfK$ with $K_{ij} = k(\bfx_i,\bfx_j,\bfpsi).$

The \emph{likelihood} $p(\bfy\mid f(\bfx), \bftheta)$ describes the probability of an observation $\bfy$ given a particular model $f$. The vector $\bftheta$ denotes additional parameters of the likelihood. 

Using Bayes' rule, the \emph{posterior} distribution $p(\bff\mid\mathcal{D})$ of the function values $\bff$ can be computed from prior and likelihood as
\begin{equation*}
  \begin{array}{lll}
    p(\bff\mid\mathcal{D},\bftheta,\bfpsi) 
      = \dfrac{p(\bfy\mid\bff,\bftheta)\,p(\bff\mid\mathcal{X},\bfpsi)}{p(\mathcal{D}\mid\bftheta,\bfpsi)}.
  \end{array}
\end{equation*}
This results in a measure for the uncertainty of the model $f$ at unobserved points $\bfx_\ast \notin \mathcal{D}$. 

Additionally, the fact that the measured quantities in the experiments are subject to measurement noise can be incorporated into the model.
The assumption 
\begin{equation*}
  \begin{array}{lll}
    y = f(\bfx) + \eps
  \end{array}
\end{equation*}
adds a normally distributed random variable of observational noise $\eps \sim \mathcal{N}(0,\sigma_n^2)$ to the formulation. The noise variance $\theta = \sigma_n^2$ is an additional parameter of the likelihood. 
It is also possible to explicitly model the mean function $m(\bfx)$. By replacing the model $f(\bfx)$ by $g(\bfx) = f(\bfx) + \bfh(\bfx)^\top \bfbeta$, i.e.
\begin{align*}
  %\begin{array}{rlrl}
    y &= f(\bfx) + \bfh(\bfx)^\top \bfbeta + \eps, \quad 
    &\text{ with } f(x) &\sim\mathcal{GP} \big(m(\bfx), k(\bfx,\bfx',\bfpsi)\big),\\[4mm]
    && \eps &\sim \mathcal{N}(0,\sigma_n^2),
  %\end{array}
\end{align*}
we allow for a global trend in the data that is formulated in terms of a vector of explicit basis functions $\bfh(\bfx)$ and corresponding coefficients $\bfbeta$.

The algorithm for Gaussian Process Regression involves estimating the following values from the given data during the training phase. The hyperparameters of the covariance function $\bfpsi$, the noise variance $\bftheta$, and the coefficients of the fixed basis functions $\bfbeta$ are determined by solving an optimization problem. The computation involves matrix inversions and has a computational complexity $\O(n^3)$, i.e. is cubic in the number of data points. For details, the reader is referred to the literature \cite{Rasmussen2005,kuss2006gaussian}.

In our study, training of the Gaussian Process of model A was performed using the ready to use implementation provided by MATLAB.
We parametrized the covariance by a squared exponential kernel and used constant basis functions, $\bfh(x) = 1$. We enabled observational noise, its variance $\bftheta = \sigma_n^2$ was found by optimization during training of the model.

%$p(\bff^\ast | \mathcal{D},\mathcal{X}^\ast)$
\subsection{Biophysical Model B}\label{sec:biophysical_model}

Extension of the elbow is governed by the triceps brachii muscle.
During elbow flexion, three muscles are involved: biceps brachii, brachialis and brachioradialis. For simplicity, only biceps brachii, which contributes most of the moment, is explicitly considered in the current study. The effects of the other two muscle are contained in the biceps brachii model in a lumped manner.

Thus, the biophysical model consists of two Hill-type muscle models, for biceps and triceps, respectively. The muscle models are arranged around a hinge joint for the elbow angle. The muscle forces contribute to the torque at the elbow over their respective moment arms.

Hill-type models describe the macroscopic, dynamic mechanical behavior of an entire muscle along a one-dimensional line of action.
The behavior is formulated by phenomenological, mathematical functions that have to be parametrized to fit experimental observations.

%Such models are often used to compute muscle forces in simulations of various movements, e.g. \cite{Siebert2003}, \cite{Kistemaker2006}.
Multiple variants of Hill-type models exist that use various configurations of mechanical elements to consider different properties and functionalities of the muscle. The original model was proposed in \cite{Hill1938}. It contains a contractile element (CE) and two elastic elements, arranged in series and in parallel to the CE.
The authors of \cite{Siebert2008} compare two different approaches using these three elements. The effect of tension in eccentric contractions is added to the Hill-type model by \cite{Till2008}. The authors of \cite{Gunther2007} add a forth, damping element to account for high-frequency damping of the muscle tissue. In \cite{Morl2012}, electromechanical delay is investigated with and without the additional damping element. 

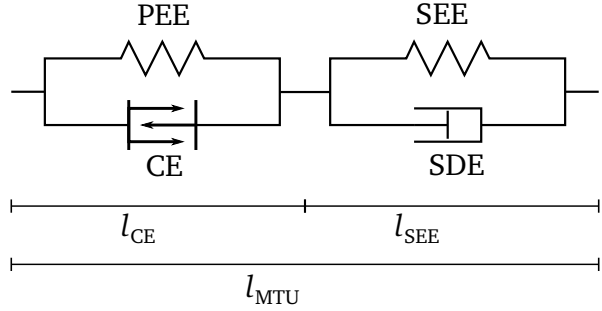
\begin{figure}%
  \centering%
  \def\svgwidth{0.5\textwidth}
  %% Creator: Inkscape inkscape 0.92.3, www.inkscape.org
%% PDF/EPS/PS + LaTeX output extension by Johan Engelen, 2010
%% Accompanies image file 'hilltype.pdf' (pdf, eps, ps)
%%
%% To include the image in your LaTeX document, write
%%   \input{<filename>.pdf_tex}
%%  instead of
%%   \includegraphics{<filename>.pdf}
%% To scale the image, write
%%   \def\svgwidth{<desired width>}
%%   \input{<filename>.pdf_tex}
%%  instead of
%%   \includegraphics[width=<desired width>]{<filename>.pdf}
%%
%% Images with a different path to the parent latex file can
%% be accessed with the `import' package (which may need to be
%% installed) using
%%   \usepackage{import}
%% in the preamble, and then including the image with
%%   \import{<path to file>}{<filename>.pdf_tex}
%% Alternatively, one can specify
%%   \graphicspath{{<path to file>/}}
%% 
%% For more information, please see info/svg-inkscape on CTAN:
%%   http://tug.ctan.org/tex-archive/info/svg-inkscape
%%
\begingroup%
  \makeatletter%
  \providecommand\color[2][]{%
    \errmessage{(Inkscape) Color is used for the text in Inkscape, but the package 'color.sty' is not loaded}%
    \renewcommand\color[2][]{}%
  }%
  \providecommand\transparent[1]{%
    \errmessage{(Inkscape) Transparency is used (non-zero) for the text in Inkscape, but the package 'transparent.sty' is not loaded}%
    \renewcommand\transparent[1]{}%
  }%
  \providecommand\rotatebox[2]{#2}%
  \newcommand*\fsize{\dimexpr\f@size pt\relax}%
  \newcommand*\lineheight[1]{\fontsize{\fsize}{#1\fsize}\selectfont}%
  \ifx\svgwidth\undefined%
    \setlength{\unitlength}{198.95853424bp}%
    \ifx\svgscale\undefined%
      \relax%
    \else%
      \setlength{\unitlength}{\unitlength * \real{\svgscale}}%
    \fi%
  \else%
    \setlength{\unitlength}{\svgwidth}%
  \fi%
  \global\let\svgwidth\undefined%
  \global\let\svgscale\undefined%
  \makeatother%
  \begin{picture}(1,0.52351682)%
    \lineheight{1}%
    \setlength\tabcolsep{0pt}%
    \put(0,0){\includegraphics[width=\unitlength,page=1]{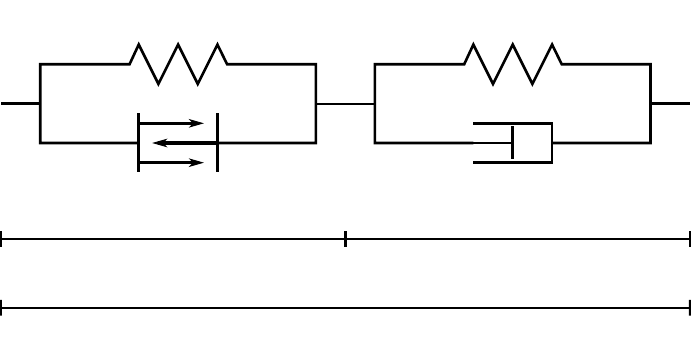}}%
    \put(0.18655677,0.12620745){\color[rgb]{0,0,0}\makebox(0,0)[lt]{\lineheight{1.25}\smash{\begin{tabular}[t]{l}$l_\text{CE}$\end{tabular}}}}%
    \put(0.65148178,0.12652906){\color[rgb]{0,0,0}\makebox(0,0)[lt]{\lineheight{1.25}\smash{\begin{tabular}[t]{l}$l_\text{SEE}$\end{tabular}}}}%
    \put(0.3982104,0.02261779){\color[rgb]{0,0,0}\makebox(0,0)[lt]{\lineheight{1.25}\smash{\begin{tabular}[t]{l}$l_\text{MTU}$\end{tabular}}}}%
    \put(0.21550776,0.48736116){\color[rgb]{0,0,0}\makebox(0,0)[lt]{\lineheight{1.25}\smash{\begin{tabular}[t]{l}PEE\end{tabular}}}}%
    \put(0.68786769,0.48868674){\color[rgb]{0,0,0}\makebox(0,0)[lt]{\lineheight{1.25}\smash{\begin{tabular}[t]{l}SEE\end{tabular}}}}%
    \put(0.70974845,0.23051932){\color[rgb]{0,0,0}\makebox(0,0)[lt]{\lineheight{1.25}\smash{\begin{tabular}[t]{l}SDE\end{tabular}}}}%
    \put(0.22986201,0.22958198){\color[rgb]{0,0,0}\makebox(0,0)[lt]{\lineheight{1.25}\smash{\begin{tabular}[t]{l}CE\end{tabular}}}}%
  \end{picture}%
\endgroup%
  \caption{Mechanical structure of the Hill-type muscle model. The force generating contractile element (CE) is parallel-connected to the parallel elastic element (PEE) and connected in series to a second parallel-connected structure consisting of the serial elastic element (SEE) and the serial damping element (SDE). The length $l_\text{MTU}$ of the whole muscle tendon unit is composed of the common length $l_\text{CE}$ of CE and PEE and the common length $l_\text{SEE}$ of SEE and SDE. The variable $l_\text{CE}$ is an internal state of the model.}
  \label{fig:hilltype}%
\end{figure}%

\newcommand{\CE}{\text{CE}}
\newcommand{\MTU}{\text{MTU}}

We employ the four-element Hill-type muscle model that is described by \cite{Hilltype2014}. Its structure is visualized in \cref{fig:hilltype}. It consists of four components: the contractile element (CE), the parallel elastic element (PEE), the serial elastic element (SEE), and the serial damping element (SDE). Inputs to the model are the muscular activation $\alpha(t)$, the length $l_\MTU(t)$ and the contraction velocity $\dot{l}_\MTU(t)$ of the muscle tendon unit (MTU).
The output of the model is the muscle force $f_\text{MTU}(t)$. The model contains one internal state variable, the length $l_\text{CE}(t)$ of the CE. The muscle dynamics determine this internal length and its time derivative, the contraction velocity $\dot{l}_\text{CE}(t)$ of the CE.

The resulting force of the MTU is given as sum of the forces of the respective parallel elements as visualized in \cref{fig:hilltype}:
\begin{equation}\label{eq:hill_type0}
  \begin{array}{lll}
    F_\MTU = F_\CE(l_\CE, \dot{l}_\CE, \alpha) + F_\text{PEE}(l_\CE) = F_\text{SEE}(l_\CE,l_\MTU) + F_\text{SDE}(l_\CE,\dot{l}_\CE,\dot{l}_\MTU,\alpha).
  \end{array}
\end{equation}
The force terms of the four elements, $F_\CE, F_\text{PEE}, F_\text{SEE}$ and $F_\text{SDE}$ are described by analytical functions that use a total of 19 parameters. A description of the detailed equations and parameters can be found in \cite{Hilltype2014}. In the following, an overview over the formulation is given with a focus on the piecewise formulated terms that contribute to the overall muscle model. In the following formulations, underlined variables designate constant parameters that either have to be specified or follow from other given parameters.

The muscle output force $F_\text{MTU}$ is computed by the second identity of \cref{eq:hill_type0}, i.e., from the forces $F_\text{SEE}$ and $F_\text{SDE}$. The force $F_\text{SEE}$ acting in the SEE is formulated as a continuous piecewise function with a constant zero, an exponential and a linear branch:
\begin{equation*}
  \begin{array}{lll}
    F_\text{SEE}(\ell_\CE,\ell_\MTU) = \begin{cases}
      0, & \ell_\text{SEE} < \underline{l_{\text{SEE},0}}\\[2mm]
      \underline{K_\text{SEE,nl}}(\ell_\text{SEE} - \underline{\ell_\text{SEE,0}})^{\underline{\nu_\text{SEE}}}, & \ell_\text{SEE} < \underline{\ell_\text{SEE,nll}}\\[2mm]
    \underline{\Delta F_\text{SEE,0}} + \underline{K_\text{SEE,l}}(\ell_\text{SEE} - \underline{\ell_\text{SEE,nll}}), & 
    \ell_\text{SEE} \geq \underline{\ell_\text{SEE,nll}}
    \end{cases}, \quad \text{with } \ell_\text{SEE} = \ell_\MTU - \ell_\CE.
  \end{array}
\end{equation*}

The damping force $F_\text{SDE}$ in the SDE is proportional to the lengthening velocity ${\dot{\ell}_\text{SEE} = \dot{\ell}_\MTU - \dot{\ell}_\CE}$ of this element. It is given by
\begin{equation}\label{eq:sde_force}
  \begin{array}{lll}
    F_\text{SDE}(l_\CE,\dot{l}_\CE,\dot{l}_\MTU,\alpha) \\ \qquad = \underline{D_\text{SDE,max}}\left( (1-\underline{R_\text{SDE}}) \dfrac{F_\text{PEE}(\ell_\CE)+F_\text{CE}(\ell_\CE, \dot{\ell}_\CE, \alpha)}{\underline{F_\text{max}} + \underline{R_\text{SDE}}} \right) (\dot{\ell}_\MTU - \dot{\ell}_\CE).
  \end{array}
\end{equation}
The amount of damping is dependent on the force $F_\MTU$ of the MTU which appears in the nominator of the fraction in \cref{eq:sde_force} as the sum of the forces $F_\text{PEE}$ and $F_\CE$. Formulas for these two forces are given in the following.

The force $F_\text{PEE}$ of the PEE is formulated piecewise as a shifted and cut off polynomial function:
\begin{equation}\label{eq:fpee}
  \begin{array}{lll}
    F_\text{PEE}(\ell_\CE) = \begin{cases}
      0, & \ell_\CE < \underline{\ell_{\text{PEE},0}}\\[4mm]
    \underline{K_\text{PEE}}(l_\CE - \underline{l_{\text{PEE},0}})^{\underline{\nu_\text{PEE}}},& \ell_\CE \geq \underline{\ell_{\text{PEE},0}}
    \end{cases}.
  \end{array}
\end{equation}

The force $F_\CE$ of the CE is the active force produced by the muscle and is given by:
\begin{equation}\label{eq:fce}
  \begin{array}{lll}
    F_\CE(l_\CE, \dot{l}_\CE, \alpha) 
    = \underline{F_\text{max}} \left(
    \dfrac{\alpha\,F_\text{isom}(\ell_\CE) + A_\text{rel}(\dot{\ell}_\CE,\ell_\CE,\alpha)}
    {1 - 
    \frac{\dot{\ell}_\CE}
    {B_\text{rel}(\dot{\ell}_\CE,\ell_\CE,\alpha)\,\ell_{\CE,\text{opt}}}}\right)
    - A_\text{rel}(\dot{\ell}_\CE,\ell_\CE,\alpha).
  \end{array}
\end{equation}
It can be seen that the active force depends on the activation level $\alpha$. 
The formulation of $F_\CE$ contains the two main characteristic curves for muscle forces, the force-length relation and the force-velocity relation.

The force-length relation is modeled by the function $F_\text{isom}(\ell_\CE)$ of isometric force, which describes the relative force for the condition $\dot{\ell}_\CE = 0$. This function is formulated piecewise for CE lengths $\ell_\CE$ smaller and larger than an optimal length $\ell_\text{CE,opt}$. 

The force-velocity relation follows from the auxiliary functions $A_\text{rel}(\dot{\ell}_\CE,\ell_\CE,\alpha)$ and $B_\text{rel}(\dot{\ell}_\CE,\ell_\CE,\alpha)$. These functions have different forms for concentric ($\dot{\ell}_\CE < 0$) and eccentric ($\dot{\ell}_\CE \geq 0$) conditions as well as for the two ranges of CE  length, $\ell_\CE < \ell_\text{CE,opt}$ and $\ell_\CE \geq \ell_\text{CE,opt}$.

\begin{figure}%
  \centering%
  \begin{subfigure}[t]{0.9\textwidth}%
    \centering%
    \includegraphics[width=\textwidth]{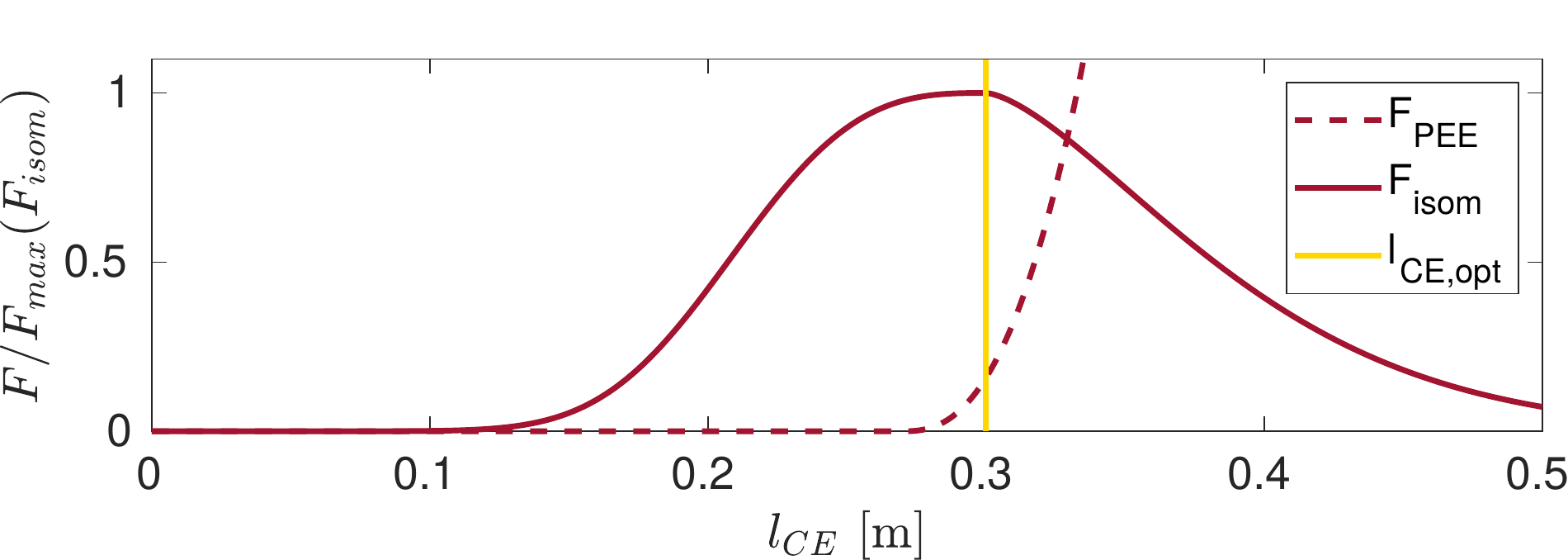}%
    \caption{Force-length curves of the PEE (dashed red line) and isometric force $F_\text{isom}$ (solid red line) for an isometric condition ($\dot{l}_\CE=0$), normalized to the maximum isometric force. The optimal length $l_\text{CE,opt}$ of the CE is shown as yellow vertical line. The force $F_\text{PEE}(l_\CE)$ of the PEE is zero for $l_\CE < 0.9\,l_\text{CE,opt}$. The isometric contraction force $F_\text{isom}$ is formulated piecewise by two branches separated by $l_\text{CE,opt}$.}%
    \label{fig:force_curves_generic_length}%
  \end{subfigure}\\[6mm]
  \begin{subfigure}[t]{0.9\textwidth}%
    \centering%
    \includegraphics[width=\textwidth]{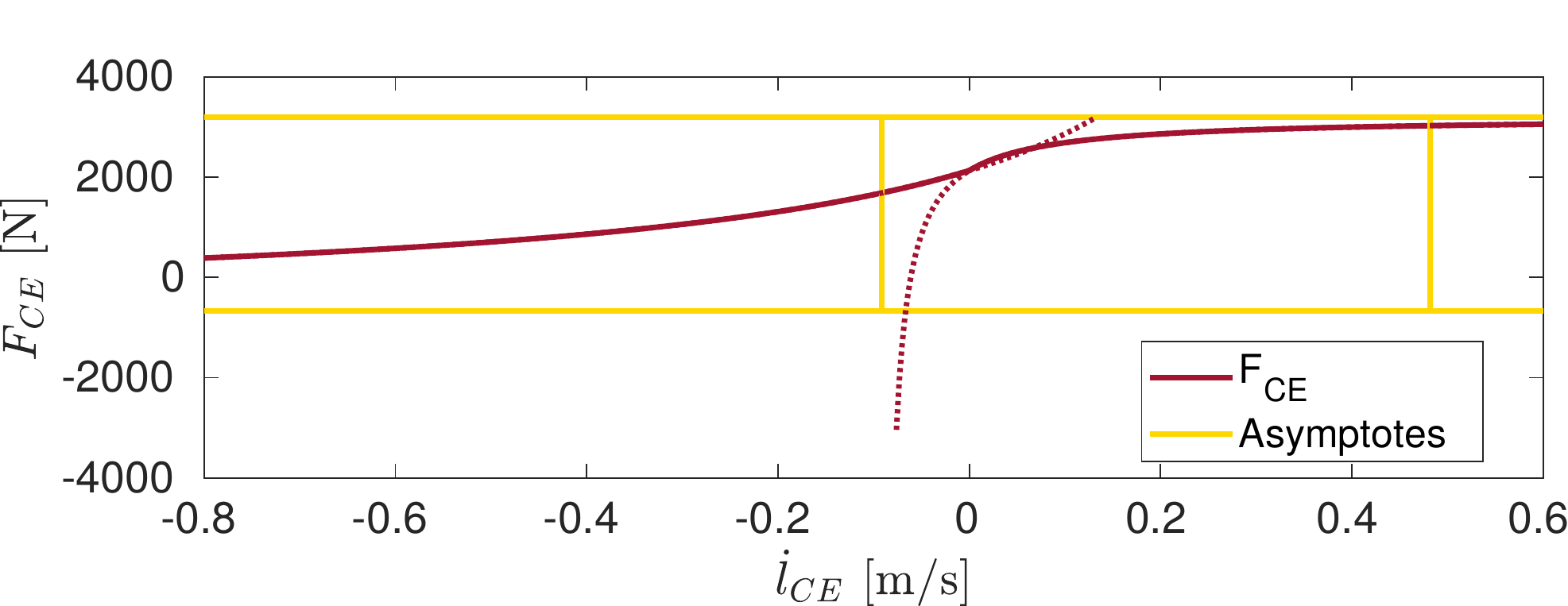}%
    \caption{Force-velocity curve $F_\CE(\dot{l}_\CE)$ of the CE at optimal length $l_\CE=l_{\CE,\text{opt}}$, and for activation level $\alpha=0.5$. The function (red solid line) is formulated piecewise, graphs of the base functions of the two branches continue as red dotted lines. Their limits and singularities are visualized by the yellow horizontal and vertical asymptotes.}%
    \label{fig:force_curves_generic_velocity}%
  \end{subfigure}%
  \caption{Force-length and force-velocity relations for the muscle model with generic parameters taken from literature \cite{Hilltype2014}.}%
  \label{fig:force_curves_generic}%
\end{figure}%

\Cref{fig:force_curves_generic} visualizes the two main characteristic curves of the model. 
\Cref{fig:force_curves_generic_length} shows how the generated force depends on the length of the CE. 
The active force $F_\CE$, given by \cref{eq:fce}, is visualized by the solid red line. It has its maximum at the optimal length $l_\text{CE,opt}$ of the CE. This can be explained by the overlap of actin and myosin filaments in the sarcomere. The overlap is lower when the actin filaments are pulled apart or pushed together. A higher overlap leads to a higher force output.

The dashed line in \cref{fig:force_curves_generic_length} represents the passive force $F_\text{PEE}$ of the elastic muscular tissue, formulated in \cref{eq:fpee}. The passive force is essentially generated by the titin proteins in the sarcomere. Only starting from a certain length, the structure exerts reaction forces against lengthening forces to avoid overstretching of the muscle.

\cref{fig:force_curves_generic_velocity} shows the force-velocity relation of the Hill-type model. The curve of $F_\CE(\dot{l}_\CE)$ is composed of two branches: 
The concentric branch for shortening contraction with $\dot{l}_\CE \leq 0$ and the eccentric branch for lengthening contraction with $\dot{l}_\CE > 0$. It can be seen that the generated force increases monotonically over the lengthening velocity. It approaches a limit for maximum positive and negative velocity. These limits can be adjusted by parameters of the model and are exemplary for how the shape of the curves of a Hill-type model can be parametrized.

In addition to the resulting muscle force $F_\MTU$, a formulation for the internal state variable $\ell_\CE$ is required.
The second identity of \eqref{eq:hill_type0} can be solved for the lengthening velocity $\dot{\ell}_\CE$ of the CE to get an evolution equation for the length $\ell_\CE$ of the CE. The derivation and the resulting formula can be found in \cite{Hilltype2014}.

To describe the activation dynamics, i.e., the evolution of the muscle activation ${\alpha \in [0,1]}$, the model of Hatze et al. \cite{Hatze1977} is used.
The activation is computed depending on the free calcium ion concentration $\gamma$ and the length $\ell_\CE$ of the CE by
\begin{equation*}
  \begin{array}{lll}
    \alpha(\ell_\CE,\gamma) = \dfrac{\underline{a_0} + \big(\rho(\ell_\CE)\,\gamma\big)^3}
    {1 + \big(\rho(\ell_\CE)\,\gamma\big)^3}.
  \end{array}
\end{equation*}
The function $\rho$ is given by
\begin{equation*}
  \begin{array}{lll}
    \rho(\ell_\CE) = \underline{c}\,\underline{\eta}\,\dfrac{(\underline{k}-1)\,\ell_\CE}{(\underline{k} - \ell_\CE/\ell_\text{CE,opt})\,l_\text{CE,opt}}.
  \end{array}
\end{equation*}
All used parameter values for the activation dynamics can be found in \cite{Bayer2017}.

In summary, we get the following coupled system of differential-algebraic equations, where $f_\CE$ and $f_\alpha$ denote the respective formulas:
\begin{align}
  F_\MTU &= F_\MTU(\ell_\MTU,l_\CE,\dot{l}_\CE,\alpha),      \label{eq:hill_type1} \\[4mm]
  \dot{l}_\CE &= f_\CE(l_\CE,\ell_\MTU,\dot{l}_\MTU,\alpha), \label{eq:hill_type2}\\[4mm]
  \alpha &= f_\alpha(\gamma, l_\CE).                      \label{eq:hill_type3}
\end{align}

To compute the joint torque in a system of an agonist and antagonist muscle pair, two instances of the presented Hill-type muscle model can be used. In our study considering the upper arm, the torque $\tau$ at the elbow is computed by multiplying the predicted forces $F_{\MTU,B}$ and $F_{\MTU,T}$ of biceps and triceps with the corresponding moment arms $\hat{r}_B$ and $\hat{r}_T$:
\begin{equation}\label{eq:muscle_torque}
  \begin{array}{lll}
    \tau = F_{\MTU,B}(l_{\MTU,B},\dot{l}_{\MTU,B},\alpha_B) \cdot \hat{r}_B - F_{\MTU,T}(l_{\MTU,T},\dot{l}_{\MTU,T},\alpha_T) \cdot \hat{r}_T.
  \end{array}
\end{equation}

\subsection{Parameter Identification for Model B}\label{sec:parameter_optimization}
The process of model identification finds the parameters that make the model B predict correct values for the specific subject, i.e., minimizes the error in the predicted outcome for the training data set.

The following minimization is performed:
\begin{align}
  &&\min\limits_{\substack{\bftheta_M,l_{\CE,M}(t), \\\forall M \in \{B,T\}, \,\forall t \in \mathcal{T}}} &\sum\limits_{t\in \mathcal{T}}|\tau(t) - \hat{\tau}(t)|^2
   \label{eq:opt_1}\\[4mm]
  &&\text{s.t. }\forall t \in \mathcal{T}:\quad &\tau(t) = F_{\MTU,B}(t,l_{\CE,B},\bftheta_B) \cdot \hat{r}_B(t) \notag\\
      &&&\qquad\quad- F_{\MTU,T}(t,l_{\CE,T},\bftheta_T) \cdot \hat{r}_T(t),                 \label{eq:opt_2}\\[4mm]
  &&& \dot{l}_{\CE,M}(t) = \dot{\hat{l}}_{\MTU,M}(t),\quad M \in \{B,T\},      \label{eq:opt_3}\\[4mm]
  &&& \bftheta_B, \bftheta_T \in \Theta,                                       \label{eq:opt_4}\\[4mm]
  &&& l_{\CE,M}(t) \in [0,l_{\MTU,M}(t)],\quad M \in \{B,T\}.                  \label{eq:opt_5}
\end{align}

The optimization variables are the parameters $\bftheta_B$ and $\bftheta_T$ for the biceps and triceps Hill-type models and the lengths $l_{\CE,B}(t)$ and $l_{\CE,T}(t)$ of the contractile elements for both models at every point in time. The variables designated as $\hat{\square}$ are the measured quantities from the training experiments. The objective function given in \cref{eq:opt_1} penalizes the difference between computed torque $\tau$ and measured torque $\hat{\tau}$ at every timestep $t \in \mathcal{T}$ of the training data. 

\Cref{eq:opt_2} computes the torque values and follows from \cref{eq:muscle_torque} of the muscle model. For every point in time, the predicted forces $F_{\MTU,B}$ and $F_{\MTU,T}$ are multiplied with the measured moment arms $\hat{r}_B$ and $\hat{r}_T$.

In \cref{eq:opt_3}, the contraction velocities are constrained to the measured values. Because the lengthening velocity $\dot{l}_\CE$ of the CE is an internal quantity and, thus, cannot be observed in experiments, we assume it to be equal to the lengthening velocity of the whole muscle: $\dot{l}_\CE \approx \dot{l}_{\MTU} = \dot{l}_\text{CE} + \dot{l}_\text{SEE}$. This requires the assumption $\dot{l}_{\text{SEE}} \approx 0$ which can be justified given the low dynamic nature of the experiments.

By \cref{eq:opt_4}, we bound each of the parameters $\bftheta_B$ and $\bftheta_T$ to a range between half and twice the generic value from literature.
The lengths of the CEs are constrained by \cref{eq:opt_5} to be positive and smaller than the length of the MTU.

All optimization variables are normalized to improve the numerical conditioning of the optimization problem. 
The parameters $\bftheta_B$ and $\bftheta_T$ are normalized with respect to generic values from literature that were taken from \cite{Gunther2007, Morl2012, Hilltype2014}. The initial values are set to one, which corresponds to the generic literature values.
The internal states $l_{\CE,B}$ and $l_{\CE,T}$, are normalized with respect to the measured MTU lengths $\hat{l}_{\MTU,B}$ and $\hat{l}_{\MTU,T}$ and initialized with zero.

%For training of model B, the optimization problem for the parameters needed to be solved. 
We implemented the Hill-type models and the constraints in MATLAB and used the nonlinear programming implementation, \code{fmincon}, to minimize the given bounded and nonlinear constrained, multivariable function.
In our study, the total number of optimization variables is computed by $2\cdot 19 + 2n = 598$, as each of the parameter vectors $\bftheta_B,\bftheta_T$ had 19 entries. 

\section{Results and Discussion}\label{sec:evaluation}

In the following, results of connecting the two model formulations, A and B, to the experimental data are presented.
At first, \cref{sec:res_feature_selection} gives details on the preprocessed data. The training phase is described in \cref{sec:res_training}. Applying the trained models to the validation data is done in \cref{sec:res_validation}. Then, \cref{sec:res_simplified_a} tests a simplified version for model A. Then, \cref{ref:res_insights_b} shows some insights into the optimized parameter values for model B.

\subsection{Feature Selection}\label{sec:res_feature_selection}
\begin{figure}%
  \centering%
  \includegraphics[width=\textwidth]{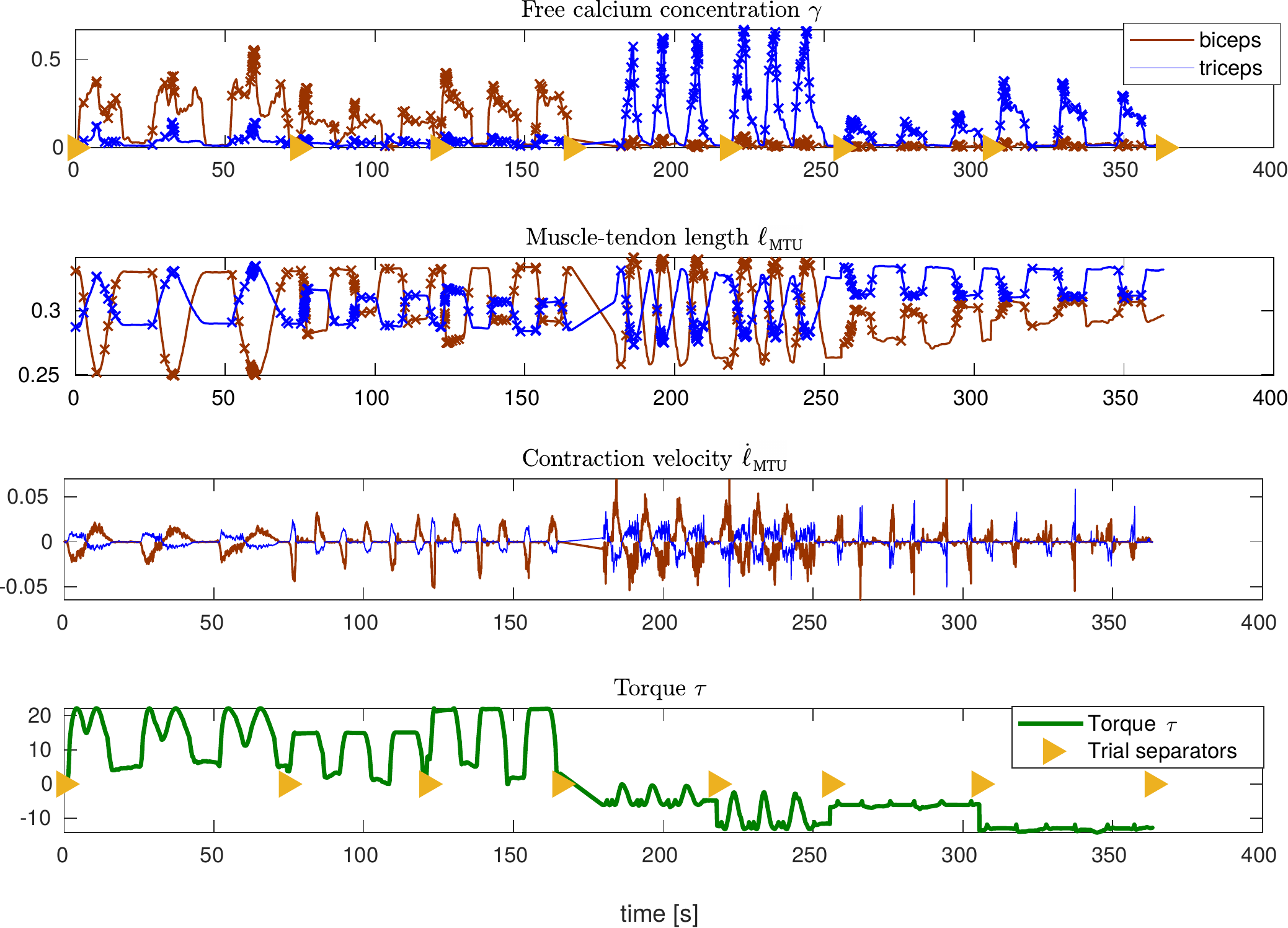}%
  \caption{Processed experimental values over time that were used for training of both models. The concatenated data of seven trials are shown, which yield an end time of $\SI{363.32}{s}$. The individual trials are separated by the yellow triangles on the $x$-axis. The three upper plots show the values of $\gamma, \ell_\MTU$ and $\dot{l}_\MTU$ for both biceps (brown) and triceps (blue), the bottom plot shows the elbow torque $\tau$. The selected feature points are visualized as crosses in the two top plots. In the upper-most plot, it can be seen that the first three trials, which correspond to elbow flexion, mainly activated the biceps muscle, whereas in the last four trials, corresponding to elbow extension, the triceps is more active.
}%
  \label{fig:selected_points}%
\end{figure}

% intro
The experimental data is split into a training and a validation dataset. \Cref{fig:selected_points} shows the processed data of the training data set. In total, we captured $N=\num{34934}$ data points for the seven experimental trials. Out of these, we select $n_\text{per\_trial}=\num{40}$ feature points in every trial, leading to a total of $n=n_\text{per\_trial}\cdot n_\text{trials}=\num{280}$ points. The selected points are visualized by crosses in the top plot of \cref{fig:selected_points}. It can be seen that the algorithm described in \cref{sec:study_feature_selection} distributes the feature points equally along the $\gamma$ axis.

\subsection{Training of the Models}\label{sec:res_training}
Training of model A consists of estimating the hyperparameters for the Gaussian Process Regression model from the training data.

For model B, the optimization problem for the biophysical parameters is solved. The resulting parameter values and their relation to the initial values are summarized in \cref{tab:model_b_parameters}. It can be seen that none of the final parameter values is limited by the constraints, which would be $\SI{-50}{\percent}$ and +\SI{100}\percent. However, it was observed that including the constraints helps the optimizer to stay in the valid range of meaningful model parameters and, thus, reach the optimum faster.

\begin{table}[t] 
  \centering
  %\begin{scriptsize}
  
    \begin{tabular}{@{}llllll@{}}
     \toprule
    \textbf{CE} 
      & $F_{\mathrm{max}}\,[\mathrm{N}]$
      & $l_{\mathrm{CE},opt}\,[\mathrm{m}]$ 
      & $\Delta W_{d} \,[\;]$ 
      & $\Delta W_{a} \,[\;]$ 
      & $\nu_{\mathrm{CE},d} \,[\;]$ \\  \midrule
    Generic & $4260$ & $0.3$ & $0.35$ & $0.35$ & $1.5$  \\ %\hline
    Biceps  & $+\SI{11.0}{\percent}$ & $+\SI{31.7}{\percent}$ & $+\SI{10.9}{\percent}$ & $+\SI{91.6}{\percent}$ & $+\SI{10.9}{\percent}$  \\ %\hline
    Triceps & $-\SI{49.0}{\percent}$ & $-\SI{25.4}{\percent}$ & $+\SI{10.9}{\percent}$ & $+\SI{5.1}{\percent}$ & $+\SI{10.9}{\percent}$  \\ %\hline 
    \addlinespace[2ex]
    \textbf{CE} 
      &  $\nu_{\mathrm{CE},a} \,[\;]$ 
      & $A_\text{rel,0} \,[\;]$  
      & $B_\text{rel,0} \,[\;]$ 
      & $S_\text{ecc} \,[\;]$ 
      & $F_\text{ecc} \,[\;]$\\  \hline
    Generic & $3.0$ & $0.25$ & $2.25$ & $2$ & $1.5$ \\ %\hline
    Biceps & $-\SI{46.4}{\percent}$ & $+\SI{14.0}{\percent}$ & $+\SI{77.5}{\percent}$ & $-\SI{4.1}{\percent}$ & $-\SI{30.1}{\percent}$ \\ %\hline
    Triceps &  $+\SI{95.9}{\percent}$ & $-\SI{20.3}{\percent}$ & $+\SI{41.4}{\percent}$ & $+\SI{22.3}{\percent}$ & $+\SI{36.8}{\percent}$ \\ %\hline
    \addlinespace[2ex]
    \textbf{PEE} 
      & $L_{\mathrm{PEE},0} \,[\;]$ 
      & $\nu_{\mathrm{PEE}} \,[\;]$ 
      & $F_{\mathrm{PEE}} \,[\;]$
      \\ \hline
    Generic & $0.9$ & $2.5$ & $2.0$  \\ %\hline
    Biceps & $+\SI{10.9}{\percent}$ & $+\SI{10.9}{\percent}$ & $+\SI{10.9}{\percent}$  \\ %\hline
    Triceps & $+\SI{10.9}{\percent}$ & $+\SI{10.9}{\percent}$ & $+\SI{10.9}{\percent}$  \\ %\hline
    \addlinespace[2ex]
    \textbf{SDE} 
      & $D_{\mathrm{SDE}} \,[\;]$ 
      & $R_{\mathrm{SDE}} \,[\;]$ 
      \\ \midrule
    Generic & $0.3$ & $0.01$    \\ %\hline
    Biceps & $+\SI{10.9}{\percent}$ & $+\SI{8.6}{\percent}$   \\ %\hline
    Triceps & $+\SI{10.9}{\percent}$ & $-\SI{11.0}{\percent}$  \\ %\hline
    \addlinespace[2ex]
    \textbf{SEE} 
      & $l_{\mathrm{SEE},0}\,[\mathrm{m}]$ 
      & $\Delta F_{\mathrm{SEE},0} \,[\;]$  
      & $\Delta U_\text{l} \,[\;]$  
      & $\Delta U_\text{nll}\,[\;]$
      \\ \hline
    Generic & $0.172$ & $0.0425$ & $0.017$ & $568$ \\ %\hline
    Biceps & $-\SI{25.0}{\percent}$ & $-\SI{42.4}{\percent}$ & $+\SI{63.3}{\percent}$ & $+\SI{59.2}{\percent}$ \\ %\hline
    Triceps & $-\SI{10.3}{\percent}$ & $+\SI{64.75}{\percent}$ & $+\SI{19.0}{\percent}$ & $+\SI{23.6}{\percent}$ \\ %\hline
    \bottomrule
  \end{tabular}

  \caption{Hill-type muscle model parameters of the four elements: CE, PEE, SDE and SEE, initial values given in literature and relative changes of the optimized values. Further explanations of the parameters and references to literature containing their initial values are given in \cite{Hilltype2014}.}
  \label{tab:model_b_parameters}
  %\end{scriptsize}
\end{table}

\begin{figure}%
  \centering%
  \begin{subfigure}[t]{0.48\textwidth}%
    \centering%
    \includegraphics[width=\textwidth]{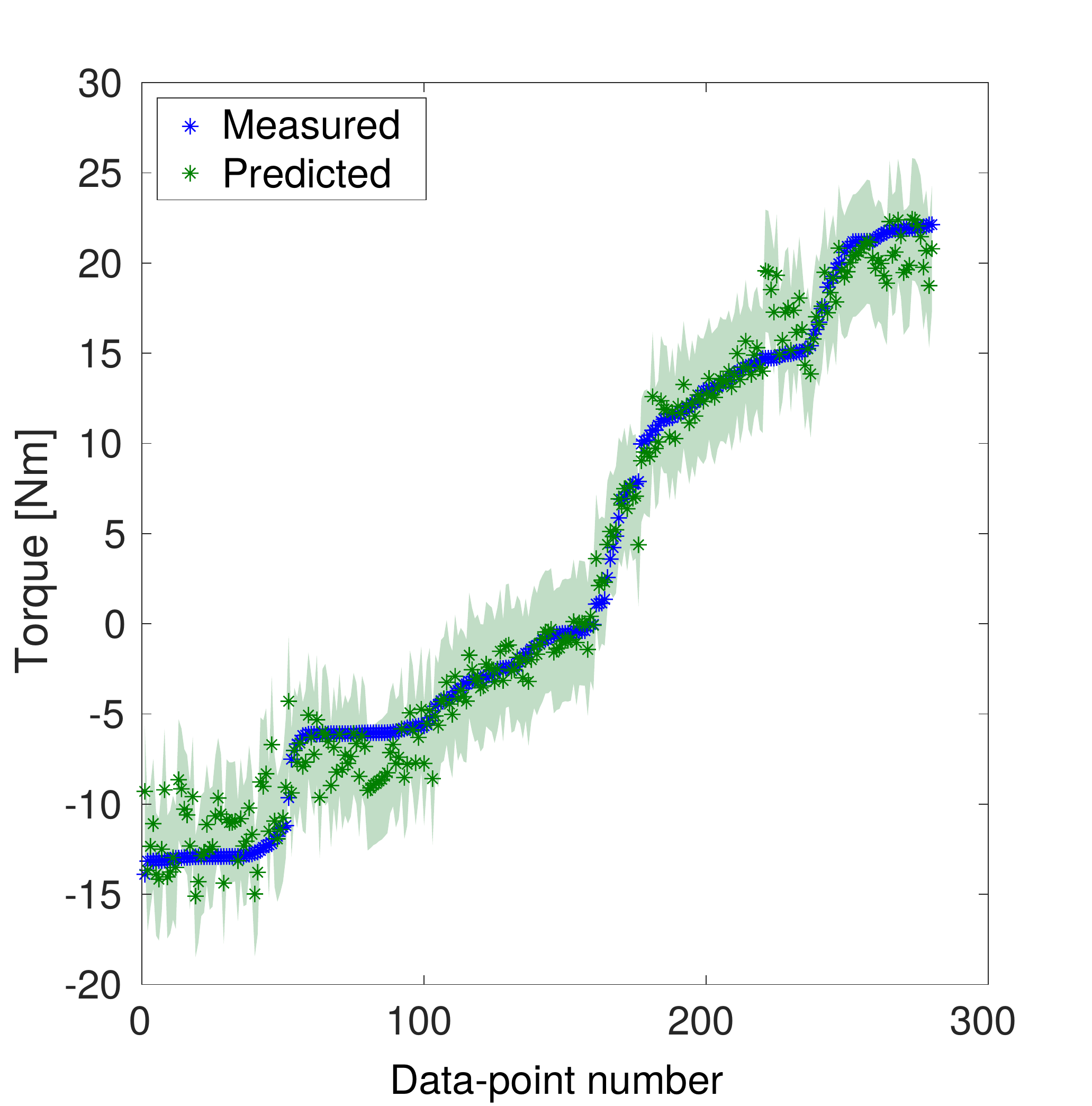}%
    \caption{Measured values (blue) and values predicted by model A (dark green), with 95\% confidence interval (light green).}%
    \label{fig:measured_optimized_torque_A}%
  \end{subfigure}%
  \quad
  \begin{subfigure}[t]{0.48\textwidth}%
    \centering%
    \includegraphics[width=\textwidth]{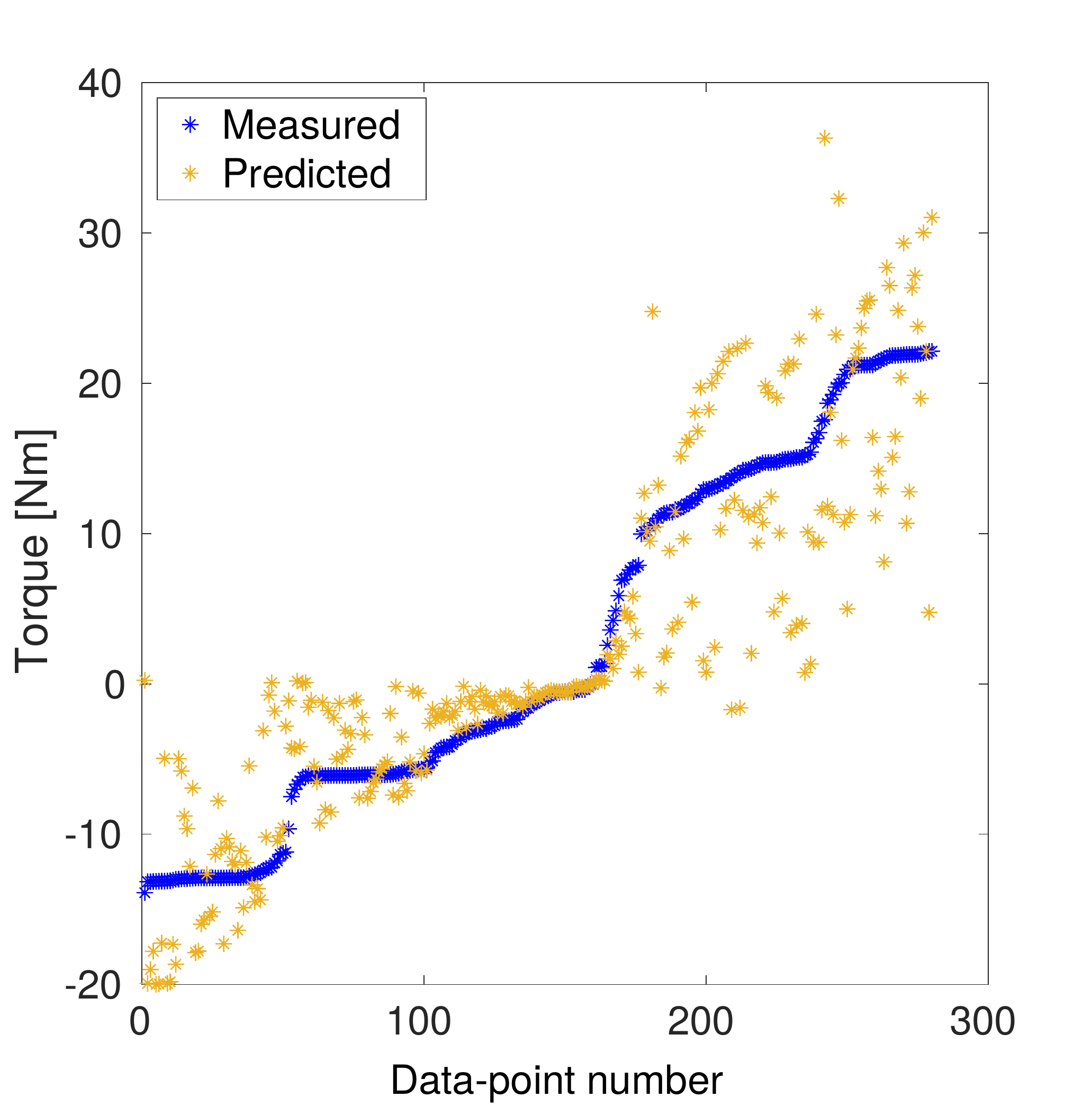}%
    \caption{Measured values (blue) and values predicted by model B (orange).}%
    \label{fig:measured_optimized_torque_B}%
  \end{subfigure}%
  \caption{Resubstitution prediction: Measured and predicted torque values for the training data set. The measured points are ordered and numbered by magnitude, the order of the predicted points matches the order of the measured points.}%
  \label{fig:measured_optimized_torque}%
\end{figure}

After training of the models A and B using the selected points of the training dataset, both models were tested by a \emph{resubstitution prediction}, i.e., predicting output from the training input data. The results are shown in \cref{fig:measured_optimized_torque_A} for model A and \cref{fig:measured_optimized_torque_B} for model B. As this evaluation only uses the subset of selected experimental values, the data points have no natural ordering. They were sorted for better visibility.

It can be seen that, for both models, the predicted values are a good fit to the measured values. For model A, the predicted 95\% confidence interval includes the actually measured values almost everywhere. For model B, the predicted values show a higher variance, especially for high torque values.

\subsection{Validation}\label{sec:res_validation}

The next evaluation uses the validation dataset and compares the predicted outputs of the models with the actual experimental values.
In contrast to the training data, where a small number $n$ of points was selected, we now use all captured values. This involves a total of \num{54e3} data points for a time span of $t=\SI{54}s$. 

\begin{figure}%
  \centering%
  \includegraphics[width=\textwidth]{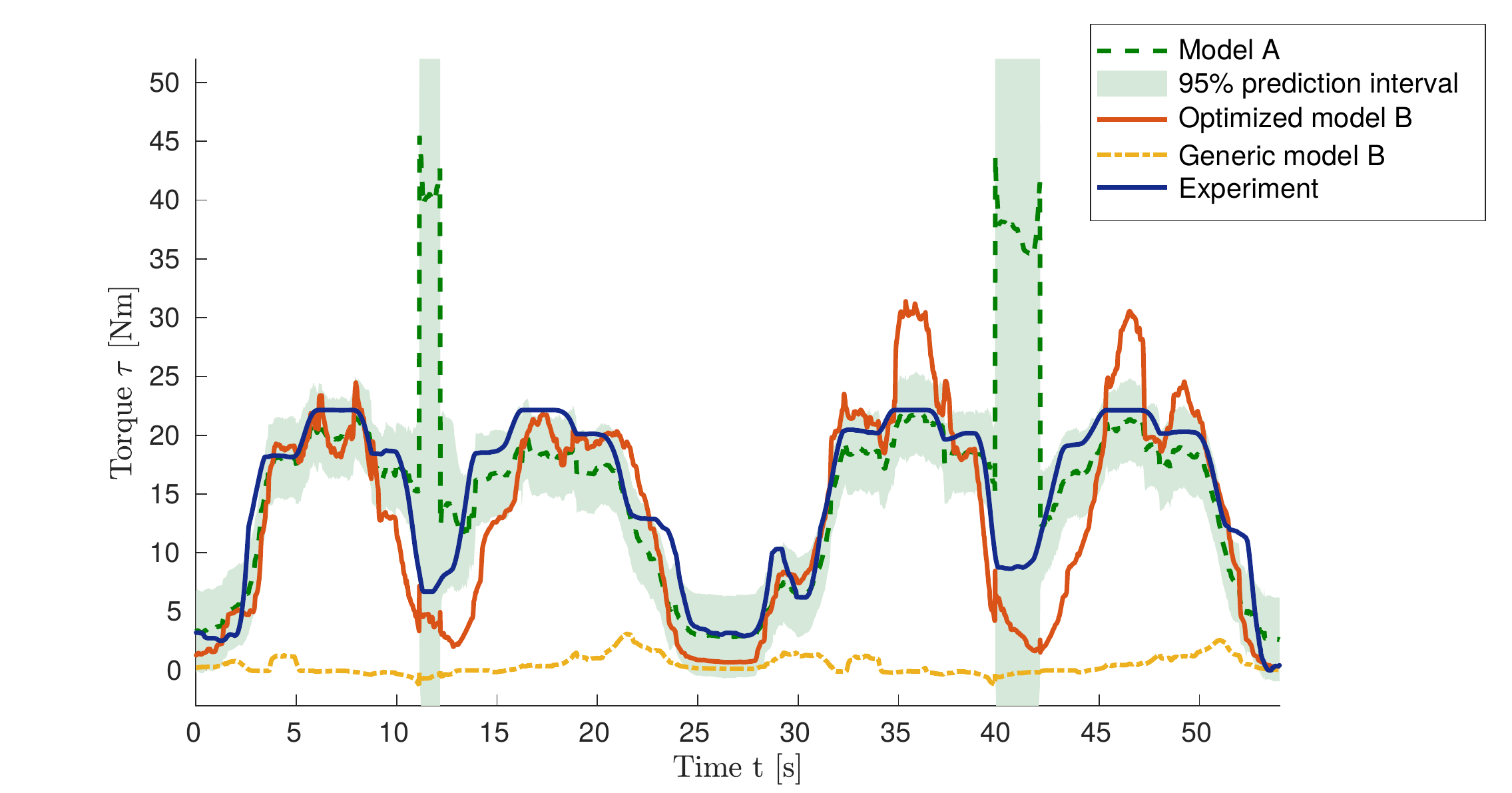}%
  \caption{Validation: Predicted torque values by model A (green), trained model B (red) and untrained model B (yellow), in comparison to the experimentally measured values (blue), for the validation data set.}%
  \label{fig:validation_bv5_40_points}%
\end{figure}%

The results are shown in \cref{fig:validation_bv5_40_points}. Comparing the green curve for model A with the blue curve for the experimental data, it can be seen that the predicted values match qualitatively for most of the time span. The predicted torque values appear consistently slightly smaller than the real values. Only for the two intervals $[\SI{11}s, \SI{12}s]$ and $[\SI{40}s, \SI{42}s]$ the predicted value is far off. The \SI{95}\percent{} confidence interval that was computed by the Gaussian Process spans a large range for these time intervals which implies that the model prediction is not to be trusted for this area. 

The biophysical model approach, model B, was tested in two variants. First, with the generic parameters from literature (yellow curve), second, with the subject-specific, optimized parameters (red curve). It can be seen that the generic model fails to predict the torque values whereas the trained model predicts reasonable values. These values are worse than most of the predictions from model A, but they succeed in giving a qualitative estimate about a low, medium or high torque output.

The match between model outputs $\tau_i$ and experimental data $\hat\tau_i$ can be quantified using the normalized root-mean-square error (NRMSE). This is a scaled version of the root-mean-square error (RMSE) and can be defined as
\begin{equation*}
  \begin{array}{lll}
    \text{RMSE} = \sqrt{\sum\limits_{i=1}^N (\tau_i - \hat{\tau_i})^2 / N},\\[4mm]
    \text{NRMSE} = \dfrac{\text{RMSE}}{\max\limits_i\{\hat\tau_i\} - \min\limits_i\{\hat\tau_i\}}.
  \end{array}
\end{equation*}
The NRMSE for model A is 0.267 which is worse than the value of 0.163 for the trained model B. The generic model B has the worst NRMSE of 0.547.

\subsection{Simplified Model A}\label{sec:res_simplified_a}
An advantage of model approach A is that it forgoes any biophysical description and the associated type of model error. It is a generic approach that does not require expert knowledge about the physiological structure. In the present study, however, some level of expert knowledge and physiological model was required in preprocessing the MoCap data, i.e. solving the inverse kinematics of the observed forearm movements to get the kinematic quantities of muscle lengths, velocities and moment arms. 

Since model A performed well in the previous validation study, we tested whether good results can also be achieved without this expert knowledge.
Consequently, the next study applies model approach A using only the elbow angle and no muscle lengths, velocities nor moment arms. Thus, the training data consists of input vectors $\bfx_i = (\phi_e(t_i), \gamma_B(t_i), \gamma_T(t_i))^\top \in \mathcal{X}$. In the following, this model is named \say{simplified model A} in contrast to the \say{full model A} that uses the complete set of input variables.

\begin{figure}%
  \centering%
  \begin{subfigure}[t]{0.48\textwidth}%
    \centering%
    \includegraphics[width=\textwidth]{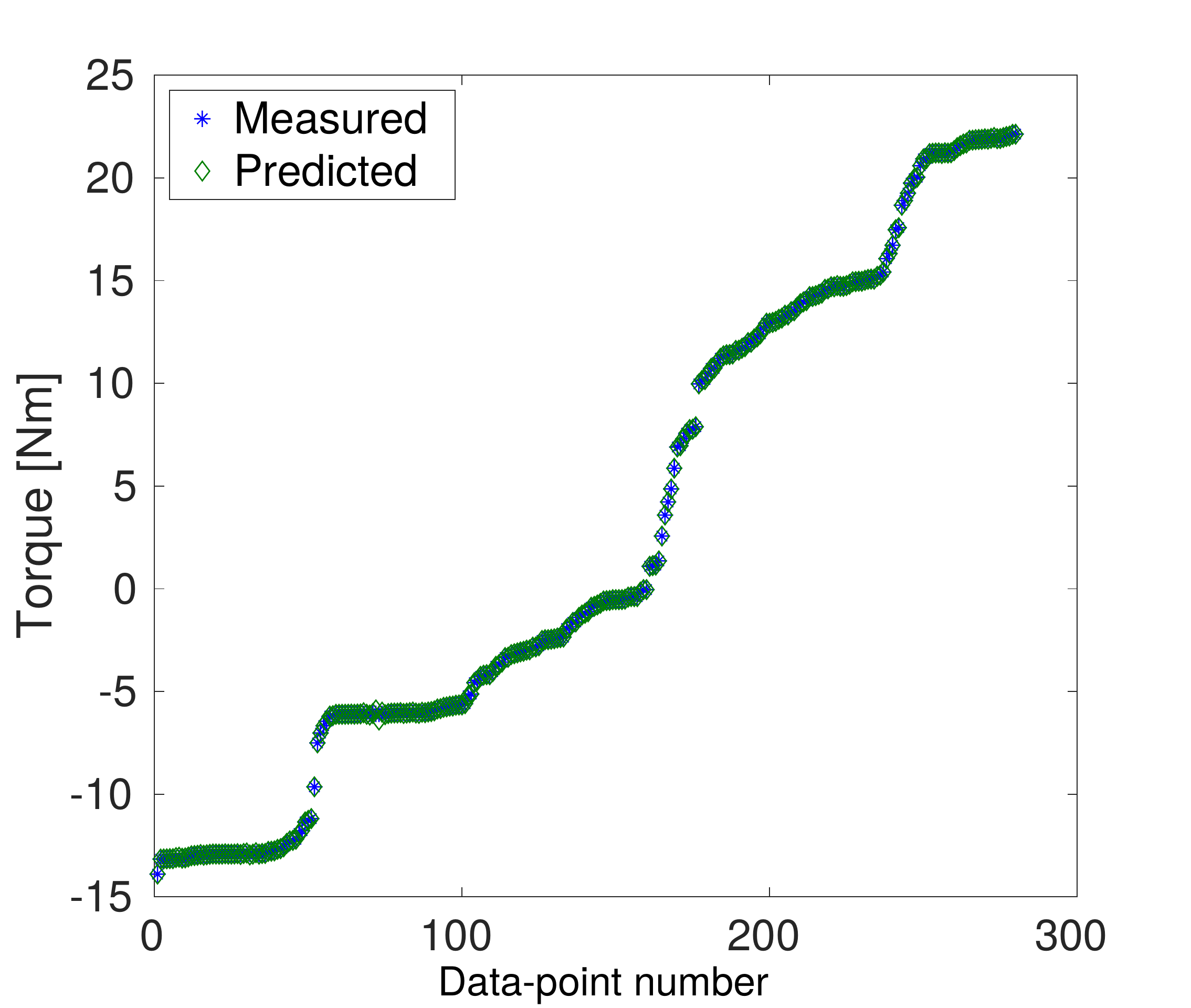}%
    \caption{Measured and predicted torque values of the training dataset, ordered and numbered by magnitude. The measured values (blue) and the values predicted by the Gaussian Process (green) lie on each other.}%
    \label{fig:measured_optimized_torque_A2}%
  \end{subfigure}%
  \quad
  \begin{subfigure}[t]{0.48\textwidth}%
    \centering%
    \includegraphics[width=\textwidth]{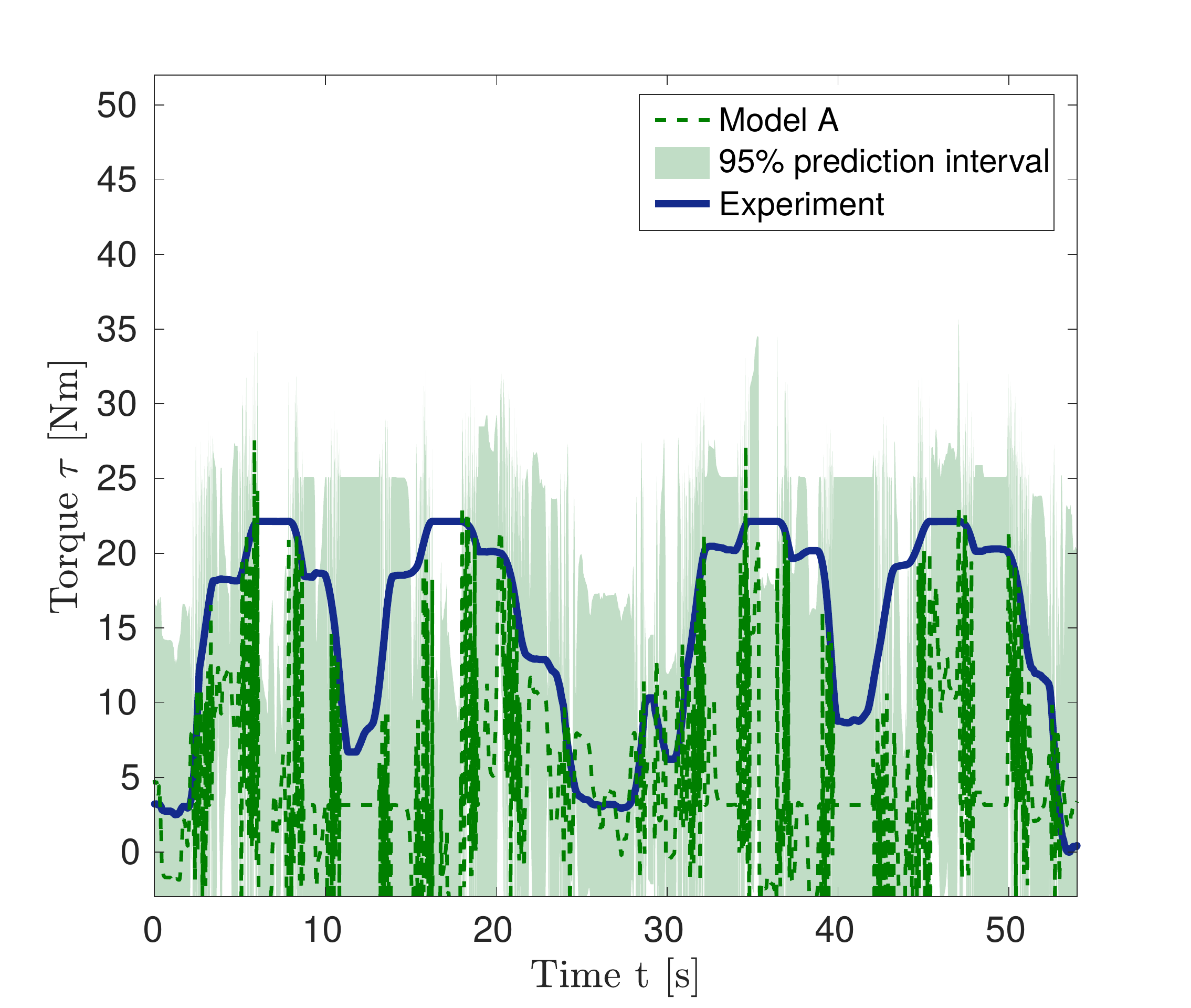}%
    \caption{Predicted torque values for the validation trials (green dotted line), 95\% confidence interval (light green) and the reference values of the experiment (blue). The plot reveals bad prediction capabilities of the simplified model A.}%
    \label{fig:measured_optimized_torque_A3}%
  \end{subfigure}%
  \caption{Result for the simplified model A, where, apart from the free calcium ion contractions, only the elbow angles, $\phi_e$, are used as training input instead of MTU lengths, velocities and moment arms.}
  \label{fig:measured_optimized_A2}%
\end{figure}

The results are shown in \cref{fig:measured_optimized_A2}. It can be seen that the resubstitution prediction in \cref{fig:measured_optimized_torque_A2} where the trained model is used to predict the training values shows a perfect fit. In contrast to the full model A, \cref{fig:measured_optimized_torque_A}, here, the learned input-output mapping shows no variance. However, the prediction for the validation dataset in \cref{fig:measured_optimized_torque_A3} shows a high error relative to the experimental data. The curve for the experimental data even lies outside the 95\% confidence interval of the prediction at some points. 

\begin{figure}%
  \centering%
  \includegraphics[width=0.35\textwidth]{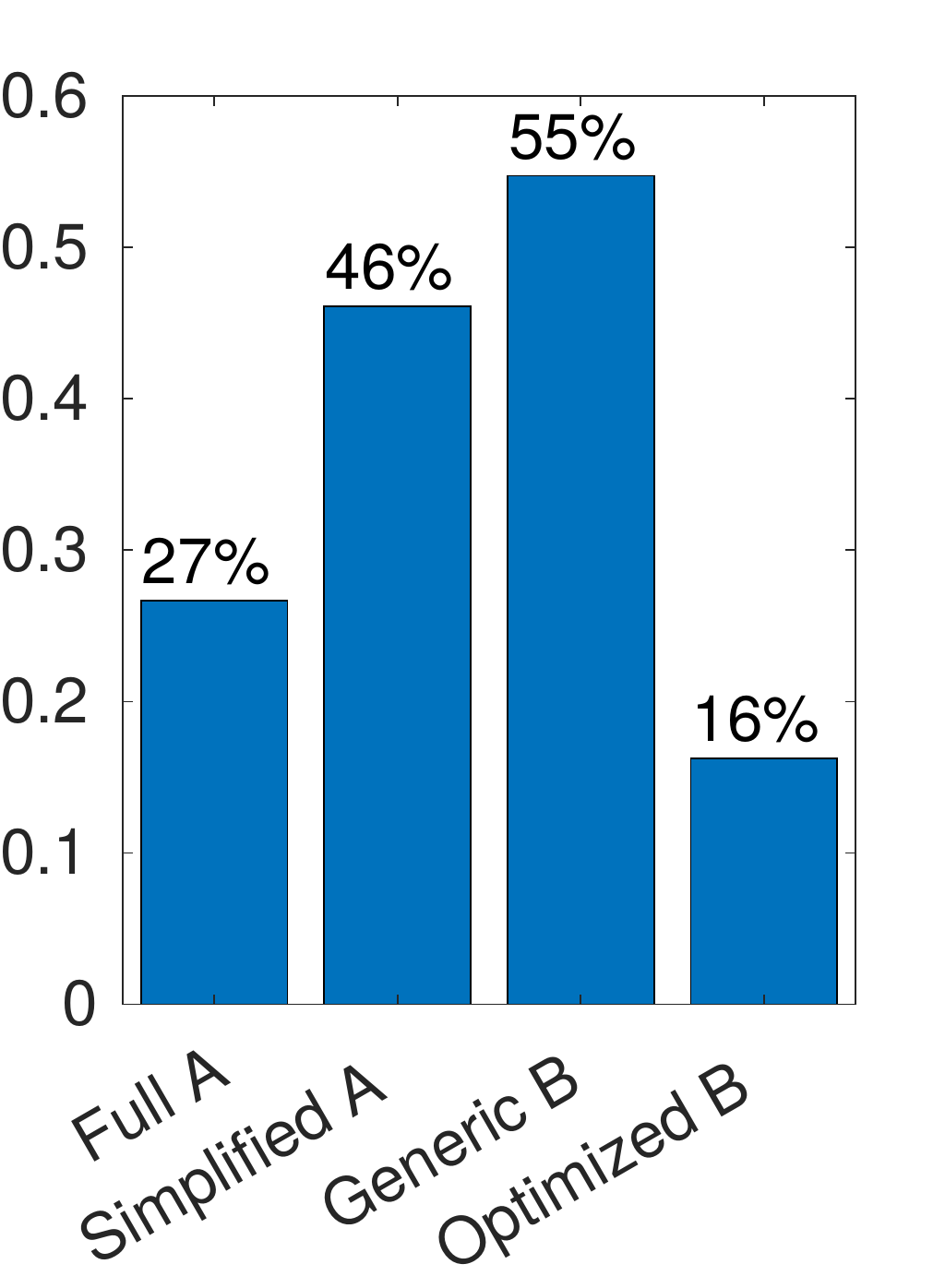}%
  \caption{Normalized Root Mean Square errors (NRMSE) of the validation trials between the respective models and the measured values. A lower error value means a better fit.}%
  \label{fig:nrmse}%
\end{figure}

The simplified model A has a NRMSE value of 0.461. For comparison, the NRMSE values of the full and simplified model A and the generic and optimized model B are summarized in \cref{fig:nrmse}. 

This evaluation shows that simplified model A gives no useful results where the training input is too scarce. Instead, preprocessing of the measurements using a subject specific geometric model, as done for the full model A, is needed to allow for a useful prediction.

\subsection{Insights of Model B}\label{ref:res_insights_b}
\begin{figure}%
  \centering%
  \begin{subfigure}[t]{0.47\textwidth}%
    \centering%
    \includegraphics[width=\textwidth]{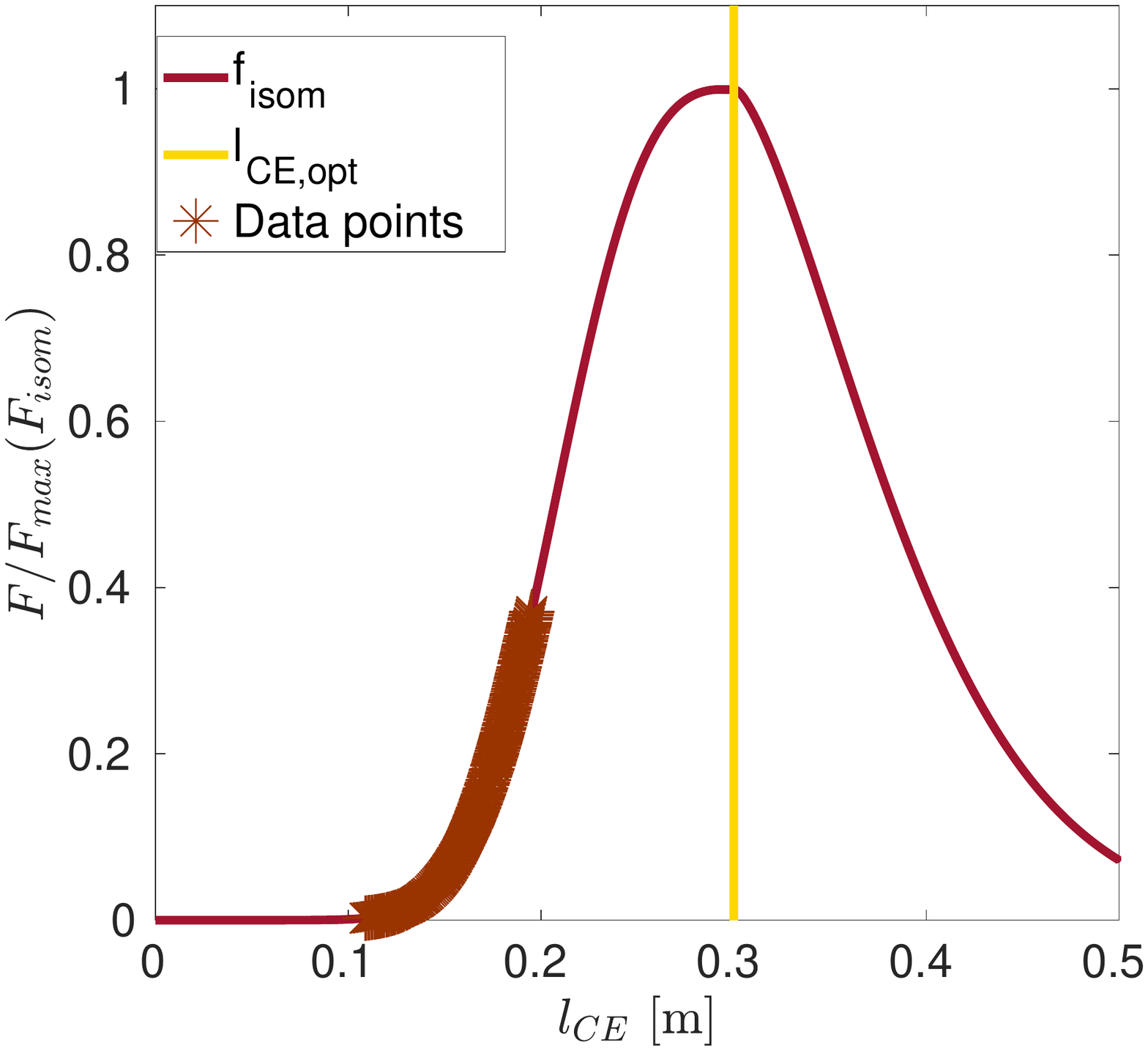}%
    \caption{Model with generic parametrization.}%
    \label{fig:biceps_a}%
  \end{subfigure}%
  \quad
  \begin{subfigure}[t]{0.47\textwidth}%
    \centering%
    \includegraphics[width=\textwidth]{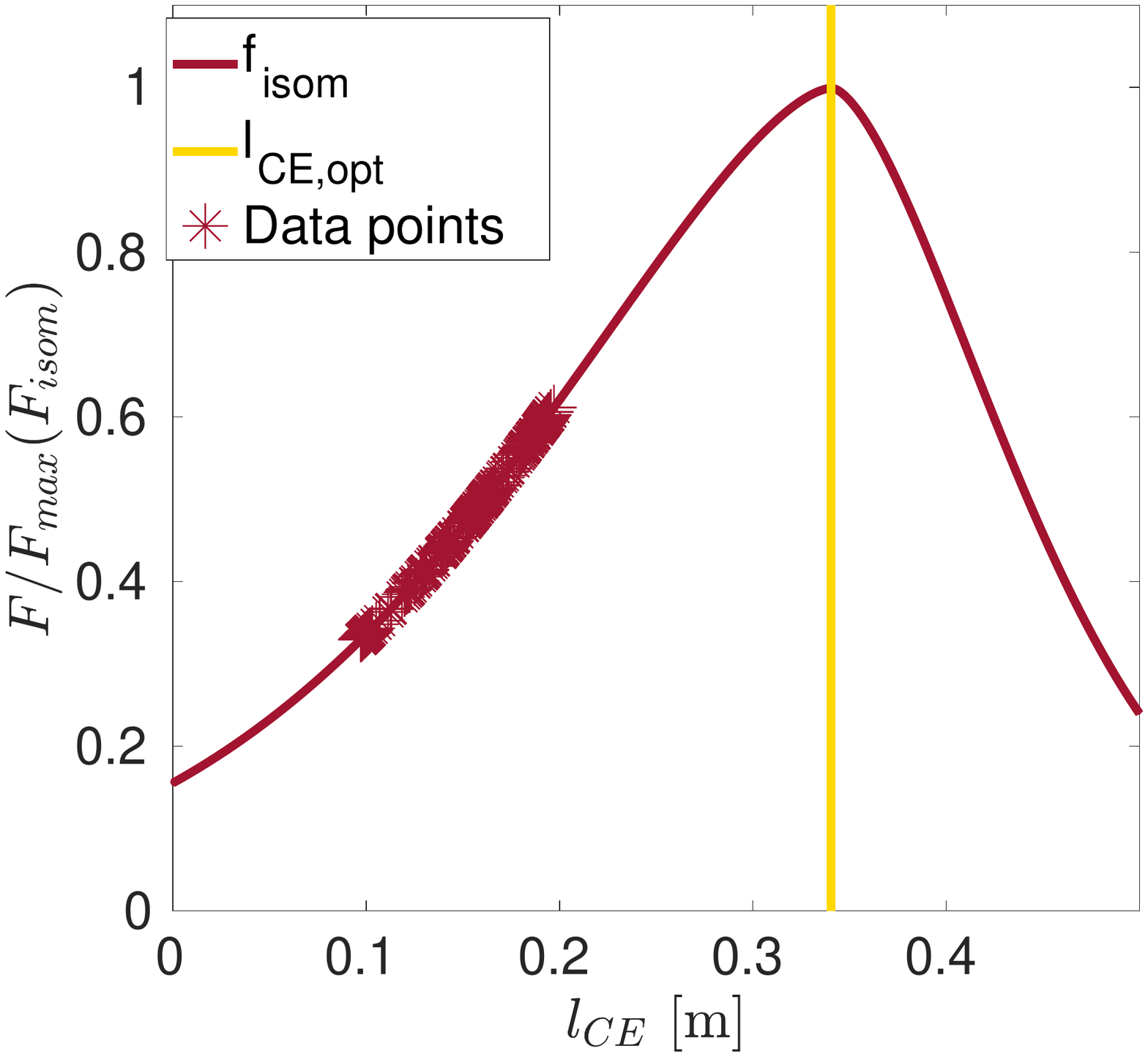}%
    \caption{Model with subject-specific parametrization.}%
    \label{fig:biceps_b}%
  \end{subfigure}%
  \caption{Isometric force-length relation of the CE for the biceps model, analogue to \cref{fig:force_curves_generic_length}, but additionally with training data points. The points are placed on the model curve and visualize the predicted relative forces for the lengths of the CE that occurred during the training trials.}%
  \label{fig:biceps_working_area}%
\end{figure}

An advantage of model approach B is that the trained parameters are physically meaningful and allow insight into the properties of the subject specific model. Furthermore, the quality of the training data can be assessed. \Cref{fig:biceps_working_area} shows the force-length relation of the biceps muscle model using the generic and the optimized parameter values. It can be seen that the subject-specific model has a smaller slope of the force curve. 
All points of the training data set are indicated by red crosses on the curves and show the operating range of the muscle in which the model has been trained. It can be seen that the experimental training data are limited to a small range of the muscle length below its optimal CE length $l_{\CE,\text{opt}}$. In order to improve the quality of the model predictions for this subject, specific additional experimental trials can be designed for model training. They can be designed to fill in values in the missing range of operation, which in this case is for larger muscle extensions.

A low computational time of the offline parameter identification and the online evaluation of the two models would be an important measure for their practical applicability. In the present study, the training phase of Model A, i.e., optimization of the quantities for the Gaussian Process Regression using 280 training data points took \SI{2.24}{\second}. The evaluation of Model A for the validation data set containing \num{54e3} points had a duration of \SI{116}{\milli\second}.

The runtimes for model B were significantly higher. The parameter optimization lasted \SI{25}{\minute} \SI{16}{\second} and the evaluation for the validation data set had a duration of \SI{13}{\second}.

The large differences in runtime between models A and B can be explained by the inefficient implementation of the biophysical model using the MATLAB programming language. During parameter identification, this model needs to be evaluated iteratively in the optimization algorithm. In contrast, the optimization within model A works with an internal implementation of the Gaussian Process which was optimized during development of the particular MATLAB functionality.
In general, the evaluation of Gaussian Process Regression has cubic time complexity whereas, for the parameter optimization of model B, iterative solvers with linear time complexity exist.

\section{Conclusion}\label{sec:study_conclusion}
In this study, elbow torques during flexion and extension of the upper arm were predicted from motion capture data and EMG measurements. Two models, A and B, were developed. Model A is non-parametric and uses Gaussian Process Regression. Model B is biophysically informed and involves two state-of-the-art Hill-type muscle models for biceps and triceps. Experiments were conducted to generate training and validation data. These training data were used for model parameter identification. Predictions from the two models were compared to real experimental values using the validation data.

Regarding the formulation and implementation, model A requires low effort and no special knowledge about the model, except where experimental motion data is preprocessed for a specific subject. In contrast, model B needs expert knowledge about the biophysical structure and the implementation of all comprised models.

Similar holds for the offline training phase. There are no parameters in model A that have to be tuned manually, which allows a quick start. Conversely, model B requires the appropriate definition of initial values and physiological constraints for the optimization problem. However, this can also be seen as an advantage for model B, as a-priori knowledge can be integrated in such a model.

On the other hand, an advantage of model A is that additional experimental data, e.g., from neighboring muscles or additional sensors, can easily be added to the model. This is not possible with model B, where the model formulation would have to be changed.

Our studies showed that both models were able to predict the levels of torque reasonably. 
\Cref{fig:nrmse} showed the best score for model A, followed by model B. It was also seen that the generic parametrization of model B does not yield a useful prediction. The same is true for a simplified version of model A, where the elbow torque was used as training input instead of derived quantities from the motion capture system that required a complex preprocessing step.

Both models provide possibilities to assess the confidence of their predictions. With model A, confidence intervals can be computed directly from the Gaussian Processes. Their usefulness was shown in the validation where regions with large errors also had a large confidence interval. 
Model B allows insight into force-length and force-velocity characteristics of the two involved muscles. The operating ranges of the muscles during the experiments can be visualized and allow assessing whether the desired model features were covered by the training phase and, thus, will yield a good prediction.

In our study, runtimes were low for model A and high for model B in both offline and online phases. However, this is due to our prototypical implementation of model B. For larger data sizes and a more sophisticated implementation, the reverse effect is expected. The runtime complexity for the training phase is better for model B (linear in time) compared to model A (cubic in time). For the online phase, costly integration over data points is needed for model A whereas model B directly provides a differential equation of the system that can be solved efficiently. 

If EMG is used to control an exoskeleton that supports the movement of the limb, it is known that the measured signals are ahead of the intended movement by a small offset. This is a result of the time delay in the neuromusculoskeletal system. This property gives the assistive exoskeleton a short time to predict the intended movement and thereby allows a seamless integration of the artificial device with human control.

When targeted at such a real-time application, both models could be considered to be integrated into the control. Model A better fits the use case of a device that could be (re\nobreakdash-)calibrated by the patient itself. Because of the built-in estimation of prediction quality, compliance and safety could be ensured more easily even for imperfect training.
Model B would need a controlled environment such as a specialist's laboratory and careful assistance for the calibration process.
After calibration, it would promise a more natural and more responsive experience because of the subject-specific model and possibly smaller compute times.

Where real-time application is not a requirement, biophysically informed models have a high potential to leverage the understanding how the human neuromusculoskeletal system operates for given tasks.
In model B of this study, the kinematics and individual muscle dynamics were described close to the current  understanding of the system. However, several aspects where not modeled as detailed as possible. The pathway from neural stimulation to excitation and activation of the muscle, the recruitment strategies including different motor units, neural feedback loops as well as effects stemming from the 3D geometry of the muscle were not considered. Therefore, this thesis develops a more detailed, biophysically informed model including these properties in the following chapters.

The presented study reproduced what similar studies in literature have shown: Subject-specific model identification for Hill-based torque prediction models can vastly improve the prediction quality compared to generic models. Our work adds to the common knowledge that this holds also for the four-element Hill-type model that was used for model B. Furthermore, a comparison with Gaussian Process Regression was given, various advantages and disadvantages of these two approaches were identified. Future work can test the two models with more subjects and increase the variety of motion in the training experiments. For example, effects resulting from high contraction velocities or eccentric contractions could be investigated to evaluate the model's potential in more complex movements.

\chapter{Generation of Meshes for the Multi-Scale Models}\label{sec:generation_of_meshes_for_multiscale}
Multi-scale models of skeletal muscles describe phenomena on different length scales and combine them into a single description. The phenomena are modeled by different sets of equations which need individual discretizations and solvers. For that, various geometrical meshes describing different physical domains are required.

The discretization considered in this work involves three-dimensional (3D) and one-dimensional (1D) meshes.
As a whole, muscles and tendons are treated as 3D domains. Muscle fascicles and myofibrils are represented by 1D fibers that are embedded in the 3D domain of the muscle.

The generation of the respective 1D and 3D meshes should be based on biomedical imaging data in order to represent actual human anatomy. The generated meshes should be of good quality such that finding numerical solutions with low error is possible. Good mesh quality usually involves mesh cells with similar lengths and angles. It should also be possible to easily partition the mesh into multiple, equally sized subdomains. This is required for efficient parallel computation.
The two requirements of good mesh quality and easy partitioning lead to the decision to employ \emph{hexahedral} elements and a \emph{structured} mesh for the 3D domains.

In this chapter, we present a workflow to construct meshes with the mentioned properties starting from biomedical data. We present novel algorithms to generate the required structured hexahedral meshes. This work contributes an implementation of the algorithms that can be used to construct all meshes needed for our biomechanical simulations.

\section{Overview and Notation of Required Meshes}\label{sec:overview_and_notation_of_required_meshes}
In the following, we summarize the meshes generated and used in this thesis and introduce their notation used in the following discussions.

The domain of the muscle belly is denoted by $\Omega_M$. A layer of fat and skin tissue is located on top of the muscle belly. It is denoted as the body domain $\Omega_B$.
The muscle belly is attached to tendons on both longitudinal ends. The tendon domains are denoted by $\Omega_{T,1}$ and $\Omega_{T,2}$. These domains are all subspaces of the 3D Euclidean space: $\Omega_M,\Omega_B,\Omega_{T,i} \subset \R^3$.

\begin{figure}%
    \centering%
    \def\svgwidth{8cm}%
    %% Creator: Inkscape inkscape 0.92.3, www.inkscape.org
%% PDF/EPS/PS + LaTeX output extension by Johan Engelen, 2010
%% Accompanies image file 'domains.pdf' (pdf, eps, ps)
%%
%% To include the image in your LaTeX document, write
%%   \input{<filename>.pdf_tex}
%%  instead of
%%   \includegraphics{<filename>.pdf}
%% To scale the image, write
%%   \def\svgwidth{<desired width>}
%%   \input{<filename>.pdf_tex}
%%  instead of
%%   \includegraphics[width=<desired width>]{<filename>.pdf}
%%
%% Images with a different path to the parent latex file can
%% be accessed with the `import' package (which may need to be
%% installed) using
%%   \usepackage{import}
%% in the preamble, and then including the image with
%%   \import{<path to file>}{<filename>.pdf_tex}
%% Alternatively, one can specify
%%   \graphicspath{{<path to file>/}}
%% 
%% For more information, please see info/svg-inkscape on CTAN:
%%   http://tug.ctan.org/tex-archive/info/svg-inkscape
%%
\begingroup%
  \makeatletter%
  \providecommand\color[2][]{%
    \errmessage{(Inkscape) Color is used for the text in Inkscape, but the package 'color.sty' is not loaded}%
    \renewcommand\color[2][]{}%
  }%
  \providecommand\transparent[1]{%
    \errmessage{(Inkscape) Transparency is used (non-zero) for the text in Inkscape, but the package 'transparent.sty' is not loaded}%
    \renewcommand\transparent[1]{}%
  }%
  \providecommand\rotatebox[2]{#2}%
  \newcommand*\fsize{\dimexpr\f@size pt\relax}%
  \newcommand*\lineheight[1]{\fontsize{\fsize}{#1\fsize}\selectfont}%
  \ifx\svgwidth\undefined%
    \setlength{\unitlength}{407.46656042bp}%
    \ifx\svgscale\undefined%
      \relax%
    \else%
      \setlength{\unitlength}{\unitlength * \real{\svgscale}}%
    \fi%
  \else%
    \setlength{\unitlength}{\svgwidth}%
  \fi%
  \global\let\svgwidth\undefined%
  \global\let\svgscale\undefined%
  \makeatother%
  \begin{picture}(1,0.28756145)%
    \lineheight{1}%
    \setlength\tabcolsep{0pt}%
    \put(0,0){\includegraphics[width=\unitlength,page=1]{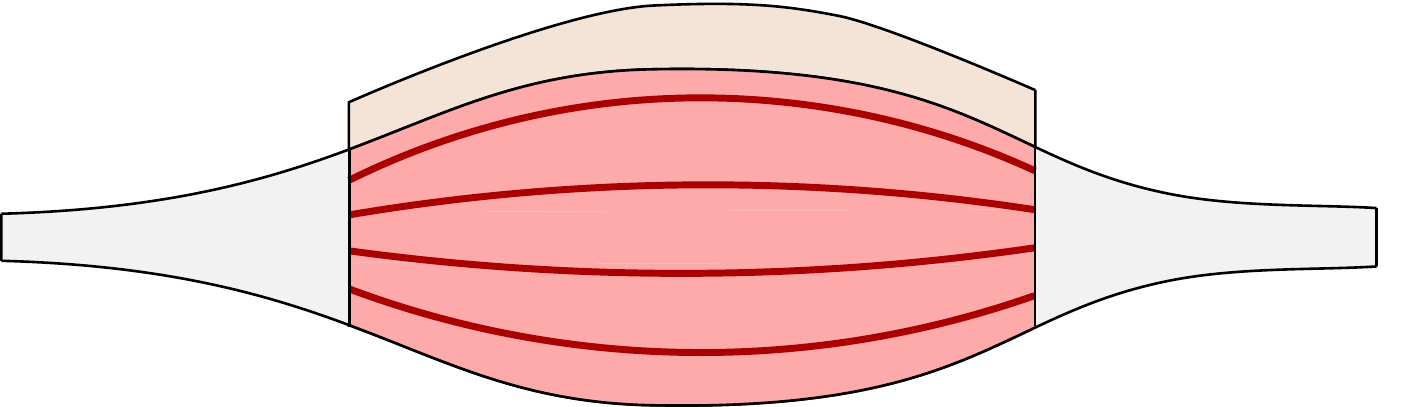}}%
    \put(0.45172276,0.17185886){\color[rgb]{0,0,0}\makebox(0,0)[lt]{\lineheight{1.25}\smash{\begin{tabular}[t]{l}$\Omega_M$\end{tabular}}}}%
    \put(0.45172276,0.24685348){\color[rgb]{0,0,0}\makebox(0,0)[lt]{\lineheight{1.25}\smash{\begin{tabular}[t]{l}$\Omega_B$\end{tabular}}}}%
    \put(0.78698257,0.10787365){\color[rgb]{0,0,0}\makebox(0,0)[lt]{\lineheight{1.25}\smash{\begin{tabular}[t]{l}$\Omega_{T,2}$\end{tabular}}}}%
    \put(0.0941124,0.10316333){\color[rgb]{0,0,0}\makebox(0,0)[lt]{\lineheight{1.25}\smash{\begin{tabular}[t]{l}$\Omega_{T,1}$\end{tabular}}}}%
    \put(0.10961439,0.26473749){\color[rgb]{0,0,0}\makebox(0,0)[lt]{\lineheight{1.25}\smash{\begin{tabular}[t]{l}$\Omega_{F,1}$\end{tabular}}}}%
    \put(0.10961439,0.21669754){\color[rgb]{0,0,0}\makebox(0,0)[lt]{\lineheight{1.25}\smash{\begin{tabular}[t]{l}$\Omega_{F,2}$\end{tabular}}}}%
    \put(0,0){\includegraphics[width=\unitlength,page=2]{domains.pdf}}%
  \end{picture}%
\endgroup%
    \caption{Visualization of the computational domains in a simulated muscle: tendons $\Omega_{T,1}, \Omega_{T,2}$, muscle belly $\Omega_M$, body domain $\Omega_B$ and fiber domains $\Omega_{F,i}$.}%
    \label{fig:fibers_domains}%
\end{figure}

Additionally, a number $n_f$ of individual muscle fibers $\Omega_{F,i} \subset \R^3$ for $i \in \{0,\dots,n_f\}$ is introduced. Each fiber is a 1D manifold embedded in the 3D domain, i.e., $\Omega_{F,i} \subset \Omega_M$. \Cref{fig:fibers_domains} summarizes the notation of the domains.

For the application of the finite element method (FEM), we create meshes for each of these domains. Formally, a 3D mesh $\Omega_\text{3D}$ is given by a number of 3D elements $\{U_{\text{3D},i}\}_{i=1,\dots,n}$ with $U_{\text{3D},i} \subset \R^3$ such that their disjoint union approximates the domain,
$\dot{\bigcup}_{i=1}^{n} U_{\text{3D},i} \approx \Omega_\text{3D}$. Similar holds for 1D meshes.

The elements are non-overlapping and can be defined by nodes and edges. In the discretizations used here, no hanging nodes are allowed, i.e., at any node all adjacent elements share the node.

\begin{figure}%
  \centering%
  \begin{subfigure}[t]{0.5\textwidth}%
    \centering%
    \includegraphics[width=\textwidth]{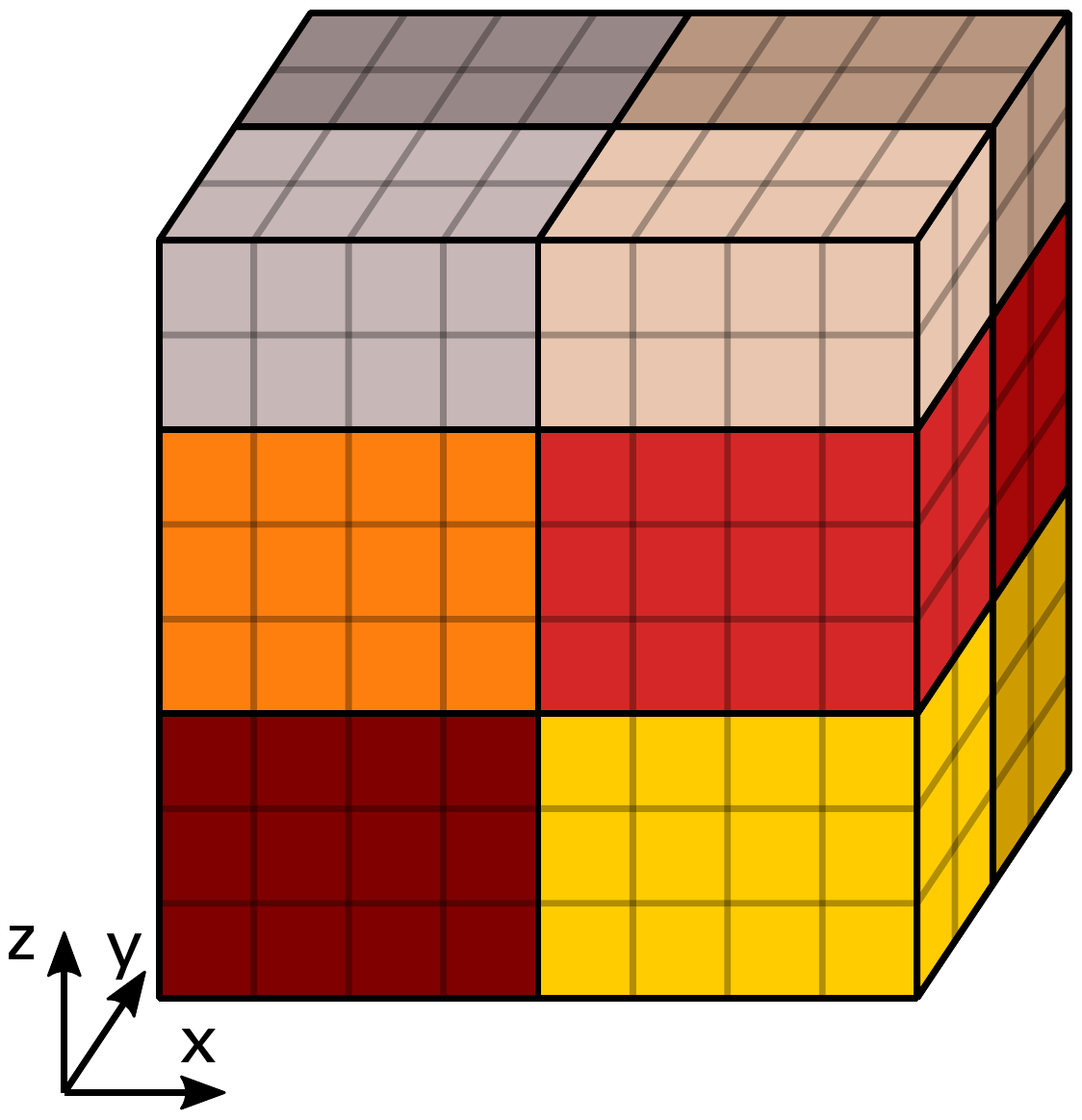}%
    \caption{Parallel decomposition of a 3D mesh with $n_x \times n_y \times n_z = 8 \times 4 \times 8$ elements into $2 \times 2 \times 3=12$ subdomains.}%
    \label{fig:fiber_creation_decomposition}%
  \end{subfigure}
  \quad
  \begin{subfigure}[t]{0.45\textwidth}%
    \centering%
    \includegraphics[width=0.6\textwidth]{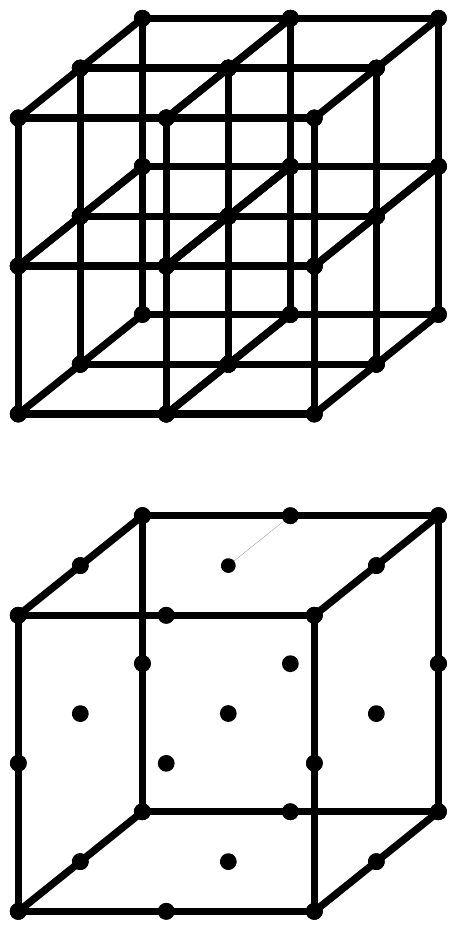}%
    \caption{Top: a linear 3D mesh with eight elements, bottom: a single quadratic element, which uses the same nodes as the linear 3D mesh at the top.}%
    \label{fig:fiber_creation_quadratic_elements}%
  \end{subfigure}    
  \caption{Structured 3D meshes that are used in the simulations: parallel partitioning and construction of quadratic elements.}%
  \label{fig:quadratic_elements_decomposition}%
\end{figure}

Furthermore, only structured, hexahedral meshes are considered in this chapter.
A 3D structured mesh is isomorphic to a 3D Cartesian grid with equidistant elements. 
This has advantages for programmatically indexing nodes and elements as well as for parallel partitioning of the domain.
\Cref{fig:fiber_creation_decomposition} shows an example of a 3D structured mesh that is partitioned into twelve subdomains. The subdomains are constructed by planar cuts through the structure of the mesh. These cuts are typically defined in a way that the resulting subdomains have similar numbers of 3D elements and, thus, every process gets a similar portion of the total computational load.

The number $n$ of 3D elements is the product of the numbers $n_i, n_j$ and $n_k$ of elements in the three coordinate directions $x,y$ and $z$ of the Cartesian grid,
 i.e., $n = n_i\,n_j\, n_k$.
Each element can be indexed by a triple $(i,j,k)$ of indices with the ranges $i \in \{0,\dots,n_i-1\}, j \in \{0,\dots, n_j-1\}$ and $k \in \{0,\dots,n_k-1\}$. 
In the simulation program, typically, consecutive indices $\iota$ are used that iterate over all elements $\iota \in \{0,\dots,n-1\}$ and are obtained from the index triples by the mapping 
$(i,j,k) \mapsto \iota = k\,n_i\,n_j + j\,n_i + i$.

The elements of such a mesh can have different numbers of \emph{degrees of freedom (dof)} depending on the desired spatial order of consistency of the finite element discretization. The number $n_{\text{dofs,}d\text{D}}$ of dofs in a $d$-dimensional element is computed from the number $n_{\text{dofs,1D}}$ of dofs along one coordinate direction of the element as $n_{\text{dofs,}d\text{D}} = n_{\text{dofs,1D}}^d$.
Consequently, linear elements have two dofs in 1D meshes and eight dofs in 3D meshes. Quadratic elements have three dofs in 1D and 27 dofs in 3D.

The dofs are located at the \emph{nodes} of the elements. In linear and quadratic elements, every node corresponds to a single dof. While the nodes form the \say{corners} of linear 3D elements, they are also located on the faces and in the interior of quadratic 3D elements. \Cref{fig:fiber_creation_quadratic_elements} shows a mesh with $2 \times 2 \times 2$ linear 3D elements at the top. The same 27 nodes can be used to define a single quadratic 3D element as shown at the bottom of \cref{fig:fiber_creation_quadratic_elements}. 

It is sufficient to develop a method for constructing structured 3D meshes with linear elements. 
Higher order elements can be geometrically constructed by using the nodes of multiple adjacent linear elements.
To generate both linear and quadratic elements, we always begin with generating a mesh with even numbers $n_i,n_j$ and $n_k$ of elements in the coordinate directions. Then, linear and quadratic meshes can be extracted from the set of nodes. Similarly, a linear 1D mesh with an even $n_i$ can be easily converted into a quadratic 1D mesh.

The next sections describe workflows and algorithms to construct 3D meshes for the domains $\Omega_M,\Omega_B,\Omega_{T,i}$ and 1D meshes for the fibers $\Omega_{F,i}$ based on anatomical information. \Cref{sec:fiber_meshes_related_works} gives an overview over available meshing software tools and existing algorithms in literature. Then, \cref{sec:preprocessing_of_the_muscle_geometry} presents a workflow to extract a smooth surface representation from anatomical imaging data. In \cref{sec:ser_alg_meshes}, two serial algorithms are presented to generate 3D meshes and 1D fibers meshes. The next section, \cref{sec:parallel_algorithm}, extends these serial algorithms formulating a parallel algorithm and shows and discusses results. Finally, \cref{sec:meshes_summary_and_conclusion} gives a summary and concludes this chapter.

\section{Related Work}\label{sec:fiber_meshes_related_works}

Generating volumetric meshes for domains enclosed by a given surface is a task that is frequently needed in computational science. It is a preprocessing step whenever spatially discretized models have to be solved numerically. In consequence, a vast amount of literature has addressed this algorithmic task and various approaches and methods have been proposed. Moreover, numerous software packages that solve this problem exist. Especially tools for Computer Aided Design and Engineering (CAD/CAE) as well as free and commercial preprocessing tools and finite element solver software include functionality to generate meshes from given surfaces.

An example from the biomechanical domain is \cite{untaroiu2013finite}. The study develops a finite element model of the lower limb of an occupant of a car with the aim to investigate injury scenarios during traffic crashes.
The lower extremity geometry was obtained by computer tomography (CT) and magnetic resonance imaging (MRI) scan data of a 50th percentile male volunteer. Different meshes of bones and ligaments were created using the three tools IA-FEMesh \cite{grosland2009ia}, TrueGrid \cite{TrueGrid} and Hyper-Mesh \cite{Hypermesh} which will be outlined in the following.

\emph{IA-FEMesh} (University of Iowa, Iowa City, USA) is an open source tool to generate hexahedral meshes \cite{grosland2009ia}. It provides an interactive environment where existing geometries can be loaded. In a visualization window, bounding boxes, called blocks, can be positioned such that they contain the whole geometry. A structured grid on the block is then projected onto the surface of the geometry. Multiple blocks can be placed to account for more complex geometries. The resulting surface mesh is improved using Laplacian smoothing which equalizes the edge lengths of the elements.
The interior nodes are generated using interpolation.
The result is a structured mesh if only one block is used or an unstructured mesh if multiple blocks are used. Further operations to manage mesh density, visually manipulate the meshes and add material properties, load and boundary conditions are available. The model can be exported in a file format for finite element analysis with ABAQUS (Dassault Systèmes, Vélizy-Villacoublay, France) \cite{ABAQUS}.

The second tool is \emph{TrueGrid} (XYZScientific Applications, Livermore, USA) \cite{TrueGrid}. It is a commercial toolkit to generate hexahedral meshes. The project was started in the early 1990s as the successor to the even older preprocessor software \emph{INGRID}. Similar to \emph{IA-FEMesh}, a projection method and a multi-block technique are used. Some effort has been put into dealing with holes and sewing together dissimilar blocks.

The third tool is \emph{Hyper-Mesh}, the commercial pre-processing and post-processing toolkit of Hyperworks (Altair HyperWorks, Troy, USA) \cite{Hypermesh}.
Altair sells infrastructure and solvers for a multitude of physics and is targeted at a wide range of industries. 
Being a commercial vendor, information about the internals of their preprocessing software are hardly provided.

More meshing software exists, such as CGALmesh \cite{Jamin2015CGALmesh} for tetrahedral meshes. The package gives quality guarantees of their generated meshes and includes four mesh optimization algorithms to further improve the mesh quality.

Another application-oriented work dedicated to the use of commercial tools is \cite{Ellankavi2018}. A workflow for patient-specific modeling, simulation and analysis of the interaction between a residual lower limb stump and the socket of a prosthesis is presented. Imaging data were taken from magnetic resonance diffusion tensor imaging where also the preferred diffusion direction of water molecules along muscle fibers is captured. The open source tool MedInria (National Institute for Research in Digital Science and Technology (Inria), France) \cite{vichot2012cardiac} was used to extract muscle fibers. The residual limb data were processed using the commercial 3D image segmentation software Simpleware ScanIP (Synopsys, Mountain View, USA). Auxiliary tasks were performed using MATLAB (MathWorks,	Natick, USA) scripts. The commercial multiphysics solver LS-DYNA (LSTC/Ansys, Canonsburg, USA) was used for the simulations.

Commercial tools usually have the advantage that more development effort was put into them, than is possible for open source codes from the scientific community. This often leads to more robust and user-friendly software. An advantage of open source software is that the used algorithms are disclosed to everyone. They are often well documented or described in a publication. This allows to assess the expected quality of the generated meshes. Conversely, commercial vendors usually have no interest in revealing their internal algorithms.

For our simulation, structured, hexahedral meshes are needed. Several of the described tools are able to generate hexahedral meshes, however the meshes are typically unstructured. For our special need of 1D muscle fibers embedded in a 3D mesh, we develop our own method that is based on the ideas of existing algorithms. In the following, an overview over the algorithmic common knowledge of creating simplex meshes and hexahedral meshes is given as a basis.

% other 3D meshing
% tets
Triangulating a 2D domain is the archetype of mesh creation. The triangulation named after B. Delaunay was formulated in 1934 \cite{delaunay1934sphere}. For a given set of points, it maximizes the minimum angle of the triangles and, thus, avoids small angles. Therefore, a guarantee on the quality of the triangulation is given.

In 1995, J. Ruppert presented the Delaunay refinement algorithm \cite{Ruppert1995}, which constructs a Delaunay triangulation conforming to prescribed connected points. This algorithm is still commonly used and also part of numerous derived meshing techniques.

In 1997, P. Chew developed an algorithm for meshing a 3D domain with tetrahedra \cite{chew1997guaranteed} and proved that the aspect ratio of the tetrahedra is bounded, i.e., degenerate, \say{flat} tetrahedra, called slivers, are avoided.

The authors of \cite{Alliez2005Variational} propose a variational approach to triangulation where a quadratic energy function is minimized. During minimization both vertex positions and connectivity are optimized. This leads to better quality meshes than by simple Delaunay triangulations.

% tets to quads:
Hexahedral meshes can be obtained from certain tetrahedral meshes by splitting up each tetrahedron into four hexahedra. This is discussed in \cite{eppstein1999linear}. A remaining issue is that the generated meshes from this procedure are highly unstructured and some hexahedra have poor quality, whereas the goal would be to construct elements that are almost equilateral.

% hexs
A different approach is to directly generate a hexahedral mesh for the given surface geometry.
The survey in \cite{owen1998survey} identifies four different strategies for generating unstructured hexahedral meshes.

The first one is a \emph{grid-based} approach. It was introduced in \cite{schneiders1996grid,schneiders1997algorithm}. The interior of a given solid is filled with a regular and Cartesian grid of as many hexahedral elements as fit into the space. Then, the gaps at the surface are filled with additional elements. This method is robust but can lead to poor quality elements near the surface. 
%The orientation of the interior grid highly depends on the orientation of the initial surfaces and may not be the natural orientation of the given volume.

The second approach for generation of hexahedral meshes are \emph{medial surface methods} \cite{price1995hexahedral, price1997hexahedral}. First, the volume is decomposed into subregions by medial surfaces such that the resulting domains are one of only 13 possible types. Predefined templates are used to fill the domains with hexahedral elements. Then, the continuity between the domains is ensured using linear programming. This approach gives good results for some geometries but has robustness issues when general geometries are considered.

The third approach is called \emph{plastering}. It was first described by \cite{blacker1993seams} and continued by \cite{staten2006unconstrained,staten2010unconstrained}.
It is a moving-front method where hexahedral elements are placed in layers starting at the boundary and moving towards the interior. Intersection of faces has to be detected when the fronts meet in the interior and rules for connecting to existing faces have to be defined.
During this process, complex shaped voids can occur in the interior. When it is no longer possible to fill the voids with hexahedra already placed elements have to be removed.
A new method, called unconstrained plastering, starts from an unmeshed volume boundary. The approach has general robustness issues and is not guaranteed to find a solution for arbitrary boundaries.

The forth approach is \emph{whisker weaving}, introduced by \cite{tautges1996whisker} and extended by \cite{ledoux2008extension,kawamura2008strategy}. Here, the dual of the hexahedral mesh is considered. The dual consists of the three surfaces per hexahedron that lie in the planes of symmetry. The surfaces of all hexahedra form topological loops. 
The principle is now to first construct the dual of the mesh, which can be determined from the given boundary surface. Then, the actual hexahedral mesh is created from the dual, using the surfaces as guides where to place the elements.
The dual forms topological loops inside the volume. One important criterion for generating good quality meshes is that self-intersections of these loops are resolved in a first step.
The approach, used with subsequent smoothing, can produce meshes of good quality. However, no guarantee is given. One problem is that the resulting mesh depends on the quality of the surface mesh and that the number of nodes can increase significantly during the method.

For the whisker weaving method and for some plastering methods, a quadrilateral mesh of the surface is required. Algorithms for creating high quality quadrangulations of closed surfaces exist \cite{dong2005quadrangulating,Kovacs2011Anisotropic,Bessmeltsev2012,Meng2016Consistent}.

Other approaches start with 3D volumetric medical imaging data instead of surfaces. In \cite{Zhang2003,Zhang20053DFiniteElementMeshing}, adaptive tetrahedral and hexahedral meshes are created from volumetric data using octree subdivision. The method avoids hanging nodes and allows a feature sensitive adaptivity. While adaptive meshing methods can reduce the number of elements in the interior of the volume, a problem is that the worst quality elements are generated at the boundary, the location where the solution in a finite element study usually is most interesting.

\nocite{Gregson2011}

Multiple reasons make the previously outlined approaches unsuited regarding the needs for our parallel muscle simulation. 

(i) The generated meshes are unstructured. When performing domain decomposition for parallel computing on unstructured meshes, graph-partitioning methods have to be used. Storing an unstructured mesh requires storage of element adjacency information. Partitioned meshes additionally require storage of the adjacent processes. In contrast, structured meshes can be trivially decomposed and stored efficiently. A decomposition can be represented in memory by a very low number of parameters.

(ii) The presented methods are designed for hexahedral meshing of arbitrary volumes. Robustness and mesh quality at the same time remain issues that are not completely solved for most of the algorithms. Often, expensive smoothing steps are needed to increase mesh quality.

(iii) In general, either no assumption can be made about orientation or alignment of hexahedra in the interior, or, for the grid-based approach, the elements at the surface have poor quality. Having a mesh that is consistently aligned with, e.g., the main diffusion direction or the preferential direction of the anisotropic material or the muscle fibers can reduce numerical errors in the finite element solution.

Consequently, a more scenario specific solution is needed that can avoid the mentioned issues. Such solutions can also be found in the literature. An example is \cite{blemker2005three}, where 3D finite element models for various complex muscle geometries around the hip are generated from magnetic resonance images. Segmentation and surface mesh generation are performed using the old, unmaintained software \emph{Nuages} (Inria, France) \cite{Nuages}.
Then, a 3D hexahedral mesh is generated using TrueGrid. A structured template mesh on a unit square is mapped to the horizontal slices of the muscle geometry that resulted from the segmentation. After mesh smoothing, the slices are connected vertically to form a 3D mesh. 
Fiber directions are described by Bézier curves in a reference volume and mapped to the muscle geometry using the same mapping. The fiber direction then are used in a transversely-isotropic material formulation. Simulations are performed using the finite element solver \emph{Nike3D} (Lawrence Livermore National Lab, Livermore, USA) \cite{Nike3D}.

We base our work on this study and use a similar mapping from a template mesh to the actual muscle volume. In comparison to \cite{blemker2005three}, we use an improved mapping based on harmonic maps, which potentially leads to better quality meshes on the slices of the muscle. Instead of the unit circle template mesh, we experiment with different reference meshes and evaluate their quality.

In the study of \cite{blemker2005three}, fiber directions are defined based on anatomically assumed directions.
However, the definition is carried out on the cuboid reference geometry. This means that the authors mentally morph the muscle geometry into the reference geometry in order to define fiber directions, using their expertise. Then, the fiber directions together with the cuboid are transformed back to the actual geometry. This approach simplifies the definition of the fibers. However, defining the fiber direction directly on the muscle geometry can lead to better results. Thus, our approach is to automatically estimate fiber directions and define fibers directly in the muscle domain. At the same time, the 3D mesh and 1D fibers are aligned in our work to allow for better numerical and data structure properties of the discretization. 

The definition of fiber directions follows a method proposed in \cite{Choi2013}. 
The directions are assumed to follow a divergence free vector field. Such a field can be created by taking the gradient of the solution of the Laplace equation. Neumann boundary conditions are defined at the attachment points of the muscle tendon complex. The solution of the Laplace equation corresponds to the pressure values of a potential flow. Its gradient corresponds to the velocity and individual fascicles or fibers can be obtain by tracing streamlines through the velocity field.
This approach is extended and validated by the studies in \cite{Inouye2015} and \cite{Handsfield2017}. We incorporate this method into our workflow.

\section{Preprocessing of the Muscle Geometry}\label{sec:preprocessing_of_the_muscle_geometry}

\begin{figure}%
  \centering%
  \includegraphics[height=15cm]{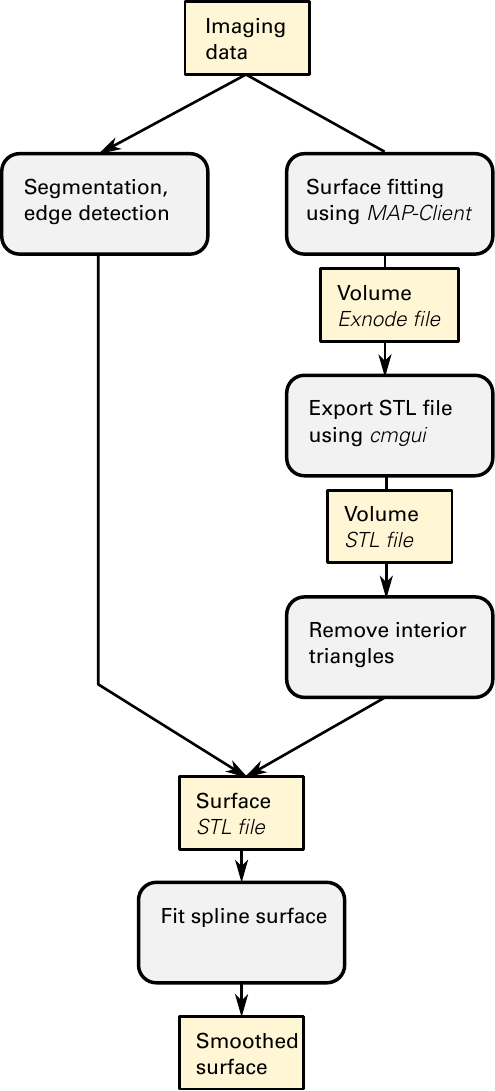}%
  \caption{Workflow of generating a surface representation of the muscle and tendons from imaging data. Operations and intermediate results are shown as gray and yellow boxes, respectively. Two alternatives are given by the two branches. On the left, the imaging data are automatically processed to directly retrieve points on the surface of the muscle. The right branch achieves the same with three steps of which the first one involves manual adjustments. At the end, a spline surface smooths the collected data from both possibilities to yield the resulting surface representation.}%
  \label{fig:scheme_preprocessing}%
\end{figure}

The first step towards creating a structured mesh is to obtain a representation of the surface of the muscle. 
Starting point is a human biomedical imaging data set. In this section, two possible workflows are presented how to extract the muscle and tendon surfaces from imaging data. 
The two workflows are visualized in \cref{fig:scheme_preprocessing}. The workflow using the branch on the left side in \cref{fig:scheme_preprocessing} is automized but only works for the particular data set and extracting the biceps muscle.
The right branch involves manual steps and is applicable for any muscle geometry.

% overview over subsections

\subsection{Data Source}
Anatomic images provide the basis for the extraction of muscle geometries.
Our used data set originates from the Visible Human Project \cite{visible_human_male} of 
the United States National Library of Medicine. 
The project has published anatomic images derived from a male corps, among other data sets.
The data, known as \say{Visible Human Male}, were published in 1994.
Colored images of transversal cross-sections were obtained by cryosectioning.
A total of \num{1871} images with dimensions of \num{2048} by \num{1216} pixels and 24 bit color depth visualize the whole human body. Parts of the upper arms are contained in approximately 500 of these images. The size of a pixel is \SI{0.33}{\milli\meter} in transversal direction and \SI{1}{\milli\meter} in axial direction. The size of the complete set of JPEG compressed images is \SI{772}{\mega\byte}. Cropping and selecting the relevant portions of the upper arm extracts a dataset with the size of \SI{35}{\mega\byte}.

\begin{figure}%
  \centering%
  \includegraphics[height=10cm]{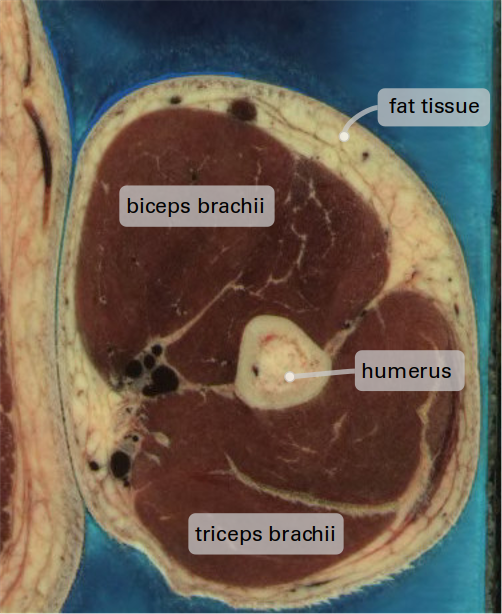}% 0483
  \caption{Exemplary extract of image number 483 from the Visible Human Male. A transversal slice of the left upper arm is shown as seen from the bottom. The biceps and triceps muscles as well as the humerus bone can be identified. }%
  \label{fig:vhp_image}%
\end{figure}

An extract of an image of the upper arm is given in \cref{fig:vhp_image}. 
The location of biceps and triceps brachii muscles can be identified in the dark red tissue. For the biceps, the two muscle heads are visible, separated by the bright diagonal line from bottom left to top right. For the triceps, at least two of the three heads can be identified. The blue background is colored frozen gelatin that was needed during cryosection to stabilize the arms.

\subsection{Automatic Surface Extraction}
This section outlines the automatic algorithm to obtain the muscle surface from the Visible Human Male data set. The scheme corresponds to the left branch in \cref{fig:scheme_preprocessing}. The algorithm was implemented in a Python script as part of the Bachelor thesis of Kusterer \cite{Kusterer} that was supervised by me.
The algorithm is capable of extracting muscle and bone geometries from the mentioned imaging dataset.

At first, the color values in the images are used to segment the pixels into muscle tissue, surrounding tissue and skeletal structure. The algorithm traverses the selected and cropped relevant parts of the images. 
%For example, to consider the biceps muscle, the image region of pixels coordinates $(x,y)$ with $x \in [1300,1720] $ and $ y\in[1030,1720]$ are considered in the images with numbers 284 to 778. 

For every such part of an image, pixels that match a certain range in the RGB color space are marked and categorized. The categories are muscle tissue and, for demonstration, also bone tissue. The corresponding color ranges are given in \cref{tab:color_ranges}.

The color based classification does not succeed everywhere as the white shade corresponds not only to bone material but also to fat and other tissue. Therefore, the algorithm removes artifacts located near the outer gelatin from the set of pixels that was categorized as bone. 

\begin{table}
  \centering%
  \begin{tabular}{|l|lll|}
    \hline
    & red & green & blue\\
    \hline
    muscle & $60 - 100$& $30-75$   & $15-60$\\
    bone   & $145-255$ & 1$35-205$ & $60-160$\\
    \hline
  \end{tabular}
  \caption{Ranges in the RGB color space to identify pixels of muscle and bone segments. The numbers correspond to 24 bit colors with the range $[0,255]$ for every color channel.}%
  \label{tab:color_ranges}%
\end{table}

\begin{figure}%
  \centering%
  \fcolorbox{black}{black}{\includegraphics[height=7cm,trim=0 0 0 6cm, clip]{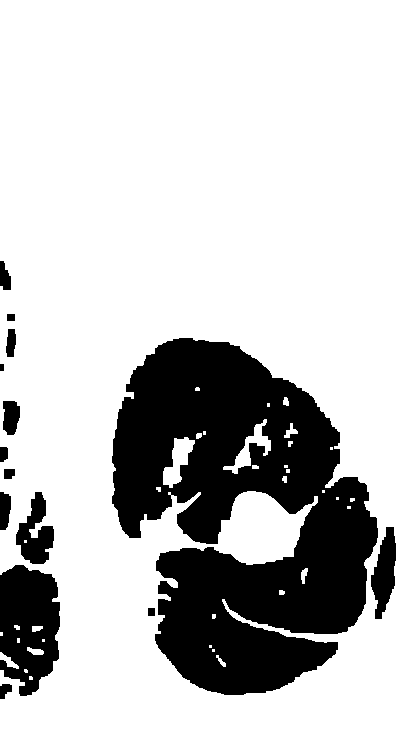}}\quad%
  \fcolorbox{black}{black}{\includegraphics[height=7cm,trim=0 0 0 4cm, clip]{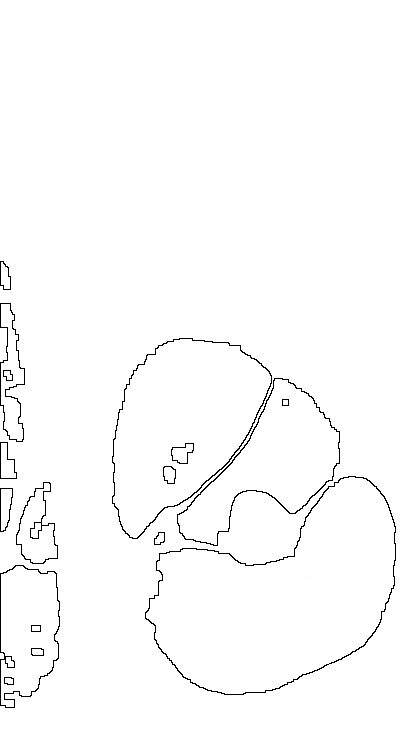}}\vspace*{5mm}\\
  \fcolorbox{black}{black}{\includegraphics[height=7cm,trim=0 0 0 6cm, clip]{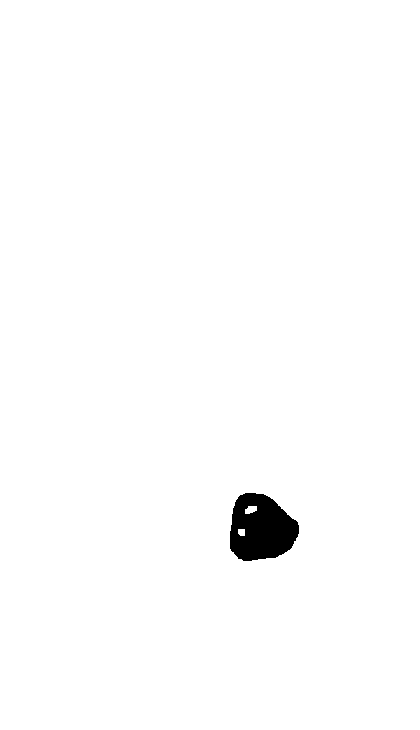}}\quad%
  \fcolorbox{black}{black}{\includegraphics[height=7cm,trim=0 0 0 6cm, clip]{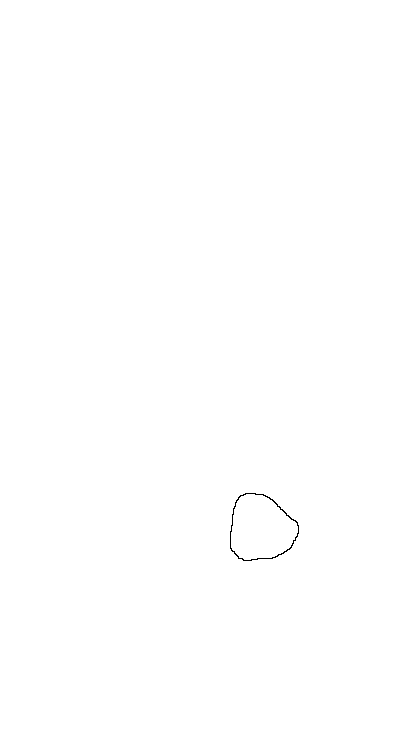}}%
  \caption{Intermediate steps of the algorithm to determine surface geometry of muscles and bones. The left column shows pixels from the image in \cref{fig:vhp_image} that were categorized to be muscle tissue (top) and bone material (bottom). The right column shows a later step in the algorithm, where the surface of muscle (top) and bone (bottom) is estimated.}%
  \label{fig:extraction}%
\end{figure}

Exemplary results for image number 483 are given in the left column of \cref{fig:extraction}.
It can be seen that the marked regions for muscle and bone have gaps in the interior resulting from differently colored tissue inside muscles and bones. On some images, the set of pixels also includes small objects outside the actual muscle and bone regions.

To reduce the gaps and small objects, the morphological operations \emph{closing} and \emph{opening} are applied on the data. These operations consist of \emph{dilation} and \emph{erosion} steps. Both are pixel based operations that traverse the dataset and for every pixel consider a window of $3\times 3$ pixels centered at the current position. Dilation picks the maximum value and erosion the minimum value from this window and assigns it as the pixel's value in a new image. In our case, values of zero and one correspond to non-categorized and categorized pixels, respectively.

Closing consists of dilation followed by erosion and closes small gaps or holes in the marked objects. Opening consists of erosion followed by dilation and removes small artifacts outside the actual bone and muscle areas. It was found effective to perform both dilation and erosion twice in sequence to yield good results containing almost no more holes nor unwanted small objects.

Next, the algorithm determines the contours of all regions with marked pixels. This leads to lines with a width of one pixel that enclose the muscle and bone areas. The right column of \cref{fig:extraction} shows the results after this step. It can be seen that numerous gaps have been closed by the morphological operations. In some images, as in the considered example, the muscle area gets split into multiple smaller enclosed regions, which is not desired. These images skipped in the processing. However, proper contours of the biceps are found in the majority of images. 

\fboxsep=0mm   % padding thickness
\fboxrule=1pt   % border thickness

In the next step, a single contour for each of muscle and bone is obtained in every image. If there are multiple contours per image, the one that is located closest to the upper right corner of the image is selected for the muscle. If all contours in an image are shorter than 20 pixels, this is an indication for bad segmentation quality and the whole image gets discarded. Because of the discarded images, the resulting surface description has a lower resolution at the respective locations. This is not a problem as the data is subsequently approximated by a smooth spline surface.

\begin{figure}%
  \centering%
  \begin{subfigure}[t]{0.48\textwidth}%
    \centering%
    \includegraphics[height=7cm]{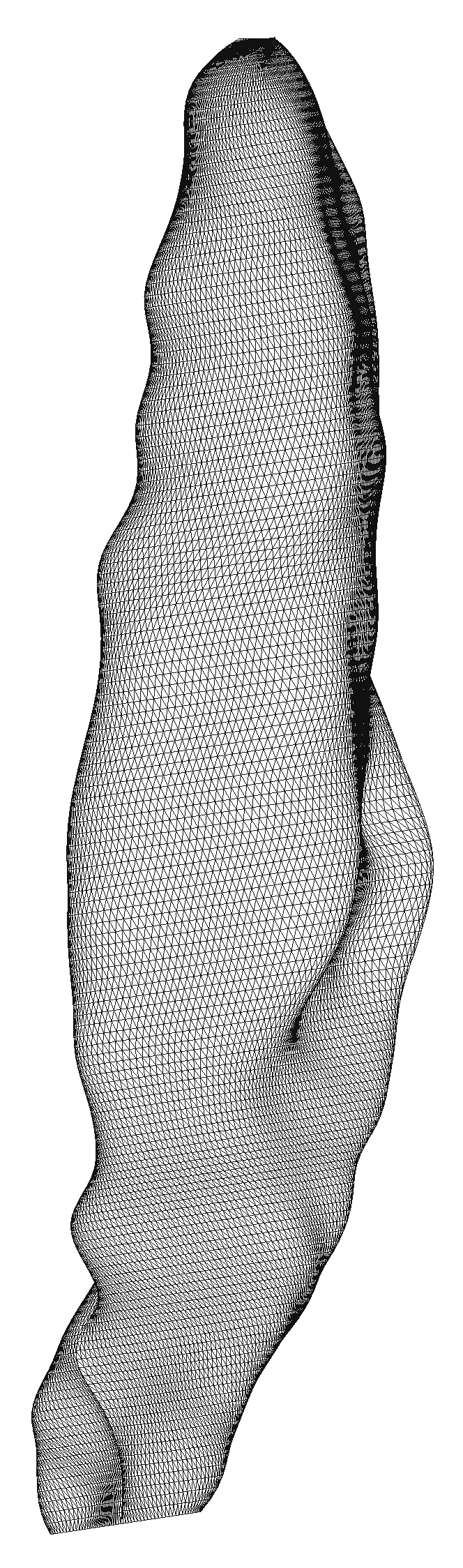}%
    \caption{Surface of the biceps brachii muscle. On the right-hand side of the muscle, the groove of the humerus bone can be seen.}%
    \label{fig:extraction_result_biceps}%
  \end{subfigure}
  \begin{subfigure}[t]{0.48\textwidth}%
    \centering%
    \includegraphics[height=7cm]{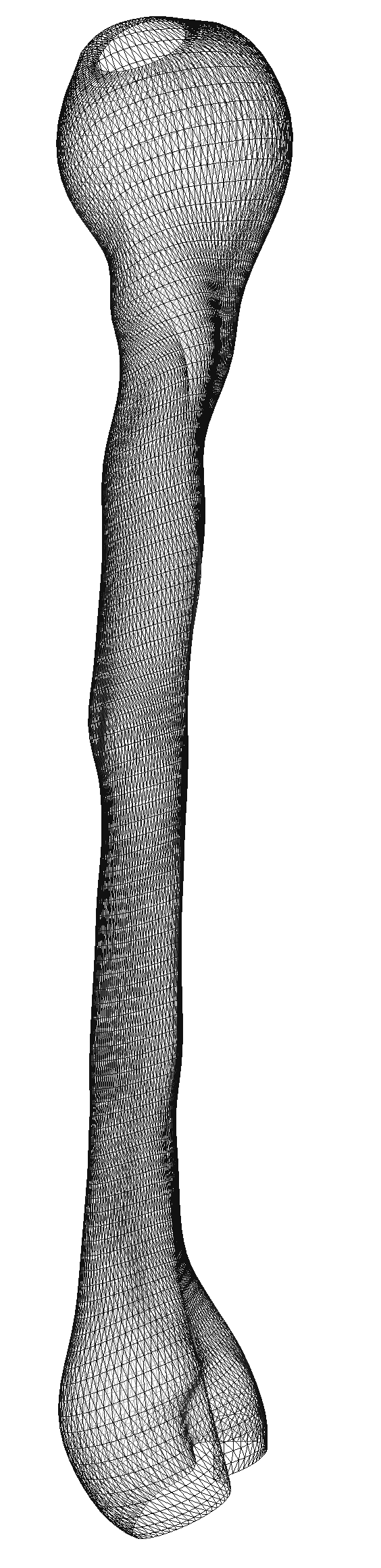}%
    \caption{Surface of the humerus.}%
    \label{fig:extraction_result_humerus}%
  \end{subfigure}    
  \caption{Surfaces of biceps and humerus bone obtained by the automatic surface extraction algorithm.}%
  \label{fig:extraction_result}%
\end{figure}

The result is a set of contours for muscle and bone in the cross-sectional planes of the images. Combining these, we get a point cloud in 3D space that approximates the surface of the biceps muscle and the surfaces of the considered bones humerus, ulna and radius. Using these points, a spline surface can be fitted and subsequently triangulated. Resulting surfaces for the biceps and humerus bones are shown in \cref{fig:extraction_result}.

The runtime for the algorithm applied on a dataset with \num{495} images and approximately \num{144e6} pixels in total was \SI{121}{\minute}. The used hardware was an AMD Ryzen 5 1600 processor with 6 cores, 3.2 GHz and \SI{16}{\giga\byte} RAM, of which a maximum of \SI{2}{\giga\byte} was used. Because processing of the images can be done in parallel, the runtime was reduced to approximately half (\SI{62}{\minute}) using 2 threads and to a quarter (\SI{30}{\minute}) using 6 threads.

The advantage of the presented algorithm is that the outcome solely depends on the imaging data and, thus, no modeling error by manual approximation of the geometry occurs. For example, the obtained surfaces of biceps and humerus geometrically fit perfectly into each other. Intermediate steps are stored as black and white images. By editing these between the steps of the algorithm, manual tweaking is possible and can be used to increase the quality of the results.

A disadvantage is that the algorithm relies on color information in the imaging data to differentiate between muscle and other tissue. Because some involved tissue types have similar colors, this approach can be error-prone. Furthermore, the color ranges need to be determined experimentally. Therefore, the algorithm is not very robust with respect to image noise and needs adjustments when it should be used to extract other muscles. Expert knowledge about the location and shape of human muscles cannot be used easily to improve the results of the algorithm.

An alternative approach is to manually segment the imaging data and construct surfaces with the help of a tool. This approach is described in the following section.

\subsection{Manually Guided Surface Extraction}\label{sec:surf_extr}

Manually guided segmentation can be done using the \emph{MAP client} of the Musculoskeletal Altas Project (MAP) \cite{mapclient}. This application allows creating and execute a workflow to achieve data processing and simulation tasks. In a graphical window, the user can place and connect various workflow steps. When executing the workflow, each step shows a dialog where the required configuration can be entered or the operations can be performed on a visual representation of the data at this workflow stage. 

Possible workflow steps include source and sink operations such as reading image data and writing meshes. Imaging data such as the 2D images from the Visual Human Male can be visualized in a 3D representation. The user can place points in the 3D space to mark boundaries of the visualized muscle and tissue structures.
Further workflow steps allow creating meshes of predefined geometrical shapes, such as cubes and cylinders and merge them into a common mesh. These meshes can be fitted to point clouds of user defined points. This is done by a least squares approach minimizing the distances between user created points and the mesh surface. Details can be found in \cite{Fernandez2018}.

The MAP client has a plugin architecture and allows creating new workflow steps. It imports features from OpenCMISS, especially data processing formats and tools from OpenCMISS Zinc. Meshes can be created with 3D cubic Hermite elements that allow for a high geometric modeling flexibility with a low number of nodes. Such meshes are stored in the OpenCMISS file format of \code{exnode} and \code{exelem} files.

\begin{figure}%
  \centering%  
  \begin{subfigure}[t]{0.48\textwidth}%
    \centering%
    \includegraphics[height=10cm]{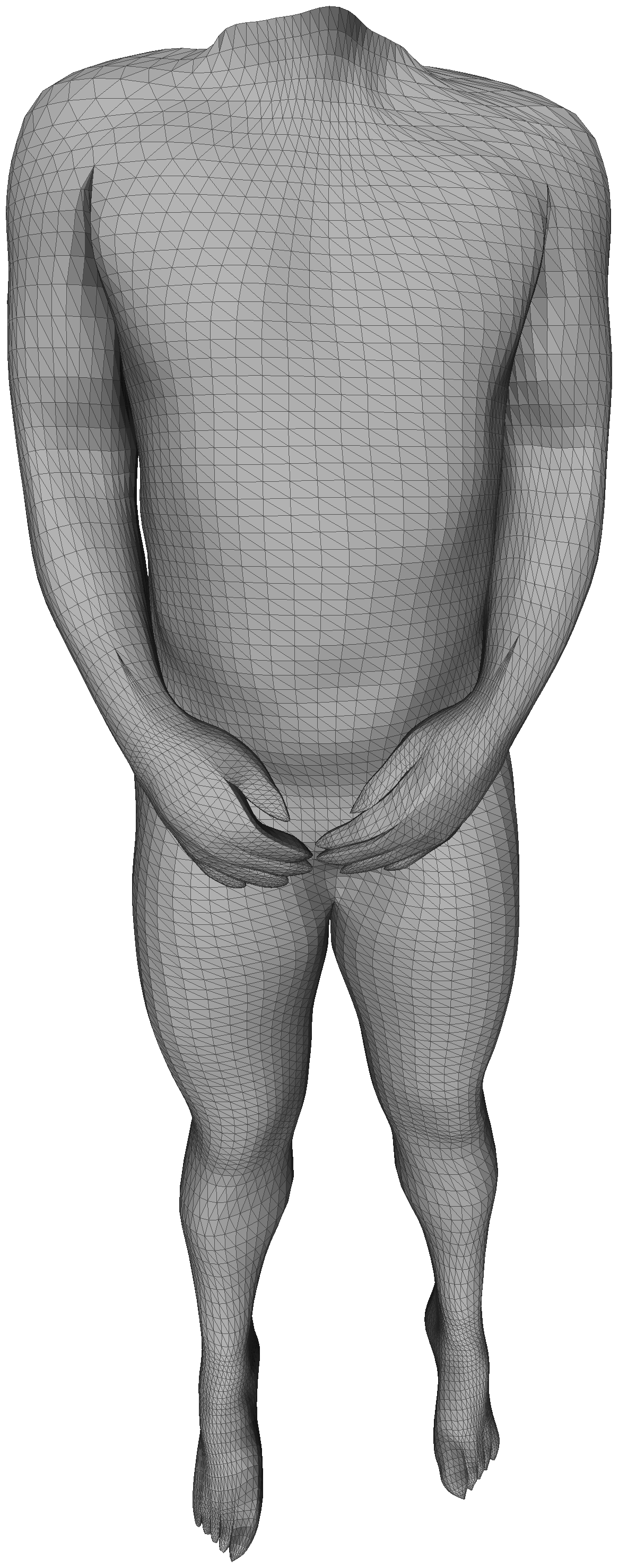}%
    \caption{Mesh of the trunk and limbs, the surface has been triangulated for visualization.}%
    \label{fig:vhp_total}%
  \end{subfigure}
  \quad
  \begin{subfigure}[t]{0.48\textwidth}%
    \centering%
    \includegraphics[height=7cm]{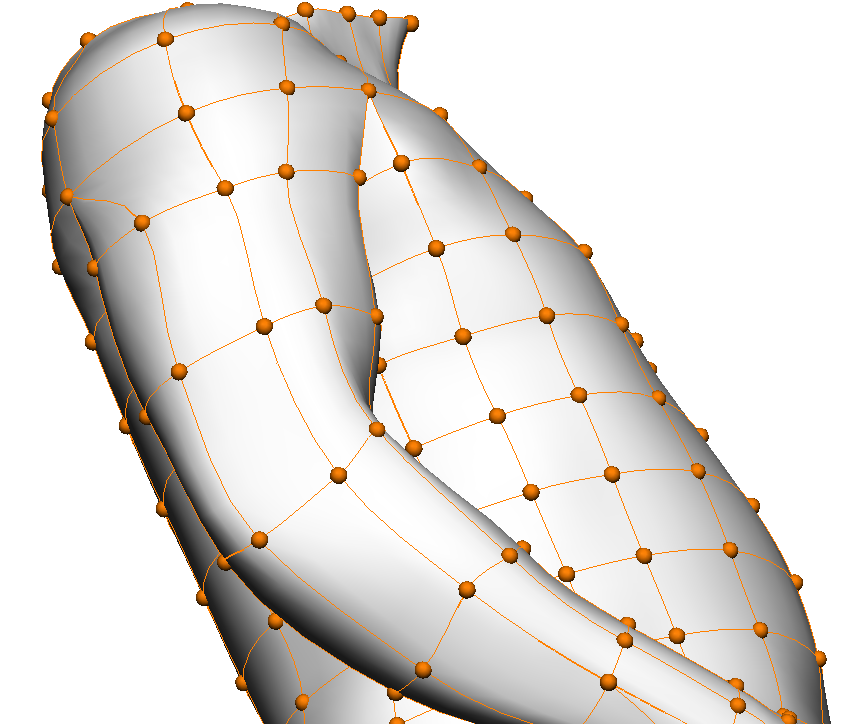}%
    \caption{Detail view of part of the right upper arm and the trunk with orange nodes and edges of a cubic Hermite element mesh.}%
    \label{fig:vhp_detail}%
  \end{subfigure} 
  \caption{Mesh of the Visible Human Male from the Visible Human Project.}%
  \label{fig:vhp_geometry}% 
\end{figure}

As a result, meshes of individual muscles or the whole human organism can be created. \Cref{fig:vhp_geometry} shows meshes that were created from the cryosectioning data of the Visible Human Male. In \cref{fig:vhp_total}, almost the whole body has been extracted. In \cref{fig:vhp_detail}, the mesh consisting of cubic Hermite elements is visualized. A relatively coarse mesh width suffices to model a smooth surface of the body. When exported in the exfiles format from the MAP client, the data can be visualized, e.g., using \emph{cmgui}, the visualization tool of OpenCMISS Zinc.

The mesh width of the meshes obtained using the MAP client was chosen such that the surface fitting yielded good results. The meshes are not necessarily ready for use in a simulation, especially if a  high mesh resolution is desired. 
Apart from the mesh width also the type of elements can be different from what is needed for a finite element simulation. Our goal is to obtain meshes with linear or quadratic Lagrange elements with configurable mesh widths for the specified upper arm muscles, such as the biceps brachii.

\begin{figure}%
  \centering%
  \includegraphics[height=10cm]{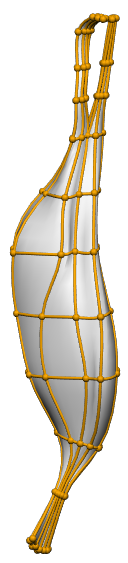}\quad%
  \includegraphics[height=10cm]{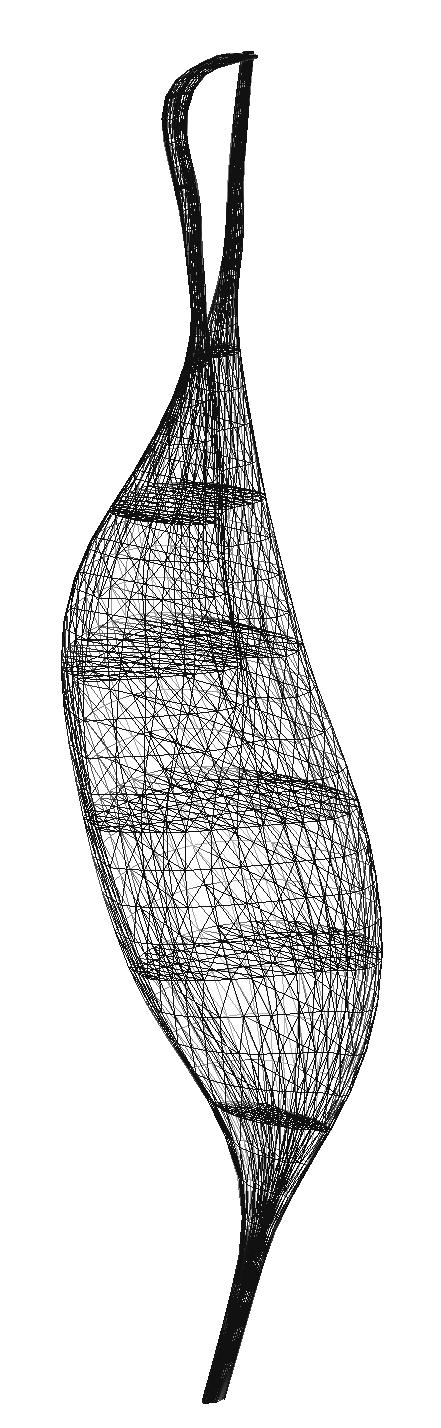}%
  \includegraphics[height=10cm]{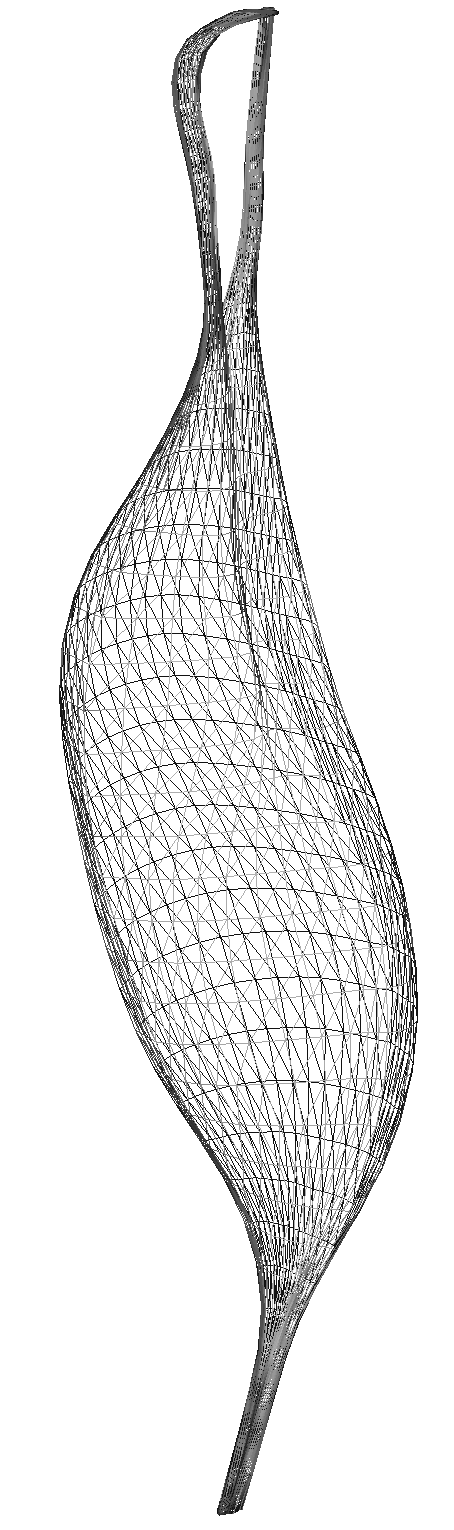}%
  \includegraphics[height=10cm,trim=-2cm 0 0 -2cm, clip]{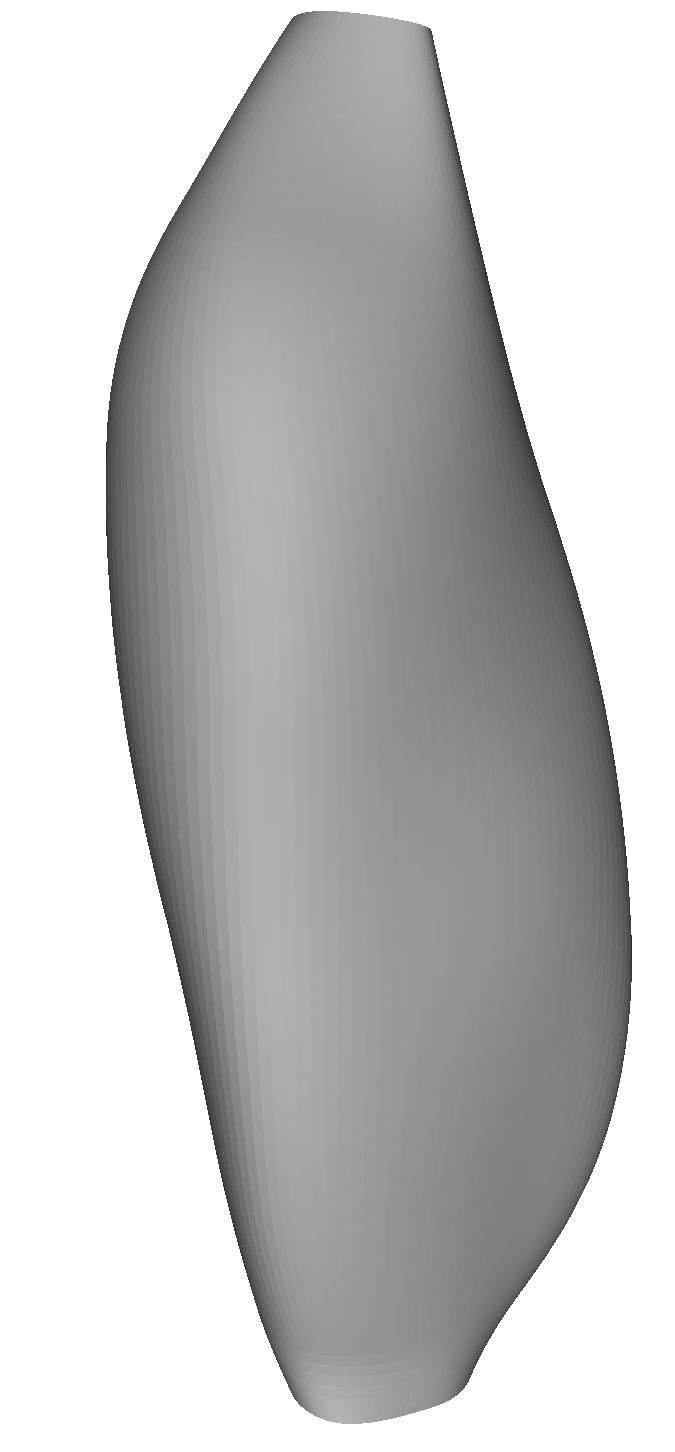}%
  \caption{Processing the geometry of the biceps brachii muscle. From left to right: mesh with cubic Hermite elements, STL mesh with inside triangles, STL surface mesh where triangles lying inside have been removed, Spline surface of the muscle belly.}%
  \label{fig:biceps_processing}%
\end{figure}

Therefore, the next step of the workflow, as visualized by the right branch of \cref{fig:scheme_preprocessing}, is to transform the volume mesh into a surface mesh which then can be used as start for further meshing. The further meshing steps are visualized in \cref{fig:biceps_processing}. The start is the Hermite mesh shown in the left-most image.

The Hermite elements can be triangulated and stored as an STL file using the tool \mbox{\code{cmgui}.} This process triangulates the non-planar faces of all Hermite elements. This leads to a dataset with triangles both on the surface and in the inside of the volume, as can be seen in the second image of \cref{fig:biceps_processing}. At this stage, the use of the MAP and OpenCMISS related tools is finished and further processing steps are performed using tools from \opendihu{} that we developed on our own.

A Python script removes the triangles inside the volume. The detection whether a triangle is inside the volume is done by casting four rays from the center of gravity of the respective triangle and determining if the rays intersect any other triangles. The rays have directions $(x,y,z) = (\pm1,\pm1,\frac13)$, where the $z$ axis is oriented along the muscle's longitudinal axis and the $x$ and $y$ axes are oriented in radial direction. The ray-triangle intersection is done using the fast Möller-Trumbore algorithm \cite{ray-triangle}. For every ray, all triangles are checked.
Only if all four rays intersect at least one more triangle, the starting triangle is considered to be inside the volume and subsequently removed from the dataset. 

This algorithm has a quadratic time complexity $\O(n^2)$ in the number of triangles $n$. It could be improved by organizing the triangles in a spatially adaptive data structure, such as an octree. However, since this preprocessing step has to be performed only once for a given geometry, the runtime is not critical and there is no need for such optimization.

The result of this operation is a triangulated surface, shown in the third image of \cref{fig:scheme_preprocessing}. The next step is to create a Spline surface of the muscle belly, as shown in the right-most image of \cref{fig:scheme_preprocessing}. This is described in the next two sections.

\subsection{Introduction of Spline Surfaces}\label{sec:nurbs}
After the surface representation of the muscle has been obtained from either the left or the right branch of the preprocessing workflow in \cref{fig:scheme_preprocessing}, the surface is given by a point cloud or a number of triangles. To remedy eventual outliers or unphysiologically sharp edges from the segmentation, a Spline surface is fitted to the data. This leads to a smooth surface representation and later to a better conditioned finite element mesh in the simulation. However, this step is optional. It is also possible to directly use the surface triangulation from \cref{sec:surf_extr} for the meshing algorithm described in \cref{sec:ser_alg_meshes}.

The surfaces use Nonuniform Rational B-splines (NURBS). A NURBS surface is a generalization of a B-spline surface. From a modeling point of view, B-spline surfaces have three advantageous properties.
First, the B-spline surface can be constructed with given smoothness properties.  
Second, the definition of a particular B-spline surface builds on intuitive geometric information, which simplifies their creation: A control polygon mesh in 3D space is defined. Its convex hull is guaranteed to contain the surface.
Third, the geometric parameters of a B-spline surface have only local impact on the shape of the surface. This allows a B-spline surface of a fixed, low polynomial degree to approximate point clouds with any number of points without loosing approximation quality.

A limitation of B-spline surfaces is that circular and spherical shapes cannot be represented. This limitation is overcome by NURBS surfaces. NURBS surfaces are defined as the perspective projection into 3D space of a B-spline surface in 4D space.

The mathematical description is given in this section, following the notation of \cite{piegl2012nurbs}. The building blocks are the B-spline basis functions of polynomial degree $p$. Given a knot vector 
\begin{align*}
  \Xi = (\xi_1, \xi_2, \dots, \xi_k) \in \R^k \quad \text{with } a=\xi_1 \leq \xi_2 \leq \cdots \leq \xi_k = b,
\end{align*}
the $i$th B-spline basis function $N_{i,n}$ of degree $n$ is defined recursively starting with the piecewise constant function
\begin{align*}
  N_{i,0}(\xi) = \begin{cases} 
    1 \, &\text{for }\xi_i \leq \xi < \xi_{i+1},\\[2mm]
    0 &\text{else},
  \end{cases}
\end{align*}
and using the following relation to define the functions of higher degree $(n > 0)$:
\begin{align*}
  N_{i,n}(\xi) = \dfrac{\xi - \xi_i}{\xi_{i+n} - \xi_i} N_{i,n-1}(\xi) + \dfrac{\xi_{i+n+1} - \xi}{\xi_{i+n+1} - \xi_{i+1}} N_{i+1,n-1}(\xi), \quad \text{for all }i > 0.
\end{align*}
Because neighboring entries in the knot vector can be equal, the fraction $0/0$ can occur. In this case, $0/0 := 0$ is defined. Note that by construction of $N_{i,n}$ a zero denominator implies that also the dividend is zero.

A B-spline curve $\bfC: \R \to \R^d$ of polynomial degree $p$ is defined as
\begin{align*}
  \bfC(u) = \s{i=1}{l}N_{i,p}(u)\,\bfP_i, \quad u \in [a,b].
\end{align*}
The coefficients $\bfP_i \in \R^d, i=1, \dots, l$, of the basis functions $N_{i,p}$ are called \emph{control points} and define the control polygon. The number $l$ of basis functions and control points is determined from the number of knots $k$ in an open knot vector and the polynomial degree $p$ as $l = k-p-1$.

The number of equal entries in series in the knot vector is the \emph{multiplicity} of the respective knot value. Usually \emph{open} knot vectors $\Xi$ are used where the first and the last knot occur with a multiplicity of $p+1$.
This makes the first and the last points of the B-spline curve coincide with the control polygon points: $\bfC(a) = \bfP_1$ and $\bfC(b) = \bfP_l$.

The multiplicities of the knots in the knot vector encode information about the smoothness of the B-spline curve. If the knot value $\hat{\xi}$ has a multiplicity of $m$, the B-spline curve will be $(p-m)$ times continuously differentiable at $\bfC(\hat{\xi})$.
% This can be seen from the fact that there exist $m$ basis functions with a support that begins at $\hat{\xi}$.

\begin{figure}%
  \centering%
  \def\svgwidth{8cm}%
  %% Creator: Inkscape inkscape 0.92.3, www.inkscape.org
%% PDF/EPS/PS + LaTeX output extension by Johan Engelen, 2010
%% Accompanies image file 'bspline_curve.pdf' (pdf, eps, ps)
%%
%% To include the image in your LaTeX document, write
%%   \input{<filename>.pdf_tex}
%%  instead of
%%   \includegraphics{<filename>.pdf}
%% To scale the image, write
%%   \def\svgwidth{<desired width>}
%%   \input{<filename>.pdf_tex}
%%  instead of
%%   \includegraphics[width=<desired width>]{<filename>.pdf}
%%
%% Images with a different path to the parent latex file can
%% be accessed with the `import' package (which may need to be
%% installed) using
%%   \usepackage{import}
%% in the preamble, and then including the image with
%%   \import{<path to file>}{<filename>.pdf_tex}
%% Alternatively, one can specify
%%   \graphicspath{{<path to file>/}}
%% 
%% For more information, please see info/svg-inkscape on CTAN:
%%   http://tug.ctan.org/tex-archive/info/svg-inkscape
%%
\begingroup%
  \makeatletter%
  \providecommand\color[2][]{%
    \errmessage{(Inkscape) Color is used for the text in Inkscape, but the package 'color.sty' is not loaded}%
    \renewcommand\color[2][]{}%
  }%
  \providecommand\transparent[1]{%
    \errmessage{(Inkscape) Transparency is used (non-zero) for the text in Inkscape, but the package 'transparent.sty' is not loaded}%
    \renewcommand\transparent[1]{}%
  }%
  \providecommand\rotatebox[2]{#2}%
  \newcommand*\fsize{\dimexpr\f@size pt\relax}%
  \newcommand*\lineheight[1]{\fontsize{\fsize}{#1\fsize}\selectfont}%
  \ifx\svgwidth\undefined%
    \setlength{\unitlength}{860.26327515bp}%
    \ifx\svgscale\undefined%
      \relax%
    \else%
      \setlength{\unitlength}{\unitlength * \real{\svgscale}}%
    \fi%
  \else%
    \setlength{\unitlength}{\svgwidth}%
  \fi%
  \global\let\svgwidth\undefined%
  \global\let\svgscale\undefined%
  \makeatother%
  \begin{picture}(1,0.7375649)%
    \lineheight{1}%
    \setlength\tabcolsep{0pt}%
    \put(0,0){\includegraphics[width=\unitlength,page=1]{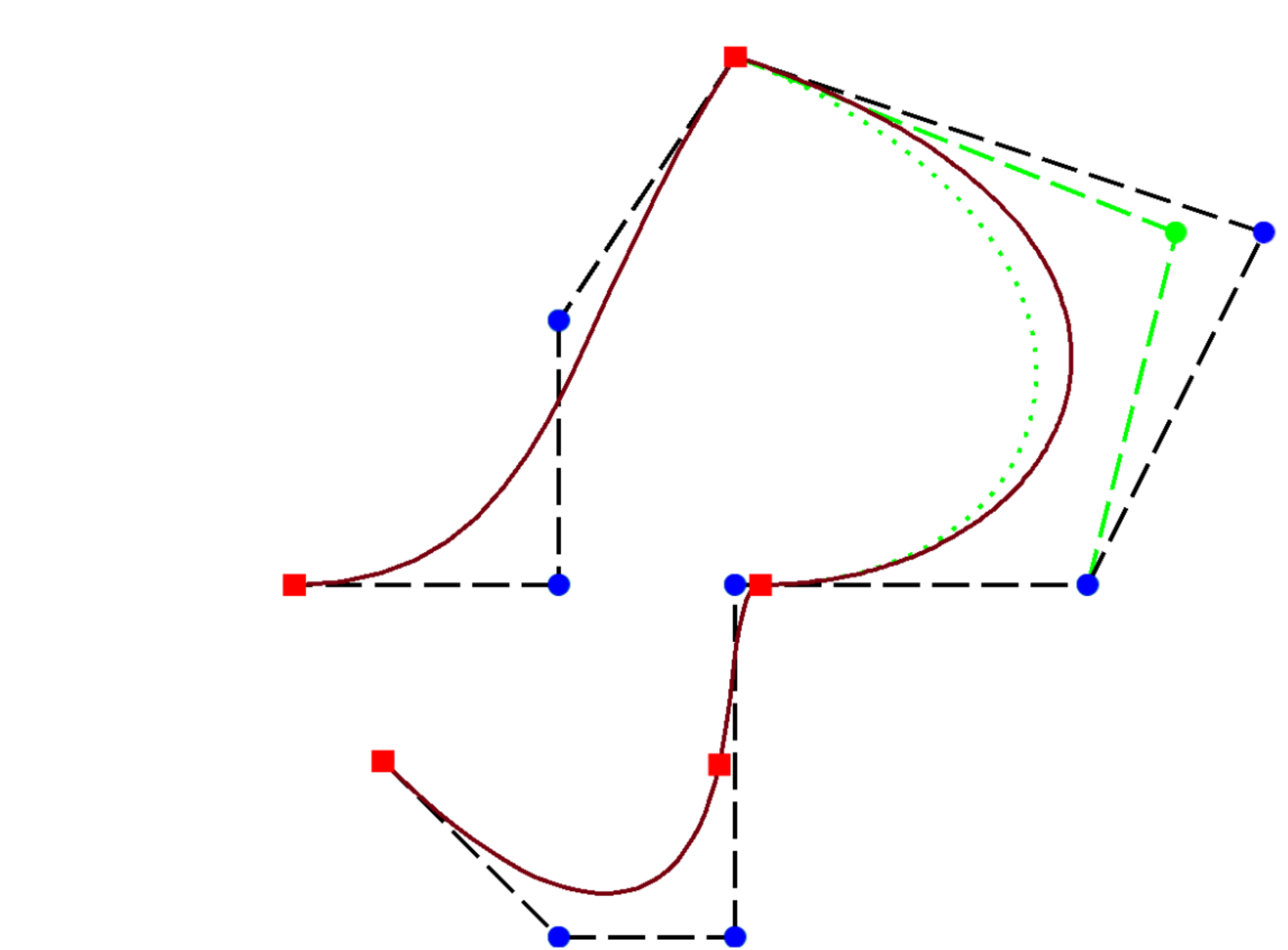}}%
    \put(-0.00039058,0.14095147){\color[rgb]{0,0,0}\makebox(0,0)[lt]{\lineheight{1.25}\smash{\begin{tabular}[t]{l}$\xi=0, m=4$\end{tabular}}}}%
    \put(0.58961371,0.13829342){\color[rgb]{0,0,0}\makebox(0,0)[lt]{\lineheight{1.25}\smash{\begin{tabular}[t]{l}$\xi=7, m=1$\end{tabular}}}}%
    \put(0.56539281,0.31799674){\color[rgb]{0,0,0}\makebox(0,0)[lt]{\lineheight{1.25}\smash{\begin{tabular}[t]{l}$m=2$\end{tabular}}}}%
    \put(0.51773233,0.71334401){\color[rgb]{0,0,0}\makebox(0,0)[lt]{\lineheight{1.25}\smash{\begin{tabular}[t]{l}$\xi=49, m=3$\end{tabular}}}}%
    \put(0.17007978,0.31565278){\color[rgb]{0,0,0}\makebox(0,0)[lt]{\lineheight{1.25}\smash{\begin{tabular}[t]{l}$m=4$\end{tabular}}}}%
    \put(0.17007978,0.36837442){\color[rgb]{0,0,0}\makebox(0,0)[lt]{\lineheight{1.25}\smash{\begin{tabular}[t]{l}$\xi=50,$\end{tabular}}}}%
    \put(0.56539281,0.36742328){\color[rgb]{0,0,0}\makebox(0,0)[lt]{\lineheight{1.25}\smash{\begin{tabular}[t]{l}$\xi=10,$\end{tabular}}}}%
  \end{picture}%
\endgroup%
  \caption{Exemplary B-spline curve (red) of degree $p=3$ for the knot vector $\Xi = (0,0,0,0,7,10,10,49,49,49,50,50,50,50)$, control points (blue) and control polygon (black).
  Positions of the curve $\bfC(\xi_i)$ at the knots $\xi_i$ are indicated by the red squares and the knot value $\xi$ and its multiplicity $m$ is given. The effect of moving one control point is shown in green.}%
  \label{fig:bspline_curve}%
\end{figure}
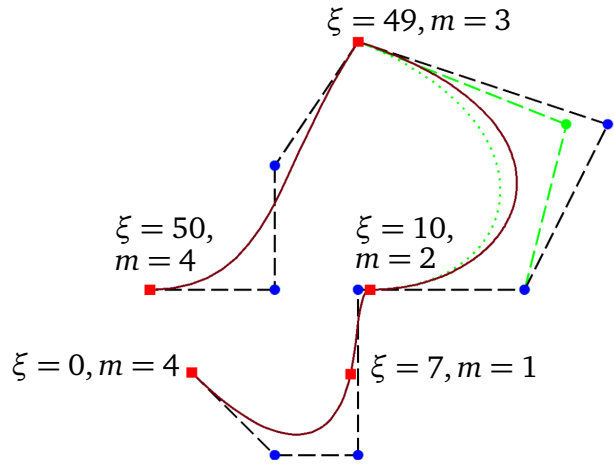

An exemplary B-spline curve is shown in \cref{fig:bspline_curve}. It uses a \emph{non-uniform} knot vector for polynomial degree $p=3$, where the differences $\xi_{i+m} - \xi_i$ between neighboring knot values vary. The effect of different multiplicities can be seen. The multiplicity $m=p=3$ places the point of the curve at the knot on the respective control point, as for $\xi=49$ in the example. The multiplicity $m=p-1=2$ places the point of the curve at the knot on the control polygon, as in the example at $\xi=10$. A lower multiplicity $m < p-1$ does not yield a higher smoothness and in turn does not force the curve to coincide with the control polygon at the respective knot. It can also be seen that the B-spline curve stays inside the convex hull of the control polygon which is a property of B-spline curves \cite{piegl2012nurbs}.

The effect of moving one of the 10 control points is visualized with green color in \cref{fig:bspline_curve}.
The B-spline basis function $N_{i,p}$ has a local support of $S=(\xi_i,\xi_{i+p+1})$. Consequently, only the corresponding part of the curve, $\bfC(\xi)$ for $\xi \in S$, changes.

A B-spline surface $\bfS : \R^2 \to \R^d$ is given by the tensor product of two B-spline curves:
\begin{equation}\label{eq:bspline_surface}
  \begin{array}{lll}
    \bfS(u,v) = \s{i=1}{l^{(1)}}\s{j=1}{l^{(2)}} N^{(1)}_{i,p^{(1)}}(u)\,N^{(2)}_{j,p^{(2)}}(v)\, \bfP_{i,j}.
  \end{array}
\end{equation}

Here, we have two polynomial degrees $p^{(1)}$ and $p^{(2)}$, the ansatz functions $N^{(1)}_{i,p^{(1)}}$ and $N^{(2)}_{j,p^{(2)}}$ with $l^{(1)}$ and $l^{(2)}$ ansatz functions per coordinate direction are constructed from the corresponding knot vectors per coordinate direction.

NURBS, B-spline curves and surfaces are formulated using \emph{homogeneous coordinates}. Every point in Cartesian coordinates $(x,y,z) \in \R^3$ has a set of homogeneous coordinates $(\tilde{x},\tilde{y},\tilde{z},w)=(x\,w,y\,w,z\,w,w)$. Thus, the Cartesian coordinates can be obtained from the homogeneous coordinates by the \emph{perspective division}, i.e., dividing all but the last coordinate by the weight $w$.

A NURBS surface is given by the same definition as the B-spline surface in \cref{eq:bspline_surface} except that the control points $\bfP_{i,j} \in \R^3$ are enriched with scalar weights $w_{i,j}$ and, thus, replaced by $(\bfP_{i,j}, w_{i,j}) \in \R^4$. The resulting surface $\bfS$ is given in homogeneous coordinates. Executing the perspective division yields the form:
\begin{align*}
  &\bfT(u,v) = \s{i=1}{l^{(1)}}\s{j=1}{l^{(2)}} R_{i,j}(u,v) \,\bfP_{i,j},\\[4mm]
  &\text{with } R_{i,j}(u,v) = \dfrac{N_{i,p^{(1)}}(u)\,N_{j,p^{(2)}}(v)\,w_{i,j}}{\s{r=1}{l^{(1)}}\s{s=1}{l^{(2)}} N_{r,p^{(1)}}(u)\,N_{s,p^{(2)}}(v)\,w_{r,s}}.
\end{align*}
The new rational basis functions $R_{i,j}$ and the possibly non-uniform knot vectors give rise to the name Non-Uniform Rational B-spline surface (NURBS).

\subsection{Fitting a Spline Surface to the Muscle Geometry}
In order to find a NURBS surface for the given triangulated surface of a muscle, at first, the part of the geometry corresponding to the tendons is removed such that the resulting triangles model only the surface of the muscle belly. In our example of the biceps muscle, the resulting belly has a length of \SI{12.8}{\mm}.

Then, twelve cross-sections are extracted from the surface triangles. As a result, we get twelve horizontal circumferential rings. On each ring, 9 equidistant points are determined. The first point is appended after the last point in every ring, such that in total we obtain a grid of $10 \times 12$ points. 

\begin{figure}%
  \centering%
  \includegraphics[width=0.35\textwidth]{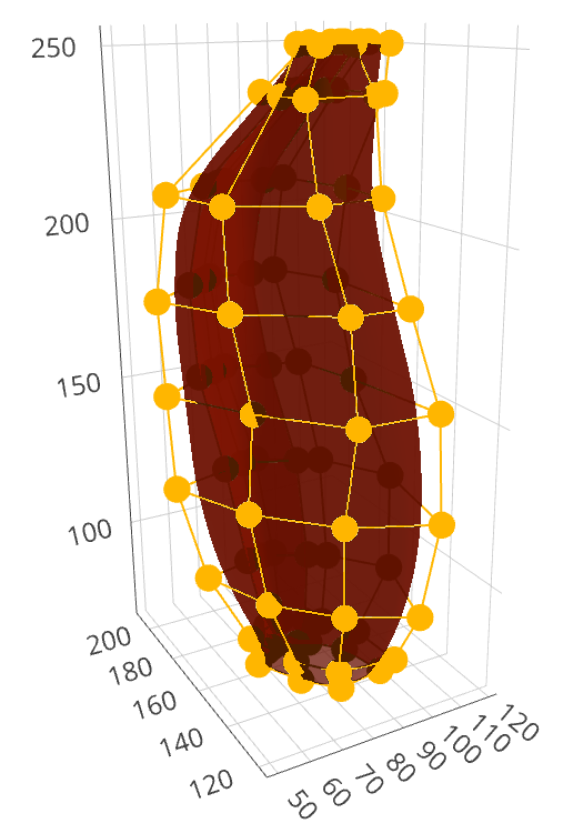}%
  \caption{Muscle surface description with Splines: Fitted NURBS surface of the biceps muscle (red) and the control polygon (orange).}%
  \label{fig:biceps_splines_control_points}%
\end{figure}

Then, the least squares surface approximation algorithm by \cite{piegl2012nurbs} is used to fit a NURBS surface to the points. The implementation of the algorithm is given by the NURBS-Python (geomdl) library. Polynomial degrees of $p^{(1)} = 3$ and $p^{(2)}=2$ are used where the first dimension corresponds to the cross-sectional direction of the muscle. The knot multiplicity is chosen as $m=1$ for both coordinate directions. We obtain a two times respective one times continuously differentiable surface in $u$ and $v$ direction. 
The resulting NURBS surface and the control polygon are visualized in \cref{fig:biceps_splines_control_points}. Note that the control polygon is different from the grid of points against which the surface is fitted.

\begin{figure}%
  \centering%
  \begin{subfigure}[t]{0.48\textwidth}%
    \centering%
    \includegraphics[height=8cm]{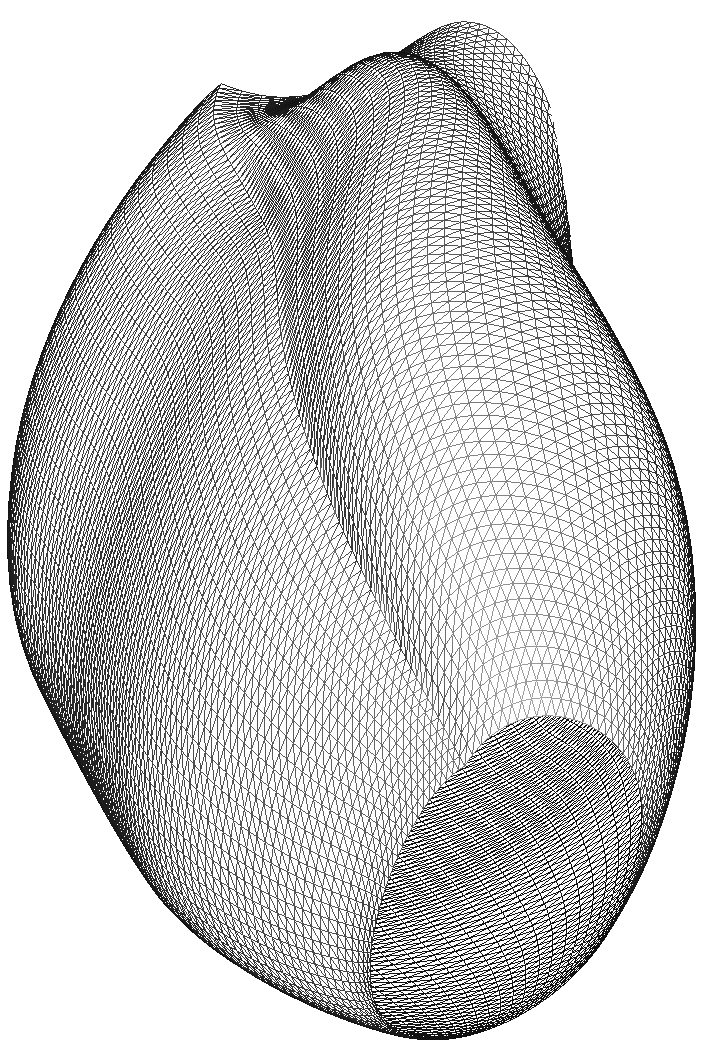}%
    \caption{First approach with $10 \times 12$ control points. The kink at the seam line along the muscle is clearly visible.}%
    \label{fig:biceps_splines_wrong}%
  \end{subfigure}
  \quad
  \begin{subfigure}[t]{0.48\textwidth}%
    \centering%
    \includegraphics[height=8cm]{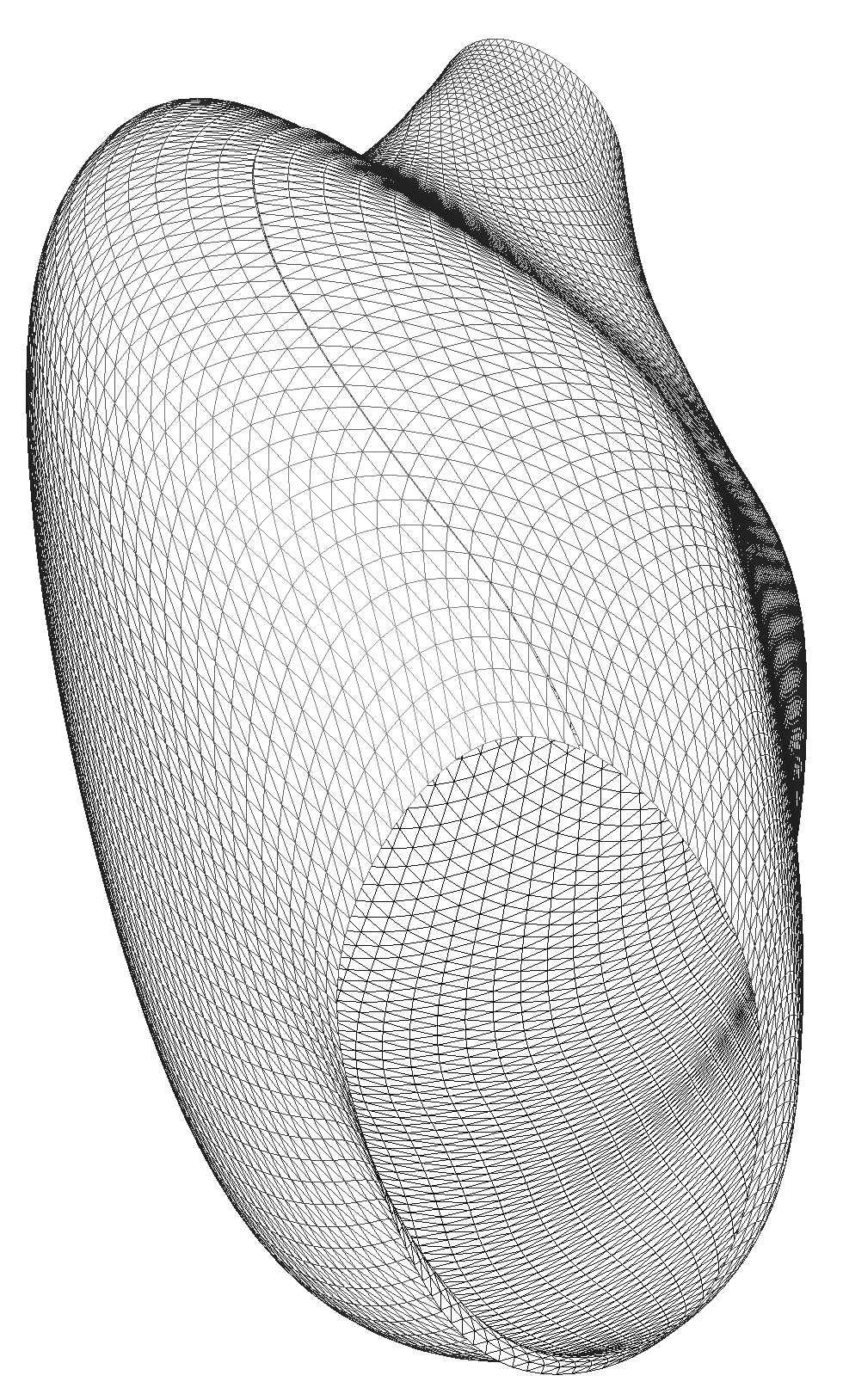}%
    \caption{Second, improved approach with $28 \times 12$ points. It can be seen that the tangents at the seam line match very well.}%
    \label{fig:biceps_splines_seam}%
  \end{subfigure}
  \caption{Muscle surface description with Splines: Fitted NURBS surface of the biceps muscle, triangulated for visualization purposes.}%
  \label{fig:biceps_splines}%
\end{figure}

\Cref{fig:biceps_splines_wrong} shows the result of this approach in more detail. We observe that the surface is non-differentiable and has a kink at the seam line where the first and last points of each ring meet. The reason for this is that the surface fitting algorithm does not pose any conditions on the tangents at the edges of the fitted NURBS surface.  Thus, the tangents mismatch.

Since no implementation of a fitting algorithm specifically for a tubular NURBS surface with periodicity in tangential direction is available, we develop a different remedy. We modify the point grid that is used for the surface fitting. The series of 9 equidistant points on each ring is replicated twice and the first point is again added as the last point. This leads to a grid of $(3\cdot 9+1) = 28 \times 12$ points which wrap around the muscle volume in circumferential direction three times. The NURBS surface fitting algorithm is applied on this grid. The resulting NURBS surface also wraps around the muscle three times with the two ends being again not properly fitting to each other. From these three wraps, the middle one is extracted. In the biceps example, this corresponds to restricting the NURBS surface $\bfT(u,v)$ from $(u,v) \in [0,1]^2$ to $(u,v) \in [0.4,0.733]\times [0,1]$.

The result is depicted in \cref{fig:biceps_splines_seam}. The tangents now match very well between the two sides of the NURBS surface. Additionally, the comparison with the initial approach in \cref{fig:biceps_splines_wrong} shows that an artificial bulge at the top of the muscle in the perspective of the visualization is removed. The overall shape of the muscle now looks smoother and more natural. Also, a comparison with the result of the automatic algorithm given in \cref{fig:extraction_result_biceps} shows that the results of our new approach are smoother.

The generated tubular surface has two holes at the top and bottom which prevent it from being an enclosing surface to the muscle belly volume. The borders of these holes each lie in a plane and, thus, the missing surfaces are treated as being planar during the subsequent creation of the 3D meshes.

For the next step, a triangulation of the tubular surface is created and stored as STL mesh file. We use the respective functionality of the NURBS-Python library that creates a structured triangle mesh using the 2D parametrization of the NURBS surface.

\section{Serial Algorithm to Create Muscle and Fiber Meshes}\label{sec:ser_alg_meshes}
Next, a 3D mesh for the muscle volume and 1D meshes for muscle fibers need to be generated from the surface representation described in the previous sections. In this section, first an algorithm for the 3D mesh is described. Then, a second algorithm that reuses results from the first algorithm is presented which generates one dimensional meshes for muscle fibers. Both algorithms are executed in serial. A derived algorithm that can run in parallel and, thus, can handle larger datasets on a distributed memory hardware is presented in \cref{sec:parallel_algorithm}.

The steps of a serial algorithm for the generation of a 3D mesh are given in \cref{alg:serial_algorithm_1}. Input is the set of triangles at the tubular surface of the muscle. The tubular surface is oriented along the $z$ axis. In the following descriptions, the muscle in considered to be oriented upright such that the $z$ axis points in vertical direction towards the top. The borders at the bottom and at the top have a constant $z$ coordinate.
\begin{algorithm}
  \begin{algorithmic}[1]%
    \Procedure{Create\_3D\_mesh}{}
    \Require Triangulated tubular surface
    \Ensure Structured 3D volume mesh
    \Statex
    \State Slice geometry           \label{alg:1.1}
    \State Triangulate 2D slices      \label{alg:1.2}
    \State Compute harmonic maps $u, v$ from the slices to a parameter space     \label{alg:1.3}
    \State Construct regular grid in parameter space and map it to slices            \label{alg:1.4}
    \State Form 3D quadrilateral elements between the 2D slices’ meshes   \label{alg:1.5}
    \EndProcedure
  \end{algorithmic}%
  \caption{Serial algorithm for the generation of 3D meshes.}%
  \label{alg:serial_algorithm_1}%
\end{algorithm}%

The idea of the algorithm is to first create 2D meshes with good quality on cross-sectional slices of the muscle volume and then combine them to get a 3D mesh. The algorithm starts with creating the horizontal 2D slices in lines \ref{alg:1.1} and \ref{alg:1.2}. The slices get vertically connected at the end of the algorithm in line \ref{alg:1.5} to create the 3D mesh. This step is visualized in \cref{fig:serial_alg_8}. 

Because the goal is to create a hexahedral mesh, the horizontal slices have to consist of quadrilaterals. 
Decomposing a 2D domain into quadrilaterals is easier for a square or circular shaped domain than for an actual cross-section of the muscle. Therefore, we introduce a separate, square or circular shaped parameter domain for creating the quadrilaterals.

\begin{figure}%
  \centering%
  \includegraphics[width=\textwidth]{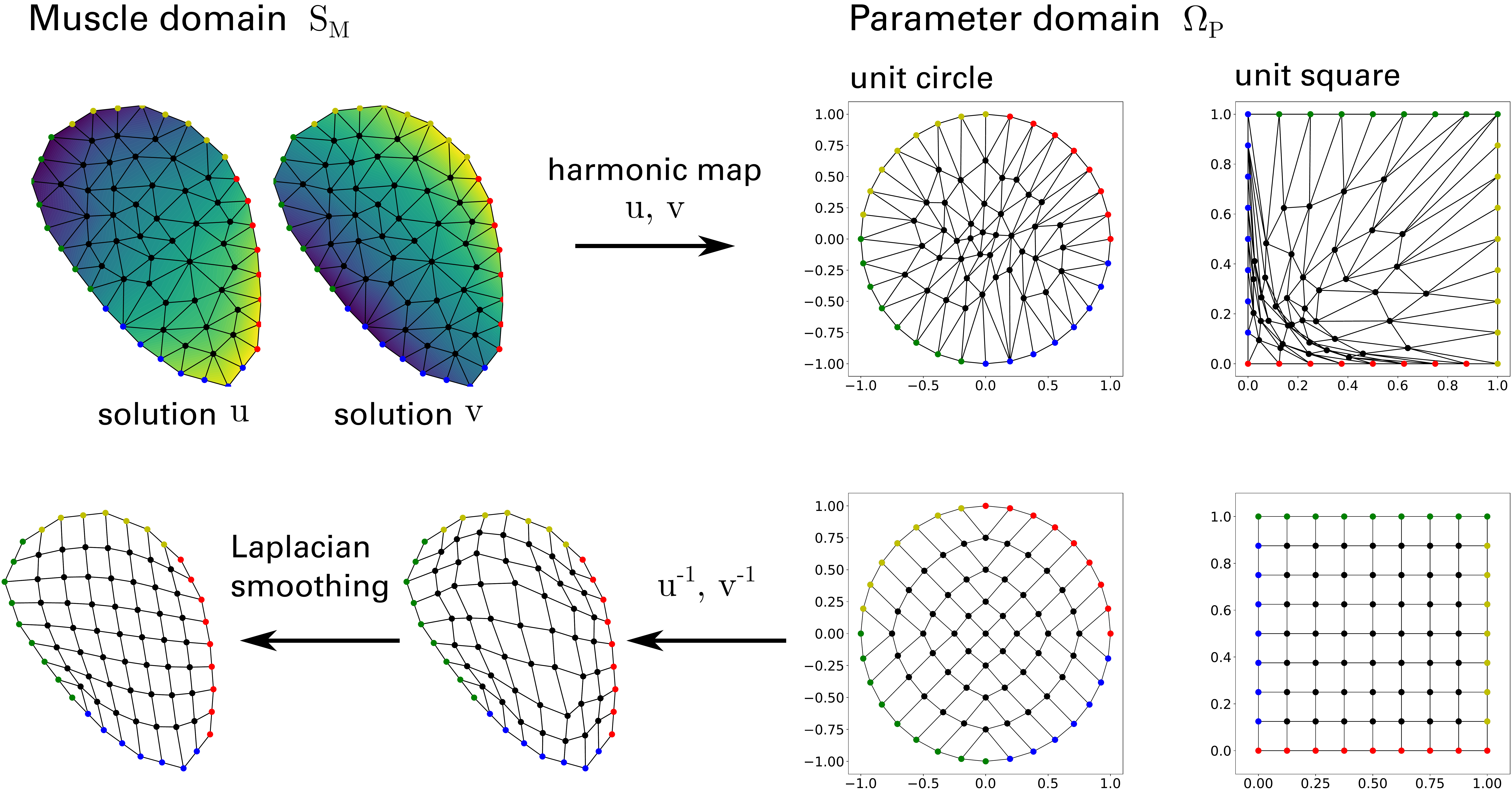}%
  \caption{Generation of the muscle meshes, overview of the mapping method between the muscle domain (left) and the parameter domain (right) using harmonic maps.}%
  \label{fig:harmonic_map_solution}%
\end{figure}

\Cref{fig:harmonic_map_solution} outlines the method.
We start at the upper left of the figure with a triangulation on the cross-sectional slices of the muscle. A mapping from the muscle slices to the parameter space at the right of the figure is computed. We use harmonic maps to ensure a smooth mapping that results in good mesh quality. This first step corresponds to line \ref{alg:1.3} in \cref{alg:serial_algorithm_1}. Different parameter domains such as unit circle and unit square are considered, as shown in \cref{fig:harmonic_map_solution}. It can also be seen at the upper right that the image of the muscle slice triangulation in the parameter domain is better in the unit circle than in the unit square. Therefore, an investigation of different parameter domains and triangulation schemes is necessary.

Next, line \ref{alg:1.4} of \cref{alg:serial_algorithm_1} defines a quadrangulation in the parameter space, shown at the lower right of \cref{fig:harmonic_map_solution}. The quadrilateral elements are mapped back to muscle slices where they are needed for the final 3D mesh. An optional smoothing step at the lower left of the figure further improves the mesh quality.
All steps of the algorithm are described in more detail in the following sections.

\subsection{Slicing of the Geometry}\label{sec:slicing_of_the_geometry}
\begin{figure}%
  \centering%
  \begin{subfigure}[t]{0.45\textwidth}%
    \centering%
    \includegraphics[height=9cm]{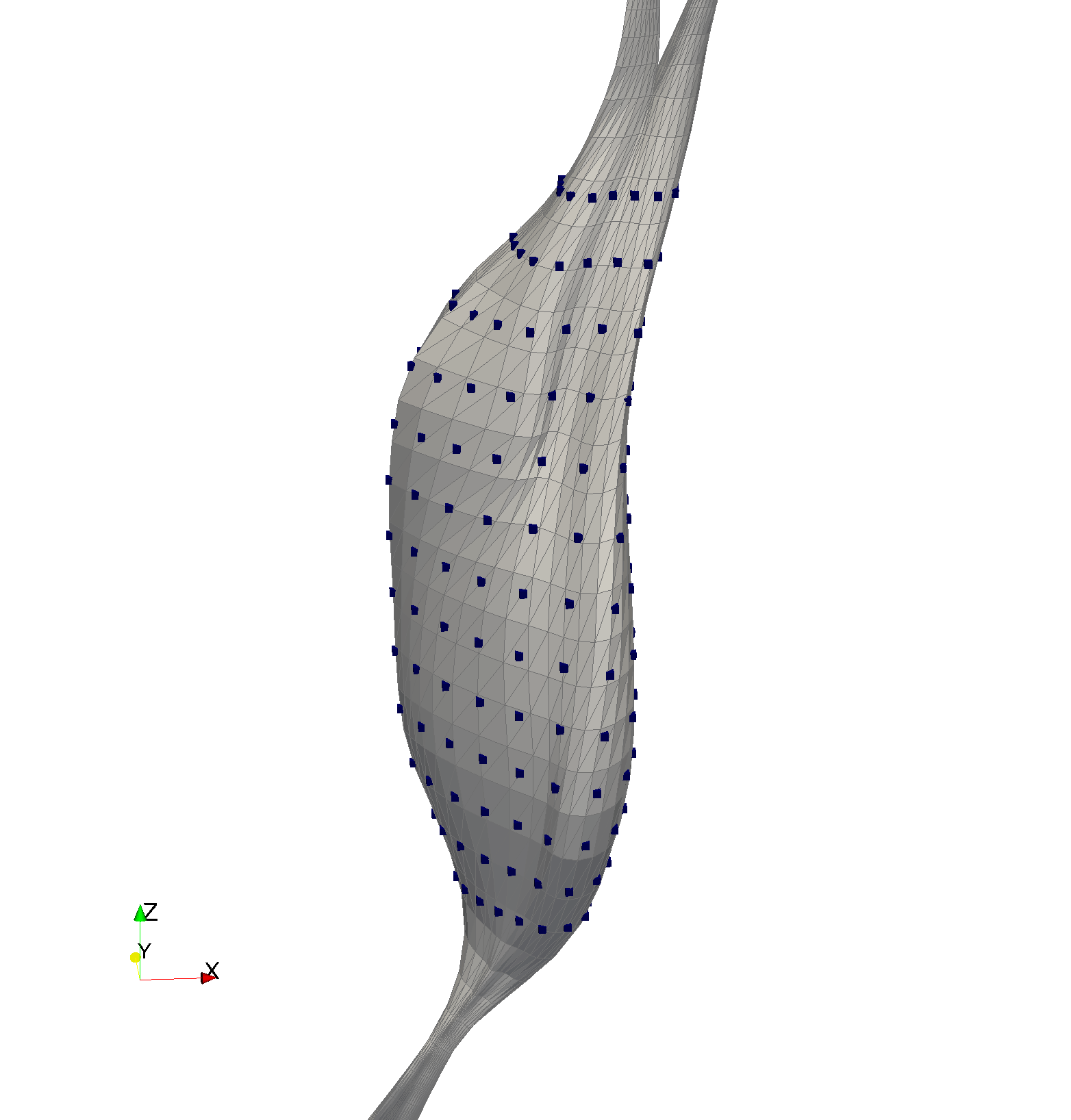}% [trim=left bottom right top, clip]
    \caption{Extracted boundary points (blue) on the biceps surface mesh (gray). This is the result of line \ref{alg:1.1} in \cref{alg:serial_algorithm_1}.}%
    \label{fig:serial_alg_0}%
  \end{subfigure}
  \quad
  \begin{subfigure}[t]{0.51\textwidth}%
    \centering%
    \includegraphics[height=9cm]{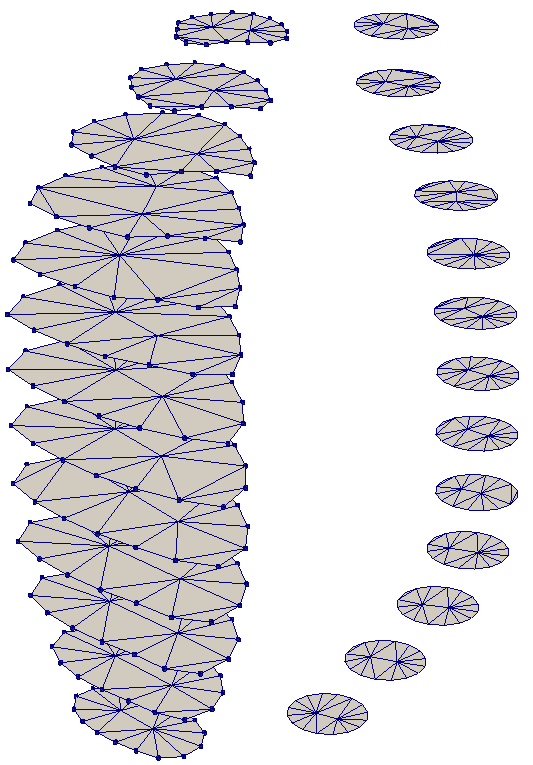}%
    \caption{The generated triangulation of the slices (left) and the image $\bfy(\bfx)$ of the triangulation under the harmonic map (right). This figure shows the result of lines \ref{alg:1.2} and \ref{alg:1.3} in \cref{alg:serial_algorithm_1}.}%
    \label{fig:serial_alg_3}%
  \end{subfigure}\\
  \centering%
  \begin{subfigure}[t]{0.49\textwidth}%
    \centering%
    \includegraphics[height=9cm]{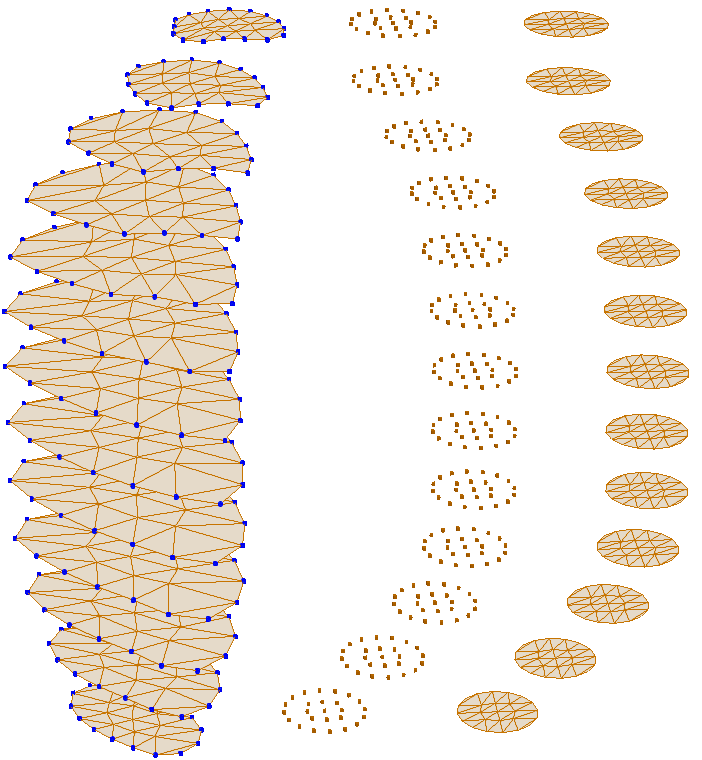}%
    \caption{Grid in parameter space (right) and muscle domain (left), result after line \ref{alg:1.4} in \cref{alg:serial_algorithm_1}.}%
    \label{fig:serial_alg_4}%
  \end{subfigure}
  \hfill{}
  \begin{subfigure}[t]{0.4\textwidth}%
    \centering%
    \includegraphics[height=9cm]{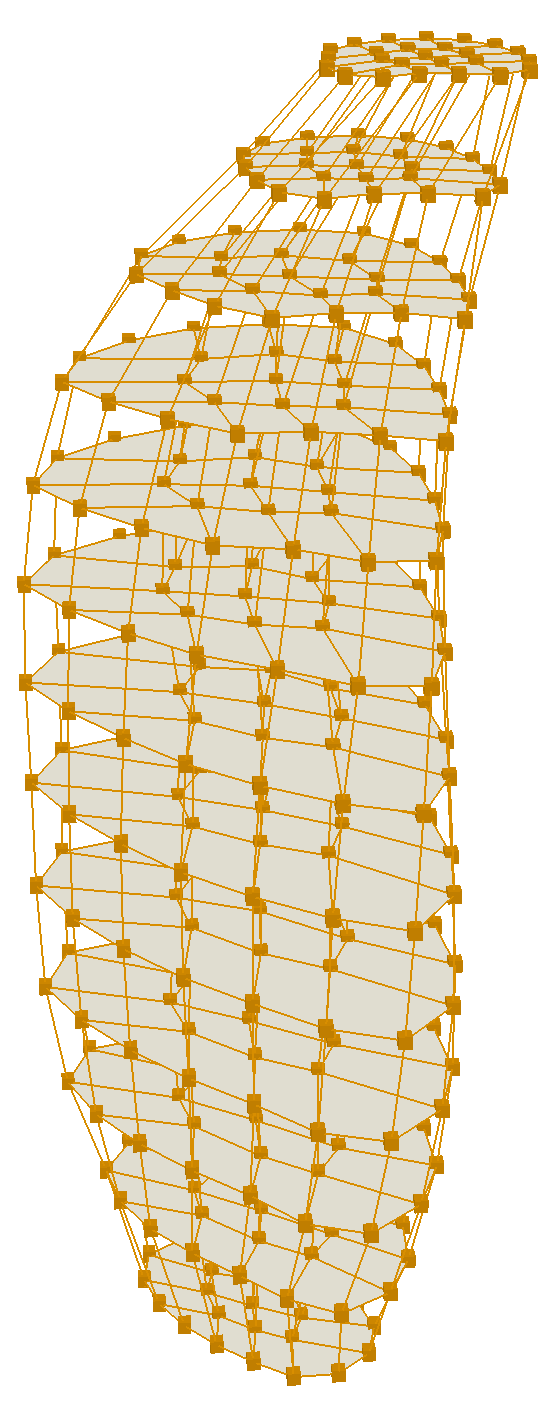}%
    \caption{3D elements formed by connecting the slices in line \ref{alg:1.5} in \cref{alg:serial_algorithm_1}.}%
    \label{fig:serial_alg_8}%
  \end{subfigure}
  \caption{Steps of the serial algorithm for 3D mesh generation, \cref{alg:serial_algorithm_1}, executed directly on the surface mesh of the biceps muscle (not the B-spline surface).}%
  \label{fig:serial_alg}%
\end{figure}

The first step in line \ref{alg:1.1} of \cref{alg:serial_algorithm_1} slices the geometry. This means that horizontal \emph{slices} of the cross-sectional area are extracted from the surface mesh. First, the muscle is divided into equidistant positions $z_i, i=1,\dots,n$ along the $z$-axis where the slices are to be extracted. As can be seen in \cref{fig:serial_alg_0}, $n=13$ $z$ coordinates are selected. Next, for every position $z_i$, all surface triangles $T_j$ that intersect the plane $Z_i = \{\bfp=(x,y,z) \mid z=z_i$\} are considered and the intersection lines $P = T_j \cap Z_i$ are computed. The method of computing plane-triangle intersection is described in the following.

Given is a triangle $T$ with points $\bfp^{1},\bfp^2,\bfp^3 ∈ \R^3$ and a value $\hat{z}$, the result is the set of points $P = T ∩ \{\bfp = (\bfp_x,\bfp_y,\bfp_z) \mid \bfp_z=\hat{z}\}$ which corresponds to a line segment $\overline{\bfp^a\bfp^b}$. 

We describe the points in the triangle by two barycentric coordinates $\xi_1$ and $\xi_2$ as
\begin{equation}\label{eq:barycentric_triangle}
  \begin{array}{lll}
    \bfp(ξ_1,ξ_2) = (1-ξ_1-ξ_2)\,\bfp^{1} + \xi_1\,\bfp^{2} + \xi_2\,\bfp^{3},  \\[4mm]
    \text{with }\xi_1+\xi_2 \leq 1, \quad 0 \leq \xi_1,\xi_2 \leq 1.
  \end{array}
\end{equation}
$\bfp_z(ξ_1,ξ_2) = \hat{z}$ defines the equation for the line through the points $\bfp^a$ and $\bfp^b$ in barycentric coordinates. The solution is given as
\begin{equation*}
  \begin{array}{lll}
    ξ_1 = m\cdot ξ_2 + c,\quad
    m = -\dfrac{\bfp_z^{3} - \bfp_z^{1}}{\bfp_z^{2} - \bfp_z^{1}}, \quad c = \dfrac{\hat{z} - \bfp_z^{1}}{\bfp_z^{2} - \bfp_z^{1}}, \quad \bfp_z^2 \neq \bfp_z^1.
  \end{array}
\end{equation*}
For $\bfp_z^1 = \bfp_z^2 \neq \bfp_z^3$ we swap $\bfp_z^2$ and $\bfp_z^3$.

%
%  # x(xi1,xi2) = (1-xi1-xi2)*x^{1} + xi1*x^{2} + xi2*x^{3},  xi1+xi2 <= 1, 0 <= xi1,xi2 <= 1
%  # x_3(xi1,xi2) = z_value  =>  (1-xi1-xi2)*x^{1} + xi1*x^{2} + xi2*x^{3} = z_value
%  #                         =>  xi1*(x^{2} - x^{1})  +  xi2*(x^{3} - x^{1})  =  z_value - x^{1}
%  #                         =>  xi2 = ((z_value - x^{1}) - xi1*(x^{2} - x^{1})) / (x^{3} - x^{1})
%  #                         =>  xi2 = (z_value - x^{1})/(x^{3} - x^{1}) - xi1 * (x^{2} - x^{1})/(x^{3} - x^{1})
%  #                         =>  xi1 = (z_value - x^{1}) / (x^{2} - x^{1})  - xi2 * (x^{3} - x^{1}) / (x^{2} - %x^{1}) 
Next, the end points of the line segment $\overline{\bfp^a\bfp^b}$ are determined.
We consider the three sides $\overline{\bfp^1\bfp^2}, \overline{\bfp^2\bfp^3}$ and $\overline{\bfp^3\bfp^1}$ of the triangle and check which of them are intersected by the $z=\hat{z}$ plane by the following three conditions:
\begin{enumerate}
\item On the triangle side $\overline{\bfp^1\bfp^2}$ the condition $ξ_2 = 0$ holds and the side intersects the plane 
at $\bfp(c,0)$ 
iff $0 \leq c \leq 1$. 
\item Similarly, on the triangle side $\overline{\bfp^1\bfp^3}$ we have the condition $ξ_1 = 0$ and the side intersects the plane 
at $\bfp(0,-c/m)$ 
iff $m\neq 0 \wedge 0 \leq -c/m \leq 1$. 
\item The third triangle side $\overline{\bfp^2\bfp^3}$ is intersected for $\hat{ξ}_1=(c+m) / (1+m)$
at $\bfp(\hat{ξ}_1,1-\hat{ξ}_1)$ 
iff ${m \neq -1 \wedge 0 \leq \hat{ξ}_1 \leq 1}$.
\end{enumerate}
If two of these three conditions for intersection of the triangle sides are met, there is an intersecting line segment $\overline{\bfp^a\bfp^b}$ with $\bfp^a \neq \bfp^b$ and the two intersection points $\bfp^a$ and $\bfp^b$ on the triangle sides are determined as stated above. The trivial cases $\bfp^a = \bfp^b$ and the case where $\bfp^a$ and $\bfp^b$ are equal to two of the triangle corners $\bfp^1, \bfp^2$ and $\bfp^3$ are handled separately in our implementation.

After the presented computations are performed for all planes $Z_i$ and all triangles $T_j$, we have a number of line segments that form a geometric \say{ring} for each $z$ plane. The line segments are ordered according to their adjacency and a counter-clockwise orientation with respect to the $z$ axis is ensured.
The length of each ring is computed. A number $m=16$ of equidistant points is selected on each ring.

Because the selected points on the rings are later used as boundary points of the resulting 3D mesh, their position relative to each other on different rings should be in a tidy manner. The positioning should enable straight connection lines in longitudinal direction of the muscle connecting the points on every ring. For illustration, \cref{fig:serial_alg_0} shows such a configuration of properly positioned ring points. Connecting the points from top to bottom is possible with smooth lines rather than zigzag lines. In result, the outer surface of the final mesh in \cref{fig:serial_alg_8} consists of a smooth quadrilateral mesh.

With given rings and number $m$ of equidistant points per ring, only the position of one point per ring is not yet fixed. To close the definition, in the following we formulate a first condition that relates the point positions of two neighboring rings and a second condition for one point at the bottom-most ring.

As mentioned, the first condition should ensure that the point positions on neighboring rings are as similar as possible. This is done by minimizing the distance between the first points on every ring.
In the algorithm, the $z$ planes are traversed from bottom to top. 
The first point $\tilde{\bfp}_{i,0}$ on a ring at $z=z_i$ is determined from the first point $\tilde{\bfp}_{i-1,0}$ of the previous ring at $z=z_{i-1}$ as the one with the minimal distance $|\tilde{\bfp}_{i,0} - \tilde{\bfp}_{i-1,0}|$. 
Thus, the searched point $\tilde{\bfp}_{i,0}$ has the property that the line between $\tilde{\bfp}_{i,0}$ and $\tilde{\bfp}_{i-1,0}$ and the tangent of the ring are perpendicular.

Given any point $\bfp$ on the ring at $z_i$ and the tangent vector $\bfu$ at this point, we can project the connection vector $\bfv$ from $\bfp$ to the start point $\tilde{\bfp}_{i-1,0}$ of the previous ring, $\bfv = \tilde{\bfp}_{i-1,0} - \bfp$, onto the tangent $\bfu$. This leads to the plumb foot point $\bfp_0$ by the computation
\begin{equation*}
  \begin{array}{lll}
    \bfp_0 = \bfp + t\,\bfu \quad \text{with }t = \dfrac{\bfv \cdot \bfu}{|\bfu|^2}.
  \end{array}
\end{equation*}
Performing this calculation for every line segment $u$ on a ring allows to select the plumb foot $\bfp_0$ with the smallest distance to the start point $\tilde{\bfp}_{i-1,0}$ of the previous ring to be the start point $\tilde{\bfp}_{i,0}$ of the current ring. This is the point that fulfills the first condition.

With this first condition, all points are only fixed relative to each other. The definition of one point, the start point $\tilde{\bfp}_{0,0}$ of the bottom-most ring, is missing. The second condition fixes this point by a prescribed plane at $x = \hat{x}$ and selects $\tilde{\bfp}_{0,0}$ such that its $x$ coordinate lies in this plane. From the (usually two) points that meet this condition, the one with lower $y$ coordinate is selected. The actual value of $\hat{x}$ is determined experimentally such that the resulting point positions are visually uniform. Not every value leads to a good result because of the shape of the biceps muscle, especially the groove where the humerus bone is located.

The resulting grid of points on the biceps surface is visualized in \cref{fig:serial_alg_0}. It can be seen that all points of the same ring have the same $z$ coordinate. By connecting neighboring points horizontally and vertically, a regular grid can be formed. This overall grid in this $x$-$z$ perspective view looks relatively uniform, e.g., compared with the gray surface triangulation mesh of the biceps geometry. The spacing between the points is lower at the top and bottom of the muscle because of the smaller circumference at these locations.

\subsection{Triangulation of the Slices}\label{sec:triangulation_of_the_slices}
The points of each ring enclose a planar, polygonal surface, a \emph{slice} $S_M$ of the muscle.
The next step in the algorithm, line \ref{alg:1.2}, is to triangulate the extracted slices, i.e., to construct triangles that decompose the polygons. The result of this step is visualized on the left side in \cref{fig:serial_alg_3}.

\begin{figure}%
  \centering%
  \begin{subfigure}[t]{0.31\textwidth}%
    \centering%
    \includegraphics[height=55mm]{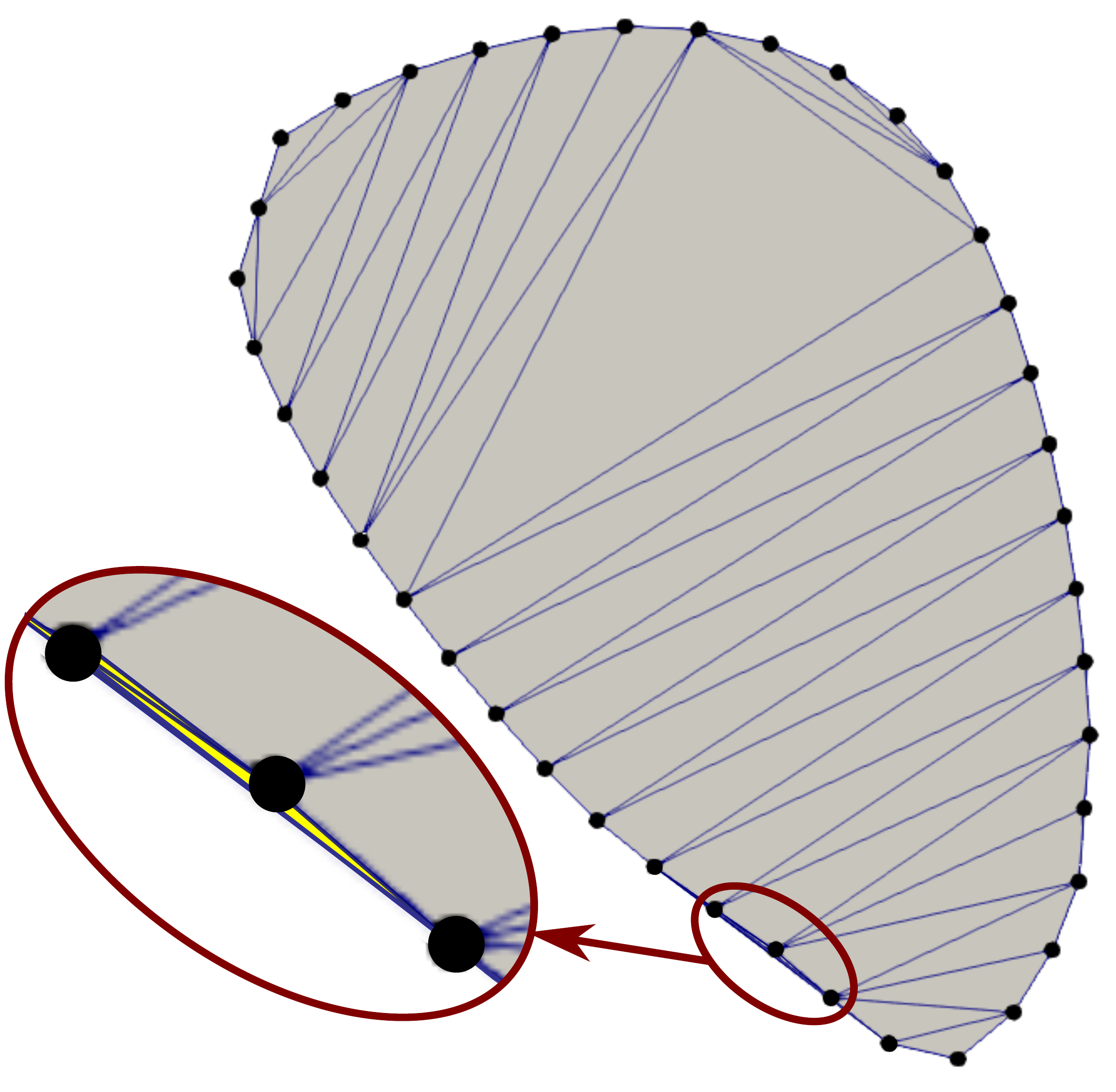}%
    \caption{First triangulation method}%
    \label{fig:triangulation_0}%
  \end{subfigure}
  \qquad
  \begin{subfigure}[t]{0.27\textwidth}%
    \centering%
    \includegraphics[height=55mm]{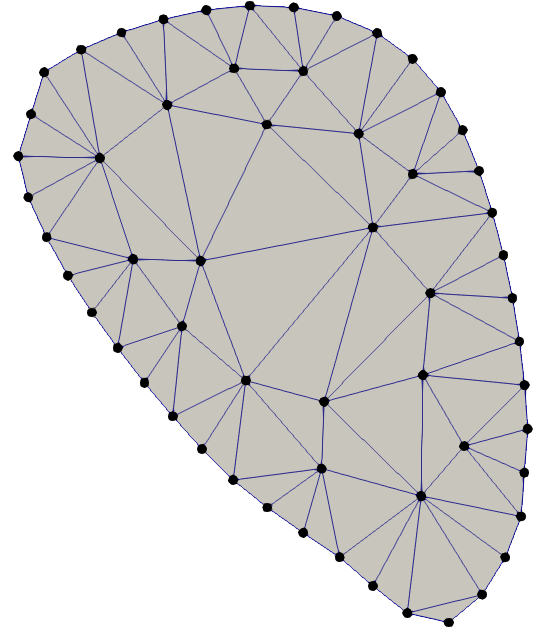}%
    \caption{Second trian\-gulation method}%
    \label{fig:triangulation_1}%
  \end{subfigure}
  \quad
  \begin{subfigure}[t]{0.32\textwidth}%
    \centering%
    \includegraphics[height=55mm]{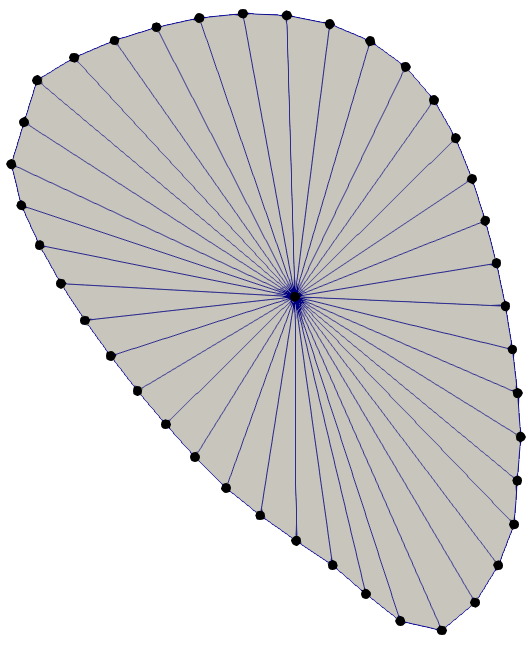}%
    \caption{Third triangulation method}%
    \label{fig:triangulation_2}%
  \end{subfigure}
  \caption{Intermediate step of 3D mesh generation: triangulation of slices, result of different triangulation methods for a slice in the center of the biceps muscle.}%
  \label{fig:triangulations}%
\end{figure}

We select three different methods to construct this triangulation. The first and second methods are based on Delaunay triangulations. The third method creates a custom triangulation using a simple construction scheme with only one additional point.
\Cref{fig:triangulations} visualizes results of the three methods for one slice.

The first method uses the tessellation algorithm from the spatial algorithms and data structures module of the Python package \emph{SciPy}. 
The Quickhull algorithm \cite{quickhull} is used which triangulates the convex hull of the points. In consequence, the triangulations of concave slices have triangles that lie outside the interior of the slice, which is a disadvantage. An advantage is that the triangulation uses all given points and no new points are added. However, this often results in meshes of lower quality than if adding additional points were allowed.
The example in \cref{fig:triangulation_0} shows such a concave slice. At the bottom of the domain, the triangles are outside the slice and almost degenerate.

The second method uses a Delaunay refinement algorithm described in \cite{Delaunay2002} and implemented in the \emph{Triangle} software \cite{shewchuk96b}. A conforming, constrained Delaunay triangulation is created. The triangulation correctly handles convex and concave domains.
Conforming means that the triangulation uses the given points at the boundary. Additional points on the boundary as well as in the interior are added. The triangulation is constraint to generate triangles with minimum angles of 20 degrees and a maximum area $A$ that is set to a value depending on the area of the bounding box. In consequence, the generated triangulations of all slices have a guaranteed mesh quality in terms of angles and about the same size and number of triangles.

Comparing the result of the second method in \cref{fig:triangulation_1} with the result of the first method in \cref{fig:triangulation_0} shows the better triangulation quality as the triangles all have larger angles.

The third method places one additional point at the center of gravity of the given points. Triangles are constructed by connecting the center point with two adjacent points on the boundary, for all given points. The resulting triangulation resembles a pie chart. For some extreme concave slices, this method also creates triangles that partly lie outside the slice, but this rarely occurs with muscle cross-sections. The advantage of this approach is its simplicity.
\Cref{fig:triangulation_2} shows the result for an exemplary slice. In contrast to the first method, the third method creates a valid triangulation despite the concave domain.

\subsection{Harmonic Maps}

Next, harmonic maps are created that allow to smoothly map a given 2D reference mesh onto an actual cross-section of the muscle. The initial application of harmonic maps to meshes used for biomedical simulations is given by \cite{marchandise2010quality} and \cite{Marchandise2_2011}. The authors improve a given, over-sampled surface mesh obtained from classical segmentation. This is done by partitioning the surface into multiple mesh partitions of zero genus (i.e., containing no holes) and transforming them to a reference space using harmonic maps. There, controlled remeshing is carried out before the transformation is reversed.

In our algorithm, harmonic maps are also used for the purpose of generating high quality meshes. In contrast to the literature, the mapping is based on the muscle slices instead of the surface. Also, different parameter domains are investigated.

A function $u: \Omega \to \R$ on a domain $\Omega \in \R^d$ is \emph{harmonic} if it is a solution of the Laplace equation $Δu = 0$.
From variational calculus, it is known that harmonic functions are extremals of the \emph{Dirichlet energy functional} \cite{weyl1940},
\begin{align*}
  E[u] = \dfrac12 \int_\Omega |\nabla u|^2 \,\d\bfx.
\end{align*}
For an intuitive understanding, the map $u$ can be seen as deforming an elastic material that is initially located tension-free in the domain $\Omega$.
Then, the Dirichlet energy $E[u]$ describes the total amount of squared stretch or elastic energy resulting from the tension that occurs in the deformed state. A harmonic map minimizes this total tension. Qualitatively, the map deforms neighborhoods of all points in $\Omega$ by a similar amount, thus, preserving geometrical structures in $\Omega$, e.g., given by a mesh. The idea of our approach is that 
applying a harmonic map on a mesh with good quality preserves the mesh quality also in the image under the map.

In \cref{alg:serial_algorithm_1}, computing the harmonic maps $u$ and $v$ is done in line \ref{alg:1.3}. For a given slice $S_M$, the functions $u$ and $v$ map from points $\bfx \in S_M$ to coordinates $u(\bfx),v(\bfx)\in \R$ of a parameter domain $\Omega_P \subset \R^2$. The parameter domain is either a unit circle or a unit square.

The vector $\bfy(\bfx) := (u(\bfx), v(\bfx))^\top$ for $\bfx \in S_M$ is interpreted as position in $\Omega_P$. The maps are constructed such that the boundary $∂S_M$ of the slice $S_M$ is mapped to the boundary $∂\Omega_P$ of the parameter domain $\Omega_P$ while preserving the distance between points on the boundary.
The mapping $\bfy: S_M \to \Omega_P$ is bijective and harmonic, i.e., the Laplacians of $u$ and $v$ are zero.
More specifically, $u : S_M \to \R$ and $v : S_M \to \R$ are solutions of
\begin{equation}\label{eq:def_harmonic_maps}
  \begin{array}{l}
    Δu(\bfx) = 0, \quad Δv(\bfx) = 0 \quad \forall \bfx \in S_M.
  \end{array}
\end{equation}
To derive suitable Dirichlet boundary conditions for these equations, we consider a uniform parametrization $\bfp:[0,1] \to ∂S_M$ of the boundary $∂S_M$ of the slice, i.e., 
\begin{align*}
  \p{l(t)}{t} = c \in \R \quad \forall t \in [0,1], \quad \text{where } l(t) := \i{0}{t} |\bfp'(\tau)| \,\d\tau.
\end{align*}
We require the image of the boundary parametrization in $\Omega_P$ to be also uniform, i.e.,%
\begin{align*}
  \p{l_P(t)}{t} = c_P \in \R \quad \forall t \in [0,1], \quad \text{where } l_P(t) := \i{0}{t} |\bfy'\big(\bfp(\tau\big)| \,\d\tau.
\end{align*}
Corresponding boundary points $\bfx_\text{M,boundary} \in S_M$ and $\bfx_\text{P,boundary} = (u_\text{P,boundary},v_\text{P,boundary})^\top$ can be defined. This leads to the following Dirichlet boundary conditions that close the definition in \cref{eq:def_harmonic_maps}:%
\begin{align}\label{eq:bc_harmonic_maps}
  u(\bfx_\text{M,boundary}) = u_\text{P,boundary}, \quad v(\bfx_\text{M,boundary}) = v_\text{P,boundary}.
\end{align}

Equations \eqref{eq:def_harmonic_maps} and \eqref{eq:bc_harmonic_maps} describe a boundary value problem of ordinary differential equations for $u$ and $v$. We solve it using the finite element method and the spatial discretization given by the triangulation of the slices.
Depending on the method of triangulation, a different number of degrees of freedom is given. For the first method with the Quickhull algorithm, no degree of freedom is present and no system of equations needs to be solved. Then, the mapping is only a FE interpolation of the boundary mapping.
For the third method, only one degree of freedom for the center point needs to be computed. The second method has as many degrees of freedom as there are additional points inserted during the Delaunay refinement.

The first step is to compute the prescribed boundary points $\bfx_\text{P,boundary}$ in parameter space. When using the first and third triangulation methods, the boundary points on the slices are  equidistant and therefore the same number of points need to be sampled equidistantly on the boundary $∂\Omega_P$ of the parameter space.
If the second triangulation method, which potentially adds additional points is used, the same number of points  are created on the boundary $∂\Omega_P$ of the slice as are given on the slice $∂S_M$. The boundary points are created such that the relations of their distances are the same on $∂\Omega_P$ as for the original points on $∂S_M$.

Using the standard procedure of the finite element method for $Δu(\bfx) = 0$ on $S_M$ and $u=f(\bfx)$ on $∂S_M$, e.g., as outlined in \cite{Remacle2010}, leads to the weak form with ansatz and test functions $\phi$,
\begin{align}\label{eq:est_fe_w}
    \i{S_M}{} (\nabla u^\top \nabla \phi + \nabla f(\bfx)^\top \nabla \phi) \,\d \bfx = 0 \quad \forall \phi \in \mathcal{H}_0^1.
\end{align}
Standard linear hat functions are used on the triangles, such that they provide the interpolation property $\phi_i(\bfx_j) = \delta_{ij}$. Using the barycentric parametrization of triangles with points $\bfp^1,\bfp^2$ and $\bfp^3$ introduced in \cref{eq:barycentric_triangle}, we define the ansatz functions and get their derivatives within the elements by:
\begin{align*}
  \phi^{(e)}_1 &= (1 - \xi_1)(1 - \xi_2), \quad&
  \phi^{(e)}_2 &= \xi_1 (1 - \xi_2), \quad &
  \phi^{(e)}_3 &= (1 - \xi_1) \xi_2,\\[4mm]
  \nabla \phi^{(e)}_1 &= (\xi_2-1, \xi_1 - 1)^\top, \quad&
  \nabla \phi^{(e)}_2 &= (1-\xi_2, -\xi_1)^\top, \quad&
  \nabla \phi^{(e)}_3 &= (-\xi_2, 1-\xi_1)^\top.
\end{align*}
The superscript $\square^{(e)}$ refers to the definition of the functions within elements. The global assembly involves composing the global nodal functions $\phi_i(\bfx)$ for nodes indexed by $i=1, \dots, n_\text{nodes}$ and using a mapping between the barycentric coordinates $\xi_1,\xi_2 \in [0,1]^2$ inside the elements to the global coordinates $\bfx \in S_M$.
Inserting the discretization
\begin{align*}
  u_h(\bfx) = \s{i=1}{n_\text{nodes}} u_i\,\phi_i(\xi_1(\bfx), \xi_2(\bfx))
\end{align*}
into \cref{eq:est_fe_w} leads to the form
\begin{equation}\label{eq:est_weak_form}
  \begin{array}{lll}
    \s{i=1}{n_\text{nodes}}u_i \i{S_M}{} \nabla_\bfx \phi_i^\top \nabla_\bfx \phi_j \,\d\bfx + \s{i=1}{n_\text{nodes}} f_i \i{S_M}{} \nabla_\bfx \phi_i^\top \nabla_\bfx \phi_j \,\d \bfx = 0\,\quad \forall j=1,\dots,n_\text{nodes}.
  \end{array}
\end{equation}
The integrations are executed element-wise and over the elemental coordinates $\xi_1,\xi_2$. 
The transformation to elemental coordinates involves the computation of the Jacobian $J=\d\bfx/\d\bfxi$ of the mapping between element coordinates $\bfxi = (ξ_1,ξ_2)$ and global coordinates $\bfx$. From the definition in \cref{eq:barycentric_triangle}, it follows that
\begin{align*}
  J = \d{\bfx}{\bfxi} = [\bfp^2-\bfp^1, \bfp^3-\bfp^1].\\[4mm]
\end{align*}
The metric tensor for this mapping is given by
\begin{align*}
  \mathcal{M} := \left(\d{x}{\bfxi}\right)^T \left(\d{x}{\bfxi}\right).
\end{align*}
The transformation of the integrals in \cref{eq:est_weak_form} introduces an additional integration factor $\sqrt{\det{\mathcal{M}}}$.
We get the following matrix equation:
\begin{align}\label{eq:est_fe_system}
  M_u \bfu = -M_f\,\bff,
\end{align}
with the vector $\bfu$ of nodal solution values, the vector $\bff$ of nodal Dirichlet boundary condition values at the boundary and the global stiffness matrices $M_u$ and $M_f$. The two global stiffness matrices are assembled from the element stiffness matrices $M^\text{(e)}$ for the degrees of freedom at all nodes respectively at the boundary nodes. The entries of the element stiffness matrices are given by
\begin{align*}
  M^\text{(e)}_{i,j} = \i{0}{1}\i{0}{1-\xi_1}   \nabla \phi_i^{(e)}(\xi_1,\xi_2)^\top \, \mathcal{M}^{-1} \, \nabla \phi_j^{(e)}(\xi_1,\xi_2) \, \sqrt{\det{\mathcal{M}}}  \,\d\xi_2\d\xi_1.
\end{align*}

By solving \cref{eq:est_fe_system} for $\bfu$, we get the discretized harmonic map $u$.
The finite element formulation and computation for $v$ is analog and uses the same global stiffness matrices.

\begin{figure}%
  \centering%
  \includegraphics[width=\textwidth]{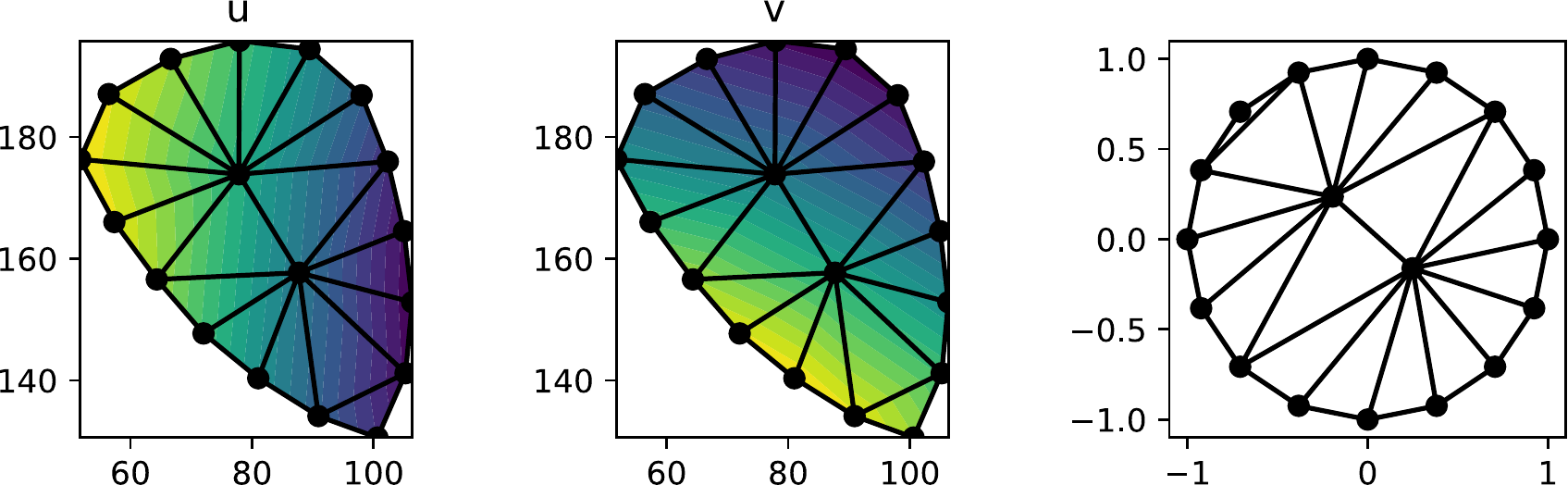}%
  \caption{Quality improvement of muscle slice meshes as a basis for 3D mesh generation: Initial triangulations and harmonic map for a slice $S_M$ of the biceps muscle. The first two plots show the solutions of $u$ and $v$ on the slice $S_M$. The third plot shows the image in $\Omega_P$ of the triangulation in $S_M$ under the harmonic map.}%
  \label{fig:harmonic_map_solution0}%
\end{figure}

\Cref{fig:harmonic_map_solution0} visualizes the triangulation of $S_M$ and the solutions $u(\bfx)$ and $v(\bfx)$ for a circular parameter domain $\Omega_P$ and an exemplary muscle slice in the first two plots. The color range from bright yellow to dark violet corresponds to increasing values of $u$ and $v$. It can be seen that the $u$ values increase from left to right whereas the $v$ values increase from bottom to top, corresponding to the horizontal and vertical coordinate axes $y_1$ and $y_2$ of $\Omega_P$.

Applying the computed harmonic map $\bfy(\bfx)$ to the triangulation of the slices results in a triangulation of the parameter domain $\Omega_P$. This is shown in the third plot of \cref{fig:harmonic_map_solution0} and in \cref{fig:serial_alg_3}.
In both figures, the triangulation of the slices was generated using the second triangulation method with the constrained Delaunay triangulation. On the right side of \cref{fig:serial_alg_3}, the image $\bfy(\bfx)$ under the harmonic map of the triangulation in the slices is shown on the unit circle parameter domain.

\subsection{Construction of a Regular Grid in the Parameter Domain}
The next step in \cref{alg:serial_algorithm_1} is the construction of a 2D structured, regular grid in the parameter domain $\Omega_P$, as stated in line \ref{alg:1.4}. This grid will then be mapped to the slices $S_M$. Creating a structured grid of quadrilateral elements in a given domain is also called \emph{quadrangulation}.

\begin{figure}%
  \centering%
  \begin{subfigure}[t]{0.48\textwidth}%
    \centering%
    \includegraphics[width=\textwidth]{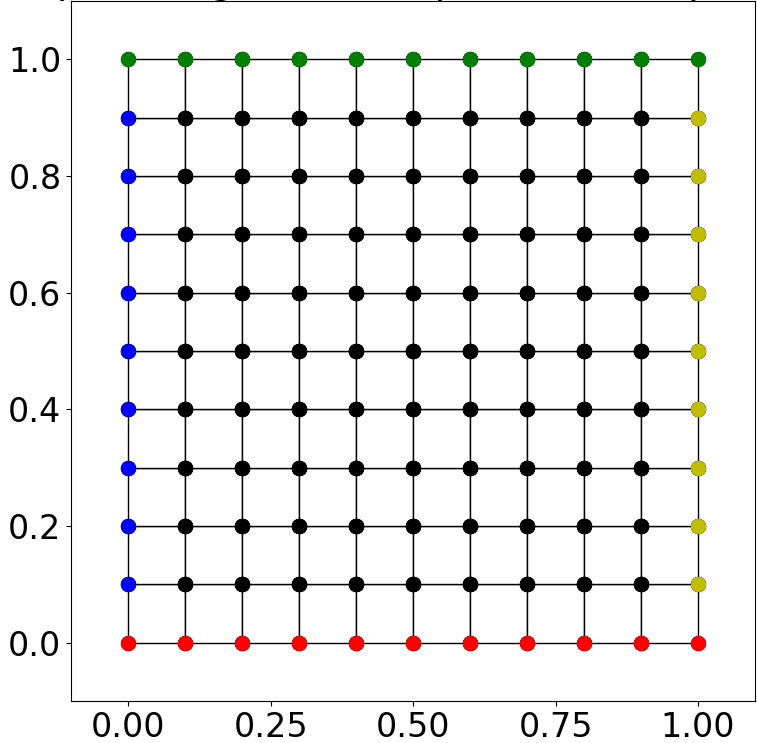}%
    \caption{Unit square, scheme 1}%
    \label{fig:quad_1}%
  \end{subfigure}
  \quad
  \begin{subfigure}[t]{0.48\textwidth}%
    \centering%
    \includegraphics[width=\textwidth]{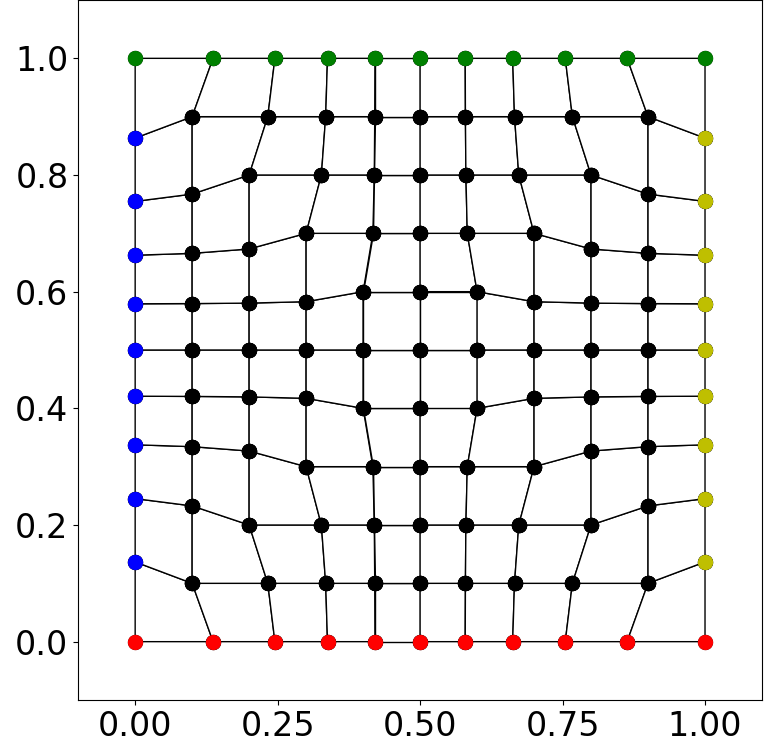}%
    \caption{Unit square, scheme 2}%
    \label{fig:quad_2}%
  \end{subfigure}
  \begin{subfigure}[t]{0.48\textwidth}%
    \centering%
    \includegraphics[width=\textwidth]{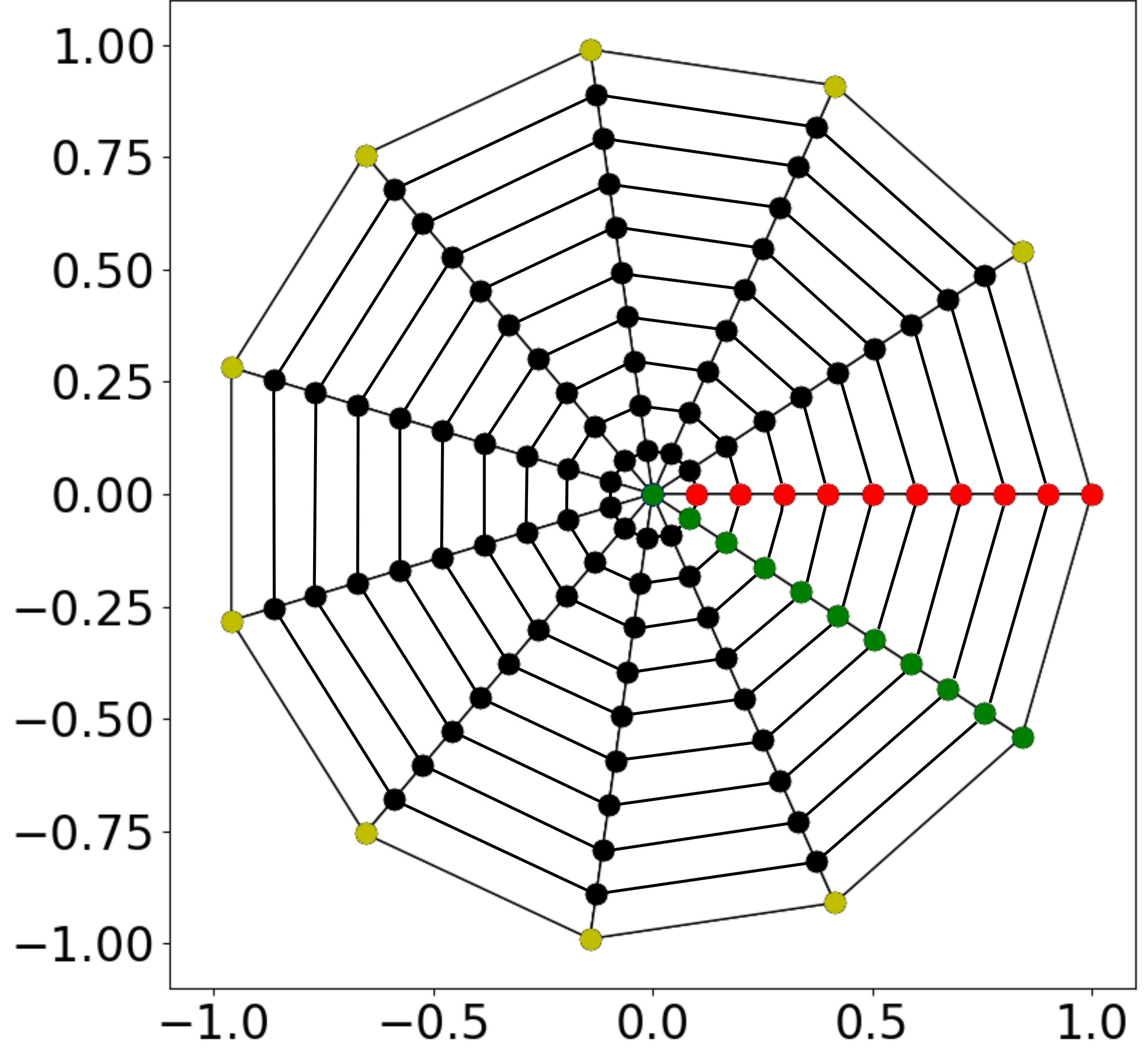}%
    \caption{Unit circle, scheme 1}%
    \label{fig:quad_0}%
  \end{subfigure}
  \quad
  \begin{subfigure}[t]{0.48\textwidth}%
    \centering%
    \includegraphics[width=\textwidth]{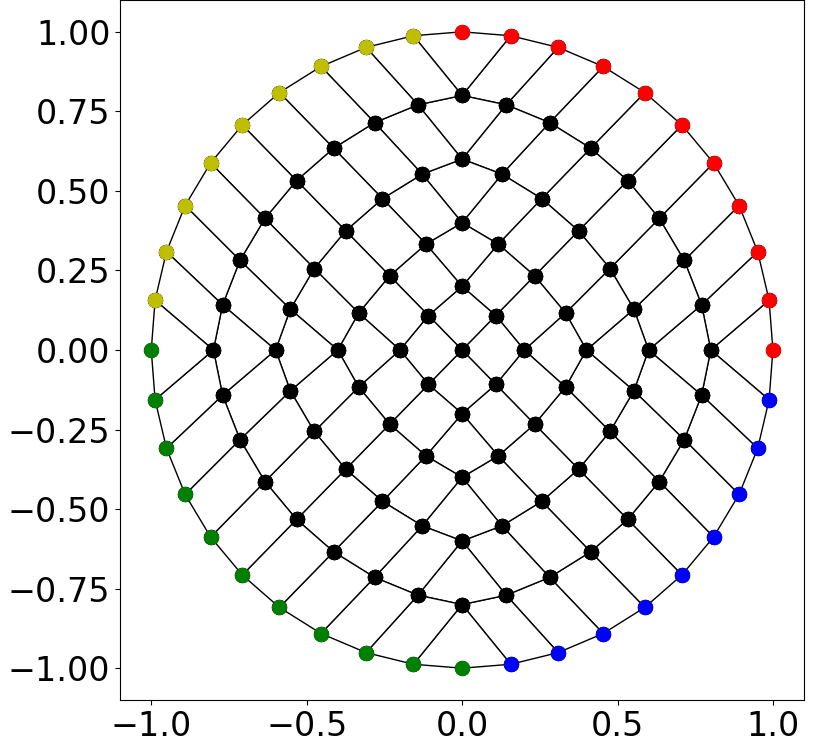}%
    \caption{Unit circle, scheme 2}%
    \label{fig:quad_3}%
  \end{subfigure}
  \caption{Four different quadrangulation schemes of the parameter domain with $11\times 11$ nodes. The boundaries of the grid are colored for better perceptibility. In (a) and (b), the parameter domain is a unit square with a uniform grid (a) and an adjusted grid (b) that tries to reduce the problem of degenerate elements at the corners of the muscle slices. In (c) and (d), quadrangulations on a unit circle parameter domain are shown. (c) shows a rotationally symmetric construction scheme whereas the approach in (d) is similar to a uniform grid.}%
  \label{fig:quads}%
\end{figure}

The parameter domain $\Omega_P$ can be selected to be either a unit square or a unit circle. For both choices, two different schemes how to generate a grid with a given number of cells can be selected.
\Cref{fig:quads} shows all four possibilities.

The first scheme, \cref{fig:quad_1}, uses an equidistant regular grid in a unit square. This is the easiest possibility to generate a quadrangulated reference domain. A possible issue is induced by the corners of the square. The grid will be mapped to a cross-section of the muscle which has no sharp corners. Therefore, the cells of the grid will be distorted at the images of the corners, usually shortening diagonals that point towards the corners and lengthening the other diagonals. This assumption motivates the second scheme in \cref{fig:quad_2}. Here, the elements are already distorted in the described manner, with increasing distortion closer to the corners. The rationale is that the mapped cells in $S_M$ will then be less distorted.

We construct our second quadrangulation scheme of the unit square as follows. The diagonals of the square divide the domain into bottom, top, left and right quarters, which are considered separately.
For example, the bottom quarter is the triangle that is formed by the corner points $(0,0), (1,0)$ and the center point $(\frac12,\frac12)$ of the square. In the bottom quarter, the horizontal $x$ coordinate of a point $(x,y)$ in a uniform grid can be described by $x = \frac12 + \textrm{tan}(\phi)\,(\frac12-y)$ where $\phi$ is the angle between a line through $(x,y)$ and the center point $(\frac12,\frac12)$ and the $y$-axis. The points of the adjusted grid in the quadrangulation scheme are constructed by altering the value of $\phi$.
On every horizontal series of points in the bottom quarter, $\phi$ is varied linearly in $[-\pi/4,\pi/4]$ instead of the nonlinear progression according to the actual angle. This leads to the larger spacing between points near the diagonals. All four quarters are treated analogously to produce the shown symmetric pattern.

A different approach is to use a unit circle, which has no corners and therefore might resemble a muscle cross-section more consistently. The first scheme of the unit circle is given in \cref{fig:quad_0}. It uses the radial and circumferential directions for the two dimensions of the grid. A disadvantage of this scheme is that the quadrilaterals at the center are degenerated to triangles. Additionally, the area of the cells varies significantly and the outer cells have unequal side lengths. 

To remedy this problem, we develop the second scheme given in \cref{fig:quad_3}. When traversing from the outer boundary towards the center point and considering the circumferential lines of grid points, the circle morphs into a square as the number of grid points decreases. This approach has the disadvantage that some cells have an inner angle of nearly \SI{180}{\degree}, especially four elements at the boundary. Apart from that, all elements have similar sized sides and angles.
The construction of this scheme is similar to the approach of scheme 2 on the unit square in that the domain is also divided into four quarters. However, different formulas for the point coordinates $(x,y)$ depending on the angle $\phi$ are used.
The detailed construction formulas of all four presented quadrangulation schemes are provided by their implementation in the code. The script \code{plot_quadrangulation_schemes.py} constructs and visualizes the four schemes with a configurable number of grid points.

Each construction scheme allows to specify the (squared) number of nodes and in consequence the number of cells. The examples in \cref{fig:quad_1,fig:quad_2,fig:quad_3} have $11 \times 11$ nodes and $10 \times 10$ cells. For \cref{fig:quad_0}, the numbers are slightly different. There are $10 \times (11 + 1)$ nodes resulting in $10 \times 11$ cells.

Next, the grid in the parameter domain is transferred to the muscle domain by applying the harmonic map $\bfy(\bfx) \in S_M$ on every point of the quadrangulation $\bfx \in \Omega_P$. This is illustrated in \cref{fig:serial_alg_4} for a parameter domain consisting of the unit circle, with quadrangulation scheme 2 and $5 \times 5$ nodes. The cells of the grid in $\Omega_P$ are shown in the right-most stack of domains. The grid points are visualized left of the grids. The resulting image of the mesh in the slices $S_M$ is shown on the left. For visualization reasons, each quadrilateral has been split into two triangles.

\subsection{Formation of Three-Dimensional Elements}

The result of the previous steps is a number of quadrangulated muscle slices. The grid on every slice has the same number of nodes and elements. The nodes on the boundary of neighboring slices are positioned similarly.

The final step of \cref{alg:serial_algorithm_1} is line \ref{alg:1.5}, the formation of 3D elements. Inserting vertical edges between all corresponding nodes on two neighboring slices creates a set of 3D hexahedral elements and, thus, an overall 3D hexahedral mesh of the muscle volume. This step is visualized in \cref{fig:serial_alg_8}.

\subsection{Generation of Fiber Meshes}\label{sec:generation_of_fiber_meshes}

1D fiber meshes are created following the approach of computing a divergence free vector field introduced in \cite{Choi2013}. The steps are given in \cref{alg:serial_algorithm_2}.

The Laplace problem to be solved can be stated as%
\begin{align}\label{eq:fiberest_laplace}
  Δp(\bfx) = 0 \quad \text{for } \bfx \in \Omega_M.
\end{align}
The vector field is given by the gradient $\nabla p$ of a solution $p$ of \cref{eq:fiberest_laplace}. The quantities can be interpreted as pressure $p$ and (negative) velocity field $\nabla p$ of a steady flow.
The muscle fibers are given as streamlines or, equivalently, pathlines in this velocity field. Every streamline $\bfx : [-c_1,c_2] \subset \R \to \Omega_M$ with $c_1,c_2 > 0$ is defined by a seed point $\bfx_0$ and the property that it is tangent to the velocity field at any point:
\begin{align*}
  \bfx(0) = \bfx_0,\quad
  \p{\bfx(s)}{s} = \nabla p\big(\bfx(s)\big).
\end{align*}

As proposed by \cite{Choi2013}, Neumann boundary conditions can be specified for the bottom and top surfaces of the muscle volume, $∂\Omega_{M, \text{bottom}}$ and $∂\Omega_{M, \text{top}}$:
\begin{equation}\label{eq:fiberest_neumann}
\begin{array}{rlrlr}
  \d{p(\bfx)}{\bfx} \cdot \bfn &= F_\text{in}, \quad &&\text{for } \bfx \in ∂\Omega_{M, \text{bottom}},\\[4mm]
  \d{p(\bfx)}{\bfx} \cdot \bfn &= F_\text{out}, \quad &&\text{for } \bfx \in ∂\Omega_{M, \text{top}}.
\end{array}
\end{equation}
The in and outflow values $F_\text{in}<0$ and $F_\text{out}>0$ are balanced such that the total inflow ${F_\text{in}\cdot \mu(∂\Omega_{M, \text{bottom}})}$ compensates the total outflow ${F_\text{in}\cdot \mu(∂\Omega_{M, \text{bottom}})}$. Here, $\mu(∂\Omega)$ is the surface area of the respective boundary.

Alternatively, Dirichlet boundary conditions can be specified:
\begin{equation}\label{eq:fiberest_dirichlet}
\begin{array}{rlrlr}
  p(\bfx) &= 0, \quad &&\text{for } \bfx \in ∂\Omega_{M, \text{bottom}},\\[4mm]
  p(\bfx) &= 1, \quad &&\text{for } \bfx \in ∂\Omega_{M, \text{top}}.
\end{array}
\end{equation}
The specification of Dirichlet boundary conditions has the same effect as Neumann boundary conditions and is easier to define. The in and outflows are still orthogonal to the boundary because the prescribed value of $p$ does not vary in the planar boundary.

\begin{figure}%
  \centering%
  \begin{subfigure}[t]{0.48\textwidth}%
    \centering%
    \includegraphics[height=10cm]{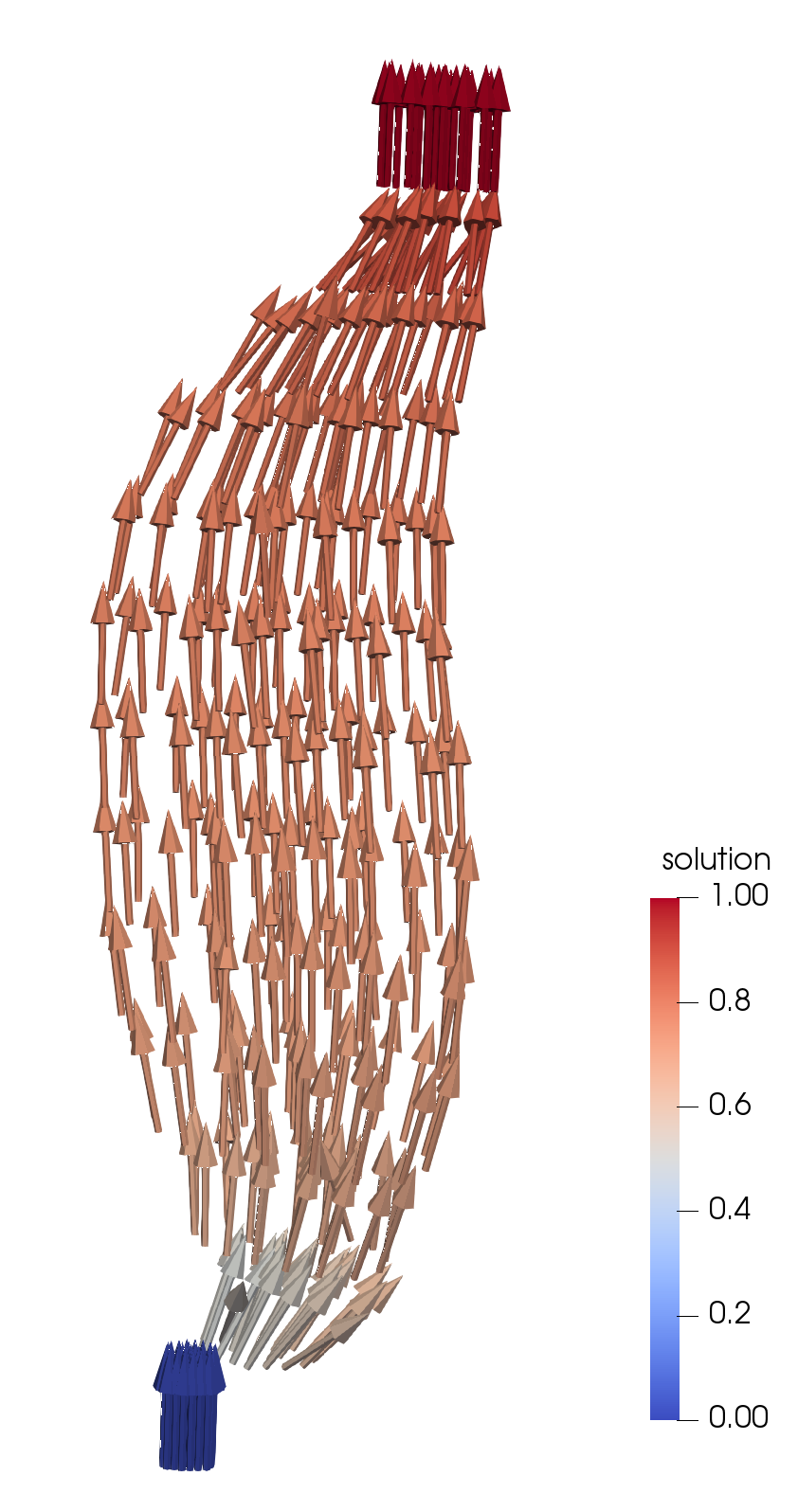}%
    \caption{Solution (color coding) and direction vectors of the gradient field for the boundary value problem \cref{eq:fiberest_laplace} with Dirichlet boundary conditions \cref{eq:fiberest_dirichlet}.}%
    \label{fig:potential_flow}%
  \end{subfigure}
  \quad
  \begin{subfigure}[t]{0.48\textwidth}%
    \centering%
    \includegraphics[height=10cm]{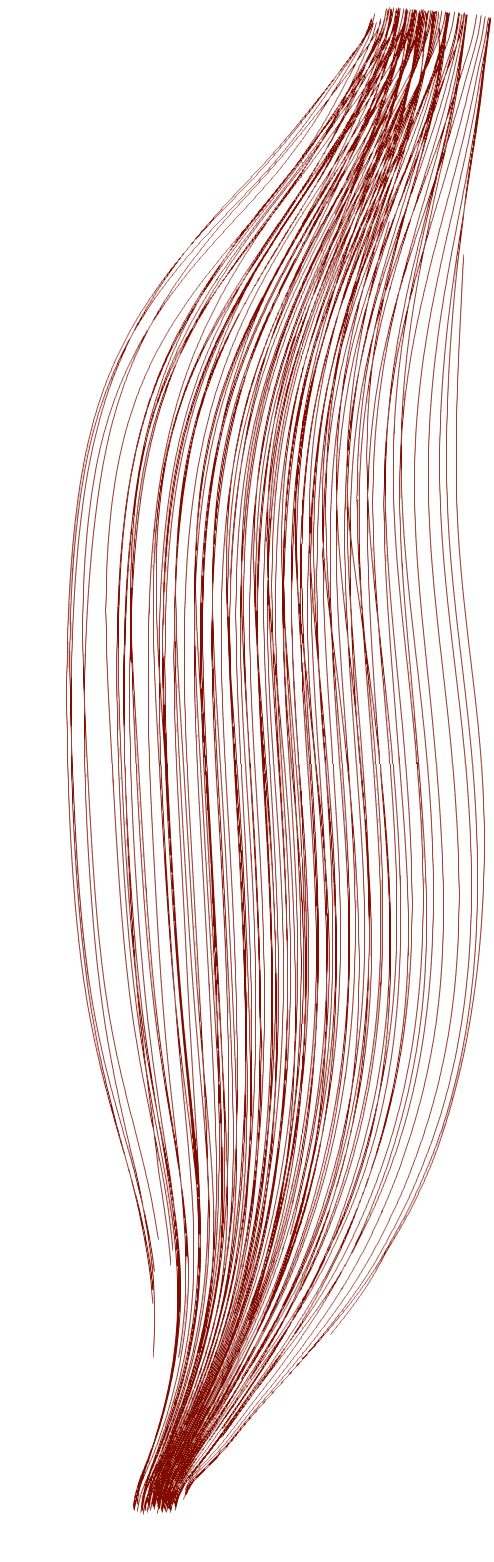}%
    \caption{Resulting streamlines that were traced through the gradient field.}%
    \label{fig:fiber_tracing_streamlines}%
  \end{subfigure}
  \caption{Setup and solution of the Laplace problem for the biceps geometry that is used to estimate muscle fibers by streamline tracing.}%
  \label{fig:potential_flow_streamlines}%
\end{figure}

The boundary value problem given by \cref{eq:fiberest_laplace,eq:fiberest_neumann}  or \cref{eq:fiberest_laplace,eq:fiberest_dirichlet} is discretized by the finite element method with linear or quadratic ansatz functions and solved by our software \opendihu{} using the 3D mesh generated from \cref{alg:serial_algorithm_1}. The divergence free gradient field is visualized in \cref{fig:potential_flow}. The gradient values are directly given by the finite element discretization. The gradient is elementwise constant for linear ansatz functions and trilinear for quadratic ansatz functions.

The next step in \cref{alg:serial_algorithm_2} is line \ref{line:2.3}, tracing streamlines through the gradient field. Seed points are selected on the 2D cross-section at the vertical center of the 3D muscle domain. The seed points are sampled regularly on the square or circular parameter domain according to the quadrangulation scheme and then mapped to the respective muscle slice.
Because the 3D mesh was created using harmonic maps, the resulting spacing between the seed points is very uniform.

For the tracing of streamlines, the semi-analytical Pollock's method \cite{Pollock1988} is often used, which was originally developed for fixed 2D finite difference grids. Extensions to irregular 3D grids and for given velocities at nodes instead of fluxes over faces have been formulated \cite{HAEGLAND2007Streamline}. Other, more accurate algorithms exist \cite{cordes1992continuous}, including higher order formulations \cite{juanes2006unified}.

Because modeling muscle fascicles is only a heuristic approach, the generated streamlines do not have to be exceptionally accurate. Therefore, we use a fully numerical method. The streamlines are generated by explicit Euler integration of the gradient vectors in top and bottom direction. A small spatial step width of $h=\num{1e-2}$ is used. Details of the algorithm are given in the next section, \cref{sec:algorithm_for_streamline_tracing}. In line \ref{line:2.4} of \cref{alg:serial_algorithm_2}, all generated streamlines are resampled to obtain the desired widths of the 1D elements.
\Cref{fig:fiber_tracing_streamlines} visualizes the resulting streamlines in the biceps muscle.

\begin{algorithm}
  \begin{algorithmic}[1]%
    \Statex\Procedure{Create\_1D\_meshes}{}
    \Require Structured 3D volume mesh
    \Ensure 1D fiber meshes
    \Statex
    \State Solve Laplacian flow problem   \label{line:2.2}
    \State Trace streamlines in the gradient field  \label{line:2.3}
    \State Resample 1D fiber meshes \label{line:2.4}
    \EndProcedure
  \end{algorithmic}%
  \caption{Serial algorithm}%
  \label{alg:serial_algorithm_2}%
\end{algorithm}%

\subsection{Algorithm for Streamline Tracing}\label{sec:algorithm_for_streamline_tracing}
The algorithm for streamline tracing uses an efficient method to traverse the mesh, which makes use of its structuredness.
At first, the element $E^{(0)}$ in the mesh that contains the first seed point $\bfp^0$ needs to be found. By construction of the mesh generation algorithms, this is always the element with the lowest index. 
However, if the seed point is not found there, the scheme is robust enough to search in all other elements.

Starting from the seed point $\bfp^0 = \bfp^{(i)}$ in element $E^{(i)}$, the next point $\bfp^{(i+1)}$ of a streamline is computed as %
\begin{align*}
  \bfp^{(i+1)} = \bfp^{(i)} + h\,\nabla p (\bfp^{(i)}).
\end{align*}
After $\bfp^{(i+1)}$ has been computed, the mesh element $E^{(i+1)}$ where it is located needs to be identified. This is needed to evaluate the gradient value $\nabla p (\bfp^{(i+1)})$ at the new point by interpolation according to the FE representation of $p$.

At first, the element $E^{(i)}$ of $\bfp^{(i)}$ is checked whether it also contains $\bfp^{(i+1)}$. If not, the neighboring element in the direction of the streamline is considered. This neighboring element is chosen among all up to 26 possible neighbors such that the direction from the previous element $E^{(i-1)}$ to the current element $E^{(i)}$ continues.

If this element is also not the right one, all other neighbors of $E^{(i)}$ are subsequently checked, ordered by their plausibility according to the previous streamline direction. If none of the 27 considered elements contains the computed point, a search among all elements of the entire mesh is performed. This case happens only for unsuited choices of the integration width $h$, i.e., if the streamline tracing skips whole elements.

The end of a streamline is detected when the streamline reaches the final $z$ plane, either at the bottom or top of the muscle volume. To make the algorithm more robust, also the case is considered where the streamline leaves the muscle domain to the side shortly before reaching the end of the muscle. This can happen due to discretization errors for streamlines that start close to the boundary of the muscle. In such a case, the missing rest of the streamline is interpolated from up to four existing neighboring parallel streamlines.

After the end of the streamline is found, tracing of the next streamline starts at the next seed point $\bfp^0_\text{next}$. The element where $\bfp^0_\text{next}$ is located can also be easily determined in the structured mesh.

The presented scheme avoids repeatedly traversing all elements of the mesh by predicting the next elements according to streamline direction and organization of seed points. This is facilitated by the structured mesh, which has well-defined element neighbor relations.
At the same time, the scheme is robust enough to also efficiently handle streamlines in other use cases. It can also be reused, e.g., in muscle fiber tracing applications of more complex shaped muscles where the fibers change directions.

\subsection{Results and Discussion}\label{sec:mesh_generation_0_results_and_discussion}
The presented \cref{alg:serial_algorithm_1,alg:serial_algorithm_2} generate a 3D mesh and 1D muscle fibers from a triangulated surface. Three different triangulation strategies for the slices and four different reference quadrangulations can be chosen. In the following, the different choices are evaluated.

\begin{figure}%
  \centering%
  \begin{subfigure}[t]{0.31\textwidth}%
    \centering%
    \includegraphics[width=\textwidth]{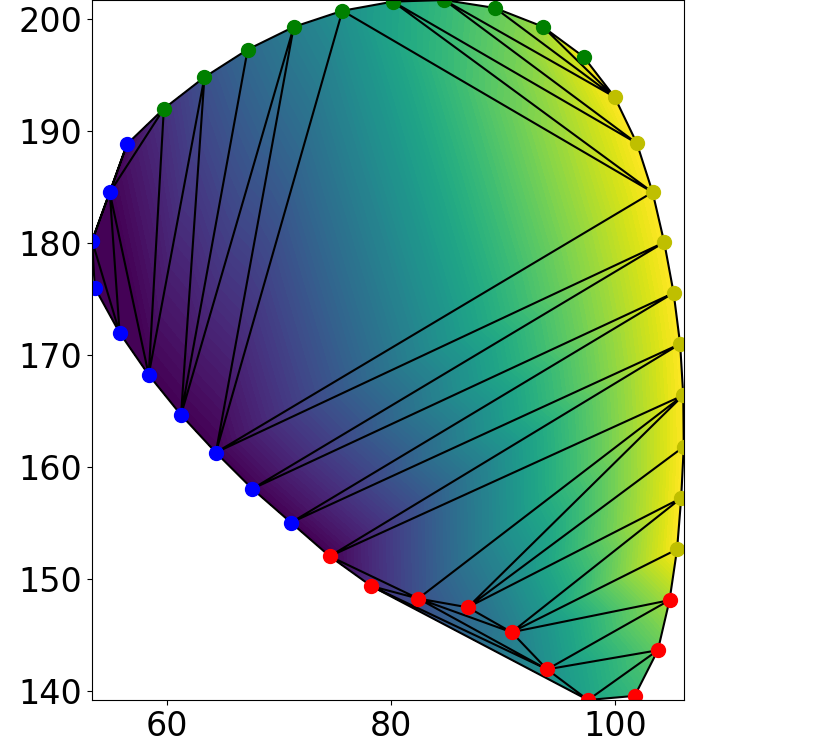}%
    \caption{First method,\\Quickhull algorithm}%
    \label{fig:triu_0}%
  \end{subfigure}
  \quad
  \begin{subfigure}[t]{0.31\textwidth}%
    \centering%
    \includegraphics[width=\textwidth]{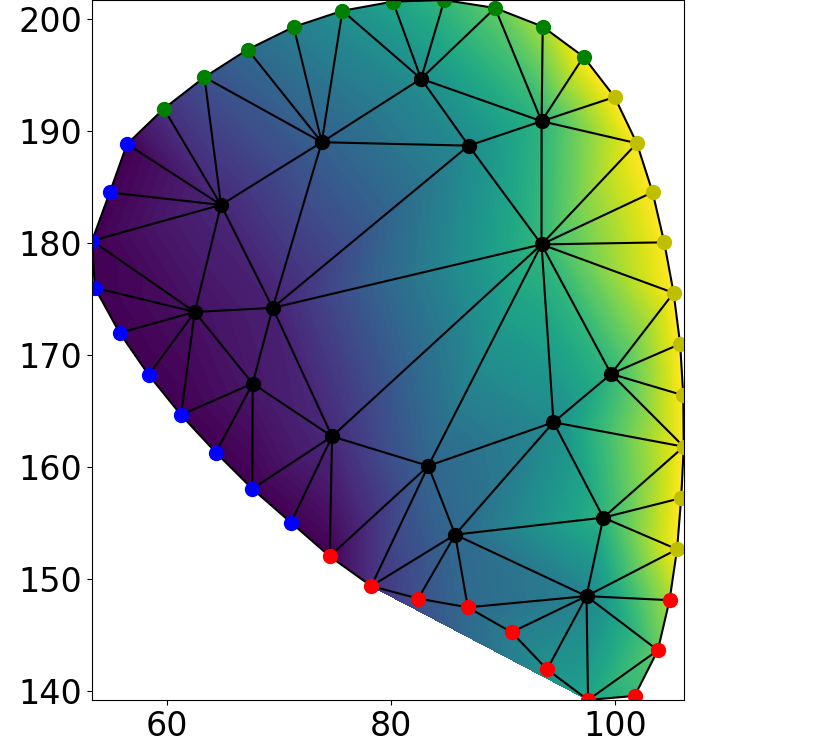}%
    \caption{Second method,\\Delaunay refinement}%
    \label{fig:triu_1}%
  \end{subfigure}
  \quad
  \begin{subfigure}[t]{0.31\textwidth}%
    \centering%
    \includegraphics[width=\textwidth]{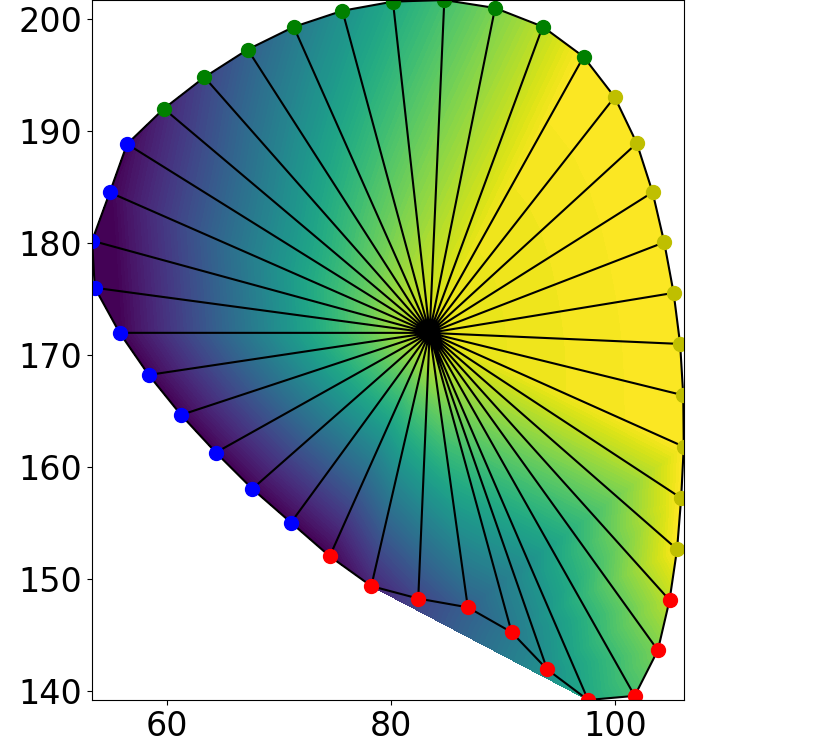}%
    \caption{Third method,\\center of gravity}%
    \label{fig:triu_2}%
  \end{subfigure}\\
  
  % square
  \begin{subfigure}[t]{0.31\textwidth}%
    \centering%
    \includegraphics[width=0.9\textwidth, trim=37mm 14mm 6mm 6mm, clip]{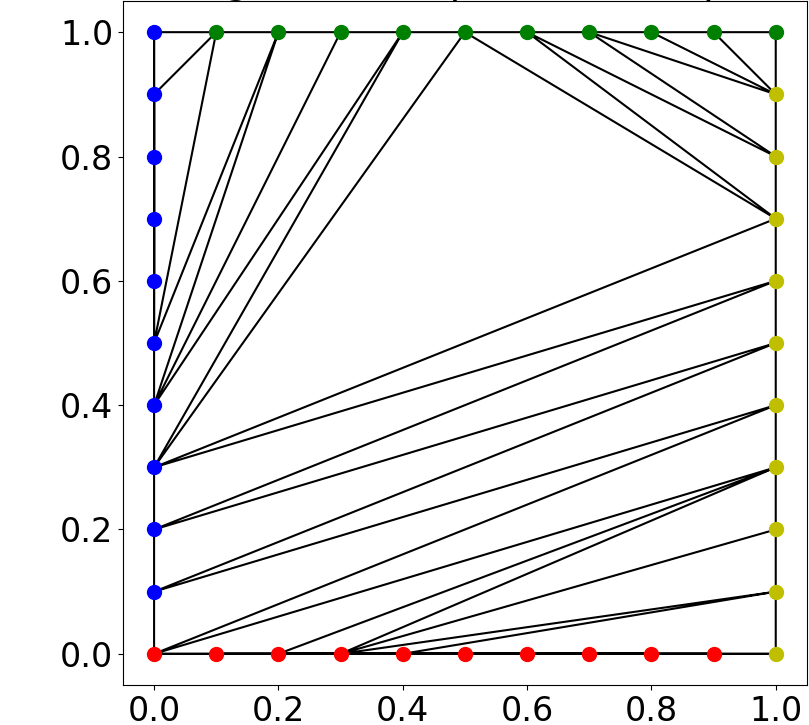}%
    %\caption{}%
    \label{fig:w_01}%
  \end{subfigure}
  \quad
  \begin{subfigure}[t]{0.31\textwidth}%
    \centering%
    \includegraphics[width=0.9\textwidth, trim=37mm 14mm 6mm 6mm, clip]{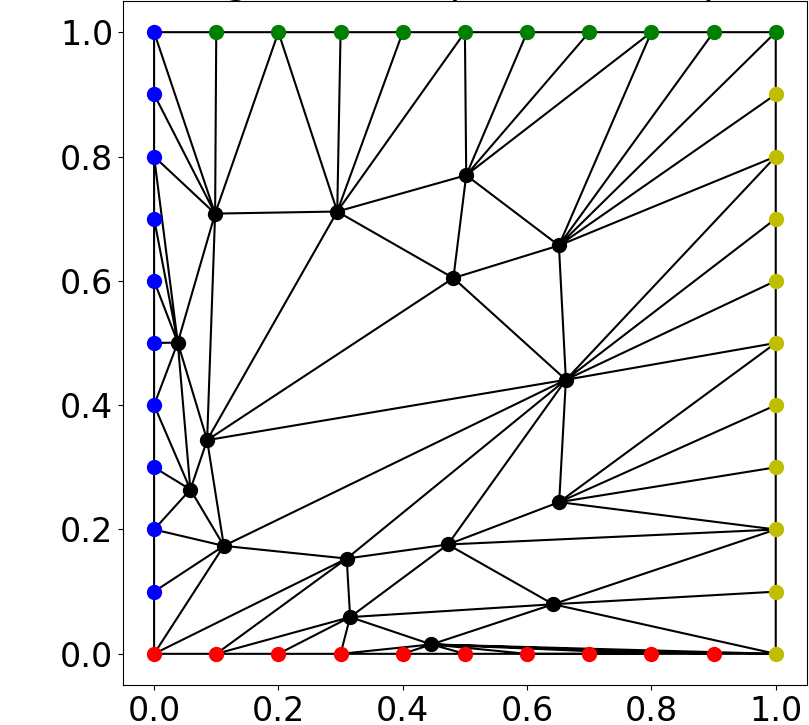}%
    %\caption{}%
    \label{fig:w_11}%
  \end{subfigure}
  \quad
  \begin{subfigure}[t]{0.31\textwidth}%
    \centering%
    \includegraphics[width=0.9\textwidth, trim=37mm 14mm 6mm 6mm, clip]{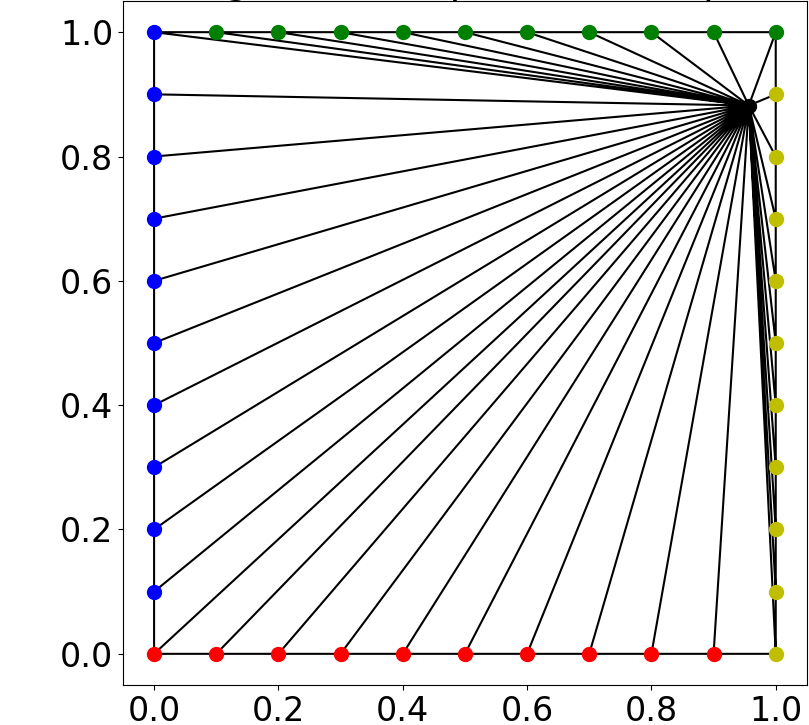}%
    %\caption{}%
    \label{fig:w_21}%
  \end{subfigure}\\
  
  % circle
  \hspace{-12mm}
  \begin{subfigure}[t]{0.31\textwidth}%
    \centering%
    \includegraphics[width=0.9\textwidth, trim=37mm 14mm 6mm 6mm, clip, angle=225,origin=c]{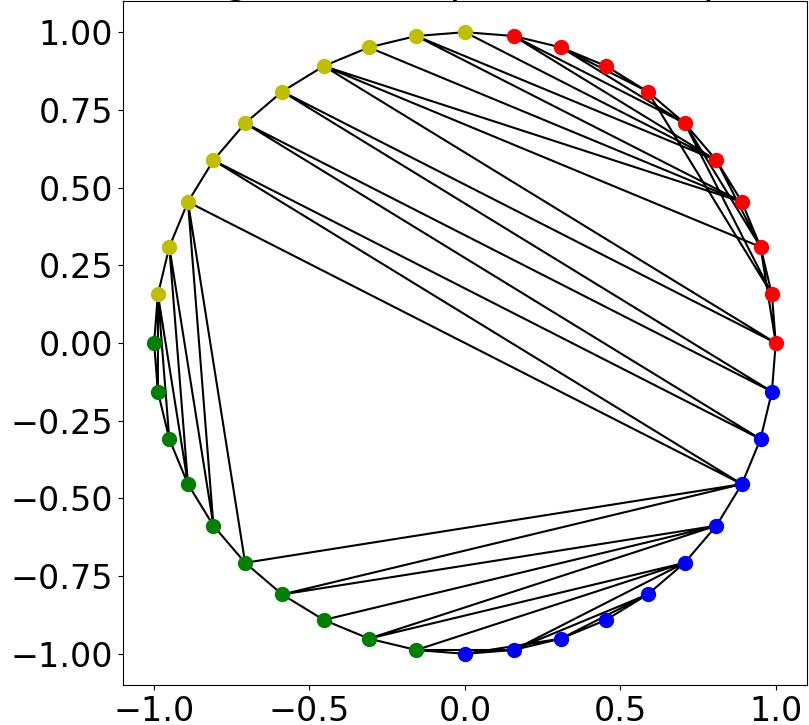}% trim=left bottom right top, clip]
    %\caption{}%
    \label{fig:w_00}%
  \end{subfigure}
  \quad
  \begin{subfigure}[t]{0.31\textwidth}%
    \centering%
    \includegraphics[width=0.9\textwidth, trim=37mm 14mm 6mm 6mm, clip, angle=225,origin=c]{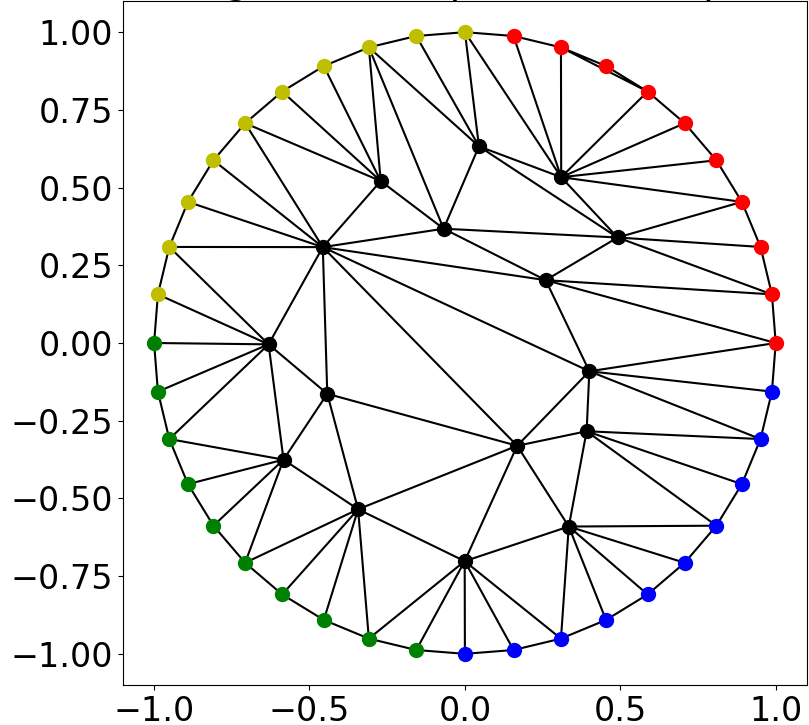}%
    %\caption{}%
    \label{fig:w_10}%
  \end{subfigure}
  \quad
  \begin{subfigure}[t]{0.31\textwidth}%
    \centering%
    \includegraphics[width=0.9\textwidth, trim=37mm 14mm 6mm 6mm, clip, angle=225,origin=c]{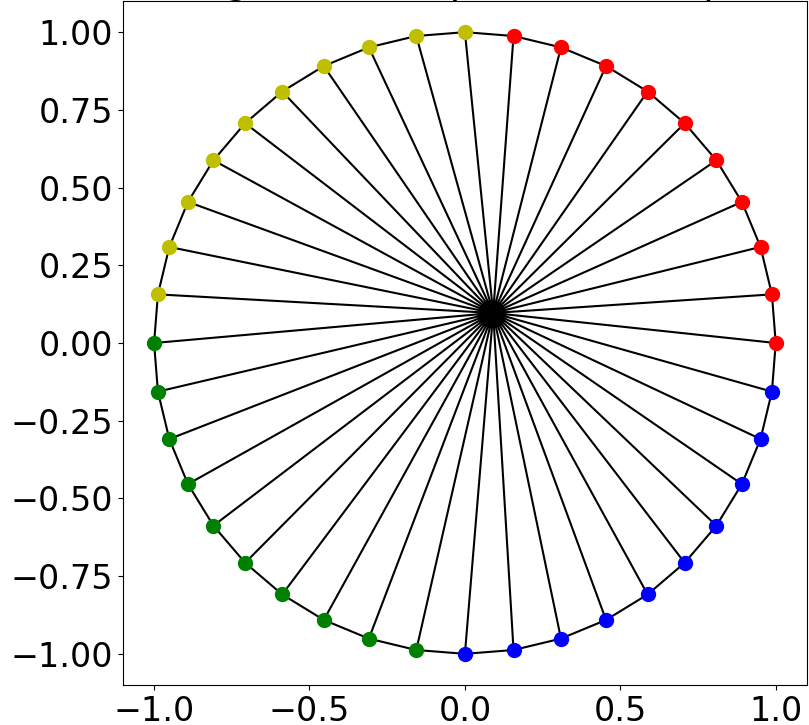}%
    %\caption{}%
    \label{fig:w_20}%
  \end{subfigure}
  
  \caption{Initial slice triangulation and harmonic maps in 3D mesh generation: Top row: Different triangulation methods for $S_M$, the color represents the solution $u$ of the harmonic map. Middle and bottom row: triangulation mapped to the parameter domain $\Omega_P$, for the unit square (middle) and the unit circle (bottom). Each column corresponds to one triangulation method.}%
  \label{fig:tri_triangulations}%
\end{figure}

The different triangulation methods for the slices discussed in \cref{sec:triangulation_of_the_slices} are visualized in the three columns of \cref{fig:tri_triangulations}. 
The top rows show the triangulation of $S_M$ and the harmonic map $u$ as color coded values from violet to yellow. $u$ is the horizontal coordinate on the reference domain. A point with violet color in $S_M$ will be mapped to the left-most point in the parameter space $\Omega_P$. Similarly, a yellow point will be mapped to a point far at the right in $\Omega_P$.

The middle and bottom rows in \cref{fig:tri_triangulations} show the image of the triangulation in $\Omega_P$ under the harmonic map, for the unit square and the unit circle, respectively. The mapping between the colored boundary points stems from the Dirichlet boundary conditions in the formulation of the harmonic map. In consequence, the boundary points are by construction equally spaced both in the muscle domain $S_M$ and in the parameter domain $\Omega_P$.

It can be seen that the triangulation appears distorted in the parameter domain. The effect is most significant for the square in \cref{fig:triu_1} and \cref{fig:triu_2}. For the latter, the center point of the muscle domain $S_M$ gets mapped far off the center of the squared parameter domain. This effect does not occur for the circle.

The reason for this lies in the triangulation of $S_M$ together with the boundary shape. In the third method, only the value at the center point is a degree of freedom in the computation of $u$ while the values at the boundary points are fixed. By the triangulation, the value of $u$ varies linearly from the boundary towards the center point.
By comparing the colored boundary points in the muscle slice in the top row with the square in the second row, it can be seen that the prescribed values for $u$ are 0 at all blue points, 1 at all yellow points and linearly increasing from 0 to 1 at the green and red points, increasing from left to right.
The yellow points of $S_M$ with the prescribed constant value of $u=1$ are approximately located on a vertical line. The first two derivatives of $u$ in vertical coordinate direction are therefore almost zero, in consequence, the Laplace equation forces the derivatives in horizontal direction to also be approximately zero. Therefore, the solution value at the center point is close to $1$. This leads to the mapped center point being close to the right boundary in the square parameter domain. The same happens for the vertical coordinate $v$ of the harmonic map.

For the circle, neighboring boundary points are not located on horizontal or vertical lines and, thus, the Dirichlet boundary conditions for $u$ and $v$ vary along the boundary. Therefore, a better mapping is obtained. The shape of the muscular slice is more similar to the circle than to the square.

Another result that can be seen in \cref{fig:tri_triangulations} is the effect of the failure of the first method to handle concave slices on the harmonic map. As the top left image shows, the first triangulation method produces triangles outside the domain. The triangles are located around the red boundary points. In the square parameter domain, these triangles are degenerated and lie on the bottom boundary. In the circle parameter domain where the respective triangles can be seen at the bottom, they even intersect other triangles. This yields an invalid triangulation.

The reasons for degenerate triangles in the square parameter domains are not solely the concave muscle slices. Also, the straight sides of the unit square lead to degenerate triangles whenever three boundary points of the same side form a triangle.
In the example in \cref{fig:tri_triangulations}, this occurs for the square in column (\subref{fig:triu_0}). As can be seen in the triangulated slice in the top row, there are three triangles that are entirely made up of blue boundary points. These triangles get mapped onto the left side of the square parameter domain where they have a vanishing surface area.

The same effect also occurs with the second triangulation method in column (\subref{fig:triu_1}) of \cref{fig:tri_triangulations} where a triangle at the bottom right comprises three red boundary points and, therefore, gets mapped to the bottom side of the square. Because of the guaranteed minimum angle in the second triangulation method, this circumstance occurs less often and only for muscle cross-sections where the boundary makes sharp turns, such as the muscle slice in this example. The third triangulation method is guaranteed to avoid this problem as all triangles are connected with the center point.

% ------------------
% parameter space triangulations

In conclusion, the second triangulation method with the unit circle and the third triangulation method with both unit square and unit circle show good behavior for use in our meshing algorithm. Next, their interplay with the quadrangulation of the parameter space needs to be investigated.

% bis hierher Änderungen eingearbeitet

\begin{figure}%
  \centering%
  \hspace{0.24\textwidth}
  \begin{subfigure}[t]{0.24\textwidth}%
    \centering%
    \includegraphics[width=\textwidth]{images/fiber_creation/u_0.png}%
    \caption{First method}%
    \label{fig:tu_0}%
  \end{subfigure}
  \begin{subfigure}[t]{0.24\textwidth}%
    \centering%
    \includegraphics[width=\textwidth]{images/fiber_creation/u_1.png}%
    \caption{Second method}%
    \label{fig:tu_1}%
  \end{subfigure}
  \begin{subfigure}[t]{0.24\textwidth}%
    \centering%
    \includegraphics[width=\textwidth]{images/fiber_creation/u_2.png}%
    \caption{Third method}%
    \label{fig:tu_2}%
  \end{subfigure}\\
  
  % square
  \begin{subfigure}[t]{0.24\textwidth}%
    \centering%
    \includegraphics[width=\textwidth]{images/fiber_creation/quad_1.png}%
    \caption{Square, scheme 1}%
    \label{fig:tquad_1}%
  \end{subfigure}
  \begin{subfigure}[t]{0.24\textwidth}%
    \centering%
    \includegraphics[width=\textwidth, trim=26mm 14mm 6mm 6mm, clip]{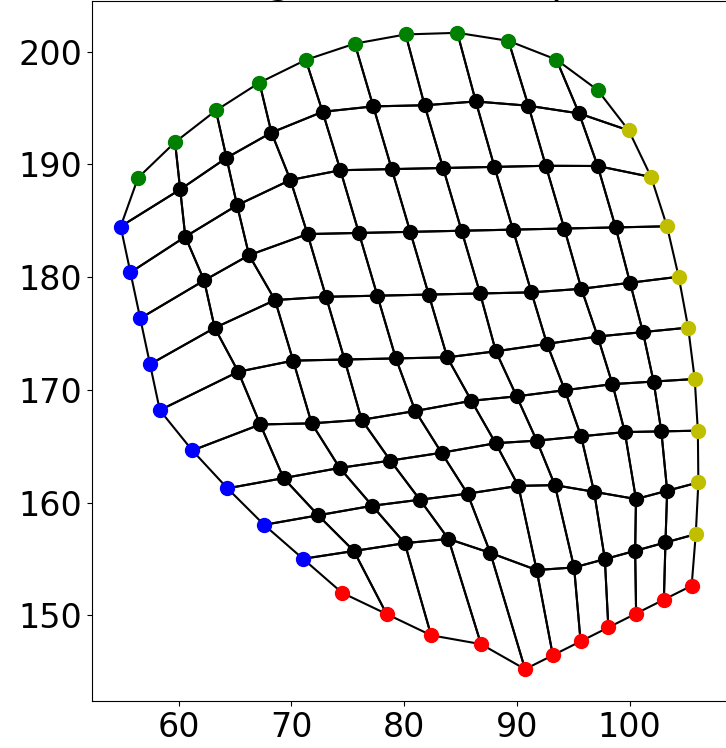}%
    %\caption{}%
    \label{fig:tri_01}%
  \end{subfigure}
  \begin{subfigure}[t]{0.24\textwidth}%
    \centering%
    \includegraphics[width=\textwidth, trim=30mm 14mm 6mm 6mm, clip]{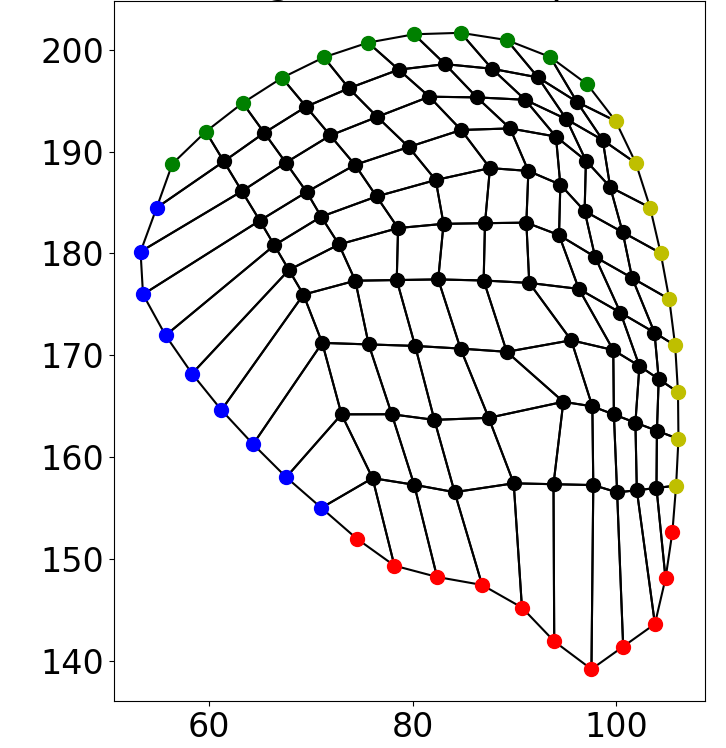}%
    %\caption{}%
    \label{fig:tri_11}%
  \end{subfigure}
  \begin{subfigure}[t]{0.24\textwidth}%
    \centering%
    \includegraphics[width=\textwidth, trim=30mm 14mm 6mm 6mm, clip]{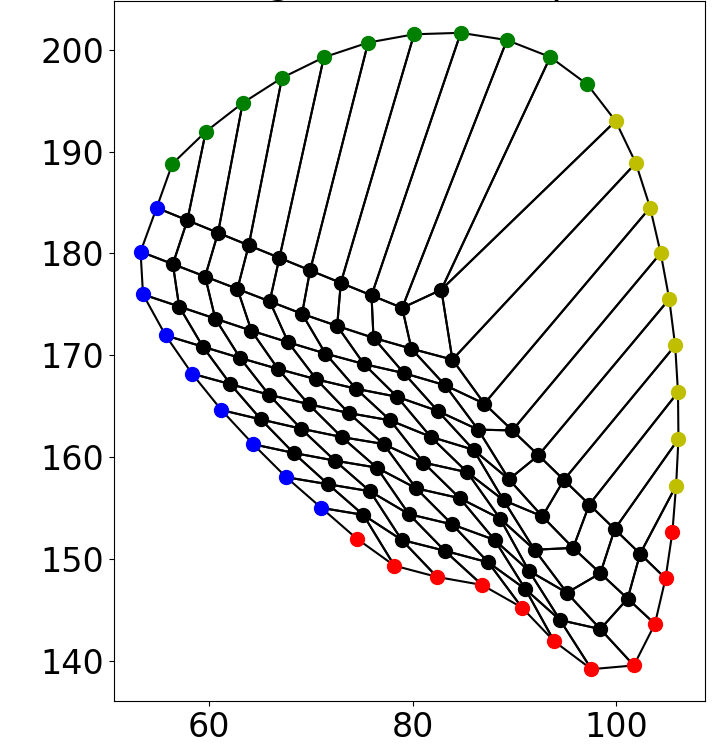}%
    %\caption{}%
    \label{fig:tri_21}%
  \end{subfigure}\\[-4mm]
  
  % square adjusted
  \begin{subfigure}[t]{0.24\textwidth}%
    \centering%
    \includegraphics[width=\textwidth]{images/fiber_creation/quad_2.png}%
    \caption{Square, scheme 2}%
    \label{fig:tquad_2}%
  \end{subfigure}
  \begin{subfigure}[t]{0.24\textwidth}%
    \centering%
    \includegraphics[width=\textwidth, trim=26mm 14mm 6mm 6mm, clip]{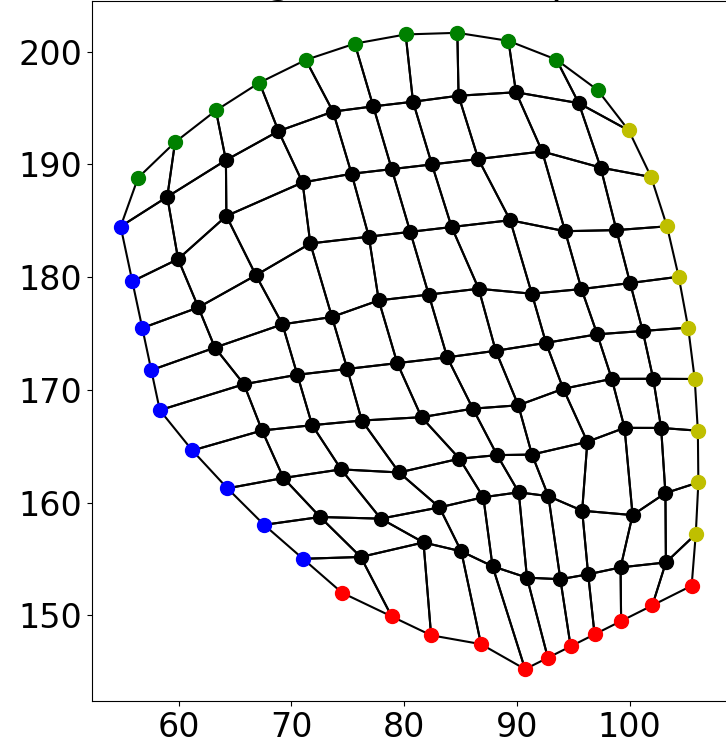}%
    %\caption{}%
    \label{fig:tri_02}%
  \end{subfigure}
  \begin{subfigure}[t]{0.24\textwidth}%
    \centering%
    \includegraphics[width=\textwidth, trim=30mm 14mm 6mm 6mm, clip]{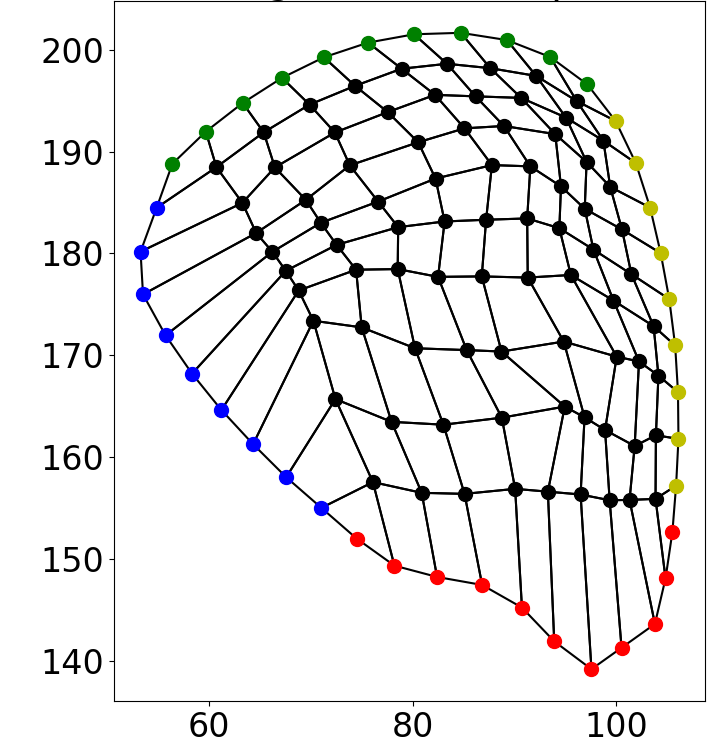}%
    %\caption{}%
    \label{fig:tri_12}%
  \end{subfigure}
  \begin{subfigure}[t]{0.24\textwidth}%
    \centering%
    \includegraphics[width=\textwidth, trim=30mm 14mm 6mm 6mm, clip]{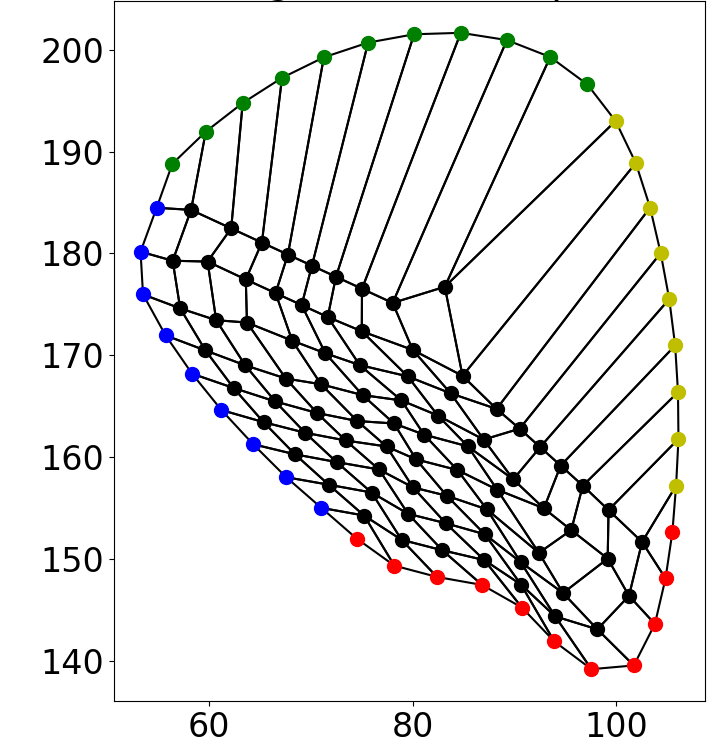}%
    %\caption{}%
    \label{fig:tri_22}%
  \end{subfigure}\\[-4mm]
  
  % circle
  \begin{subfigure}[t]{0.24\textwidth}%
    \centering%
    \includegraphics[width=\textwidth]{images/fiber_creation/quad_0.png}%
    \caption{Circle, scheme 1}%
    \label{fig:tquad_0}%
  \end{subfigure}
  \begin{subfigure}[t]{0.24\textwidth}%
    \centering%
    \includegraphics[width=\textwidth, trim=26mm 14mm 6mm 6mm, clip]{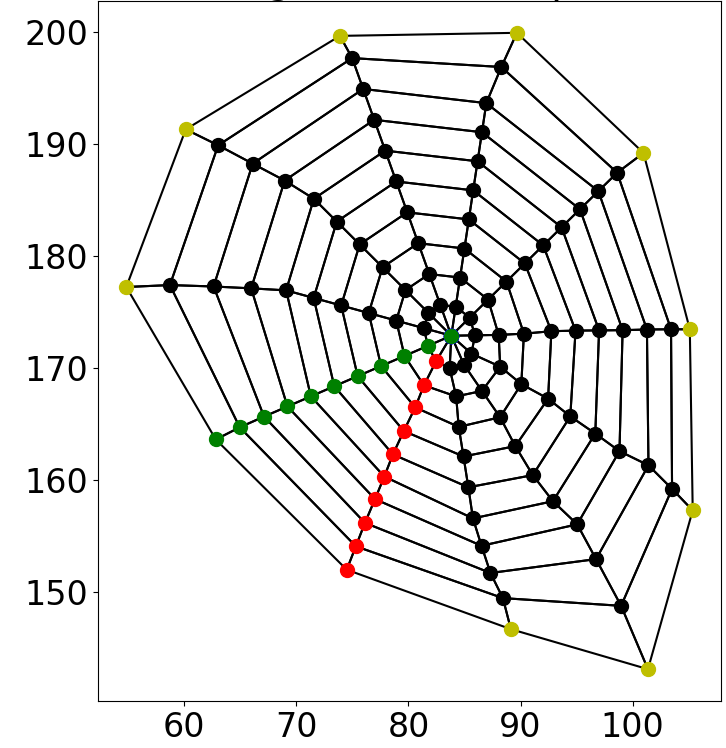}% trim=left bottom right top, clip]
    %\caption{}%
    \label{fig:tri_00}%
  \end{subfigure}
  \begin{subfigure}[t]{0.24\textwidth}%
    \centering%
    \includegraphics[width=\textwidth, trim=26mm 14mm 6mm 6mm, clip]{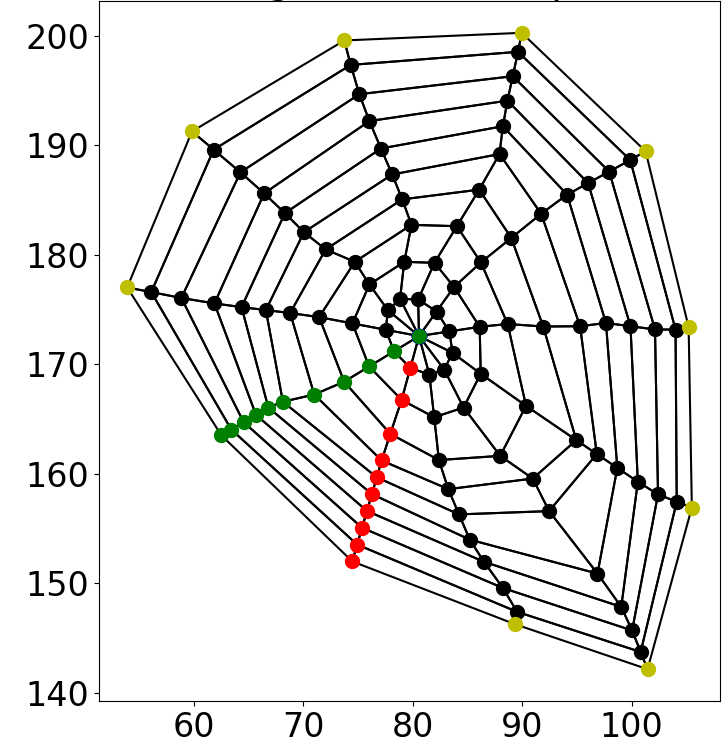}%
    %\caption{}%
    \label{fig:tri_10}%
  \end{subfigure}
  \begin{subfigure}[t]{0.24\textwidth}%
    \centering%
    \includegraphics[width=\textwidth, trim=30mm 14mm 6mm 6mm, clip]{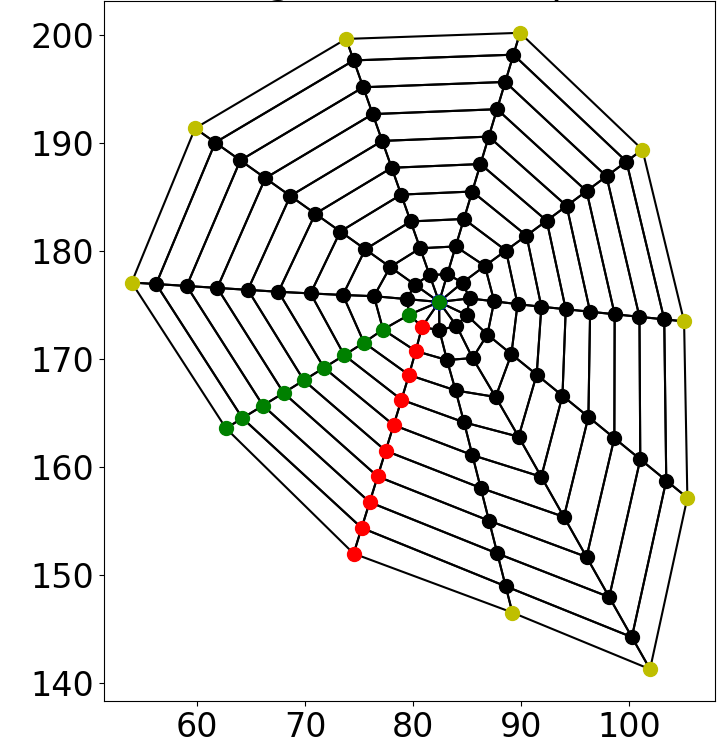}%
    %\caption{}%
    \label{fig:tri_20}%
  \end{subfigure}\\[-4mm]
  
  % circle adjusted
  \begin{subfigure}[t]{0.24\textwidth}%
    \centering%
    \includegraphics[width=\textwidth]{images/fiber_creation/quad_3.png}%
    \caption{Circle, scheme 2}%
    \label{fig:lquad_3}%
  \end{subfigure}
  \begin{subfigure}[t]{0.24\textwidth}%
    \centering%
    \includegraphics[width=\textwidth, trim=30mm 14mm 6mm 6mm, clip]{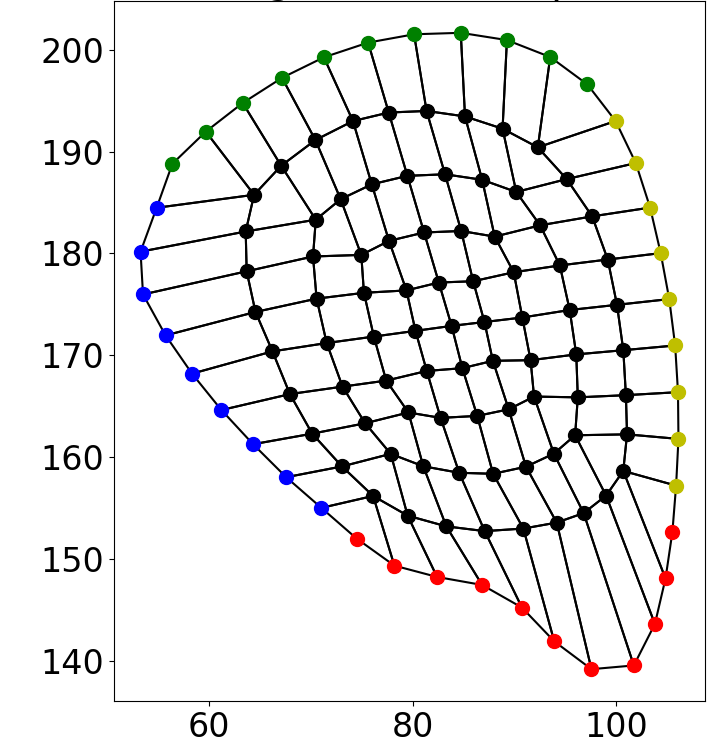}%
    %\caption{}%
    \label{fig:tri_03}%
  \end{subfigure}
  \begin{subfigure}[t]{0.24\textwidth}%
    \centering%
    \includegraphics[width=\textwidth, trim=30mm 14mm 6mm 6mm, clip]{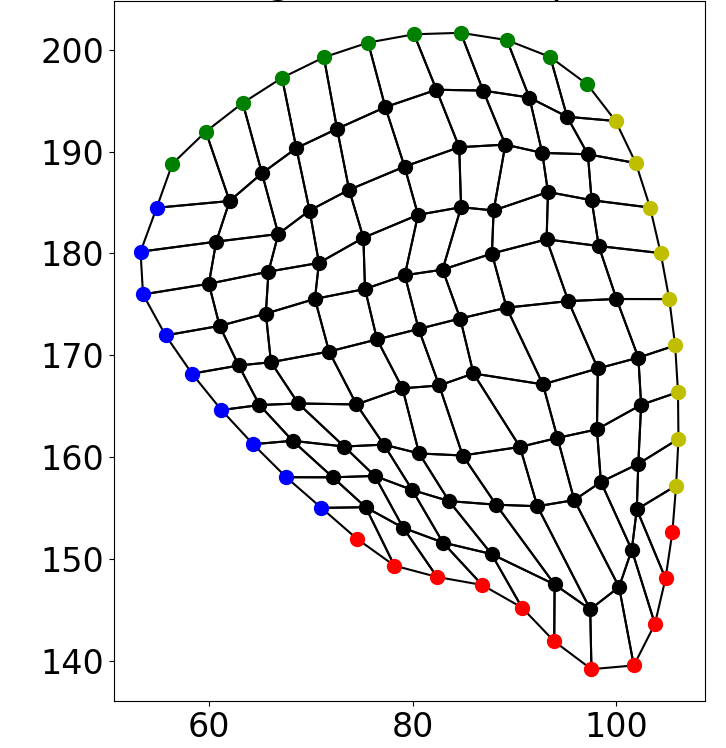}%
    %\caption{}%
    \label{fig:tri_13}%
  \end{subfigure}
  \begin{subfigure}[t]{0.24\textwidth}%
    \centering%
    \includegraphics[width=\textwidth, trim=30mm 14mm 6mm 6mm, clip]{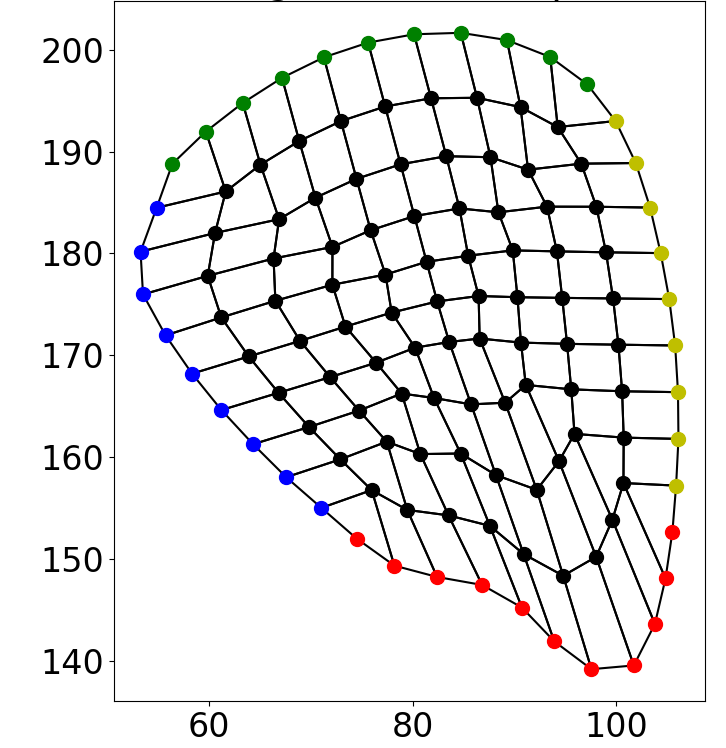}%
    %\caption{}%
    \label{fig:tri_23}%
  \end{subfigure}
  \caption{Initial triangulations, harmonic maps and final quadrangulations of muscle slices for 3D mesh generation: Meshes in the muscle slice $S_M$ for quadrangulations (rows) and triangulations (columns).}%
  \label{fig:tri_meshes}%
\end{figure}

In the next step, the algorithm creates a quadrilateral mesh in the parameter domain and computes its image in the muscle domain using the inverse of the harmonic map. The results are shown in \cref{fig:tri_meshes} for the three different initial triangulation methods (columns) and the four different schemes to create the quadrilateral mesh (rows).

In the images in column (\subref{fig:tu_2}) and rows (\subref{fig:tquad_1}) and (\subref{fig:tquad_2}), it can be seen that the previously observed effect of a bad mapping for squares and the third triangulation method also leads to a mesh in $S_M$ of poor quality.
The result for the two square schemes in column (\subref{fig:tu_1}) is better but still not satisfactory. Good results with the square reference domain are only observed for the first triangulation method in this example.

It can be seen that the approximation quality of the boundary of the domain varies. Most of all, the combination of the first triangulation method (column (\subref{fig:tu_0})) and the square parameter domain (rows (\subref{fig:tquad_1}) and (\subref{fig:tquad_2})) reproduces the shape of the slice poorly. 
The mismatch occurs at the blue and red boundary points. Additionally, the second triangulation method (column (\subref{fig:tu_1})) for the squares fails to correctly represent the round boundary at the bottom of the domain.
The degenerate triangles in the parameter domain are the cause for this effect. The harmonic map $\bfy: S_M \to \Omega_P$ is not injective and, therefore, its inverse does not exist. In our implementation, the points on the degenerate triangles in $\Omega_P$ are mapped to an arbitrarily selected location inside the corresponding triangles in $S_M$. Thus, the mapping is correctly inverted at locations of valid triangles and only creates different boundary points in the invalid areas.

Furthermore, it can be seen that an inaccurate representation of the boundary also occurs with the parameter mesh in the unit circle generated by scheme 1. In this case, the reason is the low number of elements at the boundary in the parameter domain quadrangulation.

The two schemes for the circle parameter domain in rows (\subref{fig:tquad_0}) and (\subref{fig:lquad_3}) both generate reasonable meshes for all triangulation methods, despite the different structure of the generated meshes. The best results for both schemes have been obtained with the third triangulation method.

% ------------------
% resulting meshes

% quantitative quality plots of results

%The mesh quality could easily be improved by a smoothing step, e.g., by applying Laplacian smoothing \cite{field1988laplacianSmoothingAndDelaunayTriangulations}. This technique iteratively improves the local mesh quality. Because of the local scheme, the quality of the final result depends on the quality of the initial mesh. 
%Therefore, the following study forgoes additional smoothing steps and evaluates the direct outcome of the meshing algorithm.

Next, a quantitative comparison of the resulting mesh quality for different parameters of the presented algorithm is carried out. 
The algorithm was executed for all variants with 43 slices of the biceps muscle, resulting in 43 meshes for every combination of triangulation method and quadrangulation scheme. To assess the quality of the generated meshes, the edge lengths of the elements were collected and normalized to have a mean of 1 in each mesh. 
The normalization was done to allow for a comparison between meshes with different bounding box sizes.
The standard deviation of the normalized lengths was determined in each mesh. The total mean of all standard deviations was computed. This value is a measure for the quality of the mesh. A low value means that, in every slice, the generated mesh has similar edge lengths and, in consequence, the overall mesh has good quality.

\begin{figure}%
  \centering%
  \includegraphics[width=\textwidth]{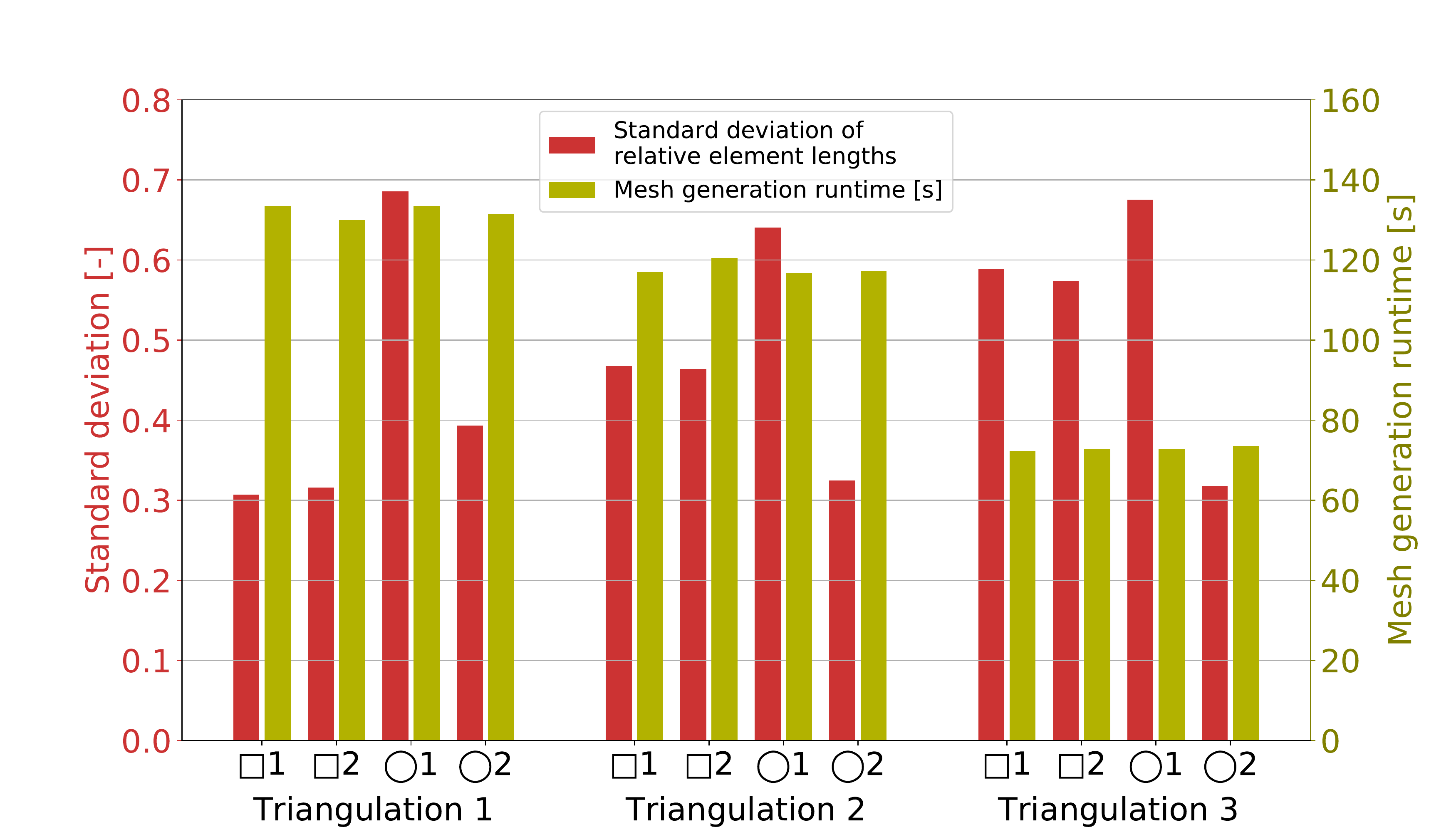}%
  \caption{3D mesh generation quality assessment: Mesh quality (red, lower is better) and generation runtime (yellow) for the three different triangulation methods and the different parameter space quadrangulation schemes. $\square 1$ and $\square 2$ designate the two quadrangulation schemes on the unit square parameter domain, $\ocircle 1$ and $\ocircle 2$ are the schemes on the unit circle parameter domain, as introduced in \cref{fig:quads}. A low value for the standard deviation of relative element lengths indicates good quality.}%
  \label{fig:mesh_quality}%
\end{figure}

\Cref{fig:mesh_quality} visualizes the results. Three groups of bars are displayed for the three triangulation methods. For every type of mesh in the parameter space, i.e., unit square ($\square$) or unit circle ($\ocircle$) and scheme 1 or 2, the standard deviation is given by the red bar and the generation runtime of the overall algorithm is given by the yellow bar. The corresponding axis labels for standard deviation and duration are given on the left and right of the diagram.

The diagram shows the lowest standard deviation of edges and therefore the best mesh quality for the first triangulation method and the square ($\square 1$ and $\square 2$), with scheme 1 having a slightly better value than scheme 2. This shows that the modified placement of the nodes in scheme 2 has no positive effect compared to scheme 1.
However, from the observations in \cref{fig:tri_triangulations}, it is known that the boundaries are not represented correctly.
This behavior does not influence the result because of the chosen metric of uniform relative edge lengths.
Similarly, good results can be seen for scheme 2 in the circular parameter domain and the second and third triangulations.

Moreover, it can be seen that certain connections between parameter domain and suited triangulation scheme exist. The square parameter domain works best with the first triangulation method. The second scheme for the parameter mesh on the unit circle works best with the triangulation methods 2 and 3.

The first scheme for the parameter mesh on the unit circle ($\ocircle 1$) shows bad results for all triangulation methods. This can be explained by looking at the generated meshes in row (\subref{fig:tquad_0}) of \cref{fig:tri_meshes}. By construction, the elements have a bad aspect ratio. This results in the high standard deviation values. However, the generated meshes still look uniform to a certain extent and can be useful in applications where such type of mesh is needed. The score could be improved by adding more nodes in circumferential direction.

The runtime of the algorithm is approximately the same for the different parameter domain meshes. It mainly depends on the triangulation of the slices. The first triangulation using the \emph{SciPy} package takes the most time, followed by the Delaunay refinement. The fastest triangulation is the custom one where only one additional point needs to be placed. In conclusion, when runtime is an issue, the third triangulation should be chosen. It achieves good quality meshes only with the second scheme of the circular parameter domain. This combination also does not suffer from the bad approximation quality of the boundary, as is the case for the unit circle with the first triangulation method.

\begin{figure}
  \centering
  \begin{tabular}{cc}
    \begin{tabular}[b]{c}
      \begin{subfigure}[b]{0.60\textwidth}%
        \centering%
        \includegraphics[width=\textwidth]{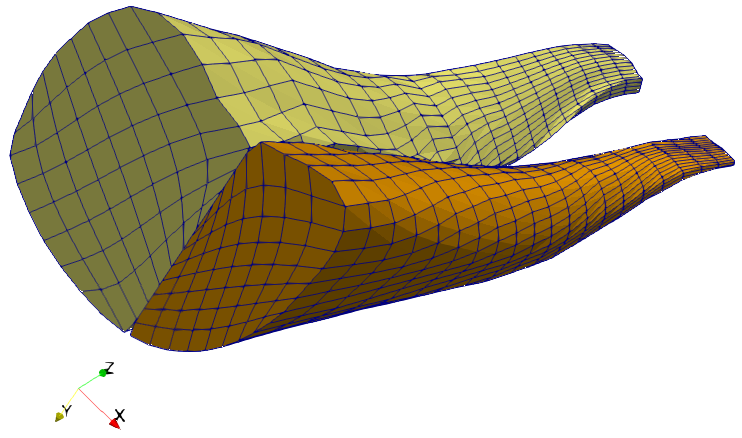}
        \caption{Two meshes for the top tendons with $9 \times 9 \times 21 = \num{1701}$ nodes each.}%
        \label{fig:tendon2}%
      \end{subfigure} \\
      \begin{subfigure}[b]{0.60\textwidth}%
        \centering%
        \includegraphics[width=\textwidth]{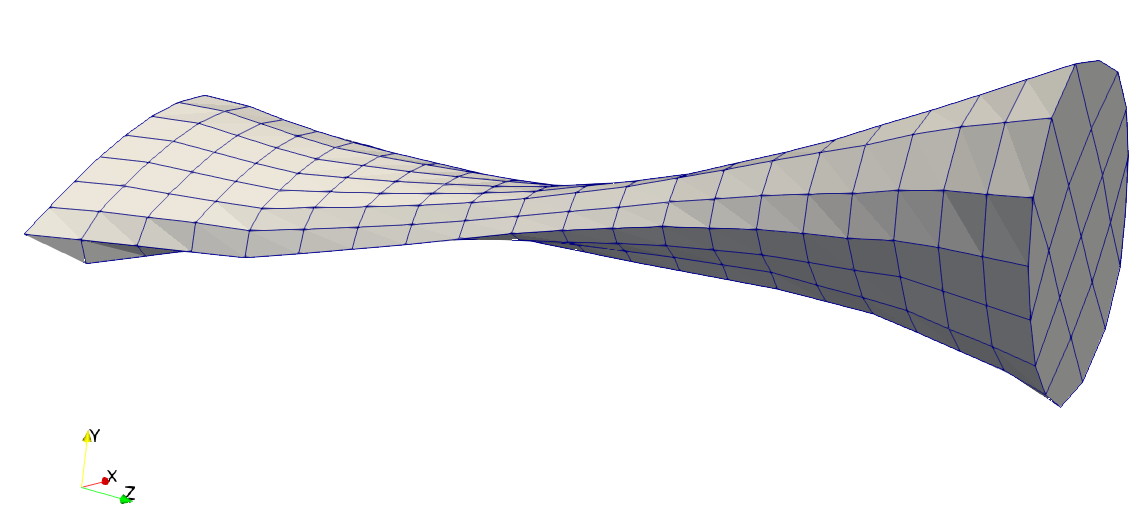}
        \caption{One mesh for the bottom tendon with $5 \times 5 \times 25 = \num{525}$ nodes.}%
        \label{fig:tendon1}%
      \end{subfigure}
    \end{tabular}
    &
    \begin{subfigure}[b]{0.30\textwidth}%
      \centering%
      \includegraphics[width=\textwidth]{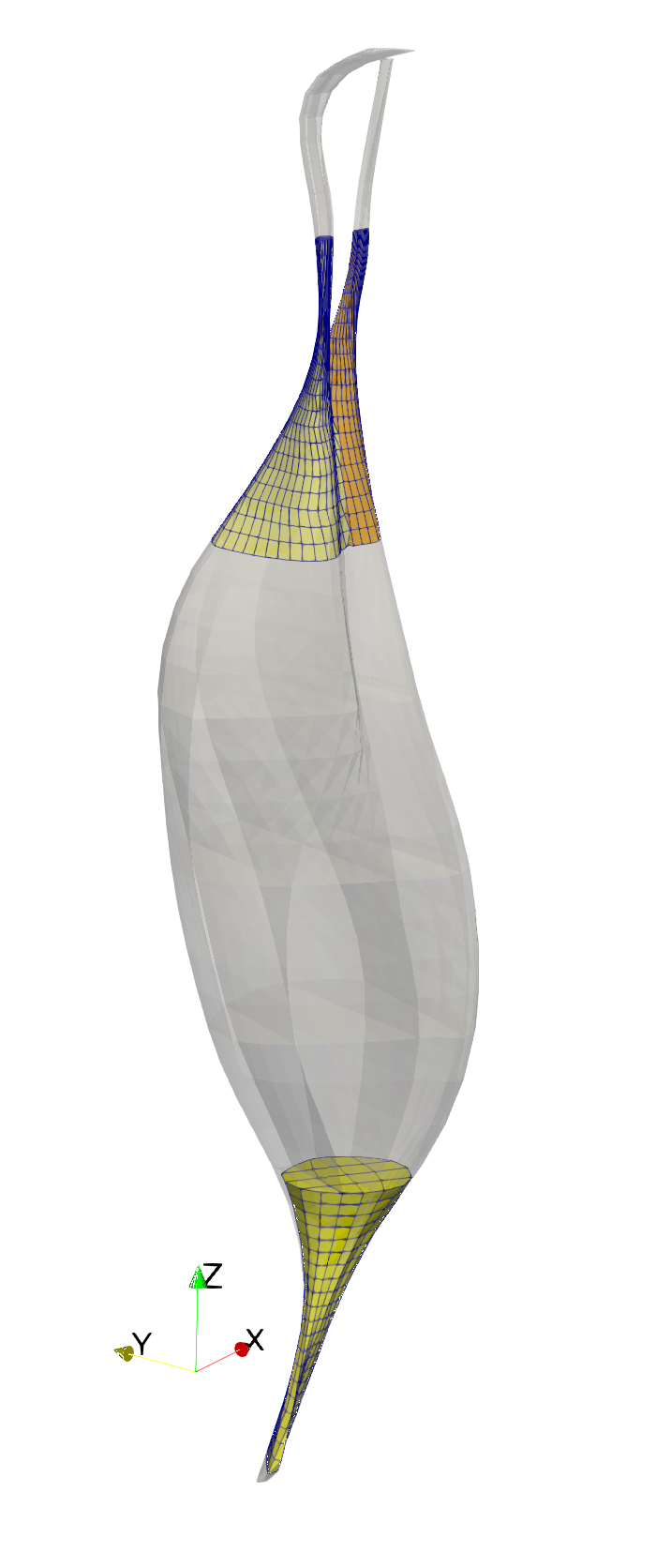}
      \caption{The tendon meshes in the volume of the whole biceps muscle.}%
      \label{fig:muscle_with_tendons}%
    \end{subfigure}
  \end{tabular}
  \caption{3D mesh generation results: Tendon meshes that were created using the serial algorithm for mesh creation.}%
  \label{fig:tendon_meshes}%
\end{figure}

\Cref{fig:tendon_meshes} shows three structured meshes $\Omega_{T,i}$ for the tendons of the biceps brachii muscle that were created using \cref{alg:serial_algorithm_1}. The tendon at the bottom of the muscle is represented by a single mesh. At the top, there are two tendons that extend the two muscle heads of the biceps. Because the meshes need to be structured, two tendon meshes are created at the top. It can be seen that the algorithm creates meshes with similar sized elements despite the difficult, wound geometry of the surfaces.

\begin{reproduce_no_break}
  The described algorithms are part of the \code{fiber_tracing} examples. Execute the following commands to get the results in this chapter:
  \begin{lstlisting}[columns=fullflexible,breaklines=true,postbreak=\mbox{\textcolor{gray}{$\hookrightarrow$}\space}]
    cd $\$$OPENDIHU_HOME/examples/fiber_tracing/streamline_tracer/scripts
    . run_evaluation.sh
  \end{lstlisting}
  Then, the visualizations will be created under \code{../processed_meshes}. Create \cref{fig:mesh_quality} with \code{plot_mesh_quality.py}.
  How to create the tendon meshes is explained at the end of \cref{sec:repro_tendon_meshes}.
\end{reproduce_no_break}

\setcounter{section}{4}
\section{Parallel Algorithm to Create Muscle and Fiber Meshes}\label{sec:parallel_algorithm}

The previously presented algorithm to create 3D and 1D meshes is not parallelized. 
%This restricts the resources that can be employed during the execution of the algorithm to those accessible by one hardware core.
Thus, the size of the handled meshes is limited by the available memory of the computer.
An algorithm that can be used with distributed memory parallelization could, in contrast, benefit from more total memory that is accessible at different compute nodes. Furthermore, the tracing of the streamlines could be performed in parallel which has the potential to reduce runtimes.

In the following, we present an extended algorithm based on the one presented in \cref{sec:ser_alg_meshes} that can be run in parallel on multiple cores. The extended algorithm employs a partitioning of the 3D volume. Every process only stores data corresponding to its own partition. This allows to run the algorithm on a distributed memory system, where data transfer between the processes occurs by sending messages using the Message Passing Interface (MPI). It is possible to create meshes with larger sizes than could fit into a single nodes' memory. This enables us to run the algorithm for meshes with very high resolution that can be used for simulations in the field of High Performance Computing. 
These meshes are partitioned into subdomains for every compute core and can be read from and written to disk concurrently.

\subsection{Overview of the Parallel Algorithm to Create Muscle and Fiber Meshes}

The steps of the algorithm and its input and output are given in \cref{alg:parallel_algorithm_1}. Input and output are the same as for the \cref{alg:serial_algorithm_1} presented in \cref{sec:ser_alg_meshes}. The input is a triangulated tubular surface of the muscle that can be obtained as described in \cref{sec:preprocessing_of_the_muscle_geometry}. A second input, the variable called \emph{boundary\_points}, is used only during recursive calls and is not set at the beginning. The output consists of the 3D mesh of the muscle volume $\Omega_M$ and embedded 1D fiber meshes $\Omega_{F,i}$. 

During execution of the algorithm, the 3D mesh of the muscle is recreated iteratively with increasing resolution and increasing number of subdomains. The algorithm is formulated recursively. 
At the finest resolution when the recursion terminates, the fiber meshes are finally generated together in all subdomains.

At first, a single process executes all the steps of \cref{alg:parallel_algorithm_1} from lines \ref{line:3.2} to \ref{line:3.11a}. This corresponds to recursion level $\ell=0$. Then, in line \ref{line:3.12}, the procedure is called again and in the first recursion executed by eight processes with eight subdomains. On the $\ell$th recursion level, the number of involved processes and subdomains is $8^\ell$. After a specified maximum recursion depth $\ell_\text{max}$ is reached, all involved processes execute the first branch of the \code{if} statement in line \ref{line:3.8} and generate the final 3D and 1D output meshes in line \ref{line:3.9}.

The steps in \cref{alg:parallel_algorithm_1} are executed concurrently by the involved processes at the respective levels.
Some of the steps only operate on the locally stored data and, thus, are independent of other processes. Other steps involve communication between processes. Whether an instruction effects only the own domain of the process or involves global communication is denoted in parentheses at the beginning of the lines in \cref{alg:parallel_algorithm_1}.

\begin{algorithm}
  \begin{algorithmic}[1]%
    \Procedure{Create\_3D\_meshes\_parallel}{}
    \Require Triangulated tubular surface
    \Require boundary\_points: $4\times 4$ points per slice
    \Ensure Structured 3D volume mesh
    \Ensure 1D fiber meshes
    \Statex
    \State (own domain) Create\_3D\_mesh(boundary\_points)        \label{line:3.2}
    \State (own domain) Fix and smooth 2D meshes  \label{line:3.3}   
    \State (global)$\hskip2.4em$     Solve Laplace problem       \label{line:3.4}
    \State (global)$\hskip2.4em$      Communicate ghost elements to neighboring subdomains      \label{line:3.5}
    \State (own domain) Trace streamlines for new subdomain boundaries      \label{line:3.6}
    \State (global)$\hskip2.4em$      new\_boundary\_points $\leftarrow$ Construct new subdomains      \label{line:3.7}
    \Statex
    \If{recursion ends}         \label{line:3.8}
    \State (own domain) Trace streamlines for fiber meshes      \label{line:3.9}
    \Else      \label{line:3.10}
    \State (global)$\hskip2.4em$      communicate boundary points      \label{line:3.11a}
    \State (global)$\hskip2.4em$      Create\_3D\_meshes\_parallel(new\_boundary\_points)      \label{line:3.11}
    \EndIf      \label{line:3.12}
    \EndProcedure
  \end{algorithmic}%
  \caption{Parallel algorithm to create muscle and fiber meshes}%
  \label{alg:parallel_algorithm_1}%
\end{algorithm}%

\subsection{Overview of the Subdomain Refinement}

The goal during the recursive calls is to determine smooth boundaries for the new subdomains. Each process splits its own subdomain into eight subdomains and then proceeds to the next recursion level.
The subdomain boundaries are determined by tracing streamlines in a divergence-free vector field through the entire muscle volume, similar to the approach in \cref{alg:serial_algorithm_2}. The divergence-free vector field is computed from the solution of a Laplace problem, which is solved in parallel on the entire mesh of the muscle in every recursion. The mesh width of this mesh gets halved in every recursion, subsequently leading to an increasingly fine mesh.

The subdomain boundaries are always aligned to streamlines in the mesh that was created last.
On each recursion level, the existing subdomain boundaries and the new boundaries for the subdomains on the next recursion level are all created anew and, thus, change slightly as the mesh refines.

As the subdomain boundaries in the interior of the volume refine, so do the outer boundaries given by the surface of the muscle. The given triangulation of the surface is sampled again on each recursion level yielding increasingly fine representations.

At the final recursion level $l_\text{max}$, the muscle is partitioned into $8^{l_\text{max}}$ subdomains and a respective fine 3D mesh in the muscle volume exists. Then, the algorithm traces the specified amount of streamlines through the whole mesh to produce 1D meshes for the muscle fibers. By construction of the subdomains, the streamlines enter and leave the subdomains through their top and bottom bounding planes. This allows parallel execution of the final streamline tracing step.

The reason that the algorithm constructs the partitioning iteratively and not once at the beginning using an initial mesh lies in the requirements for parallel streamline tracing. Each subdomain should be able to trace streamlines in longitudinal ($z$) direction of the muscle without communication to their neighbors in $x$ and $y$ directions. To ensure this property, the partitioning involves a small overlap of neighboring subdomains, i.e., a ghost layer. This ghost layer can consist of a lower number of elements if the mesh is iteratively refined than if the partitioning was created directly on a coarser mesh.

% --
\subsection{Data Structure of Boundary Points}\label{sec:data_structure_of_boundary_points}

In the following, \cref{alg:parallel_algorithm_1} is illustrated in more detail.
The execution starts with one process and the only input is the tubular muscle surface. It is given either as triangulation or in parametric form as NURBS surface.
The first step is to construct a quadrilateral mesh of this surface. This is done using the procedure explained in \cref{sec:slicing_of_the_geometry}, which creates horizontal slices of the muscle and places equidistant points on the \say{rings} of the boundaries of these slices. As explained earlier, the points are arranged such that the resulting quadrangulation of the surface has good quality. A result for the biceps muscle is visualized in \cref{fig:serial_alg_0}.

Initially, the parallel algorithm stores the points on these rings in the variable \code{boundary\_}\\\code{points}. If the procedure in \cref{alg:parallel_algorithm_1} is called recursively, the contents of this variable is passed as an argument from the previous recursion.
The set of points in \code{boundary\_points} defines the boundaries of the subdomain of the process where it is stored.

\begin{figure}%
  \centering%
  \begin{subfigure}[t]{0.48\textwidth}%
    \centering%
    \includegraphics[width=3cm]{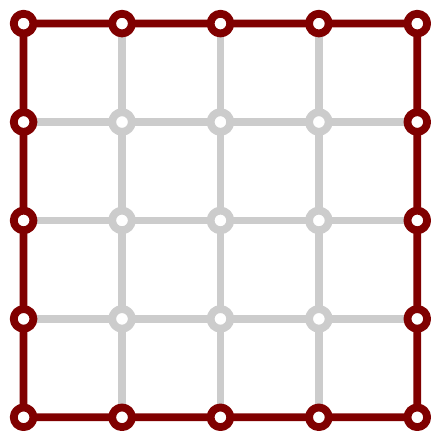}%
   \caption{Grid of $4 \times 4$ boundary points, which occurs at the beginning of the procedure of \cref{alg:parallel_algorithm_1}.}%
    \label{fig:boundary_grid_1}%
  \end{subfigure}
  \quad
  \begin{subfigure}[t]{0.48\textwidth}%
    \centering%
    \includegraphics[width=3cm]{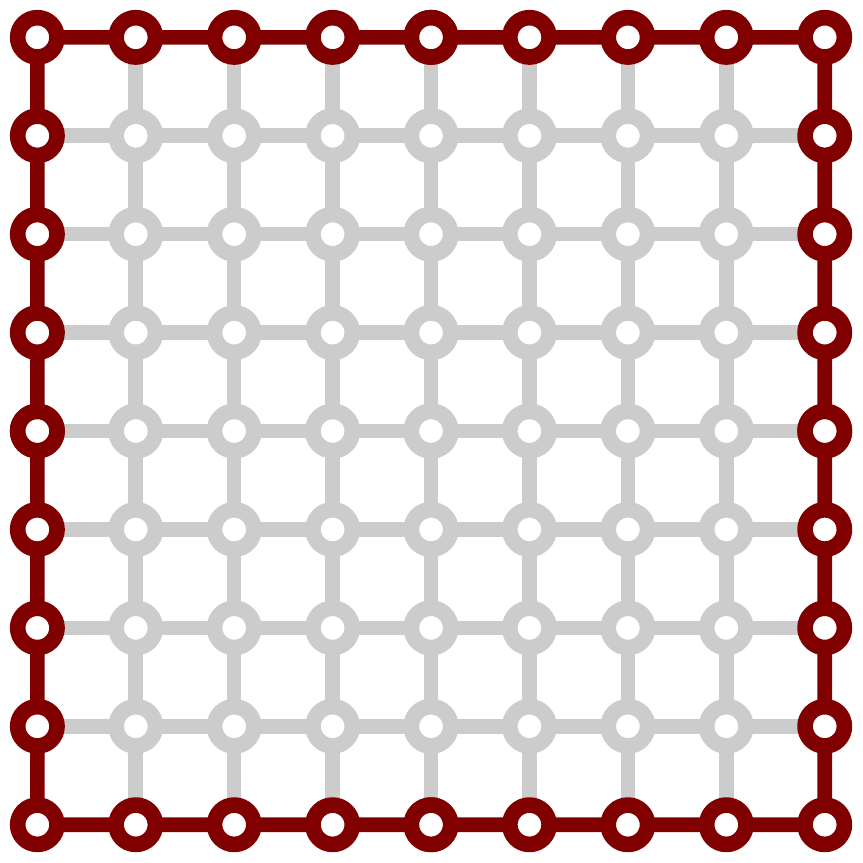}%
    \caption{Grid with $4 \times 8$ boundary points, which occurs after a refinement step at beginning of the procedure of \cref{alg:parallel_algorithm_1}.}%
    \label{fig:boundary_grid_2}%
  \end{subfigure}   
  \caption{Logical subdomain boundaries (red) and interior grid (gray) before and after the refinement at the beginning of \cref{alg:parallel_algorithm_1}.}%
  \label{fig:boundary_grid}%
\end{figure}

The points on each ring in the $x$-$y$-plane are organized such that they enclose a grid of $n_\text{el,x} \times n_\text{el,x}$ elements, where the number $n_\text{el,x}$ of elements per coordinate direction can be specified as parameter. In the following, an example with $n_\text{el,x}=4$ is considered. The grid is shown in \cref{fig:boundary_grid_1}, the $4\,n_\text{el,x}$ boundary points on the ring are visualized by red color. Note that \cref{fig:boundary_grid_1} depicts the ring as a square whereas in reality it has the potentially more irregular shape of the subdomain.

A number $(n_\text{el,z}+1)$ of these rings are stacked in $z$ direction to approximate the enclosing surface of the subdomain, where the number $n_\text{el,z}$ of elements in $z$ direction is again given by a parameter. Thus, the variable \code{boundary\_points} contains a total of $4\,n_\text{el,x}\,(n_\text{el,z}+1)$ points. Typical parameter values are $n_\text{el,x}=4$ and $n_\text{el,z}=50$. 

% refine border points
As the goal on every recursion level is to construct a mesh with half the mesh width of the mesh on the previous level, the given boundary points are refined to twice the amount by inserting new points at the centers between neighboring points. The refinement happens in all three coordinate directions. For the example with $n_\text{el,x}=4$, the resulting grid with the $4\times 8$ refined boundary points is shown in \cref{fig:boundary_grid_2}. In $z$ direction, we get $(2\,n_\text{el,z}+1)$ slices with points.

\begin{figure}%
  \centering%
  \includegraphics[height=8cm]{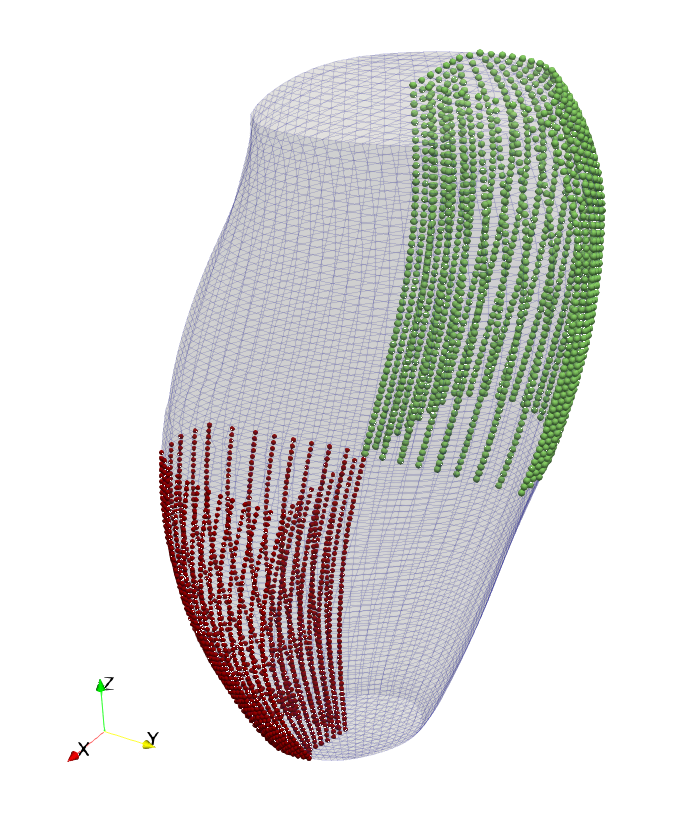}%
  \caption{Parallel 3D mesh generation: Partitioning of the muscle volume into eight subdomains during the first call to the procedure in \cref{alg:parallel_algorithm_1}. The first (red) and the eighth subdomain (green) are shown.}%
  \label{fig:subdomain}%
\end{figure}

% logical structure of new subdomains
The task in the recursive procedure is now to determine boundaries for eight subdomains. This is achieved by subdividing the given 2D slices into four 2D subdomains each. Additionally, the 3D volume is split at its vertical center. Thus, the upper and lower parts contain four subdomains each. 
\Cref{fig:subdomain} visualizes this scheme for the eight subdomains on recursion level $l=1$. The boundary points of the first and the eighth subdomain are shown. The boundary points have already been refined such that every slice in \cref{fig:subdomain} consists of $4 \times 8$ points and corresponds to the grid in \cref{fig:boundary_grid_2}

%   ^v^v^

%For this reason, the given $4 \times 4$ boundary points are refined to twice the amount of boundary points by inserting new points at the centers between neighboring points. The resulting grid is shown in \cref{fig:boundary_grid_2}. Now, it would be possible to subdivide the grid to obtain four instances of the needed grid in \cref{fig:boundary_grid_1}.
%However, this would result in constant straight connection lines between the initial boundary points. In all further recursive calls, the additional points would all be placed on these lines and thereby not properly refine the subdomain boundaries. Instead, a different approach is desired where the subdomain's boundaries in the volume follow the directions of streamlines and fibers. Thus, the approach is to define the subdomain boundaries in the interior of the global domain by traced streamlines and sample the outer boundaries from the surface triangulation with the desired mesh width. The required steps of this approach are discussed next.

% --

\subsection{Generation and Smoothing of the 3D Mesh}
% create 3D mesh

After the \code{boundary\_points} variables has been set, the next step of \cref{alg:parallel_algorithm_1} is to construct a 3D mesh in the domain.
In line \ref{line:3.2} of \cref{alg:parallel_algorithm_1}, the harmonic map  algorithm \cref{alg:serial_algorithm_1} described in \cref{sec:ser_alg_meshes} is called. Its input consists of the boundary points that define the 2D slices of the volume. This means that \cref{alg:serial_algorithm_1} does not need to construct the slice boundary rings from the surface triangulation, instead, the formulation of \cref{alg:serial_algorithm_1} can directly start with line \ref{alg:1.2} to triangulate the slices and then compute the harmonic map. For the harmonic map computation, the second triangulation method is used with a circular reference domain quadrangulated by the second scheme. The result is a set of quadrangulated 2D slices that forms a 3D mesh by vertically connecting the elements of neighbor slices.

% smoothing
Next, line \ref{line:3.3} of \cref{alg:parallel_algorithm_1} improves the mesh quality of the 2D muscle slices $S_M$ from which the 3D mesh is formed. This action consists of two steps. The first step is to ensure that no self-intersecting or degenerate quadrilaterals exist in the slice. The second step applies Laplacian smoothing to improve the mesh quality of the slice.

Theoretically, the first step should not be necessary, as the chosen quadrangulation algorithm always produces valid elements. However, in practice, small or irregularly shaped, concave domains occur and together with rounding and numerical errors in the Laplace problem computations occasionally lead to invalid meshes with self intersecting elements, especially for higher recursion depths in \cref{alg:parallel_algorithm_1}. Executing the first step therefore increases the robustness of the implementation.

\begin{figure}%
  \centering%
  \def\svgwidth{0.7\textwidth}%
  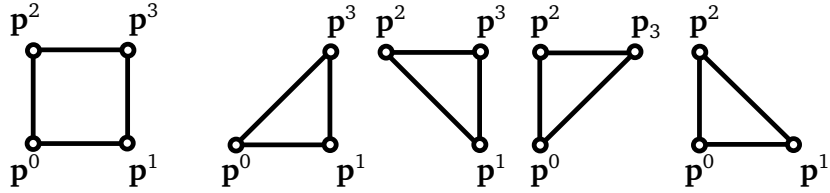%
  \caption{Decomposition of quadrilateral elements into triangles as substep of the validity check of muscle slice quadrangulations. A quadrilateral element (left) and the four triangles (right) that can be constructed from its four points. These triangles are needed for the check in \cref{alg:parallel_algorithm_1} whether the quadrilateral element is valid.}%
  \label{fig:quads_tris}%
\end{figure}%

\begin{figure}%
  \centering%
  \begin{subfigure}[t]{0.48\textwidth}%
    \centering%
    \includegraphics[width=3cm]{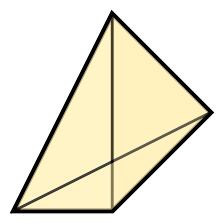}%
    \caption{Convex quadrilateral with score $s=4$ and the contained triangles, which are all oriented counterclockwise.}%
    \label{fig:triangle_score_3}%
  \end{subfigure}
  \quad
  \begin{subfigure}[t]{0.48\textwidth}%
    \centering%
    \includegraphics[width=3cm]{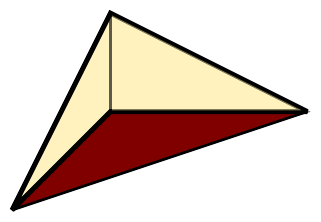}%
    \caption{Concave quadrilateral with score $s=3$ and the contained triangles. Only the red triangle is oriented clockwise.}%
    \label{fig:triangle_score_4}%
  \end{subfigure}   
  \caption{Check for valid elements in the muscle slice quadrangulations that occurs in \cref{alg:parallel_algorithm_1}: Illustration of the score of valid concave and convex quadrilaterals.}%
  \label{fig:triangle_score}%
\end{figure}%

The algorithm performs this step by repeatedly iterating over all interior mesh points in every slice $S_M$ and fixing invalid elements. To find invalid elements, for every quadrilateral the four triangles that can be formed from the points of the quadrilateral are considered, as shown in \cref{fig:quads_tris}.
For every triangle with points $\bfp^0,\bfp^1$ and $\bfp^2$, the orientation of the triangle is determined. The orientation is counterclockwise if the oriented triangle area $A_{012}$ is positive. The oriented triangle area is the determinant of the $3 \times 3$ matrix that contains the row vectors $(p^i_x,p^i_y,1)$ for the triangle points $\bfp^i=(p^i_x,p^i_y)^\top$ and can be computed by the following formula \cite{sedgewick2011algorithms}:
\begin{align*}
  A_{012} = (p^1_x-p^0_x)\,(p^2_y-p^0_y) - (p^2_x-p^0_x)\,(p^1_y-p^0_y).
\end{align*}
If the orientation is counterclockwise, a score value of the triangle is set to one, if it is clockwise, the score is set to zero. The score values of the four triangles are added up to yield a score $s$ for the quadrilateral. Only if this score is $s \geq 3$, the quadrilateral is valid. \Cref{fig:triangle_score} illustrates the cases of valid quadrilateral elements. In a valid, convex element, all four triangles lie inside the element and, thus, the score is $s=4$. If only one triangle is located outside, the quadrilateral is also valid and concave. In this case the score has the value $s=3$.

At the current mesh point in the loop over all points that are not at the boundary of the mesh, the four adjacent quadrilaterals are considered. If any of them is invalid, the algorithm tries to improve the situation by deflecting the point by a random, small vector. A maximum of 200 random deflections from the original position with exponentially increasing deflection vector sizes are tried. After each modification of the point, the scores of the four adjacent quadrilateral elements are evaluated. If the sum of the four element scores increases, the point is kept and the iteration over all interior mesh points starts anew. 

Note that this does not necessarily mean that the invalid element was fixed, only its score was improved. If it was not fixed, it will be considered again in the next iteration. For example, a convex element that initially is oriented clockwise instead of counterclockwise has a score of $s=0$. In the first iteration, one point is deflected such that the quadrilateral intersects itself but has a higher score $s\geq 0$. At least one more iteration is needed until the quadrilateral is oriented correctly.
When all elements in the slice $S_M$ are valid, this step is complete.

\subsection{Laplacian Smoothing}
\begin{figure}%
  \centering%
  \begin{subfigure}[t]{0.48\textwidth}%
    \centering%
    \includegraphics[height=7cm]{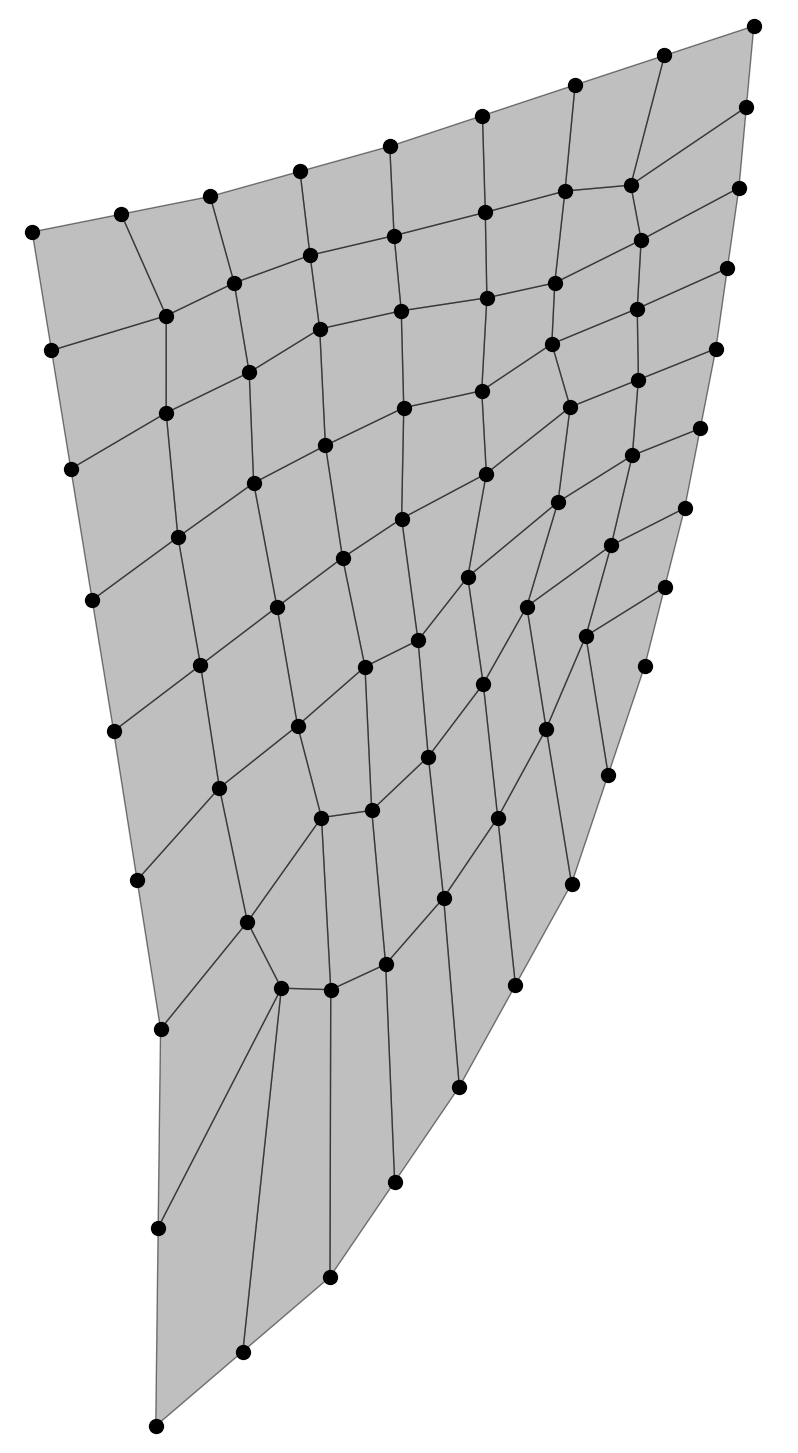}
    \caption{Initial 2D mesh of a subdomain at the boundary of the biceps muscle.}%
    \label{fig:world_mesh}%
  \end{subfigure}
  \quad
  \begin{subfigure}[t]{0.48\textwidth}%
    \centering%
    \includegraphics[height=7cm]{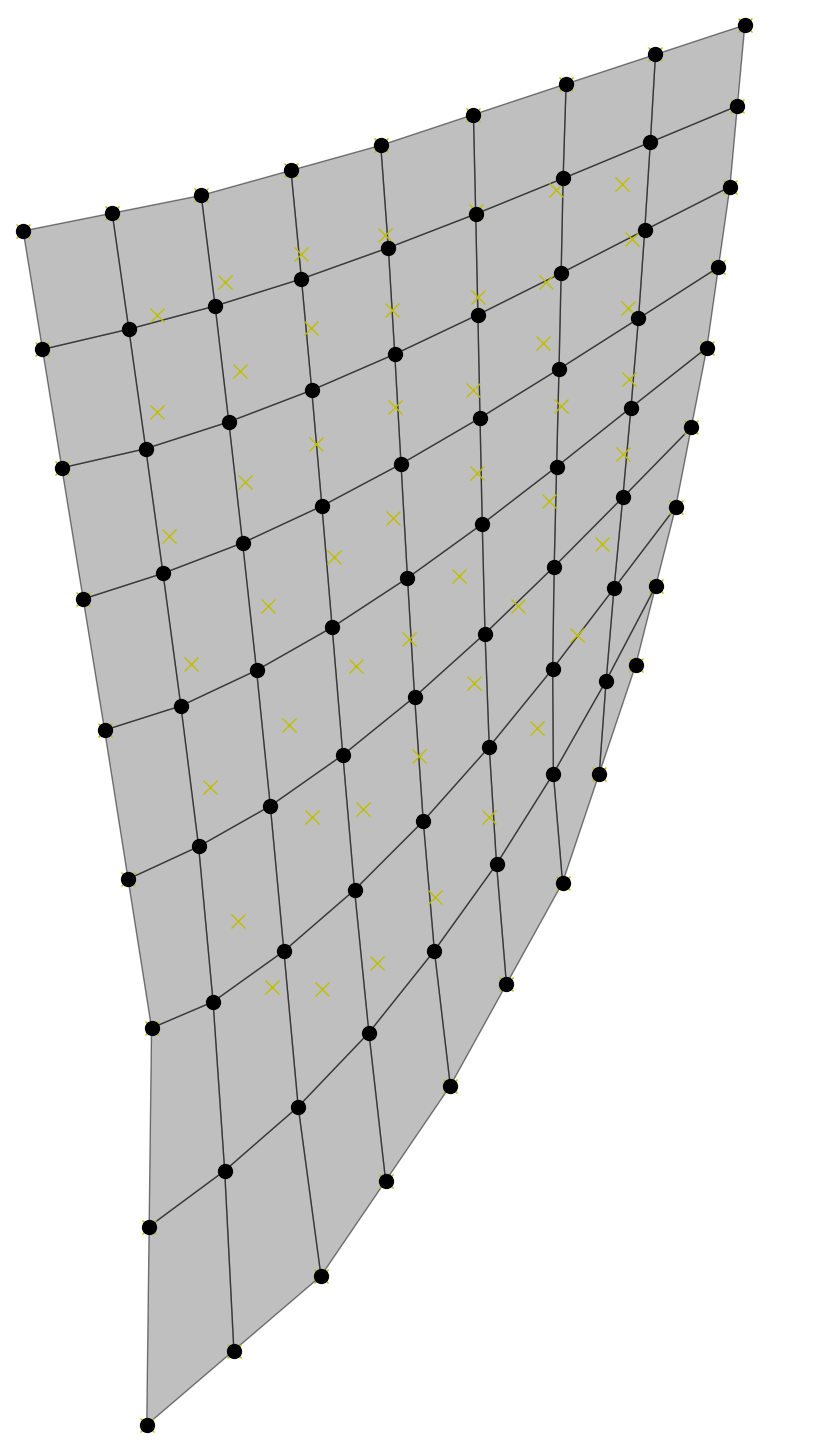}
    \caption{The mesh of (\subref{fig:world_mesh}) after 20 iterations of Laplacian smoothing.}%
    \label{fig:world_mesh_improved}%
  \end{subfigure}    
  \caption{Quality improvement of 2D muscle slice quadrangulation: Effect of Laplacian smoothing of a 2D grid which occurs in line \ref{line:3.3} of \cref{alg:parallel_algorithm_1}.}%
  \label{fig:laplace_smoothing}%
\end{figure}

The second step is the smoothing step that improves the mesh quality of the 2D slices. \num{20} iterations of Laplacian smoothing \cite{field1988laplacianSmoothingAndDelaunayTriangulations} are executed. Laplacian smoothing in our case subsequently visits all interior points of the mesh and sets the location of a point to the center of gravity of its four direct neighbors.
%The order in which the points of the mesh are traversed is changed after every iteration: In the first iteration, the points are traversed starting at the bottom left. In the second iteration, the traversal starts at the top right and moves in opposite direction compared to the first iteration. The third and fourth iterations begin at the bottom right and top left nodes of the mesh. After these four iterations the scheme is repeated. The reason for this change of the traversal is to 
\Cref{fig:laplace_smoothing} shows the effect of Laplacian smoothing for a slice in a subdomain on the first recursion level. It can be seen how the smoothing equalizes the element side lengths and angles.

\begin{figure}%
  \centering%
  \begin{subfigure}[t]{0.30\textwidth}%
    \centering%
    \includegraphics[width=\textwidth]{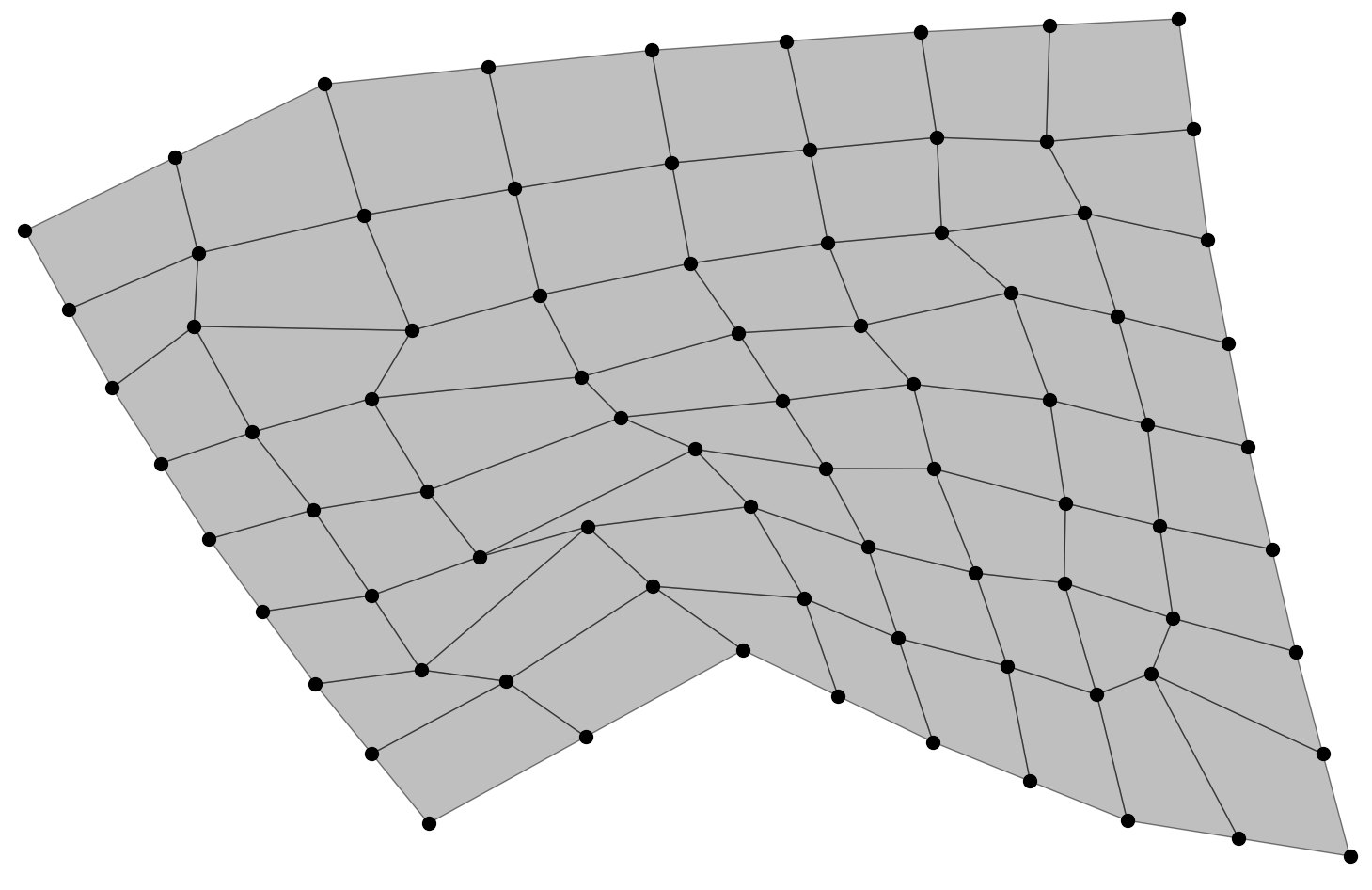}
    \caption{Initial 2D mesh.}%
    \label{fig:world_mesh_0}%
  \end{subfigure}
  \quad
  \begin{subfigure}[t]{0.30\textwidth}%
    \centering%
    \includegraphics[width=\textwidth]{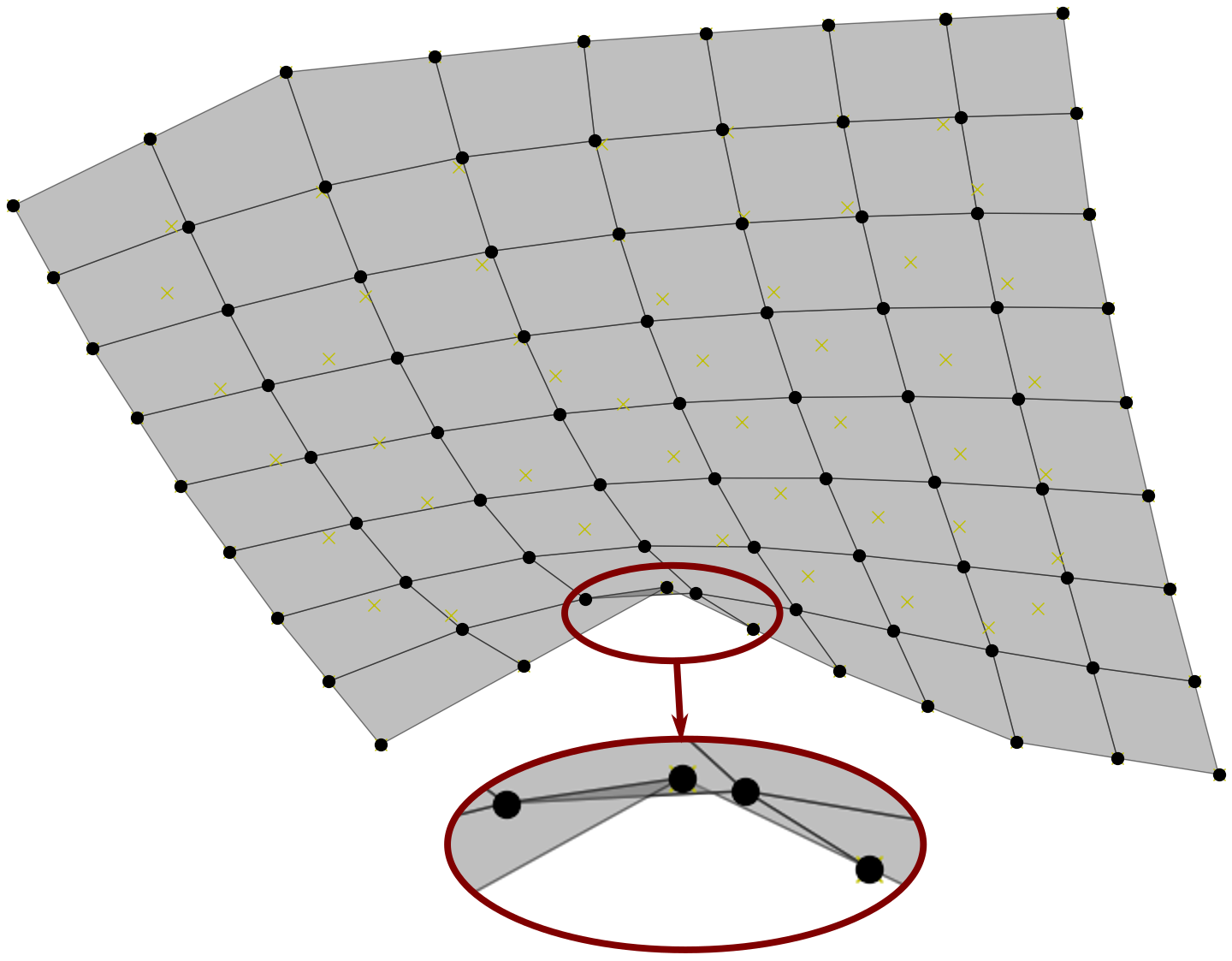}
    \caption{The mesh of (\subref{fig:world_mesh_0}) after 20 iterations of Laplacian smoothing, yielding an invalid quadrangulation.}%
    \label{fig:world_mesh_improved_0}%
  \end{subfigure}    
  \quad
  \begin{subfigure}[t]{0.30\textwidth}%
    \centering%
    \includegraphics[width=\textwidth]{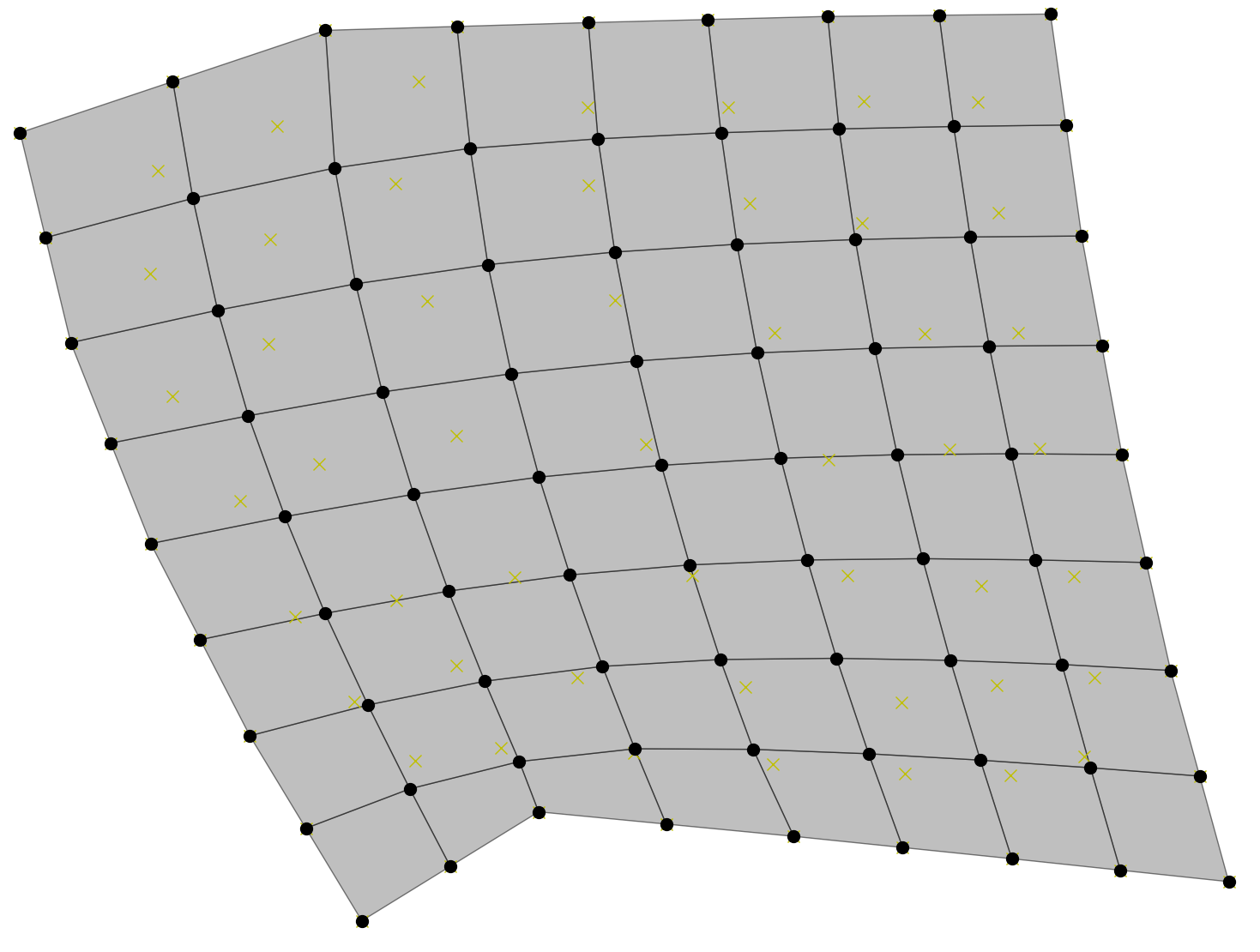}
    \caption{The mesh of (\subref{fig:world_mesh_0}) after 20 iterations of Laplacian smoothing with the validity check, yielding a valid quadrangulation.}%
    \label{fig:world_mesh_improved_1}%
  \end{subfigure}    
  \caption{Quality improvement of 2D muscle slice quadrangulation: Effects of Laplacian smoothing on concave domains.}%
  \label{fig:laplace_smoothing_0}%
\end{figure}

However, this smoothing step can invalidate a mesh by introducing overlapping quadrilaterals. An example for this case is given in \cref{fig:laplace_smoothing_0}. The initial mesh in \cref{fig:world_mesh_0} is concave and occurs during recursion level $l=2$. \Cref{fig:world_mesh_improved_0} shows the result of the smoothing, which contains one invalid element. The smoothing operation placed the fourth point of the element that also contains the three boundary points at the concavity outside the mesh. As a remedy, the smoothing method checks the validity of the adjacent elements before a point is moved. If the move results in an invalid element, the action is not carried out and the traversal continues with the next point instead.
\Cref{fig:world_mesh_improved_1} shows the resulting mesh if this check is enabled. The mesh has slightly different boundaries because the check influenced the behavior already on lower recursion levels.

% smoothing fix

\subsection{Solution of the Laplace Problem}\label{sec:solution_of_the_laplace_problem}

After the 3D mesh has been created and smoothened, the next steps are to solve the Laplace problem in the muscle domain, to trace streamlines in the gradient of the solution vector field and finally to construct the eight subdomains for the recursive calls by subdividing the own domain along the streamlines.

% refine 3D mesh
Prior to the solution of the Laplace problem, the 3D mesh gets refined further by increasing the number of elements per coordinate direction by a specified factor $r\in \N$. The rationale is to increase the number of degrees of freedom and, thus, the resolution to get a smaller numerical error in the subsequent Laplace computation.
This refinement is in addition to the refinement of the initial boundary points by a factor of two described in \cref{sec:data_structure_of_boundary_points}. The mesh with $2\,n_\text{el,x} \times 2\,n_\text{el,x} \times 2\,n_\text{el,z}$ elements gets refined to $2\,r\,n_\text{el,x} \times 2\,r\,n_\text{el,x} \times 2\,r\,n_\text{el,z}$ elements. The new points are found by interpolating in the existing mesh.

\begin{figure}%
  \centering%
  \begin{subfigure}[t]{0.45\textwidth}%
    \centering%
    \includegraphics[height=9cm]{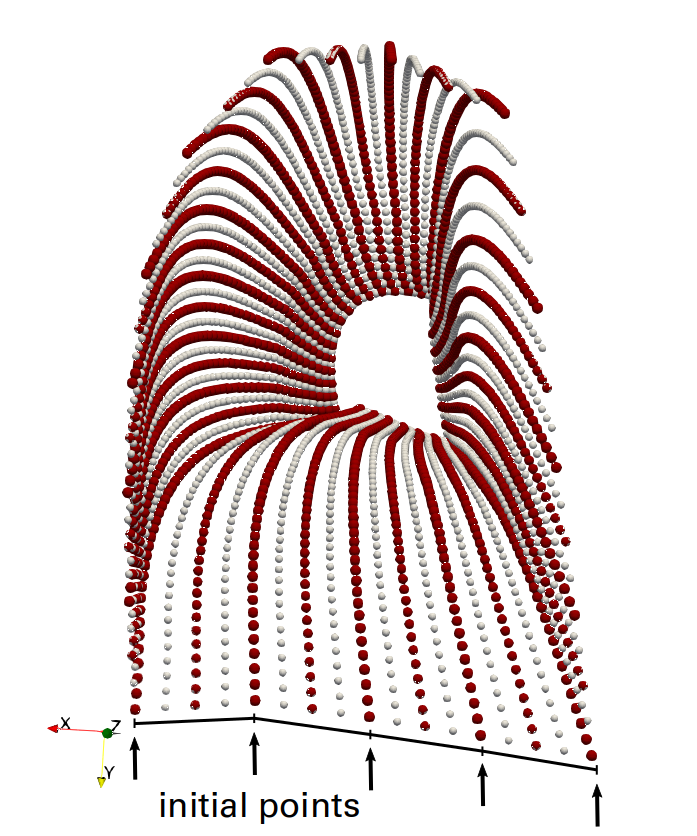}
    \caption{Initial (arrows) and refined boundary points (red) and points after additional refinement by a factor of $r=2$ (white).}%
    \label{fig:02_boundary_points}%
  \end{subfigure}   
  \quad
  \begin{subfigure}[t]{0.45\textwidth}%
    \centering%
    \includegraphics[height=9cm]{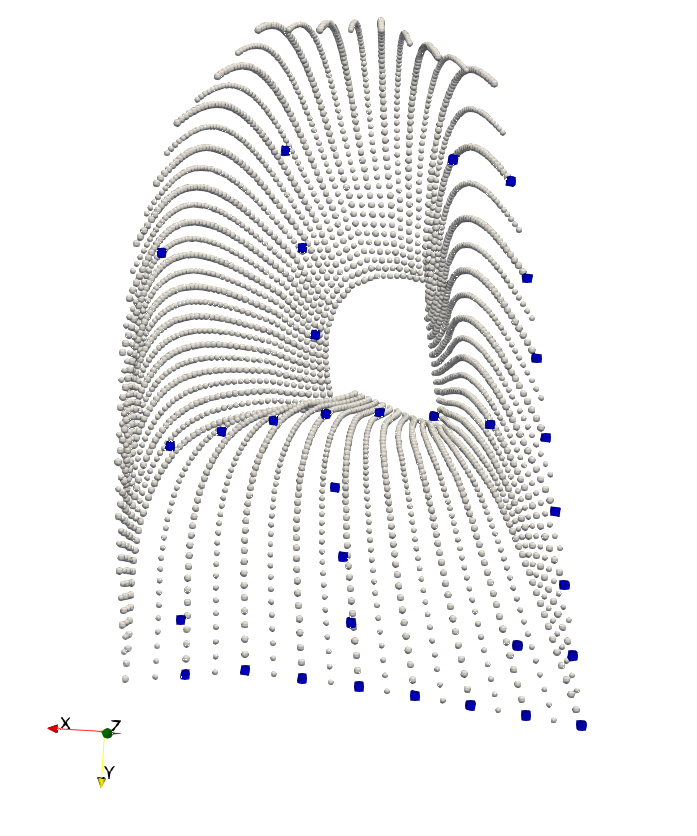}
    \caption{Seed points for the streamlines (blue).}%
    \label{fig:03_seed_points}%
  \end{subfigure}
  \\
  \begin{subfigure}[t]{0.45\textwidth}%
    \centering%
    \includegraphics[height=8cm]{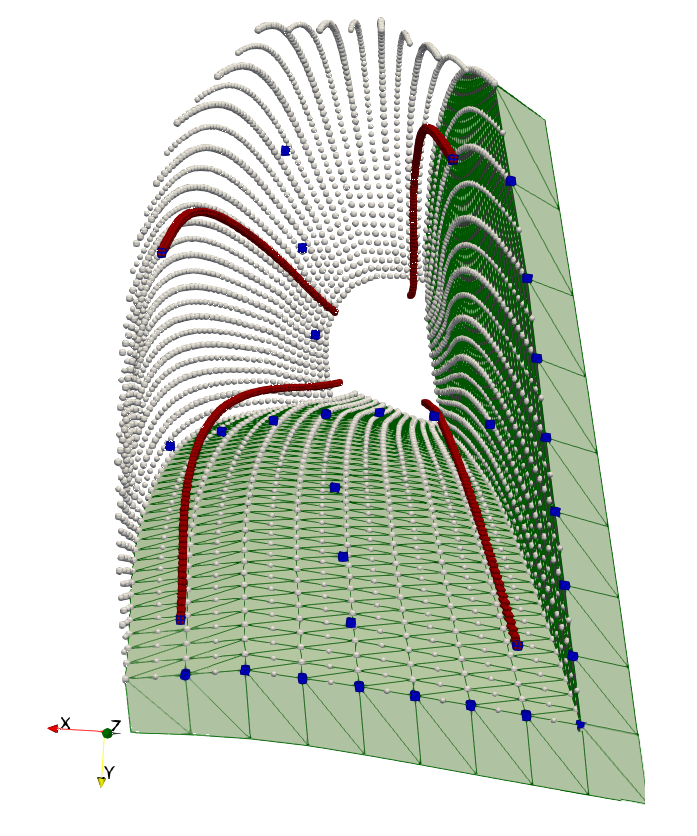}
    \caption{The four boundary streamlines (red) and the layer of ghost elements (green) at the bottom and right of the subdomain.}%
    \label{fig:05_corner_streamlines}%
  \end{subfigure}   
  \quad
  \begin{subfigure}[t]{0.45\textwidth}%
    \centering%
    \includegraphics[height=8cm]{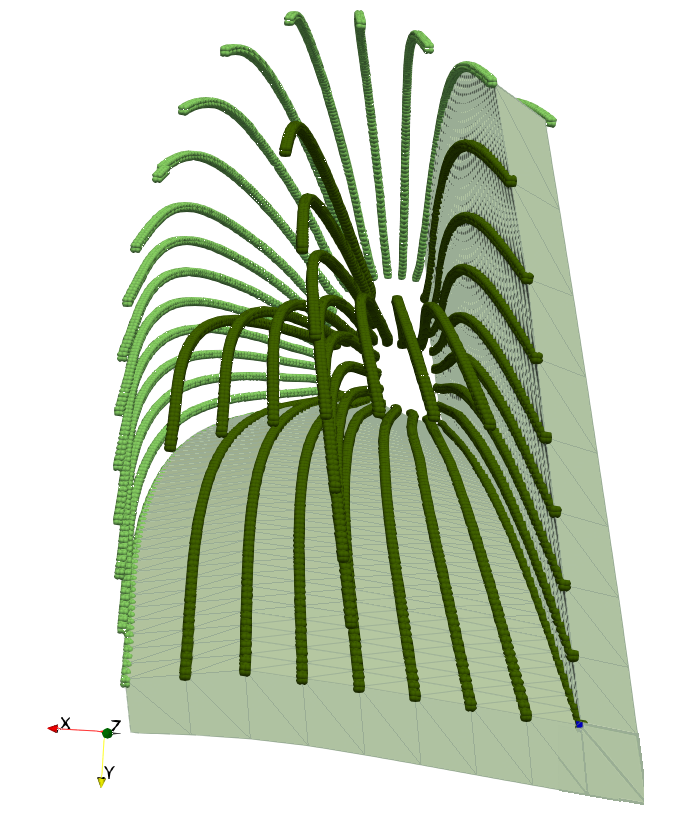}
    \caption{New boundary points on the outer (light green) and interior boundary (dark green).}%
    \label{fig:07_filled}%
  \end{subfigure}  
  \caption{Parallel generation of 3D meshes: refined boundaries, streamlines and subdomain refinement in the first subdomain for recursion level $l=1$.}%
  \label{fig:03_boundary_points_and_seed_points}%
\end{figure}

For example, the 3D mesh of \cref{fig:boundary_grid} with $2\cdot 4 \times 2\cdot 4 \times 2\cdot 50$ elements gets refined with the factor $r=2$ to $16 \times 16 \times 200$ elements.
\Cref{fig:02_boundary_points} shows the refined boundary points in this example in a view in negative $z$ direction towards the bottom of the muscle. The red points are the boundary points of the $4 \times 8$ grid, the additional white points are added in between by the refinement with $r=2$. 
Because this refinement is carried out by interpolating between the initial points, the new points are located on straight lines between the initial points.
This can especially be seen at the lower left of the figure (indicated by arrows and lines) where always five neighboring points lie on a straight line. 

\begin{figure}
  \centering
  \begin{subfigure}[t]{0.23\textwidth}%
    \centering%
    \includegraphics[height=7cm]{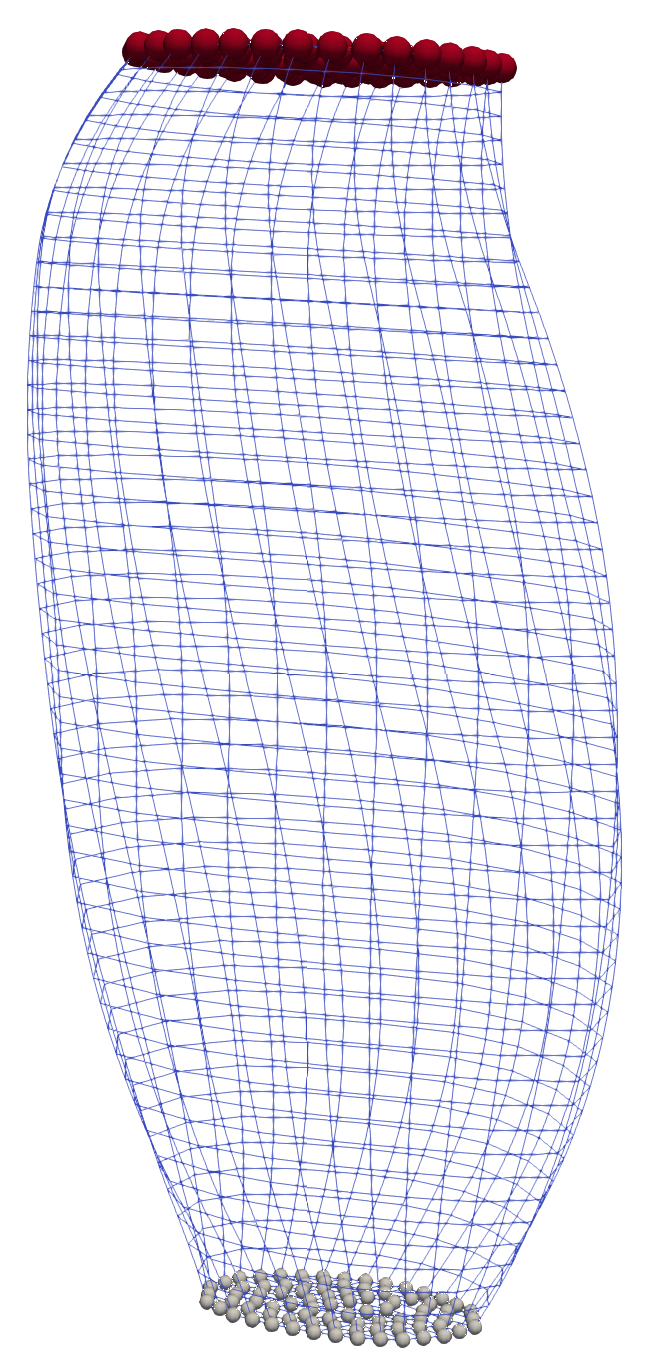}
    \caption{Location of Dirichlet boundary condition nodes at the bottom and top.}%
    \label{fig:dirichlet_bc_1}%
  \end{subfigure}
  \,
  \begin{subfigure}[t]{0.24\textwidth}%
    \centering%
    \includegraphics[height=7cm]{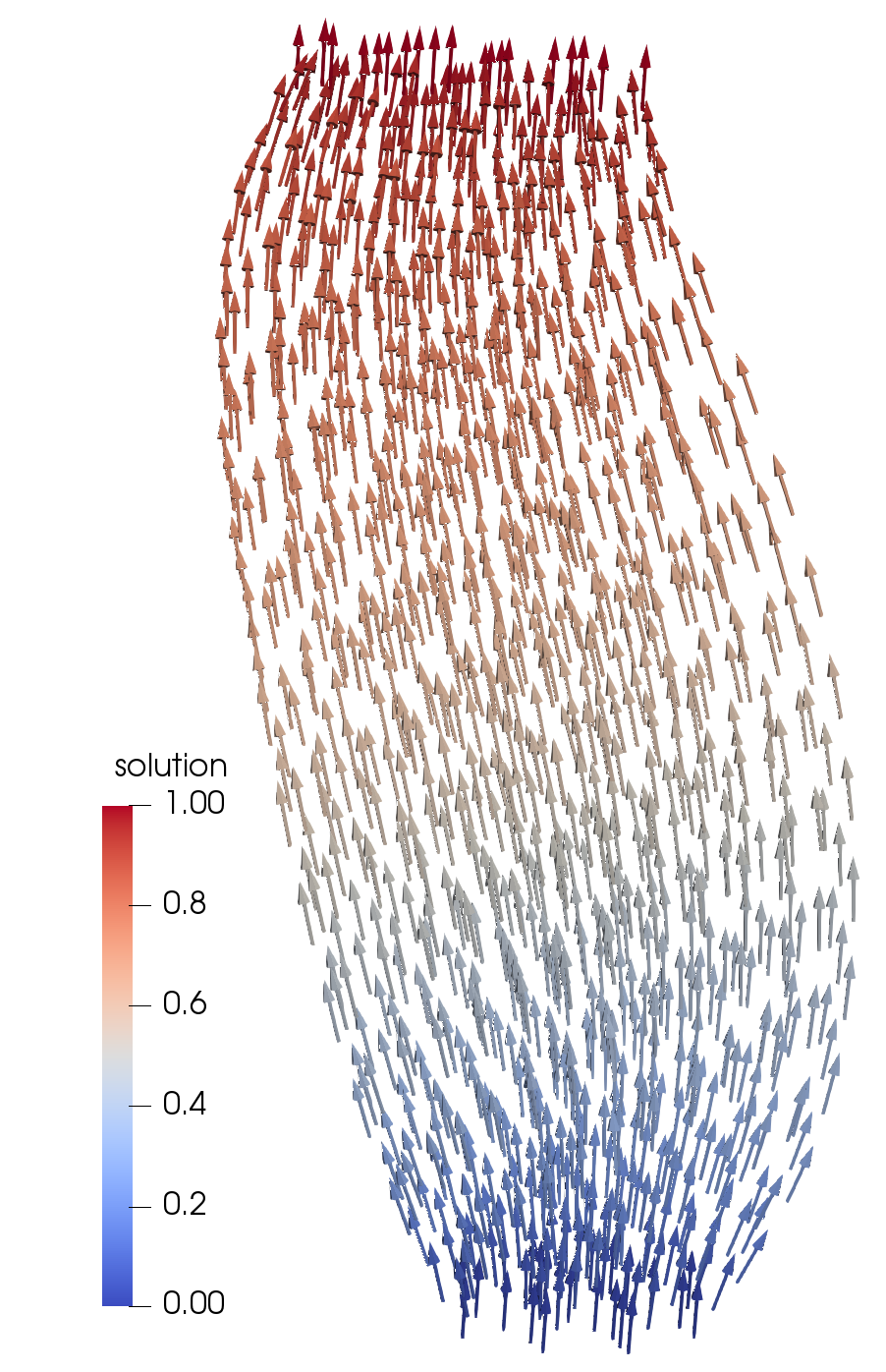}
    \caption{Solution $p$ of the Laplace problem and direction of the gradient $\nabla p$.}%
    \label{fig:laplace_1}%
  \end{subfigure}
  \qquad
  \begin{subfigure}[t]{0.19\textwidth}%
    \centering%
    \includegraphics[height=7cm]{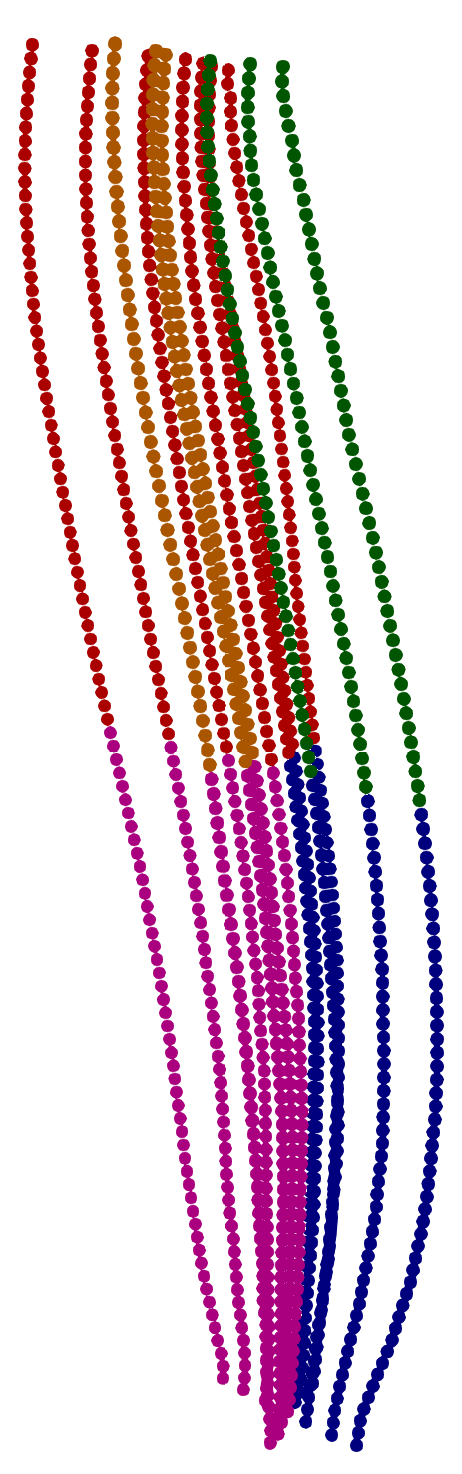}
    \caption{Traced streamlines that split the domain into eight subdomains.}%
    \label{fig:boundary_points_1}%
  \end{subfigure}
  \,
  \begin{subfigure}[t]{0.24\textwidth}%
    \centering%
    \includegraphics[height=7.2cm]{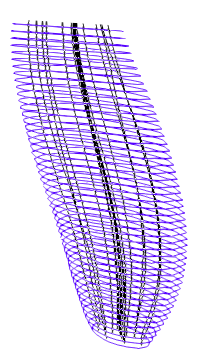}
    \caption{Rings of the slices $S_M$ and traced streamlines in the interior.}%
    \label{fig:slices_2}%
  \end{subfigure}
  \caption{Parallel 3D mesh generation: Process of subdividing the muscle volume into eight subdomains using the solution of a Laplace problem, which is an important step in the procedure of \cref{alg:parallel_algorithm_1}.}
  \label{fig:determining_subdomains}%
\end{figure}

% laplace problem
Next, in line \ref{line:3.4} of \cref{alg:parallel_algorithm_1} the Laplace problem gets solved. The same step also occurs in \cref{alg:serial_algorithm_2} and is explained in \cref{sec:generation_of_fiber_meshes}.
The equation is formulated globally and the discretization uses the existing partitioning. 
Dirichlet boundary conditions of $p(\bfx) = 0$ and $p(\bfx) = 1$ are prescribed at the bottom and top of the domain, as shown by the spheres in \cref{fig:dirichlet_bc_1}. Alternatively, Neumann boundary conditions can be used.
A parallel GMRES solver is employed to obtain the solution in a few iterations. E.g., for the biceps muscle a linear system at $l=0$ has 4131 degrees of freedom and 26 iterations are needed to obtain a residual norm below \num{1e-4}. After the solution $p(\bfx)$ is obtained, the gradient field $\nabla p(\bfx)$ is computed. The solution and the gradient directions are visualized in \cref{fig:laplace_1}.

\subsection{Communication of the Ghost Layer}

% ghost communication
Subsequently, the gradient field $\nabla p(\bfx)$ is used to trace streamlines to determine new boundaries of the subdomain. This involves tracing streamlines that start exactly on the boundary. These streamlines potentially switch between the subdomain owned by the current process and the subdomains of neighboring processes. Streamline tracing requires the gradient field values of the elements where the streamline passes through.
To avoid repeated communications in these cases, a ghost layer of a specified number $n_\text{ghost\_layer\_width}$ of elements is added to the subdomains at all parts of the boundary that touch a neighboring subdomain directly or diagonally adjacent in $x$ and $y$ direction.

The ghost layer is constructed and the node positions and values of $p$ and $\nabla p$ associated with the ghost elements are communicated between the neighboring processes after the solution of the Laplace problem.
This occurs in line \ref{line:3.5} of the algorithm. \Cref{fig:05_corner_streamlines} shows $n_\text{ghost\_layer\_width}=1$ layer of ghost elements on a subdomain at recursion level $l=1$.

\subsection{Selection of Seed Points for the Streamlines}\label{sec:selection_of_seed_points}

% seed points-1
Next, the seed points from which the streamlines start are determined on the subdomain.
All seed points are selected from the set of nodes in the structured mesh of a horizontal 2D slice. 

\begin{figure}
  \centering
  \begin{subfigure}[t]{0.30\textwidth}%
    \centering%
    \includegraphics[height=5cm]{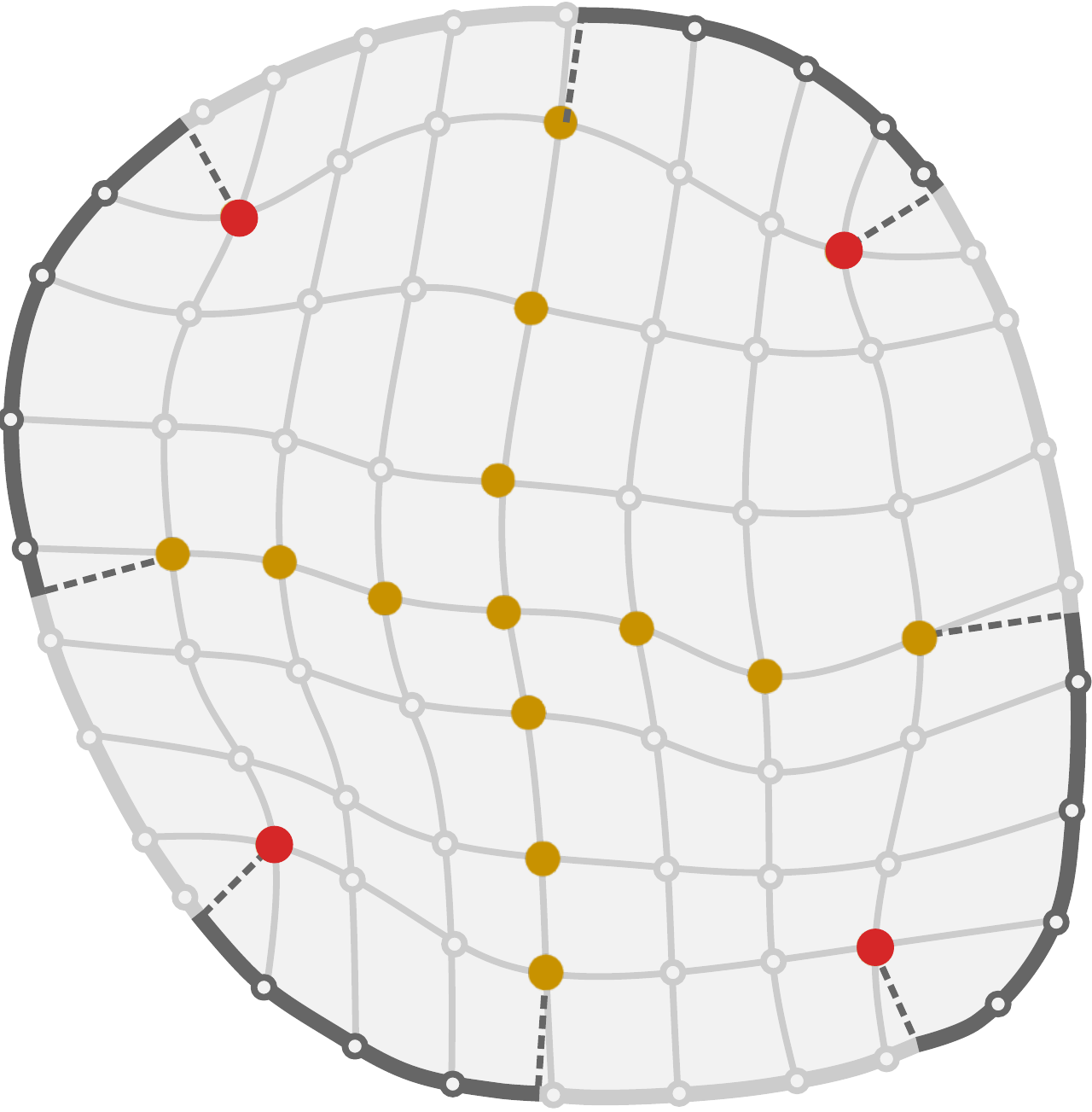}
    \caption{The seed points of the streamlines used to determine the subdomain boundaries. }%
    \label{fig:seed_points_to_send_1}%
  \end{subfigure}
  \,
  \begin{subfigure}[t]{0.30\textwidth}%
    \centering%
    \includegraphics[height=5cm]{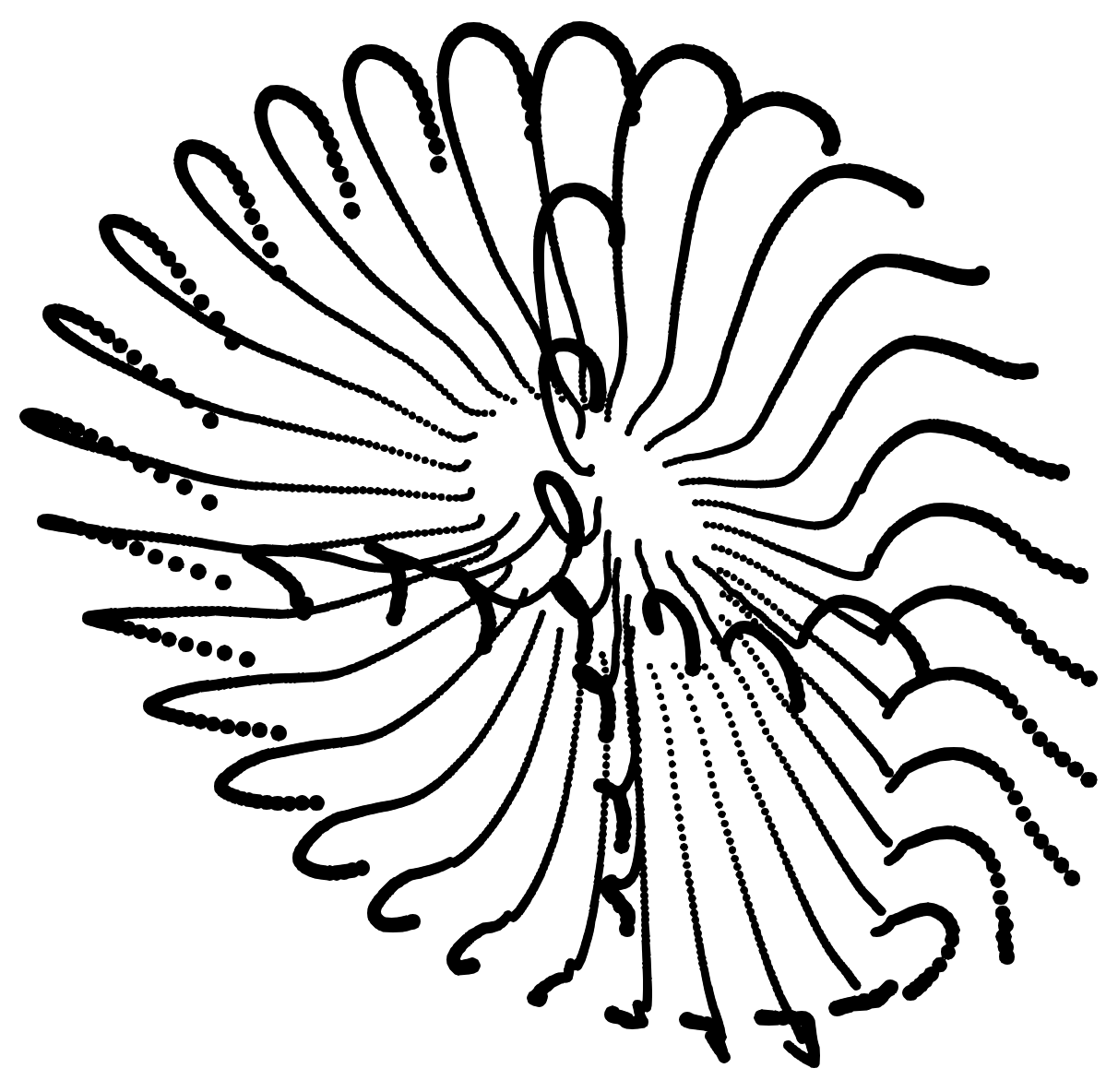}
    \caption{Streamlines and lines on the muscle surface that define the new subdomain boundaries.}%
    \label{fig:fixed_1}%
  \end{subfigure}
  \,
  \begin{subfigure}[t]{0.30\textwidth}%
    \centering%
    \includegraphics[height=5cm]{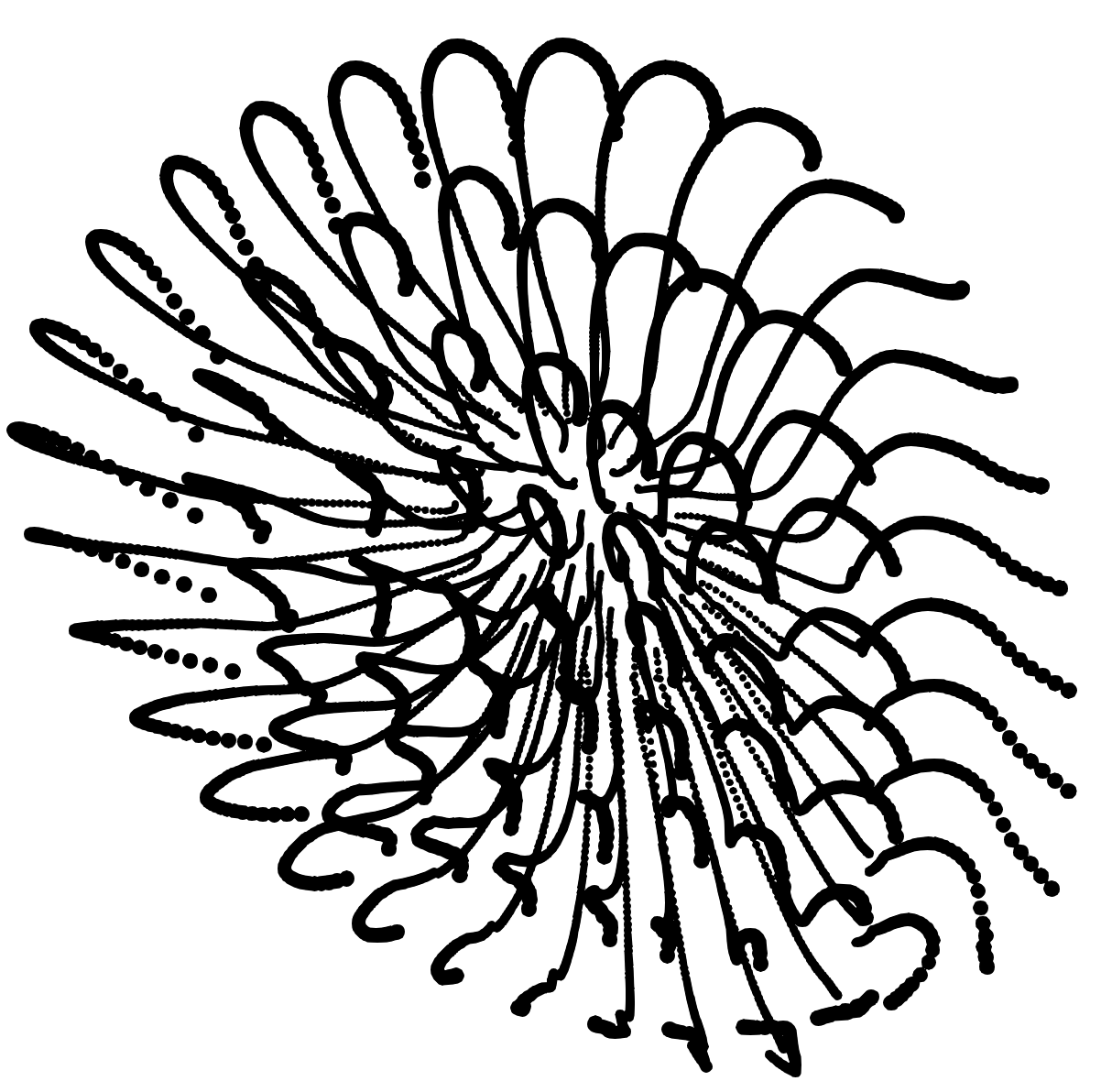}
    \caption{All streamlines and lines on the muscle surface that are created if the algorithm is run with one process and $l_\text{max}=0$.}%
    \label{fig:final_interior_1}%
  \end{subfigure}
  \caption{Parallel mesh generation: Seed points and streamlines that occur during the first call to the procedure in \cref{alg:serial_algorithm_2}, in a view from the top of the muscle.}
  \label{fig:seed_points}%
\end{figure}

\Cref{fig:seed_points_to_send_1} visualizes the structured mesh in light gray in the first call to the procedure for recursion level $l=0$ where the domain is not yet partitioned.
The selected seed points are shown by the yellow and red points.
As can be seen, the seed points consist of the nodes of the 2D mesh at the horizontal and vertical centers in this view which form the \emph{plus sign} shape given by the yellow points.
In addition, the four red points near the corners of the structured mesh are selected.
%\Cref{fig:seed_points_to_send_1} also visualizes the boundary of the slice together with a method of splitting it into eight sectors by choosing the splitting points such that they are the closest to the given outer seed points.

% seed points interior for l=0
The seed points of the \emph{plus sign} yield the streamlines that subdivide the domain into four parts in $x$ and $y$ direction. With the additional split in $z$ direction, the inner boundaries of the eight subdomain are obtained. The resulting boundaries are given in \cref{fig:fixed_1}.

The streamlines of these seed points are also depicted in \cref{fig:boundary_points_1}. The interior boundary points for the eight subdomains that partition the muscle volume at level $l=1$ are shown by different colors.
The full subdomain boundaries include also the outer surface of the muscle, which is given by the rings of the muscle slices.
\Cref{fig:slices_2} shows the streamlines in black and the circumferential rings of the muscle slices in blue that were extracted during the call to \cref{alg:serial_algorithm_1} in line \ref{line:3.2}.

% seed points interior for l>0
At higher recursion levels $l>0$, the boundaries for the new subdomains consist of those at the outer boundary of the muscle defined by the surface representation and those in the interior of the muscle.
Similar to the previously considered case at recursion level $l=0$, for $l\geq 1$ the boundaries in the interior of the muscle have to be sampled by a set of streamlines. In addition to the streamlines associated with the plus sign shaped seed points, new streamlines at the boundaries of the current subdomain have to be obtained.

\Cref{fig:03_seed_points} shows in blue all seed points that are selected in a subdomain on recursion level $l=1$ in order to create subdomain boundaries for level $l=2$.
As can be seen, in addition to the plus sign shape and the four outer seed points two lines of points in approximate $x$ and $y$ directions are selected at the lower and right edges of the image.
These are seed points for the new boundaries in the interior of the muscle. Note that the current recursion level $l=1$ also has boundaries at these locations. However, for level $l=2$ these boundaries are recreated by the new streamlines. Tracing of these streamlines potentially uses the ghost layer. The resulting streamlines and the ghost layer for $n_\text{ghost\_layer\_width}=1$ are shown in \cref{fig:05_corner_streamlines,fig:07_filled}.

\subsection{Determination of Subdomain Boundaries on the Outer Muscle Surface}

% seed points-3 outer ring
Next, the boundary points on the outer surface of the muscle are determined for the new partitioning. They are obtained by sampling the circumferential rings of the muscle surface with the resolution required in the current recursion level. In our implementation, this can be done either by sampling the original surface triangulation of the muscle or by directly evaluating the parametric form of the NURBS surface that approximates the muscle surface.

At recursion level $l=0$, the entire muscle surface is touched by the new subdomains. Thus, when traversing the circumference of the muscle four new subdomains are encountered.
In consequence, every circumferential ring needs to be split into four quarter parts for the four adjacent subdomains. For each of these new subdomains, the quarter part corresponds to two neighboring sides of the subdomain boundary in \cref{fig:boundary_grid_1}. \Cref{fig:seed_points_to_send_1} also visualizes the two neighboring sides per new subdomain as dark and light portions of the outer boundary. 
To obtain these sides, a splitting point is needed that further splits every quarter part of the circumferential ring into the two sides for the new subdomain.
In summary, the ring needs to be split into eight parts that fit to the inner subdomain boundaries.

The eight split points are determined by the eight outer streamline points. In \Cref{fig:seed_points_to_send_1}, the four outer yellow points of the plus sign and the four red points are considered. For each split location, the nearest point on the circumferential ring is determined. The employed algorithm for calculating the coordinates of the point on a ring that has the shortest distance to a given, second point is described in \cref{sec:slicing_of_the_geometry}.

After the two sections of the circumferential rings have been determined for all new subdomains, the sections are equidistantly sampled in circumferential direction with $n_\text{el,x}$ elements each to create the outer boundary points for the subdomains. 
Also in longitudinal direction of the muscle, i.e., the $z$ axis, points are sampled on each streamline and on the outer boundary surface to yield the required number of $n_\text{el,z}$ points per subdomain. The resulting boundary points obtained during recursion level $l=0$ are shown in \cref{fig:fixed_1}.

This method is also similarly required on higher recursion levels $l>0$. 
Then, however, two cases have to be considered separately. The first case involves splitting the muscle surface boundary on one process into two parts, analogously to the described method at $l=0$. The second case involves two neighboring processes that have to agree on the split point of their common part of the outer surface boundary.

In the example at recursion level $l=1$ in \cref{fig:05_corner_streamlines}, the red streamlines are used to split the boundary sides at the outer boundary of the global domain. 
The first case occurs for the upper left red streamline, which is used to bisect the shown white part of the muscle surface. 

\begin{figure}
  \centering
  \def\svgwidth{0.7\textwidth}
  %% Creator: Inkscape inkscape 0.92.3, www.inkscape.org
%% PDF/EPS/PS + LaTeX output extension by Johan Engelen, 2010
%% Accompanies image file 'fixed_1.pdf' (pdf, eps, ps)
%%
%% To include the image in your LaTeX document, write
%%   \input{<filename>.pdf_tex}
%%  instead of
%%   \includegraphics{<filename>.pdf}
%% To scale the image, write
%%   \def\svgwidth{<desired width>}
%%   \input{<filename>.pdf_tex}
%%  instead of
%%   \includegraphics[width=<desired width>]{<filename>.pdf}
%%
%% Images with a different path to the parent latex file can
%% be accessed with the `import' package (which may need to be
%% installed) using
%%   \usepackage{import}
%% in the preamble, and then including the image with
%%   \import{<path to file>}{<filename>.pdf_tex}
%% Alternatively, one can specify
%%   \graphicspath{{<path to file>/}}
%% 
%% For more information, please see info/svg-inkscape on CTAN:
%%   http://tug.ctan.org/tex-archive/info/svg-inkscape
%%
\begingroup%
  \makeatletter%
  \providecommand\color[2][]{%
    \errmessage{(Inkscape) Color is used for the text in Inkscape, but the package 'color.sty' is not loaded}%
    \renewcommand\color[2][]{}%
  }%
  \providecommand\transparent[1]{%
    \errmessage{(Inkscape) Transparency is used (non-zero) for the text in Inkscape, but the package 'transparent.sty' is not loaded}%
    \renewcommand\transparent[1]{}%
  }%
  \providecommand\rotatebox[2]{#2}%
  \newcommand*\fsize{\dimexpr\f@size pt\relax}%
  \newcommand*\lineheight[1]{\fontsize{\fsize}{#1\fsize}\selectfont}%
  \ifx\svgwidth\undefined%
    \setlength{\unitlength}{311.71094799bp}%
    \ifx\svgscale\undefined%
      \relax%
    \else%
      \setlength{\unitlength}{\unitlength * \real{\svgscale}}%
    \fi%
  \else%
    \setlength{\unitlength}{\svgwidth}%
  \fi%
  \global\let\svgwidth\undefined%
  \global\let\svgscale\undefined%
  \makeatother%
  \begin{picture}(1,0.40224075)%
    \lineheight{1}%
    \setlength\tabcolsep{0pt}%
    \put(0,0){\includegraphics[width=\unitlength,page=1]{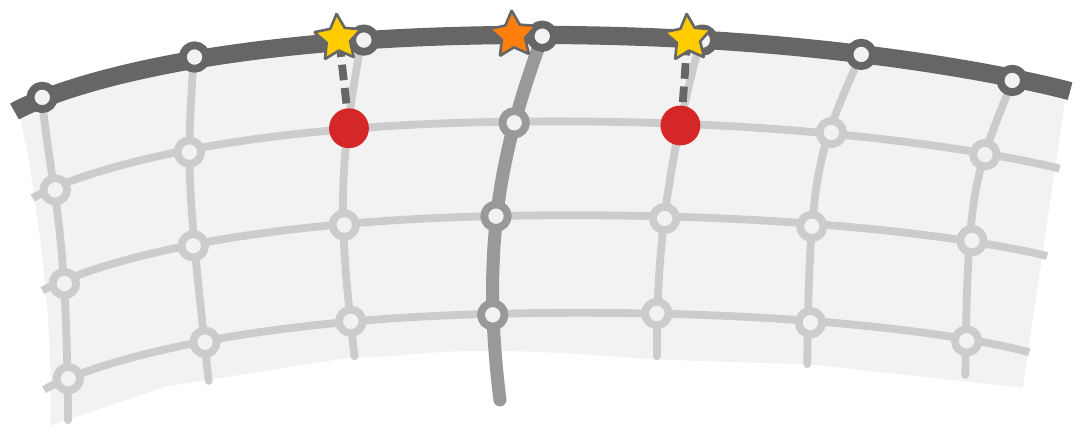}}%
    \put(0.56930606,0.12308439){\color[rgb]{0,0,0}\makebox(0,0)[lt]{\lineheight{1.25}\smash{\begin{tabular}[t]{l}Subdomain process $p_1$\end{tabular}}}}%
    \put(0.07028264,0.12308439){\color[rgb]{0,0,0}\makebox(0,0)[lt]{\lineheight{1.25}\smash{\begin{tabular}[t]{l}Subdomain process $p_0$\end{tabular}}}}%
  \end{picture}%
\endgroup%

  \caption{Parallel mesh generation: Case of partitioning the outer boundary surface that occurs for recursion level $l \geq 1$ in \cref{alg:serial_algorithm_2}. Shown are the meshes on two subdomains of processes $p_0$ and $p_1$ and the location of the streamline at the corner (red points). The orange star is the newly determined border point between the subdomain boundaries.}
  \label{fig:seed_points_case_2}%
\end{figure}

The second case occurs at the lower left and upper right borders between the shown subdomain and the neighboring subdomains of two other processes. For the case at the upper right, \cref{fig:seed_points_case_2} visualizes the following method: First, the point on the outer surface that is closest to the point of the red streamline is determined on both subdomains, visualized by the yellow stars. These points are communicated between the two processes. Each process computes the center point of these two points (orange star) and then finds the closest point to this center point on the boundary. This is done for all rings of the muscle in $z$ direction. In result, both processes have the same line on the surface in longitudinal direction of the muscle that is then used as one edge of the new subdomains.

Thus, the new subdomain boundaries on the outer boundary can be found using the red extra streamlines shown in \cref{fig:05_corner_streamlines}.
For simplicity, the algorithm always computes the four streamlines in every corner of the mesh although all of them are only required for $l=0$. In the shown example for $l=1$, the streamline in the lower right corner is not needed for the sampling of the new boundary points. 

A summary of the streamlines that are used for the new subdomain boundaries in this example is given in \cref{fig:07_filled}.
The sampled boundary points at the muscle surface are shown in light green color. These two sides of the own domain will be split into four sides for the new subdomains. In this example with $n_\text{el,x}=4$, the surface therefore gets sampled at $4\times 4=16$ lines. A comparison with the white lines in \cref{fig:05_corner_streamlines} shows that the newly sampled points are different from the initially sampled points. While in  \cref{fig:05_corner_streamlines} always five neighboring boundary points are located on a straight line, the points in \cref{fig:07_filled} follow the curved outer boundary better and, thus, refine the boundary representation.

\subsection{Parallel Algorithm for Streamline Tracing}\label{sec:parallel_streamline_tracing}
% streamline tracing method

In line \ref{line:3.6} of \cref{alg:parallel_algorithm_1}, streamlines have to be traced through the gradient field $\nabla p(\bfx)$ of the Laplace solution for all the seed points given in \cref{sec:selection_of_seed_points}. In the following, more details on the parallel method of streamline tracing is given.

This step is similar to the analog step in \cref{alg:serial_algorithm_2}. 
The same method of explicit Euler integration is used. The seed points are located at the horizontal plane at the center in vertical direction of the muscle. From there, streamlines are traced in both directions towards the ends of the muscle following the positive and negative gradient directions.
The tracing algorithm uses the efficient scheme of selecting the subsequently traversed elements described in \cref{sec:algorithm_for_streamline_tracing}. The implementation is adjusted in a way to also take into account the layers of ghost elements.

% streamline tracing in parallel-1
Since the streamlines traverse the entire muscle from the center to the bottom and top, multiple processes are involved in the computation of every streamline. To describe the scheme, all processes are numbered in $z$ direction from bottom to top by an index $i_z \in \{0, 1, \dots, n_z-1\}$ where $n_z = 2^l$ is the number of processes in $z$ direction on the current recursion level $l$.

% streamline tracing in parallel-2
The initial seed points are determined on the processes at the vertical center with index $\lfloor n_z/2\rfloor$. They are communicated to the processes below with index $\lfloor n_z/2\rfloor-1$. These two groups of processes begin with tracing the streamlines through their subdomains starting from the same seed points, the upper processes in upward and the lower processes in downward direction. Then, the end points of the traced streamlines are communicated to the next processes, which continue the tracing. The procedure repeats with further processes until the streamlines reach the bottom and top ends of the overall muscle domain.
The time complexity of this approach is $\O(n_z) = \O(\sqrt[3]{n_\text{proc}})$ with the number $n_\text{proc}$ of processes.

% streamline sampling
After the streamlines have been traced, they are sampled at equidistant positions with a distance according to the required distance between the boundary points of the subdomains.

\subsection{Recursion End: Generation of the Resulting Meshes}

% summary/überleitung
In result, one pass of \cref{alg:parallel_algorithm_1} from lines \ref{line:3.2} to \ref{line:3.7} creates boundaries for eight new subdomains. Line \ref{line:3.8} checks whether the maximum recursion $l_\text{max}$ is reached and the recursion ends. If the recursion ends, the final 3D mesh and 1D fiber meshes are constructed in line \ref{line:3.9}. In this case, the prepared boundary points are not needed for a further subdivision of the domain but are used to construct the final meshes instead.

Every resulting fiber mesh is generated by one streamline. In line \ref{line:3.9}, additional streamlines are traced starting at the remaining grid points of the 2D slice at the vertical center of the muscle that were not selected as seed points earlier. The parallel method described in \cref{sec:parallel_streamline_tracing} is used.

\begin{figure}
  \centering
  \includegraphics[width=0.5\textwidth]{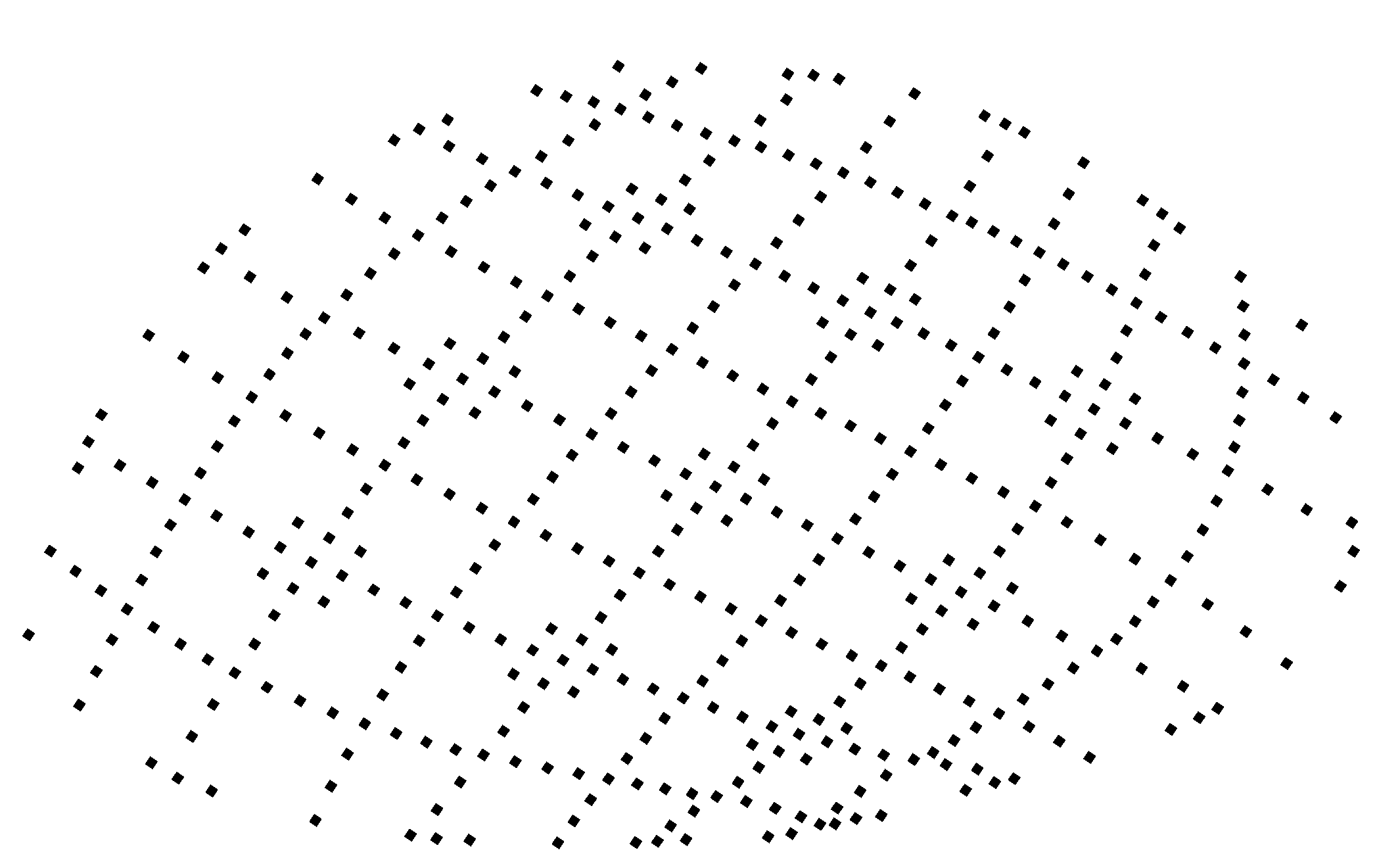}
  \caption{Parallel mesh generation: Seed points of the streamlines that are traced on processes 0 to 15 in the procedure of \cref{alg:parallel_algorithm_1} at recursion level $l=2$.}
  \label{fig:seed_points_level2}%
\end{figure}

As an example, \cref{fig:seed_points_level2} shows all seed points of streamlines for $l=2$ at the beginning of line \ref{line:3.9} in a run of \cref{alg:parallel_algorithm_1} with $n_\text{el,x}=4$, $l_\text{max}=2$ and 64 processes. 
The shown points are located at the top of the lowest 16 subdomains in the muscle, i.e., the subdomains of processes 0 to 15. The corresponding streamlines get traced in line \ref{line:3.6} to be used for the new subdomain boundary. However, since the recursion ends at $l=2$ the missing seed points of the subdomain grids are subsequently filled in and the remaining streamlines are traced. From the full grid of $31\times 31=\num{961}$ streamlines, \num{449} or $~\SI{47}{\percent}$ have already been traced at this point.

%If more fiber meshes are desired, the 2D elements of this slice can be sampled equidistantly in the parameter space used for the harmonic map computation to obtain more seed points from which additional streamlines are traced.

In summary, the 2D quadrilateral mesh at the center slice of the muscle defines the location of the resulting muscle fibers. Because the construction of this 2D mesh ensured a good mesh quality with similar element sizes, the distance between the resulting fibers is similar and a spatially homogeneous set of muscle fibers is generated.

To obtain the final 1D fiber meshes $\Omega_{F,i}$, the streamlines are sampled at equidistant $z$ intervals, specified by a parameter $\Delta z$. Because the streamlines are directed mainly along the $z$ axis, the constant $z$ interval for the sampling approximately corresponds to the resulting 1D mesh width, i.e., the distance between the points of a fiber. An advantage of this method is that the points of all fibers lie in the same $x$-$y$ planes. Thus, the total set of points can also be interpreted as a structured 3D mesh of the muscle volume $\Omega_M$. This 3D mesh is aligned with the fiber meshes and planes through the $x$ and $y$ axes. These properties are advantageous for data mapping between the 3D mesh and the 1D fiber meshes and for the numerical solution of models with anisotropic advection processes in the 3D mesh that is oriented according to the direction of the fibers.

% file output
At the end, the data is written collectively by all processes into a single file. This is done using the parallel file I/O functionality of MPI. This can be done because the absolute position in the file of every point can be calculated from the index of the point in the structured mesh.

% Examples
\Cref{fig:final_interior_1} gives an example of the resulting streamlines if the recursion ends already after one pass of the procedure at $l_\text{max}=0$. For the example with $l_\text{max}=1$, the selected seed points and the parts of the resulting streamlines in the considered subdomain are shown in \cref{fig:08_final}. Here, the dark blue streamlines on the boundary were traced as part of the refinement actions in line \ref{line:3.6}. Because the recursion ends for $l=1$, these streamlines are now reused for the final fiber meshes instead of further parallel partitioning. Additionally, the light blue streamlines in the interior were traced to obtain a full grid of fibers for the output of the algorithm.

\subsection{Continuation on the Next Recursion Level}

If line \ref{line:3.8} of \cref{alg:parallel_algorithm_1} does not detect the recursion end because the maximum recursion levels is not yet reached, the \code{else} branch in line \ref{line:3.10} is chosen.
Execution continues with the eight times higher number of processes $8^{\ell+1}$. The processes that executed the previous parts of the algorithm send their determined boundaries of the new subdomains to seven other respective processes in line \cref{line:3.11a}. 
Only the first subdomain remains on the same process. Every process stores the boundary points for its new subdomain in the variable \code{boundary\_points}. In line \cref{line:3.11}, the procedure is called recursively and the next recursion level $(l+1)$ begins.

\begin{figure}%
  \centering%
  \begin{subfigure}[t]{0.48\textwidth}%
    \centering%
    \includegraphics[height=10cm]{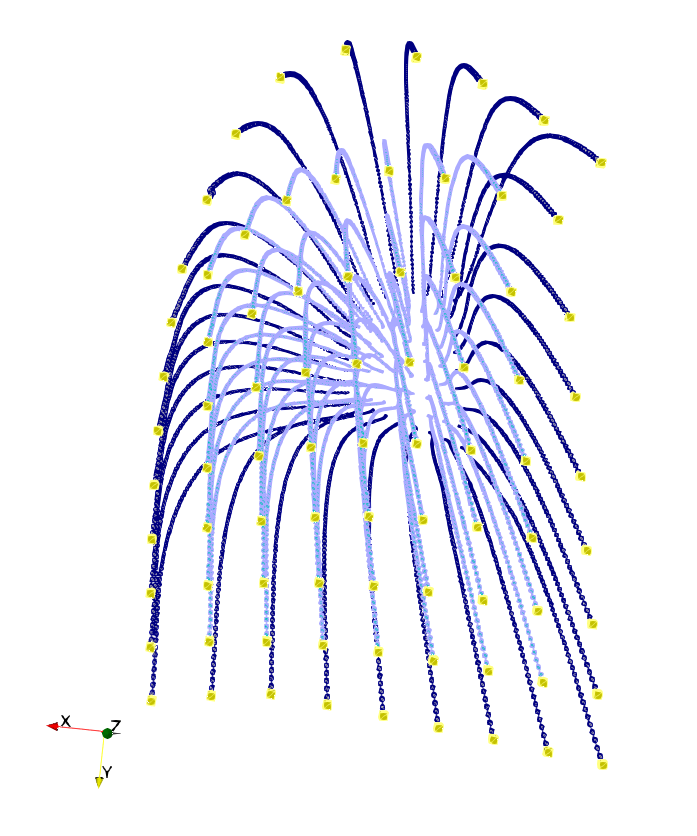}
    \caption{Seed points (yellow), traced interior streamlines (light blue) and boundary points (dark blue), generated if $l_\text{max}=1$.}%
    \label{fig:08_final}%
  \end{subfigure}
  \quad   
  \begin{subfigure}[t]{0.48\textwidth}%
    \centering%
    \includegraphics[height=10cm]{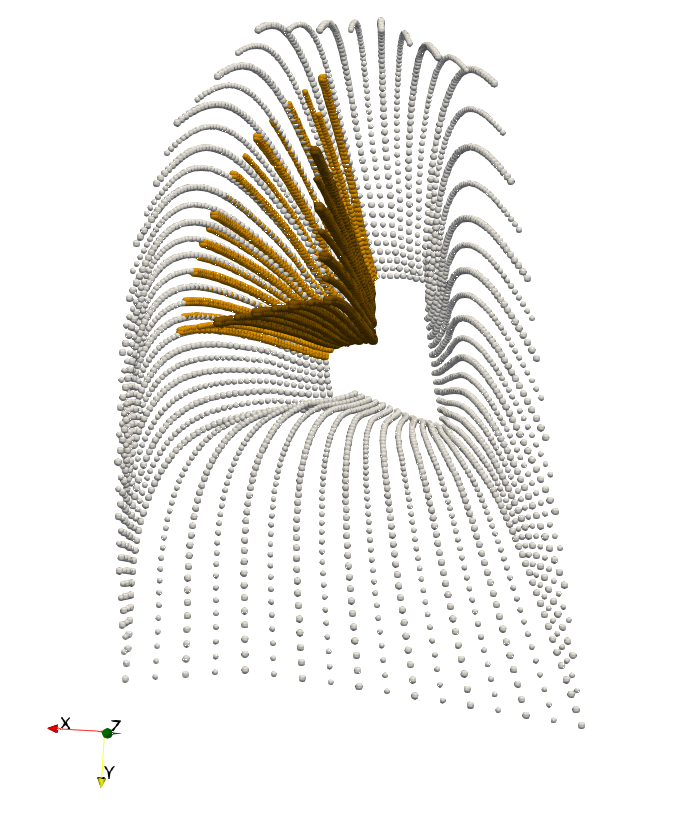}
    \caption{Boundary points of the first subdomain on level $l=2$ (dark yellow and brown) embedded in the boundary (white) of level $l=1$, generated if $l_\text{max} > 1$.}%
    \label{fig:06_subdomain}%
  \end{subfigure}
   
  \caption{Generation of 3D and 1D meshes in subdomains: Resulting streamlines after the pass of \cref{alg:parallel_algorithm_1} for recursion level $l=1$.}%
  \label{fig:improved}%
\end{figure}

\Cref{fig:06_subdomain} shows the \code{boundary\_points} of the first new subdomain on level $l=2$ for the example on recursion level $l=1$. It consists of the outer boundary (dark yellow lines) and the interior boundary (brown streamlines) and is nearly geometrically similar to the subdomain on level $l=1$. 

\subsection{Repair of Incomplete Streamlines}\label{sec:repair_of_incomplete_streamlines}

Practical tests have shown that, for irregular muscle geometries, occasionally some streamlines generated in lines \ref{line:3.6} and \ref{line:3.9} of \cref{alg:parallel_algorithm_1} can be incomplete. This means that it was not possible to obtain a streamline that runs through the entire subdomain or the entire muscle domain from top to bottom, instead points are missing for some ranges of $z$ values. This can happen if the streamlines leave the subdomains (because the ghost layer width was chosen too small) or due to numerical errors in irregularly shaped elements mainly on high recursion levels where the system matrix is badly conditioned.

\begin{figure}%
  \centering%
  \def\svgwidth{0.8\textwidth}
  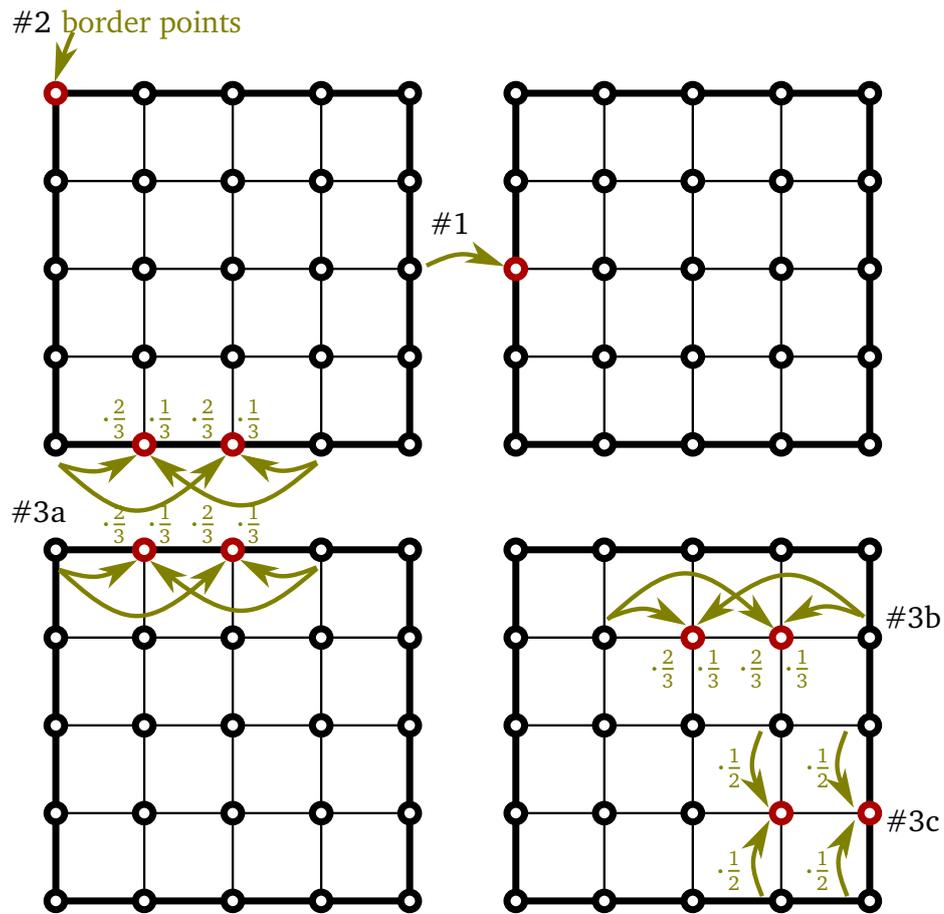
  \caption{Repair of streamlines used during partitioning and fiber approximation in the 3D and 1D mesh generation: Examples of the four repair mechanisms for estimating incomplete streamlines during the parallel algorithm. Invalid streamlines are indicated by red circles, valid streamlines by black circles. The brown arrows show the direction of data transfer.}%
  \label{fig:fix_invalid}%
\end{figure}

To obtain meaningful results even in these cases, three different repair mechanisms are introduced that interpolate the missing data from valid streamlines. \Cref{fig:fix_invalid} visualizes the cases by examples in a setting of four subdomains with grids of $5 \times 5$ fibers each. The repair mechanisms $\#1$ to $\#3a$ only apply to boundary points. They are executed in line \ref{line:3.6} of the algorithm after the local portions of the streamlines have been traced and before the end points of the streamlines are sent to the neighbor processes below and above that continue the streamline tracing. Mechanismn $\#3b$ and $\#3c$ repair invalid streamlines in the final result and are executed during line \ref{line:3.9} of \cref{alg:parallel_algorithm_1}.

Mechanism \#1 checks all streamlines at subdomain boundaries in the interior, which are shared between neighboring processes. If a streamline is incomplete on one process but complete on the neighbor process, the data of the complete side are transferred such that both processes have the same valid points for this streamline. In the example in \cref{fig:fix_invalid}, the valid streamline data are sent from the top left to the top right subdomain.

Mechanism \#2 checks streamlines at the outer corners of the subdomains. Incomplete streamlines at these locations are recreated from the given boundary points. Because the set of boundary points is twice as coarse as the required number of sample points at these streamlines, every second point gets interpolated from the top and bottom neighbor points.

Mechanism \#3a is concerned with streamlines at interior subdomain boundaries that could not be fixed by mechanism \#1 because the streamlines are incomplete on both sharing processes. In this case, the streamlines are interpolated from the two complete neighboring streamlines that are located next along the boundary as shown in the example in \cref{fig:fix_invalid}. Instead of the factors $\frac13$ and $\frac23$, the actual relation of distances between the seed points of the streamlines is used. The same interpolation is executed independently on both involved processes. Because the valid streamlines have the same data on both subdomains, the resulting fixed streamlines will also be identical.

Mechanisms \#3b and \#3c follow the same approach. They are applied to the interior fibers of the final result and can repair any number of incomplete fibers that are located between complete fibers. This case rarely occurs, a cause can be errors in the numerical solution of the Laplace problem.  In example \#3b in \cref{fig:fix_invalid}, the two invalid streamlines are interpolated from their left and right valid neighbors. In example \#3c, no valid right neighbor exists. Instead, the streamlines are interpolated by using valid positions from the upper and lower neighbors.

\subsection{Post-processing and Output of the Generated Streamlines}\label{sec:postprocessing_of_the_generated_streamlines}

After repairing invalid streamlines, the final result of the algorithm is a grid with $(2\,n_\text{el,x}\,n_x+1) \times (2\,n_\text{el,x}\,n_x+1)$ fibers in the $x$-$y$ plane and a configurable number of points in $z$ direction, where $n_x = 2^{\ell_\text{max}}$ is the number of subdomains per coordinate direction on the last recursion level. 

If a higher number of fibers is desired than is naturally generated by the parallel algorithm, additional fibers can be created by interpolation in the existing grid of fibers, which is parallel partitioned. The implementation of the presented algorithm in \opendihu{} includes this post-processing functionality as part of the mesh generation program. Alternatively, the step can be applied separately on any binary output file that contains a grid of fibers. 

The action of increasing the number of fibers proceeds as follows. The initial grid contains the fibers that were created from the streamlines, called \emph{key fibers}.
A specified number $m$ of additional fibers is placed between the key fibers in both $x$ and $y$ coordinate directions.
The additional fibers together with the key fibers form a grid of fibers in the muscle cross-sections with an $m$ times finer mesh width. In the grid of key fibers, every portion bounded by $2 \times 2$ key fibers contains $(2+m)^2 - 4$ additional fine fibers. The total number of fibers depending on $n_x$ and $m$, therefore, is $N=(2\,n_\text{el,x}\,n_x\,(1+m)+1)^2$. Due to construction, this number is always odd. This is a desired property because it yields an even number of elements per coordinate direction and this allows to construct a mesh with quadratic ansatz functions.

The new fibers are computed by barycentric interpolation. The location of every new point $\bfp$ is calculated from the nearest points $\bfp_0, \bfp_1, \bfp_2$ and $\bfp_3$ of key fibers in the $x$-$y$ plane, numbered according to  \cref{fig:quads_tris}, by%
\begin{align}\label{eq:mesh_barycentric_interpolation}
  \bfp = (1-\alpha_x)\,(1-\alpha_y)\,\bfp_0 + \alpha_x\,(1-\alpha_y)\,\bfp_1 
        + (1-\alpha_x)\,\alpha_y\,\bfp_2 + \alpha_x\,\alpha_y\,\bfp_3.
\end{align}
Here, the factors $\alpha_x,\alpha_y \in [0,1]$ are chosen in a way to create the fine grid of fibers:
\begin{align*}
  \alpha_x = i / (m+1), \quad \alpha_y = j / (m+1)\quad \text{ for }i,j = 0, \dots,m, \quad (i,j) \neq (0,0).
\end{align*}
As a result, we can generate a 3D mesh where the number of points in $x$ and $y$ directions can be adjusted by the parameter $m$. 

\begin{figure}[H]
  \centering%
  \begin{subfigure}[t]{0.48\textwidth}%
    \centering%
    \includegraphics[width=\textwidth]{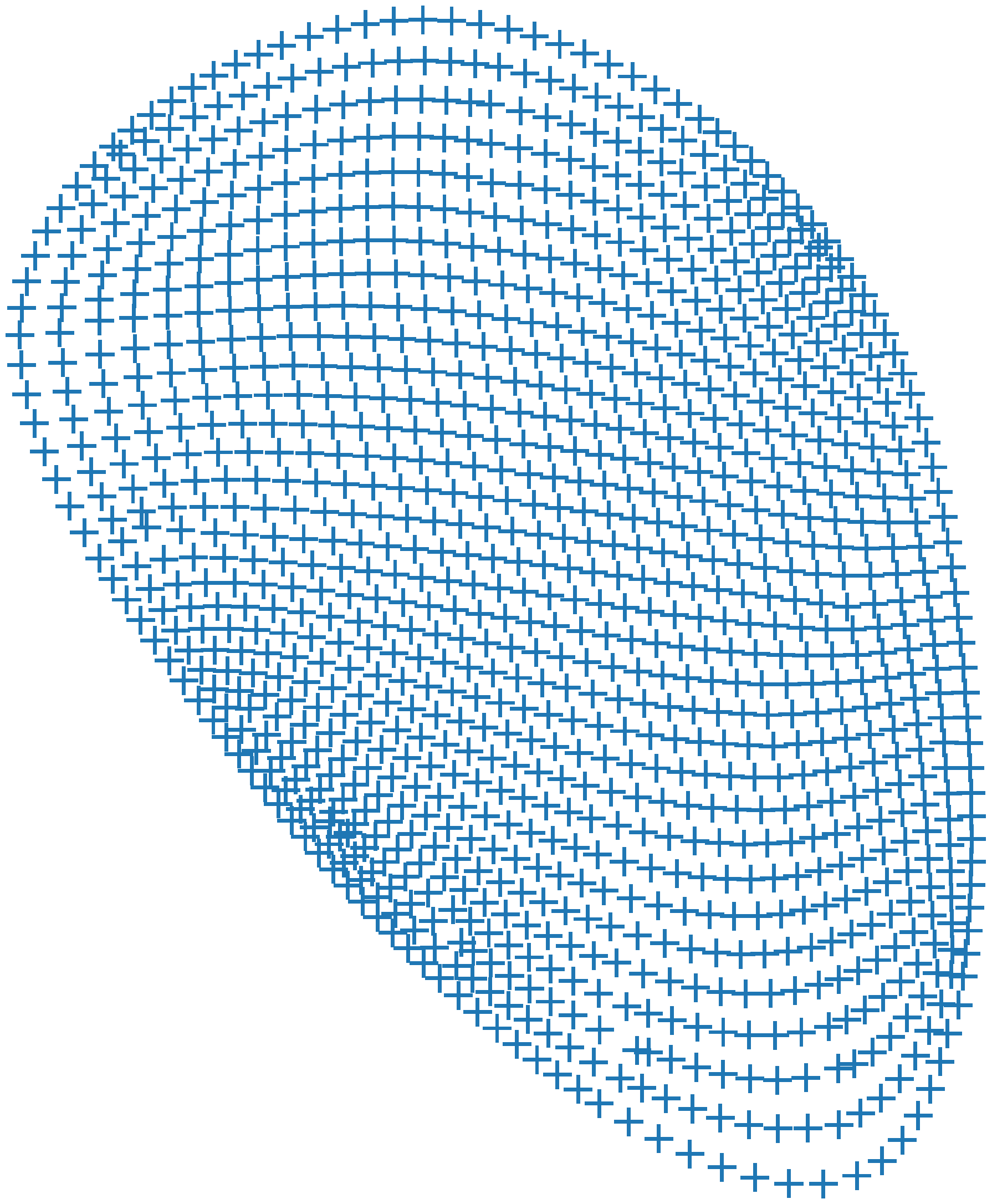}%
    \caption{Mesh points in a $33\times 33$ grid at the center cross-section of the biceps muscle.}%
    \label{fig:left_biceps_brachii_33x33fibers_bin_csv}%
  \end{subfigure}
  \quad
  \begin{subfigure}[t]{0.48\textwidth}%
    \centering%
    \includegraphics[width=\textwidth]{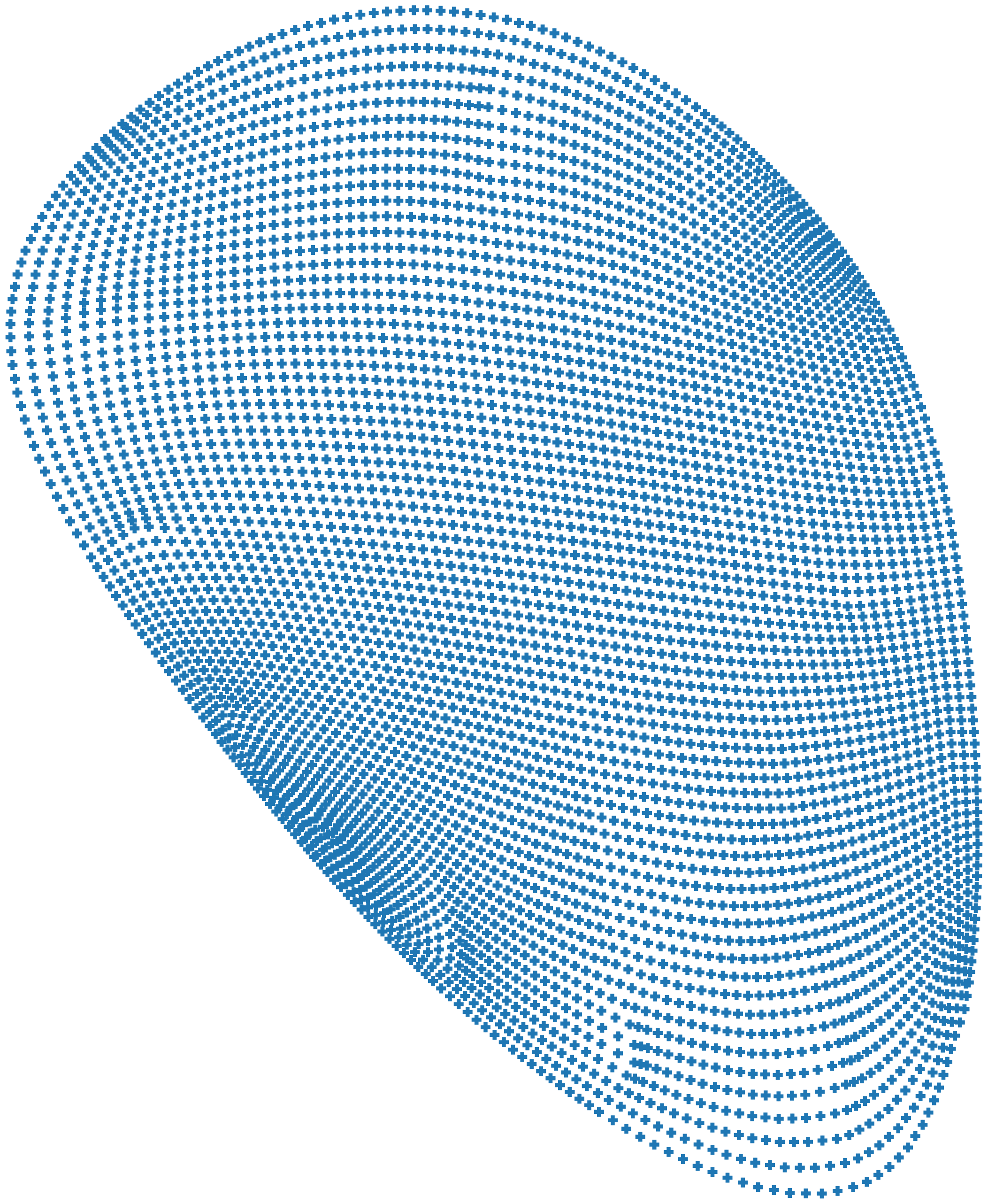} % also png
    \caption{Refined mesh points in a $71\times 71$ grid that were obtained from (a) by barycentric interpolation.}%
    \label{fig:left_biceps_brachii_71x71fibers_bin_csv}%
  \end{subfigure}   
  \caption{Refinement of existing meshes to obtain derived meshes with any number of nodes.}%
  \label{fig:left_biceps_brachii_33x33fibers_refined}%
\end{figure}

An advantage of this algorithm is that each process only has to keep the data of its own subdomain in memory at any time. This allows parallel processing of very large meshes. For small-enough meshes that do not fall under this restriction, the utility script \code{resample_bin_fibers.py} can be used to create meshes of any resolution from any other mesh using the barycentric interpolation in \cref{eq:mesh_barycentric_interpolation}. An example is given in \cref{fig:left_biceps_brachii_33x33fibers_refined}, where a mesh of  $33\times 33$ fibers is refined by interpolation to a mesh with $71\times 71$ fibers.

% seed points 33x33, 71x71 from scaling script

The resulting points are stored in a binary file format. The contents of this output file can either be interpreted as grid points of a 3D mesh or as points of individual 1D fibers. This is an advantage in a multi-scale simulation where both a 3D muscle mesh and multiple embedded 1D fiber meshes occur: First, all mesh information of both $\Omega_M$ and $\Omega_{F,i}$ can be given by a single file. And second, the 3D mesh is aligned with the 1D fibers and all 3D mesh points are also 1D mesh points. 

The spacing in $z$ direction between points on a fiber is typically chosen as $\Delta z = \SI{0.01}{\cm}$. This value was found to ensure a low error in the model for propagation of electric stimuli along the muscle. The value leads to 1481 points per fiber on the belly of the biceps muscle. 

Every point coordinate is stored in the output file as double precision value with eight bytes. The file contains a header of 72 bytes with descriptive information such as the number of fibers, some parameter values and a time stamp. The total file size therefore can be calculated by $72+N\cdot 1481\cdot 3\cdot 8$ bytes.

Often, the spatial resolution of the 3D mesh does not need to be as high as those of the fibers. The relation of the 3D and 1D mesh widths as well as the number of 1D meshes should be chosen such that the numerical error of the simulation in both domains is balanced.
In case the 3D mesh should be coarser than the output of the algorithm, we can use only a subset of the points contained in the output file. Then, a stride in $x$, $y$, and $z$ direction is specified in the settings for the simulation. The corresponding coarse grid of points is extracted and used to construct the 3D mesh that is then used for the simulation.
%The 1D fiber meshes use all given points in the file.
% When a 3D mesh with a smaller spatial resolution in $x$ and $y$ direction than the number of fibers is used in a simulation together with 1D fiber meshes from the same file, some of the fibers at the outer layer can be located outside of the 3D mesh. A way to avoid this is to not use the outer layer of fibers. 

\begin{figure}%
  \centering%
  \includegraphics[width=\textwidth]{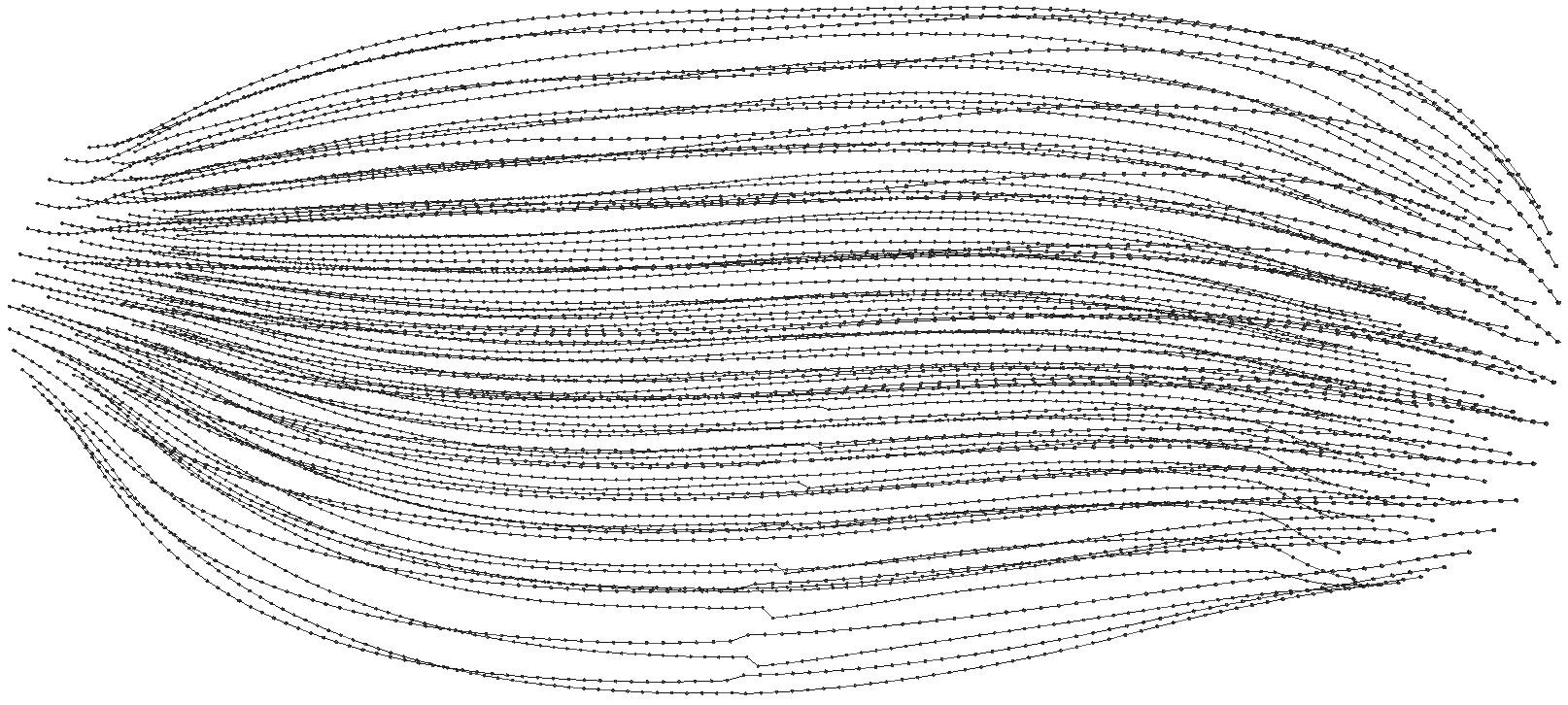}%
  \caption{Evaluation of the parallel mesh generation algorithm, \cref{alg:parallel_algorithm_1}: Resulting fibers and points on the fibers created with the parallel algorithm, $9\times 9$ fibers with $1481$ nodes each. Irregularities in the outer surface can be seen in the  center at the bottom of the image.}%
  \label{fig:poor_boundary_33x33}%
\end{figure}

A remaining issue concerns the mesh quality on the outer boundary. In general, the 3D mesh created by \cref{alg:parallel_algorithm_1} has good quality because the interior points result from smooth streamlines that were traced through a divergence free vector field. 
The points at the boundary, however, are either sampled from a triangulation of a tubular surface of the muscle or computed from the NURBS formulation. This surface is derived from imaging data, as described in \cref{sec:preprocessing_of_the_muscle_geometry}. If the triangulation is used, the quality of the boundary points of the created mesh depends on the quality of the muscle surface and its triangulation. In a case where this quality is poor, only the outer layer of elements of the created 3D mesh is affected. \Cref{fig:poor_boundary_33x33} shows an example for this effect in a grid of $9 \times 9$ fibers. It can be seen that only the fibers at the bottom of the image have an irregularity at their center. Such an irregularity potentially occurs at every $z$ coordinates where a new subdomain begins. The cause is that, at these locations, the points on the rings are slightly shifted relative to each other.

A remedy in such a case is to discard the outer layer of fibers and construct the mesh only from points of the inner streamlines.
Accordingly, our implementation of the presented algorithm \cref{alg:parallel_algorithm_1} always creates two different output files. The first output file contains all fibers, the second contains all except the outer layer of fibers. The second file contains only $N=(2\,n_\text{el,x}\,n_x\,(1+m)-1)^2$ instead of $N=(2\,n_\text{el,x}\,n_x\,(1+m)+1)^2$ fibers.

\section{Results and Discussion}

The following section presents results of the parallel algorithm for mesh generation, \cref{alg:parallel_algorithm_1}. In addition, the effect of various parameters is investigated.

Two types of parameters can be distinguished. Parameters of the first type influences the number of nodes in the resulting mesh. These parameters have to be set such that the desired mesh resolution is achieved. Often, multiple, different parameter combinations are possible to achieve a given mesh resolution.
Parameters of the second type have no effect on the mesh resolution but on the quality of the mesh. Usually, the parameter combination that gives the highest mesh quality should be chosen.

In the following, \cref{sec:mesh_generation_resulting_meshes} shows results of the algorithm. Then, \cref{sec:mesh_generation_mesh_size_parameters} outlines how parameters of the first type affect the mesh resolution. A specific parameter, the recursion width, is discussed in \cref{sec:mesh_generation_recursion_width}. Subsequently, \cref{sec:mesh_generation_mesh_quality_parameters} evaluates and discusses parameters of the second type, which affect the mesh quality.

\subsection{Resulting Meshes}\label{sec:mesh_generation_resulting_meshes}

At first, results of the whole workflow described in \cref{sec:overview_and_notation_of_required_meshes} to \cref{sec:parallel_algorithm} are presented.
The input for the mesh generation algorithm is a geometry representation, which is typically extracted from biomedical imaging. The output of the parallel algorithm, \cref{alg:parallel_algorithm_1}, comprises a 3D mesh with hexahedral elements as well as multiple, embedded 1D fiber meshes. 

\begin{figure}%
  \centering%
  \begin{subfigure}[t]{0.30\textwidth}%
    \centering%
    \includegraphics[height=7cm]{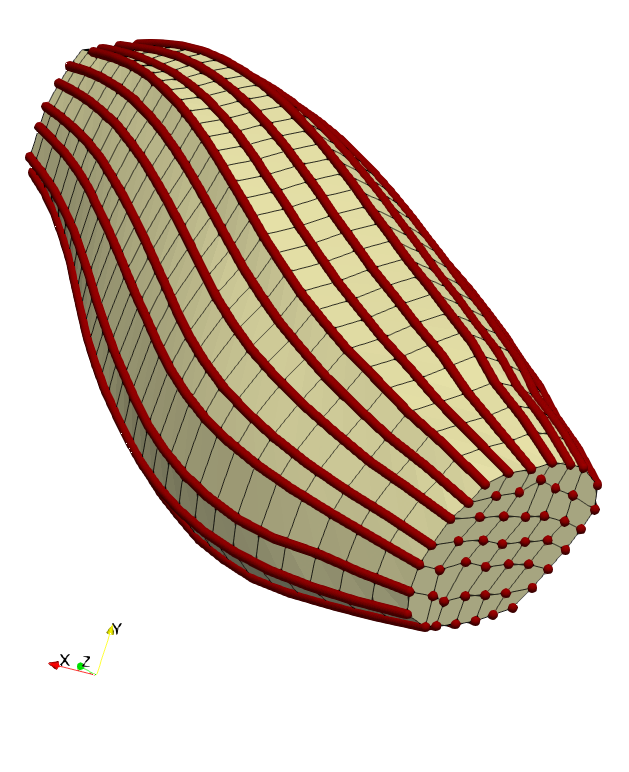}
    \caption{Grid of $7 \times 7$ fibers (red) and the aligned 3D mesh with $7 \times 7 \times 30$ nodes (yellow).}%
    \label{fig:muscle_mesh_0}%
  \end{subfigure}
  \quad
  \begin{subfigure}[t]{0.30\textwidth}%
    \centering%
    \includegraphics[height=7cm]{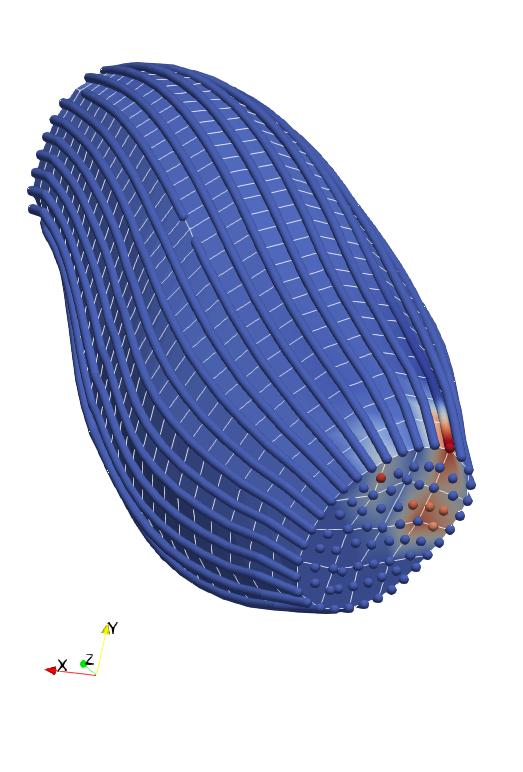}
    \caption{Grid of $9 \times 9$ fibers and 3D mesh with the solution of an electrophysiology simulation.}%
    \label{fig:muscle_mesh_1}%
  \end{subfigure}  
  \quad 
  \begin{subfigure}[t]{0.30\textwidth}%
    \centering%
    \includegraphics[height=7cm]{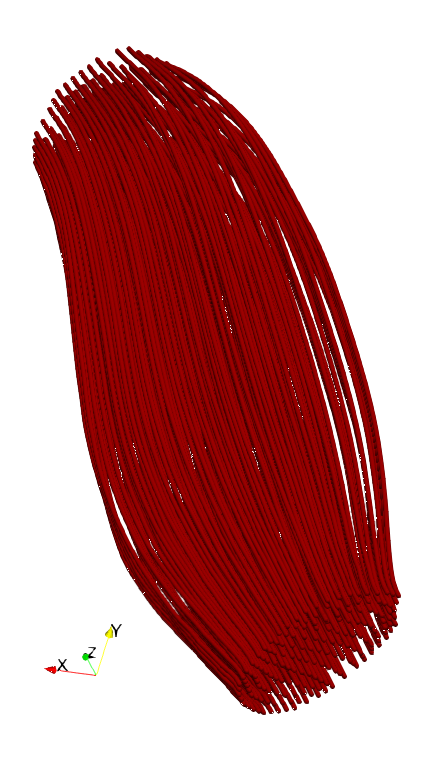}
    \caption{Grid of $13 \times 13$ fibers.}%
    \label{fig:muscle_mesh_2}%
  \end{subfigure}
  \\
  \begin{subfigure}[t]{0.25\textwidth}%
    \centering%
    \includegraphics[height=5cm]{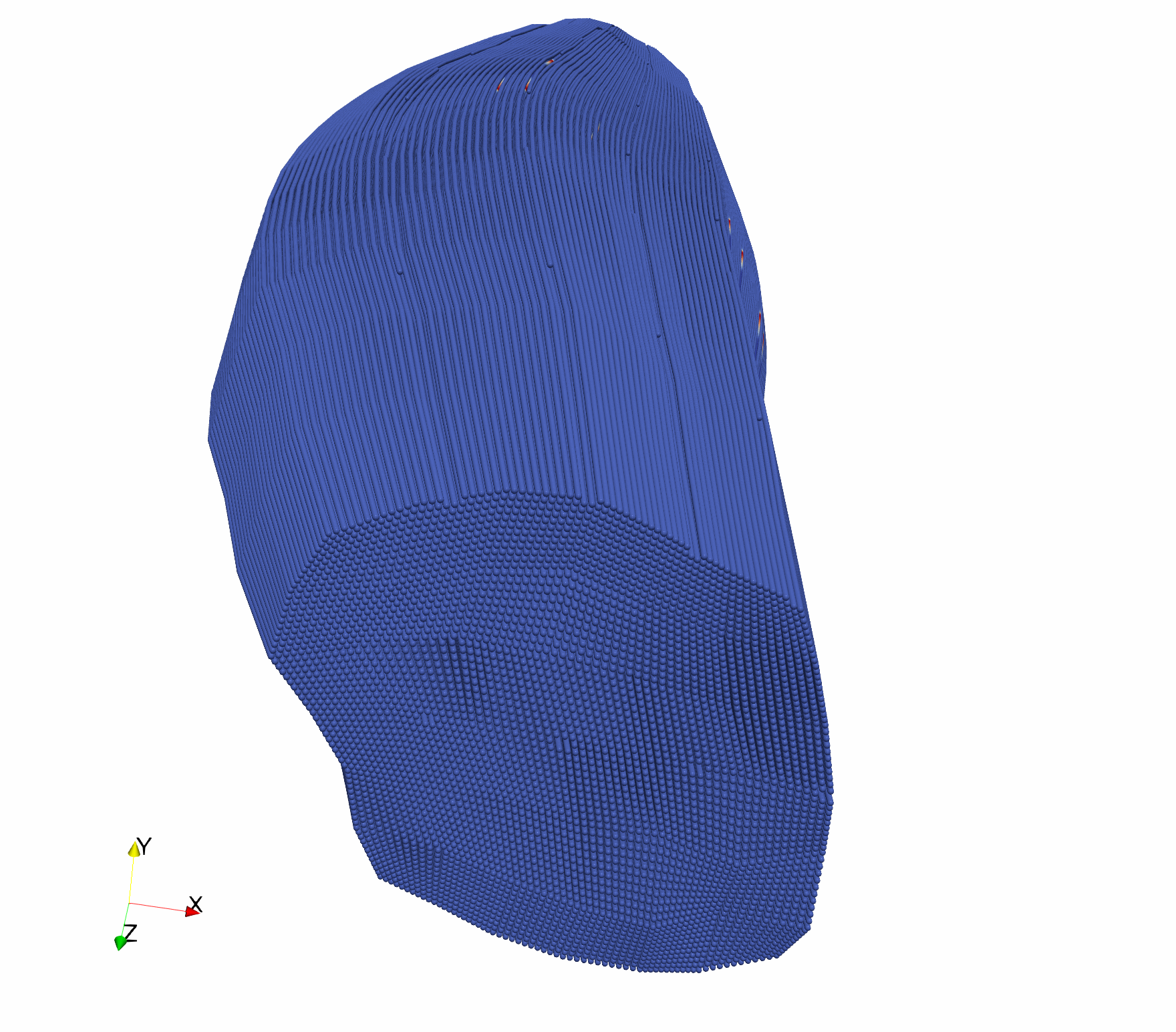}
    \caption{Grid $67 \times 67$ muscle fibers for the biceps geometry}%
    \label{fig:muscle_mesh_3}%
  \end{subfigure} 
  \,
  \begin{subfigure}[t]{0.15\textwidth}%
    \centering%
    \includegraphics[height=5cm]{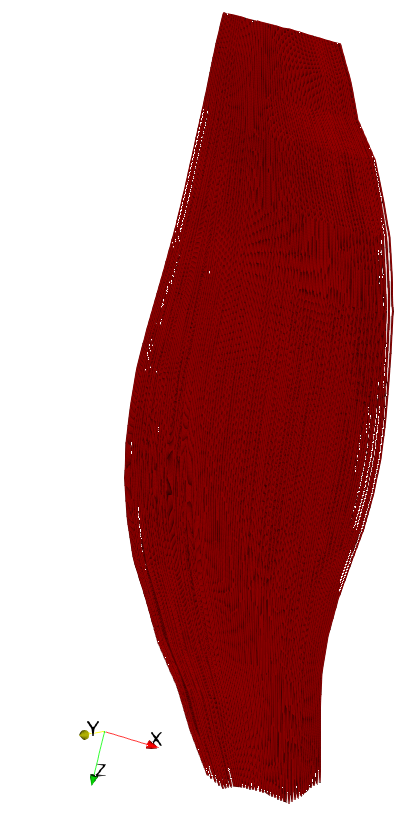}
    \caption{$25 \times 25$ fibers of triceps}%
    \label{fig:muscle_mesh_5}%
  \end{subfigure} 
  \hfill
  \begin{subfigure}[t]{0.55\textwidth}%
    \centering%
    \includegraphics[width=\textwidth]{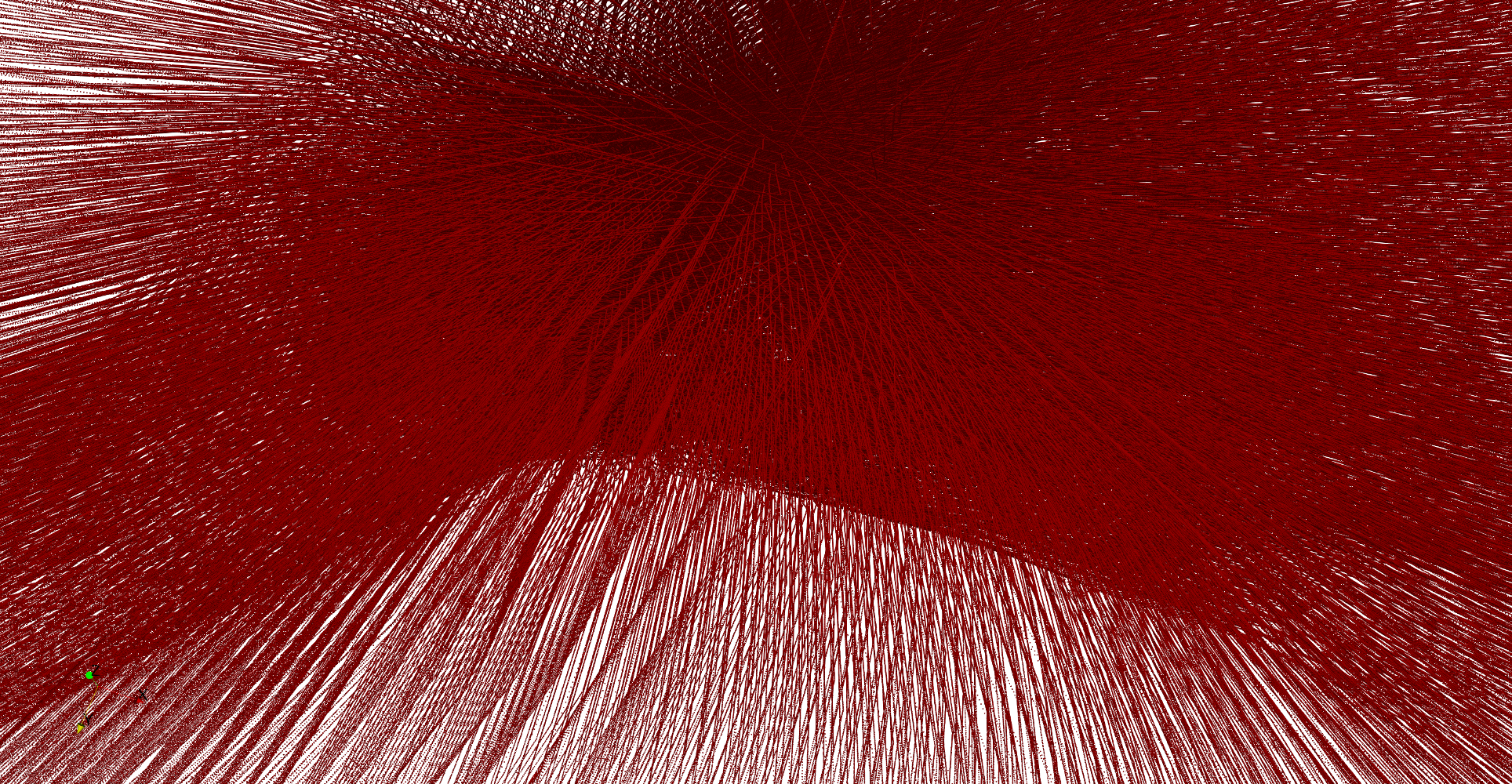}
    \caption{Grid of $67 \times 67$ fibers for the triceps geometry as seen from within the muscle. The total number of points is \num{8982489}.}%
    \label{fig:muscle_mesh_4}%
  \end{subfigure}  
   
  \caption{Evaluation of the parallel mesh generation algorithm, \cref{alg:parallel_algorithm_1}: 1D fiber meshes and corresponding 3D meshes. The biceps geometry is used in (a)-(d), the triceps geometry in (e) and (f). The fibers have 1481 nodes each in the biceps muscle and 2001 nodes each in the triceps muscle.}%
  \label{fig:muscle_meshes}%
\end{figure}

\Cref{fig:muscle_meshes} visualizes some results for the biceps and triceps muscles.
The parameter values $n_\text{el,x}=4$, $n_\text{el,z}=50$ and $m=0$ are chosen. If the recursion level is set to $l_\text{max}=0$, the algorithm generates meshes with the smallest possible number of fibers, which is a grid of $7 \times 7$ fibers.  
\Cref{fig:muscle_mesh_0} shows a grid of $7\times 7$ fibers and the corresponding 3D mesh that was sampled from the fiber data using every 50th point in $z$ direction of the fiber meshes. It can be seen that the generated fibers traverse all nodes of the generated 3D mesh and, thus, the 3D mesh is aligned with the fiber direction.

\Cref{fig:muscle_mesh_1} shows a similar result with $9 \times 9$ fibers. Here, the colors correspond to the solution of an electrophysiology simulation. Blue regions indicate that the fiber membranes have an electric potential equal to their resting potential, which indicates no activation. Orange and red colors correspond to activated regions. It can be seen that the activation is present at the same locations on both the fibers and the 3D mesh. In the simulation, this requires data mapping from the fiber meshes to the 3D mesh. Because all nodes of the 3D mesh are located on the fibers, this data transfer becomes trivial.

\Cref{fig:muscle_mesh_2,fig:muscle_mesh_3} present grids with $13 \times 13$ and $67 \times 67$ fibers of the biceps muscle, respectively. Results with larger numbers of fibers are not shown here because in such visualizations the fibers become less distinguishable. \Cref{fig:muscle_mesh_4,fig:muscle_mesh_5} show fibers for the triceps geometry.

\subsection{Effect of Mesh Size Parameters}\label{sec:mesh_generation_mesh_size_parameters}

Next, the type of parameters that affect the resulting mesh resolution is discussed.
The choices of the maximum recursion level $\ell_\text{max}$, the number $n_\text{el,x}$ of elements in $x$ direction of the subdomains and the fine grid parameter $m$ determine the resulting number $N$ of fibers and, thus, the file size of the binary output file. The formulas for these numbers were given in \cref{sec:postprocessing_of_the_generated_streamlines}. \Cref{tab:file_sizes} lists exemplary numbers of fibers and file sizes for $n_\text{el,x}=4$ and different values of $\ell_\text{max}$ and $m$. The number $n_\text{proc}$ of required processes to reach the maximum recursion level is also listed, it depends on $\ell_\text{max}$ by $n_\text{proc}=8^{\ell_\text{max}}$.

Two different numbers of fibers and corresponding file sizes are listed for every parameter combination. The two variants correspond to the two files that include respectively omit the fibers at the boundary.

The table shows that meshes with different sizes can be constructed by appropriate choices of parameters. A realistic biceps muscle contains about \num{200000} to \num{400000} muscle fibers \cite{MacDougall1984}. The table shows that constructing a mesh in this range yields a file with a size of $\approx$\SI{10}{\gibi\byte}.\footnote{
In this work, file sizes are given using multiples of bytes (B) and the prefixes defined in the ISO/IEC International System of Quantities \cite{ISOmebi}. The prefixes are: 1 kibibyte (\SI{1}{\kibi\byte})=$2^{10}$ bytes, 1 mebibyte (\SI{1}{\mebi\byte})=$2^{20}$ bytes, 1 gibibyte (\SI{1}{\gibi\byte})=$2^{30}$ bytes}
A mesh that contains \SI{1}{\percent} of the realistic number of fibers can be stored in a file with size of $\approx$\SI{100}{\mebi\byte}.

The binary files to store the generated meshes are small compared to ASCII-based file formats as each point coordinate is represented by only eight bytes. For comparison, the ASCII-based \emph{exnode} format defined within the OpenCMISS framework uses 24 characters, i.e., 24 bytes to store one point coordinate. Additionally, a larger memory overhead for the description of the data is needed such that \emph{exnode} files are more than three times larger than the binary files used in \opendihu{}.

The binary file format uses no compression that could further reduce the file size. The reason is that no extra effort should be needed when writing programs that parse these files. Thus, they can easily be handled by codes in different programming languages. For example, within \opendihu{} the file format is understood by various Python scripts and C++ programs.

\begin{table}
  \centering%
  \begin{tabular}{|l|l|l| r@{\,=\,}l r|rr|}
    \hline
    max.                      &fine grid     &\# proc. &  \multicolumn{2}{c}{\begin{tabular}{lrr} \# fibers && \end{tabular}}&file size \\
    level $\ell_\text{max}$   & $m$          & $n_\text{proc}$         & \multicolumn{2}{c}{\begin{tabular}{lll} &&\end{tabular}}&\\
    \hline
    0     & 0      & 1     & $9\times 9$      & \num{81} & \SI{2.7}{\mebi\byte} \\
          &        &       & $7\times 7$      & \num{49} & \SI{1.7}{\mebi\byte}\\
    0/1   & 1/0    & 1/8   & $17\times 17$    & \num{289} & \SI{9.8}{\mebi\byte}\\
          &        &       & $15\times 15$    & \num{225} & \SI{7.6}{\mebi\byte}\\
    0/1/2 & 3/1/0  & 1/8/64 & $33\times 33$    & \num{1089} & \SI{36.9}{\mebi\byte}\\
          &        &       & $31\times 31$    & \num{961} & \SI{32.6}{\mebi\byte}\\\hline
    0     & 7      & 1     & $65\times 65$    & \num{4225} & \SI{143.2}{\mebi\byte}\\
          &        &       & $63\times 63$    & \num{3969} & \SI{134.5}{\mebi\byte}\\
    2     & 7      & 64     & $257 \times 257$ & \num{66049} & \SI{2.2}{\gibi\byte}\\
          &        &       & $255 \times 255$ & \num{65025} & \SI{2.2}{\gibi\byte}\\
    2     & 15     & 64    & $513 \times 513$ & \num{263169} & \SI{8.7}{\gibi\byte}\\
          &        &       & $511 \times 511$ & \num{261121} & \SI{8.6}{\gibi\byte}\\
    \hline
  \end{tabular}
  \caption{Parallel 1D and 3D mesh generation: Different parameter choices of $l_\text{max}$ and $m$ and the resulting number $n_\text{proc}$ of processes, number of fibers and file size. Some results can be achieved with different parameter combinations, e.g., both $\ell_\text{max}=0, m=1$ and $\ell_\text{max}=1, m=0$ result in $17\times 17$ fibers. These combinations are separated by slashes.}%
  \label{tab:file_sizes}%
\end{table}

\subsection{Effect of the Recursion Width}\label{sec:mesh_generation_recursion_width}

Some numbers of fibers can be achieved with multiple, different parametrizations that use different recursion widths. \Cref{tab:file_sizes} contains such alternatives for $\ell_\text{max}$ and $m$ separated by slashes in the second and third row. For example, the three combinations $(\ell_\text{max} = 0, m=3)$, $(\ell_\text{max} = 1, m=1)$, and $(\ell_\text{max} = 2, m=0)$ all lead to a grid of $31 \times 31$ fibers (without boundary layer). However, the spatial location of the fibers in the muscle is not identical for these alternatives, because the intermediate mesh used for the streamline tracing of the fibers is differently resolved.
In the case with recursion depth $\ell_\text{max} = 0$ and fine grid interpolation parameter $m=3$, numerous of the resulting fibers are interpolated from a coarse grid whereas in the case with $\ell_\text{max} = 2$ and $m=0$ all fibers are key fibers and are obtained by streamlines tracing through a fine mesh.

\begin{figure}%
  \centering%
  \begin{subfigure}[t]{0.45\textwidth}%
    \centering%
    \includegraphics[width=0.9\textwidth,trim=0 0 8mm 1cm, clip]{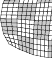}
    \caption{Result for parameters $l_\text{max}=0, m=3$, i.e., with three interpolated fibers between every two traced fibers.}%
    \label{fig:31x31_l0_center}%
  \end{subfigure}
  \quad
  \begin{subfigure}[t]{0.45\textwidth}%
    \centering%
    \includegraphics[width=0.9\textwidth,trim=0 0 8mm 1cm, clip]{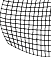}
    \caption{Result for parameters $l_\text{max}=2, m=0$, i.e., without interpolation.}%
    \label{fig:31x31_l2_center}%
  \end{subfigure}   
   
  \caption{Comparison of generated meshes of the biceps with different maximum recursion levels $l_\text{max}$. A lower left portion of the full mesh with $31 \times 31$ fibers is shown.}%
  \label{fig:different_recursion_levels}%
\end{figure}

\Cref{fig:different_recursion_levels} shows parts of the resulting meshes at the longitudinal center of the muscle for these two cases.
In \cref{fig:31x31_l0_center}, the mesh obtained with $l_\text{max}=0$ consists of a grid of traced key fibers and an interpolated finer grid of fibers. The key fiber grid  is given by the corners of the gray checkerboard pattern in the image. It can be seen that the mesh consists of patches with $4 \times 5$ or $5 \times 5$ fibers that each have equal element lengths and angles. In comparison, the mesh in \cref{fig:31x31_l2_center} that was obtained with $l_\text{max}=2$ consists only of key fibers. Here, the change in shape going from one element to its neighbors occurs more smoothly than in \cref{fig:31x31_l0_center}. This qualitatively implies a higher mesh quality.

To quantify this effect, we introduce a measure for mesh quality and compare the scores of the three alternatives in the present example. We consider all angles that occur in an element in the $x$-$y$ plane. The mean value of all angles is obviously $\pi/2$. The variance of all angles can be used as the measure for mesh quality. If the variance is low, this indicates similar elements and, thus, good mesh quality. 

\begin{figure}%
  \centering%
  \includegraphics[height=6cm]{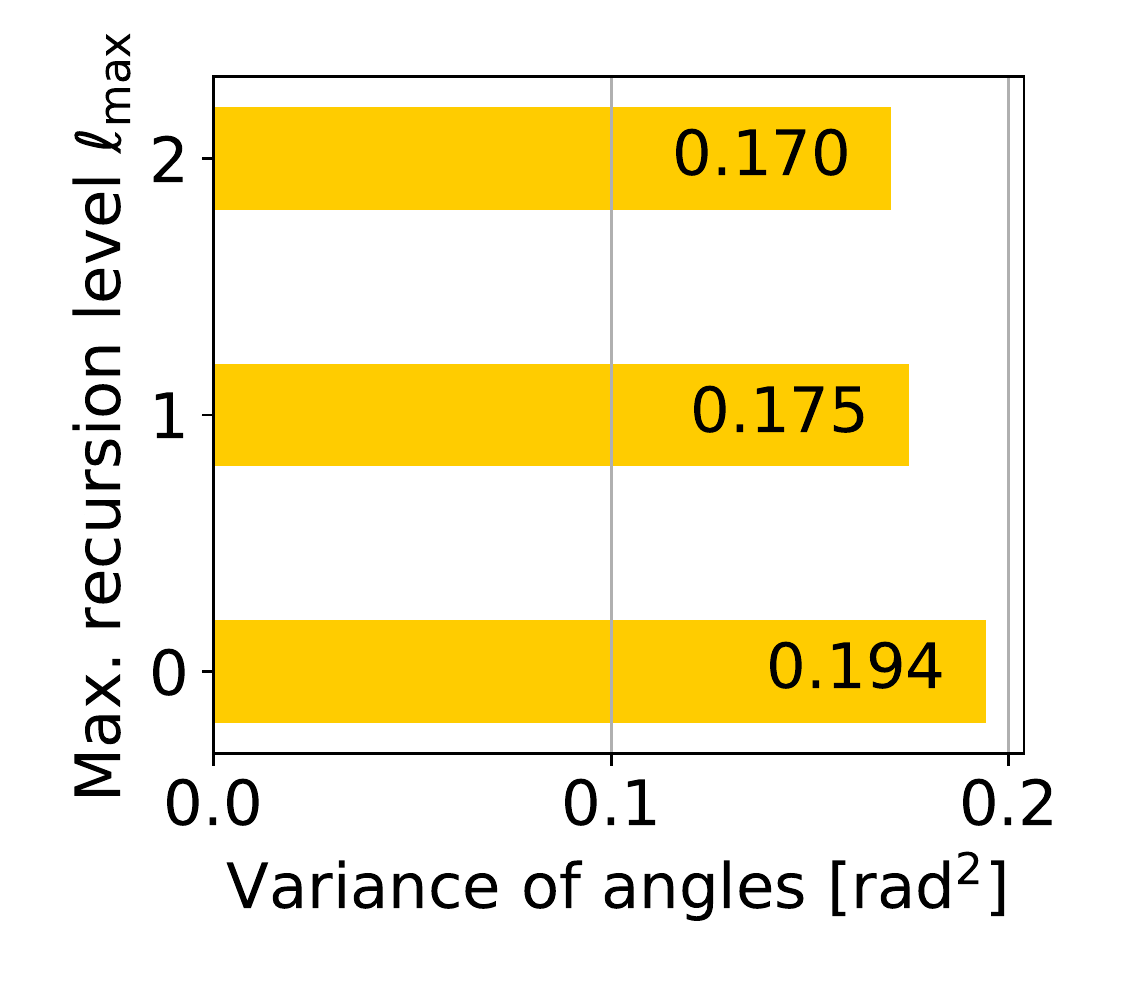}%
  \caption{Variance of the element angles for meshes with the same number of $31 \times 31$ fibers, but created by different recursion levels $\ell_\text{max}$. The parameters correspond to the third row of \cref{tab:file_sizes}. A lower variance means better mesh quality.}%
  \label{fig:2mesh_quality}%
\end{figure}

For the present example, the variance was computed for the mesh with $31 \times 31$ fibers and \num{1481} nodes per fiber and, thus, $\num{1332000}$ 3D elements in total.
\Cref{fig:2mesh_quality} plots the variance for the three cases given in the third row of \cref{tab:file_sizes}, i.e., parameter combinations $(l_\text{max}=0,m=3)$, $(l_\text{max}=1,m=1)$ and $(l_\text{max}=2,m=0)$. The 3D mesh corresponding to the lowest bar contains the 2D mesh shown in \cref{fig:31x31_l0_center} and the 3D mesh corresponding to the upper-most bar contains \cref{fig:31x31_l2_center}. 

It can be seen that the quality of meshes on higher recursion levels with less interpolation increases as expected.
This emphasizes the benefit of the parallel algorithm that uses finer meshes compared to the mesh used during serial execution of the algorithm.

\subsection{Effect of Mesh Quality Parameters}\label{sec:mesh_generation_mesh_quality_parameters}

In addition to the parameters that affect the resulting mesh sizes, $n_\text{el,x}$, $l_\text{max}$ and $m$, further options exist to tune the behavior of \cref{alg:parallel_algorithm_1} and in result lead to meshes with different quality. These options are described in the following.

The surface that is the input to \cref{sec:parallel_algorithm} can be represented either as triangulation or in parametric form as NURBS surface. The triangulation can either be the result of the image segmentation step or it can be obtained by triangulating a NURBS surface. Thus, if the approximation of the geometry by a smooth spline surface is desired 
it is possible to choose between both options. 

One difference is the resulting runtime. To sample a point on the surface using the NURBS representation, the nonlinear equation has to be inverted using a Newton scheme for each point. This is slower than using the triangulation where rings on $x$-$y$ planes are extracted initially and then equidistantly sampled, as explained in \cref{sec:slicing_of_the_geometry} and \cref{sec:data_structure_of_boundary_points}.

The Laplace problem $Δp=0$ that is solved in every recursion depends on the discretization and mesh resolution on every subdomain.
In addition to the number $n_\text{el,x}$ of elements in $x$ and $y$ directions, the mesh resolution also follows from the number $n_\text{el,z}$ of elements in $z$ direction.

The number of elements in this intermediate mesh is also influenced by the factor $r\in \N$ of the refinement described in \cref{sec:solution_of_the_laplace_problem}. While $r=1$ corresponds to no refinement, for $r>1$ the number of elements is increased by the factor $r^3$.
Note that the output meshes of the algorithm depend on $n_\text{el,x}$ but not on $n_\text{el,z}$ nor $r$ as they are generated later after the process of streamline tracing.

Furthermore, the finite element discretization of the Laplace problem can either use linear or quadratic ansatz functions, leading to the respective linear or quadratic elements in the mesh.
The type of boundary conditions for the Laplace problem in \cref{eq:fiberest_laplace} can be selected among the Neumann boundary conditions given by \cref{eq:fiberest_neumann} or the Dirichlet boundary conditions given by \cref{eq:fiberest_dirichlet}.

After the Laplace problem is solved, the gradient direction $\nabla p(\bfx)$ at a point $\bfx$ in the domain needed for streamline tracing can be determined by two different methods.
Either the gradient vector field is precomputed using finite differences and the nodal values of the solution field $p$ and then evaluated at $\bfx$.
Or the gradient value is directly interpolated at $\bfx$ in the 3D element using a linear combination of the solution values and derivatives of the ansatz functions of the element.

During parallel streamline tracing, the width $n_\text{ghost\_layer\_width}$ of the ghost layer is important. If it is too small, streamlines leave the domain of the process and have to be repaired, i.e., approximated by neighboring streamlines as described in \cref{sec:repair_of_incomplete_streamlines}. We found that a value of $n_\text{ghost\_layer\_width}=2\,r$ is enough and minimizes the number of invalid streamlines leaving the domain. With some parameter combinations, invalid streamlines still occur occasionally. Those result from badly conditioned elements and gradient values with high numerical errors that cannot be fixed by a larger ghost layer.
By including the factor $r$ in the ghost layer width, the actual sizes of the ghost layer is always the same independent of the chosen refinement.

To investigate the effect of all options, a parameter study is conducted in the following. We fix the values of $n_\text{el,x}=4$, $n_\text{el,z}=50$, $m=1$, $l_\text{max}=1$, and $n_\text{ghost\_layer\_width}=2\,r$ and vary all other parameters. The resulting meshes consist of $33 \times 33$ fibers and $31 \times 31$ fibers if the boundary layer is omitted, as described in \cref{sec:postprocessing_of_the_generated_streamlines}. We compare the mesh quality of the 3D meshes that result from the $31 \times 31$ fibers.
As before, the variance of the element angles is used to rate the quality of each resulting mesh.

To identify a parameter combination in the study, a scenario name is composed of one character each for the various options, as explained in the following.
The linear or quadratic formulation of the Laplace problem is indicated by the characters \say{$\ell$} or \say{q}. Neumann and Dirichlet boundary conditions are indicated by \say{N} and \say{D}. The refinement level $r$ is specified by the respective integer value. Finally, \say{g} or \say{s}  indicates whether the precomputed gradient field (\say{g}) is used in  streamline tracing or the solution values (\say{s}) and derivatives of the ansatz functions.

For example, the scenario considered in \cref{fig:different_recursion_levels,fig:2mesh_quality} can be specified as \say{$\ell$D2s}, as it uses the linear mesh with Dirichlet boundary conditions, a refinement factor $r=2$ and the solution values to compute the gradient.

The following study was performed for the biceps geometry in two variants, firstly using the approximated NURBS surface directly and secondly using a triangulation obtained from the NURBS surface. These two variants are indicated by \say{splines} for the NURBS surface and \say{stl} for the STL file containing the triangulation.

\begin{figure}%
  \centering%
  \begin{subfigure}[t]{0.48\textwidth}%
    \centering%
    \includegraphics[width=\textwidth]{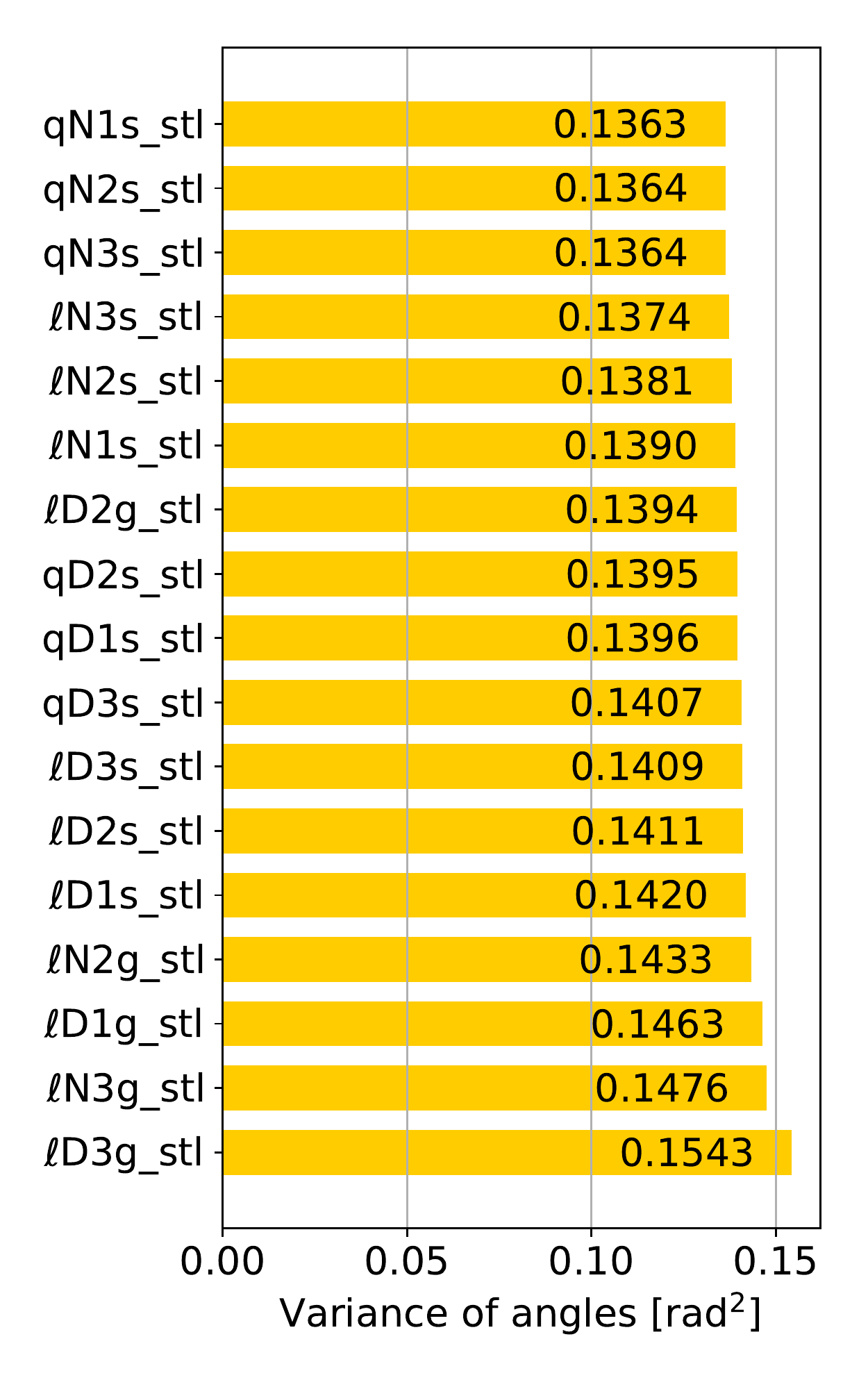}%    
    \caption{Scenario using the surface triangulation.}%
    \label{fig:mesh_quality_options_0}%
  \end{subfigure}
  \begin{subfigure}[t]{0.48\textwidth}%
    \centering%
    \includegraphics[width=\textwidth]{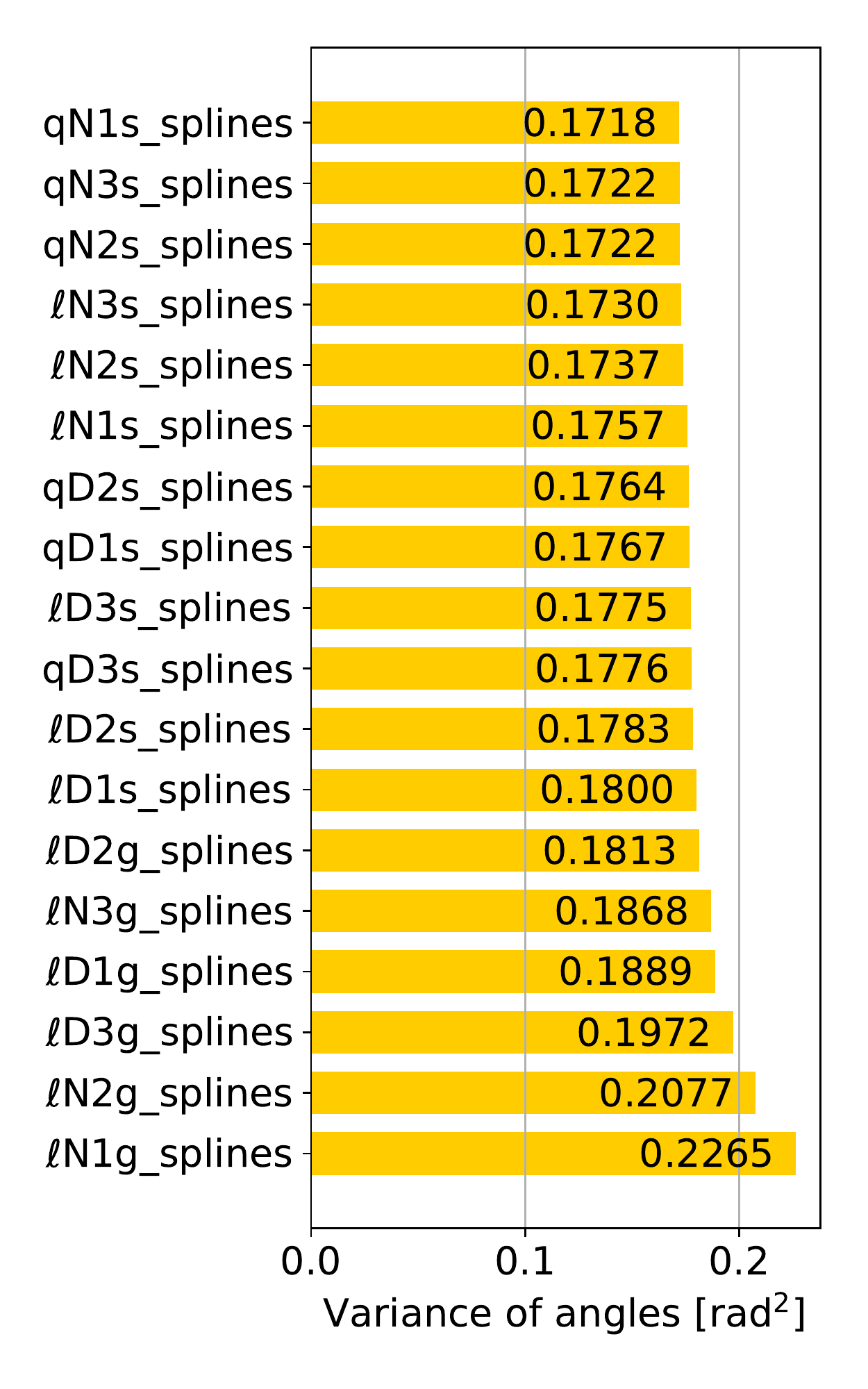}%
    \caption{Scenario using the spline surface.}%
    \label{fig:mesh_quality_options_1}%
  \end{subfigure}
  \caption{Parallel mesh generation algorithm: Comparison of the mesh quality that results from different options in the mesh generation algorithm. A lower variance means better mesh quality.}%
  \label{fig:3mesh_quality}%
\end{figure}

%Additionally, also the variance of relative element lengths in the $x$-$y$ planes was computed, using the calculation explained in \cref{sec:mesh_generation_0_results_and_discussion}. Most scenarios yielded a value of \num{2.2e-2}. Scenarios with significantly different values were discarded, as their results contained incomplete streamlines.
\Cref{fig:3mesh_quality} presents the resulting variances of the element angles.
The scenarios are sorted according to their mesh quality score, i.e., the variance of their element angles. This means the results are ordered by improving mesh quality from bottom to top.

\begin{figure}%
  \centering%
  \includegraphics[width=0.8\textwidth]{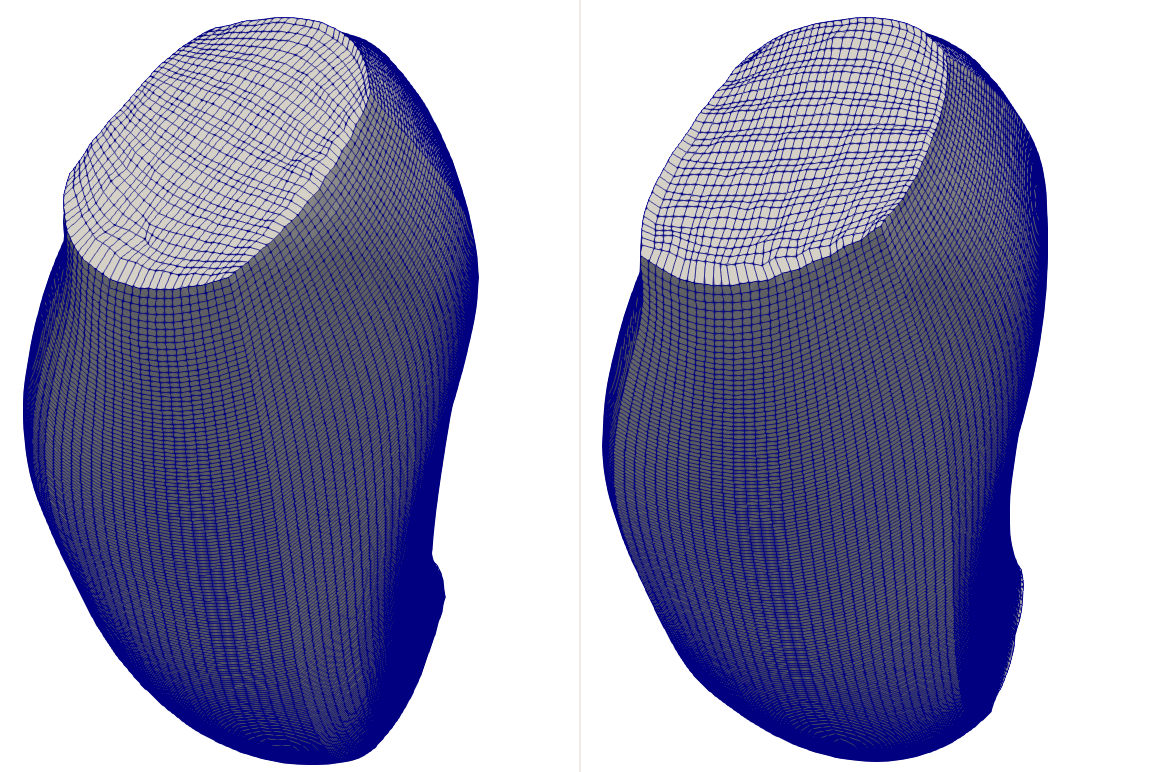}%
  \caption{Parallel mesh generation algorithm: Results of the scenarios using a surface triangulation (stl, left) and a NURBS surface (splines, right).}%
  \label{fig:stl_splines_results}%
\end{figure}

Two separate plots  for the \say{stl} and \say{splines} scenarios are shown in \cref{fig:mesh_quality_options_0} and \cref{fig:mesh_quality_options_1}.
The two resulting meshes of the best options, \say{qN1s\_stl} and \say{qN1s\_splines} are visualized left and right in \cref{fig:stl_splines_results}.
In the top plane of the muscle belly, it can be seen that the orientation of the mesh is slightly different. This explains the large difference of the angle variance values between the two scenarios in \cref{fig:3mesh_quality}, which are higher for \say{splines} than for \say{stl}. The scores of parameter combinations should therefore only be compared among the same surface representation. A statement regarding which of the two options is better is not reasonable from this data set.

The results in \cref{fig:3mesh_quality} show that almost all values are close together, which indicates similar good mesh qualities for different parameter combinations.
Nevertheless, the ranking reveals some differences between the options. 
A comparison of the rankings in the columns for \say{stl} or \say{splines} shows that some parameter choices consistently resulted in better meshes. 
A better result was achieved if Neumann boundary conditions were used (\say{N}) compared to Dirichlet boundary conditions (\say{D}). 
Similarly, quadratic ansatz functions (\say{q}) performed better than linear ansatz functions (\say{$\ell$}). This is reasonable as quadratic ansatz functions yield a higher spatial consistency in the finite element formulation. 

A higher refinement factor $r$ of the internal mesh was beneficial for the cases with linear ansatz functions. For quadratic ansatz functions, the effect of $r$ is less clear. The study shows that no refinement ($r=1$) is often better in this case.
The variant without the precomputed gradient field (\say{s}) performed generally better than the variant with gradient field computation (\say{g}).

However, further studies with different recursion depths showed that the effect of some options also depended on the scenario. For a higher recursion depth, Dirichlet boundary conditions turned out to be more robust in the sense that fewer incomplete streamlines occurred.
%Larger refinement factors led to smaller mesh widths and in consequence to a smaller ghost layer for the streamline tracing, which always has a thickness of one element. Thus, more streamlines left their subdomain and became invalid. Therefore, e.g., a smaller value of $r$ yielded better results for $\ell_\text{max}=2$.

In summary, the quadratic formulation (\say{q}) and the streamline tracing using solution values (\say{s}) could be shown to be better options than their alternatives. For a maximum recursion level $\ell_\text{max}=1$, the best parameter combination among the tested combination was \say{qN1s}. In a separate study for $\ell_\text{max}=2$, the combination \say{qD2s} was found to be as robust as \say{qN1s}.

\subsection{Post-processing of the Meshes}

To further improve the mesh quality on every cross-section, we apply two more post-processing steps, one local and one global transformation.
As can be seen in the cross-section of \cref{fig:stl_splines_results}, some rows of elements in the generated mesh have zigzag lines, and not all elements are equally sized. Furthermore, there are almost degenerate elements with small interior angles. 

Thus, we apply a first, local transformation on the mesh. This operation consists of Laplacian smoothing and randomly deflecting points, where interior element angles are smaller than \SI{20}{\degree}. If such deflections result in invalid self-intersecting elements, the self-intersection is resolved, which potentially again introduces small interior angles. The total transformation consists of 25 iterations of alternatingly applying the smoothing step and the improvement step of small interior angles.

\begin{figure}
  \centering%
  \begin{tabular}{cc}
    \begin{subfigure}[t]{0.30\textwidth}%
      \centering%
      \includegraphics[width=\textwidth]{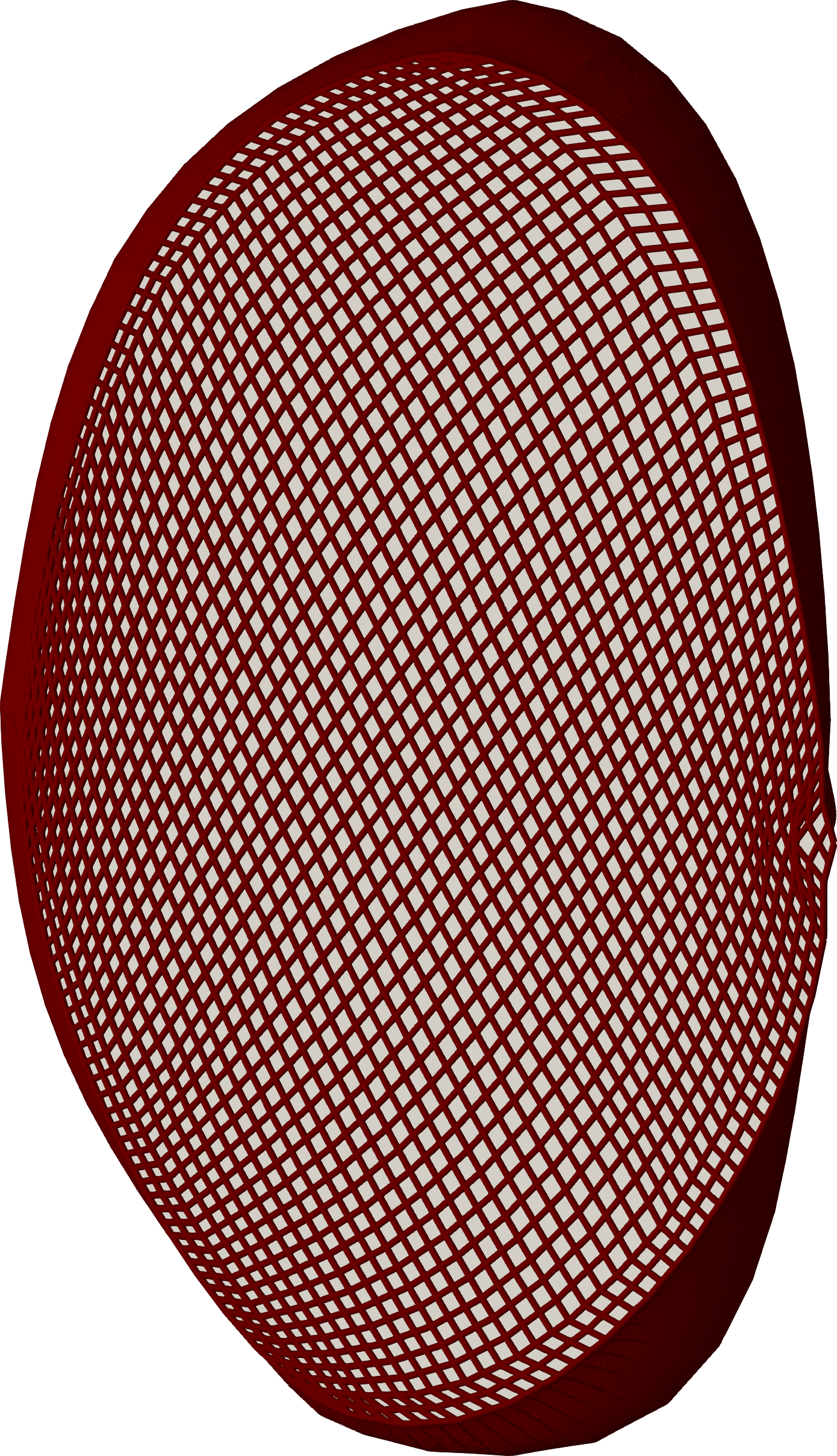}%
      \caption{Original mesh at the upper-most cross-section of the muscle.}%
      \label{fig:mesh47_a}%
    \end{subfigure}
    &
    \begin{subfigure}[t]{0.30\textwidth}%
      \centering%
      \includegraphics[width=\textwidth]{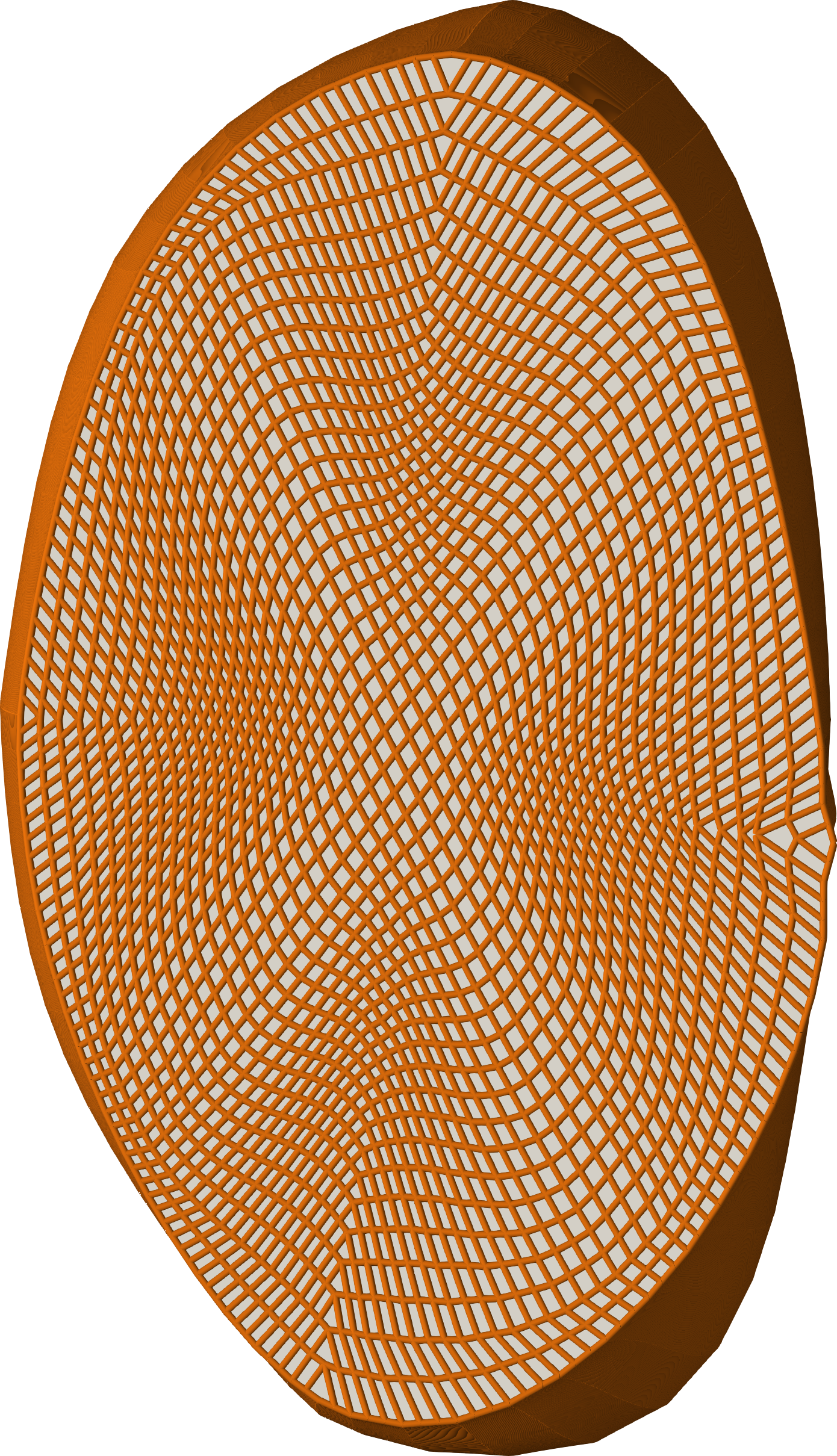}
      \caption{Transformed mesh with better mesh quality.}%
      \label{fig:mesh47_b}%
    \end{subfigure}  
    \\
    \begin{subfigure}[t]{0.48\textwidth}%
      \centering%
      \includegraphics[width=\textwidth]{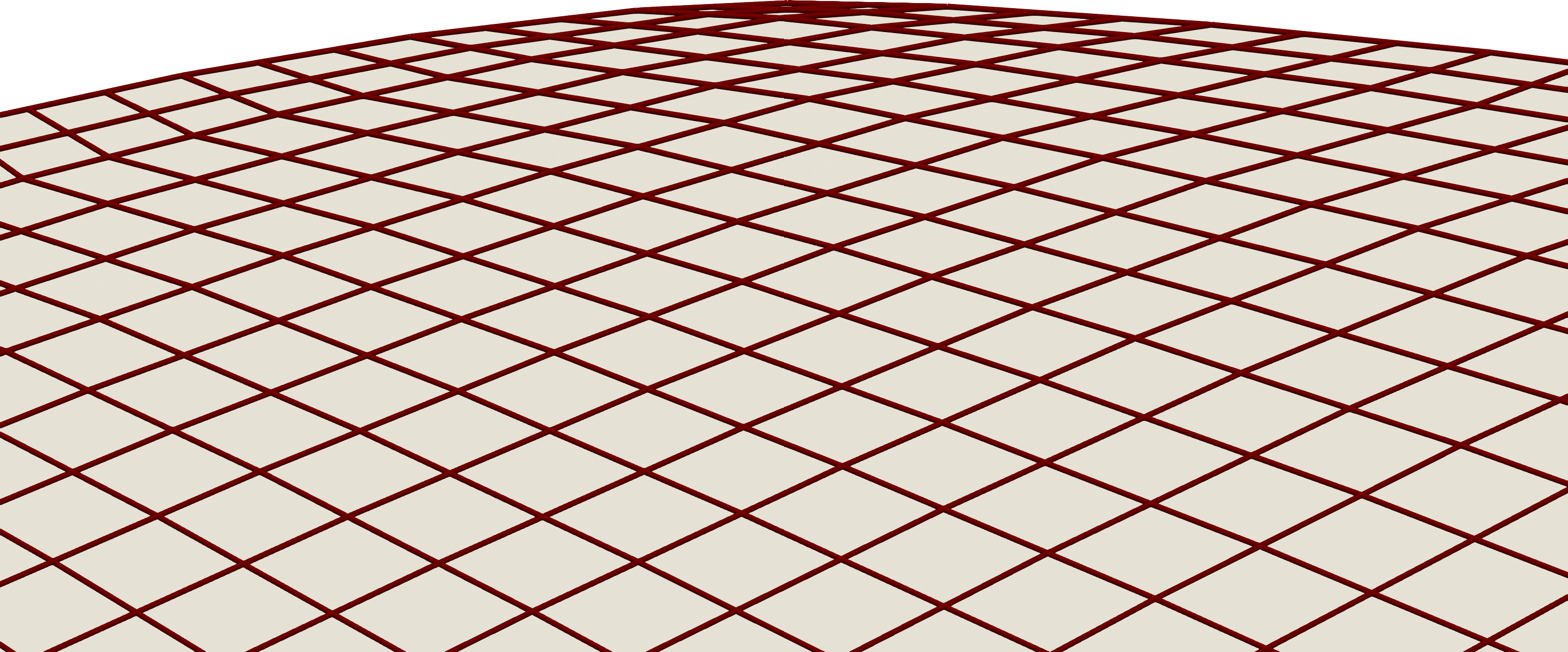}
      \caption{Extract at the left boundary of the original mesh in (a), rotated clockwise by \SI{90}{\degree}.}%
      \label{fig:mesh47a_}%
    \end{subfigure}
    &
    \begin{subfigure}[t]{0.48\textwidth}%
      \centering%
      \includegraphics[width=\textwidth]{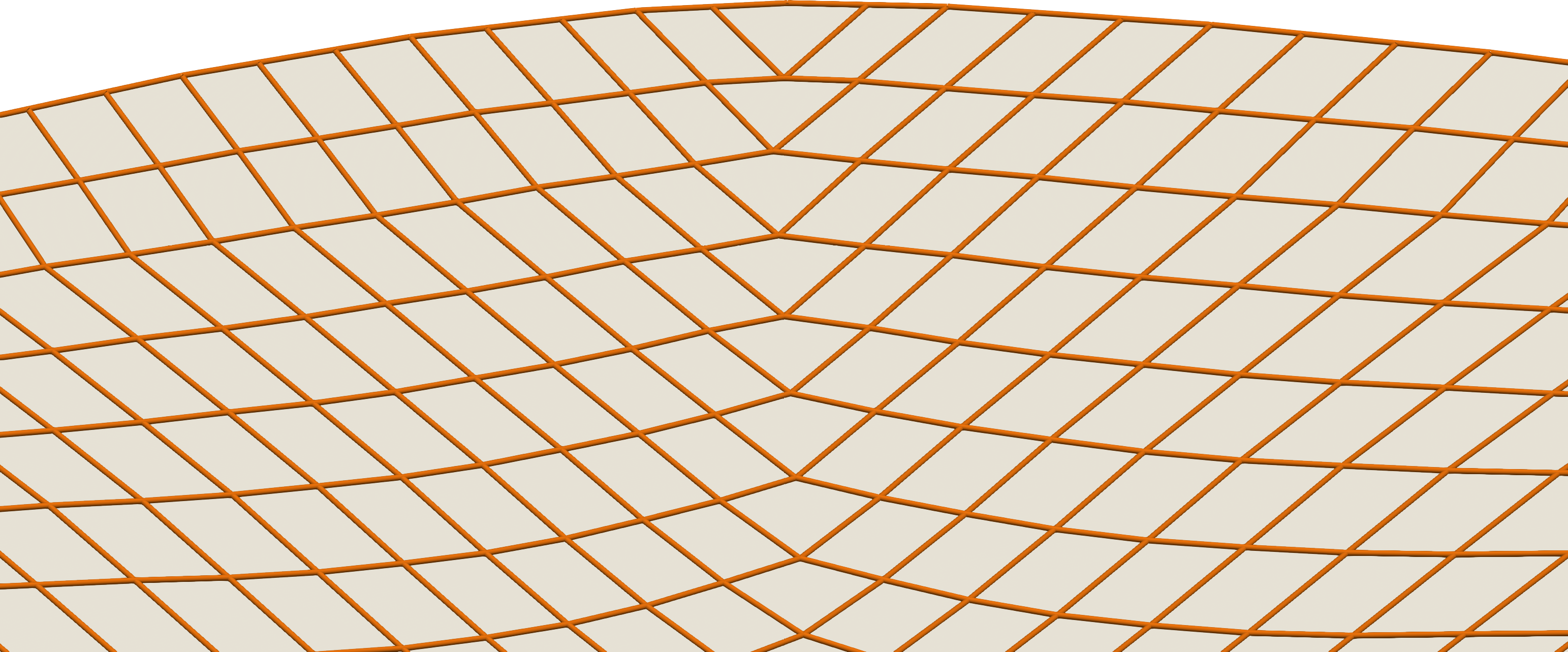}
      \caption{Analog extract to (c) of the transformed mesh (b).}%
      \label{fig:mesh47b_}%
    \end{subfigure}    
  \end{tabular}
  \begin{subfigure}[t]{0.7\textwidth}%
    \centering%
    \includegraphics[width=\textwidth]{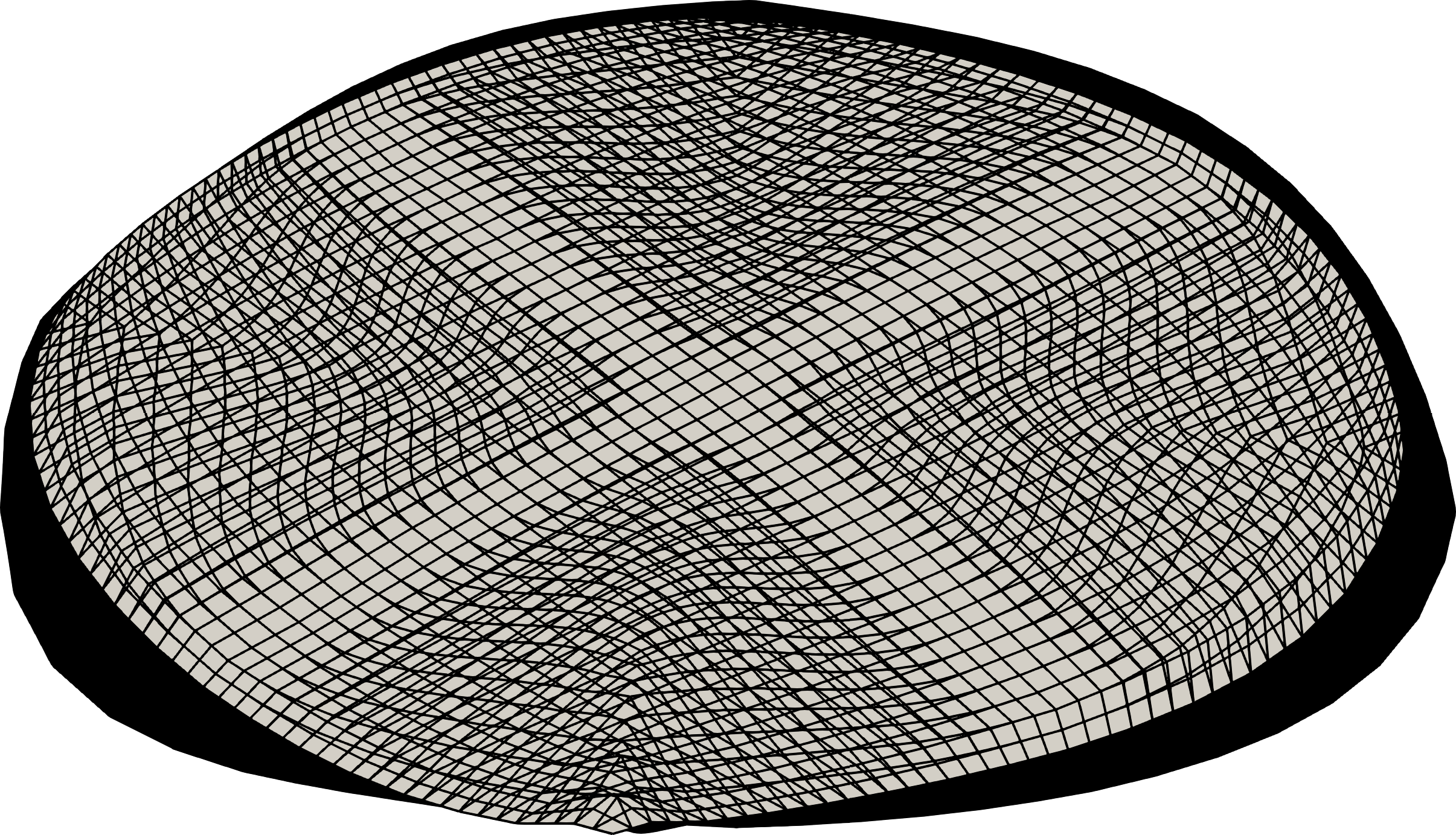}
    \caption{Overlay of the original mesh (a) and the transformed mesh (b), rotated clockwise by \SI{90}{\degree}. The unmodified regions can be seen.}%
    \label{fig:mesh_47ab}%
  \end{subfigure}
  \caption{Transformation of a biceps mesh with $47\times 47$ points to improve the mesh quality.}%
  \label{fig:mesh47}%
\end{figure}

An exemplary resulting mesh after this transformation is shown in \cref{fig:mesh47_a}. It can be seen that all lines in the mesh are smooth and almost straight. At the right center of the shown mesh, the effects of the deflection step, which improves small interior angles, can be seen. However, at the left, top and bottom boundary of the mesh, degenerate elements with small interior angles remain. This is especially true for the four mesh points on the boundary that correspond to the corners of the quadrangulation. At these points, elements are present that have two sides that are part of the mesh boundary, forming an interior angle of almost \SI{180}{\degree}. Three points of these elements are located in an almost straight line. As a consequence, the Laplacian smoothing step moves the fourth point of these elements close to this straight line, which adds another large interior angle and degenerates the element. This effect also occurs for interior elements of the mesh that are close to these points on the boundary.

\Cref{fig:mesh47_b} shows the detail of the left boundary of the mesh in \cref{fig:mesh47_a}, where this effect can be seen. The area of the elements decreases towards the boundary and the elements get more degenerate in this direction.

As a remedy, we perform the second, global transformation step to counteract this tendency. In this step, most of the points in every cross-section of the mesh are translated by a fixed mapping, such that the small elements at the boundary are transformed into elements of better quality.

% plot of function f(r)
\begin{figure}
  \centering%
  \includegraphics[width=0.5\textwidth]{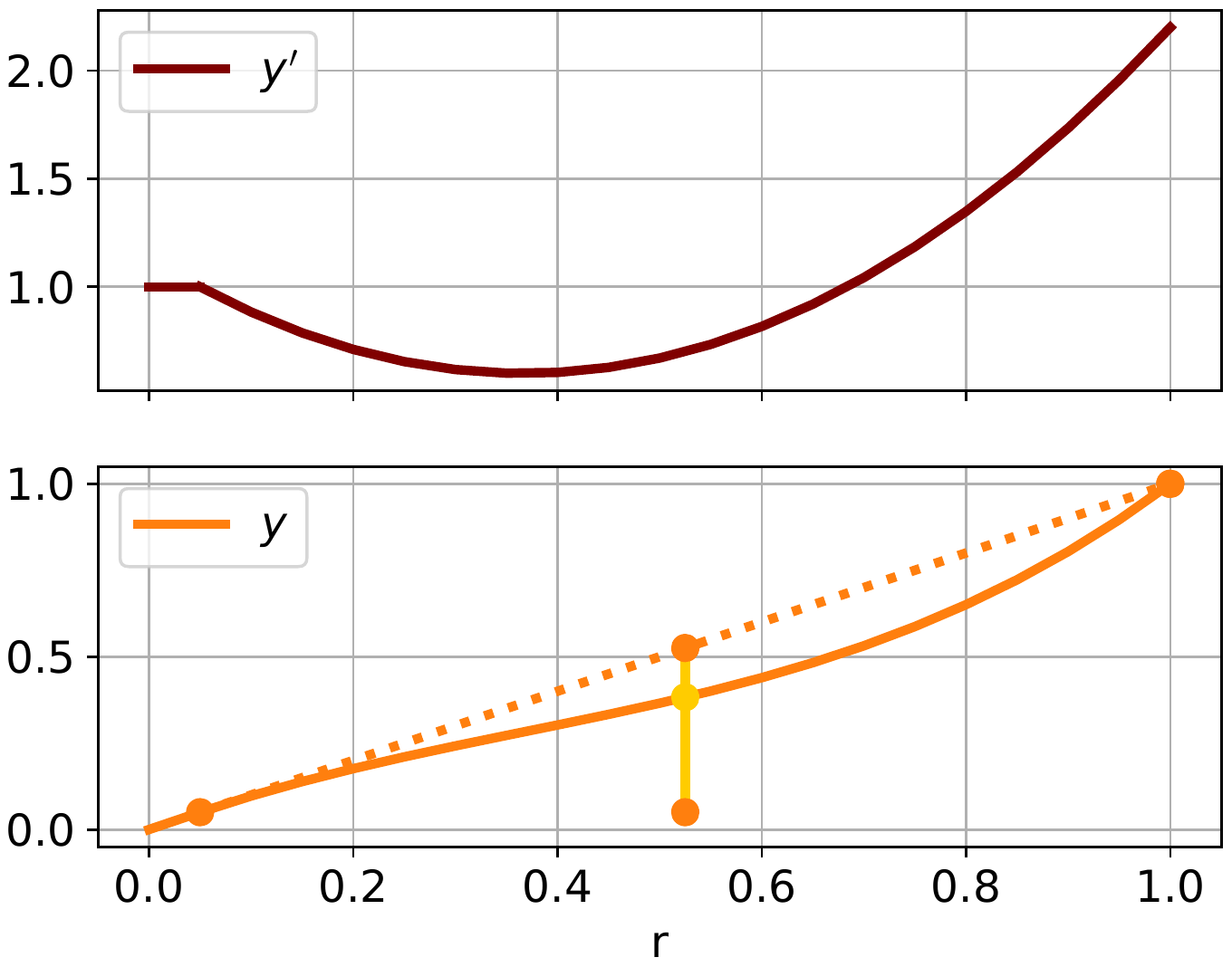}%
  \caption{Transformation of a mesh to improve the mesh quality. The depicted function $f$ transforms the elements in radial direction.}%
  \label{fig:extend_mesh_plot}%
\end{figure}%

Each mesh point on a cross-section of the mesh is represented by polar coordinates $(r,\varphi)$. The radius $r$ is transformed according to the function $r_\text{new} = y(r_\text{old})$ that is depicted in the lower plot of \cref{fig:extend_mesh_plot}. This piecewise defined function $f\in \CC^1([0,1]\to [0,1])$ is linear for $x\in [0,s]$ and a polynomial of degree 3 for $x \in [s,1]$. It passes through the yellow point in \cref{fig:extend_mesh_plot}, which in $x$ direction is at the center of the interval $[s,1]$ and in $y$ direction is at fraction $\alpha$ of the shown yellow line.
The parameter $\alpha$ controls the extent, to which the mesh is transformed in radial direction. The parameter $s$ specifies the size of a region around the center of the mesh where no transformation occurs. We obtain good results by choosing $s=0.05$ and $\alpha=0.7$. The resulting function takes the form $f(r) = 1.330\,r^3 - 1.463\,r^2 + 1.136\,r - 0.003$.

The first derivative $f'(x)$ of this function is shown in the upper plot of \cref{fig:extend_mesh_plot}. It quantifies the amount of extension or compression of the mesh elements in radial direction. The right side of the plot corresponds to the outer boundary of the mesh. There, the elements are extended, since $f'(r) > 0$. To compensate this extension, the elements have to be compressed towards the interior of the mesh  where $f'(r) < 0$. The range of $r \in [0,s]$ corresponds to the region in the interior of the mesh that is not transformed.

The points of the mesh are transformed in radial direction by adjusting their coordinate $r$ and not transformed in circumferential direction, i.e., the angle $\varphi$ remains constant. However, the application of the function $f$ on $r$ is additionally modulated by a piecewise sine function depending on $\varphi$. The transformation $f$ is only fully applied at the four radii corresponding to the described special points on the boundary, around which the degenerated elements occur. In between these lines, the transformation is reduced and some points of the mesh are not transformed at all.

\Cref{fig:mesh47_b} shows the mesh of \cref{fig:mesh47_a} after this transformation has been applied. \Cref{fig:mesh47b_} shows the extract of the mesh from the left boundary that corresponds to the same extract of the original mesh in \cref{fig:mesh47a_}.
It can be seen that the quality of the elements is improved close to the boundary. The area of the rectangles is now approximately equal and no small interior angles occur.

In \cref{fig:mesh_47ab}, the previous and the transformed mesh are overlaid to show the regions that remain unmodified. The unmodified parts form a \say{cross} shape that touches the boundary at the regions where the mesh quality is also good in the original mesh.

% original and transformed mesh

\subsection{Usage of the Generated Meshes in Simulations}

In some biomechanical simulations, also a body fat and skin layer on top of the muscle is considered. In this case, an appropriate mesh is required that is attached to the muscle mesh and seamlessly matches the elements of the muscle mesh. The construction of such a mesh is discussed later in \cref{sec:construction_and_partitioning_of_the_mesh} together with the parallel partitioning.

\begin{figure}
  \centering%
  \includegraphics[width=\textwidth]{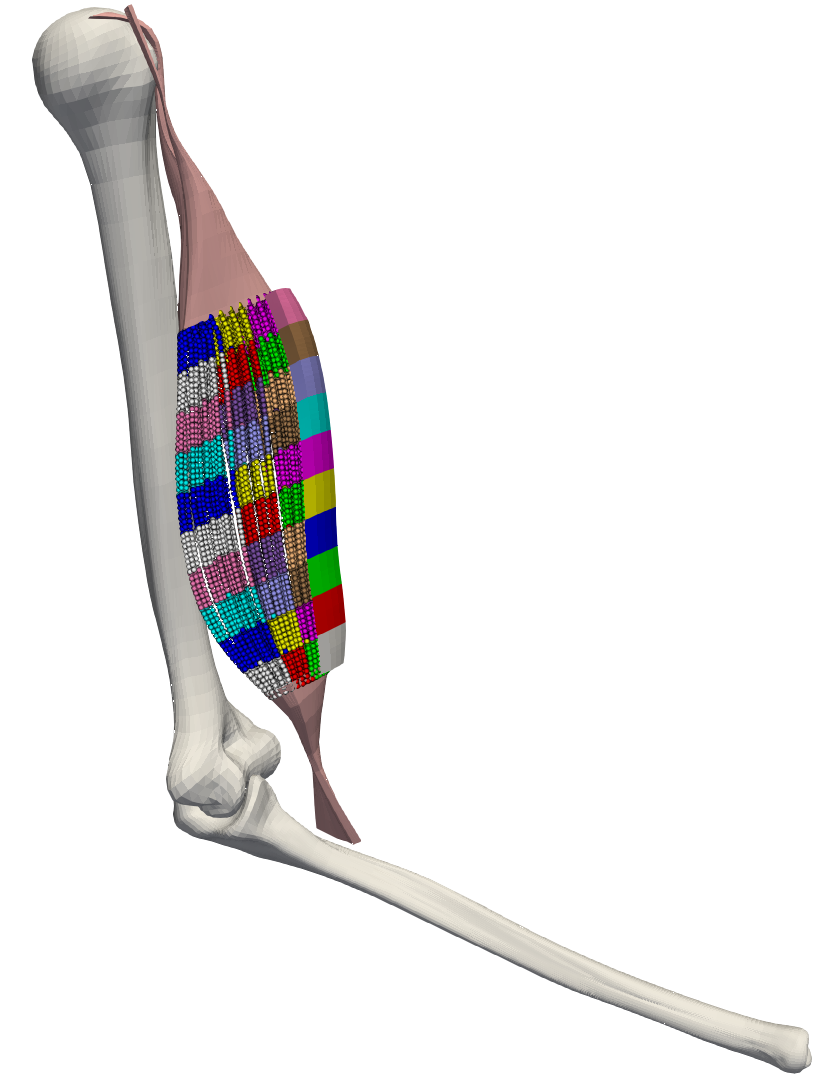}%
  \caption{Summary visualization of the simulation setup in this work (I): Biceps muscle with tendons and bones, parallel partitioned fibers and a fat layer mesh.}%
  \label{fig:partitioning_biceps}%
\end{figure}

A visualization of such a parallel setting is given in \cref{fig:partitioning_biceps}. Tendons connect the muscle belly of the biceps brachii muscle to the humerus and ulna bones at the top left and bottom, respectively. The muscle mesh consists of fibers that are colored according to a parallel partitioning. In a parallel partitioning, every process contributes calculations only on its associated spatial subdomain to the overall computation. On the right-hand side of the muscle, a layer of adipose tissue is attached to the muscle belly. This layer is needed, if electromyography on the skin surface is simulated. 

% fat layer mesh and partitioning

\begin{figure}
  \centering%
  \includegraphics[width=\textwidth]{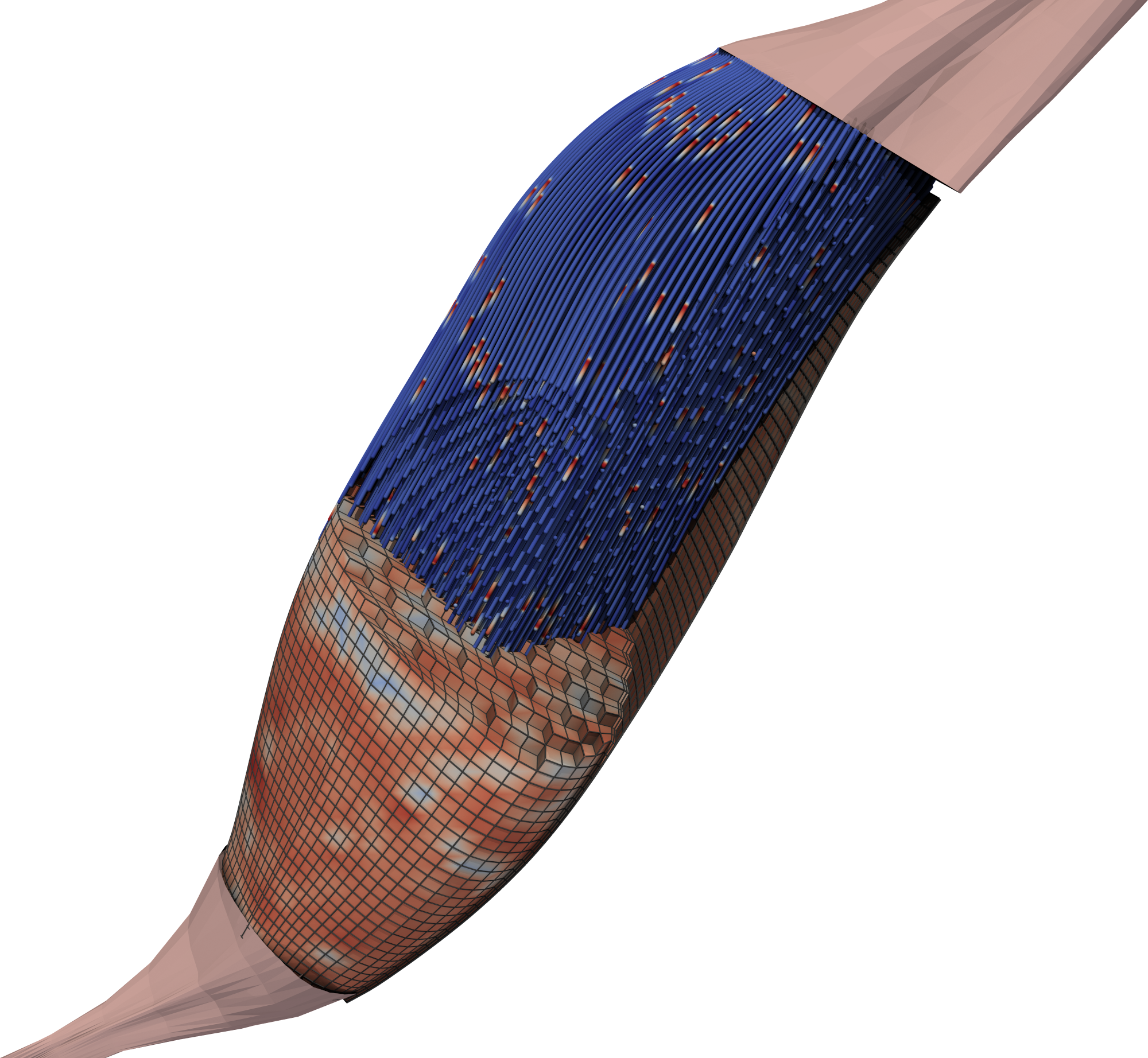}%
  \caption{Summary visualization of the simulation setup in this work (II): Muscle fibers in the upper part and 3D mesh elements in the lower part of the muscle belly.}%
  \label{fig:muscle_meshes_raytrace}%
\end{figure}

\Cref{fig:muscle_meshes_raytrace} shows another use case of various meshes in a multi-scale simulation model. The muscle is cut open for visualization purposes. The figure depicts numerous muscle fibers in the upper part of the muscle belly. The fibers are colored according to simulation results of the electric membrane potential, which  is responsible for the activation of the muscle. The lower part of the muscle shows elements of the 3D mesh. The coloring corresponds to the electric potential, that is measured during intramuscular electromyography in the interior of the muscle or during surface electrophysiology on the outside of the muscle.

% 3D and 1D meshes
\clearpage

%19G     resulting_meshes
%The created meshes have been written to "resulting_meshes". Now you can copy these files to the examples/electrophysiology/input directory

%real    1047m29.662s
%user    3970m33.404s
%sys     4m6.717s

\label{sec:repro_tendon_meshes}
\begin{reproduce}
  The parallel algorithm is implemented in the example \code{parallel_fiber_estimation}. 
  Numerous parameters can be set on the command line. After compilation, run the program as follows to get a description of all available options.
  \begin{lstlisting}[columns=fullflexible,breaklines=true,postbreak=\mbox{\textcolor{gray}{$\hookrightarrow$}\space}]
    cd $\$$OPENDIHU_HOME/examples/fiber_tracing/parallel_fiber_estimation/build_release
    ./generate ../settings_generate.py --help
  \end{lstlisting}
  Running the program without options and \code{--help} uses sensible default values. A given surface triangulation of the biceps muscle gets used by default. To compute the examples shown in this section, use and adjust the following script that runs the \code{generate} program and computes the mesh quality:
  \begin{lstlisting}[columns=fullflexible,breaklines=true,postbreak=\mbox{\textcolor{gray}{$\hookrightarrow$}\space}]
    cd $\$$OPENDIHU_HOME/examples/fiber_tracing/parallel_fiber_estimation/build_release
    ../run.sh
  \end{lstlisting}
  Computation of mesh and file sizes as shown in \cref{tab:file_sizes} can be done using the \emph{compute\_sizes.py} script.
  
  While the previously given commands are good for exploring the algorithms, generation of the meshes used for the simulation involves some more steps. Dedicated scripts exist that perform all steps and call the algorithms with the proper parameters.
  Starting from the STL file extracted from cmgui, as explained in \cref{sec:surf_extr}, the next steps are: 
  \begin{itemize}[leftmargin=1cm]
  \item[(i)] Scale the points from millimeters (used in the Visible Human dataset) to centimeters (used in the simulation), 
  \item[(ii)] remove the interior triangles, 
  \item[(iii)] translate the mesh such that the bounding box begins at $z=0$, this is needed for the programs used in the next steps, 
  \item[(iv)] create the spline surface representation as explained in \cref{sec:nurbs}, 
  \item[(v)] compile and run the \opendihu{} programs to create the binary files of the 3D mesh and the 1D fibers meshes, the algorithm in \cref{sec:parallel_algorithm} is used, 
  \item[(vi)] adjust the indexing and undo the translation in (iii), \item[(vii)] refine the created meshes of key fibers by different numbers $m$ of fine grid fibers, in total 10 different mesh sizes are created for differently refined simulations, 
  \item[(viii)] create meshes for the fat layer $\Omega_B$ on top of the muscle surface, also in 10 different resolutions.
  \end{itemize}
  
  Two scripts are given for the biceps brachii and triceps brachii muscles. They perform all listed steps and also create intermediate output files that can be used to understand the process. Some steps are automatically skipped if the resulting output file already exists from a previous run. This is especially helpful for the removal of the interior triangles from the initial file which takes nearly a full day.
  
  A third scripts creates three meshes $\Omega_{T,i}$ for the tendons of the biceps muscle, as visualized in \cref{fig:tendon_meshes}. At the bottom, a single tendon mesh is created whereas at the top, two separate tendons exist. The script involves numerous rotation and cropping operations of the initial surface, before the algorithm of \cref{sec:ser_alg_meshes} is executed. The three output files of the tendon meshes have the same file format as the muscle meshes. The files have the extension \emph{.bin} for \say{binary}. The script \code{examine_bin_fibers.py} can be used to debug the created binary files.
  
  The three scripts can be executed as follows:
  \begin{lstlisting}[columns=fullflexible,breaklines=true,postbreak=\mbox{\textcolor{gray}{$\hookrightarrow$}\space}]
    cd $\$$OPENDIHU_HOME/examples/electrophysiology/meshes
    ./process_meshes_biceps.sh
    ./process_meshes_triceps.sh
    ./process_meshes_tendons.sh
  \end{lstlisting}
  The output can be found in the subdirectory \code{processed_meshes}. For the total output about \SI{68}{\gibi\byte} of drive space is required, however, the resulting meshes have a size of only \SI{19}{\gibi\byte}. A total runtime of more than a day is to be expected.
  
\end{reproduce}

\section{Conclusion and Future Work}\label{sec:meshes_summary_and_conclusion}

This chapter presented algorithms for creating muscle meshes that are needed for multi-scale simulations of the musculoskeletal system.
For the biceps muscle, 3D meshes for tendons on both ends and the muscle were created. Additionally, 1D fibers meshes were generated that are embedded in the mesh of the muscle. The 3D mesh and the 1D meshes resulting from the parallel algorithm are aligned with each other. This facilitates data mapping between the meshes and reduces numerical errors. All generated meshes are structured, which allows an efficient parallelization.

First, an overview of available meshing software and known algorithms in the literature was given. Very little software tools were capable of generating structured meshes and none fitted our special needs. Therefore, own algorithms were developed to generate meshes starting from medical imaging data.

A workflow was presented to generate a smooth surface triangulation from imaging data. Our base data was the male dataset from the Visible Human Project. Two alternatives within this workflow were presented, where the first alternative executed automatic image segmentation based on morphological operations and the second alternative used semi-automatic segmentation tools from the Physiome project. Then, smooth NURBS surfaces were fitted to the extracted boundaries of the muscle volumes.

Next, a novel algorithm to create structured meshes from a triangulated muscle surface was presented. The algorithm used harmonic maps on 2D slices in combination with regular grids in a parameter space to achieve good mesh quality. A method of computing streamlines in a divergence-free vector field to estimate muscle fibers, which is established in the literature, was used. It allowed embedding 1D meshes for muscle fibers in the created 3D meshes of the muscle. Numerical experiments tested and evaluated different choices of triangulation and quadrangulation schemes for the 2D cross-section and reference domains in our algorithm.

Next, a parallelized algorithm was introduced that was based on our first, serial algorithm. The algorithm used distributed memory parallelism and provided the same features as the serial algorithm, having the same formats for input and output. The difference was that it constructed a fine, partitioned mesh for streamline tracing that was distributed over all employed processes. Thus, it was possible to create finer meshes using more compute nodes. Differently resolved meshes of the biceps and triceps muscle volumes and muscle fibers were created using this algorithm. The superiority of the parallel algorithm using a higher number of processes compared to the serial execution was explained and demonstrated in a numerical experiment. Several options to fine-tune the algorithm were evaluated. Post-processing methods were described that improved the mesh quality of the resulting meshes.

The presented algorithms and their implementation in \opendihu{} are the basis for further computations within this work. They are used to generating structured hexahedral meshes with good mesh quality. These meshes are required for efficient, parallel finite element simulations of various aspects of the neuromuscular system.

The presented algorithms are specialized for fusiform muscles and require the muscle geometry to be oriented along one coordinate axis (the $z$ axis) in order to generate a structured mesh  that comprises planar slices that are normal to that direction.
The algorithms can also be applied to any tubular surface geometry of more complex muscles and will construct the corresponding structured 3D mesh. The generated 1D fiber meshes, however, are only valid for muscles, where the approach of streamline tracing through the solution of the Laplacian potential flow problem with boundary conditions at the bottom and top ends of the muscle can be applied.
In literature, this approach has been successfully used for various muscles with more complex fiber architectures, such as the tibialis anterior, gluteus maximus and deltoid muscles \cite{Choi2013}. However, the locations where boundary conditions were prescribed was not always at the bottom and top ends of the muscle.

If in future work muscles with more complex layouts should be simulated, the approach could be as follows. Depending on the complexity of the outer geometry, first the presented algorithms (either \cref{alg:serial_algorithm_1,alg:serial_algorithm_2} or \cref{alg:parallel_algorithm_1}) can be used to create a structured 3D mesh. Then, a potential flow simulation can be manually setup in \opendihu{} using the 3D mesh and boundary conditions defined at proper locations. Seed points have to be defined and the streamline tracer of \opendihu{} can be used to create fiber meshes. In consequence, the resulting 1D fibers will not be aligned with the 3D mesh. Algorithmically, this poses no problem to the simulations in \opendihu{} as the data mapping functionality can handle arbitrarily positioned meshes. However, the parallel partitioning gets more involved as the combined domain of 1D and 3D meshes has to be partitioned equally for both mesh types.

\chapter{Muscle Fibers and Motor Units}\label{sec:muscle_fibers_and_motor_units}
The activation of muscle fibers is governed by the functional organization of the fibers in motor units (MU). An MU is the set of fibers that are innervated by the same $\alpha$-motor neuron, together with the neuron. If a motor neuron fires, all muscle fibers within the MU are activated. The association of the muscle fibers with MUs needs to be specified for electrophysiology simulations that consider activated muscle fibers. This chapter describes algorithms to achieve an MU-fiber association based on biophysical principles.

\section{Introduction}\label{sec:mu_intro}
Given a number of muscle fibers, the goal is to assign each fiber to one MU out of a set of given MUs. A muscle with a fusiform geometry, such as the biceps brachii is considered. Because muscle fibers do not branch or interrupt within the belly of such a muscle, the task can be reduced to the 2D problem on a cross-section of the muscle.

Properties of MUs have been subject to various investigations in literature. The number of MUs in a human muscle can be estimated by anatomical and physiological methods \cite{MacIntosh2006}. Anatomical methods include counting large-diameter fibers in postmortem tissue. The morphological studies of \cite{Feinstein1955} revealed high variations between different muscles. For example, the brachioradialis muscle has an estimated number of \num{333} MUs with \num{410} muscle fibers on average whereas the external rectus muscles in the eye have \num{2970} MUs with an average of only 9 muscle fibers. 

Physiological methods involve comparing the electrical and mechanical responses of artificially activated muscles, e.g., as in \cite{Milner-Brown1973b,Thomas1990b}. Typically, a high number of MUs with a smaller force or electric response is observed and a smaller number of MUs with a higher response. 
The review of \cite{Enoka2001} collects available experimental results and concludes an exponential distribution of the number of fibers per MU over all the MUs in a muscle.
% Enoka and Fuglevand (2001): Motor unit physiology: some unresolved issues

The spatial arrangement of the fibers of an MU can be revealed by a histochemical method \cite{brandstater1969histochemical}. It was found that the fibers of an MU appear at random positions but are grouped in a subregion of the muscular cross-section. The size of the subregion varies among muscles and fiber types and can be as large as one quarter of the cross-section, as in the tibialis anterior of the rat \cite{Edstrom1968}. Although the fibers of an MU are located in proximity they usually do not touch each other, i.e., there are always fibers of other MUs in between the fibers associated with an MU.

In our algorithm for assigning fibers to MUs we incorporate the following properties that are founded on biophysical experiments. 
\begin{itemize}
\item[(a)] The number of fibers per MU is  exponentially distributed. 
\item[(b)] The fibers of an MU are spatially distributed around a center point of the MU territory.
\item[(c)] The MU center points are reasonably separated from each other. However, the MU territories intermingle. 
\item[(d)] The spatial extents of the MU territories are proportional to the number of fibers of the MUs. 
\item[(e)] The exact locations of the fibers are random, but the overall density of fibers in the muscle is approximately constant. 
\item[(f)] Neighboring fibers are not innervated by the same motor neuron and therefore belong to different MUs.
\end{itemize}

Further physiological properties of fibers such as their fast or slow-twitch type as well as the distribution of electrical and mechanical properties are not subject to the fiber assignment algorithm. They are considered during configuration of the simulations of electrophysiology or muscular contraction.

\subsection{Related Works}
Simulations involving individually resolved muscle fibers are scarce in the literature. Therefore, not much previous work exists regarding methods to algorithmically assign fibers to MUs. The chemo-electro-mechanical skeletal muscle framework of \cite{Heidlauf2013} uses a method introduced in \cite{Roehrle2012} where center points of MU territories are positioned normally distributed around two distinct weighting centers for fast- and slow-twitch fibers. 
The algorithm randomly selects from certain sets of fibers and assigns them to MUs with exponentially increasing MU sizes. 
The method is applied to determine up to 50 MUs in the tibialis anterior (TA) muscle.

This algorithm fulfills the previously formulated properties (a)-(e). Among those, the fulfillment of (c) is not guaranteed but may be given by the random nature of the algorithm. An assumed issue regarding property (b) is that no predictions can be made about the fiber locations of the larger MUs. The larger MUs get assigned to previously unassigned fibers in a late stage of the algorithm, after most of the fibers have been selected for smaller MUs. The largest MU simply gets associated with all remaining fibers that were not selected for other MUs. In the worst case, these fibers can accumulate at multiple different regions, e.g., at boundaries of the muscle which is not physiological.

Instead of the 3D setting in \cite{Roehrle2012} that was needed for the complex anatomy of the TA muscle, we restrict our problem to a 2D cross-section of a fusiform muscle such as the biceps brachii. In comparison, our method creates MU territories of equal quality for all MUs and additionally fulfills property (e). Slow- and fast-twitch fibers are not treated differently by our algorithm, their properties are considered later in the simulation settings.

3D Simulations of skeletal muscle exist that treat MU association as homogenized property in the muscle volume. The approach in \cite{harry2018} assigns volume fractions of MUs to every spatial point of a 3D FEM discretization grid. MU center points are selected randomly ensuring a minimal distance. Prescribed volume fractions are sequentially assigned for each MU. The degree of intermingling can be adjusted by a parameter. It is shown that the algorithm highly depends on the order in which the MUs are traversed. The algorithm fulfills the properties (b)-(e), fulfilling (a) is possible by using appropriate parameter values.

In comparison, our method is targeted at MU assignment to individual fibers. However, distributing volume fractions is also the first step in our method. The volume fractions are interpreted as probabilities of the fibers being assigned to the respective MU. Therefore, our method can also be used as generator for homogenized formulations. A difference is that our method does not depend on a traversing order of MUs and ensures the exponential distribution of MU sizes.

\subsection{Two Alternative Premises for Motor Unit Assignment}
% Röhrle2012: TA 12 elements, MU association
% Heidlauf2013: 400 fibers, TA, OpenCMISS, mechanics

We identify two different sets of requirements which lead to two different methods.
Both methods fulfill the properties (a)-(f) listed in \cref{sec:mu_intro}.
The first premise is to assign motor units to a given number of fibers such that every fiber is associated with one MU. The second, alternative premise is to assign motor units only to a portion of the given set of fibers, discarding unassigned ones and thereby reducing the resulting set of fibers.

With the first setup, simulations of muscles that are gradually activated by motor neurons are possible. Activating the full muscle corresponds to activating all MUs and, in consequence, all muscle fibers. In reality, it is hardly possible to voluntarily activate all fibers in a muscle. This approach is also chosen in the presented literature, \cite{Roehrle2012} and \cite{harry2018}.

The second setup corresponds to modeling only a part of the muscle. The discarded fibers can be seen as belonging to other MUs that are not part of the simulation. When running highly parallelized simulations containing a large number of fibers, the missing fibers can introduce load imbalances, if they are computationally treated equally to the fibers with MUs. Even if no extra computational time is spent for the discarded fibers, a parallel domain decomposition becomes more involved than with the first setup where all fibers in a grid are present.

An advantage of the second setup is that the MU assignment to the fibers is generally easier. It also has its analog in volume fraction methods, where scalar fields of factors $f_k: \Omega \subset \R^3 \to [0,1]$ representing multiple MU territories, $k=1, \dots, n_\text{MU}$, can be easily defined. With this setup it is possible, for example, to perform analogous EMG simulations with the Multidomain model of \cite{Klotz2020} and the fiber based model of \cite{Mordhorst2015}.

In the remainder of this chapter, \cref{sec:method1_assignment} presents method 1, which fulfills the first premise where all fibers are associated to the MUs. Next, \cref{sec:method2_selection} introduces method 2, which only associates some fibers to MUs. Methods 1 and 2 fulfill the properties (a)-(e). Two derived methods 1a and 2a are subsequently constructed in order to also fulfill property (f). They are presented in \cref{sec:method3_modification}. Results and a discussion is given in \cref{sec:mu_results_and_discussion} before the chapter ends with a conclusion in \cref{sec:mu_conclusion}.

\section{Method 1: Assignment of Motor Units to a Given Set of Fibers}\label{sec:method1_assignment}

\begin{figure}%
  \centering%
  \begin{subfigure}[t]{0.32\textwidth}%
    \centering%
    \includegraphics[width=\textwidth]{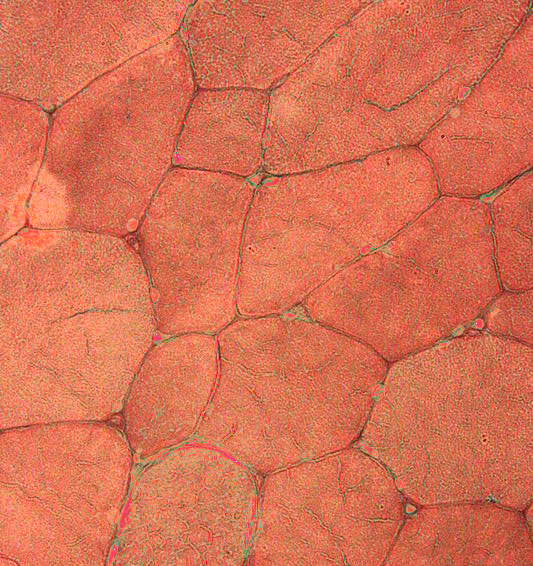}%
    \caption{Muscle fibers in the cross-section of a muscle visualized by Gömöri trichrome stain\footnotemark}%
    \label{fig:mu_grid0}%
  \end{subfigure}
  \quad
  \begin{subfigure}[t]{0.28\textwidth}%
    \centering%
    \includegraphics[width=\textwidth]{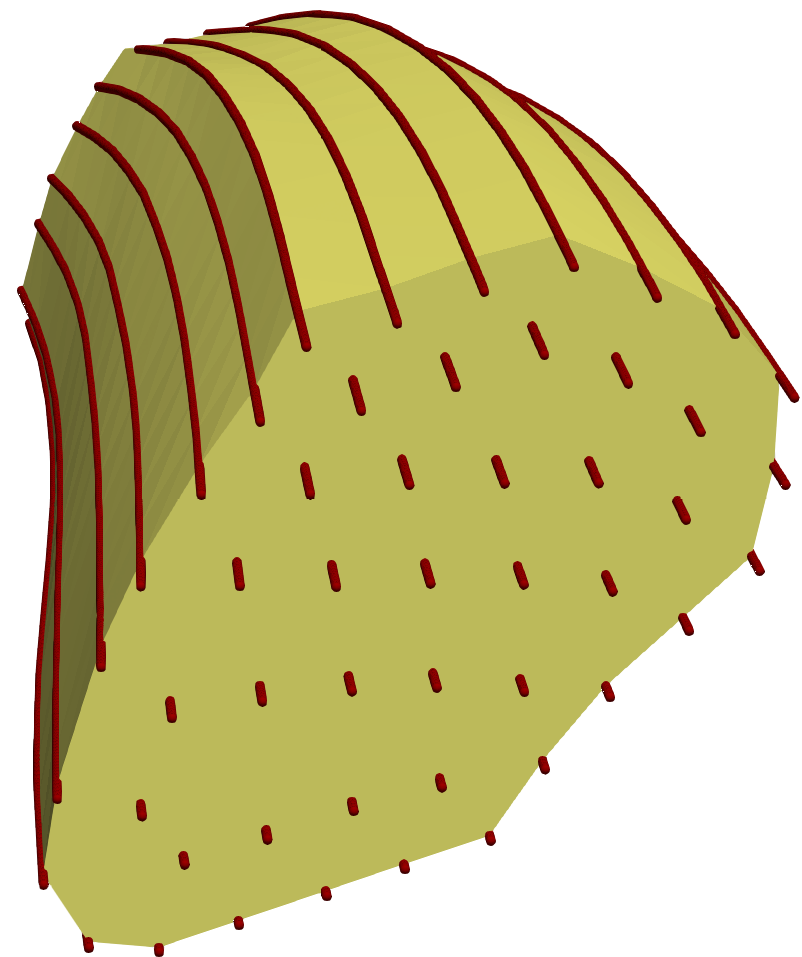}%
    \caption{A cut open muscle with $49$ muscle fibers in the simulation domain}%
    \label{fig:mu_grid1}%
  \end{subfigure}
  \quad
  \begin{subfigure}[t]{0.33\textwidth}%
    \centering%
    \includegraphics[width=\textwidth]{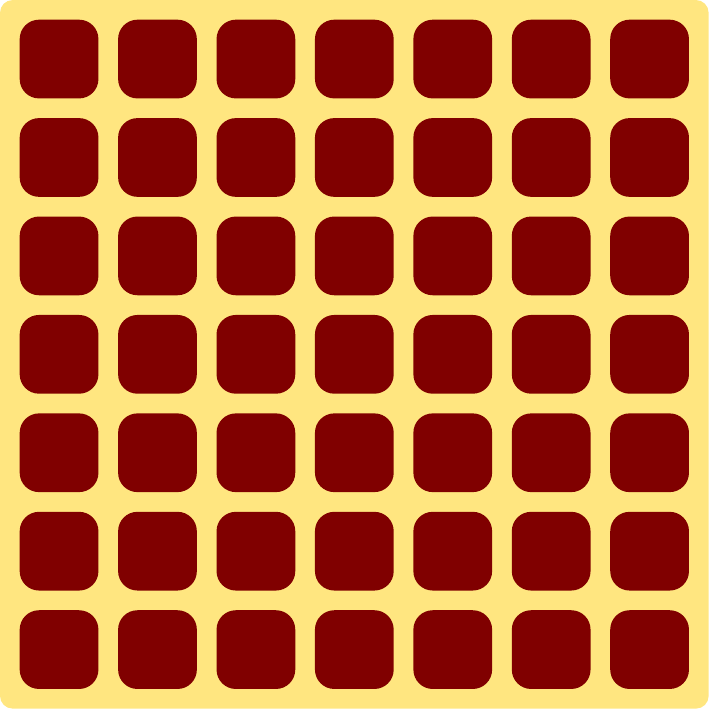}%
    \caption{A quadratic grid of $7 \times 7$ fibers, used for the methods to assign MUs to fibers.}%
    \label{fig:mu_grid2}%
  \end{subfigure}
  \caption{Representation of muscle fibers for the methods to associate fibers with MUs: From the real muscle to a quadratic grid.}%
  \label{fig:mu_grid}%
\end{figure}

In our methods to assign MUs to muscle fibers, the considered set of muscle fibers is organized in a regular grid. \Cref{fig:mu_grid0} shows 
a part of the cross-section of a skeletal muscle, the domains of individual fibers are visible. \Cref{fig:mu_grid1} visualizes the representation of muscle fibers in our simulations. In this figure, a relatively low number of 49 fibers was modeled. The fibers are approximated by 1D lines with uniform spacing in radial direction. For the algorithms to assign MUs to fibers, the muscle cross-section is considered as a logical 2D grid with a quadratic number of $n \times n$ fibers. Such a grid is visualized in \cref{fig:mu_grid2}.

\footnotetext{Image copyright © 25/12/2010 Michael Bonert (\url{https://commons.wikimedia.org/wiki/User:Nephron}), CC BY-SA 3.0 (\url{https://creativecommons.org/licenses/by-sa/3.0/legalcode}). The picture shows mitochondrial myopathy, it was cropped and the color was adjusted to make the \say{ragged red fibers} less prominent.}
\stepcounter{footnote}

The first method for the assignment of MUs to a given set of fibers associates the $n \times n$ fibers to a set of $n_\text{MU}$ motor units.
First, for each fiber ${(i,j)}$ in the grid, the probabilities $p(i,j,k_\text{MU})$ to be assigned to MU $k_\text{MU}$ are computed. This computation involves the solution of an optimization problem. Second, the sampling step assigns the actual MU indices to the fibers.

\subsection{Stochastic Formulation of Motor Unit Assignment}\label{sec:stochastic_formulation_and_algorithm}

In order to fulfill the formulated properties, the following three conditions are enforced on the probabilities. 
\begin{itemize}
\item[(i)] The probabilities at every fiber  have to be valid, i.e., positive and sum up to 1 for all MUs.
\item[(ii)] The portions fibers associated to MUs have to approximately follow an exponential progression $q$, with MU 1 containing the least and MU $n_\text{MU}$ containing the most fibers.
\item[(iii)] For any given MU, the spatial arrangement of its fibers in the cross-sectional plane is described by a radial kernel function $\hat{p}$. The fiber density of the MU increases when moving closer to the center point of the MU. This condition approximates the fact that the fibers of an MU are located in proximity, forming the MU territory.
\end{itemize}

The exponential progression $q$ in condition (ii) is defined as follows,
\begin{align}\label{eq:mus_q}
  q(k_\text{MU}) = b^{k_\text{MU}} / \s{\ell=1}{n_\text{MU}} b^\ell.
\end{align}
%
% f(x) = 1/(1+a*x^2),  ∫f(x)dx = arctan(sqrt(x)*x)/sqrt(a) + c
% stddev = sqrt(variance) = sqrt(∫{-∞,+∞} (f(x)-0)^2 dx) = sqrt(pi/(2*sqrt(a)))
% => a = pi^2 / (4*\sigma^4)
%
% 
The basis $b$ is a parameter and should be set to a value greater than one, e.g., $b=\num{1.2}$. \Cite{Enoka2001} formulate the function to be proportional to $\exp(\log(R)/n_\text{MU}\cdot k_\text{MU})$ where $R$ is the constant ratio between the sizes of the largest and smallest MUs. Our form is equivalent with $b = R^{1/n_\text{MU}}$.

The value of $q$ in \cref{eq:mus_q} is always positive. The division by the scaling factor ensures that the probabilities for all MUs sum up to one. Thus, condition (i) is fulfilled. The construction with the exponential function fulfills condition (ii).

For condition (iii), center positions $\bfx_{k_\text{MU}}, k_\text{MU}=1, \dots, n_\text{MU}$ of the MU territories are defined. 
The center positions are quasi-randomly selected inside the inner 80\% of the $n \times n$ grid of fibers. A band at the boundary with width of 10\% is not considered because the MU center points should not be at the border of any MU territory but rather at their center.

\begin{figure}%
  \centering%
  \includegraphics[width=0.4\textwidth]{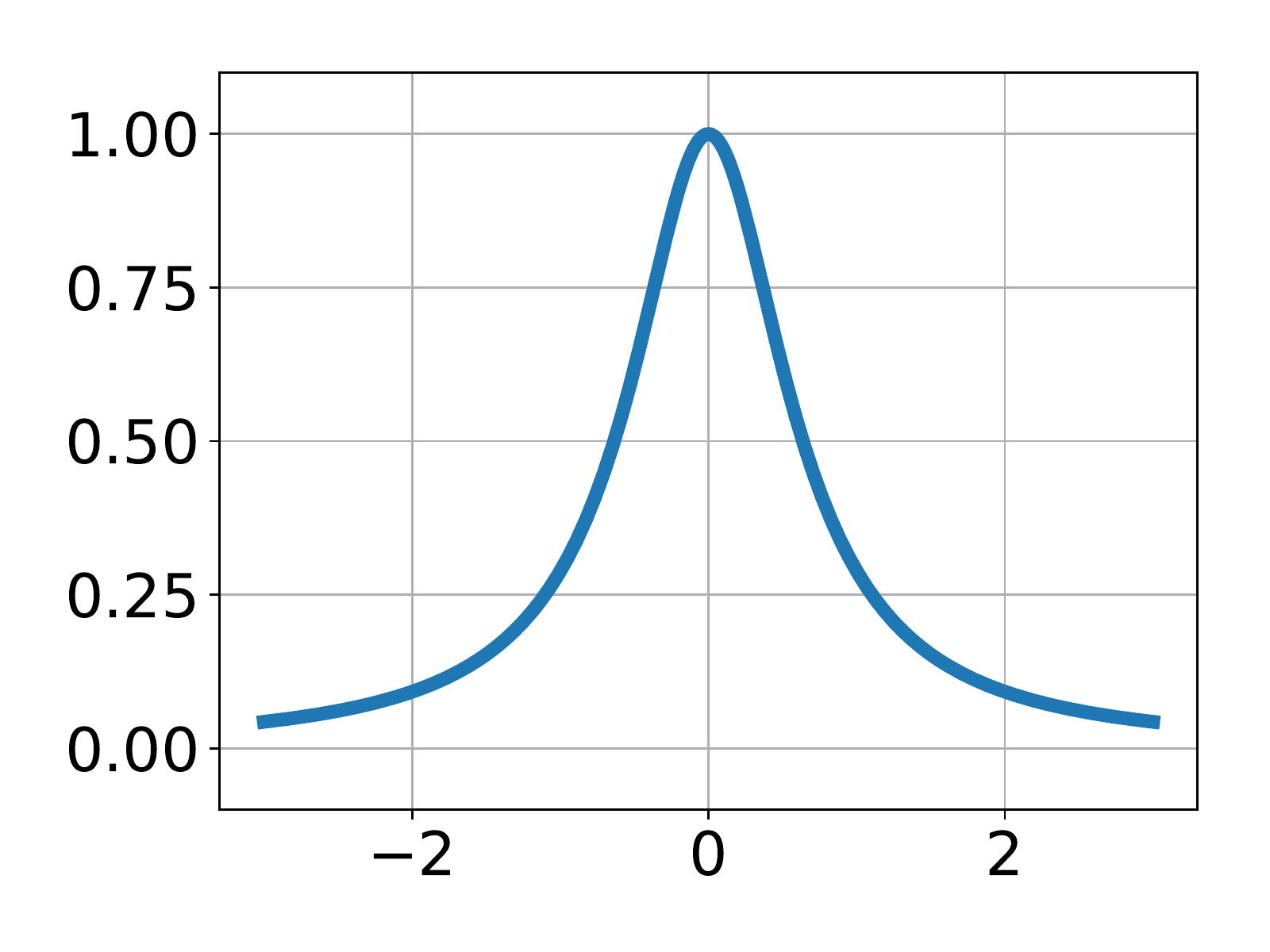}%
  \caption{Function $1/(1+a\,|x|^2)$, similar to $\hat{p}$ of \cref{eq:phat_kernel}, for $\sigma=1$.}%
  \label{fig:mu_phat}%
\end{figure}

The used quasi-random sequence is the following Weyl low-discrepancy sequence \cite{Weyl1916}:
\begin{equation}\label{eq:weyl}
  \begin{array}{lll}
    x_0 = 0.5, \qquad &y_0 = 0.5,\\[4mm]
    x_{i} = x_0+(i\cdot \alpha_1) \mod \num{1.0}, \qquad\qquad 
    &y_{i} = y_0+(i\cdot \alpha_2) \mod \num{1.0},\\[4mm]
    \text{with }\alpha_1 = \num{0.5545497}, \qquad &\alpha_2 = \num{0.308517}.
  \end{array}
\end{equation}
It is known that the sequences $x_i$ and $y_i$ are equidistributed in $[0,1)$ for any irrational $\alpha_1$ and $\alpha_2$ \cite{Weyl1916}. The chosen values lead to a sequence of 2D points $(x,y)\in $$[0,1)^2$ with low discrepancy and a good coverage of the domain for any number of sequence elements. Accordingly, the MU territory center points are defined as 
\begin{align*}
  \bfx_{k_\text{MU}} = \big((0.1 + 0.8\,x_{k_\text{MU}})\,n,\, (0.1 + 0.8\,y_{k_\text{MU}})\,n\big)^\top
\end{align*}
The radial kernel function $\hat{p}$ that describes the spatial probability distribution for a given MU $k_\text{MU}$ according to condition (iii) is defined as follows,
\begin{align}\label{eq:phat_kernel}
  \hat{p}(i,j,k_\text{MU}) = \dfrac{1}{1 + a\,\vert\bfx_{k_\text{MU}} - \bfx_{i,j}\vert^2}, \quad \text{with }a = \dfrac{\pi^2}{4\sigma^4}.
\end{align}
The coordinates $i$ and $j$ specify the grid point $\bfx_{i,j} = (i,j)^\top$ of the fiber. The factor $a$ is computed from the given standard deviation $\sigma$ of the spatial distribution of the MU territory around the center point. A lower value of $\sigma$ leads to smaller and \say{sharper} MU territories, for higher values of $\sigma$, the MU territories intermingle more with each other.
\Cref{fig:mu_phat} shows the graph of the function for $\sigma=1$. 

This kernel function was chosen because it can be computed efficiently with a low number of basic operations unlike, e.g., a Gaussian kernel function which requires costly evaluation of an exponential function.

If the kernel function $\hat{p}$ in \cref{eq:phat_kernel} is used to describe the probability of fiber $(i,j)$ to be in MU $k_\text{MU}$, then condition (iii) is fulfilled but conditions (i) and (ii) will not automatically be fulfilled. Instead of $\hat{p}$, a derived term $p(i,j,k_\text{MU})$ is introduced in the following that satisfies all requirements.

To ensure condition (ii), additional scalar factors $\lambda_k, k=1\dots n_\text{MU}$ for the MUs are introduced that yield the required exponential distribution.
To ensure condition (i), the term is normalized by a respective division. The resulting formulation is given as follows:
\begin{align}\label{eq:mu_p}
  p(i,j,k_\text{MU};\{\lambda_k\}_{1\dots n_\text{MU}}) = \dfrac{\hat{p}(i,j,k_\text{MU}) \cdot \lambda_{k_\text{MU}}}{\s{\ell_\text{MU}=1}{n_\text{MU}} \hat{p}(i,j,\ell_\text{MU}) \cdot \lambda_{\ell_\text{MU}}}.
\end{align}

Now, the factors $\{\lambda_k\}_{1\dots n_\text{MU}}$ have to be determined accordingly. Setting $\lambda_{k_\text{MU}} = q(k_\text{MU})$ would not yield the required exponential distribution of probabilities, because the MU center points $\bfx_{k_\text{MU}}$ have varying distances between each other.
Therefore, the accumulated total probability of all fibers to be associated to a particular MU is different for each MU. This is the case even before scaling with any factors $\{\lambda_k\}$.

Instead, the values of the factors have to be determined by solving a global optimization problem.
The objective function to be minimized is given by%
\begin{align}\label{eq:mus_objective}
  F(\{\lambda_k\}_{1\dots n_\text{MU}}) = \s{k_\text{MU}=1}{n_\text{MU}} \left(q(k_\text{MU}) - \s{i=1}{n}\s{j=1}{n}p(i,j,k_\text{MU};\{\lambda_k\}_{1\dots n_\text{MU}}) / n^2\right)^2.
\end{align}
It sums up the quadratic error for every MU between the desired, exponentially distributed probability $q(k_\text{MU})$ per fiber and the achieved probability per fiber under the current set of the scaling factors $\lambda_k$. 
The achieved probability is computed by a sum over all fibers $(i,j)$ and the formulated radial probability density function $p$ divided by $n^2$ to get the value per fiber.
After solving the optimization problem and plugging the factors $\{\lambda_k\}$ into \cref{eq:mu_p} we get every probability for a fiber to be in an MU by \cref{eq:mu_p}. The optimization problem is described in more detail in the following section.

\subsection{Algorithm to Solve the Optimization Problem}
The optimization problem to be solved in order to compute the scaling factors in \cref{eq:mu_p} can be stated as:
\begin{align}\label{eq:mus_opt}
  \text{``Find}\quad \{\lambda_k\}_{1\dots n_\text{MU}} \text{ with } \lambda_k > 0 \quad \text{ s.t. } \quad F(\{\lambda_k\}_{1\dots n_\text{MU}}) \quad \text{ is minimal''.}
\end{align}
The objective function $F$ was given in \cref{eq:mus_objective}. The solution is obtained by a Quasi-Newton method, more specifically the limited-memory version of the \emph{Broyden}-\emph{Fletcher}-\emph{Goldfarb}-\emph{Shanno} algorithm with box constraints by the authors of \cite{byrd1995limited}. Their Fortran implementation is made accessible in Python by the \emph{SciPy Optimize} package.

With increasing number $n^2$ of fibers and increasing number $n_\text{MU}$ of MUs, the evaluation duration for the objective function and the number of optimization parameters increases. For numbers about $n^2 > 1000$ and $n_\text{MU} > 25$, the solution times become unfeasible.

As a remedy we develop an algorithm to split the large optimization problem into multiple smaller ones which reduces the total runtime. 
The set of factors $\{\lambda_k\}_{1\dots n_\text{MU}}$ is partitioned into chunks, i.e., subsets of given size $n_\text{per\_chunk}$ leading to a total of $n_\text{chunks} = \lceil n_\text{MU} / n_\text{per\_chunk} \rceil$ chunks.
Remainder chunks towards the end potentially get one set element less. 
The factors for chunk number $c$ are selected in a strided manner as $\{\lambda_k\}$ with indices ${k=c, c+n_\text{chunks}, c+2\,n_\text{chunks}, \dots}$. For example, for $n_\text{MU}=13$ and $n_\text{per\_chunk}=4$ we get chunks of sizes $4,3,3,3$ and subsequently solve for $\{\lambda_k\}$ with $k \in \{1,5,9,13\}$, $\{2,6,10\}$, $\{3,7,11\}$, $\{4,8,12\}$.

A number of $n_\text{chunks}$ optimization problems is solved subsequently where the optimization parameters are each time given by the next chunk. During this loop, more and more scalar factors are determined.
Initially, all scalar factors $\lambda_k$ are set to one. 
After each solved optimization problem, the respective $\lambda_k$ values are updated with the values of the found minimizer.
The number $n_\text{factors\_up\_to\_chunk}$ of already solved scalar factors up to the current iteration starts with zero and increases by $n_\text{per\_chunk}$ after each iteration, finally arriving at $n_\text{MU}$ after the last iteration.

These smaller optimization problems have a similar formulation as the overall problem with different values for some variables.
The formulation of the optimization problem involves \cref{eq:mus_q,eq:phat_kernel,eq:mu_p,eq:mus_objective}. 
The number $n_\text{MU}$ of motor units in \cref{eq:mu_p,eq:mus_objective} is replaced by the number $(n_\text{factors\_up\_to\_chunk}+n_\text{per\_chunk})$ of factors that will have been solved after the current iteration. 
%This means that only so many MUs are considered at this iteration of the algorithm.

As each of the small optimization schemes only solves for $n_\text{per\_chunk}$ factors, the argument of the objective in \cref{eq:mus_objective} contains only the factors of the current chunk.
The other $n_\text{factors\_up\_to\_chunk}$ factors in the definition that are not given by the argument of the objective are set to the solutions obtained in previous iterations.

The indexing of the MU center positions $\bfx_{k_\text{MU}}$ in \cref{eq:phat_kernel} is adjusted such that only the MUs up to and including the current chunk are considered. The indexing in the exponential progression formulated by \cref{eq:mus_q}, however, stays the same, here the argument $k_\text{MU}$ of the function $q$ refers to the full set of MUs.

In summary, the iteratively considered settings contain an increasing number of MUs. The number of optimization parameters and, thus, new MUs is kept constant while the number of summands in the objective function increases. In consequence, the evaluation of the objective function gets more expensive with increasing iteration number while the size of the optimization problem stays constant. The latter has more influence on the optimizer duration.
By increasing the chunk size $n_\text{per\_chunk}$, the size of the small optimization problems can be decreased to any value. This makes the presented algorithm applicable for any large number $n_\text{MU}$ of MUs.

The result in this approach is not exactly the same as if one big optimization problem including all scalar factors at once would be solved. However, the error is small because of the interleaved MUs indices that are considered in every iteration. Because the resulting MU distribution finally gets drawn from the computed random probability distribution the error is hardly noticable in the result.

\subsection{Sampling Motor Unit Indices from the Given Probabilities}

The next step is to assign actual MU numbers to every fiber. Drawing samples $Y$ with the given probabilities $p(i,j,k_\text{MU})$ is done using inverse transform sampling. The inverse cumulative distribution function (CDF) $F_p^{-1}(k_\text{MU})$ is applied onto a random value $X$ drawn from a continuous uniform distribution $\mathcal{U}$:
\begin{align*}
  Y = F_p^{-1}(X)\quad \text{with }X\sim \mathcal{U}\big(0,F_p(n_\text{MU})\big), \quad F_p(k_\text{MU}) = \s{\ell_\text{MU}=1}{k_\text{MU}} p(i,j,\ell_\text{MU}).
\end{align*}
Computing the inverse of the CDF is computationally cheap as the probabilities are discrete and, thus, a loop over the values of the CDF suffices.

To reduce outliers during the random sampling where some MUs get exceptionally little fibers (such as none) or exceptionally many fibers assigned, the sampling procedure is performed five times.
Each time, a histogram with bins for the MUs is computed and provides the number of fibers per MUs. For all MUs $k$ that have zero fibers assigned, the one fiber where the probability for the respective MU $k$ is highest is determined. This is usually the fiber closest to the center point $\bfx_k$ of the MU $k$. The MU assignment of this fiber is changed to $k$, such that the MU $k$ is no longer empty but is associated to one fiber.

In each of the five iterations, the squared error between the sampled MU sizes and the expected sizes according to the probabilities, given by $q(k_\text{MU})\cdot n^2$ is computed. The MU assignment of the iteration that yielded the smallest error is used for the final result of method 1.

\section{Method 2: Assignment of Motor Units to a Selection of Fibers}\label{sec:method2_selection}

The second method proceeds similar to the first method in that at first the probability for a specific MU is defined for any fiber $(i,j)$ in the $n \times n$ grid. Then the actual MU assignments are sampled from the probability distributions. The difference to the first method is that any fiber is also allowed to not be assigned to any MU. This makes the definition of the probabilities easier and no optimization is required.

The three conditions defined in \cref{sec:stochastic_formulation_and_algorithm} are also imposed for the second method. The definition of the MU center positions $\bfx_{k_{MU}}$ follows the same low-discrepancy series. Also, the radial kernel function \cref{eq:phat_kernel} can be reused to describe the spatial distribution of probability for a given MU. Instead of \cref{eq:mu_p}, the probability is formulated directly as the product of the kernel function $\hat{p}$ and the exponential progression $q$:
\begin{align*}
  \tilde{p}(i,j,k_\text{MU}) = \hat{p}(i,j,k_\text{MU}) \cdot q(k_\text{MU}).
\end{align*}
To ensure that the function is within the bounds of a probability, $p \leq 1$, the result is divided by the maximum occuring value,%
\begin{align*}
  p(i,j,k_\text{MU}) = \dfrac{\tilde{p}(i,j,k_\text{MU})}{\max\limits_{\substack{\bar{i},\bar{j} = 1,\dots,n\\\bar{k}_\text{MU}=1,\dots,n_\text{MU}}} \tilde{p}(\bar{i},\bar{j},\bar{k}_\text{MU})}.
\end{align*}

In the sampling step, for every fiber $(i,j)$ the probabilities $p(i,j,k_\text{MU}), k_\text{MU}=1,\dots,n_\text{MU}$ for the MUs and the remaining probability $\bar{p} = 1-\sum_{k_\text{MU}} p(i,j,k_\text{MU})$ are computed. Then, the MU index is randomly drawn from the set of numbers $\{1,\dots,n_\text{MU}\}$ and a \say{null} event that corresponds to no assigned MU, according to the computed probabilities $p$ and $\bar{p}$. 

As a result, we get MUs that satisfy all three conditions formulated in \cref{sec:stochastic_formulation_and_algorithm}, including the approximate exponential progression of the MU sizes. However, the resulting number of fibers with assigned MU is an outcome of the algorithm and cannot be prescribed.

\section{Assignment of Different Motor Units for Neighboring Fibers}\label{sec:method3_modification}

As mentioned in \cref{sec:mu_intro}, one observation in staining experiments was that the fibers of an MU typically do not touch each other, i.e., neighboring fibers always belong to different MUs.
However, the presented methods 1 and 2 assign neighboring fibers with a high probability to the same MU. To create a grid with an MU assignment that avoids this behavior, the idea is to interleave four smaller grids where the MU assignments were obtained independently of each other but with the same parameters. In the following, this method is named \say{method 1a} and \say{2a} depending on whether the partial grids were handled with method 1 or 2.

\begin{figure}%
  \centering%
  \includegraphics[width=0.4\textwidth]{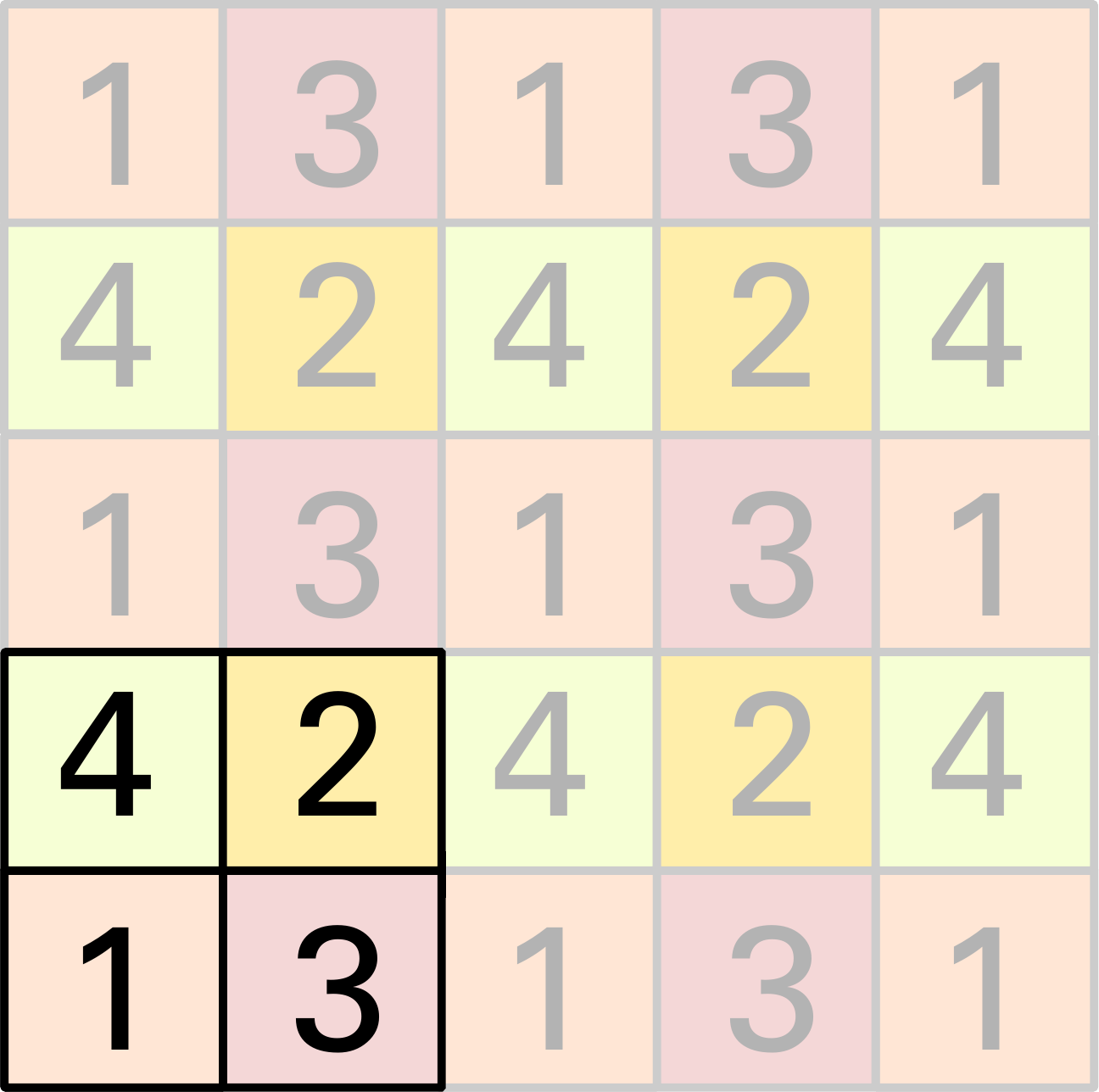}%
  \caption{Repeating scheme for interleaving the four partial grids. The partial grids are indicated by the numbers and have different colors. The pattern is highlighted at the bottom left of the figure.}%
  \label{fig:interleaving_scheme}%
\end{figure}

First, either method 1 or method 2 are applied four times to smaller grids of fibers, the \emph{partial grids}, as visualized in \cref{fig:interleaving_scheme}. The partial grids contain $n_\text{part} \times n_\text{part} = n/2 \times n/2$ fibers. In each partial grid, MU assignments with $n_\text{MU,part} = n_\text{MU}/4$ MUs are created. The basis $b$ is changed to $b_\text{part} = b^4$ and the standard deviation of the kernel function is changed to $\sigma_\text{part} = \sigma/2$. In result, we get the same exponential distribution of number of fibers per MU on every partial grid.

The four smaller grids are then merged according to the scheme shown in \cref{fig:interleaving_scheme}. Fibers of the first partial grid directly touch only fibers of the third and fourth grids and touch fibers of the second partial grid diagonally. By using this scheme, neighboring fibers in any of the partial grids are always separated by fibers of other grids. 

The MU indices that are assigned in the partial grids are mapped to the resulting, large grid also in an interleaved manner. MU $k$ of the $l$\emph{th} grid is mapped to the resulting MU $(4\,(k-1)+\ell)$. For illustration, the MUs 1,2,3 of the first grid are mapped to MUs $1,5,9$, MUs 1,2,3 of the second grid are mapped to MUs $2,6,10$, etc. Since the MU sizes in the partial grids follow the defined exponential progression, this also holds for the final MUs. 

The location of the MU center points is determined by contiguous elements of the same Weyl sequence given in \cref{eq:weyl} for all partial grids. First, all MU center points for the first partial grid are assigned, then for the second, third and forth. By this construction, all MU center points are distributed with similar spacing between each other and the placement of similar sized MUs close to each other is avoided.

\section{Results and Discussion}\label{sec:mu_results_and_discussion}

\begin{figure}%
  \centering%
  \begin{subfigure}[t]{0.48\textwidth}%
    \centering%
    \includegraphics[width=\textwidth]{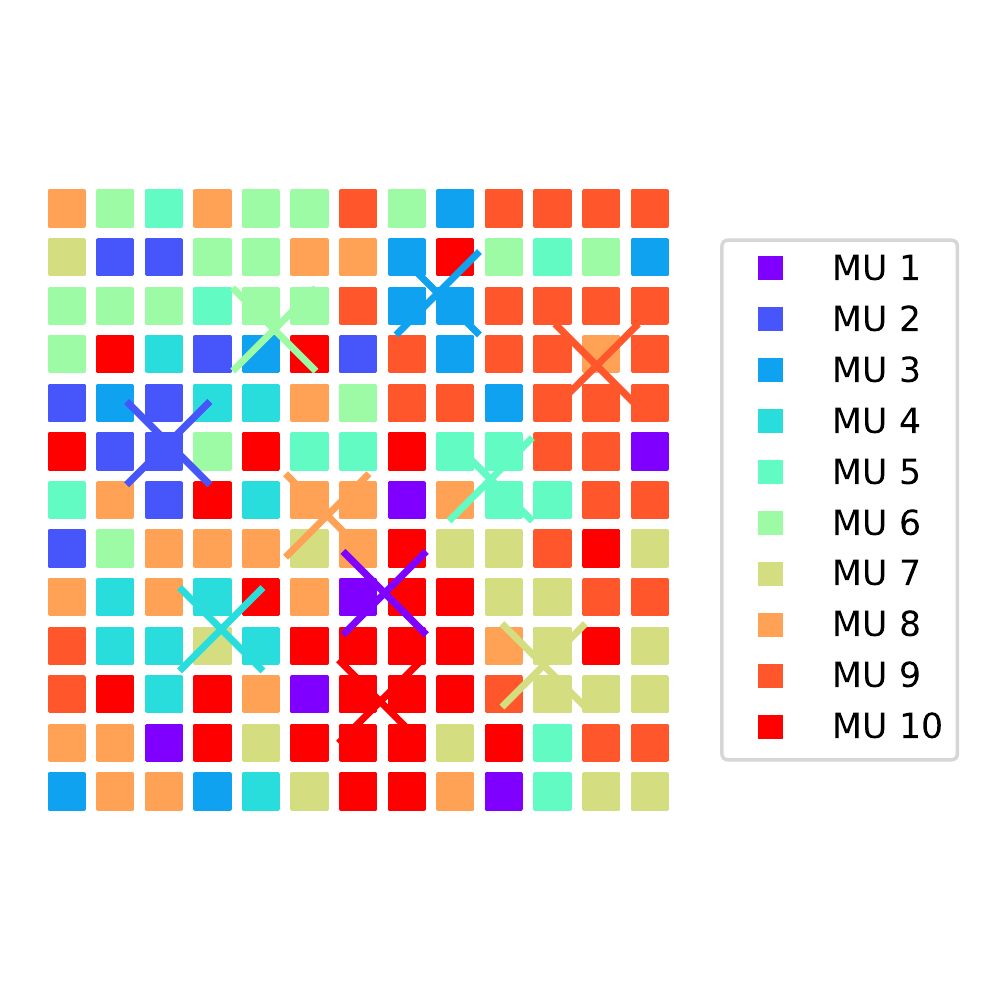}%
    \caption{Result for method 1 with $\sigma = n/10 = 1.3$}%
    \label{fig:MU_fibre_distribution_13x13_10_2d_fiber_distribution}%
  \end{subfigure}
  \quad
  \begin{subfigure}[t]{0.48\textwidth}%
    \centering%
    \includegraphics[width=\textwidth]{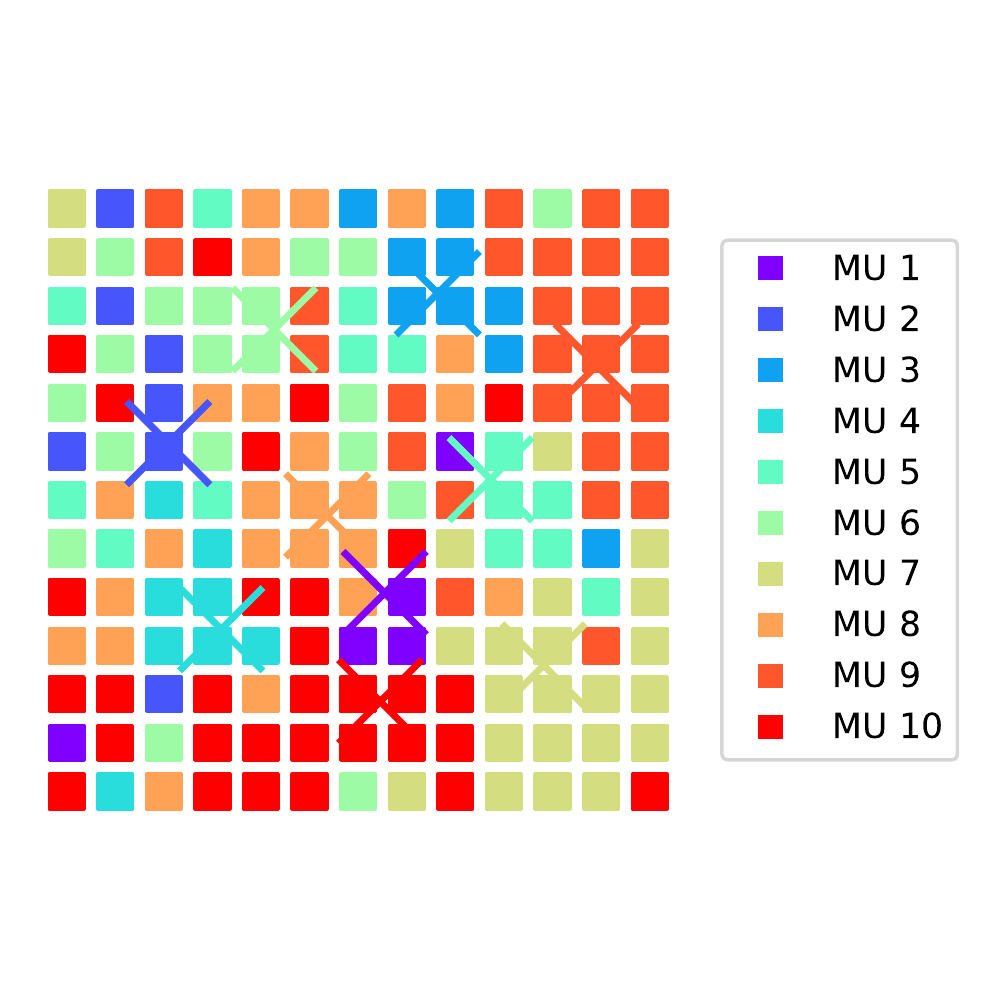}%
    \caption{Result for method 1 with $\sigma = n/100 = 0.13$}%
    \label{fig:MU_fibre_distribution_13x13_10_2d_fiber_distribution_sigma}%
  \end{subfigure}
  \begin{subfigure}[t]{0.48\textwidth}%
    \centering%
    \includegraphics[width=\textwidth]{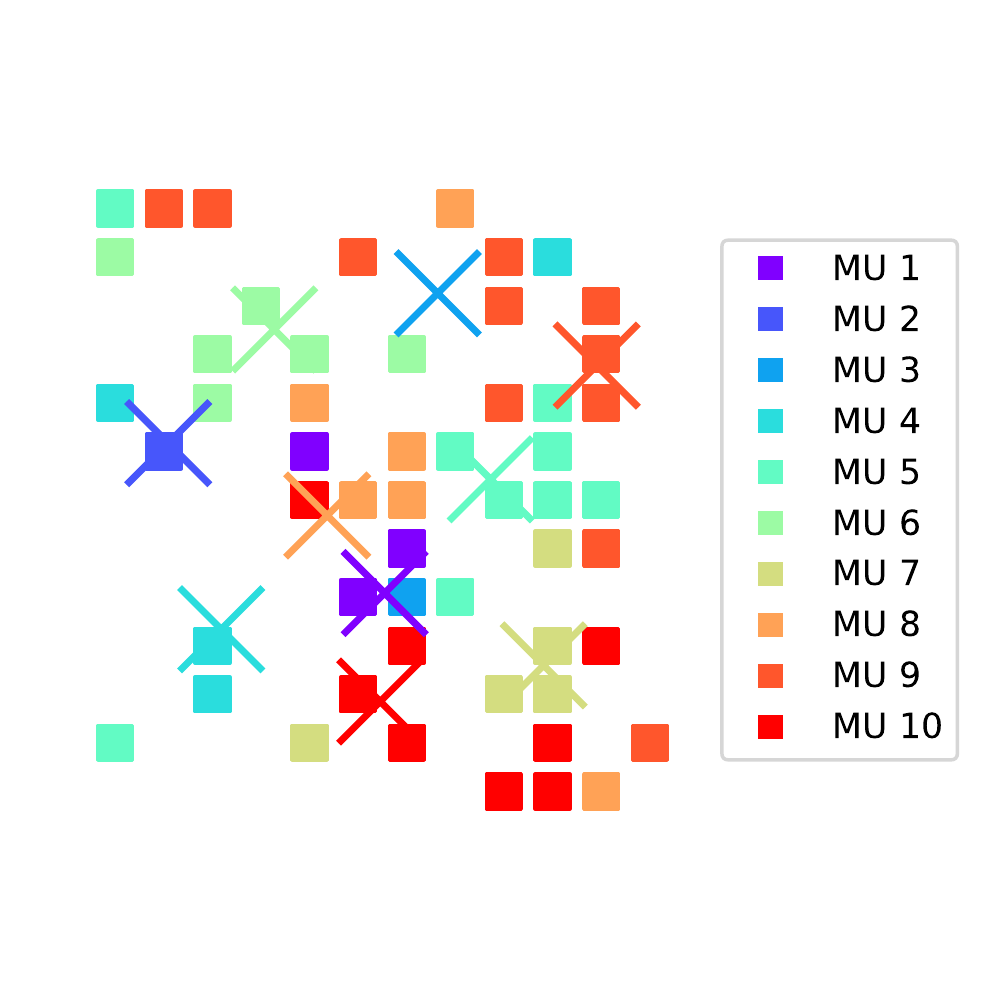}%
    \caption{Result for method 2 with $\sigma = n/10 = 1.3$, only 54 out of 169 fibers, i.e., \SI{32}{\percent} have an assigned motor unit.}%
    \label{fig:MU_fibre_distribution_sparse_13x13_10_2d_fiber_distribution}%
  \end{subfigure}
  \quad
  \begin{subfigure}[t]{0.48\textwidth}%
    \centering%
    \includegraphics[width=\textwidth]{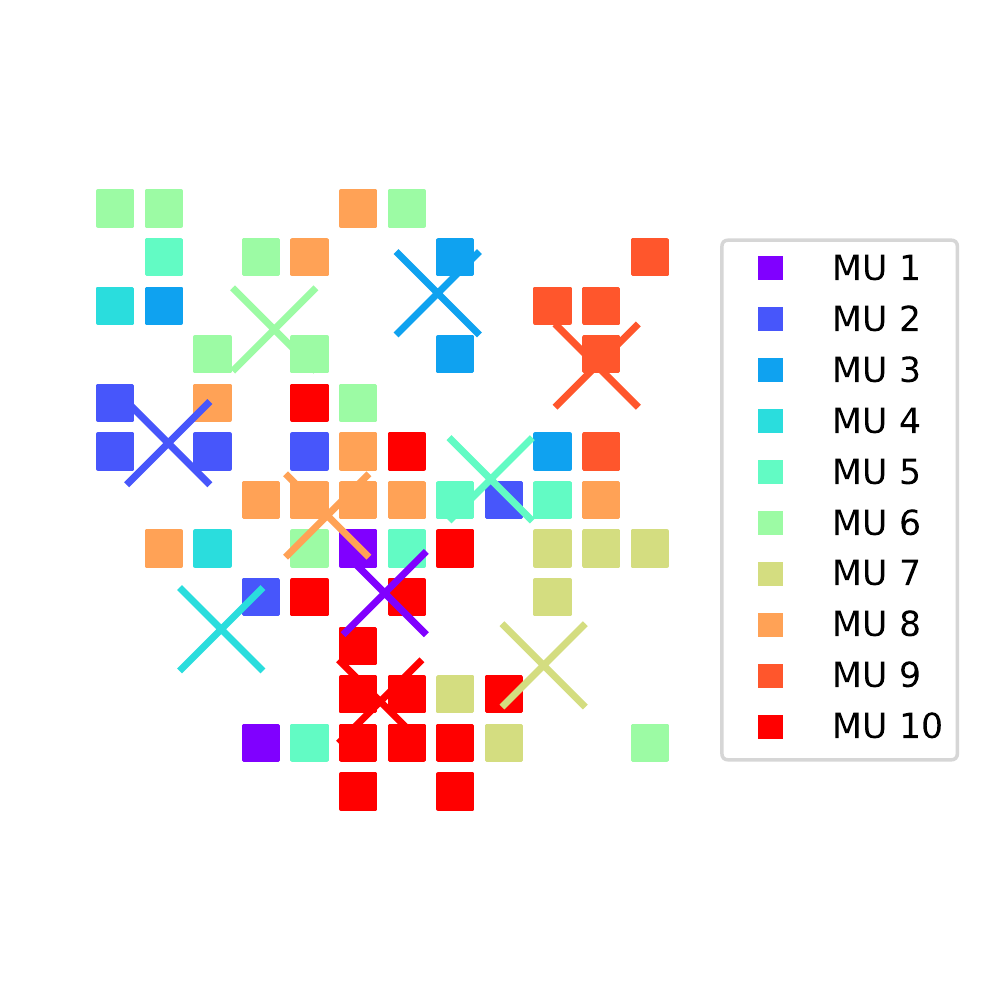}%
    \caption{Result for method 2 with $\sigma = n/100 = 0.13$, only 63 out of 169 fibers, i.e., \SI{37}{\percent} have an assigned motor unit.}%
    \label{fig:MU_fibre_distribution_sparse_13x13_10_sigma_2d_fiber_distribution}%
  \end{subfigure}
  \caption{Resulting MU assignments to a grid of $n\times n = 13 \times 13 = 169$ fibers. Each MU is represented by a color, the MU center points $\bfx_{k_\text{MU}}$ are the same for all scenarios and are shown by the colored crosses.}%
  \label{fig:mus_results1}%
\end{figure}

In the following, results of the methods described in \cref{sec:method1_assignment,sec:method2_selection,sec:method3_modification} with different parameters are presented.
\Cref{fig:mus_results1} shows the resulting assignment of MUs to fibers for methods 1 and 2. The number of fibers per coordinate direction is $n=13$, a number of $n_\text{MU} = 10$ MUs is considered and two different values for the kernel function parameter $\sigma$ are used.

In \cref{fig:MU_fibre_distribution_13x13_10_2d_fiber_distribution}, method 1 is used with basis $b=1.2$ and a kernel function with standard deviation of a tenth of the grid, $\sigma=n/10$. Each square represents one fiber, their colors refer to the MU index as indicated by the legend. Colored crosses visualize the center points $\bfx_{k_\text{MU}}$ of the respective MUs.

It can be seen that the MU territories, i.e., the regions of the fibers of an MU are located around the center points of the MUs. Because of the random sampling, the fibers of an MU are not all located closely together but spread over a larger area. Especially for MU 8, depicted by light orange color, some fibers are located further away from the center point, which is approximately at the center of the grid. On the other hand, for MU 10 most of the red marked fibers are located close to the center point of the MU, which can be found in the center of the lower third of the grid.

\begin{figure}%
  \centering%
  \includegraphics[width=0.8\textwidth]{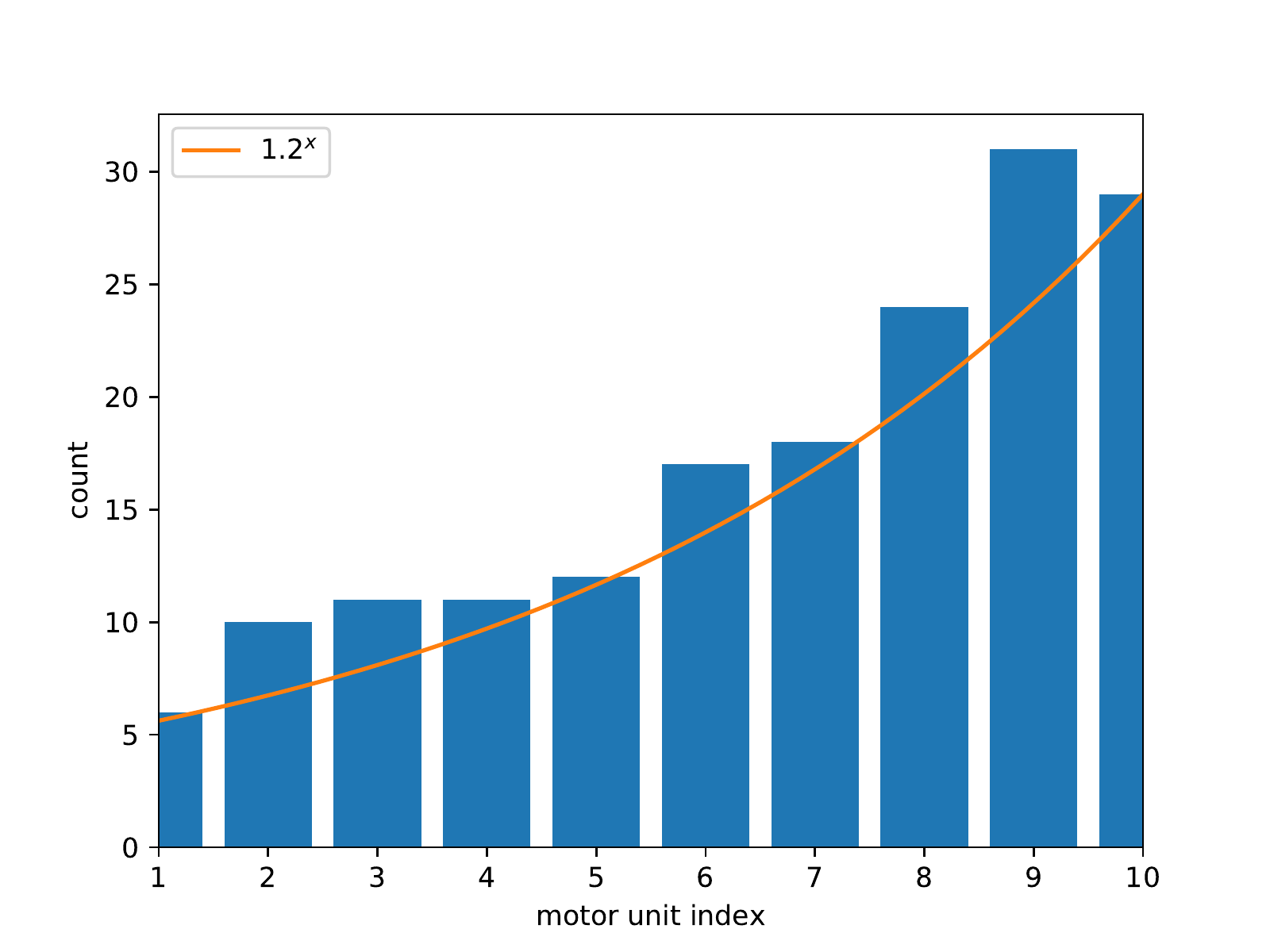}%
  \caption{Histogram of the number of fibers per MU in \cref{fig:MU_fibre_distribution_13x13_10_2d_fiber_distribution}. The orange line corresponds to the ideal exponential distribution $y=c\cdot 1.2^x$.}%
  \label{fig:MU_fibre_distribution_13x13_10_fiber_distribution}%
\end{figure}

The histogram for the setting considered in \cref{fig:MU_fibre_distribution_13x13_10_2d_fiber_distribution} is shown in \cref{fig:MU_fibre_distribution_13x13_10_fiber_distribution}. It can be seen that the number of fibers per motor unit approximately follows the prescribed exponential function with basis $b=1.2$. The observed deviation is due to the random sampling.

\begin{figure}%
  \centering%
  \begin{subfigure}[t]{0.31\textwidth}%
    \centering%
    \includegraphics[width=\textwidth]{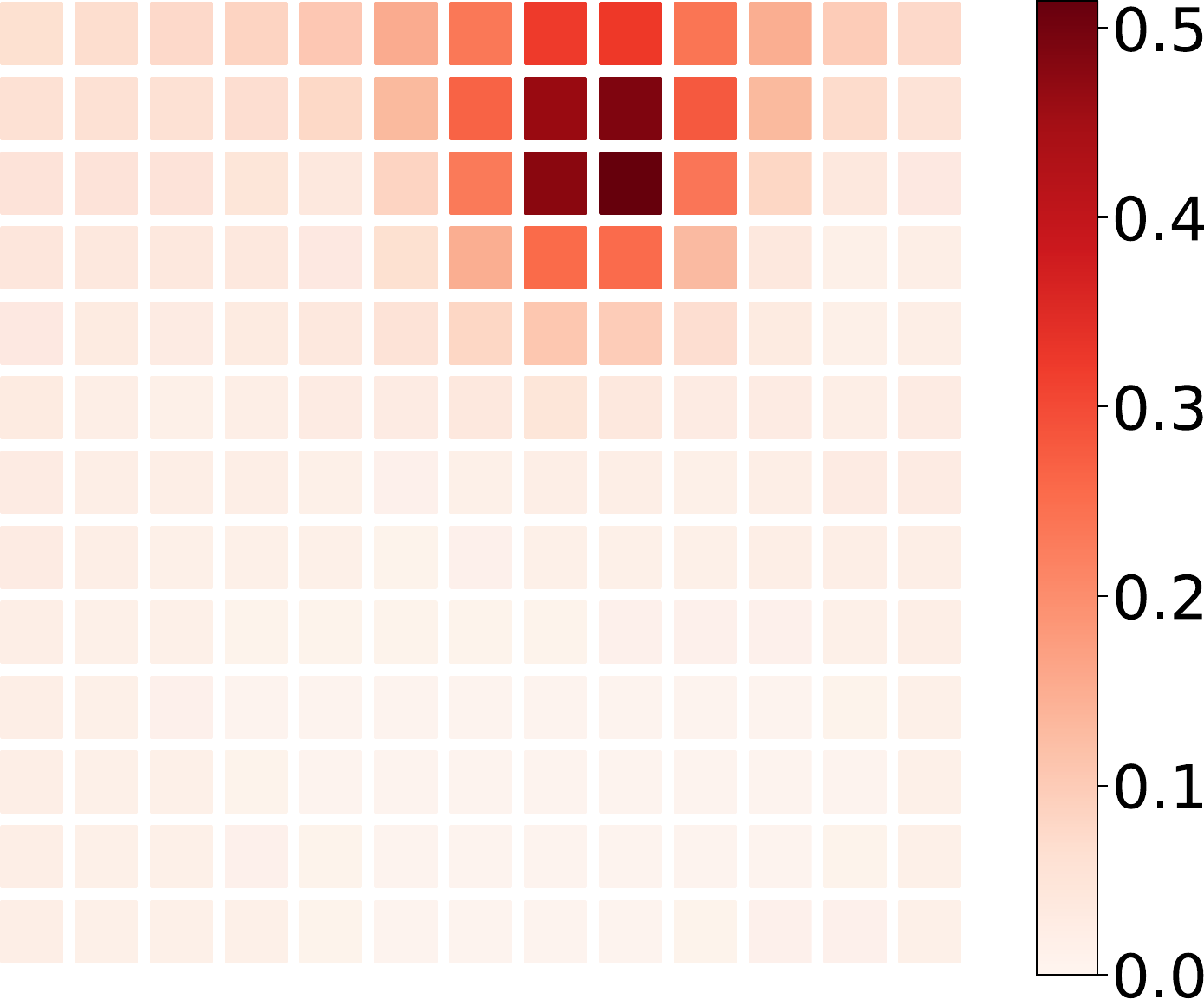}%
    \caption{$n=13, \sigma = n/10 = 0.13$, MU 3}%
    \label{fig:MU_fibre_distribution_13x13_10_fibers_mu3}%
  \end{subfigure}
  \quad
  \begin{subfigure}[t]{0.31\textwidth}%
    \centering%
    \includegraphics[width=\textwidth]{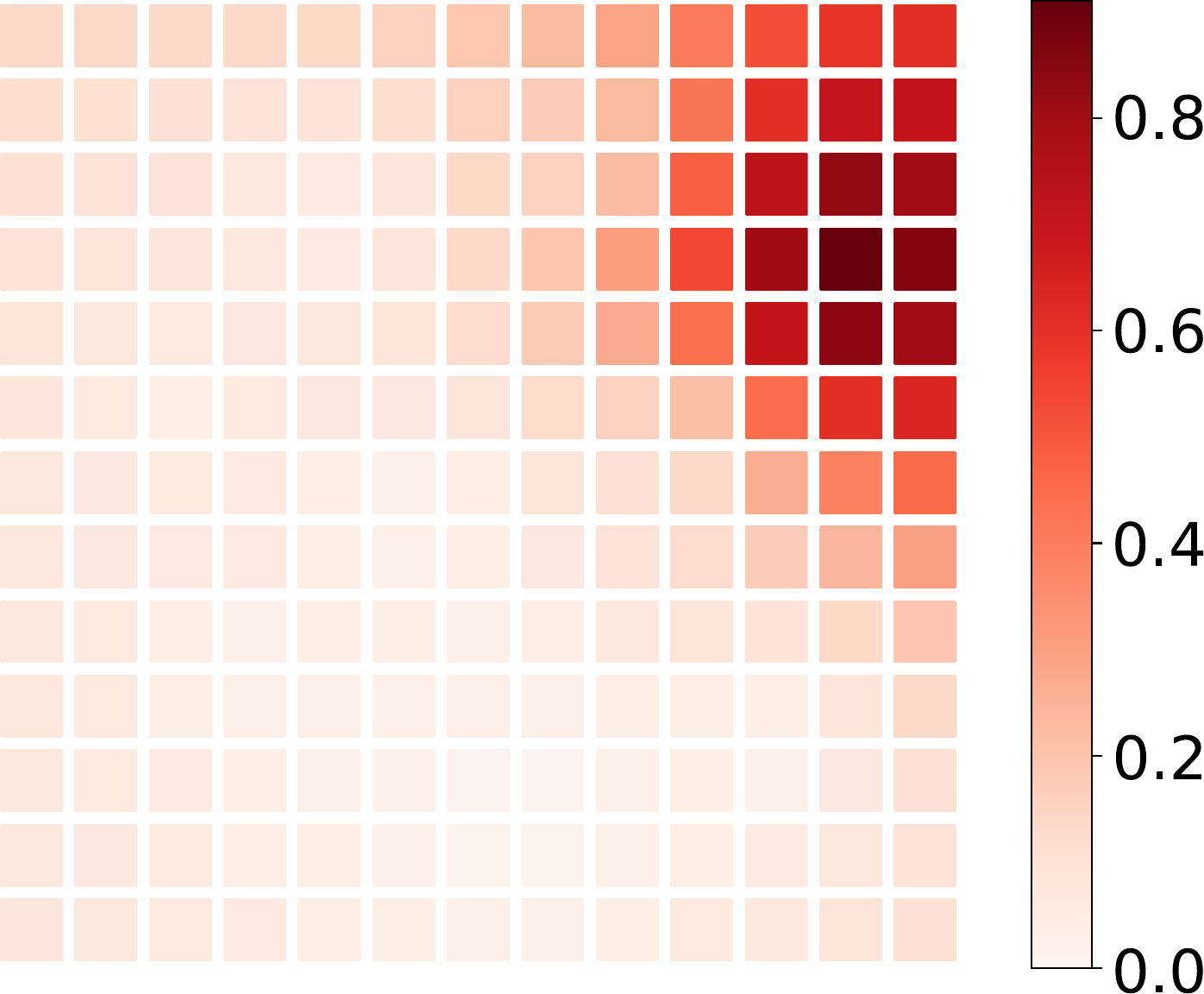}%
    \caption{$n=13, \sigma = n/10 = 0.13$, MU 9}%
    \label{fig:MU_fibre_distribution_13x13_10_fibers_mu9}%
  \end{subfigure}
  \quad
  \begin{subfigure}[t]{0.31\textwidth}%
    \centering%
    \includegraphics[width=\textwidth]{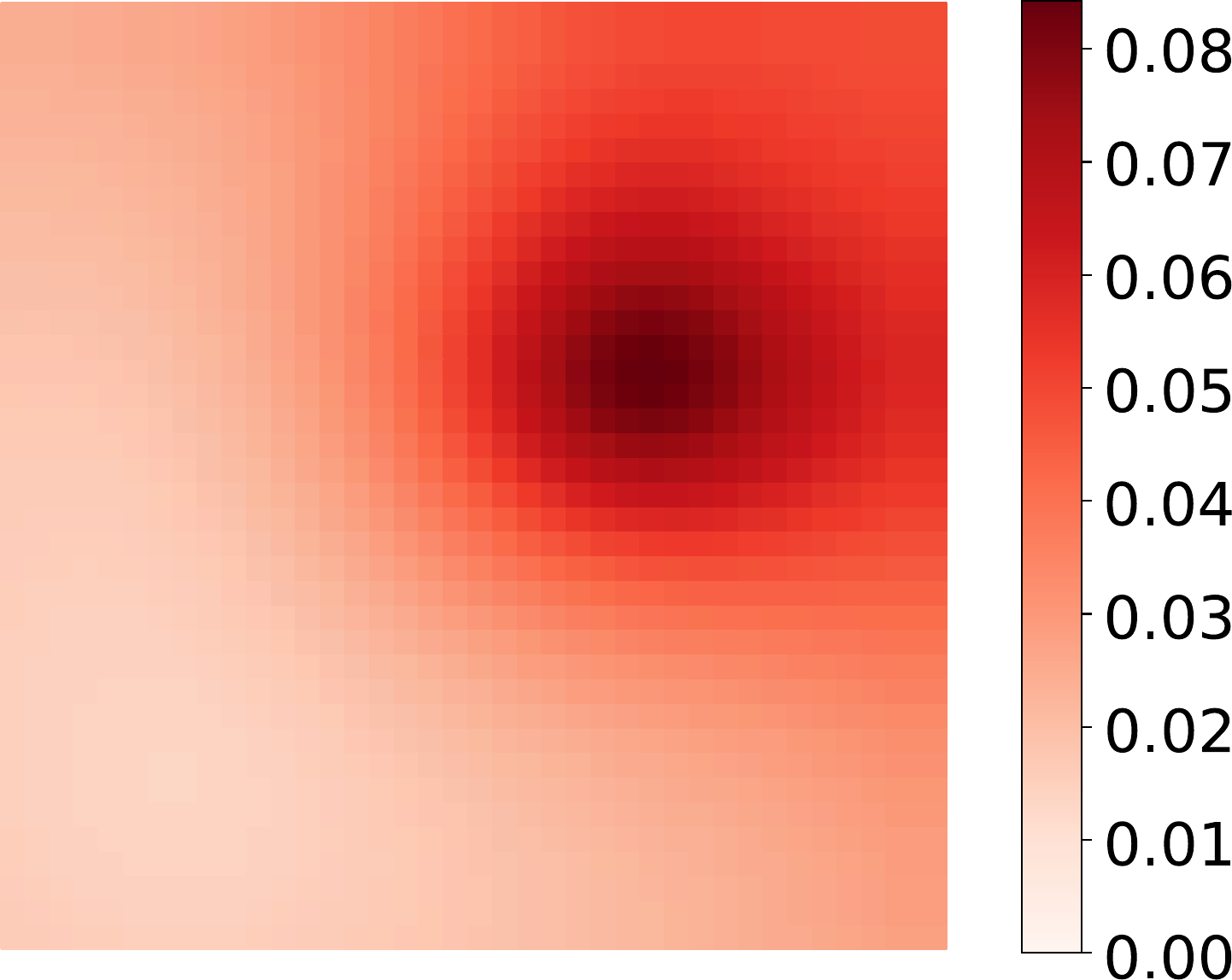}%
    \caption{$n=37, \sigma = n/10 = 0.37$, MU 41}%
    \label{fig:MU_fibre_distribution_37x37_50_fibers_mu41}%
  \end{subfigure}
  \caption{Probability at every fiber to be in a given MU, for different grid sizes and number of MUs.}%
  \label{fig:mu_kernel_fkt}%
\end{figure}

\Cref{fig:mu_kernel_fkt} shows the values of the probability function of a specific MU for all fibers, formulated in \cref{eq:mu_p}. \Cref{fig:MU_fibre_distribution_13x13_10_fibers_mu3,fig:MU_fibre_distribution_13x13_10_fibers_mu9} correspond to the scenario considered in \cref{fig:MU_fibre_distribution_13x13_10_2d_fiber_distribution}. The comparison shows that the probability is the highest around the center of MU 3 and MU 9, respectively. When moving away from the MU centers, the probability follows approximately the shape of the radial kernel function in \cref{eq:phat_kernel}.
An image of the radial kernel function in higher resolution is given by \cref{fig:MU_fibre_distribution_37x37_50_fibers_mu41}, where the probability function is depicted for MU 41 in a scenario with 50 MUs in a grid of $37 \times 37$ fibers.

It can be observed, however, that the probability distribution in \cref{fig:MU_fibre_distribution_13x13_10_fibers_mu3,fig:MU_fibre_distribution_13x13_10_fibers_mu9} does not entirely follow the kernel function. The effects of the scaling factors $\{\lambda_k\}$ in \cref{eq:mu_p} are visible, e.g., at the top-most and right-most fibers. There, the probability increases again compared to the interior of the grid. The purpose of the scaling factors is to enforce the exponential distribution of MU sizes.

\begin{figure}%
  \centering%
  \includegraphics[width=0.8\textwidth]{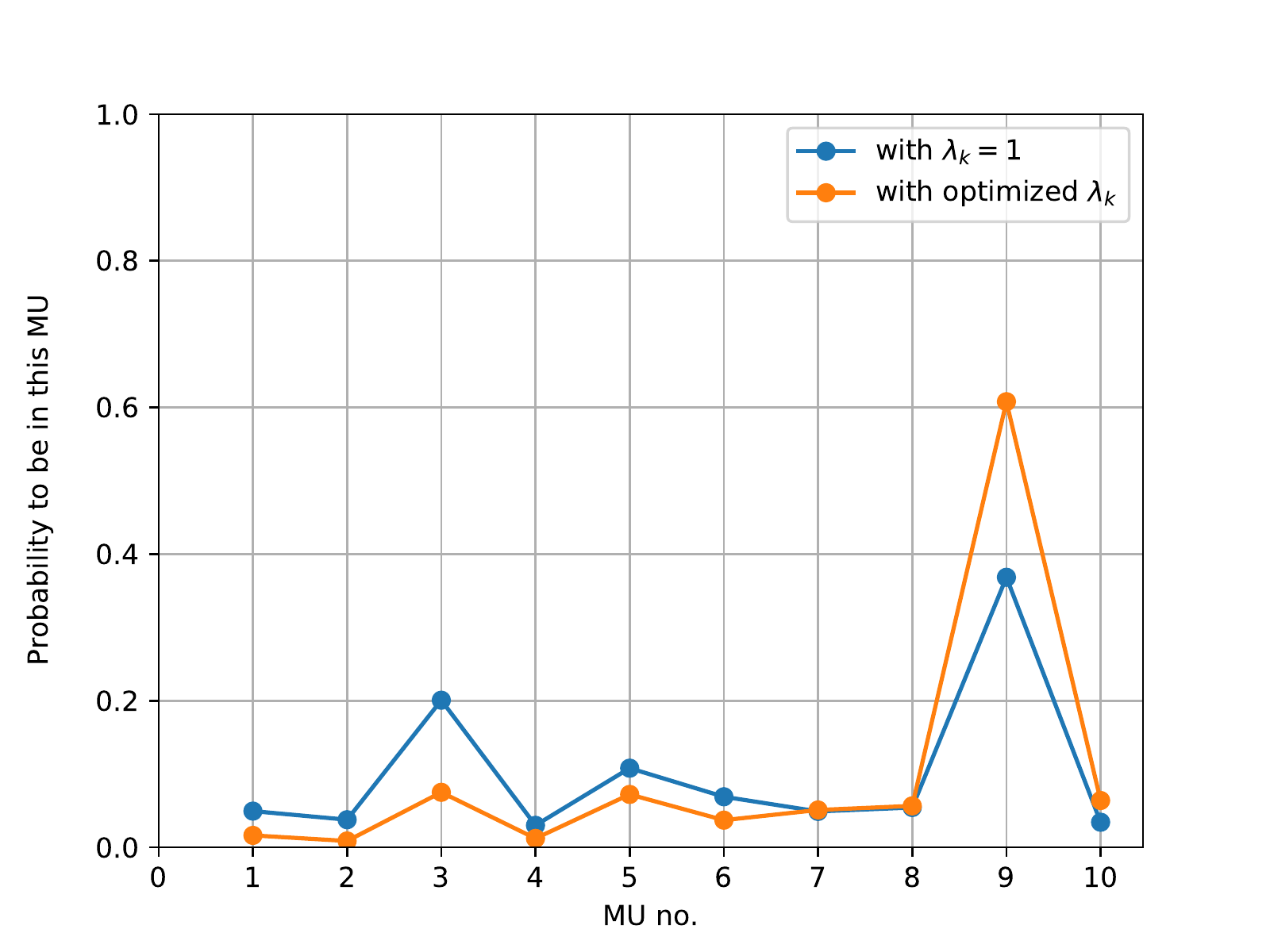}%
  \caption{Probability of the top right fiber in the $13 \times 13$ grid to be in a given MU, without considering the scaling factors $\{\lambda_k\}_{1\dots n_\text{MU}}$ (blue) and including the scaling factors (orange).}%
  \label{fig:MU_fibre_distribution_13x13_10_pdf}%
\end{figure}

This effect is illustrated more clearly in \cref{fig:MU_fibre_distribution_13x13_10_pdf}. It shows the value of $p(i,j,k_\text{MU})$ for the top right fiber in the grid, $(i,j)=(13,13)$, for all values of $k_\text{MU}$. The blue curve indicates the probability that results from the kernel functions, only according to the distance of the top right fiber to the respective MU centers. In other words, the scaling factors $\{\lambda_k\}_{1\dots n_\text{MU}}$ are removed or equivalently set to one. By inspecting again \cref{fig:MU_fibre_distribution_13x13_10_2d_fiber_distribution}, it can be seen that the MU centers of MUs 9, 3 and 5 are---in this order---closest to the top right fiber whereas MU 4 and 10 are the furthest away. Consequently, the blue curve in \cref{fig:MU_fibre_distribution_13x13_10_pdf} has peaks at 9, 3 and 5 and low values for 4 and 10.

When incorporating the scaling factors $\{\lambda_k\}_{1\dots n_\text{MU}}$ that were found by the optimization problem in \cref{eq:mus_opt}, the probabilities change to the orange curve. It can be seen that the probability for the fiber to be in MU 9 increases. MU 9 which is expected to have a rather high number of fibers according to the exponential progression.
It gets more fibers from the top right corner. The areas left to and below the center of MU 9 are at the same time close to the centers of MU 3 and 5 and therefore can also be occupied by fibers of MUs 3 and 5. Thus, the optimization performs an exchange where MU 9 forgoes the bottom and left fibers and, conversely, obtains portions of the fibers in the top right area from MUs 3 and 5. Consequently, the probability in \cref{fig:MU_fibre_distribution_13x13_10_pdf} decreases for MUs 3 and 5. The shape of the final probability distribution for MU 9 in \cref{fig:MU_fibre_distribution_13x13_10_fibers_mu9} is the kernel function stretched to the top and right.
By looking at \cref{fig:MU_fibre_distribution_13x13_10_2d_fiber_distribution}, it can be seen that, by chance, the top right fiber indeed gets assigned to MU 9.

The influence of the kernel width $\sigma$ is demonstrated by comparing \cref{fig:MU_fibre_distribution_13x13_10_2d_fiber_distribution} with \cref{fig:MU_fibre_distribution_13x13_10_2d_fiber_distribution_sigma}. In \cref{fig:MU_fibre_distribution_13x13_10_2d_fiber_distribution_sigma} the value of $\sigma$ is only a tenth of the value in \cref{fig:MU_fibre_distribution_13x13_10_2d_fiber_distribution}. All other parameters are the same such that a similar exponential distribution of MU sizes is obtained. It can be seen that the MU territories are less interleaved and have clearer borders. For example, the territory of MU 7 at the bottom left of the domain has a cohesive shape in \cref{fig:MU_fibre_distribution_13x13_10_2d_fiber_distribution_sigma} whereas the respective fibers are more scattered in \cref{fig:MU_fibre_distribution_13x13_10_2d_fiber_distribution}.

In comparison, the results for method 2 with the same two values of $\sigma$ are shown in \cref{fig:MU_fibre_distribution_sparse_13x13_10_2d_fiber_distribution,fig:MU_fibre_distribution_sparse_13x13_10_sigma_2d_fiber_distribution}. All other parameters are kept the same. It can be seen how method 2 only associates some fibers with MUs. For the larger standard deviation $\sigma$ in \cref{fig:MU_fibre_distribution_sparse_13x13_10_2d_fiber_distribution}, only \SI{32}{\percent} of the fibers get assigned to a MU, for the smaller value of $\sigma$, the fraction is slightly higher with \SI{37}{\percent}. Similar to method 1, the effect of more cohesive MU territories for smaller $\sigma$ values can also be observed in the results of method 2.

Next, the two methods 1 and 2 are investigated for a higher number of $n_\text{MU}=100$ motor units and a grid of $n \times n=67 \times 67 = \num{4489}$ fibers.

\begin{figure}%
  \centering%
  \begin{subfigure}[t]{0.48\textwidth}%
    \centering%
    \includegraphics[width=\textwidth]{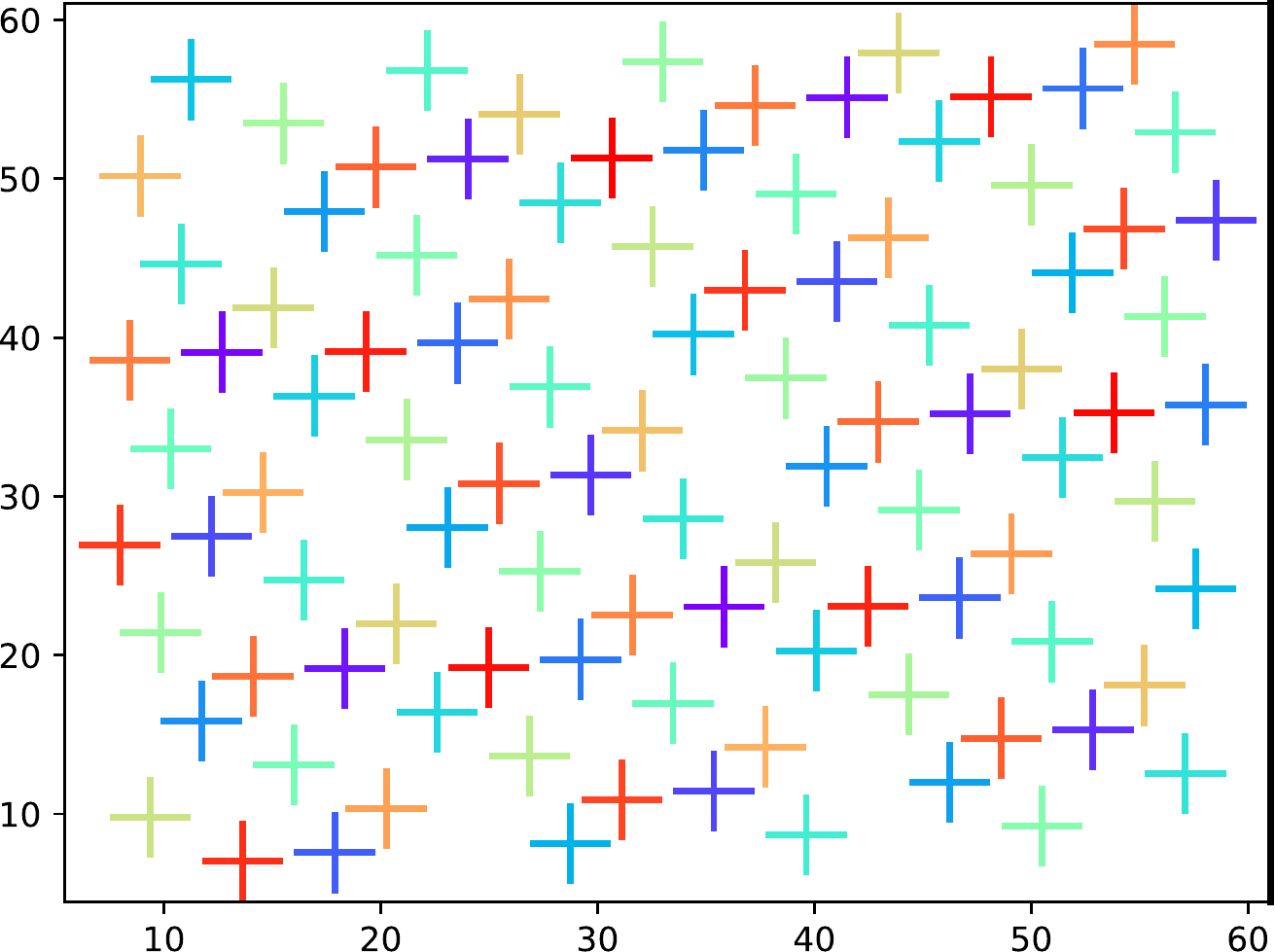}%
    \caption{MU center points.}%
    \label{fig:MU_fibre_distribution_67x67_100_mu_positions}%
  \end{subfigure}
  \quad
  \begin{subfigure}[t]{0.48\textwidth}%
    \centering%
    \includegraphics[width=\textwidth]{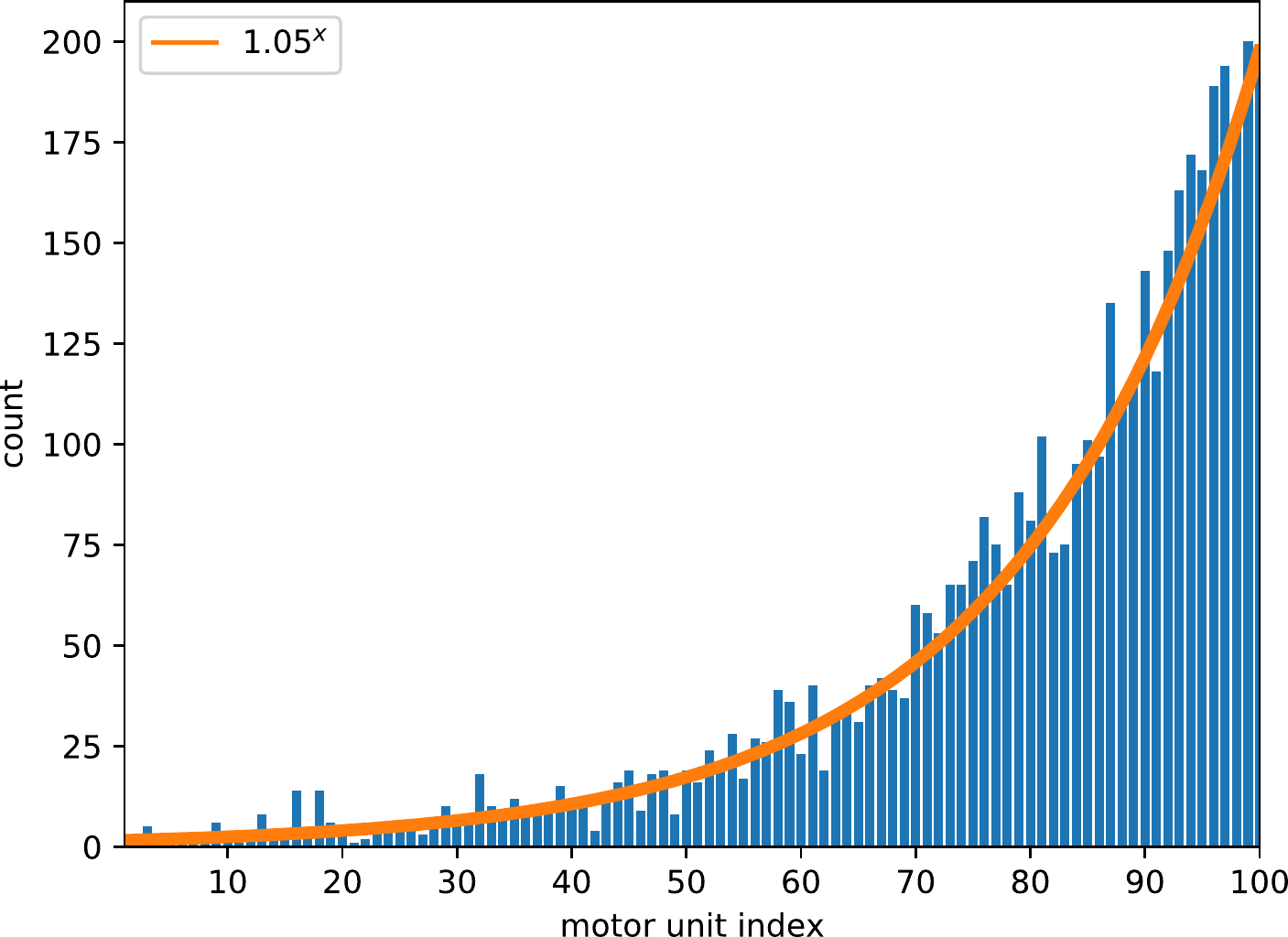}%
    \caption{Histogram of number of fibers assigned to MUs (blue) and the prescribed exponential progression $y=c\cdot 1.05^x$ (orange).}%
    \label{fig:MU_fibre_distribution_67x67_100_fiber_distribution}%
    %total number of fibers: 4489
  %  MU 11, n. fibers optimal: 2.802, n. fibers expected: 0.945, realized: 1
  %MU 41, n. fibers optimal: 12.108, n. fibers expected: 8.809, realized: 15
  %MU 61, n. fibers optimal: 32.126, n. fibers expected: 30.439, realized: 29
  % MU 81, n. fibers optimal: 85.241, n. fibers expected: 127.908, realized: 118
  %MU 91, n. fibers optimal: 138.849, n. fibers expected: 137.036, realized: 123
  
  \end{subfigure}
  \begin{subfigure}[t]{0.48\textwidth}%
    \centering%
    \includegraphics[width=\textwidth]{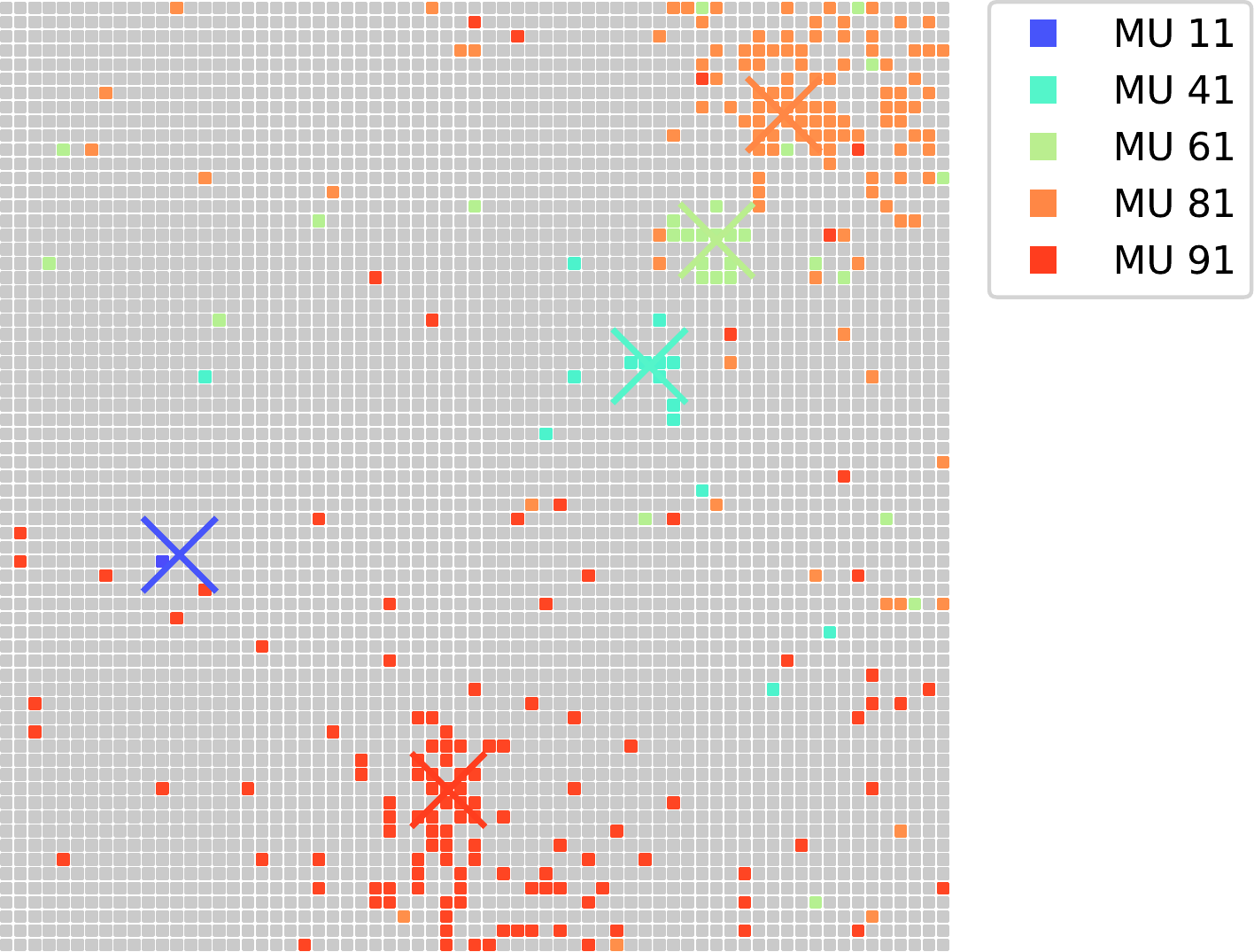}%
    \caption{Result of method 1 with $\sigma = n/100 = \num{0.67}$. The colored fibers are assigned to one of five selected MUs: 11, 41, 61, 81 and 91. The MU sizes are: 
    MU 11: 1 fiber,
    MU 41: 2 fibers,
    MU 61: 15 fibers,
    MU 81: 32 fibers,
    MU 91: 72 fibers}%
    \label{fig:MU_fibre_distribution_67x67_100_some_2d_fiber_distribution.pdf}%
  \end{subfigure}
  \quad
  \begin{subfigure}[t]{0.48\textwidth}%
    \centering%
    \includegraphics[width=\textwidth]{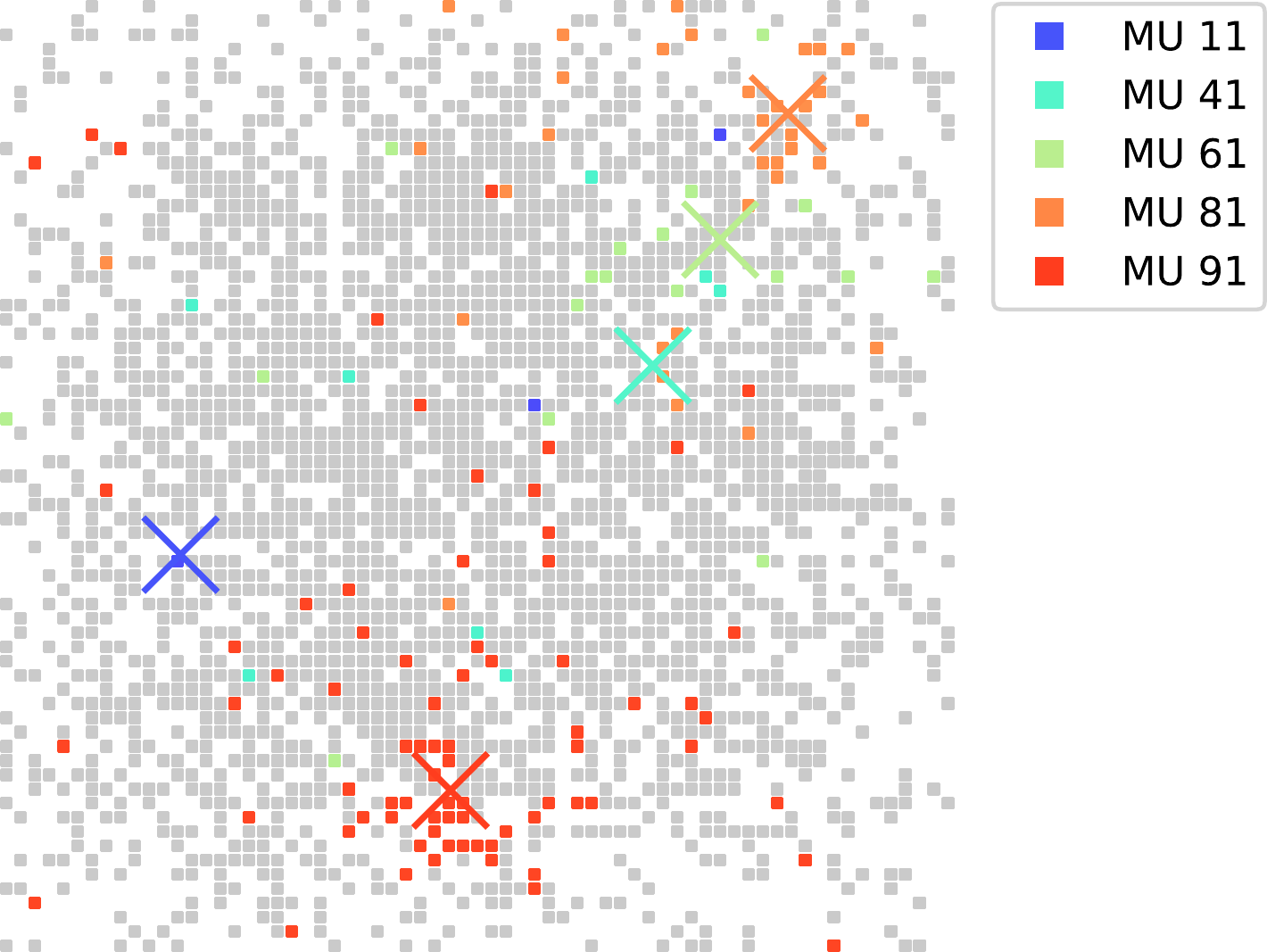}%
    \caption{Result of method 2 with $\sigma = 0.04\cdot n = \num{2.68}$. The colored fibers are assigned to one of five selected MUs: 11, 41, 61, 81 and 91. The MU sizes are:  
    MU 11: 3 fibers, MU 41: 8 fibers, MU 61: 18 fibers, MU 81: 34 fibers, MU 91: 75 fibers}%
    \label{fig:MU_fibre_distribution_sparse2_67x67_100_2d_fiber_distribution}%
  \end{subfigure}
  \caption{Results of the presented algorithm to assign MUs to fibers, using a grid of $67 \times 67$ fibers and 100 MUs.}%
  \label{fig:100mus_results}%
\end{figure}

\Cref{fig:MU_fibre_distribution_67x67_100_mu_positions} shows the MU center points $\bfx_{k_\text{MU}}$. The color corresponds to the MU index and follows the same rainbow color scheme as in \cref{fig:mus_results1}. Since the construction scheme is the deterministic Weyl sequence in \cref{eq:weyl}, the first 10 MU center positions are the same as for the scenario with $n_\text{MU}=10$. It can be seen that the MU centers have similar distances throughout the grid and that, in general, MUs located next to each other have different colors and therefore are differently sized.

\Cref{fig:MU_fibre_distribution_67x67_100_fiber_distribution} shows the histogram of the MUs, i.e., the number of fibers per MU. Following \cite{Enoka2001} the prescribed basis for the exponential progression was reduced because of the higher number of fibers. It was set to $b=1.05$. It can be seen that the resulting MU size distribution closely matches the prescribed function. Because the ratio of fibers to MUs (4489/100) is higher than in the previous setting (169/10), the deviation of the realized MU sizes from the prescribed curve appears smaller than in \cref{fig:MU_fibre_distribution_13x13_10_fiber_distribution}.

\Cref{fig:MU_fibre_distribution_67x67_100_some_2d_fiber_distribution.pdf} shows the result for method 1. The width of the radial kernel function was chosen as $\sigma = n/100 = \num{0.67}$. Only the fibers of five selected MUs and their center points are visualized for better clarity.

The algorithm for the 4489 fibers and 100 MUs was performed with a  chunk size of $n_\text{per\_chunk}=10$, yielding a total number of $n_\text{chunks}=10$ chunks. The runtime was \SI{45}{\min} \SI{53.5}{\sec} on a single core of an Intel Core i5-6300U CPU with base frequency of 2.40GHz and 19.5 GiB of RAM.

\Cref{fig:MU_fibre_distribution_sparse2_67x67_100_2d_fiber_distribution} shows the result for method 2. All resulting fibers that were associated to an MU are shown as gray or colored squares, leaving white spaces for unassigned fibers. Again, only the fibers of five selected MUs are colored. The kernel parameter was set to $\sigma = 0.04\cdot n = \num{2.68}$ which resulted in 2328 of 4489 fibers or \SI{52}{\percent} of the fibers being assigned an MU. When the parameter is instead set to $\sigma = n/100  = \num{0.67}$ as in the study with 10 MUs before, the result assigns only 136 fibers or \SI{3}{\percent}. This shows that method 2 is very sensitive to the choice of the standard deviation parameter $\sigma$.

The comparison with \cref{fig:MU_fibre_distribution_67x67_100_some_2d_fiber_distribution.pdf} shows that the resulting MU territories are more dispersed than for method 1. This can be explained with the higher value of $\sigma$. Obtaining \say{sharper} MU territories would require a smaller $\sigma$, however, this results in less fibers being assigned to MUs.

Furthermore, \cref{fig:MU_fibre_distribution_sparse2_67x67_100_2d_fiber_distribution} shows that the fiber density decreases towards the outer border of the domain. In reality, staining studies on skeletal muscles do not find this effect.
%In an electrophysiology simulation with muscle fibers where the aim is to compute EMG signals on the surface of a muscle, this effect is not desired, as the border of the domain is the important part that contributes to the measurements of surface EMG.

An advantage of method 2 over method 1 is that we do not have to solve any optimization problem. In consequence, the algorithm for the scenario in \Cref{fig:MU_fibre_distribution_sparse2_67x67_100_2d_fiber_distribution} was completed in \SI{8}{\sec} on the same hardware as before.

%chunksize 10, Total duration: 352.44 s 
% MU_fibre_distribution_combined_67x67_100.txt_2d_fiber_distribution_.pdf
% MU_fibre_distribution_combined_67x67_100.txt_fiber_distribution_.pdf
% MU_fibre_distribution_combined_67x67_100_0_2d_fiber_distribution_.pdf
% MU_fibre_distribution_combined_67x67_100_1.txt_2d_fiber_distribution_.pdf
% MU_fibre_distribution_combined_67x67_100_2.txt_2d_fiber_distribution_.pdf
% MU_fibre_distribution_combined_67x67_100_3.txt_2d_fiber_distribution_.pdf

%chunksize 5, Total duration: 293.03 s

% sparse:
% MU_fibre_distribution_combined_sparse_251x251_100_2d_fiber_distribution.pdf

\begin{figure}%
  \centering%
  \begin{subfigure}[t]{0.48\textwidth}%
    \centering%
    \includegraphics[width=\textwidth]{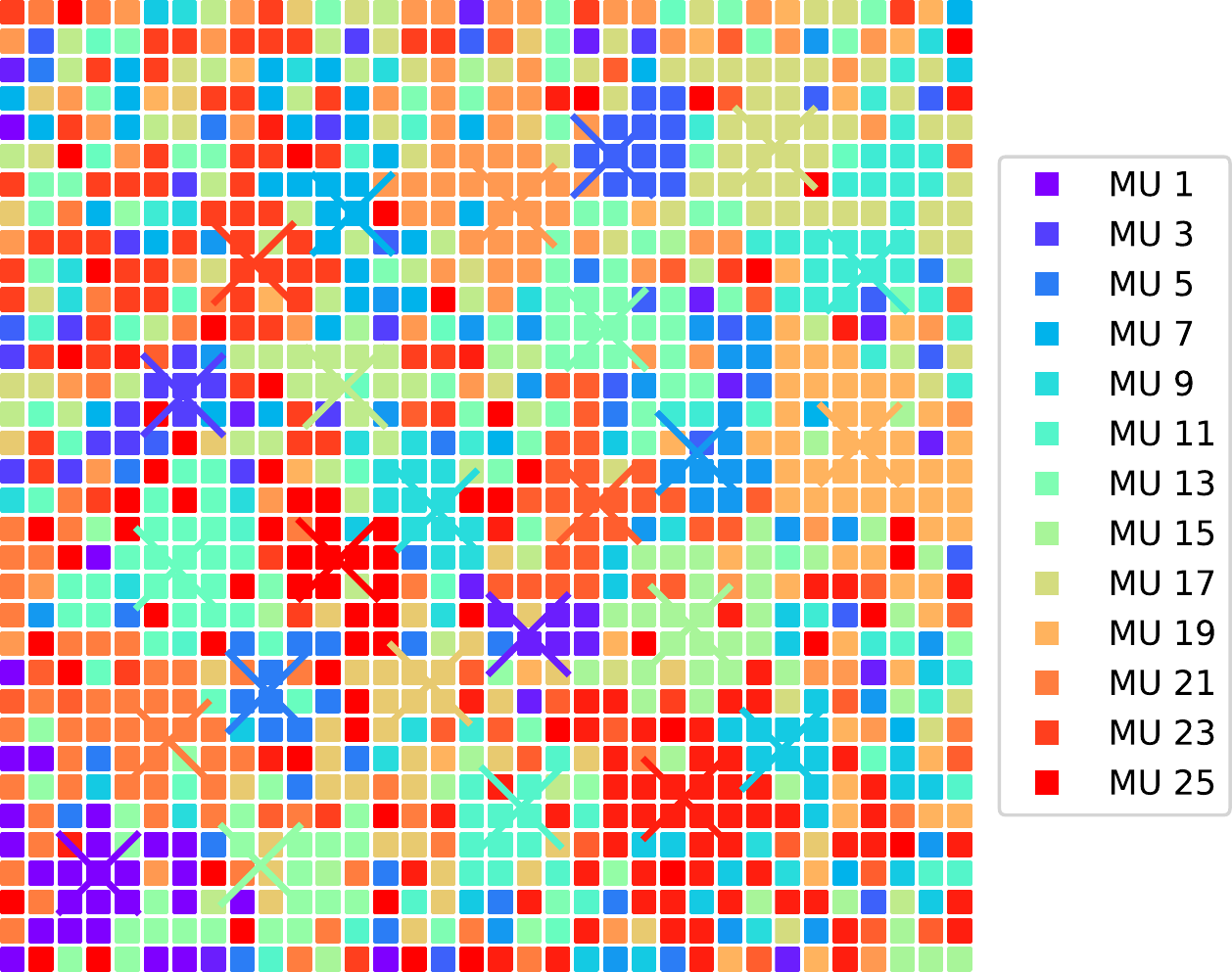}%
    \caption{First partial grid, $n=34$.}%
    \label{fig:mu_3partial_1}%
  \end{subfigure}
  \,
  \begin{subfigure}[t]{0.48\textwidth}%
    \centering%
    \includegraphics[width=\textwidth]{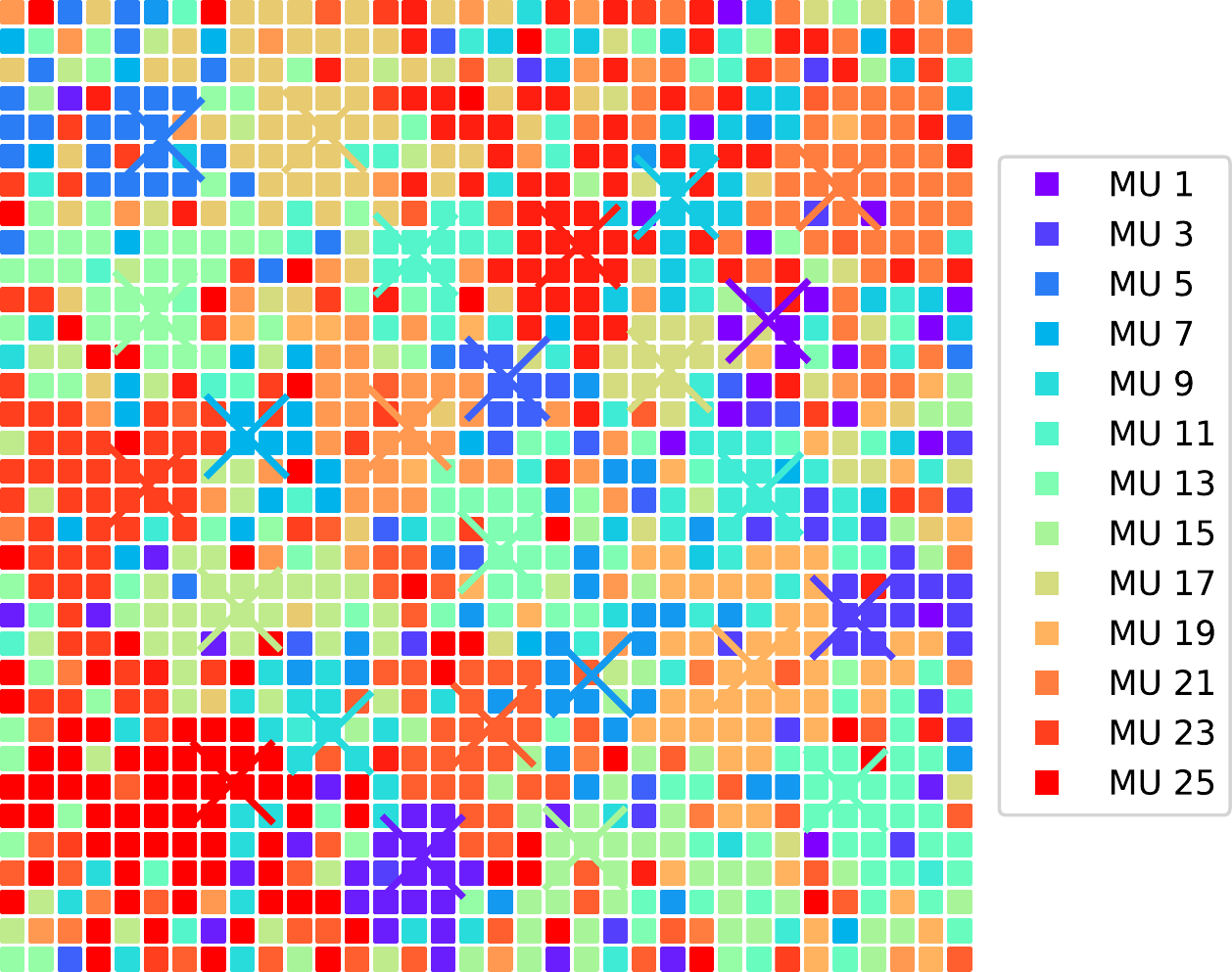}%
    \caption{Second partial grid, $n=34$.}%
    \label{fig:mu_3partial_2}%
  \end{subfigure}
  \,
  \begin{subfigure}[t]{0.48\textwidth}%
    \centering%
    \includegraphics[width=\textwidth]{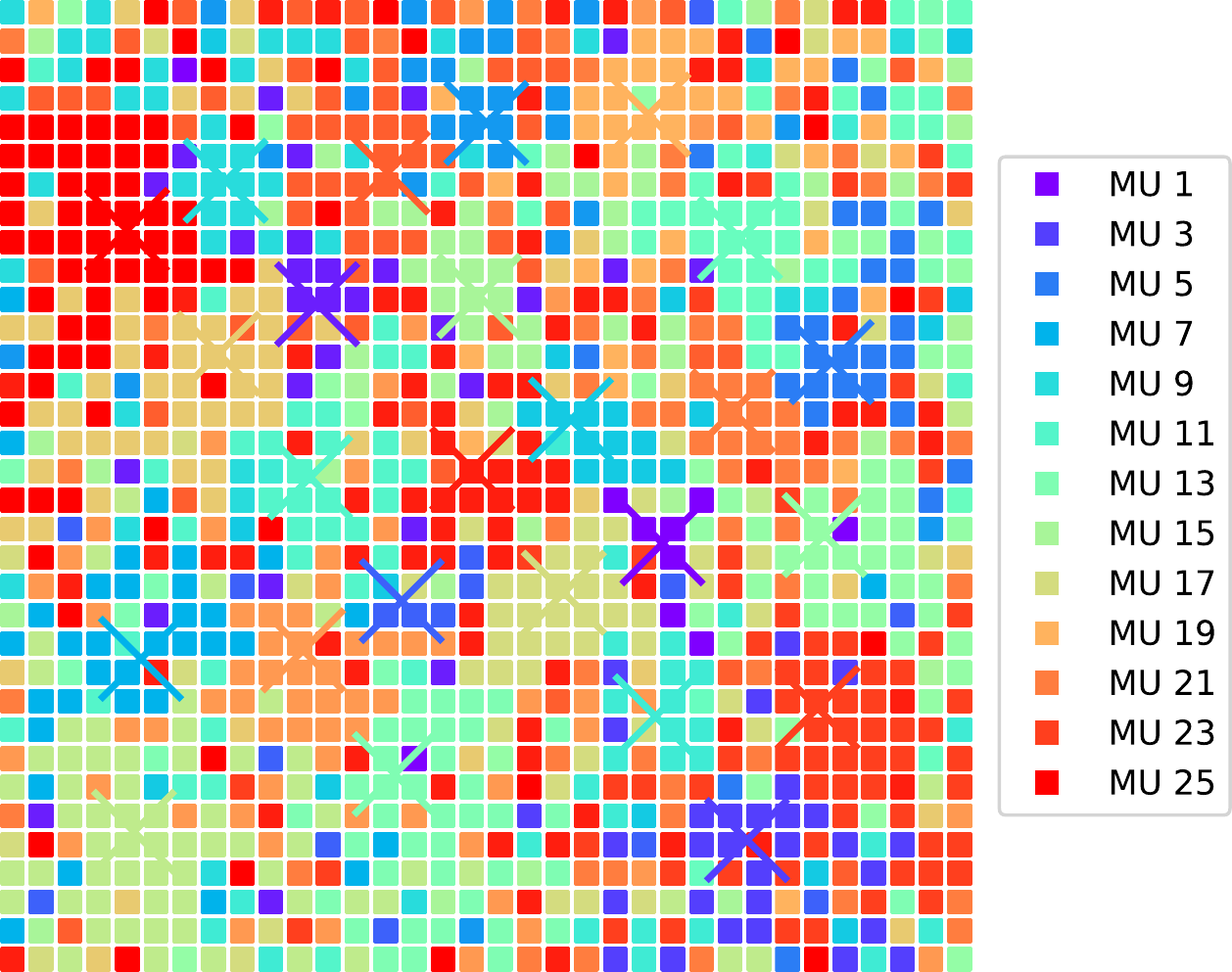}%
    \caption{Third partial grid, $n=34$.}%
    \label{fig:mu_3partial_3}%
  \end{subfigure}
  \,
%  \begin{subfigure}[t]{0.48\textwidth}%
%    \centering%
%    \includegraphics[width=\textwidth]{images/motor_unit_assignment/MU_fibre_distribution_combined_67x67_100_3_2d_fiber_distribution_.pdf}%
%    \caption{Forth partial grid.}%
%    \label{fig:mu_3partial_4}%
%  \end{subfigure}
  \begin{subfigure}[t]{0.48\textwidth}%
    \centering%
    \includegraphics[width=\textwidth]{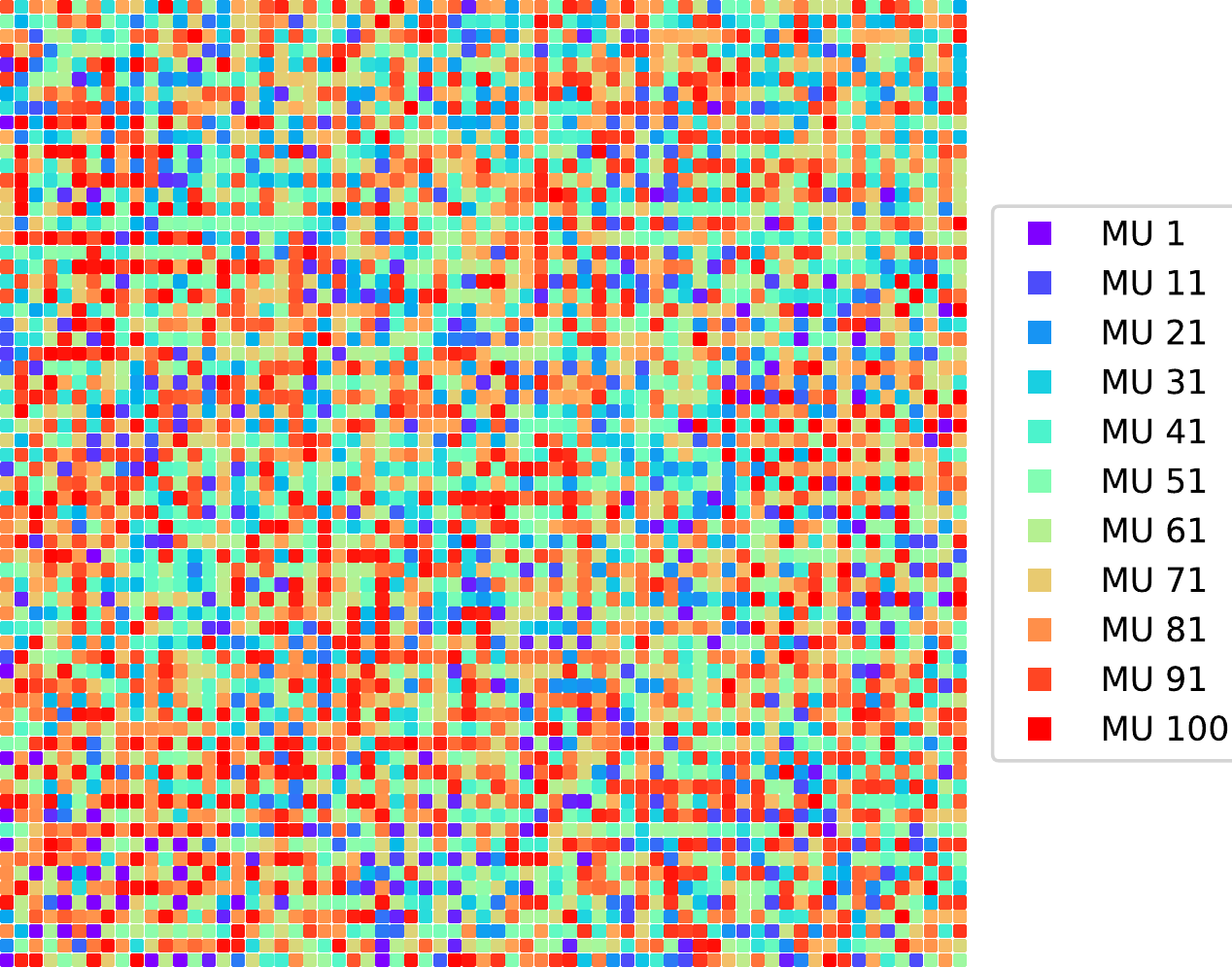}%
    \caption{Resulting interleaved grid, $n=67$.}%
    \label{fig:mu_3_1}%
  \end{subfigure}
  \caption{Assignment of motor units to fibers using method 1a. Shown are the first three partial grids (a)-(c) to be interleaved and the result (d), parameters $n=67, n_\text{MU}=100, \sigma = n/10 = 6.7, b=1.05$.}%
  \label{fig:mu_method3_partial}%
\end{figure}

\begin{figure}%
  \centering
  \includegraphics[width=0.9\textwidth]{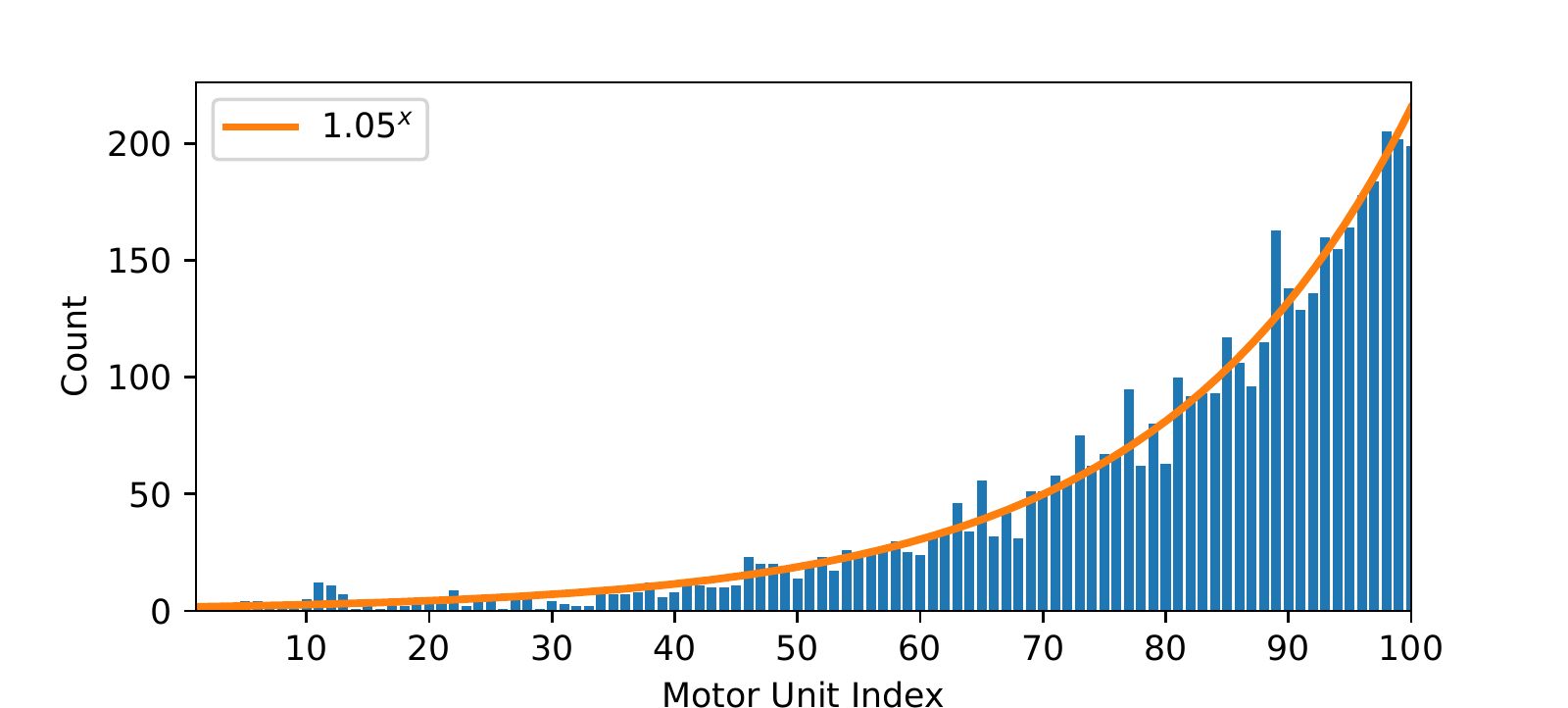}%
  \caption{Histogram of number of fibers assigned to MUs for method 1a, for the scenario that is shown in \cref{fig:mu_method3_partial}.}%
  \label{fig:mu_method3_distribution}%
\end{figure}

Next, the extension of methods 1 and 2, called 1a and 2a, are investigated that ensure that neighboring fibers are not associated to the same MU.
\Cref{fig:mu_method3_partial} shows results for method 1a for $n_\text{MU}=100$ MUs. In \cref{fig:mu_3partial_1,fig:mu_3partial_2,fig:mu_3partial_3}, three of the four partial grids with $n=34$ are shown. 
Because parameters are the same for those smaller grids, the generated MU assignments look similar for all partial grids, except for different MU center positions. In \cref{fig:mu_3_1}, the resulting grid with $n=67$ is shown that is obtained by interleaving the four partial grids. In this MU association, all neighboring fibers belong to different MUs. The resulting distribution of MU sizes is shown in \cref{fig:mu_method3_distribution}. It can be seen that the algorithm for method 1a achieves the approximate, prescribed exponential progression.

An advantage of method 1a is also that the runtime decreases compared to method 1. The result in \cref{fig:mu_3_1} could be computed in \SI{5}{\min} \SI{52.4}{\sec} with $n_\text{per\_chunk}=10$ or in \SI{4}{\min} \SI{53.0}{\sec} with $n_\text{per\_chunk}=5$, compared to the \SI{45}{\min} \SI{53.5}{\sec} of method 1.

\begin{figure}%
  \centering
  \includegraphics[width=0.9\textwidth]{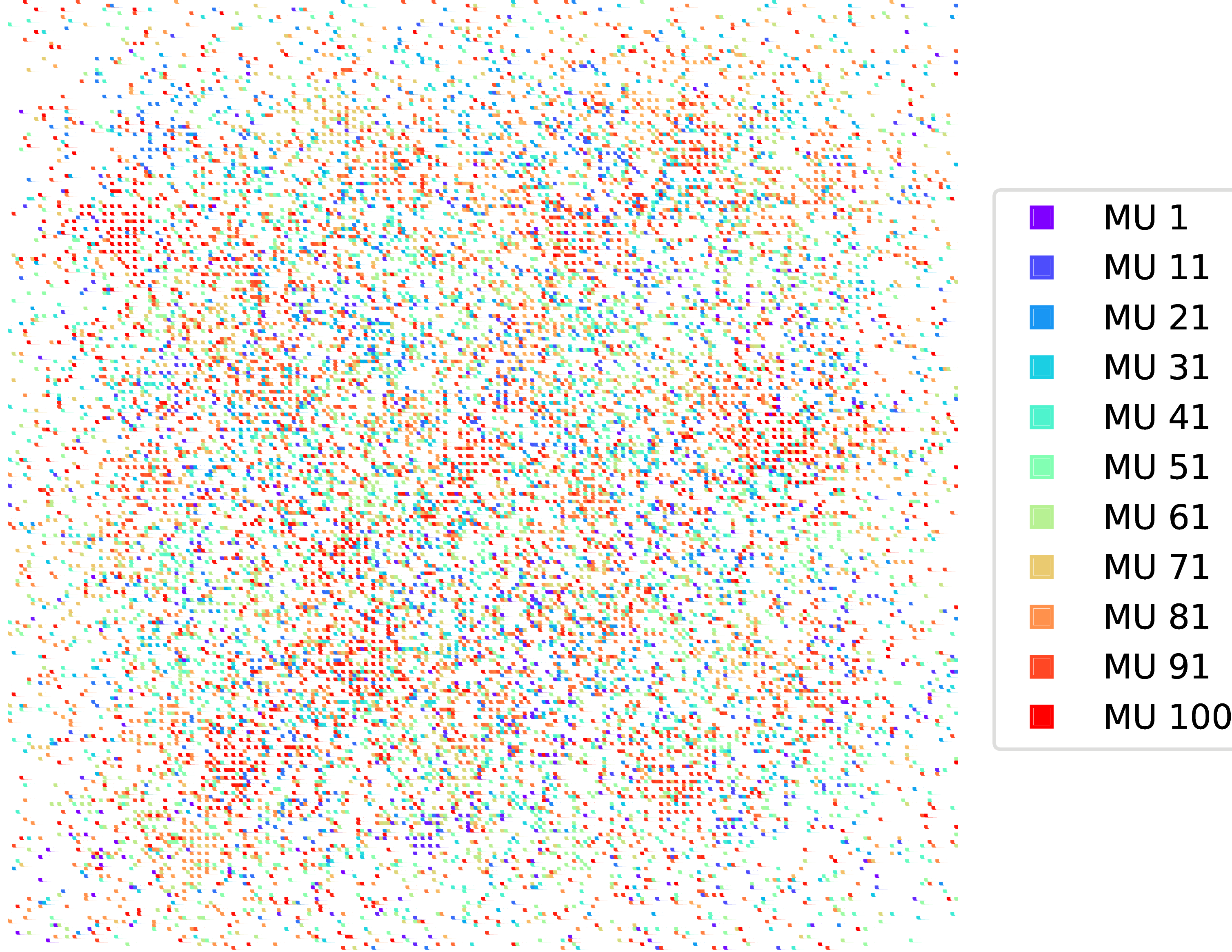}%
  \caption{Assignment of motor units to fibers using method 2a and parameters $n=251, n_\text{MU}=100, \sigma = n/100 = 2.51, b=1.05$.}%
  \label{fig:mu_method3_2}%
\end{figure}

Method 2a cannot be reasonably used with the same parameters as method 1a. If it is used to generate fibers assigned to 100 MUs, a grid of $67 \times 67$ leads to the majority of MUs having only 1 fiber. Therefore, a larger grid is needed. \Cref{fig:mu_method3_2} shows the result for $n=251$. The result assigns \num{13618} of the $n^2 = \num{63001}$ initial fibers to MUs, i.e. \SI{22}{\percent}. The number of fibers per MU varies between \num{41} and \num{244}. As can be seen, fibers of the same MU are separated by either a fiber of a different MU or by an unassigned fiber, i.e., a hole in the grid.

%It can be seen in \cref{fig:mu_method3_2} that the fibers are dispersed over the grid and holes are present throughout the domain. The resulting set of fibers can be viewed as being arbitrarily positioned in the domain instead of following the grid. When this viewpoint is adopted, the property that neighboring fibers are of different MUs no longer holds as the separation is often given by a hole, which does not prevent the fibers touching each other.

% resulting fibers: 13618 / 63001 = 22\%

\begin{reproduce}
  Run the script \code{generate_fiber_distribution.py} without arguments to get usage information.
  The script contains the implementation for all three presented methods. For example, to run method 1a to get the result of \cref{fig:mu_method3_partial}, use:
  \begin{lstlisting}[columns=fullflexible,breaklines=true,postbreak=\mbox{\textcolor{gray}{$\hookrightarrow$}\space}]
    generate_fiber_distribution.py MU_fibre_distribution_combined_67x67_100 100 3 1 67 1.05 100 10
  \end{lstlisting}
  Then, existing fiber distribution files can be visualized by the following script:
  \begin{lstlisting}[columns=fullflexible,breaklines=true,postbreak=\mbox{\textcolor{gray}{$\hookrightarrow$}\space}]
    $\$$OPENDIHU_HOME/examples/electrophysiology/input/plot_fibre_distribution_2d.py MU_fibre_distribution_combined_67x67_100.txt 67
  \end{lstlisting}
\end{reproduce}

\section{Summary and Conclusion}\label{sec:mu_conclusion}
In the beginning of this chapter, two methods 1 and 2 for associating MUs with fibers in a given 2D grid were presented. The methods were constructed based on biophysical properties of MU distribution. The fibers were located in intermingling MU territories that were each centered at different MU center points.
The density of fibers belonging to an MU decreased with higher distance from the center and was described by a radial kernel function.
The number of fibers assigned to the motor units approximately followed an exponential progression where the first MU contained the lowest number of fibers and the last MU contained the largest amount. 
Whereas method 1 assigned MUs to all available fibers, method 2 only assigned MUs to some fibers yielding a lower number of fibers in the result.
Evaluation of the literature showed that no comparable method with these properties existed previously.

Next, the methods 1a and 2a were introduced that built upon methods 1 and 2. They ensured that neighboring fibers were assigned to different MUs, another behavior that was known from anatomical studies.

Steps of the algorithms and their final satisfaction of the design criteria were demonstrated with various visualizations. Results were shown for different parameter values. The influence of the kernel width on the \say{sharpness} or intermingledness of MU territories was pointed out. 

It was found that methods 2 and 2a typically produced results where only 20-50\% of the fibers get an MU assigned. This ratio highly depended on the problem size and the kernel function width and no direct predictions about the number of resulting fibers was possible. Since the kernel parameter at the same time also influenced the sharpness of the MU territories, adjusting parameters to the desired outcome was an issue for these methods. Reasonable results were only achieved for a higher number of initial fibers. In contrast, methods 1 and 1a robustly produced exponentially distributed MU assignments for all parameter values.

In method 2a for large grid sizes the fibers were dispersed over the grid and \say{holes}, i.e., unassigned fibers, were present throughout the domain. The fiber density decreased towards the boundaries of the domain. No experimental evidence exists that this behavior occurs in reality. It might also be unfavorable when EMG simulations are performed where the boundary layers contribute most to the measured EMG signal on the muscle surface. In contrary, method 1a did not show this behavior.

The runtime for \num{4489} fibers and 100 MUs was over \SI{45}{\min} for method 1 and below \SI{10}{\sec} for method 2. The large difference could be explained with the optimization problem that needed to be solved for method 1. To handle large runtimes for a high number of MUs, an algorithm was presented that splits the optimization problem in smaller chunks that could be solved faster.

With the use of the extended methods 1a and 1b, runtime decreased. For method 1a the runtime was under \SI{6}{\min}. These runtimes are all considered acceptable since the task occurs only once during preprocessing. 

In conclusion, the developed method 1a proved to be robust for all tested parameter combinations and fulfilled all considered biophysical properties of MU distributions. The exponential distribution of MU sizes and the sharpness of MU territories are adjustable through parameters. If the condition that neighboring fibers belong to different MUs is not desired, method 1 can be used instead.

The presented methods are implemented and made available as Open Source software within \opendihu{}.
The program stores the resulting MU assignments in a plain text file format that is compatible with both OpenCMISS Iron and \opendihu{}. Thus, it can and will be used in simulations of the multi-scale chemo-electromechanical model.

\chapter{Models and Discretization}\label{chap:models_and_discretization}
In this chapter, the mathematical description of the multi-scale model and its discretization is presented. 
We use the multi-scale chemo-electro-mechanical model that was introduced in literature \cite{Roehrle2012,Heidlauf2013,Heidlauf2015Diss,Mordhorst2015}. Additional models known from literature are incorporated that  were previously only simulated in isolation: The multidomain description for electrophysiology \cite{Klotz2020}, a model of neural stimulation \cite{Cisi2008} and sensory organ models such as the muscle spindle model of Mileusnic et al. \cite{Mileusnic2006Spindle}. Similarly, models of Golgi tendon organs can be added \cite{Mileusnic2006Golgi}.

% components of the multi-scale model
\begin{figure}%
  \centering%
  \includegraphics[width=\textwidth]{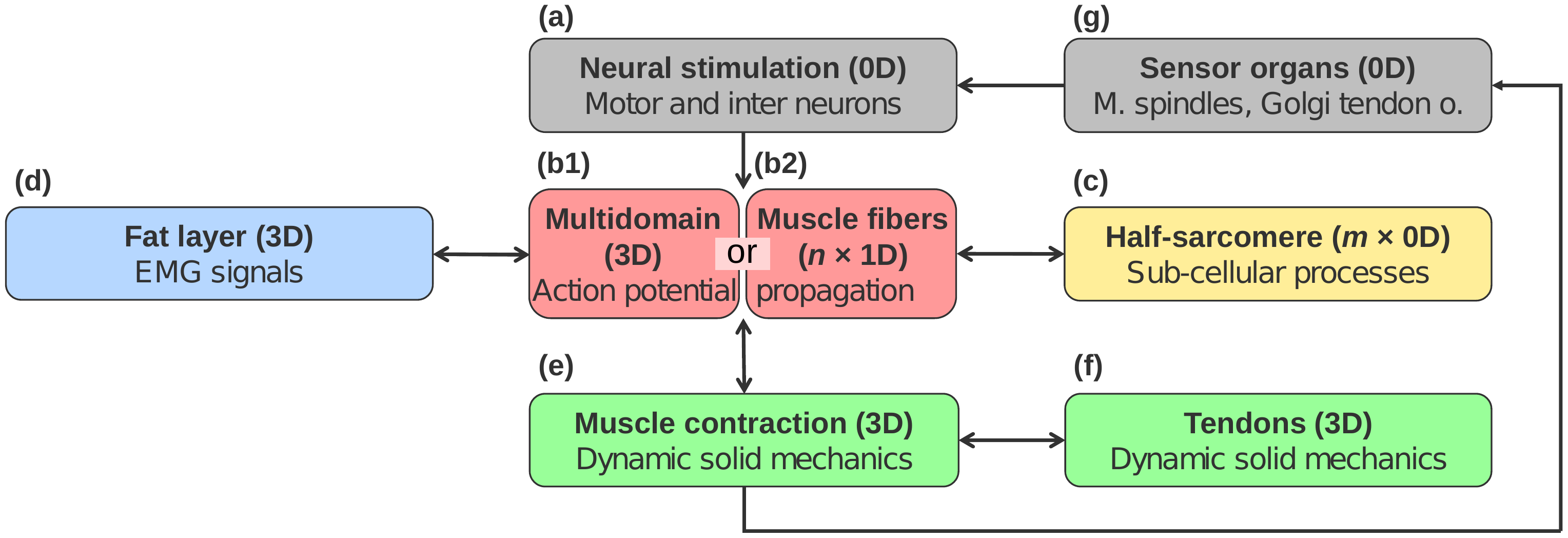}%
  \caption{Interacting components of the multi-scale model.}%
  \label{fig:multi-scale-model}%
\end{figure}

\Cref{fig:multi-scale-model} shows an overview of the components of the implemented multi-scale model.
A pool of motor neurons drives the stimulation of the muscular system in \cref{fig:multi-scale-model} (a). 
The axons of each motor neuron innervate the muscle fibers corresponding to the same MU and transmit rate-encoded stimulation signals.

In the muscle tissue, action potentials propagate starting at the neuromuscular junctions and subsequently reach the whole length of the muscle.
In our multi-scale model, two different formulations are available to describe this phenomenon. The multidomain description (\cref{fig:multi-scale-model} (b1)) models the MUs from a  homogenized 3D perspective. The description with muscle fibers (\cref{fig:multi-scale-model} (b2)) models action potential propagation explicitly with $n$ 1D muscle fibers. 

Both of these descriptions of electrophysiology involve a subcellular model (\cref{fig:multi-scale-model} (c)). This model describes the ionic processes involving the fiber membranes and taking place within one half of a sarcomere as the smallest unit to generate muscle forces. A large number $m$ of instances of this model has to be computed. 

In addition to the physiology of the muscle, a layer of body fat and skin on top of the muscle belly can be added to the model. This 3D fat layer (\cref{fig:multi-scale-model} (d)) is used to simulate EMG recordings on the skin surface. The model for the fat layer is unidirectionally coupled with the muscle fiber model (\cref{fig:multi-scale-model} (b2)) or bidirectionally coupled with the multidomain model (\cref{fig:multi-scale-model} (b1)). Using the multidomain model, it is, thus, possible to simulate external stimulation by electrodes on the skin, which is subject to research in neuroprosthetics.

The activated muscle generates force by subcellular processes on a molecular scale. They are computed on the cellular level by the half-sarcomere model (c). On the macroscopic scale, stresses lead to strains and contraction of the muscle. This effect is described by the muscle contraction model on a 3D domain (\cref{fig:multi-scale-model} (e)). 
The description is coupled with the electrophysiology models (b1),(b2) by the geometry of the contracting muscle and fibers. It is coupled with the subcellular model by the generated active stresses of the half-sarcomere. Displacements and stresses can be computed for the muscle belly itself, but also for the connected body fat layer and for elastic tendons (\cref{fig:multi-scale-model} (f)). Depending on the research questions, the contraction model is either formulated quasi-static or fully dynamic taking into account inertia effects.

Sensory organs such as muscle spindles and Golgi tendon organs sense fiber stretch and contraction velocity (\cref{fig:multi-scale-model} (g)). They are connected with the motor neuron pool by layers of interneurons and modulate the stimulation in \cref{fig:multi-scale-model} (a).

In this chapter, \cref{sec:model_equations} presents mathematical descriptions of the electrophysiology model components in the multi-scale framework and \cref{sec:model_muscle_contraction} derives the solid mechanics models. Then, \cref{sec:discretization,sec:discretization_mechanics} address the spatial and temporal discretizations of the electrophysiology and mechanics descriptions, respectively.

\section{Electrophysiology Model Equations}\label{sec:model_equations}
In the following, more details and mathematical descriptions are given for the outlined models. The section begins with the 0D half-sarcomere model in \cref{sec:subcelullar_model}, followed by the bidomain and monodomain models in \cref{sec:bidomain_model,sec:monodomain_model}, which constitute the muscle fiber based model of electrophysiology. \Cref{sec:multidomain_model} continues with the multidomain model. Electric conduction in the body fat layer is described in \cref{sec:electric_conduction_body_domain}. An overview of the continuum mechanics model used for muscle contraction is given in \cref{sec:model_muscle_contraction}.
% ----------
\subsection{Subcellular Model}\label{sec:subcelullar_model}

Propagation of electric stimuli along muscle fibers involves activation and deactivation of ion channels and ion pumps in the fiber membrane  (the sarcolemma) and in the transverse tubules.
Functioning of these processes on the subcellular scale have first been suggested in 1952 by Hodgkin and Huxley after their studies of the squid giant axons \cite{Hodgkin1952,hodgkin1952propagation}. To date, their mathematical model still serves as the basis for electrophysiology models and some of their predictions, e.g., on gating currents that occur during opening of channels, were experimentally confirmed later.

The fiber membrane separates intra- and extracellular space and can be locally described by an electric circuit. The membrane voltage $V_m=\phi_i-\phi_e$ is the difference between the intra and extracellular potentials $\phi_i$ and $\phi_e$. The membrane stores charges $Q$, quantifiable by its electric capacitance $C_m$:
\begin{align}\label{eq:subcellular_model_helper1}
  Q = C_m\cdot V_m.  
\end{align}
A change in the transmembrane potential, e.g., induced by an action potential leads to a change in $Q$, which is accounted for by an electric current $I$ over the membrane. This can be formally obtained by the derivative of \cref{eq:subcellular_model_helper1} with respect to time:%
\begin{align}\label{eq:subcellular_model_helper2}
  \d{Q}{t} = C_m \cdot \d{V_m}{t}.
\end{align}
The current $I=\d Q / \d t$ is realized by ions passing through the membrane.
Significant ions in this process are sodium $(\text{Na}^{+})$ and potassium ions $(\text{K}^{+})$.
Considering a particular point on the fiber, these ions diffuse through ion-specific channels in the membrane.
The diffusion is driven by an interplay of the ion concentration gradient and the electric field that is caused by action potentials.

Without any electric field imposed by action potentials, the equilibrium state of the diffusion process for sodium and potassium ions is given by their Nernst potentials $E_{\text{Na}^{+}}$ and $E_{\text{K}^{+}}$. These voltage levels depend on logarithmic relations between extra- and intracellular concentrations scaled by constants describing the thermal energy and the number of electrons.
In thermodynamic equilibrium, the membrane voltage is equal to the Nernst potential $E_i$ of the involved ions $i$. 
At a higher membrane voltage $V_m$, the remainder ($V_m - E_i$) is the part of the electric field that drives the
ion fluxes and electric currents through the membrane. The currents depend on the conductivity $g_i$ of
the membrane for ion $i$. 

Apart from sodium and potassium ions, the diffusion of less frequent ions and ionic pumps can be lumped by a leakage current $I_L$ that is modeled by a channel with constant conductivity $\bar{g}_L$.
With this, the total ionic membrane current $I_\text{ion}$ is formulated as
\begin{subequations}
\begin{align}\label{eq:subcellular_model_helper4}
  I_\text{ion}(V_m)  &= I_{\text{Na}^{+}} + I_{\text{K}^{+}} + I_L \\
  & = g_{\text{Na}^{+}}\,(V_m - E_{\text{Na}^{+}}) + g_{\text{K}^{+}}\,(V_m - E_{\text{K}^{+}}) + \bar{g}_L\,(V_m - E_L). \label{eq:subcellular_model_helper5}
\end{align}
\end{subequations}
The conductivities $g_{\text{Na}^{+}}$ and $g_{\text{K}^{+}}$ for the sodium and potassium channels depend on the transmembrane voltage $V_m$ and its history.

In addition to the ionic current $I_\text{ion}$, an externally driven current $I_\text{ext}$ can be modeled that occurs as a result of neural stimulation at the neuromuscular junctions. Substituting the current $I=\d Q / \d t$ in \cref{eq:subcellular_model_helper2}, we get the following differential equation for the membrane voltage $V_m$:
\begin{align}\label{eq:subcellular_model_helper3}
  C_m \cdot \d{V_m}{t} = -I_\text{ion}(V_m) + \dfrac{I_\text{ext}}{A}.
\end{align}
The negative sign of the ionic current $I_\text{ion}$ is in accordance with the definition of the membrane voltage as $V_m=\phi_i-\phi_e$. The external current $I_\text{ext}$ is divided by the surface area $A$ of the stimulating electrode or neuromuscular junction, as the description considers an infinitesimal area on the membrane.

Hodgkin and Huxley suggested that ion channels can be activated and deactivated. This molecular process requires independent \say{gating} particles to move to a new position in order for a channel to be activated. For the potassium channel, four of these independent events have to occur, each modeled by a probability $n$. The resulting probability for the channel to open is, thus, $n^4$. For the sodium channel, three such events are assumed for activation and another one for the deactivation of the channel, described by the probabilities $m$ and $h$, respectively. The values of the probabilities change over time and modulate the conductivities of the ion channels:
\begin{align*}
  g_{\text{Na}^{+}} &= \bar{g}_{\text{Na}^{+}} \cdot m(t)^3 \cdot h(t), \quad &g_{\text{K}^{+}} &= \bar{g}_{\text{K}^{+}} \cdot n(t)^4.
\end{align*}
Here, $\bar{g}_{\text{Na}^{+}}$ and $\bar{g}_{\text{K}^{+}}$ are channel specific constants. The gating variables $n,m$ and $h$ can be interpreted as probabilities for the events to occur or as the amount of occurred events related to all available gating particles. 
The evolution of the activation probability $n$ is modeled by the following ordinary differential equation (ODE):%
\begin{align*}
  \d{n}{t} = \alpha_n(V_m) \cdot (1-n) + \beta_n(V_m) \cdot n,
\end{align*}
analogously for $h$ and $n$. The transition rates between activation probability $n$ and deactivation probability $(1-n)$ are nonlinearly dependent on the membrane voltage $V_m$.

\begin{figure}%
  \centering%
  \def\svgwidth{0.4\textwidth}
  %% Creator: Inkscape inkscape 0.92.3, www.inkscape.org
%% PDF/EPS/PS + LaTeX output extension by Johan Engelen, 2010
%% Accompanies image file 'ode_solution.pdf' (pdf, eps, ps)
%%
%% To include the image in your LaTeX document, write
%%   \input{<filename>.pdf_tex}
%%  instead of
%%   \includegraphics{<filename>.pdf}
%% To scale the image, write
%%   \def\svgwidth{<desired width>}
%%   \input{<filename>.pdf_tex}
%%  instead of
%%   \includegraphics[width=<desired width>]{<filename>.pdf}
%%
%% Images with a different path to the parent latex file can
%% be accessed with the `import' package (which may need to be
%% installed) using
%%   \usepackage{import}
%% in the preamble, and then including the image with
%%   \import{<path to file>}{<filename>.pdf_tex}
%% Alternatively, one can specify
%%   \graphicspath{{<path to file>/}}
%% 
%% For more information, please see info/svg-inkscape on CTAN:
%%   http://tug.ctan.org/tex-archive/info/svg-inkscape
%%
\begingroup%
  \makeatletter%
  \providecommand\color[2][]{%
    \errmessage{(Inkscape) Color is used for the text in Inkscape, but the package 'color.sty' is not loaded}%
    \renewcommand\color[2][]{}%
  }%
  \providecommand\transparent[1]{%
    \errmessage{(Inkscape) Transparency is used (non-zero) for the text in Inkscape, but the package 'transparent.sty' is not loaded}%
    \renewcommand\transparent[1]{}%
  }%
  \providecommand\rotatebox[2]{#2}%
  \newcommand*\fsize{\dimexpr\f@size pt\relax}%
  \newcommand*\lineheight[1]{\fontsize{\fsize}{#1\fsize}\selectfont}%
  \ifx\svgwidth\undefined%
    \setlength{\unitlength}{581.6692915bp}%
    \ifx\svgscale\undefined%
      \relax%
    \else%
      \setlength{\unitlength}{\unitlength * \real{\svgscale}}%
    \fi%
  \else%
    \setlength{\unitlength}{\svgwidth}%
  \fi%
  \global\let\svgwidth\undefined%
  \global\let\svgscale\undefined%
  \makeatother%
  \begin{picture}(1,0.75243665)%
    \lineheight{1}%
    \setlength\tabcolsep{0pt}%
    \put(0,0){\includegraphics[width=\unitlength,page=1]{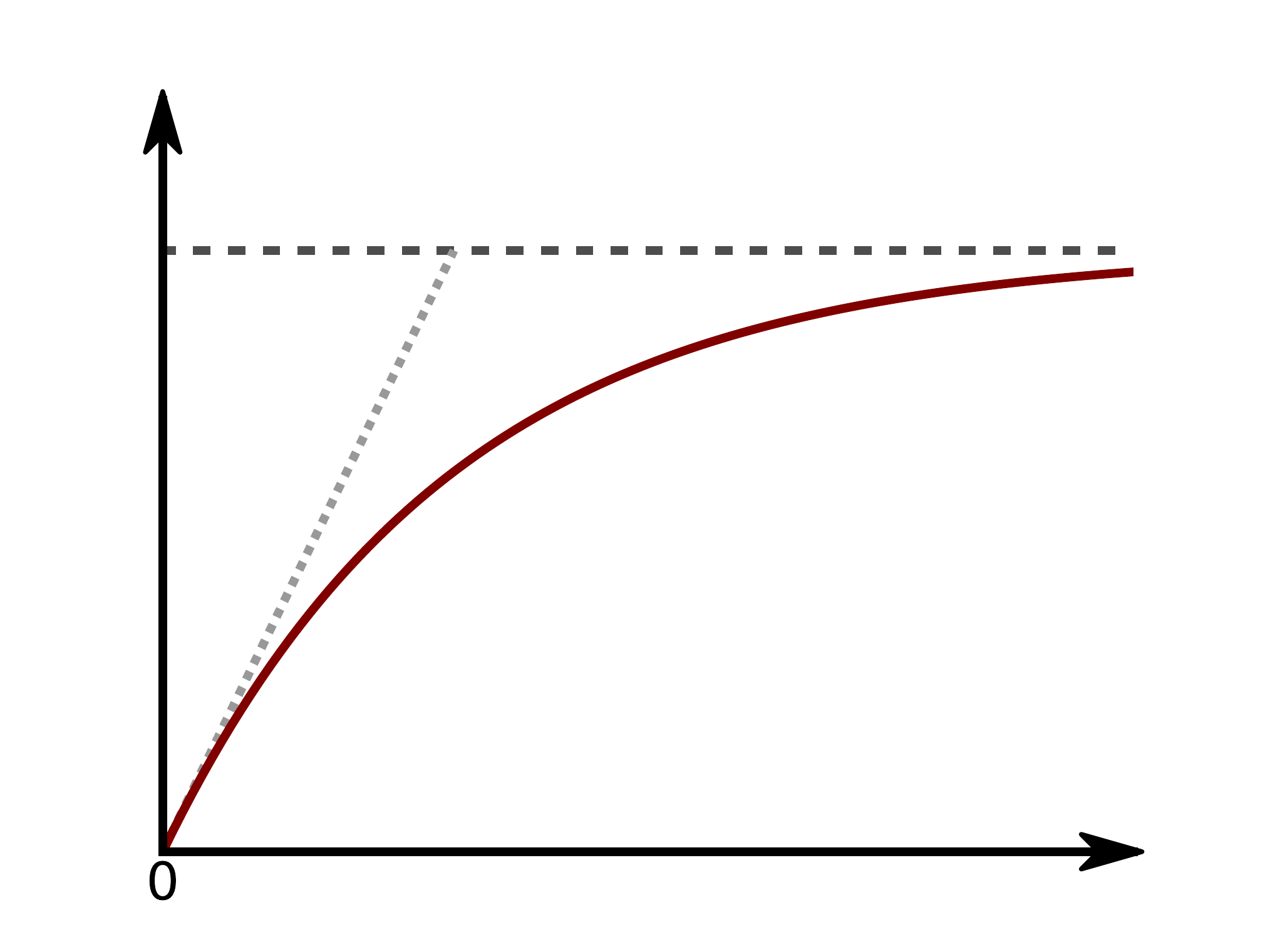}}%
    \put(0.3315636,-0.00911793){\color[rgb]{0,0,0}\makebox(0,0)[lt]{\lineheight{0}\smash{\begin{tabular}[t]{l}$\tau_n$\end{tabular}}}}%
    \put(0,0){\includegraphics[width=\unitlength,page=2]{ode_solution.pdf}}%
    \put(-0.03301438,0.51298673){\color[rgb]{0,0,0}\makebox(0,0)[lt]{\lineheight{0}\smash{\begin{tabular}[t]{l}$n_\infty$\end{tabular}}}}%
    \put(0.86382485,-0.00911793){\color[rgb]{0,0,0}\makebox(0,0)[lt]{\lineheight{0}\smash{\begin{tabular}[t]{l}$t$\end{tabular}}}}%
    \put(0.52521248,0.37586131){\color[rgb]{0.50196078,0,0}\makebox(0,0)[lt]{\lineheight{0}\smash{\begin{tabular}[t]{l}$n(t)$\end{tabular}}}}%
  \end{picture}%
\endgroup%
  \caption{Subcellular model: Graph of the analytic solution (red) of the ordinary differential equation that is part of the activation model of ion channels for constant transmembrane voltage and initial condition $n(0)=0$, given in \cref{eq:subcellular_model_helper0}. The variables $n_\infty$ and $\tau_n$ can be interpreted as the equilibrium value and a characteristic timescale, respectively.}%
  \label{fig:ode_solution}%
\end{figure}

For a constant $V_m$, this ODE has an analytical solution
\begin{align}\label{eq:subcellular_model_helper0}
  n(t) = n_\infty\big(1 - \exp(1 - \dfrac{1}{\tau_n}t)\big),
\end{align}
which for $t\to \infty$ converges to the equilibrium value $n_\infty := \alpha_n\,\tau_n$ as shown in \cref{fig:ode_solution}. The time constant $\tau_n := 1/(\alpha_n + \beta_n)$ indicates how fast the solution approaches the equilibrium, e.g., when starting from $n(0)=0$, half of the value of the equilibrium is reached after $t_{1/2}=\log(2)\,\tau_n$. The smaller $\tau_n$, the stiffer is the ODE, which needs to be considered in the choice of a suitable numerical solution scheme.

Because the transmembrane voltage $V_m$ changes over time, the ODEs for $n, m$ and $h$ have to be solved numerically. Then, the dependent ionic current $I_\text{ion}$ can be calculated. Thus, the model is a system of differential-algebraic equations (DAE).

The internal states in this model can be combined into a state vector $\bfy = (n,m,h)^\top$. 
The combined right-hand side for all states is formulated as a vector-valued function $G(V_m,\bfy)$.
In summary, the system of DAEs for the subcellular model on a subcellular domain $\Omega_s$ can be written in the following form:
\begin{align}
  \dfrac{\partial\bfy}{\partial t} &= G(V_m,\bfy),& I_\text{ion}&=I_\text{ion}(V_m,\bfy) \quad \text{on }\Omega_s\label{eq:subcellular}.
\end{align}
For an exemplary solution that shows how the membrane potential changes over time, see \cref{fig:action_potentials}. 

The system of equations in \cref{eq:subcellular} together with \cref{eq:subcellular_model_helper3} describe the subcellular processes on a single point $\Omega_s \subset \Omega_f$ on a muscle fiber $\Omega_f$. 
It does not model the interaction between neighboring points that leads to propagation of action potentials.
To account for action potential propagation, ionic currents $I_\text{ion}$ on multiple points are coupled within the multidomain or fiber models that are formulated in the multi-scale framework. This is described in the following sections, \cref{sec:bidomain_model,sec:monodomain_model,sec:multidomain_model}.
Using these models, the system of ODEs in \cref{eq:subcellular} has to be solved for multiple subcellular points $\Omega_s^i$ in the muscle domain.

After Hodgkin and Huxley proposed this model in 1952, more detailed models were formulated that take into account more ion channels, ion pumps and more advanced biochemical processes within the cell. One particular model is the one proposed by Shorten et al. \cite{Shorten2007}, which adds the full pathway from activation to excitation-contraction coupling in the sarcomere. It has a state vector of $\bfy \in \R^{56}$ and is used to compute active stresses for simulations of muscle contraction. 
It can also be written in the form given in \cref{eq:subcellular}.
Apart from $I_\text{ion}$, another value $\gamma = H(\bfy,\dot{\lambda}_f)$ is computed by an additional equation from the vector of states $\bfy$ and the fiber contraction velocity $\dot{\lambda}_f$, which is given to the model as a parameter.
The value $\gamma$ is a lumped activation parameter in the range $\gamma \in [0,1]$ that describes the amount of active stress generated in the sarcomere and can be linked to the continuum mechanics model of muscle contraction.

%and alter the ion concentration gradients across the membrane.
%ion-specific channels
%electrochemical gradient
%channels respond to a change in transmembrane potential 
%altered ion concentration gradients

% ----------
\subsection{Bidomain Model}\label{sec:bidomain_model}

A description of electrophysiology on a general 3D muscle tissue is given by the bidomain model formulated by \cite{tung1978bi,peskoff1979electric}. The bidomain model considers the intra (index $i$) and extracellular spaces (index $e$) in a homogenized setting, such that the two domains coexist at every spatial point $\bfx \in \Omega\subset \R^3$. Similar to the setting of the subcellular model, the two domains in the bidomain model have locally varying electric potential fields $\phi_i$ and $\phi_e$ that yield a locally varying transmembrane voltage $V_m=\phi_i - \phi_e$. Electric conduction within the two domains is governed by conductivity tensors $\bfsigma_i$ and $\bfsigma_e$. 

Assuming static conditions, a spatially varying electric potential $\phi$ induces the electric field $E=-\grad \phi$.
According to Ohm's law, the resulting current density $j$ is given by 
\begin{align}\label{eq:bidomain_helper1}
  j = \bfsigma\,E = -\bfsigma\,\grad \phi \quad \text{in }\Omega.
\end{align}
This holds for both intra and extracellular domain, yielding expressions for $j_i$ and $j_e$.

The intracellular and the extracellular domain are electrochemically coupled. Thus, one assumption is that currents are preserved and a change in current density on one domain corresponds to the opposite change in current density in the other domain. This is expressed by the divergence of the current densities, which in one domain equals to the negated value in the other domain:
\begin{align}\label{eq:bidomain_helper2}
  \div(j_i) = -\div(j_e) \quad \text{in }\Omega.
\end{align}
This change in current density directly corresponds to a current flow over the membrane:%
\begin{align*}
  \div(j_i) = A_m\,I_m \quad \text{in }\Omega.
\end{align*}
Here, the factor $A_m$ describes the membrane area to domain volume relationship. It is needed to convert the units between current per volume and current per area. The membrane current $I_m$ is given by the subcellular model of Hodgkin and Huxley in \cref{eq:subcellular_model_helper3}. Neglecting the external current $I_\text{ext}$ in \cref{eq:subcellular_model_helper3} and using the formulation of the intracellular current density $j_i$ in \cref{eq:bidomain_helper1}, we get:
\begin{align*}
  \div\big(\bfsigma_i\,\grad(\phi_i)\big) = A_m\,\big(C_m\,\p{V_m}{t} + I_\text{ion}(V_m)\big) \quad \text{in }\Omega.
\end{align*}
The ionic current $I_\text{ion}$ can be computed by \cref{eq:subcellular_model_helper5}. 
By plugging \cref{eq:bidomain_helper1} also into \cref{eq:bidomain_helper2} and rewriting the equations in terms of the extracellular potential $\phi_e$ and the transmembrane voltage $V_m = \phi_i-\phi_e$, we get the bidomain equations:%
\begin{subequations}
  \begin{align}
    \div\big((\bfsigma_i + \bfsigma_e)\,\grad(\phi_e)\big) + \div(\bfsigma_i\,\grad(V_m)\big) &= 0,\label{eq:bidomain1} \\[4mm]
    \div\big(\bfsigma_i\,\grad(V_m)\big) + \div(\bfsigma_i\,\grad(\phi_e)\big) &= A_m\,\big(C_m\,\p{V_m}{t} + I_\text{ion}(V_m)\big).  \label{eq:bidomain2}
  \end{align}
\end{subequations}
With appropriate boundary conditions, these equations are often used to model cardiac electrophysiology. They also serve as a basis for the fiber models in our multi-scale setting, which will be described in the next section.

% ----------
\subsection{Monodomain Model}\label{sec:monodomain_model}
% ----------
One approach to modeling skeletal muscle electrophysiology is to explicitly resolve muscle fibers and compute propagating action potentials on these spatial domains.
Propagation of action potentials can be described by the monodomain equation, which is a specialization of the bidomain equations for a one-dimensional intracellular space.

We assume a muscle domain $\Omega_M \subset \R^3$ with a number of embedded 1D manifolds $\Omega_f^j\subset \R^3$ for $j=1,\dots,n$ that represent muscle fibers. The domain $\Omega_M$ represents the extracellular space and each fiber domain $\Omega_f^j$ represents a separate  intracellular space.
It is further assumed that electric conduction in the extracellular space is directed equally to the embedded fibers. This can be stated as%
\begin{align}\label{eq:monodomain_helper1}
  \bfsigma_i = k\cdot \bfsigma_e.  
\end{align}
The intracellular conductivity tensor $\bfsigma_i$ (here prolonged from the scalar value $\sigma_i$ on a fiber with tangent $\bfa \in \R^3$ to the 3D domain by $\bfsigma_i = \sigma_i \, \bfa \otimes \bfa$) and the extracellular conductivity $\bfsigma_e$ are multiples of each other with a scaling factor $k\in\R$.

Plugging \cref{eq:monodomain_helper1} into the first bidomain equation, \cref{eq:bidomain1}, and restricting the domain to a 1D fiber $\Omega_f^j$ allows to combine the terms related to $\phi_e$:
\begin{align*}
  \div\big(\sigma_i\,\grad(\phi_e)\big) = -\dfrac{k}{k+1} \div\big(\sigma_i\,\grad(V_m)\big) \quad \text{on }\Omega_f^j.
\end{align*}
Using the second bidomain equation, \cref{eq:bidomain2}, we get the expression
\begin{align*}
  \div\big(\sigma_\text{eff} \,\grad(V_m)\big) = A_m\big(C_m\,\p{V_m}{t} + I_\text{ion}(V_m,\bfy)\big) \quad \text{on }\Omega_f^j.
\end{align*}
The effective conductivity $\sigma_\text{eff}$ combines the intra and extracellular conductivities, $\sigma_i$ and $\sigma_e$, analog to a parallel circuit:
\begin{align*}
  \sigma_\text{eff} := \sigma_i \parallel \sigma_e = \dfrac{\sigma_i\,\sigma_e}{\sigma_i + \sigma_e}.
\end{align*}
Rearranging the terms yields the classical form of the monodomain equation:
\begin{align} %eq:monodomain
  &\dfrac{\partial V_m}{\partial t} = \dfrac{1}{A_m\,C_m}\big(\sigma_\text{eff}\dfrac{\partial^2 V_m}{\partial x^2} - A_m\,I_\text{ion}(V_m,\bfy)\big) \quad && \text{for }x \in \Omega_f^j. \label{eq:monodomain}
\end{align}

The multi-scale framework uses multiple instances of the monodomain equation \\\cref{eq:monodomain} together with the first bidomain equation \cref{eq:bidomain1} to model electrophysiology in the fibers and the extracellular domain \cite{Mordhorst2015}. In addition to the fiber domains $\Omega_f^j$, two instances of the muscle domain $\Omega_M$ are needed for the bidomain equation, one for the intracellular and one for the extracellular space. The transmembrane potential $V_m$ is unidirectionally coupled from the fiber meshes to the intracellular space of the first bidomain equation. The extracellular potential $\phi_e$ corresponds to the signals that are measured during intramuscular EMG recording.

Within the multi-scale framework, it is also possible to couple a model for electric conduction in an additional layer of body fat tissue. This is subsequently described in \cref{sec:electric_conduction_body_domain}. 
Then, electric current fluxes between the muscle and body fat domains have to be modeled.

If no such additions should be made to the model, the following Neumann boundary conditions are used to close the description:
\begin{subequations}\label{eq:monodomain_bc}
\begin{align}
  \p{V_m}{x} &= 0 & \quad \text{on }∂\Omega_f^j, \label{eq:monodomain_bc1}\\[4mm]
  \big(\bfsigma_i\,\grad(V_m)\big) \cdot \bfn_m &= -\big(\bfsigma_i\,\grad(\phi_e)\big)\cdot \bfn_m & \quad\text{ on }∂\Omega_M, \label{eq:monodomain_bc2}\\[4mm]
  \big(\bfsigma_e\,\grad(\phi_e)\big) \cdot \bfn_m &=0  & \quad\text{ on }∂\Omega_M, \label{eq:monodomain_bc3}
\end{align}
\end{subequations}
with the outward normal vector $\bfn_m$. 
\Cref{eq:monodomain_bc1} defines homogeneous Neumann boundary conditions for the monodomain equation \cref{eq:monodomain} at the two ends of each 1D muscle fiber domain. 
The boundary conditions on $∂\Omega_M$ are related to the bidomain equations given in \cref{eq:bidomain1,eq:bidomain2}.
\Cref{eq:monodomain_bc2} is equivalent to a homogeneous Neumann boundary condition on the intracellular current density $j_i$ (cf. \cref{eq:bidomain_helper1}) and is expressed in terms of the transmembrane voltage $V_m$ and the extracellular potential $\phi_e$.
Another homogeneous Neumann boundary condition on $\phi_e$ as given by \cref{eq:monodomain_bc3} is required.

% ----------
\subsection{Multidomain Model}\label{sec:multidomain_model}

The multidomain model is an alternative approach to the description based on the monodomain and bidomain equations described in \cref{sec:bidomain_model,sec:monodomain_model}. It was proposed in \cite{Klotz2020} and describes the same physics. However, the muscle fibers are homogenized and all equations are formulated using a single 3D muscle domain $\Omega_M$.

The multidomain equations generalize the two bidomain equations and allow taking into account multiple MUs by defining a separate intracellular space for each MU. Thus, at every spatial point $\bfx \in \Omega_M$ one extracellular and  $N_\text{MU}$ intracellular domains or compartments coexist, where $N_\text{MU}$ is the number of MUs. As before, the extracellular domain has the electric potential $\phi_e$ and  conductivity tensor $\bfsigma_e$. For each compartment $k = 1, \dots, N_\text{MU}$, a separate electric potential $\phi_i^k$, transmembrane voltage $V_m^k = \phi_i^k-\phi_e$, conductivity tensor $\bfsigma_i^k$, surface to volume ratio of the membrane $A_m^k$ and membrane capacitance $C_m^k$ are defined.

Analog to the fibers of a MU that exhibit different densities at different locations in the muscle, each compartment occupies different locations within the domain to a different extent. This is described by the relative occupancy factor $f_r^k: \Omega_M \to [0,1]$ for MU $k$. The factors have different values in the domain according to the presence of the MU at the respective location. At every point, their sum is one, $\sum_{k=1}^{N_\text{MU}} f_r^k = 1$, if all MUs should be considered or less than one if the effect of remainder MUs that will not be activated in the simulation scenario is neglected.

The first multidomain equation is similar to the first bidomain equation \cref{eq:bidomain1} and balances the current flow between the extracellular space and the weighted sum of all intracellular spaces:%
\begin{align}\label{eq:multidomain_helper1}
  \div\big(\bfsigma_e\,\grad(\phi_e)\big)  + \ds\sum\limits_{k=1}^{N_\text{MU}} f_r^k\,\div\big(\bfsigma_i^k\,\grad(V_m^k + \phi_e)\big) = 0.
\end{align}
By defining a total intracellular conductivity tensor $\bfsigma_i = \sum_{k=1}^{N_\text{MU}}f_r^k\,\bfsigma_i^k$, \cref{eq:multidomain_helper1} can be restated as
\begin{align}\label{eq:multidomain1}
  \div\big((\bfsigma_e + \bfsigma_i)\,\grad(\phi_e)\big) + \sum\limits_{k=1}^{N_\text{MU}} f_r^k\,\div\big(\bfsigma_i^k\,\grad(V_m^k)\big) = 0.
\end{align}

The second multidomain equation equals the second bidomain equation \cref{eq:bidomain2}. It describes the current over the membrane and holds for every compartment:%
\begin{align*}
  \div(\bfsigma_i^k\,\grad(V_m^k + \phi_e)\big) = A_m^k\,\Big(C_m^k\,\p{V_m^k}{t} + I_\text{ion}(V_m^k)\Big) && \forall k \in \{1, \dots, N_\text{MU}\}.
\end{align*}
It is convenient to rearrange it for $\partial{}V_m^k / \p t$:
\begin{align}\label{eq:multidomain2}
  \p{V_m^k}{t} = \dfrac{1}{A_m^k\,C_m^k} \Big(\div\big(\bfsigma_i^k\,\grad(V_m^k + \phi_e)\big) - A_m^k I_\text{ion}(V_m^k)\Big) && \forall  k \in \{1, \dots, N_\text{MU}\}.
\end{align}
The current $I_\text{ion}$ over the membrane is again computed by the subcellular model given by \cref{eq:subcellular_model_helper5}.

The resulting system of \cref{eq:multidomain1,eq:multidomain2} constitutes the first and second multidomain equations and can be used to compute muscle electrophysiology.
The boundary conditions are defined analogously to \cref{eq:monodomain_bc2,eq:monodomain_bc3}:
\begin{subequations}\label{eq:multidomain_bc}
\begin{align}
  \big(\bfsigma_i^k\,\grad(V_m^k)\big) \cdot \bfn_m &= -\big(\bfsigma_i^k\,\grad(\phi_e)\big)\cdot \bfn_m && \quad\text{on }∂\Omega_M\quad && \forall k \in \{1,\dots,N_\text{MU}\},  \label{eq:multidomain_bc1}\\[4mm]
  \big(\bfsigma_e\,\grad(\phi_e)\big) \cdot \bfn_m &= 0, && \quad\text{on }∂\Omega_M \label{eq:multidomain_bc2}
\end{align}
\end{subequations}
where $\bfn_m$ is the outward normal vector on $∂\Omega_M$.

% ----------
\subsection{Electric Conduction in the Body Domain}\label{sec:electric_conduction_body_domain}

Surface EMG signals are the result of electric conduction in the electrically active muscle tissue as well as in surrounding inactive tissue such as adipose tissue and skin or connective tissue such as tendons and ligaments. This surrounding tissue is summarized by the body domain $\Omega_B$, which partly shares its boundary with the muscle domain $\Omega_M$. 

\begin{figure}[t]
  \centering
  \includegraphics[width=50mm]{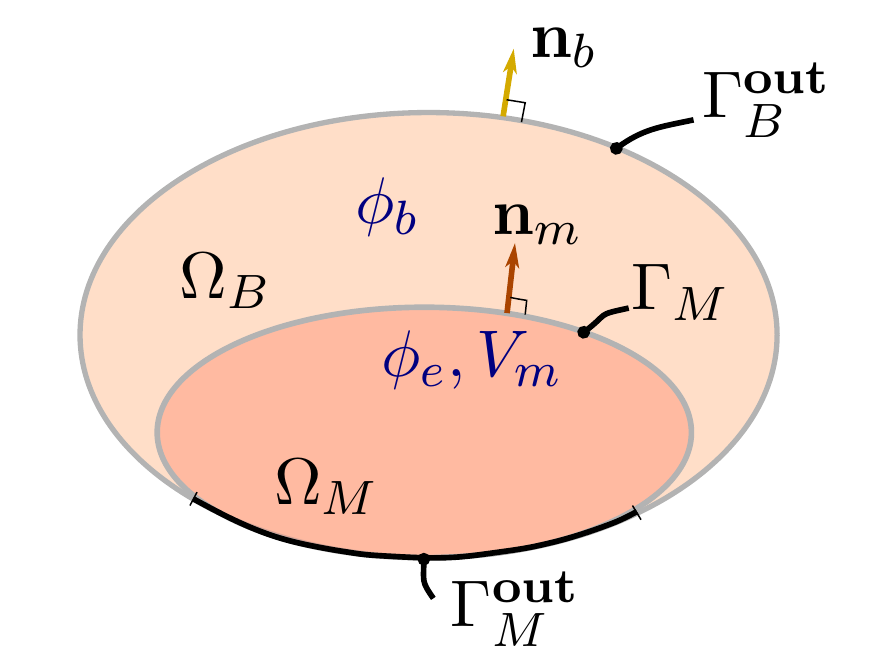}
  \caption{Computational domains for the simulation of surface EMG. The body domain $\Omega_B$ and the muscle domain $\Omega_M$ share a part of their boundary, $\Gamma_M$, which has a normal vector $\bfn_m$. The outer boundary is composed of $\Gamma_B^\text{out}$ and $\Gamma_M^\text{out}$ and has the outward normal vector $\bfn_b$. }
  \label{fig:body_domain_visualization}
\end{figure}

\Cref{fig:body_domain_visualization} visualizes these domains and defines their names: The domains $\Omega_M$ and $\Omega_B$ have outward normals $\bfn_m$ and $\bfn_b$, the outer boundary is composed of $\Gamma_B^\text{out}$ and $\Gamma_M^\text{out}$ and the variables $\phi_e,V_m$ and $\phi_b$ are defined as shown within the domains $\Omega_M$ and $\Omega_B$.

The work of \cite{Mordhorst2015} proposes an isotropic conductivity $\bfsigma_b$ and a harmonic electric potential $\phi_b$ in the body domain $\Omega_B$:
\begin{align}\label{eq:body}
  \div \big(\bfsigma_b\,\grad (\phi_b)\big) = 0 \qquad \text{on } \Omega_B.
\end{align}

The electric potentials $\phi_e$ and $\phi_b$ of the neighboring domains $\Omega_M$ and $\Omega_B$ as well as the current densities are continuous on the shared boundary $\Gamma_M$. This is described by the following two coupling conditions:
\begin{subequations}\label{eq:body_domain_coupling}
  \begin{align}
    \phi_e &= \phi_b  \qquad &&\text{on } \Gamma_M, \label{eq:body_domain_bc1}   \\[4mm]
    \big(\bfsigma_e\, \grad( \phi_e)\big)\cdot \bfn_m &= \big(\bfsigma_b\, \grad(  \phi_b)\big)\cdot \bfn_m \qquad &&\text{on } \Gamma_M.\label{eq:body_domain_bc2}
  \end{align}
\end{subequations}
On the outer boundary $\Gamma_B^\text{out}$, homogeneous Neumann boundary conditions are assumed:
\begin{align}% eq:body_domain_bc3
  \big(\bfsigma_b\, \grad( \phi_b)\big)\cdot \bfn_b &= 0 \qquad &&\text{on } \Gamma^\text{out}_B.\label{eq:body_domain_bc3}
\end{align}

The description of the body domain has to be combined either with the fiber based description in \cref{sec:monodomain_model} or the multi-domain description in \cref{sec:multidomain_model}. In the literature, this combination was mathematically described for the fiber based model in \cite{Mordhorst2015} and for the multidomain model in \cite{Klotz2020}. Correspondingly, additional boundary conditions either given by \cref{eq:monodomain_bc} or \cref{eq:multidomain_bc} are assumed: For the fiber based description, which uses the bidomain equation for volume conduction, the boundary conditions are:%
\begin{subequations}
\begin{align}
  \big(\bfsigma_i\,\grad(V_m)\big) \cdot \bfn_m &= -\big(\bfsigma_i\,\grad(\phi_e)\big)\cdot \bfn_m && \quad\text{on }∂\Omega_M=\Gamma_M \cup \Gamma_M^\text{out}, \label{eq:monodomain_fat_bc1}\\[4mm]
  \big(\bfsigma_e\,\grad(\phi_e)\big) \cdot \bfn_m &=0  && \quad\text{on }∂\Gamma_M^\text{out}. \label{eq:monodomain_fat_bc2}
\end{align}
\end{subequations}
For the multidomain description with fat layer, the boundary conditions are:
\begin{subequations}
\begin{align}
  \big(\bfsigma_i^k\,\grad(V_m^k)\big) \cdot \bfn_m &= -\big(\bfsigma_i^k\,\grad(\phi_e)\big)\cdot \bfn_m && \quad\text{on }∂\Omega_M=\Gamma_M \cup \Gamma_M^\text{out},  \label{eq:multidomain_fat_bc1}\\[4mm]
  \big(\bfsigma_e\,\grad(\phi_e)\big) \cdot \bfn_m &= 0 && \quad\text{on }∂\Gamma_M^\text{out}. \label{eq:multidomain_fat_bc2}
\end{align}
\end{subequations}
The first condition in \cref{eq:multidomain_fat_bc1} is enforced for all compartments $k=1,\dots,N_\text{MU}$.
%

% ----------

\section{Model of Muscle Contraction}\label{sec:model_muscle_contraction}

%Because of possibly large strains, a nonlinear hyperelastic formulation is used. 
%For mathematical foundations in continuum mechanics, we refer to basic literature such as the books of Holzapfel \cite{holzapfel2000nonlinear} and Marsden and Hughes \cite{marsden1994mathematical}, as well as literature on the application of the finite element method in continuum mechanics \cite{zienkiewicz1977finite,SUSSMAN1987357,zienkiewicz2005finite}.

%The following section provides a more profound introduction of solid mechanics to complement the overview given in \cref{sec:model_muscle_contraction}. It also describes the finite element discretization of the solid mechanics model and the algorithms used to obtain a numeric solution. 

Muscle contraction is described on the organ level by a solid mechanics model. The goal is to describe the deformation of the tissue caused by the internal forces that are generated by sarcomeres and as a response to outer constraints such as applied forces, the attachment to tendons and inertia effects.

Different modeling approaches exist to describe the mechanical muscle behavior. Dynamic \emph{finite elasticity} methods for large strains exist that use hyperelastic materials, both compressible and incompressible. Further, \emph{linear elasticity} descriptions with linearizations at various levels are used in appropriate applications where small strains can be assumed. The whole range from simplifying linearized models to accurate nonlinear approaches can be found in the literature, sometimes with varying conventions and symbols.
In this section, we introduce consistent notation and formulate the model equations for both approaches. The discretization and solution is discussed later in \cref{sec:discretization_mechanics}.

The derivation largely follows the book of Holzapfel \cite{holzapfel2000nonlinear} and the discretization follows the work of Zienkiewicz, Taylor et al. \cite{zienkiewicz1977finite,zienkiewicz2005finite}. Further details can be found also in the book of Marsden and Hughes \cite{marsden1994mathematical}.

%  The implementation of a solver for such generic descriptions exploiting parallel execution and integrating a multi-scale biomechanics model, being a contribution of this work, is an interdisciplinary endeavour. Therefore, we introduce consistent notation and summarize the required basics and the derivation up to the final algorithm  such that it may serve also readers that are not specialized in the field of continuum mechanics. 

% -v-

\subsection{Geometric Description}\label{sec:geometric_description}

% introduce quantities
% F, C, E, E(u), variation δE
%  S, P, sigma

We begin with the geometric description of the material body and define the basic quantities that are subsequently used to describe the physics. We consider the 3D muscle domain $\Omega_0=\Omega_M \subset \R^3$ in reference configuration at time $t=0$ that deforms into a current configuration $\Omega_t$ at time $t$. The material points are given by $\bfX \in \Omega_0$. The corresponding points $\bfx \in \Omega_t$ in the current configuration are defined by the function $\bfx = \bfvarphi_t(\bfX)$. 

In the following, capital letters refer to quantities in material or Lagrangian description, i.e., defined in the reference configuration and small letters refer to quantities in spatial or Eulerian description, i.e., defined in the current configuration.

The relation of point coordinates in the current configuration with respect to the reference configuration can also be described by the displacements field $\bfU$:
\begin{align*}
  \bfx(\bfX) = \bfX + \bfU(\bfX).
\end{align*}
The symbol $\bfu$ with $\bfu(\bfx(\bfX))=\bfU(\bfX)$ denotes the displacements formulated in current configuration.
The current velocity $\bfv$ is the time derivative of the displacements, $\bfv := \dot{\bfu}$.

The deformation gradient $\bfF$ is the second order tensor that is obtained by differentiating the function $\bfvarphi_t$. It is given using the unit vectors $\bfe_i$ and components $F_{aA}$:
\begin{align*}
  \bfF &= F_{aA}\,\bfe_a \otimes \bfe_A, \quad && F_{aA} = \p{x_a}{X_A}.
\end{align*}
Capital and small indices refer to reference and current configuration, respectively. The deformation gradient can also be expressed using the displacement field $\bfU$:
\begin{align}\label{eq:solid1}
  \bfF = \bfI + ∇\bfU.
\end{align}
Here and in the following, the gradient symbol $∇$ refers to differentiation with respect to material coordinates $\bfX$. 
We assume Cartesian coordinates.

\begin{figure}
  \centering%
  \def\svgwidth{0.7\textwidth}
  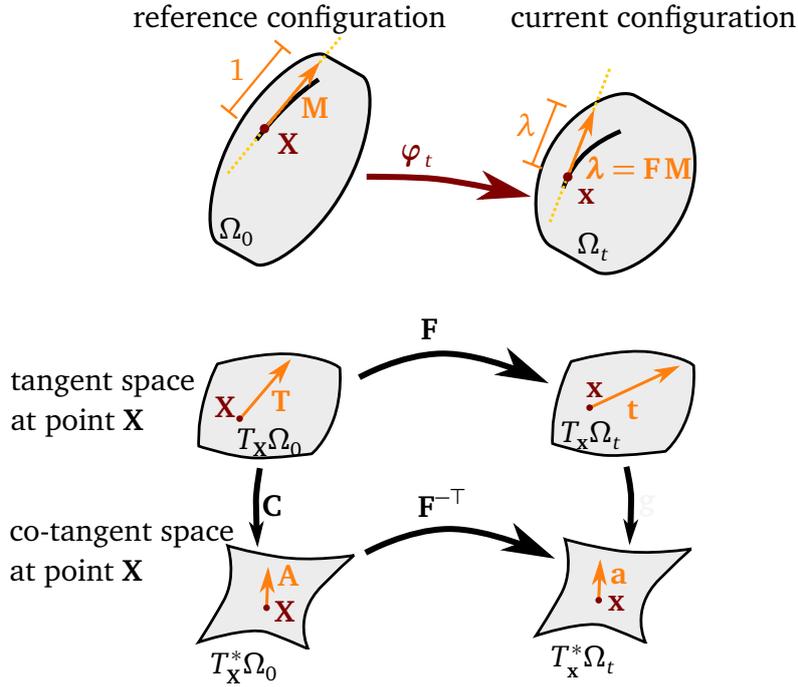%
  \caption{Vector spaces and variables used in the geometric description of the solid mechanics model. The left side shows the reference configuration with tangent and co-tangent space of point $\bfX$. The right side shows tangent and co-tangent space for the current domain and a point $\bfx$. The spatial stretch $\lambda$ is defined by mapping a material element $\bfM$ to the current configuration. The maps $\bfvarphi_t$, $\bfF$ and $\bfF^{-\top}$ map tangents $\bfT,\bft$ and normals $\bfA,\bfa$ between the configurations.}%
  \label{fig:geometric_quantities}%
\end{figure}

% tangent, normal, volume map
The determinant of the deformation gradient is $J:= \det \bfF>0$. It is positive for any physically valid transformation.
The deformation gradient is used to map geometric quantities from the reference to the current configuration:
\begin{subequations}\label{eq:geometry_maps}
  \begin{align}
    \bft &= \bfF\,\bfT, & \text{(tangent map)} \label{eq:tangent_map}\\[4mm]
    \bfa &= \cof(\bfF)\,\bfA, & \text{(normal map)} \label{eq:normal_map}\\[4mm]
    v &= J\,V. & \text{(volume map)}\label{eq:volume_map}
  \end{align}
\end{subequations}
As given in \cref{eq:tangent_map} and visualized in \cref{fig:geometric_quantities}, the tensor $\bfF$ maps material tangents $\bfT$ in $\Omega_0$ to the corresponding spatial line elements $\bft$ in $\Omega_t$. 
Accordingly, the spatial stretch at a point $\bfx \in \Omega_t$ in a certain direction is given by $\lambda=\sqrt{\bflambda^\top\,\bflambda}$ with $\bflambda = \bfF\,\bfM$, where $\bfM$ is a material line element with unit length pointing in the respective Lagrangian direction.

In \cref{eq:normal_map}, the cofactor of $\bfF$ given by $\cof(\bfF) = J\,\bfF^{-\top}$ maps normals $\bfA$ and surface areas $|\bfA|$ from $\Omega_0$ to the corresponding values $\bfa$ and $|\bfa|$ in $\Omega_t$. Nanson's formula, $\d \bfa = \cof(\bfF)\,\d\bfA$, is used to transform surface integrals from Eulerian to Lagrangian description.
Note that tangents at a point $\bfX$ live in the tangent space $T_\bfX\Omega_0$ and normals live in the co-tangent space $T^\ast_\bfX\Omega_0$.

\Cref{eq:volume_map} describes the volume map from $\Omega_0$ to $\Omega_t$, which simply scales the reference volume $V$ by the determinant $J$ to obtain the volume $v$ in the current configuration.

% geometric quantities

Furthermore, the deformation gradient $\bfF$ is used to define the right Cauchy Green tensor $\bfC = \bfF^\top\bfF$, which maps from tangent to co-tangent space in reference configuration, and subsequently the Green-Lagrange strain tensor:
\begin{align*}
  \bfE = \dfrac12(\bfC - \bfI).
\end{align*}
This strain measure can be interpreted as comparing the current Lagrangian metric $\bfC$, a measure for the symmetric part of the current deformation, with the reference metric which is the identity. Using \cref{eq:solid1}, the Green-Lagrange strain tensor can be formulated in terms of derivatives of the displacements:%
\begin{align}\label{eq:green_lagrange_u}
  \bfE &= \dfrac12\big((∇\bfU)^\top + ∇\bfU + ∇\bfU^\top ∇\bfU\big).
\end{align}

In case of small displacements, a simplification is to not distinguish between reference and current configuration.
The strain expression given in \cref{eq:green_lagrange_u} can be linearized by neglecting products of the derivatives and using the spatial displacements $\bfu$ instead of $\bfU$. As a result, the linearized strain tensor $\eps$ is given by:
\begin{align}\label{eq:linearized_helper3}
  \bfeps = \dfrac12\big((∇\bfu)^\top + ∇\bfu\big).
\end{align}
It can be used together with linear material models to derive a completely linear model.

\subsection{Stress Metrics}

Continuum mechanical models establish equations for the unknown displacement function $\bfu$ and its evolution in time via relations between stresses and strains. In the following, we introduce the required stress metrics.

% stress measures 
The Cauchy stress tensor $\bfsigma$ results from Euler's cut principle: we consider the mechanical action on an arbitrary, virtual cut out of the body in current configuration. The contact forces on the cut surface at a point $\bfx$ are described by the traction force $\bft$.
The traction vector acts on the current configuration and is a function of the position $\bfx \in \Omega_t$ and the local orientation of the cut given by the normal vector $\bfn$. Cauchy's theorem states that this relation is linear and can be described by the second order Cauchy stress  tensor $\bfsigma$:
\begin{align}\label{eq:cauchy_theorem}
  \bft = \bfsigma \cdot \bfn.
\end{align}

Thus, the Cauchy stress describes the \say{true stress} of contact forces per deformed surface area. Both slots of the second order tensor are associated with the current configuration. More specifically, $\bfsigma$ is contravariant and maps from a normal $\bfn$ in co-tangent space $T^\ast_\bfx\Omega_t$ to the traction $\bft$ in tangent space $T_\bfx\Omega_t$. 

While the physical description is natural in this Eulerian setting, the numerical treatment is more convenient in the Lagrangian setting, where we can integrate over a non-deforming domain. 
Moreover, a two-point setting, where surface areas are measured in the undeformed configuration and traction forces are measured in the deformed configuration, is often useful in engineering. This is the natural setting, e.g., in tension tests. Therefore, other stress measures involving the reference configuration are defined.

% numerics -> physics
% pull-back, push-back
Using the mappings presented in \cref{eq:geometry_maps}, all quantities can be transformed between both configurations. 
The physical derivation can be carried out equivalently in a Lagrangian or Eulerian setting and switching between them is possible at any point in the derivation. For this purpose, two operations are defined: the pull-back $\varphi^\ast(\bfa) = \bfF^\top\bfa\,\bfF$ and push-forward operations $\varphi_\ast(\bfA) = \bfF^{-\top}\bfA\,\bfF^{-1}$, which bring tensors from Eulerian to Lagrangian description and vice-versa.

% sigma, 1st PK, 2nd PK
The first Piola-Kirchhoff stress tensor $\bfP$ measures contact forces in the current configuration with regard to the area of the reference configuration and relates to the Cauchy stress as $\bfP = \bfsigma\,\cof(\bfF)$. The second Piola-Kirchhoff tensor $\bfS$ is a fully Lagrangian field given as the pull-back of the Cauchy stress scaled by $J$:%
\begin{align*}
  \bfS = \varphi^\ast(J\,\bfsigma) = J\,\bfF^{-1}\bfsigma\,\bfF^{-\top}.
\end{align*}
%It appears in the summary of the muscle contraction model in \cref{eq:contraction} in connection with the strain energy function $\Psi$.

% stress tensors
\begin{figure}
  \centering%
  \def\svgwidth{0.5\textwidth}
  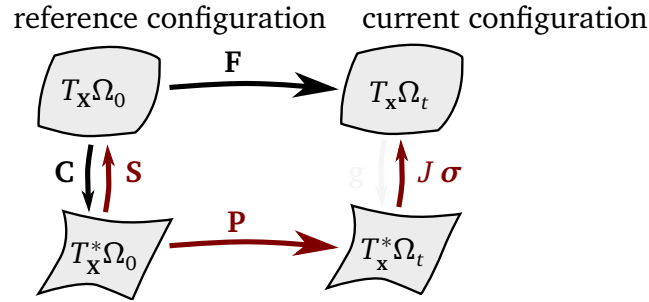%
  \caption{Stress tensors and geometric maps that can be used together in a solid mechanics formulation. The right Cauchy-Green tensor $\bfC$  and the second Piola-Kirchhoff stress $\bfS$ are dual Eulerian tensors and map between tangent space $T_\bfX\Omega_0$ and co-tangent space $T^\ast_\bfX\Omega_0$ in the reference domain. The deformation gradient $\bfF$ and the first Piola-Kirchhoff stress $\bfP$ are dual two-point tensors mapping from the reference to the current configuration. The Cauchy stress $\bfsigma$ is defined entirely in the Eulerian setting.
  %The Eulerian metric $\bfg$, which is the identity in cartesian coordinates and the Kirchhoff stress $J\,\bfsigma$ (where $\bfsigma$ is the Cauchy stress) are the dual objects in the Eulerian setting. All three pairs of dual tensors are linked together by transformations such as pull-back and push-forward.
  }%
  \label{fig:stress_tensors}%
\end{figure}

\Cref{fig:stress_tensors} summarizes the geometric maps by black arrows and the stress measures by red arrows. 
%To relate strains and stresses in a material model, the corresponding tensors have to be dual objects. Here, three settings are possible: The Eulerian setting uses the metric $\bfg$ and the dual Kirchhoff stress $J\,\bfsigma$. The two-point setting uses the deformation gradient $\bfF$ and the dual first Piola-Kirchhoff stress $\bfP$. The Lagrangian setting uses the right Cauchy-Green tensor $\bfC$ and the dual second Piola-Kirchhoff stress $\bfS$. 
%Different matching pairs of strain and stress tensors can be used to describe physical relations. 
The Lagrangian setting defines the right Cauchy-Green tensor $\bfC$ and the second Piola-Kirchhoff stress tensor $\bfS$.
We use these quantities in the derivation of the discretized equations, because the Lagrangian formulation is natural for this task and allows integrating over the non-deforming domain $\Omega_0$.

The Cauchy stress $\bfsigma$ is completely defined in the Eulerian setting. It is used to formulate physical balance principles.

% ---
\subsection{Overview of the Physical Relations}

\begin{figure}
  \centering%
  \def\svgwidth{\textwidth}
  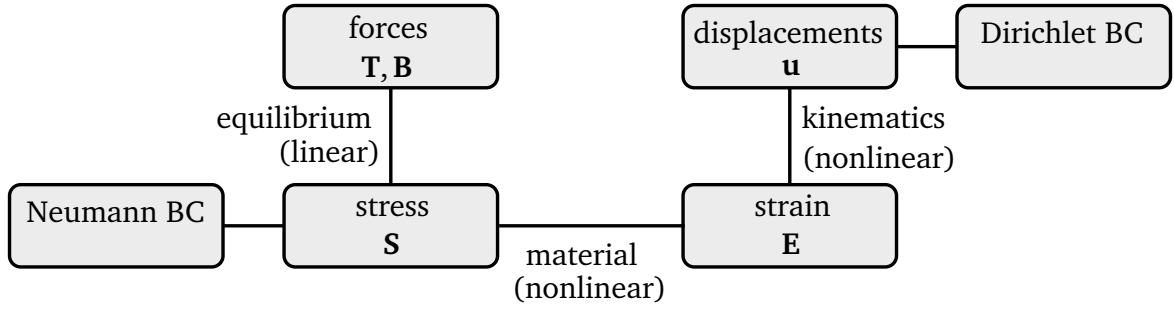%
  \caption{The three relations between various quantities that compose the solid mechanics model: Equilibrium links traction and body forces $\bfT$ and $\bfB$ to the stresses $\bfS$. A material model connects them to strains $\bfE$. The kinematic relations yield the resulting displacement field $\bfu$. Note that all quantities in this diagram are given in Lagrangian formulation.}%
  \label{fig:tonti_diagram}%
\end{figure}

The previously introduced quantities are linked together by various relations, which are summarized in the diagram in \cref{fig:tonti_diagram}. The goal is to find the relationship between given forces (top left in \cref{fig:tonti_diagram}) and the resulting deformation of the body described by the displacements (top right in \cref{fig:tonti_diagram}).
Prescribed external traction forces $\bfT$ and external or inertial body forces $\bfB$ act on the body and result in stresses $\bfS$ satisfying the \emph{equilibrium} relation. A \emph{material law} connects stresses $\bfS$ and strains $\bfE$. The \emph{kinematics} of the body determine the relationship between displacements $\bfu$ and strains $\bfE$. Geometric Dirichlet boundary conditions prescribe displacements and Neumann boundary conditions such as traction forces contribute to the stress field. 

Whereas the equilibrium relation is linear, the material and kinematic descriptions can both be chosen to be linear or nonlinear. 
In cases of small strains, geometric and material linearity can be assumed. We derive two such formulations: a linear formulation where all relations are linear and a nonlinear formulation with nonlinear material and kinematic relations.

% Tonti diagram
% linear, static and dynamic

\subsection{Assumptions and Model Equations}\label{sec:assumptions_and_model_equations}
% ----

The foundation of continuum mechanics usually builds on three balance principles: conservation of mass, of momentum and of angular momentum. In the following, these principles are presented in their Eulerian forms.

First, we assume \emph{conservation of mass} in terms of the densities $\rho_0(\bfX)$ and $\rho(\bfx)$ in reference and current configurations:
\begin{align*}
  \ds\int\limits_{V_0} \rho_0\,\d V = \int\limits_{V_t} \rho \,\d v.
\end{align*}
The equation holds for all corresponding subdomains $V_0\subset \Omega_0$ and $V_t \subset \Omega_t$. With the intermediate step of deducing $\d/\d t \int_{\Omega_t} \rho \,\d v=0$, we get the following differential equation:%
\begin{align}\label{eq:contraction_helper1}
  \dot{\rho}(\bfv,t) + \rho(\bfx,t)\,\div\big(\bfv(\bfx,t)\big) = 0.
\end{align}

As muscle tissue largely consists of water, it is typically assumed to be an incompressible domain. This is equivalent to a constant density, $\dot{\rho}=0$, and, thus, \cref{eq:contraction_helper1} reduces to
\begin{align}\label{eq:assumption_1_local}
  \div(\bfv(\bfx,t)) = 0.
\end{align}

The second assumption is the \emph{balance of momentum}, which is expressed as %
\begin{align}\label{eq:assumption_2_integral}
  \d{t} \ds\int\limits_{V_t} \rho\,\bfv\, \d v = \ds\int\limits_{V_t} \rho\,\bfb \,\d v + \ds\int\limits_{∂V_t} \bft \,\d a.
\end{align}
Here, $\bfb$ describes a body force and $\bft$ describes a traction force that acts on the surface of the domain $V_t$. Using the Cauchy theorem \cref{eq:cauchy_theorem}, it can be replaced by the Cauchy stress $\bfsigma$. The corresponding differential form is given by the following differential equation:%
\begin{align}\label{eq:assumption_2_local}
  \rho\,\dot{\bfv}(\bfx,t) = \rho\,\bfb(\bfx,t) + \div\bfsigma(\bfx,t).
\end{align}
It relates external forces to the internal stress field and describes the \emph{equilibrium} relation in \cref{fig:tonti_diagram} in Eulerian form.
For the discretization, a Lagrangian form is typically used.

For hyperelastic materials, which we consider in the muscle model, the equilibrium relation can also be formulated in terms of the \emph{Hellinger-Reissner energy functional} $\Pi_L(\bfu,p)$, which describes the potential energy of the system depending on the displacement and pressure functions $\bfu$ and $p$. Analog to the local form in \cref{eq:assumption_2_local}, it contains terms for the external loads and for the internal response of the body.
The functional is additively composed of external and internal potential energy:
\begin{align}\label{eq:hellinger_reissner}
  \Pi_L(\bfu,p) = \Pi_\text{ext}(\bfu) + \Pi_\text{int}(\bfu,p).
\end{align}
The external energy functional is formulated by
\begin{align}\label{eq:pi_ext}
  \Pi_\text{ext}(\bfu) = -\ds\int_{\Omega_0} \bfB\, \bfu\,\d V - \ds\int_{∂\Omega_0^t}\bar{\bfT}\,\bfu\,\d S,
\end{align}
with body force $\bfB$ in reference configuration and prescribed surface traction $\bar{\bfT}$ on the traction boundary $∂\Omega_0^t$. The body force term $\bfB$ also includes the inertial forces of mass density times acceleration, $\rho\,\dot{\bfv}$, in case of a dynamic model.
The internal energy functional $\Pi_\text{int}(\bfu,p)$ describes the strain energy of the system depending on the displacement field $\bfu$ and the hydrostatic pressure $p$. The term is defined in \cref{sec:section_with_pi_int}.

The \emph{principle of stationary potential energy} demands that the potential energy functional $\Pi_L$ is stationary.
Variational calculus and differentiation of \cref{eq:hellinger_reissner} lead to the local Eulerian description given in \cref{eq:assumption_2_integral,eq:assumption_2_local}.

The third assumption is the \emph{balance of angular momentum} and can be formulated using the 3D cross-product:%
\begin{align*}
  \d{t} \ds\int\limits_{V_t} \bfx \times (\rho\,\bfv)\, \d v = \ds\int\limits_{V_t} \bfx \times (\rho\,\bfb) \,\d v + \ds\int\limits_{∂V_t} \bfx \times \bft\,\d a.
\end{align*}
This can be shown to be equivalent to the symmetry of the Cauchy stress tensor, $\bfsigma = \bfsigma^\top$.

A further assumption in the multi-scale muscle framework is to only consider isothermal conditions. 
An activated muscle performs work and energy is added to the system by metabolism. Further, the muscle is not thermodynamically isolated. The system is not closed regarding conversion and transfer of energy and, thus, the balance of energy cannot be modeled easily.

Regarding the required relations to obtain the deformation of the body from external loads given in \cref{fig:tonti_diagram}, the \emph{equilibrium} relation is given by \cref{eq:assumption_2_local} and the nonlinear \emph{kinematic} relation is given by \cref{eq:green_lagrange_u}.
The \emph{material} relation has yet to be defined. The mathematical description is closed by defining a constitutive relation between stresses and strains in the next sections.

\Cref{sec:material_linear_model} defines a linear material model that can be used together with linearized kinematics to formulate a fully linear model. \Cref{sec:material_nonlinear_model} presents the nonlinear material model to proceed with the fully nonlinear description. 

\subsection{Linear Material Model}\label{sec:material_linear_model}

For a linear constitutive relation between strain and stress, the linearized strain tensor $\bfeps$, defined in \cref{eq:linearized_helper3} is used together with the Eulerian Cauchy stress $\bfsigma$. The generic linear material model is \emph{Hooke}'s law, given by 
\begin{align}\label{eq:linearized_helper2}
  \bfsigma = \C:\bfeps
\end{align}
with the fourth order material tensor%
\begin{align*}
  \C_{abcd} = K\,δ_{ab}\,δ_{cd} + μ\,(δ_{ac}\,δ_{bd} + δ_{ad}\,δ_{bc} - \dfrac23 δ_{ab}\,δ_{cd}).
\end{align*}
The bulk modulus $K$ is a measure for the (in-)compressibility and the shear modulus $\mu$ specifies the elastic shear stiffness. $δ_{ab}$ is the Kronecker delta.
The material tensor $\C$ exhibits the following major and minor symmetries:%
\begin{subequations}\label{eq:symmetries}
\begin{align}
  \C_{abcd} &= \C_{cdab}, \quad &\text{(major symmetries)}\\[4mm]
  \C_{abcd} &= \C_{bacd} = \C_{abdc} = \C_{badc}, \quad & \text{(minor symmetries)}
\end{align}
\end{subequations}
effectively reducing the number of independent entries from 81 to 21 for 3D domains.

To incorporate force generation in the muscle, the stress can be additively composed of the passive stress $\bfsigma$ and an additional active stress term $\bfsigma_\text{active}$:%
\begin{align}\label{eq:active_stress_linear}
  \bfsigma^\text{total} = \bfsigma + \bfsigma^\text{active}.
\end{align}

\subsection{Nonlinear Material Modeling}\label{sec:material_modeling}
Next, we present the derivation of a nonlinear model that does not make any linearization assumptions of small strains as in the previous section.
We begin with the description of the material law, which links strains and stresses.

The scalar strain energy function $\Psi$ describes the elastic energy of the material depending on the deformation.
The definition of $\Psi$ suffices to describe the behavior of a hyperelastic material.
The strain energy function links the right Cauchy Green tensor $\bfC$ to the second Piola-Kirchhoff stress tensor $\bfS$ by the relation%
\begin{align}\label{eq:material_model_helper1}
  \bfS = 2\p{\Psi(\bfC)}{\bfC}.
\end{align}

The \emph{principle of material objectivity} requires that material properties are invariant under a change of observer. As a result, the \emph{representation theorem for isotropic materials} states that the stress tensor can be represented using three strain invariants $I_1, I_2$ and $I_3$. For a transversely isotropic material, two invariants $I_4$ and $I_5$ that depend on the anisotropy direction $\bfa_0$ (corresponding to a fiber direction) are added.
Consequently, we can formulate the strain energy function $\Psi=\Psi(I_1,I_2,I_3,I_4,I_5)$ in terms of these invariants. The principle strain invariants $I_1$ to $I_3$ of the right Cauchy-Green tensor $\bfC$ and the additional anisotropic invariants $I_4$ and $I_5$ are defined as:
\begin{align*}
  &I_1(\bfC) = \tr(\bfC),  &
  &I_2(\bfC) = \dfrac12\big(\tr(\bfC)^2 - \tr(\bfC^2)\big), &
  I_3(\bfC) = \det(\bfC) = J^2,\\[4mm]
  &I_4(\bfC,\bfa_0) = \bfa_0 \cdot \bfC \, \bfa_0, &
  &I_5(\bfC,\bfa_0) = \bfa_0 \cdot \bfC^2 \, \bfa_0. &
\end{align*}
The fiber stretch is related to the fourth invariant by $\lambda_f = \sqrt{I_4}$. Note that requiring incompressibility is equivalent to enforcing $J=1$, and, in this case, we get ${I_3(\bfC) = 1}$. 

It is convenient to use a decoupled description, where the deformation gradient $\bfF$ and the right Cauchy-Green tensor $\bfC$ are multiplicatively decomposed into volume-changing (volumetric) and volume-preserving (isochoric) parts:%
\begin{align*}
  \bfF &= (J^{1/3}\bfI)\,\bar{\bfF},  & \bfC &= (J^{2/3}\bfI)\,\bar{\bfC}.
\end{align*}
Here, the volumetric parts are the identity tensors scaled by a power of the determinant $J$ of the deformation gradient. The isochoric or distortional parts $\bar{\bfF}$ and $\bar{\bfC}$ are given by%
\begin{align}\label{eq:reduced_fc}
  \bar{\bfF} &= J^{-1/3}\,\bfF,  & \bar\bfC &= J^{-2/3}\,\bfC.
\end{align}
The reduced invariants $\bar{I}_1$ to $\bar{I}_5$ of the reduced right Cauchy-Green tensor $\bar\bfC$ are defined accordingly.
Similarly, the strain energy function has a decoupled representation with volumetric part $\Psi_\text{vol}$ and isochoric part $\Psi_\text{iso}$:
\begin{align}\label{eq:psi_iso}
  \Psi = \Psi_\text{vol}(J) + \Psi_\text{iso}(\bar{\bfC}) = \Psi_\text{vol}(J) + \Psi_\text{iso}(\bar{I}_1,\bar{I}_2,\bar{I}_4,\bar{I}_5).
\end{align}

Using the decoupled form, any incompressible material can be modeled with the \emph{penalty method} as follows. 
The material behavior is given by the isochoric strain energy $\Psi_\text{iso}(\bar{\bfC})$, e.g., by employing the Mooney-Rivlin model in \cref{eq:mooney_rivlin}. The volumetric part is defined as
\begin{align*}
  \Psi_\text{vol}(J) &= \kappa\,G(J) \qquad \text{with } G(J) = \dfrac12 (J-1)^2,
\end{align*}
with the incompressibility parameter $\kappa$ and the penalty function $G(J)$. This function is strictly convex and approaches zero as $J$ approaches 1. For large values of $\kappa$, the behavior is nearly incompressible. A disadvantage of this method is, that the resulting system becomes singular for $J \to 1$.

A better approach in this regard is to use a mixed formulation, where incompressibility is enforced exactly using a Lagrange multiplier. This approach is also implemented in OpenDiHu and is the preferred method for incompressible materials. 

In OpenDiHu, the strain energy function of a new material can be given using the following four terms:
\begin{align}\label{eq:definition_psi}
  \Psi = \Psi_\text{vol}(J) + \Psi_\text{iso}(\bar{I}_1,\bar{I}_2,\bar{I}_4,\bar{I}_5) + \Psi_1(I_1,I_2,I_3) + \Psi_2(\bfC,\bfa_0).
\end{align}
The decoupled form is available with $\Psi_\text{vol}$ and $\Psi_\text{iso}$, the coupled form for isotropic materials can be used via $\Psi_1$. The term $\Psi_2$ gives the most flexibility, as the constitutive model can be directly formulated using the right Cauchy-Green tensor $\bfC$ and the fiber direction $\bfa_0$. The unused terms among $\Psi_\text{vol},\Psi_\text{iso},\Psi_1$ and $\Psi_2$ can be defined as constant zero. The incompressibility constraint using Lagrange multipliers can be switched on or off such that both incompressible and compressible materials can be computed.
%
% ---

\subsection{The Nonlinear Material Model for Muscle Contraction}\label{sec:material_nonlinear_model}

In the muscle contraction model of \cite{Heidlauf2013}, the strain energy function is additively composed of two passive terms, one isotropic, one anisotropic, and one additional active term:
\begin{align}\label{eq:transiso_mooney_rivlin}
  \Psi(\bfC) = \Psi_\text{isotropic}(I_1,I_2) + \Psi_\text{anisotropic}(\lambda_f) + \Psi_\text{active}(\gamma).
\end{align}
The isotropic term $\Psi_\text{isotropic}$ is formulated in terms of the strain invariants $I_1=\tr(\bfC)$ and $I_2=\big(\tr(\bfC)^2 - \tr(\bfC^2)\big)/2$. The anisotropic term $\Psi_\text{anisotropic}$ depends on the fiber stretch $\lambda_f$. The active term $\Psi_\text{active}$ yields the active stress that results from muscular activation, which is described by the activation parameter $\gamma$.
%Note the missing dependency on the third invariant $I_3$, which is constant because of the enforced incompressibility.

The passive behavior of muscle tissue is modeled by a transversely isotropic Mooney-Rivlin material.
The isotropic part is given by the Mooney-Rivlin formulation:%
\begin{align}\label{eq:mooney_rivlin}
  \Psi_\text{isotropic}(I_1,I_2) = c_1\,(I_1 - 3) + c_2\,(I_2-3).
\end{align}
The values of the two material parameters $c_1$ and $c_2$ can be determined by compression tests and are summarized in the work of \cite{Heidlauf2013}.

The anisotropic behavior depends only on the fiber stretch $\lambda_f$. The formulation in \cite{Heidlauf2013} uses two material parameters $b$ and $d$ and the following function:
\begin{align*}
  \Psi_\text{anisotropic}(\lambda_f) = \dfrac{b}{d}(\lambda_f^d - 1) - b\,\log(\lambda_f).
\end{align*}

The active contribution is directly formulated in terms of the second Piola-Kirchhoff stress $\bfS$. The relation between the active stress $\bfS_\text{active}$ and the active contribution $\Psi_\text{active}$ of the strain energy function as well as the definition of $\bfS_\text{active}$ is given as follows:
\begin{align}\label{eq:active_stress_term}
  \bfS_\text{active} = \dfrac{1}{\lambda_f}\p{\Psi_\text{active}}{\lambda_f} \bfA \otimes \bfA = \dfrac{1}{\lambda_f} \cdot S_\text{max,active}\cdot f_\ell(\lambda_f)\cdot\bar{\gamma}\, \bfA \otimes \bfA.
\end{align}
Here, the resulting active stress tensor $\bfS_\text{active}$ is the second order tensor oriented according to the material fiber direction $\bfA: \Omega_0 \to \R^3$ and given by the dyadic product $\bfA \otimes \bfA = A_{i}\,A_{j}\,\bfe_i \otimes \bfe_j$, scaled by the maximum active stress parameter $S_\text{max,active}$, a function $f_\ell$ that models the force-length relation, and the 3D homogenized value $\bar{\gamma}$ of the activation parameter $\gamma \in [0,1]$ following from the half-sarcomere model.

In the deforming body fat layer, the active stress contribution is disregarded. For simulating tendons, different material models can be used such as the model proposed by Carniel et al. \cite{Carniel2017}, which describes microstructural interactions between collagen fibers and their matrix in addition to the elastic response of the fibers themselves. To alter the material model, the definition of $\Psi$ can simply be changed while all other equations of the solid mechanics model remain intact. 
%Similarly, other material models can be defined using the framework of the strain energy function.

\subsection{Summary of the Solid Mechanics Model Equations}
In summary, the model of solid mechanics for muscle contraction is solved for the unknown displacements $\bfu$ and additionally the velocities $\bfv$ if a dynamic formulation is considered.

The model equations follow from the following balance principles:
\begin{subequations}\label{eq:contraction}
  \begin{align}
    \div(\bfv) &= 0, \qquad &&\text{(incompressibility)} \label{eq:contraction_1}\\[4mm]
    \rho\,\dot{\bfv} &= \rho\,\bfb + \div\bfsigma, && \text{(balance of linear momentum)}\label{eq:contraction_2}\\[4mm]
    \bfsigma &= \bfsigma^\top, && \text{(balance of angular momentum)}\label{eq:contraction_3}
  \end{align}
\end{subequations}
with the constant density $\rho$, external body forces $\bfb$ and the Cauchy stress tensor $\bfsigma$.

Additionally, geometric relations between displacements $\bfu$ and strains $\bfE$ or $\bfeps$ are assumed, either fully nonlinear in \cref{eq:green_lagrange_u} or with corresponding linearization assumptions in \cref{eq:linearized_helper3}.
Furthermore, a material model is given that relates strains and stresses. A linear model is described in \cref{sec:material_linear_model}. The framework for nonlinear hyperelastic models uses a strain energy function $\Psi$ as described in \cref{sec:material_modeling}. A particular nonlinear material model for muscle contraction from the literature is described in \cref{sec:material_nonlinear_model}.

The description of the multi-scale model \cite{Roehrle2012,Heidlauf2013} assumes quasi-static conditions, which means that the velocities are set to zero, $\bfv=\bfzero$, and inertial terms are neglected. As a consequence, the incompressibility constraint in \cref{eq:contraction_1} has to be formulated differently and the balance of momentum in \cref{eq:contraction_2} reduces to $\rho\,\bfb + \div \bfsigma = 0$.
Our implementation extends the model to the fully dynamic formulation given in \cref{eq:contraction_1,eq:contraction_2,eq:contraction_3}. 

Initial conditions for the displacements $\bfu$ and velocities $\bfv$ define the initial pose of the muscle tissue:
\begin{align*}
  \bfu(\bfx,0) &= \bfu_0(\bfx), & \bfv(\bfx,0) &= \bfv_0(\bfx) \quad &&\text{for } \bfx \in \Omega_M.
\end{align*}
Dirichlet boundary conditions for $\bfu$ and $\bfv$ can fix certain parts of the muscle, e.g., at the attachment points of the tendons:
\begin{align*}
  \bfu(\bfx,t) &= \bar{\bfu}(t), & \bfv(\bfx,t) &= \bar{\bfv}(t) \quad &&\text{for } \bfx \in ∂\Omega_\text{Dirichlet}.
\end{align*}
Additionally, Neumann boundary conditions can be used to prescribe traction forces on the surface.

The derivation of the finite element formulation and the resulting numerical scheme to obtain the solution functions $\bfu$ and $\bfv$ are discussed in \cref{sec:discretization_mechanics}.

%\subsection{Sensory Organs and Motor Neurons}
%text of paper:
%Muscle spindles and Golgi tendon organs are located in the muscle and sense fiber stretch, contraction velocity, contraction acceleration and muscle forces. 
%They are connected via layers of inter neurons to the motor neurons, which reside in the spinal cord.
%In turn, the motor units innervate and activate the muscle.

%We use CellML models to compute the dynamics of muscle spindles \cite{Mileusnic2006} and motor neurons \cite{CisiKohn2008}.

\section{Discretization of the Electrophysiology Models}\label{sec:discretization}

The partial and ordinary differential equations described in the last section contain spatial and temporal derivatives that have to be discretized to be solved numerically. For temporal derivatives, we use timestepping schemes, for spatial derivatives, we employ the finite element method.

In this section, we describe the discretization of the subcellular and electrophysiology models that were presented in the last section. A description of the discretization of the solid mechanics model follows in \cref{sec:discretization_mechanics}.

We begin with the discretization in time in \cref{sec:discretization_monodomain}, followed by the spatial discretization for the monodomain (\cref{sec:discretization_diffusion,sec:mass_lumping}) and multidomain models (\cref{sec:discretization_multidomain,sec:discretization_body_domain}).

\subsection{Discretization of the Monodomain Model}\label{sec:discretization_monodomain}

Electrophysiology models typically consist of a reaction-diffusion equation. The diffusion term describes the electric conduction in the tissue and the reaction term includes the subcellular model. In our model, the monodomain equation  \cref{eq:monodomain} used in the fiber based description and the second multidomain equation \cref{eq:multidomain2} are equations of this type.

This type of partial differential equation is often solved using operator splitting schemes. A first order operator splitting scheme is Godunov splitting \cite{Godunov2003}. It was used for the solution of the chemo-electro-mechanical model in \cite{Roehrle2012}. In addition to Godunov splitting, we also employ the second order accurate Strang splitting scheme \cite{Strang1968}.

In the following, the application of these two schemes is illustrated for the monodomain equation \cref{eq:monodomain}. The right-hand sides of the diffusion and reaction terms are denoted in short as $\mathcal{L}_1$ and $\mathcal{L}_2$:%
\begin{align*}
  \mathcal{L}_1(V_m) := \dfrac{1}{A_m\,C_m} \sigma_\text{eff}\dfrac{\partial^2 V_m}{\partial x^2}, &&
  \mathcal{L}_2(V_m) := -\dfrac{1}{C_m} I_\text{ion}(V_m,\bfy).
\end{align*}
Then, the monodomain equation takes the form:
\begin{align}\label{eq:monodomain_operator_splitting}
  \p{V_m}{t} = \mathcal{L}_1(V_m) + \mathcal{L}_2(V_m).
\end{align}

A timestepping scheme is constructed that starts with a given initial value $V_m^{(0)}$ and computes solution values $V_m^{(i)}$ at discrete points in time $t^{(i)} = i\cdot\dt$ with a fixed timestep width $\dt$.
Godunov splitting proceeds by alternatingly performing steps in the two directions of the right-hand sides $\mathcal{L}_1$ and $\mathcal{L}_2$. In the first substep per iteration, an intermediate value $V_m^{\ast}$ is calculated, which is used as starting point for the second substep. Each of the substeps are performed using independent timestepping scheme, e.g., the explicit Euler scheme:
\begin{subequations}\label{eq:monodomain_godunov}
  \begin{align}
    V_m^{\ast} &= V_m^{(i)} + \dt \mathcal{L}_1(V_m^{(i)},t^{(i)}),\\[4mm]
    V_m^{(i+1)} &= V_m^{\ast} + \dt \mathcal{L}_2(V_m^{\ast},t^{(i)})
  \end{align}
\end{subequations}

Strang splitting uses a similar approach with three substeps per timestep and two intermediate values $V_m^{\ast}$ and $V_m^{\ast\ast}$:
\begin{subequations}\label{eq:monodomain_strang}
  \begin{align}
    V_m^{\ast} &= V_m^{(i)} + \dfrac{\dt}{2} \mathcal{L}_1(V_m^{(i)},t^{(i)}),\\[4mm]
    V_m^{\ast\ast} &= V_m^{\ast} + \dt \mathcal{L}_2(V_m^{\ast},t^{(i)}),\\[4mm]
    V_m^{(i+1)} &= V_m^{\ast\ast} + \dfrac{\dt}{2} \mathcal{L}_1(V_m^{\ast\ast},t_{(i)}+\dfrac12 \dt).
  \end{align}
\end{subequations}
%
% stuff needed to create the plot
%\definecolor{vred}{RGB}{170,0,0}
%\definecolor{vyellow}{RGB}{212,170,0}
%\definecolor{vgreen}{RGB}{0,170,0}
%\begin{equation*}
%  \begin{array}{lll}
%    \frac{\partial}{\partial t} V_m  = \textcolor{vred}{c_1 \frac{\partial^2}{\partial x^2} V_m }\\[4mm]
%    \frac{\partial}{\partial t} \bfy = \textcolor{vyellow}{G(V_m,\bfy)}\\[4mm]
%    \frac{\partial}{\partial t} V_m = \textcolor{vyellow}{c_2\,I_\text{ion}(V_m,\bfy)}\\[4mm]
%    \textcolor{vyellow}{\dt_\text{0D}}\quad
%    \textcolor{vred}{\dt_\text{1D}}\quad
%    \dt_\text{splitting}\quad
%    \textcolor{vgreen}{\dt_\text{3D}}\quad
%    t^{(i)}, t^{(i)} + \dt_\text{splitting}/2, t^{(i)} + \dt_\text{splitting} ...\\[4mm]
%    t^{(i)} + \textcolor{vgreen}{\dt_\text{3D}}\\[4mm]
%    t^{(i+1)}
%  \end{array}
%\end{equation*}
%\textcolor{vyellow}{0D:}
%\textcolor{vred}{1D:}
%\textcolor{vgreen}{3D:}

Note that each substep can either be executed as a single timestep of the chosen method as in \cref{eq:monodomain_godunov,eq:monodomain_strang} or divided into several steps with timestep widths $\dt_{0D}$ (for the 0D subcellular model represented by $\mathcal{L}_1$) and $\dt_{1D}$ (for the diffusion equation represented by $\mathcal{L}_2$).

\begin{figure}%
  \centering%
  \begin{subfigure}[t]{0.48\textwidth}%
    \centering%
    \def\svgwidth{\textwidth}
    \includegraphics[width=\textwidth]{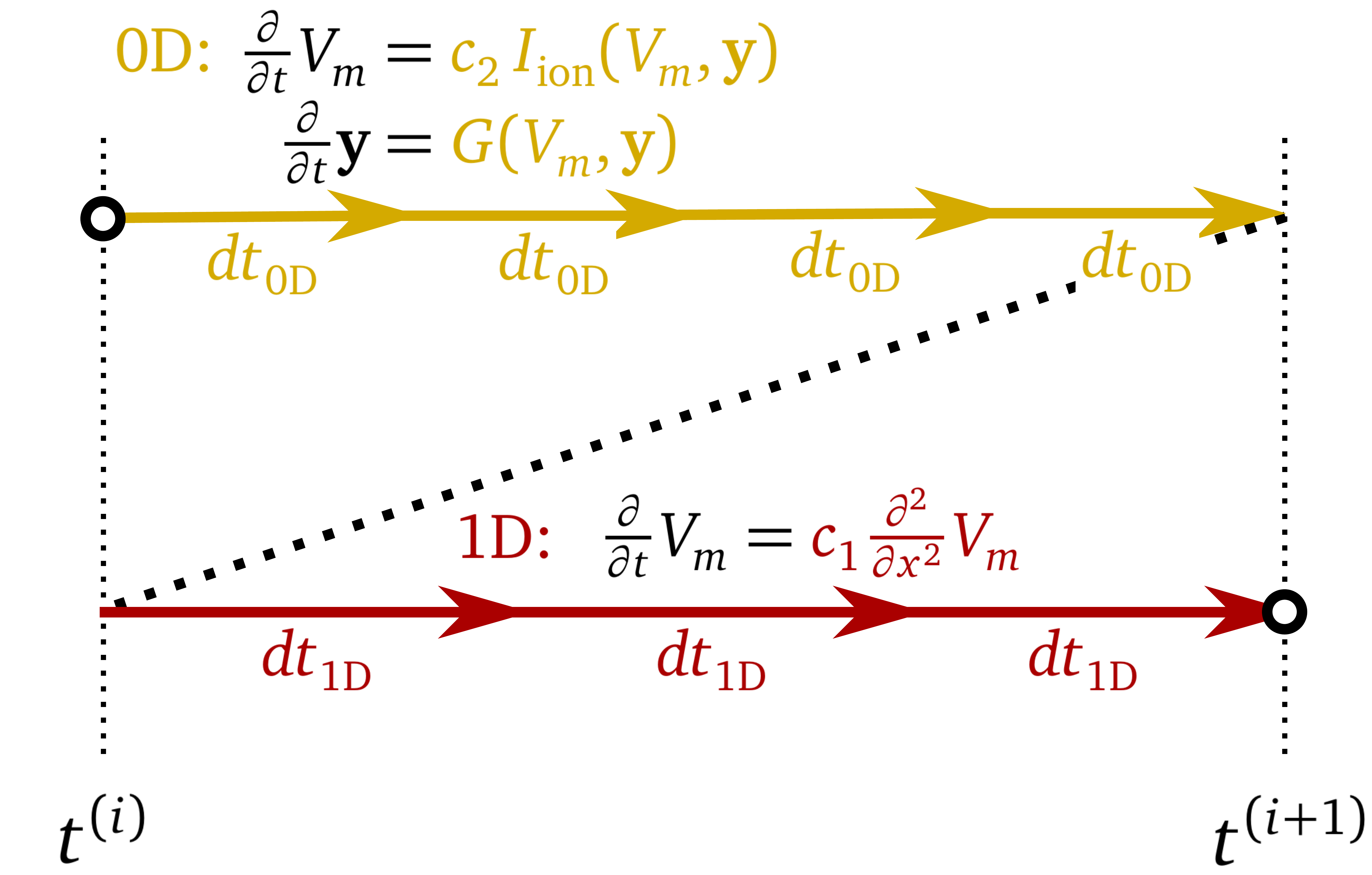}
    \caption{The Godunov splitting uses two substeps: 0D, 1D.}%
    \label{fig:godunov_splitting}%
  \end{subfigure}
  \quad
  \begin{subfigure}[t]{0.48\textwidth}%
    \centering%
    \includegraphics[width=\textwidth]{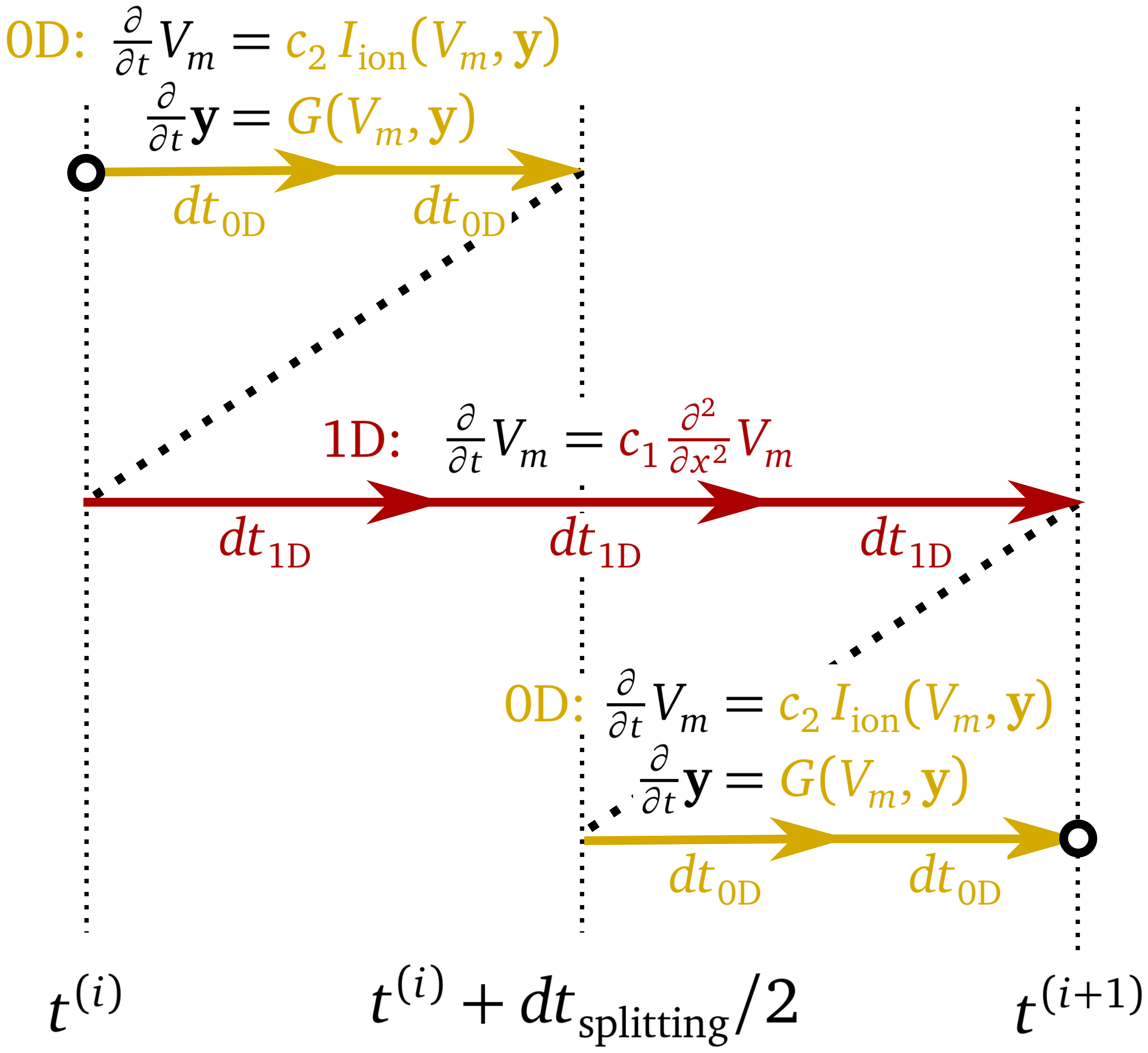}
    \caption{The Strang splitting uses three substeps: 0D, 1D, 0D.}%
    \label{fig:strang_splitting}%
  \end{subfigure}
  \caption{Godunov and Strang splitting schemes that are used to solve the monodomain equation. The equation is split into a reaction part (0D,yellow) and a diffusion part (1D,red) and these parts are solved alternatingly. The visualizations show one splitting timestep starting at the left circle and completing at the right circle.}%
  \label{fig:splitting_schemes}%
\end{figure}

\Cref{fig:splitting_schemes} visualizes both splitting schemes applied to the monodomain equation. 
The yellow arrows correspond to the solution of the 0D subcellular model. The red arrows correspond to the solution of the 1D diffusion equation. The timestep width of one splitting step is $\dt_\text{splitting}$. Depending on how the timestep widths are chosen in relation to each other, different numbers of subcycles are used in the solution of the 0D and 1D problems.

Instead of the explicit Euler method in \cref{eq:monodomain_godunov,eq:monodomain_strang}, other timestepping methods can be used for the substeps. 
We use the following schemes, which are listed as single steps for the generic ODE ${\p V_m / \p t = \mathcal{L}(V_m,t)}$:
\begin{subequations}\label{eq:ode_solver_schemes}
  \begin{align}
    V_m^{(i+1)} &= V_m^{(i)} + \dt \mathcal{L}(V_m^{(i)},t^{(i)}), \label{eq:explicit_euler}\\[4mm]
    V_m^{(i+1)} &= V_m^{(i)} + \dfrac{\dt}{2}\Big(
      \mathcal{L}(V_m^{(i)},t^{(i)}) + \mathcal{L}\big(V_m^{(i)} + \dt \mathcal{L}(V_m^{(i)},t^{(i)}),t^{(i+1)}\big)\Big), \label{eq:heun}\\[4mm]
    V_m^{(i+1)} &= V_m^{(i)} + \dt \mathcal{L}(V_m^{(i+1)},t^{(i+1)}), \label{eq:implicit_euler}\\[4mm]
    V_m^{(i+1)} &= V_m^{(i)} + \dfrac{\dt}{2}\Big(
      \theta\,\mathcal{L}(V_m^{(i+1)},t^{(i+1)}) + (1-\theta)\,\mathcal{L}(V_m^{(i)},t^{(i)})\Big).\label{eq:crank_nicolson}
  \end{align}
\end{subequations}
Here, \cref{eq:explicit_euler} is the first-order accurate explicit Euler scheme, \cref{eq:heun} is the second-order accurate Heun scheme, \cref{eq:implicit_euler} is the first order accurate implicit Euler scheme, and \cref{eq:crank_nicolson} is the Crank-Nicolson scheme \cite{CrankNicolson1947}, which for $\theta=0$ equals the explicit Euler and for $\theta=1$ equals the implicit Euler scheme. For $\theta=\frac12$, it is second order accurate. An advantage of the implicit schemes in \cref{eq:implicit_euler,eq:crank_nicolson} is, that, for our considered diffusion problems, they are unconditionally stable. A disadvantage is, that a linear equation has to be solved in every timestep.

A second order accurate timestepping scheme yields a faster decrease of the numerical error with decreasing step size and, thus, in many cases allows a larger step size than a first order scheme.
To obtain a second order scheme for the monodomain equation, we use Strang splitting (\cref{eq:monodomain_strang}) with the Crank-Nicolson scheme (\cref{eq:crank_nicolson}) for the diffusion term $\mathcal{L}_1$ and Heun's method (\cref{eq:heun}) for the reaction term $\mathcal{L}_2$. In the subcellular model, the system of ODEs with state vector $\bfy$ given in \cref{eq:subcellular} is solved with Heun's method along with the equation in terms of $V_m$.

Next, the spatial derivatives in the diffusion part $\mathcal{L}_2$ of the split equation have to be discretized. Then, both the multidomain and the fiber based models can be solved using the splitting scheme.

\subsection{Discretization of the Diffusion and Laplace Equations}\label{sec:discretization_diffusion}

For the spatial discretization, we first derive the finite element formulation for a generic parabolic diffusion equation in a domain $\Omega\subset\R^d$ of arbitrary dimensionality $d$. Then, specialization to 1D yields the formulation for the monodomain equation. Considering a 3D domain, the formulation is an important building block for the discretization of the multidomain model. This is shown in more detail in a later section, \cref{sec:discretization_multidomain}

We consider the following diffusion problem in the variable $u: \Omega \times [0,t_\text{end}] \to \R$ with Neumann boundary conditions on a part of the boundary $\Gamma_f \subset ∂\Omega$ with normal vector $\bfn$:
\begin{align*}
  \p{u}{t} &= \div(\bfsigma \grad u), &(\bfsigma\,\grad u) \cdot \bfn &= f \quad \text{on }\Gamma_f, & (\bfsigma\,\grad u) \cdot \bfn &= 0 \quad \text{on } ∂Ω\backslash \Gamma_f.
\end{align*}
We discretize the temporal derivative using the Crank-Nicolson scheme as in \cref{eq:crank_nicolson}. Following the procedure of the Galerkin finite element formulation with the Hilbert space $H^1_0(\Omega)$ of test functions $\phi$ that are zero on the boundary, we arrive at the following weak form:
\begin{align*}
  \ds\int_Ω \big(\theta\,∇\cdot(\bfsigma ∇ \bfu^{(i+1)})  + (1-\theta)\,∇\cdot(\bfsigma ∇ u^{(i)})\big)\,\phi \,\d\bfx &&\\
    \qquad = \dfrac{1}{\dt} \ds\int_Ω(u^{(i+1)} - u^{(i)})\,\phi\,\d\bfx, &&\qquad \forall \phi \in H^1_0(\Omega).
\end{align*}
For brevity, we express divergence and gradient using the nabla operator. 

To discretize the weak form in space, we choose a function space $V_h = \text{span}\{\varphi_j \mid j = 1, \dots, N\}$ to represent the solution as $u = \sum_{j=1}^N u_j \phi_j$. Applying the divergence theorem, we obtain:
\begin{equation}\label{eq:diffusion_helper1}
  \begin{array}{l}
    \ds\sum\limits_{j=1}^{N} \big(\theta\,u_j^{(i+1)} + (1-\theta)\,u_j^{(i)}\big)  
    \left(-\ds\int_Ω \bfsigma\,∇\varphi_j\cdot ∇\varphi_k \,\d\bfx + \ds\int_{∂Ω} \big(\bfsigma\,∇\varphi_j\cdot \bfn\big)\,\varphi_k \,\d\bfx  \right) \\
      \quad = \dfrac{1}{\dt} \sum\limits_{j=1}^{N} \big(u_j^{(i+1)} - u_j^{(i)}\big) \ds\int_Ω \varphi_j\,\varphi_k\,\d\bfx, \qquad \forall k = 1,\dots, N.
  \end{array}
\end{equation}
This iteration step can be written in matrix notation in terms of the vectors of unknowns $\bfu^{(i)}=(u^{(i)}_0,\dots,u^{(i)}_N)^\top$ at timestep $i$:%
\begin{align*}
  \bfA\,\bfu^{(i+1)} = \bfb(\bfu^{(i)}).
\end{align*}
The system matrix $\bfA$ and the right-hand side $\bfb$ are given by:
\begin{align*}
  \bfA &= \theta\,(\bfK_{\bfsigma} + \bfB_{\bfsigma}) -\dfrac{1}{\dt}\bfM, &
  \bfb &= \big((\theta-1)\,(\bfK_{\bfsigma} + \bfB_{\bfsigma}) - \dfrac{1}{\dt} \bfM \big)\,\bfu^{(i)}.
\end{align*}
The formulation uses the standard stiffness matrix $\bfK_{\bfsigma}$, the matrix $\bfB_{\bfsigma}$ of the boundary integral and the mass matrix $\bfM$, whose components are defined as%
\begin{align}\label{eq:diffusion_matrices}
  \bfK_{\bfsigma,kj} &= -\ds\int_Ω (\bfsigma\, ∇\varphi_j)\cdot ∇\varphi_k \,\d\bfx,&
     \bfB_{\bfsigma,kj} &= \ds\int_{\Gamma_f} \big((\bfsigma\,∇\varphi_j)\cdot \bfn\big)\,\varphi_k \,\d\bfx,&
     \bfM_{kj} &= \ds\int_Ω \varphi_j\,\varphi_k\,\d\bfx.
\end{align}
Note that, after applying the divergence theorem, the definition of the stiffness matrix has a minus sign.

Next, we take into account the Neumann boundary condition $\bfsigma∇u\cdot \bfn = f$ on the boundary $\Gamma_f$. The flux $f$ over the boundary is discretized by $M$ separate ansatz functions $\psi_j$ on $\Gamma_f$ as $f = \sum_{j=1}^M f_j\, \psi_j$.
The flux values are summarized in a vector $\bff=(f_1,\dots,f_M)^\top$.
Plugging this into \cref{eq:diffusion_helper1} yields the following equation in matrix notation:%
\begin{align}\label{eq:diffusion_helper2}
  \tilde{\bfA}\,\bfu^{(i+1)} = \tilde{\bfb}(\bfu^{(i)}),  
\end{align}
with the system matrix $\tilde{\bfA}$ and right-hand side $\tilde{\bfb}$: 
\begin{align*}
  \tilde{\bfA} &= \theta\,\bfK_{\bfsigma} -\dfrac{1}{\dt}\bfM, &
    \tilde{\bfb} &= \big((\theta-1)\,\bfK_{\bfsigma} - \dfrac{1}{\dt} \bfM \big)\,\bfu^{(i)} - \bfB_{\Gamma_f}\,\big(\theta\,\bff^{(i+1)} + (1-\theta)\,\bff^{(i)}\big),
\end{align*}
and the boundary matrix $\bfB_{\Gamma_f}$ given by:
\begin{align}\label{eq:definition_boundary_matrix}
  \bfB_{\Gamma_f,kj} &= \ds\int_{\Gamma_f} \psi_j\,\varphi_k \,\d\bfx.
\end{align}
Note that incorporating the Neumann boundary conditions in the weak form corresponds to the following exchange of the boundary matrices $\bfB_\sigma$ and $\bfB_{\Gamma_f}$:%
\begin{align}\label{eq:boundary_relation}
  \bfB_\sigma\,\bfu = \bfB_{\Gamma_f}\,\bff.
\end{align}

\Cref{eq:diffusion_helper2} is used to solve the diffusion part of the monodomain equation given in \cref{eq:monodomain} after inserting the corresponding constant prefactors.

When deriving or implementing new models or optimizing solver code, it is often beneficial to study certain effects in isolation. It can help to use a toy problem such as the simple Laplace problem $Δu = 0$, possibly with Neumann boundary condition $\partial u/\partial \bfn = f$. 
By specializing the formulation in \cref{eq:diffusion_helper2} accordingly, we obtain the system
\begin{align*}
  (\bfK_\bfI + \bfB_\bfI)\,\bfu = \bfzero
\end{align*}
for the case without boundary condition (set $\bfB_\bfI$ to zero to assume homogeneous Neumann boundaries) or
\begin{align}\label{eq:discretization_laplace}
  \bfK_\bfI\,\bfu = -\bfB_{\Gamma_f}\,\bff
\end{align}
to include the formulated Neumann boundary condition.

\subsection{Using Mass Lumping for Implicit Timestepping}\label{sec:mass_lumping}
Implicit timestepping schemes such as implicit Euler or the Crank-Nicolson scheme for $\theta=\frac12$ need to solve a linear equation in every timestep.
Assuming homogeneous Neumann boundary conditions for simplicity, the iteration step of the canonical Crank-Nicolson scheme follows from \cref{eq:diffusion_helper2}:
\begin{subequations}\label{eq:lumping_crank_nicolson}
  \begin{align}
    \big(\dfrac1{2}\bfK-\dfrac{1}{\dt}\bfM\big)\, \bfu^{(i+1)} &= \big(-\dfrac12{\bfK} - \dfrac{1}{\dt}\bfM\big)\, \bfu^{(i)}\label{eq:lumping_crank_nicolson_1}\\[4mm]
    \Leftrightarrow \quad (\bfI - \dfrac{\dt}{2}\,\bfM^{-1}\bfK)\,\bfu^{(i+1)}&= (\bfI + \dfrac{\dt}{2}\,\bfM^{-1}\bfK)\,\bfu^{(i)}.\label{eq:lumping_crank_nicolson_2}
  \end{align}
\end{subequations}
For the implicit Euler method, we obtain:%
\begin{subequations}\label{eq:lumping_implicit_euler}
  \begin{align}
    \ds(\bfK-\frac{\bfM}{\dt})\,\bfu^{(i+1)} &=\,\ds -\frac{\bfM}{\dt}\bfu^{(i)}\label{eq:lumping_implicit_euler_1}\\[4mm]
    \Leftrightarrow \quad (\bfI - \dt\,\bfM^{-1}\bfK)\,\bfu^{(i+1)}&= \,\bfu^{(i)}.\label{eq:lumping_implicit_euler_2}
  \end{align}
\end{subequations}
Both iteration steps in \cref{eq:lumping_crank_nicolson_1,eq:lumping_crank_nicolson_2} and in \cref{eq:lumping_implicit_euler_1,eq:lumping_implicit_euler} are equivalent, as the second equation follows from the first one by left multiplication of $(-\dt \bfM^{-1})$. In the second equations, the matrices to be multiplied are created by a sum of the unity matrix $\bfI$ and another matrix term that is scaled by the potentially small timestep width $\dt$. For the implicit Euler in \cref{eq:lumping_implicit_euler_2}, the matrix on the right-hand side even reduces to the identity matrix. This is preferred over the first iteration steps in \cref{eq:lumping_crank_nicolson_1,eq:lumping_implicit_euler_1} as it leads to better conditioned matrix-vector multiplications.

The required inversions of the mass matrix cannot be carried out explicitly as the inversion would fill in numerous matrix entries and eliminate the sparse structure. This is not feasible for highly resolved meshes with many degrees of freedom. Instead, \emph{mass lumping} is used, where the mass matrix $\bfM$ is approximated by a diagonal matrix with diagonal entries equal to the row sums in $\bfM$ \cite{Hinton1976}. Thus, multiplication with the inverse mass matrix corresponds to a rescaling of columns by the inverse lumped diagonal entries.

\subsection{Discretization of the Multidomain Model}\label{sec:discretization_multidomain}
With the prerequisites of temporal discretization in \cref{sec:discretization_monodomain} and the finite element formulation of a diffusion equation in \cref{sec:discretization_diffusion}, we can now discretize the multidomain model. Since this has not been previously done in literature using the finite element method, the subsequent derivation is more detailed.

The first multidomain equation given in \cref{eq:multidomain1} yields the following form after applying the finite element derivation in \cref{eq:diffusion_helper2}:
\begin{align}\label{eq:multidomain_discretization_helper_multidomain1}
  \big(\bfK_{\bfsigma_e + \bfsigma_i} + \bfB_{\bfsigma_e + \bfsigma_i}\big)\,\bfphi_{e} +  \s{k=1}{N_\text{MU}} f_r^k \big(\bfK_{\bfsigma_i^k} + \bfB_{\bfsigma_i^k}\big)\,\bfV_m^k = 0.  
\end{align}
Here, $\bfphi_{e}$ and $\bfV_m^k$ are the vectors of degrees of freedom for the extracellular potential $\phi_e$ and membrane voltage $V_m^k$ of compartment $k$. The matrices are defined by \cref{eq:diffusion_matrices} and do not yet include the boundary conditions.
The subscripts of the stiffness matrices $\bfK$ and boundary integral matrices $\bfB$ refer to the anisotropy tensors that occur in their definitions.

The diffusion part of the second multidomain equation, \cref{eq:multidomain2}, discretized with Crank-Nicolson, yields the system%
\begin{align}\label{eq:multidomain_discretization_helper_multidomain2}
  \bfA\,\mat{\bfV_m^{k,(i+1)}\\ \bfphi_{e}^{(i+1)}} = \bfb,  
\end{align}
with the $1 \times 2$ block system matrix $\bfA$ and right-hand side vector $\bfb$ given by:%
\begin{subequations}\label{eq:multidomain_discretization_helper_multidomain3}
\begin{align}
   \bfA &= \matt{
      \dfrac{\theta}{A_m^k\,C_m^k}(\bfK_{\bfsigma_i^k} + \bfB_{\bfsigma_i^k}) -\dfrac{1}{\dt}\bfM & \quad
      \dfrac{\theta}{A_m^k\,C_m^k}(\bfK_{\bfsigma_i^k} + \bfB_{\bfsigma_i^k})
    }, \\[4mm]
    \bfb &= \Big( \dfrac{\theta-1}{A_m^k\,C_m^k}(\bfK_{\bfsigma_i^k} + \bfB_{\bfsigma_i^k}) - \dfrac{1}{\dt}\bfM\Big)\,\bfV_m^{k,(i)} 
      + \dfrac{\theta - 1}{A_m^k\,C_m^k}(\bfK_{\bfsigma_i^k} + \bfB_{\bfsigma_i^k})\,\bfphi_e^{(i)}.
\end{align}
\end{subequations}

A separate instance of this equation holds for every compartment $k$. Again, the integrals over the boundary are still present in the $\bfB_{\bfsigma_i^k}$ matrices.
To resolve this and to close the formulation, we have to consider the fluxes over the boundary of all involved unknowns and to replace them using the boundary conditions.

One required boundary conditions to solve the multidomain model without body domain is given in \cref{eq:multidomain_bc1}. The boundary condition for compartment $k$ in terms of the intracellular potential $\phi_i^k$,
\begin{align}\label{eq:multidomain_discretization_helper1}
  (\bfsigma_i^k\,∇\phi_i^k) \cdot \bfn_m = 0 \qquad \text{on } ∂\Omega_M,
\end{align}
is expressed in terms of the unknowns $V_m^k$ and $\phi_e$ to yield the condition
\begin{align}\label{eq:multidomain_discretization_helper2}
  (\bfsigma_i^k\,∇V_m^k) \cdot \bfn_m &= -(\bfsigma_i^k\,∇\phi_e)\cdot \bfn_m =: p^k \qquad \text{on }∂\Omega_M.
\end{align}
We define the value of this flux to be equal to a helper variable $p^k$.
A second flux is formulated for the extracellular potential $\phi_e$. We assign its value to the helper variable $q$:
\begin{align}\label{eq:definition_q}
  (\bfsigma_e ∇ \phi_e)\cdot \bfn_m =:q \qquad \text{on }∂\Omega_M.
\end{align}

We can now express the flux value $\big((\bfsigma_e + \bfsigma_i)\,∇\phi_e\big) \cdot \bfn_m$, which occurs in the discretized first multidomain equation, \cref{eq:multidomain_discretization_helper_multidomain1}, in terms of the variables $p^k$ and $q$. Using \cref{eq:multidomain_discretization_helper1,eq:multidomain_discretization_helper2} and the relation $\phi_e = \phi^k_i - V_m^k$, we derive:
\begin{align}\label{eq:multidomain_discretization_helper3}
   \big((\bfsigma_e + \bfsigma_i)\,∇\phi_e\big) \cdot \bfn_m
    &= (\bfsigma_e\,∇\phi_e)\cdot \bfn_m + (\bfsigma_i\,∇\phi_e)\cdot \bfn_m = q - \s{k=1}{N_\text{MU}} f_r^k\,p^k.
\end{align}

We discretize the flux values $p^k$ and $q$ analogously to the Neumann boundary condition flux $f$ in \cref{sec:discretization_diffusion} and summarize the degrees of freedoms in vectors $\bfp^k$ and $\bfq$.

Next, we combine the flux values with the first and second multidomain equation.
Plugging the generic relation \cref{eq:boundary_relation} for boundary integral terms into the discretization of the first multidomain equation, \cref{eq:multidomain_discretization_helper_multidomain1}, and using the derived flux values in \cref{eq:multidomain_discretization_helper2,eq:multidomain_discretization_helper3} leads in a first step to the following equation:
\begin{align*}
  \bfK_{\bfsigma_e + \bfsigma_i}\,\bfphi_{e} + \bfB_{\Gamma_M}\,\big(\bfq - \s{k=1}{N_\text{MU}} f_r^k\,\bfp^k\big) +  \s{k=1}{N_\text{MU}} f_r^k \big(\bfK_{\bfsigma_i^k}\,\bfV_m^k + \bfB_{\Gamma_M}\,\bfp^k\big) = 0.  
\end{align*}

It can be seen that the terms involving $\bfp^k$ cancel out, such that we get:
\begin{align}\label{eq:multidomain_discretization_helper4}
    \bfK_{\bfsigma_e + \bfsigma_i}\,\bfphi_{e} + \s{k=1}{N_\text{MU}} f_r^k \bfK_{\bfsigma_i^k}\,\bfV_m^k = -\bfB_{\Gamma_M}\,\bfq.
\end{align}

If the multidomain description is used without body fat domain, the boundary condition in \cref{eq:multidomain_bc2} is used and the right-hand side in \cref{eq:multidomain_discretization_helper4} vanishes. If a body domain is considered, the right-hand side interacts with the body domain model, which is considered in the next section.

Adding boundary conditions to the discretization of the second multidomain equation proceeds using \cref{eq:multidomain_discretization_helper_multidomain2,eq:multidomain_discretization_helper_multidomain3}.
Carrying out the analog procedure to the first multidomain equation, we plug in \cref{eq:boundary_relation} to yield the matrix equation
\begin{align}\label{eq:multidomain_discretization1}
  \bfA\,\mat{\bfV_m^{k,(i+1)}\\ \bfphi_{e}^{(i+1)}} = \bfb
\end{align}
with system matrix $\bfA$ and right-hand side vector $\bfb$ given by
\begin{align}
 \bfA &= \matt{
    \dfrac{\theta}{A_m^k\,C_m^k}\bfK_{\bfsigma_i^k} -\dfrac{1}{\dt}\bfM & \quad
    \dfrac{\theta}{A_m^k\,C_m^k}\bfK_{\bfsigma_i^k},
  },\label{eq:multidomain_discretization2} \\[4mm]
  \bfb &= \Big( \dfrac{\theta-1}{A_m^k\,C_m^k}\bfK_{\bfsigma_i^k} - \dfrac{1}{\dt}\bfM\Big)\,\bfV_m^{k,(i)} 
    + \dfrac{\theta - 1}{A_m^k\,C_m^k}\bfK_{\bfsigma_i^k}\,\bfphi_e^{(i)}\nonumber \\[4mm]
  & +\dfrac{\theta-1}{A_m^k\,C_m^k}\bfB_{\Gamma_M}\,\bfp^{k,(i)} - \dfrac{\theta-1}{A_m^k\,C_m^k}\bfB_{\Gamma_M}\,\bfp^{k,(i)}
  -\dfrac{\theta}{A_m^k\,C_m^k}\bfB_{\Gamma_M}\,\bfp^{k,(i+1)} + \dfrac{\theta}{A_m^k\,C_m^k}\bfB_{\Gamma_M}\,\bfp^{k,(i+1)}.\nonumber 
\end{align}
Again, the boundary terms involving $\bfp^k$ vanish to yield the following expression for $\bfb$:%
\begin{align}\label{eq:multidomain_discretization3}
    \bfb &= \Big( \dfrac{\theta-1}{A_m^k\,C_m^k}\bfK_{\bfsigma_i^k} - \dfrac{1}{\dt}\bfM\Big)\,\bfV_m^{k,(i)} 
      + \dfrac{\theta - 1}{A_m^k\,C_m^k}\bfK_{\bfsigma_i^k}\,\bfphi_e^{(i)}.
\end{align}
In summary, \cref{eq:multidomain_discretization_helper4} with $\bfq=\bfzero$ coupled with $N_\text{MU}$  instances of \cref{eq:multidomain_discretization1,eq:multidomain_discretization2,eq:multidomain_discretization3} comprises the discretization for the multidomain model without body domain. Definitions of the involved stiffness and mass matrices are given in \cref{eq:diffusion_matrices}.

\subsection{Discretization of the Multidomain Model for Surface EMG}\label{sec:discretization_body_domain}

To discretize the multidomain model with the electric potential $\phi_b$ in the body domain, we extend the formulation without body domain in \cref{sec:discretization_multidomain}.
The body domain adds the electric potential $\phi_b$ to the vector of unknowns, for which the system has to be solved. As before, we discretize the field using finite element ansatz functions and solve for the vector $\bfphi_b$ of degrees of freedom.

The model for $\phi_b$ is the Laplace equation given in \cref{eq:body} with homogeneous Neumann boundary conditions given in \cref{eq:body_domain_bc3}. According to \cref{eq:discretization_laplace}, the discretized equation is given by
\begin{align}\label{eq:discretized_body}
  \bfK_{\bfsigma_b}\,\bfphi_b = 0.
\end{align}

In addition, the value of the body potential $\phi_b$ is coupled to the extracellular potential $\phi_e$ in the muscle domain $\Omega_M$ via the coupling conditions on the boundary $\Gamma_M$ given in \cref{eq:body_domain_coupling}.

We write the discretized and coupled multidomain equations as a linear system of equations in generic block-matrix form:
\begin{align}\label{eq:discretized_multidomain_body}
  \left[
  \begin{array}{@{}c|c|c|c@{}}
    \bfA_{V_m,V_m}^k & \bfB_{V_m,\phi_e}^k & &\\[2mm]
    \bfB_{\phi_e,V_m}^k & \bfB_{\phi_e,\phi_e} & &\bfB_{\Gamma_M} \\[2mm] \hline
    &&\bfC_{\phi_b,\phi_b} & -\bfB_{\Gamma_M}\\[2mm]\hline
    & \bfI_{\Gamma_M,\phi_e} & -\bfI_{\Gamma_M,\phi_b} &\\[2mm]
  \end{array}
  \right]
  \left[
  \begin{array}{@{}c@{}}
    \bfV_{m}^{k,(i+1)}  \\[2mm]\hline 
    \bfphi_{e}^{(i+1)} \\[2mm]\hline
    \bfphi_{b}^{(i+1)}  \\[2mm]\hline
    \bfq^{(i+1)}
  \end{array}\right]
  = 
  \left[\begin{array}{@{}c@{}}
    \bfb_{V_m}^{k,(i)} \\[2mm]
    \bfzero\\\hline
    \bfzero\\\hline 
    \bfzero
  \end{array}\right].
\end{align}

The vector of unknowns consists of the degrees of freedom in the finite element formulation at the next timestep $(i+1)$ of the transmembrane voltage $\bfV_m^{k,(i+1)}$, the extracellular potential $\bfphi_{e}^{(i+1)}$, the body potential $\bfphi_{b}^{(i+1)}$, and additionally the flux $\bfq^{(i+1)}$ over the shared boundary $\Gamma_M$ of the muscle and the body domain, which was defined in \cref{eq:definition_q}. For illustration purposes, only one compartment, $k=1$, for one MU, $N_\text{MU}=1$, is considered.

We refer to parts of the matrix in \cref{eq:discretized_multidomain_body} as block rows and block columns according to the given block-structure.

The first block row in the matrix equation is given by the discretized second multidomain equation. Following \cref{eq:multidomain_discretization2,eq:multidomain_discretization3}, the matrices and the right-hand side are given by
\begin{align*}
  &\bfA^k_{V_m,V_m} = \dfrac{\theta}{A_m^k\,C_m^k}\bfK_{\bfsigma_i^k} -\dfrac{1}{\dt}\bfM, \qquad
  \bfB^k_{V_m,\phi_e} = \dfrac{\theta}{A_m^k\,C_m^k}\bfK_{\bfsigma_i^k},\\[4mm]
  &\bfb_{V_m}^{k,(i)} = \Big( \dfrac{\theta-1}{A_m^k\,C_m^k}\bfK_{\bfsigma_i^k} - \dfrac{1}{\dt}\bfM\Big)\,\bfV_m^{k,(i)} 
      + \dfrac{\theta - 1}{A_m^k\,C_m^k}\bfK_{\bfsigma_i^k}\,\bfphi_e^{(i)}.
\end{align*}

The second block row describes the first multidomain equation that was derived in \cref{eq:multidomain_discretization_helper4}. The flux term $\bfq$ has been brought to the left-hand side and is incorporated by the boundary matrix $\bfB_{\Gamma_M}$ defined in \cref{eq:definition_boundary_matrix}. The other matrices are formulated as follows:
\begin{align*}
  \bfB_{\phi_e,V_m}^k &= f_r^k \bfK_{\bfsigma_i^k}, & 
  \bfB_{\phi_e,\phi_e} &= \bfK_{\bfsigma_e + \bfsigma_i}.
\end{align*}

The third block row is the formulation of the harmonic body potential $\phi_b$ and the matrix $\bfC_{\phi_b,\phi_b}$ equals the system matrix $\bfK_{\bfsigma_b}$ in \cref{eq:discretized_body}. The coupling condition on the flux $q$ in  \cref{eq:body_domain_bc2} is accounted for by including the boundary matrix $\bfB_{\Gamma_M}$ in the last column. The minus sign comes from the fact that the outward normal vector on $\Gamma_M$ as the boundary of $\Omega_B$ has the opposite direction to the normal vector on $\Gamma_M$ that is used for the models in the muscle domain $\Omega_M$. Using the helper variable $\bfq^{(i+1)}$, the second and third row of \cref{eq:discretized_multidomain_body} are coupled according to the prescribed condition in \cref{eq:body_domain_bc2}.

The other coupling condition, \cref{eq:body_domain_bc1}, is accounted for by the last block row in \cref{eq:discretized_body}. The degrees of freedom for the extracellular potential $\bfphi_e^{(i+1)}$ and the body potential $\bfphi_b^{(i+1)}$ have equal values on the boundary $\Gamma_M$. The matrices $\bfI_{\Gamma_M,\phi_e}$ and $\bfI_{\Gamma_M,\phi_b}$ are identity matrices that only have nonzero entries on the diagonal for the boundary degrees of freedom in the meshes of muscle domain and body domain, respectively.

Because the vector $\bfq^{(i+1)}$ is not an unknown in the system, the respective values in \cref{eq:discretized_multidomain_body} have to be eliminated.
As a result, we obtain the following system, which is formulated for a generic number $N_\text{MU}$ of MUs:
\begin{align}\label{eq:discretized_multidomain_body2}
  \left[\begin{array}{@{}ccc|c|c@{}}
    \bfA_{V_m,V_m}^1 &&& \bfB_{V_m,\phi_e}^1 &\\[2mm]
    &\ddots&&\vdots&\\[2mm]
    &&\bfA_{\phi_e,V_m}^{N_\text{MU}} & \bfB_{V_m,\phi_e}^{N_\text{MU}}&\\[2mm]
    \bfB_{\phi_e,V_m}^1 & \dots & \bfB_{\phi_e,V_m}^{N_\text{MU}} & \bfB_{\phi_e,\phi_e} & \bfD \\[2mm] \hline
    &&&\bfE & \tilde{\bfC}_{\phi_b,\phi_b}
  \end{array}\right]
  \left[\begin{array}{@{}c@{}}
    \bfV_{m}^{1,(i+1)}  \\[2mm]
    \vdots\\[2mm]
    \bfV_{m}^{N_\text{MU},(i+1)}\\[2mm]\hline 
    \bfphi_{e}^{(i+1)} \\[2mm]\hline
    \tilde{\bfphi}_{b}^{(i+1)}
  \end{array}\right]
  = 
  \left[\begin{array}{@{}c@{}}
    \bfb_{V_m}^{1,(i)} \\[2mm]
    \vdots \\[2mm]
    \bfb_{V_m}^{N_\text{MU},(i)}\\[2mm]
    \bfzero\\[2mm]\hline
    \bfzero
  \end{array}\right].
\end{align}

Formally, the elimination step is carried out by adding the equations of the third block row in \cref{eq:discretized_multidomain_body}, that correspond to the boundary degrees of freedom on $\Gamma_M$, to the corresponding equations of the same degrees of freedom in the second block row. This eliminates the last block column, which corresponds to $\bfq^{(i+1)}$. Next, the duplicate boundary degrees of freedom, that appear in both the $\Omega_M$ and $\Omega_B$ meshes, get unified. The corresponding matrix columns in the third block column are removed. To preserve the entries in the third block row, they are added in the sub matrix of block row three and block column two.

Now considering the updated matrix equation in \cref{eq:discretized_multidomain_body2}, all sub blocks are equal to \cref{eq:discretized_multidomain_body}, except for the former matrix $\bfC_{\phi_b,\phi_b}$ and the new matrices $\bfD$ and $\bfE$. The new matrix $\tilde{\bfC}_{\phi_b,\phi_b}$ is obtained from $\bfC_{\phi_b,\phi_b}$ by removing all rows and columns of boundary degrees of freedom. The removed entries are contained in the new matrices $\bfD$ and $\bfE$.

The size of the system matrix in \cref{eq:discretized_multidomain_body2} equals $a\times a$, where the number $a$ is composed of $N_\text{MU}+1$ times the number of degrees of freedom in the muscle mesh plus the number of degrees of freedom in the fat layer mesh without the boundary degrees of freedom on $\Gamma_M$. Accordingly, the vector $\tilde{\bfphi}_b^{(i+1)}$ is the same as $\bfphi_b^{(i+1)}$ except that it does not contain the boundary degrees of freedom, which are already included in $\bfphi_e^{(i+1)}$.

\Cref{eq:discretized_multidomain_body2} describes one iteration of the Crank-Nicolson scheme that is used to solve the multidomain model. This iteration is carried out alternatingly with the subcellular model according to the chosen operator splitting scheme. 

The first $N_\text{MU}$ block rows in \cref{eq:discretized_multidomain_body2} contain the second multidomain equation for every MU. The second-to-last block row contains the first multidomain equation and the last block row  contains the body fat layer model.

Because of the implicit formulation, electric conduction in the intracellular and extracellular space and the body domain are bidirectionally coupled. Therefore, the model can be used to simulate the effects of natural activation in the muscle on EMG signals on the skin surface as well as the reverse effect of external stimulation on the surface on the electrophysiology.

\subsection{Discretization of the Fiber Based Electrophysiology Model}

% Nach zwei Abschnitten über das Multidomain-Modell wäre es hier glaub ich gut, noch einen Abschnitt über das fiber-based Modell einzufügen, mit entsprechender Überschrift. Der Abschnitt darf gern kurz sein und Gemeinsamkeiten und Unterschiede zu den vorausgehenden Abschnitten erklären. Eine kurze Zusammenfassung in generischer Matrix-Block-Schreibweise wie oben mit Erklärung uu den Blöcken wäre sehr hilfreich. Ich denke, man versteht dann auch die Parallelisierung und Implementierung besser, wenn man hier schonmal ganz klar die unabhängigen Blöcke für die Fasern gesehen hat. 

The fiber based electrophysiology model consists of multiple independent 1D fiber domains, where the monodomain equation \cref{eq:monodomain} is solved. The transmembrane voltage $V_m$ is then mapped to a 3D mesh of the muscle domain and unidirectionally coupled to the first bidomain equation \cref{eq:bidomain1}. The first bidomain equation is solved for the extracellular potential $\phi_e$ and possibly the electric potential $\phi_b$ in the body fat domain, which corresponds to EMG signals on the skin surface.

The temporal discretization of the monodomain equation was described in \cref{sec:discretization_monodomain}. The diffusion term within the operator splitting requires a spatial discretization for which we use the finite element method. This 1D diffusion equation is given as
\begin{align}\label{eq:discretization_diffusion_term}
  \p{V_m}{t} = \dfrac{\sigma_\text{eff}}{A_m\,C_m} \p{V_m}{x}{2}.
\end{align}
It can be solved using a timestepping scheme such as the implicit Euler method to obtain time-discrete values $V_m^{(i)}, i=1,2,\dots$ for the transmembrane potential. The discretization leads to the matrix equation given in \cref{eq:diffusion_helper2} and to the variants presented in \cref{sec:mass_lumping} if mass lumping is used. In the stiffness and mass matrices, the anisotropic conduction tensor is replaced by the constant scalar prefactor $c := \sigma_\text{eff}/(A_m\,C_m)$ of the spatial second derivative in \cref{eq:discretization_diffusion_term}.

The first bidomain equation \cref{eq:bidomain1} is a 3D Poisson problem in terms of the unknown extracellular potential $\phi_e$. According to \cref{eq:discretization_laplace}, the finite element discretization is given by%
\begin{align*}
  \bfK_{\bfsigma_i + \bfsigma_e} \bfphi_e^{(i+1)} &= - \bfB_{\Gamma_f}\bff + \textbf{rhs},
\end{align*}
where the right-hand side $\textbf{rhs}$ of the Poisson problem is the transmembrane flow and is given by
\begin{align}\label{eq:static_bidomain_rhs}
  \textbf{rhs} &= -\bfK_{\bfsigma_i} \bfV_{m,3D}^{(i+1)}.
\end{align}
Here, $\bfphi_e^{(i+1)}$ and $\bfV_{m,3D}^{(i+1)}$ are the vectors of degrees of freedom on the 3D mesh for the extracellular potential $\phi_e$ and the membrane potential $V_m$ at timestep $(i+1)$. With the homogeneous Neumann boundary conditions for $V_m$ and $\phi_e$ given in \cref{eq:monodomain_bc}, the boundary term $\bfB_{\Gamma_f}$ vanishes.

In summary, the following matrix equations are solved for the fiber based electrophysiology model with $n$ fibers:
\begin{subequations}\label{eq:discretized_fibers}
  \begin{align}
    \left[\begin{array}{@{}ccc@{}}
      \bfA &&\\
      &\ddots&\\
      &&\bfA
    \end{array}\right]
    \left[\begin{array}{@{}c@{}}
      \bfV_{m}^{1,(i+1)}  \\
      \vdots\\
      \bfV_{m}^{n,(i+1)}
    \end{array}\right]
    &= 
    \left[\begin{array}{@{}c@{}}
      \bfV_{m}^{1,(i)}  \\
      \vdots\\
      \bfV_{m}^{n,(i)}
    \end{array}\right],\label{eq:discretized_fibers_1} \\[4mm]
    \bfV_{m,3D}^{(i+1)} &= \bfP \left[\begin{array}{@{}c@{}}
      \bfV_{m}^{1,(i+1)} \label{eq:discretized_fibers_2} \\
      \vdots\\
      \bfV_{m}^{n,(i+1)}
    \end{array}\right],\\[4mm]
    \bfK_{\bfsigma_i + \bfsigma_e} \bfphi_e^{(i+1)} &= -\bfK_{\bfsigma_i} \bfV_{m,3D}^{(i+1)} \label{eq:discretized_fibers_3}
  \end{align}
\end{subequations}
with the system matrix $\bfA$ for a single fiber given according to \cref{eq:lumping_implicit_euler_2} by
\begin{align*}
  \bfA &= \bfI - \dt\,\bfM_c^{-1}\bfK_c.
\end{align*}
\Cref{eq:discretized_fibers_1} solves the diffusion part of the operator splitting in \cref{eq:monodomain_operator_splitting}. After the values $\bfV_{m}^{j,(i+1)}$ for the timestep $(i+1)$ are computed on the 1D fiber meshes, the homogenized vector $\bfV_{m,3D}^{(i+1)}$ in the 3D mesh of the muscle domain $\Omega_M$ is obtained by the prolongation operation $\bfP$ in \cref{eq:discretized_fibers_2}. The homogenized vector is used in the right-hand side of the bidomain model in \cref{eq:discretized_fibers_3}, which computes the discretized extracellular potential $\bfphi_e^{(i+1)}$.

\Cref{eq:discretized_fibers_3} can be extended by adding a body fat layer $\Omega_B$ and the corresponding model for the electric potential $\phi_b^{(i+1)}$. Then, the vector of unknowns contains the degrees of freedom for both $\phi_e^{(i+1)}$ and $\phi_b^{(i+1)}$. The stiffness matrix $\bfK_{\bfsigma_i + \bfsigma_e}$ is obtained by integrating over both meshes in $\Omega_M \cup \Omega_B$. Only in the elements of the finite element mesh for $\Omega_B$, the conduction tensors are redefined as $\bfsigma_i = \bfzero$ and $\bfsigma_e = \bfsigma_b$. This sets the right-hand side of \cref{eq:discretized_fibers_3} to zero in $\Omega_B$ and the solution $\phi_b$ in harmonic according to the model in \cref{eq:body}. The coupling conditions \cref{eq:body_domain_coupling} between $\phi_e$ and $\phi_b$ and the outer Neumann boundary conditions \cref{eq:body_domain_bc3} for $\phi_b$ are satisfied automatically by this approach.

The comparison between the discretized multidomain model in \cref{eq:discretized_multidomain_body2} with the discretized fiber based model in \cref{eq:discretized_fibers} reveals several differences. Whereas the multidomain description consists of a single coupled linear system for electric conduction in the intracellular, extracellular and body domains, the formulations are only unidirectionally coupled in the fiber based description. While the multidomain model always computes the EMG signals on the skin surface in every timestep, the corresponding model in the fiber based description can be solved with larger timestep widths, using subcycling for the action potential propagation model.

As can be seen in \cref{eq:discretized_fibers_1}, the system matrix is decoupled and contains independent problems for every fiber. This is an advantage compared to the multidomain model, where a system describing the whole muscle domain has to be solved. On the downside, separate representations of the transmembrane voltage $V_m$ exist in the fiber based description. The representation in the 3D mesh has to be computed by interpolation from the representation on the fibers. The multidomain description has a single vector of degrees of freedom for $V_m$ with fewer entries than in the fiber-based description.

\subsection{Summary of Domains and Meshes}

Various finite element meshes occur in the formulation of the multi-scale model.
If the fiber based description is used, the description requires finite element meshes for the 1D fiber domains $\Omega_f^j$ for ${j=1,\dots,n}$. Further meshes are needed for the 3D muscle domain $\Omega_M$ and for the 3D body domain $\Omega_B$. The meshes for $\Omega_M$ and $\Omega_B$ share nodes on their common boundary $\Gamma_M$. The fiber meshes are embedded in the muscle domain. Their nodes do not necessarily have to coincide with the nodes of the muscle mesh.

The subcellular model is solved at locations $\Omega_s^i$ for $i=1,\dots,m$. These locations are the nodes of the fiber meshes for the fiber based description and the nodes of the muscle mesh for the multidomain description. We therefore have the inclusion $\Omega_s^i \subset \Omega_f^j \subset \Omega_M$.

For the solid mechanics model, the unified 3D domain $\Omega = \Omega_M \cup \Omega_B$ is used. The mesh for the continuum mechanics formulation can be different from the meshes used for the electrophysiology model. In fact, the continuum mechanics mesh has special requirements in order to yield a consistent formulation. Our implementation uses two overlaid meshes of quadratic and linear hexahedral elements for displacements and the hydrostatic pressure. 

Often, the required accuracy of the electrophysiology model is higher than for the continuum mechanics model, such that differently resolved meshes can be used. To facilitate data mapping, the nodes of the mechanics mesh should be chosen as subset of the nodes of the electrophysiology meshes.

%We spatially discretize the variables in \cref{eq:multi-domain1,eq:multi-domain2,eq:body,eq:bc1,eq:bc2,eq:bc3,eq:monodomain,eq:bidomain,eq:subcellular} using the finite element method with linear ansatz functions. The subcellular points $\Omega_s^i$ are placed at the nodes of the muscle domain mesh $\Omega_M$ for the multi-domain model and at the nodes of the fiber domain meshes $\Omega_f^j$ for the fiber based model. The fibers $\Omega_f^j$ are embedded in the muscle domain $\Omega_M$, however, the nodes of their meshes do not necessarily coincide. The nodes on the internal boundary $\Gamma_M$ between $\Omega_M$ and $\Omega_B$ are shared between the meshes of $\Omega_M$ and $\Omega_B$.

% ---

\section{Discretization and Solution Approach for the Solid Mechanics Model}\label{sec:discretization_mechanics}

After the formulation of linear and nonlinear models for solid mechanics in \cref{sec:model_muscle_contraction}, this section discusses their discretization and derives finite element formulations for the linearized model and the nonlinear model, both static and dynamic.
We also describe the algorithms used to obtain a numerical solution.

The implementation of a solver for generic hyperelastic descriptions is an interdisciplinary endeavor, if parallel execution is exploited and the model is integrated in a multi-scale biomechanics model.
Therefore, we give a comprehensive derivation of the formulas used to numerically solve the equations matching the implementation in OpenDiHu, such that the implementation is also comprehensible for readers that are not specialized in the field of continuum mechanics. More details on finite element discretizations for solid mechanics models can be found in the literature \cite{zienkiewicz1977finite,SUSSMAN1987357,zienkiewicz2005finite}.

% ---x---
\subsection{Discretization of the Linear Model}\label{sec:linearized_mechanics_model}

In this section, we discuss the linearized and static model. Besides the nonlinear model, our software OpenDiHu also implements the linearized description. The linear model exhibits better numerical properties and can be solved faster than the generic model. Thus, it can serve as a toy problem or can be used for mechanical systems, where the linearization assumptions are valid.

By assuming small strains, we can use the linearized kinematic relation in \cref{eq:linearized_helper3} to express the linear strain tensor $\bfeps$. The material model is Hooke's law formulated in \cref{eq:linearized_helper2}. It relates the strain to the Cauchy stress by $\bfsigma = \C:\bfeps$.

% linear, static and dynamic

Using variational calculus, the system response of external forces and infinitesimal, compatible, virtual displacements $δ\bfu$ is studied. 
We start with the \emph{principle of virtual work}, which states that in equilibrium the virtual work $δW$ performed by external forces along any virtual displacements $δ\bfu$ is zero. Equivalently, the external virtual work $δW_\text{ext}$ is equal to the internal virtual work $δW_\text{int}$:
\begin{align}
  δW_\text{int}(\bfu,δ\bfu) &= δW_\text{ext}(δ\bfu) \qquad && \forall δ\bfu \in H^1_0(\Omega)\label{eq:linearized_helper1}.
\end{align}
Here, the external virtual work $δW_\text{ext}$ is given by the product of external forces $\bft$ and the virtual displacements $δ\bfu$ at the same location. The internal virtual work $δW_\text{int}$ is the body's response in terms of stresses $\bfsigma$ and virtual strains $\bfeps$.
In summary, \cref{eq:linearized_helper1} is equivalent to the following equilibrium equation:
\begin{align}
  \ds\int_\Omega \bfsigma(\bfu) : δ\bfeps\,\d\bfx &= \ds\int_{∂\Omega} \bft : δ\bfu\,\d \bfx &&\forall δ\bfu \in H^1_0(\Omega).\label{eq:linearized_helper1b}
\end{align}
The vectors contain the degrees of freedom of a finite element discretization. The operator \say{:} denotes the component-wise product. 

Often, it is easier to write the equations in component form. Indices $a,b,c,\dots$ are used to specify a dimension index in $\{1,\dots,d\}$. The letters $L,M \in \{1,\dots,N\}$ denote indices over degrees of freedom in a mesh with $N$ nodes. The Einstein sum convention is used where repeated indices implicitly indicate summation, except when the indices are in parentheses.
Thus, the right-hand side $\bff$ of \cref{eq:linearized_helper1b} with ansatz functions $\phi^L$ and the degrees of freedom $δu_a^L$ of $δ\bfu$ can be written as:
\begin{align}\label{eq:linearized_mechanics_rhs}
  \bff_a = \ds\int_{∂\Omega} t_{(a)}\,δu_{(a)}^L\,\phi^L \,\d \bfx.
\end{align}

By combining the kinematic relation between displacements $\bfu$ and linearized strains $\bfeps$ in \cref{eq:linearized_helper3}, the material relation between $\bfeps$ and the stress $\bfsigma$ in \cref{eq:linearized_helper2}, the equilibrium relation between $\bfsigma$ and the right-hand side vector $\bff$ in \cref{eq:linearized_helper1b} and after discretizing displacements and virtual displacements, we get the following linear matrix equation:
\begin{align}\label{eq:linearized_helper4}
  \bfK\,\bfu = \bff.
\end{align}
The stiffness matrix $\bfK$ has rows and columns for every combination of degree of freedom $L,M \in \{1,\dots,N\}$ and dimension indices $a,b \in \{1,2,3\}$. The entries are given by:
\begin{align*}
  \bfK_{LaMb} = \ds\int_{\Omega} \mathbb{C}_{adbc}\p{\phi^L(\bfx)}{x_{d}}\p{\phi^M(\bfx)}{x_{c}}\,\d \bfx.
\end{align*}
The resulting model in \cref{eq:linearized_helper4} describes the passive behavior of a body under the linearization assumptions and can be used in an appropriate biomechanical application.

However, for contracting muscle tissue, we also need to incorporate active stresses that are generated at the sarcomeres in the muscle. As described in \cref{eq:active_stress_linear}, an active stress term $\bfsigma^\text{active}$  can be considered. Because this active term is prescribed by the activation dynamics and the subcellular model, it has to appear on the right-hand side of the linear model.
We add the active stress term $\bfsigma^\text{active}$ to the external virtual work in \cref{eq:linearized_helper1}, yielding the equation:
\begin{align}\label{eq:linearized_helper5}
  δW_\text{int}(\bfu,δ\bfu) &= \bff + \ds\int_\Omega \bfsigma^\text{active} : δ\bfeps_{-}\,\d\bfx &&\forall δ\bfu \in H^1_0(\Omega).
\end{align}
The active stress is associated with compression, i.e., negative virtual strains $δ\bfeps < 0$. Therefore, we use $δ\bfeps_{-}$ which is defined equal to $δ\bfeps$ for $δ\bfeps < 0$ and zero otherwise.
From \cref{eq:linearized_helper5}, we get the same discretized linear system as in \cref{eq:linearized_helper4}, but with an additional term $\bff^\text{ active}$ in the right-hand side that contains the discretized prescribed active stress field $\bfsigma^\text{active}_{ab}(\bfx)$:
\begin{align}\label{eq:linearized_helper6}
  \bff^\text{ active}_{La} = \ds\int_{Ω}\bfsigma^\text{active}_{ab}(\bfx)\,\p{\phi^L(\bfx)}{x_{b}} \,\d\bfx.
\end{align}
\subsection{Discretization of the Nonlinear Static Hyperelastic Model}\label{sec:static_hyperelastic_fe_model}

Next, we discuss the discretization of the nonlinear solid mechanics model, which uses the model equations introduced in \cref{sec:model_muscle_contraction}.
We begin with the discretization of a static, incompressible problem, where no velocities have to be considered.
The discretization is extended to the dynamic model in \cref{sec:solver_dynamic_hyperelasticity_fe_model}.

%In \cref{sec:linearized_mechanics_model}, the ingredients of a solid mechanics model derivation consisting of equilibrium, material and kinematic equations were outlined and used to derive a linearized description. For the nonlinear model, the material equations were discussed in \cref{sec:material_modeling} and the stress and elasticity tensors were derived in \cref{sec:stress_and_elasticity}. This section uses these building blocks and presents the full derivation for the generic hyperelastic finite element model.
As described in \cref{sec:assumptions_and_model_equations}, the equilibrium equation can be formulated in terms of the \emph{Hellinger-Reissner energy functional} $\Pi_L(\bfu,p) = \Pi_\text{int}(\bfu,p) + \Pi_\text{ext}(\bfu)$
given in \cref{eq:hellinger_reissner}.
It consists of the external energy functional, given in \cref{eq:pi_ext} as
\begin{align*}
  \Pi_\text{ext}(\bfu) = -\ds\int_{\Omega_0} \bfB\, \bfu\,\d V - \ds\int_{∂\Omega_0^t}\bar{\bfT}\,\bfu\,\d S,
\end{align*}
with body force $\bfB$ and surface traction $\bar{\bfT}$, and of the internal energy functional
\label{sec:section_with_pi_int}
\begin{align}\label{eq:mechanics_helper1}
  \Pi_\text{int}(\bfu,p) = \ds\int_{\Omega_0} \Psi_\text{iso}\big(\bar\bfC(\bfu)\big)\,\d V
    + \ds\int_{\Omega_0} p\,\big(J(\bfu) - 1\big)\,\d V.
\end{align}
Here, $\Psi_\text{iso}$ is the isochoric strain-energy density function introduced in \cref{eq:psi_iso} in terms of the reduced right Cauchy-Green tensor $\bar{\bfC}$ defined in \cref{eq:reduced_fc}.
The first term in \cref{eq:mechanics_helper1} describes the isochoric elastic response of the material, the second term adds the incompressibility constraint $J=1$ with the Lagrange multiplier $p$. The value of $p$ is computed as part of the model and can be identified as the hydrostatic pressure. Therefore, the second term is interpreted as the elastic response to compression and is included in the internal energy functional $\Pi_\text{int}$.

According to the \emph{principle of stationary potential energy}, the system is in equilibrium, if the potential energy functional is stationary.
This is the case, if the first variation $δ\Pi_L$ is zero.
Using the additive structure of $\Pi_L$, we can express the principle of stationarity as
\begin{subequations}
  \begin{align}
    D_{δ\bfu}\Pi_L(\bfu, p) &= D_{δ\bfu}\Pi_\text{int}(\bfu,p) + D_{δ\bfu}\Pi_\text{ext}(\bfu) \overset{!}{=} 0, & \forall δ\bfu  \label{eq:variations_functional_zero_a}\\[4mm]
    D_{δp}\Pi_L(\bfu, p) &= D_{δp}\Pi_\text{int}(\bfu,p) \overset{!}{=} 0 & \forall δp. \label{eq:variations_functional_zero_b}
  \end{align}
\end{subequations}
The variations of the internal and external energy functionals are defined as
\begin{align}\label{eq:def_variation}
  D_{δ\bfu}\Pi(\bfu) &= \d{\eps} \Pi(\bfu + \epsδ\bfu)\big|_{\eps=0}, & 
  D_{δp}\Pi(p) &= \d{\eps} \Pi(p + \epsδp)\big|_{\eps=0}.
\end{align}
They can be identified as the internal and external virtual work,
\begin{align*}
  D_{δ\bfu}\Pi_\text{int}(\bfu,p) &= δW_\text{int}, & D_{δ\bfu}\Pi_\text{ext}(\bfu) &= -δW_\text{ext}.
\end{align*}
Thus, \cref{eq:variations_functional_zero_a} can be expressed as 
\begin{align*}
  δW_\text{int} - δW_\text{ext} &= 0,
\end{align*}
which is the form of the equilibrium equation that was used in \cref{eq:linearized_helper1} in the derivation of the linearized model in \cref{sec:linearized_mechanics_model} . The Euler-Lagrange equations corresponding to the variational problem are the local incompressibility constraint and the partial differential equation of balance of momentum presented in \cref{eq:contraction_1,eq:contraction_2}.

Executing the derivative in the definitions of the variations in \cref{eq:def_variation} yields the following terms:
\begin{align*}
  &D_{δ\bfu}\Pi_\text{int}(\bfu,p)  = \ds\int_{\Omega_0} \bfS(\bfu,p): δ\bfE(δ\bfu)\,\d V,
  \qquad D_{δp}\Pi_\text{int}(\bfu,p) =\ds\int_{\Omega_0} \big(J(\bfu) - 1\big)δp\,\d V, \\[4mm]
  &D_{δ\bfu}\Pi_\text{ext}(\bfu) = -\ds\int_{\Omega_0} \bfB\cdot δ\bfu\,\d V - \ds\int\limits_{∂\Omega^t_0} \bar{\bfT}\cdot δ\bfu\,\d S,
\end{align*}
where the variational variables $δp,δ\bfu$ and $δ\bfE$ are the virtual pressure, virtual displacements, and virtual strains.

% discretization
We discretize the solutions of the functional for the displacements $\bfu(\bfx)$ and pressure $p(\bfx)$ and their variations using different ansatz functions $\phi^L$, $L=1,\dots,N_u$ and $\psi^L$, $L=1,\dots,N_p$:
\begin{align*}
   u_a &= \hat{u}_a^L \phi_{(a)}^L, & δu_a &= δ\hat{u}_a^L \phi_{(a)}^L,   & p &= \hat{p}^L \psi^L, & δp &= δ\hat{p}^L \psi^L.
\end{align*}
Again, Einstein summation over repeated indices, in this case the index $L$, is used. The displacements function is vector-valued and given by $\bfu(\bfx) = (u_1(\bfx), u_2(\bfx), u_3(\bfx))^\top$. The vectors containing the degrees of freedom are denoted by $\hat{\bfu} = (\hat{u}^L)_{L=1,\dots,N_u}$ and $\hat{\bfp} = (\hat{p}^L)_{L=1,\dots,N_p}$.

The kinematics equation to compute virtual strains from virtual displacements follows from \cref{eq:green_lagrange_u} in Lagrangian description and is given by $δ\bfE = \sym(\bfF^\top ∇\bfu)$. Its discretized form is given as follows, where the subscript comma $\square_{,A}$  indicates the derivative with respect to the indexed coordinate $\bfX_A$:
\begin{align*}
  δE_{AB} &= \dfrac12\left(F_{aB}\, \phi_{(a),A}^M + F_{aA}\, \phi_{(a),B}^M\right)δ\hat{u}_{a}^M.
\end{align*}
%
% summarize equations
In summary, the resulting set of discretized nonlinear equations can be formulated as:
\begin{align}\label{eq:mechanics_static_system}
  δW_\text{int}(\bfu,p) - δW_\text{ext} &= 0 \qquad \forall\,δ\bfu, \\[4mm]
  D_{δp}\Pi_L(\bfu) &= 0 \qquad \forall\,δp,
\end{align}
with the following discretized terms:
\begin{subequations}\label{eq:mechanics_static_system2}
  \begin{align}
    δW_\text{int}(\hat{\bfu},\hat{\bfp})  = \ds\int_{\Omega}\dfrac12  S_{AB}(\hat{\bfu},\hat{\bfp})\, \left(F_{aB}\, \phi_{(a),A}^M + F_{aA}\, \phi_{(a),B}^M\right)δ\hat{u}_{a}^M \,\d V,\\[4mm]
    δW_\text{ext}  = \ds\int_{\Omega} B_a \phi_{(a)}^M\,δ\hat{u}^M_a \,\d V +\ds\int_{∂\Omega}  \bar{T}^L_a\,\phi_{(a)}^L\, \phi_{(a)}^M\,δ\hat{u}^M_a\,\d S, \\[4mm]
    D_{δp}\Pi_L(\hat{\bfu}) = \ds\int_\Omega \big(J(\hat{\bfu}) - 1\big)\,δp\,\d V .
  \end{align}
\end{subequations}

The nonlinear system of equations in \cref{eq:mechanics_static_system,eq:mechanics_static_system2} can now be solved for the unknown vectors  $\hat{\bfu}$ and $\hat{\bfp}$ of degrees of freedom using a Newton scheme.

\subsection{Discretization of the Nonlinear Dynamic Hyperelastic Model}

We extend the discretization of the static model in the last section for the dynamic model.
The vector of unknowns is extended by a velocity function $\bfv: \Omega_t \to \R^3$. The additional equation $\dot{\bfu} = \bfv$ relates the displacements and the velocity.

As noted in the derivation of the equilibrium equation in \cref{sec:assumptions_and_model_equations}, the body force term $\bfB$ in the external energy functional also includes the inertial forces $\bfB_\text{inertial} = \rho_0\,\dot{\bfv}$ to describe the dynamic behavior.

The resulting nonlinear system of equations is given as follows:
\begin{subequations}\label{eq:mechanics_dynamic0}
  \begin{align}
    δW_\text{int}(\bfu,p) - δW_\text{ext}(\bfv) &= 0 \qquad \forall\,δ\bfu,\label{eq:mechanics_dynamic_system1}\\[4mm]
    \bfv &= \dot{\bfu},\label{eq:mechanics_dynamic_system2}\\[4mm]
    D_{δp}\Pi_L(\bfu) &= 0 \qquad \forall\,δp.\label{eq:mechanics_dynamic_system3}
  \end{align}
\end{subequations}

\subsection{Computation of the Stress Tensor and the Elasticity Tensor}\label{sec:stress_and_elasticity}

In the Newton solver, we need to compute the stress tensor $\bfS$ and its derivative $\C$, called the elasticity tensor, given the current displacement field $\bfu$. The relations are defined by the material model given by the strain energy function. This section presents the algorithm how to obtain the values of $\bfS$ and $\C$ from the displacements $\bfu$. While the derivation is formulated in terms of the displacement function $\bfu$, it is also valid for the finite element discretization, i.e., using the vector $\hat{\bfu}$  of degrees of freedom instead.

Following \cref{eq:material_model_helper1}, the second Piola-Kirchhoff stress $\bfS$ is given by the derivative of the strain energy function $\Psi$ with respect to $\bfC$.
For the representation using the invariants, the chain rule has to be used:%
\begin{align*}
   \bfS &= 2\,\p{\Psi(\bfC)}{\bfC} = \p{\Psi}{I_a}\p{I_a}{\bfC}.
\end{align*}
Using the decoupled form, the resulting stresses are also decoupled as $\bfS = \bfS_\text{vol}+\bfS_\text{iso}$. The volumetric stress $\bfS_\text{vol}$ describes the elastic response to compression, the isochoric stress $\bfS_\text{iso}$ describes the response to the deviatoric deformation. In the following, all steps to compute these stresses are listed. The rationale is to give a condensed reference of the implemented algorithm in OpenDiHu to facilitate further development.
For the derivation of all intermediate steps, we refer to the literature \cite{holzapfel2000nonlinear}.

At first, the reduced stress tensor $\bar{\bfS}$ that neglects the volumetric change is formulated as:%
\begin{align*}
  \bar{\bfS} = 2\p{\Psi_\text{iso}(\bar{I}_1,\bar{I}_2,\bar{I}_4,\bar{I}_5)}{\bar{\bfC}} &= \bar{\gamma}_1\,\bfI + \bar{\gamma}_2\,\bar{\bfC}
  + \bar{\gamma}_4\, \bfa_0 \otimes \bfa_0 + \bar{\gamma}_5\,(\bfa_0 \otimes \bar{\bfC}\,\bfa_0 + \bfa_0\bar{\bfC}\otimes \bfa_0).
\end{align*}
In case of an isotropic material, the terms involving $\bfa_0$ are not needed. The prefactors are given by derivatives of the strain energy function with respect to the reduced invariants:
\begin{align*}
  \bar{\gamma}_1 &= 2\left(\p{\Psi_\text{iso}(\bar{I}_1, \bar{I}_2)}{\bar{I}_1} + \bar{I}_1\,\p{\Psi_\text{iso}(\bar{I}_1, \bar{I}_2)}{\bar{I}_2}\right),
  &\bar{\gamma}_2 &= -2\p{\Psi_\text{iso}(\bar{I}_1, \bar{I}_2)}{\bar{I}_2},
  &\bar{\gamma}_4 &= 2\p{\Psi_\text{iso}}{\bar{I}_4}\\[4mm]
  \bar{\gamma}_5 &= 2\p{\Psi_\text{iso}}{\bar{I}_5}
\end{align*}
Using the fourth order identity tensor $\mathbb{I}$ and the projection tensor $\mathbb{P}$,%
\begin{align*}
  (\mathbb{I})_{abcd} &= \delta_{ac}\,\delta_{bd}, &
  \mathbb{P} &= \mathbb{I} - \dfrac13 \bfC^{-1} \otimes \bfC,
\end{align*}
the stress tensors can finally be computed as
\begin{align*}
  \bfS_\text{iso} &= J^{-2/3}\mathbb{P}:\bar{\bfS}, &
  \bfS_\text{vol} &= J\,p\,\bfC^{-1}, &
  \bfS &= \bfS_\text{iso} + \bfS_\text{vol}.
\end{align*}
In the compressible case including the penalty method, the value of $p$, that is needed for $\bfS_\text{vol}$, is given by the constitutive model as $p = \d \Psi_\text{vol}(J)/\d J$. In the incompressible case, $p$ is the unknown Lagrange multiplier that gets computed as part of the numerical solution. In that case, $p$ has the physical meaning of the hydrostatic pressure.

Using the present algorithm, the stress tensor $\bfS$ can, thus, be computed from derivatives of the strain energy function $\Psi$ and the right Cauchy Green tensor $\bfC$, which can be calculated from the displacement field $\bfu$.

Another important quantity for the numerical solution of the nonlinear system is the fourth order elasticity tensor $\C$, which is defined as
\begin{align*}
  \C = 2\p{\bfS(\bfC)}{\bfC} = 4\dfrac{\partial^2 \Psi(\bfC)}{\partial\bfC\partial\bfC}.
\end{align*}
It is the derivative of the stress tensor and is required in the Jacobian matrix of an iteration of the nonlinear Newton solver. Like the material tensor in \cref{eq:symmetries}, it shows major and minor symmetries and has 21 independent entries.

Like the stress tensor, the elasticity tensor is also additively composed into a volumetric term $\C_\text{vol}$ and an isochoric term $\C_\text{iso}$. The volumetric term can be computed by:%
\begin{align*}
  \mathbb{C}_\text{vol} &= J\,\tilde{p}\,\bfC^{-1} \otimes \bfC^{-1} - 2\,J\,p\,\bfC^{-1} \odot \bfC^{-1}, &
  \big(\bfC^{-1} \odot \bfC^{-1}\big)_{abcd} &= \dfrac12\big(C^{-1}_{ac}\,C^{-1}_{bd} + C^{-1}_{ad}\,C^{-1}_{bc}\big).
\end{align*}
The term includes two pressure variables $\tilde{p}$ and $p$. In the incompressible formulation, both variables equals the Lagrange multiplier $p$. For the compressible formulation, $\tilde{p}$ is derived as $\tilde{p} = p + J\,\d p/\d J$ and $p$ is computed from the volumetric strain energy function as stated above.

The isochoric term $\mathbb{C}_\text{iso}$ of the elasticity tensor follows from the following algorithm listing the quantities to compute:%
\begin{align*}
  &\bar{\delta}_1 = 4\left(\dfrac{∂^2\Psi_\text{iso}}{∂\bar{I}_1\,∂\bar{I}_1} + 2\,\bar{I}_1\dfrac{∂^2\Psi_\text{iso}}{∂\bar{I}_1\,∂\bar{I}_2} +\dfrac{∂\Psi_\text{iso}}{∂\bar{I}_2} + \bar{I}_1^2\,\dfrac{∂^2\Psi_\text{iso}}{∂\bar{I}_2\,∂\bar{I}_2}\right), \,
  \bar{\delta}_2 = -4\left(\dfrac{∂^2\Psi_\text{iso}}{∂\bar{I}_1\,∂\bar{I}_2} + \bar{I}_1\,\dfrac{∂^2\Psi_\text{iso}}{∂\bar{I}_2\,∂\bar{I}_2}\right),\\[4mm]
  &\bar{\delta}_3 = 4\dfrac{∂^2\Psi_\text{iso}}{∂\bar{I}_2\,∂\bar{I}_2}, \quad
  \bar{\delta}_4 = -4\dfrac{∂\Psi_\text{iso}}{∂\bar{I}_2}, \quad
  \bar{\delta}_5 = 4\left(\dfrac{∂^2\Psi_\text{iso}}{∂\bar{I}_1\,∂\bar{I}_4} +\bar I_1 \dfrac{∂^2\Psi_\text{iso}}{∂\bar{I}_2\,∂\bar{I}_4}\right),\\[4mm]
  &\bar{\delta}_6 = -4\dfrac{∂^2\Psi_\text{iso}}{∂\bar{I}_2\,∂\bar{I}_4}, \,\,\,\,
  \bar{\delta}_7 = 4\dfrac{∂^2\Psi_\text{iso}}{∂\bar{I}_4\,∂\bar{I}_4}, \,\,\,\,
  \mathbb{I}_{abcd} = δ_{ac}\,δ_{bd}, \,\,\,\,
  \bar{\mathbb{I}}_{abcd} = δ_{ad}\,δ_{bc}, \,\,\,\,
  \mathbb{S} = (\mathbb{I} + \bar{\mathbb{I}}) / 2, \\[4mm]
  &\p{\bar I_5}{\bar\bfC} = \bfa_0 \otimes \bar\bfC\,\bfa_0 + \bfa_0\,\bar\bfC \otimes \bfa_0, \quad
  \dfrac{∂^2\bar{I}_5}{∂\bar{\bfC}∂\bar{\bfC}} = \p{\bar{\bfC}}(\bfa_0 \otimes \bar\bfC\,\bfa_0 + \bfa_0\,\bar\bfC \otimes \bfa_0),
\end{align*}
\begin{align*}
  &\bar{\mathbb{C}} = J^{-4/3}\bigg(\bar{\delta}_1\,\bfI \otimes \bfI + \bar{\delta}_2\,\big(\bfI \otimes \bar{\bfC} + \bar{\bfC} \otimes \bfI\big) + \bar{\delta}_3\bar{\bfC} \otimes \bar{\bfC} + \bar{\delta}_4\,\mathbb{S}
  +\bar{δ}_5\,(\bfI \otimes \bfa_0 \otimes \bfa_0 + \bfa_0 \otimes \bfa_0 \otimes \bfI)\\[4mm]
  &\hspace*{1cm} +\bar{δ}_6\,(\bar{\bfC} \otimes \bfa_0 \otimes \bfa_0 + \bfa_0 \otimes \bfa_0 \otimes \bar{\bfC})
  +\bar{δ}_7\,(\bfa_0 \otimes \bfa_0 \otimes \bfa_0 \otimes \bfa_0) \\[4mm]
  &\hspace*{1cm} + \bar{δ}_8\,\Big(\bfI \otimes \p{\bar{I}_5}{\bar{\bfC}} + \p{\bar{I}_5}{\bar{\bfC}} \otimes \bfI \Big)
  + \bar{δ}_9\,\Big(\bar{\bfC} \otimes \p{\bar{I}_5}{\bar{\bfC}} + \p{\bar{I}_5}{\bar{\bfC}} \otimes \bar{\bfC} \Big) + \bar{δ}_{10}\Big(\p{\bar{I}_5}{\bar{\bfC}} \otimes \p{\bar{I}_5}{\bar{\bfC}}\Big) \\[4mm]
  &\hspace*{1cm}+ \bar{δ}_{11} \Big(\bfa_0 \otimes \bfa_0 \otimes \p{\bar{I}_5}{\bar{\bfC}} + \p{\bar{I}_5}{\bar{\bfC}} \otimes \bfa_0 \otimes \bfa_0 \Big) + \bar{δ}_{12} \dfrac{∂^2\bar{I}_5}{∂\bar{\bfC}∂\bar{\bfC}}\bigg)\\[4mm]
  &\tilde{\mathbb{P}} = \bfC^{-1} \odot \bfC^{-1} - \dfrac13 \bfC^{-1} \otimes \bfC^{-1} \\[4mm]
  &\mathbb{C}_\text{iso} = \mathbb{P} : \bar{\mathbb{C}} : \mathbb{P}^\top + \dfrac23 J^{-2/3} \bar{\bfS} : \bfC\,\tilde{\mathbb{P}} - \dfrac23\big(\bfC^{-1}\otimes \bfS_\text{iso} + \bfS_\text{iso}\otimes \bfC^{-1}\big)
\end{align*}
Then, $\C = \C_\text{vol} + \C_\text{iso}$ can be calculated.
%
%

% invariants: I1-I5
% transversely isotropic
% reduced invariants, reduced quantities for compressible materials
% strain energy function, derivative
% elasticity tensor
% -> computation of S and C

\subsection{Nonlinear Solver for the Solid Mechanics Model}\label{sec:solver_static_hyperelastic_fe_model}

% Newton solver
The governing nonlinear system of equations is solved by a Newton scheme. We define the vector of the unknown degrees of freedom as $(\hat{\bfu},\hat{p}) =: \bfz$. Then, the nonlinear equation takes the general form $\bfW(\bfz) = 0$. By linearization around a value $\bfz$, we get%
\begin{align*}
  \bfW(\bfz+Δ\bfz) = \bfW(\bfz) + \bfJ\,Δ\bfz + o(\bfz + Δ\bfz),
\end{align*}
with the increment $Δ\bfz = (Δ\hat\bfu, Δ\hat{p})$ and the Jacobian matrix $\bfJ = \partial {\bfW}/\partial {\bfz}$.
Neglecting the sublinear error term $o(z + Δz)$, we can start from an initial guess $\bfz^{(0)}$ and proceed to find the root of $\bfW$ using the following iterative Newton scheme:%
\begin{subequations}\label{eq:newton_scheme}
  \begin{align}
    \bfJ\,Δ\bfz^{(n)} = -\bfW(\bfz^{(n)}),\label{eq:mechanics_linear_system}\\[4mm]
    \bfz^{(n+1)} = \bfz^{(n)} + Δ\bfz^{(n)}.
  \end{align}
\end{subequations}
\Cref{eq:mechanics_linear_system} is a linear system of equations with the system matrix given by $\bfJ$, which has to be solved in every iteration step $n$. The linear system of equations can be expressed as follows:
\begin{align}\label{eq:static_newton_iteration}
  \matt{\bfk_{δ\bfu,Δ\bfu} & \bfk_{δp,Δ\bfu}^\top \\[2mm]
  \bfk_{δp,Δ\bfu} & \bfzero} \, \matt{Δ\hat{\bfu} \\[2mm] Δ\hat{p}} 
  =
  \matt{-\bfR_{δ\bfu} \\[2mm] -\bfR_{δp}}.
\end{align}
The definition of the right-hand sides $\bfR_{δ\bfu} = δW_\text{int} - δW_\text{ext}$ and $\bfR_{δp}=D_{δp}\Pi_L$ is given in \cref{eq:mechanics_static_system}. The system matrix is composed as follows. The upper left part consists of 3 times 3 blocks of submatrices, each with size $N_u \times N_u$ and the entries given by:
\begin{align*}
  \bfk_{δ\bfu,Δ\bfu,(L,a),(M,b)} &= \ds\int_\Omega \phi_{(a),B}^L\tilde{k}_{abBD}\phi_{(b),D}^M\,\d V &\text{with}\quad 
  \tilde{k}_{abBD} &= δ_{ab}\,S_{BD} + F_{aA}\,F_{bC}\,\mathbb{C}_{ABCD}.
\end{align*}
Here, $S_{BD}$ and $\mathbb{C}_{ABCD}$ are entries of the second Piola-Kirchhoff stress tensor $\bfS$ and the elasticity tensor $\mathbb{C}$. The computation of these terms uses the description in \cref{sec:stress_and_elasticity}.

The lower left part of the system matrix in \cref{eq:static_newton_iteration} is given by 1 times 3 blocks of submatrices, each with size $N_p \times N_u$ and entries given by:
\begin{align*}
  \bfk_{δp,Δ\bfu,L,(M,a)} = \ds\int_\Omega J\,\psi^L\,(F^{-1})_{Ba}\,\phi_{(a),B}^M \,\d V.
\end{align*}
The upper right part equals the transposed lower left block such that the system matrix is symmetric. Solving the system in \cref{eq:static_newton_iteration} in every iteration of the Newton scheme in \cref{eq:newton_scheme} converges to the solution of the static solid mechanics problem.

\subsection{Discretization and Solution of the Dynamic Hyperelastic Model}\label{sec:solver_dynamic_hyperelasticity_fe_model}
% dynamic hyperelasticity (6.9.2)

The dynamic model is given by the system of nonlinear equations in \cref{eq:mechanics_dynamic}. In addition to the spatial discretization with finite elements, we need to discretize the temporal derivatives of the displacement field $\bfu$ and the velocity field $\bfv$.
The time derivatives are discretized to timesteps $t=i\cdot \dt$ with an implicit Euler scheme:
\begin{align*}
  \dot{\bfu} &\leadsto \dfrac1{\dt}(\bfu^{(i+1)} - \bfu^{(i)}), & \dot{\bfv} &\leadsto \dfrac1{\dt}(\bfv^{(i+1)} - \bfv^{(i)}).
\end{align*}

Because of the added inertial body force, the external virtual work now depends on the vector of unknowns.
In consequence, we split the external virtual work $δW_\text{ext}$ into a dead part $δW_\text{ext,dead}$ that solely depends on external forces and an inertial part:%
\begin{align*}
  δW_\text{ext} = δW_\text{ext,dead} + \ds\int_{\Omega} \rho_0\,\dfrac{v^{(i+1),L}_{(a)} - v^{(i),L}_{(a)}}{\dt}\,\phi_{(a)}^L\, \phi_{(a)}^M\,δ\hat{u}^M_a \,\d V = 0.
\end{align*}
In summary, the system of equations to proceed from timestep $i$ to $(i+1)$ is given as:
\begin{subequations}\label{eq:mechanics_dynamic}
  \begin{align}
    δW_\text{int}({\bfu^{(i+1)}},p^{(i+1)}) - δW_\text{ext}(\bfv^{(i)},\bfv^{(i+1)}) &= 0 \qquad &&\forall\,δ\bfu,\label{eq:mechanics_dynamic1}\\[4mm]
    \dfrac1{\dt}(\bfu^{(i+1)} - \bfu^{(i)}) - \bfv^{(i+1)} &= 0,\label{eq:mechanics_dynamic2}\\[4mm]
    D_{δp}\Pi_L(\bfu^{(i+1)}) &= 0 \qquad &&\forall\,δp.\label{eq:mechanics_dynamic3}
  \end{align}
\end{subequations}
Here, \cref{eq:mechanics_dynamic1} is the principle of virtual work, \cref{eq:mechanics_dynamic2} relates displacements $\bfu$ and velocities $\bfv$ and \cref{eq:mechanics_dynamic3} is the incompressibility constraint.

The system is again solved using the Newton scheme presented in \cref{sec:solver_static_hyperelastic_fe_model}.
The linear system for each Newton iteration takes the following form:
\begin{align*}
  \matt{
    \bfk_{δ\bfu,Δ\bfu} & \bfl_{δ\bfu,Δ\bfv} & \bfk_{δp,Δ\bfu}^\top \\[2mm]
    \bfl_{δ\bfv,Δ\bfu} & \bfl_{δ\bfv,Δ\bfv} & \bfzero \\[2mm]
    \bfk_{δp,Δ\bfu} & \bfzero & \bfzero
  } \, 
  \matt{Δ\hat{\bfu} \\[2mm] Δ\hat{\bfv} \\[2mm] Δ\hat{p}} 
  =
  \matt{-\bfR_{δ\bfu} \\[2mm] -\bfR_{δ\bfv} \\[2mm] -\bfR_{δp}}.
\end{align*}
The entries $\bfk_{δ\bfu,Δ\bfu}$ and $\bfk_{δp,Δ\bfu}$ are the same as in the static case in \cref{eq:static_newton_iteration}.
The other non-zero entries are given by 
\begin{align*}
  \bfl_{δ\bfu,Δ\bfv,(L,a),(M,b)} &= \dfrac1{\dt}\delta_{ab} \ds\int_{\Omega} \rho_0\,\,\phi_{(b)}^M \,\phi_{(a)}^L \,\d V, & 
  \bfl_{δ\bfv,Δ\bfu,(L,a),(M,b)} &= \dfrac{1}{\dt}\delta_{ab}\,\delta^{LM},\\[4mm]
  \bfl_{δ\bfv,Δ\bfv,(L,a),(M,b)} &= -\delta_{ab}\,\delta^{LM}.
\end{align*}

Note that in the dynamic problem, the system matrix is unsymmetric. It would be symmetric if the entries $\bfl_{δ\bfu,Δ\bfv}$ and $\bfl_{δ\bfv,Δ\bfu}^\top$ were the same. This would be the case for a density of one, $\rho_0 = 1$, and if the term $\int_{\Omega} \phi_{b}^M \phi_{a}^L \,\d V$ would be replaced by $\delta_{ab}\delta^{LM}$. The second condition means that a lumped mass matrix would be used where the diagonal entries are set to the row sums of the original matrix.

We discretize the finite element solution in space by \emph{Taylor-Hood} elements. This type of element uses quadratic ansatz functions $\phi$ for the displacements and velocities and linear ansatz functions $\psi$ for the Lagrange multiplier or hydrostatic pressure $p$ on a 3D hexahedral mesh. This choice was proven to exhibit no locking \cite{zienkiewicz2005finite}. Locking is a phenomenon of degraded convergence of the finite element method for solid mechanics problems and occurs for improper discretization schemes.

For a compressible material, the incompressibility constraint which is the last equation in the systems \cref{eq:mechanics_static_system} or \cref{eq:mechanics_dynamic} is removed. Instead of solving for the pressure $p$ as a Lagrange multiplier, the value is given by the constitutive model as described in \cref{sec:material_modeling}. In consequence, the system matrix of the linear system of equations that is solved in the Newton iterations has a smaller size for compressible materials.

Moreover, the size varies depending on whether the static or the dynamic problem given in \cref{sec:solver_static_hyperelastic_fe_model,sec:solver_dynamic_hyperelasticity_fe_model} is solved. Assuming a linear mesh with $N_p$ degrees of freedom and a quadratic mesh with $N_u$ degrees of freedom, the square system matrix has $3\,N_u$ rows and columns for a static compressible formulation, $3\,N_u + N_p$ for a static incompressible formulation, $6\,N_u$ for a dynamic compressible model, and $6\,N_u+N_p$ for a dynamic incompressible model.

% static compressible:    3*N_u
% static incompressible:  3*N_u + N_p
% dynamic compressible:   3*N_u + 3*N_u
% dynamic incompressible: 3*N_u + 3*N_u + N_p

In any case, the mechanics model can be linked to the subcellular model by defining the active stress as given in \cref{eq:active_stress_term}. Since the active stress does not depend directly on the passive behavior, the active stress term can be added as a constant to the passive stress term. This constant also has no influence on the Jacobian matrix $\bfJ$. As the subcellular model depends on the fiber stretch $\lambda_f = \sqrt{I_4}$, there is a feedback loop between the subcellular and the solid mechanics model.

Details on the connection to the subcellular model as well as details on the numerical solution schemes are given in \cref{sec:solid_mechanics_solver}.

%, including the solver schemes, how initial values are chosen and measures to speed up convergence such as load stepping are discussed in the implementation and result sections.
%To speed up the computation, the initial guess of the vector of unknowns in every timestep is linearly extrapolated from the two previous timesteps.

\chapter{Usage of the Software OpenDiHu}\label{chap:usage}

\Opendihu{} is an open source software framework for static and dynamic multi-physics problems that can be solved with the finite element method. 
It was developed essentially by the author as part of this work with some code contributions given by Aaron Krämer, Nehzat Emamy and Felix Huber from the \emph{Institute for Parallel and Distributed Systems} and the \emph{Institute of Applied Analysis and Numerical Simulation}.
We use it to simulate the multi-scale models presented in the last chapter: biophysical problems describing biomechanics and neurophysiology of the musculoskeletal system.

In this chapter, we introduce basic concepts and present details on the usage of our software. The next chapter \cref{sec:implementation} continues with a discussion of internal software aspects and gives details on how the different algorithms and solvers are implemented.

We begin this chapter with an explanation of the design goals in \cref{sec:design_goals}. Next, \cref{sec:usage} showcases how to set up simulation programs for various scenarios based on examples. Some biochemical models are conveniently formulated and shared in the bioengineering community using the CellML description language \cite{Cellml2003}. \Cref{sec:usage_cellml} gives more details on the usage of CellML models in our software. Finally, \cref{sec:output_file_formats} discusses various output formats and compatible tools for visualization.

\section{Design Goals}\label{sec:design_goals}

Simulations of complex multi-scale models require the combination of tailored numerical solution schemes. Spatial mesh resolutions and timestep widths should be chosen carefully to avoid instabilities and to allow the completion of useful simulation time spans in feasible runtimes.
Numerical solvers on 1D, 2D and 3D meshes have to be coupled and the data should be mapped between these meshes. 

The simulations should be parallelized to efficiently exploit today's hardware and reduce the runtime to a minimum. Parallel runs should be performant on small workstations, on larger compute servers and on supercomputers. 

Scenarios for different use cases should be possible, ranging from convenient debugging with simple physics and small problem sizes to large runs with highly resolved meshes and comprehensive models. Schemes for input and output of the data should be available for all of these use cases. Established community standards for models, such as the CellML description, and output file formats should be considered.
The configuration of models and solver parameters should be flexible, well organized and properly documented to allow an efficient workflow.

%Input and output have to be processed in proper data formats for different purposes, ranging
% efficient storage and visualization of large datasets

In addition to this feature list from a user perspective, further requirements can be formulated from a developer perspective.
The program code should be modular, structured and well documented to allow discovery, reuse and extension in the future. The implemented solvers should compute correct results, which should be testable to preserve correctness during code changes.

OpenDiHu aims to fulfill these requirements.
%With these user and developer requirements in mind we develop our software framework named \opendihu{}.
%In the different contexts of reporting and software development, the presented capitalized name and an all lower case version of the name are used, respectively. 
The name originates from the Digital Human project that aimed to advance the field of biomechanics by \say{providing new possibilities to improve the understanding of the neuromuscular system by switching from small-sized cluster model problems to realistic simulations on HPC clusters} \cite{DihuWeb}. The software framework contributes to this goal.
%The software has been introduced in a publication \cite{Maier2019}.

In the following, we concretize the requirements and formulate design goals to guide the software development.
The design goals can be summarized under the keywords \emph{usability}, \emph{performance} and \emph{extensibility}. They span the field of requirements from user-centric properties to developer centric properties.

\subsection{Usability}
Usability is defined in ISO 9241-11 \cite{ISO9241} as \say{the extent to which a product can be used by specified users to achieve specified goals with effectiveness, efficiency, and satisfaction in a specified context of use.}  
We target at users with a basic understanding of biophysics, numerics, programming and command line usage in Linux. 
The specified goals include---in increasing complexity---to reproduce results of existing studies,  analyze the simulation results,  adjust parameters of existing simulations to achieve different model behavior, conduct studies over a set of different parameters,  exchange numerical schemes to improve stability or efficiency, combine implemented parts of models to a new multi-physics model, and implement new solvers for completely new physics. The context of use lies in scientific and educational studies. 
%The program can be run in serial or in parallel on a small-scale compute server or compute cluster or on a supercomputer.

We base the usage of the framework on command line programs and scripts and do not include a graphical user interface (GUI). A GUI would need to present an abstract, simplified layer of the simulation setup, that reduces the understanding of the actual process. Furthermore, it would be difficult to keep a GUI up to date with all functionality that gets implemented in the software over time. The advantage of a command-line-only-program is, that it can be easily used in automated studies with different parameter combinations. Furthermore, it simplifies usage on remote computers such as compute clusters and supercomputers.

In this context, good usability is ensured by using the Python programming language for the configuration of the simulation.
The computational code of every simulation is written in C++ and compiled to a hardware specific executable, which enables good computational performance. The user can configure all parameters using Python scripting. The Python3 interpreter is linked into the C++ program such that the configuration script can be parsed at runtime when the simulation program is executed.

Thus, users can organize the simulation settings using their own variable names.
Users can compute derived parameter values within the settings script, they have the flexibility to organize the settings in multiple files, and define own command line parameters for every example. Input data and results of the simulation can be preprocessed and post-processed directly in the Python settings script.

\subsection{Perfomance}
The second design goal is to achieve good performance.
OpenDiHu satisfies this goal by supporting parallel execution on the one hand and by providing efficient algorithms on the other hand. 

Simulations can be run on distributed and shared memory systems. The computational domains are mainly discretized with structured meshes, that can easily be partitioned into subdomains for multiple processes. For large scale simulations on multiple cores, the input data such as node positions can be specified in a distributed way. Hence, every process only needs to know its own portion of the whole simulation and the total amount of data can exceed the storage capabilities of a single compute core.

Efficient algorithms involve efficient numerical solvers such as multigrid or conjugate gradient schemes and optimized data handling within the software framework.
We use the external library \emph{PETSc} \cite{petsc-efficient1997} for the parallel storage of vectors and matrices and for linear algebra operations. PETSc provides a large collection of preconditioners, linear and nonlinear solvers that can be chosen at runtime. At the same time, it offers low-level access to the locally stored data, which, e.g., allows us to optimize data transfer between different arrays.

For multi-scale models, good performance can only be achieved, when the data transfer between different solvers is also efficient. Using profiling tools, runtime hotspots in various simulation programs were regularly identified and evaluated. The portions of code, that use  most runtime should be the actual computations and not memory allocations or data transfer operations.
Using these insights, the framework was efficiently constructed, e.g., by avoiding repeated memory initialization and expensive copy operations whenever possible.

\subsection{Extensibility}
By extensibility, we refer to the possibility to add solvers for new physical processes to the existing framework.
This is facilitated when existing components of the framework are documented and can be reused. 

On the highest level, existing simulation programs using models in the CellML format can be altered to solve different physics by exchanging the CellML model. For example, model extensions of the active mechanical behavior of the half-sarcomeres can be implemented in the corresponding CellML model and without changing the C++ code.

It is also possible to use the adapters in OpenDiHu for the numerical coupling library \emph{preCICE} \cite{precice} to numerically couple  software packages to OpenDiHu. The surface coupling adapter allows coupling external mechanics solvers and exchange displacements and traction forces. The volume coupling adapter allows, e.g., to use the electrophysiology solver in OpenDiHu and couple it with external models of the muscle or other organs.

On the next level, which still does not require changing the C++ core, the modular building blocks of model solvers such as timestepping schemes for the solution of ODEs, operator splittings or coupling schemes can be newly combined for different behavior.

Solving other models, for which no solver has been designed, involves adding new code to the software framework.
The solvers in OpenDiHu use structures like function spaces consisting of meshes and basis functions, linear system solvers and output writers for data output, which are self-contained and get reused at different locations in the framework. A completely new model, e.g., an electro-magnetic description of electrophysiology would require a dedicated new solver class. 
Two template classes exist in OpenDiHu that can be copied and adjusted to create such a new solver for either transient or static problems.

Polymorphism concepts of the C++ programming language such as object orientation (OO) and template meta-programming (TMP) allow writing generic algorithmic code that gets specialized for the particular use-cases at runtime (OO) or at compile time (TMP). For example, most of the solvers are independent of the type of mesh they operate on. Similarly, an explicit timestepping scheme has the same definition regardless whether it solves a 0D subcellular model with a high-dimensional state vector or a 3D linear elasticity formulation discretized by finite elements, where the solution is a vector field with three components.
These concepts help to reuse existing structures in OpenDiHu.

%\subsection{Overview of this Chapter}
%Usage of the software and its implementation reflects the properties presented in the previous sections. More %details are given in the remainder of this chapter. TODO

%Details are given on the implementation, the underlying algorithms, and the application from a user's perspective. For selected topics, a comparison to other existing simulation frameworks for biomechanical problems such as OpenCMISS is given.
%The aim of this chapter is to describe how the software is used and how it is implemented. We begin with an introduction of its usage in \cref{sec:usage}. The remainder of this chapter addresses the implementation. \Cref{sec:data_handling_with_petsc} presents details on the data handling based on the library PETSc. \Cref{sec:fem_matrices_and_bc} describes the assembly of finite element matrices and shows an algorithm for Dirichlet boundary conditions.  TODO

\section{Usage of OpenDiHu}\label{sec:usage}
We begin with aspects of OpenDiHu that are relevant from a user's perspective. This section outlines the basic organization of the repository in \cref{sec:organization_of_the_directory}, the installation procedure in \cref{sec:installation} and demonstrates the usage of given example scenarios in \cref{sec:exemplary_usage_1,sec:exemplary_usage_2,sec:exemplary_usage_3}. \Cref{sec:summary_of_existing_solver_classes} summarizes all available solver classes.

\subsection{Organization of the Repository}\label{sec:organization_of_the_directory}
The complete OpenDiHu is contained in a single git repository which is hosted on GitHub at \url{https://github.com/maierbn/opendihu}. In addition, the documentation is hosted on the \say{Read the Docs} website under \url{https://opendihu.readthedocs.io} \cite{opendihuWeb}. This documentation is split into a user and a developer documentation. The user part includes introductory information such as installation instructions, a description of most of the existing simulation scenarios with images of the results, instructions how to build and run them and a complete reference of the settings of all available solvers.

\begin{figure}
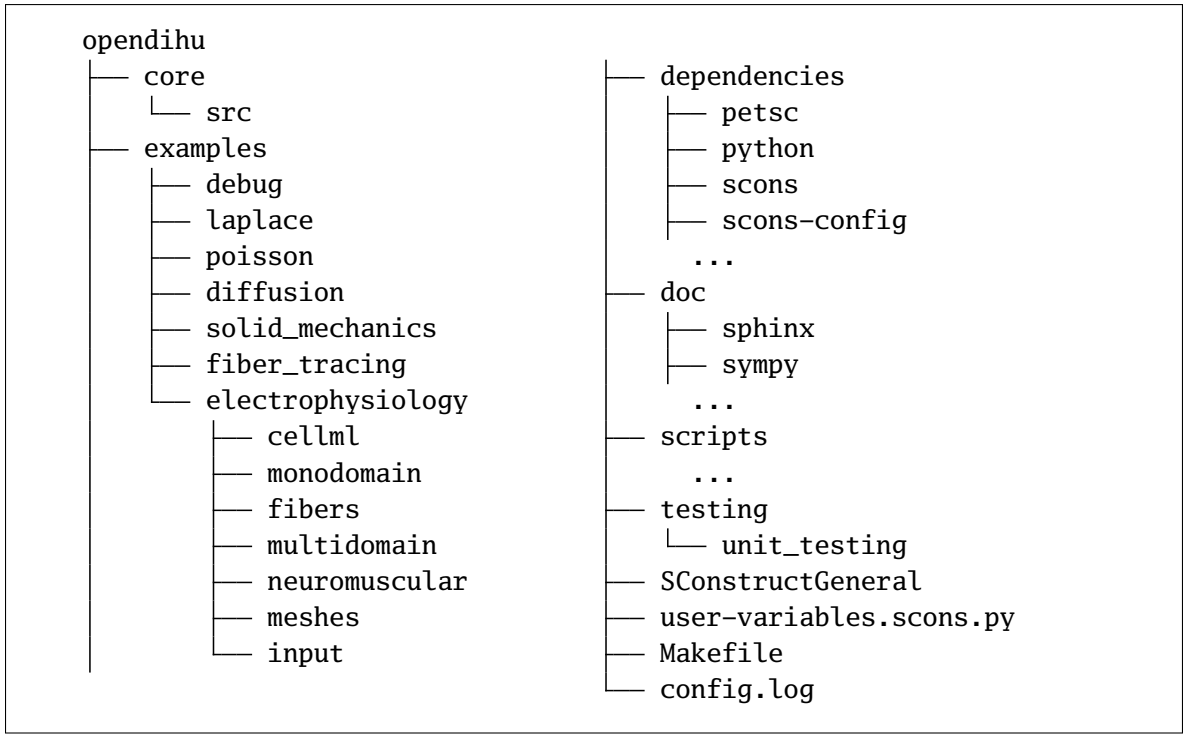

\centering
%\begin{lstlisting}[columns=fullflexible,breaklines=true,postbreak=\mbox{\textcolor{gray}{$\hookrightarrow$}\space}]

\begin{framed}
\begin{subfigure}[t]{0.45\textwidth}%
\begin{Verbatim}[fontsize=\small]
opendihu
├── core
│   └── src
├── examples
│   ├── debug
│   ├── laplace
│   ├── poisson
│   ├── diffusion
│   ├── solid_mechanics
│   ├── fiber_tracing
│   └── electrophysiology
│       ├── cellml
│       ├── monodomain
│       ├── fibers
│       ├── multidomain
│       ├── neuromuscular
│       ├── meshes
│       └── input

\end{Verbatim}
\end{subfigure}
\begin{subfigure}[t]{0.45\textwidth}%
\begin{Verbatim}[fontsize=\small]

├── dependencies
│   ├── petsc
│   ├── python
│   ├── scons
│   ├── scons-config
│     ...
├── doc
│   ├── sphinx
│   ├── sympy
│     ...
├── scripts
│     ...
├── testing
│   └── unit_testing
├── SConstructGeneral
├── user-variables.scons.py
├── Makefile
└── config.log
\end{Verbatim}
\end{subfigure}
\end{framed}
\caption{Contents of the main opendihu directory.}%
\label{fig:directory_structure}%
\end{figure}

After cloning the git repository, the directory has the contents shown in \cref{fig:directory_structure}, which will be explained in the following.
The software consists of a core library, that provides all functionality such as solvers and data handling. In addition, examples are created, that set up specific simulation scenarios and import the required solvers by linking to the core library. 

The subdirectory \code{core/src} contains all C++ code that is compiled into the core library. This source code consists of approximately \num{90000} code lines, \num{24000} blank lines and \num{19000} comment lines contained in approximately 700 files and structured in a directory tree with approximately 70 total subdirectories.

The \code{examples} directory contains all simulation scenarios that are packaged with Open-\break DiHu. Each of the approximately 65 examples demonstrates how to solve a different model, often in several variations with different parameters and numerical schemes. 
%Each example is a collection of one or several small C++ source code files that define the solvers and the corresponding Python settings files. 
The examples are grouped by the subdirectories shown in \cref{fig:directory_structure} in different categories: technical examples for debugging, scenarios for solving the Laplace, Poisson and diffusion equations, various solid mechanics models, fiber tracing examples that can be used to generate meshes as described in \cref{sec:generation_of_meshes_for_multiscale}, and the electrophysiology models. 
The electrophysiology examples are further structured as given in \cref{fig:directory_structure} with increasing complexity: 
subcellular CellML model solvers (0D) in the subdirectory \code{cellml}, 
solvers for the monodomain equation (1D), i.e., electrophysiology on a single fiber in \code{monodomain}, 
models with multiple 1D fibers also coupled with the 3D EMG model or muscle contraction model in \code{fibers}, 
the same but with the multidomain model in \code{multidomain}, 
and models of motor neurons coupled with fibers and multidomain models in the \code{neuromuscular} directory. 

The directory \code{meshes} contains scripts and raw data to generate all meshes needed by the simulations. The directory \code{input} collects all input files that are used in any example, e.g., cellml models, meshes, and text files that specify MU assignments and firing times. Because this directory contains large files, it is not included in the git repository but hosted on a separate file server.

The \code{dependencies} directory contains the source files and installations of all external packages, such as PETSc in \code{petsc}, the Python3 interpreter in \code{python}, and the SCons related packages in \code{scons} and \code{scons-config}. This directory will be automatically filled with more subdirectories during the installation procedure.

The \code{doc} directory collects various documents and mathematical derivations that help to understand certain solvers. For example, the directory \code{doc/sphinx} contains the whole online documentation, which is hosted on the \say{Read The Docs} website \cite{opendihuWeb} and built using the reStructuredText markup language and the Sphinx generation system. The \code{doc/sympy} directory contains Python scripts with the derivation of various equations using the symbolic math package SymPy.

Various utility Python scripts are stored in the directory \code{scripts}. Users should add this directory to the \code{$\$$PATH} environment variable in their system such that the scripts can be invoked from the command prompt. For example, the \code{catpy} and \code{plot} scripts list and visualize Python-based output files of simulations, other scripts can be used to inspect and manipulate binary mesh files.

The directory \code{testing/unit_testing} contains the code for all unit tests. Furthermore, the files that exist directly under the top level \code{opendihu} directory are relevant for the build system, which will be explained in the next section.

\subsection{Installation}\label{sec:installation}
% scons

The installation procedure involves three steps: First, the dependencies, i.e., all required external packages have to be located. Second, the OpenDiHu core library is compiled and linked. Third and optionally, unit tests are compiled, linked and executed.

The first step consists of finding the location of each dependency, determining the corresponding header and library files that are needed for inclusion and linking, and potentially determining special compiler or linker flags. The step can be configured to fit the individual system setup and use case by taking into account already existing dependencies and enabling or disabling optional packages.

The second step compiles the source code and links it to all dependencies that were collected in the first step. The result is a static library, which contains the functionality of OpenDiHu in executable form. To run a particular simulation, an additional program with a small source code file has to be written, compiled and linked to this library.

For the compilation of unit tests in the third step, a similar action is performed. The step builds and links three unit testing executables that are subsequently run with one, two and six processes and conduct various functional tests of the implementation.

Currently, the following fifteen dependencies are used with OpenDiHu: the standard for shared-memory parallelisation \emph{MPI} \cite{openmpi}, the data handling and numerics library \emph{PETSc} \cite{petsc-efficient1997,petsc-user-ref,petsc-web-page}, the \emph{Python interpreter} \cite{python3}, a set of \emph{Python packages} including \emph{NumPy} \cite{numpy}, \emph{SciPy} \cite{scipy} and \emph{Matplotlib} \cite{matplotlib} among others, a \emph{Base64} compression library \cite{base64}, the unit testing framework \emph{Googletest} \cite{googletest}, the compile-time differentiation toolbox \emph{SEMT} \cite{gutterman2004symbolic,semt}, the parallel file I/O library \emph{ADIOS2} \cite{adios2}, the vectorization toolbox \emph{Vc} \cite{vc2012,Kretz2015} and its newer version \emph{std-simd} \cite{hoberock2016working}, the library for parallel time integration with multigrid \emph{XBraid} \cite{xbraid-package}, the solver and converter for CellML models \emph{OpenCOR} \cite{OpenCOR2015}, the XML parser \emph{libxml2} \cite{libxml2}, the coupling library \emph{preCICE} \cite{precice} and the logging library \emph{Easylogging++} \cite{easyloggingpp}. The build system has to install these dependencies and possibly cope with different sets of available versions and prerequisites.

% This action usually relies on configuration files or command line options to manually enable or disable certain dependecies and to manually specify their locations.

Popular build systems exist that facilitate the three mentioned installation step, e.g., CMake, GNU Autoconf and SCons. CMake uses a three-step process of configuration, generation and building, which requires users to have the corresponding know-how. Autoconf creates a configure script that relies on command line options instead of a configuration file for all settings. 

Considering the usability goal for OpenDiHu, we chose the build system SCons as it requires little previous knowledge and can read its settings from a Python based configuration file. SCons performs the three steps of the installation procedure by a single command. Packages that are not yet installed are downloaded and installed automatically, using transparent bash commands.
Furthermore, the SCons build system itself is packaged along with our software and, thus, no additional installation steps are required to begin the build process (apart from checking that some essential packages such as a compiler and an MPI implementation are available).

SCons allows to both specify the installation configuration and extend the functionality by using the Python scripting language. Based on the scons-config package \cite{sconsconfig}, we added functionality to detect and automatically download the dependencies that are required for OpenDiHu. For some dependencies, multiple versions are tried if the first attempt fails, e.g., for Python and PETSc.
This adds robustness for different systems and typically allows to set up OpenDiHu on a new Linux computer by only executing a single \say{\code{scons}} command.

The top-level files listed in \cref{fig:directory_structure} are related to the build system. The file \code{SConstruct}\code{General} contains Python code that defines various flags for the usual build targets: the \code{release} targets creates optimized and hardware-specific binaries, the \code{debug} target disables optimization and adds debugging symbols to the executables, other debugging targets produce intermediate outputs after the  preprocessing, assembly or optimization stages. All options are documented in the help text of the build system.

The \code{user-variables.scons.py} file contains the configuration and can be adjusted by the user to enable or disable certain packages or features. The \code{Makefile} contains convenient shortcuts for longer build command, e.g., the command \say{\code{make}} builds the debug and release targets and runs the unit tests. During installation, all text output and progress information is appended to the log file \code{config.log}. If the installation fails, this file contains all information that is required to track down the respective issue.

After the installation and build step of the core library, individual simulation scenarios can be developed and executed. The place for the code of these simulations is in the \code{examples} subdirectory, where numerous predefined simulation scenarios are given. In the following, we demonstrate how to use OpenDiHu and, more specifically, we present the structure and configuration of a simulation program by considering three of these examples in increasing complexity: a Laplace problem in \cref{sec:exemplary_usage_1}, a simulation of muscle contraction in \cref{sec:exemplary_usage_2} and a simulation of the neuromuscular system in \cref{sec:exemplary_usage_3}.

\subsection{Exemplary Usage: Laplace Problem}\label{sec:exemplary_usage_1}
% Laplace example
Every simulation consists of a single C++ source file that gets compiled to an executable program and a Python file that defines all settings for the simulation. The example considered in the following solves a 2D Laplace problem and is located in the directory \code{examples/laplace/laplace2d}. The considered C++ source file is \code{src/laplace_structured.cpp} and the corresponding Python settings file is \code{settings_lagrange_quadratic.py}. The contents of the two files are listed in \cref{fig:laplace_example_source,fig:laplace_example_settings}. The directory additionally contains code for other scenarios that have different parameters and discretizations.

After compilation by the \say{\code{scons}} command, an executable is created in the \code{build_release} subdirectory of the example. In this directory, the simulation can be run with the following command:
\begin{lstlisting}[columns=fullflexible,breaklines=true,postbreak=\mbox{\textcolor{gray}{$\hookrightarrow$}\space}]
  ./laplace_structured ../settings_lagrange_quadratic.py
\end{lstlisting}
Here, the first item is the program name. We pass the filename of the settings file as the first command line argument. 
%For some examples, additional arguments can be given, those are accessible from within the Python settings script and can be used to adjust parameters.

\Cref{fig:laplace_example_source} lists the full C++ source code of this example. \Cref{code:c0l1} includes the main header file of opendihu, which makes all functionality available. The rest of the source file contains the definition of the \code{main} function. \Cref{code:c0l9} defines the context object \code{settings}, which uses the command line arguments given by \code{argc} and \code{argv}. This line invokes the Python interpreter on the Python settings file and stores all parameters in the \code{settings} object.

\Cref{code:c0l12,code:c0l13,code:c0l14,code:c0l15,code:c0l16,code:c0l17} define an object named \code{equationDiscretized}, which is of the \code{Finite}\code{ElementMethod} class located in the \code{SpatialDiscretization} namespace. The new object uses the settings object that was defined before.

The \code{FiniteElementMethod} class takes several class template arguments enclosed in angle brackets. The first in \cref{code:c0l13} specifies the mesh type, which, in this 2D example, is a structured deformable mesh of dimension two. Furthermore, \cref{code:c0l14} specifies quadratic Lagrange basis functions and \cref{code:c0l15} specifies Gauss quadrature with three Gauss points per dimension. The argument in \cref{code:c0l16} defines the equation that is discretized by this finite element method class, which, in this case, is the static Laplace equation $Δ\bfu = 0$.

In \cref{code:c0l20}, the solver is executed, performs the computation and writes the configured output files. The program finally returns in \cref{code:c0l22}.

% laplace2D c++ source
\begin{figure}
\centering
\begin{framed}
%\begin{Verbatim}[fontsize=\small]
\begin{lstlisting}[basicstyle=\small\ttfamily,commentstyle=\color{gray},numbers=left]
  #include <cstdlib>
  #include "opendihu.h" $\label{code:c0l1}$

  int main(int argc, char *argv[])
  {
    // 2D Laplace equation 0 = du^2/dx^2 + du^2/dy^2
    
    // initialize and parse settings from input file
    DihuContext settings(argc, argv);   $\label{code:c0l9}$
    
    // define the tree of solvers (here only one FEM solver)
    SpatialDiscretization::FiniteElementMethod<   $\label{code:c0l12}$
      Mesh::StructuredDeformableOfDimension<2>,   $\label{code:c0l13}$
      BasisFunction::LagrangeOfOrder<2>,          $\label{code:c0l14}$
      Quadrature::Gauss<3>,                       $\label{code:c0l15}$
      Equation::Static::Laplace                   $\label{code:c0l16}$
    > equationDiscretized(settings);              $\label{code:c0l17}$
    
    // run the simulation
    equationDiscretized.run();  $\label{code:c0l20}$
    
    return EXIT_SUCCESS;    $\label{code:c0l22}$
  }
\end{lstlisting}
%\end{Verbatim}
\end{framed}
\caption{Example source file of an OpenDiHu solver for the 2D Laplace problem.}%
\label{fig:laplace_example_source}%
\end{figure}

The problem to be solved is parametrized by the settings file \code{settings_lagrange_}\code{quadratic.py}, which is listed in \cref{fig:laplace_example_settings}. The code in this file can use all features of the Python scripting language. For example, in \cref{code:c1l5}, the \emph{NumPy} numerics packages is imported and its sine function is used in \cref{code:c1l11,code:c1l14}. The print statement in \cref{code:c1l16} is executed in the for loop in \cref{code:c1l7} and produces informational output about boundary condition values during execution. 

The settings file has to define the variable \code{config} to be a Python dictionary, i.e., an associative container data structure. This dictionary contains the parameter names and values that are required by the solver in the C++ program. 
In \cref{fig:laplace_example_settings}, the \code{config} dictionary is defined in \cref{code:c1l18,code:c1l19,code:c1l20,code:c1l21,code:c1l22,code:c1l23,code:c1l24,code:c1l25,code:c1l26,code:c1l27,code:c1l28,code:c1l29,code:c1l30,code:c1l31,code:c1l32,code:c1l33,code:c1l34,code:c1l35,code:c1l36,code:c1l37,code:c1l38,code:c1l39,code:c1l40,code:c1l41,code:c1l42,code:c1l43,code:c1l44,code:c1l45,code:c1l46,code:c1l47,code:c1l48,code:c1l49,code:c1l50,code:c1l51,code:c1l52}. It contains global options such as filenames of log files in \cref{code:c1l19,code:c1l20,code:c1l21,code:c1l22} followed by specific options for the finite element method object in \cref{code:c1l24,code:c1l25,code:c1l26,code:c1l27,code:c1l28,code:c1l29,code:c1l30,code:c1l31,code:c1l32,code:c1l33,code:c1l34,code:c1l35,code:c1l36,code:c1l37,code:c1l38,code:c1l39,code:c1l40,code:c1l41,code:c1l42,code:c1l43,code:c1l44,code:c1l45,code:c1l46,code:c1l47,code:c1l48,code:c1l49,code:c1l50}. The exact meaning of all parameters is documented in the online documentation \cite{opendihuWeb} and also sketched by the comments in the file. Some parameters will be presented in the following.

The parameter set consists of mesh parameters in \cref{code:c1l25,code:c1l26,code:c1l27,code:c1l28}, problem parameters in \cref{code:c1l31,code:c1l32,code:c1l33,code:c1l34}, 
solver parameters in \cref{code:c1l37,code:c1l38,code:c1l39,code:c1l40,code:c1l41,code:c1l42,code:c1l43,code:c1l44} and output writers in \cref{code:c1l47,code:c1l48,code:c1l49,code:c1l50}. 
The mesh in this scenario is a cartesian grid on the unit square. The number of elements is specified by the parameter \code{`nElements`}. The number of elements in $x$ and $y$-directions is given by the variables \code{nx} and \code{ny}, which are defined in \cref{code:c1l1} of the settings file. 

The parameter \code{`inputMeshIsGlobal`} is relevant for parallel execution. Its value specifies, whether all parameters apply to the global problem (\code{True}) or to a local subdomain (\code{False}). If the given example is executed by four processes, the mesh will have $10\times 10$ elements, as specified by \code{nx} and \code{ny}. 
However, if \code{`inputMeshIsGlobal`} is set to \code{False}, each of the $2\times 2$ subdomains would create a mesh of this size, yielding a total mesh of $20 \times 20$ elements. Instead of the Cartesian grid, it is also possible to define the node positions of every element. Then, it is beneficial to only specify the data for the own subdomain on each process for meshes with large numbers of elements and large numbers of processes.

This example problem uses Dirichlet boundary conditions. Values of a sine curve are prescribed at the boundaries $y=0$ and $y=1$ of the unit square. \Cref{code:c1l31} of the settings file sets the boundary conditions to the variable \code{bc}. This variable is defined in the loop before the \code{config} dictionary in \cref{code:c1l6,code:c1l7,code:c1l8,code:c1l9,code:c1l10,code:c1l11,code:c1l12,code:c1l13,code:c1l14,code:c1l15,code:c1l16}. The \code{bc} variable itself is a dictionary that specifies the prescribed values for every degree of freedom. 

\Cref{code:c1l38,code:c1l39} specify the employed preconditioner and solver by strings that are given to the solver library PETSc. Thus, all linear solvers available in PETSc can be used.
Error tolerances on the residual norm and a maximum number of iterations can be specified. It is also possible to dump the system matrix, right-hand side and solution vectors to a text file or a MATLAB readable file using the options in \cref{code:c1l42,code:c1l43}.

Output of the simulation results is configured by specifying a list of output writers in \cref{code:c1l47,code:c1l48,code:c1l49,code:c1l50}. The considered example has the two output writers with formats \code{`Paraview`} and \code{`PythonFile`}. The former writes files that can be visualized by the software \emph{ParaView}, the latter outputs files that can be easily parsed with a Python script. Both output writers either generate binary or human-readable files, depending on the \code{`binary`} option. Binary files have smaller file sizes and are used for large datasets. The human-readable text files make it is easier to debug the output.

\begin{figure}[H]
  \centering%
  \includegraphics[width=0.7\textwidth]{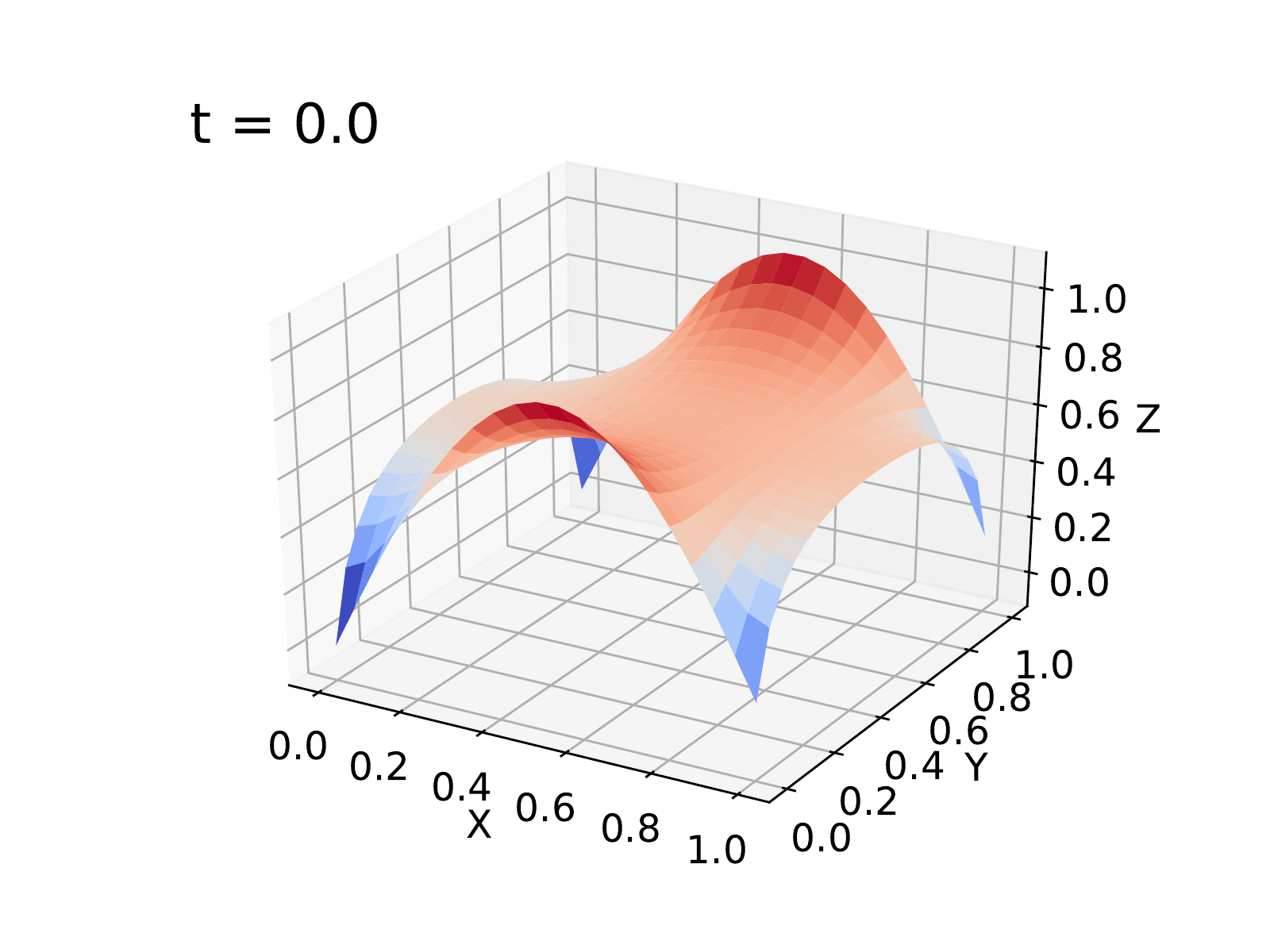}%
  \caption{Visualization of the solution of the exemplary 2D Laplace problem. The plot was created using the \code{plot} command on the output of the \code{PythonFile} output writer.}%
  \label{fig:laplace_example_plot}%
\end{figure}

After the program has been run, the \code{out} subdirectory contains the two output files created by the output writers. The Python based file can be visualized using the command \say{\code{plot}}, which is also provided by OpenDiHu. \Cref{fig:laplace_example_plot} shows the resulting \emph{Matplotlib} visualization. The figure shows that the Dirichlet boundary conditions for $y=0$ and $y=1$ are met and the solution is a harmonic function.

% laplace2D settings
\begin{figure}
\centering
\begin{framed}
\begin{lstlisting}[basicstyle=\footnotesize\ttfamily,language=Python,commentstyle=\color{gray},numbers=left]
  nx = 10;       ny = nx          # number of elements   $\label{code:c1l1}$
  mx = 2*nx + 1; my = 2*ny + 1    # number of nodes    $\label{code:c1l2}$
                                                        $\label{code:c1l3}$
  # specify boundary conditions                         $\label{code:c1l4}$
  import numpy as np       $\label{code:c1l5}$
  bc = {}                    $\label{code:c1l6}$
  for i in range(mx):        $\label{code:c1l7}$
    x = i/mx                 $\label{code:c1l8}$
                                $\label{code:c1l9}$
    # bottom line                   $\label{code:c1l10}$
    bc[i] = np.sin(x*np.pi)       $\label{code:c1l11}$
                                   $\label{code:c1l12}$
    # top line                        $\label{code:c1l13}$
    bc[(my-1)*mx + i] = np.sin(x*np.pi)       $\label{code:c1l14}$
                                              $\label{code:c1l15}$
    print("{}, bc: {}, {}".format(i, bc[i], bc[(my-1)*mx + i]))       $\label{code:c1l16}$
                                                                             $\label{code:c1l17}$
  config = {                                                                      $\label{code:c1l18}$
    "solverStructureDiagramFile":     "solver_structure.txt",      # diagram file $\label{code:c1l19}$
    "logFormat":                      "csv",     # "csv" or "json", format of log $\label{code:c1l20}$
    "scenarioName":                   "laplace", # scenario name for log file     $\label{code:c1l21}$
    "mappingsBetweenMeshesLogFile":   None,      # a log file about mappings      $\label{code:c1l22}$
    "FiniteElementMethod": {                                                      $\label{code:c1l23}$
      # mesh parameters                                                           $\label{code:c1l24}$
      "nElements":                   [nx, ny],   # number of elements in x and y  $\label{code:c1l25}$
      "inputMeshIsGlobal":           True,       # if nElements is a global number$\label{code:c1l26}$
      "physicalExtent":              [1.0, 1.0], # physical domain size           $\label{code:c1l27}$
      "physicalOffset":              [0, 0],     # physical location of origin    $\label{code:c1l28}$
                                                                                  $\label{code:c1l29}$
      # problem parameters                                                        $\label{code:c1l30}$
      "dirichletBoundaryConditions": bc,         # Dirichlet BC as dict           $\label{code:c1l31}$
      "dirichletOutputFilename":     None,       # output file for Dirichlet BC   $\label{code:c1l32}$
      "neumannBoundaryConditions":   [],         # Neumann BC                     $\label{code:c1l33}$
      "prefactor":                   1,          # constant prefactor c in $\textcolor{gray}{c\Delta u}$   $\label{code:c1l34}$
                                                                                  $\label{code:c1l35}$
      # linear solver parameters                                                         $\label{code:c1l36}$
      "solverType":                  "gmres",    # linear solver scheme           $\label{code:c1l37}$
      "preconditionerType":          "none",     # preconditioner scheme          $\label{code:c1l38}$
      "relativeTolerance":           1e-15,      # stopping criterion, rel. tol.  $\label{code:c1l39}$
      "absoluteTolerance":           1e-10,      # stopping criterion, abs. tol.  $\label{code:c1l40}$
      "maxIterations":               1e4,        # maximum number of iterations   $\label{code:c1l41}$
      "dumpFormat":                  "default",  # format for data dump           $\label{code:c1l42}$
      "dumpFilename":                None,       # filename for dump              $\label{code:c1l43}$
      "slotName":                    None,       # connector of solver            $\label{code:c1l44}$
                                                                                  $\label{code:c1l45}$
      # output writers                                                            $\label{code:c1l46}$
      "OutputWriter" : [                                                          $\label{code:c1l47}$
        {"format": "Paraview",   "filename": "out/laplace", "binary": False},     $\label{code:c1l48}$
        {"format": "PythonFile", "filename": "out/laplace", "binary": False},     $\label{code:c1l49}$
      ]                                                                           $\label{code:c1l50}$
    }                                                                             $\label{code:c1l51}$
  }                                                                               $\label{code:c1l52}$
\end{lstlisting}
\end{framed}
\caption{Python settings file of the Laplace example corresponding to the source file in \cref{fig:laplace_example_source}.}%
\label{fig:laplace_example_settings}%
\end{figure}

% plot of Laplace solution

\subsection{Exemplary Usage: Multidomain Model With Solid Mechanics}\label{sec:exemplary_usage_2}
% with multidomain, first show multidomain excerpt, then full solver_structure

The next example to be studied is a simulation of electrophysiology and muscle contraction. It uses the multidomain model on the muscle and body fat domains and bidirectionally couples the nonlinear solid mechanics model. The example is located in the directory \code{examples/electrophysiology/multidomain/multidomain_contraction}.

% multidomain_contraction C++ source
\begin{figure}
\centering
\begin{framed}
%\begin{Verbatim}[fontsize=\small]
\begin{lstlisting}[basicstyle=\scriptsize\ttfamily,commentstyle=\color{gray},numbers=left]
  #include <cstdlib>
  #include "opendihu.h"     $\label{code:c2l2}$

  int main(int argc, char *argv[])
  {
    // 3D multidomain coupled with contraction
    
    // initialize everything, handle arguments and parse settings from input file
    DihuContext settings(argc, argv);       $\label{code:c2l9}$
    
    typedef Mesh::StructuredDeformableOfDimension<3> MeshType;    $\label{code:c2l11}$

    Control::Coupling       $\label{code:c2l13}$
    <
      OperatorSplitting::Strang<          $\label{code:c2l15}$
        Control::MultipleInstances<     // subcellular model  $\label{code:c2l16}$      
          TimeSteppingScheme::Heun<       $\label{code:c2l17}$
            CellmlAdapter<                $\label{code:c2l18}$
              57,71,  // nStates,nAlgebraics: 57,71 = Shorten, 4,9 = Hodgkin Huxley     $\label{code:c2l19}$
              FunctionSpace::FunctionSpace<MeshType,BasisFunction::LagrangeOfOrder<1>>  $\label{code:c2l20}$
            >  
          >
        >,
        TimeSteppingScheme::MultidomainWithFatSolver<       // multidomain    $\label{code:c2l24}$
          SpatialDiscretization::FiniteElementMethod<       // FEM for initial potential flow    $\label{code:c2l25}$
            MeshType,                                       $\label{code:c2l26}$
            BasisFunction::LagrangeOfOrder<1>,
            Quadrature::Gauss<3>,
            Equation::Static::Laplace                            $\label{code:c2l29}$
          >,
          SpatialDiscretization::FiniteElementMethod<       // anisotropic conduction      $\label{code:c2l31}$
            MeshType,                                            $\label{code:c2l32}$
            BasisFunction::LagrangeOfOrder<1>,
            Quadrature::Gauss<5>,
            Equation::Dynamic::DirectionalDiffusion
          >,
          SpatialDiscretization::FiniteElementMethod<       // isotropic conduction in fat layer     $\label{code:c2l37}$
            MeshType,                                             $\label{code:c2l38}$
            BasisFunction::LagrangeOfOrder<1>,
            Quadrature::Gauss<5>,
            Equation::Dynamic::IsotropicDiffusion
          >
        >   $\label{code:c2l43}$
      >,
      MuscleContractionSolver<          // solid mechanics $\label{code:c2l45}$
        Mesh::CompositeOfDimension<3>     $\label{code:c2l46}$
      >
    > problem(settings);     $\label{code:c2l48}$
    
    problem.run();  $\label{code:c2l50}$
    
    return EXIT_SUCCESS;
  }
\end{lstlisting}
%\end{Verbatim}
\end{framed}
\caption{Source code of the simulation program that computes the multidomain model coupled with the solid mechanics model.}%
\label{fig:example_multidomain_source}%
\end{figure}

\Cref{fig:example_multidomain_source} shows the source code of the C++ file. The overall structure of the code is the same as in the previous Laplace example: \Cref{code:c2l2} includes the OpenDiHu header, the main function consists of the definition of a settings object in \cref{code:c2l9}, the definition of the solver in \crefrange{code:c2l13}{code:c2l48} and its execution in \cref{code:c2l50}. The only difference is the definition of the solver, which contains more nested class templates.

The problem is numerically solved by computing the multidomain model, transferring the activation parameter $\gamma$ from the multidomain mesh to the elasticity mesh, computing the solid mechanics model, and then mapping the deformed geometry back to the multidomain mesh. This compute cycle repeats in every timestep. This coupling between two models is performed by the \code{Control::Coupling} class defined as the outer-most solver in \cref{code:c2l13}. It nests the two solvers of the model parts: The first is the class named \code{OperatorSplitting::Strang} in \cref{code:c2l15}. It computes the multidomain electrophysiology model. The second class is the \code{MuscleContractionSolver} in \cref{code:c2l45}. It calls the solid mechanics solver and incorporates the activation and active stress term.

The multidomain model itself is computed using two coupled solvers. As formulated in \cref{sec:discretization}, a Strang operator splitting is used that alternates between solving the subcellular model and the electric conduction part of the multidomain model. In the code, these two parts are defined in \cref{code:c2l16} and \cref{code:c2l24}. As can be seen, the first part that solves the subcellular model consists of the three nested classes in \cref{code:c2l16,code:c2l17,code:c2l18}. The inner-most is the \code{CellmlAdapter}, which loads and executes a DAE model description from a CellML file. Its template arguments are the number of states and number of algebraic variables in \cref{code:c2l19} and the type of the function space in \cref{code:c2l20} used for spatial discretization.
The CellML model is solved by the enclosing Heun timestepping scheme in \cref{code:c2l17}. Because we need to solve the subcellular model for every compartment $k \in 1,\dots,N_\text{MU}$, a \code{MultipleInstances} class is used in \cref{code:c2l16}, which encloses the timestepping scheme and applies it on the domains for every compartment.

The second part of the multidomain model is the electric conduction in the intracellular, extracellular and body fat domains. It corresponds to solving the linear system of equations given in \cref{sec:discretization_body_domain}. This is done in OpenDiHu by the \code{MultidomainWithFatSolver} defined in \crefrange{code:c2l24}{code:c2l43}. It can be seen, that it nests three classes of type \code{FiniteElement}\\\code{Method}. 

The first one in \cref{code:c2l25} is used to initially solve a potential flow problem, from which the fiber direction can be estimated. This approach \cite{Choi2013} is also used in the fiber generation algorithms described in \cref{sec:generation_of_fiber_meshes}. As a result, we get the anisotropy direction in the 3D domain, which is needed to define the anisotropic intracellular conduction tensors $\bfsigma_i^k$. As the problem to be solved is a Laplace problem, the equation to be discretized by the class is defined accordingly in \cref{code:c2l29}.

The second and third nested finite element classes are defined in \cref{code:c2l31,code:c2l37}. They define the isotropic electric conduction in the muscle domain and the anisotropic electric conduction in the fat domain and are used to set up the stiffness and mass matrices for these subproblems.

Several meshes are involved in the definition of this example. As described in \cref{sec:electric_conduction_body_domain}, the computational domain consists of a muscle domain and a fat domain. Both domains are discretized by a 3D structured mesh, which is given as \code{MeshType} in \cref{code:c2l11}. This type is referenced for the subcellular model in \cref{code:c2l20} and for the conduction parts in \cref{code:c2l26,code:c2l32,code:c2l38}. 
%Only in the anisotropic conduction class, it refers to the fat domain, in the other uses it specifies the muscle domain.
For the muscle contraction solver in \cref{code:c2l46}, we use a different, \say{composite} type of mesh. This type is a combination of the two structured meshes for the body and for the fat domain, as the whole tissue should be considered in the computation of the deformation.

% multidomain_contraction solver_structure
\begin{figure}
\centering
\begin{framed}
\begin{lstlisting}[basicstyle=\scriptsize\ttfamily,language=Python,commentstyle=\color{gray},numbers=left]
  config = {
    "scenarioName": "multidomain_contraction",
    "Solvers": {
      "potentialFlowSolver": {...},   $\label{code:c3l4}$
    },
    "Coupling": {                $\label{code:c3l6}$
      "timeStepWidth": 1e-3,
      "Term1": {            # multidomain     $\label{code:c3l8}$
        "StrangSplitting": {                  $\label{code:c3l9}$
          
          "Term1": {        # subcellular model      $\label{code:c3l11}$
            "MultipleInstances": {                    $\label{code:c3l12}$
              "nInstances": variables.n_compartments,    $\label{code:c3l13}$
              "instances": [        # settings for each motor unit     $\label{code:c3l14}$
              {
                "ranks": list(range(n_ranks)),    $\label{code:c3l16}$
                "Heun": {
                  "CellML" : {
                  }
                }
              } for compartment_no in range(variables.n_compartments)] $\label{code:c3l21}$
            },   $\label{code:c3l122}$
          }, 
          "Term2": {        # conduction term of multidomain      $\label{code:c3l24}$
            "MultidomainSolver": {
              
              "PotentialFlow": {                                   $\label{code:c3l27}$
                "FiniteElementMethod": {  
                  "solverName": "potentialFlowSolver",            $\label{code:c3l29}$
                },
              },                                                    $\label{code:c3l31}$
              
              "OutputWriter": [ $\label{code:c3l33}$
              ]
            },
          }
        },
      },
      "Term2": {        # solid mechanics    $\label{code:c3l39}$
        "MuscleContractionSolver": {
          
          # the actual solid mechanics solver
          "DynamicHyperelasticitySolver": {
           }
        }
      }
    }
  }
\end{lstlisting}
\end{framed}
\caption{Excerpt of the settings file for the multidomain and solid mechanics solver.}%
\label{fig:example_multidomain_settings}%
\end{figure}

The Python settings script, that corresponds to the C++ source file, is given in \cref{fig:example_multidomain_settings}. Only the main structure of the \code{config} dictionary is outlined and the details are left out. It can be seen, that the definition in this file has a hierarchical structure. It is the same tree-like structure as in the C++ source.

The settings for the top-level coupling scheme start in \cref{code:c3l6}. First, some settings related to the timestepping scheme itself are specified of which one, the timestep width, is shown. Then, the settings of the two nested solvers are listed under \code{`Term1`} in \cref{code:c3l8} and \code{`Term2`} in \cref{code:c3l39}. Similarly, also the nested Strang splitting scheme in \cref{code:c3l9} defines its two nested solvers under \code{`Term1`} (\cref{code:c3l11}) and \code{`Term2`} (\cref{code:c3l24}).

\Crefrange{code:c3l12}{code:c3l122} define settings for the \code{MultipleInstances} class, which holds separate instances of the subcellular solver for all motor units.
The number of instances is specified by the \code{`nInstances`} parameter. A list containing the particular parameters for each instance is given under the keyword \code{`instances`} in \crefrange{code:c3l14}{code:c3l21}. This construct is a Python list comprehension, an inline definition of list entries defined by the for loop in \cref{code:c3l21}. The nested specifications of parameters of the Heun and the CellML methods, not shown in \cref{fig:example_multidomain_settings}, depend on the iteration index \code{compartment_no} of this loop. 
Each of these instance gets computed by a defined set of processes, specified under the parameter \\\code{`ranks`} in \cref{code:c3l16}. In this multidomain example, all processes take part in the computation of all multidomain compartments and, thus, all instances of the \code{MultipleInstances} class are computed by all processes. The expression in \cref{code:c3l16} expands to a list \code{[0,1,2,...]} indicating all available processes.

% solvers
Specifications of the parameters for a \code{FiniteElementMethod} class, similar to the Laplace example considered in \cref{sec:exemplary_usage_1}, also appear in the example of this section, once for each of the three occurrences of this class. The excerpt of the settings file in \cref{fig:example_multidomain_settings} shows one of these specifications, the \code{PotentialFlow} finite element method, in \crefrange{code:c3l27}{code:c3l31}. 
This \code{FiniteElementMethod} class shares its mesh and its linear solver with other classes. The mesh and the linear solver both have specific parameters that were listed as blocks in the settings file of the Laplace problem in \cref{fig:laplace_example_settings}. To avoid duplication of this information and to share linear solvers and meshes, these parameters are not repeated for every class, by which they are required. Instead, parameters for linear solvers and meshes can be specified globally at the beginning of the settings file and referenced at the locations in inner classes, when they are used. In the example settings in \cref{fig:example_multidomain_settings}, this is indicated for the linear solver. Its parameters are defined under the global \code{`Solvers`} keyword  in the beginning and the name \code{`potentialFlowSolver`} in \cref{code:c3l4}. These settings are referenced in the finite element method in \cref{code:c3l29} using the \code{`solverName`} keyword. Internally, only one solver object with the related data structures of PETSc is created and reused where ever the solver is referenced by its solver name.

The meshes use an analog approach, in which all meshes can be defined under a global \code{`Meshes`} keyword (not shown in \cref{fig:example_multidomain_settings}) and referenced in the solver objects by their \code{`meshName`}. This is helpful especially for meshes with many node positions that can be specified once and reused throughout all solvers.

% output writers
The output of the results is written to files by output writers that are defined as shown in the previous Laplace example in \cref{fig:laplace_example_settings}. Almost all solver classes allow configuring associated output writers. In \cref{fig:example_multidomain_settings}, such output writer settings are listed in \cref{code:c3l33} within the multidomain solver. Additional output writers can be defined in the Heun scheme of the subcellular model and in the \code{MuscleContractionSolver} for the solid mechanics models. Each output writer outputs files with the solution variables of the respective solver. Different time intervals can be set for the writers to allow for different output frequencies of large data such as all subcellular model states and of smaller data such as the solid mechanics outputs.

% command
The following exemplary command can be used to run the program for this example:
\begin{lstlisting}[columns=fullflexible,breaklines=true,postbreak=\mbox{\textcolor{gray}{$\hookrightarrow$}\space}]
  mpirun -n 2 ./multidomain_contraction ../settings_multidomain_contraction.py very_coarse.py --end_time=10
\end{lstlisting}
Similar to the Laplace example in \cref{sec:exemplary_usage_1}, the program \code{./multidomain_contraction} is called with the settings file \code{../settings_multidomain_contraction.py} as its first argument. In addition, a second script, \code{very_coarse.py}, is given as second argument. This script gets loaded from within the Python settings script and defines a number of high-level parameters in a separate \code{variables} namespace. These parameters are then used in the settings file. For example, \cref{code:c3l13} of \cref{fig:example_multidomain_settings} uses the variable \code{n_compartments}, which is defined in the so-called \emph{variables} file \code{very_coarse.py}. The file name refers to the coarse discretization that is chosen in the particular scenario.

The rationale of this second script is to summarize important parameter values in a smaller and easier readable file. Whereas the full settings file corresponding to \cref{fig:example_multidomain_settings} contains approximately 500 lines and a complex nested structure, the variables script only contains about 200 lines, mainly value assignments to parameters and descriptive comments. 

Several of these variables files exist in the \code{variables} subdirectory of the example. They define different scenarios for the given simulation such as different mesh resolutions or CellML model files. By exchanging the filename in the second argument of the command line, these different scenarios can be easily executed.

The last argument in the command, \say{\code{--end_time=10}}, gets also parsed by the Python script. It allows to set the end of the simulation time span to the specified value via the command line. Other options are available to alter various parameters in this scenario. These command line arguments take precedence over the parameter values that are specified in the Python scripts. This type of command line argument makes it possible to easily conduct parameter studies, e.g., from bash scripts, where the program can be called with different parameter values.
The architecture involving a main settings file with all parameters in the hierarchical solver structure, a set of small variables files with dedicated parameter choices and the possibility to override all parameters from the command line is present in most of the advanced examples in OpenDiHu.

Another thing to note is that the given command begins with \say{\code{mpirun -n 2}}, which instructs MPI to launch the program using two processes. Here, any other number is possible and a corresponding domain decomposition is computed automatically. The parallelism is only bounded by the number of available elements in the meshes.

\subsection{Data Connections in the Example of a Multidomain Model with Solid Mechanics}\label{sec:exemplary_usage_2b}

The control flow of a simulation program with nested solvers such as the coupled electrophysiology and muscle contraction model studied in the previous section is defined by the tree of solvers in the C++ source file. This structure is reflected in the Python settings file.
The corresponding data flow, which connects the solvers, is another important property that has to be specified.
To help with this step, the program generates a diagram of its data connections whenever the program stops (either after completing or when interrupted by the shell).

\begin{figure}
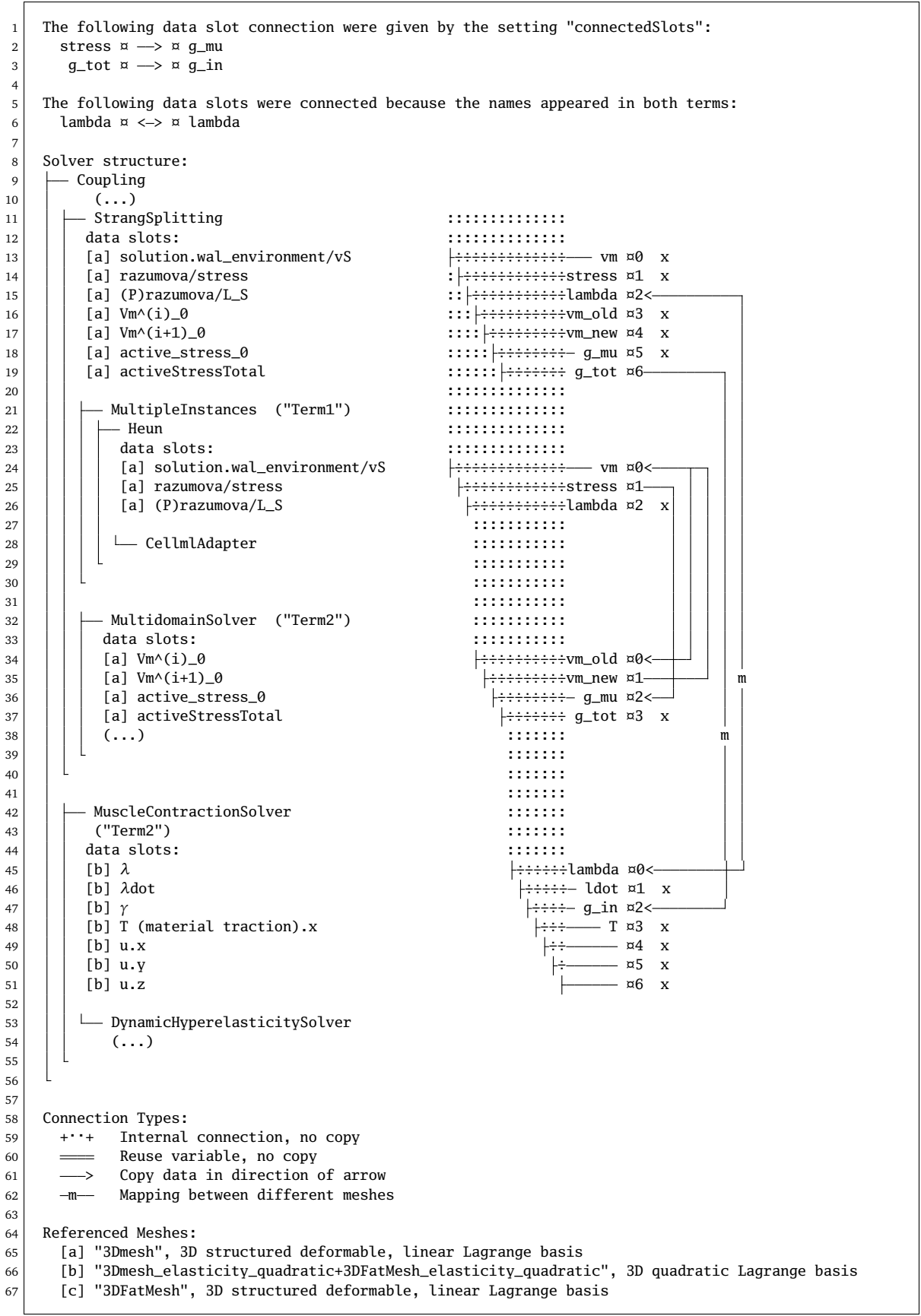

\centering
\begin{framed}
\begin{Verbatim}[fontsize=\scriptsize\ttfamily,numbers=left]
The following data slot connection were given by the setting "connectedSlots":
  stress ¤ ──> ¤ g_mu  
   g_tot ¤ ──> ¤ g_in  

The following data slots were connected because the names appeared in both terms:
  lambda ¤ <─> ¤ lambda

Solver structure: 
├── Coupling  
│     (...)
│ ├── StrangSplitting                          ::::::::::::::           
│ │  data slots:                               ::::::::::::::           
│ │  [a] solution.wal_environment/vS           ├÷÷÷÷÷÷÷÷÷÷÷÷÷─── vm ¤0  x
│ │  [a] razumova/stress                       :├÷÷÷÷÷÷÷÷÷÷÷÷stress ¤1  x
│ │  [a] (P)razumova/L_S                       ::├÷÷÷÷÷÷÷÷÷÷÷lambda ¤2<──────────┐
│ │  [a] Vm^(i)_0                              :::├÷÷÷÷÷÷÷÷÷÷vm_old ¤3  x        │
│ │  [a] Vm^(i+1)_0                            ::::├÷÷÷÷÷÷÷÷÷vm_new ¤4  x        │
│ │  [a] active_stress_0                       :::::├÷÷÷÷÷÷÷÷─ g_mu ¤5  x        │
│ │  [a] activeStressTotal                     ::::::├÷÷÷÷÷÷÷ g_tot ¤6─────────┐ │
│ │                                            ::::::::::::::                  │ │
│ │ ├── MultipleInstances  ("Term1")           ::::::::::::::                  │ │
│ │ │ ├── Heun                                 ::::::::::::::                  │ │
│ │ │ │  data slots:                           ::::::::::::::                  │ │
│ │ │ │  [a] solution.wal_environment/vS       ├÷÷÷÷÷÷÷÷÷÷÷÷÷─── vm ¤0<────┬─┐ │ │
│ │ │ │  [a] razumova/stress                    ├÷÷÷÷÷÷÷÷÷÷÷÷stress ¤1───┐ │ │ │ │
│ │ │ │  [a] (P)razumova/L_S                     ├÷÷÷÷÷÷÷÷÷÷÷lambda ¤2  x│ │ │ │ │
│ │ │ │                                           :::::::::::            │ │ │ │ │
│ │ │ │ └── CellmlAdapter                         :::::::::::            │ │ │ │ │
│ │ │ └                                           :::::::::::            │ │ │ │ │
│ │ └                                             :::::::::::            │ │ │ │ │
│ │                                               :::::::::::            │ │ │ │ │
│ │ ├── MultidomainSolver  ("Term2")              :::::::::::            │ │ │ │ │
│ │ │  data slots:                                :::::::::::            │ │ │ │ │
│ │ │  [a] Vm^(i)_0                               ├÷÷÷÷÷÷÷÷÷÷vm_old ¤0<──┼─┘ │ │ │
│ │ │  [a] Vm^(i+1)_0                              ├÷÷÷÷÷÷÷÷÷vm_new ¤1───┼───┘ │ m
│ │ │  [a] active_stress_0                          ├÷÷÷÷÷÷÷÷─ g_mu ¤2<──┘     │ │
│ │ │  [a] activeStressTotal                         ├÷÷÷÷÷÷÷ g_tot ¤3  x      │ │
│ │ │  (...)                                          :::::::                  m │
│ │ └                                                 :::::::                  │ │
│ └                                                   :::::::                  │ │
│                                                     :::::::                  │ │
│ ├── MuscleContractionSolver                         :::::::                  │ │ 
│ │   ("Term2")                                       :::::::                  │ │
│ │  data slots:                                      :::::::                  │ │
│ │  [b] λ                                            ├÷÷÷÷÷÷lambda ¤0<────────┼─┘
│ │  [b] λdot                                          ├÷÷÷÷÷─ ldot ¤1  x      │
│ │  [b] γ                                              ├÷÷÷÷─ g_in ¤2<────────┘
│ │  [b] T (material traction).x                         ├÷÷÷──── T ¤3  x
│ │  [b] u.x                                              ├÷÷────── ¤4  x
│ │  [b] u.y                                               ├÷────── ¤5  x
│ │  [b] u.z                                                ├────── ¤6  x
│ │                                                                     
│ │ └── DynamicHyperelasticitySolver    
│ │     (...)                                
│ └                                                                     
└                                                                       
                                                                        
Connection Types:
  +··+   Internal connection, no copy
  ════   Reuse variable, no copy
  ───>   Copy data in direction of arrow
  ─m──   Mapping between different meshes

Referenced Meshes:
  [a] "3Dmesh", 3D structured deformable, linear Lagrange basis
  [b] "3Dmesh_elasticity_quadratic+3DFatMesh_elasticity_quadratic", 3D quadratic Lagrange basis
  [c] "3DFatMesh", 3D structured deformable, linear Lagrange basis
\end{Verbatim}
\vspace{-5mm}
\end{framed}
\caption{Solver structure diagram that shows the data connections of the solvers.}%
\label{fig:example_multidomain_solver_structure}%
\end{figure}

\Cref{fig:example_multidomain_solver_structure} shows such a solver structure diagram for the current example. It is given as a text file and visualizes data connections using only Unicode characters. This is advantageous for computers that only provide remote access, such as compute clusters. 
On the left side, the tree of nested solvers is given. On the right side, lines indicate the corresponding data connections.

% multidomain_contraction solver_structure

Each solver has a fixed number of \emph{data connector slots}. A data connector slot is a scalar field variable or one component of a vector-valued field variable on a certain mesh. In coupling or operator splitting schemes, values can be transferred from a data connector slot of the first solver to a data connector slot of the second solver. This data transfer either reuses the internal data structure, if possible or it involves a copy operation. Either way, after the transfer, the second solver knows the corresponding values of the first solver and can use them in subsequent computations.

Each field variable has a given, fixed name defined by the solver. The corresponding data connector slot can have a custom name with a maximum length of six character, which is assigned from the Python settings.
The diagram shows the field variable names on the left under the \say{data slots} lists of the solvers. The corresponding data connector slots are marked by the \say{¤} symbol and a slot number on the right. The custom name of the slot is written before the \say{¤} symbol.

For example, the \code{MuscleContractionSolver} listed in line 42 has field variables for the fiber stretch \code{$\lambda$}, contraction velocity \code{$\lambda$dot}, muscle activation \code{$\gamma$}, traction in material description \code{T} and displacements in $x$, $y$ and $z$-direction, \code{u.x}, \code{u.y} and \code{u.z}. As can be seen in \cref{fig:example_multidomain_solver_structure}, the first four data  slots correspondingly have the names \code{lambda}, \code{ldot}, \code{g_in} and \code{T}. The fiber stretch \code{$\lambda$} is a quantity that is computed by the solver, and the activation parameter \code{$\gamma$} is a field variable that is an input to the solver and used for the computation of the active stress. However, data connector slots make no distinction between input and output slots, they simply expose the corresponding field variable to be connected to other slots.

The Heun solver in line 22 that solves the subcellular model has three slots: the slot \code{vm} of the transmembrane voltage, the slot named \code{stress} of the active stress parameter $\gamma$ and the slot \code{lambda}, which is the input of the relative half-sarcomere length of the subcellular model.

The multidomain solver in line 32 has four slots: the slot \code{vm_old} exposes the field variable for $V_m^{(i)}$, the transmembrane voltage at the previous timestep. After solving the linear system of equations, the field variable $V_m^{(i+1)}$, which is connected to the slot \code{vm_new}, holds the transmembrane voltage for the next timestep. Another slot used for data input is \code{g_mu}, which retrieves the muscle activation parameter $\gamma$ from each compartment. The multidomain solver computes the resulting activation parameter at slot \code{g_tot} by the weighted sum over the $\gamma$ values at the intracellular compartments. 

In case of the multidomain solver, separate field variables exist for every compartment at the same data connector slot. The solver structure diagram in \cref{fig:example_multidomain_solver_structure} shows the field variables for slots 0 to 2 ending in \say{\code{_0}} for the compartment $k=0$. Similar field variables exist for $k=1,\dots,N_\text{MU}$, however, those are not shown in the diagram. Similarly, the Heun scheme in line 22 is nested in the \code{MultipleInstances} scheme in line 21. Here, the field variables that connect to the slots \code{vm}, \code{stress} and \code{lambda} also have different instances for every compartment. To resolve the ambiguity of multiple field variables of the same kind being associated with a single slot, every exposed field variable for data transfer has to be identified by its slot and potentially an array index within this slot.

In case of nested solvers, the parent solver class always exposes data connector slots of its children. For example, the \code{StrangSplitting} class in line 11 has no own slots, but exposes the slots of its two children. The slots with indices 0 to 2 are the same as the slots of its \code{`Term1`} in line 21, the slots 3 to 6 are identical to the \code{`Term2`}, the \code{MultidomainSolver} in line 32. These connections are indicated in the diagram by the dotted vertical connection lines.
Note that the outer-most solver always contains the slots of all nested solvers. In the example in \cref{fig:example_multidomain_solver_structure}, this is the outer \code{Coupling} scheme. The slot listing has been omitted in the visualization.

The actual connections between the data slots of different solvers are indicated by the arrows on the right-hand side of the slots. Unconnected slots are marked by an \say{x}.
The data transfer behavior is as follows. Each coupling and operator splitting scheme has two nested solvers. The coupling scheme executes the first solver, transfers the data over the connected data slots from the first to the second solver, executes the second solver, and then transfers the data according to the connected slots from the second to the first solver. For the Strang splitting scheme, this data transfer happens twice per timestep, as defined by the splitting algorithm (cf. \cref{fig:strang_splitting}).

The interaction between the subcellular model and the multidomain model is given by the arrows between the Heun scheme in line 22 and the \code{MultidomainSolver} in line 32. After the solution of the subcellular model, the transmembrane voltage is transferred from slot 0 (\code{vm}) of the Heun scheme to slot 0 (\code{vm_old}) of the multidomain solver. At the same time, the stress is transferred from slot 1 (\code{stress}) to slot 2 (\code{g_mu}). After the linear system has been solved, the values for the new timestep are transferred back from slot 1 (\code{vm_new}) to slot 0 (\code{vm}).

At the outer \code{Coupling} scheme, after the electrophysiology model consisting of the subcellular and multidomain model parts have been solved, the active total stress is transferred from slot 6 (\code{g_tot}) in line 19 to the slot 2 (\code{g_in}) of the \code{MuscleContractionSolver}. Note that the starting slot \code{g_tot} is shared between \code{StrangSplitting} and \code{Multidomain}\code{Solver}, shown by the dotted vertical lines. Then, the solid mechanics model uses the activation value, computes new displacements and updates the slots \code{lambda} and \code{ldot}. The value in \code{lambda} is transferred back to slot 2 of the \code{StrangSplitting}, where it is shared with the subcellular model. The value of the slot \code{ldot}, which is the contraction velocity, is not used here, however, some subcellular CellML models make use of this value. In such a case, the corresponding connection line can be added.

In addition to the \code{lambda} slot, the \code{MuscleContractionSolver} updates the muscle geometry with the new deformed configuration. This occurs outside of data connector slots using defined relationships or mappings between the elasticity and electrophysiology meshes.
Reasons for this exception are, first, that a mesh is not owned by a solver class in the same way as other data, e.g., as a solution vector. And second, the geometry information is different from normal field variables. Changing the geometry of a mesh, e.g, invalidates finite element system matrices.

% meshes
Each field variable is associated with a mesh, which is referenced by \code{[a]} and \code{[b]} in front of the field variable names. The referenced meshes are listed at the bottom of the diagram in line 64. The reference \code{[a]} corresponds to the mesh in the muscle domain used for the subcellular and multidomain models. The reference \code{[b]} is the composite mesh of both muscle and fat domain used for the solid mechanics problem. 

If data connector slots of different meshes are connected, the values get mapped between the slots. This is indicated by an \say{m} on the connection line. In the presented example, the activation value $\gamma$ gets mapped from the multidomain mesh \code{[a]} to the elasticity mesh \code{[b]} and the fiber stretch value $\lambda$ gets mapped in the opposite direction.

The different connection types are also listed in the legend in line 58. The dotted connection lines of shared slots between nested solvers refer to internal connections where the slots are reused and no data copy operation is necessary. The solid arrows indicate a copy operation. The legend shows also double connection lines, which indicate that the field variable of two slots can be reused and no copy is required. This type of connection is not present in the current example, but occurs for example in most of the fiber based electrophysiology models. The last connection type is the mapping, indicated by a line with an \say{m} character.

Which connection type to use is determined by OpenDiHu. In case of matching meshes, the double line connection that reuses the field variable is preferred. However, it is not always possible because changes in a reused field variable also influence the field variable at its original point of use, which may not be desired. In the current example, the reason why the \say{copy} connections are used between the subcellular and multidomain solvers lies in the number of compartments. The subcellular models holds the data of all compartments in an array-of-vectorized-struct memory layout, such that the order of the compartments' variables in memory is different than the required order for the slot. Thus, the data have to be copied during transfer between connected slots.

The specification of which slots to connect with each other is given in the settings file. Three possibilities how to define slot connections exist: First, the slot numbers of connected slots can be given in the settings of coupling and operator splitting schemes. Second, the names of connected slots can be specified under the global keyword \code{`connectedSlots`}. In the given example, this is the case for the slots listed in \cref{fig:example_multidomain_solver_structure} in lines 1 to 3. Third, slots with the same name are connected automatically. In the considered example, this is the case for the \code{lambda} slot, which is named identically in the \code{CellmlAdapter} (line 26) and the \code{MuscleContractionSolver} (line 45). Slots connected by the third possibility are also listed at the top of the diagram, here in lines 5 and 6.

\subsection{Exemplary Usage: Neuromuscular System}\label{sec:exemplary_usage_3}

\begin{figure}
  \centering%
  \includegraphics[width=0.9\textwidth]{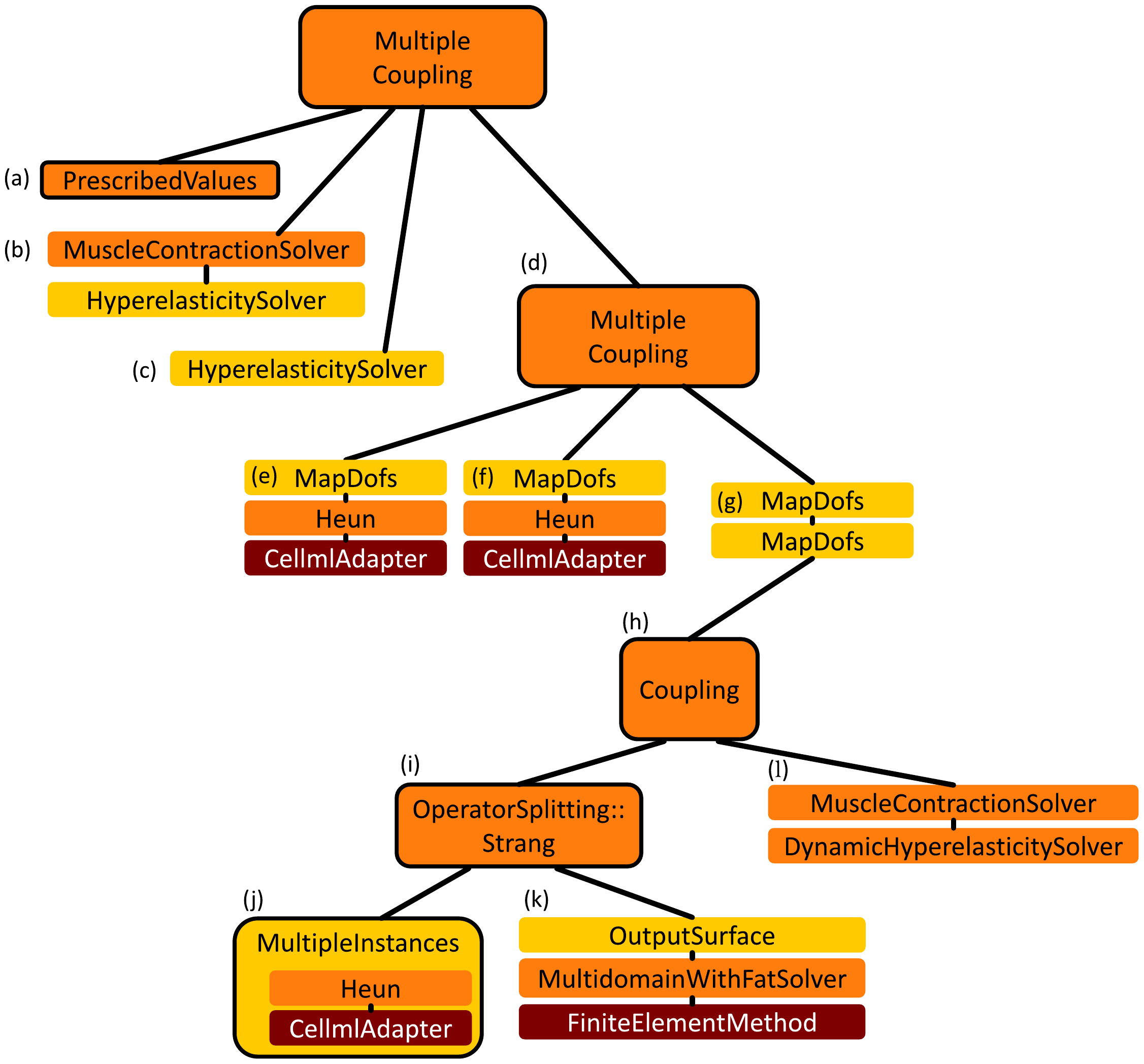}%
  \caption{Solver tree for a simulation of the neuromuscular system. Classes of type \code{DiscretizableInTime} are shown as red boxes, \code{TimeSteppingScheme}s are given by orange boxes.}%
  \label{fig:solver_tree_multidomain_spindles}%
\end{figure}

In the example in the last sections \cref{sec:exemplary_usage_2,sec:exemplary_usage_2b}, the nested solver structure was a binary tree. However, also scenarios with a more general tree structure exist. The solver tree for a simulation of the neuromuscular system including sensory feedback is shown in \cref{fig:solver_tree_multidomain_spindles}. The tree corresponds to the example in the directory \code{examples/electrophysiology/neuromuscular/spindles_multidomain}.

% solver tree

The following solver classes are involved in this example. For executing multiple solvers in series, the top-level \code{MultipleCoupling} class exists, which calls its nested solvers one by one in every timestep.
The \code{PrescribedValues} class (a) can be configured to set any of its field variables to prescribed values. The values can be set by callback functions in the Python settings that get frequently called by the solver to update the values over time.

A \code{MuscleContractionSolver} combines either a static or a dynamic hyperelasticity model with the active stress term used in the muscle contraction model. The class in (b) uses the static \code{HyperelasticitySolver}. The solvers in (a) and (b) compute the static contraction of the muscle under a prescribed constant activation level. Then, another \code{HyperelasticitySolver} (c) stretches the muscle tissue again by a prescribed external force to yield a prestretched muscle. The actual transient simulation is then performed under the subtree at (d). 
Again, a \code{MultipleCoupling} class is used to run all nested solvers in every timestep.

In (e) and (f), two CellML models are solved using a Heun scheme, the first one for the muscle spindles and the second one for the motor neurons. A filter step is applied on the resulting signals and the values are copied to the destination field variables using the \code{MapDofs} class in (e), (f) and (g). A \code{MapDofs} class is able to copy degrees of freedom between two field variables, apply custom Python functions on the values and communicate values between processes in parallel execution.

The subtree under (h) is identical to the example presented in \cref{sec:exemplary_usage_2}. It solves the electrophysiology model using the multidomain equations in (j) and a subcellular model in (k), coupled to the solid mechanics model in (l).

The tree in \cref{fig:solver_tree_multidomain_spindles} consists of solver classes of different types. The orange boxes indicate timestepping schemes. Internally, these classes derive from a \code{TimeSteppingScheme} interface class. They have a common set of parameters such as the timestep width and end time.
The boxes with dark red background color are classes of type \code{DiscretizableInTime}. They represent a term or equation that can be nested in a timestepping scheme. There only exist two different classes of this type: The \code{CellmlAdapter}, which contains a system of DAEs given by a CellML model and the \code{FiniteElementMethod}, which discretizes the generalized Laplace operator $\div(\bfsigma \grad u)$.

Besides the classes of the presented example shown in \cref{fig:solver_tree_multidomain_spindles}, further solver classes are available in OpenDiHu. A comprehensive list of all available solver classes is given in the following section.

% summary of existing solvers additional solvers exist
\subsection{Summary of Existing Solver Classes}\label{sec:summary_of_existing_solver_classes}

All the timestepping schemes introduced in \cref{eq:ode_solver_schemes} are available to solve ODEs given by \code{Discretizable}\code{InTime} objects:
The explicit schemes are the explicit Euler and Heun's method. The available implicit schemes are the implicit Euler and Crank-Nicolson method.
Implemented operator splitting schemes are the Godunov and Strang splittings. The implementation of the \code{Coupling} class is identical to Godunov splitting.
As mentioned in \cref{sec:exemplary_usage_3}, \code{DiscretizableInTime} objects are either given by the \code{CellmlAdapter} or the \code{FiniteElementMethod.}

Some classes are special solvers for dedicated models: A \code{StaticBidomainSolver} is used to solve the first bidomain equation \cref{eq:bidomain1}. The \code{Multidomain}\code{Solver} and \code{Multidomain}\code{WithFatSolver} classes solve the multidomain models \cref{eq:multidomain1,eq:multidomain2} without and with body fat domain. A class \code{FastMonodomainSolver} exists that improves the parallel performance of the fiber based electrophysiology solver using the monodomain equation \cref{eq:monodomain}.

Solid mechanics models can be computed by a series of specialized solvers. The \code{QuasiStaticLinearElasticitySolver} class uses a \code{FiniteElementMethod} object to compute 3D linear elasticity using Hooke's Law with an additional active stress term, as derived in \cref{sec:linearized_mechanics_model}. The \code{HyperelasticitySolver} class solves the static hyperelasticity formulation for any material model, as presented in \cref{sec:static_hyperelastic_fe_model}. The \code{DynamicHyperelasticity}\code{Solver} class inherits from the \code{HyperelasticitySolver} class and adds functionality to solve the dynamic hyperelasticity formulation shown in \cref{sec:solver_dynamic_hyperelasticity_fe_model}. 
Both the static and the dynamic hyperelastic solvers do not incorporate the active stress term that is present in the muscle contraction model. This is handled by another class, the \code{MuscleContractionSolver}. It uses either a \code{HyperelasticitySolver} or a \code{DynamicHyperelasticitySolver} object and adds the functionality accordingly.

Instead of solving a model numerically, also precalculated analytic solutions can be used. This can be done using the \code{PrescribedValues} class, which uses a Python function to set the solution values. Further auxiliary classes exist that are no numerical solver: The \code{MapDofs} class gives flexibility to transfer certain degrees of freedom between field variables. The \code{Dummy} class can be used as a placeholder. The \code{OutputSurface} class extracts a 2D mesh at the surface of a 3D mesh and writes it to an output file using the normal output writers. This can be used to reduce the amount of data output for finely resolved EMG simulations, where only the values at the surface are of interest.

Moreover, adapters to external software tools are implemented. The class \code{Nonlinear}\code{ElasticitySolverFebio} allows to use the solver \emph{FEBio} \cite{Maas2012,maas2017febio} for solving a continuum mechanics model and couple it to an electrophysiology model in OpenDiHu.
Two adapters to the numerical coupling library preCICE \cite{precice}, \code{PreciceAdapter} and \code{PreciceAdapterVolumeCoupling} exist for surface and volume coupling. They can be configured to implicitly or explicitly couple any field variables to external solvers or to couple two separate instances of OpenDiHu.
% summary structure, how to define the settings, help with gui program
For more details on the solver classes and their configuration, we refer to the online documentation \cite{opendihuWeb}.

\subsection{Graphical Helper Program}
For the creation of new examples from scratch, a helper program with a graphical user interface exists. The program was created by Matthias Tompert in his Bachelor thesis and is included in the OpenDiHu repository under \code{scripts/gui/gui.py}.

Graphical widgets allow selecting and nesting compatible solver classes for OpenDiHu. The corresponding Python settings are automatically shown with their default values and can be adjusted in the graphical user interface. For every option, explanatory comments are displayed and buttons allows to open the corresponding page of the online documentation in a web browser.

The program also features a horizontally split code editor, which shows both the C++ code and the corresponding Python code for the settings, either for a single node in the solver tree or as a global view of the whole example. 
After the user completes the adjustments of the solver structure and the settings, the C++ source file and the Python settings can be exported and used with OpenDiHu.

The program is also capable of parsing existing C++ and Python files and, thus, loading an existing example into the graphical representation to be extended by the user. The program is able to parse a large portion of the solvers and options that are available in OpenDiHu. However, some more advanced examples, e.g., where the Python settings contain complex code constructs are not fully supported.

\begin{figure}%
  \centering%
  \begin{subfigure}[t]{0.71\textwidth}%
    \centering%
    \includegraphics[height=8cm]{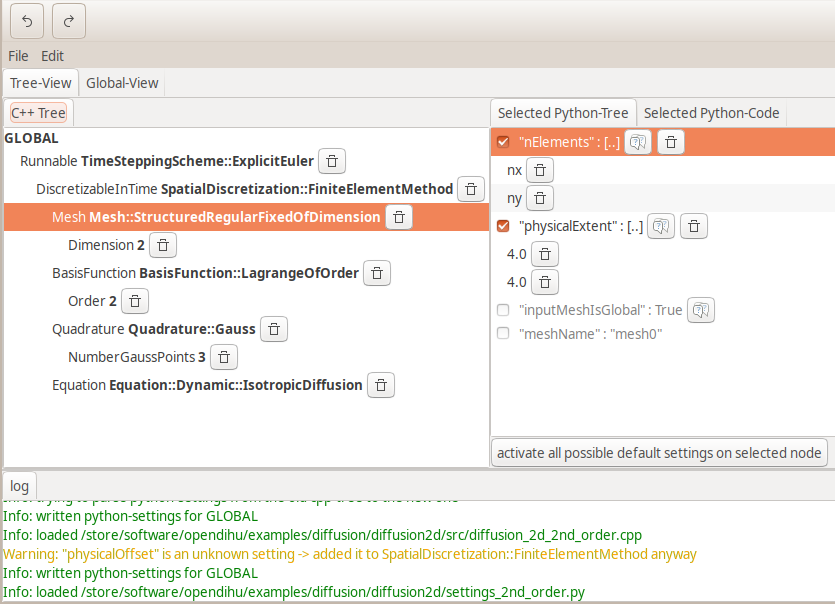}
    \caption{User interface with the tree of nested solvers on the left and the Python settings for the selected mesh on the right.}%
    \label{fig:gui}%
  \end{subfigure}
  \,
  \begin{subfigure}[t]{0.27\textwidth}%
    \centering%
    \includegraphics[height=8cm]{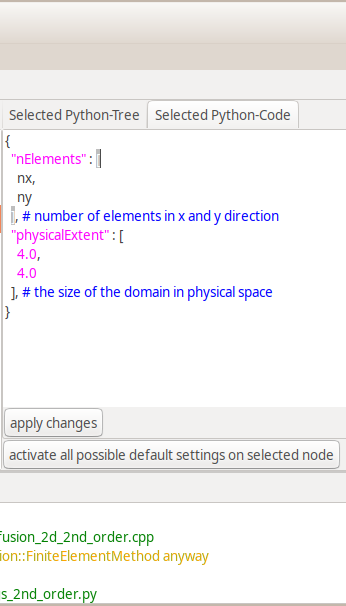}
    \caption{Alternative view of the right pane that shows the Python settings editor.}%
    \label{fig:gui3}%
  \end{subfigure}
  \caption{Graphical helper program to create and adjust the C++ and Python codes of OpenDiHu examples.}%
  \label{fig:gui_a}%
\end{figure}

\Cref{fig:gui_a} shows the user interface after the 2D diffusion example of OpenDiHu has been loaded. The left pane in \cref{fig:gui} represents the solver tree as it appears in the C++ file. The right pane in \cref{fig:gui} displays the settings for the selected item in the left pane, which, in this case, is the mesh. Additional options that are not yet present in the Python settings are grayed out and can be added by clicking the checkboxes.
\Cref{fig:gui3} shows an alternative view in the right pane, which displays the corresponding Python code. The user can also directly edit the settings there.

\section{Usage of CellML Models}\label{sec:usage_cellml}

The CellML description language can be used to describe mathematical models of a wide range of physiological processes. Arbitrary systems of differential-algebraic equations (DAE) can be represented.
We use it for incorporating and exchanging subcellular models, which describe the electrophysiology on a muscle fiber, and for models of motor neurons or sensory organs.
The CellML infrastructure is popular in the bioengineering community. The CellML website of the Physiome project hosts over 600 curated CellML models from different areas. Each model can be downloaded in CellML format or as source code containing the expressions of the equations in various programming languages such as MATLAB, Python and C.

\subsection{Integration of CellML in OpenDiHu and Comparison with Other Framework}

Mathematically, a CellML model describes the functions $G$ and $H$ of the following DAE:
\begin{align}\label{eq:cellml_generic_dae}
  \p{\bfy(t)}{t} &= G\big(t,\bfy(t),\bfh(t),\hat{\bfc},\hat{\bfp}(t)\big), & \bfh(t) &= H\big(\bfy(t),\hat{\bfc},\hat{\bfp}(t)\big).
\end{align}
Here, $\bfy$ is the state vector and $\bfh$ is a vector with additional values that are derived from the state vector. The vectors of constants $\hat{\bfc}$ and parameters $\hat{\bfp}$ are prescribed and fixed over time for $\hat{\bfc}$ or varying over time for $\hat{\bfp}$.

Various open source tools exist to create or manipulate CellML models and to solve them and visualize the results \cite{pmid18579471}. A comprehensive list is given on the CellML website \cite{cellmlWebsite} and some of them, which are relevant to our work, are outlined in the following.

There exist two application programming interfaces (APIs), the \emph{CellML API} and the newer \emph{libCellML}, which allow direct access to the structures of the CellML model from, e.g., C++ code \cite{pmid20377909}. 

\begin{figure}%
  \centering%
  \begin{subfigure}[t]{0.45\textwidth}%
    \centering%
    \includegraphics[width=\textwidth]{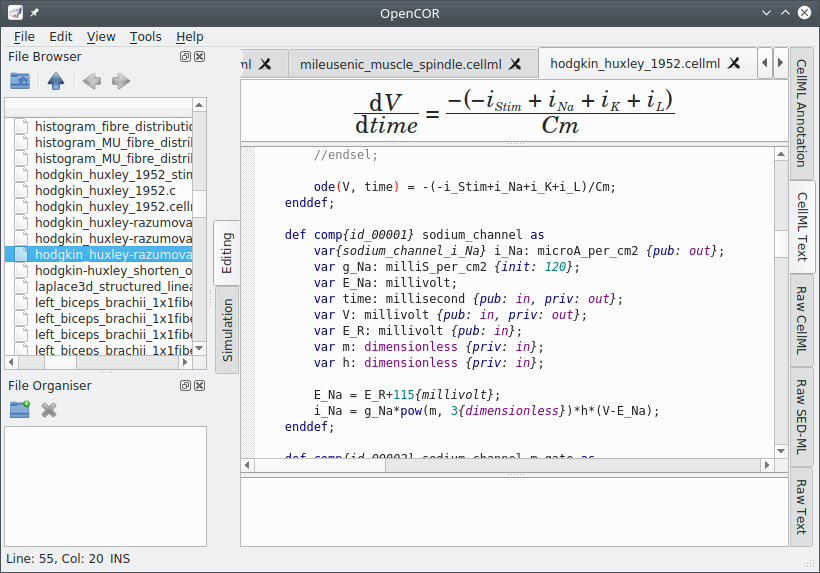}
    \caption{CellML editor with the ODE for the membrane voltage \say{V} in the Hodgkin-Huxley cellular model, which corresponds to \cref{eq:subcellular_model_helper4} inserted into \cref{eq:subcellular_model_helper3}.}%
    \label{fig:opencor1}%
  \end{subfigure}
  \quad
  \begin{subfigure}[t]{0.45\textwidth}%
    \centering%
    \includegraphics[width=\textwidth]{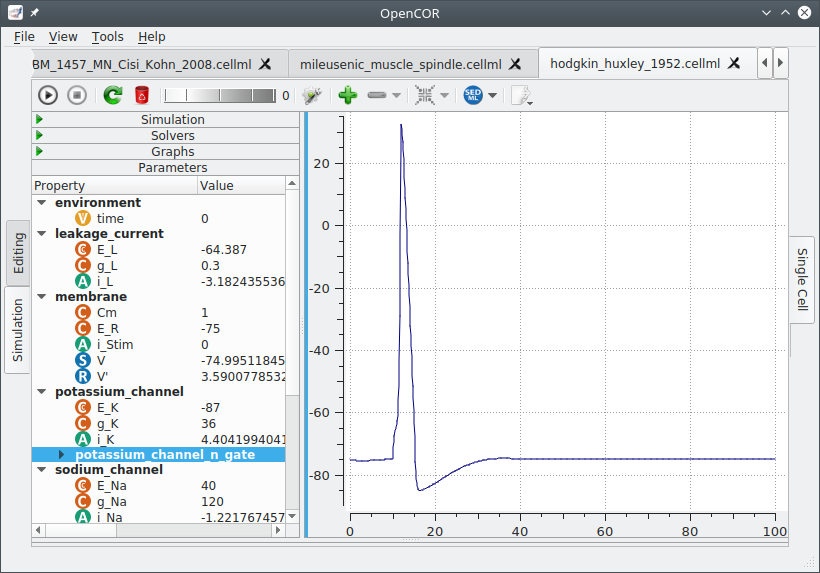}
    \caption{Visualization of a simulated action potential $V_m$ over time.}%
    \label{fig:opencor2}%
  \end{subfigure}
  \caption{The CellML modeling environment OpenCOR.}%
  \label{fig:opencor}%
\end{figure}%

\emph{OpenCOR} \cite{OpenCOR2015} provides a modeling environment in a graphical user interface, where models can be edited. 
\Cref{fig:opencor1} shows the interface with the editor on the right. Mathematical equations are described in a declarative language and can be rendered to mathematical notation, as seen in the upper part in \cref{fig:opencor1}. OpenCOR automatically transfers the equations to the XML-based MathML syntax and integrates them in the XML-based CellML description.
OpenCOR can also be used to solve the system of DAEs using implicit solvers such as backward differentiation formulas. The solver parameters can be adjusted and the solver can be started from the graphical user interface. \Cref{fig:opencor2} shows the interface that lists all variables with their current values on the left and a visualization of the result, in this case an action potential, on the right.

OpenCOR also provides command line functionality to convert CellML files into C code. This generated C code can evaluate all model equations but not solve the DAE system. Because OpenCOR is robust and well established in the bioengineering modeling community, we decide to use it in OpenDiHu. The installation procedure of OpenDiHu downloads and installs OpenCOR automatically.

During execution of a simulation, our framework parses the C code of CellML models, compiles a shared library and executes the functions, all at runtime. Thus, the CellML model can be directly given as a C file. Otherwise, if the model file is in XML format, it is assumed to be a CellML description and automatically converted to the required C code using the OpenCOR command line interface.

If a CellML model is manually simulated in the OpenCOR graphical user interface with time-varying input signals, these signals have to be hard-coded in the model, e.g., as a piecewise defined function. This is acceptable for getting insight into the models, but counteracts the idea of modular models that can be shared and recombined. 
As a remedy, we design our framework in a way that simulations of CellML models with configurable time-varying input signals are possible without the need to change the CellML description.

CellML models are limited to single-cell systems of DAEs and are not designed for PDEs that, e.g., involve multiple instances of a DAE system on a given geometry. Thus, the monodomain equation cannot be solved with OpenCOR and a multi-scale software framework is needed for this task. Two such frameworks with CellML support, which were described in the introduction in \cref{sec:intro_related_software}, are Chaste and OpenCMISS Iron. In the following, we relate and compare the approaches of CellML integration in OpenDiHu and these existing frameworks.

\begin{table}
  \centering%
  \begin{tabular}{|c|l|l|l|l|l|}
    \hline
    Symbol        & \multicolumn{3}{l|}{Name}            & Computed          & Initial values\\
    \cline{2-4}
                  & OpenCOR    & OpenCMISS & OpenDiHu   & by model?          & can be set?\\
    \hline
    $\bfy$        & \code{state}     & \code{STATES}    & \code{state}     & by timestepping   & yes \\[2mm]
    $\partial\bfy/\partial t$ & \code{rate}      & \code{RATES}     & \code{rate}      & yes               & no  \\[2mm]
    $\hat{\bfc}$  & \code{constant}  & \code{CONSTANTS} & \code{constant}  & no                &in CellML  \\[2mm]
    $\bfh$        & \code{algebraic}  & \code{WANTED}    & \code{algebraic} & yes               & no  \\[2mm]
    $\hat{\bfp}$  & -  & \code{KNOWN}     & \code{parameter} & no                & yes \\
    \hline
  \end{tabular}
  \caption{The different CellML quantities and their properties and names in various tools.}%
  \label{tab:cellml_names}%
\end{table}

The variables in the generic DAE in \cref{eq:cellml_generic_dae} have different names in the different software packages. \Cref{tab:cellml_names} compares the symbols and their names in OpenCOR, OpenCMISS in OpenDiHu and summarized how their values are determined. 

All three software packages have the concept of \code{state} and \code{rate} vectors, where the states $\bfy$ are the input and the rates $\partial\bfy/\partial t$ are the output of the CellML formulas. Similarly, the constants $\hat{\bfc}$ are always a set of predefined values that are fixed during the computations.

The algebraic formulas lead to the values in $\bfh$, independently of the timestepping scheme. These algebraic values can be considered as the resulting quantities of interest of the model and are typically written to output files or transferred to coupled solvers.

Moreover, OpenCMISS and OpenDiHu define parameters $\hat{\bfp}$, which influence the behavior of the model. Their values can be changed by a coupled solver or prescribed from the settings. 
In OpenDiHu, any constant or algebraic variable in a CellML model can be converted into a parameter. All occurrences of the constant or algebraic variable in the CellML description get replaced by the parameter variable. For former algebraic variables, this replacement step overrides the equations that would have defined the algebraic value. Exemplary use cases are to set the external stimulation current $I_\text{ext}$ in \cref{eq:subcellular_model_helper3} or to set the fiber stretch in a strain-dependent subcellular model.

OpenCMISS uses a similar concept, where some algebraics in the CellML description can be declared as \code{WANTED} to be read by the framework. Some of the constants can be declared as \code{KNOWN} such that OpenCMISS sets their values from other computations within OpenCMISS. (Assigning new values to algebraics as in OpenDiHu is not possible.) Because the terms \code{WANTED} and \code{KNOWN} can be ambiguous if either seen from within the CellML model or from the framework, we decide to use the terms \code{algebraics} and \code{parameters} instead.

The last two columns of \cref{tab:cellml_names} summarize the purpose of the different quantities. The CellML description defines formulas for the states, rates and algebraics. Rates and algebraics are directly calculated by the code that is generated from the CellML model, the vector of states is then computed from the vector of rates by the timestepping scheme. The initial values of the states are either explicitly specified in the OpenDiHu settings, e.g., to allow different values for different instances of a model. Or, if this specification is omitted,  the initial values are set according to the specification in the CellML file. The parameter values always have to be specified in the Python settings. By definition, the constants cannot be set from OpenDiHu, but are given in the CellML model. If the value of a constant should be specified from the settings, the variable should instead be configured to be a parameter.

From a computational point of view, a CellML model computes the following function in terms of the introduced variable names:
\begin{equation}
  \left(
    \begin{array}{cc}
      \texttt{rates} \\ \texttt{algebraics} 
    \end{array}
  \right) = \texttt{cellml}\left(\texttt{states}, \texttt{constants}\right).
  \label{eq:cellml_generic}
\end{equation}

In the fiber based electrophysiology model, CellML is needed to formulate the reaction term in the monodomain equation \cref{eq:monodomain}, which is repeated here:
\begin{align}\label{eq:monodomain_2}
  \p{V_m}{t}  = \dfrac{\sigma_\text{eff}}{A_m\,C_m} \p{V_m}{x}{2} - \dfrac{1}{C_m} I_\text{ion}(V_m,\bfy).
\end{align}
The \code{states} vector in \cref{eq:cellml_generic} includes both $V_m$ and $\bfy$ in \cref{eq:monodomain_2}. In consequence, the computed \code{rates} contain $\partial V_m/\partial t$ and $\partial\bfy/\partial t$. The right-hand side of \cref{eq:cellml_generic}, i.e., the \code{cellml} function calculates the term $(-1/C_m \cdot I_\text{ion})$, which is the reaction part of the monodomain equation in \cref{eq:monodomain_2}. Thus, the CellML computation can be directly used in the operator splitting approach in \cref{sec:discretization_monodomain}.

For the solution of CellML models, OpenCMISS implements the explicit forward Euler scheme or allows to use the backward differentiation formula (BDF) schemes with adaptive order of convergence of SUNDIALS. Recently, an implementation of the second order explicit Heun scheme was added by Aaron Krämer. Accuracy and runtimes were investigated for Euler, Heun and BDF solvers for the subcellular model within the monodomain equation. Because of the operator splitting scheme, only very small timespans have to be solved by those solvers, which does not redeem the overhead of advanced schemes such as the BDF solver, ultimately yielding the best performance for the Heun solver. Based on these investigations, we choose to implement the forward Euler and Heun schemes for the solution of CellML models in OpenDiHu.

% comparison to Chaste
The following differences exist between the approaches to support CellML models in Chaste, OpenCMISS Iron and OpenDiHu:
Chaste tries to automatically determine the CellML variable names of standard quantities such as the membrane voltage and the stimulation current. This requires potentially less user intervention when CellML models are exchanged.
In OpenDiHu, the step of identifying the CellML variables to be connected to the coupled solvers is done manually to give the user complete control over the setup. It can be achieved in a clear way with the Python settings script. Another difference in OpenDiHu is that all computational code is guaranteed to invoke vector instructions, i.e., following the single-instruction multiple data (SIMD) paradigm. Chaste only relies on the optimization behavior of the Intel compiler, which is not guaranteed to be optimal, e.g., for non-Intel hardware.

% comparison to Iron
The computational core Iron from the OpenCMISS package employs the CellML API and also requires manual connections of CellML variables to the solver code. These variable mappings have to be hard-coded in the main Fortran program (if the Python wrappers are not used) and are compiled into the program. Thus, a CellML model is a fixed part of a compiled simulation program. In contrast, OpenDiHu allows to configure the CellML model at runtime.
Another difference in the implementation is that Iron uses a non-optimal memory layout for the state vector, which prohibits vectorization and slows down the solution compared to OpenDiHu.

\subsection{Mapping of CellML Variables to Slots and Parameters}

Preparing the OpenDiHu solver for use with a CellML model consists of the two steps of adjusting the C++ template parameters and setting up the variable mappings in the Python settings. The two C++ template parameters have to be set to the sizes of the state vector $\bfy$ and the algebraics vector $\bfh$. The code snipped in \cref{fig:example_shorten_cellml} belongs to the example program in \code{examples/electrophysiology/cellml/shorten}, which solves a single-cell CellML model:
\begin{figure}[H]
\centering
\begin{framed}
\begin{lstlisting}[basicstyle=\small\ttfamily,commentstyle=\color{gray},numbers=left]
  TimeSteppingScheme::ExplicitEuler<
    CellmlAdapter<56,71>
  >
\end{lstlisting}
\end{framed}
\caption{C++ code snipped that solves a CellML model with an explicit Euler scheme. The two template parameters 56 and 71 correspond to the number of states and algebraics, respectively.}%
\label{fig:example_shorten_cellml}%
\end{figure}
In this case, the model contains 56 states and 71 algebraics. The reason that these numbers have to be fixed at compile-time is that this allows the data structures in the implementation to have a fixed layout and be allocated on the stack instead of the heap, which improves the performance.

If the given numbers are not matching the variables in the CellML file, appropriate warnings or errors are generated, containing the correct C++ code to be copied to the C++ file. If the numbers are too high, the solver still works correctly, however, some memory and computation time is wasted for the excess variables.

The other step is configuring the connections between the CellML computation and input data or coupled solvers. This involves defining a \code{mappings} parameter. \Cref{fig:example_mapping} shows such a definition for the multidomain example with fat layer and a contraction model, which was presented in \cref{sec:exemplary_usage_2}.

\begin{figure}
\centering
\begin{framed}
%\begin{Verbatim}[fontsize=\small]
\begin{lstlisting}[basicstyle=\small\ttfamily,commentstyle=\color{gray},numbers=left,language=python]
  mappings = {
    # function in OpenDiHu      name in CellML model    # comment
    
    ("parameter", 0):           "wal_environment/I_HH", # I_stim (constant)   $\label{alg:5.3}$
    ("parameter", 1):           "razumova/L_S",         # $\textcolor{gray}{\lambda}$ (constant)      $\label{alg:5.4}$
    
    ("connectorSlot", "vm"):    "wal_environment/vS",   # $\textcolor{gray}{V_m}$ (state)            $\label{alg:5.7}$
    ("connectorSlot", "stress"):"razumova/stress",      # $\textcolor{gray}{\gamma}$ (algebraic) $\label{alg:5.8}$
    ("connectorSlot", "lambda"):"razumova/L_S",         # $\textcolor{gray}{\lambda}$ (constant)        $\label{alg:5.9}$
  }
  
  parameters_initial_values = [0.0, 1.0]                # I_stim=0, $\textcolor{gray}{\lambda}$=1                    $\label{alg:5.12}$
\end{lstlisting}
%\end{Verbatim}
\end{framed}
\caption{Specification of parameters and connector slots in a CellML model. The listed settings define two CellML variables to be parameters and specify three connector slots to transfer values between coupled solvers.}%
\label{fig:example_mapping}%
\end{figure}

The \code{mappings} define which CellML constants or algebraics are treated as parameters. Lines \ref{alg:5.3} and \ref{alg:5.4} make the stimulation current and fiber stretch constants accessible from outside the CellML model by making them parameters. The variables are identified by their names and the model components they are defined in in the CellML model. In this example, the first parameter is the stimulation current \code{I_HH} within the \code{wal_environment} model component and the second parameter is the fiber stretch or half-sarcomere length \code{L_S} in the \code{razumova} component.
The initial values for these parameters are given in line \ref{alg:5.12}, which sets the stimulation current to zero and the fiber stretch to one.

The second information in the \code{mappings} parameter is which variables from the CellML model are exposed to coupled solvers in OpenDiHu. This happens by defining connector slots that can be connected between the solvers as shown in \cref{fig:example_multidomain_solver_structure}.
Three slots are defined in lines \ref{alg:5.7} to \ref{alg:5.9} with slot names \code{`vm`}, \code{`stress`} and \code{`lambda`}. The corresponding CellML variables are again specified by their model component name and their own name. 

CellML variables of all three different types are connected in the example. The membrane voltage $V_m$ in slot \code{`vm`} is part of the state vector $\bfy$. In this example, it is used in a bidirectional coupling with the diffusion solver. The second slot, \code{`stress`}, connects to the activation parameter $\gamma$, which is part of the algebraic vector $\bfh$. It is an output of the model. The slot \code{`lambda`} refers to a constant in the CellML description, which has been transformed to a parameter in line \ref{alg:5.4}. It is used as an  input and the received values at these slots are moved to the corresponding locations in the CellML formulation.

\subsection{Consistent Physical Units in CellML Models and the Multi-Scale Framework}

The variables in a CellML model describe physical quantities. CellML handles their physical units and computes the appropriate conversions when combining model components within a CellML description.
For the integration of a CellML model in external solvers such as OpenDiHu, we have to take care that the units are consistent.

\sisetup{retain-unity-mantissa = false}
The subcellular models that we use are formulated with the following units for length, time, electric current and capacitance:%
\begin{align*}
   \SI{1}{\centi\meter} &= \SI{1e-2}{\meter}, &
   \SI{1}{\milli\second} &= \SI{1e-3}{\second}, &
   \SI{1}{\micro\ampere} &= \SI{1e-6}{\ampere}, &
   \SI{1}{\micro\farad} = \SI{1e-6}{\farad}.
\end{align*}
These basic units also fix derived units such as \SI{1}{\kilo\hertz} for frequencies and \SI{1}{\milli\volt} for voltages. For example, the membrane capacitance $C_m$ has to be specified in units \SI{1}{\micro\farad\per\square\centi\meter} and the stimulation current $I_\text{stim}$ in the units \SI{1}{\micro\ampere\per\square\centi\meter}.

With this system of units, values are in a similar scale when computing subcellular models. However, these units are less suitable for organ-scale computations, as the derived mass and density units are \SI{1e-14}{\kilogram} and \SI{1e-8}{\kilogram\per\meter\cubed} and the derived force and stress units are \SI{1e-10}{\newton} and \SI{1e-6}{\pascal}. For the dynamic solid mechanics model, where these quantities play a role, we use the following different system of units:
\begin{align*}
   \SI{1}{\centi\meter} &= \SI{1e-2}{\meter}, &
   \SI{1}{\milli\second} &= \SI{1e-3}{\second}, &
   \SI{1}{\newton}.
\end{align*}
The length and time scales are identical to the subcellular model and allow for consistent coupling. The coupling of active stresses from the subcellular model to the solid mechanics model uses the unit-less activation parameter $\gamma \in [0,1]$, which is transferred to stress units by multiplication with a maximum active stress value in the solid mechanics model.

Derived units in the solid mechanics system of units are \SI{1e2}{\kilogram\per\meter\cubed} for the density, \SI{1e4}{\meter\per\square\second} for the acceleration and $\SI{1}{\newton\per\square\centi\meter} = \SI{10}{\kilo\pascal}$ for the stress. The values of material parameters and boundary conditions have to be given with respect to these units.
The units allow for smaller values in the solid mechanics computation than in the unit system of the subcellular model. Moreover, it is  convenient to specify forces directly in terms of \SI{1}{\newton}.

\subsection{Specification of Stimulation Times Using Callback Functions}\label{sec:stimulation_times_callbacks}

A muscle fiber is activated by impulse trains that are generated from a motor neuron and stimulate the fiber at its neuromuscular junction. At the synaptic terminal, neurotransmitters are released and open certain ion channels, which results in depolarization of the muscle fiber membrane.
This process can either be modeled by adding an external stimulation current $I_\text{stim}$ through the dedicated ion channels or by directly prescribing the transmembrane voltage $V_m$ to reflect the resulting depolarized state. The first approach is more accurate as it also describes the depolarization process at the stimulated parts of the fiber. The electric \say{far field} away from the stimulation point, however, is the same for both approaches.

In OpenDiHu, it is possible to configure either approach. Setting the stimulation current is more involved as the actual value of  $I_\text{stim}$ has to be chosen depending on the mesh width. Furthermore, multiple adjacent nodes have to be stimulated such that the electric current that is added to the system balances with the amount that is carried away by the diffusion term. The nonlinear subcellular model fails to compute a valid solution, if too much current is present. With too little current, the membrane potential stays below the activation threshold and no action potential is triggered. 

Prescribing the transmembrane voltage to a value above the depolarization threshold at multiple adjacent nodes leads to equivalent action potentials independent of the mesh width. However, a suitable value for the prescribed voltage also has to be chosen in accordance with the employed subcellular model.

The stimulation current $I_\text{stim}$ is a CellML parameter and the transmembrane voltage $V_m$ corresponds to a state in the CellML model. The values of both parameters and states can be adjusted during the simulation. This feature is implemented by means of callback functions in the Python settings. A callback is a user defined function that gets called in regular intervals during the simulation, receives various information about the current state of the simulation and can alter some values such as the states vector $\bfy(t)$ or the parameters vector $\hat{\bfp}(t)$.

\begin{figure}
\centering
\begin{framed}
%\begin{Verbatim}[fontsize=\small]
\begin{lstlisting}[basicstyle=\footnotesize\ttfamily,commentstyle=\color{gray},numbers=left,language=python]

  # callback function that can set parameters, i.e. stimulation current
  def $\textcolor{Maroon}{\text{\ttfamily set\_specific\_parameters}}$(n_nodes_global, time_step_no, current_time, $\label{alg:6.3}$
                              parameters, fiber_no):    
    
    # determine if fiber gets stimulated at the current time
    if fiber_gets_stimulated(fiber_no, current_time):    $\label{alg:6.6}$
      stimulation_current = 40.
    else:
      stimulation_current = 0.

    innervation_node_global = int(n_nodes_global / 2)    $\label{alg:6.11}$
    parameters[(innervation_node_global),0,0] = stimulation_current    $\label{alg:6.12}$

  # callback function that can set states, e.g., prescribe $\textcolor{gray}{V_m}$ for stimulation
  def $\textcolor{Maroon}{\text{\ttfamily set\_specific\_states}}$(n_nodes_global, time_step_no, current_time,       $\label{alg:6.15}$
                          states, fiber_no):            

    # determine if fiber gets stimulated at the current time
    if fiber_gets_stimulated(fiber_no, current_time):    $\label{alg:6.18}$
      innervation_node_global = int(n_nodes_global / 2)    $\label{alg:6.19}$
      states[(innervation_node_global),0,0] = 40.0      $\label{alg:6.20}$

  config = {        $\label{alg:6.22}$
    (...)          
    
    # callback to adjust parameters
    "setSpecificParametersFunction":         $\textcolor{Maroon}{\text{\ttfamily set\_specific\_parameters}}$,    $\label{alg:6.28}$
    "setSpecificParametersCallInterval":     1e3,
    "setSpecificStatesFrequencyJitter":      0,                          
                    
    # callback to alter values of states
    "setSpecificStatesFunction":             $\textcolor{Maroon}{\text{\ttfamily set\_specific\_states}}$,      $\label{alg:6.33}$
    "setSpecificStatesCallInterval":         2*int(1/stimulation_frequency/dt_0D),     $\label{alg:6.34}$
    
    "setSpecificStatesCallFrequency":        stimulation_frequency,    $\label{alg:6.36}$
    "setSpecificStatesCallEnableBegin":      0,                        $\label{alg:6.37}$
    "setSpecificStatesRepeatAfterFirstCall": 0.01,                     $\label{alg:6.38}$
    "setSpecificStatesFrequencyJitter":      [0.1,-0.2,0.0],           $\label{alg:6.39}$
                         
    # callback to postprocess the result
    "handleResultFunction":                  $\textcolor{Maroon}{\text{\ttfamily handle\_result}}$,    $\label{alg:6.41}$
    "handleResultCallInterval":              1e4,         
     
    "additionalArgument":                    fiber_no,        $\label{alg:6.43}$
  }

\end{lstlisting}
%\end{Verbatim}
\end{framed}
\caption{Settings that define neural spike trains activating muscle fibers. The definition of the two callback functions \code{set_specific_parameters} and \code{set_specific_states} is demonstrated.}%
\label{fig:example_callback_functions}%
\end{figure}

\Cref{fig:example_callback_functions} defines two such callback functions used in the fiber based electrophysiology model to add electric stimulation to the monodomain model. Either suffices to implement the stimulation. The function \code{set_specific_parameters} in line \ref{alg:6.3} and the function \code{set_specific_states} in line \ref{alg:6.15} both receive similar information from the simulation as their function arguments: the total number \code{n_nodes_global} of nodes in the current fiber, the current integer timestep number \code{time_step_no}, the corresponding floating-point number \code{current_time} of the current simulation time, and the number \code{fiber_no} that identifies the current fiber.

The variables \code{parameters} and \code{states} are the output of the callback functions that alter the parameter and state values, respectively. Both callbacks determine, whether the current fiber should be stimulated at the current time, in lines \ref{alg:6.6} and \ref{alg:6.18}.
If yes, the parameter or state at the center point of the fiber, computed in lines \ref{alg:6.11} and \ref{alg:6.19}, gets changed accordingly. In the real scenario, three adjacent points get stimulated instead of a single point.

Because the conversion of transferred data between the Python code and the C++ code costs some runtime, the number of transferred values is reduced to a minimum. Only the parameters and states that should be changed are indicated in the \code{parameters} and \code{states} variables in lines \ref{alg:6.12} and \ref{alg:6.20}. These variables are Python dictionaries, i.e., key-value pairs. The key is a tuple of three items: First, the global coordinates $(x,y,z)$ of the node where the parameter or state change is applied. In case of a 1D fiber mesh, this is only a single coordinate. Second, the dof index on this node. This is different from zero only for Hermite ansatz functions, which have multiple dofs per node. And third, the index of the parameter or state that should be set. Parameter 0 corresponds to the stimulation current as defined in line \ref{alg:5.3} of \cref{fig:example_mapping}, and state 0 corresponds to the transmembrane voltage $V_m$. The new value to set is the value of the key-value pair.

The comparison of the two callbacks functions shows one difference: In the callback for the parameters, the stimulation current is set to zero when there is no stimulation. In the callback for the states, nothing is done during this time. The reason for this is that the state values will be continuously updated from the rates by the timestepping scheme, whereas the parameters keep their values until they are changed from the callback. This has consequences on the times at which the callback functions have to be called from the simulation, which are described in the following.

Invoking the Python interpreter on a callback requires some time. Calling the callback after every small timestep of the simulation is, thus, not performant. We model the stimulation of a fiber by a piecewise constant function with two possible values for on and off.
In the approach that sets the stimulation current, the callback \code{set_specific_parameters} has to be called at the onset and at the end of every stimulation spike. If the approach with the prescribed membrane voltage is used, the callback \code{set_specific_states} has to be called after every stimulation onset in every subsequent timestep until the stimulation is over.

The requirements for both approaches can be satisfied by defining a small constant interval of timesteps after which the callback functions are invoked. This call interval can be specified in the \code{config} dictionary of the Python file in \cref{fig:example_callback_functions} , which is shown in excerpts from line \ref{alg:6.22} onwards. The \code{config} variable references the callback functions in lines \ref{alg:6.28} and \ref{alg:6.33} and the parameters for the call interval in the next lines. Note that this configuration is only shown for demonstration, a real configuration should either specify the states callback or the parameters callback function, not both.

Line \ref{alg:6.34} in \cref{fig:example_callback_functions} shows how the call interval can be computed to correspond to a given stimulation frequency \code{stimulation_frequency}, given the timestep width \code{dt_0D}. The prefactor of two occurs because the callback would be called twice per timestep in the Strang splitting scheme.

Real impulse trains from the motor neuron pool typically follow a base frequency with some added jitter that offsets the exact firing times from the base frequency by a small random time. Furthermore, studies are often designed to start with a completely inactive muscle and switch on certain MUs after specified times. To efficiently account for these two demands, we add another way to specify the times when the \code{set_specific_states} callback gets invoked. In this second way of specification, the \code{setSpecificStatesCallInterval} parameter is disabled by setting it to zero. Then, the three options \code{CallFrequency}, \code{CallEnableBegin}, \code{RepeatAfterFirstCall} and \code{Frequency}\code{Jitter} (prefixed by \code{set}\code{Specific}\code{States}) given in lines \ref{alg:6.36} to \ref{alg:6.39} are significant. 

\begin{figure}%
  \centering%
  \includegraphics[width=0.8\textwidth]{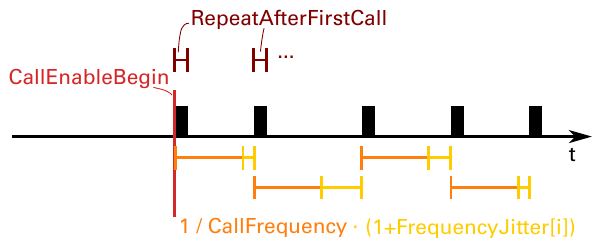}%
  \caption{Parametrization of stimulation times in electrophysiology simulations. The neuronal impulse train is given by the black spikes. The parameters \code{CallEnableBegin}, \code{RepeatAfterFirstCall}, \code{CallFrequency} and \code{FrequencyJitter} (in the settings all prefixed by \code{setSpecificStates}) specify the shape of the spike train.}%
  \label{fig:stimulation_times}%
\end{figure}%

Their meaning is illustrated in \cref{fig:stimulation_times}. \code{CallEnableBegin} specifies the time when the callback should be called for the first time. Then, it is called with a frequency that is additively composed of the base frequency given by \code{CallFrequency} and one entry of the parameter \code{FrequencyJitter}. This parameter is a ring buffer of relative factors by which the regular time span between subsequent firing events is prolonged. For example, if \code{FrequencyJitter} contains the list \code{[0.1,-0.2,0.0]}, the time span $T_{01}$ between the first two firing events is \SI{10}{\percent} longer than according to the base frequency $f$, the next timespan $T_{12}$ is \SI{20}{\percent} shorter and the next time span $T_{23}$ exactly equals the inverse base frequency, $T_{23} = 1/f$. Subsequently, the scheme repeats. Typically, this parameter is set to a randomly generated list with a large number of entries.
After each onset of a stimulation, the \code{setSpecificStatesCallInterval} function is called repeatedly in every subsequent timestep for a time span given by \code{RepeatAfterFirstCall}. 

In the fiber based electrophysiology example, every fiber has its own instance of the Python settings, and it is possible to specify different parameter values for different fibers or motor units, e.g., to set a different beginning time of the stimulations. The fibers can be distinguished by the last parameter of the callbacks, which receives the custom value that is defined by the \code{`additionalArgument`} parameter in line \ref{alg:6.43}. In the given example, the current fiber number is used here, but any other Python variable is possible.

\begin{figure}%
  \centering%
  \includegraphics[width=\textwidth]{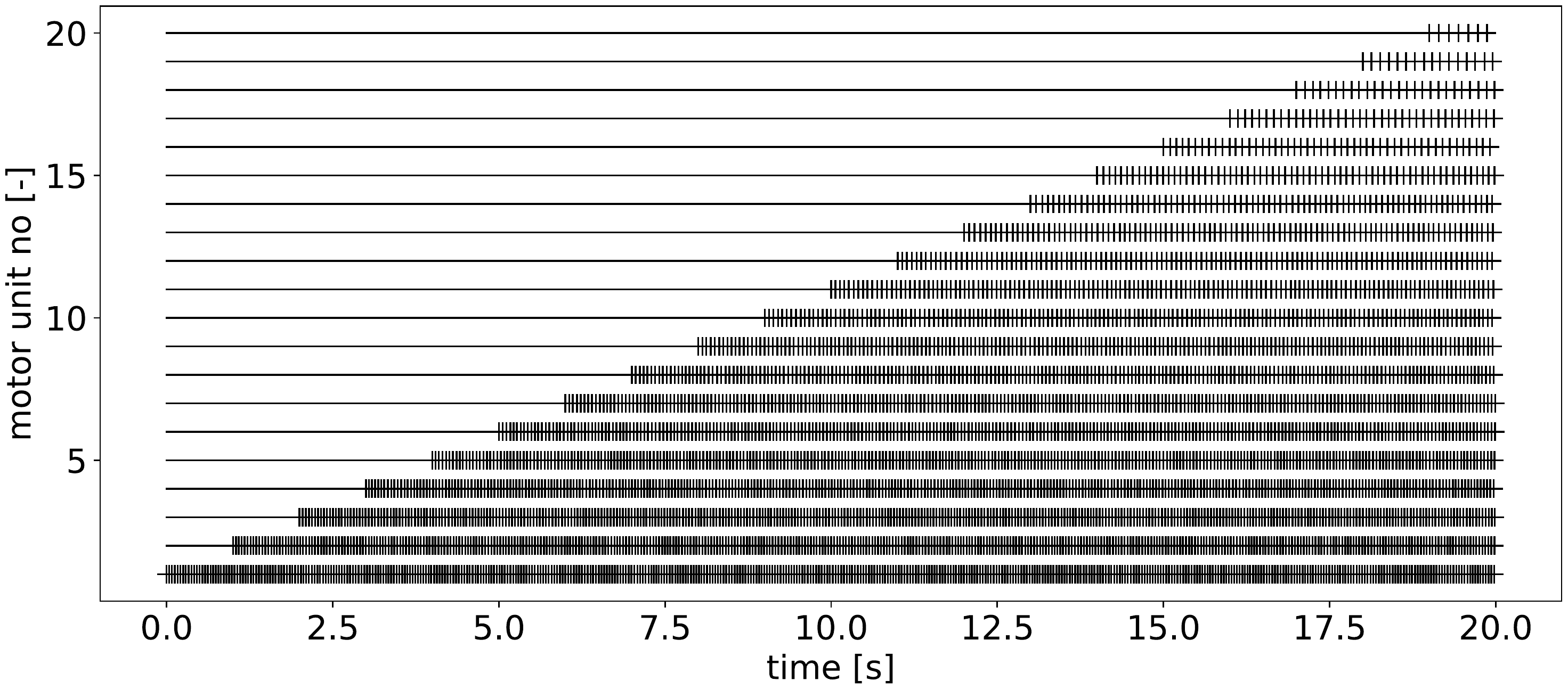}%
  \caption{Firing times for a scenario with 20 motor units with ramp-like activation and different stimulation frequencies.}%
  \label{fig:firing_times_ramp}%
\end{figure}

\Cref{fig:firing_times_ramp} shows a scenario, where different parameters are set for different MUs. The figure shows the firing times of fibers grouped to 20 MUs, which are activated in a ramp in the first $t=\SI{19}{\second}$. The base frequency decreases from \SI{23.92}{\hertz} to \SI{7.66}{\hertz} for MUs 1 to 20, which reproduces a scenario in literature \cite{Klotz2020}. The frequency jitter parameter is a list of 100 randomly chosen values between \SI{-10}{\percent} and \SI[retain-explicit-plus]{+10}{\percent}. The \code{CallEnableBegin} parameter enables the stimulation of the next MU every second.

Similar to the two presented callbacks, which set parameters and states, a third callback \code{handle_result} can be defined as given in line \ref{alg:6.41} of \cref{fig:example_callback_functions}. This callback function gets called in a fixed interval specified by \code{`handleResultCallInterval`}. It receives the complete vectors of states $\bfy$ and intermediates $\bfh$ and can be used to perform custom post-processing or to output custom data files from the Python script.

In summary, variables of CellML models can be coupled to other solvers. Their parameters and values can be adjusted from the settings file. Callback functions are used to alter values during the simulation. This flexibility comes at the runtime cost of invoking the Python interpreter, therefore the times when to call the callback functions have to be specified appropriately. Special methods exist to model steady stimulation with frequency jitter, which occurs in typical neural stimulation of muscle fibers.

% --------------------------------------------
\section{Output File Formats}\label{sec:output_file_formats}

After the simulation program completes, the computed results can be visualized using external tools.
As mentioned in the previous sections, output writers are used to generate output files in various formats. The formats of the output writers and additional options are configured in the Python settings under the parameter \code{OutputWriter}. The following formats are supported:\code{`ParaView`}, \code{`ExFile`}, \code{`PythonFile`}, \code{`PythonCallback`} and\break\code{`MegaMol`}.
The corresponding output can be visualized and post-processed using different tools, which will be presented in the following. We use simulation results of a fiber-based electrophysiology scenario with 49 1D fibers and a 3D muscle mesh to showcase the different output data formats.

\subsection{Output of VTK Files for the Use with ParaView}
The canonical way to visualize simulation results computed by OpenDiHu is to use the software ParaView \cite{paraview}. 
The required output file formats are defined by the Visualization Toolkit (VTK) specification \cite{vtk}. Depending on the mesh type in OpenDiHu, different file types are generated:
\emph{RectilinearGrid} files (with file ending \code{.vtr}) for the output of \say{regular fixed} meshes that represent a cartesian grid,
\emph{StructuredGrid} files (\code{.vts}) for the output of \say{structured deformable} meshes, i.e., structured meshes that can deform over time, 
\emph{UnstructuredGrid} files (\code{.vtu}) for the output of unstructured meshes, and
\emph{PolyData} files (\code{.vtp}) containing connected points are used to represent multiple muscle fibers in a single file.
ParaView can be used to load and visualize all of these file types.

All of these files are XML based and their payload data can be configured to be either written in ASCII representation or in Base64 encoding. Base64 encoding also translates the raw data into ASCII characters. The data stream is split into pieces of 6 bits, which are each represented by an 8-bit-ASCII character. Thus, the required memory is $4/3$ of the raw data. Compared to a full ASCII representation containing the digits of all numerical values, this leads to a significant reduction of file sizes.

The VTK file format specifies parallel file output, where each process writes its local data to a separate file and one additional master file references the pieces in all files. This parallel file output scheme is implemented in OpenDiHu. However, it can lead to an impractically large number of small output files for high degrees of parallelism. 

Therefore, we additionally implement a second approach, where non-parallel VTK files, which contain the whole dataset, are written. The same type of output files is generated during serial and parallel execution of the program. 
Writing the data to such a file is done using the parallel output capabilities of MPI. The respective MPI functions allow to collectively write data to the same file from different processes at different locations in the file. For parallel execution, every process only writes its own local data and no communication of the payload data to a master process is necessary. 

Because the byte boundaries in a Base64 encoded data stream coincide with multiples of 8 bits only every three bytes, the processes that write neighboring parts in the output file have to coordinate the bit offsets of their data streams. For this, a small amount of data has to be communicated between these processes. However, the cost of this communication is negligible.

With this improved output scheme, one file is generated per mesh and output timestep of the simulation. The frequency of output timesteps can be configured in the Python settings.
It is also possible and useful to combine all 1D fiber meshes into a single output file per timestep to reduce the number of output files.

Different meshes can be written with different frequency. For example, for a simulation of fiber based electrophysiology with EMG signals, it is reasonable to output the comprehensive dataset of all fibers less frequently than the smaller dataset of EMG signals on the 2D skin surface. To associate the output files with the correct times, a timestamp of the current simulation time is added to every file. Furthermore, partitioning information is added, i.e., which part of the mesh is computed by which process.

To synchronize output files of different meshes with different output frequencies in the visualization tool, additional \emph{series files} (with file ending \code{.series}) are automatically created for every mesh. Such a file references all available output files of a mesh with their simulation times in JSON format. These files can be opened in ParaView to get a time-series of the simulation result.

Using the series files is also convenient, if the simulation is run in a directory, where old simulation results from previous runs exist. Because the series files are updated every timestep and only reference the newly created files, opening these files in ParaView only visualizes newly created simulation output, in contrast to opening a whole directory, which would potentially also load old results.

\subsection{Visualization With ParaView}
\begin{figure}%
  \centering%
  \includegraphics[width=\textwidth]{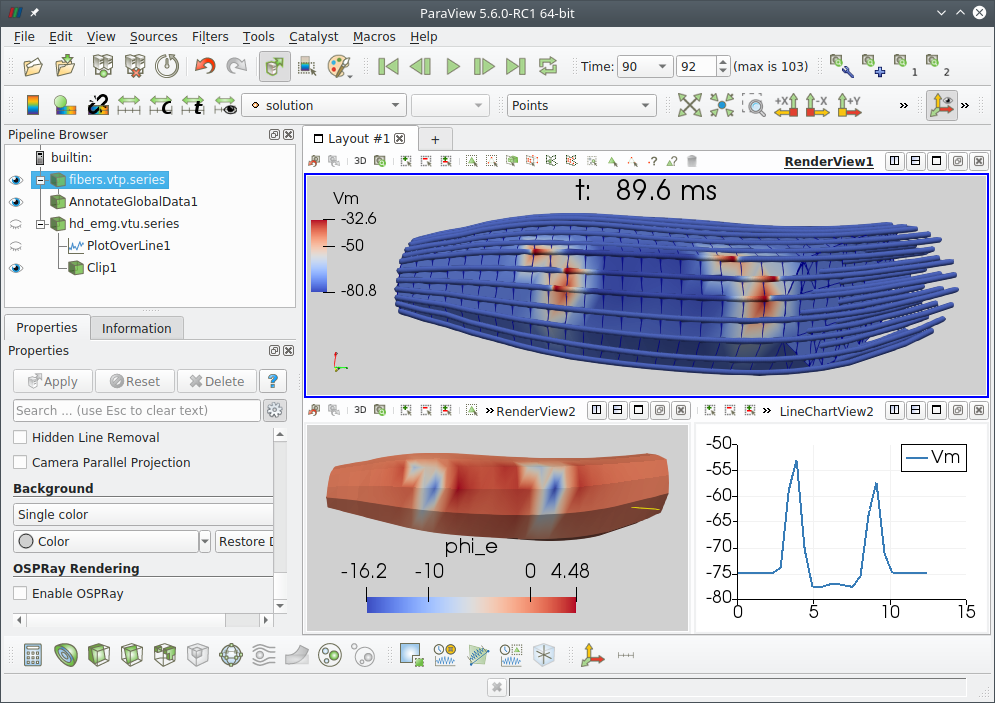}%
  \caption{Visualization of simulation results with ParaView: ParaView window with a visualization of muscle fibers and a 3D muscle mesh.}%
  \label{fig:paraview_output}%
\end{figure}

ParaView allows various manipulations and types of visualization of the loaded data. \Cref{fig:paraview_output} shows the ParaView window with simulation data of a fiber-based electrophysiology scenario. The loaded data are organized in a tree of datasets with applied filters, which can be seen in the \say{Pipeline Browser} in the top left. The center view shows a visualization of the muscle fibers and the 3D mesh at simulation time $t=\SI{89.6}{\milli\second}$. An animation of the transient data can be shown by using the playback controls in the top bar.

The visualization in the center top view displays the membrane voltage $V_m$ at the fibers and in the 3D mesh, colored by the scheme shown at the left in the view. The 3D mesh is sliced on the right-hand side of the muscle to make the fiber dataset better visible.

The view on the bottom left depicts the extra-cellular potential $\phi_e$ on the 3D mesh. The view on the bottom right shows a plot of the value of $\phi_e$ along a horizontal line on the surface of the muscle.

It can be seen that three fibers near the surface are activated, and that the action potentials effect the EMG value given by $\phi_e$ on the surface.

For larger datasets, a head-less render server of ParaView also can be run in parallel on a remote server and the graphical user interface shown in \cref{fig:paraview_output} can be used as the client to interactively control the visualization.

ParaView also supports ray tracing using the OSPRay ray tracing engine. With ray tracing, more advanced lighting and the computation of shadows are possible.

\subsection{ExFiles and Visualization with CMGUI}

Another option in OpenDiHu is to output files in the \say{ExFile} format. This format originates from the software environment of OpenCMISS. Output of results in simulations with OpenCMISS Iron relies on this type of files. The visualization toolbox of OpenCMISS Zinc is able to create various visualizations of the data given in this format. The program \emph{CMGUI} provides a graphical user interface to visualize the data.

The output consists of corresponding \code{.exelem} and \code{.exnode} files containing information at element and node level, respectively. The mesh is assumed to be unstructured and, thus, the information which nodes correspond to a particular element has to be explicitly stored. It is stored in the \code{.exelem} file. The payload data are contained in the \code{.exnode} file. The file format supports parallel output to separate files. However, only serial output is supported in OpenDiHu.
ExFiles are ASCII-based and, thus, only usable up to a certain problem size.

An advantage of the \say{ExFile} format is that also higher order elements can be represented. The visualization tools are capable of representing the geometric data accordingly, e.g., it is possible to visualize the correct shape of cubic Hermite 3D hexahedral elements. In contrast, ParaView only visualizes linear elements and linearly interpolates the data between the nodes of an element.

The program CMGUI can be used to visualize the output files of OpenDiHu in ExFile format.
In the graphical user interface, the \code{.exelem} and \code{.exnode} files can be loaded. Representations of loaded points, lines and elements can be added to the visualization in the scene editor. Various options such as coordinate frames and parameters for shading and tessellation can be set. The visualizations can be colored using predefined appearances or according to the loaded solution values.

For larger datasets, these manual adjustments are tedious. For example, for a dataset with 49 fibers, the user would have to load 49 \code{.exelem} and 49 \code{.exnode} files one by one. Instead, the Perl scripting interface of CMGUI can be used. Every command in the GUI corresponds to a Perl command and CMGUI can load and execute those commands from a given Perl script.

OpenDiHu automatically creates such Perl scripts. The generated script for a mesh loads all generated output files into CMGUI, adds a  corresponding visualization depending on the mesh dimensionality and opens the required CMGUI windows such that the data are immediately visible. This is an improvement to OpenCMISS Iron, where all steps have to be done manually. By using the generated Perl script, less expert knowledge on the usage of CMGUI is required, and it is also possible to visualize datasets with a large number of fibers.

\begin{figure}%
  \centering%
  \begin{subfigure}[t]{0.62\textwidth}%
    \centering%
    \includegraphics[width=\textwidth]{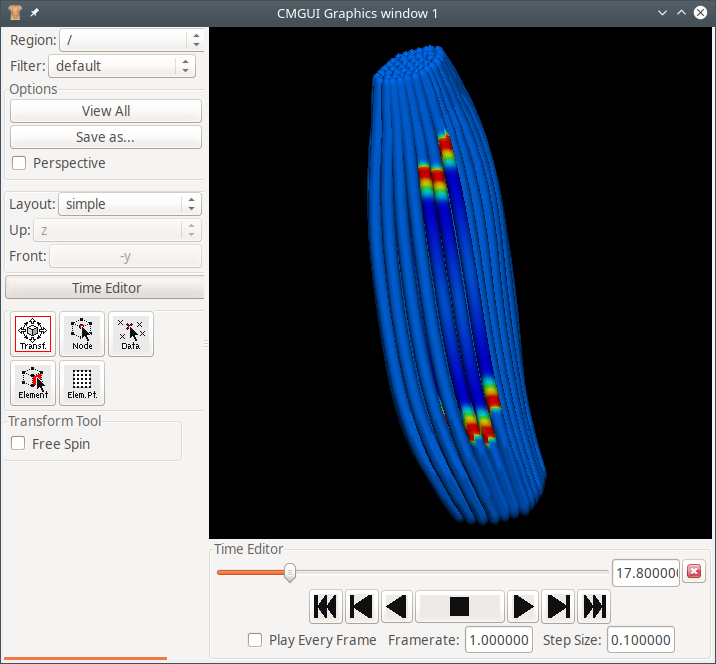}
    \caption{The main graphics window that displays the visualization and allows to control the current view and the current timestep.}%
    \label{fig:current_configuration_1}%
  \end{subfigure}
  \begin{subfigure}[t]{0.363\textwidth}%
    \centering%
    \includegraphics[width=\textwidth]{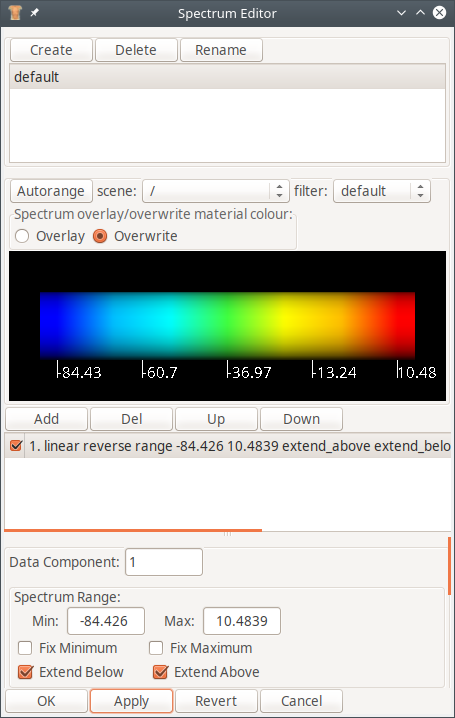}
    \caption{The spectrum editor that can be used to adjust the coloring according to the loaded solution values.}%
    \label{fig:cmgui_spectrum}%
  \end{subfigure}
  \caption{Visualization of the results of an electrophysiology simulation with CMGUI involving 49 muscle fibers.}%
  \label{fig:cmgui_output1}%
\end{figure}

\Cref{fig:cmgui_output1} shows two windows of CMGUI. In \cref{fig:current_configuration_1}, the main graphics window can be seen with a visualization of 49 muscle fibers. The membrane potential is visualized by varying colors, and action potentials can be seen on three of the shown fibers. Similar to ParaView, the transient data can be animated by using the controls at the bottom.
\Cref{fig:cmgui_spectrum} shows the spectrum editor, where the color scheme can be adjusted to the range of the loaded data.

\begin{figure}%
  \centering%
  \begin{subfigure}[t]{0.48\textwidth}%
    \centering%
    \includegraphics[width=\textwidth]{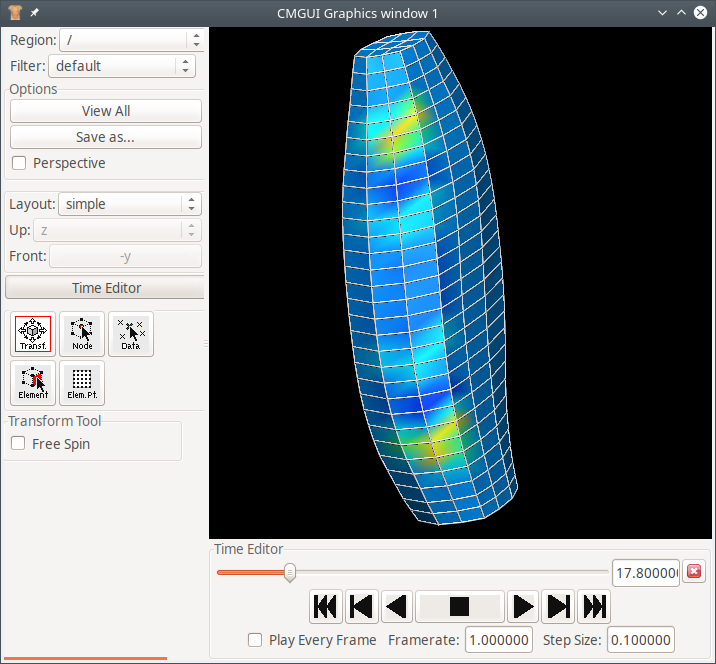}%
    \caption{Graphics window with the visualization of a 3D mesh.}%
    \label{fig:cmgui_emg}%
  \end{subfigure}
  \begin{subfigure}[t]{0.48\textwidth}%
    \centering%
    \includegraphics[width=\textwidth]{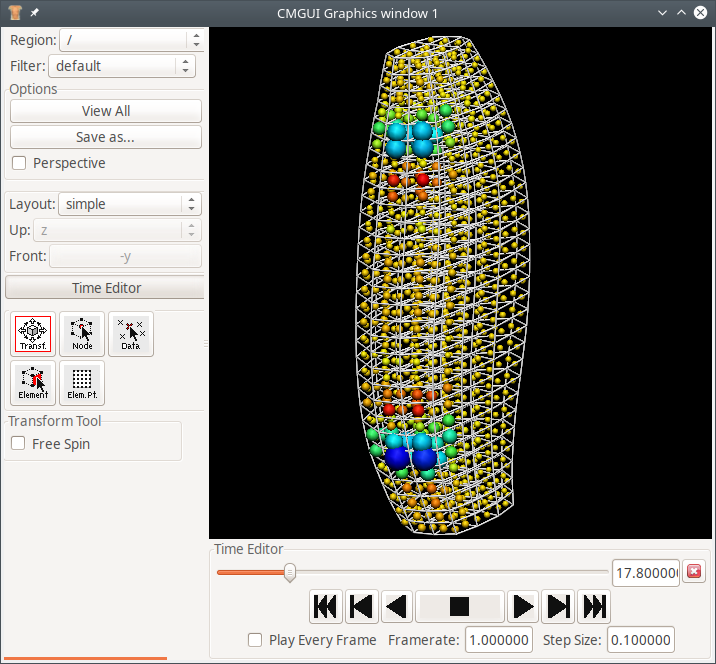}
    \caption{Visualization of the same data as in (a), but using sphere glyphs at every node.}%
    \label{fig:cmgui_phie}%
  \end{subfigure}
  \caption{Visualization of data on a 3D mesh with CMGUI.}%
  \label{fig:cmgui_output2}%
\end{figure}

The other Perl script besides the one used in \cref{fig:cmgui_output1} to visualize the muscle fibers addresses the 3D mesh of the muscle.
\Cref{fig:cmgui_output2} shows the graphics windows with the resulting visualizations of this dataset. In \cref{fig:cmgui_emg}, the extracellular potential $\phi_e$ is visualized on the muscle surface. The visualization contains the colored 3D representation for the mesh and a 1D representation of the mesh consisting of white tubes.

\Cref{fig:cmgui_phie} demonstrates the feature of visualizing nodal data using glyphs. The $V_m$ values at every node are represented by colored circles with a radius that corresponds to the value. With this representation, it is possible to also show the data inside the muscle volume.

\subsection{Python Output Files}
Another option in OpenDiHu is to output data in a Python-friendly format, which can easily be parsed from within a python script.
The data can then be used, e.g., for error analysis or to convert them to other custom formats.

If the format \code{PythonFile} is specified in the output writer, the data get written to output files. If the format \code{PythonCallback} is specified, the same data are passed to a callback function and can directly be used in the Python settings script during the running simulation. 

For output, the data are organized in a Python dictionary. The output files either contain the plain Python code of this dictionary or a binary representation obtained by the \emph{pickle} package of Python. In parallel execution, every process writes its own file containing the data of the corresponding subdomain. OpenDiHu provides a Python module to parse these output files. The data representation, whether the data are stored in binary or in human-readable format and whether it is composed of multiple files resulting from parallel execution is abstracted and transparent in the call to this module.

The utility program \code{plot} can be used to quickly visualize the simulation results in such Python output files. It creates plots and animations of 1D and 2D structured meshes and chooses different layouts for the type of data, e.g, a plot over time for single-cell CellML models or an animation with multiple plots for subcellular models with multiple ion channels. This script is useful mainly for 1D and 2D toy problems, such as the Laplace, Poisson and Diffusion problems.

\begin{figure}%
  \centering%
  \includegraphics[width=\textwidth]{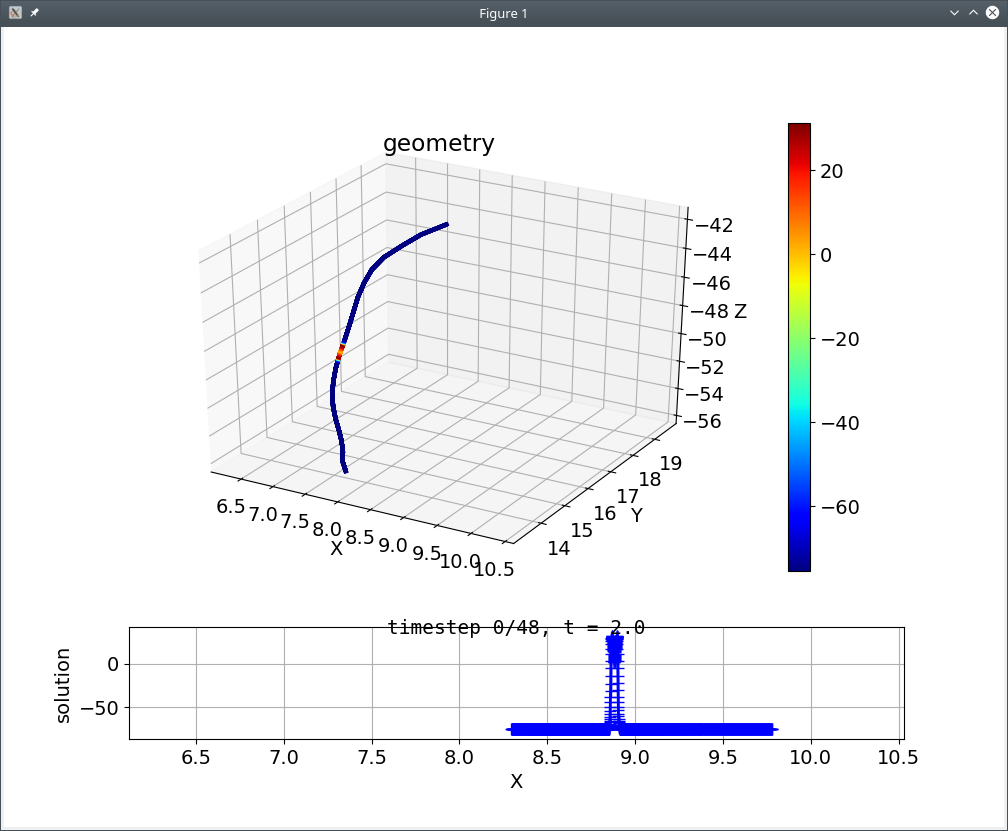}%
  \caption{Visualization of Python based simulation results using the \code{plot} utility.}%
  \label{fig:python_output}%
\end{figure}

\Cref{fig:python_output} shows the output of the \code{plot} script for one muscle fiber. The top plot visualizes the geometry in 3D space, colored by the membrane potential $V_m$. The plot below shows the spatial progression of the $V_m$ value along the $x$-axis. However, for the visualization of 3D data, other options such as ParaView or CMGUI are better suited and should be used instead.

\subsection{ADIOS output files and MegaMol}
Another output format is the binary-pack file format defined by the Adaptable Input Output System library (ADIOS2). This type of output is selected by the OpenDiHu output writers for the \code{`MegaMol`} format. ADIOS2 provides a framework for high-performance computing data management \cite{adios2}. ADIOS2 manages self-describing data that allows rapid metadata extraction also from large data sets.

Output files in this format can be loaded into the visualization software MegaMol by experts. MegaMol has been successfully used together with OpenDiHu to implement in-situ visualization, where OpenDiHu shares the computed simulation data with MegaMol using the ADIOS2 format and triggers updates of the visualization by sending asynchronous messages to MegaMol during the runtime of the simulation. As both OpenDiHu and MegaMol can run in parallel, the partitioned data needs to be merged from all processes only at the stage of rendering the visualization image. For highly parallel runs on supercomputers, the local data that are generated by the OpenDiHu processes on the same compute node can be shared in memory with one instance of MegaMol per compute node. Then, all MegaMol instances collectively render the resulting visualization. This approach bypasses the costly file output operation on the highly distributed file system of a supercomputer.

\begin{figure}
\centering
\begin{framed}
%\begin{Verbatim}[fontsize=\small]
\begin{lstlisting}[basicstyle=\footnotesize\ttfamily,commentstyle=\color{gray},numbers=left,language=python]
    string   config                         (...)
    string   meta                           $\text{"}$current time: 2021/3/30 19:48:05,$\textcolor{gray}{\hookleftarrow}$
                                             hostname: lapsgs05, n ranks: 4$\text{"}$
    string   version                        $\text{"}$opendihu 1.2, built $\textcolor{gray}{\hookleftarrow}$
                                             Mar 27 2021, C++ 201402, GCC 7.5.0$\text{"}$
    double   localBoundingBox               10*{4, 6} = -56.3 / 19.7732   $\label{alg:7.6}$
    double   globalBoundingBox              10*{6} = -56.3 / 19.7732     $\label{alg:7.7}$
    double   global_radius                  10*scalar = 0.1 / 0.1
    int32_t  nPointsPerCoordinateDirection  10*{3} = 4 / 31
    int32_t  nodeOffsetOnOwnComputeNode     10*{4} = 0 / 368
    int32_t  node_count                     10*scalar = 496 / 496
    int32_t  rankNo                         10*{496} = 0 / 3                 $\label{alg:7.12}$
    double   emg                            10*{496} = -12.0536 / 4.89757    $\label{alg:7.13}$
    double   transmembraneFlow              10*{496} = -125.014 / 226.428    $\label{alg:7.14}$
    double   vm                             10*{496} = -81.3198 / -27.4762   $\label{alg:7.15}$
    double   xyz                            10*{1488} = -56.3 / 19.7732      $\label{alg:7.16}$
\end{lstlisting}
%\end{Verbatim}
\end{framed}
\caption{Contents of the output file created by ADIOS2.}%
\label{fig:adios_output}%
\end{figure}             

The generated output files can be inspected using the \code{bpls} utility. \Cref{fig:adios_output} shows a description of the 3D dataset extracted from the binary-pack format that was written by a simulation with four processes.
Each line corresponds to one variable in the file. The first column specifies the variable type and the second column is the name of the variable. The third column contains structural information with minimum and maximum values for numeric types. 

The first three shown variables are of type \code{string} and contain metadata for the simulation run. The \code{config} variable contains the Python settings code of the scenario and, thus, accurately describes the settings of the simulation run. 
The values of the \code{meta} and \code{version} variables are fully listed in \cref{fig:adios_output} and contain meta information about the simulation program and the particular run. 

For the numeric values, the third column specifies the dimension of the stored data. The file contains the simulation output for 10 different timesteps, which can be seen in the third column. For example, the \code{localBoundingBox} variable in line \ref{alg:7.6} stores 10 instances of a matrix with dimension $4\times 6$. The four rows of this matrix correspond to the four processes and the columns store the six values of the geometric bounding box of the subdomain on the respective process. This information is required by MegaMol to constrain the volume that has to be rendered on each process. Further structural information is contained in the variables in lines \ref{alg:7.7} to \ref{alg:7.12}. 
The remaining variables contain the payload data.  The variables \code{emg}, \code{transmembraneFlow} and \code{vm} correspond to $\phi_e$, the right-hand side of the first bidomain equation in \cref{eq:static_bidomain_rhs}, and $V_m$, respectively. The variable  \code{xyz} holds the geometry information for all nodes.

\begin{reproduce_no_break}
  The visualizations in this section are based on outputs of the following simulation:
  \begin{lstlisting}[columns=fullflexible,breaklines=true,postbreak=\mbox{\textcolor{gray}{$\hookrightarrow$}\space}]
    cd $\$$OPENDIHU_HOME/examples/electrophysiology/fibers/fibers_emg/build_release
    ./fast_fibers_emg ../settings_fibers_emg.py output_demo.py
  \end{lstlisting}
  
  Output files for ADIOS2, CMGUI, ParaView and Python will be generated in corresponding subdirectories under \code{out/}.
  The following commands invoke the respective visualization tool in the corresponding output directory:
  \begin{lstlisting}[columns=fullflexible,breaklines=true,postbreak=\mbox{\textcolor{gray}{$\hookrightarrow$}\space}]
    paraview fibers.vtp.series              # ParaView
    cmgui fibers.com                        # CMGUI
    cmgui hd_emg.com                        # CMGUI
    plot fibers_0000001_MeshFiber_*.py      # Python
    bpls hd_emg.bp -la                      # ADIOS2
  \end{lstlisting}
  In the graphical user interfaces of CMGUI and ParaView, more settings have to be adjusted to obtain the results shown in \cref{fig:paraview_output,fig:cmgui_output1,fig:cmgui_output2}.
  
  The listing shown in \cref{fig:adios_output} was obtained by a simulation with 4 processes. 
  Because the ExFile output writer does not work for parallel execution, the corresponding option has to be disabled in the \code{output_demo.py} variables file prior to execution:
  \begin{lstlisting}[columns=fullflexible,breaklines=true,postbreak=\mbox{\textcolor{gray}{$\hookrightarrow$}\space}]
    mpirun -n 4 ./fast_fibers_emg ../settings_fibers_emg.py output_demo.py
  \end{lstlisting}
  Afterwards, the shown listing can be obtained by \code{bpls -la hd_emg.bp}.
\end{reproduce_no_break}

\chapter{Implementation of the Software OpenDiHu}\label{sec:implementation}

After the usage of OpenDiHu has been described in the last chapter, we now discuss the implementation of the algorithms and solvers that are available in the framework. This chapter begins with the basic data organization in \cref{sec:data_handling_with_petsc} and generic algorithms to set up finite element discretizations and parallel partitionings of a problem in \cref{sec:fem_matrices_and_bc,sec:parallel_partitioning_and_sampling_of_the}. Then, details are given on the implementation of particular solvers. \Cref{sec:parallel_partitioning_for_fiber_based} discusses the solvers for the fiber based electrophysiology model, \cref{sec:parallel_solver_multidomain} addresses the multidomain solver and \cref{sec:computation_cellml_models} presents optimizations for the solver of the subcellular model. The chapter closes in \cref{sec:data_mapping_between_meshes} with a discussion of the data mapping required in coupling schemes.

\section{Data Handling with PETSc}\label{sec:data_handling_with_petsc}

% introduction Petsc
OpenDiHu processes various types of data: geometry data, the discretized solution data, system matrices and vectors in the specification of the mathematical model such as right-hand sides and prescribed values in boundary conditions.
All these data need to be organized in accordance with the parallel partitioning. Linear system solvers need to be applied on matrices and vectors to obtain the solution.
The result of the simulation has to be invariant under a change of the number of processes that execute the program.

For parallel data handling and solvers of linear and nonlinear systems, the \emph{Portable, Extensible Toolkit for Scientific Computation (PETSc)} \cite{petsc-web-page,petsc-user-ref,petsc-efficient1997} is used. PETSc provides a large collection of solvers and preconditioners that can be selected and configured at runtime. More solvers are accessible through interfaces to external software, such as the \emph{Multifrontal Massively Parallel Sparse Direct Solver (MUMPS)} \cite{mumps2001,mumps2019} and the preconditioner library \emph{HYPRE} \cite{falgout2002hypre}. PETSc natively supports MPI parallelism and provides parallel data structures for vectors and matrices. Numerous operations on the data are provided including value communication and access, housekeeping, arithmetical operations, and more advanced calculations in the field of linear algebra.

Since MPI is used, processes can be identified by their \emph{rank} $r$ within the used \emph{MPI communicator}. An MPI communicator is a subset of processes that can communicate with each other. The rank of a process is its number in this communicator, i.e., a consecutive number starting with zero.

\subsection{Organization of Parallel Partitioned Data}\label{sec:oragnization_of_parallel_partitioned_data}

% partitioned 1D vector, local, global numbering
Basic building blocks in the implementation of \opendihu{} are \emph{field variables} that represent scalar fields.
A scalar field $v : \Omega \to \R$ defined on a domain $\Omega \subset \R^3$ is represented in the program by its finite element discretization. It comprises, on the one hand, the specification of the mesh of $\Omega$, i.e, the node positions, elements and ansatz functions and on the other hand the values of the coefficients of the ansatz functions. The values of the coefficients are called \emph{degrees of freedom (dof)}. Meshes with linear ansatz functions have one dof on every node. In the following, regular Cartesian meshes with linear ansatz functions are considered.

The partitioning of a regular, $d$-dimensional mesh is constructed as follows. A partitioning in terms of number of processes is given in the form $n_x \times n_y \times n_z = n_\text{proc}$, where $n_x,n_y$ and $n_z$ are the number of processes or subdomains in $x$, $y$ and $z$ direction, respectively. For 2D meshes, $n_z$ is set to one, for 1D meshes, $n_y$ and $n_z$ are set to one. The given mesh is partitioned on the level of elements. In every coordinate direction $i \in \{x,y,z\}$, the number $N^\text{el}_i$ of elements is equally distributed to the specified number $n_i$ of processes. Every process gets either $\lfloor N^\text{el}_i/n_i+1\rfloor$ or $\lfloor N^\text{el}_i/n_i \rfloor$ elements, where the larger number of elements is assigned to the processes with lower ranks. Thus, the subdomains with smaller index in $x$, $y$ and $z$ direction potentially have one layer of elements more than other subdomains.

\begin{figure}%
  \centering%
  \includegraphics[width=0.6\textwidth]{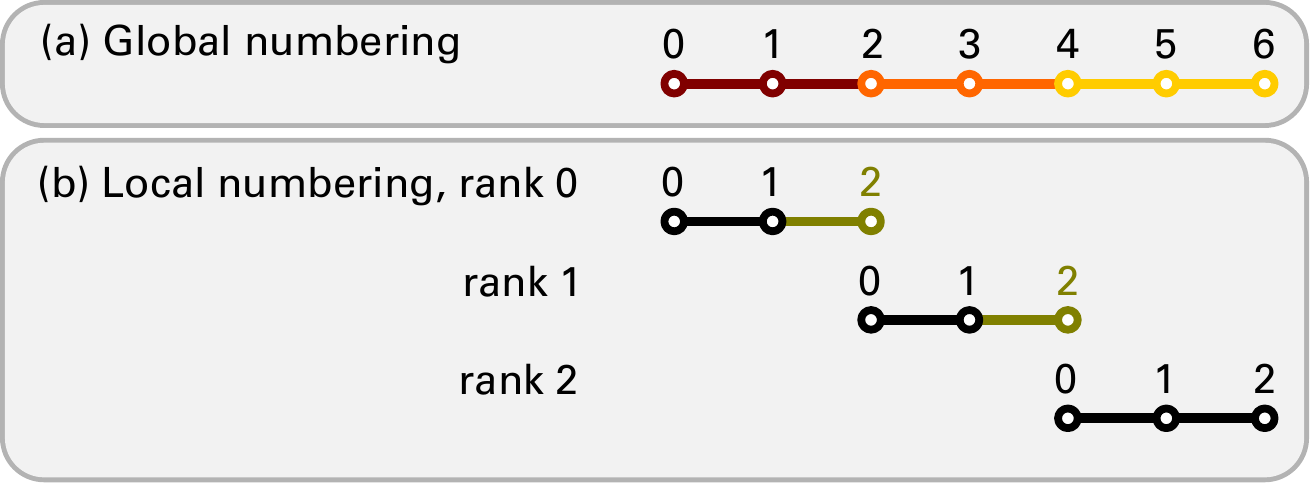}%
  \caption{Partitioning and local and global numbering of a 1D mesh with $N^\text{el}_x=6$ elements partitioned to $n_x=3$ processes. Ghost nodes are marked in green in (b)}%
  \label{fig:1d_nodes}%
\end{figure}

\begin{figure}%
  \centering%
  \includegraphics[width=\textwidth]{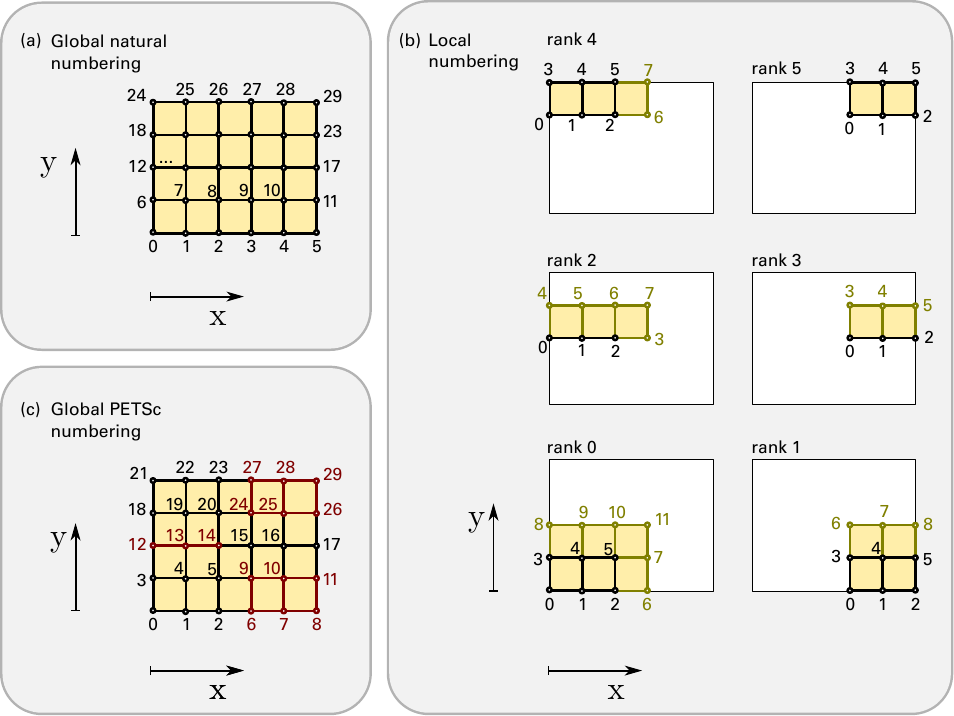}%
  \caption{Subdomains and numberings of a 2D mesh with $N^\text{el}_x \times N^\text{el}_y=5\times 4$ elements partitioned to $n_x\times n_y = 2 \times 3$ processes. (a)-(c) show the different numberings needed for (a) boundary condition specification, (b) identification of local non-ghost dofs (black) and local ghost dofs (green), and (c) identification of global dofs.}%
  \label{fig:dof_numbering}%
\end{figure}

For example, in \cref{fig:1d_nodes}, a 1D mesh with $N^\text{el}_x=6$ elements is partitioned into three subdomains with two elements each. 
\Cref{fig:dof_numbering} (a) shows a 2D mesh with $N^\text{el}_x \times N^\text{el}_y = 5 \times 4$ elements, a partitioning to $n_x \times n_y = 2 \times 3$ processes is given in \cref{fig:dof_numbering} (b).

The nodes of the mesh are assigned to the same subdomains as their adjacent elements. 
The assignment of the nodes that lie on the cutting planes between the subdomains remains to be specified. These nodes are assigned to the subdomain of the adjacent element in positive $x$, $y$ and $z$ direction such that each of these nodes is also owned by a single rank. On all other adjacent ranks, the node is stored as so called \emph{ghost} node. In contrast, the other local nodes are called \emph{non-ghost} nodes in the following.

The assignment of nodes to processes leads to the situation, that subdomains with the highest index in $x$, $y$ and $z$ direction (i.e., the subdomains at the \say{right}, \say{top}, or \say{back} end of the domain) potentially have one layer of nodes more than other subdomains. This effect is
intentionally chosen to be balanced out by our strategy to assign a higher number of elements to 
subdomains with lower index. Therefore, the total number of nodes and dofs is  approximately equally distributed. Moreover, in the limit for $N^\text{el}_x,N^\text{el}_y,N^\text{el}_z \to \infty$, the imbalance vanishes totally. 

In the exemplary partitionings in \cref{fig:1d_nodes} (b) and \cref{fig:dof_numbering} (b), owned nodes are represented by black circles and numbers, ghost nodes are represented by green circles and numbers. In the 1D example in \cref{fig:1d_nodes}, the three subdomains with two elements each have three, two and two nodes. In the 2D example in \cref{fig:dof_numbering}, the six subdomains have either three (ranks 2 and 3) or six (ranks 0,1,4 and 5) nodes while the number of elements varies between six (rank 0) and two (rank 5). This demonstrates the construction of nearly equally sized subdomains in terms of the number of assigned nodes.

%1D subdomain and dof numbering

%2D dof numbering local and global

\renewcommand{\Vec}{\code{Vec}}

On the partitioned meshes, field variables can be defined to represent the scalar and vector fields in the FEM computations. A field variable in \opendihu{} manages its values using the basic PETSc data type for storing scalar fields: the \emph{\Vec{}}. It represents a vector $\tilde{\bfv} \in \R^{n_\text{global}}$ with $n_\text{global}$ values. The vector is distributed to $n_\text{proc}$ processes according to the partitioning of the mesh, such that every value is owned by exactly one process. 

In a PETSc \Vec{}, every rank $r$ locally stores a distinct portion of $n_\text{local\_without\_ghosts} \leq n_\text{global}$ values of the global vector of dofs. Therefore, every dof is \emph{owned} by exactly one rank. These dofs correspond to the local nodes in the partitioning.
Additionally, the process maintains storage for $n_\text{ghosts}$ ghost dofs that are owned by other ranks. 
%Again, the ghost dofs correspond to the ghost nodes of the partitioning. 
PETSc is able to communicate corresponding values between all ranks where the dof is present either as ghost or non-ghost dof. 

In total, the local buffer of a \Vec{} stores $n_\text{local\_with\_ghosts} = n_\text{local\_without\_ghosts} + n_\text{ghosts}$ values. The non-ghost dofs are located at array positions $0,\dots,n_\text{local\_without\_ghosts}-1$, the ghost dofs follow at positions $n_\text{local\_without\_ghosts}, \dots, n_\text{local\_with\_ghosts}-1$. This array is consecutive in memory. The latter part for the ghost dofs is called the \emph{ghost buffer}. 

The local dofs in every subdomain are numbered according to the layout of this buffer. \Cref{fig:1d_nodes} (b) shows the local dof numbering on the three ranks. It proceeds through all non-ghost dofs followed by the ghost dofs. A global numbering of all dofs is given in \Cref{fig:1d_nodes} (a). It is needed, if global operations have to be performed with the \Vec{}, e.g., computing matrix vector products.

In the following, we outline how the finite element stiffness and mass matrices are assembled in parallel.
The algorithm proceeds by iterating over the elements of the mesh. The contributions to the matrix, i.e., the \say{element matrices}, are computed at the dofs of every element. 
Additional material data, such as values of a diffusion tensor, may be stored at the dofs and are used in these computations.

The step of assembling the global matrix entries adds up the contributions of all elements that are adjacent to every dof.
This includes a parallel reduction operation for the ghost dofs, which contribute to matrix entries that are owned by a different subdomain.

% Petsc ghosts, Petsc communication functions
PETSc provides specific functionality for the two required communication operations: (i) gathering data into the ghost buffers on every rank, i.e., communicating the values from the owning rank to the ghost buffers at all other ranks, where the respective dofs are ghosts, and (ii) the global reduction of values between ghost and non-ghost dofs, i.e., communicating the values from the ghost buffers back to the one rank, where they are non-ghosts and adding their values to the values present at the respective rank.

%First, communicating the values from the owning rank to the ghost buffers at all other ranks where the respective dofs are ghosts. 
%Second, . \Cref{fig:1d_comm1} and \cref{fig:1d_comm2} visualize the data flow in the two operations. 

\begin{figure}%
  \centering%
  \begin{subfigure}[t]{0.45\textwidth}%
    \centering%
    \includegraphics[height=6cm]{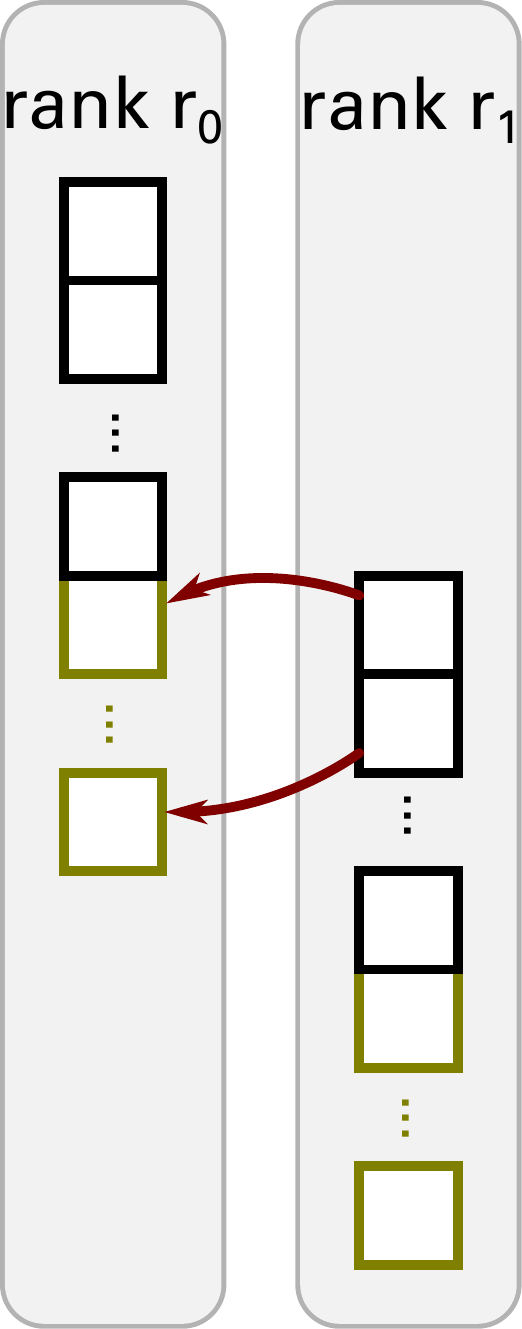}
    \caption{\code{startGhostManipulation()}: Communication of values at $r_1$ to the ghost buffer at $r_0$.}%
    \label{fig:1d_comm1}%
  \end{subfigure}
  \quad
  \begin{subfigure}[t]{0.45\textwidth}%
    \centering%
    \includegraphics[height=6cm]{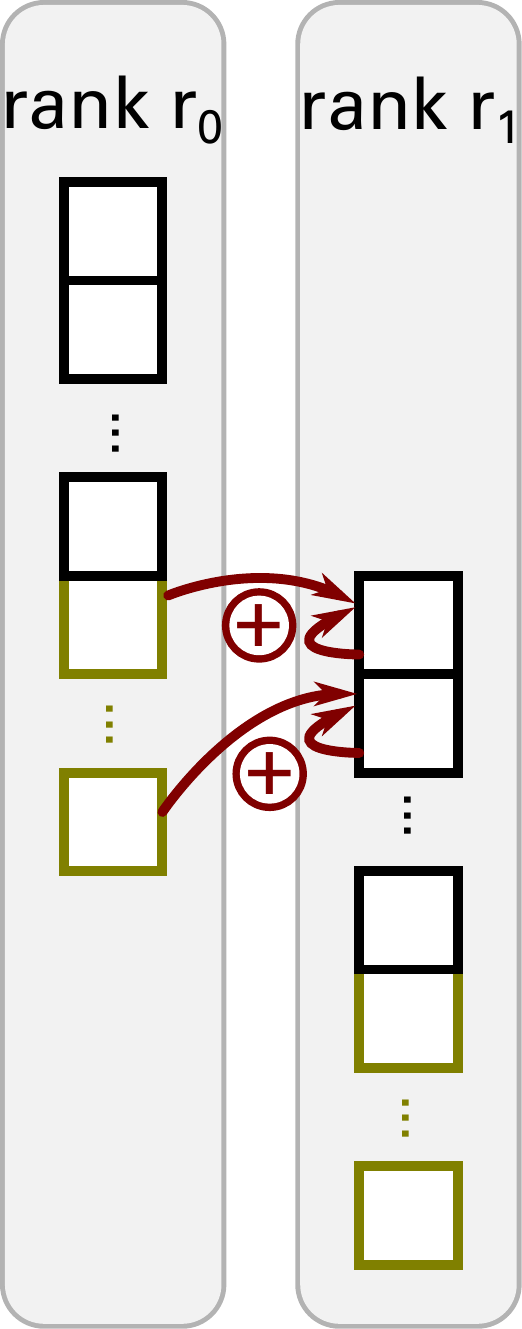}
    \caption{\code{finishGhostManipulation()}: Communication and addition of ghost values at $r_0$ to the non-ghost dofs at $r_1$.}%
    \label{fig:1d_comm2}%
  \end{subfigure}
  \caption{Communication operations for ghost values in an example with two ranks $r_0$ and $r_1$. Depicted are the vectors of local storage for non-ghost (black) and ghost values (green). The red arrows indicate the data transfer. The visualized operations are needed, e.g., in the assembly of finite element stiffness and mass matrices and their application on the vector of unknowns.}%
  \label{fig:1d_comm}%
\end{figure}%
The \opendihu{} code wraps the two operations in the methods \code{startGhostManipulation()} and \code{finishGhostManipulation()}. After the call to \code{startGhostManipulation()}, the vector can be accessed using the local dof numbering. Values of the local dofs including ghosts can be retrieved, inserted or added as needed, e.g., during FEM matrix assembly. 
After a concluding call to \code{finishGhostManipulation()}, the vector is in a valid global state. Then, global operations such as adding or scaling the whole vector, computing a norm or a matrix vector product can be performed by using the respective PETSc routines. For these operations, the partitioning is transparent, i.e., the calls are the same for serial and parallel execution. Individual entries of the vector can now be accessed using a global numbering. However, every process can still only access the non-ghost dofs owned by its subdomain. The two operations can be interpreted as switching between a local and a global view on the vector object.

One thing to note is that calling \code{startGhostManipulation()} and \code{finishGhostManipu-}\\\code{lation()} directly in sequence changes the values of the vector. The reason is that during the call to \code{startGhostManipulation()}, the ghost buffers get filled with the ghost values from other subdomains. Then, by \code{finishGhostManipulation()} the values in every ghost buffer get summed up and added to the value at the corresponding non-ghost dof. Thus, these dof values finally have a multiple of their initial value.  This is usually not intended. Thus, between the calls to the two methods either all ghost values have to be set, such as during computation of the stiffness matrix. Or, if the ghost values were only needed for reading instead of updating them, the ghost buffers have to be cleared to zero. For the latter, a helper method \code{zeroGhostBuffer()} exists. A typical usage is therefore to call \code{startGhostManipulation()}, then operate on the local dof values including ghosts, and then finish with \code{zeroGhostBuffer()} and \code{finishGhostManipulation()}.

\subsection{Numbering Schemes for Nodes and Degrees of Freedom}\label{sec:numbering_schemes_for}

% partitioned 2D vector, domain numbering, for structured grid with linear ansatz functions
PETSc's definition of the local value buffer used by \Vec{} objects dictates the local numbering scheme of dofs on meshes of any dimensionality. While, for 1D meshes, the numbering as given in \cref{fig:1d_nodes} seems natural, for 2D and 3D meshes, a more complex ordering of local dofs is needed.

\begin{figure}%
  \centering%
  \includegraphics[width=0.5\textwidth]{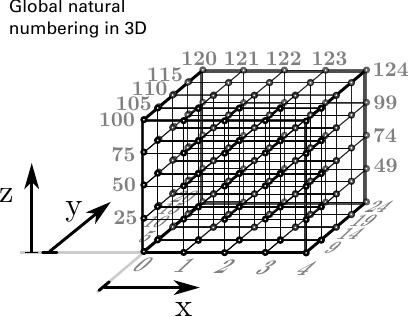}%
  \caption{Global natural numbering of nodes in a mesh with $4 \times 4 \times 4$ linear elements.}%
  \label{fig:dof_numbering_global_3D}%
\end{figure}

Three different numbering schemes for nodes and dofs exist within \opendihu{}. They are visualized in \cref{fig:dof_numbering} for a 2D mesh. The first is the \emph{global natural} numbering scheme, which numbers all $n_\text{global} = N^\text{dofs}_x \times N^\text{dofs}_y \times N^\text{dofs}_z$ global dofs in the structured mesh. It starts with zero and iterates through the mesh using the triple of coordinate indices $(i,j,k)$ for the $x$, $y$ and $z$ axis with the ranges $i \in \{0,\dots,N^\text{dofs}_x-1\}$, $j \in \{0,\dots,N^\text{dofs}_y-1\}$ and $k \in \{0,\dots,N^\text{dofs}_z-1\}$. The numbering proceeds fastest in $x$ or $i$ direction, then in $y$ or $j$ direction and then in $z$ or $k$ direction. Examples are shown in \cref{fig:dof_numbering} (a) for a 2D mesh  and in \cref{fig:dof_numbering_global_3D} for a 3D mesh.

The intention of this first numbering is to facilitate the problem description by the user. If values for a variable in the whole computational domain should be specified, the order of the given value list will be interpreted according to this numbering. Boundary conditions can be given for some dofs by simply specifying the corresponding dof numbers in global natural numbering. The advantage is that this numbering scheme is easily understandable from a users' perspective and independent of the partitioning.

The second numbering scheme is the \emph{local} numbering. An example is given in \cref{fig:dof_numbering} (b). It specifies the order of dofs in the local PETSc \Vec{} and is defined locally on every subdomain for the non-ghost and ghost dofs. At first, all non-ghost dofs are numbered with the order equal to the one in the global natural scheme. Then, all ghost dofs are numbered, again in the order of the global natural scheme. This numbering has the counter-intuitive property of jumps between some neighboring nodes.

The third numbering scheme is called \emph{global PETSc} numbering and is defined by PETSc. It is the numbering used to access global \Vec{}s. It is also the ordering of the rows and columns of matrices. The numbering starts with all local non-ghost numbers on rank 0, then proceeds over all non-ghost numbers of rank 1 and continues like this for all remaining ranks. An example for this numbering is given in \cref{fig:dof_numbering} (c). The portions of local dofs for the different ranks are indicated by the grid of red and black colors. This numbering depends on the partitioning and, thus, on the number of processes. For serial execution it is identical to the global natural numbering.

\subsection{Parallel Data Structures in \Opendihu{}}\label{sec:paralel_data_structures_in_opendihu}

\begin{figure}%
  \centering%
  \includegraphics[width=0.8\textwidth]{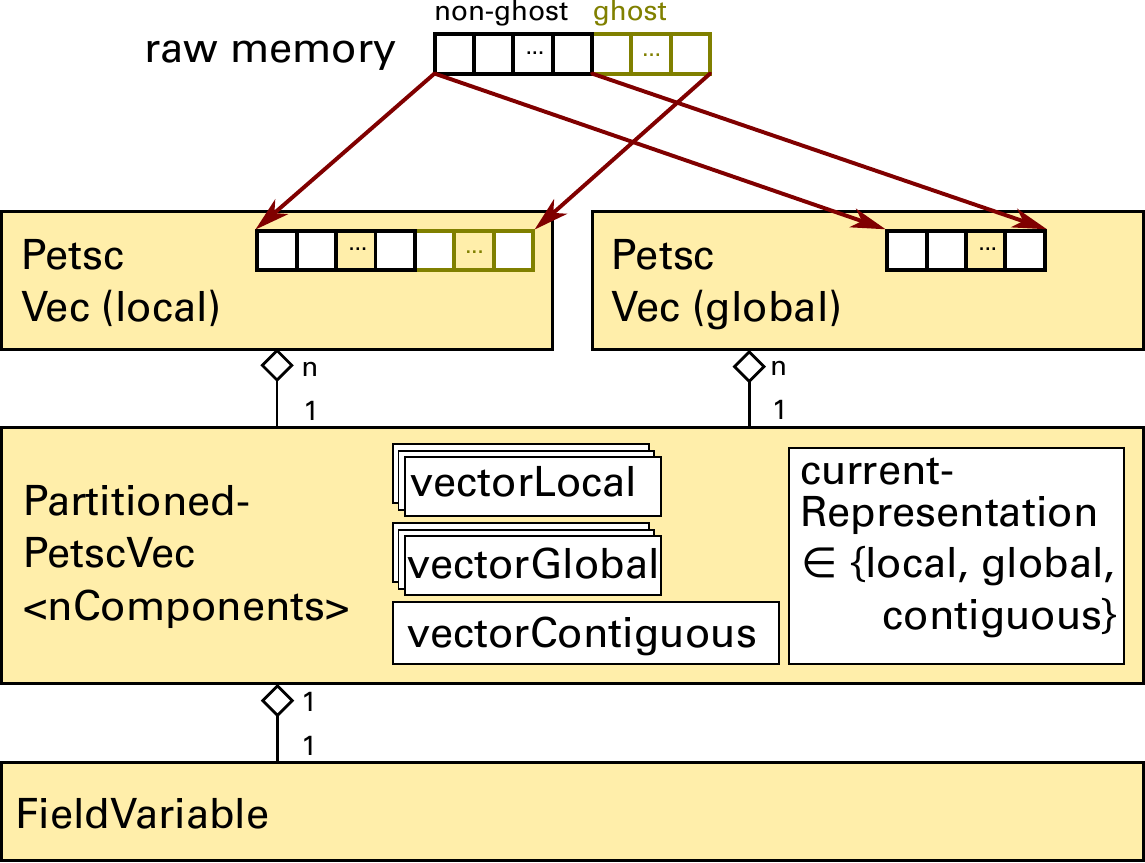}%
  \caption{Classes in \opendihu{} that represent vectors in parallel execution. The abstraction layer increases from raw memory at the top to the \code{FieldVariable} at the bottom.}%
  \label{fig:partitioned_petsc_vec}%
\end{figure}

All operations on scalar and vector fields in the simulation break down to manipulating variables of the \Vec{} type provided by PETSc. 
Because this involves low level operations such as working with different numbering schemes and communicating ghost values, an abstraction layer on a higher level is implemented in \opendihu{}. The data handling classes are visualized in \cref{fig:partitioned_petsc_vec} with the data representation in raw memory at the top and increasing abstraction towards the bottom of the figure.

Depending on whether the local or the global PETSc numbering scheme describes the data, two different objects of the \Vec{} type are used: one local and one global \Vec{}. In \cref{fig:partitioned_petsc_vec}, these two \Vec{}s are represented by the \emph{PETSc Vec (local)} and \emph{PETSc Vec (global)} boxes. At any time, only one of these is in a valid state and allows to manipulate the data. Internally, both PETSc \Vec{}s use the same memory to store their data. However, as shown at the top of \cref{fig:partitioned_petsc_vec}, the memory range of the local \Vec{}'s buffer includes the ghost buffer, which is never accessed by the global \Vec{}. As mentioned, PETSc functions are available to switch the valid state between the two \Vec{}'s, involving communication of ghost values. 
%Because of the shared memory, no costly value copy operation is needed for this action.

The next abstracting class is \emphcode{PartitionedPetscVec<nComponents>}. It represents a discretized vector field $\bfv : \Omega \to \R^c$ with a given number of components $c$. The number of components $c$ is a template parameter to the class, which has to be specified at compile time. 
An example for such vector fields is the \emph{geometry field} with $c=3$, which is defined for every mesh and specifies the node positions. Another example is the solution variable of the considered problem. For scalar problems such as the Laplace equation, it has $c=1$ component, for vector-valued problems, e.g., static elasticity, it has $c=3$ components, namely the displacements in $x$, $y$ and $z$ direction. For the subcellular model of Shorten \cite{Shorten2007}, the solution variable, i.e., the vector of states has $c=57$ components.

For each component, a separate pair of local and global \Vec{}s is stored in the variables \emphcode{vectorLocal} and \emphcode{vectorGlobal}. The global number of entries in each of the \Vec{}s is given by the number $n_\text{global}$ of dofs in the mesh that discretizes the domain $\Omega$. Thus, the memory layout of such a multi-component vector is struct-of-array (SoA). 

Besides \code{vectorLocal} and \code{vectorGlobal}, a third variable \emphcode{vectorContiguous} of type \Vec{} exists in the class \code{PartitionedPetscVec}. It contains the concatenated values of all component vectors in \code{vectorGlobal}. Its size is therefore $c\cdot n_\text{global}$ and the layout is again SoA but stored in a single \Vec{}.

This representation is chosen when a timestepping scheme operates on a state vector with multiple components. An example is the solution of multiple instances of a subcellular problem. Here, the dofs in the mesh correspond to the individual instances and the components are the state variables of the system of ODEs. Thus, the contiguous vector begins with the values of the first state for all instances, then stores the values of the second state for all instances, etc. If the right-hand side of the system of ODEs is evaluated together for all instances, this memory layout is very efficient as it leads to a cache aware access pattern.

Only one of the three vectors \code{vectorLocal}, \code{vectorGlobal} and \code{vectorContiguous} is valid at any time and can be used to retrieve or update the vector values. 
A state variable \emphcode{currentRepresentation} in \code{PartitionedPetscVec<nComponents>} indicates which one that is. The state and the \Vec{} variables are encapsulated and hidden in the class, i.e., not directly accessible from outside. Instead, the class provides data access methods and ways to change the internal representation. For example, calls to \code{startGhostManipulation()} and \code{finishGhostManipulation()} change the representation from global to local and from local to global, respectively.
Thus, it is ensured that only the current valid representation gets accessed at any time.

As noted before, the change between local and global representation does not involve data copying because of the shared physical data structures.
When the representation is changed from local to contiguous, the $c$ sets of values of the \code{vectorLocal} variables have to be copied into the buffer of \code{vectorContiguous}. This operation is performed by copying memory blocks  \mbox{(\code{memcpy})} instead of the slower iteration over all values and the value-wise copy. The reverse change from contiguous back to local representation happens analogously. Thus, the change between all representations is fast. Despite occurring often during transient simulations, profiling of simulations has shown negligible runtime for the action of switching between these representations.

The top level class in the value storage hierarchy as shown in \cref{fig:partitioned_petsc_vec} is the \emphcode{Field Variable}, which contains a \code{PartitionedPetscVec} and adds numerous methods to facilitate access to the data container. Model formulations use this class to manipulate scalar and vector fields. At the same time, the underlying global PETSc \Vec{} can still be obtained from a \code{FieldVariable}. Vector operations such as addition, norms and matrix-vector products are performed using the low-level PETSc functions on the global \Vec{} obtained from the \code{FieldVariable}s.

\subsection{Discussion of Several Design Decisions}
% design decision compile time c and separate vectors, comparison with OpenCMISS iron
In the following, some of the design decisions in \crefrange{sec:oragnization_of_parallel_partitioned_data}{sec:paralel_data_structures_in_opendihu} are discussed.
% numbering, not using DMDA
In the present code, PETSc functionality is used for value storage and organization of ghost values transfer. 
The employed PETSc data model naturally corresponds to a 1D mesh.
The representation of arbitrary dimensional meshes is added by \opendihu{} and involves the presented local and global PETSc numberings. 

PETSc also provides the management of abstract 2D and 3D mesh objects in the \code{DM} (data management) module. It allows to automatically create a partitioning with local numberings and data vectors. However, the mesh always has a symmetric ghost node layout, where ghost layers are present on all faces of a subdomain (box stencil) or also at diagonal neighbors (star stencil). This partitioning layout is based on distributing the nodes of the mesh to all processes. It is needed, e.g., for Finite Difference computations. For the finite element method, however, we need an element based partitioning with ghost layers only on one end of the mesh per coordinate direction. Therefore, we do not use this functionality of PETSc and instead implemented the numberings for 2D and 3D meshes on our own.

% layout with multiple vectors, not one long vector
Another choice was made regarding the data layout in the \code{PartitionedPetscVec} class. Instead of an interleaved storage of the component values in one long \Vec{} in array-of-struct (AoS) memory layout, one separate \Vec{} for each component is stored, which corresponds to SoA layout. 
%The reason for this is better performance as discussed earlier. 
Thereby, the implementation differs from OpenCMISS Iron, which is also based on PETSc, but uses the AoS  approach. 

In Iron, not only the values of multiple components, but actually the values of multiple field variable are combined into a single \Vec{}. A local numbering is defined that enumerates all components, all dofs, and all field variables. Differences to our code are, that Iron uses unstructured meshes, which additionally are allowed to contain different types of elements in a single mesh. Field variables can be defined with dofs either  associated with nodes or with elements. 
All these possible variations are accounted for by the local numbering. 
The construction of the numbering is, thus, a complex process. Iron implements it by a loop over all $n_\text{global}$ dofs of the domain. The same loop is executed in parallel by all $n_\text{proc}$ processes. The runtime complexity of this approach is $\O(n_\text{global})$ regardless of the partitioning. 
In contrast, \opendihu{} constructs its local numberings separately on each process and only iterates over the $n_\text{local\_with\_ghosts}$ dofs, leading to a runtime complexity of $\O(n_\text{local\_with\_ghosts}) = \O(n_\text{global} / n_\text{proc})$. In a weak scaling experiment with constant relation $n_\text{global} / n_\text{proc}$, the approach of Iron yields infinite runtime in the limit for $n_\text{global} \to \infty$, whereas the runtime in the approach of \opendihu{} stays constant.

For \opendihu{}, the AoS approach with separate \Vec{}s was chosen for three reasons. First, it is more cache efficient than the alternative during the computation of the subcellular model, as explained in \cref{sec:paralel_data_structures_in_opendihu}.

Second, the AoS structure is easier, and it allows to treat the components separately, which makes modular code possible. Only a single local dof numbering has to be constructed per mesh, and it can be reused for all components of all field variables.

Third, it is possible to extract one component of a vector-valued field variable and place it into another, scalar field variable without copying. This is used during the solution of the monodomain equations given in \cref{eq:monodomain}. There, the subcellular models have a vector-valued solution variable and the diffusion problem needs a scalar solution variable that consists of the first component of the vector-valued variable of the subcellular model. This first component is the transmembrane voltage $V_m$. The program needs to switch between these two required vectors in every timestep of the splitting scheme. Only with the chosen representation by multiple \Vec{}s, the \Vec{} for the particular component can be efficiently exchanged between the two field variables without an expensive copy operation.

% nComponents compile time
Another design decision was to make the number $c$ of components fixed at compile time. 
Upon construction of a new \code{FieldVariable}, its number of components needs to be known. Typically, this is the case and does not pose any restriction.
The main advantage is that local variables that hold all components for a given dof can be allocated on the stack instead of a much slower dynamic allocation on the heap. For example, in a dynamic solid mechanics problem, the solution \code{FieldVariable} contains three components each for displacements and velocities plus one component for the pressure, in total $c=7$ components. The program can use static arrays with seven entries as temporary variables to handle these values in various computations. If the number of components was not fixed at compile time, a costly dynamic allocation of the seven components would be needed wherever values of the \code{FieldVariable} are retrieved. 
In addition, with a compile-time fixed $c$ the compiler knows the size of the arrays and can perform automatic optimizations such as vectorization and loop unrolling.

The C++ implementation of \code{FieldVariable}s and all other constructs that depend on the number of components is generic, as the  $c$ value is a template argument. Specializations for particular numbers of components such as for the scalar case $c=1$ are possible using \emph{template specialization}.
This flexibility while using object orientation is an advantage over codes using procedural programming languages such as the Fortran standard used by OpenCMISS Iron. It contributes to the extensibility design goal of \opendihu.

\subsection{Implemented Basis Functions}
% different mesh types and ansatz functions, numberings

In the FEM, the number of dofs and nodes per element depends on the chosen ansatz functions or basis functions. \Opendihu{} supports linear and quadratic Lagrange as well as cubic Hermite basis functions. 
\Cref{tab:ansatz_functions_mesh} shows these three sets of functions and the resulting node configuration of an element in a 1D, 2D and 3D mesh.
Profiling showed that evaluation of the basis functions contributes most to the runtime during calculation of the stiffness matrix. Therefore, care was taken to choose the formulations of the basis functions among different factorizations that need the least operations. Those are listed in \cref{tab:ansatz_functions_mesh}.

\begin{table}
  \centering%
  \begin{tabular}{|ll|lll|}
    \hline
    \multicolumn{2}{|l|}{Ansatz functions} \hspace*{43.82mm} & \multicolumn{3}{l|}{Element shapes}  \\
                                         && 1D \hspace*{11mm} & 2D \hspace*{11mm}& 3D\hspace*{11.75mm}\\
    \hline
  \end{tabular}
  \begin{tabular}{|ll|lll|}
    \vspace*{-5mm}
      &&&&\\
      \includegraphics[width=2cm]{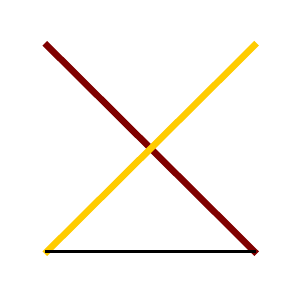}&
      \begin{minipage}{5cm}
        \vspace{-2cm}
        $\phi_0(\xi) = 1-\xi$,\\[2mm] 
        $\phi_1(\xi) = \xi$
      \end{minipage}
       &
      \includegraphics[width=15mm]{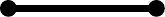} & 
      \includegraphics[width=15mm]{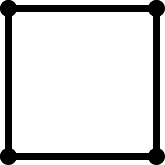} &
      \includegraphics[width=2cm]{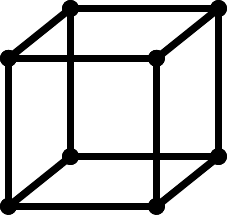} \\
    \hline
      \includegraphics[width=2cm]{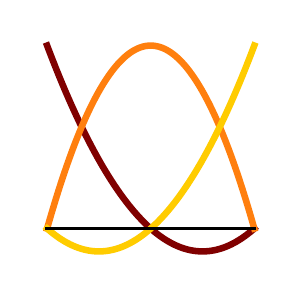}&
      \begin{minipage}{5cm}
        \vspace{-15mm}
        $\phi_0(\xi) = (2\,\xi - 1) \,(\xi-1), $ \\[2mm]
        $\phi_1(\xi) = 4\,(\xi - \xi^2),$    \\[2mm] 
        $\phi_2(\xi) = 2\,\xi^2 - \xi$
      \end{minipage}
       &
      \includegraphics[width=15mm]{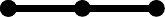} & 
      \includegraphics[width=15mm]{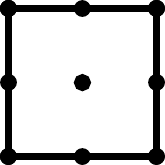} &
      \includegraphics[width=2cm]{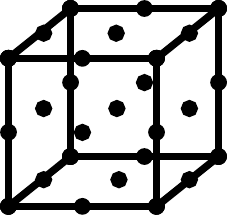} \vspace{2mm}\\
    \hline
      \includegraphics[width=2cm]{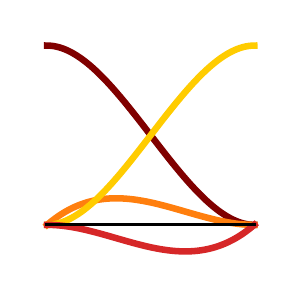}&
      \begin{minipage}{5cm}
        \vspace{-1cm}
        $\phi_0(\xi) = 2\,\xi^3- 3\,\xi^2 + 1, $\\[2mm]
        $\phi_1(\xi) = \xi\,(\xi-1)^2,         $\\[2mm]
        $\phi_2(\xi) = \xi^2 \, (3 - 2\,\xi),  $\\[2mm]
        $\phi_3(\xi) = \xi^2 \, (\xi-1)  $
      \end{minipage}
       &
      \includegraphics[width=15mm]{images/implementation/ansatz_mesh_1.pdf} & 
      \includegraphics[width=15mm]{images/implementation/ansatz_mesh_2.pdf} &
      \includegraphics[width=2cm]{images/implementation/ansatz_mesh_3.pdf} \vspace{2mm}\\
    \hline
  \end{tabular}
  \caption{Finite element ansatz functions and resulting element shapes of hexahedral meshes in 1D, 2D and 3D. From top to bottom: Linear Lagrange, quadratic Lagrange and cubic Hermite ansatz functions.}%
  \label{tab:ansatz_functions_mesh}%
\end{table}

In the program, every basis function is defined by a class that specifies the constant, static numbers $n_\text{dofs\_per\_basis}$ of dofs per 1D element and $n_\text{dofs\_per\_node}$ of dofs per node. Furthermore, the actual functions and their first derivatives are implemented. 
All algorithms working with meshes or ansatz functions only use this information given in the basis function class. Therefore, it is easily possible to introduce new nodal ansatz functions as needed, e.g., a cubic Lagrange basis, by accordingly defining a new class.

If any Lagrange basis is used, every node has exactly one dof, i.e., $n_\text{dofs\_per\_node}=1$. With the 1D Hermite basis, every node has $n_\text{dofs\_per\_node}=2$ dofs, one that describes the function value and one that defines the derivative at the particular node. For higher dimensional meshes, the bases are constructed by the tensor product approach. For 2D meshes, this results in four and for 3D meshes in eight dofs per node for the Hermite basis. For example, at a node at location $\bfx$ in a 2D mesh, the first dof describes the value $f(\bfx)$ of a scalar field $f:\Omega \to \R$ and the others relate to the derivatives $\partial_x f(\bfx)$, $\partial_y f(\bfx)$ and $\partial_{xy} f(\bfx)$. 

Note that the dof values for derivatives only match the real derivatives of $f$ in meshes with unity mesh widths. In a general, the derivatives are scaled by the element lengths. In general meshes with varying element sizes, the represented FE solution $f$ is not continuously differentiable at element boundaries, i.e., $f\in\CC^0(\Omega,\R)$.

For quadratic Lagrange and cubic Hermite basis functions, the numbering schemes presented in \cref{sec:numbering_schemes_for} have to be adjusted, such that, at every node, all dofs are enumerated in sequence before the numbering continues at the next node.

\subsection{Implemented Types of Meshes}
\begin{figure}%
  \centering%
  \def\svgwidth{0.5\textwidth}
  %% Creator: Inkscape inkscape 0.92.3, www.inkscape.org
%% PDF/EPS/PS + LaTeX output extension by Johan Engelen, 2010
%% Accompanies image file 'meshes.pdf' (pdf, eps, ps)
%%
%% To include the image in your LaTeX document, write
%%   \input{<filename>.pdf_tex}
%%  instead of
%%   \includegraphics{<filename>.pdf}
%% To scale the image, write
%%   \def\svgwidth{<desired width>}
%%   \input{<filename>.pdf_tex}
%%  instead of
%%   \includegraphics[width=<desired width>]{<filename>.pdf}
%%
%% Images with a different path to the parent latex file can
%% be accessed with the `import' package (which may need to be
%% installed) using
%%   \usepackage{import}
%% in the preamble, and then including the image with
%%   \import{<path to file>}{<filename>.pdf_tex}
%% Alternatively, one can specify
%%   \graphicspath{{<path to file>/}}
%% 
%% For more information, please see info/svg-inkscape on CTAN:
%%   http://tug.ctan.org/tex-archive/info/svg-inkscape
%%
\begingroup%
  \makeatletter%
  \providecommand\color[2][]{%
    \errmessage{(Inkscape) Color is used for the text in Inkscape, but the package 'color.sty' is not loaded}%
    \renewcommand\color[2][]{}%
  }%
  \providecommand\transparent[1]{%
    \errmessage{(Inkscape) Transparency is used (non-zero) for the text in Inkscape, but the package 'transparent.sty' is not loaded}%
    \renewcommand\transparent[1]{}%
  }%
  \providecommand\rotatebox[2]{#2}%
  \newcommand*\fsize{\dimexpr\f@size pt\relax}%
  \newcommand*\lineheight[1]{\fontsize{\fsize}{#1\fsize}\selectfont}%
  \ifx\svgwidth\undefined%
    \setlength{\unitlength}{199.66479204bp}%
    \ifx\svgscale\undefined%
      \relax%
    \else%
      \setlength{\unitlength}{\unitlength * \real{\svgscale}}%
    \fi%
  \else%
    \setlength{\unitlength}{\svgwidth}%
  \fi%
  \global\let\svgwidth\undefined%
  \global\let\svgscale\undefined%
  \makeatother%
  \begin{picture}(1,1.54592749)%
    \lineheight{1}%
    \setlength\tabcolsep{0pt}%
    \put(0,0){\includegraphics[width=\unitlength,page=1]{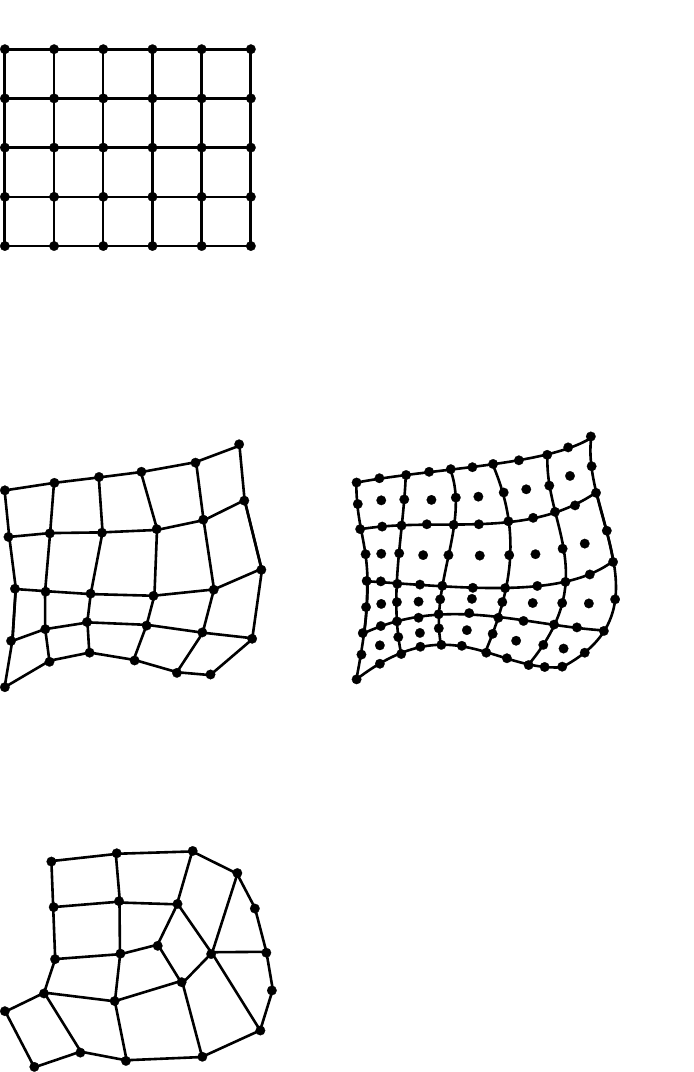}}%
    \put(-0.00173729,1.51282513){\color[rgb]{0,0,0}\makebox(0,0)[lt]{\lineheight{1.25}\smash{\begin{tabular}[t]{l}\textbf{\code{Mesh::}\textbf{RegularFixed}\textbf{OfDimension<2>}}\end{tabular}}}}%
    \put(-0.00173729,0.93430057){\color[rgb]{0,0,0}\makebox(0,0)[lt]{\lineheight{1.25}\smash{\begin{tabular}[t]{l}\textbf{\code{Mesh::}\textbf{StructuredDeformable}\textbf{OfDimension<2>}}\end{tabular}}}}%
    \put(-0.00173729,0.35565382){\color[rgb]{0,0,0}\makebox(0,0)[lt]{\lineheight{1.25}\smash{\begin{tabular}[t]{l}\textbf{\code{Mesh::}\textbf{UnstructuredDeformable}\textbf{OfDimension<2>}}\end{tabular}}}}%
    \put(0,0){\includegraphics[width=\unitlength,page=2]{meshes.pdf}}%
  \end{picture}%
\endgroup%
  \caption{The three implemented mesh types in \opendihu{}, each time for 2D linear Lagrange or Hermite ansatz functions (left) and for 2D quadratic Lagrange ansatz functions (right).}%
  \label{fig:meshe_types}%
\end{figure}

Meshes of different types can be selected independently of the choice of basis functions. Three types are supported. \Cref{fig:meshe_types} visualizes meshes of these types with linear and quadratic elements.

The first type is \code{Mesh::RegularFixedOfDimension<D>} where \code{D} $\in \{1,2,3\}$ is a compile-time constant of the dimension. This type describes a rectilinear, regular structured mesh that is defined by a fixed mesh width $h$ in all coordinate directions. This mesh is \say{fixed}, which means that the positions of the nodes cannot change after the mesh object was created. 
Regular fixed meshes describe a line (1D), a rectangular (2D) or a cuboid domain (3D). This mesh type exists, because such domains are often used in exemplary problems to study certain effects independently of the shape of the domain. A regular fixed mesh can be easily configured by specifying origin point coordinates, mesh widths and number of elements. For this mesh type, matrix assembly in the FEM is simplified and more efficient by using precomputed stencils.

The second mesh type is \code{Mesh::StructuredDeformableOfDimension<D>}. The structured deformable mesh is a generalization of the regular fixed mesh. The mesh again has a structure of $N^\text{el}_x \times N^\text{el}_y \times N^\text{el}_z$ elements. Contrary to the regular fixed mesh, the nodes can now have arbitrary positions. In the name of this mesh, \say{deformable} indicates that the node locations can be changed over time. Thus, this mesh type is usable in dynamic solid mechanics problems, where the domain deforms over time. If the user wants to configure a mesh of this type, they either have to provide the same information as for regular fixed meshes---then, a mesh with fixed mesh width will be created---or they provide the positions of all nodes, yielding an arbitrarily shaped domain as shown in \cref{fig:meshe_types}.

The third mesh type is \code{Mesh::UnstructuredDeformableOfDimension<D>}. In contrast to the two other types, this mesh is unstructured implying that element adjacency is no longer given implicitly.
The example at the lower third of \cref{fig:meshe_types} shows capabilities of this mesh type: The overall shape of the domain is not restricted to resemble a rectangle. Protruding parts like the element at the bottom left are possible. Furthermore, not every node needs to be adjacent to exactly four elements in 2D. The example shows nodes with three and five adjacent elements that allow to properly approximate the round shape of the right side of the domain. The mesh is again \say{deformable}, which means that it can be used for elasticity problems.
In order to configure such a mesh, the node positions have to be specified, similar to a structured deformable mesh. Additionally, the elements with links to their corresponding nodes have to be given.
\Opendihu{} implements a second possibility to specify these meshes. A pair of \code{exelem} and \code{exnode} files, which are common in the OpenCMISS community, can be loaded.

A disadvantage of unstructured meshes is that the simple parallel partitioning scheme of subdividing the domain according to element index ranges is not applicable. Instead, the set of elements for every subdomain needs to be computed individually. Typically, this is done using graph partitioning methods in order to minimize subdomain border lengths while ensuring equal subdomain sizes. Another disadvantage is that information about neighbor elements and neighbor subdomains has to be stored explicitly, while it is given implicitly in structured meshes.
For these reasons, unstructured meshes can be used in \opendihu{} only for serial computation. The construction of parallel partitionings is only possible with the other two, structured mesh types.

The choice, which mesh type to use in a simulation, has to be made at compile time. A simulation program can be easily compiled for different meshes by substituting the type in the main C++ source file.
By proper abstraction in the code, all implemented algorithms are independent of the used mesh type when run in serial. Some algorithms, e.g., streamline tracing, are specialized for structured meshes to exploit the structure and lead to more efficient code. Unit tests ensure the correct solution of a Laplace problem with all combinations of mesh type, dimensionality and ansatz function.

\subsection{Composite Meshes}\label{sec:composite_meshes}

\begin{figure}%
  \centering%
  \begin{subfigure}[t]{0.45\textwidth}%
    \centering%
    \def\svgwidth{0.7\textwidth}%
    %% Creator: Inkscape inkscape 0.92.3, www.inkscape.org
%% PDF/EPS/PS + LaTeX output extension by Johan Engelen, 2010
%% Accompanies image file 'meshes_composite.pdf' (pdf, eps, ps)
%%
%% To include the image in your LaTeX document, write
%%   \input{<filename>.pdf_tex}
%%  instead of
%%   \includegraphics{<filename>.pdf}
%% To scale the image, write
%%   \def\svgwidth{<desired width>}
%%   \input{<filename>.pdf_tex}
%%  instead of
%%   \includegraphics[width=<desired width>]{<filename>.pdf}
%%
%% Images with a different path to the parent latex file can
%% be accessed with the `import' package (which may need to be
%% installed) using
%%   \usepackage{import}
%% in the preamble, and then including the image with
%%   \import{<path to file>}{<filename>.pdf_tex}
%% Alternatively, one can specify
%%   \graphicspath{{<path to file>/}}
%% 
%% For more information, please see info/svg-inkscape on CTAN:
%%   http://tug.ctan.org/tex-archive/info/svg-inkscape
%%
\begingroup%
  \makeatletter%
  \providecommand\color[2][]{%
    \errmessage{(Inkscape) Color is used for the text in Inkscape, but the package 'color.sty' is not loaded}%
    \renewcommand\color[2][]{}%
  }%
  \providecommand\transparent[1]{%
    \errmessage{(Inkscape) Transparency is used (non-zero) for the text in Inkscape, but the package 'transparent.sty' is not loaded}%
    \renewcommand\transparent[1]{}%
  }%
  \providecommand\rotatebox[2]{#2}%
  \newcommand*\fsize{\dimexpr\f@size pt\relax}%
  \newcommand*\lineheight[1]{\fontsize{\fsize}{#1\fsize}\selectfont}%
  \ifx\svgwidth\undefined%
    \setlength{\unitlength}{78.11014911bp}%
    \ifx\svgscale\undefined%
      \relax%
    \else%
      \setlength{\unitlength}{\unitlength * \real{\svgscale}}%
    \fi%
  \else%
    \setlength{\unitlength}{\svgwidth}%
  \fi%
  \global\let\svgwidth\undefined%
  \global\let\svgscale\undefined%
  \makeatother%
  \begin{picture}(1,1.12066202)%
    \lineheight{1}%
    \setlength\tabcolsep{0pt}%
    \put(0,0){\includegraphics[width=\unitlength,page=1]{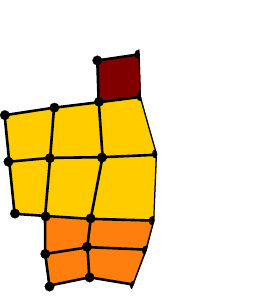}}%
    \put(-0.04395108,1.01176876){\color[rgb]{0,0,0}\makebox(0,0)[lt]{\lineheight{1.25}\smash{\begin{tabular}[t]{l}\textbf{\code{Mesh::}CompositeOfDimension<2>}\end{tabular}}}}%
    \put(0,0){\includegraphics[width=\unitlength,page=2]{meshes_composite.pdf}}%
  \end{picture}%
\endgroup%

    \caption{A composite mesh that is created from three structured meshes (different colors) and a possible subdivision for parallel partitioning (white vertical line).}%
    \label{fig:meshes_composite}%
  \end{subfigure}
  \quad
  \begin{subfigure}[t]{0.45\textwidth}%
    \centering%
    \includegraphics[width=\textwidth]{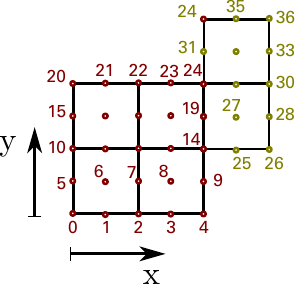}
    \caption{Numbering scheme of dofs for a composite mesh, which is created from two quadratic meshes.}%
    \label{fig:dof_numbering_structured_composite}%
  \end{subfigure}
  \caption{Examples for composite meshes that combine the advantages of structured and unstructured meshes.}%
  \label{fig:composite_meshes}%
\end{figure}

To overcome the limitations of structured meshes regarding possible domain shapes and, at the same time, preserving the advantage of efficient parallel partitioning, \emph{composite} meshes are introduced. These meshes of type \code{Mesh::CompositeOfDimension<D>} are built using multiple meshes of type \code{Mesh::StructuredDeformableOfDimension<D>}, called \emph{submeshes} in this context. The structured submeshes are positioned next to each other to form a combined single mesh on the union of the domains of all meshes. \Cref{fig:meshes_composite} shows a 2D example where three structured meshes are combined to a composite mesh. As can be seen, the submeshes can have different numbers of elements.
The nodes on the borders between touching structured meshes are shared between the individual meshes. Thus, these nodes contain only a single set of dofs like every other node in the mesh.

In the code, composite meshes reuse the implementation of structured meshes by defining different numbering schemes for nodes and dofs over the whole composite domain. The numbering of nodes starts with all nodes of the first submesh, then proceeds over all remaining nodes of the second submesh and so on, until all nodes are numbered. The numbering of dofs is analog. \Cref{fig:dof_numbering_structured_composite} shows an example with two quadratic submeshes with four and two elements. The resulting composite mesh has six elements. The node numbers in the first structured mesh are identical to the corresponding nodes in the composite mesh. The numbering continues in the set of remaining nodes of the second structured mesh and the shared nodes on the border between the meshes are skipped in the numbering, as they already have a number assigned. The shared nodes have the numbers 14, 19 and 24.

In parallel execution, this scheme is executed first on the non-ghost and then on the ghost nodes of the subdomains of all submeshes. Thus, the local numbering of the composite scheme visits the non-ghost nodes of all subdomains first before iterating over the ghost dofs on all subdomains. Thus, the ghost buffer is consecutive in memory as required by the parallel PETSc \Vec{}s.

For the construction of this numbering, the shared nodes of different submeshes, which lie at the same position, have to be determined.
The identification of shared nodes occurs according to their position in the physical domain. The distance in every coordinate direction has to be lower than the tolerance of \num{1e-5} for a pair of nodes to be considered identical and shared. The shared nodes are determined on every local subdomain of the underlying structured meshes. To correctly number ghost nodes that are shared between submeshes, communication between processes is necessary.

Using the set of shared nodes, mappings in both directions between the local numberings of the submeshes and the local and global PETSc numberings of the composite mesh are constructed. These mappings are used to transfer operations on the composite mesh to operations on the structured submeshes. Thus, every implemented algorithm can transparently work also on composite meshes.

The creation of the numbering schemes requires that neighboring elements on different submeshes are located on the same process. If this was not the case, submeshes would potentially have ghost nodes at their outer border, which does not occur in normal structured meshes and would disallow reusing their implementation.
Furthermore, the MPI communicator of the submeshes has to be the same and no subdomain can be empty. This means that a composite mesh has to be partitioned, such that every submesh is subdivided into the same number of partitions involving all processes. If these requirements are fulfilled, the parallel implementation of any algorithm on structured meshes can be reused for composite meshes.
\Cref{fig:meshes_composite} shows a valid partitioning of the exemplary composite mesh to two subdomains.

To configure composite meshes in the settings, their submeshes have to be specified as usual for structured meshes. Then, a list of all submeshes is given for the composite mesh. In parallel execution, a proper partitioning that fulfills the requirements has to be constructed in the Python script of the settings as well.

\begin{figure}%
  \centering%
  \includegraphics[width=\textwidth]{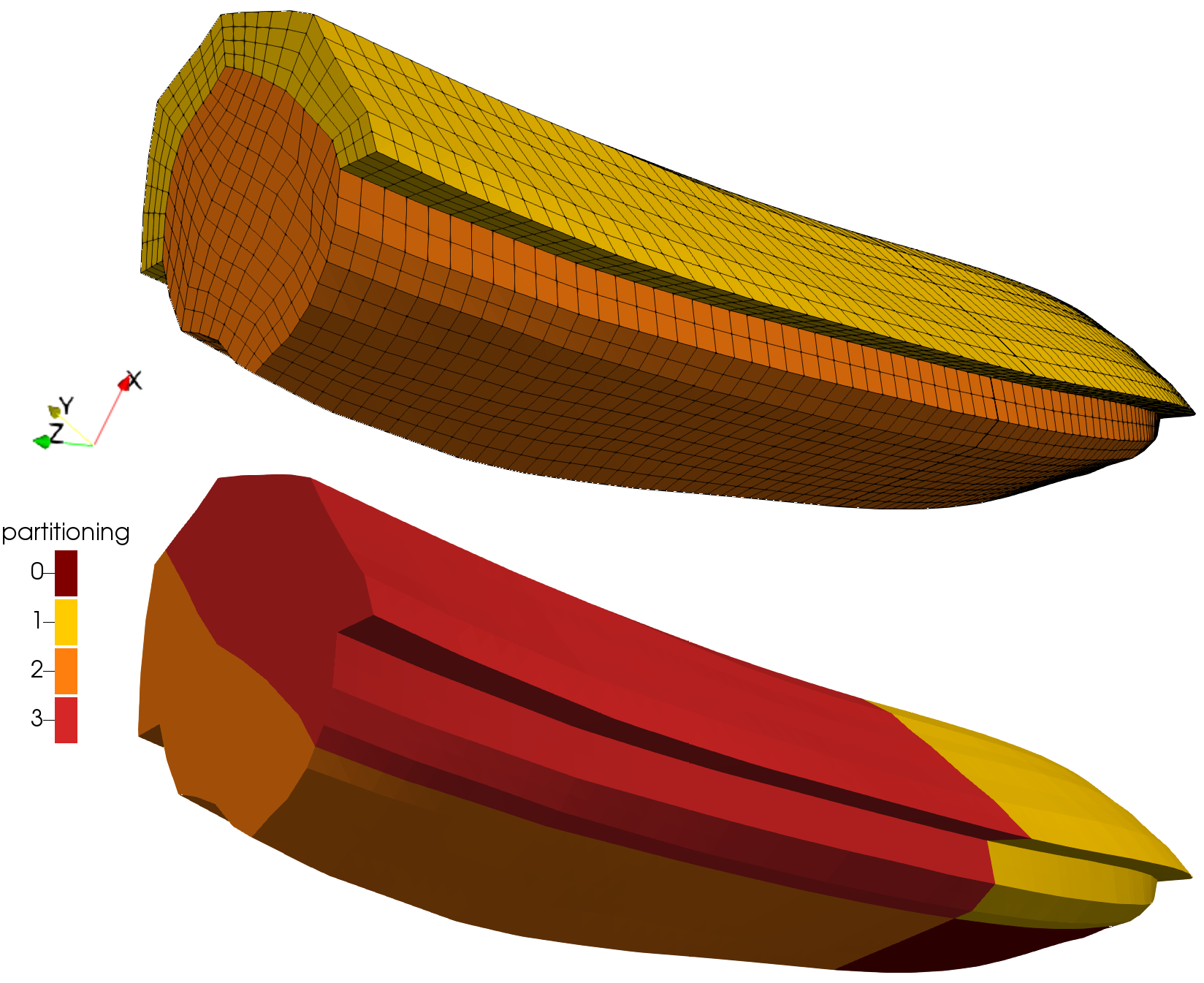}%
  \caption{Composite mesh of the biceps muscle. Top: the two structured meshes from which the composite mesh is created, bottom: partitioning to four processes.}%
  \label{fig:composite_muscle_mesh0}%
\end{figure}

An application of composite meshes is the biceps muscle with a fat and skin layer. \Cref{fig:composite_muscle_mesh0} visualizes the composite mesh. It consists of two structured submeshes for the muscle belly and the body layer on top, as visualized in the top image. The bottom image shows a partitioning to four processes. As can be seen, the domain can be split along the $x$ and $z$ coordinate axes to produce valid partitionings. Using this decomposition strategy, any number of subdomains (limited by the number of elements, though) is possible.

\section{Finite Element Matrices and Boundary Conditions}\label{sec:fem_matrices_and_bc}
Another important mathematical object besides the vector, which has to be represented in finite element simulation programs, is the matrix. Matrices are mainly needed to store the linear system of equations that results from the discretized weak formulation within the FEM. Dirichlet boundary conditions can be enforced by adjusting the system matrix.

In the following sections, the storage of matrices is discussed, an efficient, parallel algorithm to assemble the FEM system matrix is presented and evaluated and a second parallel algorithm for handling Dirichlet boundary conditions is given.

\subsection{Storage of Matrices}
% matrix storage in Petsc
The storage of matrices is delegated to PETSc, like the storage of vectors. The default sparse  matrix format of PETSc, \emph{compressed row storage (CRS)} or \say{AIJ} in PETSc notion, is used. 
The representation stores the non-zero locations and their values for every row of the matrix.

The system matrices in the FEM have as many rows and columns as there are global dofs in the system. The typical linear system of equations can be expressed as:
\begin{align*}
  \bfK\,\bfu = \bff,
\end{align*}
with system matrix $\bfK$ and the parallel vectors $\bfu$ and $\bff$ of the solution and right-hand side, respectively. The partitioning of the rows of the matrix corresponds to the partitioning of the right-hand side vector $\bff$. Thus, every rank has the complete information of a subset of lines in this matrix equation.

Every rank stores a submatrix of size $n_\text{local\_without\_ghosts} \times n_\text{global}$. In PETSc, this submatrix is composed of two blocks. The \emph{diagonal} block is a square matrix of size $(n_\text{local\_without\_ghosts})^2$ and holds only the columns of the local dofs. The rest of the columns are stored in the \emph{off-diagonal} block which is a non-square matrix in general.

The memory of these two storage blocks needs to be preallocated prior to the assignment of matrix entries. This allows PETSc to allocate the whole data storage in one chunk instead of potential reallocations for every new matrix entry. According to the documentation of PETSc, this can speed up the assembly runtime by a factor of 50 \cite{petsc-web-page}. For the preallocation, the numbers of non-zero entries per row in the two storage blocks need to be estimated. The estimated numbers need to be equal to or greater than the actual number of non-zeros per row.

The stiffness and mass matrices in the FEM have a banded non-zero structure that implies a maximum number of non-zero entries per matrix row. The value can be computed as follows:
\begin{equation}\label{eq:nonzero_estimates}
  \begin{array}{ll}
    n_\text{1D\_overlaps} = (2\,n_\text{dofs\_per\_basis} - 1)\cdot n_\text{dofs\_per\_node},\\[4mm]
    n_\text{non-zeros} = \left(n_\text{1D\_overlaps}\right)^d.
  \end{array}
\end{equation}
Here, the number $n_\text{dofs\_per\_basis}$ of dofs per 1D element is 2 and 3 for linear and quadratic Lagrange bases and 4 for cubic Hermite basis functions. The number $n_\text{dofs\_per\_node}$ of dofs per node is 1 for Lagrange basis functions and 2 for Hermite basis functions. The value $n_\text{1D\_overlaps}$ describes the number of basis functions in a 1D mesh that have overlapping support with a given basis function. By the tensor product approach, the resulting estimate $n_\text{non-zeros}$ of non-zero entries per row is computed by exponentiation of $n_\text{1D\_overlaps}$ with the dimensionality $d$.

Because no assumption can be made about how the bands of non-zero entries in the matrix are distributed to the diagonal and off-diagonal storage parts, the same value of $n_\text{non-zeros}$ is used as estimate to preallocate both the diagonal and the off-diagonal part of the local matrix storage.

\begin{figure}%
  \centering%
  \includegraphics[width=\textwidth]{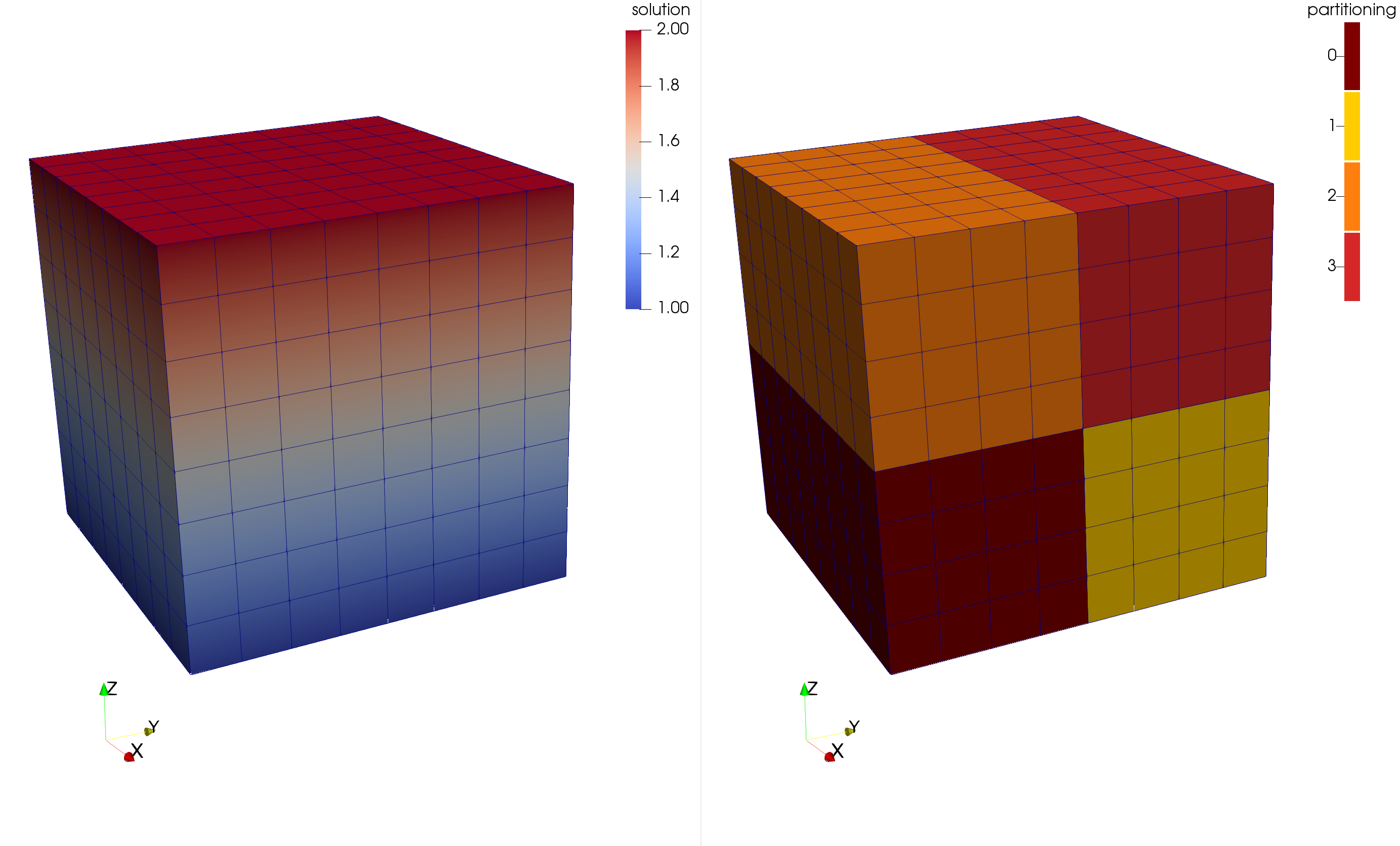}%
  \caption{Solution of the Laplace equation $ \Delta\bfu = \bff$ with prescribed values $\bfu|_\text{bottom}=1$ and $\bfu|_\text{top}=2.$ The mesh consists of $4\times 4\times 4$ quadratic elements and, thus, $9^3$ nodes. Left: Solution, right: Partitioning to four processes.}
  \label{fig:3d_laplace}%
\end{figure}

In the following, the non-zero structure of an exemplary stiffness matrix is shown. A 3D regular fixed mesh of $4\times 4 \times 4$ elements with quadratic Lagrange basis functions is considered. The Laplace equation is solved with Dirichlet boundary conditions at the bottom and top planes of the volume. The prescribed values are 1 at the bottom and 2 at the top. The solution is visualized in the left of \cref{fig:3d_laplace}. The computation is performed with four processes. \Cref{fig:3d_laplace} shows the partitioning on the right.

\begin{figure}%
  \centering%
  \includegraphics[width=0.7\textwidth]{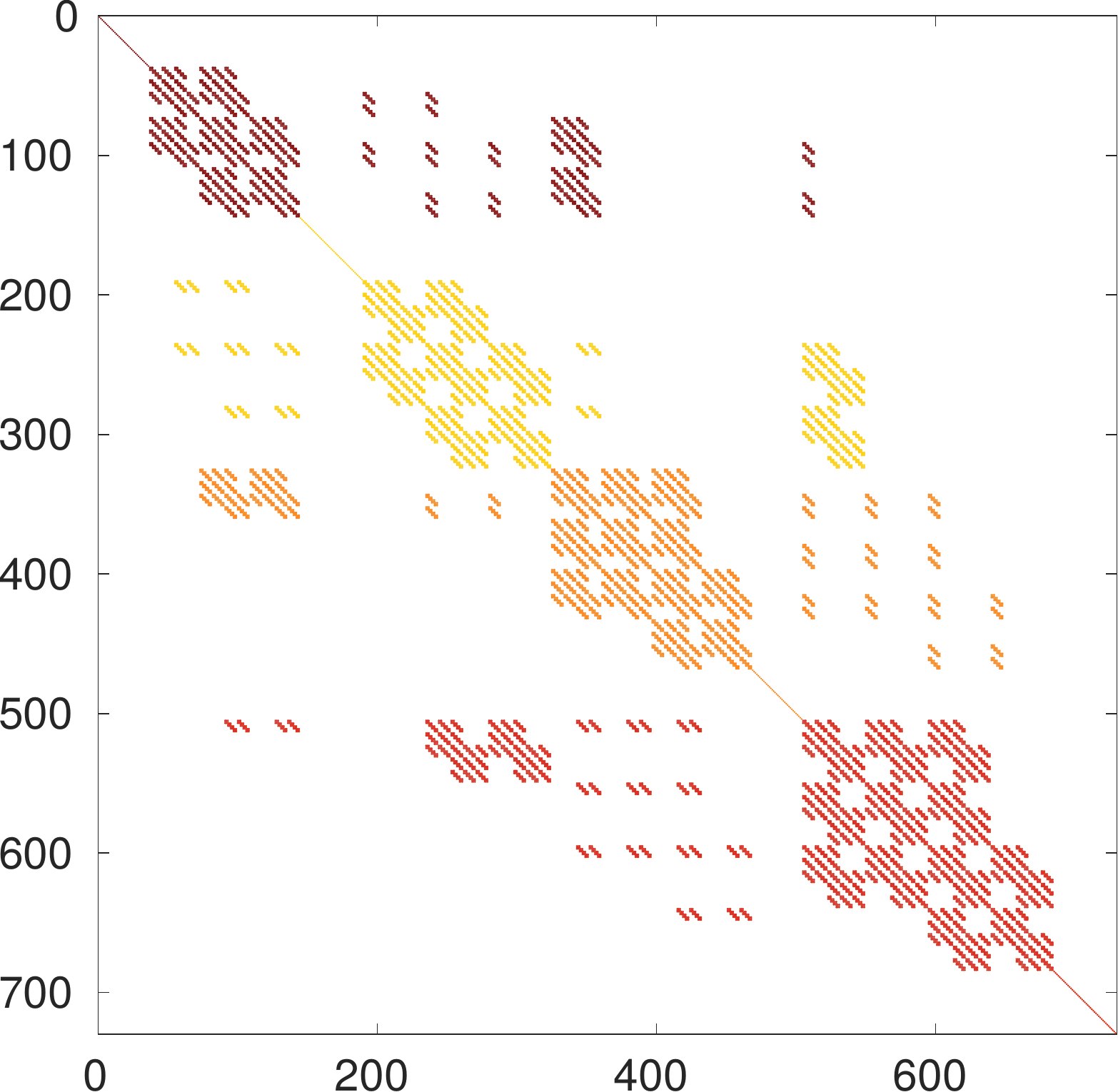}%
  \caption{Sparsity pattern, i.e., locations of non-zero entries of the $729 \times 729$ stiffness matrix $\bfK$ for the example problem in \cref{fig:3d_laplace}. The rows for the four processes are given by different colors matching the partitioning in \cref{fig:3d_laplace}. The processes have 144, 180, 180 and 225 local dofs.}%
  \label{fig:sparsity_pattern}%
\end{figure}

The non-zero estimates computed by \cref{eq:nonzero_estimates} are $n_\text{1D\_overlaps}=5$ and $n_\text{non-zeros} = 125$. The mesh has 4 elements, thus, 9 nodes per coordinate direction, and, therefore, $n_\text{global} = 9^3 = 729$ dofs.
\Cref{fig:sparsity_pattern} shows the resulting sparsity pattern of the stiffness matrix $\bfK$. The portions of the four processes are indicated by different colors. The maximum number of non-zeros per row and column is indeed 125, as calculated. Some rows have less non-zero entries. These correspond to dofs that lie on the boundary of the domain. The rows with only one non-zero entry on the diagonal enforce the Dirichlet boundary conditions. 
The total size of preallocated memory for the diagonal and off-diagonal blocks on all processes is $2\,n_\text{global}\,n_\text{non-zeros} = \num{182250}$. The actual number of non-zero entries is $\num{35937}$, which is approximately $\SI{20}{\percent}$ of the preallocated values.

% This will be explained in more detail in the next section.

\begin{reproduce_no_break}
  In any example, the system matrix can be written to a MATLAB compatible file by specifying the settings \code{'dumpFormat': 'matlab', 'dumpFilename': 'out'}.
   %\code{"dumpFormat": "matlab", "dumpFilename": "out"}. 
  To get the non-zero structure for the example in \cref{fig:sparsity_pattern}, compile the \emph{laplace3d} example and run the following:
  \begin{lstlisting}[columns=fullflexible,breaklines=true,postbreak=\mbox{\textcolor{gray}{$\hookrightarrow$}\space}]
    cd $\$$OPENDIHU_HOME/examples/laplace/laplace3d/build_release
    mpirun -n 4 ./laplace_quadratic ../settings_quadratic_matrix_output.py -ksp_view
  \end{lstlisting}
  The flag \code{-ksp\_view} is parsed by PETSc and outputs matrix statistics such as the number of preallocated and actual non-zeros. 
  A file \code{out\_matrix\_000.m} is created that can be loaded in MATLAB. Use \code{spy(stiffnessMatrix)} to plot the non-zero structure.
\end{reproduce_no_break}

\subsection{Assembly of Finite Element Matrices}\label{sec:assembly_of_finite_element_matrices}
% stiffness matrix assembly/integration

Next, the algorithm to compute stiffness and mass matrices in parallel for the application of the $d$-dimensional FEM is discussed.
The matrix entries to be computed are given by%
\begin{align}\label{eq:alg_assembly_m}
  m_{i,j} = \i{\Omega}{} I(\bfx) \,\d\bfx,
\end{align}
where the integrand $I$ is derived from the respective FEM formulation in weak form.

A generic algorithm for the evaluation of this integral and parallel assembly to a global matrix is presented in \cref{alg:fe_matrix_assembly}. Multiple variants of this algorithm, which only differ in their achieved performance, have been implemented for evaluation purposes. They are discussed in \cref{sec:performance_of_the_algorithm_for_pma}. The listed algorithm in \cref{alg:fe_matrix_assembly} shows the fastest variant.

\begin{algorithm}
  \begin{algorithmic}[1]%
    \Procedure{Assemble FE system matrix}{}
    \For{elements $e=\{e_1,e_2,e_3,e_4\}$ in all elements}      \label{line:4.2}
      \For{sampling point $\bfxi$}                              \label{line:4.3}
        \State Compute Jacobian $J_e(\bfxi)$                    \label{line:4.4}
        \State Evaluate integrand $I_{e,i,j}(\bfxi) = c\cdot I(J_e,\bfxi)$ \Comment{for all elements $e$/dofs $(i,j)$ at once}                    \label{line:4.5}
      \EndFor
      \State matrix\_entries[$i$,$j$] = Quadrature($I_{e,i,j}(\bfxi)$)   \Comment{for all el. $e$/dofs $(i,j)$ at once}                    \label{line:4.6}
      
      \For{dof $i$ = $0,\dots,n_\text{dofs\_per\_element}-1$}                      \label{line:4.7}
        \For{dof $j$ = $0,\dots,n_\text{dofs\_per\_element}-1$}                    \label{line:4.8}
          \State rows \hspace*{4.1mm} = dofs $i$ of elements $e_1,e_2,e_3,e_4$                    \label{line:4.9}
          \State columns = dofs $j$ of elements $e_1,e_2,e_3,e_4$  \label{line:4.10}
          \State matrix[rows,columns] = matrix\_entries[$i$,$j$]   \label{line:4.11}
        \EndFor
      \EndFor
    \EndFor
    \State Call PETSc final matrix assembly                    \label{line:4.12}
    \EndProcedure
  \end{algorithmic}%
  \caption{Finite element matrix assembly}%
  \label{alg:fe_matrix_assembly}%
\end{algorithm}%

The main loop in line \algref{alg:fe_matrix_assembly}{line:4.2} iterates over the local elements of the subdomain. 
The shown implementation iterates over sets of four elements $e_1,e_2,e_3$ and $e_4$. A simpler variation of the algorithm is to instead visit every single local element in its own iteration.
However, the more efficient variant is the presented one that always considers the set $e$ of four elements at once. 
Explicit vectorization is employed on all following operations on these four elements, such that 
the four sequences of calculations for the elements are performed by identical instructions. This adheres to the single-instruction-multiple-data (SIMD) paradigm. The vectorization is explicit since the C++ library \emph{Vc} \cite{vc2012,Kretz2015} is used. Vc provides zero-overhead C++ types for explicitly data-parallel programming and directly employs the respective vector instructions where these types are used.

To compute the integral in \cref{eq:alg_assembly_m}, a node based quadrature rule is used. In our code, the quadrature rule has to be chosen at compile time among Gauss, Newton-Cotes and Clenshaw-Curtis quadrature rules. All three schemes are implemented for different numbers $n_\text{sampling\_points}$ of sampling points. The loop in lines \algref{alg:fe_matrix_assembly}{line:4.3}---\algref{alg:fe_matrix_assembly}{line:4.5} iterates over the respective sampling points $\bfxi \in [0,1]^d$ in the element coordinate system. In line \algref{alg:fe_matrix_assembly}{line:4.4}, the Jacobian matrix of the mapping from element to world coordinate frame is computed at the given coordinate $\bfxi$ for all elements in the set $e$. The Jacobian is needed in the integrand for the transformation of the integration domain.

In line \algref{alg:fe_matrix_assembly}{line:4.5}, the integrand $I$ is evaluated for all elements in $e$ and also for all pairs $(i,j)$ of local dofs in each of these elements. The indices $i$ and $j$ are in the range $i,j \in \{0,1,\dots,n_\text{dofs\_per\_element}-1\}$ with the number $n_\text{dofs\_per\_element}$ of dofs per element. The set of $4\,(n_\text{dofs\_per\_element})^2\cdot (n_\text{sampling\_points})^d$ computed values is passed to the implementation of the $d$-dimensional quadrature rule in line \algref{alg:fe_matrix_assembly}{line:4.6}. The numerical values of the integrals get computed for all considered elements in $e$ and dof pairs $(i,j)$, yielding $4\,(n_\text{dofs\_per\_element})^2$ quadrature problems to be solved at once. This means that the result of the quadrature rule is a linear combination of quadrature weights and vector-valued function evaluations instead of scalar function values.

Next, the two loops in lines \algref{alg:fe_matrix_assembly}{line:4.7}---\algref{alg:fe_matrix_assembly}{line:4.11} assign the computed values stored in the variable \code{matrix_entries} to the actual matrix. The loops iterate over all dof pairs $(i,j)$ per element. The corresponding rows and columns are determined in lines \algref{alg:fe_matrix_assembly}{line:4.9} and \algref{alg:fe_matrix_assembly}{line:4.10} and the respective computed value is assigned in line \algref{alg:fe_matrix_assembly}{line:4.11}. 
The values are added to the matrix entry indicated by the row and column index. Since all dofs including ghosts are considered on every local domain, the same matrix entry can get contributions on multiple processes.

Thus, the last step in line \algref{alg:fe_matrix_assembly}{line:4.12} is a PETSc call that communicates and sums all matrix entry contributions to the respective processes where the dof is non-ghost. Additionally, the call frees the residual preallocated memory that was not needed for non-zero entries and finalizes the internal data structure of the CRS storage format.

In the last iteration over local elements of the main loop in line \algref{alg:fe_matrix_assembly}{line:4.2}, the remaining number of elements is potentially less than four. Nevertheless, the computations proceed as normal. The spare entries of the SIMD vectors get computed using dummy values and are discarded at the end.

For the case of vector-valued finite element problems, e.g., linear elasticity with a solution vector of vector-valued displacements, two more inner loops over the components of the vector are inserted. As a result, the presented algorithm can be used to assemble any FEM matrix on any mesh type and for any formulation given by the term $I$ in \cref{eq:alg_assembly_m}.
Examples are stiffness and mass matrices for the Laplace operator with and without diffusion tensor or stiffness and mass matrices for the linear equations that have to be solved during the solution of nonlinear, dynamic elasticity problems.

% algorithms all local, there is no globalToLocal map
Note that the algorithm operates in parallel execution entirely on data stored in the local subdomain and does not need any global information. The loop iterates over local elements. For every element, the indices in the local numbering of the nodes that are adjacent to the element are needed. In structured meshes, this information is determined from the numbers $N^\text{el}_x \times N^\text{el}_y \times N^\text{el}_z$ of local elements in the coordinate directions. In unstructured meshes, these indices are explicitly stored in the elements. To assemble the global matrix, PETSc uses the mapping from local to global numbering, which it can maintain by storing the constant offset in the global numbering on every subdomain. Mappings from global dof or node numbers to local numbers are not needed in this algorithm. 
In general, storing global information, which would require memory of $\O(n_\text{global})$, is avoided in all algorithms to ensure good parallel weak scaling properties.

%A different approach is to iterate over the global matrix entries rather than the elements. Then, 

\subsection{Performance of the Algorithm for Parallel Matrix Assembly}\label{sec:performance_of_the_algorithm_for_pma}

In the following, the performance of two variations of the algorithm in \cref{alg:fe_matrix_assembly} will be examined. The first variation is to not use explicit vectorization and, thus, iterate over the elements one by one instead of the groups of four elements in line \algref{alg:fe_matrix_assembly}{line:4.2}.

The second variation is to not accumulate multiple values for the application of the quadrature scheme in line \algref{alg:fe_matrix_assembly}{line:4.6}. Instead, the loop over the sampling points in line \algref{alg:fe_matrix_assembly}{line:4.3} is made the inner-most loop and placed inside  the loop in line \algref{alg:fe_matrix_assembly}{line:4.8}. Then, the quadrature scheme only computes a single value at once. In consequence, this value can directly be stored in the resulting matrix, and the temporary variable \code{matrix\_entries} is not needed. This loop reordering requires the evaluations of the Jacobian and the integrand in lines \algref{alg:fe_matrix_assembly}{line:4.4} and \algref{alg:fe_matrix_assembly}{line:4.5} to also be located in the new inner-most loop over the sampling points. 

The algorithm with these two variations corresponds to the naive way of implementing matrix assembly because iterating first over elements, then over dof pairs and then performing the quadrature directly mirrors the mathematical description.

\begin{figure}%
  \centering%
  \includegraphics[width=\textwidth]{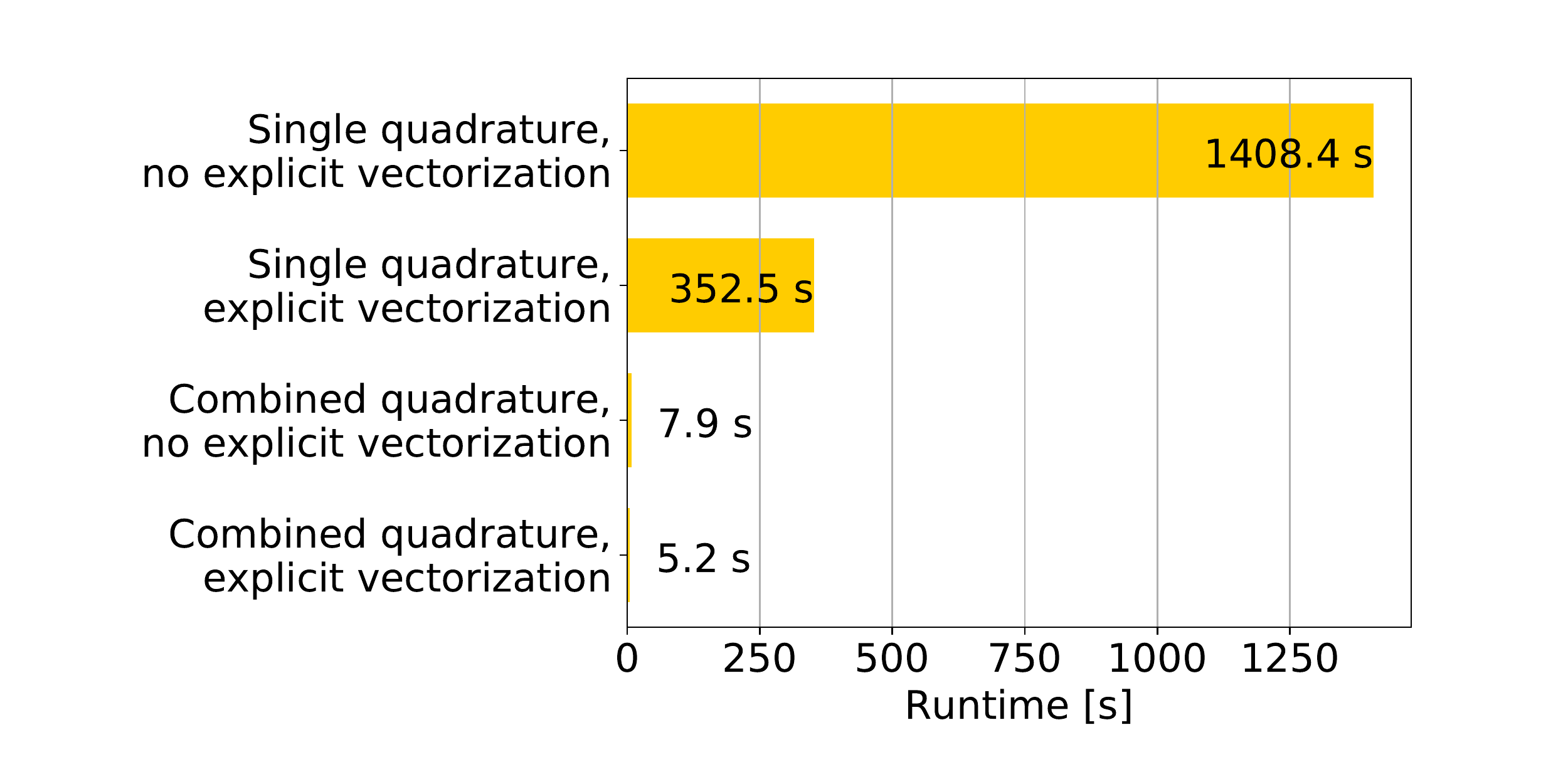}%
  \caption{Runtimes of different optimizations for the algorithm to assemble the FEM stiffness matrix.}%
  \label{fig:matrix_runtimes}%
\end{figure}

Different combinations of these two variations result in four variants of the algorithm. A study was conducted to measure their effects on the runtimes. A simulation with the same settings as in \cref{fig:3d_laplace} was run except for a larger number of $50 \times 50 \times 50$ elements. This setup lead to a total number of $n_\text{global}=\num{1030301}$ dofs. Gauss quadrature with three sampling points per coordinate direction and, thus, $3^3=27$ sampling points in total was used. The program was executed with four processes on an AMD EPYC 7742 processor with base frequency of \SI{2.25}{\giga\hertz}, maximum boost frequency of \SI{3.4}{\giga\hertz}, \SI{2}{\tera\byte} of memory and a memory bandwidth of \SI{204.8}{\giga\byte\per\second} per socket.
The runtime for the assembly of the stiffness matrix with dimensions $n_\text{global} \times n_\text{global}$ was measured for all four variants. \Cref{fig:matrix_runtimes} presents the resulting runtimes.

It can be seen that a large difference in runtime exists between the variants with quadrature of single values compared to the combined quadrature. In the case of no explicit vectorization (first and third bar from the top in \cref{fig:matrix_runtimes}), the runtime reduces to less than \SI{0.6}{\percent}. In the case of explicit vectorization (second and fourth bar from the top in \cref{fig:matrix_runtimes}), the runtime reduces to less than \SI{1.5}{\percent}. The reason for this enormous gain in performance is three-fold. First, the values of the Jacobian can be reused for the same element and sampling point. Second, the combined quadrature for multiple values yields more cache-efficient memory access, because the vector of values is stored consecutively in memory and can be fetched from the cache by less load operations. For the single quadrature, the individual values are fetched at different times from different memory locations.
Third, the compiler is able to employ SIMD instructions for the combined quadrature, a process called auto-vectorization.

The performance improvements from the second variation, the use of explicit vectorization by simultaneously computing the entries for four elements at once can be seen by comparing the first and second bars and the third and forth bars in \cref{fig:matrix_runtimes}. The runtime reduction of explicit vectorization with single quadrature from \SI{1408.4}{\second} to \SI{352.5}{\second} is exactly by the expected factor of four. This shows that explicit vectorization works as expected, and that no auto-vectorization could be performed by the compiler for the single quadrature. The runtime reduction of explicit vectorization from \SI{7.9}{\second} to \SI{5.2}{\second} during combined quadrature corresponds to a speedup of only approximately \num{1.5}. This shows that combined quadrature without explicit vectorization already allows the compiler to employ some auto-vectorization. However, using the explicit vectorization approach on the level of different elements instead of the level of quadrature values still has a positive effect. 

In total, the performance gain from the most naive implementation (top bar in \cref{fig:matrix_runtimes}) to the most optimized version (bottom bar in \cref{fig:matrix_runtimes}) equals a speedup of more than \num{270}. Together with the solution of the linear system using an algebraic multigrid preconditioner and a GMRES solver with a residual norm tolerance of \num{1e-10}, the total runtime of the program to solve the Laplace problem with over a million degrees of freedom using a modest parallelism of four processes takes \SI{28}{\second}.

\begin{reproduce_no_break}
  The results of \cref{fig:matrix_runtimes} can be reproduced as follows. The explicit vectorization can be turned on and off with the variable \code{USE_VECTORIZED_FE_MATRIX_ASSEMBLY} in the configuration of the SCons build system in the file \code{$\$$OPENDIHU_HOME/user-variables.scons.py} (ca. line 75). Normally, only the variant with combined quadrature is implemented. To test the single quadrature, checkout the git branch \code{fem_assembly_measurement}. The single quadrature is on by default, to change back to the combined quadrature, edit the following line:
  \begin{lstlisting}[columns=fullflexible,breaklines=true,postbreak=\mbox{\textcolor{gray}{$\hookrightarrow$}\space}]
    vi $\$$OPENDIHU_HOME/core/src/spatial_discretization/finite_element_method/01_stiffness_matrix_integrate.tpp +17
  \end{lstlisting}
  For all variants of the algorithm, compile and run the following example: 
  \begin{lstlisting}[columns=fullflexible,breaklines=true,postbreak=\mbox{\textcolor{gray}{$\hookrightarrow$}\space}]
    cd $\$$OPENDIHU_HOME/examples/laplace/laplace3d/build_release
    mpirun -n 4 ./laplace_quadratic ../settings_quadratic.py
  \end{lstlisting}
  The duration of the algorithm for stiffness matrix assembly will be printed.
\end{reproduce_no_break}

% stencils
\subsection{Assembly of Finite Element Matrices for Regular Meshes}

For equidistant meshes of type \code{Mesh::StructuredRegularFixedOfDimension<D>}, all elements are similar through the uniform grid resolution and, thus, all elements matrices equal the same constant matrix. In consequence, the integral terms in \cref{eq:alg_assembly_m} can be precomputed analytically and no numerical quadrature at runtime is needed. This speeds up the determination of the FEM matrices.

We implement matrix assembly using precomputed values for the stiffness and mass matrices of the Laplace operator for linear Lagrange basis  functions. For the stiffness matrix of the Laplace operator, the integral term $(-\int_{\Omega^\text{el}} ∇\phi_i\cdot ∇\phi_j \,\d\bfxi)$ is calculated analytically. The result is a value for every combination of the dofs $i$ and $j$ in the element. 
Thus, the contribution of one representative element in the mesh to the values at adjacent dofs is known. 
To get the matrix entry for a particular dof, the element contributions of all elements that are adjacent to the node need to be summed up. For this process, it is convenient to represent the precomputed values in \emph{stencil notation}.

\Cref{tab:stencils_laplace} shows the stencils for element contributions in the left column and the resulting stencils for the dofs in the right column.
In the element contribution stencils, dof $i$ is chosen as the first dof in the local dof numbering. Values are calculated for all choices of dof $j$ in the element and the values are noted in the stencil. The location of dof $i$ is marked by the underlined number. Stencils for all other locations of dof $i$ follow by symmetry.

The node stencils describe the values of the term $(-\int_{\Omega} ∇\phi_i\cdot ∇\phi_j \,\d\bfxi)$ with the integration over the whole domain. 
The node $i$ is fixed and marked in the stencil notation by the underlined number. Values for all neighboring nodes $j$ are computed and listed in the stencils.
For a given node $i$, the integral over the whole domain $\Omega$ is the sum of integrals over all elements $\Omega^\text{el}$ adjacent to node $i$. These have been computed in the element contribution stencils. As can be seen in \cref{tab:stencils_laplace}, the node stencils follow by adding up mirrored variants of the element stencils centered around the underlined node.

\begin{table}
  \centering
  \begin{tabular}{|c|c|c|c}
      \hline
      Dim. & Element contribution & Node stencil\\
      \hline
      1D &
  \begin{minipage}{4cm}
    \begin{equation*}
       \left[\begin{array}{ccc}
          \underline{-1} & 1\\
      \end{array}\right] \quad 
    \end{equation*}
  \end{minipage} 
      &
  \begin{minipage}{4cm}
    \begin{equation*}
      \left[\begin{array}{ccc}
          1 & \underline{-2} & 1\\
      \end{array}\right]
    \end{equation*}
  \end{minipage} 
       \\[4mm]
       \hline
      2D &
  \begin{minipage}{5cm}
    \begin{equation*}
      \dfrac16\left[
        \begin{array}{ccc}
          1 & 2 \\
          \underline{-4} & 1
        \end{array}
      \right]
    \end{equation*}
  \end{minipage}  &
  \begin{minipage}{5cm}
    \begin{equation*}
        \dfrac13\left[
          \begin{array}{ccc}
            1 & 1 & 1\\
            1 & \underline{-8} & 1 \\
            1 & 1 & 1
          \end{array}
        \right]
    \end{equation*}
  \end{minipage}  \\[4mm]
      \hline
      3D &
  \begin{minipage}{6cm}
    \begin{equation*}
      \begin{array}{ll}
        \text{center:} &
        \dfrac1{12}\left[\begin{array}{ccc}
            0 & 1\\
            \underline{-4} & 0\\
        \end{array}\right] \\[4mm]
        \text{bottom:}& 
        \dfrac1{12}\left[\begin{array}{ccc}
            1 & 1\\
            0 & 1\\
        \end{array}\right]
      \end{array}
    \end{equation*}
  \end{minipage} &
  \begin{minipage}{6cm}
    \begin{equation*}
      \begin{array}{ll}
        \text{top:} &
        \dfrac1{12}
        \left[\begin{array}{ccc}
            1 & 2 & 1\\
            2 & 0 & 2\\
            1 & 2 & 1
        \end{array}\right] \\[4mm]
        \text{center:} &
        \dfrac1{12}
        \left[\begin{array}{ccc}
            2 & 0 & 2\\
            0 & \underline{-32} & 0\\
            2 & 0 & 2
        \end{array}\right] \\[4mm]
        \text{bottom:}& \text{same as top}
      \end{array}  
    \end{equation*}
  \end{minipage}
  \\\hline
  \end{tabular}
  \caption{Stencils of the finite element stiffness matrix of $Δu$ for a regular mesh with mesh width $h=1$ and linear ansatz functions. The stiffness matrix entries can be computed by multiplication with a mesh width dependent factor.}
  \label{tab:stencils_laplace}
\end{table}

\begin{table}
  \centering
  \begin{tabular}{|c|c|c|c}
      \hline
      Dim. & Element contribution & Node stencil\\
      \hline
      1D &
  \begin{minipage}{4.5cm}
    \begin{equation*}
       \dfrac1{6}\left[\begin{array}{ccc}
          \underline{2} & 1\\
      \end{array}\right] \quad 
    \end{equation*}
  \end{minipage} 
      &
  \begin{minipage}{4.5cm}
    \begin{equation*}
      \dfrac1{6}\left[\begin{array}{ccc}
          1 & \underline{4} & 1\\
      \end{array}\right]
    \end{equation*}
  \end{minipage} 
       \\[4mm]
       \hline
      2D&
  \begin{minipage}{5cm}
    \begin{equation*}
      \dfrac1{36}\left[
        \begin{array}{ccc}
          2 & 1 \\
          \underline{4} & 2
        \end{array}
      \right]
    \end{equation*}
  \end{minipage}  &
  \begin{minipage}{5cm}
    \begin{equation*}
        \dfrac1{36}\left[
          \begin{array}{ccc}
            1 & 4 & 1\\
            4 & \underline{16} & 4 \\
            1 & 4 & 1
          \end{array}
        \right]
    \end{equation*}
  \end{minipage}  \\[4mm]
      \hline
      3D &
  \begin{minipage}{6cm}
    \begin{equation*}
      \begin{array}{ll}
        \text{center:} &
        \dfrac1{216}\left[\begin{array}{ccc}
            4 & 2\\
            \underline{8} & 4\\
        \end{array}\right] \\[4mm]
        \text{bottom:}& 
        \dfrac1{216}\left[\begin{array}{ccc}
            2 & 1\\
            4 & 2\\
        \end{array}\right]
      \end{array}
    \end{equation*}
  \end{minipage} &
  \begin{minipage}{6cm}
    \begin{equation*}
      \begin{array}{ll}
        \text{top:} &
        \dfrac1{216}\left[\begin{array}{ccc}
            1 & 4 & 1\\
            4 & 16 & 4\\
            1 & 4 & 1
        \end{array}\right] \\[4mm]
        \text{center:} &
        \dfrac1{216}\left[\begin{array}{ccc}
            4 & 16 & 4\\
            16 & \underline{64} & 16\\
            4 & 16 & 4
        \end{array}\right] \\[4mm]
        \text{bottom:}& \text{same as top}
      \end{array}  
    \end{equation*}
  \end{minipage}
  \\\hline
  \end{tabular}
  \caption{Stencils of the finite element mass matrix for a regular mesh with mesh width $h=1$ and linear ansatz functions. The mass matrix entries can be computed by multiplication with a mesh width dependent factor.}
  \label{tab:stencils_mass_laplace}
\end{table}

The entries in the stiffness matrix are computed from the node stencils by a multiplication with a mesh dependent prefactor.
For 1D, 2D and 3D meshes with mesh width $h$, these prefactors are $1/h$, 1 and $h$, respectively. Thus, e.g., the 1D stiffness matrix has the entries $-2/h$ on the diagonal, $1/h$ on the secondary diagonals above and below the main diagonal and zero everywhere else.

A similar computation is possible for the mass matrix, where the term $\int \phi_i\,\phi_j \,\d\bfxi$ can be precalculated. The element and node stencils for the mass matrix are given in \cref{tab:stencils_mass_laplace}.

The precalculated values can only be used for meshes with uniform mesh width and linear Lagrange basis functions. In \opendihu{}, the type of the mesh and basis function is fixed at compile time. The stencil based approach to set the entries of stiffness and mass matrix is implemented as partial template specialization of the template, which otherwise uses the numerical algorithm presented in \cref{sec:assembly_of_finite_element_matrices}. Thus, the stencil based implementation is instantiated automatically by the compiler for regular fixed meshes of all dimensionalities with linear basis functions. 

The conditions for the stencil based approach are fulfilled whenever regular fixed meshes and linear bases are used, e.g., for toy problems or studies where the shape of the domain is irrelevant and, e.g., a cuboid cutout of muscle tissue is sufficient. Mathematical models involving a Laplace operator, such as Laplace, Poisson or diffusion problems can benefit from the faster system matrix setup.

Another purpose of the stencil based approach in \opendihu{} besides runtime reduction is to verify the implementation of the numerical integration method of \cref{alg:fe_matrix_assembly}. Because of the regular mesh and linear ansatz functions, the numerical method computes the exact result with proper quadrature schemes and, thus, can be compared to the stencil based approach. Several unit tests ensure that the generated system matrices of the two approaches are equal.

\subsection{Algorithm for Dirichlet Boundary Conditions}
% algorithm for Dirichlet boundary conditions

The assembled system matrix needs to be adjusted when Dirichlet boundary conditions are specified.
Dirichlet boundary conditions are ensured by replacing equations involving the prescribed dofs by the definition of the boundary conditions. This involves changes in the system matrix and right-hand side of the finite element formulation.

Consider the following matrix equation resulting from a FE discretization with four dofs $u_1$ to $u_4$:
\begin{equation*}
  \begin{array}{llll}
    \mat{m_{11} & m_{12} & m_{13} & m_{14} \\ m_{21} & m_{22} & m_{23} & m_{24} 
       \\ m_{31} & m_{32} & m_{33} & m_{34} \\ m_{41} & m_{42} & m_{43} & m_{44} }
    \mat{u_1 \\ u_2 \\ u_3 \\ u_4} = \mat{f_1 \\ f_2 \\ f_3 \\ f_4}.
  \end{array}
\end{equation*}
We assume a Dirichlet boundary condition for the last dof, $u_4 = \hat{u}_4$. Enforcing this condition is accomplished by the following adjusted system of equations:
\begin{equation*}
  \begin{array}{llll}
    \mat{m_{11} & m_{12} & m_{13} & 0\\ m_{21} & m_{22} & m_{23} & 0 \\ m_{31} & m_{32} & m_{33} & 0 \\ 0 & 0 & 0 & 1}
    \mat{u_1 \\ u_2 \\ u_3 \\ u_4} 
    = \mat{f_1 - m_{14}\,\hat{u}_4 \\
           f_2 - m_{24}\,\hat{u}_4  \\ 
           f_3 - m_{34}\,\hat{u}_4 \\
          \hat{u}_4}.
  \end{array}
\end{equation*}
The last equation has been replaced by $u_4 = \hat{u}_4$, all summands in the other equations where $u_4$ occurred have been brought to the right-hand side and $u_4$ has been substituted by the prescribed value $\hat{u}_4$.

Thus, setting a dof $u_i$ to a prescribed value $\hat{u}$ corresponds to 
subtracting the column vector of column $i$ of the system matrix multiplied with $\hat{u}$ from the right-hand side, replacing the right-hand side entry at $i$ by $\hat{u}$ and setting row $i$ and column $i$ in the matrix to all zero and the diagonal entry $m_{ii}$ to one.

This method also works for more prescribed values as demonstrated with the additional Dirichlet boundary condition $u_2 = \hat{u}_2$. Executing the scheme results in the following system:
\begin{equation*}
  \begin{array}{llll}
    \mat{m_{11} & 0 & m_{13} & 0\\ 0 & 1 & 0 & 0 \\ m_{31} & 0 & m_{33} & 0 \\ 0 & 0 & 0 & 1}
    \mat{u_1 \\ u_2 \\ u_3 \\ u_4} 
    = \mat{f_1 - m_{12}\,\hat{u}_2 - m_{14}\,\hat{u}_4 \\
          \hat{u}_2 \\ 
           f_3 - m_{32}\,\hat{u}_2 - m_{34}\,\hat{u}_4 \\
          \hat{u}_4}.
  \end{array}
\end{equation*}

During parallel execution, only a distinct subset of rows of the matrix equation is accessible on every rank. 
However, for the subtractions at the right-hand side, the full vector of prescribed boundary conditions values is needed on every rank. Additionally, the corresponding matrix entries are required. While the needed matrix entries are all stored on the local rank, the vector of prescribed values is partitioned to all ranks and only the local subdomain is accessible. 
%However, most of the prescribed values are not needed for the subtraction, because the corresponding matrix entries are zero. 
Some of the non-accessible prescribed values correspond to a non-zero matrix entry, though, and are not needed for the subtraction. In consequence, some data transfer between processes is required. In the following, the identification of the values that have to be communicated is illustrated with an example.

\begin{figure}%
  \centering%
  \includegraphics[width=0.6\textwidth]{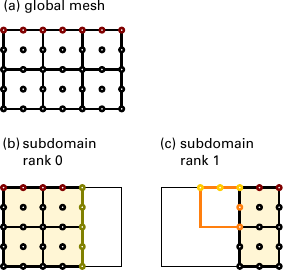}%
  \caption{Example to illustrate ghost element transfer that is needed for handling Dirichlet boundary conditions in parallel. A mesh with Dirichlet boundary conditions on the red nodes is given in (a). The subdomains for two processes are given in (b) and (c). (c) shows a ghost element in orange that is sent from rank 0 to rank 1.}%
  \label{fig:dirichlet_bc_example}%
\end{figure}%

\Cref{fig:dirichlet_bc_example} (a) shows a 2D quadratic mesh with $3 \times 2$ elements. The top layer of nodes has prescribed Dirichlet boundary conditions, marked by the red circles. The elements are partitioned to two processes with subdomains containing four and two elements, as shown in \cref{fig:dirichlet_bc_example} (b) and \cref{fig:dirichlet_bc_example} (c). 

On rank 0, the right-most layer of nodes consists of ghost nodes. All other nodes are non-ghost and correspond to the matrix rows and right-hand side entries that have to be manipulated by this rank.
In every of these matrix rows $i$, only entries in columns $j$ that correspond to nodes in the same element as $i$ are non-zero. For these columns $j$, the prescribed Dirichlet boundary condition values $\hat{u}_j$ need to be known, such that the product of matrix entry $m_{ij}$ and prescribed value $\hat{u}_j$ can be subtracted from the right-hand side at the corresponding row $i$. This is fulfilled for rank 0 since the required boundary condition values are all part of the subdomain. The top right boundary condition node in the top right element of rank 0's subdomain is stored as ghost value, the other four are non-ghosts.

As can be seen in \cref{fig:dirichlet_bc_example} (c), the subdomain of rank 1 has no ghost nodes. The rank owns three boundary condition nodes in the top layer. It has to perform right-hand side subtractions for the twelve rows of the matrix equation that correspond to the $3\times 4$ other nodes, which have no prescribed boundary condition. For the bottom two horizontal layers of nodes, the subtraction terms are zero, because the bottom element of the subdomain has no boundary conditions and, thus, these matrix entries are all zero.
The upper three horizontal layers of nodes that all belong to the upper element, however, lead to non-zero right-hand side subtraction terms because of the boundary conditions at the top. The terms can be computed for all but the two orange nodes. 
At the corresponding rows $i$, the prescribed values $\hat{u}_j$ for all five yellow and red boundary condition nodes $j$ are needed. However, the left two yellow nodes are not stored on the subdomain of rank 1. They have to be communicated from the subdomain of rank 0. As rank 1 has no topology knowledge of rank 0's subdomain, the information that the missing boundary condition nodes are in the same element as the two orange nodes has to be also transferred. 

In total, the information indicated in \cref{fig:dirichlet_bc_example} (c) by the orange element with the two orange nodes and the three yellow boundary condition nodes and values have to be communicated from rank 0 to rank 1. This element is called \emph{ghost element}.
Rank 0 knows that rank 1 will need this information because it stores the right-most yellow node and the orange nodes in subdomain 1 as ghost nodes in its own subdomain. Therefore, no request from rank 1 is necessary.

In general, every rank constructs ghost elements from own elements that contain both at least one boundary condition node and at least one ghost node without boundary condition. The global indices of all nodes of these two kinds and the corresponding boundary condition values are packaged as ghost element and sent to the rank of the neighboring subdomain. Every process potentially sends multiple ghost elements to multiple neighboring ranks. 

Because a rank does not know the number of ghost elements it will receive a-priori, one-sided communication is employed, which was introduced with the MPI 2.0 standard. More specifically, \emph{passive target} communication is used where only the sending rank is explicitly involved in the transfer.

After the data are received, the proper matrix entries can be retrieved from the local matrix storage and 
the subtraction operations on the right-hand side of the formulation can be performed. The algorithm has to ensure that the same subtraction is not executed multiple times, when the particular pair of nodes is obtained once from the local subdomain and once from a received ghost element.
After solving the linear system with the updated stiffness matrix and right-hand side, the dofs on nodes with Dirichlet boundary conditions will have the prescribed values.

Considering the overhead for ensuring Dirichlet boundary conditions, the question may arise whether the partitioning scheme should be designed in a better way to simplify the presented algorithm.
However, applying Dirichlet boundary conditions is the only process where the subdomains including ghost nodes, which were created by the parallel partitioning of the mesh, do not provide all required local information. All other algorithms such as matrix assembly successfully operate on the given partitioning. 
Therefore, designing the ghost information of subdomains differently, e.g., by storing a full ghost layer of elements or nodes around the local subdomains seems not beneficial. In fact, our presented approach is minimal with respect to stored local mesh information. Furthermore, the communicated information for the Dirichlet boundary conditions only involves a few elements depending on the number of boundary conditions. Of these elements, only a subset of nodes is actually communicated.

Another alternative approach would be to store all Dirichlet boundary condition information globally on all processes, such as done in OpenCMISS Iron. This approach is not chosen, because the required total storage would increase linearly with the number of processes. Thus, the possible number of boundary conditions would be limited by available memory.

In summary, the presented algorithm fits our design goals of good performance. It is used in our implementation to enforce Dirichlet boundary conditions for static and dynamic problems. The algorithm is executed after assembling the stiffness matrix. In consequence, for static problems the linear system solver sets the prescribed values at the respective dofs and the boundary conditions are fulfilled. For dynamic problems with constant stiffness matrices, Dirichlet boundary conditions have to be ensured in every timestep. After running the algorithm on the system matrix once, the operation on the right-hand side vector needs to be repeated in every timestep on the updated right-hand side.

% extension of representation in PartitionedPetscVecWithDirichletBc
% -> PartitionedPetscVecForHyperelasticity, no do not explain, because too detailed

% Neumann boundary conditions
\subsection{Neumann Boundary Conditions}

The other supported boundary conditions besides Dirichlet boundary conditions are the Neumann type boundary conditions.
In general, they are formulated on a subset $\Gamma_f \subset ∂\Omega$ of the boundary $∂\Omega$ with outwards normal vector $\bfn$ as follows:
\begin{align}\label{eq:impl_neumann}
  \big(\bfsigma\,\grad u(\bfx)\big) \cdot \bfn = f(\bfx) \quad \text{for }\bfx \in \Gamma_f.
\end{align}
Here, $\bfsigma$ is a conductivity tensor, which describes the anisotropy of the problem. For problems with a scalar solution function $u: \Omega \to \R$, the Neumann boundary condition is interpreted as a flux $f$ over the boundary of the quantity described by $u$. For elasticity problems, the solution function $\bfu: \Omega \to \R^d$ is vector-valued with $d=2$ or $d=3$ and describes the displacement field. Then, the value $\bff$ corresponds to a traction force $\bff$ per area that acts on the surface $\Gamma_f$.

In a finite element formulation, we use Neumann boundary conditions to resolve the boundary integrals that appear after applying the divergence theorem on the weak form. In the derivation in \cref{sec:discretization_diffusion}, these boundary integrals were summarized by the matrix $\bfB_{\bfsigma}$.
By using the definition of the Neumann boundary condition in \cref{eq:impl_neumann}, we can derive the following equation for the boundary integral:
\begin{subequations}
\begin{align}
  -\ds\sum\limits_{j=1}^{M} u_j\, \ds\int_{\Gamma_f} ( \bfsigma\,\grad \varphi_j \cdot \bfn)\, \varphi_k\,\d\bfx 
  &= -\ds\sum\limits_{j=1}^{M} \ds\int_{\Gamma_f} f_j\,\psi_j\,\varphi_k \,\d\bfx \label{eq:impl_neumann_a} \\[4mm] 
    &=: \text{rhs}_k, \label{eq:impl_neumann_b}
\end{align}
\end{subequations}
where the dofs $u_j$ and the ansatz functions $\varphi_j$ for $j=1,\dots,M$ discretize the solution function $u(\bfx)$, the dofs $f_j$ and the ansatz functions $\psi_j$ discretize the flux $f(\bfx)$ and $\varphi_k$ is the test function, which is chosen from the same function space as the ansatz functions.

\Cref{eq:impl_neumann_a} is equivalent to the matrix equation \cref{eq:boundary_relation}, and \cref{eq:impl_neumann_b} defines the final right-hand side vector $\textbf{rhs}$ of the linear system of equations to be solved. Analogously, the discretization of the Laplace problem in \cref{eq:discretization_laplace} contains a right-hand side of $\textbf{rhs} = -\bfB_{\Gamma_f}\,\bff$ with the vector $\bff$ of dofs of the discretized flux function $f$.

For elasticity problems where the solution $\bfu(\bfx)$ and the traction $\bff(\bfx)$ are vector-valued, the definition of the right-hand side, which is equivalent to \cref{eq:impl_neumann_b}, can be formulated as
\begin{align}\label{eq:impl_neumann_c}
  \text{rhs}_{aM}:= -\ds\int\limits_{∂Ω} T_a^L \, \psi^L(\bfx)\, δu_M  \,\d \bfx.
\end{align}
Here, the right-hand side vector $\textbf{rhs}$ consists of the given coefficients $\text{rhs}_{aM}$, where $a=1,\dots,d$ is the index over the dimension and $M=1,\dots,N$ iterates over the dofs in the discrete function space. $T_a^L$ is the dof of the discretized traction force using ansatz functions $\psi^L$ and summation over the repeated index $L$. $δu_M$ is the virtual displacement, which is equivalent to the test function in the Galerkin finite element formulation.

In OpenDiHu, a class exists that parses Neumann boundary conditions from the Python settings and computes the negative right-hand side vector $-\textbf{rhs}$, either for the scalar case in \cref{eq:impl_neumann_b} or the vector-valued case in \cref{eq:impl_neumann_c}.

If no Neumann boundary conditions are specified for parts of the boundary $∂\Omega$, the right-hand side vector is set to zero for the corresponding dofs. This means that specifying no Neumann boundary conditions is equivalent to specifying homogeneous Neumann boundary conditions, i.e., setting $f\equiv 0$.

\begin{figure}
  \centering%
  \includegraphics[width=\textwidth]{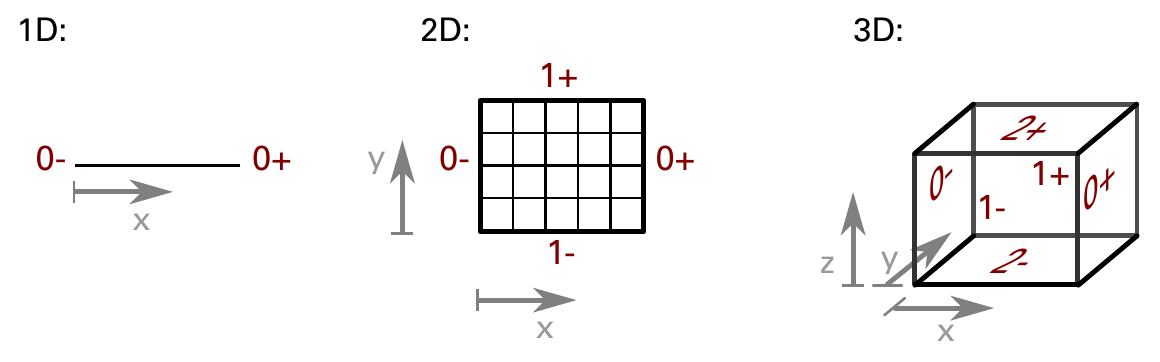}%
  \caption{Notion of the faces of 1D, 2D and 3D elements, as used in the definition of Neumann boundary conditions.}%
  \label{fig:faces}%
\end{figure}

Neumann boundary conditions are specified in the Python settings as a list of elements with associated flux or traction values. This is in contrast to the Dirichlet boundary conditions, which are defined per dof or node.
In every element with Neumann boundary conditions, the boundary face that is part of $\Gamma_f$ has to be specified. The face is identified by one of the strings \code{`0-`}, \code{`0+`}, \code{`1-`}, \code{`1+`}, \code{`2-`} or \code{`2+`}, which describe the positive or negative coordinate directions of the element coordinate system, as given in \cref{fig:faces}.

% solver structure of contraction

For elasticity problems, where the function $f(\bfx)$ is interpreted as traction force, two more options can be set.
The first option is \code{`divideNeumannBoundaryConditionValues}\code{ByTotalArea`}. If set to \code{True}, the traction force vector is interpreted as a total force on the whole surface. The value of $T$ in \cref{eq:impl_neumann_c} is computed by scaling down the given value by the total surface area. This allows to conveniently specify a total force, which, e.g., acts on the lower end of the muscle. Without this option, the traction force is interpreted as force per area unit.

The second option \code{`isInReferenceConfiguration`} allows switching between reference and current configuration to specify the traction force. The mapping between the traction $\bfT$ in reference configuration and the traction $\bft$ in current configuration is given by the inverse deformation gradient $\bfF$:%
\begin{align}\label{eq:impl_transformation}
  \bfT = \bfF^{-1}\,\bft.
\end{align}
Because the implemented model uses the Lagrangian formulation with the right-hand side term defined in \cref{eq:impl_neumann_c}, the transformation in \cref{eq:impl_transformation} and subsequently the right-hand side have to be computed in every timestep of a dynamic problem.

\begin{figure}%
  \centering%
  \begin{subfigure}[t]{0.31\textwidth}%
    \centering%
    \includegraphics[width=\textwidth]{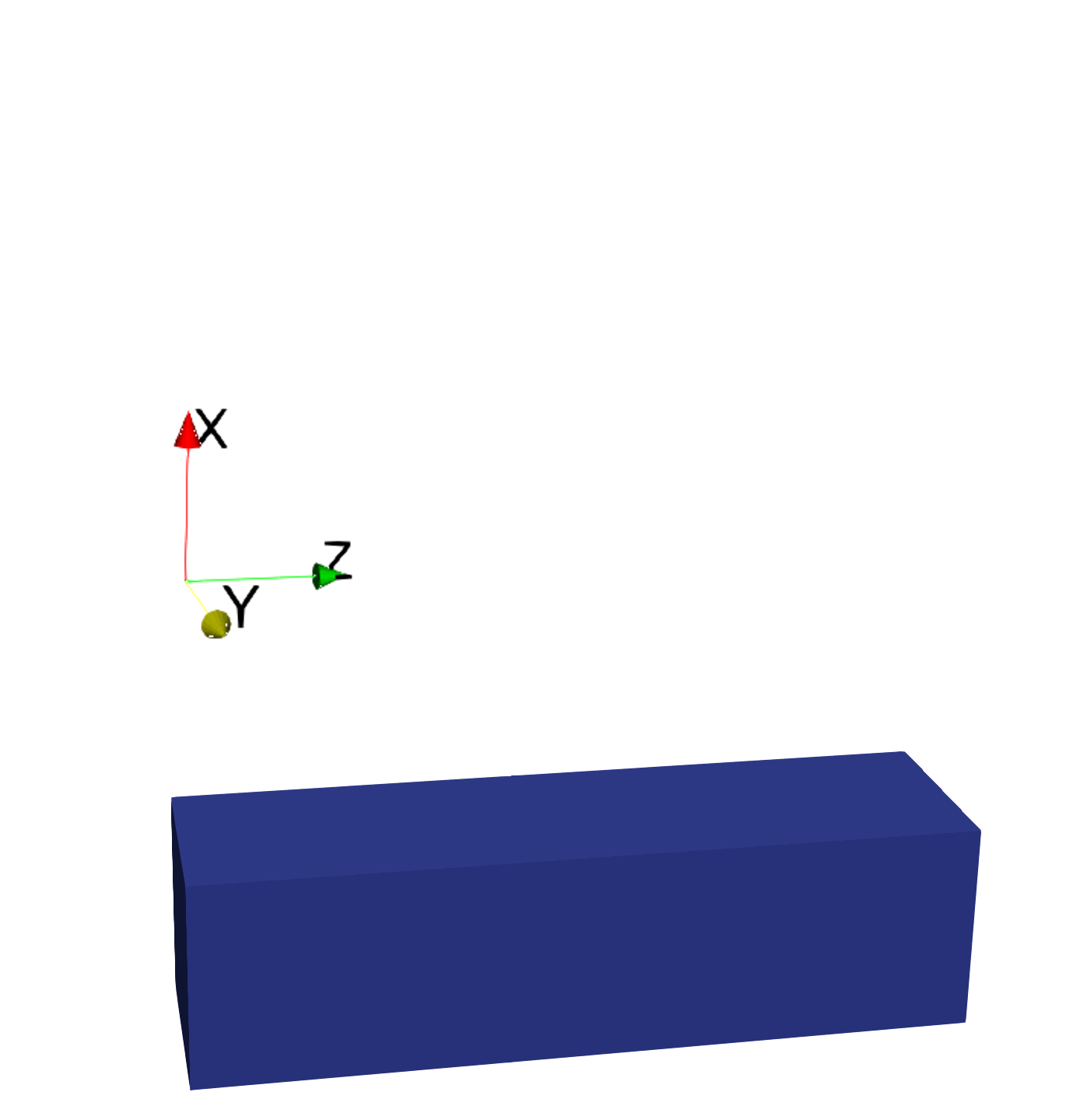}
    \caption{Boundary condition given in reference configuration, $t=\SI{1}{\milli\second}$}%
    \label{fig:reference_configuration_1}%
  \end{subfigure}
  \quad
  \begin{subfigure}[t]{0.31\textwidth}%
    \centering%
    \includegraphics[width=\textwidth]{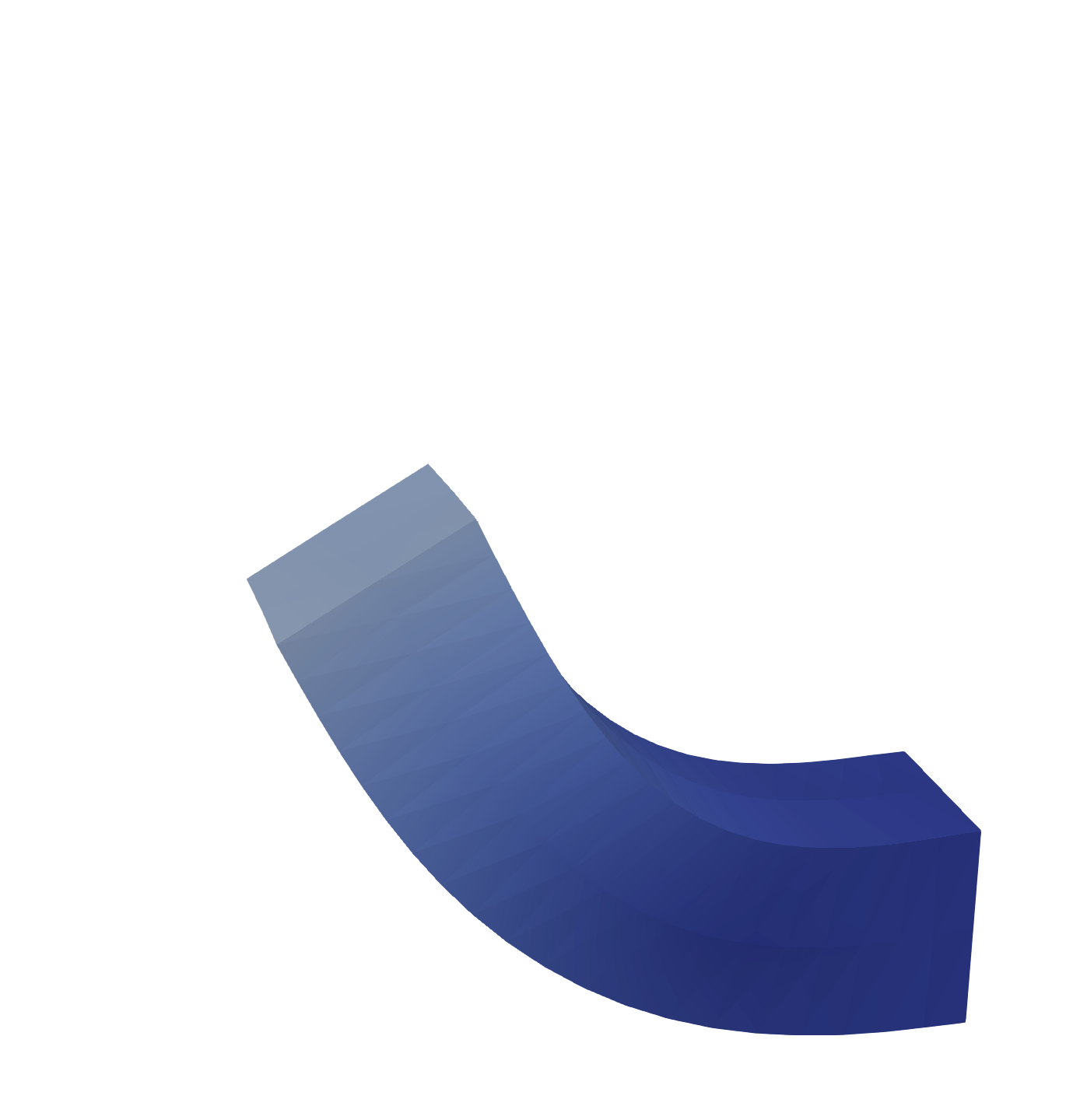}
    \caption{Boundary condition given in reference configuration, $t=\SI{30}{\milli\second}$}%
    \label{fig:reference_configuration_2}%
  \end{subfigure}
  \quad
  \begin{subfigure}[t]{0.31\textwidth}%
    \centering%
    \includegraphics[width=\textwidth]{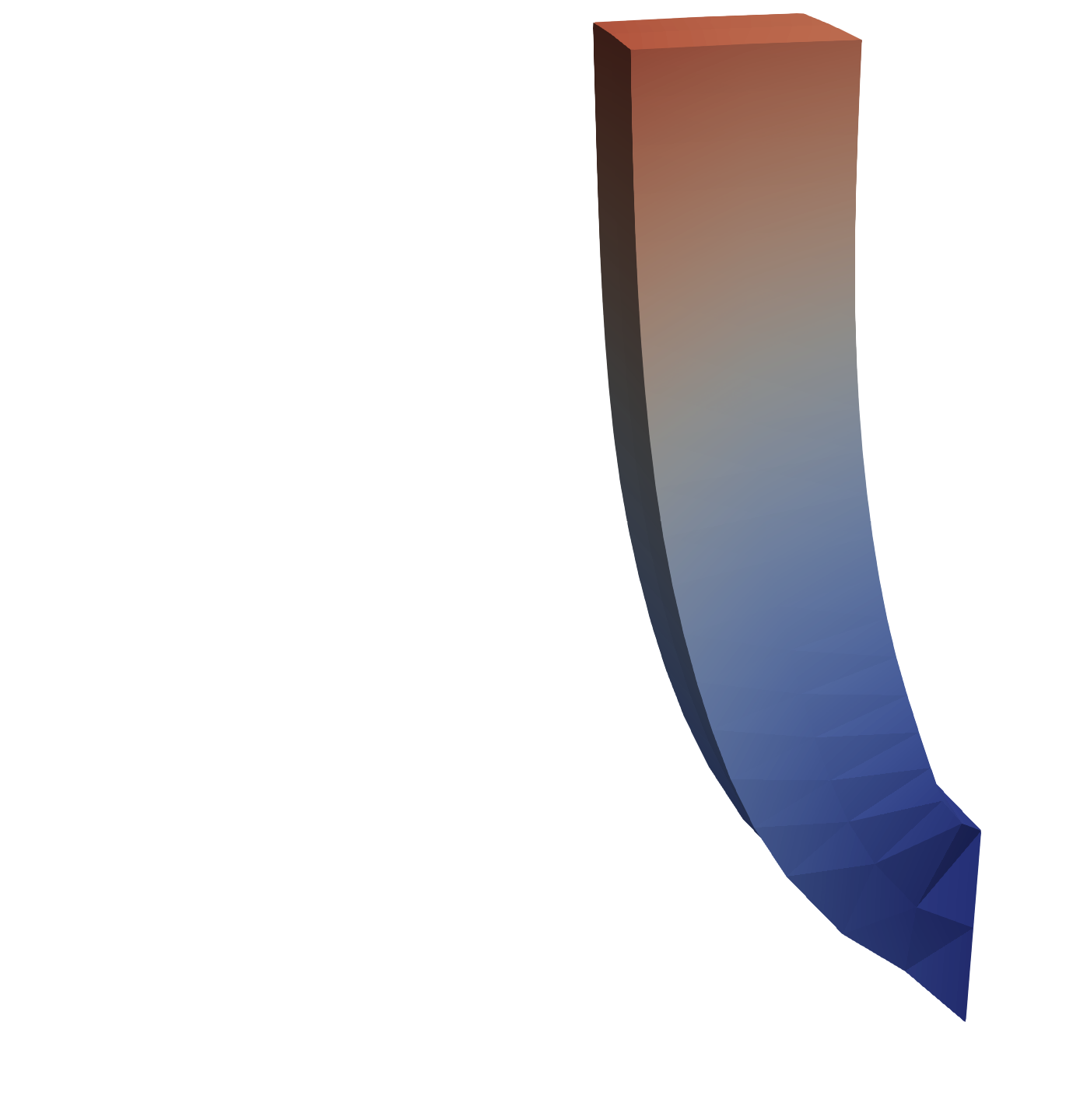}
    \caption{Boundary condition given in reference configuration, $t=\SI{60}{\milli\second}$}%
    \label{fig:reference_configuration_3}%
  \end{subfigure}
  \\
  \begin{subfigure}[t]{0.31\textwidth}%
    \centering%
    \includegraphics[width=\textwidth]{images/implementation/current_configuration_1_cropped.png}
    \caption{Boundary condition given in current configuration, $t=\SI{1}{\milli\second}$}%
    \label{fig:current_configuration_1}%
  \end{subfigure}
  \quad
  \begin{subfigure}[t]{0.31\textwidth}%
    \centering%
    \includegraphics[width=\textwidth]{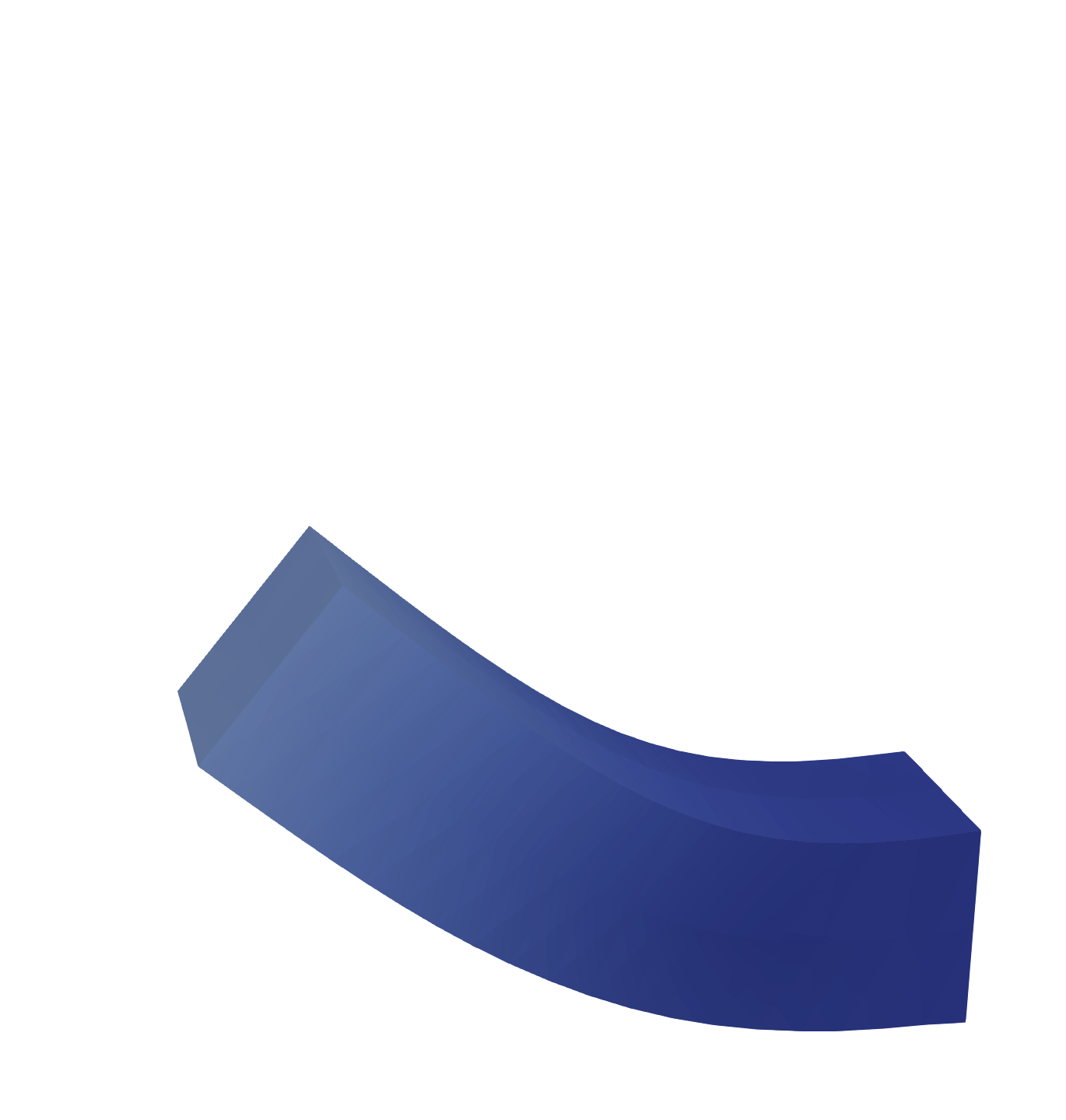}
    \caption{Boundary condition given in current configuration, $t=\SI{30}{\milli\second}$}%
    \label{fig:current_configuration_2}%
  \end{subfigure}
  \quad
  \begin{subfigure}[t]{0.31\textwidth}%
    \centering%
    \includegraphics[width=\textwidth]{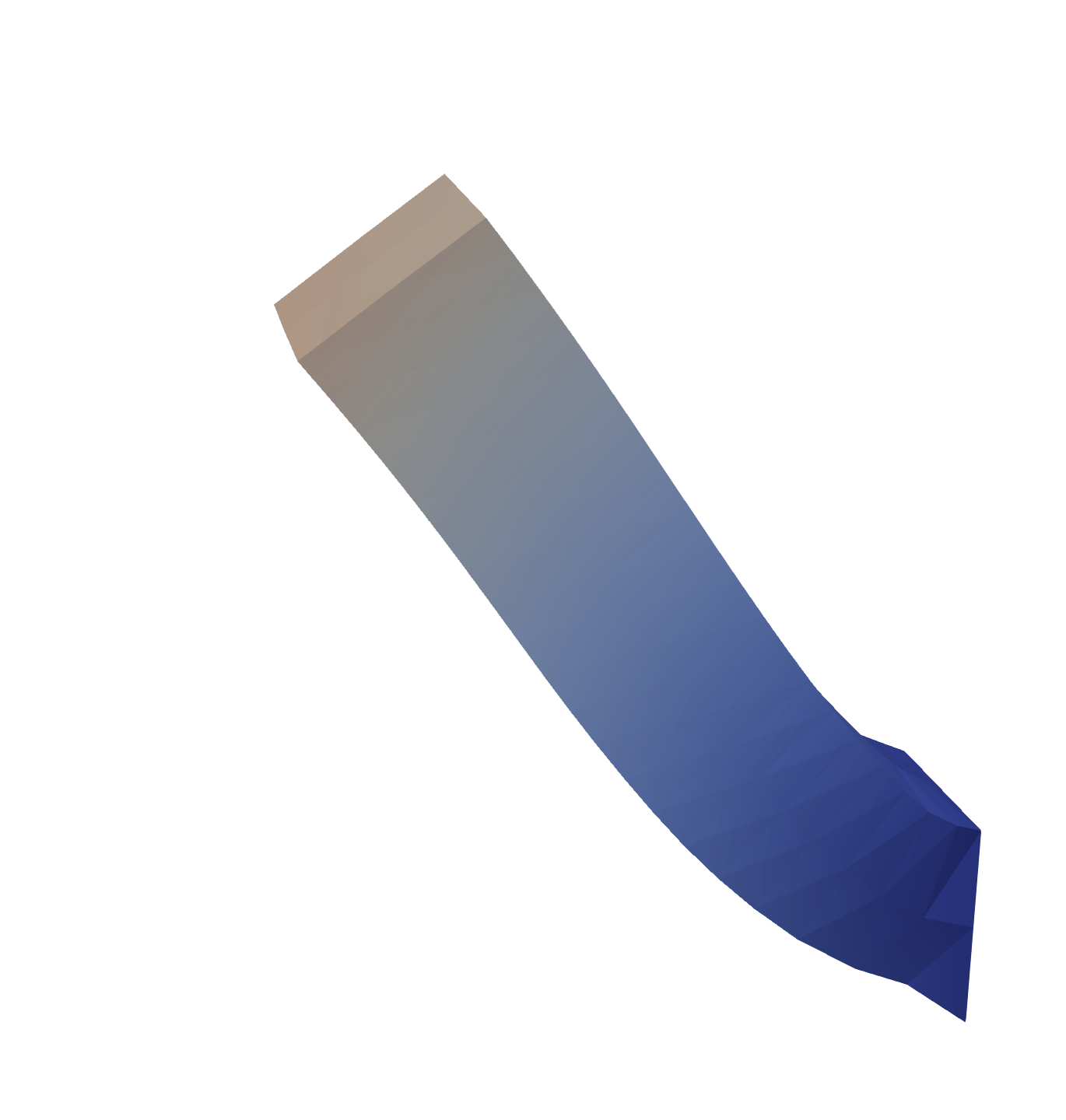}
    \caption{Boundary condition given in current configuration, $t=\SI{60}{\milli\second}$}%
    \label{fig:current_configuration_3}%
  \end{subfigure}
  \caption{Influence of whether external traction forces are defined in the current or in the reference configuration. Simulation of a dynamic solid mechanics problem with a Mooney Rivlin material model and a constant traction force in $x$ direction.}%
  \label{fig:traction_current_reference_configuration}%
\end{figure}

\Cref{fig:traction_current_reference_configuration} illustrates the difference between Neumann boundary conditions that are specified in the  reference configuration and the current configuration. A horizontal, cuboid rod is fixed at its right end and a traction force in positive $x$ direction acts on the surface on its other end.
A dynamic hyperelasticity model with isotropic Mooney-Rivlin material is solved.

\Cref{fig:reference_configuration_1,fig:reference_configuration_2,fig:reference_configuration_3} show the simulation results, where the traction force is specified in reference configuration, \cref{fig:current_configuration_1,fig:current_configuration_2,fig:current_configuration_3} show results of the same simulation, but with the traction force specified in the current configuration.
It can be seen that the rod bends more to the right, if the traction force is specified in reference configuration. In this case, the force is always acting perpendicular to the rod, whereas, in the other version, the direction of the applied force in the global coordinate system stays constant.

\begin{reproduce}
  Use the following commands to create the results in \cref{fig:traction_current_reference_configuration}.
  \begin{lstlisting}[columns=fullflexible,breaklines=true,postbreak=\mbox{\textcolor{gray}{$\hookrightarrow$}\space}]
    cd $\$$OPENDIHU_HOME/examples/electrophysiology/fibers/fibers_contraction/with_tendons_precice/meshes
    ./create_cuboid_meshes.sh     # create the cuboid mesh
    cd $\$$OPENDIHU_HOME/examples/electrophysiology/fibers/fibers_contraction/with_tendons_precice/traction_current_or_reference_configuration
    mkorn && srr       # build
    ./muscle_precice settings_current_configuration.py ramp.py
    ./muscle_precice settings_reference_configuration.py ramp.py
  \end{lstlisting}
\end{reproduce}

% matrices? how they are stored in PETSc, different approaches in elasticity and multidomain -> no
% slot connector data and transfer, no already described

\section{Parallel Partitioning and Subsampling of Meshes}\label{sec:parallel_partitioning_and_sampling_of_the}

% dedicated solver

The derivation of increasingly detailed models in the domain of biomechanics has to be complemented by engineering of efficient software that is used to solve these models. Using proper parallelization allows to increase the amount of computational load that is possible to handle. In turn, this allows to simulate more complex models with higher resolution and ultimately enables physiological and pathological insights on a new level.

For detailed multi-scale model solvers, parallelization is a complex task. 
The paradigm has to be regarded  during the whole setup process of the system. Different descriptions for the same physical behavior have to be evaluated with respect to their solvability in parallel. For a given model, suitably parallelizable numerical solution schemes have to be selected. The implementation of individual solvers and their coupling have to take into account the parallel environment. 
Discretization schemes enabling parallel domain decomposition are required. Their representation on compute hardware with distributed memory has to be taken into account as well as ensuring acceptable conditioning of large scale problems. To ensure fast runtimes, load balancing between compute nodes and parallel scalability are important.

All these fundamental considerations potentially depend on each other and require a comprehensive solution. 
Thus, it is often difficult to port existing, isolated solver software that was designed for serial or moderately parallel execution to efficiently fit into a highly-parallel, multi-scale solution framework. To not (re-)create this kind of isolated solvers for individual model components, we focus on their parallel design from the ground up in the current and following sections.

In this section, we introduce algorithms for the generation of parallel partitioned meshes, which are fundamental ingredients to all our solvers. The parallel organization of the data and their indexing using various numbering schemes
has already been discussed in \cref{sec:oragnization_of_parallel_partitioned_data} and \cref{sec:numbering_schemes_for}, respectively. 
In the following, we consider the parallel partitioning problem on a higher level and provide algorithms to construct the domain decomposition for various meshes in the multi-scale model discretization. Meshes with different mesh widths are obtained by subsampling a finely resolved initial mesh, which is the outcome of the algorithms described in \cref{sec:generation_of_meshes_for_multiscale}, and, in practice, is given to a particular simulation program by the respective mesh input file.

\Cref{sec:algorithm_for_partitioning_and_sampling,sec:partitioning_requirements} set the scene and define our requirements for well-behaved parallelized meshes. \Cref{sec:partitioning_alg1,sec:partitioning_alg2} give details on the implemented algorithms and \cref{sec:partitioning_user_options} addresses the configuration for the user. \Cref{sec:partitioning_results} concludes by comparing the resulting partitionings for different parameters.
%The subsequent sections present the parallel solvers for various parts of the multi-scale model.

%\subsection{Algorithm for Partitioning and Sampling the 3D Mesh}\label{sec:algorithm_for_partitioning_and_sampling}
\subsection{Specification of the Partitioning}\label{sec:algorithm_for_partitioning_and_sampling}

Structured meshes of the types \code{RegularFixed}\code{OfDimension<D>} or \code{Structured}\code{Deformable}\code{OfDimension<D>} are partitioned for parallel execution by distributing the elements to all processes. As mentioned in \cref{sec:oragnization_of_parallel_partitioned_data}, planar cuts in the space of the element indices separate the subdomains. For example, in computations on a structured 3D mesh with $N_x^\text{el} \times N_y^\text{el} \times N_z^\text{el}$ global elements, the process with rank $r$ owns a subdomain with
$N_x^{\text{el,local,}r} \times N_y^{\text{el,local,}r} \times N_z^{\text{el,local,}r}$ local elements.
The sizes of the local subdomains depend on the specified total number of subdomains $n_i$ in each coordinate direction $i \in \{x,y,z\}$.
Given $n_i$, the number of local elements in every subdomain along the coordinate axis $i$ can be set to either $N_i^{\text{el,local}} = \lfloor N^\text{el}_i/n_i+1\rfloor$ or $N_i^{\text{el,local}} = \lfloor N^\text{el}_i/n_i \rfloor$ to allow for good load balancing.

A prerequisite to construct such a partitioning for $n_\text{proc}$ processes is to fix the numbers of subdomains $n_x \times n_y \times n_z = n_\text{proc}$. In OpenDiHu, the Python settings file can either specify the global numbers $N_i^\text{el}$ of elements or separate local numbers $N_i^{\text{el,local,}r}$ of elements for every rank $r$. This step involves setting the option \code{inputMeshIsGlobal} to either \code{True} or \code{False} as explained in \cref{sec:exemplary_usage_1}.

Specifying the global numbers of elements is often useful for toy problems, when the total element count is small and the actual partitioning is not important. In this case, PETSc is used to determine optimal subdomain sizes for all processes and, subsequently, constructing the partitioning. Because the partitioning is not yet known at the time of parsing of the Python settings, spatial information such as node positions or boundary conditions have to be specified on every rank for the whole domain.

Most of the electrophysiology examples, however, use the specification of local numbers of elements. Thus, every rank only needs to specify the local data of its subdomain, such as node positions and boundary conditions. This is a prerequisite for good parallel weak scaling behavior, as the amount of data processing on each process stays constant when simultaneously increasing problem size and total process counts.

In the electrophysiology examples, the partitioning into $n_x \times n_y \times n_z$ subdomains can be specified by the command line parameter \code{--n_subdomains n_x n_y n_z}, where \code{n_x}, \code{n_y} and \code{n_z} are replaced by the actual numbers. Their product has to match the process count $n_\text{proc}$ that is given to MPI to start the program.
If this option is not specified, the values are determined automatically by the following algorithm: For all partitions of the number $n_\text{proc}$ into three integer factors, a performance value $p$ is computed as follows:
\begin{align*}
  p = (n_x-n_\text{opt})^2 + (n_y-n_\text{opt})^2 + (n_z-n_\text{opt})^2.
\end{align*}
The optimal value is given by $n_\text{opt} = n_\text{proc}^{1/3}$, which, in general, is not an integer. The partitioning with the lowest value of $p$ is selected among all partitions, as it leads to nearly cuboid subdomains with the best volume-to-surface ratio. An advantage of this method is that it is independent of the mesh size.

\subsection{Requirements for Partitioning and Sampling of the 3D Mesh Based on 1D Fiber Meshes}\label{sec:partitioning_requirements}
%\subsection{Requirements for Partitioning and Sampling of the 3D Mesh}

Next, we specify desired properties of the parallel partitioned 3D meshes, which are used together with 1D muscle fiber meshes in the discretization of fiber based multi-scale models. Subsequently, we construct an algorithm to generate the accordingly partitioned 3D meshes in parallel by sampling a finer dataset based on 1D fiber meshes.

Simulation scenarios with fiber based electrophysiology use a 3D muscle mesh and embedded 1D fiber meshes, which are generated from the same node positions as described in \cref{sec:postprocessing_of_the_generated_streamlines}. The binary input file contains a structured grid of points, which can be either interpreted as 1D fibers by connecting the points in $z$-direction or as 3D mesh by additionally connection points in $x$ and $y$-directions.

Usually, all points in such a file are used to define the 1D fiber meshes and the 3D mesh is constructed from only a subset of the available points. To obtain a 3D mesh with approximately equal mesh widths in all coordinate directions, the point data are sampled by constant strides in $x$, $y$ and $z$ direction.  The stride in fiber direction ($z$ direction) is typically chosen larger than the strides in transverse directions as the distance between the given points is smaller in this direction.

In the following, we discuss the sampling procedure that generates the partitioned 3D mesh from the fiber data in more detail.
Given a structured hexahedral fine 3D mesh, numbers of subdomains $n_i$ and sampling stride parameters \code{sampling_stride_$i$} for the three coordinate directions $i\in\{x,y,z\}$, we have to determine the nodes that should be part of each subdomain in the resulting coarser hexahedral 3D mesh. 

\begin{figure}
  \centering%
  \includegraphics[width=\textwidth]{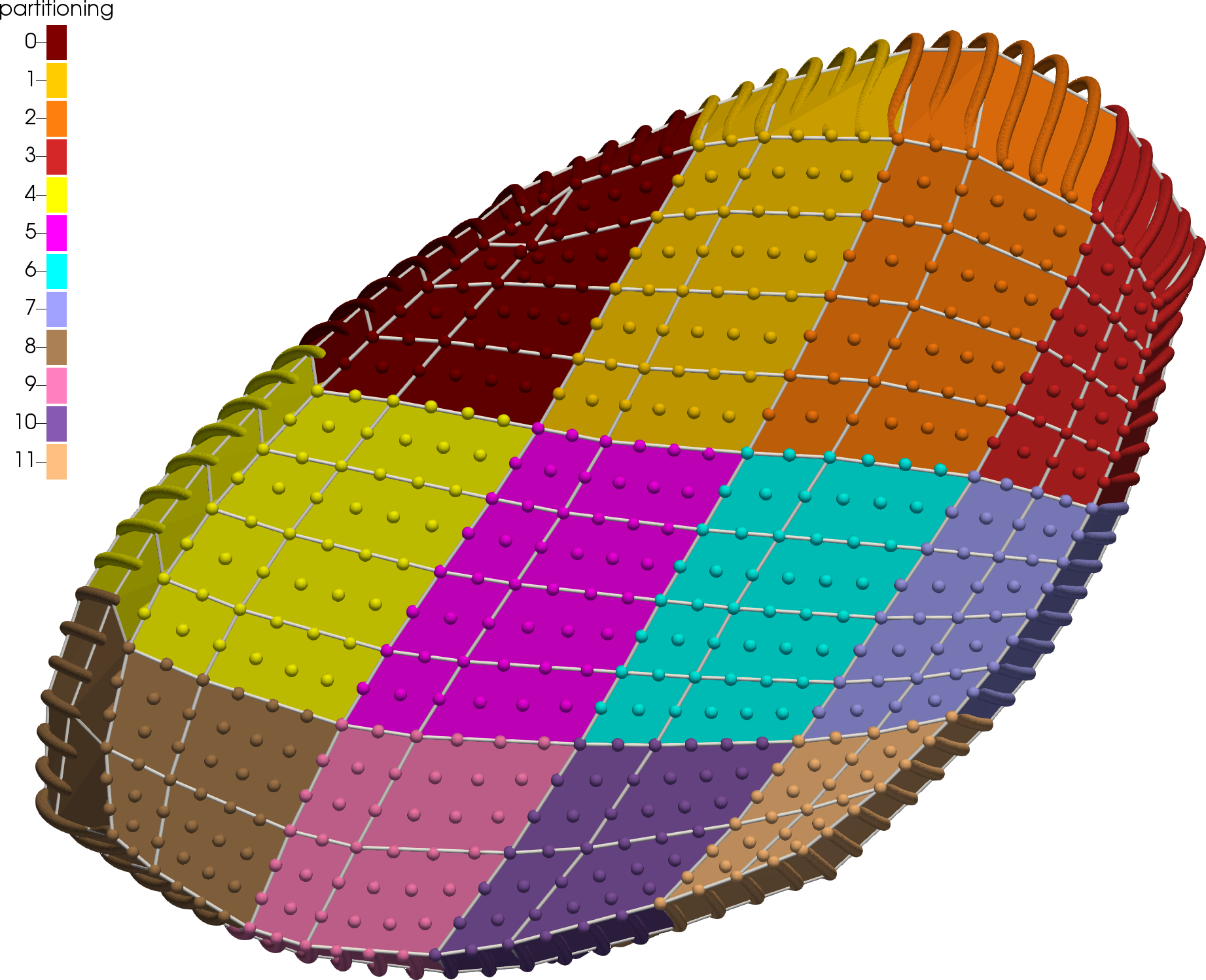}%
  \caption{Partitioning and subsampling of a fiber mesh to twelve processes. The fiber data indicated by the spheres are sampled with a stride parameter of two to obtain the partitioned quadratic coarse mesh given by the white elements. The subdomains are indicated by different colors. The image shows a perspective view on the top 2D face of the 3D muscle mesh.}%
  \label{fig:partitioning1}%
\end{figure}

For illustration, \cref{fig:partitioning1} shows the initial fine mesh visualized by spheres that are arranged in fibers, that run from the shown cross-section to the back. The resulting sampled mesh is given by the white elements and uses a subset of the nodes in the fine mesh. The sampled mesh is partitioned into the colored subdomains. 
Furthermore, the coarse mesh consists of quadratic elements that are formed from two by two white standard elements, in the cross-section each. Hence, every subdomain contains an even number of the white elements in horizontal and vertical directions.

% quadratic partitioning

The requirements for the sampling and partitioning algorithm are as follows: 
\begin{enumerate}[label=(\roman*)]
\item The resulting coarser 3D mesh should use every $k$th node, where $k$ is adjustable by the parameter \code{sampling_stride_$i$} in the settings.
\item The number of nodes in every subdomain should be approximately equal to allow for a good load balancing in the computation.
\item There should be as little \say{remainder elements} that have a different mesh width than the majority of the elements as possible.
\item If a quadratic shape functions are required, e.g., for solid mechanics models, the number of (standard) elements in every subdomain in every coordinate direction has to be even to allow for the generation of quadratic hexahedral elements.
\end{enumerate}

Clearly, not all requirements can be fulfilled exactly for all given input meshes. For combinations of given input mesh sizes and sampling strides that lead to an even number of sampled nodes, requirement (iv) cannot be fulfilled. 
Exact fulfillment of requirement (ii), i.e., an equal number of nodes in every subdomain is also only possible for suited parameter choices. Therefore, we relax requirement (i) and also occasionally allow different step widths between the selected nodes on the fine grid. Having varying distances between the nodes leads to elements with different mesh widths, which is unfavorable in terms of the numerical conditioning of the problem. Therefore, the number of such elements should be as low as possible, which is also stated by requirement (iii).

To avoid differently sized elements as far as possible, we work with a granularity parameter. This parameter specifies the amount of nodes to summarize and treat as an indivisible unit. For example, a value of \code{granularity_x=2} specifies that pairs of two neighboring points are in the same element. Then, subdomain boundaries and element boundaries can only occur at every second node.

\subsection{Algorithm for Partitioning and Sampling of the 3D Mesh Based on 1D Fiber Meshes}\label{sec:partitioning_alg1}

Important steps in the algorithm for sampling the fine mesh and constructing the partitioning are, first, to determine the locations of the new subdomains in the original fine grid, second, to determine the number of sampled points in each subdomain and third, to determine which points from the fine grid will be sampled in every subdomain of the coarse grid. The steps have to be carried out independently for all three coordinate directions. Thus, it suffices to only consider the algorithm for the partitioning along one axis.
In the following, we present the algorithms of the first two steps for the $x$-axis. 

The algorithm for the first step is given in \cref{alg:n_fibers_in_subdomain}. Input to the function \code{n_fibers_in_}\code{subdomain_x} is a subdomain coordinate in the range $[0,n_x-1]$ that identifies the subdomain. The output to be computed is the number of grid points in the fine grid or, equivalently, the number of fibers that are contained in the subdomain. Calling this function for all subdomains defines the partitioning of the fine grid.

% partitioning algorithm
\begin{figure}
  \centering%
  \includegraphics[width=\textwidth]{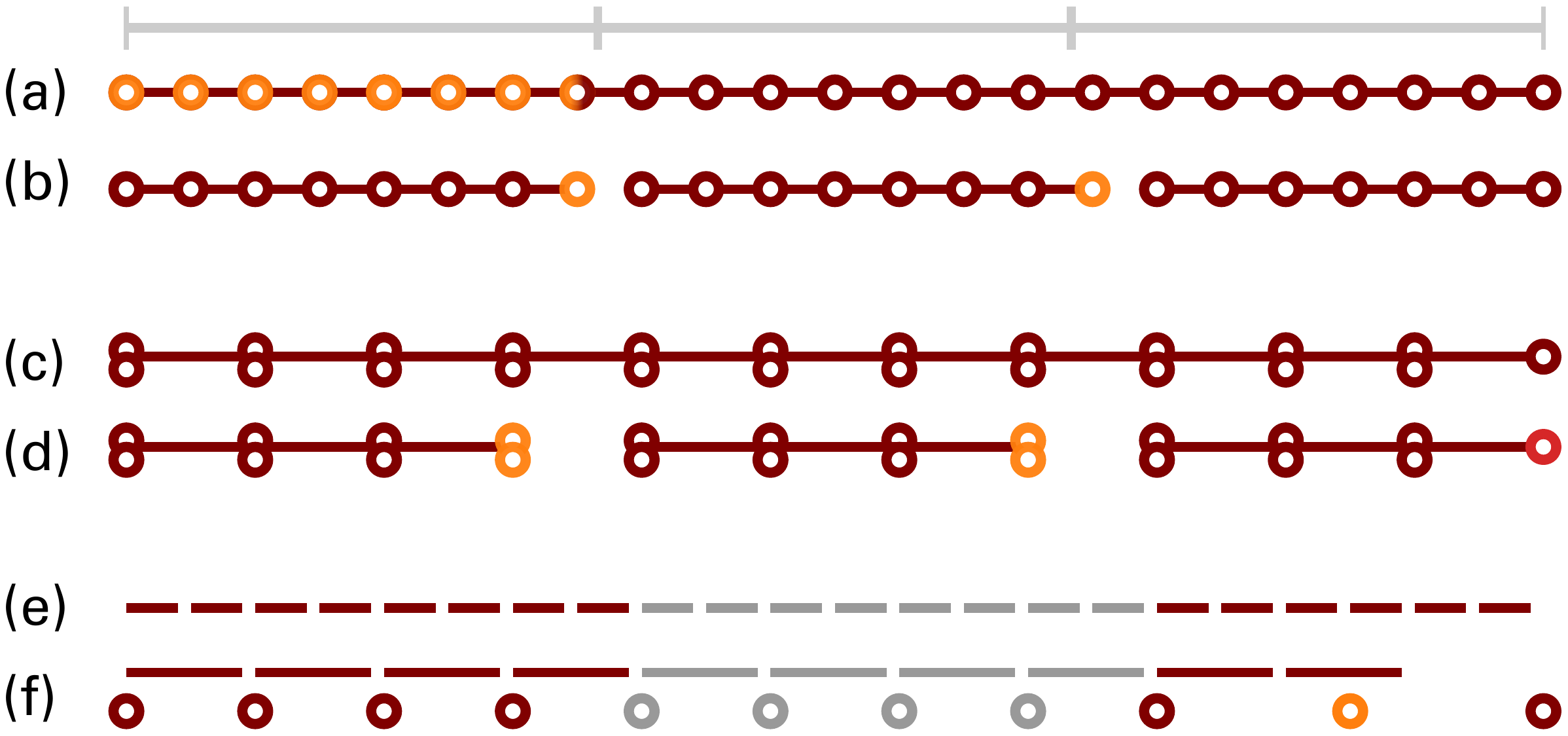}%
  \caption{Visualization of the steps of the partitioning algorithms given by \cref{alg:n_fibers_in_subdomain,alg:n_sampled_points_in_subdomain} that yield the partitioning shown in \cref{fig:partitioning1}.}%
  \label{fig:partitioning_algorithm}%
\end{figure}%

\begin{algorithm}
  \begin{algorithmic}[1]%
    \Procedure{n\_fibers\_in\_subdomain\_x}{subdomain\_coordinate\_x}
    \Require Index of a subdomain in $x$-direction
    \Ensure Number of fibers that are contained in this subdomain
    \Statex
    \State   $\alpha$ = $\lfloor$ n\_fibers\_x / $n_x$ / granularity\_x $\rfloor$ * granularity\_x   \label{alg:3.2}
    \Statex
    \State a1 = $\lfloor$(n\_fibers\_x - $n_x$ * $\alpha$) / granularity\_x $\rfloor$ \label{alg:3.3}  \Comment{subdomains with $>\alpha$ nodes}
    \State a2 = $n_x$ - a1                        \label{alg:3.4}              \Comment{subdomains with $\alpha $ nodes}
    \Statex
    \If{subdomain\_coordinate\_x < a1} \Comment{first a1 subdomains} \label{alg:3.5}
      \State \textbf{return} $\alpha$ + granularity\_x \label{alg:3.6}
    \ElsIf{subdomain\_coordinate\_x < $n_x$ - 1}\label{alg:3.7}
      \State \textbf{return} $\alpha$                        \label{alg:3.8}
    \Else  \Comment{last subdomain}  \label{alg:3.9}
      \State \textbf{return} $\alpha$ + n\_fibers\_x \% granularity\_x  \label{alg:3.10}
    \EndIf
    \EndProcedure
  \end{algorithmic}%
  \caption{Computation of subdomain sizes, needed for the construction of a parallel partitioning.}%
  \label{alg:n_fibers_in_subdomain}%
\end{algorithm}%

\Cref{fig:partitioning_algorithm} provides a visualization of the algorithmic steps, corresponding to the partitioning in vertical direction of the mesh shown in \cref{fig:partitioning1}. \Cref{fig:partitioning_algorithm} (a) shows a 1D mesh with \code{n_fibers_x=23} nodes or fibers.
% By comparing with \cref{fig:partitioning1}, it can be seen that nodes and fibers are equivalent in this point of view.
The goal is to partition them to $n_x=3$ subdomains. According to requirement (ii), the nodes should be distributed equally to the subdomains. Dividing 23 nodes by 3 subdomains yields an average number of $7\frac23$ nodes per subdomain, which is indicated by the orange color in \cref{fig:partitioning_algorithm} (a). 

For now, we neglect the granularity parameter and set \code{granularity_x=1}. 
Line \ref{alg:3.2} of the algorithm computes the rounded down value $\alpha$ of the average number fraction. Every subdomain should obtain either $\alpha$ or $(\alpha+1)$ nodes. We specify that the first \code{a1} subdomains obtain $(\alpha+1)$ nodes and the remaining subdomains obtain $\alpha$ nodes. 
The amount of nodes that remain after we fill every subdomain with $\alpha$ nodes is the difference between all nodes \code{n_fibers_x} and  $n_x \cdot \alpha$. This difference is equal to \code{a1} and the formula in line \ref{alg:3.3} of the algorithm computes the value of \code{a1} accordingly. The remainder number of subdomains \code{a2} follows as given in line \ref{alg:3.4}.
The visualization in \cref{fig:partitioning_algorithm} (b) shows that, in the example, \code{a1=2} subdomains obtain $\alpha+1=8$ nodes and only the last subdomain, i.e., \code{a2=1}, obtains $\alpha=7$ nodes.

The rest of \cref{alg:n_fibers_in_subdomain} checks whether the given subdomain coordinate \code{subdomain}\code{_coor}\code{dinate_x} refers to a subdomain with $(\alpha+1)$ or with $\alpha$ nodes by comparing the coordinate with \code{a1} in line \ref{alg:3.5}. The first branch of the \code{if} statement returns the high number of nodes $(\alpha+1)$, the other branches return the low number $\alpha$, as far as the granularity parameter is neglected.

Next, we discuss the algorithm with a granularity value that is different from 1. 
Assuming a value of, e.g., \code{granularity_x=2}, always two neighboring nodes are grouped and the algorithm acts on these groups instead of individual nodes. The visualization in \cref{fig:partitioning_algorithm} (c) shows this grouping. Because the considered example has an odd total number of 23 nodes, only a single nodes remains for the last group.

The number of nodes per subdomains should now be a multiple of the granularity. This is ensured in line \ref{alg:3.2} of \cref{alg:n_fibers_in_subdomain} by dividing by the granularity, rounding down and multiplying again with the granularity. The subdomains obtain either \code{$\alpha$} or \code{($\alpha$ + granularity_x)} nodes. The computation of the number \code{a1} of subdomains with the higher number of nodes in line \ref{alg:3.3} requires a division by \code{granularity_x} as every subdomain with the higher number takes \code{granularity_x} extra nodes. The rounding down in line \ref{alg:3.3} is needed to obtain an integer value even if the total number of nodes is not a multiple of the granularity.

In the example in \cref{fig:partitioning_algorithm} (d), the subdomains obtain either $\alpha=6$ or $\alpha +$ \code{granularity_x}$=8$ nodes. In fact, for the last subdomain, only seven nodes remain, as the total number of 23 nodes is not divisible by the granularity of two.
In the algorithm, this is accounted for by the last branch of the \code{if-else} construct in line \ref{alg:3.10}, where only the remaining nodes are added to the last subdomain.

\subsection{Algorithm for Sampling Points from a Fine Fiber Mesh}\label{sec:partitioning_alg2}

Next, we can sample points from the nodes that were assigned to each subdomain. The sampling process is parametrized by the value of \code{sampling_stride_x}, which specifies the step width of the nodes from the fine mesh to select for the coarse mesh.
\Cref{alg:n_sampled_points_in_subdomain} lists the function that determines the number of sampled points in a given subdomain. Similar to \cref{alg:n_fibers_in_subdomain}, the input is a 1D subdomain coordinate. The output is the number of sampled points in this subdomain.

\begin{algorithm}
  \begin{algorithmic}[1]%
    \Procedure{n\_sampled\_points\_in\_subdomain\_x}{subdomain\_coordinate\_x}
    \Require Index of a subdomain in $x$-direction
    \Ensure Number of points in the subdomain for the coarse 3D mesh
      \Statex
      \State n = n\_fibers\_in\_subdomain\_x(subdomain\_coordinate\_x)  \hypertarget{alg:4.2}
    \State \textbf{if} subdomain\_coordinate\_x == $n_x$ - 1 \textbf{then}      \hypertarget{alg:4.3}                      
      \State \hspace{0.8em} n -= 1                                                     \hypertarget{alg:4.4}
    \Statex
    \If{linear 3D elements}                                                     \hypertarget{alg:4.5}
      \State result = $\lfloor$ n / sampling\_stride\_x $\rfloor$               \hypertarget{alg:4.6}
    \Else                                                       \hypertarget{alg:4.7}
      \State result = $\lfloor$ n / (sampling\_stride\_x * 2) $\rfloor$ * 2              \hypertarget{alg:4.8}
    \EndIf
    \Statex
    \If{subdomain\_coordinate\_x == $n_x$ - 1}              \hypertarget{alg:4.9}
      \State result += 1              \hypertarget{alg:4.10}
    \EndIf              \hypertarget{alg:4.11}
    \State \textbf{return} result
    \EndProcedure
  \end{algorithmic}%
  \caption{Algorithm for sampling the fine mesh to obtain the coarser 3D mesh}%
  \label{alg:n_sampled_points_in_subdomain}%
\end{algorithm}%

First, line \hyperlink{alg:4.2}{2} of \cref{alg:n_sampled_points_in_subdomain} calls \cref{alg:n_fibers_in_subdomain} to obtain the number of fine grid points in the subdomain. The number of elements \code{n} is equal to the number of points for all except the last 1D subdomain, which has one element less. This can be seen, e.g., in \cref{fig:partitioning1}, where the first process with rank 0 (dark brown at the upper left) does not own the nodes on its subdomain boundary, whereas the last process with rank 11 (light brown at the lower right) owns all nodes on its subdomain boundary.
Thus, lines \hyperlink{alg:4.3}{3} and \hyperlink{alg:4.4}{4} of \cref{alg:n_sampled_points_in_subdomain} decrement the value of \code{n} to yield the correct number of elements. 

The corresponding visualization in \cref{fig:partitioning_algorithm} (e) assumes \code{granularity_x=2} and shows $n=8$ elements for both the first and the second subdomain and $n=6$ elements for the last subdomain. 

The resulting number of sampled points is obtained from the number of elements by a division by the sampling stride parameter and rounding down in lines \hyperlink{alg:4.5}{5} to \hyperlink{alg:4.8}{8}. For the last subdomain, line \hyperlink{alg:4.10}{10} increments the result by one to account for the additional node on the boundary.

Depending on whether the sampled mesh should contain linear or quadratic elements, the number of elements obtained from the algorithm has no restriction, or it has to be even. This is checked in the \code{if} statement in line \hyperlink{alg:4.5}{5}. In case of quadratic elements, an even number of elements is enforced by the formula in line \hyperlink{alg:4.8}{8}.

In the considered example, we require quadratic elements and set \code{sampling_stride_x=2}. The visualization in \cref{fig:partitioning_algorithm} (f) shows the number of elements as long bars, which equals the \code{result} variable before line \hyperlink{alg:4.9}{9} in the algorithm. The resulting number of nodes is given in \cref{fig:partitioning_algorithm} (f) by the circles below.

The actual selection of the nodes from the fine grid according to the stride parameter and using the determined subdomains and their numbers of contained nodes is a straight-forward task and not part of the algorithms listed here. For quadratic elements in the last subdomain, the potentially different mesh widths are resolved by selecting the second-last node in the middle between the third-last and the last node. In \cref{fig:partitioning_algorithm} (f), this case occurs in the last subdomain. The orange node is sampled at the middle between the two neighboring dark red nodes. This behavior can also be observed in the corresponding partitioning in \cref{fig:partitioning1} for the elements given by white lines in the lowest row. These elements have a larger vertical mesh width of three sampled points than the other elements, which have a vertical mesh width of two sampled points.

\subsection{User Options for the Algorithms}\label{sec:partitioning_user_options}

By adjusting the sampling stride and granularity parameters, it is possible to tune the outcome of the partitioning algorithms.
The trade-off between the two requirements given in \cref{sec:partitioning_requirements} by the numbers (ii) and (iii), i.e., that each subdomain obtains the same number of nodes, and that the least possible number of remainder elements is generated, can also be managed in the settings by enforcing either of the two requirements.

Moreover, we set the granularity parameters to the same value as the sampling parameters by default and additionally ensure  for quadratic finite elements that the granularities are a multiple of two. This setting typically yields partitionings with equally sized elements. However, the number of nodes per subdomain is not always optimal.

To allow users to enforce a partitioning, where every rank gets the exact same number of nodes, except for the last subdomains in each coordinate direction, which potentially gets one layer of nodes less, we provide the option \code{distribute_nodes_equally}, which can be set in the variables files. If this option is set to \code{True}, the granularity values are internally fixed to one for \say{linear} meshes and to two for \say{quadratic} meshes, i.e., discretizations with quadratic finite element ansatz functions.

\subsection{Results}\label{sec:partitioning_results}

\begin{figure}%
  \centering%
  \begin{subfigure}[t]{0.48\textwidth}%
    \centering%
    \includegraphics[width=\textwidth]{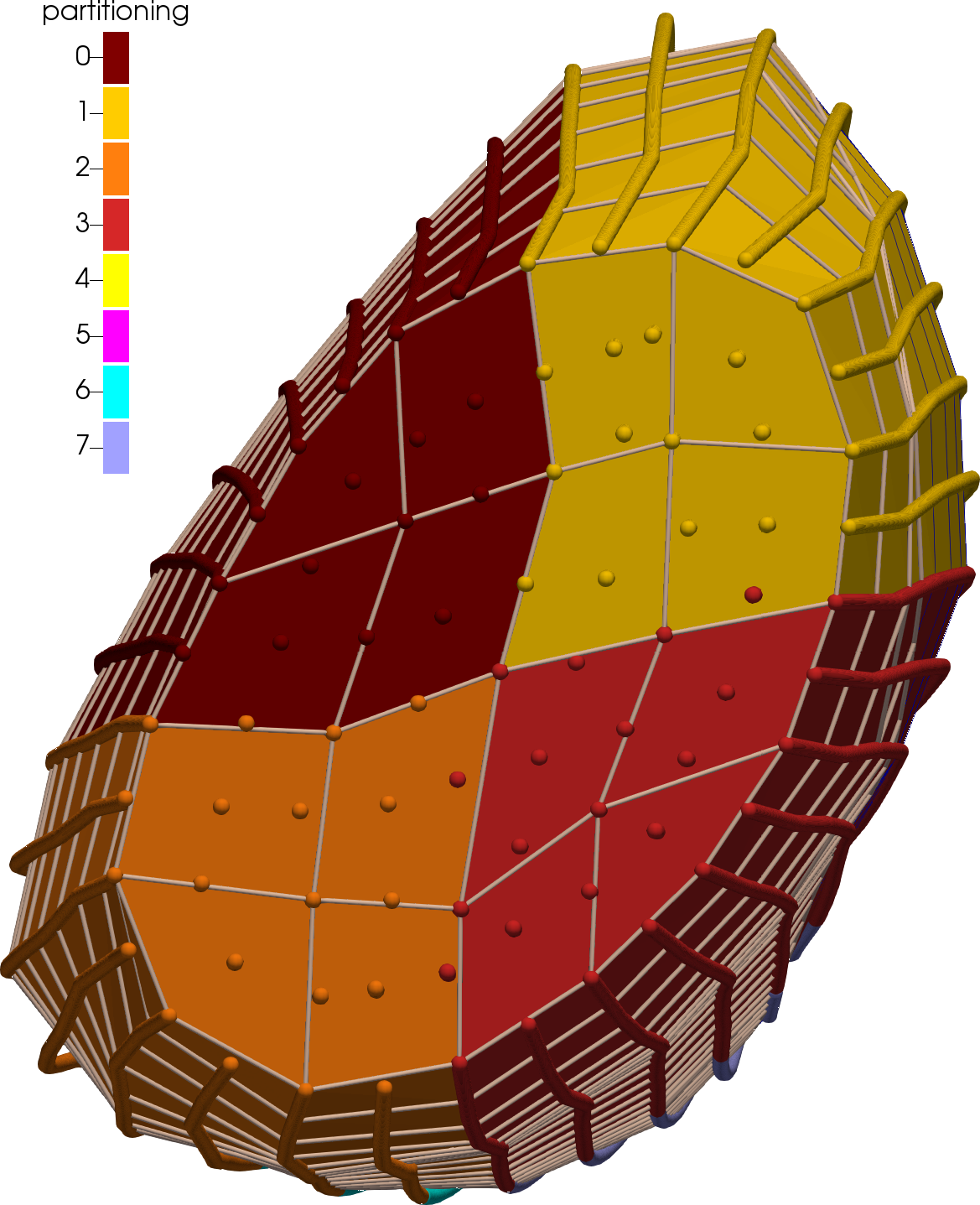}
    \caption{Resulting sampled mesh with the option \code{distribute_nodes_equally=False}.}%
    \label{fig:partitioning4}%
  \end{subfigure}
  \quad
  \begin{subfigure}[t]{0.48\textwidth}%
    \centering%
    \includegraphics[width=\textwidth]{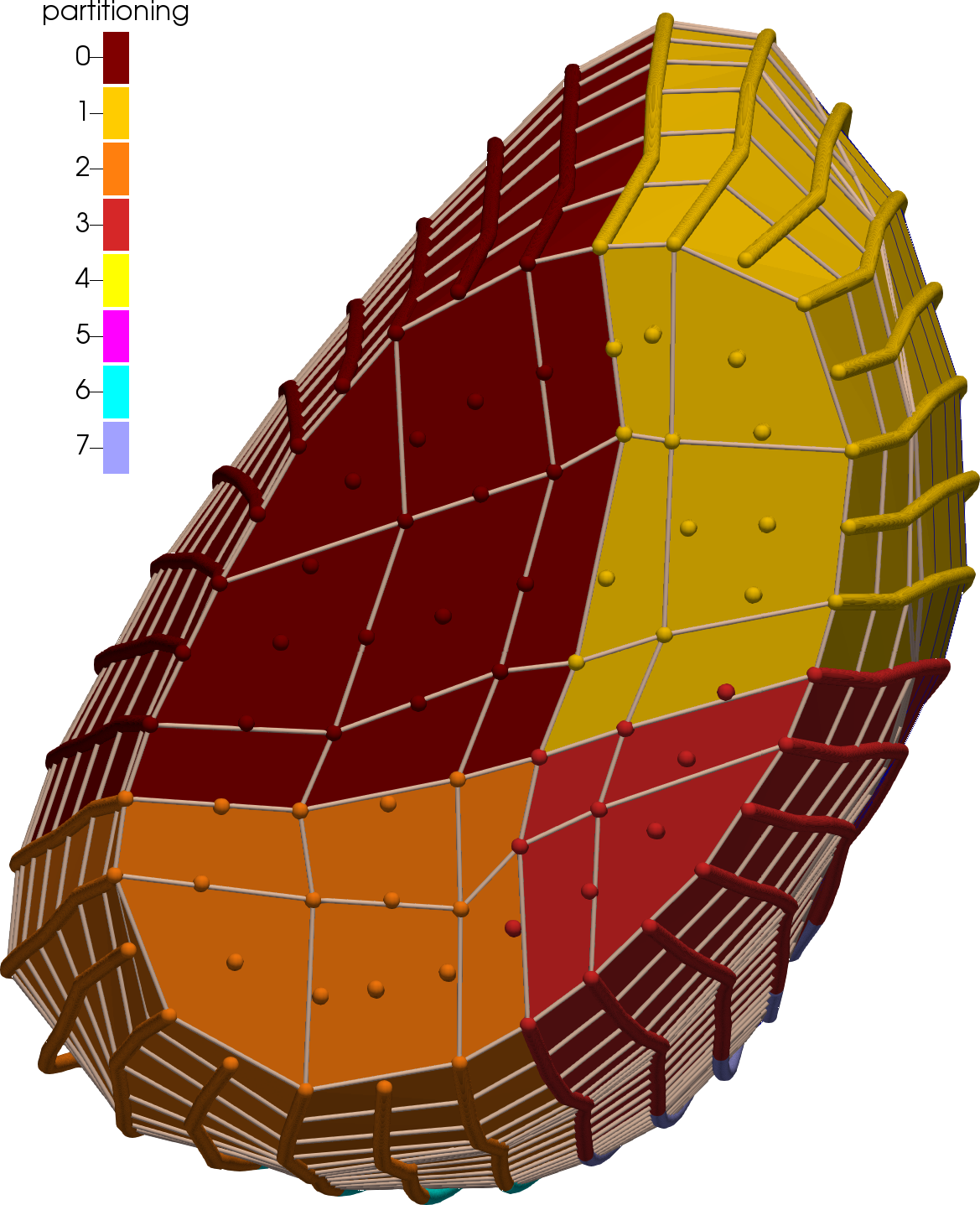}
    \caption{Resulting sampled mesh with the option \code{distribute_nodes_equally=True}.}%
    \label{fig:partitioning3}%
  \end{subfigure}
  \caption{Mesh partitions generated by the sampling algorithm with different settings. A fine mesh with 49 fibers is sampled with a stride parameter of two and partitioned to eight processes.}%
  \label{fig:partitioning3_4}%
\end{figure}

\begin{figure}%
  \centering%
  \begin{subfigure}[t]{0.48\textwidth}%
    \centering%
    \includegraphics[width=\textwidth]{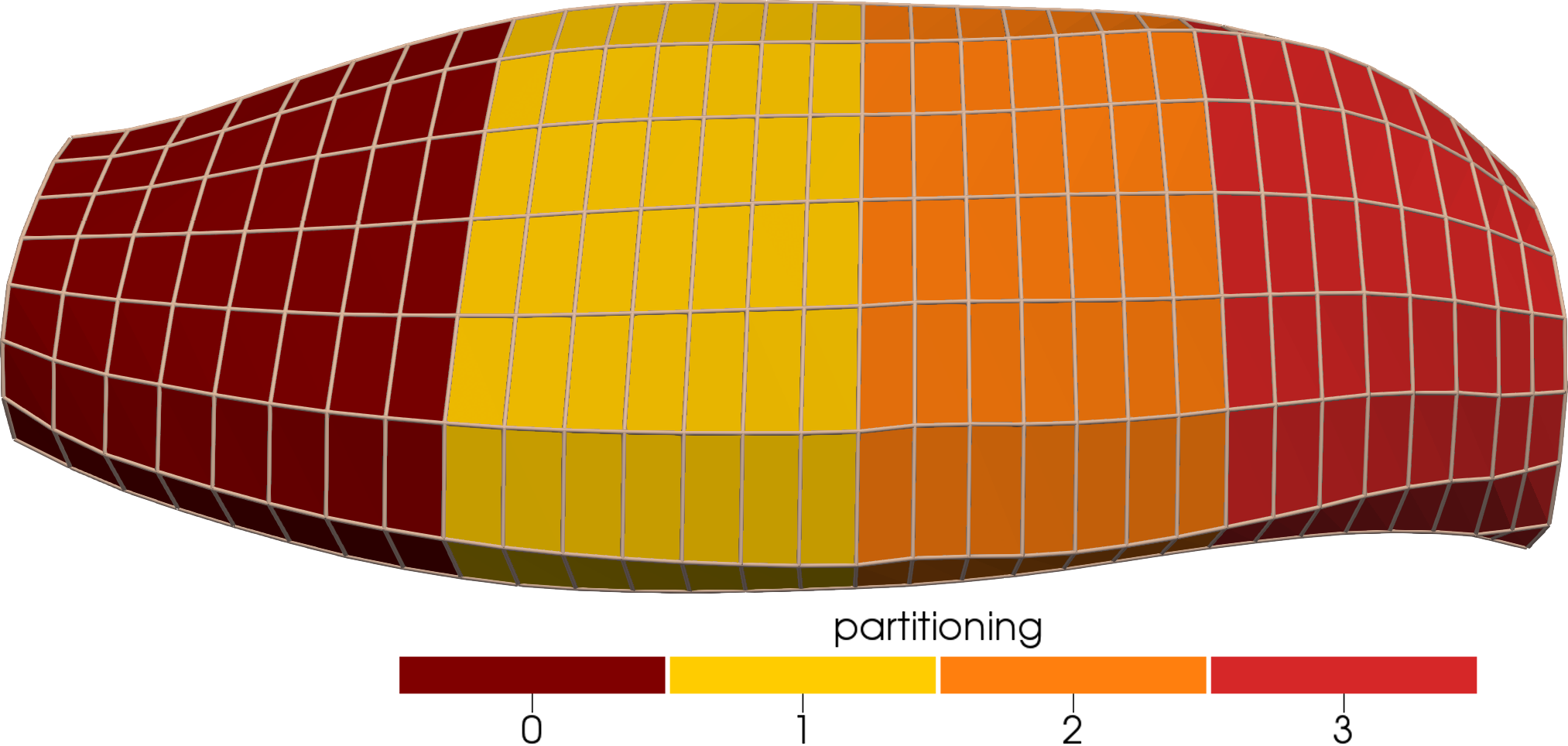}
    \caption{Resulting sampled mesh with option \code{distribute_nodes_equally=False}. The mesh width is constant, but the partitioning is not perfectly balanced.}%
    \label{fig:partitioning6}%
  \end{subfigure}
  \quad
  \begin{subfigure}[t]{0.48\textwidth}%
    \centering%
    \includegraphics[width=\textwidth]{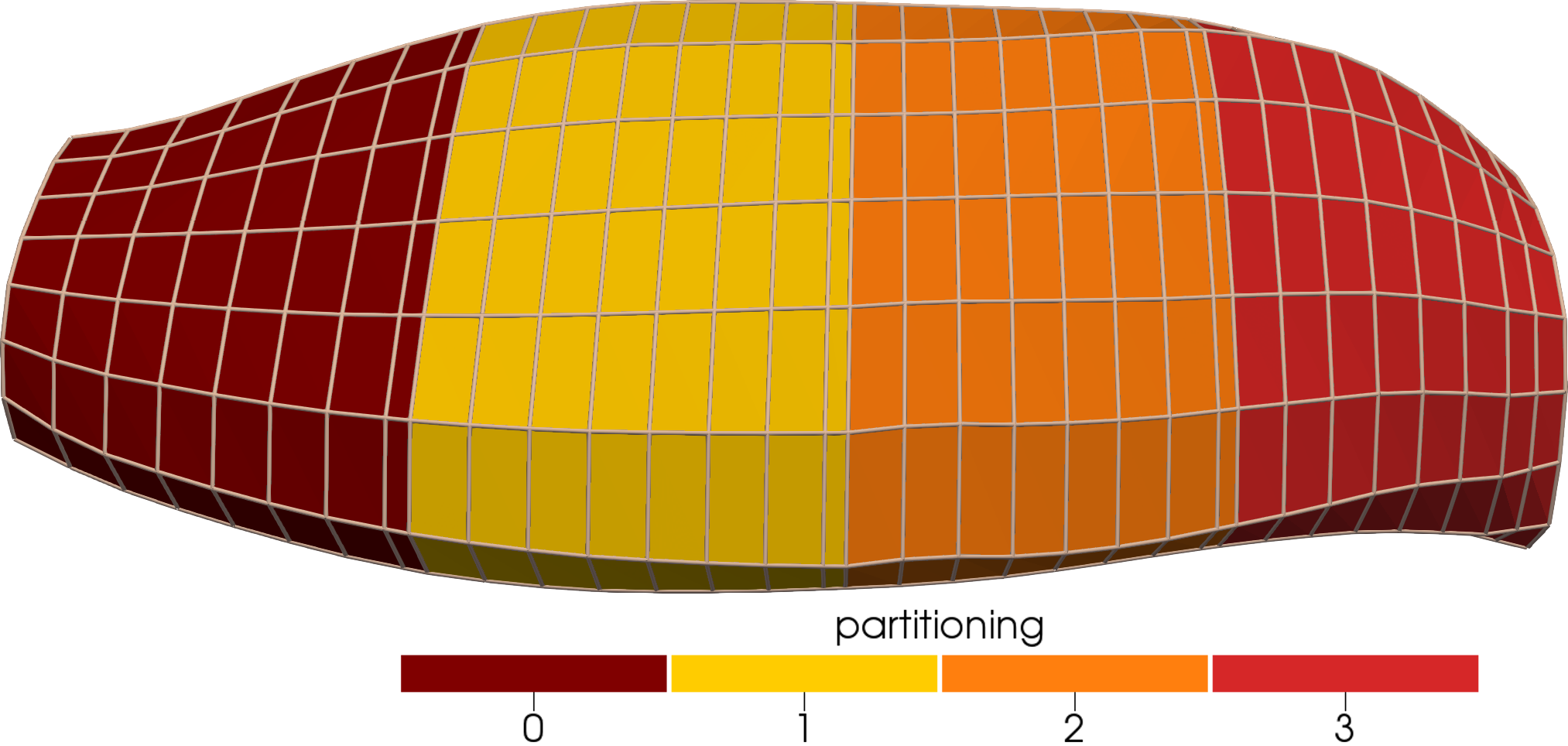}
    \caption{Resulting sampled mesh with option \code{distribute_nodes_equally=True}. The partitioning is perfectly balanced, but the mesh width is not constant.}%
    \label{fig:partitioning5}%
  \end{subfigure}
  \caption{Sampling a mesh along the fiber direction. The original mesh has 1481 nodes and is sampled with a stride value of 50.}%
  \label{fig:partitioning56}%
\end{figure}

The different results for the \code{distribute_nodes_equally} option are demonstrated in \cref{fig:partitioning3_4,fig:partitioning56}. \Cref{fig:partitioning3_4} shows the automatic partitioning, where a simulation of fiber based electrophysiology with a grid of $9 \times 9$ fibers is executed with eight processes and the stride values \code{sampling_stride_x} and \code{sampling_stride_y} are set to two.
By default, a linear mesh of $4\times 4$ elements in $x$ and $y$-directions is created with $2\times 2 \times 2=8$ subdomains, as shown in \cref{fig:partitioning4}. Only the first four subdomains can be seen in the visualization, the other four are located behind and hidden in the background.

The distribution of the fibers to the two 1D subdomains along both $x$ and $y$ directions yields four fibers for the first and five fibers for the second 1D subdomain. Thus, the total 3D subdomains of the first four processes contain $16,20,20$ and 25 fibers.

\Cref{fig:partitioning3} shows the same scenario, except that the option \code{distribute_nodes_equally} has been set. The resulting partitioning is different and the fiber distribution is reversed, five and four fibers are assigned to the two 1D subdomains in both $x$ and $y$ directions. As a result, we get $25,20,20$ and $16$ fibers for the first four 3D subdomains. Note that this is the best balanced partitioning of a structured mesh that is possible for $9 \times 9$ fibers.
The subdomain sizes are the same as in \cref{fig:partitioning4}, except for a different order. However, for larger examples using more processes, the respective partitioning with the \code{distribute_nodes_equally} option is always optimal, whereas the balance rapidly degrades without this option.

While, in this example, there is no difference between \cref{fig:partitioning4}  and \cref{fig:partitioning3} in terms of load balancing, the 3D mesh quality of the generated partitioning is worse for \cref{fig:partitioning3}. As can be seen in \cref{fig:partitioning3}, the first and the third subdomain have one layer of elements more in both $x$ and $y$ direction, and these elements have half the mesh width of the normal elements. Additionally, the second and fourth subdomain also contain elements of different mesh widths.

Similar effects can also be studied in the scenario of \cref{fig:partitioning56}, where the same mesh is partitioned to four processes in $z$-direction. The number of nodes in $z$-direction is \num{1481} and the sampling stride is chosen as \code{sampling_stride_z=50}. \Cref{fig:partitioning6,fig:partitioning5} show the resulting partitioning without and with the \code{distribute_nodes_equally} option. Again, the second scenario shows \say{remainder} elements with smaller mesh widths at the boundaries of every subdomain. The distribution of nodes is $400,350,350$ and $381$ nodes per subdomain in \cref{fig:partitioning6} and $371,370,370$ and $370$ nodes per subdomain for the scenario in \cref{fig:partitioning5}, where the \code{distribute_nodes_equally} option has been set. The first case has the better 3D mesh quality, whereas only the second case yields the perfect load balancing.

In summary, it is possible to tweak the created partitioning by adjusting the sampling stride and deciding between mesh quality and perfect load balancing. For electrophysiology simulations, which impose high computational load because of the subcellular model, the load balancing aspect is more important and the option \code{distribute_nodes_equally}  should be set to \code{True}. In simulations with elasticity models, the quality of the 3D meshes is more important and the partitioning for the corresponding meshes should be parametrized with the  \code{distribute_nodes_equally} option set to \code{False}.

% partitioning with and without distribute_nodes_equally

% partitioning with and without distribute_nodes_equally

\begin{reproduce_no_break}
  The partitioning in \cref{fig:partitioning1} is obtained by the following simulation:
  \begin{lstlisting}[columns=fullflexible,breaklines=true,postbreak=\mbox{\textcolor{gray}{$\hookrightarrow$}\space}]
    cd $\$$OPENDIHU_HOME/examples/electrophysiology/fibers/fibers_contraction/no_precice/build_release
    mpirun -n 12 ./biceps_contraction ../settings_biceps_contraction.py partitioning_demo.py --n_subdomains 4 3 1
  \end{lstlisting}
  The partitionings in \cref{fig:partitioning3_4,fig:partitioning56} are created by the following simulations. For \cref{fig:partitioning4,fig:partitioning6}, edit the variables file \code{partitioning_demo.py} and set \code{distribute_nodes_equally = False}. For \cref{fig:partitioning3,fig:partitioning5}, set \code{distribute_}\code{nodes_equally = True}.
  \begin{lstlisting}[columns=fullflexible,breaklines=true,postbreak=\mbox{\textcolor{gray}{$\hookrightarrow$}\space}]
    cd $\$$OPENDIHU_HOME/examples/electrophysiology/fibers/fibers_emg/build_release
    mpirun -n 8 ./fast_fibers_emg ../settings_fibers_emg.py partitioning_demo.py
    mpirun -n 4 ./fast_fibers_emg ../settings_fibers_emg.py partitioning_demo.py --n_subdomains 1 1 4
  \end{lstlisting}
\end{reproduce_no_break}

\section{Parallel Solver for the Fiber Based Electrophysiology Model}\label{sec:parallel_partitioning_for_fiber_based}

After discussing the general partitioning and sampling of 3D and 1D meshes in the last section, we now focus on the concrete application for the fiber based electrophysiology model.
We describe our basic solver and algorithmic improvements that yield lower runtimes.

\begin{figure}
  \centering%
  \includegraphics[width=\textwidth]{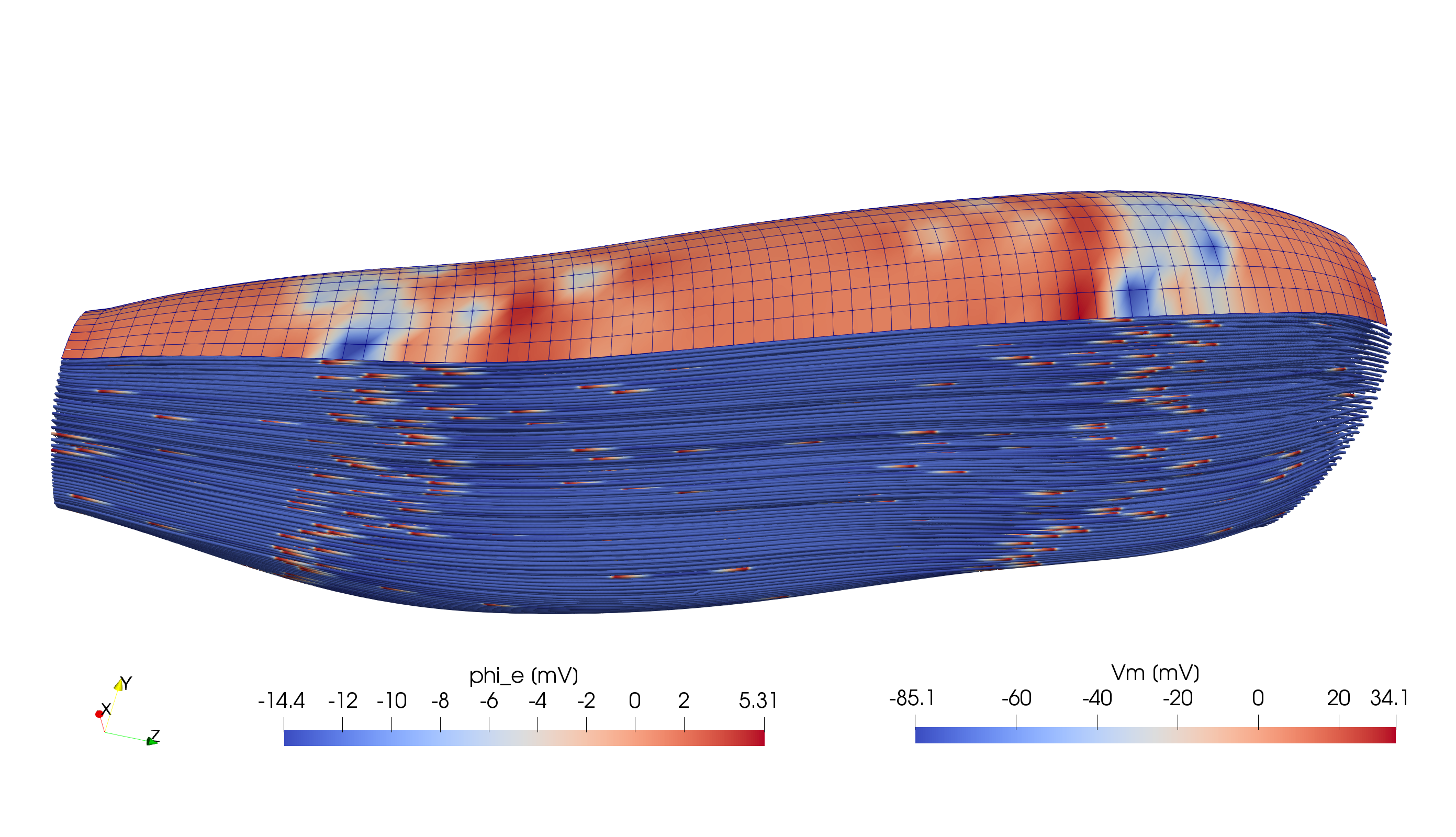}
  \caption{Simulation result of the fiber based electrophysiology model with 1369 muscle fibers and a 2D surface mesh on top of the muscle. The fibers are colored according to the transmembrane potential $V_m$, the surface is colored according to the EMG values given by the extracellular potential $\phi_e$.}%
  \label{fig:solver_fibers_3}%
\end{figure}

The fiber based electrophysiology model consists of the action potential propagation model given by the 1D monodomain equation \cref{eq:monodomain} and a 0D subcellular model as described in \cref{sec:subcelullar_model}. The 0D and 1D problems are solved on the 1D fiber meshes. They are coupled to the 3D bidomain problem given in \cref{eq:bidomain1}, which computes the EMG values. In summary, the components (b2),(c) and (d) of the diagram in \cref{fig:multi-scale-model} are involved in this computation.
\Cref{fig:solver_fibers_3} shows a simulation result of this model, where the 1D fibers and the surface of the 3D mesh can be seen.

% fibers mesh

In the following, \cref{sec:parallel_partitioning_for_fiber_based_solver} begins with a description of the solver structure and the parallelization. Subsequently, performance improvements considering the parallel execution of the solver are discussed. \Cref{sec:improved_parallel_solver_for_fiber_based} presents a variant, where a faster solver is employed for the 1D part of the computation. \Cref{sec:adaptive_computation_for_fiber_based} shows how the computational load can be reduced by only computing activated parts of the muscle.

\subsection{Parallel Solver Structure}\label{sec:parallel_partitioning_for_fiber_based_solver}
\begin{figure}%
  \centering%
  \begin{subfigure}[t]{0.30\textwidth}%
    \centering%
    \includegraphics[width=\textwidth]{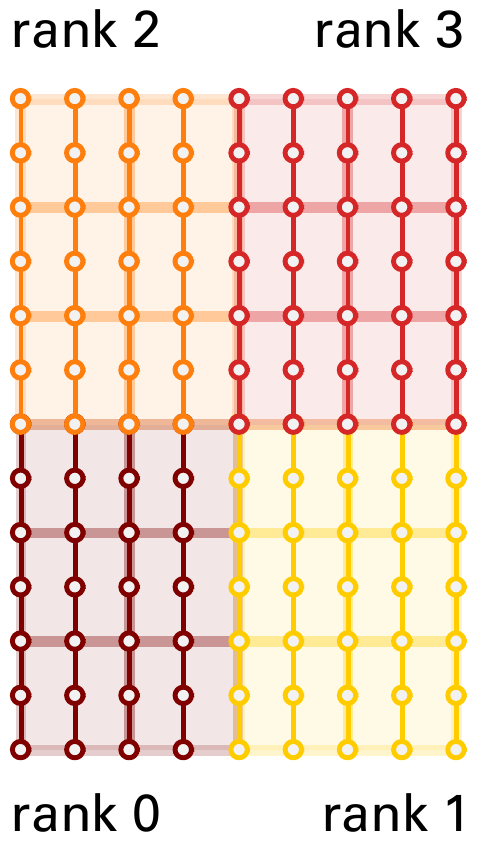}
    \caption{Visualization of the 3D mesh with embedded 1D fibers, partitioned to four ranks.}%
    \label{fig:mesh_structure}%
  \end{subfigure}
  \qquad
  \begin{subfigure}[t]{0.45\textwidth}%
    \centering%
    \includegraphics[width=0.9\textwidth]{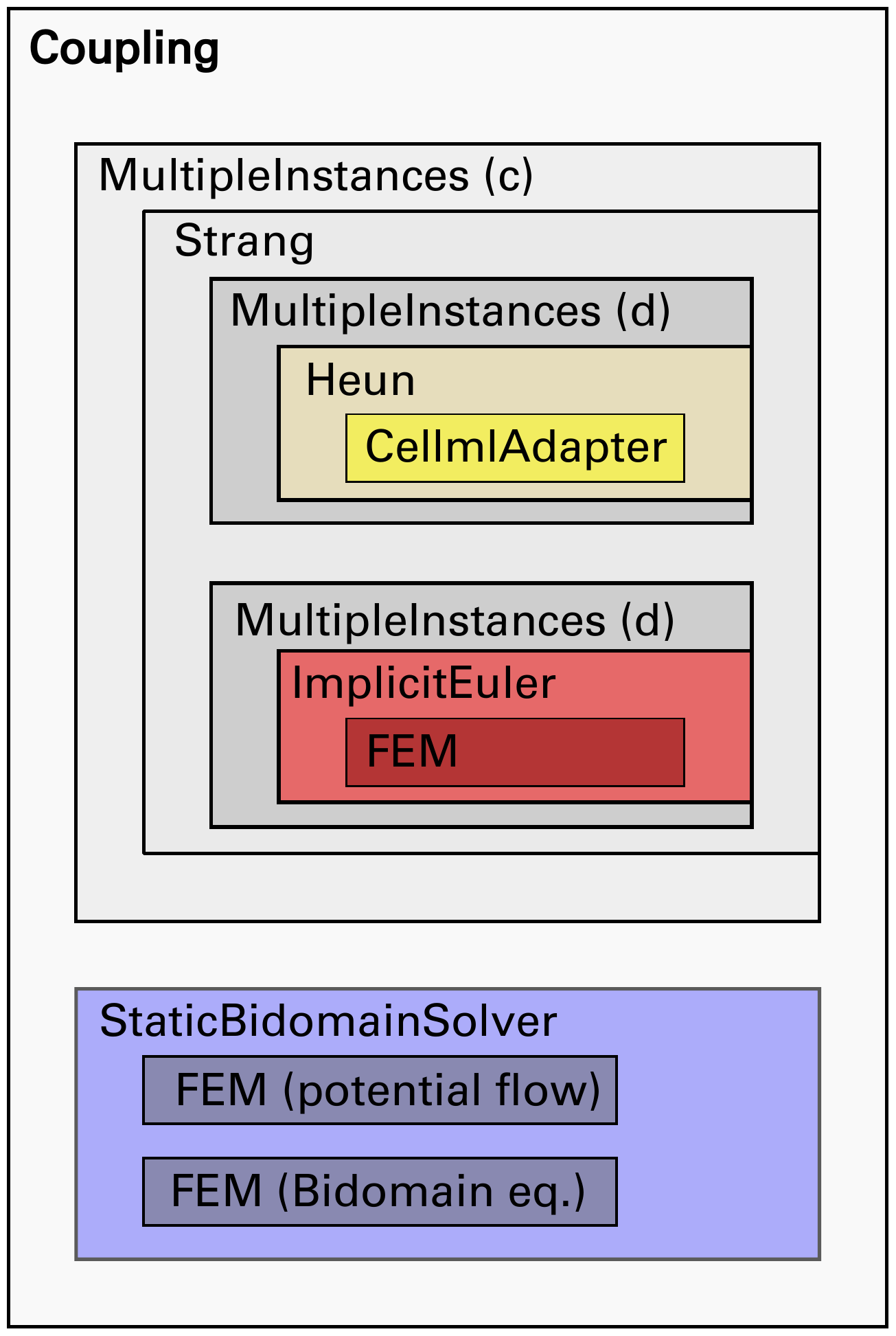}
    \caption{Structure of the OpenDiHu example program to solve the fiber based electrophysiology model. The colors match the scheme introduced in the overview chart in \cref{fig:multi-scale-model}.}%
    \label{fig:program_structure}%
  \end{subfigure}
  \\[8mm]
  \begin{subfigure}[t]{0.48\textwidth}%
    \centering%
    \includegraphics[height=8cm]{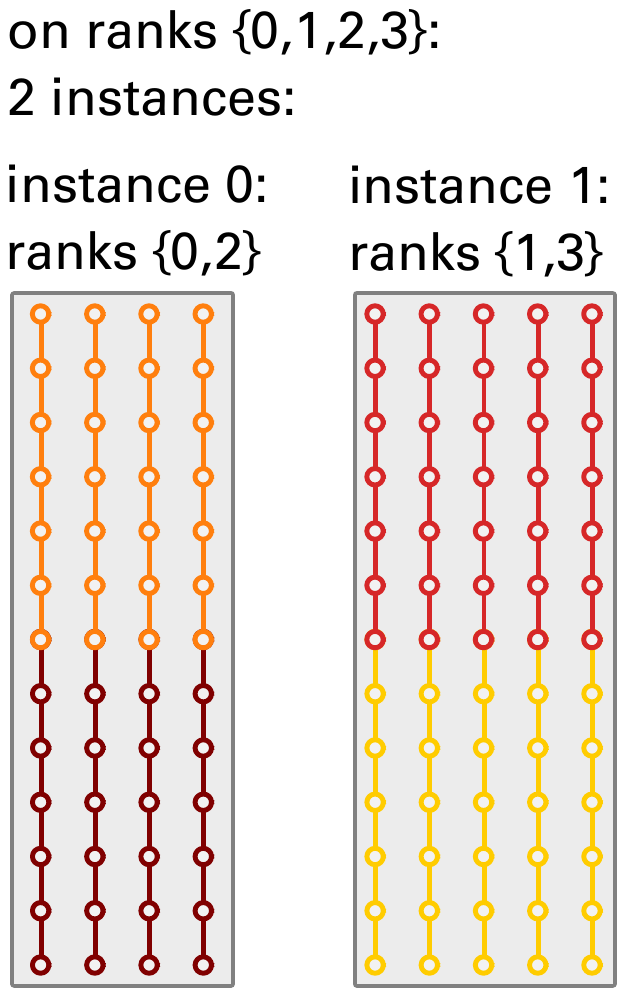}
    \caption{Instances of the outer \code{MultipleInstances} class in \cref{fig:program_structure}.}%
    \label{fig:fiber_partitioning1}%
  \end{subfigure}
  \,
  \begin{subfigure}[t]{0.48\textwidth}%
    \centering%
    \includegraphics[height=7.5cm]{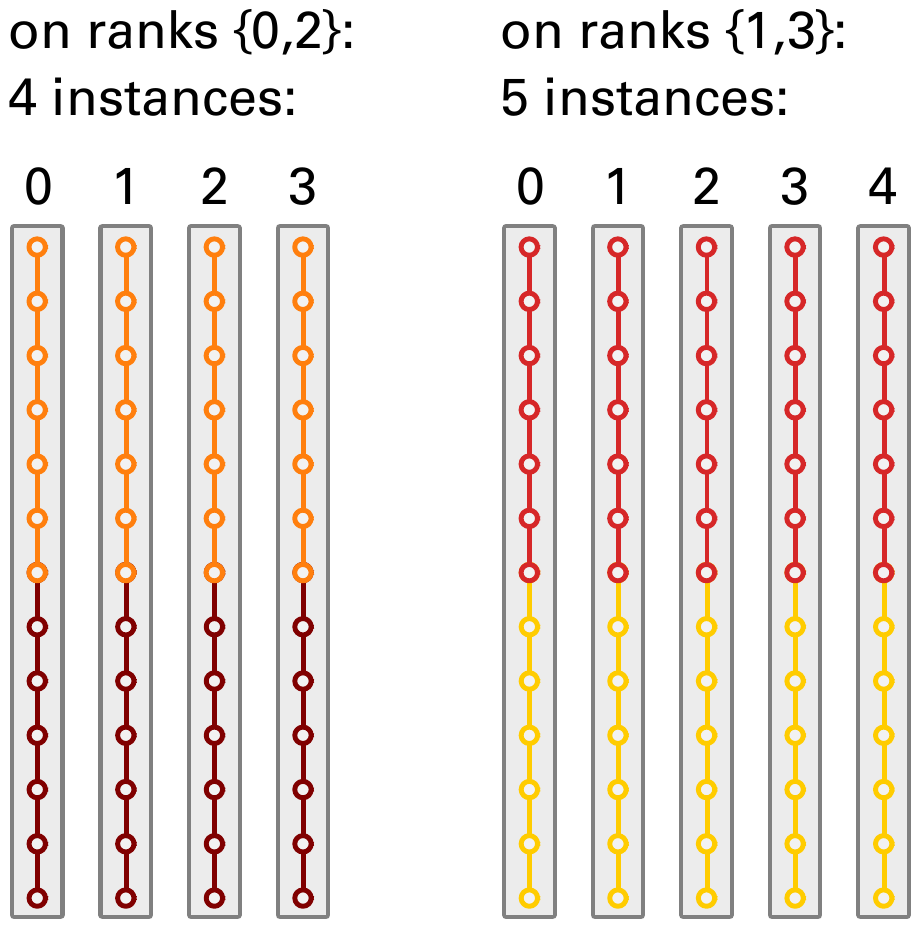}
    \caption{Instances of the inner \break\code{MultipleInstances} classes in \cref{fig:program_structure}.}%
    \label{fig:fiber_partitioning2}%
  \end{subfigure}
  \caption{Visualizations for the discussion of the program structure and partitioning used for fiber based electrophysiology simulations. The circles and lines represent the 1D meshes, their coloring indicates the MPI rank.}%Etwas mehr Information, z.B. Kreise sind Datenpunkte, Linien deuten Fasern an, Farben die MPI-Ranks?? Nee, kein Platz!
  \label{fig:partitioning_program}%
\end{figure}

For better visualization, we consider the 2D setting of a mesh and embedded 1D fibers partitioned to $2\times 2$ processes as shown in \cref{fig:mesh_structure} by different colors. However, all discussions are also valid for the real 3D setting shown in the last section and for arbitrary partitionings to $n_x \times n_y \times n_z$ processes.

% program structure and partitioning

\Cref{fig:program_structure} shows the program structure of the example that solves the fiber based electrophysiology model. The outer class is a \code{Coupling} that alternates between computing the monodomain equation \cref{eq:monodomain} on the 1D fibers and computing the static bidomain equation \cref{eq:bidomain1} on the 3D domain. The second part, the bidomain solver, is given in \cref{fig:program_structure} by the class \code{StaticBidomainSolver}, which includes two \code{FiniteElementMethod} classes. The first class solves the potential flow to obtain the fiber direction for the anisotropic conduction tensor, the second class is used to discretize the spatial derivatives in the bidomain equation.

The first part of the coupling scheme in \cref{fig:program_structure} consists of a \code{Multiple}\code{Instances} class, which encloses the Strang operator splitting. The splitting has two child solvers for the subcellular model and the diffusion or conduction term. The first child consists of another \code{MultipleInstances} class with a \code{Heun} scheme and the \code{CellmlAdapter}\nolinebreak.
The second child of the Strang splitting also consists of a \code{MultipleInstances} class and a combination of an \code{ImplicitEuler} scheme (alternatively a \code{CrankNicolson} scheme can be used) and a \code{FiniteElementMethod}.

A \code{MultipleInstances} class can be used to apply a solver to more than one problem of the same kind. The class allows to specify a number of instances of its nested solver. Each instance can be given a subset of processes that will take part in the computation of the instance. Each process then iterates over all instances, for which it is part of the subset. Thus, the nested solver of a \code{MultipleInstances} class is called in series for all instances that share a process/MPI rank, and it is called in parallel and independently for 1D model instances that have disjoint subsets of ranks.

Furthermore, the class provides a common output writer, which collectively writes the data of all instances. This allows, e.g., to create a single output file in every timestep containing the data of all fibers. Especially for large scenarios, this is more practical than having as many output files as fibers.

The settings that have to be specified in the Python file for a \code{MultipleInstances} class comprise the number of instances and a list with the according number of entries, which further configure the instances. Each list entry can be \code{None} if the rank does not take part in the computation of the corresponding instance. 
Otherwise, the list entry consists of (i) a specification of all ranks that should collectively compute the corresponding instance and (ii) the settings of the corresponding nested solver. 

The own MPI rank of a process is known in the Python settings file. This allows to specify different settings for different ranks in the same file. By omitting the configuration of irrelevant instances and setting their list entry to \code{None}, the amount of data is reduced and parsing of the script is sped up, especially for large problem sizes.

The settings and corresponding subdomains of the \code{MultipleInstances} classes that are indicated by (c) and (d) in \cref{fig:program_structure} are shown in \cref{fig:fiber_partitioning1,fig:fiber_partitioning2}, respectively.
As can be seen in \cref{fig:fiber_partitioning1}, the outer \code{MultipleInstances} class separates the subdomains that are not connected by any fibers, such that they can be computed in parallel and independently of each other. In the example of \cref{fig:mesh_structure}, the subdomains of ranks 0 and 2 can be computed independently of the subdomains of ranks 1 and 3. 
As a consequence, all processes specify that their \code{MultipleInstances} class has two instances. 
At rank 0, the list of instance settings contains the settings of the nested Strang solver with all information of rank 0's subdomain (in the first item) and the value \code{None}, as rank 0 has no information about fibers outside its subdomain (in the second item). Ranks 1, 2 and 3 specify their subdomain accordingly, as shown in \cref{fig:fiber_partitioning1}.

During computation, ranks 0 and 2 as well as ranks 1 and 3 enter the \code{Strang} solver class collectively with a shared MPI communicator.
The inner \code{MultipleInstances} classes employ the 0D subcellular and the 1D electric conduction solver on multiple fibers. As shown in \cref{fig:fiber_partitioning2}, ranks 0 and 2 specify four instances with the settings of the four shared fibers. At the same time and concurrently, ranks 1 and 3 specify five instances with settings for their five shared fibers. 

Note that the multiplicity of the 0D instances on a fiber is not achieved by another \code{MultipleInstances} class, but the model is solved for all points on the mesh together, using parallelism on the lower, instruction-based level.

These different splits of the geometry allow to compute the electrophysiology model on the fibers in parallel. The partitioning of the domain has to be the same for the 3D mesh and the embedded fibers to allow value mapping from the fibers to the 3D mesh without communication. The fibers are oriented along the $z$-direction in the 3D setting. This explains, why the ranks for a particular fiber, e.g., $\{0,2\}$ or $\{1,3\}$ are not direct successors of each other but increasing with a stride equal to the number of subdomains in $x$ and $y$ directions, $n_x \cdot n_y$.

\subsection{Improved Parallel Solver Scheme using the Thomas Algorithm}\label{sec:improved_parallel_solver_for_fiber_based}
% FastMonodomainSolver

The monodomain model, which is solved on each fiber, consists of a reaction-diffusion equation, which is solved using the Strang operator splitting.
The diffusion part uses an implicit timestepping scheme, which leads to a linear system of equations to be solved in every timestep.
As the finite element method with linear ansatz functions is used for spatial discretization, this linear system has a tridiagonal system matrix.

In the solver tree structure in \cref{fig:program_structure}, this solution step occurs in the solvers under the second inner \code{MultipleInstances} class.  As can be seen in \cref{fig:fiber_partitioning2}, the dofs of each fiber that are part of this linear system are partitioned to multiple processes. Hence, this linear system is solved using a parallel conjugate-gradient solver of PETSc.

However, there is the possibility to improve the performance by exploiting the tridiagonal matrix structure. The \emph{Thomas algorithm} is the specialization of Gaussian elimination for this matrix type and is known to efficiently solve such a system in linear time complexity. More specifically, it only requires a first downwards sweep through the matrix entries for forward substitution and a second upwards sweep for back substitution to compute the solution. It is stable for diagonally dominant matrices and this condition is met for the governing system matrix.

\begin{figure}
  \centering%
  \includegraphics[width=0.6\textwidth]{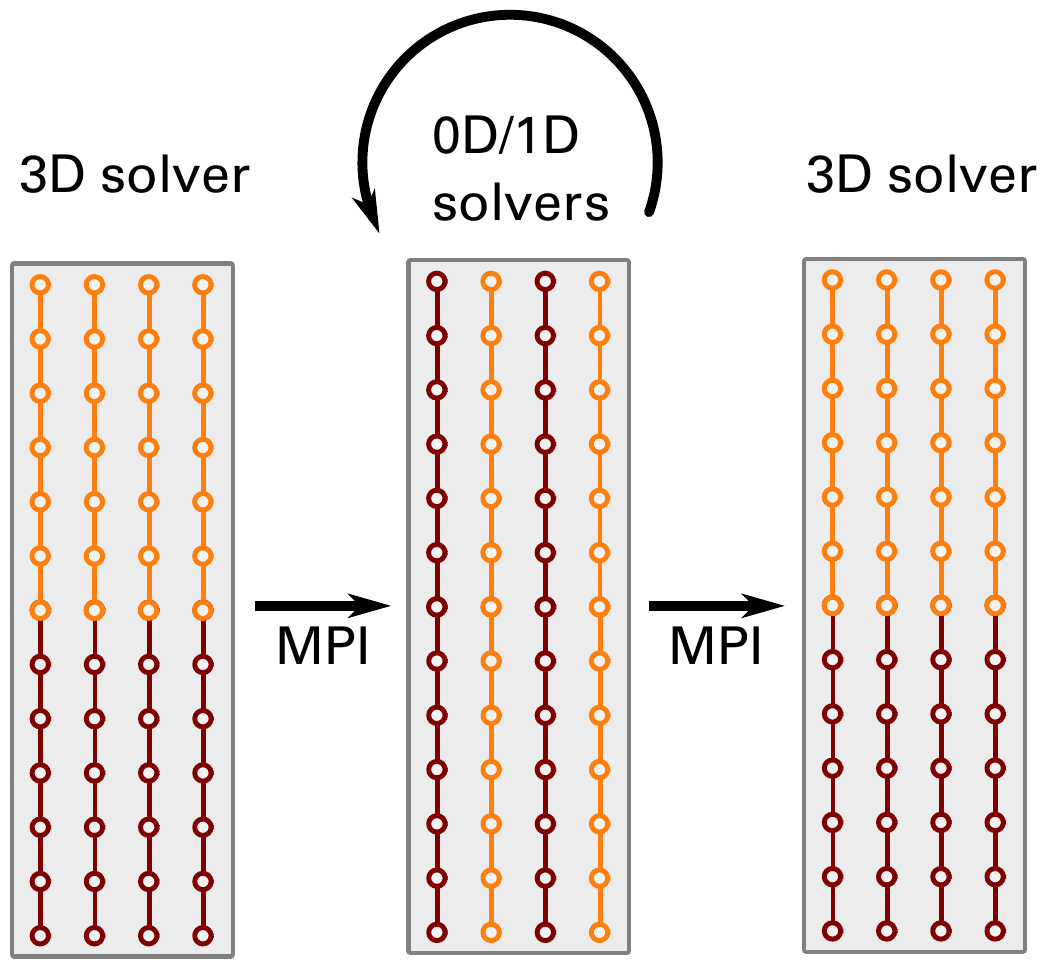}
  \caption{Algorithmic steps of the \code{FastMonodomainSolver} to efficiently solve the 0D/1D problems. After a timestep of the 3D solver (left), the partitioned fibers, visualized by different colors for the MPI ranks, are communicated using MPI, such that every fiber is accessible on a single MPI rank (middle). The 0D/1D solvers compute multiple subsequent timesteps until the next 3D coupling step. Then, the original partitioning is restored by a second communication step and the 3D solver can continue with the next timestep (right).}%
  \label{fig:fast_monodomain_scheme}%
\end{figure}

As the Thomas algorithm is not parallel, we have to gather the matrix data on a single process in order to employ the algorithm. In OpenDiHu, the \code{FastMonodomainSolver} class is tailored to the parallel solution of fiber based electrophysiology using the Thomas algorithm. 
\Cref{fig:fast_monodomain_scheme} outlines the steps performed by the \code{FastMonodomainSolver} class.

During initialization, the \code{FastMonodomainSolver} class initializes its nested solver tree as normal and the parallel partitioning of the fibers is carried out as described in \cref{sec:parallel_partitioning_for_fiber_based}. This is visualized on the left in \cref{fig:fast_monodomain_scheme} for four fibers and two ranks. At the beginning of the first timestep, the communication to gather complete fiber data on single processes is carried out. The fiber data are communicated, such that every fiber is completely accessible at a single processes. The assignment of the fibers to processes occurs in a round-robin fashion, i.e., the first fiber is sent to rank 0, the second to the next rank, etc. As a result, every process has approximately the same number of complete fibers. 
This is shown in the middle image of \cref{fig:fast_monodomain_scheme}, where the red colored rank has all values of the first and third fiber and the orange colored rank has all values of the second and fourth fiber.

The processes then each compute the full monodomain model consisting of the Strang splitting with the subcellular model on the nodes of each fiber and the diffusion part using the Thomas algorithm. 
This is done in a separate serial implementation for the now locally owned fibers, i.e., not using the nested solvers. The solution is obtained for as many subsequent timesteps as were specified in the settings. When the end time of the enclosing coupling scheme is reached, the fiber data are communicated back to the original partitioned fibers, as shown on the right in \cref{fig:fast_monodomain_scheme}.
Then, the coupling scheme continues with the data mapping from the partitioned fibers to the 3D domain and with the \code{StaticBidomainSolver}. Afterwards, the \code{FastMonodomainSolver} is called again and performs its computation anew starting with the communication step.

% fibers mesh

In the C++ file, the \code{FastMonodomainSolver} class is inserted as a wrapper to the outer \code{MultipleInstances} class that is indicated by (c) in the solver structure in \cref{fig:mesh_structure}. In the Python settings, the class does not add an additional nesting level such that the same settings file can be used for programs with and without the \code{FastMonodomainSolver} class and yields the same simulation results.

In summary, the efficient serial computation of the monodomain model in the \code{Fast}\code{MonodomainSolver} is wrapped by communication steps of  the partitioned fiber data. The frequency of this communication step is determined by the timestep width of the coupling scheme. 
The scenario solves the bidomain equation to simulate EMG signals. A typical sampling frequency of EMG capture devices is $f=\SI{2}{\kilo\hertz}$, which corresponds to a coupling timestep width of $\dt_\text{3D}=\SI{0.5}{\milli\second}$. The timestep widths $\dt_\text{0D}$ of the subcellular model and $\dt_\text{1D}$ of the diffusion term have to be set at maximum to $\SI{1e-3}{\milli\second}$, yielding \num{500} timesteps of computations on the fiber between subsequent communication steps. As a result, the communication cost is negligible.

\subsection{Adaptive Computation of the Subcellular Model}\label{sec:adaptive_computation_for_fiber_based}
% adaptive solution of cells and whole fibers

During simulations of the fiber based electrophysiology model, often only a small fraction of the given fibers is activated.
The reason is, that, in physiological conditions, the smaller MUs are activated first and the larger MUs only get activated when the full force of the muscle is required. As the majority of the fibers belongs to larger MUs, a high portion of fibers is less frequently activated, also depending on the scenario.
But even if the scenario specifies a tetanic stimulation of all MUs, the larger MUs have lower stimulation frequencies, which again leads to less action potentials on large MUs than on smaller MUs in the same time span.

A naive solver of the monodomain models always computes all 1D electric conduction problems on the fiber meshes and all 0D subcellular models on the nodes of the fiber meshes, regardless of their activation state. In the following, we present a method in OpenDiHu that exploits the infrequent activation events on most of the fibers while obtaining the same solution as the naive solver.

We assume that the subcellular models are initialized in their equilibrium state, where the temporal derivative of the state vector $\bfy$ vanishes, $∂\bfy/∂t = 0$. The first algorithmic improvement is to only consider those fibers in the solver that have yet been stimulated. This improves the performance especially for \say{ramp like} motor recruitment, where more and larger MUs are activated over time. However, after all MUs have been activated at least once, all fibers are computed again and no more performance improvement is obtained.

The second improvement is to only compute instances of the subcellular model at those points, where it is not in equilibrium. To determine, whether an instance of the subcellular model is in equilibrium, we compare the solution before and after one integration step by the Heun method. Only if the relative change of any component of the state vector $\bfy$ is larger than \num{1e-5}, we consider the model to be not in equilibrium.

This check requires to compute the solution of the subcellular model, the avoidance of which is subject of the improved scheme. Therefore, we use the property of the 1D diffusion problem discretized by linear finite elements that the value at one spatial point can only influence its two neighbors in a single timestep. This allows us to avoid checking the equilibrium condition at points that are surrounded by other points in equilibrium. This means that the subcellular model does not have to be solved at most points in equilibrium, which drastically reduces the runtime. The 1D electric conduction problem, however, has to be solved for the whole fiber mesh if at least one point it is not in equilibrium.

\begin{figure}
  \centering
  \includegraphics[width=0.7\textwidth]{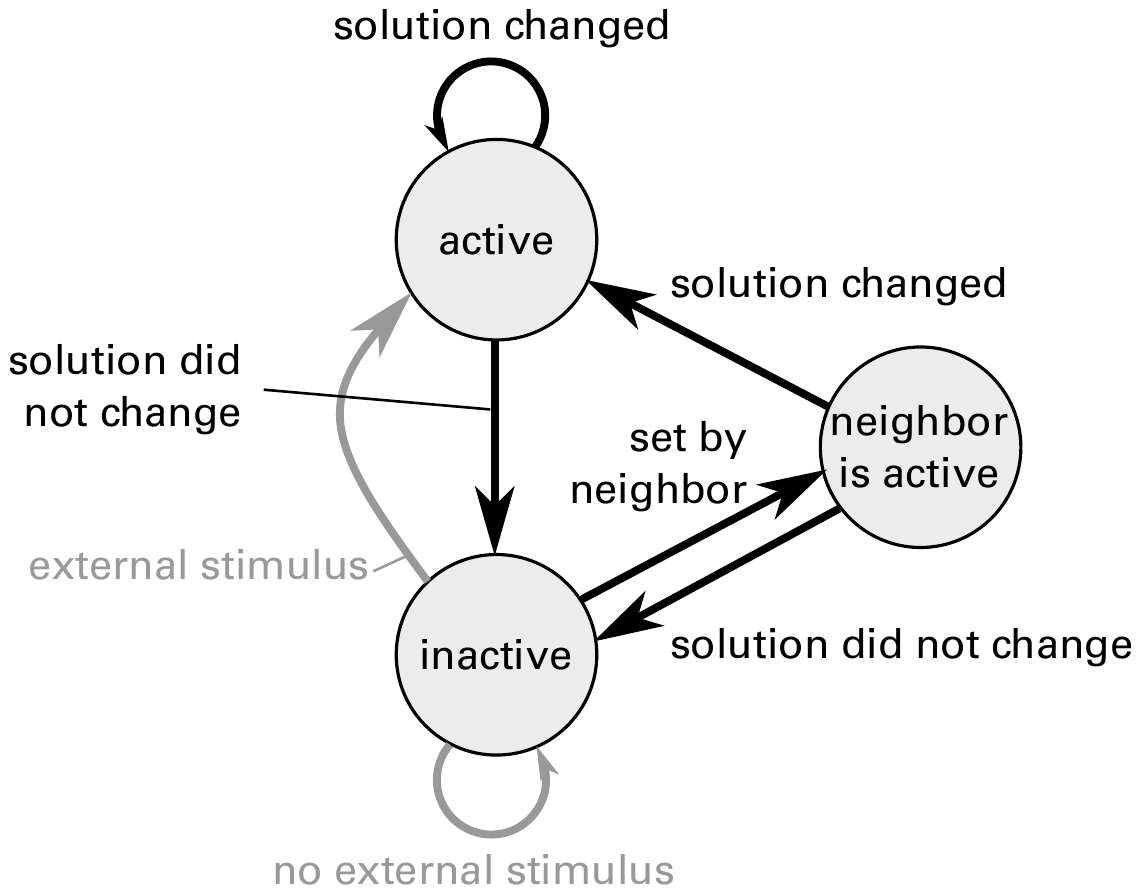}%
  \caption{Transition diagram for the adaptive computation of the subcellular model. The diagram shows the transition between local states of points on the fibers. Points in inactive state do not perform the computation of the 0D subcellular model.}
  \label{fig:state_chart}
\end{figure}

In our method, each subcellular point can be in one of the three states \say{active}, \say{inactive} and \say{neighbor is active}.
If the subcellular model is not in equilibrium, the point is in the state \say{active} and has to be solved in the next timestep. If the subcellular model is in equilibrium and does not have to be solved because the solution vector stays constant, the point is in the state \say{inactive}. The state \say{neighbor is active} occurs for a previously inactive point, of which at least one neighbor became active and, thus, the check if the point is still in equilibrium has to be performed and the subcellular model has to be solved in the next timestep. After each solution step, the state of a point changes according to the transitions given in \cref{fig:state_chart}.

An active point stays active, if the solution has changed in the last numerical integration step. It transitions to inactive, if the solution did not change. The same applies to points in the state \say{neighbor is active}, which also change to \say{active} or \say{inactive} after one timestep. 
An inactive state cannot be activated by a check on the point itself, as this state implies that no computation and no subsequent equilibrium check are carried out. The only transition for a point $A$ from an inactive state occurs, when a neighbor point $B$ reaches the state \say{active} (or for external stimulation). Then, point $A$ changes to \say{neighbor is active}.
For propagating action potentials along a fiber that is in the \say{inactive} state, this leads to a propagating front of points in the \say{neighbor is active} state.

Initially, all states are set to \say{active}. If no stimulation occurs and the subcellular model is in equilibrium, they momentarily change to \say{inactive}. Upon external stimulation, the stimulated points are automatically set to \say{active} and their neighbors are set to \say{neighbor is active} such that the effect of the stimulation can be considered in subcellular model computations.

\begin{figure}%
  \centering%
  \includegraphics[width=\textwidth]{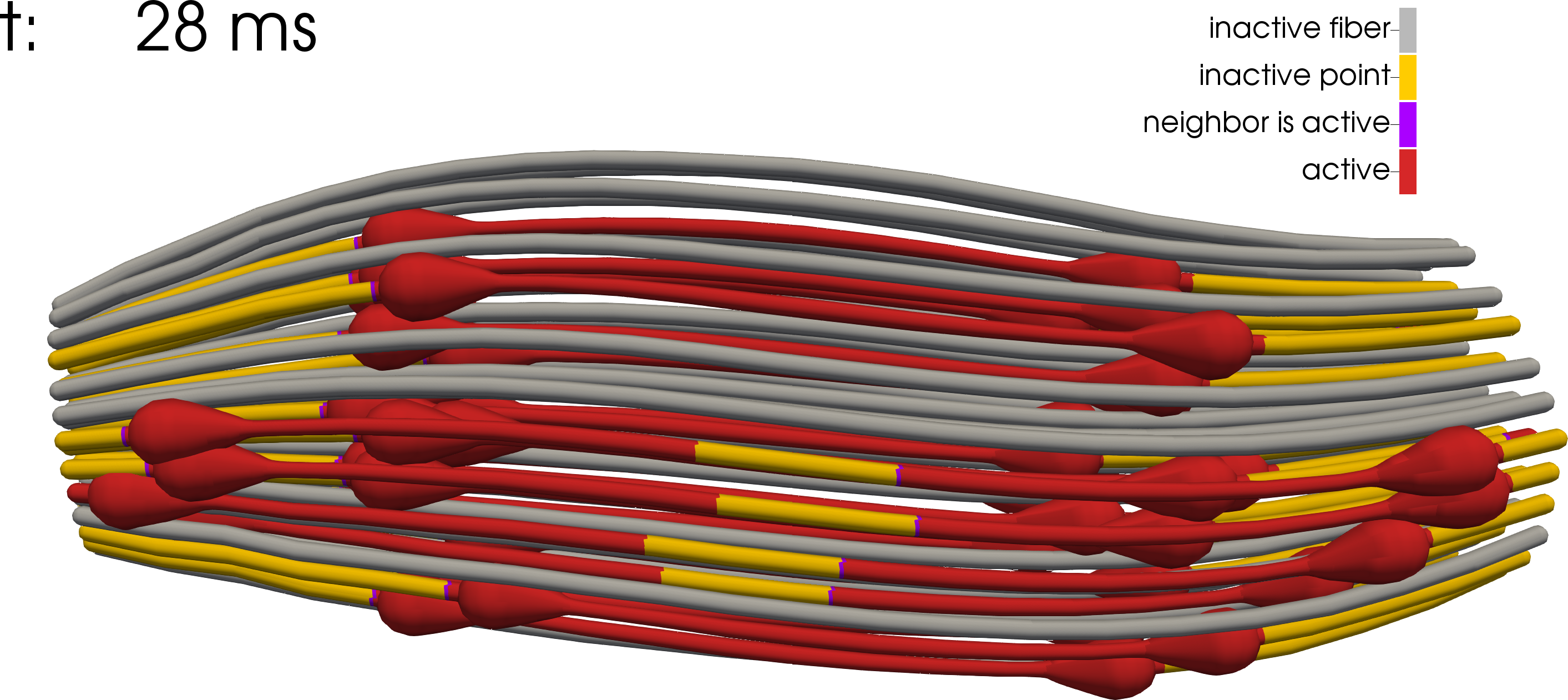}%
  \caption{Simulation scenario that demonstrates the adaptive computation method of fibers and subcellular points. A simulation of the monodomain equation on a set of 49 fibers with the subcellular model of Hodgkin and Huxley is shown. The transmembrane potential is visualized by the fiber radius. The states of the points used in the algorithm are given by the different colors.}%
  \label{fig:compute_state3}%
\end{figure}

\Cref{fig:compute_state3} shows a simulation, where the effect of both improvements is visible. The Hodgkin-Huxley subcellular model has been solved on a set of 49 fibers. At the displayed time of $t=\SI{28}{\milli\second}$, two MUs have been activated. The value of the membrane potential $V_m$ is visualized by the radius of the fibers. The active or inactive state of the improved scheme is indicated by the colors.

% compute state

It can be seen that several fibers have gray color which indicates that they have not yet been stimulated and, thus, are not part of the computation. The other fibers have been stimulated either by the first or the second MU. Action potentials at two different distances from the center corresponding to the two MUs can be identified by the bulbous shapes. The red parts of the fibers contain the active points, where the subcellular model is not in equilibrium. At the yellow regions, the subcellular models are in equilibrium, and no computational work is performed there. The yellow regions are at the outer ends of the fibers that were not yet reached by the action potentials as well as around the center for fibers of the first MU. This demonstrates the repolarisation effect, after which the model reaches its equilibrium state again. 

The purple colored points are in the state \say{neighbor is active} and can be found between active and inactive points. As the algorithm iterates over all points of a fiber from left to right, these purple points only occur at the left boundaries of active regions. At their right boundaries, the initial \say{neighbor is active} points transition to \say{active} or \say{inactive} directly after the computation step within this iteration.

Instead of individual nodes on the fiber mesh, our implementation treats SIMD vectors of four or eight such adjacent nodes (depending on the hardware capabilities) as one point in the algorithm.  If one of these nodal instances is not in equilibrium, the whole SIMD vector is considered not in equilibrium and transitions to the \say{active} state. This coarser granularity of the model instances allows to solve the subcellular problem in chunks according to the SIMD lane width using SIMD instructions.

\begin{reproduce_no_break}
  The scenario of \cref{fig:compute_state3} can be run as follows:
  \begin{lstlisting}[columns=fullflexible,breaklines=true,postbreak=\mbox{\textcolor{gray}{$\hookrightarrow$}\space}]
    cd $\$$OPENDIHU_HOME/examples/electrophysiology/fibers/fibers_emg/build_release
    mpirun -n 4 ./fast_fibers_emg ../settings_fibers_emg.py compute_state_demo.py
  \end{lstlisting}
  Instead of four processors, you can use as many as you have to speed up the computation.
\end{reproduce_no_break}
\section{Parallel Solver for the Multidomain Electrophysiology Model}\label{sec:parallel_solver_multidomain}

After the details on the parallel partitioning and solvers for the fiber based electrophysiology model have been discussed in \cref{sec:parallel_partitioning_and_sampling_of_the,sec:parallel_partitioning_for_fiber_based}, we now consider the multidomain based model of electrophysiology, which includes electric conduction in the body fat layer. The class for the implicit solver within the operator splitting is the \code{MultidomainWithFatSolver} class, which has been introduced in \cref{sec:exemplary_usage_2}.

\begin{figure}
  \centering
  \includegraphics[width=0.7\textwidth]{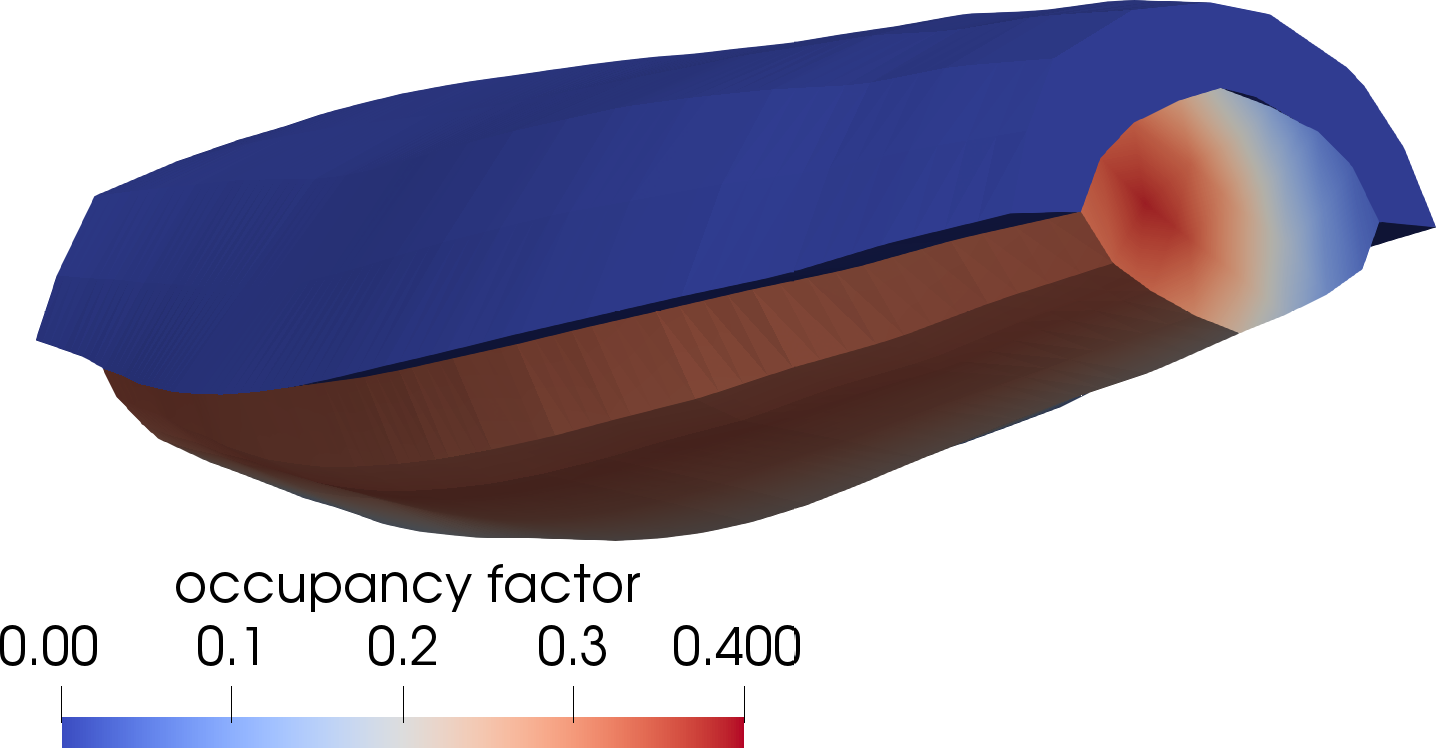}%
  \caption{Visualization of the domains for the multidomain electrophysiology model: The body fat domain is shown in blue, the muscle domain is colored according to the values of one occupancy factor $f_r^k$, which specifies the territory of a MU.}
  \label{fig:solver_multidomain_mesh}
\end{figure}

The multidomain based electrophysiology model contains the two multidomain equations, \cref{eq:multidomain1,eq:multidomain2}, which are solved on the 3D domain. The model leads to a large linear system of equations that is solved in every timestep, described in \cref{sec:discretization_body_domain}. 
\Cref{fig:solver_multidomain_mesh} visualizes the body fat and muscle domains, on which the multidomain model is solved. The coloring of the muscle domain also gives an example for the occupancy factor $f_r^k$, which specifies to which extend every point in the domain is occupied by a particular MU.

% multidomain mesh

\subsection{Construction and Partitioning of the Mesh}\label{sec:construction_and_partitioning_of_the_mesh}

\begin{figure}
  \centering%
  \def\svgwidth{0.6\textwidth}
  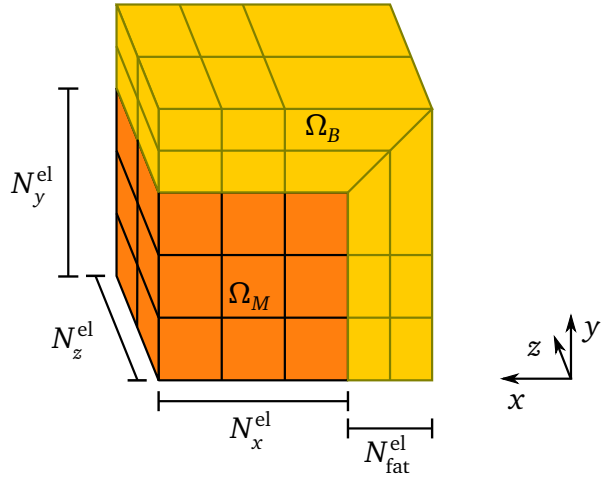%
  \caption{Layout of the composite 3D mesh for the multidomain model with fat layer. The orange elements belong to the mesh of the muscle domain $\Omega_M$, the yellow elements are added on top to represent the body fat domain $\Omega_B$.}%
  \label{fig:structured_grid_n_nodes}%
\end{figure}

The mesh used in this solver is a composite mesh of type \code{Mesh::CompositeOfDimension<D>}, as introduced in \cref{sec:composite_meshes}. \Cref{fig:structured_grid_n_nodes} shows the layout, in particular how the mesh of the body fat domain $\Omega_B$ is connected with the mesh of the muscle domain $\Omega_M$. The muscle and body fat meshes have $N_x^\text{el} \times N_y^\text{el} \times N_z^\text{el}$ and $(N_x^\text{el}+ N_y^\text{el}) \times N_\text{fat}^\text{el} \times N_z^\text{el}$ elements, respectively. Only the muscle mesh has been generated from medical imaging data by the pipline given in \cref{sec:generation_of_meshes_for_multiscale}. The fat mesh is created on top of the muscle mesh geometry and has to use the same number of elements as the muscle mesh along the muscle surface for compatibility in the composite mesh. Only the physical thickness of the adipose tissue layer and the corresponding number $N_\text{fat}^\text{el}$ of elements in radial direction have to be specified.  (The mesh generation step is implemented in the script \code{create_fat_layer.py}.) 

% fat layer mesh and partitioning

\Cref{fig:multidomain_matrix_mesh} shows such as composite mesh.
The muscle mesh is based on a dataset of $13\times 13$ fibers with \num{1481} nodes per fiber. This fine mesh is sampled as described in \cref{sec:algorithm_for_partitioning_and_sampling} with stride values of $3,3$ and $20$ in $x,y$ and $z$ directions and \code{distribute_nodes}\code{_equally=True}. As a result, we get $N_x^\text{el} \times N_y^\text{el} \times N_z^\text{el} = 5 \times 4 \times 75$ elements.
The fat mesh consists of a $\SI{1}{\centi\meter}$ adipose tissue layer with $N_\text{fat}^\text{el}=4$ elements. The muscle and fat meshes have \num{2280} and \num{3800} dofs.

% comment about parallelization
The partitioning of the composite mesh into $n_x \times n_y \times n_z$ subdomains cannot be chosen arbitrarily. The reason is that both the muscle and the body fat mesh have to be partitioned into the same number of subdomains. 
If, e.g., a partitioning of $n_x=n_y=2$ is chosen, the cube in \cref{fig:structured_grid_n_nodes} gets divided by one horizontal planar cut and one vertical planar cut. This divides the orange muscle mesh into four subdomains as expected. The yellow body fat mesh, however, is only partitioned to three of the four processes as there are no yellow elements below the horizontal cut and left of the vertical cut.

Thus, a valid partitioning can only be created if either $n_x$ or $n_y$ is set to one. Because there is no restriction on $n_z$, the total mesh can still be partitioned in two dimensions to a product of subdomains, either as $1 \times n_y \times n_z$ or as $n_x \times 1 \times n_z$.
The example mesh in \cref{fig:multidomain_matrix_mesh} is partitioned to $2 \times 1 \times 2$ subdomains as shown by the different colors.

% scenario_name: matrix,  n_subdomains: 2 1 2,  n_ranks: 4,  end_time: 0.002
% dt_0D:           1e-03    multidomain solver:         1000 it. of gmres (10000 it. of gmres), lumped mass matrix: False, initial guess: previous solution
% dt_multidomain:  1e-03    multidomain preconditioner: euclid (euclid), symmetric precond.: True
% dt_splitting:    1e-03    theta: 1.0, solver tolerances, abs: 1e-15, rel: 1e-15
% fiber_file:              ../../../input/left_biceps_brachii_13x13fibers.bin
% fat_mesh_file:           ../../../input/left_biceps_brachii_13x13fibers.bin_fat.bin
% cellml_file:             ../../../input/hodgkin_huxley_1952.c
% firing_times_file:       ../../../input/MU_firing_times_always.txt
% ********************************************************************************
% 4 ranks, partitioning: x2 x y1 x z2
%   sampling 3D mesh with stride 3 x 3 x 20 
%   distribute_nodes_equally: True
%     linear 3D mesh    nodes global: 6 x 5 x 76 = 2280, local: 3 x 5 x 38 = 570
%     linear 3D mesh elements global: 5 x 4 x 75 = 1500, local: 3 x 4 x 38 = 456
%     fat mesh, n points total:    3800 (10 x 5 x 76), (per process: 3 x 5 x 38 = 570)
% 
\begin{figure}
  \centering%
  \includegraphics[width=\textwidth]{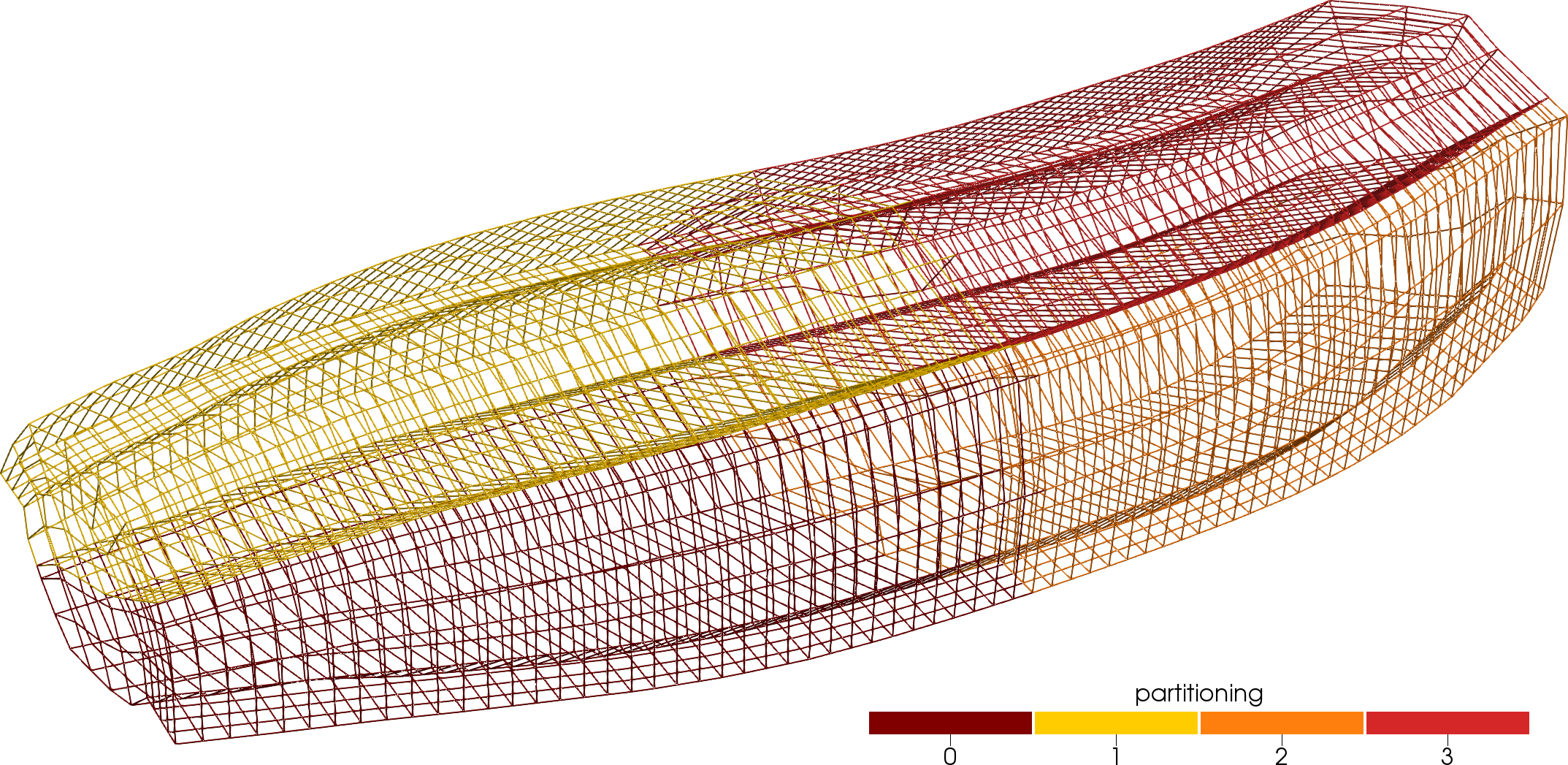}%
  \caption{Composite mesh of the multidomain example, partitioned into four parallel subdomains.}%
  \label{fig:multidomain_matrix_mesh}%
\end{figure}%

\subsection{Structure of the System Matrix}\label{sec:structure_multidomain_system_matrix}

\begin{figure}
  \centering%
  \includegraphics[width=0.3\textwidth]{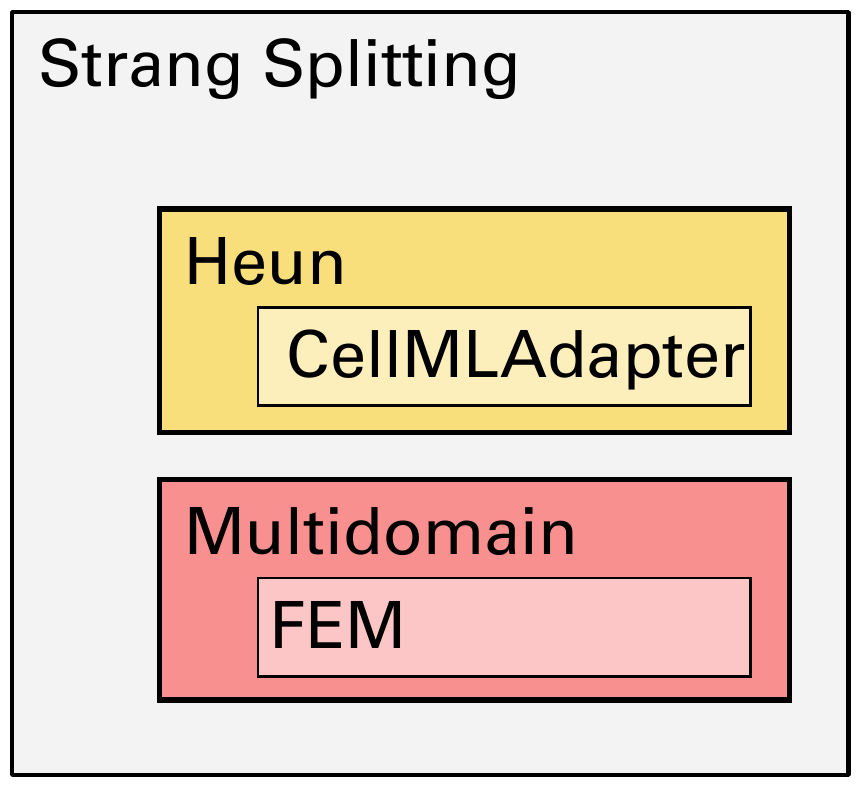}%
  \caption{Solver structure of the multidomain solver, consisting of the Strang operator splitting, which contains the two nested solvers for the subcellular model (Heun scheme with CellML adapter) and the solver for the diffusion part of the multidomain problem. The colors match the scheme introduced in the overview chart in \cref{fig:multi-scale-model}.}%
  \label{fig:multidomain_matrix_mesh}%
\end{figure}%

\Cref{fig:multidomain_matrix_mesh} shows the solver structure of a simulation of the multidomain model. The Strang operator splitting couples the Heun scheme of the 0D subcellular model with the multidomain solver, which is given by the \code{MultidomainWithFatSolver} class. The \code{MultidomainWithFatSolver} class uses nested \code{FiniteElementMethod} classes to describe the anisotropic electric conduction in the muscle domain and the isotropic electric conduction in the fat domain.
The system matrix for the system of equations is given in \cref{eq:discretized_multidomain_body2} in \cref{sec:discretization_body_domain}. 
The solver calculates the matrix block entries using the stiffness and mass matrices computed by the nested \code{FiniteElementMethod} classes, i.e., the unknowns are organized in blocks in the system matrix for each MU.

\begin{figure}%
  \centering%
  \begin{subfigure}[t]{0.49\textwidth}%
    \centering%
    \includegraphics[width=\textwidth]{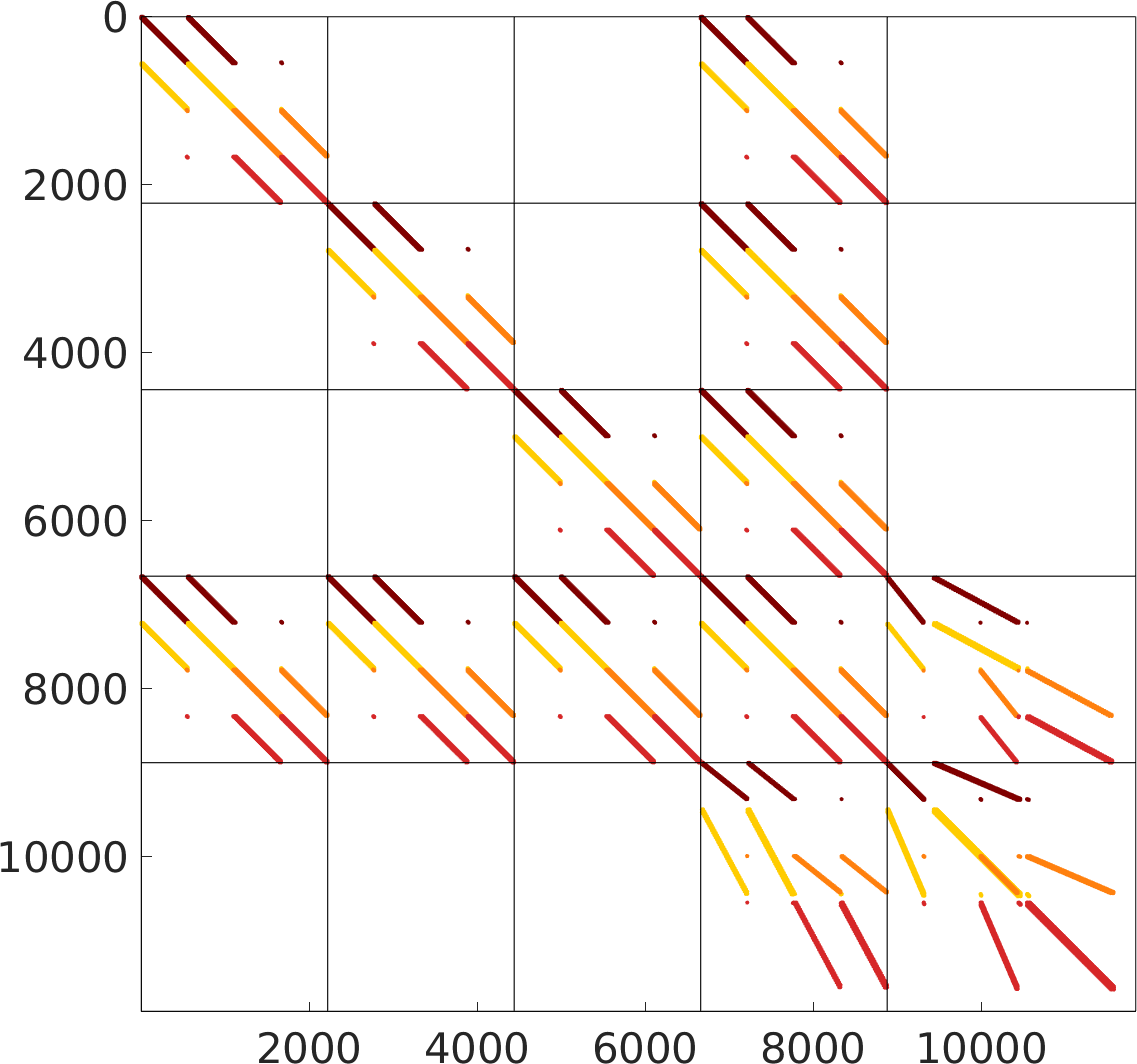}
    \caption{Original matrix layout.}%
    \label{fig:original_matrix}%
  \end{subfigure}
  \,
  \begin{subfigure}[t]{0.49\textwidth}%
    \centering%
    \includegraphics[width=\textwidth]{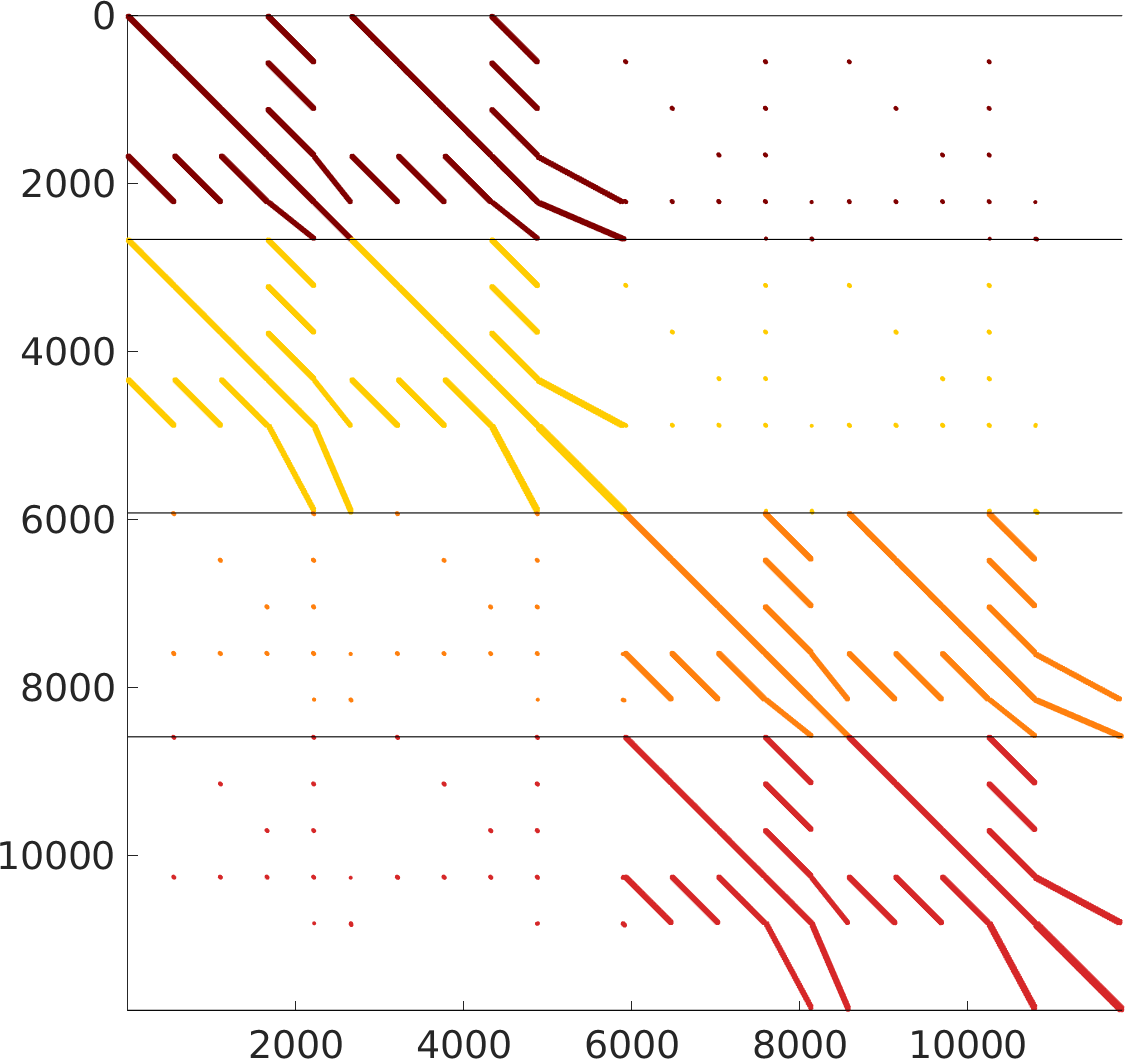}
    \caption{Reordered matrix layout.}%
    \label{fig:reordered_matrix}%
  \end{subfigure}
  \caption{Nonzero structure of the system matrix of the multidomain problem for three MUs. The five blocks in every row and column in (a) correspond to the dofs of the three MUs, the extracellular potential in the muscle domain, and the body potential.}%
  \label{fig:original_reordered_matrix}%
\end{figure}

\Cref{fig:original_matrix} shows the location of non-zeros in the resulting sparse matrix for three MUs. The matrix blocks are indicated by boxes and correspond to the symbolic formulation given in \cref{eq:discretized_multidomain_body2}. The first three blocks correspond to the electric conduction problems of the 3 MUs given by the second multidomain equations in \cref{eq:multidomain2}, the fourth row and column of blocks corresponds to the first multidomain equation \cref{eq:multidomain1}, and the last block corresponds to the electric conduction problem in the fat domain. It can be seen that the dimension of the last block is different, corresponding to the number of dofs in the fat mesh.

In this visualization, it may seem that most of the blocks only have three non-zero entries per row, however, the actual number is higher (the \say{lines} consist of multiple diagonals of non-zero entries) with a maximum of 27 entries, as the finite element ansatz function of a node in the 3D mesh has overlapping support with the ansatz functions of other nodes in a $3\times 3 \times 3$ grid.
The actual non-zero structure per block is close to the example shown in \cref{fig:sparsity_pattern}.

The colors in \cref{fig:original_matrix} correspond to the four processes, as defined in the partitioned mesh in \cref{fig:multidomain_matrix_mesh}. 
The entries in every block are all partitioned in the same way to the four processes, as given by the partitioning of the nested \code{FiniteElementMethod} classes. The data structure for this layout is the \code{MATNEST} type of PETSc. 

However, to be able to apply the multitude of PETSc solvers to this linear system, the matrix has to be transferred to the canonical parallel matrix layout of PETSc, which groups all dofs of the subdomains together. As this conversion is not available in PETSc, it is done in OpenDiHu by reordering the dofs and, as a consequence, the matrix entries. The same permutation is applied to the rows and to the columns of the matrix. The result of this operation is shown in \cref{fig:reordered_matrix}. It can be seen that the portions for each process are now consecutive matrix rows. The non-zero structure within each process resembles the global matrix structure of the original matrix.

% multidomain uses two fem objects, parallel partitioning
% reorder matrix entries
% for 16 processes scenario

\subsection{Properties of a Diagonal Block-Matrix for the Preconditioner}\label{sec:multidomain_diagonal_matrix}
With the reordered matrix, the linear system can now be solved using almost any preconditioner and linear solver of the PETSc framework. 
For the construction of the preconditioner $\mathcal{P}$ with left preconditioning matrix $P=\mathcal{P}(A)$, we can either use the system matrix $A$ or provide a different matrix $A'$. The preconditioned linear system $P^{-1}A$ should have a smaller condition number than $A$ and, thus, solving the preconditioned system iteratively should be significantly faster than the original $A$ system.

\begin{figure}
  \centering%
  \includegraphics[width=0.7\textwidth]{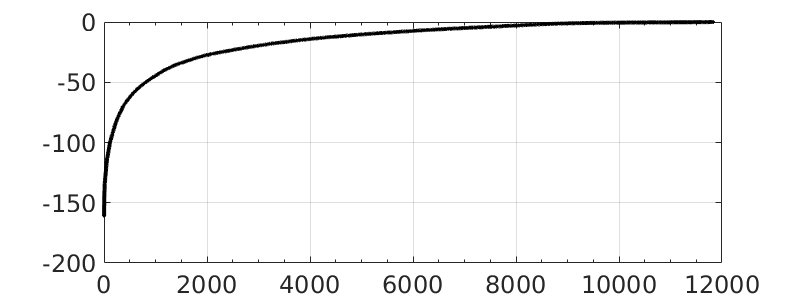}%
  \caption{Real parts of the eigenvalues sorted by magnitude and corresponding to the example in \cref{fig:multidomain_matrix_mesh}. The non-zero eigenvalue with largest and smallest absolute values are $\lambda_\text{max} = \num{-161.2576}$ and $\lambda_\text{min} = \num{-0.0116}$.}%
  \label{fig:eigenvalues}%
\end{figure}

To compute the condition number of the system matrix $A$, we determine its spectrum. \Cref{fig:eigenvalues} shows the real parts of all eigenvalues of $A$. The imaginary parts vanish for almost all eigenvalues. The matrix is singular with one zero-eigenvalue. This property corresponds to the fact that the membrane potential in the problem is abitrary with respect to a constant offset. The singular problem can be solved using appropriate iterative solvers.

% eigenvalues, spectrum, condition number is bad -> preconditioning
The real parts of the eigenvalues are all negative, which is in line with the fact that the model consists of a combination of several diffusion problems. The progression in \cref{fig:eigenvalues} shows a large difference between the largest and the smallest eigenvalues. The condition number of $A$ can be computed by $\textrm{cond}(A) = |\lambda_\text{max}| / |\lambda_\text{min}| = 161.2576 / 0.0116 \approx \num{1.4e5}$. Thus, the problem is ill-conditioned and can benefit from for preconditioning. The condition number is also dependent on the spatial mesh resolution and increases for larger problem sizes.

% PETSc tutorial on preconditioning: https://www.mcs.anl.gov/petsc/meetings/2016/slides/tutorial1.pdf

% the ten smallest eigenvalues of the matrix:
%     0.0000
%    -0.0116
%    -0.0183
%    -0.0193
%    -0.0204
%    -0.0211
%    -0.0222
%    -0.0229
%    -0.0235
%    -0.0239
% 
% the highest eigenvalues:
%  -161.2576
%  -157.7988
%  -153.7242
%  -151.2197
%  -149.2269
%  -144.5006
%  -143.2441
%  -142.4697
%  -139.5421
%  -139.2625

\begin{figure}%
  \centering%
  \begin{subfigure}[t]{0.49\textwidth}%
    \centering%
    \includegraphics[width=\textwidth]{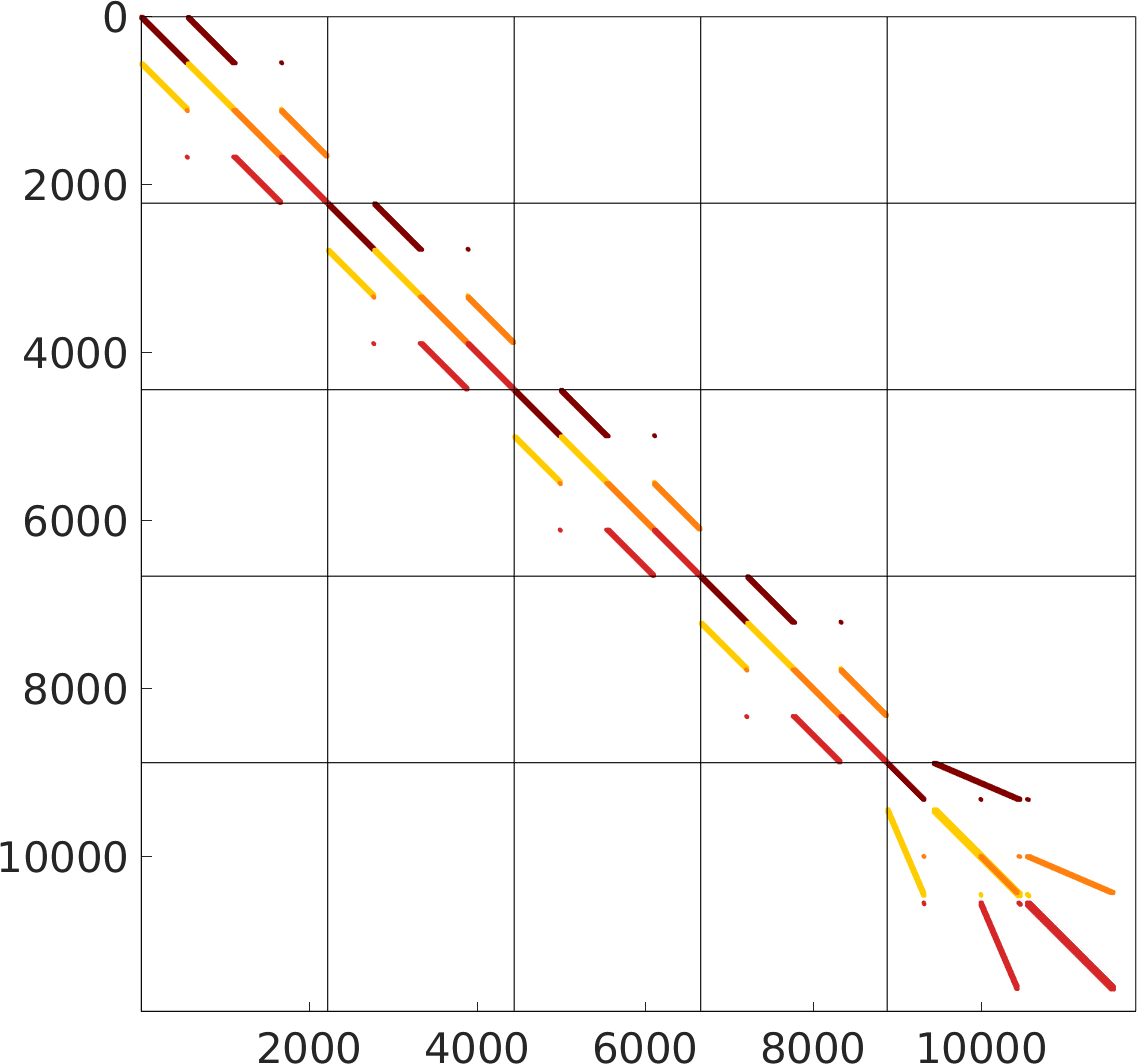}
    \caption{Original matrix layout.}%
    \label{fig:original_diagonal_matrix}%
  \end{subfigure}
  \,
  \begin{subfigure}[t]{0.49\textwidth}%
    \centering%
    \includegraphics[width=\textwidth]{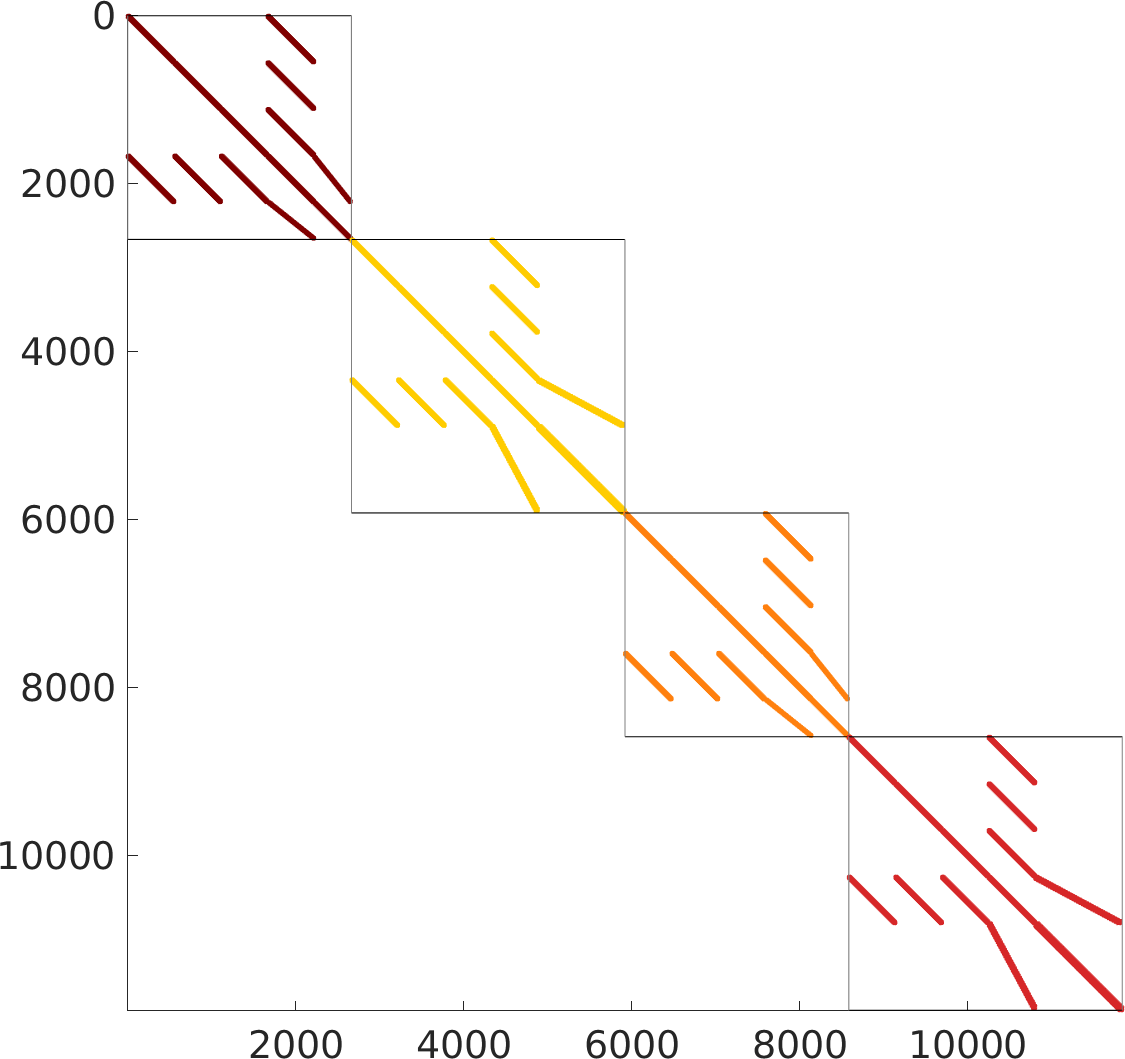}
    \caption{Reordered matrix layout.}%
    \label{fig:reordered_diagonal_matrix}%
  \end{subfigure}
  \caption{Nonzero structure of the symmetric preconditioner matrix of the multidomain problem. The symmetric matrices are obtained from the full matrices in \cref{fig:original_reordered_matrix} by removing all blocks outside the main diagonal.}%
  \label{fig:original_reordered_diagonal_matrix}%
\end{figure}

We experiment with a preconditioning matrix that only uses the diagonal blocks of the system matrix in reordered matrix layout, as shown in \cref{fig:reordered_matrix}. 
\Cref{fig:reordered_diagonal_matrix} shows the non-zero structure of the resulting matrix and compares it with the non-zero structure of the diagonal blocks of the matrix in original ordering in \cref{fig:original_diagonal_matrix}. 
As all diagonal blocks are symmetric matrices on both orderings, the resulting block-diagonal matrices $A'$ are also symmetric in contrast to the original matrix $A$.

Furthermore, it can be seen that the matrices in \cref{fig:original_reordered_diagonal_matrix} are different. The reordered layout depends on the parallel partitioning and contains only matrix entries within one subdomain, i.e., decouples the problems for different subdomains. In contrast, the diagonal blocks of the original system matrix are independent of the partitioning and contain dependencies between dof in different subdomains.
We use the reordered diagonal matrix for the preconditioner, as this approach is compatible with the parallel matrix storage in PETSc and allows to use the preconditioners and solvers of PETSc.
The decoupled entries on every rank potentially allows for a faster computation in the application of the preconditioner.

\subsection{Mesh and Matrices for Higher Degrees of Parallelism}
\begin{figure}
  \centering%
  \includegraphics[width=\textwidth]{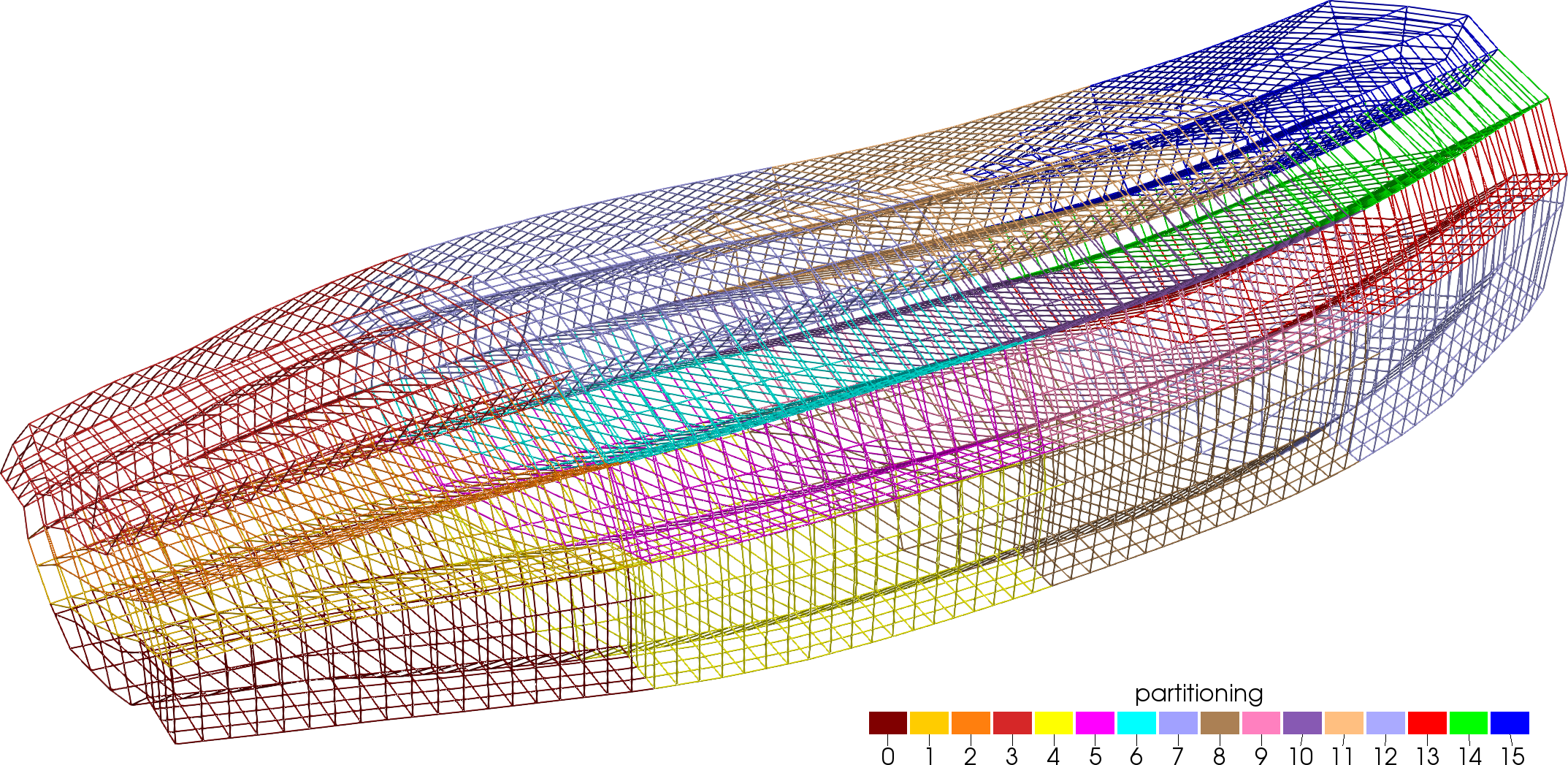}%
  \caption{Partitioning of the mesh in the multidomain example into 16 subdomains.}%
  \label{fig:16_multidomain_matrix_mesh}%
\end{figure}

\begin{figure}%
  \centering%
  \begin{subfigure}[t]{0.49\textwidth}%
    \centering%
    \includegraphics[width=\textwidth]{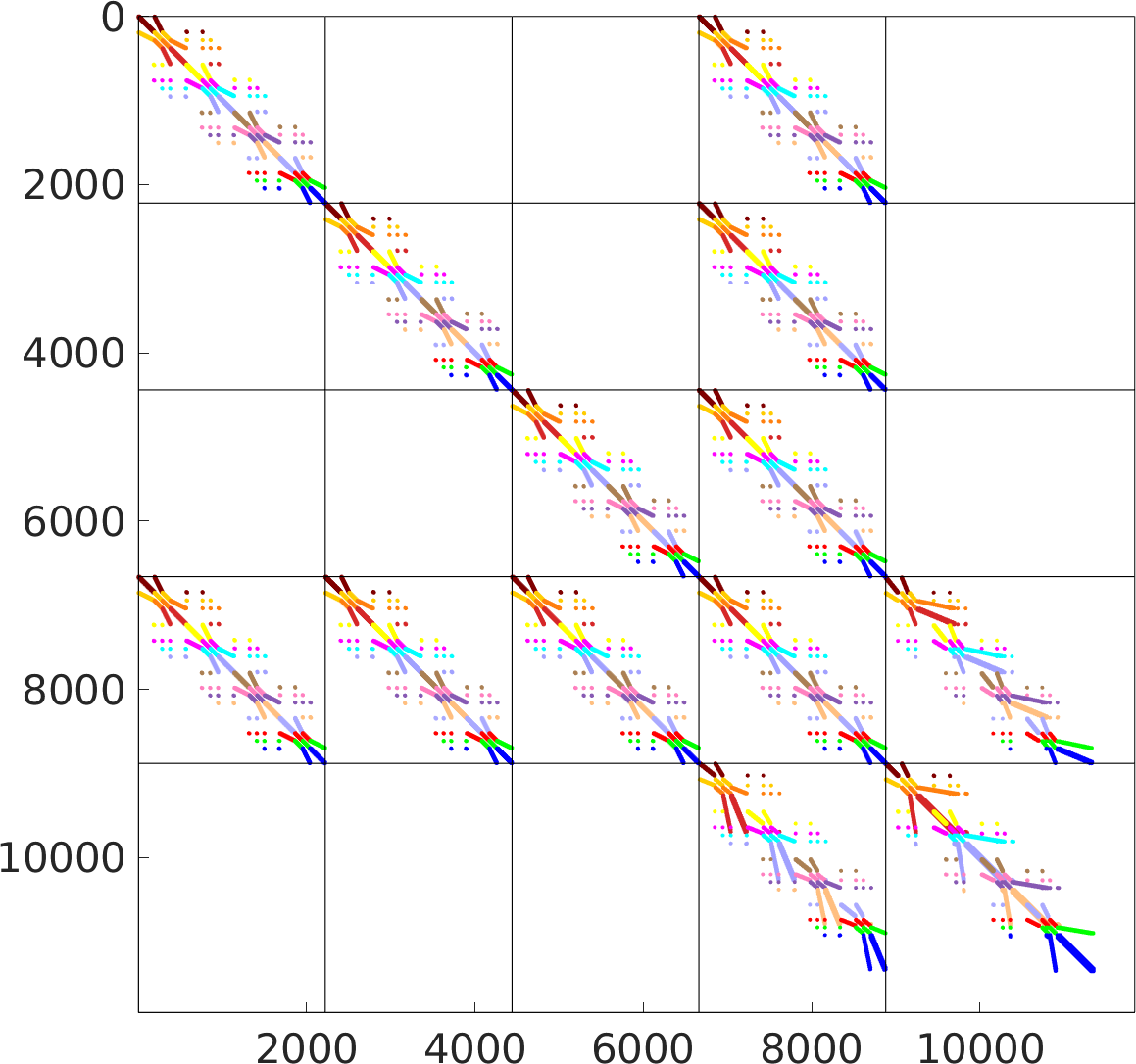}
    \caption{Full matrix in original matrix layout.}%
    \label{fig:16_original_matrix}%
  \end{subfigure}
  \,
  \begin{subfigure}[t]{0.49\textwidth}%
    \centering%
    \includegraphics[width=\textwidth]{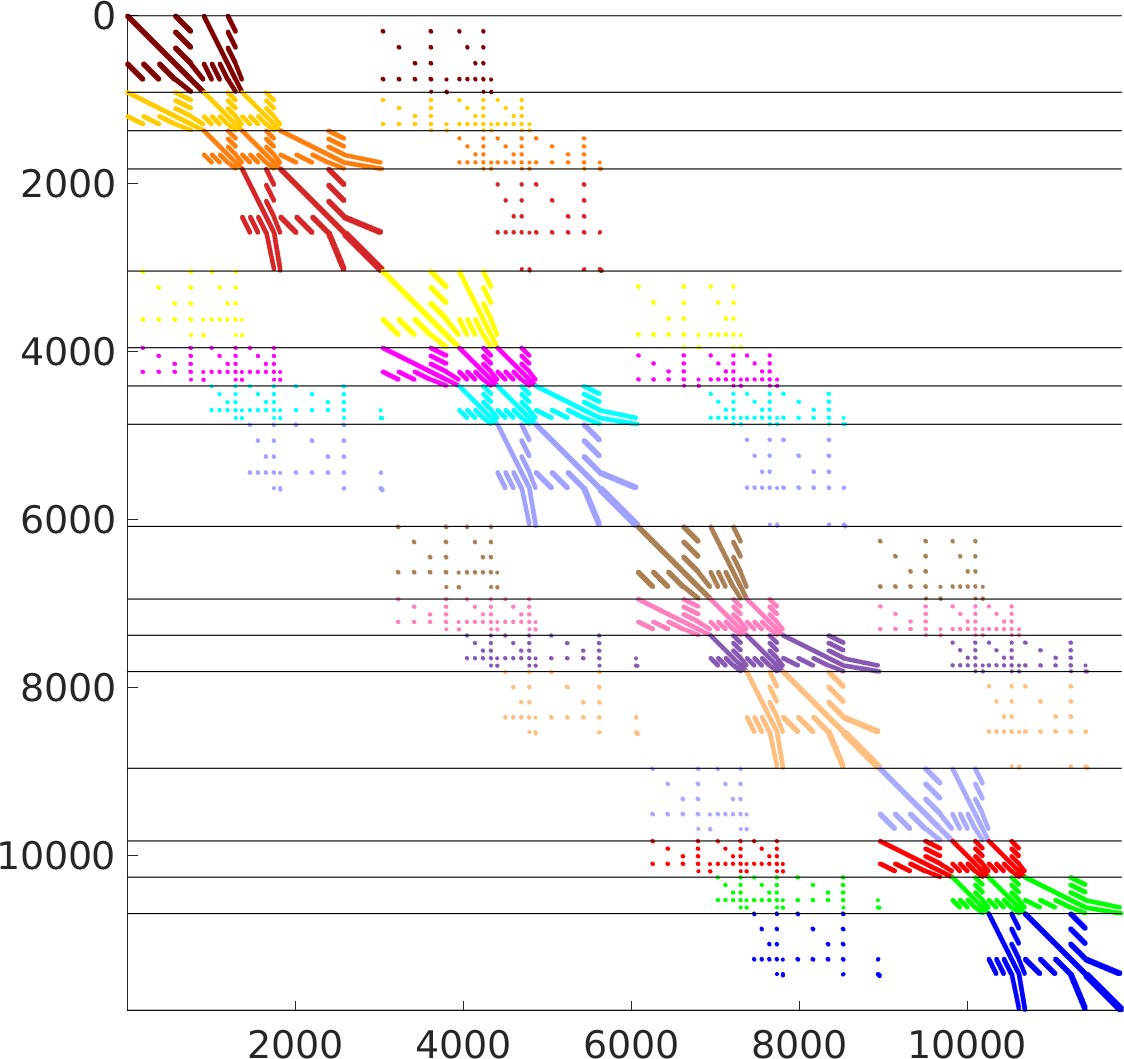}
    \caption{Full matrix in reordered matrix layout.}%
    \label{fig:16_reordered_matrix}%
  \end{subfigure}
  \\
  \begin{subfigure}[t]{0.49\textwidth}%
    \centering%
    \includegraphics[width=\textwidth]{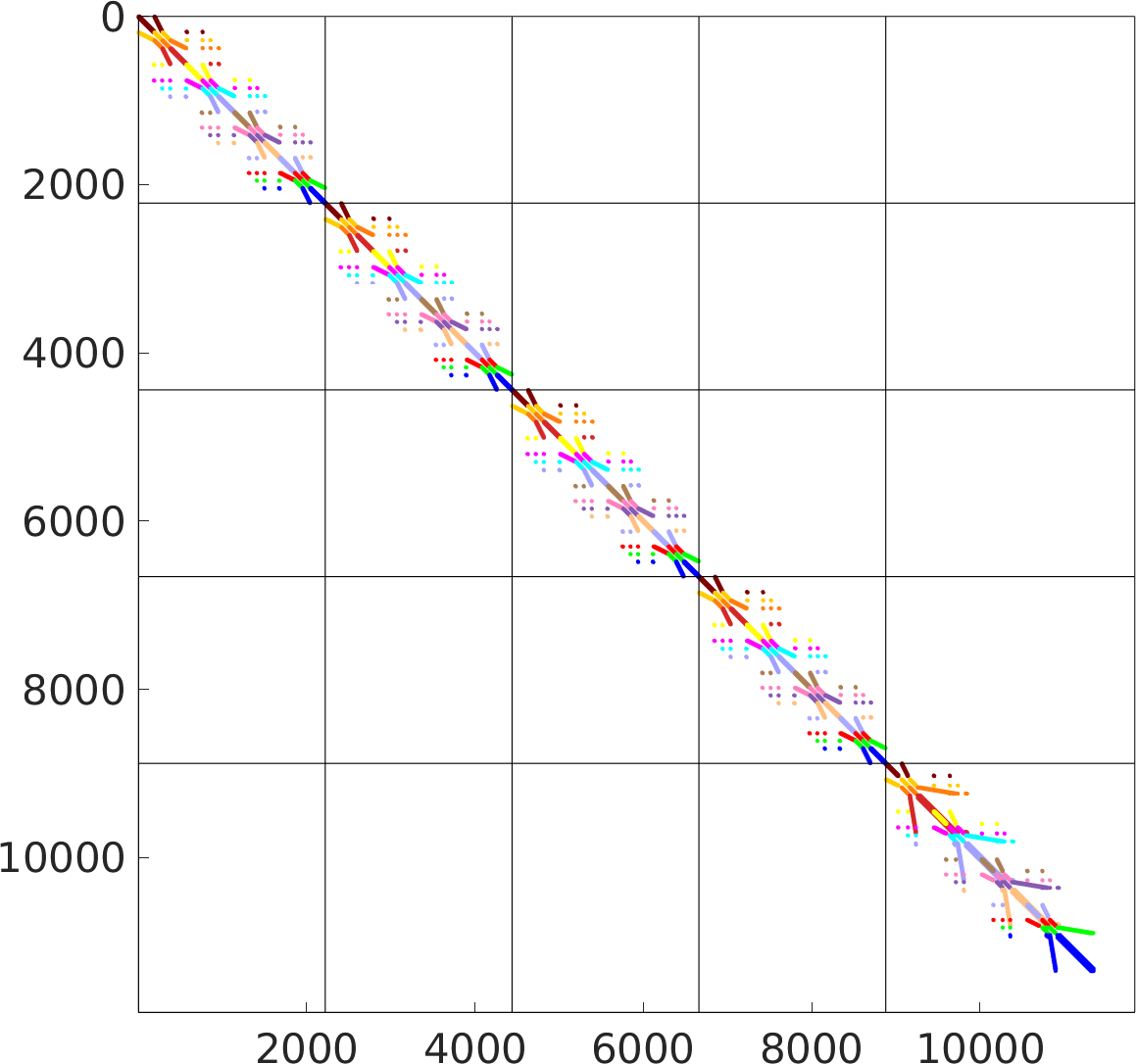}
    \caption{Block-diagonal submatrix in original matrix layout.}%
    \label{fig:16_original_diagonal_matrix}%
  \end{subfigure}
  \,
  \begin{subfigure}[t]{0.49\textwidth}%
    \centering%
    \includegraphics[width=\textwidth]{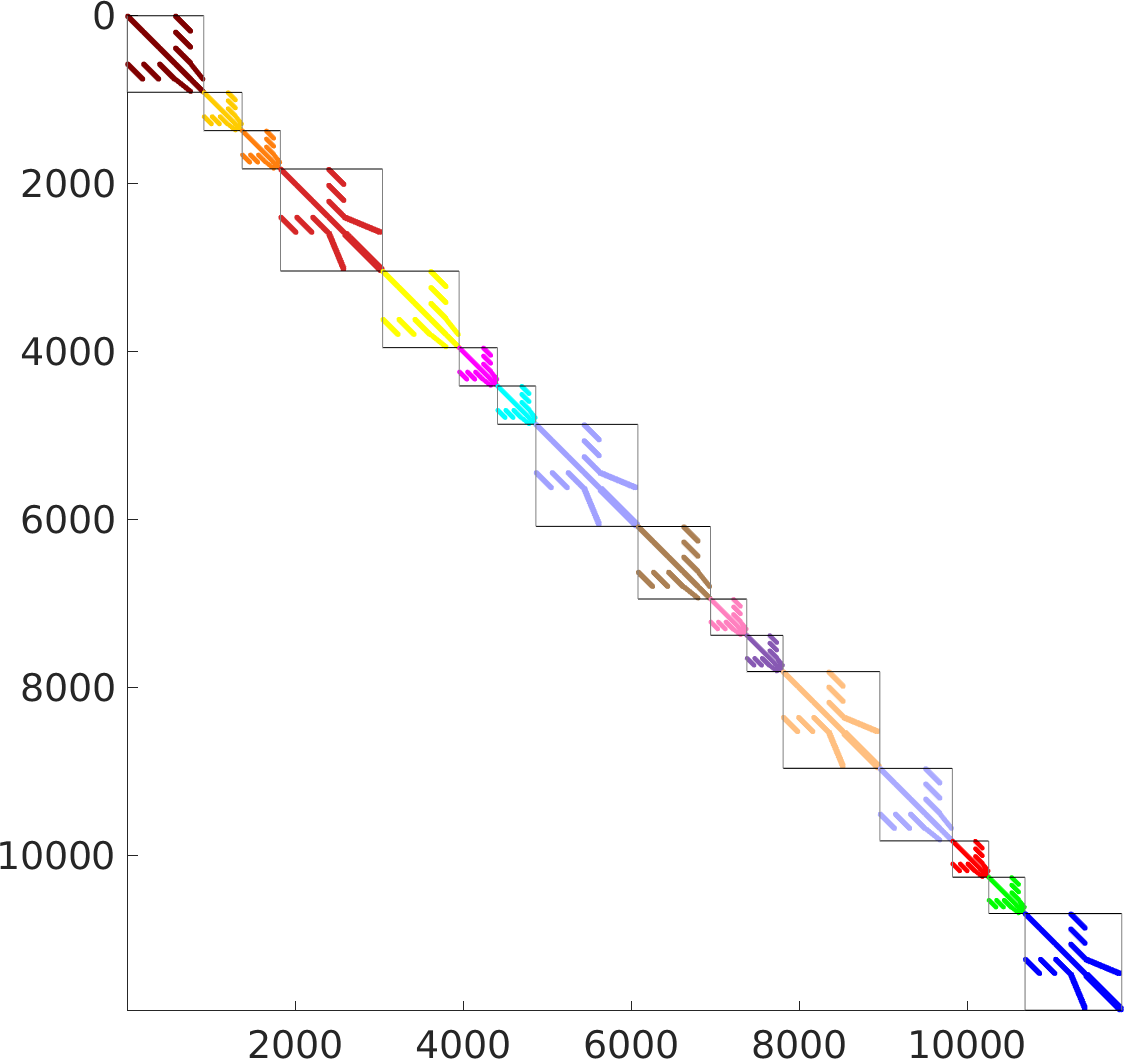}
    \caption{Block-diagonal submatrix in reordered matrix layout.}%
    \label{fig:16_reordered_diagonal_matrix}%
  \end{subfigure}
  \caption{Nonzero structure of the full system matrices in (a) and (b) and the symmetric preconditioner matrices in (c) and (d) of the multidomain problem, partitioned into 16 subdomains. The comparison with \cref{fig:original_reordered_matrix} reveals smaller relative diagonal band widths for the larger number of subdomains in this example.}%
  \label{fig:16_original_reordered_diagonal_matrix}%
\end{figure}

To show the effect of a higher degree of parallelism on the matrix structure, we also partition the same mesh as in \cref{fig:multidomain_matrix_mesh} to 16 processes. The resulting partitioning of the mesh is given in \cref{fig:16_multidomain_matrix_mesh}.
\Cref{fig:16_original_reordered_diagonal_matrix} shows the non-zero structure of the system matrix and the diagonal matrices for the preconditioner. \Cref{fig:16_original_matrix} contains the original matrix structure that is permuted to the structure in \cref{fig:16_reordered_matrix}. 
The symmetric matrices for the preconditioner are shown in \cref{fig:16_original_diagonal_matrix,fig:16_reordered_diagonal_matrix}. 
A comparison with \cref{fig:original_reordered_diagonal_matrix} shows that the width of the non-zero band decreases for higher parallelizations.

\begin{reproduce_no_break}
  The following commands run one timestep of the multidomain simulation with fat layer:
  \begin{lstlisting}[columns=fullflexible,breaklines=true,postbreak=\mbox{\textcolor{gray}{$\hookrightarrow$}\space}]
    cd $\$$OPENDIHU_HOME/examples/electrophysiology/multidomain/multidomain_with_fat/build_release
    mpirun -n 4 ./multidomain_with_fat ../settings_multidomain_with_fat.py matrix.py
    mpirun -n 16 ./multidomain_with_fat ../settings_multidomain_with_fat.py matrix.py
  \end{lstlisting}
  To inspect the system matrix, define a directory where the matrix should be stored. This can be done by setting the parameter \code{config[`Solvers`][`multidomainLinear}\code{Solver`][`dumpFilename`]}, e.g., to \code{`out/matrix/m`}. Then, the directory \code{out/matrix} will contain MATLAB files with the system matrix. To create the plots, open MATLAB, load the system matrix from the  respective file and open the script \code{display_matrix}\code{_entries.m}. Adjust the name of the matrix variable in the first code block, the run the desired steps of the Live Script to produce various plots.\\
  The saved file contains the system matrix already in the reordered layout shown in \cref{fig:reordered_matrix,fig:16_reordered_matrix}. The MATLAB script reverses the permutation that was applied in OpenDiHu to generate the plots of \cref{fig:original_matrix,fig:16_original_matrix}.
\end{reproduce_no_break}

% scenario_name: matrix,  n_subdomains: 4 1 4,  n_ranks: 16,  end_time: 0.002
% dt_0D:           1e-03    multidomain solver:         1000 it. of gmres (10000 it. of gmres), lumped mass matrix: False, initial guess: previous solution
% dt_multidomain:  1e-03    multidomain preconditioner: euclid (euclid), symmetric precond.: True
% dt_splitting:    1e-03    theta: 1.0, solver tolerances, abs: 1e-15, rel: 1e-15
% fiber_file:              ../../../input/left_biceps_brachii_13x13fibers.bin
% fat_mesh_file:           ../../../input/left_biceps_brachii_13x13fibers.bin_fat.bin
% cellml_file:             ../../../input/hodgkin_huxley_1952.c
% firing_times_file:       ../../../input/MU_firing_times_always.txt
% ********************************************************************************
% 16 ranks, partitioning: x4 x y1 x z4
%   sampling 3D mesh with stride 3 x 3 x 20 
%   distribute_nodes_equally: True
%     linear 3D mesh    nodes global: 6 x 5 x 77 = 2310, local: 2 x 5 x 19 = 190
%     linear 3D mesh elements global: 5 x 4 x 76 = 1520, local: 2 x 4 x 19 = 152
%     fat mesh, n points total:    3850 (10 x 5 x 77), (per process: 2 x 5 x 19 = 190)
% 

% details multidomain solver, reordering of matrix entries

% solver structure of contraction
%\begin{figure}
%  \centering%
%  \includegraphics[width=0.5\textwidth]{images/implementation/solver_structure_contraction.pdf}%
%  \caption{solver structure}%
%  \label{fig:prestrech1b}%
%\end{figure}%

% ----

\section{Computation of CellML Models}\label{sec:computation_cellml_models}

In the following, we consider the computation of models that are given in CellML description, such as the subcellular model in the multi-scale muscle model.

The subcellular model is a system of DAEs that is solved at every node of the meshes in the discretized muscle.
For the fiber based electrophysiology description, instances of the 0D subcellular model are computed on every node of every 1D fiber mesh. The 0D instances are coupled by the monodomain equation on every fiber. For the multidomain description, 0D model instances are solved at every node of the 3D muscle mesh for every compartment.

The subcellular model is provided as a CellML file and can be configured in the Python settings as described in \cref{sec:usage_cellml}.
The class in OpenDiHu that computes CellML model instances for all nodes of a given mesh is the \code{CellMLAdapter}. It computes the expression $G$ of the right-hand side of the ODE system, to obtain the vector of rates $\partial \bfy / \partial t = G(\bfy)$ and the expression $H$ for the algebraics $\bfh=H(\bfy)$. The new state vector $\bfy$ is computed from the previous vector by a timestepping scheme, which uses the computed rates $\partial \bfy / \partial t$ as right-hand side. In the solver tree, the timestepping solver class has to be the parent node of the \code{CellMLAdapter}.

CellML models can be obtained as C source files, which can be compiled to shared libraries, loaded and accessed by the solver program. This approach is used in both OpenCMISS and OpenDiHu.
The operation of computing multiple instances of a CellML model at once can be done more efficiently than in the naive way of repeatedly executing the model function, as done in OpenCMISS. To exploit the structure of computing multiple model instances together, dedicated C code has to be generated from the CellML model at runtime for a given number of model instances.
In the following, we describe our code generation functionality for this purpose.

\subsection{Data Flow for the Computation of CellML Models}

\begin{figure}%
  \centering%
  \includegraphics[width=0.7\textwidth]{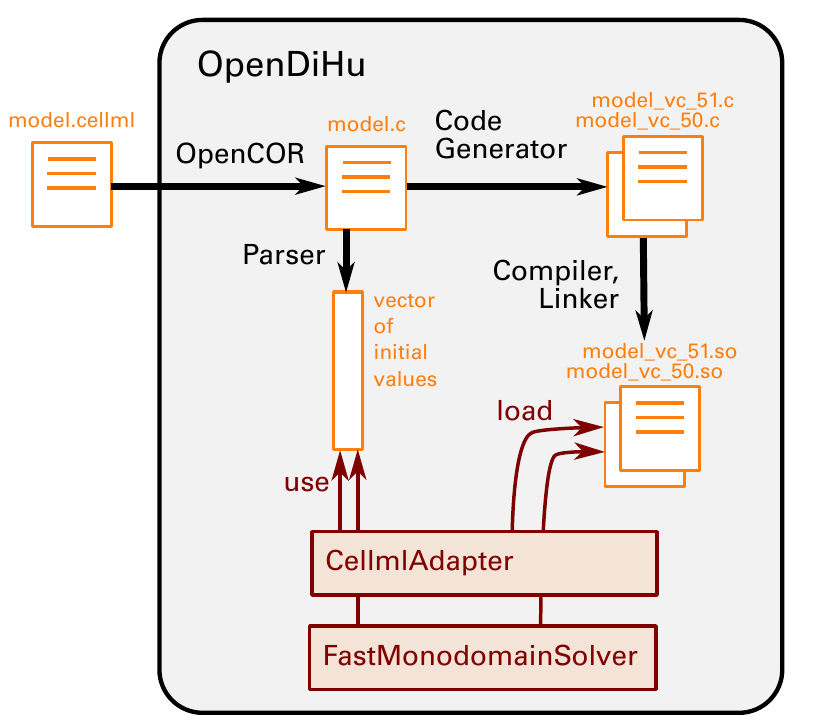}%
  \caption{Processing of the given CellML model prior to solution. The CellML description is converted to C code using OpenCOR. The parser loads the C source code file and determines the contained initial values. Additionally, it parses the compute instructions into an internal syntax tree. The code generator produces optimized C code that can solve as many instances of the model as needed on every process according to the global partitioning of the domain. The generated C code is compiled, linked to a shared library and accessed from the solver code.}
  \label{fig:cellml_scheme}%
\end{figure}%

\Cref{fig:cellml_scheme} shows the information flow for the CellML subsystem in OpenDiHu. On the left, a subcellular model is specified in CellML format in a file \code{model.cellml}. OpenDiHu uses the command line interface of OpenCOR to generate corresponding C code in the file \code{model.c}. The C code computes the functions $G(\bfy)$ and $H(\bfy)$ for the right-hand side and algebraics vector, respectively. A parser traverses the generated C source file and stores all instructions in an internal syntax tree data structure. The parser also determines the initial values for the state vector $\bfy$ from the code that initializes the variables.
Next, certain constants and algebraics in the compute instructions are replaced by parameter variables, as configured in the settings.

Then, a code generator outputs new C code that is optimized for a given number of CellML instances according to the number of nodes in the processes' subdomain within the global domain decomposition. 
This step is executed in parallel by different processes, but only once for every required number of model instances.

For example, if two fibers with 100 elements each are computed by $2 \times 2$ processes, the 101 nodes on each fiber are equally distributed to two different processes. As a result, each MPI rank has to compute either 51 or 50 CellML instances. Thus, the code generators on two of the ranks produce source code files for 51 and 50 model instances, named \code{model_vc_51.c} and \code{model_vc_50.c} in \cref{fig:cellml_scheme}, respectively. After generation, the source files are compiled and linked to a shared library, resulting in the shared object files \code{model_vc_51.so} and \code{model_vc_50.so} in \cref{fig:cellml_scheme}.

The generation, compilation and linking steps are performed only by one process per source file.
If a source file or shared library with the required name already exists from a previous run, the respective code generation and compilation steps are omitted.
In the example, only two processes generate and compile the code. All four processes synchronize after all shared libraries have been generated, before proceeding to execute the computations.

The generated shared libraries contain machine-code to compute $G(\bfy)$ and $H(\bfy)$. They are loaded into the simulation program and executed by the \code{CellmlAdapter} class with the corresponding values, as indicated in \cref{fig:cellml_scheme}. Furthermore, the \code{Cellml}\code{Adapter} uses the previously inferred vector of initial values to initialize the state vector before the first timestep. Also, the \code{FastMonodomainSolver} class presented in \cref{sec:improved_parallel_solver_for_fiber_based}, which efficiently solves the monodomain equation, makes use of the code generator and the shared libraries to evaluate the operators G and H of the subcellular model.

\subsection{Optimizations in the Generated Code}\label{sec:optimizations_in_the_generated}

The code generator can be configured to employ various types of optimizations in the generated code. These optimizations can be selected in the settings by the parameter \code{`optimizationType`}.

\begin{figure}%
  \centering%
  \begin{subfigure}[t]{0.6\textwidth}%
    \centering%
    \includegraphics[width=\textwidth]{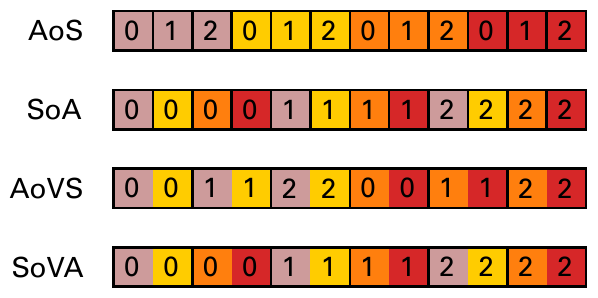}
    \caption{Data in different memory layouts.}%
    \label{fig:memory_layouts}%
  \end{subfigure}\\[4mm]
  \begin{subfigure}[t]{0.7\textwidth}%
    \centering%
    \includegraphics[width=\textwidth]{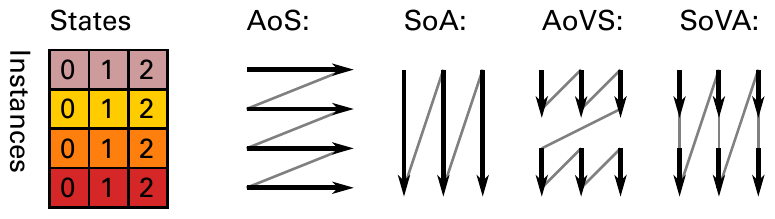}%
    \caption{Schemes how to construct the memory layouts. On the left, the entries are organized in a 2D field according to the state index and instance index. On the right, the traversal schemes for the different layouts are shown.}%
    \label{fig:memory_layouts1}%
  \end{subfigure}
  \caption{Different memory layouts for the CellML model: Array-of-Struct (AoS), Struct-of-Array (SoA), Array-of-Vectorized-Struct (AoVS), and Struct-of-Vectorized-Array (SoVA). The entries for four instances of the CellML model are shown by different colors, where each contains the three state variables 0,1 and 2.}%
  \label{fig:memory_layouts_both}%
\end{figure}

The naive way to solve multiple CellML model instances leads to storing the state vectors
in an Array-of-Struct (AoS) memory layout. The \say{struct} containing all components of the state vector for a single CellML model is stored at consecutive locations in memory and multiple structs for all computed instances are lined up next to each other. \Cref{fig:memory_layouts} shows the AoS layout in the top row for four model instances given by different colors. Each instance contains the three state variables $0,1$ and $2$. 

The transposed memory layout is Struct-of-Array (SoA), where the same state components for all model instances are close in memory. In the example in the second row of \cref{fig:memory_layouts}, always four states of the same kind are stored contiguously. 
\Cref{fig:memory_layouts1} shows the construction schemes for the memory layouts. Comparing the scheme for SoA with AoS, it can be seen that the traversal in the 2D field of values is now vertical instead horizontal.

Such a vertical layout is a prerequisite for employing single-instruction-multiple-data (SIMD) parallelism. SIMD instructions perform the same calculations on multiple components of SIMD vectors simultaneously. In the visualization of SoA in \cref{fig:memory_layouts}, always four operands could be loaded simultaneously from memory to the vector registers in the CPU. Modern processors support the AVX2 instruction set with a SIMD lane width of $\mathcal{W}_T=4$ double values or the AVX-512 instruction set with $\mathcal{W}_T=8$ double values.

\begin{figure}
\centering
  \begin{subfigure}[t]{\textwidth}%
  \centering%
\begin{framed}
\begin{lstlisting}[basicstyle=\footnotesize\ttfamily,commentstyle=\color{gray},numbers=left]
  ALGEBRAIC[1] = ( - 0.100000*(STATES[0]+50.0000))/(exp(- (STATES[0]+50.0000)/10.0
  RATES[1] =  ALGEBRAIC[1]*(1.00000 - STATES[1]) -  ALGEBRAIC[5]*STATES[1];
  ...
\end{lstlisting}
\end{framed}
    \caption{Original C code for one CellML model instance generated by OpenCOR.}%
    \label{fig:cellml_codes_original}%
  \end{subfigure}\\[4mm]
  \begin{subfigure}[t]{\textwidth}%
  \centering%
\begin{framed}
\begin{lstlisting}[basicstyle=\footnotesize\ttfamily,commentstyle=\color{gray},numbers=left]
  #pragma omp for simd
  for (int i = 0; i < 1481; i++)
    algebraics[1481+i] = ( - 0.100000*(states[0+i]+50.0000))/(exp(- (states[0+i]+5
    
  #pragma omp for simd
  for (int i = 0; i < 1481; i++)
    rates[1481+i] =  algebraics[1481+i]*(1.00000 - states[1481+i]) -  algebraics[7
  ...
\end{lstlisting}
\end{framed}
    \caption{Generated code for optimization type \code{`simd`}.}%
    \label{fig:cellml_codes_simd}%
  \end{subfigure}\\[4mm]
  \begin{subfigure}[t]{\textwidth}%
  \centering%
\begin{framed}
\begin{lstlisting}[basicstyle=\footnotesize\ttfamily,commentstyle=\color{gray},numbers=left]
  // fill input vectors of states and parameters
  for (int stateNo = 0; stateNo < nStates; stateNo++)
    for (int i = 0; i < nVcVectors; i++)  // Vc vector no
      for (int k = 0; k < $\mathcal{W}_T$; k++)  // entry no in Vc vector 
        statesVc[i*nStates + stateNo][k] = states[stateNo*nInstances + i*$\mathcal{W}_T$+k]; $\label{alg:st_aovs}$
     // statesVc[stateNo*nVcVectors + i][k] = states[stateNo*nInstances + i*$\textcolor{gray}{\mathcal{W}_T}$+k]  $\label{alg:st_sova}$

  for (int i = 0; i < nVcVectors; i++)
  {
    algebraicsVc[i*nAlgebraics + 1] = ( - 0.100000*(statesVc[i*nStates + 0]+50.000 $\label{alg:b_aovs}$
  //algebraicsVc[371+i] = ( - 0.100000*(statesVc[0+i]+50.0000))/(exponential(- (st $\label{alg:b_sova}$
    ...
  }
\end{lstlisting}
\end{framed}
    \caption{Generated code for optimization type \code{`vc`}.}%
    \label{fig:cellml_codes_vc}%
  \end{subfigure}\\[4mm]
  \begin{subfigure}[t]{\textwidth}%
  \centering%
\begin{framed}
\begin{lstlisting}[basicstyle=\footnotesize\ttfamily,commentstyle=\color{gray},numbers=left]
  #pragma omp parallel for
  for (int i = 0; i < 1481; i++)
  {
    algebraics[1481+i] = ( - 0.100000*(states[0+i]+50.0000))/(exp(- (states[0+i]+5
    rates[1481+i] =  algebraics[1481+i]*(1.00000 - states[1481+i]) -  algebraics[7
    ...
  }
\end{lstlisting}
\end{framed}
    \caption{Generated code for optimization type \code{`openmp`}.}%
    \label{fig:cellml_codes_openmp}%
  \end{subfigure}
\caption{Output of the CellML code generator in OpenDiHu for 1481 model instances and different optimization types.  The model is the subcellular model of Hodgkin and Huxley, and the code shows only two formulas of this model. Furthermore, the lines are truncated.}%
\label{fig:cellml_codes}%
\end{figure}

\Cref{fig:cellml_codes} demonstrates the code generation and presents different approaches to efficiently evaluate the operators of a CellML model for multiple instances. \Cref{fig:cellml_codes_original} shows the original code for a single model instance, which can be obtained from the CellML website or exported from a CellML model using OpenCOR. The listing shows the computation of the algebraic variable with index one and the rate with index one. The formulas typically use other states, algebraics and constant variables and consist of basic arithmetic such as additions, multiplications, potentiations to integer exponents and exponential functions. Some models such as the subcellular model of Shorten et al. \cite{Shorten2007} also involve piecewise definitions that include \say{inline if} branching operations.

Calling the code in \cref{fig:cellml_codes_original} for multiple model instances is associated with the AoS memory layout.
An improvement is the generated code with optimization type \code{`simd`} in \cref{fig:cellml_codes_simd}, which assumes the data to be organized in the SoA memory layout. The code is generated specifically to solved 1481 instances of the model. The array indexing for the \code{algebraics} and \code{rates} variables sums the constant offset according to the memory layout and the number of the model instance. For example, for the second algebraic (with former index 1), the offset is 1481 because so many memory locations are filled with values of the first algebraic (with former index 0).

Furthermore, every formula is enclosed in a loop over all 1481 instances of the model. The loops have OpenMP pragmas that instruct the compiler to use SIMD instructions for the loop body, if possible. Because of the consecutive storage, $\mathcal{W}_T$ loop iterations can be combined into a single computation using vector instructions. For the remainder iterations at the end of the loop, the compiler automatically adds different instructions with corresponding smaller SIMD vector lengths.

The approach of using OpenMP pragmas has the advantage that it is independent of the actual hardware capabilities and does not fix the SIMD vector size $\mathcal{W}_T$. If vectorization is disabled at compile-time, sequential CPU code is generated and the same valid solution is computed. A disadvantage is that the performance of the generated code depends on the vectorization ability of the compiler and its detection that the variables have the proper memory layout. For some constructs such as exponential functions or branching instructions, no vectorization is employed and the particular loop falls back to serial code. Such behavior is observed when inspecting the vectorization reports, which are emitted by the compiler.

Thus, we implement another optimization type \code{`vc`} in the code generator that guarantees usage of vector instructions for all formulas. We use the C++ library \emph{Vc}, which provides a wrapper to hardware-specific vector instructions and abstracts the SIMD lane width \cite{vc2012,Kretz2015}. Using the data types of this library also allows writing hardware independent code and to achieve performance portability, like with the \code{`simd`} optimization type. 
As \emph{Vc} only supports vectorization up to the AVX2 instruction set, we also use the \code{std::experimental::simd} specification, which is currently considered by the International Organization for Standardization (ISO) and the International Electrotechnical Commission (IEC) for inclusion in the C++ standard library \cite{hoberock2016working}. Switching between these two libraries is transparent in the code and depends on whether the compiler supports C++17.

Similar to the \code{`simd`} optimization type, the \code{`vc`} optimization type also uses a memory layout where consecutive memory entries correspond to different instances of the model, and the traversal direction in \cref{fig:memory_layouts1} is vertical for at least $\mathcal{W}_T$ entries. \Cref{fig:memory_layouts_both} shows two such memory layouts for a SIMD vector length of $\mathcal{W}_T=2$: Array-of-Vectorized-Struct (AoVS) and Struct-of-Vectorized-Array (SoVA). Both are implemented in the code generator. 

The SoVA memory layout is very similar to SoA, the only difference is, that, in SoVA, entries are always accessed in multiples of the SIMD vector length $\mathcal{W}_T$. The advantage of SoVA is that the array indices are given by the sum of a constant offset with the loop index, whereas, with the AoVS layout, a multiplication is required for every access.

AoVS resembles more the AoS layout. Its advantage over SoVA is that the complete state vector $\bfy$ for any model instance is located more locally in memory. As the total computation iterates over model instances, the accessed memory is more coherent than for the same iteration scheme with the AoVS layout. This possibly leads to more cache hits, however, for set-associative caches, the effect is reduced. Due to the complexity of today's cache architectures, only measurements can decide which of the two memory layouts leads to a faster execution. Our measurements show that the SoVA layout leads to \SI{2}{\percent} shorter runtimes than the AoVS memory layout and, thus, is the preferred choice.

% hodgkin huxley
%   vc-sova 216.6 s
%   vc-aovs 217.3 s

% shorten
%   vc-sova 6873.6 s
%   vc-aovs 7000.3 s

% other run:
%                     subdomains        user  total comp.         0D         1D  bidomain  duration_init     write       mem    n
% scenarioName nRanks                                                                                                            
% vc           18      [3, 2, 3]  186.429444   183.516278  45.524128  95.482744  2.693811       3.053507  1.707781  0.224 GB   18
% vc-aovs      18      [3, 2, 3]  178.800859   177.067879  44.365866  92.375215  2.323516       1.917626  1.673526  0.223 GB  198
% vc-sova      18      [3, 2, 3]  180.776722   179.158511  45.464952  93.226485  2.329576       1.810070  1.685208  0.223 GB  180
% ------------------------------------------------------------------------------------------------------------------------
% 

\Cref{fig:cellml_codes_vc} shows the resulting code using the AoVS memory layout. The commented lines \ref{alg:st_sova} and \ref{alg:b_sova} show the corresponding code for the SoVA memory layout.
At the beginning of the generated program code, the given data in the \code{states} variable are copied to the \code{statesVc} variable in the new memory layout. Nested loops over all states, over the SIMD vectors and over the scalar values within the SIMD vector are used for this operation. For comparison, lines \ref{alg:st_aovs} and \ref{alg:st_sova} show the corresponding indexing of the \code{statesVc} variables for the AoVS and SoVA memory layouts, respectively.

For the evaluation of the model operators, we iterate over the number \code{nVcVectors} of SIMD vectors instead of the number of model instances as for \code{`simd`}. In the example with 1481 instances, we have \code{nVcVectors=$\lceil 1481/\mathcal{W}_T \rceil$=371} SIMD vectors for $\mathcal{W}_T=4$. 
Accordingly, the offsets for indexing the variables in the SoVA layout are smaller, e.g., in line \ref{alg:b_sova}, the offset for indexing the \code{algebraicsVc} variables is 371 instead of 1481 for the non-vectorized variable in the previously considered \code{`simd`} code. 
Comparing the statements for AoVS and SoVA in lines \ref{alg:b_aovs} and \ref{alg:b_sova}, it can be seen that the AoVS memory layout involves an additional multiplication during the indexing of the array.

In case of branching instructions in the CellML formulas, the Vc library provides an implementation of the \say{inline if} statement for SIMD vectors, which checks the condition, potentially executes both branches and merges the components from the active branches into the resulting SIMD vector.

Profiling the execution of the \code{`vc`} code for different subcellular models shows that about half of the runtime is spent in evaluating the exponential function. Therefore, we use the following approximation:
\begin{align}\label{eq:apx-e-function}
  \textrm{exp}(x) \approx \textrm{exp}^\ast(x) = \left( 1 + \dfrac{x}{n}\right)^n.
\end{align}
The series converges to the exact value for $n\to \infty$. We choose $n=1024$ and are able to compute the approximate value by only one addition and 11 multiplications using the following formula:
\begin{align*}
    \textrm{exp}^\ast(x) = \left( 1 + \dfrac{x}{1024}\right)^{2^{10}} = \left(\cdots{{\left({\left(\left( 1 + \dfrac{x}{1024}\right)^2\right)}^2\right)}^{\scriptsize\iddots}}\right)^2.
\end{align*}
In the subcellular models of Hodgkin and Huxley \cite{Hodgkin1952} and Shorten et al. \cite{Shorten2007}, the values for $x$ are bounded by $|x| < x_\text{max} = 12$, and we get a relative error of the approximation of $|(\textrm{exp}^\ast - \textrm{exp})(x_\text{max}) / \textrm{exp}(x_\text{max})| < 0.07.$
This approximation can be enabled or disabled in the code generation.

Another optimization is implemented for exponentiation $a^b$. In the considered CellML models, only integer exponents $b\in \mathbb{Z}$ occur. We add a recursive implementation of the power function that requires a logarithmic number of multiplications. 

The code generator with the \code{`vc`} optimization type is also used by the \code{FastMonodomain}\code{Solver} class described in \cref{sec:improved_parallel_solver_for_fiber_based}. The generated codes for the \code{FastMonodomainSolver} class additionally contain the Heun scheme to solve the model, integrate code for the stimulation of muscle fibers and  export certain algebraic values that were declared as parameters in the settings.

Another possiblity to improve the performance besides instruction-level parallelism is thread-level parallelism. The \code{`openmp`} optimization type generates code containing OpenMP pragmas that distribute the computations to multiple OpenMP threads with shared memory. \Cref{fig:cellml_codes_openmp} shows the generated code for this optimization type. A loop iterates over all model instances and the variables are stored in SoA memory layout. The loop iterations are independent of each other as they correspond to different instances of the CellML model. OpenMP distributes the workload to a predefined number of threads that can be specified by environment variables.

\subsection{Code Generation for GPUs}

Besides instruction-level and thread-level parallelism, which were discussed in the last section, accelerator hardware such as GPUs can be considered to reduce the runtime of solving a CellML model.
Our code generator features the \code{`gpu`} optimization type to generate code that is called on the CPU and then offloads the main computations to a GPU.

We use OpenMP 4.5 to instrument the generated code for device offloading. At the time of writing, only an experimental version of GCC 11 is fully capable of compiling this code. In our studies, the code is compiled for the \emph{nvptx} target, which generates and compiles device-specific CUDA code using the NVIDIA parallel thread execution (PTX) instruction set architecture. We successfully run the computation on various NVIDIA GPUs, including a GeForce RTX 3080. However, the approach is device-agnostic and other accelerator hardware can also be used.

\Cref{fig:cellml_codes_gpu} shows an excerpt of the generated code. It resembles the code of the\break\code{`openmp`} optimization type, except that the OpenMP pragma in lines \ref{alg:pragma_line} and \ref{alg:pragma_line2} is different. The lines specify the variables to be mapped to and from the target device: The vectors of states and parameters as well as the current simulation time \code{t} are sent to the GPU and, after computation, the rates and algebraics are transferred back to the CPU.

\begin{figure}
\centering
\begin{framed}
\begin{lstlisting}[basicstyle=\footnotesize\ttfamily,commentstyle=\color{gray},numbers=left]
  #pragma omp target parallel for \ $\label{alg:pragma_line}$
                     map(to:states,t,parameters) map(from:rates,algebraics) $\label{alg:pragma_line2}$
  for (int i = 0; i < 1481; i++)
  {
    algebraics[1481+i] = ( - 0.100000*(states[0+i]+50.0000))/(exp(- (states[0+i]+5
    rates[1481+i] =  algebraics[1481+i]*(1.00000 - states[1481+i]) -  algebraics[7
    ...
  }
\end{lstlisting}
\end{framed}
\caption{Generated code for optimization type \code{`gpu`} corresponding to the scenario in \cref{fig:cellml_codes}.}%
\label{fig:cellml_codes_gpu}%
\end{figure}

Using the \code{CellmlAdapter}, it is, thus, possible to run any CellML model on the GPU. 
However, for the fiber based electrophysiology model uploading and downloading the data of all model instances between CPU to GPU in every timestep is clearly not the most efficient way to utilize the GPU. Therefore, we add efficient GPU integration with proper memory management to the \code{FastMonodomainSolver} class, which is specialized to solve the monodomain equation for multiple fibers. The class allows computing multiple timesteps in series on the GPU between subsequent points of synchronization with the CPU. This synchronization is only required, e.g., for writing output files or coupling to a solid mechanics solver.

The generated GPU source code for the \code{FastMonodomainSolver} contains the full algorithm for solving multiple timesteps of the electrophysiology model for multiple fibers with a given number of nodes each. The Strang splitting scheme is used, which solves the 0D subcellular part and the 1D electric conduction part in the scheme 0D-1D-0D.
The 0D part is solved by the Heun scheme. The 1D part is computed either with the implicit Euler method or the Crank-Nicolson method. The linear system of equations is solved using the linear complexity Thomas algorithm.

The parallelization on the GPU uses a fixed number of thread teams, where all threads in a team execute the same code.
For the 0D problem, the iterations of the two nested loops over fibers and model instances per fiber are distributed to all thread teams, such that the iterations are \emph{workshared}. Thus, the 0D subcellular models are computed concurrently for all instances. Between the computations of the 0D and 1D parts, synchronization occurs as the data on all instances on a fiber are accessed in the solution of the 1D problem. The 1D computations are distributed on the fiber level, before the second 0D computation in the Strang splitting is again distributed on the model instance level.
Another synchronization occurs after each timestep of the whole Strang splitting.

The data transfer in both directions between CPU and GPU is reduced to a minimum. Initially, all required parameters and initial values have to be transferred to GPU memory. The initial state vector $\bfy$ is only sent once to the GPU and all model instances of all fibers get initialized to these same values. Further data to be sent includes parameters that describe the stimulation times as presented in \cref{sec:stimulation_times_callbacks}, locations of the neuromuscular junction and the distribution of fibers to motor units. Instead of the callback functions described in \cref{sec:stimulation_times_callbacks}, the stimulation times can be altered by an input file. For details, we refer to the online documentation \cite{opendihuWeb}.

During computation, smaller amounts of data are transferred before and after each set of consecutive timesteps on the GPU. The data to be sent to the GPU before the computations consist of the CellML parameter values and the lengths of all elements in the 1D mesh, which change, if muscle contraction is computed on the CPU. The data to be transferred back to the CPU after the computations on the GPU consist of a subset of the state vector for every model instance. This subset contains only those components of $\bfy$ that should be written to an output file on the CPU or are required for coupling to another solver. Thus, the majority of the data stay on the GPU.

% ---
\section{Solid Mechanics Solver}\label{sec:solid_mechanics_solver}
Next, we discuss details on the solver of the solid mechanics models, which is needed for the muscle contraction part of the multi-scale model, described in \cref{sec:model_muscle_contraction}.

\Cref{sec:solver_linear_model_elasticity} gives details on the solver for the linear solid mechanics model. \Cref{sec:specification_of_nonlinear_ma} addresses the nonlinear model and describes how the material model is specified. \Cref{sec:convergence_improvements_for_the_nonlinear_solver} presents the timestepping method for the dynamic problem and describes the implemented measures to improve the convergence.

\subsection{Solver for the Linear Model}\label{sec:solver_linear_model_elasticity}
% linear fem
As noted in \cref{sec:summary_of_existing_solver_classes}, the \code{QuasiStaticLinearElasticitySolver} class can be used to solve the linearized solid mechanics model described in \cref{sec:material_linear_model} and discretized in \cref{sec:linearized_mechanics_model}. 
Within this solver class, the matrix equation \cref{eq:linearized_helper4} is assembled and solved by an object of the \code{FiniteElementMethod} class, which is the same class that is used to solve  Laplace problems.

If the solver is explicitly coupled with an electrophysiology model, we obtain a quasi-static formulation of muscle contraction. The activation parameter $\bar{\gamma}$ on the 3D mesh is transferred from the electrophysiology model to the elasticity model.
Then, the linear system of equations of the elasticity model is solved using the new muscle activation values in the right-hand side. The system matrix stays constant in all timesteps. After the new displacements have been computed, the geometries of the 3D mesh and the embedded 1D fiber meshes are updated accordingly.

In this scenario, the active stress tensor $\bfsigma^\text{active}$ in \cref{eq:linearized_helper6} is computed as the product of the activation parameter $\bar{\gamma}$ with a scalar maximum active stress parameter $\sigma_\text{max,active}$ and an anisotropy tensor $\bfa$:
\begin{align}\label{eq:solid_mechanics_solver_1}
  \bfsigma^\text{active} = \sigma_\text{max,active} \,\bar{\gamma}\,\bfa.
\end{align}
The tensor $\bfa$ can be specified in the Python settings by a $3 \times 3$ matrix and allows to specify the anisotropic active behavior of the muscle tissue. In this specification, the first unit vector $\bfe_1=(1,0,0)^\top$ designates the fiber direction, $\bfe_2$ and $\bfe_3$ specify the transverse direction. Prior to the computation in \cref{eq:solid_mechanics_solver_1}, the basis of the given matrix is changed, such that $\bfe_1$ in the old basis maps to the fiber direction in the new basis and the new basis is orthonormal. This change of basis is performed at every point in the muscle with the respective fiber direction. Thus, it is possible to specify transversely isotropic material behavior with contraction in fiber direction.

\subsection{Specification of Nonlinear Material Models}\label{sec:specification_of_nonlinear_ma}
% load steps, initalization in the dynamic case

To compute the nonlinear model, the \code{HyperelasticitySolver} class is used for the static formulation of a passive material, the \code{DynamicHyperelasticitySolver} class is used for the dynamic passive behavior, and the \code{MuscleContractionSolver} is used for either the static or the dynamic model with active stress contribution.

These solver classes can be coupled to the electrophysiology model in the same way as described in \cref{sec:solver_linear_model_elasticity}.
Similar to \cref{sec:solver_linear_model_elasticity}, the \code{MuscleContractionSolver} adds an active stress term to the formulation according to the formula in \cref{eq:active_stress_term}. The force-length relation $f_\ell(\lambda_f)$ can either be added by the \code{MuscleContractionSolver}  or specified in the CellML description as part of the subcellular model for the activation parameter $\gamma$.

To specify the passive material behavior, the strain energy function $\Psi$ has to be defined.
This definition has to be available at compile-time and is specified in the C++ code. 

Four different terms can be defined to describe the material model in different forms such as the coupled or decoupled representation. The four terms are introduced in \cref{sec:material_modeling} and given in \cref{eq:definition_psi} as follows:
\begin{align}\label{eq:definition_psi1}
  \Psi = \Psi_\text{vol}(J) + \Psi_\text{iso}(\bar{I}_1,\bar{I}_2,\bar{I}_4,\bar{I}_5) + \Psi_1(I_1,I_2,I_3) + \Psi_2(\bfC,\bfa_0).
\end{align}
Formulas for these terms can be specified using C++ expressions with a syntax specified by the \emph{SEMT} library \cite{semt,gutterman2004symbolic} (and also described in the online documentation of OpenDiHu \cite{opendihuWeb}). Mathematical functions such as power and log functions are available, intermediate variables can be defined and reused, and constants for material parameters can be used, whose values can be specified in the Python settings.

The implementation uses the SEMT library to symbolically differentiate the given terms with respect to their function arguments. Thus, all values used in the Newton solver including the Jacobian matrix can be computed automatically. Using this technology, OpenDiHu provides the flexibility to add new material models at compile-time without the need for manual differentiation.

Additionally, three options, which alter the computation and efficiency, have to be set in the C++ description: The first option specifies, whether the material is considered incompressible. If this option is set to true, the solution approach with Lagrange multiplier $p$ is used, otherwise the unknowns only contain the displacements $\bfu$ and possibly the velocities $\bfv$. The second option specifies, if the active stress term $\bfS^\text{active}$ should be added to the material. This option is only relevant for the \code{MuscleContractionSolver} class, disabling it allows computing passive tissue.

The third option determines, if the fiber direction $\bfa_0$ appears in the description of the material model. Only if this option is enabled, the corresponding invariants $I_4$ and $I_5$ are available for the definition of the $\Psi_\text{iso}$ term in \cref{eq:definition_psi1}. If disabled, all terms in the formulas in \cref{sec:stress_and_elasticity} that involve $\bfa_0$ are left out of the computation, which speeds up the computations in the solver.

\subsection{Convergence Improvements for the Nonlinear Solver}\label{sec:convergence_improvements_for_the_nonlinear_solver}

The nonlinear equation is solved using the Scalable Nonlinear Equations Solvers (SNES) component of PETSc, which provides Newton-type and quasi-Newton methods for solving systems of nonlinear equations. The method to use and other parameters such as the line-search type can be configured in the Python settings file.

Fast convergence of a Newton-based nonlinear solver is facilitated with a good initial guess for the vector of unknowns. Therefore, we predict the solution functions $\bfu$ and $\bfv$ for the next timestep in a dynamic problem using the following computations:
\begin{align*}
  \bfu^{(i+1),\text{predicted}} &= \bfu^{(i)} + \dt\,\bfv^{(i)}, & 
  \bfa^{(i)} &= \dfrac1{\dt}(\bfv^{(i)} - \bfv^{(i-1)}), & 
  \bfv^{(i+1),\text{predicted}} &= \bfv^{(i)} + \dt\,\bfa^{(i)}.
\end{align*}
The predicted displacements $\bfu^{(i+1),\text{predicted}}$ for the next timestep $(i+1)$ are estimated by a forward Euler scheme from the displacements $\bfu^{(i)}$ and velocities $\bfv^{(i)}$ of the current timestep $i$. The current acceleration $\bfa^{(i)}$ is estimated by finite differences from the current and previous velocities, $\bfv^{(i)}$ and $\bfv^{(i-1)}$. The predicted velocities $\bfv^{(i+1),\text{predicted}}$ for the next timestep again use a forward Euler method with the estimated acceleration values $\bfa^{(i)}$.
Using the initial guess $(\bfu^{(i+1),\text{predicted}},\bfv^{(i+1),\text{predicted}},p^{(i)})^\top$, the solution vector $(\bfu^{(i+1)},\bfv^{(i+1)},p^{(i+1)})^\top$  for the next timestep can be obtained by the nonlinear system solver.

Independently of the predictions of initial values from previous timesteps, the convergence of the nonlinear solver within a timestep can be  improved by employing load stepping. This approach involves solving $N>1$ sub problems with increasing load steps. In each step $i$, the problem is solved with the right-hand side $\bff_i = \alpha_i\,\bff$, scaled by the load factor $\alpha_i \in [0,1]$. The obtained solution in iteration $i$ is used as the initial guess for the subsequent load step $(i+1)$. Increasing values of $\alpha_i$ are used until the final solution is found for $\alpha_N=1$. Typical load factors are $(\alpha_i)_{i=1,\dots,N} = (b^{-(N-1)}, b^{-(N-2)}, \dots, b^{0})$ for a basis $b>0$.

The list of load factors can be specified in the settings. If the nonlinear solver diverges or fails because an unphysical negative determinant $J$ of the deformation gradient occurs, the current load factor is automatically reduced and the solution processes is started again, using the last valid solution as initial guess. If the last successful solution was found for load factor $\alpha_i$ and the current load factor $\alpha_{i+1}$ fails, a new load factor $\alpha^\ast_{i+1} = (\alpha_i + \alpha_{i+1})/2$ is inserted in the list of load factors between $\alpha_i$ and $\alpha_{i+1}$ and the solution of the nonlinear problem with this new factor is attempted. 

In case of a poorly conditioned problem, it can happen that no more solution can be found, regardless of how far the load factor gets decreased. If the difference between two load factors falls below a configurable threshold, the nonlinear solution process for the current timestep is aborted.

Practical tests with the dynamic incompressible problem have shown that the convergence sometimes degrades only for a single timestep and returns to normal in the next timestep. Thus, we allow a single timestep $i$ to diverge and, in this case, continue with the next timestep $(i+1)$ using the (diverged) solution with the lowest residual norm from timestep $i$ to predict the initial guess for timestep $(i+1)$.

\section{Data Mapping Between Meshes}\label{sec:data_mapping_between_meshes}

% introduction, mapping required for operator splitting, different dimensionalities, source to target mesh
% goal is to construct a mapping in the order of the target mesh
% parallel partitioning -> treat source mesh as point cloud

% target->source is interpolation, construct transposed mapping
After the implementation of various solvers for specific parts of the multi-domain model has been described in the previous sections, we now focus on the data mapping between different meshes that occurs in the coupling schemes between the execution of the coupled solvers. 

Data mapping between meshes is required in scenarios that involve both a finely resolved 3D mesh for the electrophysiology model and a coarse 3D mesh for the solid mechanics model. Moreover, data are mapped between the 3D muscle mesh and the embedded 1D fiber meshes in the fiber based electrophysiology model. In these two cases, the mapping has to be carried out in both directions between the involved meshes. The operation can be characterized as \emph{volume mapping}.

We implement a generic mapping scheme between two meshes of any dimensionality and with any relative orientation with respect to each other. 
Given is a finite element interpolant on a \emph{source} mesh, defined by the dof values at the nodes. The goal is to set the dof values of the \emph{target} mesh, such that the error between the finite element representations on the common domain of source and target mesh is as low as possible. By using the ansatz functions of the target mesh, the constructed mapping has the same order of accuracy as the target mesh interpolant. For a linear target mesh, the mapping operation is second order accurate, for a quadratic target mesh, the mapping operation is third order accurate.

The considered source and target meshes are possibly partitioned. In order to perform the data mapping directly between two such meshes without communication between the processes, the partitioning of the volumes would have to be identical. However, this is not practical for different meshes and would disallow different orientations of source and target meshes. The only way to allow such a mapping is to consider either the source or the target mesh as a point cloud and construct the mapping between individual points and a mesh.

Considering the case of mapping the activation parameter value $\gamma$, which is stored on multiple 1D fibers, to the value $\bar{\gamma}$ on the 3D muscle mesh, it is natural to consider the source mesh as a point cloud. Then, instead of multiple fibers, we have a set of points, where $\gamma$ is known. This set of source points is mapped to the enclosing 3D target mesh. On every process, the source points have to be located inside the local subdomain of the target mesh.

The reverse mapping in this example is also required: The geometry of the 3D muscle mesh has to be mapped to the fibers points, such that a deformation of the muscle also affects the embedded fibers. This reverse mapping from the target mesh to the source fiber meshes or points is trivial: The source values can be interpolated in the target mesh using the finite element discretization.
We construct the mapping from source to target mesh to be the transpose operation to this interpolation. In the following section, we introduce the method with a graphical example.

\subsection{Construction of the Parallel Data Mapping}

\begin{figure}%
  \centering%
  \def\svgwidth{0.4\textwidth}
  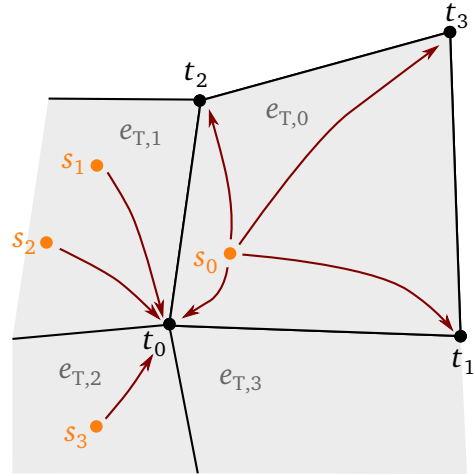%
  \caption{Data mapping scheme from source points (orange) to a target mesh (black).}%
  \label{fig:mapping_between_meshes_2}%
\end{figure}

\Cref{fig:mapping_between_meshes_2} shows a scenario, where data mapping is performed from a source mesh to a target mesh. The source mesh is given by the orange points $s_0$ to $s_3$. The target mesh is visualized by the black and gray elements $e_{\text{T},0}$ to $e_{\text{T},3}$  and nodes $t_0$ to $t_3$.

The value at $s_0$ contributes to all nodes $t_0$ to $t_3$ of the target element $e_{\text{T},0}$ as indicated by the red arrows. The relations between the contributions to $t_0,t_1,t_2$ and $t_3$ are determined by the values of the respective target element ansatz functions $\phi_0$ to $\phi_3$, evaluated at the location of $s_0$.
Similarly, the source points $s_1$ to $s_3$ contribute to the nodes of their enclosing target elements $e_{\text{T},1}$ and $e_{\text{T},2}$. 

% mapping scheme

As a consequence, all shown source points $s_0$ to $s_3$ influence the value of the target node $t_0$, as indicated by the red arrows.
The value $\hat{t}_0$ at point $t_0$ is computed using the values $\hat{s}_i$ at the points $s_i$ for $i=1,\dots,4$ as follows:
\begin{align}\label{eq:mapping_source_target}
  \hat{t}_0 = \s{i=0}{4} \alpha_i\,\hat{s}_i, \quad \text{with }\alpha_i = \dfrac{\phi_{\noexpand\mkern-4mu t_0}(\bfxi_{s_i})}{\s{i=0}{4} \phi_{\noexpand\mkern-4mu t_0}(\bfxi_{s_i})}.
\end{align}
Here, $\phi_{\noexpand\mkern-4mu t_0}$ is the finite element ansatz function for the node $t_0$. It is evaluated at the locations $\bfxi_{s_i}$ of the source points $s_i$ in the respective target elements. The factors $\alpha_i$ specify the fractions, with which the different contributions to $\hat{t}_0$ are scaled. Their construction ensures the property $\sum_{i=1}^4 \alpha_i = 1$.
Note that the number of summands in the sum over the source points can be different from 4 for other target points.

% reverse mapping scheme
\begin{figure}%
  \centering%
  \begin{subfigure}{0.4\textwidth}
    \centering
    \def\svgwidth{\textwidth}
    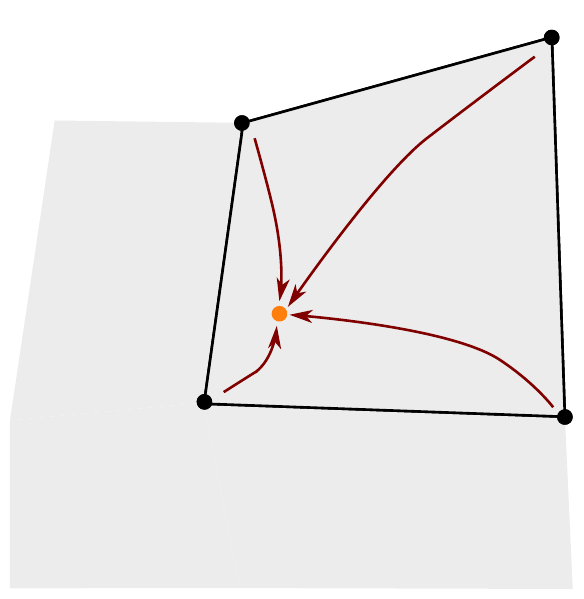%
    \caption{Mapping from target to source points using the reverse scheme of \cref{fig:mapping_between_meshes_2}.}%
    \label{fig:mapping_between_meshes_3}%
  \end{subfigure}
  \quad
  \begin{subfigure}{0.4\textwidth}
    \centering
    \def\svgwidth{\textwidth}
    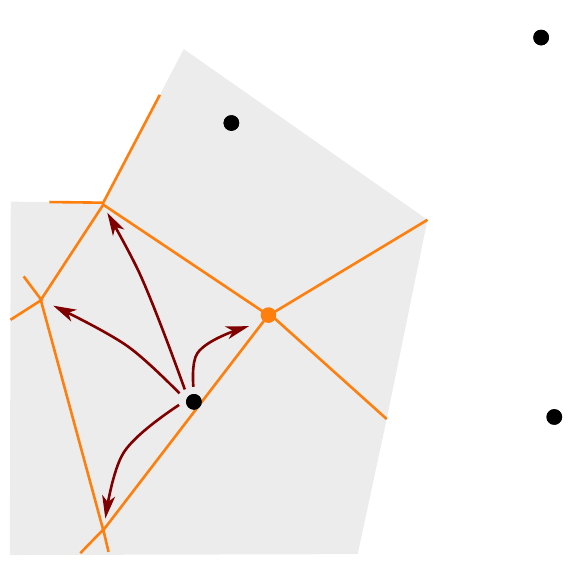%
    \caption{Inverse mapping scheme where the roles of source and target mesh are swapped.}%
    \label{fig:mapping_between_meshes_4}%
  \end{subfigure}
  \caption{Data mapping from target mesh to source mesh.}%
  \label{fig:mapping_between_meshes_34}%
\end{figure}%

The reverse mapping from target nodes to source points uses the same data dependencies between dofs in the source and in the target meshes. \Cref{fig:mapping_between_meshes_3} shows the scheme for the reverse mapping. It is the same as in \cref{fig:mapping_between_meshes_2}, except that the direction of the arrows has been flipped. As noted earlier, the mapping from the target to the source mesh is simply an interpolation in the target mesh. The value $\hat{s}_0$ is computed from $\hat{t}_0$ to $\hat{t}_3$ using the finite element interpolation formula in element $e_{\text{T},0}$:%
\begin{align*}
  \hat{s}_0 = \s{i=0}{4}\hat{t}_i\,\phi_{\noexpand\mkern-4mu t_i}(\bfxi_{s_0}).
\end{align*}
Again, the contribution factors sum up to one, $\sum_{i=0}^4 \phi_{\noexpand\mkern-4mu t_i}(\bfxi_{s_0}) = 1$.

This mapping scheme has the advantage that it requires no communication between the involved processes to determine the target dofs, to which a source dof contributes to. Considering the example in \cref{fig:mapping_between_meshes_2} and assuming that the four target elements are located on four different subdomains, it can be seen that each source point $s_i$ only has to access the target element, where it is contained, to determine the respective element coordinates $\bfxi_{s_i}$.

For the computation of the factors $\alpha_i$ in \cref{eq:mapping_source_target} and for the computation of the target dofs, communication is required. This communication step is the same exchange of ghost dof values, which is also needed for the assembly of finite element stiffness and mass matrices. In the implementation, it is available by the respective functionality of PETSc as described in \cref{sec:oragnization_of_parallel_partitioned_data}.
The reverse mapping from target to source meshes, i.e., the interpolation scheme, works without any communication as all required data are local to the processes.

Instead of reversing the source to target mapping as described, it is often also possible to change the roles of source and target mesh and construct a new mapping in this way. This is only possible, if the two meshes have the same dimensionality, as in the considered example visualizations with two 2D meshes. \Cref{fig:mapping_between_meshes_4} shows the presented mapping scheme with the roles of source and target meshes reversed.
By comparing with \cref{fig:mapping_between_meshes_3}, it can be seen that, in this case, the node $t_0$ contributes to the same nodes $s_0$ to $s_3$ in both approaches. However, the contribution factors are different. In general, the dependent nodes in both meshes are not necessarily the same in the two mapping directions. This means that, in general, the reversed or transposed mapping is not equal to the inverse mapping that is created by interchanging source and target meshes.

For mappings between different dimensionalities, only the approach of reversing the mapping in one direction is possible. For example, mapping from a 1D mesh to a 3D mesh allows no interpolation in the 1D mesh to get the 3D mesh data, as the 1D mesh occupies only a subset of the domain of the 3D mesh. This is a reason for implementing the presented mapping scheme, where the mapping direction can be reversed. Another advantage is that, once the mapping is constructed, both mapping directions are available and the expensive operation of locating the points of one mesh inside the elements of the other mesh has only be performed once.

\subsection{Special Treatment of Coarse Meshes}
% special case with coarse meshes

% reverse mapping scheme
\begin{figure}%
  \centering%
  \begin{subfigure}{0.4\textwidth}
    \centering
    \def\svgwidth{\textwidth}
    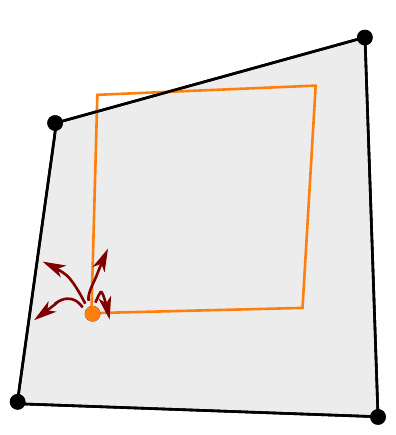%
    \caption{Mapping scheme from source to target mesh.}%
    \label{fig:mapping_between_meshes_5}%
  \end{subfigure}
  \quad
  \begin{subfigure}{0.4\textwidth}
    \centering
    \def\svgwidth{\textwidth}
    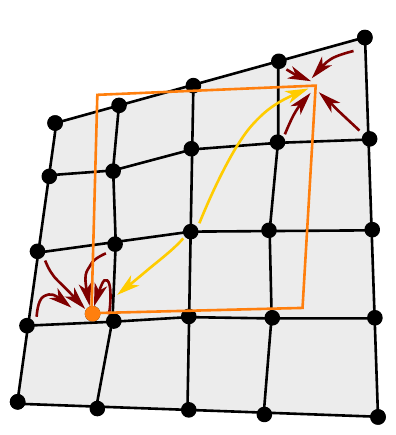%
    \caption{Reverse mapping scheme from target mesh to source points.}%
    \label{fig:mapping_between_meshes_6}%
  \end{subfigure}
  \caption{Data mapping scheme with additional data dependencies for a coarse source mesh.}%
  \label{fig:mapping_between_meshes_56}%
\end{figure}%

While the mapping error in the described scheme converges to zero, when the mesh widths approach zero, an issue occurs, if one of the meshes is significantly coarser than the other. \Cref{fig:mapping_between_meshes_5} depicts the case of a coarse source mesh in orange color that is mapped to a finer target mesh in black and gray colors. According to the presented scheme, the source points $s_0$ and $s_3$ contribute to target nodes as visualized by the red arrows. The analog contributions for $s_1$ and $s_2$ are not shown in \cref{fig:mapping_between_meshes_5}. Some target mesh nodes in this example have large distances to the source points and, as a result, do not get contributions from any source point. For example, this is the case for the target points $t_1$ and $t_2$.

To define the value at $t_2$ depending on the source data, we add new contributions from all nodes of the source element in which $t_2$ is located. These contributions are visualized by the yellow arrows in \cref{fig:mapping_between_meshes_5}.
The contributions use the ansatz functions of the source element, and the operation is equivalent to interpolating the value for $t_2$ in the source mesh:
\begin{align*}
  \hat{t}_2 = \s{i=0}{4}\hat{s}_i\,\phi_{\noexpand\mkern-4mu s_i}(\bfxi_{t_2}).
\end{align*}
The location $\bfxi_{t_2}$ of $t_2$ in element coordinates of the source element is required for this computation. Analogously, corresponding contributions are added for the other target nodes that do not yet get any contribution from the source data.

These additional contributions are also present in the reversed mapping scheme from the target to the source mesh. As can be seen in \cref{fig:mapping_between_meshes_6}, the value at $t_2$ contributes to the source nodes $s_0$ to $s_3$. At any source node, the number of contributions increases accordingly. The visualization in \cref{fig:mapping_between_meshes_6} shows five incoming arrows with contributions for $s_0$ and $s_3$. The actual number is higher, since not all target nodes with additional contributions are visualized. At the target nodes, the contribution factors get rescaled, such that they add up to 1 and their relations are preserved.

\subsection{Computation of Element Coordinates For Mapped Points}

During the setup of the mapping between the source mesh and the target mesh, we need to find, for every source point $s_i$, the target element $e_{\text{T},j}$ that contains $s_i$. Furthermore, we need to determine the local element coordinates $\bfx_{s_i} = (\xi_1,\xi_2,\xi_3)^\top$ of the point in this element. To check, if the point is inside a particular element, we compute its coordinates in the element coordinate system. If the coordinates $\bfxi$ are inside the range of $[0,1]^d$, the point is considered inside this element, and the coordinates are determined.

The source point is given by coordinates $\bfx=(x_1,x_2,x_3)^\top \in \R^3$ in the world coordinate frame. The point is related to the $d$-dimensional element coordinate frame $(\xi_1,\dots,\xi_d)$ of its containing target element by the following map:
\begin{align}\label{eq:mapping_parameter_world}
  \bfx(\bfxi) = \s{i=1}{n_\text{dofs}} \phi_i(\bfxi)\, \bfx^i.
\end{align}
Here, $\phi_i$ for $i=1,\dots,n_\text{dofs}$, are the nodal ansatz functions of the target element. The element  has $n_\text{dofs}$ dofs and is given by its node positions $\bfx^i$.

The computation of the element coordinates $\bfxi$ from the world coordinates $\bfx$ consists of inverting the mapping in \cref{eq:mapping_parameter_world}. In the following, several approaches are presented to perform this inversion for different mesh types.

For meshes of type \code{StructuredRegularFixedOfDimension<D>}, the inversion can be performed analytically. The mesh is a Cartesian grid with a fixed mesh width $h$. For quadratic elements, $h$ denotes the side length of an element, not the distance between adjacent nodes.
The computation of the element coordinates $\bfxi$ for the point $\bfx$ uses the position $\bfx^1$ of the first node and is given by:
\begin{align*}
  \bfxi = (\bfx - \bfx^1) / h.
\end{align*}
This formula is also used for 1D meshes of any type.

For non-Cartesian 2D meshes with linear ansatz functions, i.e., meshes of type \code{Struc}\code{tured}\code{DeformableOfDimension<2>}, the inversion of the mapping from element to world coordinate frame in \cref{eq:mapping_parameter_world} can also be done analytically. We consider this problem in a generic way, where both the point $\bfx$ and the nodes $\bfx^1$ to $\bfx^4$ of the 2D element are embedded in 3D space, $\bfx,\bfx^1,\dots,\bfx^4 \in \R^3$. The computation determines the element coordinates $(\xi_1,\xi_2)$ of the projection of $\bfx$ onto the plane of the triangle $(\bfx^1,\bfx^2,\bfx^3)$. 

\begin{figure}%
  \centering%
  \includegraphics[width=\textwidth]{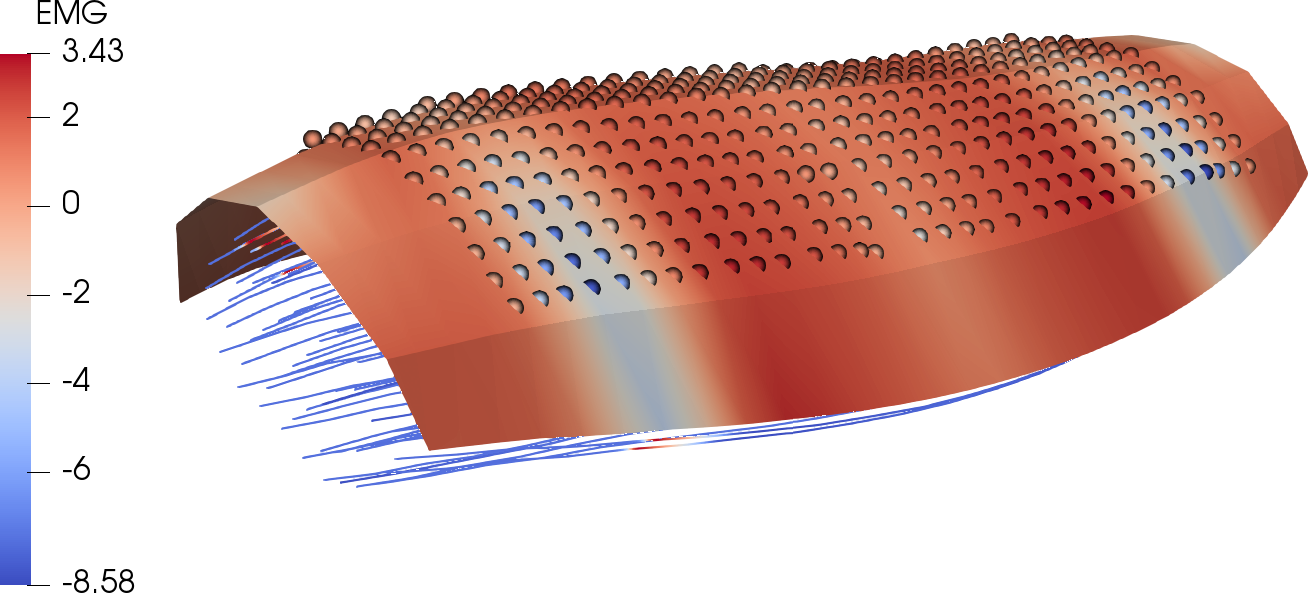}%
  \caption{Simulation of surface EMG using the fiber based electrophysiology model with body fat layer. Only the top surface of the fat layer mesh is shown. The spheres correspond to the position of surface electrodes that are used to sample the simulation result in a spatial grid.}%
  \label{fig:electrodes}%
\end{figure}

This functionality is used for specifying electrodes on the skin surface. The electrode positions are specified as a 2D grid in 3D space above the muscle. The mapping automatically projects the points of this grid onto the surface of the 3D mesh. \Cref{fig:electrodes} shows such a use case. A simulation of surface EMG is shown, the coloring corresponds to the potential $\phi_b$ in millivolts in the body domain. A grid of electrode points, visualized by spheres, is mapped onto the surface of the muscle mesh and simulates electrode patches that capture high density surface EMG.

% electrodes

After calculating the mentioned projection on the 2D plane, the computation has to invert the map in \cref{eq:mapping_parameter_world}. The ansatz functions $\phi_i$ are bilinear in the coordinates $\xi_1$ and $\xi_2$. This quadratic equation has two solutions for the unknown coordinates $\bfxi$. The formulas for those solutions have been determined using the symbolic mathematics toolbox \emph{SymPy} \cite{meurer2017sympy} and the solution, where the point is inside the element or closer to its center is chosen.

For generic hexahedral 3D meshes, \cref{eq:mapping_parameter_world} is a cubic equation in $\bfxi$, and the analytic inversion is not feasible. However, for simplex elements, i.e., tetrahedra given by points $\bfx^1$ to $\bfx^4$, it is possible. The ansatz in this case is given by:
\begin{align*}
  \bfx &= (1-\xi_1-\xi_2-\xi_3)\,\bfx^1 + \xi_1\,\bfx^2 + \xi_2\,\bfx^3 + \xi_3\,\bfx^4.
\end{align*}
This can be reformulated as:
\begin{align}\label{eq:simplex_ansatz}
   \bfx-\bfx^1 &= (\bfx^2 - \bfx^1)\,\xi_1 + (\bfx^3  - \bfx^1)\,\xi_2 + (\bfx^4 - \bfx^1)\,\xi_3.
\end{align}
This linear system of three equations can be solved for the three unknowns $\xi_1,\xi_2$ and $\xi_3$.

To invert the mapping for hexahedral elements, we proceed as follows.
A hexahedral element can be subdivided into five simplex elements. 
Four outer simplex elements share their faces with parts of the hexahedral's surface. One interior simplex element only touches the hexahedron surface by its edges. 

In each of the four outer simplex elements, we define a coordinate system $(\xi_1,\xi_2,\xi_3)$ with the origin located at a corner of the hexahedron. 
In these elements, the coordinates $\bfxi$ for the point $\bfx$ can be computed  using the ansatz in \cref{eq:simplex_ansatz}. The computed coordinate values can be transformed to the hexahedral coordinate system by applying the appropriate mirror operations $\xi \mapsto (1-\xi)$ on some coordinates. Using the average values of the hexahedral coordinates resulting from all four outer simplex elements gives a good approximation for the correct hexahedral element coordinates $\bfxi$ of the point $\bfx$.

To obtain the correct element coordinates, these approximate values are used as initial guess in a Newton scheme, which subsequently tries to find the root of $\bfr = (\bfxi - \bfx(\bfxi))$ and, thus, invert the mapping in \cref{eq:mapping_parameter_world}. 

Runtime measurements have shown that the lower number of Newton iterations resulting from the heuristic with the four simplex elements to compute an initial guess outweighs the additional runtime for the heuristic and, in total, leads to a faster computation.

The Newton scheme uses the inverse Jacobian matrix of the mapping in \cref{eq:mapping_parameter_world}. If the residual norm $\Vert\bfr\Vert_2$ cannot not be brought under the threshold of \num{1e-8} in 16 iterations, this indicates that the problem of inverting the Jacobian is badly conditioned and the Jacobian has a large numerical error. In this case, the optimization is restarted using the derivative-free Nelder-Mead algorithm.

Before applying the Newton and Nelder-Mead algorithms, our implementation performs two basic checks that can directly terminate the computation of the coordinates: First, the coordinates of the point $\bfx$ are compared with the bounding box of all nodes of the element. If the point is outside the bounding box, the element coordinates do not have to be computed, and a different element, which contains the point $\bfx$, is searched. Second, all node positions are checked for equality with the point $\bfx$. If $\bfx$ is the same as one of the node positions, the element coordinates are directly known. This case frequently occurs, if one of source and target mesh is a subset of the other.

\subsection{Conditioning of the Problem and Mapping Tolerances}

\begin{figure}%
  \centering%
  \includegraphics[width=0.3\textwidth]{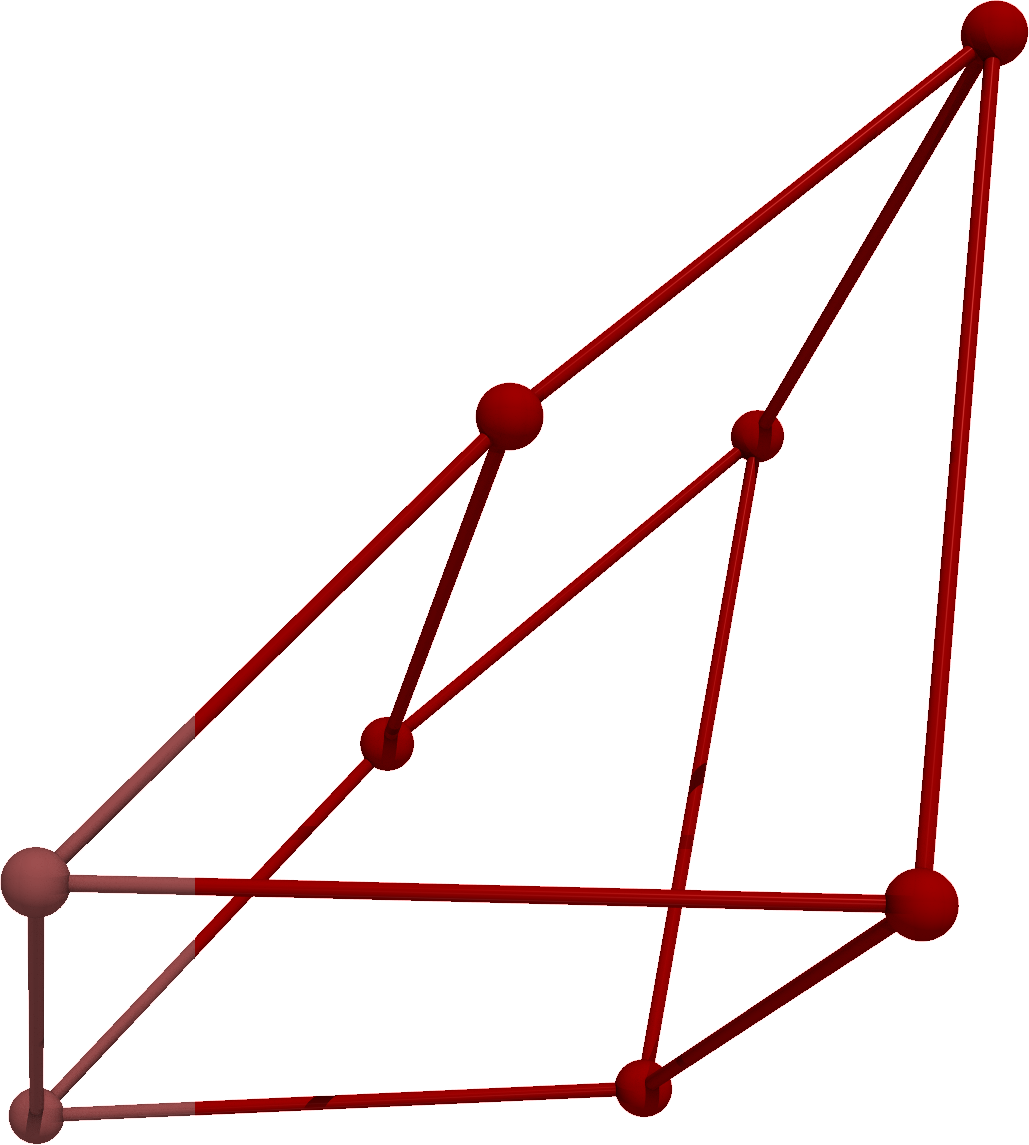}%
  \caption{Hexahedral element of the muscle mesh with an interior angle of nearly \SI{180}{\degree}. Such elements lead to poor conditioning of the Jacobian matrix inversion problem and, as a consequence, require numerous Newton iterations in the setup of the data mapping between meshes.}%
  \label{fig:bad_element}%
\end{figure}

As mentioned in the last section, the Newton scheme that solves the inverse problem of mapping a point from elemental coordinates to world coordinates uses the inverse Jacobian matrix of the mapping. The inversion of this matrix has a high numerical error, if the condition number of the Jacobian matrix is large. A large condition number can be found for 3D hexahedral elements, where the two element coordinate directions for $\xi_1$ and $\xi_2$ are almost linearly dependent. This is the case for elements with an interior angle of nearly \SI{180}{\degree}. \Cref{fig:bad_element} shows such an element, which occurs at the outer boundary of the muscle mesh.

% bad_element

If the conditioning is too bad and the computed inverse Jacobian has a large numerical error, the Newton scheme fails to find a solution in the given maximum number of iterations and the Nelder-Mead algorithm is used instead. This algorithm usually succeeds. However, it requires significantly more compute time than the Newton scheme. The case where the Nelder-Mead algorithm is needed, however, only occurs for a small number of elements and only in highly-resolved meshes.

\begin{figure}%
  \centering%
  \begin{subfigure}{0.9\textwidth}
    \centering
    \includegraphics[width=\textwidth]{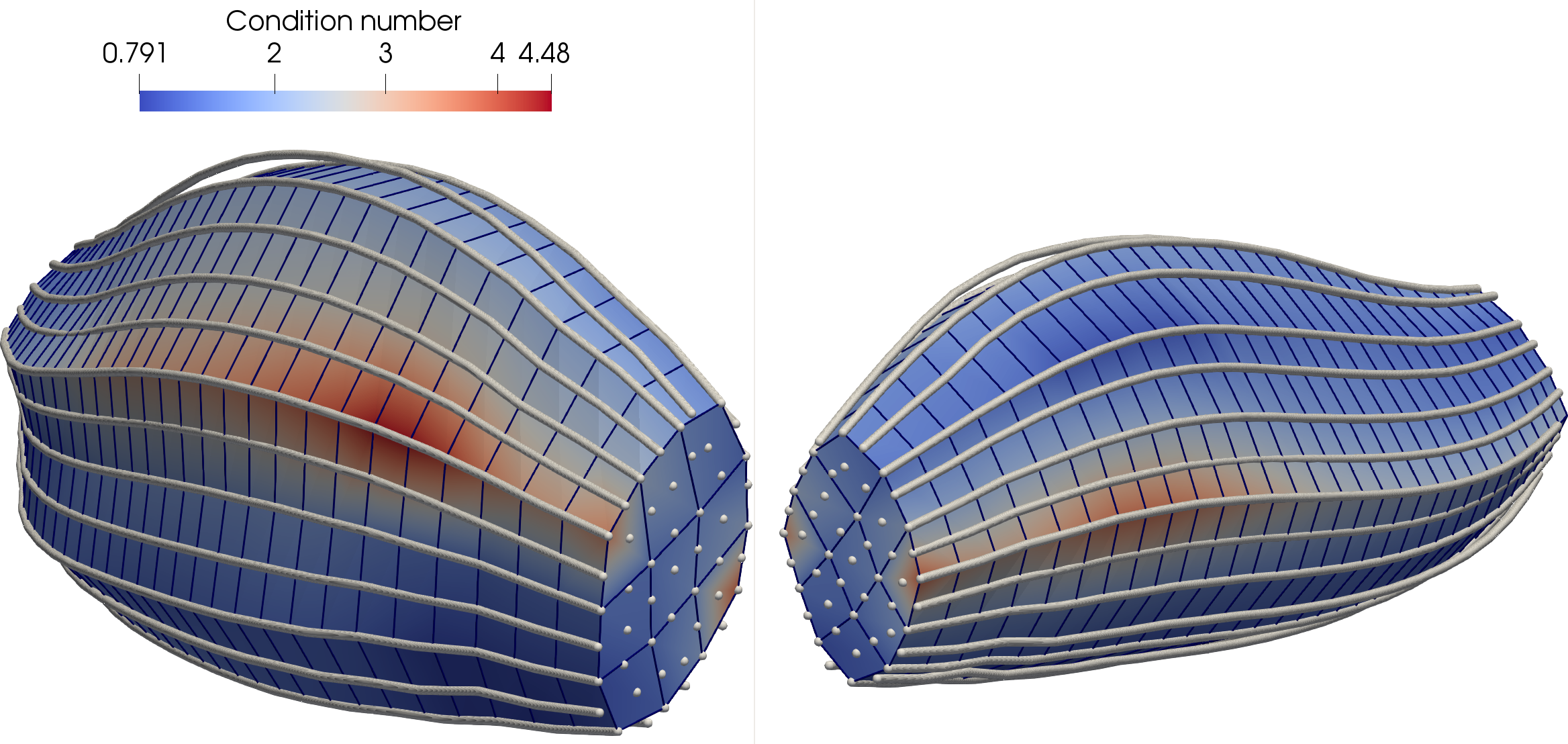}%
    \caption{Scenario with $7\times 7$ fibers.}%
    \label{fig:condition_number7x7}%
  \end{subfigure}\\[4mm]
  \begin{subfigure}{0.9\textwidth}
    \centering
    \includegraphics[width=\textwidth]{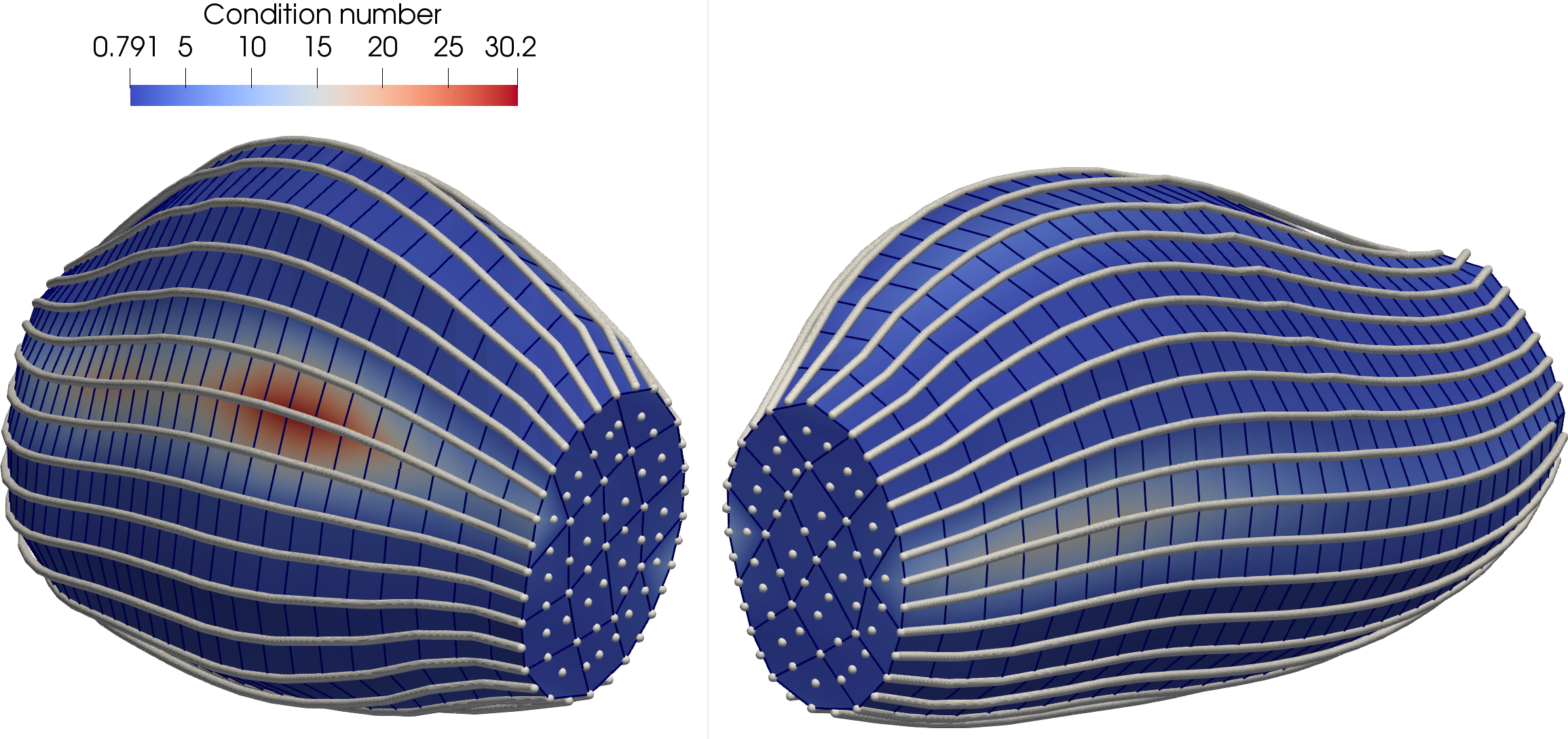}%
    \caption{Scenario with $9\times 9$ fibers}%
    \label{fig:condition_number9x9}%
  \end{subfigure}
  \caption{Fiber meshes and corresponding muscle mesh, obtained with a sampling stride of two. The left and right view show both sides of the muscle mesh. The 3D mesh is colored by the condition number of the Jacobian matrix.}%
  \label{fig:condition_number2}%
\end{figure}

\Cref{fig:condition_number2} shows the condition number of the Jacobian matrix of the mapping from element to world coordinates per element. The condition number is numerically approximated using the \emph{von Mises} power iteration algorithm to obtain the largest eigenvalue of the Jacobian and its inverse.
It can be seen in \cref{fig:condition_number7x7} that the elements with the highest condition number are located along longitudinal lines on the outer surface of the muscle mesh. Two such lines exist on both sides of the muscle. The cross-sectional mesh at the top of the muscle shows that the elements along these lines have large interior angles at the respective positions. 

% condition number 7x7

\Cref{fig:condition_number9x9} shows the same information for a different mesh with $9\times 9$ fibers instead of the subset of $7\times 7$ fibers in \cref{fig:condition_number7x7}. In \cref{fig:condition_number9x9}, the outer surface of the muscle mesh is smoother and the interior angle of the elements along the respective longitudinal lines is even closer to \SI{180}{\degree}. Thus, the resulting maximum condition number has a higher value of \num{30.2} compared to \num{4.48} in the example of \cref{fig:condition_number7x7}.

Another effect can be seen in the visualization in \cref{fig:condition_number7x7}. The muscle mesh was generated from the fiber data with sampling strides of two in the cross-sectional directions. As a consequence, the nodes of the 3D mesh are part of every second fiber. This results in some outer fibers being located outside the domain of the 3D mesh. Such a case can be seen for the upper-most fiber in \cref{fig:condition_number7x7}. 

To also involve such fibers in the computation, we enable data mapping between the 3D mesh and fibers that are outside but close to the 3D mesh. We add a tolerance parameter $\xi_\text{tolerance}$ to the implementation that specifies, how far outside the mesh fibers can be located to still be included in the mapping. On the element level, a point is considered to be part of an element, if its element coordinates $(\xi_1,\xi_2,\xi_3)$ are no further than $\xi_\text{tolerance}$ off the element domain, i.e., for %
\begin{align*}
  -\xi_\text{tolerance} \leq \xi_i \leq 1 + \xi_\text{tolerance}\quad \forall i \in \{1,2,3\}.
\end{align*}
This treats the outside fibers as if they were located inside the 3D mesh. For the fibers in the interior, the threshold leads to potentially multiple neighboring elements claiming ownership of a point. In this case, the element that contains the point without this tolerance value is chosen.
By default, the tolerance value is set to $\xi_\text{tolerance}=0.1$, but it can be adjusted to different values in the Python settings file if needed.

In summary, OpenDiHu can map data between any two overlapping meshes. The inversion of the mapping from elemental to world coordinates is an important task of this problem, which is non-trivial for 3D hexahedral elements and is solved numerically. The combination of fiber meshes with a 3D muscle mesh leads to specific effects such as degraded condition numbers or fibers outside the 3D mesh that have to be considered in the mapping.

\begin{reproduce_no_break}
  The visualizations in \cref{fig:condition_number2} were obtained using the \code{electrophysiology/fibers/fibers_emg} example. The condition number of the Jacobian is computed by the \code{StaticBidomainSolver} if the parameter \code{`enableJacobianConditionNumber`} is set to \code{True}.
\end{reproduce_no_break}

\chapter{Numerical Results and Discussion}\label{sec:results}

After various numerical models and methods for biophysical simulations of the neuromuscular system were described in the previous chapters, the remainder of this work deals with their application and discusses the newly obtained insights.
The current chapter presents numerical results and demonstrates the use of OpenDiHu for all major components of the multi-scale models. 

\Cref{sec:poisson_diffusion} begins with the simulation of toy problems such as Poisson and diffusion equations, which are used as building blocks for the more advanced simulations. Subsequently, dedicated solvers for the solid mechanics problem, the CellML models, the fiber based electrophysiology and the multidomain model are presented in \cref{sec:solver_solid_mechanics,sec:results_cellml_models,sec:results_fiber_based_electrophysiology,sec:solver_multidomain_model}. Finally, \cref{sec:coupled_electrophysiology_and_solid_mechanics} combines the fiber based electrophysiology model and the multidomain model with the solid mechanics solver to yield a comprehensive multi-physics simulation of muscle contraction.

% ==================
%
% =-------------------

\section{Solution of Poisson and Diffusion Problems}\label{sec:poisson_diffusion}

Setting up a composite multi-scale simulation, where multiple equations are coupled, requires a profound understanding of the model components. 
Thus, it can help to first simulate isolated models. We provide simple examples with our software, such as Laplace and Diffusion problems, as prototypes for elliptic and parabolic partial differential equations. The examples with analytic solutions are also used to validate the basic finite element solvers.

\begin{figure}
  \centering%
  \begin{subfigure}[t]{0.38\textwidth}%
    \centering%
    \includegraphics[width=\textwidth]{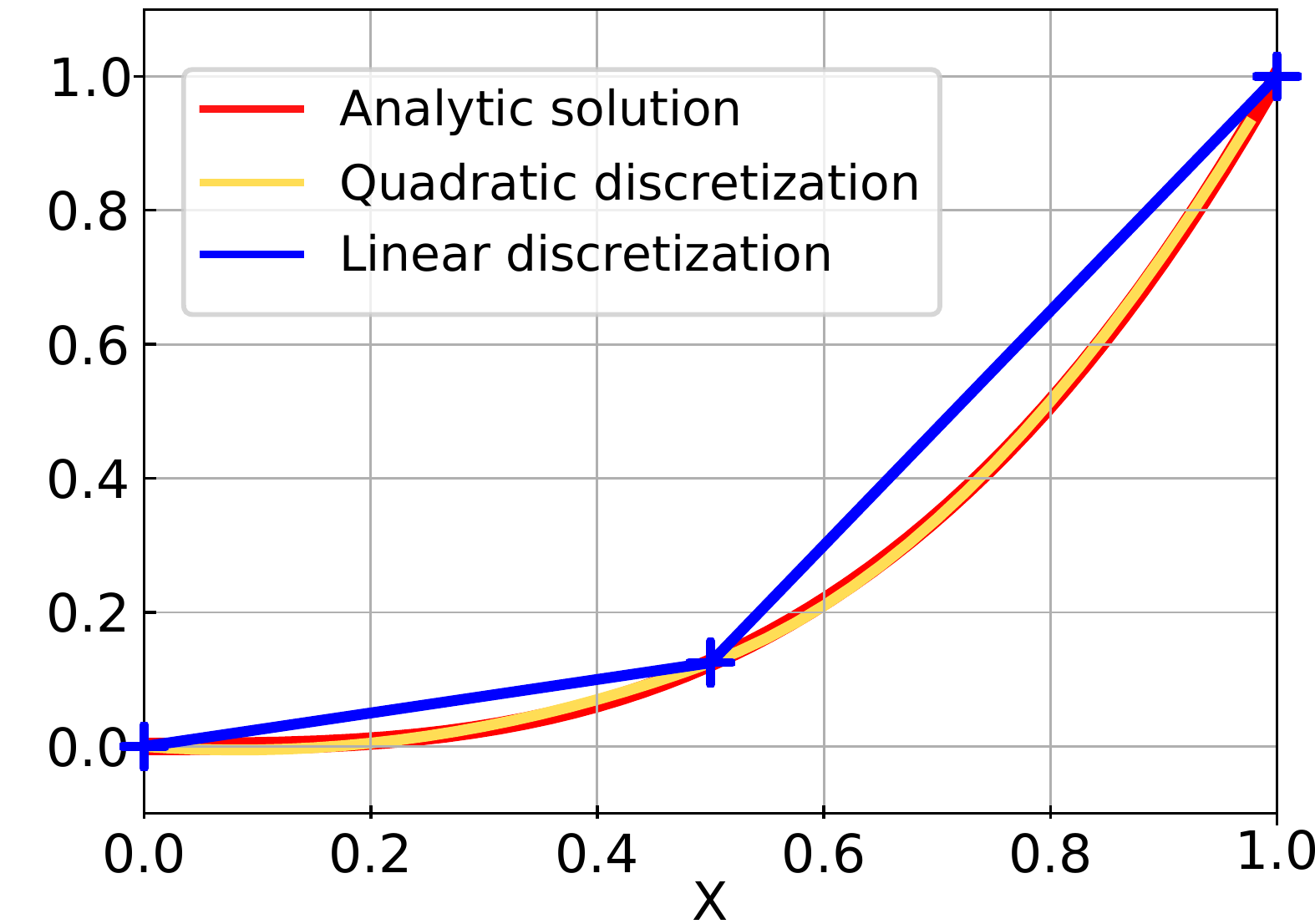}%
    \caption{Solution of a 1D Poisson problem for linear and quadratic ansatz functions.}%
    \label{fig:poisson}%
  \end{subfigure}\quad
  \begin{subfigure}[t]{0.58\textwidth}%
    \centering%
    \includegraphics[width=\textwidth]{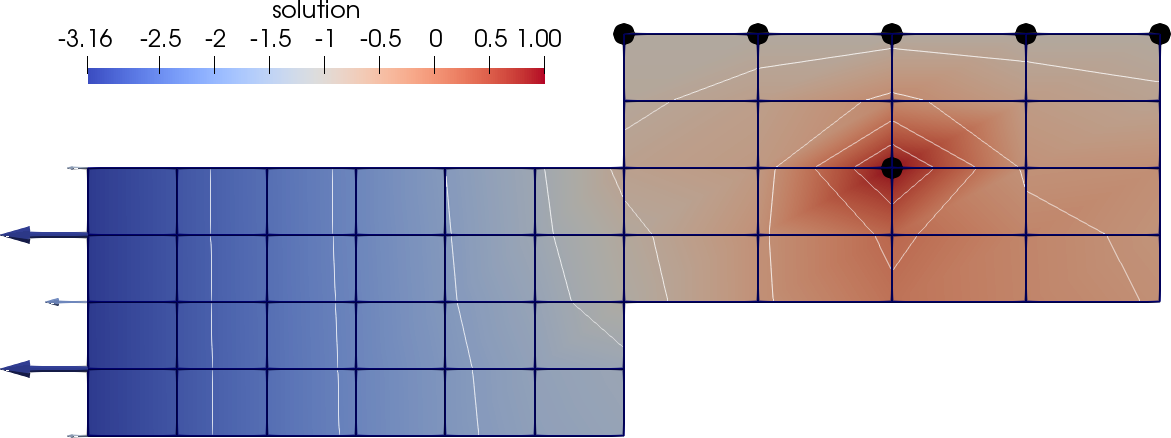}%
    \caption{Finite element mesh and solution of a 2D electric conduction problem .}%
    \label{fig:laplace_composite_1}%
  \end{subfigure}
  \caption{Exemplary problems that can be solved with OpenDiHu and are part of the multi-scale problem.}%
  \label{fig:poisson_laplace_composite_1}%
\end{figure}

In this section, we showcase three of these simple problems. First, we consider the 1D Poisson problem $u''(x) = f$ on $\Omega=[0,1]$ with Dirichlet boundary conditions $u(0)=0$ and $u(1)=1$ and right-hand side $f(x)=6\,x$. The analytic solution is $y(x)=x^3$. \Cref{fig:poisson} shows the analytic solution and the results of the finite element computation with linear and quadratic ansatz functions for two elements. Both linear and quadratic finite element solutions cannot exactly represent the cubic function, however, yield the best possible approximation. 
The first bidomain equation given in \cref{eq:bidomain1} is a 3D version of this Poisson problem and is needed in the multi-domain model to simulate EMG signals on the muscle surface.

The second example is a 2D Laplace problem $c(x)\,\Delta u(x) = 0$. The solution is given in \cref{fig:laplace_composite_1} and can be interpreted as a static electric potential field. 
The discretization uses quadratic Lagrange ansatz functions and is composed of two joined rectangular parts, each given by a structured mesh. The conductivity is set as $c=1$ in the left part and as $c=2$ in the right part. Dirichlet boundary conditions prescribe the electric potential at the five upper points in the right mesh to $u = -1$ and at the center of the right mesh as $u=1$. In addition, Neumann boundary conditions $\partial u / \partial \bfn = -1$ corresponding to an outward electric current are set on the left boundary of the left mesh with the normal vector $\bfn$ pointing to the left. \Cref{fig:laplace_composite_1} visualizes the values of the degrees of freedom of the right-hand side contribution of the Neumann boundary conditions by the arrows. 
A 3D Laplace problem is also part of the multi-scale model and describes volume conduction in the adipose tissue domain as formulated in \cref{eq:body}.

% poisson - laplace

\begin{figure}
  \centering%
  \begin{subfigure}[t]{0.4\textwidth}%
    \centering%
    \includegraphics[width=\textwidth]{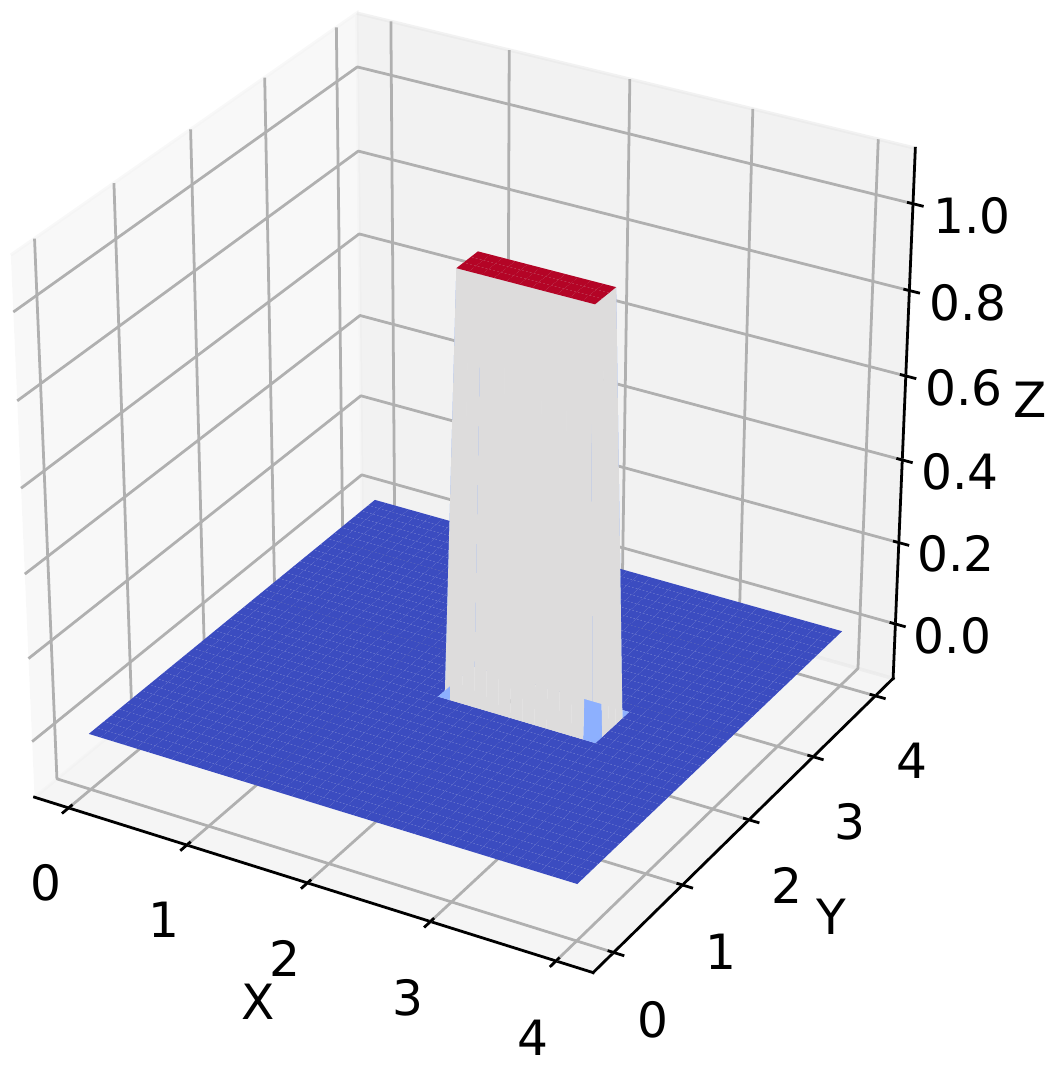}%
    \caption{Initial charge distribution, $t=0$.}%
    \label{fig:diffusion1}%
  \end{subfigure}\quad
  \begin{subfigure}[t]{0.4\textwidth}%
    \centering%
    \includegraphics[width=\textwidth]{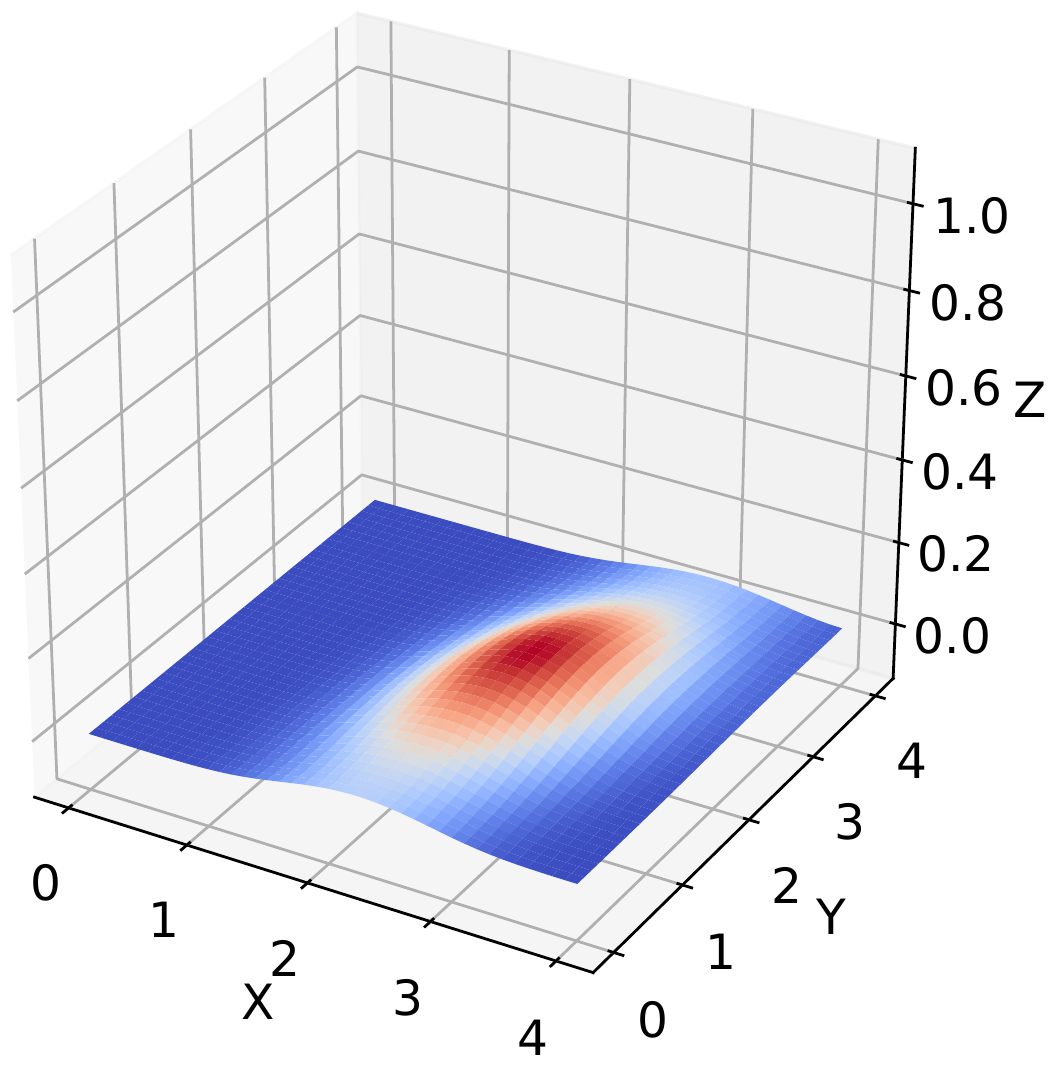}%
    \caption{Solution at $t=5$.}%
    \label{fig:diffusion2}%
  \end{subfigure}
  \caption{A 2D electric conduction problem as a demonstrator for the solution of transient problems in OpenDiHu.}%
  \label{fig:diffusion}%
\end{figure}

The third presented example solves the 2D diffusion equation $\partial u/\partial t - \div(\bfsigma\,\grad u) = 0$ with homogeneous Neumann boundary conditions. The equation can be interpreted as a transient electric conduction problem. As shown in \cref{fig:diffusion1}, the initial charge distribution is $u=1$ in a rectangle in the inner of the domain and $u=0$ everywhere else. The anisotropic diffusion or conductivity tensor $\bfsigma$ is constant in the domain and set to %
\begin{align*}
  \bfsigma = \frac15\,\mat{1 & 1\\
              1 & 6}.
\end{align*}
A regular mesh with $40\times 40$ elements and linear finite element ansatz functions is used. \Cref{fig:diffusion2} shows the solution at time $t=5$, where the initially discontinuous charge distribution has smoothed out and has expanded mainly in $y$ direction, which is the preferential direction of electric conduction in this example. A 3D version of this equation is part of the multidomain model and given by \cref{eq:multidomain1}.

% diffusion

\begin{reproduce_no_break}
  The three presented simulations can be executed and visualized as follows:
  \begin{lstlisting}[columns=fullflexible,breaklines=true,postbreak=\mbox{\textcolor{gray}{$\hookrightarrow$}\space},language=python]
    cd $\$$OPENDIHU_HOME/examples/poisson/poisson1d_2/build_release
    ./linear ../settings_1d.py && plot out/*.py
    ./quadratic ../settings_1d.py && plot out/*.py
    ./hermite ../settings_1d.py && plot out/*.py

    cd $\$$OPENDIHU_HOME/examples/laplace/laplace_composite/build_release/
    ./laplace_composite_2d ../settings_2d_2.py && paraview paraview_state.pvsm

    cd $\$$OPENDIHU_HOME/examples/diffusion/anisotropic_diffusion/build_release
    ./anisotropic_diffusion2d ../settings2d.py && plot out/*.py
  \end{lstlisting}
\end{reproduce_no_break}

%-----
\section{Simulation of Solid Mechanics Models}\label{sec:solver_solid_mechanics}

Next, we demonstrate the solid mechanics solvers, which can be used to compute muscle contraction. In this section, we focus on the passive material behavior.
As described in \cref{sec:material_linear_model,sec:material_modeling}, the mechanics equations can be computed in linearized or in nonlinear form within OpenDiHu.
In the following, \cref{sec:comparison_linear_nonlinear} applies both model approaches in a simulation of an externally stretched muscle and compares the results. Then, \cref{sec:validation_nonlinear} validates the implementation of the nonlinear hyperelasticity solver in OpenDiHu. Finally, \cref{sec:simulation_hyperelastic_tendon} showcases, using the simulation of a tendon, how more complex material models can be computed.

%-----
\subsection{Comparison of Linear and Nonlinear Mechanics Models}\label{sec:comparison_linear_nonlinear}

We demonstrate the use of linear and nonlinear mechanics models in a simulation of an externally stretched biceps muscle. The muscle belly is fixed at its lower end and an upwards pulling force acts on the upper end, effectively stretching the muscle tissue in vertical direction.

We solve two scenarios with the same geometry and boundary conditions but different material models. The first scenario uses the linearized mechanics model given in \cref{sec:material_linear_model}. We use material parameters obtained from porcine in vitro indentor tests in literature \cite{schock1982vivo} and set the bulk modulus to $K=\SI{39}{\kilo\pascal}$ and the shear modulus to $\mu=\SI{48}{\kilo\pascal}$.

The second scenario uses the incompressible transversely isotropic hyperelastic muscle material based on the Mooney-Rivlin description without active stress, which is defined in \cref{sec:material_nonlinear_model}. The material parameters are set to the values given \cite{Heidlauf2016}.

\begin{figure}
  \centering%
  \hfill
  \begin{subfigure}[t]{0.4\textwidth}%
    \centering%
    \includegraphics[height=12cm]{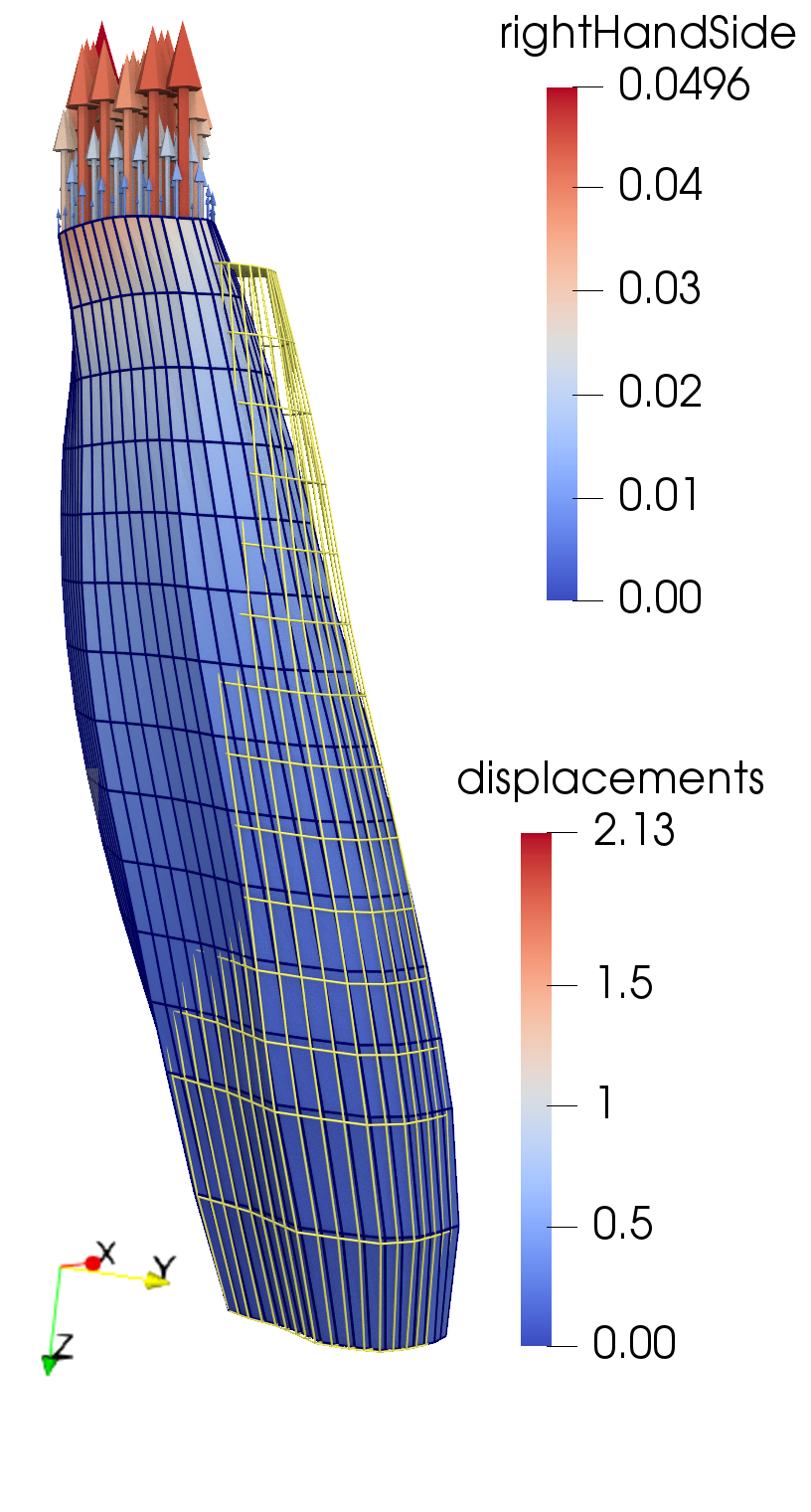}%
    \caption{Solution of the linear model. The arrows at the top visualize the (negated) right-hand side of the finite element formulation, with the absolute values indicated by the arrow lengths and color. The surface of the muscle mesh is colored according to the values of the displacements.}%
    \label{fig:lin_nonlin_muscle_mechanics_b}%
  \end{subfigure}\hfill
  \begin{subfigure}[t]{0.4\textwidth}%
    \centering%
    \includegraphics[height=12cm]{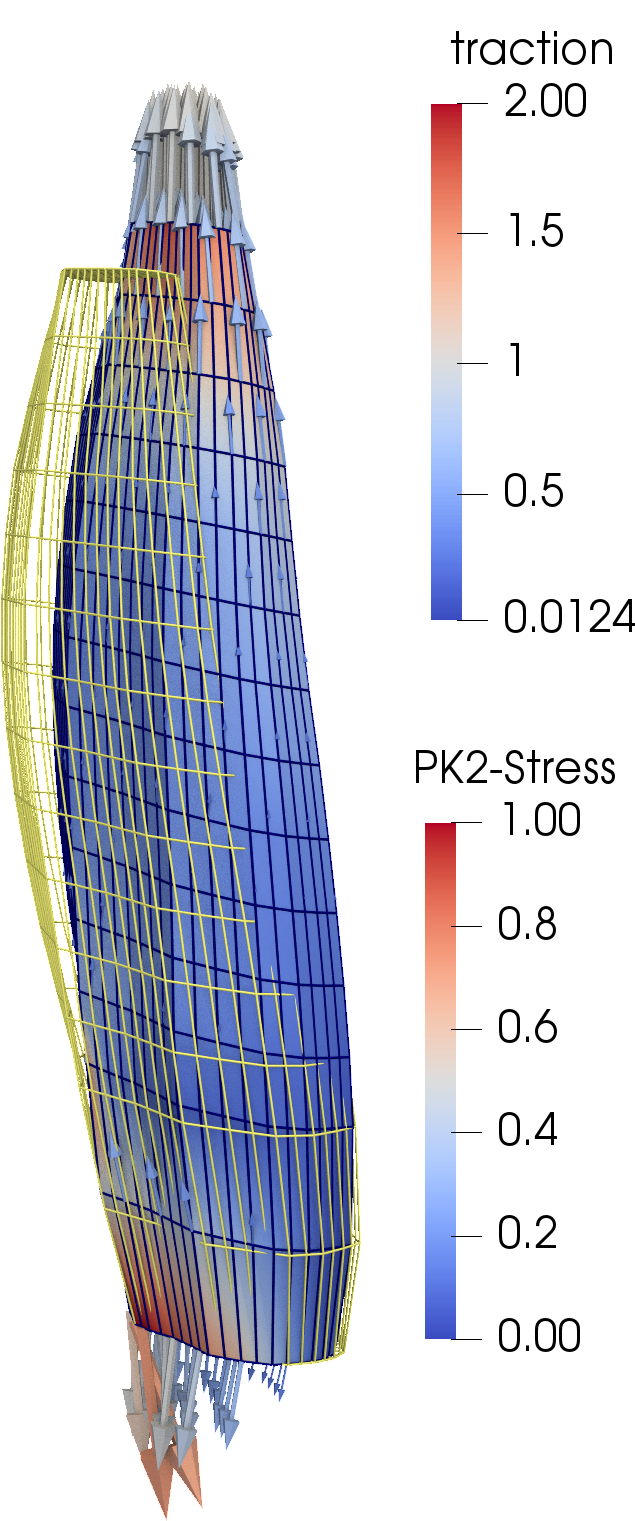}%
    \caption{Solution of the nonlinear model. The arrows specify the traction vectors in current configuration.  Their absolute values are indicated by the arrow sizes and the color. The surface of the muscle mesh is colored according to the second Piola-Kirchhoff stress.}%
    \label{fig:lin_nonlin_muscle_mechanics_a}%
  \end{subfigure}
  \hfill
  \caption{Solid mechanics solver example: Comparison of linear and nonlinear mechanics models. A biceps muscle is stretched by an applied force. The yellow mesh specifies the identical reference configuration in both scenarios.}%
  \label{fig:lin_nonlin_muscle_mechanics}%
\end{figure}

\Cref{fig:lin_nonlin_muscle_mechanics} shows the geometric setup of the model. We discretize the biceps geometry by a 3D mesh with 252  elements, quadratic finite element ansatz functions,  and a total of $13 \times 13 \times 15 = 2535$ nodes.
The linearized material model uses this mesh to construct the stiffness matrix and to solve the linear system.
For the nonlinear model, an additional coarser linear mesh is constructed, and linear-quadratic Taylor-Hood elements are used for the discretization. 

A total force of $(F_x,F_y,F_z) = (0,\SI{-0.4}{\newton},\SI{-3}{\newton})$ is applied, which points in negative $z$-direction, i.e., upwards in \cref{fig:lin_nonlin_muscle_mechanics}, and slightly in negative $y$-direction, i.e, to the left in \cref{fig:lin_nonlin_muscle_mechanics}.
Instead of a single force vector acting on a point, the equivalent constant surface load is applied on the whole top face of the muscle geometry.

We consider a static problem where no timestepping is required. In the linear model, the resulting displacements are obtained by a GMRES solver, which solves the linear system of equations \cref{eq:linearized_helper4} corresponding  to the finite element formulation.
The nonlinear model uses increasing load steps as described in \cref{sec:convergence_improvements_for_the_nonlinear_solver}, which are adaptively refined in case the solver diverges at one load step. The scheme solves a system of nonlinear equations for every load step, and the contained linear system is solved by a direct solver.

% linear and nonlinear mechanics solvers

The results of the linear and nonlinear models are shown in \cref{fig:lin_nonlin_muscle_mechanics_b,fig:lin_nonlin_muscle_mechanics_a}. 
In both images, the identical reference configuration is given by the yellow wireframe and the deformed muscle is given by the solid body with colored mesh. 
The deformed body in the linear model in \cref{fig:lin_nonlin_muscle_mechanics_b} is colored according to the resulting vector of unknowns, which contains the displacements.  Arrows on the upper end of the geometry indicate the negative right-hand side of the linear system as formulated in \cref{eq:linearized_mechanics_rhs}. The arrows correspond to the applied Neumann boundary conditions in the weak form of the finite element formulation and point in the direction of the applied surface load.

In the visualization of the nonlinear model in \cref{fig:lin_nonlin_muscle_mechanics_a}, the deformed muscle body is colored according to the second Piola-Kirchhoff (PK2) stress. It can be seen that the stress is highest at the bottom bearing and at the top end, where the muscle cross-section is smaller. The arrows visualize the traction forces $\bft$ on virtual horizontal cuts. As a result, the arrows that can be seen on top of the muscle geometry correspond to the applied external force, and the arrows at the bottom indicate the forces on the bearing.

A comparison of the two obtained results from the linear and nonlinear models shows a qualitatively different outcome. With the linear model, the muscle bends to the left, whereas, with the nonlinear model, it bends to the right. This effect is a result of the different material behavior. The linear model is isotropic and the deformation follows the direction of the applied force, which points to the upper left. The nonlinear model has an anisotropy and is stiffer in fiber direction. As a consequence, the muscle deforms less in longitudinal direction and therefore moves to the right.
Thus, the material models influence the bending direction in this scenario.

In a second example, we compare the muscle stretches that results from different external forces acting in $z$-direction.  We use the same scenario as before and increase the applied force from 0 to \SI{15}{\newton}. We measure the displacement of one node in the top face of the muscle, for both the linear model and the nonlinear model. While the stress-strain relations  in 1D extension tests  can be derived analytically for linear and nonlinear models, our examples considers a real 3D setting where this relation is influenced by the geometry, e.g., by non-parallel fiber directions, as the force is not applied exactly in fiber direction.

\begin{figure}
  \centering%
  \includegraphics[width=0.7\textwidth]{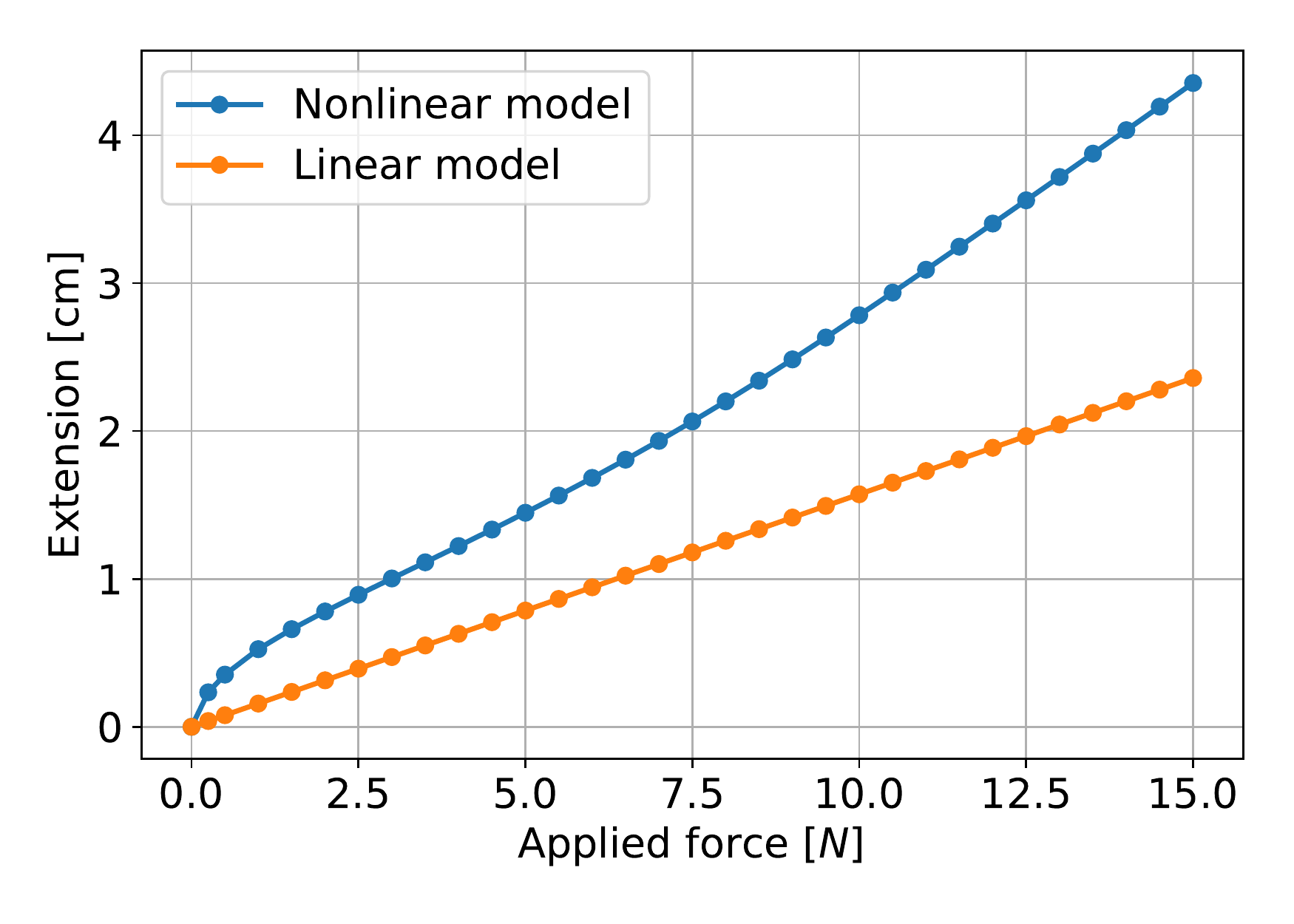}%
  \caption{Solid mechanics example: Quantitative comparison of the relation between applied force and extension of the muscle for a linear and a nonlinear solid mechanics model.}%
  \label{fig:linear_nonlinear_displacements}%
\end{figure}

\Cref{fig:linear_nonlinear_displacements} shows the resulting muscle extensions for different applied external forces for the linear and nonlinear models.
It can be seen that the stretch of the muscle increases nonlinearly for the transversely isotropic hyperelastic model, in contrast to the linear progression of the linear model.
The slopes of the two curves are qualitatively different, which is a result of the chosen material parameters from different experimental origins. It would be possible to scale the linear model to better match the nonlinear model behavior by simply reducing the value of the bulk modulus accordingly.

% study of displacements, linear  nonlinnear

The two presented studies show that a linear isotropic material model can give significantly different results than a more accurate nonlinear transversely isotropic model. Therefore, simulations of muscle contraction that target high accuracy should use the according nonlinear models. Nevertheless, both approaches are implemented and can be used with OpenDiHu.

\begin{reproduce_no_break}
  The two simulations for \cref{fig:lin_nonlin_muscle_mechanics} can be run as follows:
  \begin{lstlisting}[columns=fullflexible,breaklines=true,postbreak=\mbox{\textcolor{gray}{$\hookrightarrow$}\space}]
    cd $\$$OPENDIHU_HOME/examples/solid_mechanics/linear_elasticity/muscle/build_release
    ./linear_elasticity ../settings_linear_elasticity.py
    cd $\$$OPENDIHU_HOME/examples/solid_mechanics/mooney_rivlin_transiso/build_release
    ./3d_hyperelasticity ../settings_3d_muscle.py --njacobi=1
  \end{lstlisting}
  The study in \cref{fig:linear_nonlinear_displacements} can be run and plotted using the scripts in the repository at \href{https://github.com/dihu-stuttgart/performance}{github.com/dihu-stuttgart/performance}
  in the directory \code{opendihu/}\\\code{23_linear_nonlinear_mechanics}.
\end{reproduce_no_break}

%-----
\subsection{Validation of the Nonlinear Solid Mechanics Solver}\label{sec:validation_nonlinear}

Next, we perform tests to validate our implementation of the nonlinear hyperelasticity solvers.
We simulate the same scenario with our software and with the nonlinear finite element analysis tool \emph{FEBio} \cite{Maas2012}.
FEBio is developed at the University of Utah and the Columbia University in the USA. FEBio contains solid mechanics solvers that can be run from the command line or a graphical user interface model. An extensive model library contains material models also from the domain of biomechanics. The mechanics solver uses the PARDISO linear solver \cite{pardiso2020}, which exploits shared memory parallelism.

An adapter in OpenDiHu exists, which can output the required configuration file for FEBio, run the solver, and parse the computed solution from the text files that are output by FEBio. Thus, we can conduct our validation studies fully in OpenDiHu by using similar Python settings files and the same meshes for the computation in OpenDiHu and the reference solution computed by FEBio.

Apart from the present study, the FEBio adapter in OpenDiHu can also be used to solve quasi-static coupled problems with the electrophysiology part solved in OpenDiHu and the mechanics part solved in FEBio. However, test have shown that the interfacing method of generating configuration files and parsing result files in every timestep leads to higher runtimes than directly using the mechanics solver of OpenDiHu.

\begin{figure}
  \centering%
  \hfill
  \begin{subfigure}[t]{0.45\textwidth}%
    \centering%
    \includegraphics[height=8cm]{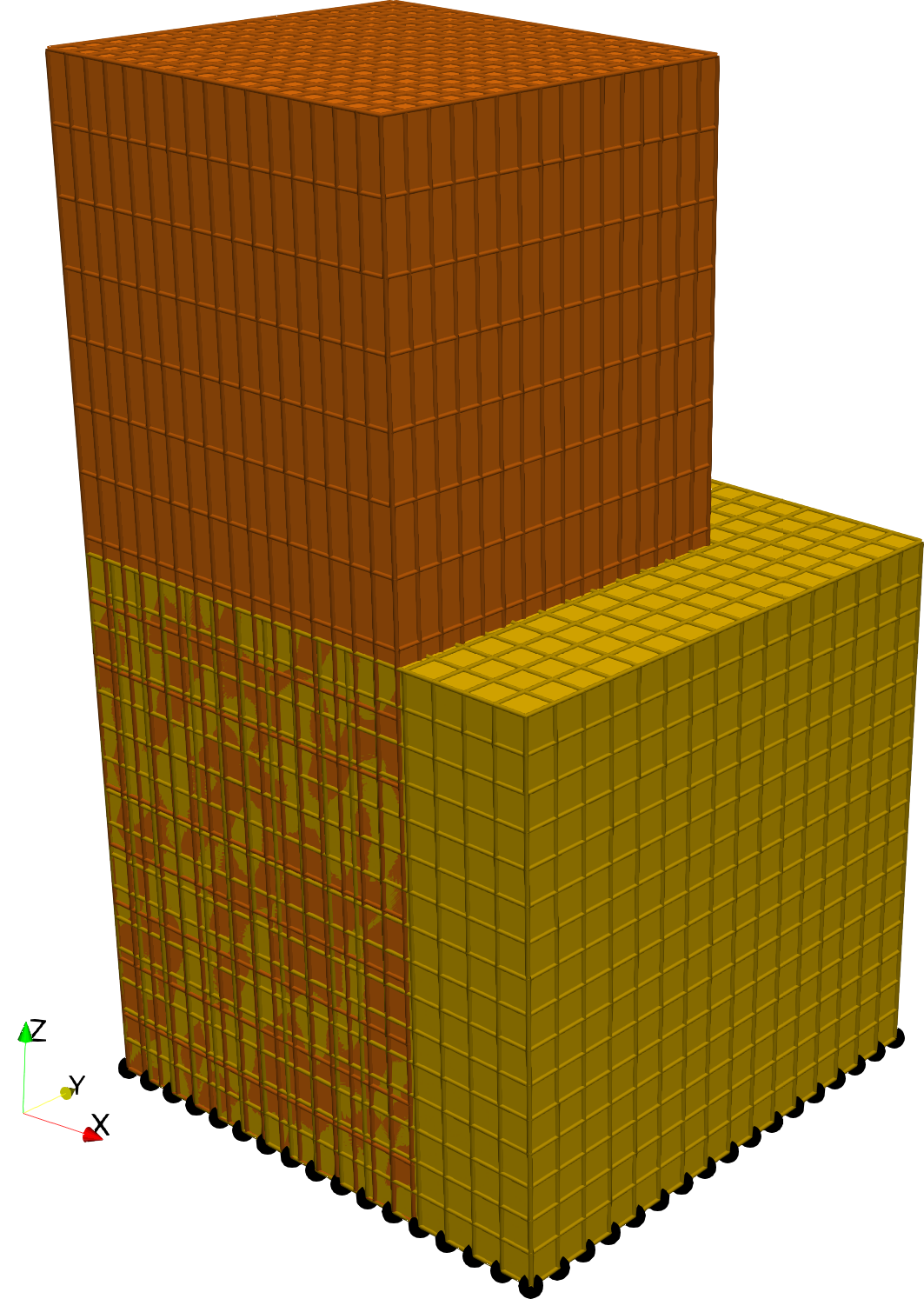}%
    \caption{Tensile test scenario used in the first validation experiment.}%
    \label{fig:tensile_test_img}%
  \end{subfigure}\hfill
  \begin{subfigure}[t]{0.45\textwidth}%
    \centering%
    \includegraphics[height=8cm]{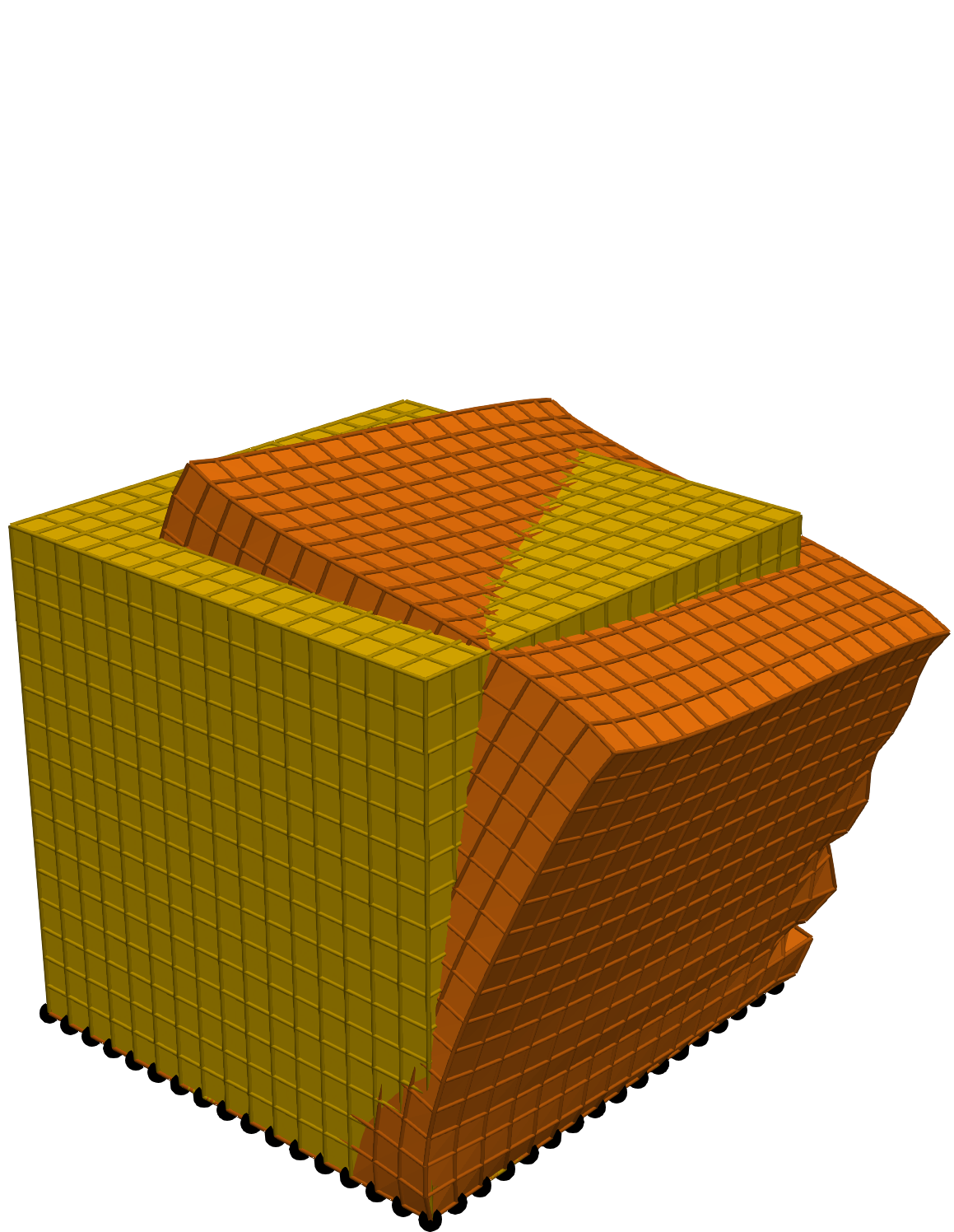}%
    \caption{Shear test scenario used in the second validation experiment.}%
    \label{fig:shear_test_img}%
  \end{subfigure}
  \hfill
  \caption{Scenarios used for validation of the solid mechanics solver. The reference and the current configuration are given by the yellow and orange meshes, respectively.}%
  \label{fig:tensile_shear_test_img}%
\end{figure}

In our validation studies, we consider a unit cube  that is discretized by $8\times 8 \times 8$ quadratic elements and 4913 degrees of freedom. \Cref{fig:tensile_shear_test_img} shows the discretized cube in yellow color. Its orientation is given by the coordinate frame in the lower left of \cref{fig:tensile_test_img}. 
The following Dirichlet boundary conditions are prescribed: All points of the lower face are fixed at $z=0$. The points of the two edges ($y=0 \wedge z=0$) and ($x=0 \wedge z=0$) are additionally fixed in $y$ and $x$ directions, respectively. The corner at $x=y=z=0$ is fixed completely. Thus, the cube can freely deform in its bottom plane, but not move nor rotate as a whole. 

% visualization of the scenarios

The first study is a tensile test, where a uniform surface load pointing in positive $z$ direction is applied on the top face of the cube. We increase the force from 1 to \SI{50}{\newton}. For the largest force, the cube deforms as shown by the orange geometry in \cref{fig:tensile_shear_test_img}. Note that the volume is preserved due to the incompressibility constraint in the material model.

We use an incompressible and isotropic Mooney-Rivlin material with parameters $c_1=c_2=1$. The material can be simulated in three different forms in OpenDiHu. In the following, we list all model formulations in OpenDiHu and the reference formulation in FEBio,
expressed by the strain energy functions $\Psi$, $\Psi_\text{iso}$ and $\Psi_\text{vol}$ introduced in the modeling chapter in \cref{sec:material_modeling}:

\begin{enumerate}[label=(\roman*)]
\item the \say{fully incompressible}, mixed $u$-$p$ formulation, which ensures incompressibility using the Lagrange multipliers,
\begin{align}
    \Psi_\text{iso}(\bar{I}_1,\bar{I}_2) &= c_1\,(\bar{I}_1 - 3) + c_2\,(\bar{I}_2 - 3), \quad J=1, \label{eq:validation_incompressible}
\end{align}
\item the \say{nearly incompressible} formulation in terms of the invariants $I_1$ to $I_3$,
\begin{subequations}\label{eq:validation_1b}
  \begin{align}      
      \Psi(I_1,I_2,I_3) &= c_1\,(I_1 - 3) + c_2\,(I_2 - 3) + \kappa\,(\sqrt{I_3} - 1)^2 - d\,\log(\sqrt{I_3}),\label{eq:validation_nearly_incompressible_1} \\
         d &= 2\,(c_1 + 2\,c_2) \label{eq:validation_nearly_incompressible},
\end{align}
\end{subequations}
\item the nearly incompressible formulation given in decoupled form, in terms of the reduced invariants $\bar{I}_1$ and $\bar{I}_2$,
\begin{subequations}\label{eq:validation_1c}
  \begin{align}
      \Psi_\text{iso}(\bar{I}_1,\bar{I}_2) &= c_1\,(\bar{I}_1 - 3) + c_2\,(\bar{I}_2 - 3),\label{eq:validation_nearly_incompressible_decoupled_1} \\ 
      \Psi_\text{vol}(J) &= \kappa\,G(J) \quad\text{with }  G(J) = \dfrac14\big(J^2 - 1 - 2\,\log(J)\big), \label{eq:validation_nearly_incompressible_decoupled}
  \end{align}
\end{subequations}
\item and the one used in FEBio, which also describes a nearly incompressible material in decoupled form, but with a different penalty function $G(J)$,
\begin{subequations}\label{eq:validation_1d}
  \begin{align}
      \Psi_\text{iso}(\bar{I}_1,\bar{I}_2) &= c_1\,(\bar{I}_1 - 3) + c_2\,(\bar{I}_2 - 3),\label{eq:validation_nearly_incompressible_decoupled_febio_1} \\ 
      \Psi_\text{vol}(J) &= \kappa\,G(J) \quad\text{with } G(J) = \dfrac12\big(\log(J)\big)^2 \label{eq:validation_nearly_incompressible_decoupled_febio}.
  \end{align}
\end{subequations}
\end{enumerate}
For the three nearly incompressible descriptions in \cref{eq:validation_1b,eq:validation_1c,eq:validation_1d}, we set the incompressibility parameter to $\kappa=\num{1e3}$.

\begin{figure}
  \centering%
  \includegraphics[width=\textwidth]{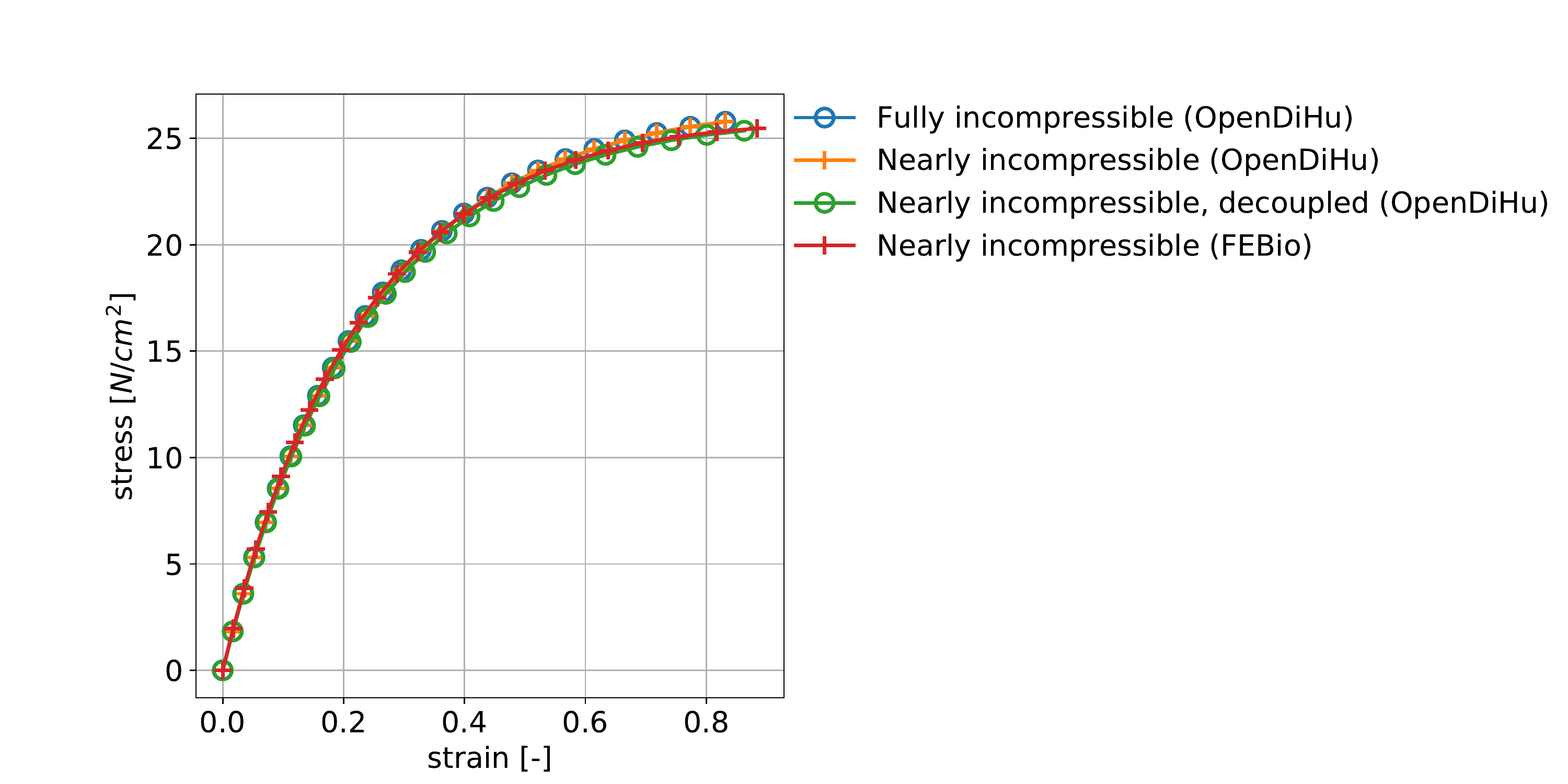}%
  \caption{Solid mechanics solver validation: Results of the tensile test validation experiment. The stress-strain curve for three different formulations in OpenDiHu and the computation in FEBio match.}%
  \label{fig:validation_tensile_test}%
\end{figure}

We compare the resulting normal stress value $S_{33}$ in $z$-direction of the second Piola-Kirchhoff stress tensor $\bfS$ for all formulations listed in \crefrange{eq:validation_incompressible}{eq:validation_1d}. For the tensile test, this stress value is constant throughout the domain. \Cref{fig:validation_tensile_test} shows the computed stresses over the computed strain values.
It can be seen that the three formulations in OpenDiHu yield approximately the same results as the reference solution given by FEBio over the whole range of applied forces.

% results tensile test

As the previous tensile test only validates stress and strain in one direction, we additionally conduct a numerical shear experiment. A shear force $\bfF=(0.1\alpha, 0.05\alpha, 0)^\top$ is applied on the top face of the cube and $\alpha$ is again varied between 1 and $\SI{50}{\newton}$. \Cref{fig:shear_test_img} shows the deformed configuration for the highest force by the orange colored body.

% results shear test
\begin{figure}
  \centering%
  \includegraphics[width=0.7\textwidth]{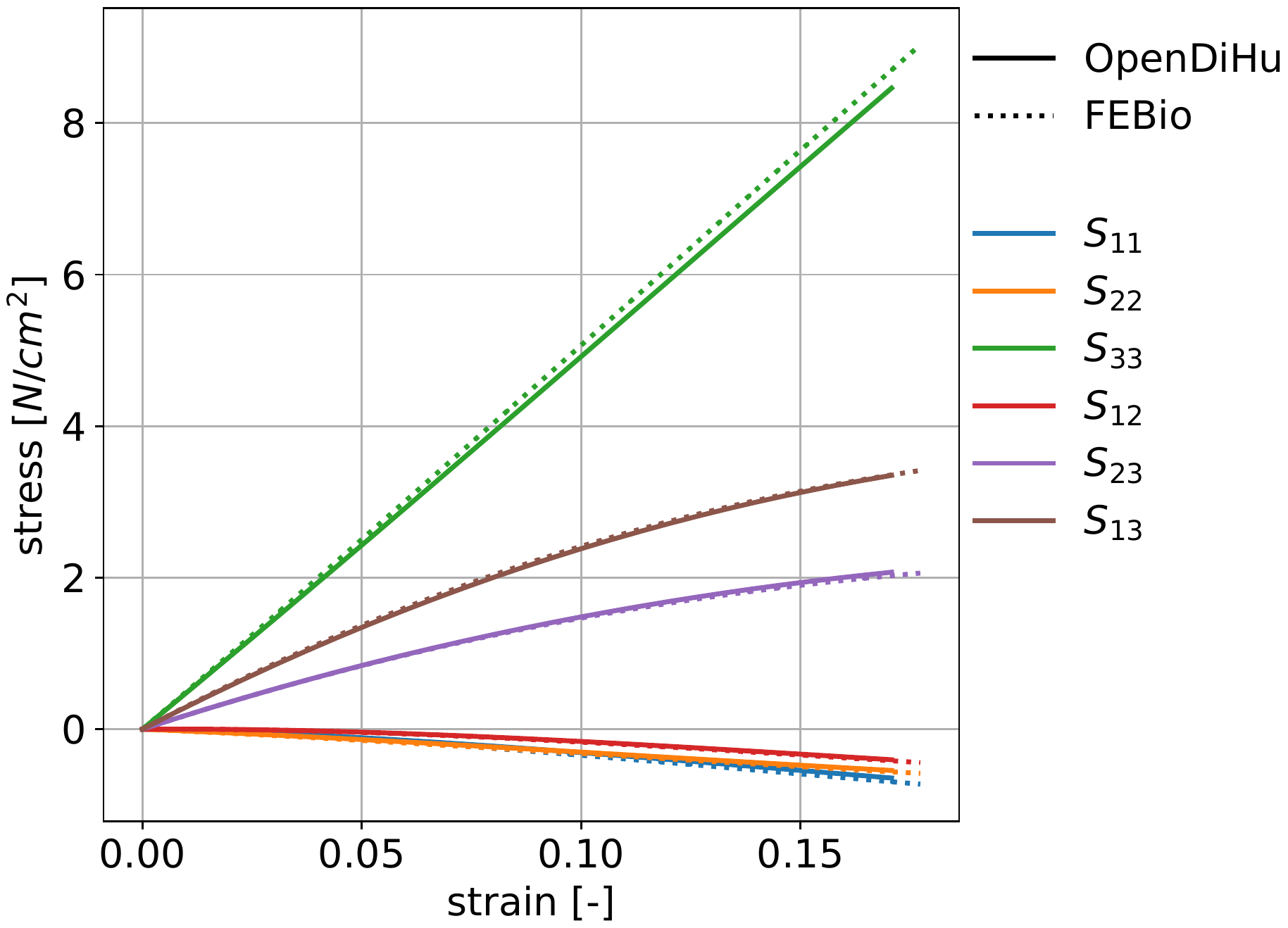}%
  \caption{Solid mechanics solver validation:Results of the shear test validation experiment. The values of the second Piola-Kirchhoff tensor computed by OpenDiHu (solid lines) and FEBio (dotted lines) closely match.}%
  \label{fig:validation_shear_test}%
\end{figure}

In this second study, we consider one point in the interior of the domain, which is 3 elements below the top face of the mesh. We compare all six distinct entries of the symmetric second Piola-Kirchhoff tensor $\bfS$ between the fully incompressible model in OpenDiHu and the nearly incompressible model in FEBio. 

\Cref{fig:validation_shear_test} shows the computed values in a stress-strain diagram. The solutions of OpenDiHu and FEBio are given by solid and dotted lines, respectively. It can be seen that the curves coincide, which validates the implementation in OpenDiHu.

\begin{reproduce_no_break}
  The tensile test validation experiment can be reproduced by the following commands:
  \begin{lstlisting}[columns=fullflexible,breaklines=true,postbreak=\mbox{\textcolor{gray}{$\hookrightarrow$}\space}]
    cd $\$$OPENDIHU_HOME/examples/solid_mechanics/tensile_test/build_release
    ../run_force.sh
    cd $\$$OPENDIHU_HOME/examples/solid_mechanics/tensile_test
    ./plot_validation.py
  \end{lstlisting}
  The shear test can be executed analogously by replacing \code{tensile_test} by \code{shear_test} in the given paths.
\end{reproduce_no_break}

%-----
\subsection{Simulation of a Hyperelastic Tendon Material}\label{sec:simulation_hyperelastic_tendon}

Next, we demonstrate the use of a more complex constitutive material model, which represents tendon tissue. The material is formulated in \cite{Carniel2017}. The model describes microstructural interactions between collagen fibers and their matrix. It consists of a transversely isotropic model, which describes the high stiffness in fiber direction, and a coupled model for the compressive response. The model is formulated in terms of a logarithmic strain measure.

%from the paper: The high stiffness of tendons under tensile tests is handled by a transversely isotropic model while the coupled compressive response is modeled by means of a Fung-type potential in terms of Seth-Hill’s generalized strain tensors. In present study the logarithm strain measure is used instead of the usually employed Green-Lagrange strain

\begin{figure}
  \centering%
  \begin{subfigure}[t]{\textwidth}%
    \centering%
    \includegraphics[width=0.9\textwidth]{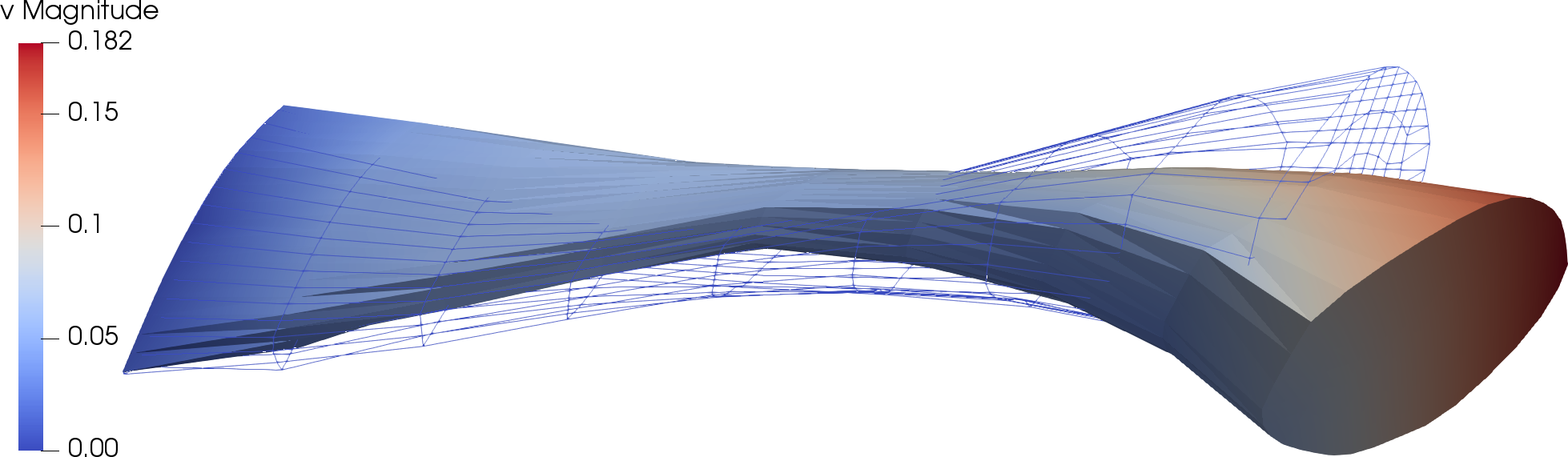}%
    \caption{Dynamic simulation of the lower tendon of a biceps brachii. The attachment to the ulna bone is at the left end. The free right end bends due to the applied surface traction.}%
    \label{fig:dynamic_mooney_rivlin_6}%
  \end{subfigure}\\[4mm]
  \begin{subfigure}[t]{\textwidth}%
    \centering%
    \includegraphics[width=0.9\textwidth]{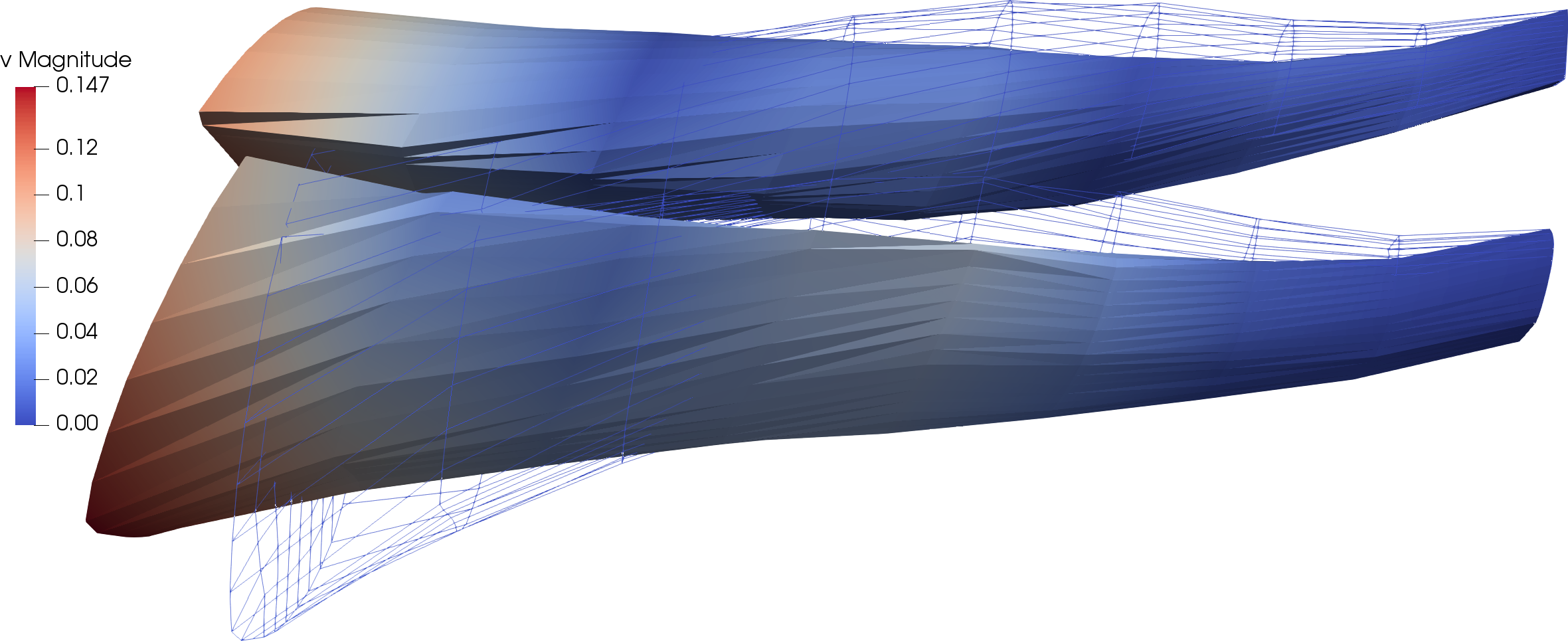}%
    \caption{Simulation of the two upper tendons of the two biceps heads.}%
    \label{fig:dynamic_mooney_rivlin_7}%
  \end{subfigure}
  \caption{Simulation of tendons as a showcase of dynamic simulations with complex material models. The color coding indicates the velocity.}%
  \label{fig:tendon_material_simulation}%
\end{figure}

\Cref{fig:tendon_material_simulation} shows the geometries of the tendons of the biceps brachii and the results of the simulations. The lower tendon in \cref{fig:dynamic_mooney_rivlin_6} is fixed at its left end and a constant surface traction of \SI{1}{\newton} in total pulls to the right. The image shows the initial configuration by the wireframe mesh and the current configuration after $t=\SI{10}{\ms}$, colored according to the resulting velocity.
Similarly, the upper tendons in \cref{fig:dynamic_mooney_rivlin_7} are fixed at the right ends and stretched to the left resulting from the applied force at the left end.

In summary, this section demonstrated the capabilities of the solid mechanics solvers in OpenDiHu. \Cref{sec:comparison_linear_nonlinear,sec:simulation_hyperelastic_tendon} simulated extension of the biceps muscle and tendons due to external forces. The comparison of results from a linear and a nonlinear model showed that a linear isotropic material cannot always accurately predict the behavior of muscle tissue and, thus, a nonlinear model is required. The validation experiments in \cref{sec:validation_nonlinear} demonstrated that OpenDiHu correctly computes deformation and stresses of incompressible materials.

The solid mechanics solvers can also be coupled to solvers of electrophysiology to simulate muscle contraction resulting from the spatially heterogeneous activation and considering the neuronal stimulation dynamics. Moreover, coupled simulations of the muscle and tendons are possible. Such simulations are described in \cref{sec:fiber_based_contraction} and \cref{sec:surface_coupling_contraction}, respectively.

% ------------
%
% f===========

% ==================
%
% =-------------------

% --------------------------------
% application
\section{Simulation of CellML Models}\label{sec:results_cellml_models}
The subcellular models used in the multi-scale model are given in CellML description and can be solved in OpenDiHu using the \code{CellmlAdapter} class.
In the following, we show simulation results of the most commonly used CellML models in this work.

\subsection{Simulation of Subcellular Models}
\begin{figure}
  \centering%
  \includegraphics[width=0.8\textwidth]{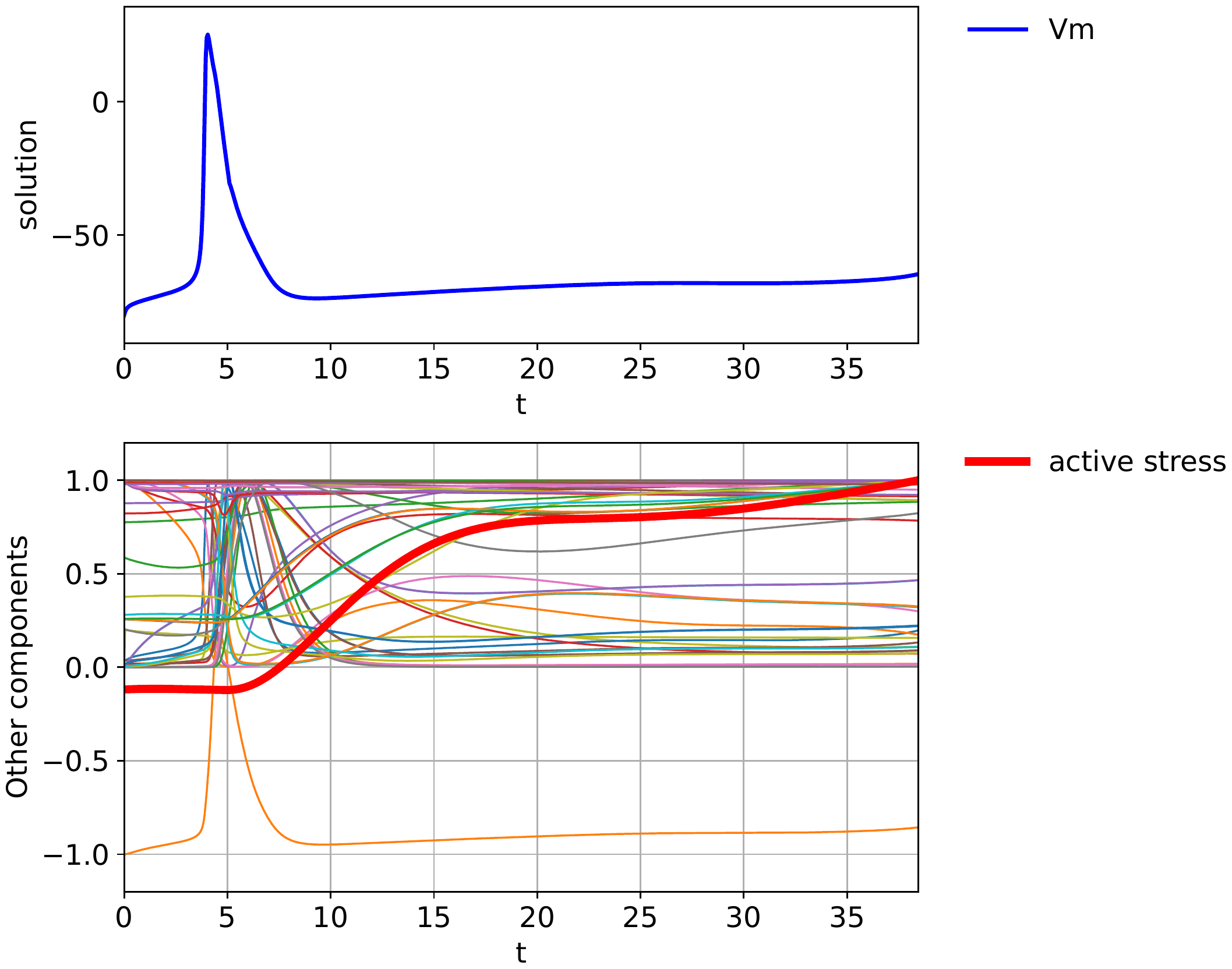}%
  \caption{Simulation of the Shorten subcellular model over time $t$ in milliseconds. The cell is stimulated at $t=\SI{5}{\ms}$. The upper plot shows the membrane voltage $V_m$, the lower plot show all other variables, some of which are listed in the legend on the right.}%
  \label{fig:shorten_over_time}%
\end{figure}

First, we consider a single instance of the subcellular model of Shorten et al. \cite{Shorten2007}. We solve the model with Heun's method with a timestep width of $\dt_\text{0D} = \num{1e-5}$. A stimulation current of $I_\text{stim}=\SI{40}{\micro\ampere\per\square\centi\meter}$ is applied during the time range $[\SI{5}{\ms},\SI{5.1}{\ms}]$. \Cref{fig:shorten_over_time} shows the resulting values of the membrane voltage $V_m$ over time in the upper plot and the temporal evolution of all other variables in the lower plot.
%\footnote{Note that the purpose of the lower plot in \cref{fig:shorten_over_time} is not to identify the individual variables, but to give an overview impression of the activation dynamics.}.

% Shorten over time

In the upper plot, the depolarization and repolarization of the membrane can be seen upon the stimulation at $t=\SI{5}{\ms}$, exhibiting the characteristic action potential shape. The membrane voltage $V_m$ reaches its equilibrium value approximately $\SI{5}{\ms}$ after the stimulation. 
The lower plot in \cref{fig:shorten_over_time} shows longer durations for several other variables to return to the equilibrium state. As a consequence, the generated force or active stress of the sarcomeres, indicated by the thick red line in \cref{fig:shorten_over_time}, does not directly decrease after the stimulation is over. For frequent stimulation patterns, the force level would show a smooth progression over time.

\begin{figure}
  \centering%
  \includegraphics[width=0.8\textwidth]{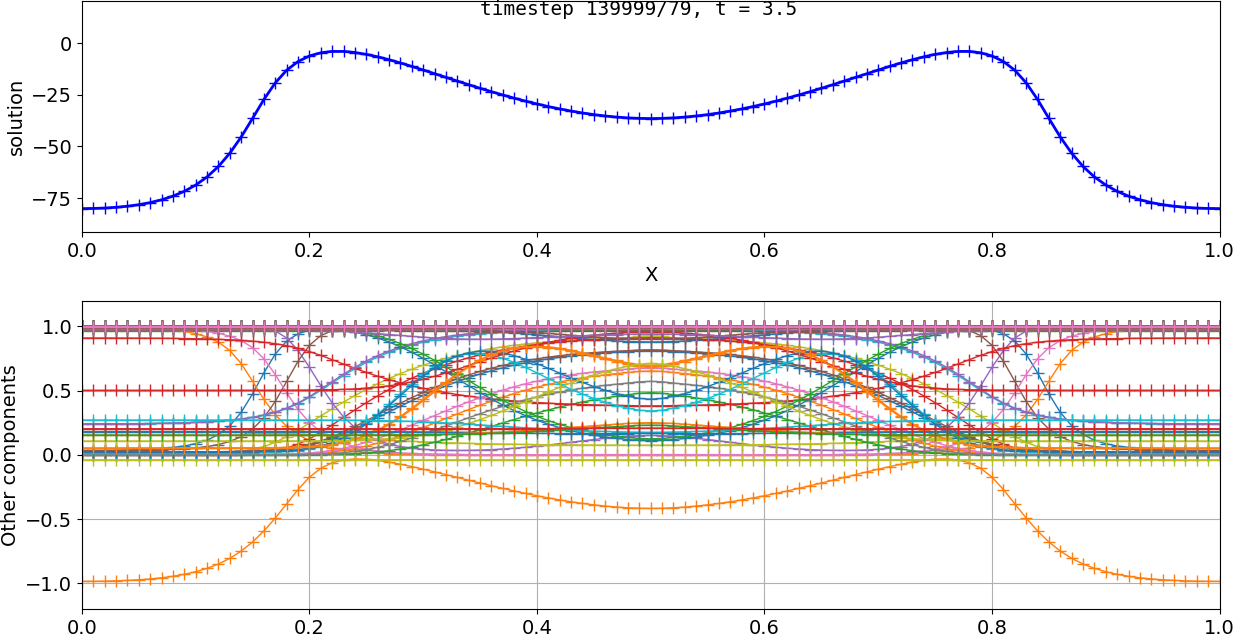}%
  \caption{Simulation of action potential propagation on a 1D mesh with the Shorten subcellular model. The upper plots shows the membrane voltage $V_m$ at time $t=\SI{3.5}{\ms}$ over the fiber along the $x$ axis. The lower plot shows all other variables of the model.}%
  \label{fig:shorten_03_50}%
\end{figure}

The propagation of the action potentials of the Shorten subcellular model can be simulated with the monodomain equation (\cref{eq:monodomain}). \Cref{fig:shorten_03_50} shows simulation results on a 1D muscle fiber mesh with length \SI{1}{\cm}, discretized to 100 elements. The muscle fiber is stimulated at its center at $t=0$. The upper plot in \cref{fig:shorten_03_50} displays the membrane voltage $V_m$ at time $t=\SI{3.5}{\ms}$. Two action potentials, which move towards both ends of the fiber can be identified. The lower plot shows all other variables, normalized to the value range $[-1,1]$. At the outer ends of the fiber, the variables are still in equilibrium, whereas towards the center, their values change dynamically as the action potentials propagate. 

% Shorten over mesh

\begin{figure}
  \centering%
  \includegraphics[width=0.9\textwidth]{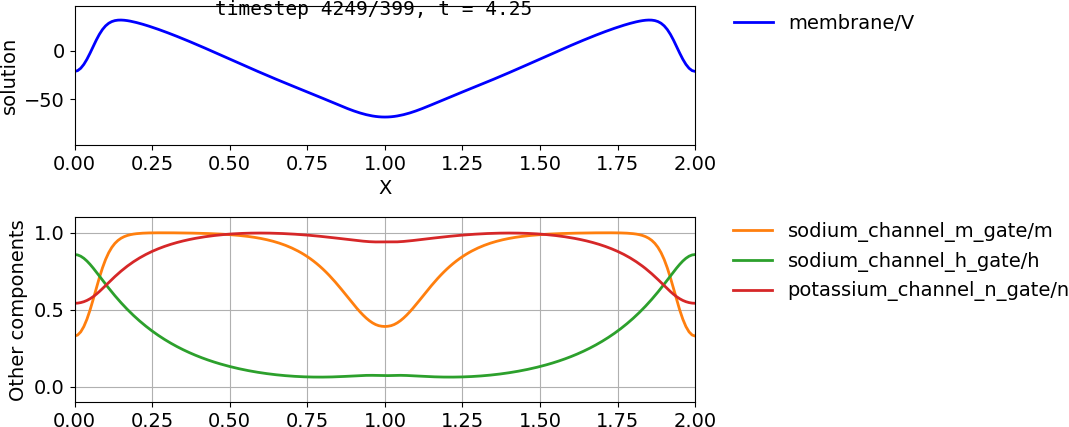}%
  \caption{Simulation of action potential propagation on a 1D mesh with the Hodgkin-Huxley subcellular model for time $t=\SI{4.25}{\ms}$, analog to \cref{fig:shorten_03_50}.}%
  \label{fig:hodgkin_huxley_over_mesh}%
\end{figure}

\Cref{fig:hodgkin_huxley_over_mesh} shows analog results for a simulation with the subcellular model of Hodgkin and Huxley \cite{Hodgkin1952}. Two action potentials can be seen  at time $t=\SI{4.25}{\ms}$ on a fiber mesh with 200 nodes and of \SI{2}{\cm} length. The system state is fully described by the values of the four variables that are plotted in \cref{fig:hodgkin_huxley_over_mesh}.

% Hodgkin_Huxley over mesh

\begin{reproduce_no_break}
  The simulation of a single instance of the Shorten model and the plot of \cref{fig:shorten_over_time} can be obtained as follows:
  \begin{lstlisting}[columns=fullflexible,breaklines=true,postbreak=\mbox{\textcolor{gray}{$\hookrightarrow$}\space}]
    cd $\$$OPENDIHU_HOME/examples/electrophysiology/cellml/shorten/build_release
    ./cellml ../settings_cellml.py
    cd out && plot
  \end{lstlisting}
  The simulation of the monodomain equation for the Shorten model shown in \cref{fig:shorten_03_50} can be executed and visualized as follows: 
  \begin{lstlisting}[columns=fullflexible,breaklines=true,postbreak=\mbox{\textcolor{gray}{$\hookrightarrow$}\space}]
    cd $\$$OPENDIHU_HOME/examples/electrophysiology/monodomain/new_slow_TK_2014_12_08/build_release
    ./shorten_implicit ../settings_new_slow_TK_2014_12_08.py
    cd out && plot
  \end{lstlisting}
\end{reproduce_no_break}

\subsection{Simulation of Motor Neuron Models}

If an activation model with a pool of motor neurons is considered in the neuromuscular multi-scale model, the transient behavior of motor neurons has to be simulated as well. In our simulations, we use the motor neuron model of Cisi and Kohn \cite{Cisi2008}. The drive parameter of the model is set to a constant value of \SI{0.01}{\volt\per\second}. As a consequence, the motor neuron fires with a frequency that depends on the input drive, which in the presented scenario is approximately \SI{25}{\hertz}.

\begin{figure}
  \centering%
  \includegraphics[width=\textwidth]{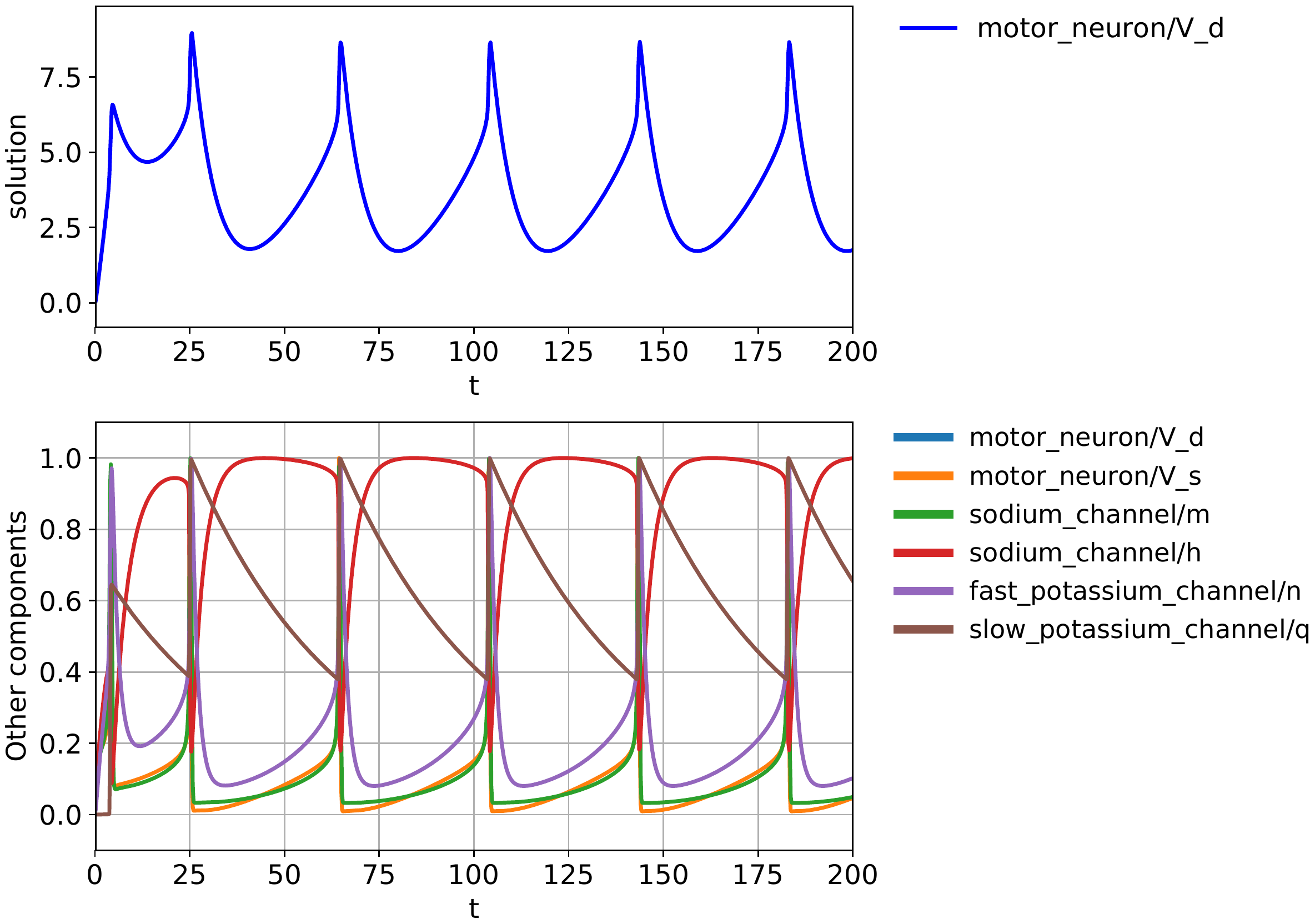}%
  \caption{Simulation of the motor neuron model of Cisi and Kohn \cite{Cisi2008} for a constant input drive of  \SI{0.01}{\volt\per\second}. The evolution of all variables of the model is plotted over time $t$ in milliseconds.}%
  \label{fig:motoneuron_plot}%
\end{figure}

\Cref{fig:motoneuron_plot} shows the evolution of the different variables of the model over time. Six firing times can be identified. 

% Motoneuron over time

To connect the motor neuron with the fibers of the MU in the simulation, we stimulate the muscle fibers whenever the $V_s$ value of the motor neuron reaches a certain threshold. It is possible to configure a pool of several motor neurons with different model parameters and different input drive values. Each motor neuron can be connected to a different set of fibers, according to the MU to fiber association. As a result, the MUs get physiologically activated according to the different firing frequencies of the motor neurons.

In summary, we showed simulations of the 0D subcellular models of Shorten et al. and Hodgkin and Huxley, simulations of these models together with the 1D conduction model in the monodomain equation, and a simulation of motor neurons. All of these models are given in CellML description and can be further combined with other parts of the multi-scale model, e.g., to simulate surface EMG signals.

\begin{reproduce_no_break}
  The simulation and visualization for \cref{fig:motoneuron_plot} can be executed with the following commands:
  \begin{lstlisting}[columns=fullflexible,breaklines=true,postbreak=\mbox{\textcolor{gray}{$\hookrightarrow$}\space}]
    cd $\$$OPENDIHU_HOME/examples/electrophysiology/monodomain/motoneuron_cisi_kohn/build_release
    ./motoneuron_cisi_kohn ../settings_motoneuron_cisi_kohn.py
    cd out && plot motoneuron*
  \end{lstlisting}
  This simulation also computes the monodomain equation for one muscle fiber that gets activated whenever the motor neuron fires.
\end{reproduce_no_break}
%-----

% --------
%
% f==============

% ==================
%
% =-------------------

\section{Simulation of Fiber Based Electrophysiology}\label{sec:results_fiber_based_electrophysiology}
% ==================
%
% =-------------------

In this section, we consider surface EMG signals on the upper arm by simulating the activation of the biceps brachii muscle. We use the fiber based multi-scale model consisting of 1D action potential propagation on muscle fibers, potentially involving a 0D subcellular model, and the 3D bidomain model.
In \cref{sec:overview_emg_simulation}, we introduce the setting of the simulation and present an exemplary scenario to compute EMG signals.
Subsequently, we simulate various scenarios to investigate the effects of different model parameters and numerical settings on the resulting EMG signal. \Cref{sec:simfiber_mu} considers the effects of single motor units, \cref{sec:simfiber_fat} the fat layer, and \cref{sec:effects_of_the_mesh_width_emg} shows effects of the mesh width. \Cref{sec:simfiber_electrodes} presents a way to simulate realistic EMG electrodes and \cref{sec:simfiber_decomposition} deals with the decomposition of EMG signals.
In \cref{sec:sim_rosenfalck}, we describe our simulations with a phenomenological model for action potential propagation.

\subsection{Overview of the EMG Simulation}\label{sec:overview_emg_simulation}

% full_muscle_emg_raytrace_1.png
\begin{figure}
  \centering%
  \includegraphics[width=0.6\textwidth]{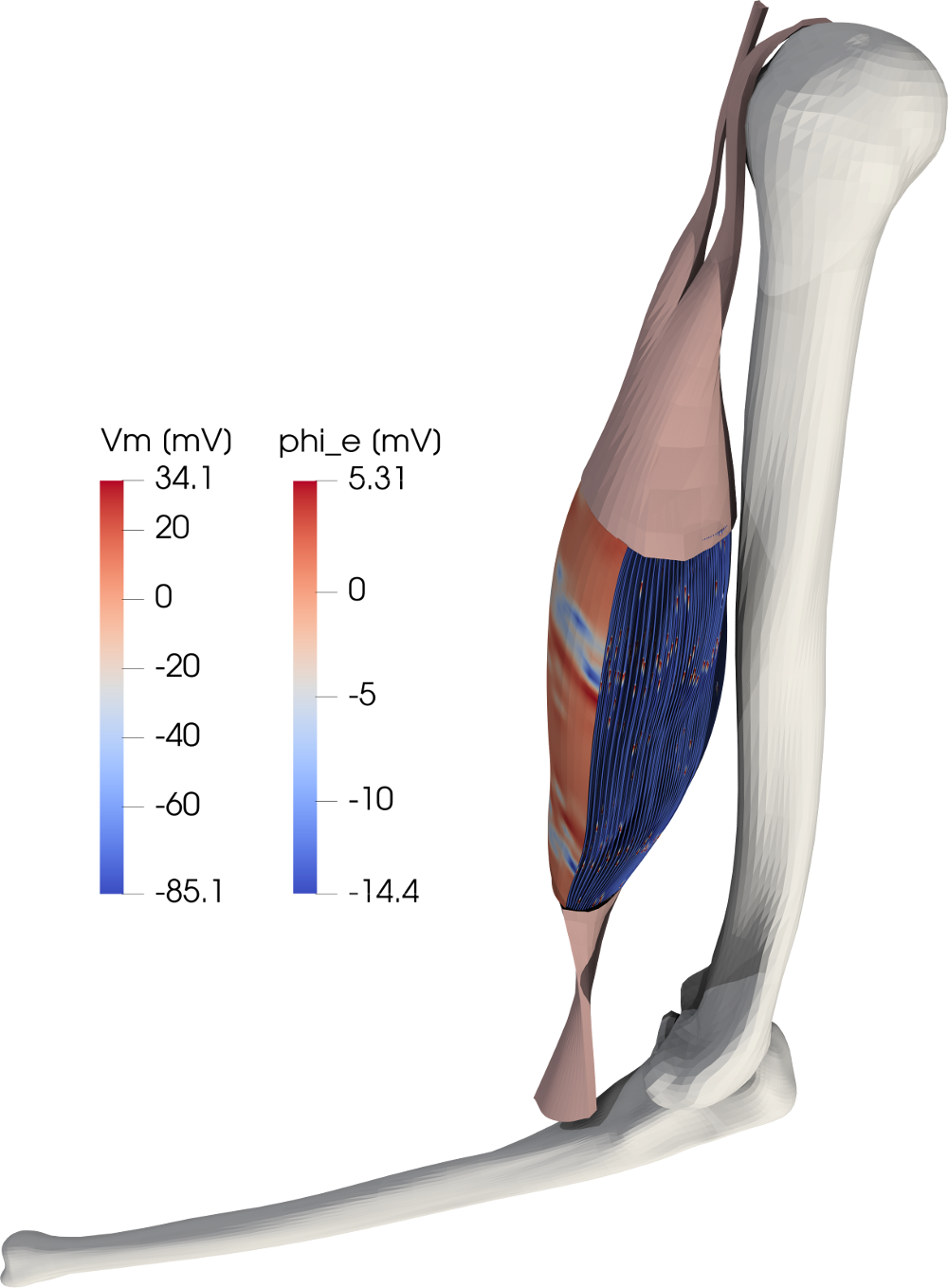}%
  \caption{Considered setting for simulations of surface EMG for the upper arm, consisting of the biceps brachii muscle, tendons and bones. A simulation result of the membrane voltage $V_m$ on the muscle fibers and the extracellular potential $\phi_e$ on the surface is shown.}%
  \label{fig:full_muscle_emg_raytrace_1}%
\end{figure}

\Cref{fig:full_muscle_emg_raytrace_1} shows the setting of the biceps muscle and the tendons, which attach to the skeleton near the shoulder and to the ulna bone in the forearm. For the simulation of EMG, we only consider the muscle belly of the biceps muscle. \Cref{fig:full_muscle_emg_raytrace_1} shows muscle fibers inside the muscle, which run in longitudinal direction between the tendons at both ends. The image also visualizes the results of an EMG simulation. The fibers are colored according to the transmembrane potential $V_m$. On some fibers, action potentials can be seen.

The surface of the muscle is colored by the  extracellular electric potential $\phi_e$. In a reasonable approximation, the value of $\phi_e$ corresponds to the measured EMG signals on the skin surface. 
Additionally, we consider volume conduction in a layer of adipose tissue on top of the muscle in the following section.

For the EMG simulations, we solve the multi-scale model of fiber based electrophysiology. We solve the monodomain equation \cref{eq:monodomain} independently on all 1D muscle fiber meshes. 
After a fixed number of timesteps, we map the membrane voltage $V_m$ from the 1D meshes to the 3D mesh. Subsequently, we solve the static bidomain equation \cref{eq:bidomain1} on the muscle domain and potentially the body fat domain to obtain the $\phi_e$ values on the skin surface.

% fibers mesh
\begin{figure}
  \centering%
  \includegraphics[width=\textwidth]{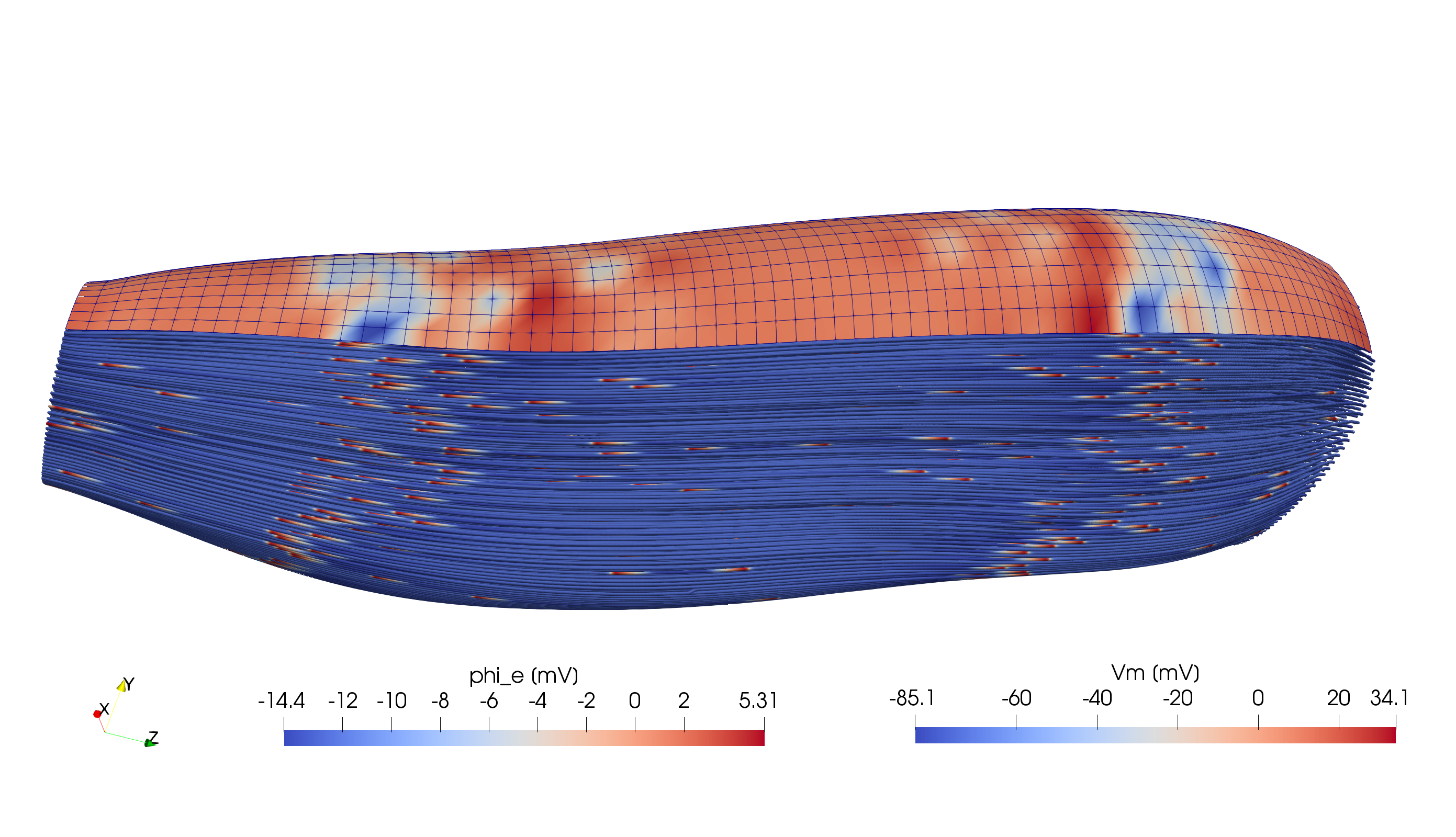}
  \caption{Overview of some of the meshes in an electrophysiology simulation: 1369 muscle fibers are located in the muscle belly. A 2D surface mesh on top of the muscle describes the computed EMG values. The visualized simulation is the same as in \cref{fig:full_muscle_emg_raytrace_1}.}%
  \label{fig:fibers_4}%
\end{figure}

\Cref{fig:fibers_4} shows a close-up view of the active muscle fibers and the resulting EMG signals on the upper surface, which are identical to \cref{fig:full_muscle_emg_raytrace_1}. The scenario considers 961 muscle fibers, each described by a 1D mesh with 1481 nodes. It can be seen that they are approximately equally spaced as a result of the meshing algorithms described in \cref{sec:generation_of_meshes_for_multiscale}.

\Cref{fig:fibers_4} also shows the mesh of the muscle surface, which is colored according to the extracellular potential $\phi_e$. It can be seen that the values correlate with the activation state of the underlying fibers. At the two blue colored regions at the surface near the left and right end of the muscle, the $\phi_e$ value is close to its minimum, while the majority of fibers exhibits its maximum positive $V_m$ value. Towards the center of the muscle, the value of $\phi_e$ increases to its maximum, which reflects the hyperpolarization of the muscle fibers behind the propagating action potentials, i.e., the overshoot of the membrane voltage before it reapproaches the equilibrium level.

The lower left corner in \cref{fig:fibers_4} shows the coordinate frame that is used in all simulations. The $z$ axis is approximately  oriented in fiber direction, the $x$ and $y$ axes are oriented in transverse direction and describe cross-sectional planes of the muscle.

The scenario uses the subcellular model of Hodgkin and Huxley \cite{Hodgkin1952}. The shown image corresponds to the simulation time of $t=\SI{200}{\milli\second}$. The simulation also considers a body fat mesh layer on top of the muscle, which is not visualized in \cref{fig:fibers_4}.

We run the simulation with 128 processes on a two-socket shared-memory node comprising two AMD EPYC 7742 64-core processors with \SI{2.87}{\giga\hertz} clock frequency and \SI{1.96}{\tebi\byte} RAM. The total runtime for a simulation end time of one second is \SI{8}{\hour} \SI{38}{\minute}.

\subsection{Effects of Single Motor Units on the Electromyography Signal}\label{sec:simfiber_mu}

Next, we investigate how the surface EMG signals are influenced by several parameters of the simulation. We begin by studying EMG of only a single activated MU in the muscle.

\begin{figure}[H]
  \centering%
  \begin{subfigure}[t]{\textwidth}%
    \centering%
    \includegraphics[width=0.6\textwidth]{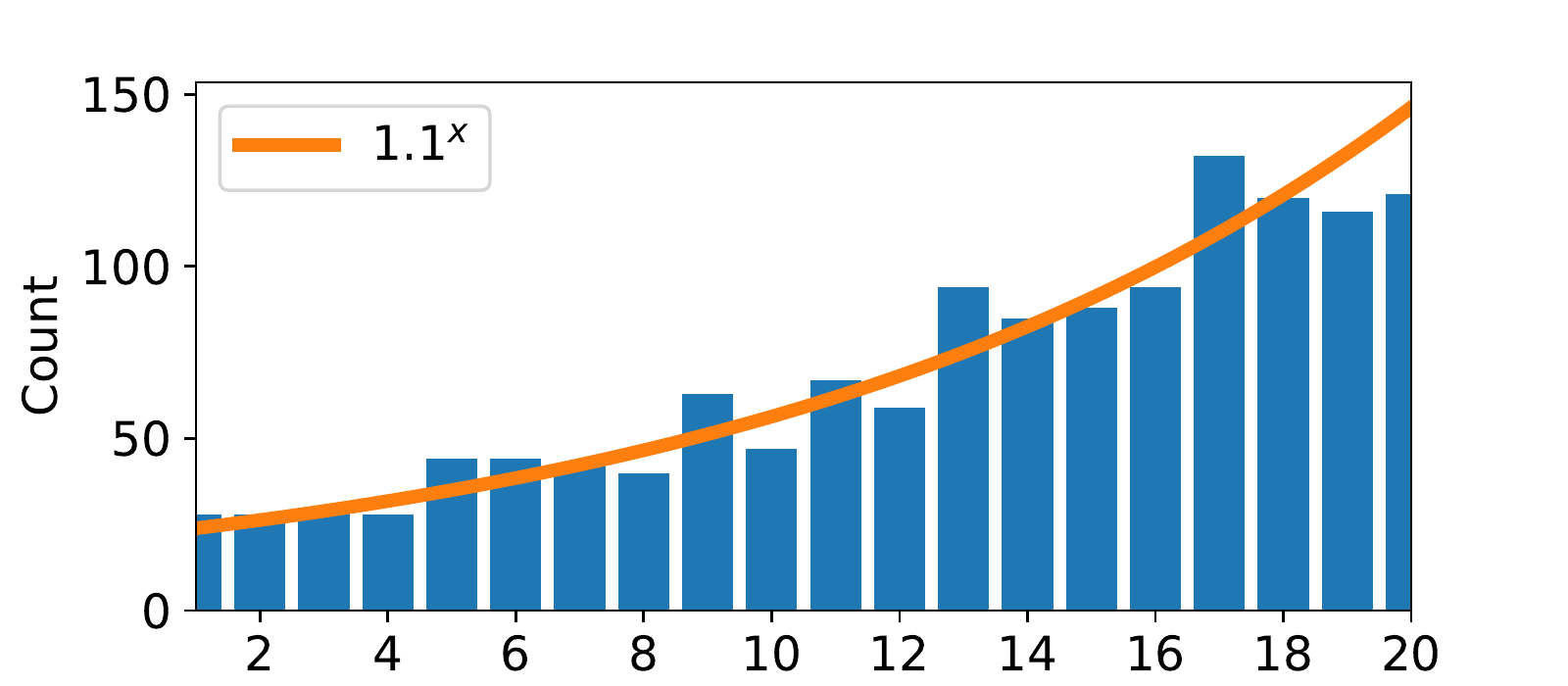}%
    \caption{Exponential distribution of motor unit sizes. The diagram shows the motor unit numbers with the corresponding sizes or fibers counts of the MUs.}%
    \label{fig:MU_fibre_distribution_37x37_20c_txt_fiber_distribution}%
  \end{subfigure}\\
  \begin{subfigure}[t]{\textwidth}%
    \centering%
    \includegraphics[width=0.6\textwidth]{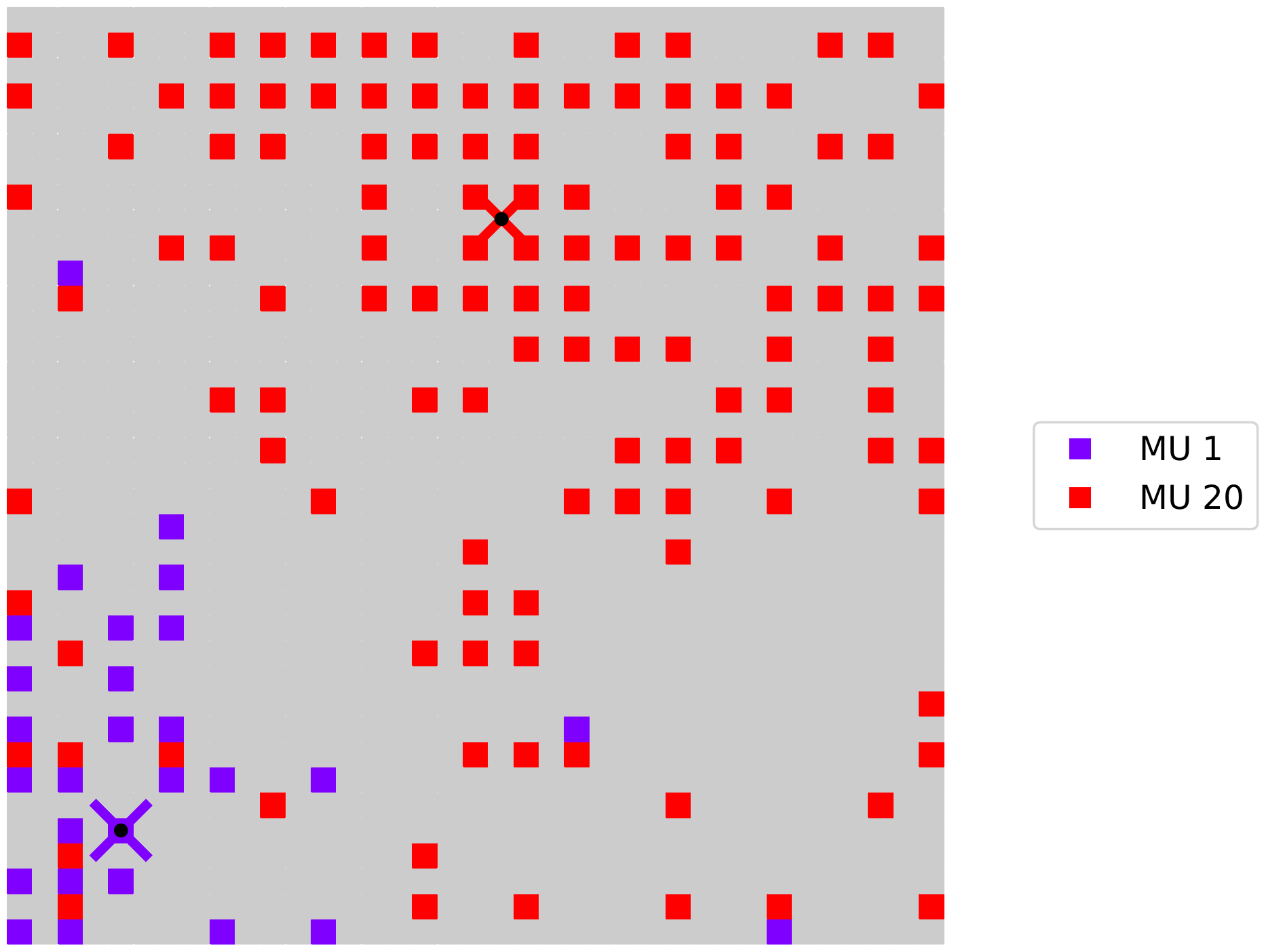}%
    \caption{Fibers that belong to motor units 1 and 20. The crosses are the center points around which the MU territories are generated by our algorithm.}%
    \label{fig:MU_fibre_distribution_37x37_20c_txt_2d_fiber_distribution}%
  \end{subfigure}
  \caption{Fiber based upper arm EMG simulation: Assignment of the 1369 fibers to 20 motor units used in the simulation scenario for fiber based electrophysiology.}%
  \label{fig:MU_fibre_distribution_37x37_20c}%
\end{figure}

The first scenario contains 20 MUs that connect to an exponentially increasing number of fibers as shown in \cref{fig:MU_fibre_distribution_37x37_20c_txt_fiber_distribution}. The progression follows the function $y=c\,1.1^x$ for an appropriate constant $c>0$.
The MU assignment is created using method 1a of the algorithm described in \cref{sec:muscle_fibers_and_motor_units}, where the MU territories are centered around given points, and neighboring fibers are never part of the same MU. 

\Cref{fig:MU_fibre_distribution_37x37_20c_txt_2d_fiber_distribution} shows the fibers that are assigned to the smallest and to the largest MU, MU 1 and MU 20. For this visualization, the muscle cross-section is mapped to the large gray square and every colored small square corresponds to one fiber. The purple and red crosses indicate the center of the MU territories for MU 1 and 20, respectively. As a consequence, the fibers of MU 1 are mostly located at the bottom left of the cross-section and the fibers of MU 20 are mostly located in the upper right region of the muscle cross-section.
The visualization shows that the fibers of the same MU always have some spacing between them, which is due to the construction of the MU assignment algorithm.
\newpage

% MU fiber sizes Vm

We begin with a simulation scenario, where only a single MU is stimulated, and study the effect on the surface EMG. The fibers of the respective MU are stimulated with a frequency $f=\SI{24}{\hertz}$ starting at time $t=\SI{0}{\milli\second}$. 
Each of the $13\times 13=\num{1369}$ fibers consists of a mesh with 1481 nodes, the 3D mesh of the muscle contains $19 \times 19 \times 38 = \num{13718}$ nodes and the 3D mesh of the fat layer contains $37 \times 5 \times 38 = \num{7030}$ nodes. The domains are partitioned into 27 subdomains associated to 27 MPI ranks. The subcellular model of Hodgkin and Huxley is used, yielding a total number of more than $\num{8.1e6}$ degrees of freedom. 
The timestep widths are $\dt_\text{0D}=\dt_\text{splitting} = \SI{2.5e-3}{\milli\second}$, $\dt_\text{1D} = \SI{6.25e-4}{\milli\second}$ and $\dt_\text{3D} = \SI{5e-1}{\milli\second}$, leading to 4 subcycles for the 1D model in each splitting step and 200 splitting steps per solution of the bidomain equation.

We compute the linear systems for the initial potential flow problem to estimate fiber directions in the 3D domain, \cref{eq:fiberest_laplace}, and for the bidomain equation \cref{eq:bidomain1}, which is solved in every timestep  using a conjugate gradient solver. 
The program uses the \code{FastMonodomainSolver} class for the electrophysiology model. The Thomas algorithm solves the linear system of the diffusion problem. We use the \code{`vc`} optimization type and employ the scheme to only compute active fibers and the subcellular problems that are not in equilibrium.

The computation of a simulated time span with $t_\text{end}=\SI{100}{\milli\second}$  on an AMD EPYC 7742 64-core processor with \SI{2492}{\mega\hertz} base frequency and \SI{1.96}{\tera\byte} RAM takes approximately $\SI{100}{\second}$ in the scenario that activates only the smallest MU, and $\SI{126}{\second}$ in the scenario that activates only the largest MU.

\begin{figure}[H]
  \centering%
  \begin{subfigure}[t]{0.7\textwidth}%
    \centering%
    \includegraphics[width=\textwidth]{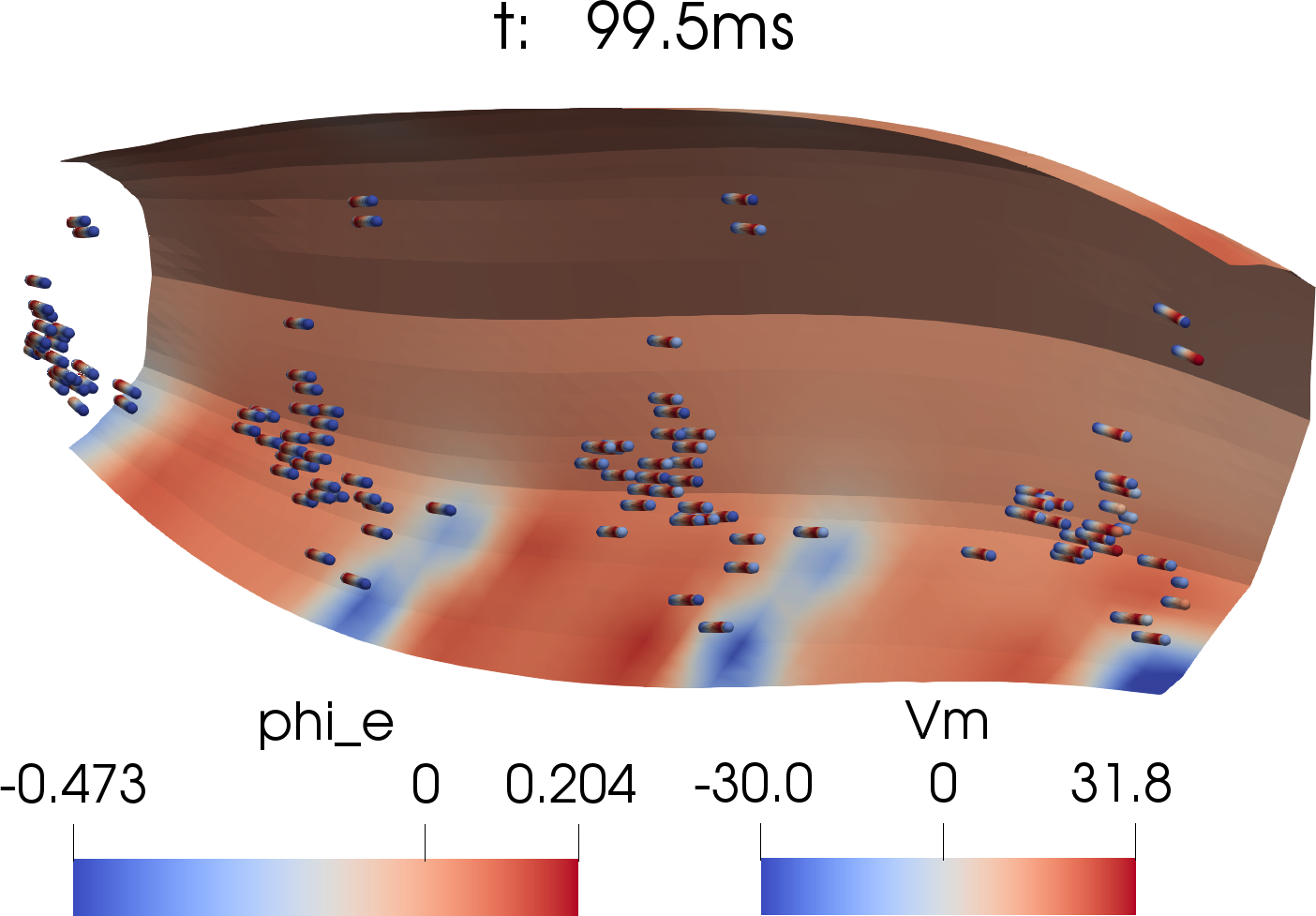}%
    \caption{Membrane voltage $V_m$ at active parts of the fibers (foreground) and EMG signals $\phi_e$ on the skin surface (background).}%
    \label{fig:mu01a}%
  \end{subfigure} \,
  \begin{subfigure}[t]{0.25\textwidth}%
    \centering%
    \includegraphics[width=\textwidth]{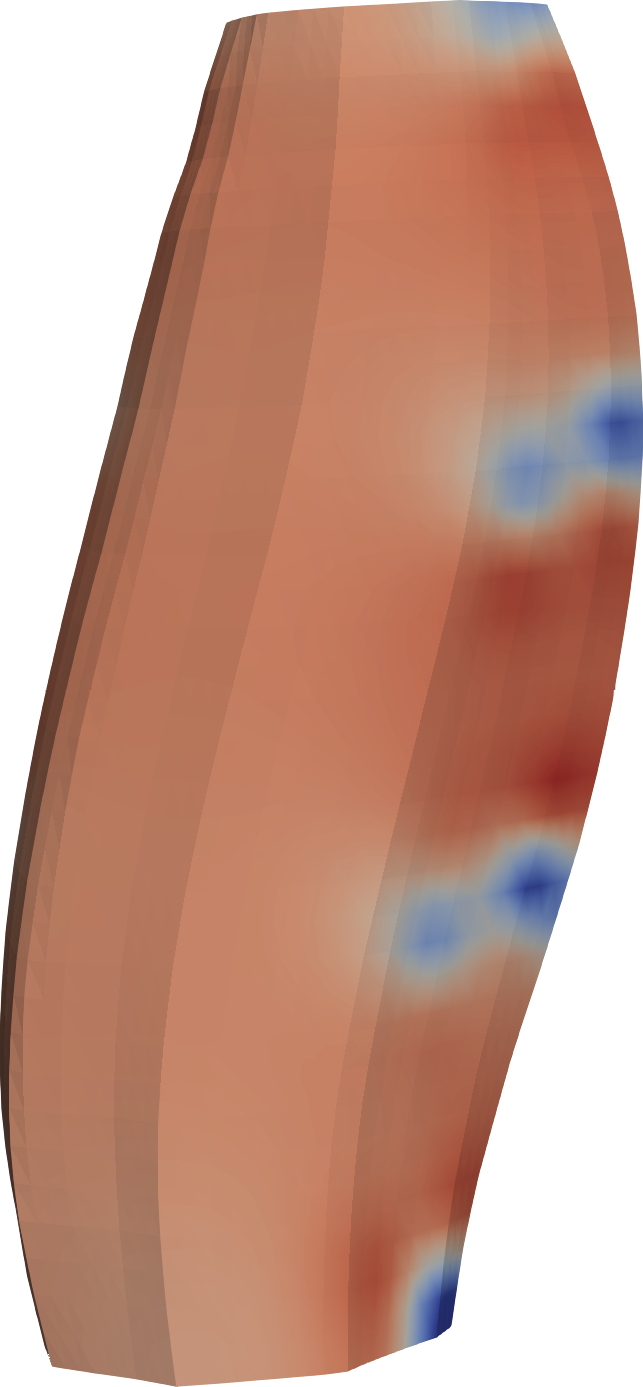}%
    \caption{Resulting surface EMG.}%
    \label{fig:mu01b}%
  \end{subfigure}   
  \caption{Fiber based EMG simulation for the upper arm (biceps) model: Simulation result at $t=\SI{99.5}{\milli\second}$ where only MU 1 is activated.}%
  \label{fig:result_mu1}%
\end{figure}

\Cref{fig:result_mu1} shows the result for the scenario of activating the smallest MU, MU 1. In \cref{fig:mu01a}, the surface is shown in the background and colored according to the extracellular potential $\phi_e$, which represents the EMG signal. The muscle volume is not shown. Instead, the active parts of the respective fibers are displayed as tubes in the 3D domain. Their color visualizes the value of the transmembrane voltage $V_m$. In every of these small tube segments, the rising and declining shape of an action potential can be observed by the color progression from blue over orange to red for the rising part and back to blue for the declining part.

In this scenario, the fibers of MU 1 are stimulated three times within the first $\SI{100}{\milli\second}$ at $\SI{0}{\milli\second}$, $\SI{41.6}{\milli\second}$ and $\SI{83.3}{\milli\second}$.  The innervation zone contains the starting points for the propagating stimulus on every fiber. The scenario positions the neuromuscular junctions randomly with a uniform distribution within the central $\SI{10}{\percent}$ of every muscle fiber. The activated parts of the fibers visible in \cref{fig:mu01a} correspond to the propagated action potentials of the last two stimulations in this scenario.

By comparing the results in \cref{fig:mu01a} with the fiber distribution in \cref{fig:MU_fibre_distribution_37x37_20c_txt_2d_fiber_distribution}, it can be seen that fibers of MU 1 are located opposite of the outer arm surface, which is at the upper side of the cross-sectional square diagram in \cref{fig:MU_fibre_distribution_37x37_20c_txt_2d_fiber_distribution}. The left side of the diagram in \cref{fig:MU_fibre_distribution_37x37_20c_txt_2d_fiber_distribution} corresponds to the lower part of the skin in \cref{fig:mu01a}. This part of the skin is closer to the activated fibers and, thus, the effect on the surface EMG is highest for this region.

\Cref{fig:mu01b} shows the skin surface as seen from the inside of the arm in \cref{fig:mu01a}. The active region is located on the right-hand side in this image.
It can be seen that the active region on the skin surface, which results from fibers of the activated MU 1, only spans a small portion of the surface.

% Mu no. 1

% Mu no. 20
\begin{figure}[H]
  \centering%
  \begin{subfigure}[t]{0.7\textwidth}%
    \centering%
    \includegraphics[width=\textwidth]{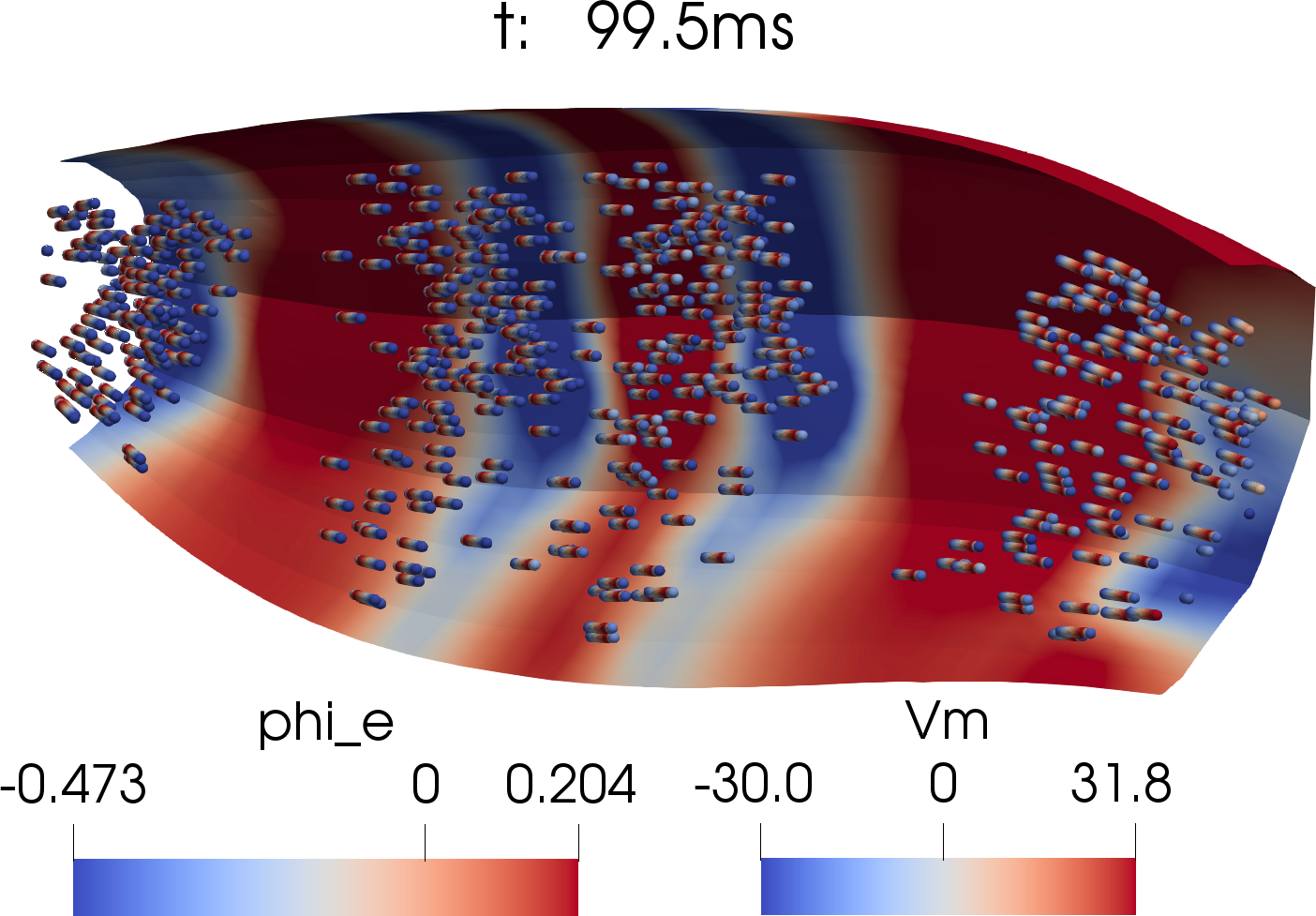}%
    \caption{Membrane voltage $V_m$ at active parts of the fibers (foreground) and EMG signals $\phi_e$ on the skin surface (background).}%
    \label{fig:mu20a}%
  \end{subfigure} \,
  \begin{subfigure}[t]{0.25\textwidth}%
    \centering%
    \includegraphics[width=\textwidth]{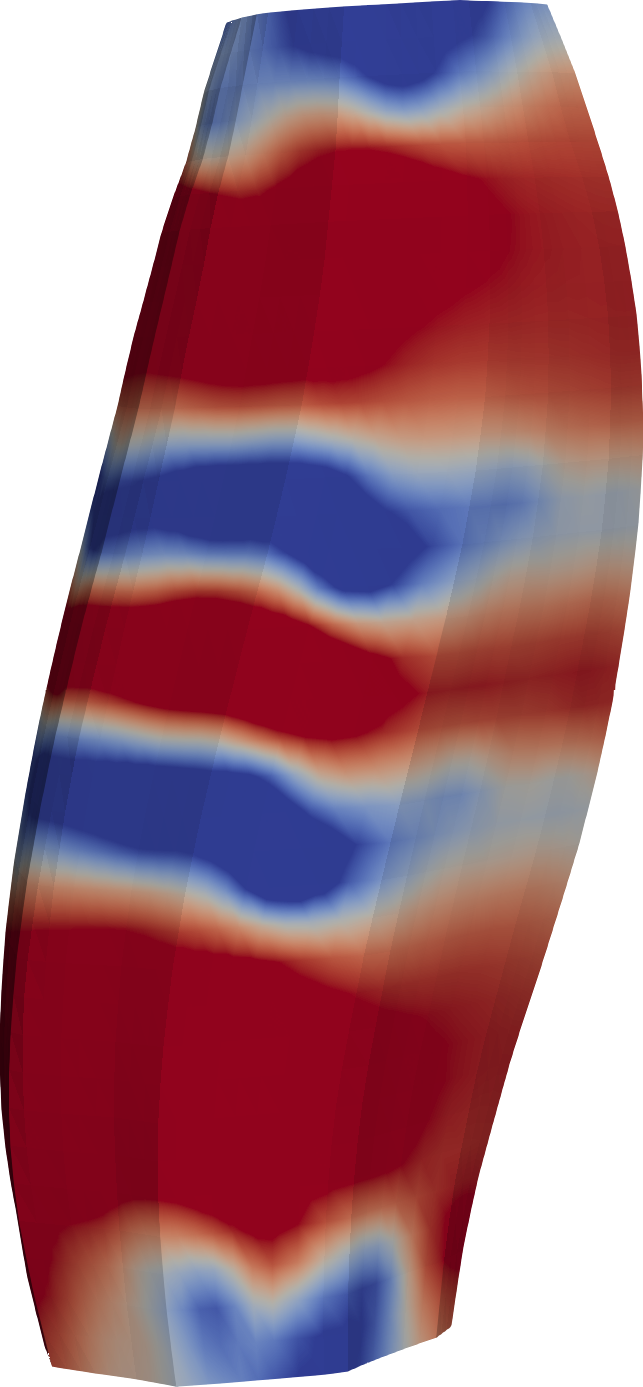}%
    \caption{Resulting surface EMG.}%
    \label{fig:mu20b}%
  \end{subfigure}   
  \caption{Fiber based EMG simulation for the upper arm (biceps) model: Simulation result at $t=\SI{99.5}{\milli\second}$ where only motor unit 20 is activated, analog to \cref{fig:result_mu1}.}%
  \label{fig:result_mu20}%
\end{figure}%

\Cref{fig:result_mu20} shows the analogous scenario that activates MU 20 instead of MU 1. \Cref{fig:mu20a} shows that, now, more fibers are activated as MU 20 is larger than MU 1. According to the MU layout in \cref{fig:MU_fibre_distribution_37x37_20c_txt_2d_fiber_distribution}, the active fibers are also located closer to the skin surface. This layout results in a stronger EMG signal compared to the previous scenario. 

The color coding in the two scenarios in \cref{fig:result_mu1,fig:result_mu20} is identical, and it can be seen that the absolute value of the extracellular potential $\phi_e$ is larger in the scenario for MU 20. For the scenario with MU 1 in \cref{fig:result_mu1}, the value range of the extracellular potential $\phi_e$ is $[\SI{-0.473}{\milli\volt}, \SI{0.204}{\milli\volt}]$. For the scenario with MU 20 in \cref{fig:result_mu20}, it is $[\SI{-0.834}{\milli\volt}, \SI{0.579}{\milli\volt}]$, which is more than twice the range.

\Cref{fig:mu20b} shows the overall EMG signal on the skin surface for MU 20. Compared to the result of MU 1 in \cref{fig:mu01b}, nearly the inverse region is activated. It can, thus, be observed that the EMG signal is highly influenced by the location and size of the MUs. MUs with territories closer to the skin surface have a larger effect on the EMG signals than MUs that are located further away. As seen in \cref{fig:mu01b}, the influence of fibers completely vanishes if the distance is larger than a certain value. The effects of several close fibers add up, such that large MUs located near the surface have the largest impact on the resulting EMG signal.

\subsection{Effects of the Fat Layer on the Electromyography Signal}\label{sec:simfiber_fat}

% different fat meshes
\begin{figure}[H]
  \centering%
  \begin{subfigure}[t]{0.3\textwidth}%
    \centering%
    \includegraphics[width=\textwidth]{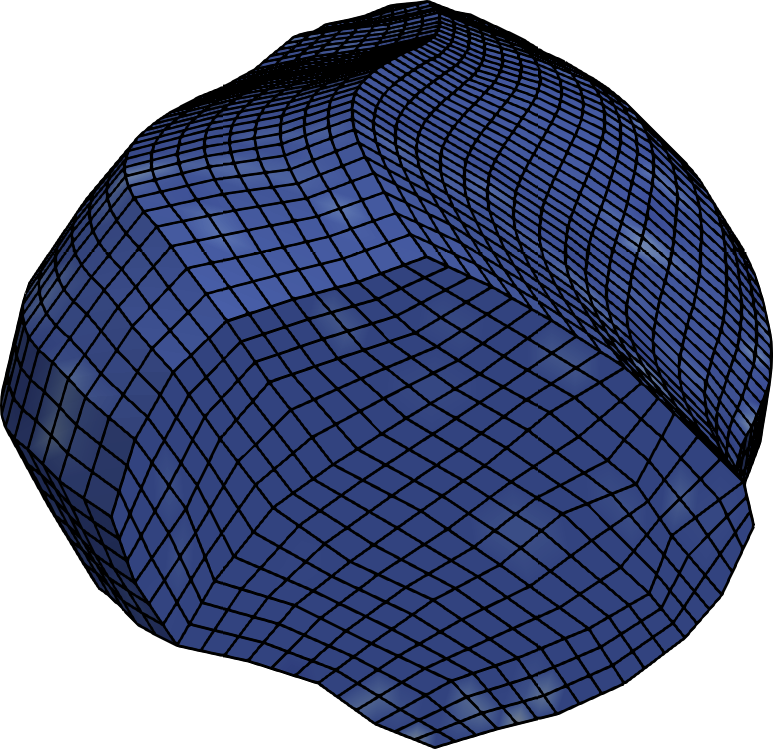}%
    \caption{Muscle mesh without fat layer.}%
    \label{fig:fibers_emg_mesh_no_fat}%
  \end{subfigure} \quad
  \begin{subfigure}[t]{0.3\textwidth}%
    \centering%
    \includegraphics[width=\textwidth]{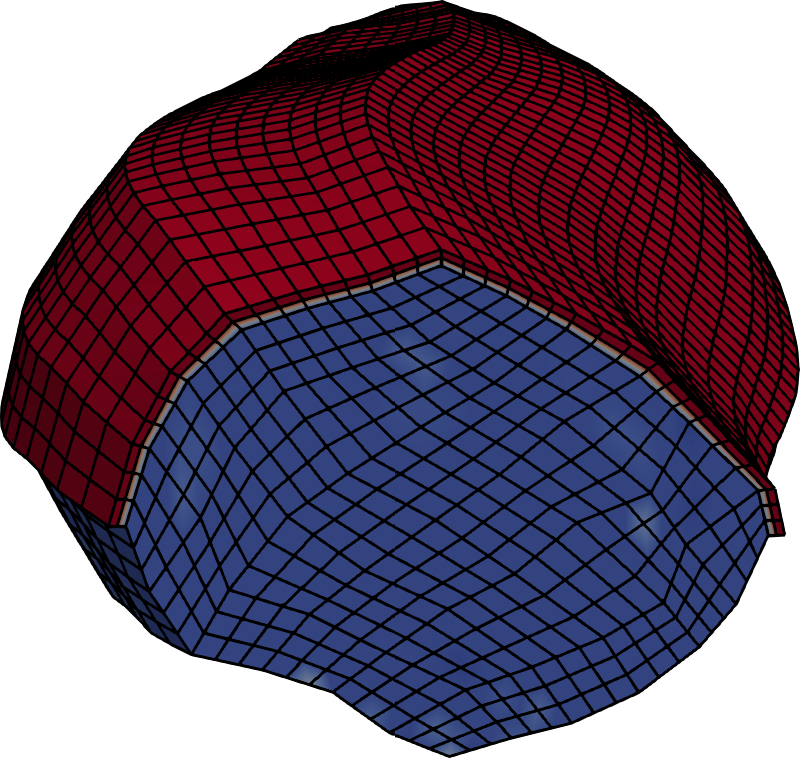}%
    \caption{Muscle with thin fat layer.}%
    \label{fig:fibers_emg_mesh_thin_fat}%
  \end{subfigure}  \quad
  \begin{subfigure}[t]{0.3\textwidth}%
    \centering%
    \includegraphics[width=\textwidth]{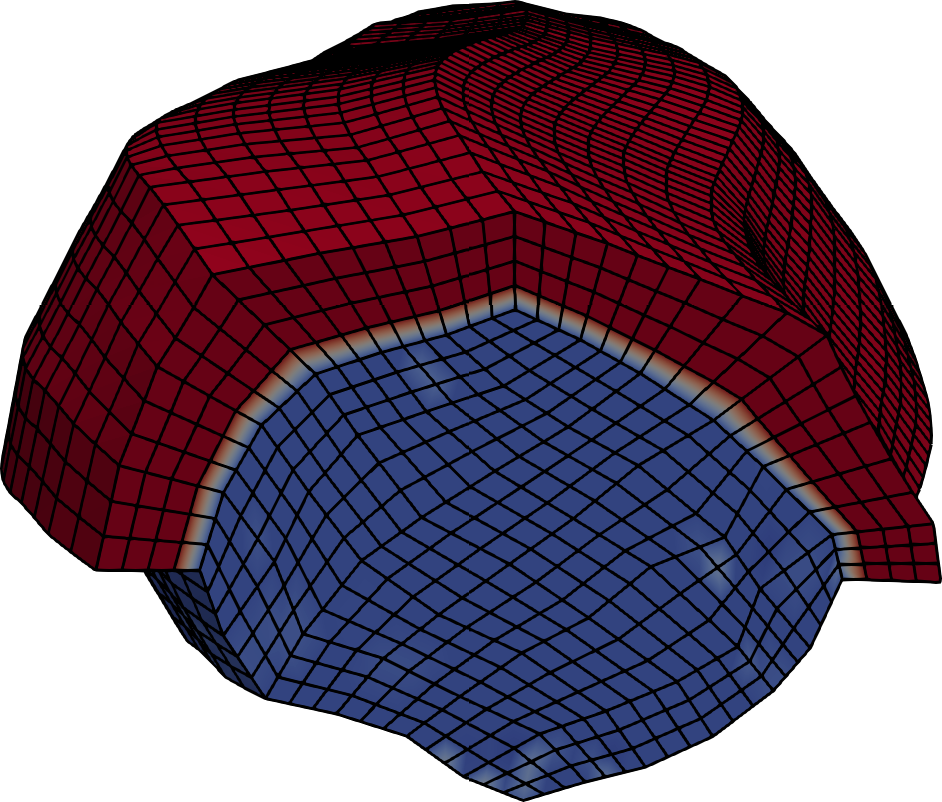}%
    \caption{Muscle with thick fat layer.}%
    \label{fig:fibers_emg_mesh_thick_fat}%
  \end{subfigure}   
  \caption{Fiber based EMG simulation for the upper arm (biceps) model: Meshes for the muscle domains (blue) and the layer of adipose tissue (red) used in the study to compare different fat layer widths.}%
  \label{fig:fibers_emg_mesh_fat}%
\end{figure}%

In the next study, we investigate the effect of the fat layer on the resulting EMG signals. The same scenario as in the previous section is used, except that the size of the body fat domain is varied and the activated MUs are chosen differently. We consider the domains and meshes shown in \cref{fig:fibers_emg_mesh_fat}: Scenario (a) only considers the muscle domain without additional  fat layer. Scenario (b) adds a thin fat layer with thickness of $\SI{2}{\milli\meter}$, discretized by two layers of finite elements. Scenario (c) considers a fat layer with thickness of $\SI{1}{\centi\meter}$ and four layers of elements. The scenario in the previous section also used this thick fat layer.

In this series of experiments, the first 10 MUs are activated with different stimulation frequencies ranging from \SI{7}{\hertz} for the smallest MU to \SI{15.15}{\hertz} for MU 10. The runtime of the simulation for one scenario on the same hardware as in the previous section is approximately $\SI{9}{\minute}$.

\begin{figure}[H]
  \centering%
  \begin{subfigure}[t]{0.36\textwidth}%
    \centering%
    \includegraphics[width=\textwidth]{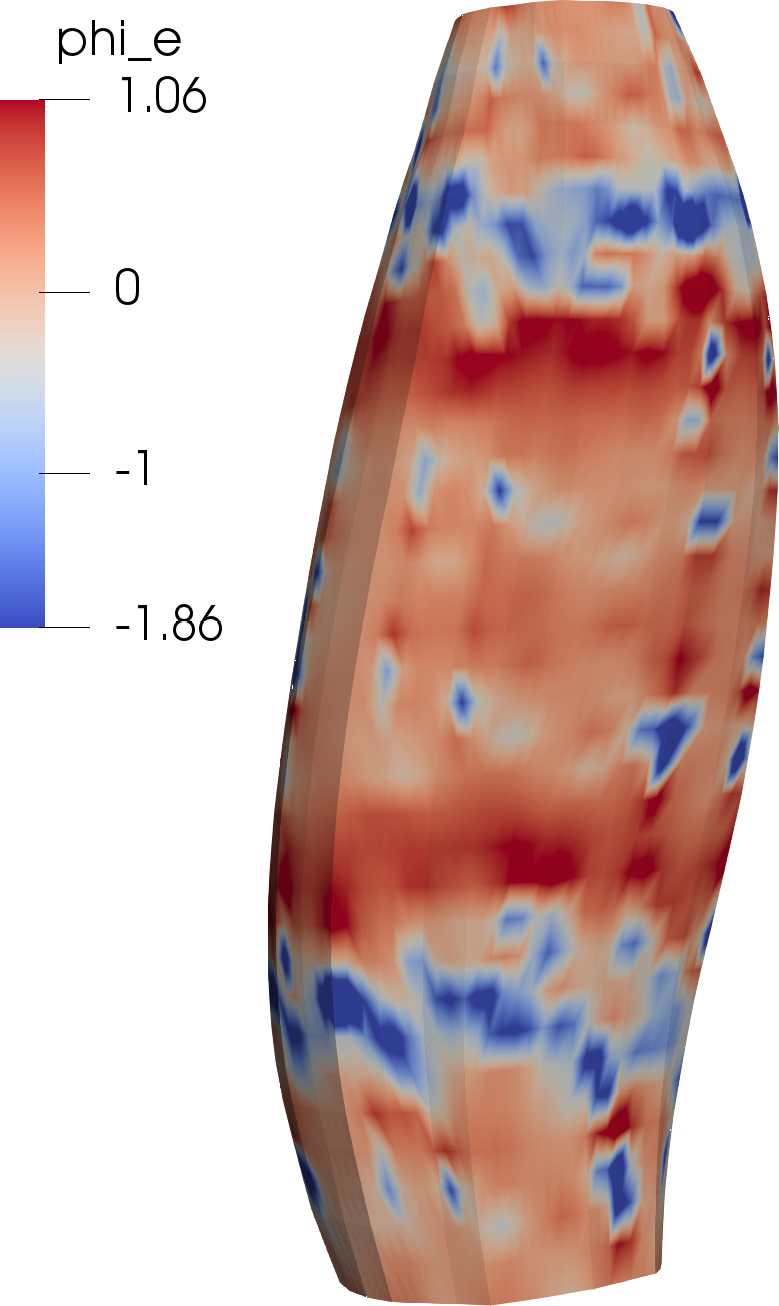}%
    \caption{Simulation without fat layer.}%
    \label{fig:fibers_emg_no_fat}%
  \end{subfigure} \quad
  \begin{subfigure}[t]{0.25\textwidth}%
    \centering%
    \includegraphics[width=\textwidth]{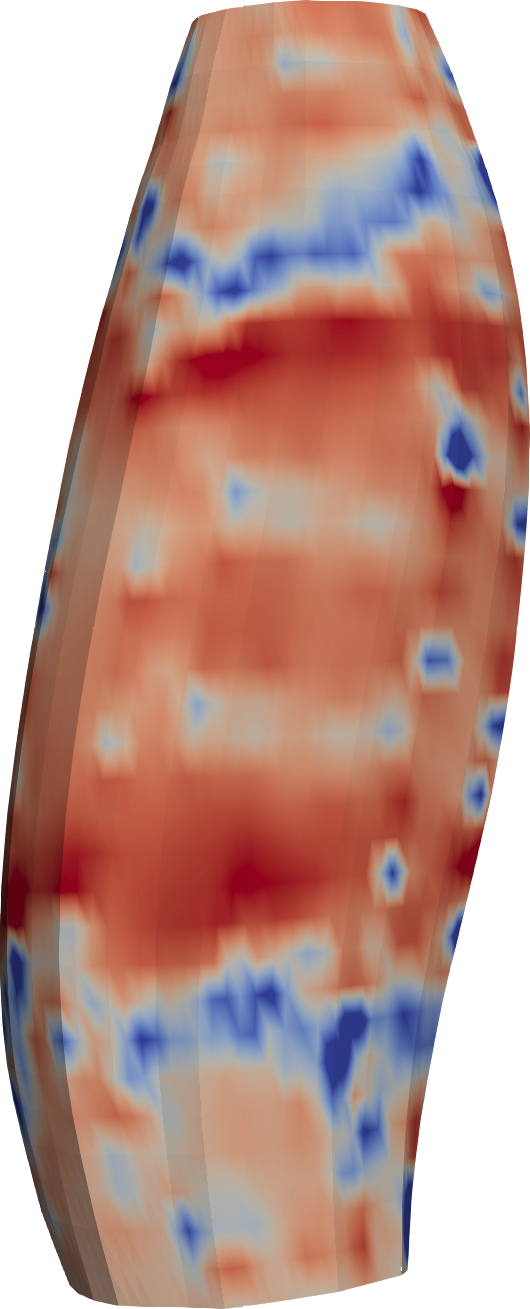}%
    \caption{Simulation with a thin fat layer.}%
    \label{fig:fibers_emg_thin_fat}%
  \end{subfigure}  \quad
  \begin{subfigure}[t]{0.28\textwidth}%
    \centering%
    \includegraphics[width=\textwidth]{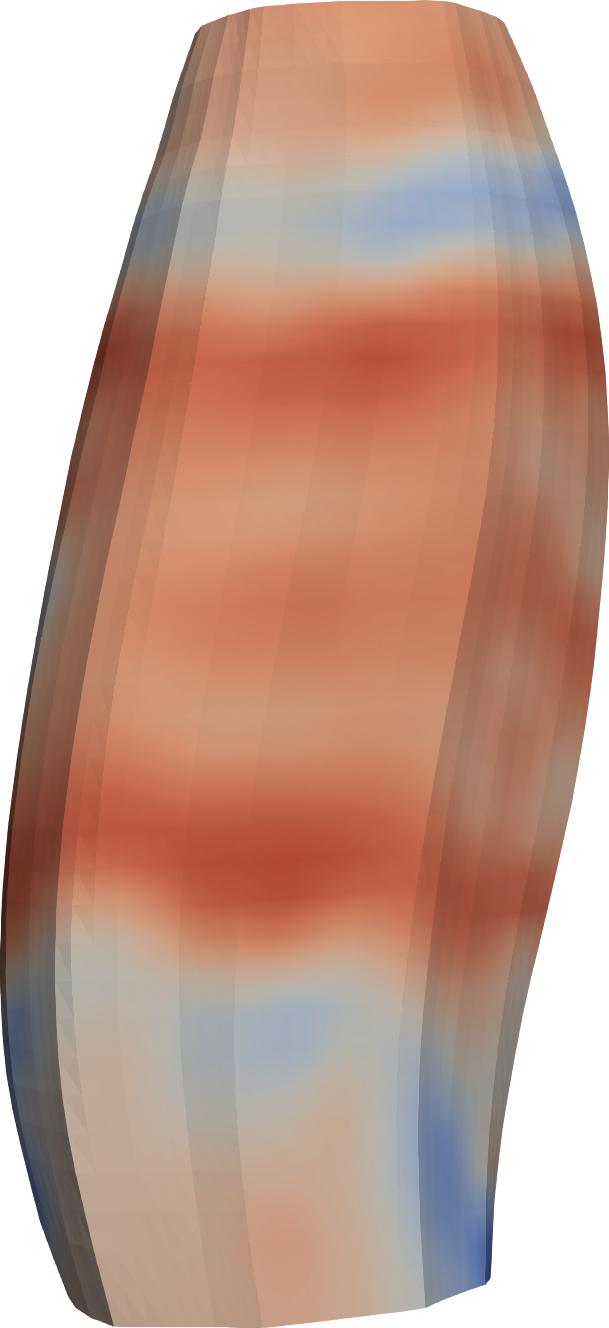}%
    \caption{Simulation with a thick fat layer.}%
    \label{fig:fibers_emg_thick_fat}%
  \end{subfigure}   
  \caption{Fiber based EMG simulation for the upper arm (biceps) model: Simulated surface EMG signals for the different fat layers shown in \cref{fig:fibers_emg_mesh_fat}.}%
  \label{fig:fibers_emg_fat}%
\end{figure}

\Cref{fig:fibers_emg_fat} shows the simulation results at $t=\SI{100}{\milli\second}$ for the three scenarios with different fat layers. The figure uses the same color coding for the extracellular potential $\phi_e$ in all three scenarios. It can be seen that the volume conduction in the fat layer significantly smooths the resulting EMG signal, especially for the thick fat layer. The scenarios with no fat layer and the thin fat layer also exhibit a small difference.
This effect has implications for experimental studies, where the EMG recordings capture the less resolved spatial information, the more tissue is located between the muscle and the surface electrodes.

% results for different fat meshes

\begin{reproduce}
  The simulations in this section use the examples \code{examples/electrophysiology/fibers/fibers_emg} and \code{examples/electrophysiology/fibers/fibers_fat_emg} with the variables file \code{20mus_fat_comparison.py}.

  The scenario data that are necessary to run the simulations are given in the repository at \href{https://github.com/dihu-stuttgart/performance}{github.com/dihu-stuttgart/performance}
  in the directory \code{opendihu/}\code{18_fibers_emg}. The main scripts that runs the simulations for the two sections are the following:
  \begin{lstlisting}[columns=fullflexible,breaklines=true,postbreak=\mbox{\textcolor{gray}{$\hookrightarrow$}\space}]
    ./run_single_MUs.sh
    ./run_compare_fat_layer.sh
  \end{lstlisting}
\end{reproduce}

\subsection{Effects of the Mesh Width on the Electromyography Signal}\label{sec:effects_of_the_mesh_width_emg}

% Hier herausarbeiten, dass das eine sehr wichtige Frage für Deine Arbeit ist, quasi die Rechtfertigung, dass man hohe Auflösungen braucht oder: Ziel der Arbeit war unter anderem, hetauszufinden, wieviel qualitativ und quantitativ neue Effekte sichtbar werden, wenn die Auflösung hoch genug ist. 
% Evtl. sollte man hier auch nochmal erwähnen, wie viele Fasern ein durchschittlicher Bizeps in der Realität hat?

One goal of our simulation studies is to evaluate the required mesh width and the necessary number of fibers to obtain accurate simulation results of surface EMG signals.
Experimental studies reveal a large variation in the number of muscle fibers in a real biceps brachii muscle. MacDougall et al. estimate in vivo numbers for elite and intermediate bodybuilders and untrained control subjects and find comparable numbers for these groups \cite{MacDougall1984}. They determine $278.5 \pm 60.7 \times \num{1e3}$ muscle fibers for the group untrained subjects.
Thus, we simulate scenarios with different 3D mesh resolutions and numbers of fibers up to the realistic number of \num{273529}. By comparing the obtained simulation results, we can determine if certain effects are only visible for high resolutions.

We consider a scenario with 100 MUs and increase the spatial resolution and the number of processes that execute the computation on the supercomputer Hawk at the High Performance Computing Center Stuttgart. Each compute node consists of two AMD EPYC 7742 processors with 64 cores each, a clock frequency of \SI{2.25}{\giga\hertz} and \SI{256}{\giga\byte} memory per node.

% fibers mesh
\begin{figure}
  \centering%
  \includegraphics[width=0.5\textwidth]{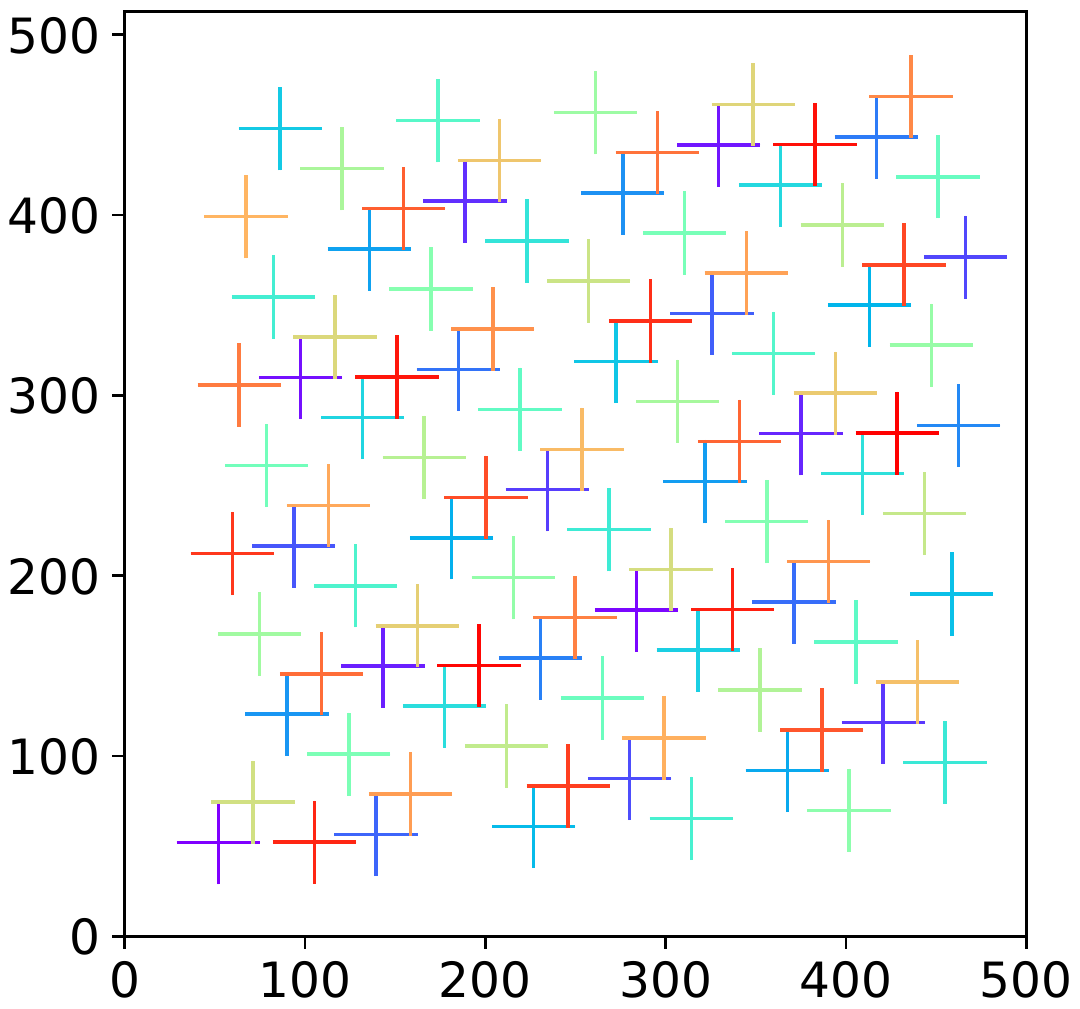}%
  \caption{Fiber based EMG simulation for the upper arm (biceps) model; study of different mesh widths, MU territory center points. The shown center points of the 100 motor units are used in all different scenarios within the study of different mesh widths.}%
  \label{fig:MU_fibre_distribution_523x523_100mus_txt_mu_positions}%
\end{figure}

Our simulated scenarios consider between \num{1369} and \num{273529} fibers. The specified number of 100 MUs has to be assigned to these numbers of fibers for each scenario.
We use the method 1a of the algorithm described in \cref{sec:muscle_fibers_and_motor_units}. The MU territories are centered around quasi-randomly generated center points, as shown in \cref{fig:MU_fibre_distribution_523x523_100mus_txt_mu_positions}. It can be seen that the MU territory center points are homogeneously distributed in space.

% MU distributions
\begin{figure}
  \centering%
  \begin{subfigure}[t]{0.45\textwidth}%
    \centering%
    \includegraphics[width=\textwidth]{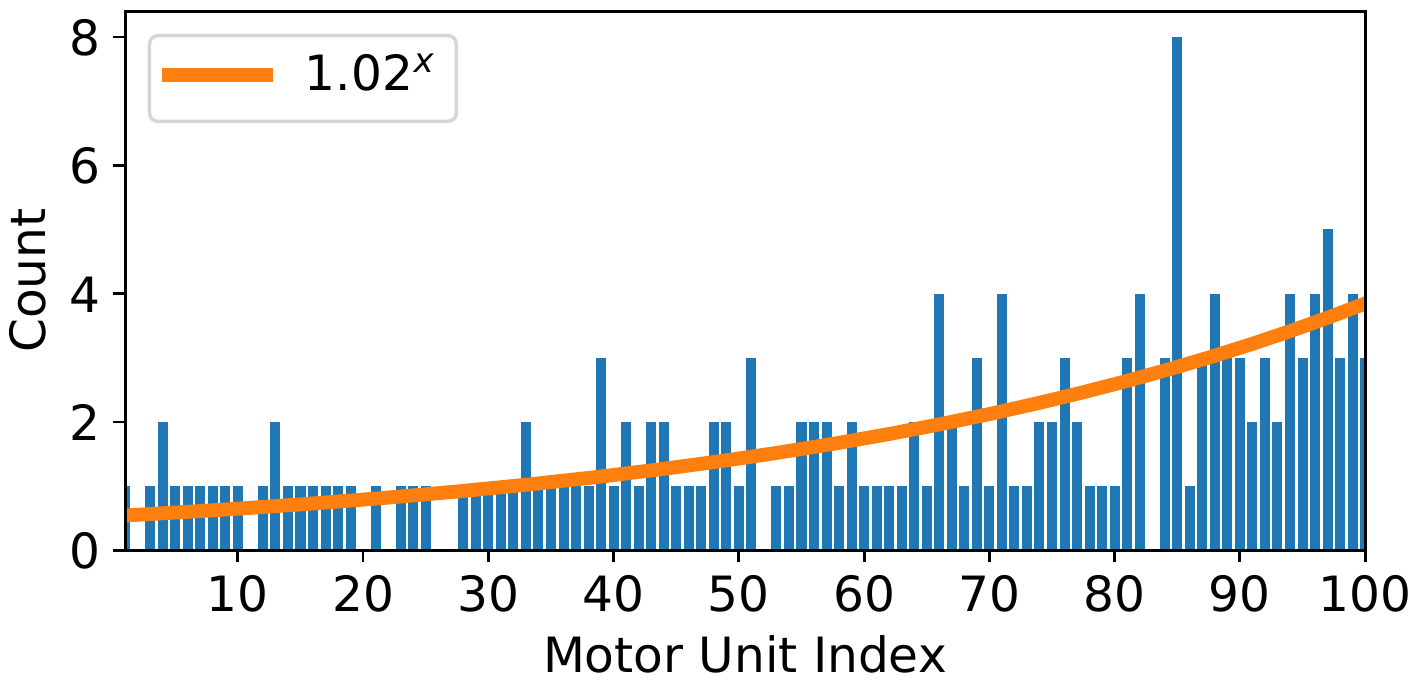}%
    \caption{Size distribution for $13\times 13 = 169$ fibers.}%
    \label{fig:mus_13}%
  \end{subfigure} \quad
  \begin{subfigure}[t]{0.45\textwidth}%
    \centering%
    \includegraphics[width=\textwidth]{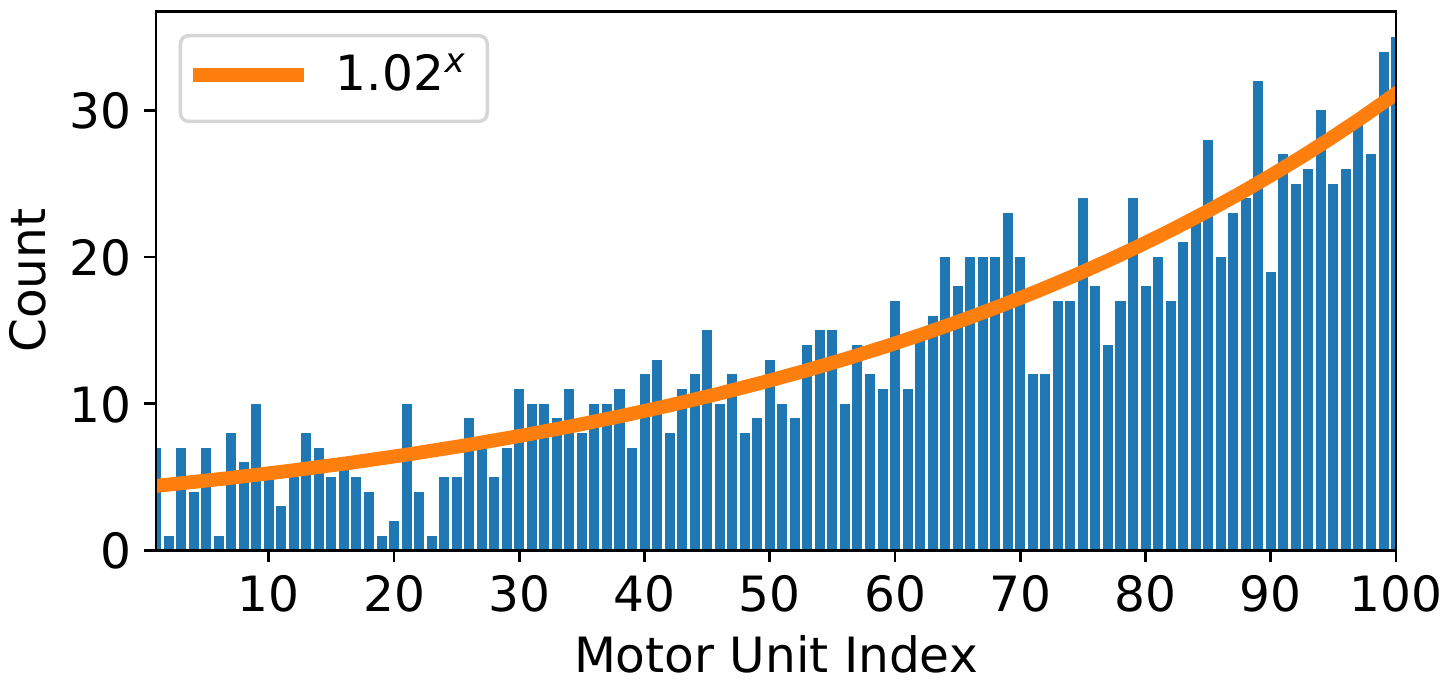}%
    \caption{Size distribution for $37\times 37 = 1369$ fibers.}%
    \label{fig:mus_37}%
  \end{subfigure}  \\
  \begin{subfigure}[t]{0.45\textwidth}%
    \centering%
    \includegraphics[width=\textwidth]{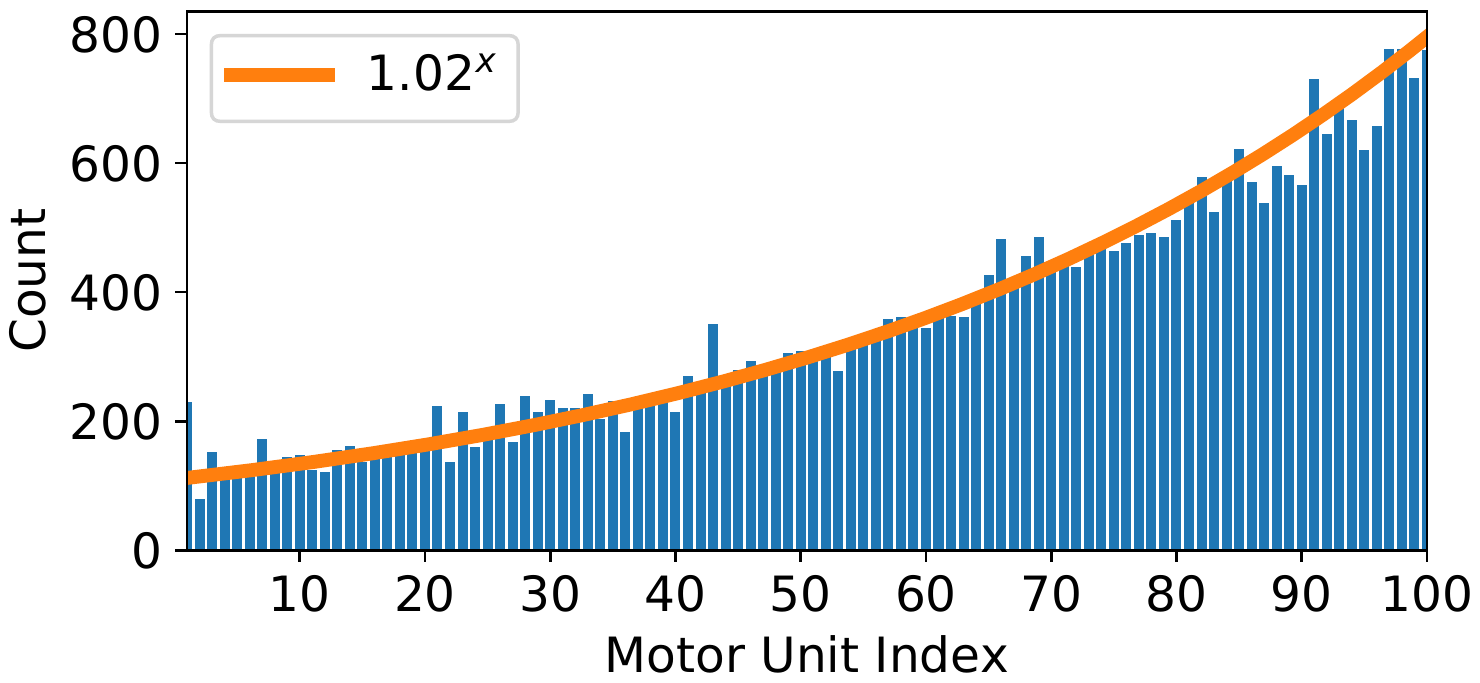}%
    \caption{Size distribution for $187\times 187 = \num{34969}$ fibers.}%
    \label{fig:mus_187}%
  \end{subfigure} 
  \begin{subfigure}[t]{0.45\textwidth}%
    \centering%
    \includegraphics[width=\textwidth]{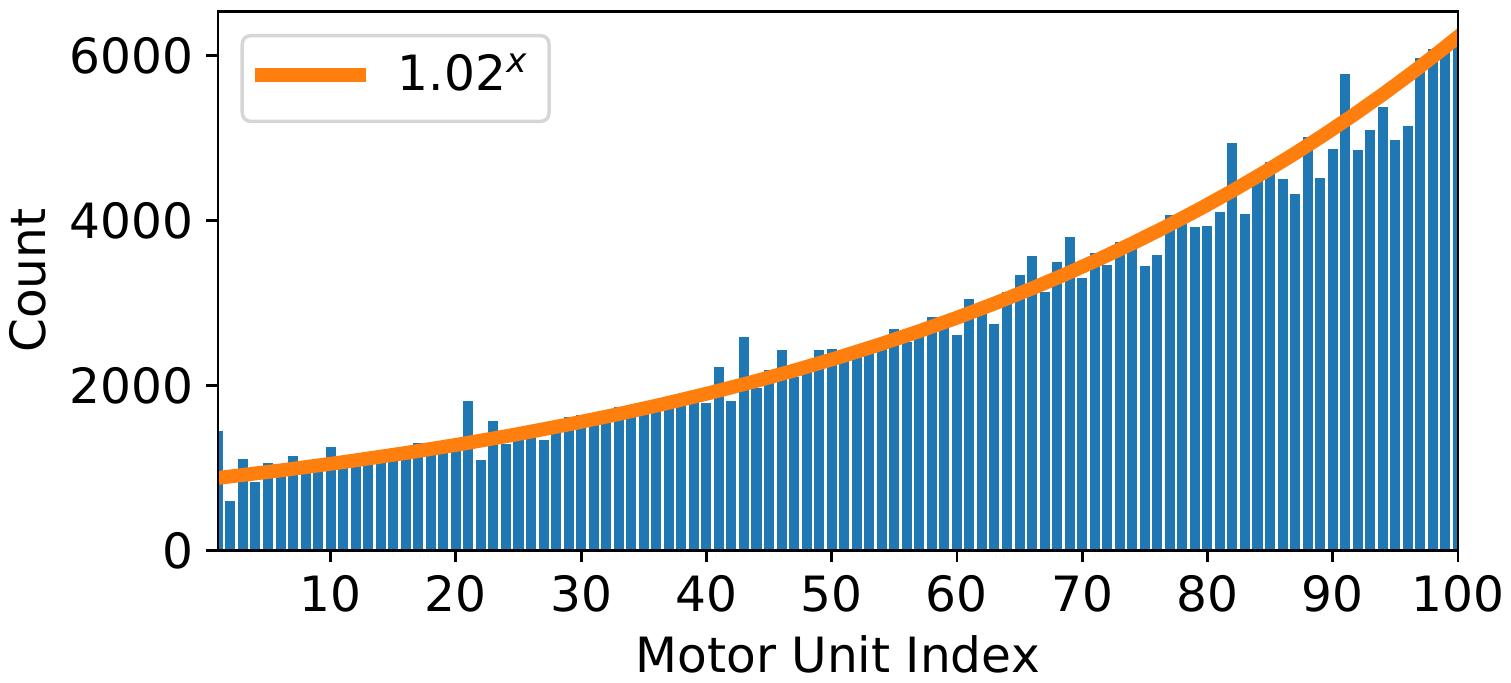}%
    \caption{Size distribution for $523\times 523 = \num{273529}$ fibers.}%
    \label{fig:mus_523}%
  \end{subfigure}   
  \caption{Fiber based EMG simulation for the upper arm (biceps) model; Distribution of the sizes of the 100 MUs in the scenarios with different number of fibers.}%
  \label{fig:mu_sizes_100mus}%
\end{figure}%

For every fiber, the algorithm assigns a MU with a close center point with higher probability than a MU whose center is located further away. The total number of fibers per MU is progressing exponentially for the MUs from 1 to 100. The progression is described by an exponential function with basis $1.02$. \Cref{fig:mu_sizes_100mus} shows the MU size distributions for four scenarios with increasing numbers of fibers from \num{169} to \num{273529}. For 169 fibers in \cref{fig:mus_13}, not all 100 MUs get associated with a fiber. Further, it can be seen that the error of the actual size distribution to the exponential function decreases with increasing number of fibers. For the largest scenario in \cref{fig:mus_523}, the MU sizes range from 602 to 6097 fibers.

The number of approximately $\num{3e6}$ fibers in the largest scenario matches the realistic number in a biceps muscle \cite{MacDougall1984}. The number of MUs can be higher in reality, approximately by a factor of 5 \cite{Feinstein1955,MacIntosh2006}. Thus, the modeled MUs in this scenario can be seen as a combination of multiple real MUs. Especially the smallest MUs, which in reality can consist of only some dozens of fibers, are lumped by the first few MUs in our scenario. We restrict the number of MUs to 100 to be able to simulate the same problem also with smaller resolutions, e.g., with only 169 fibers.

\begin{figure}
  \centering%
  \begin{subfigure}[t]{0.47\textwidth}%
    \centering%
    \includegraphics[width=\textwidth]{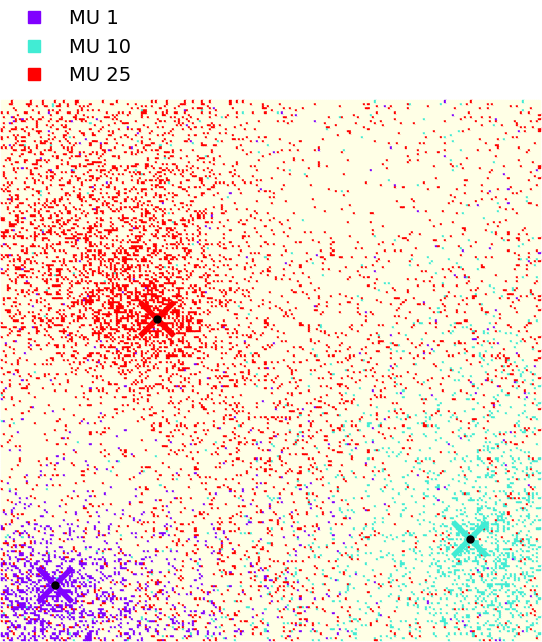}%
    \caption{Partial problem with a quarter of the whole set of fibers and 25 MUs, which occurs in the algorithm 1a described in \cref{sec:muscle_fibers_and_motor_units}. Only three MUs are shown. The MU territory centers are indicated by crosses.}%
    \label{fig:mu_assignment_part0}%
  \end{subfigure} \quad
  \begin{subfigure}[t]{0.47\textwidth}%
    \centering%
    \includegraphics[width=\textwidth]{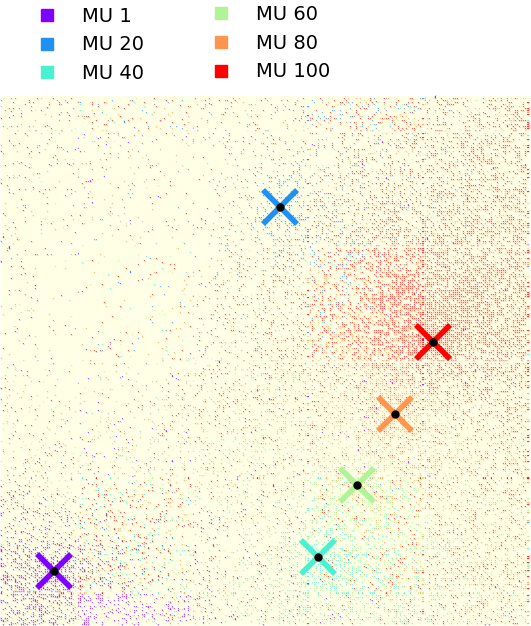}%
    \caption{The resulting assignments of MUs to fibers after four parts similar to (a) have been combined, here only shown for six MUs.}%
    \label{fig:mu_assignment_total}%
  \end{subfigure} 
  \caption{Fiber based EMG simulation for the upper arm (biceps) model; association of MUs to the fibers. The square domain corresponds to a cross-section in the muscle, every colored point is one fiber, and the color corresponds to the MU. }%
  \label{fig:mu_assignment_100}%
\end{figure}%

As described in \cref{sec:muscle_fibers_and_motor_units}, the MU assignment algorithm ensures that neighboring fibers are part of different MUs, by splitting the assignment problem for the given set of fibers into four smaller problems and then interleaving the results of the four parts. \Cref{fig:mu_assignment_part0} shows the first of these four parts, where 25 MUs are associated to a subset of the fibers for the largest scenario with \num{273529} fibers. It can be seen that the three visualized MUs are largely clustered around their MU territory centers.

The final association of fibers and MUs is given in \cref{fig:mu_assignment_total}. Six selected MUs are shown, of which the first, MU 1, corresponds to the first MU in \cref{fig:mu_assignment_part0}. The figure shows that the fibers, especially the ones of the larger MUs, are distributed far across the muscle. Comparing the smallest MU, MU 1, with the largest MU, MU 100, gives an impression of the MU size differences in this scenario.

\begin{table}
  \centering%
  \begin{tabular}{|r|c|c|c|r|r|c|c|}
    \hline
    \#fibers        & \multicolumn{2}{c|}{3D stride} & 2D surface  & 3D dofs    & 0D dofs & \#proc. & \#comp.\\
    \cline{2-3}
                         & $x,y$ & $z$                & mesh       & (k=1000)      &  && nodes\\
    \hline
    $37^2=\num{1369}$    & 2     & 8  & $19 \times 186$  & \num{67}\,k     & \num{8109}\,k  & \num{144} & 3\\[2mm]  % 6x6x4
    $67^2=\num{4489}$    & 2     & 4  & $34 \times 371$  & \num{428}\,k     & \num{26592}\,k  & \num{448} & 7\\[2mm]  % 7x8x8
    $109^2=\num{11881}$  & 2     & 3  & $55 \times 495$  & \num{1497}\,k     & \num{70383}\,k  & \num{1152} & 18\\[2mm]  % 12x12x8
    $187^2=\num{34969}$  & 2     & 2  & $94 \times 741$  & \num{6547}\,k     & \num{207156}\,k  & \num{3600} & 57\\[2mm] % 15x15x16
    $277^2=\num{76729}$  & 2     & 1  & $139 \times 1481$  & \num{28614}\,k  & \num{454542}\,k  & \num{7744} & 121\\[2mm] % 22x22x16
    $523^2=\num{273529}$ & 2     & 1  & $262 \times 1481$  & \num{101661}\,k  & \num{1620}\,M  & \num{26912} & 421\\ % 29x29x32
    \hline
  \end{tabular}
  \caption{Fiber based EMG simulation for the upper arm (biceps) model; Parameters of spatial discretization and parallel partitioning. The 3D stride refers to the stride with which the 3D mesh is generated from the 0D points. The 2D surface is the output of the EMG and corresponds to one face of the 3D mesh.}%
  \label{tab:emg_study_parameters}%
\end{table}

The numerical parameters of the simulations are the same as in the last section. The scenario is computed for a simulation time span of $\SI{1}{\s}$. The MUs are activated in a ramp every $\SI{2}{\ms}$ such that all MUs are active after $\SI{200}{\ms}$. The fiber radius
and the stimulation frequency for the MUs are exponentially distributed with basis $1.02$, similar to the MU size. The fiber radius increases from \SI{40}{\micro\meter} to \SI{55}{\micro\meter}, and the stimulation frequency decreases from \SI{24}{\hertz} to \SI{7}{\hertz} for MUs 1 to 100. A random frequency jitter of \SI{10}{\percent} is assumed.

The surface to volume ratio $A_m$ of the membrane is determined by assuming a cylindrical shape and can be computed from the fiber radius $r$ as $A_m = 2/r$ \cite{Klotz2020}. We model \SI{70}{\percent} slow twitch and \SI{30}{\percent} fast twitch fibers. Accordingly, the membrane capacitance $C_m$ is set to $C_m = \SI{0.58}{\micro\farad\per\centi\meter\squared}$ for the 70 smallest MUs and to $C_m = \SI{1}{\micro\farad\per\centi\meter\squared}$ for the 30 largest MUs.

\Cref{tab:emg_study_parameters} lists the spatial discretization and parallel partitioning parameters. The first column shows the number of fibers. Their number increases, however, the mesh resolution of every 1D fiber mesh stays constant at 1480 elements per fiber. The stride that defines the 3D mesh is given in the second and third columns. The stride in radial direction of the muscle, i.e., in the $x$ and $y$ coordinate directions, stays constant. Because the fiber density increases, the 3D mesh is refined accordingly. The stride along the fibers, i.e., in $z$ direction is reduced, such that the mesh widths of the 3D mesh in all three coordinate directions remain balanced.

The resulting EMG recordings of each simulation are described by 2D meshes, which contain the values of the 3D muscle meshes without fat layer on the surface at one side of the muscle.  The fourth column in \cref{tab:emg_study_parameters} lists the dimensions of these surface meshes. 

The next two columns list the number of dofs in the 3D mesh and the number of dofs in all fibers. For these scenarios, it is not practical to output the 3D mesh or the 1D fiber meshes in regular time intervals, because this would produce large amounts of data that could hardly be processed. Instead, we only output the 2D surface mesh in the ParaView format every \SI{10}{\milli\second}.

The last two columns in \cref{tab:emg_study_parameters} show the numbers of processes and compute nodes that are used on Hawk. One compute node contains 128 physical cores, four cores share a $\SI{16}{\mebi\byte}$ level three (L3) cache. However, we decide to use only 64 cores per compute node, i.e., two cores per L3 cache, because measurements showed that this reduces the overall computation times more than it increases the runtime due to the decreased parallelism. The total computation time of this scenario with a timespan of $\SI{1}{\s}$ is $\SI{2}{\hour}$ $\SI{20}{\minute}$ for the scenario with \num{76729} fibers and 7744 processes.

\begin{figure}
  \centering%
  \begin{subfigure}[t]{0.19\textwidth}%
    \centering%
    \includegraphics[height=12cm]{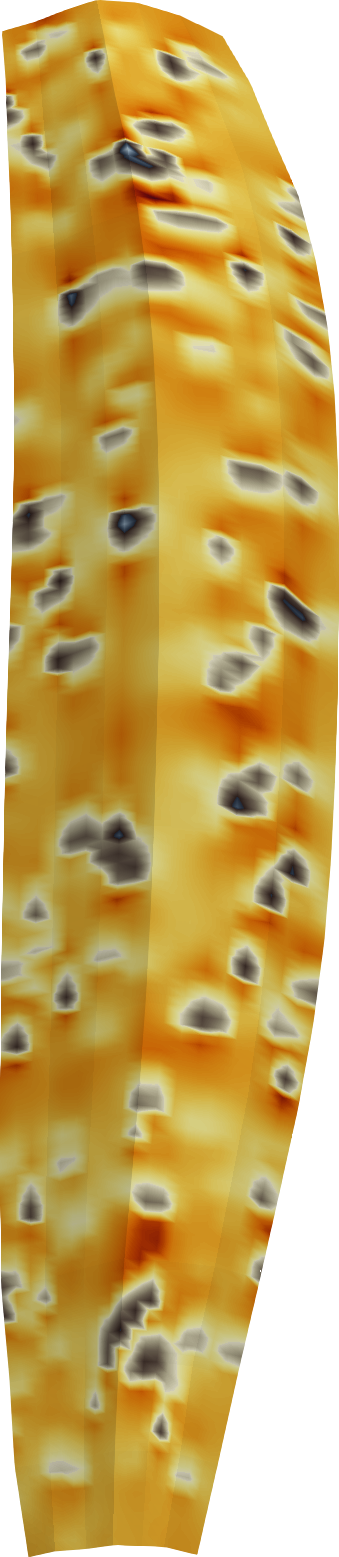}%
    \caption{$1369$ fibers.}%
    \label{fig:emg37}%
  \end{subfigure} \,
  \begin{subfigure}[t]{0.19\textwidth}%
    \centering%
    \includegraphics[height=12cm]{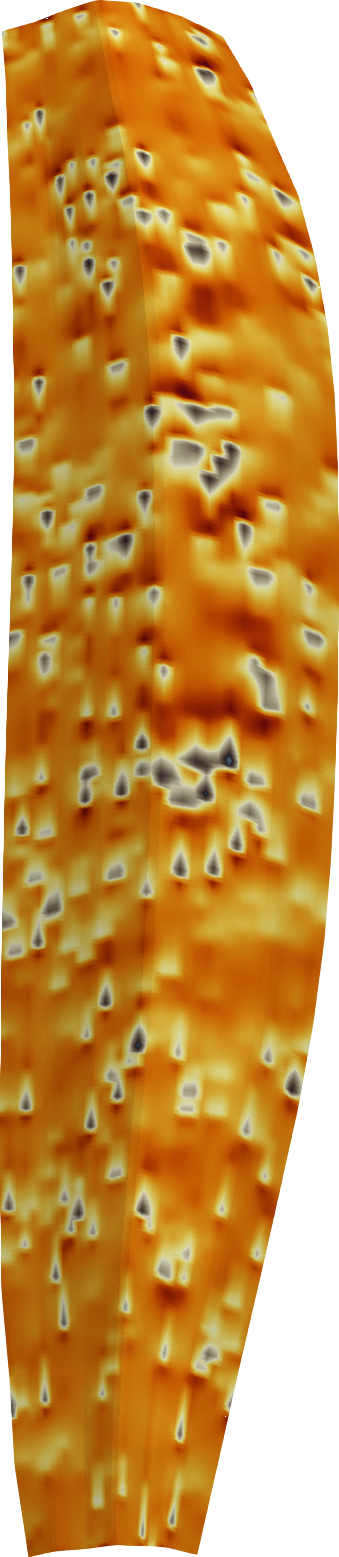}%
    \caption{$4489$ fibers.}%
    \label{fig:emg67}%
  \end{subfigure}  \,
  \begin{subfigure}[t]{0.19\textwidth}%
    \centering%
    \includegraphics[height=12cm]{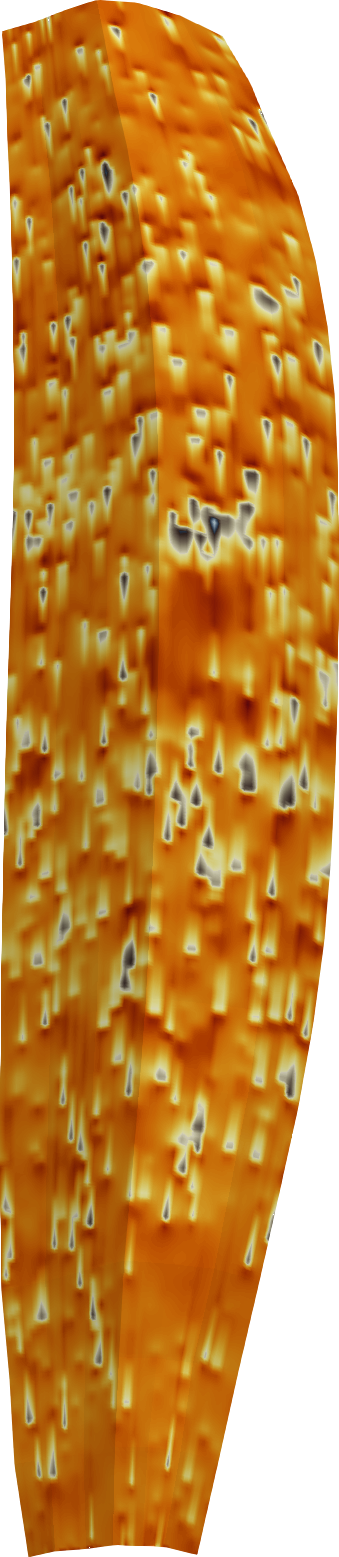}%
    \caption{$\num{11881}$ fibers.}%
    \label{fig:emg109}%
  \end{subfigure}    \,
  \begin{subfigure}[t]{0.35\textwidth}%
    \centering%
    \includegraphics[height=12cm]{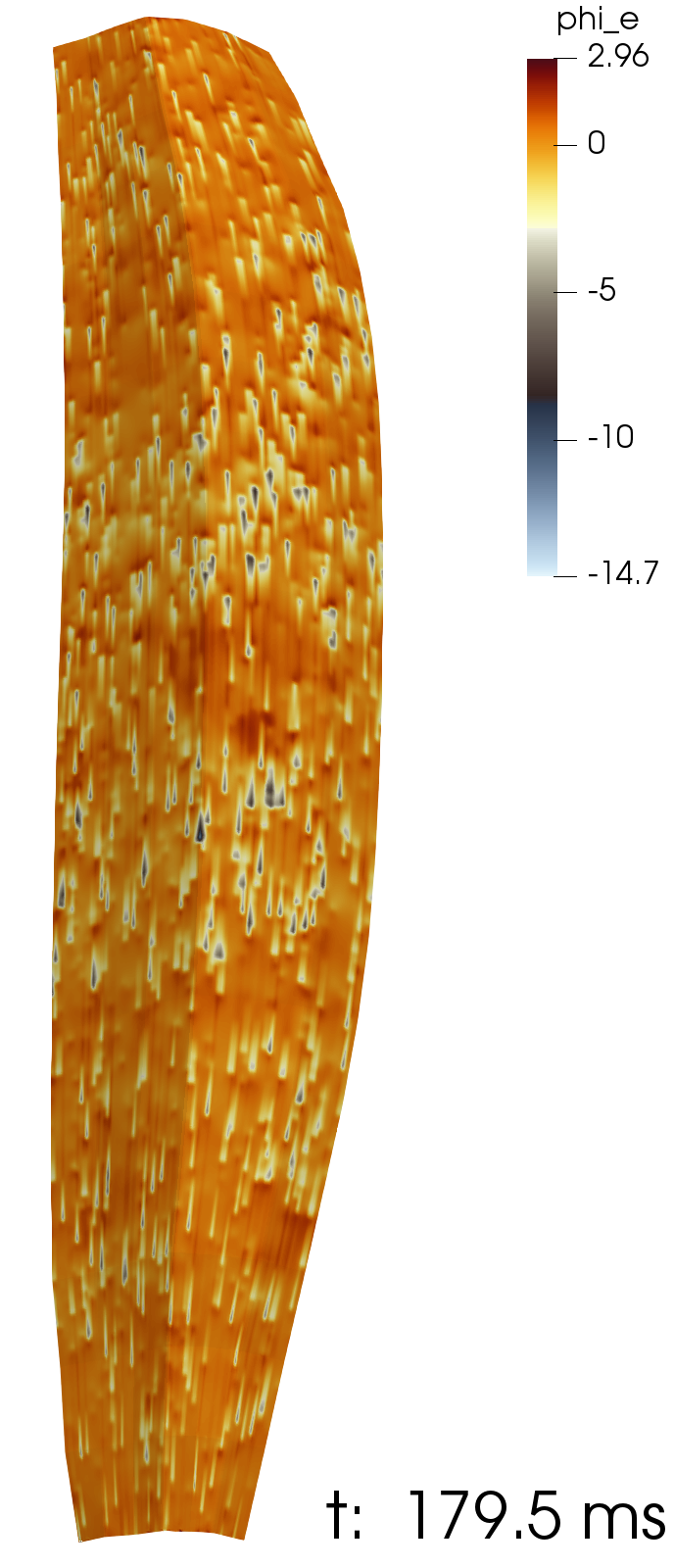}%
    \caption{$\num{34969}$ fibers.}%
    \label{fig:emg187}%
  \end{subfigure}   
  \caption{Fiber based EMG simulation for the upper arm (biceps) model; simulated surface EMG signals for different numbers of fibers and different mesh widths of the 3D mesh, see \cref{tab:emg_study_parameters} for details. The same color coding of the EMG signal $\phi_e$ is used in the four scenarios.}%
  \label{fig:emg_hpc}%
\end{figure}

\begin{figure}
  %\centering%
  \begin{subfigure}{0.30\textwidth}%
    \centering%
    \includegraphics[height=125mm]{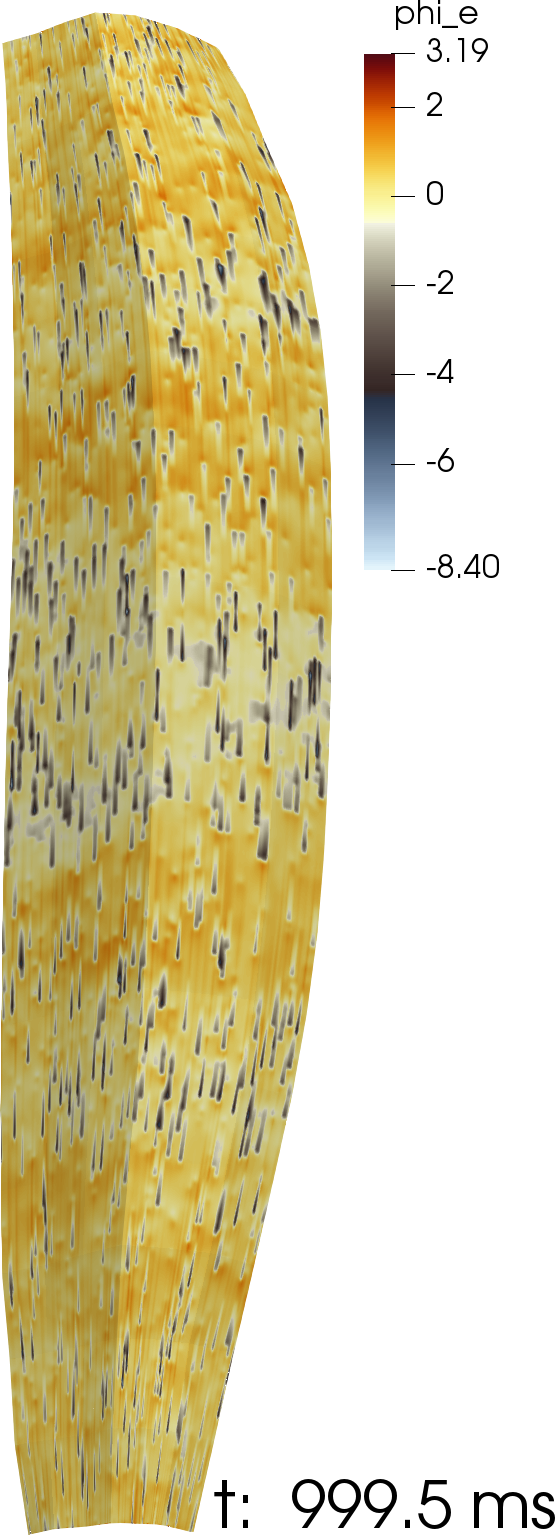}%
    \caption{\num{76729} fibers.}%
    \label{fig:emg277}%
  \end{subfigure}\,
  \begin{subfigure}{0.30\textwidth}%
    \centering%
    \includegraphics[height=125mm]{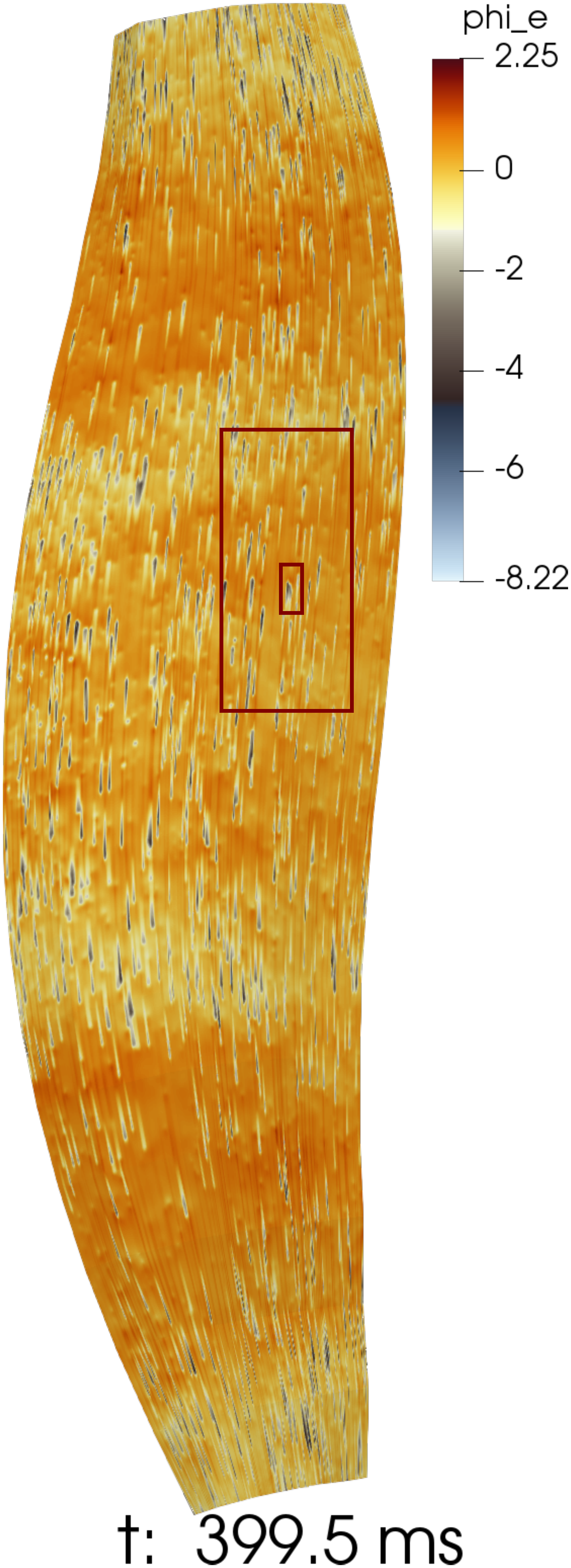}%
    \caption{\num{273529} fibers.}%
    \label{fig:emg273529b}%
  \end{subfigure}\hspace{-1cm}
  \begin{minipage}{0.5\textwidth}
    \vspace{48mm}
    \begin{subfigure}[t]{0.5\textwidth}%
      \centering%
      \includegraphics[height=77mm]{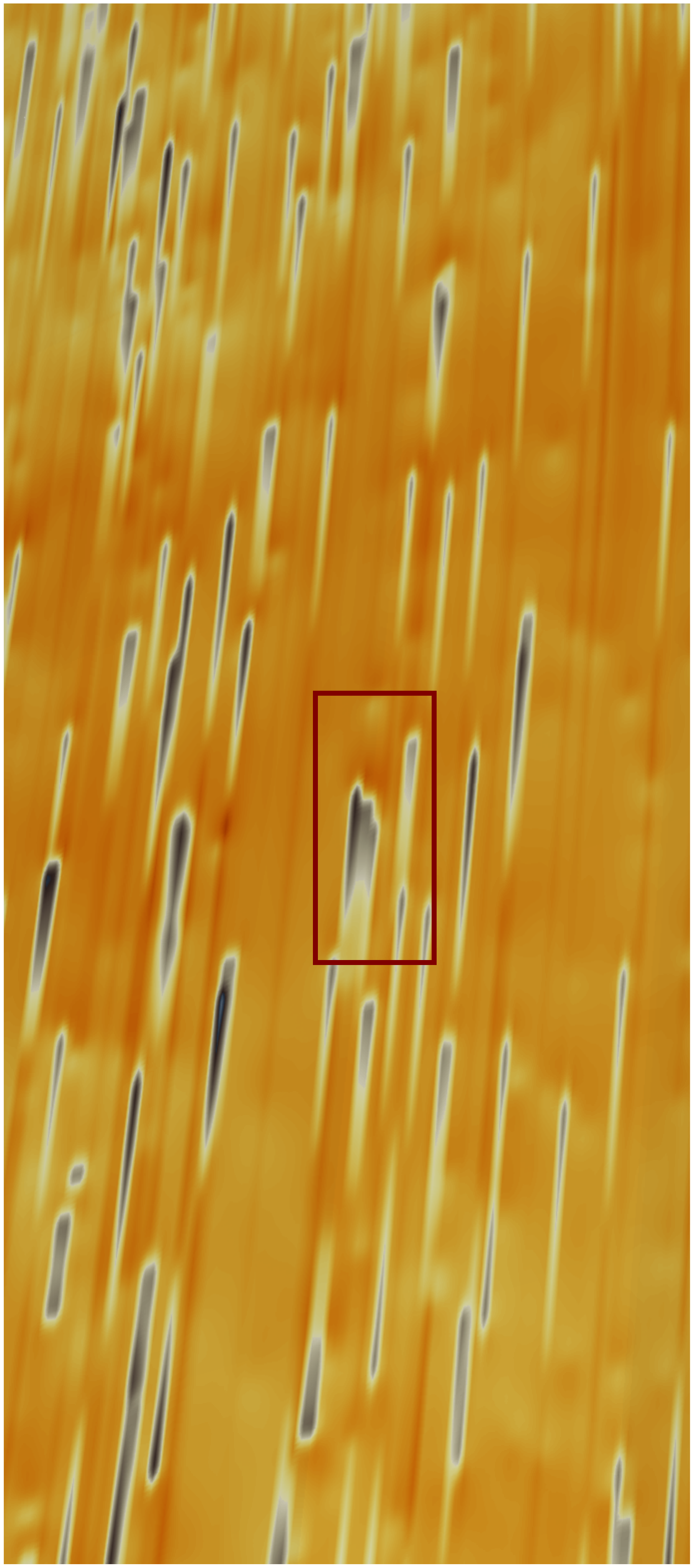}%
      \caption{Detail view.}%
      \label{fig:emg273529c}%
    \end{subfigure}\hspace{-5mm}
    \begin{subfigure}[t]{0.49\textwidth}%
      \centering%
      \includegraphics[height=77mm]{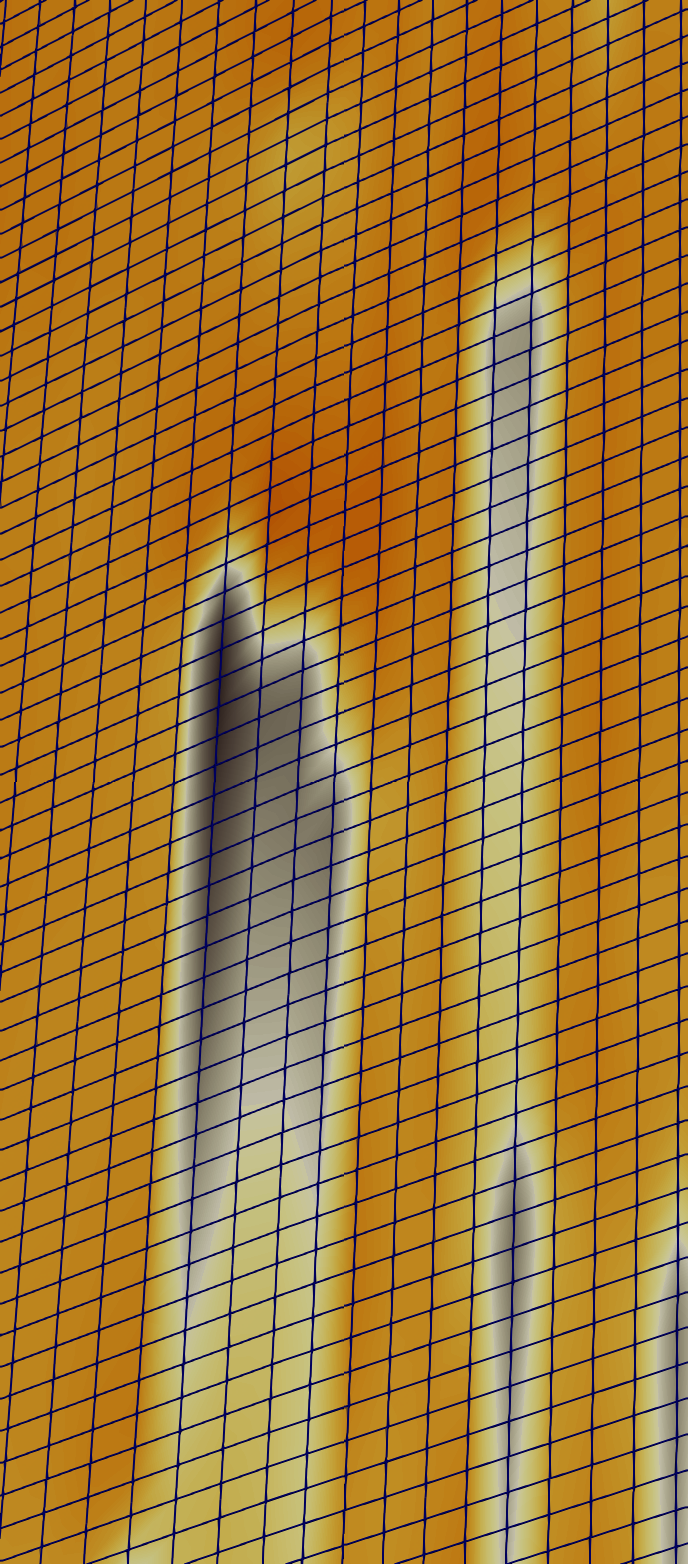}%
      \caption{Detail view.}%
      \label{fig:emg273529d}%
    \end{subfigure}
  \end{minipage}
  \caption{Fiber based EMG simulation for the upper arm (biceps) model; simulated surface EMG signals with realistic fiber counts, continued from \cref{fig:emg_hpc}. Figure (a) shows a scenario with \num{76729} fibers, which is approximately a third of the realistic number for the biceps muscle. Figures (b)-(d) show the result with a realistic number of \num{273529} fibers. (c) shows a detail view of (b) indicated by the outer red box. (d) shows another zoomed in view of (b) and (c), also indicated by the red boxes.}%
  \label{fig:emg_hpc2}%
\end{figure}

\Cref{fig:emg_hpc,fig:emg_hpc2} show the resulting surface EMG signals for different resolutions. The color visualizes the value of the extracellular potential $\phi_e$ according to the shown color bar. Because of sign conventions in the definitions of the electric potentials, the spikes in the EMG signals, which result from the action potentials, are negative.

The resulting electric potential in \cref{fig:emg_hpc} exhibits different regions of higher activation that move over time from the center of the muscle towards its ends. The size of these regions at time ${t=\SI{179.5}{\ms}}$ decreases from \cref{fig:emg37} to \cref{fig:emg187} as the mesh width decreases. 
Dark-colored strong signals can be seen, which mainly correspond to fibers that are located directly underneath the shown muscle surface.
Apart from these strong signals, also weaker artifacts occur, which are shown in yellow and orange colors. They result from the superposition of several fibers of the same or different MUs. The number of recognizable weak signals is higher for the simulations with higher numbers of fibers and finer mesh resolution.

The four scenarios in \cref{fig:emg_hpc} share the material parameters, territory centers and relative size distributions of the 100 MU and the activation scheme. However, the location of the neuromuscular junctions is determined randomly and varies between the scenarios. Therefore, the resulting EMG signals are not refined images of each other.  However, a similarity of activated regions on a coarse scale can be observed in all scenarios.

% results for different mesh widths

\Cref{fig:emg_hpc2} shows two more scenarios with many fibers at times of $t\approx \SI{1}{\s}$ and $t\approx \SI{0.4}{\s}$. The scenario in \cref{fig:emg277} simulates approximately \num{76e3} fibers, which is in the order of a third of the realistic number of fibers in the biceps muscle. 
This scenario can also be interpreted as only activating a third of the available fibers in the muscle, resulting in the respective lower percentage of maximum voluntary contraction force.

\Cref{fig:emg273529b} shows the scenario where a realistic number of \num{273e3} fibers was simulated. The computational effort for these two scenarios can only be tackled with High Performance Computing. The last two rows of \cref{tab:emg_study_parameters} show that 121 and 421, respectively, compute nodes were used for the computations.

\Cref{fig:emg273529c,fig:emg273529d} present details of the simulated surface EMG of the scenario in \cref{fig:emg273529b}. The cut-outs are indicated by the red boxes in \cref{fig:emg273529b,fig:emg273529c}. \Cref{fig:emg273529d} also visualizes the elements of the 2D and 3D meshes. In both scenarios of \cref{fig:emg_hpc2}, the 3D mesh is as finely resolved in $z$ direction (vertical in \cref{fig:emg273529d}) as the muscle fibers. In transversal direction (horizontal in \cref{fig:emg273529d}), the element sizes are twice as large as the spacing between the fibers. 

The right side of \cref{fig:emg273529d} shows two almost aligned action potentials that propagate towards the top of the image. The upper action potential originates from a fiber that is located at the center between the element boundaries. Its electric potential is distributed to two adjacent nodes on the surface mesh, having the same activated values.
In contrast, the lower action potential results from a fiber that is directly located on the element boundaries. It can be seen that the activation on the surface decays rapidly in transverse (horizontal) direction, as the neighbor elements already almost exhibit the same electric potential as the background level. This underlines the importance to use finely resolved meshes to accurately represent surface EMG signals.

% results for different mesh widths

\begin{reproduce_no_break}
  Use the following commands to run the EMG simulation of the biceps muscle with fat layer and electrodes:
  \begin{lstlisting}[columns=fullflexible,breaklines=true,postbreak=\mbox{\textcolor{gray}{$\hookrightarrow$}\space}]
    cd $\$$OPENDIHU_HOME/examples/electrophysiology/fibers/fibers_fat_emg/build_release
    mpirun -n 16 fibers_fat_emg ../settings_fibers_fat_emg.py 50mus.py
    cd out/50mus
    plot_emg.py ./electrodes.csv ./stimulation.log 25900 26000    # plot the result, here for time span 25.9s - 26s
  \end{lstlisting}
\end{reproduce_no_break}

\subsection{Simulation of EMG electrodes}\label{sec:simfiber_electrodes}
% fibers_fat_emg with electrodes

While the surface EMG simulation results as presented in the last section in \cref{fig:emg_hpc} are suited for insights into the temporal and spatial variation of the electric potential, real experiments are constraint to capture values at the discrete locations of electrodes.
For some applications such as the evaluation of EMG decomposition algorithms, it is beneficial to obtain simulated values at electrode locations.

One possibility would be to extract nodal values from the simulated surface meshes to simulate electrodes. However, the distance between the nodes in the mesh is not constant in the whole mesh, whereas EMG electrode arrays have a fixed inter-electrode spacing. We, therefore, follow a different approach and allow to directly specify a grid of electrodes close to the muscle surface. These points are then mapped onto the surface of the muscle and the respective values are calculated by evaluating the finite element interpolant at the respective locations.

In OpenDiHu, a 2D grid of surface electrodes can be defined in the Python settings file by specifying the grid parameters and inter electrode distances. As a result, the simulation distributes the electrodes to the processes according to the parallel partitioning of the 3D mesh, evaluates the computed EMG values at the respective locations and outputs them in a single text file of comma separated values.

\begin{figure}
  \centering%
  \includegraphics[width=0.7\textwidth]{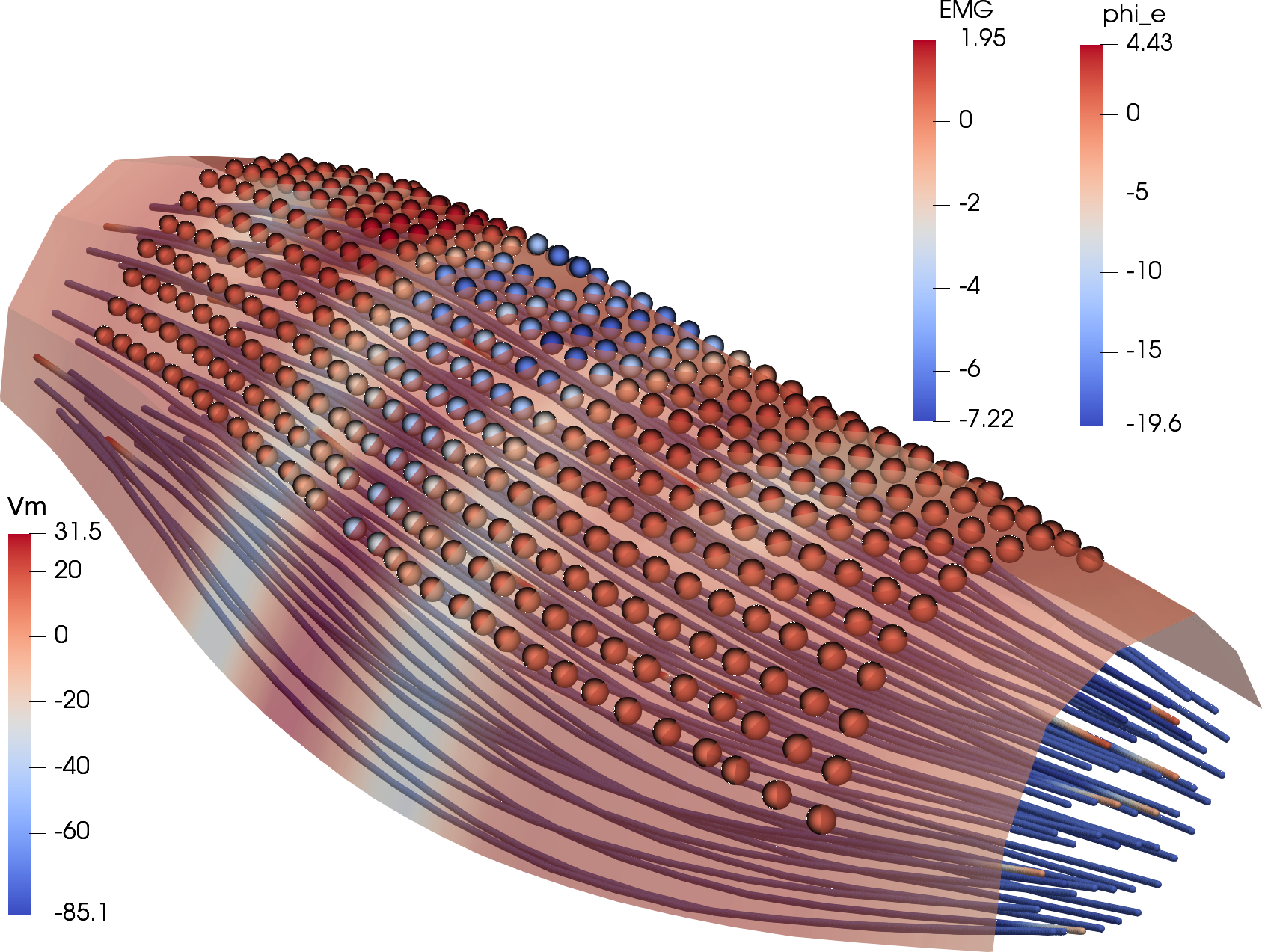}%
  \caption{Fiber based EMG simulation for the upper arm (biceps) model; simulation of surface EMG and capturing electrodes. The scenario contains 49 muscle fibers, a fat layer, of which only the surface is shown, and a grid of $12 \times 32$ equidistant electrodes.}%
  \label{fig:fibers_fat_emg2_electrodes}%
\end{figure}

\Cref{fig:fibers_fat_emg2_electrodes} shows simulation results of the fiber based electrophysiology model with 49 fibers, fat layer and an array of $12\times 32$ electrodes. The electrodes are visualized as spheres. The muscle fibers below the fat layer are colored according to the transmembrane voltage $V_m$. Only the upper surface of the fat layer is shown and colored according to the extracellular potential $\phi_e$. The EMG electrodes capture the values of the scalar field $\phi_e$ at their locations. The color coding for the electrodes has a different \emph{EMG} color scale to make the resulting signals more distinguishable. Two activated bands across the muscle surface can be seen, which are also present in the electrode values.

\begin{figure}
  \centering%
  \begin{subfigure}[t]{0.6\textwidth}%
    \centering%
    \includegraphics[width=\textwidth]{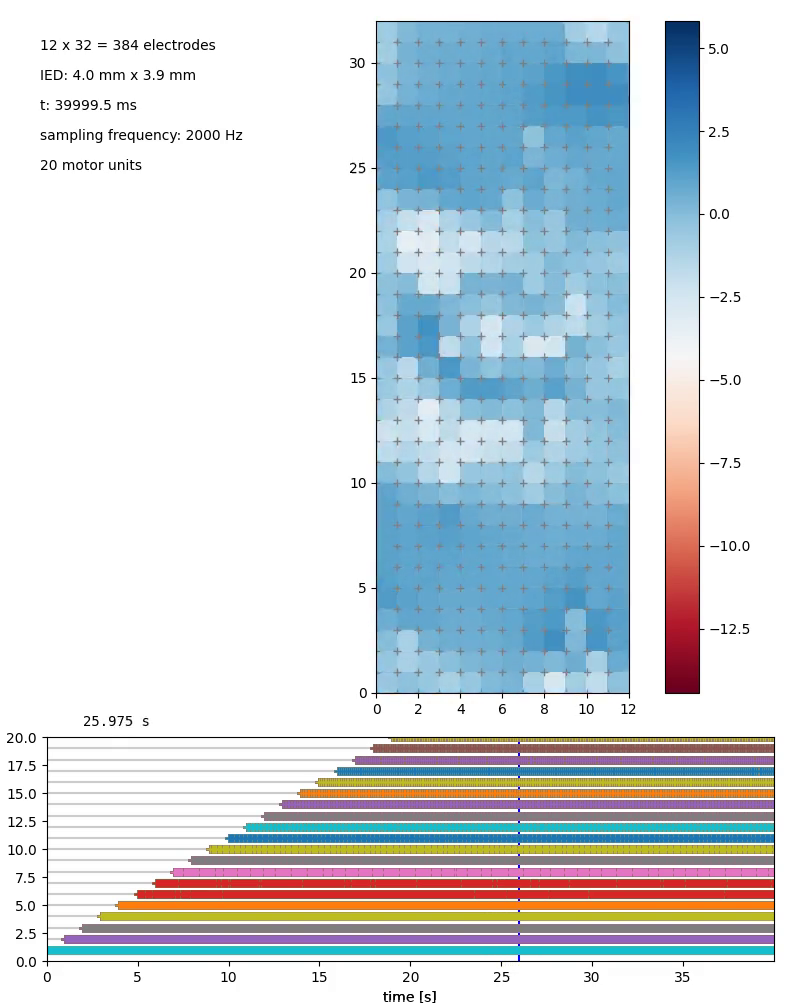}%
    \caption{Snapshot of an animation of the simulation results that can be generated with the utility of OpenDiHu. The top part visualizes the current values of the EMG electrodes at time $t=\SI{25.975}{\s}$. The bottom part shows the MU activation ramp, each colored line shows to the firing time range of one MU.}%
    \label{fig:emg_video_37}%
  \end{subfigure}  \,
  \begin{subfigure}[t]{0.38\textwidth}%
    \centering%
    \includegraphics[width=\textwidth]{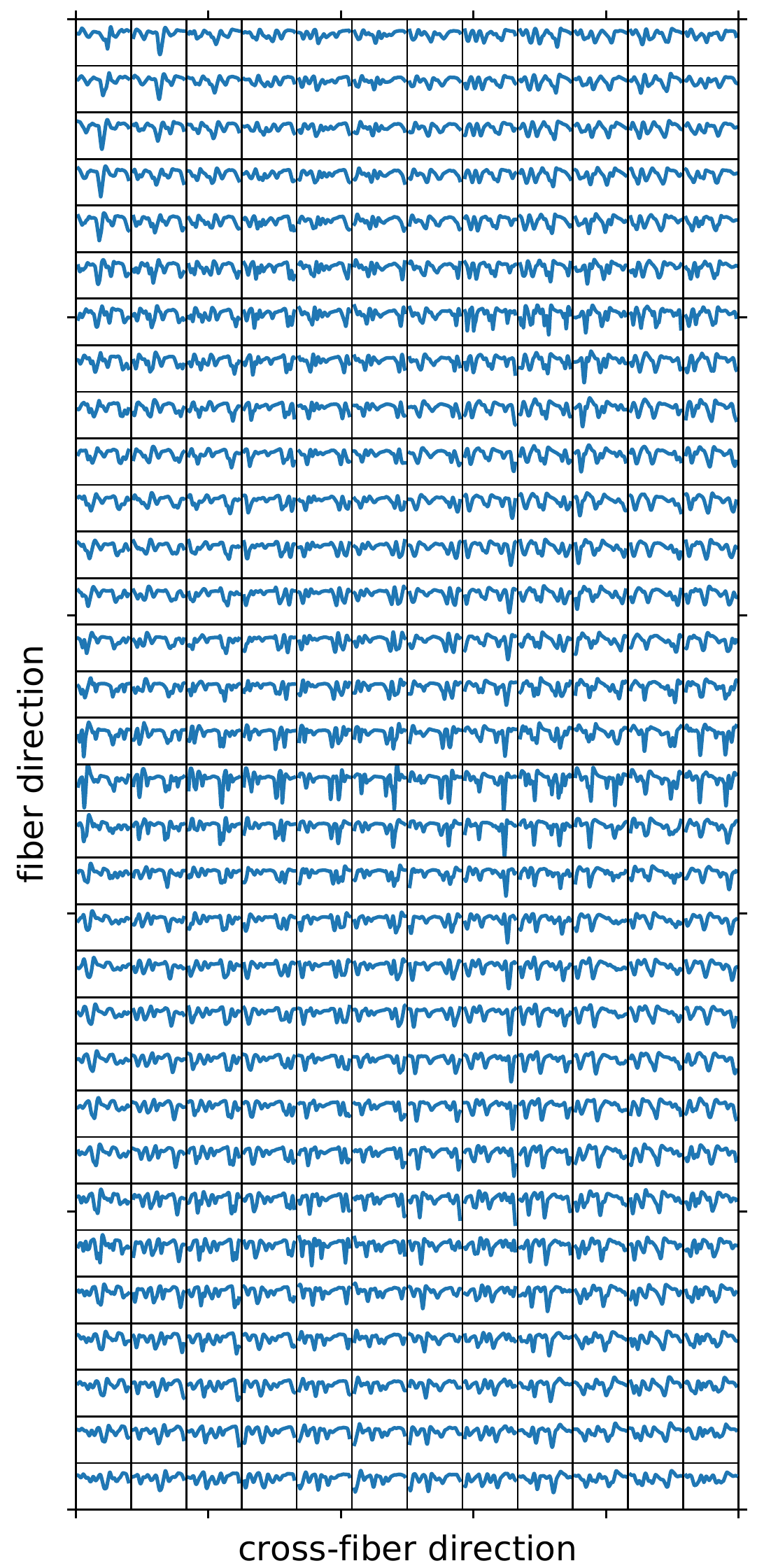}%
    \caption{Surface EMG at $12 \times 32$ electrodes in the time range $[\SI{25.9}{\s},\SI{26}{\s}]$.}%
    \label{fig:emg_video_plot}%
  \end{subfigure}
  \caption{Fiber based EMG simulation for the upper arm (biceps) model; results of surface EMG obtained at simulated electrodes.}%
  \label{fig:emg_video}%
\end{figure}

To visually evaluate the simulated EMG signals at the electrodes, OpenDiHu provides utilities to create the visualizations shown in \cref{fig:emg_video}. \Cref{fig:emg_video_37} shows a single frame from an animation. On the upper right, the grid of electrodes is displayed. The EMG signal at the electrodes is given by the colored tiles and changes over time. At the bottom of the image, the activation times of the MUs are visualized. Every horizontal line corresponds to one MU. The colored markers indicate when the respective MU fires. As the shown example visualizes data for \SI{40}{\s}, the individual firing times are not distinguishable. In the animation, a vertical bar moves over the time axis and indicates the current simulation time. The picture displays the EMG values at time $t=\SI{25.975}{\s}$. The upper left of the image shows a text with static information about the dataset, containing the electrode grid size, the inter electrode distance (IED), the end time, the sampling frequency of the electrodes, i.e., the frequency with which the computed EMG signals values are stored to the output file, and the number of MUs.

\Cref{fig:emg_video_plot} shows another, static visualization of simulated EMG data. The diagram contains boxes for all electrodes in the $12\times 32$ grid. The value of the EMG signal is plotted over time in every box for the respective electrode. \Cref{fig:emg_video_plot} visualizes the data of \cref{fig:emg_video_37} for the time interval $[\SI{25.9}{\s},\SI{26}{\s}]$.
The diagram enables experts to visually identify propagating action potentials from the tile columns. The propagation velocity of the action potentials can be estimated from the time shift of matching spikes in vertically adjacent boxes.

\subsection{Decomposition of Surface EMG Signals}\label{sec:simfiber_decomposition}

Surface EMG recordings are a valuable tool to gain insights into the neuromuscular system. They are used, e.g., for the diagnosis of muscular disorders and in clinical studies that aim to advance biomedical understanding.

% why, how are EMG signals generated -> fig
As described earlier, the EMG signals on the skin surface originate from the activated muscle fibers. Effects from volume conduction of action potentials on all muscle fibers are superpositioned and contribute to the EMG signal. The scaling of the contributions to the overall signal depends on several factors such as the distance of the fibers to the skin surface. As all fibers in the same MU get activated simultaneously, each MU`s contribution shows a characteristic \say{shape} in the resulting surface EMG signal. This shape is influenced by the number and location of the muscle fibers relative to the electrodes and the location of the neuromuscular junctions.

\begin{figure}
  \centering%
  \includegraphics[width=0.85\textwidth]{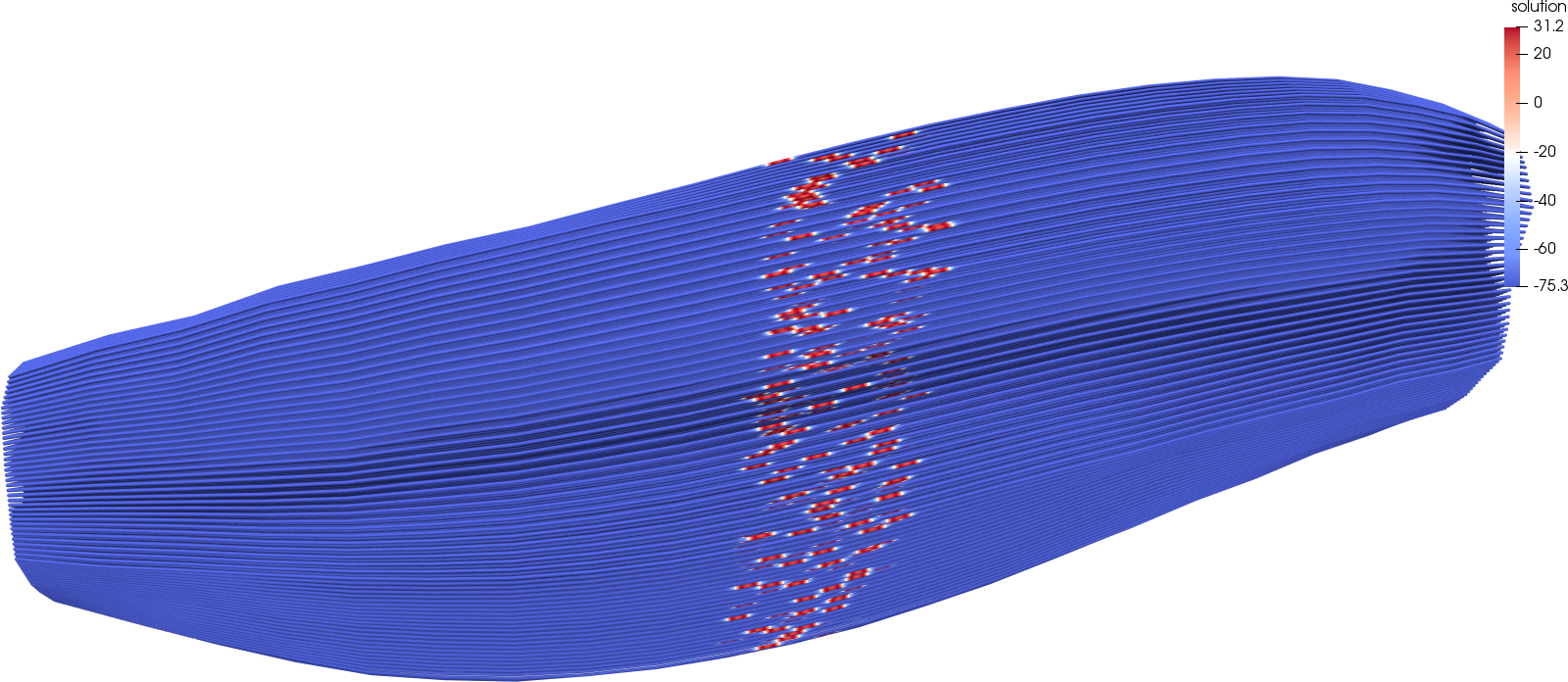}%
  \caption{Fiber based EMG simulation for the upper arm (biceps) model; a simulation result that reveals the locations of the neuromuscular junctions. The figure depicts 1369 fibers after $\SI{1}{\ms}$, which have initially been stimulated at the neuromuscular junction. The color coding corresponds to the membrane potential $V_m$, which has a positive value near the points of stimulation.}%
  \label{fig:emg_video_37_junctions}%
\end{figure}

In our simulation, the location of the neuromuscular junctions is chosen pseudo-randomly (but deterministic) during initialization in the central \SI{10}{\percent} of every muscle fiber. \Cref{fig:emg_video_37_junctions} shows the state of a simulation with $1369$ fibers at $t=\SI{1}{\ms}$, where all fibers have been activated at $t=\SI{0}{\ms}$. The color coding indicates the potential $V_m$ of the membrane, which at the shown time has only depolarized near the locations of the neuromuscular junctions.

% gCKC method
% Evtl. auch hier in 1-2 Sätzen etwas ausholen, warum man das macht: Die Simulationsdaten enthalten alle Informationen über die tatsächlich dem simulierten EMG-Signal zugrunde liegende Stimulation. Man kann sie also nutzen, um die decomposition methods zu validieren.
Methods exist that seek to decompose the surface EMG signal into the contributions of the individual MUs. Given a surface EMG recording, such methods output a number of recovered MUs and their firing times.
In our simulation studies, all relevant information is available that determines the EMG signal resulting from MU activity: the location of the fibers and their association to MUs, the positions of the neuromuscular junctions and the innervation pulses for each MU.
Thus, our simulation can be used to validate and evaluate EMG decomposition methods.

One popular EMG decomposition method is \emph{Gradient Convolution Kernel Compensation} (gCKC) \cite{Holobar2007b,Holobar2007}, which, in the following, will be outlined and then applied on simulated data.

Most decomposition methods, including the gCKC algorithm, assume that the EMG signal at an electrode is composed of the convolutional mixture of the activity of $N$ MUs.
The activity of each MU $k\in \{1,\dots,N\}$ is described by the innervation pulse trains, which activate the fibers of MU $k$, given as a point process of neural inputs at stimulation times $\varphi_r$. The source signal $s_k$ in the muscle, which represents the effect of MU $k$ is described as a spike $s_k(t) = \sum_r\delta(t - \varphi_r)$, where $\delta$ is the dirac delta function.

The vector of observed EMG value $\bfx \in \R^m$ at a time $t$ is composed of the temporal convolution over $L$ time-shifted sources $\bfs$ and a term $\bfomega$ of additive Gaussian noise:
\begin{align*}
  \bfx(t) = \s{l=0}{L-1}\bfH(l)\,\bfs(t-l) + \bfomega(t).
\end{align*}
Here, $\bfH$ is the $m\times n$ mixing matrix for $m$ observations and $n$ MU sources and $\bfs = (s_k)_{1,\dots,n}$ is the vector of source signals. The sum over $L$ previous values in this convolutive mixture can be reformulated by moving the summation into the matrix-vector product. The dimensions of the matrix $\bfH$ and the vector $\bfs$ are extended accordingly. 
An optimization problem yields the separation vectors, with which the innervation pulse trains $\varphi_r$ of the MUs can be recovered from the recorded EMG signals $\bfx$. The gCKC algorithm determines the inverse effect of applying the unknown mixing matrix by solving a derived optimization problem using a gradient descent scheme.

The gCKC decomposition algorithm is implemented in the DEMUSE software, a commercial, MATLAB based tool that allows automatic and semi-automatic EMG decomposition \cite{demuse}. In collaboration with Lena Lehmann from the \emph{Institute of Signal Processing and System Theory} and the \emph{Institute for Modelling and Simulation of Biomechanical Systems}, we evaluated the performance of gCKC decomposition on simulated surface EMG signals.

% different performance depending on parameters
We simulate fiber based electrophysiology scenarios with fat layer and 1369 fibers using the same model parameter as in \cref{sec:effects_of_the_mesh_width_emg}. In the first scenario, a fat layer with thickness of $\SI{1}{\cm}$ is modelled. The simulated EMG signal is sampled in an electrode array with a frequency of $\SI{2}{\kilo\hertz}$ and a grid size of $12 \times 32$ fibers, as shown in \cref{fig:emg_video}.

\begin{figure}
  \centering%
  \includegraphics[width=\textwidth]{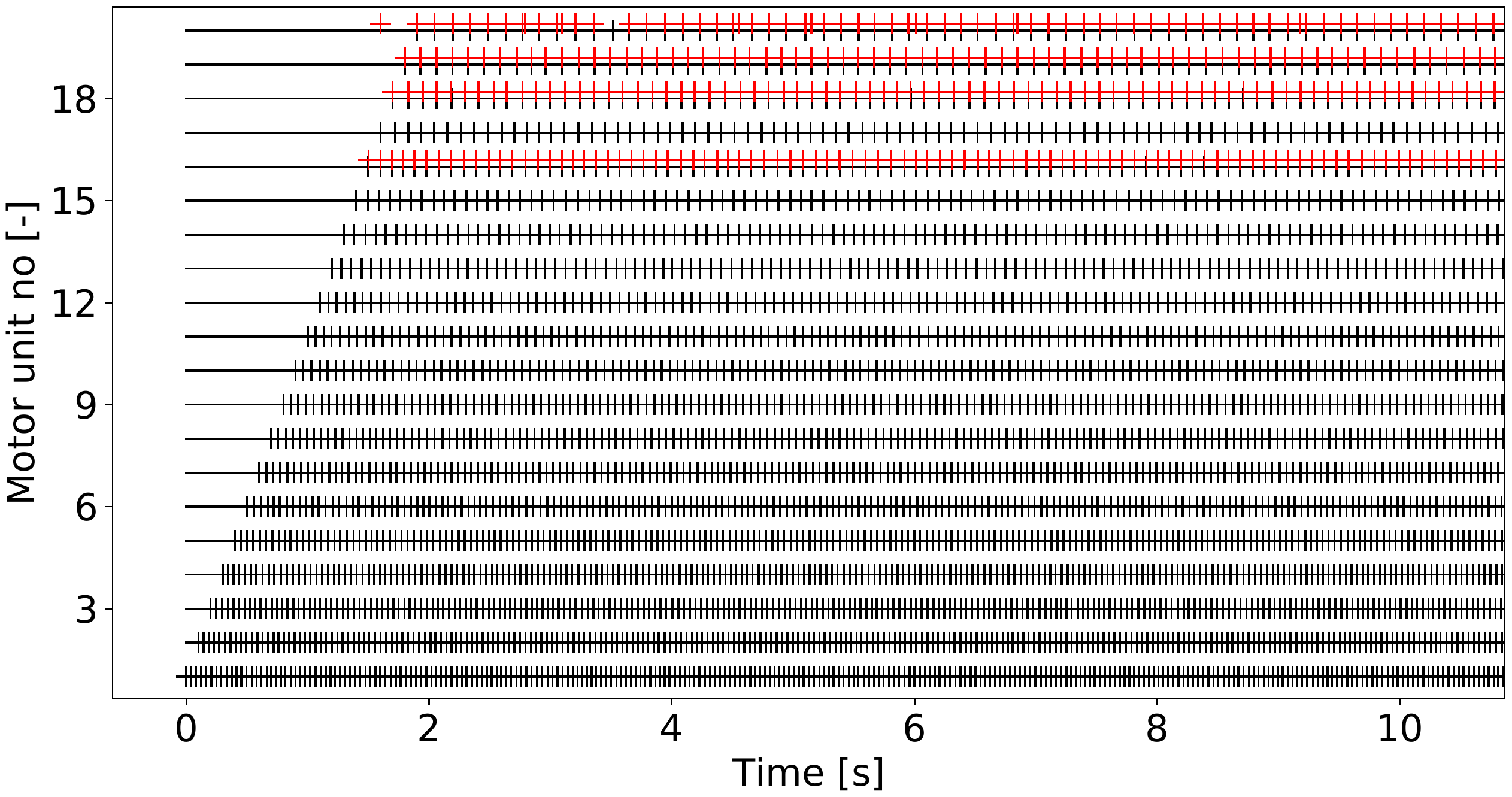}%
  \caption{Validation experiment for EMG decomposition based on fiber-based simulations of the biceps brachii muscle: Match of EMG decomposition results with simulated data. The true firing pattern over time for the 20 MUs in the simulation is shown by black markers. The recovered firing times of the gradient convolution kernel compensation algorithm are given by the red markers. The algorithm detected the four MUs 16, 18, 19 and 20.}%
  \label{fig:emg_20mus-50s-old2}%
\end{figure}

\Cref{fig:emg_20mus-50s-old2} shows the firing times of the 20 MUs in the first \SI{10}{\second}. The different MUs are initially activated every \SI{100}{\ms} to generate the shown \say{ramp} activation pattern, which later helps to identify the recovered MUs from the decomposition.  From $t=\SI{1.8}{\second}$ on, all MUs fire with their respective constant frequency, subject to jitter values of $\SI{10}{\percent}$.

In this first scenario, the gCKC decomposition algorithm was applied on the first $t=\SI{40}{\second}$ of simulated EMG data. The preconfigured algorithm in DEMUSE was used without manual intervention. While the simulated EMG recording consisted of an electrode grid of $12 \times 32$ fibers, only a rectangular subset of $5\times 13$ channels at the lower center of the grid was used for the decomposition to mimic a realistic electrode array size. 

In reality, some of the recorded channels may measure invalid data due to inappropriate surface contact of the electrodes, noisy signals at the particular measurement location or other experimental difficulties. The DEMUSE tool can automatically detect such channels and discard the corresponding data from the decomposition scheme. Despite our simulation does not contain invalid channels, the DEMUSE software discarded four of the 65 simulated channels.

% old 20MUs scenario

\Cref{fig:emg_20mus-50s-old2} shows the innervation pulses that were detected by DEMUSE as red vertical markers. A time span of \SI{50}{\s} was simulated of which only the first \SI{11}{\s} are visualized in \cref{fig:emg_20mus-50s-old2}. DEMUSE found four MUs in this scenario, i.e., \SI{20}{\percent} of the 20 simulated MUs. 
The recovered MUs were identified in the set of simulated MUs by matching the  average firing frequency and the activation onset time in the ramp scheme. A first visual comparison with the original stimulation times given by the black markers shows a good agreement.

% old fiber distribution: Smallest MU: 2, Largest MU: 256
\begin{figure}
  \centering%  \,
  \begin{subfigure}[t]{0.45\textwidth}%
    \centering%
    \includegraphics[width=\textwidth]{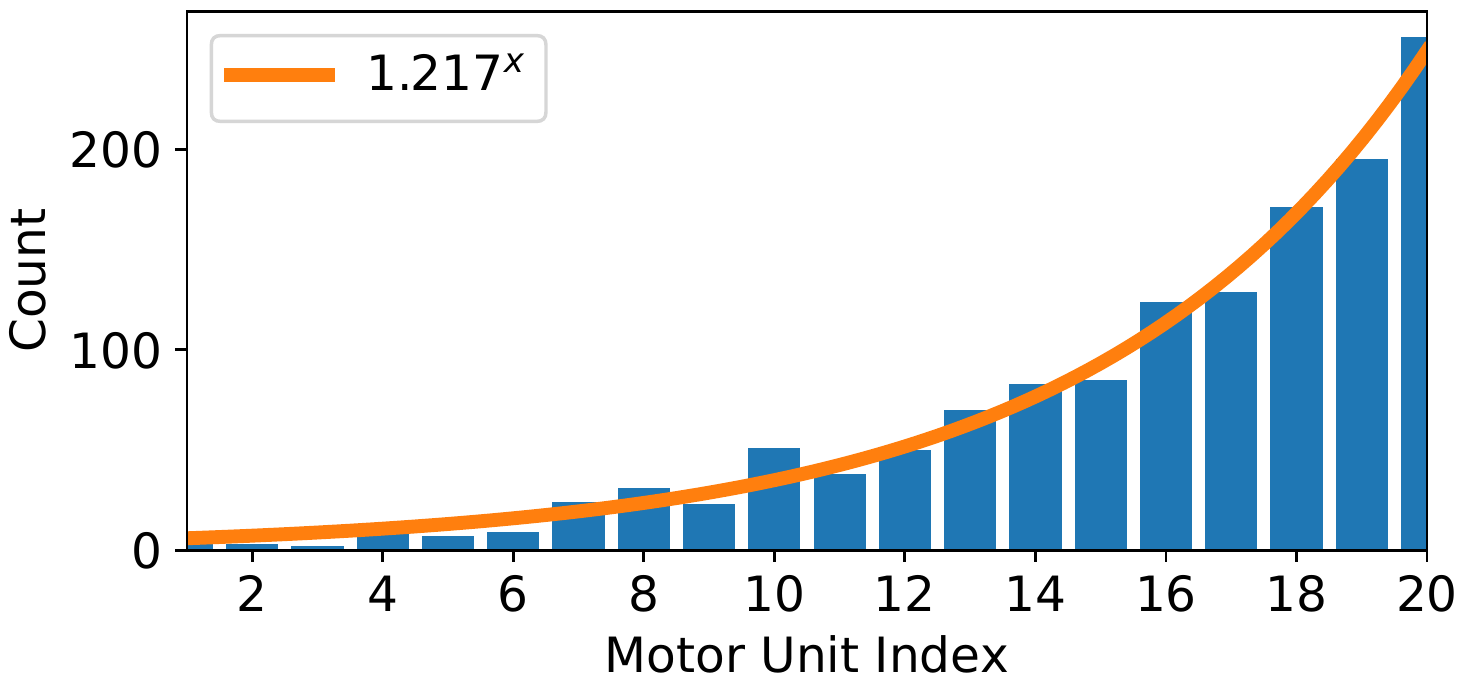}%
    \caption{Number of fibers of the 20 MUs in this scenario, following an exponential progression with basis $1.217$.}%
    \label{fig:oldmus_progression}%
  \end{subfigure}\hfill
  \begin{subfigure}[t]{0.45\textwidth}%
    \centering%
    \includegraphics[width=\textwidth]{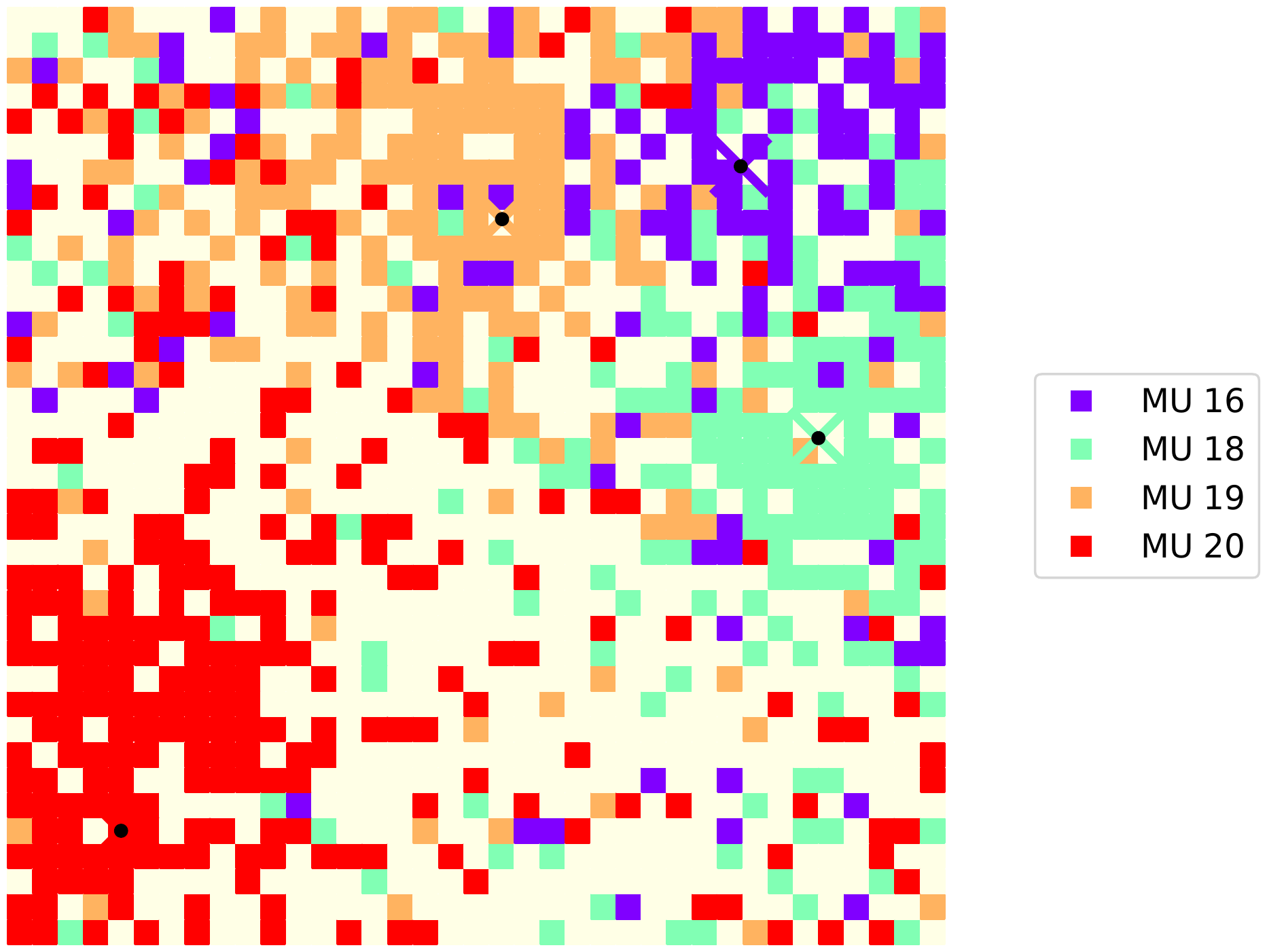}%
    \caption{Spatial location of the fibers of the four MUs that were recovered by the EMG decomposition algorithm, indicated by different colors. The territory center points of the MUs used in the generation algorithm are indicated by black dots. The yellow background area corresponds to other MUs, which were not detected.}%
    \label{fig:oldmus_2d}%
  \end{subfigure}
  \caption{Validation experiment for EMG decomposition based on fiber-based simulations of the biceps  muscle: Association of the fibers with motor units for the first scenario with 20 MUs, given in \cref{fig:emg_20mus-50s-old2}.}%
  \label{fig:oldmus}%
\end{figure}

% new fiber distribution: Smallest MU: 42, Largest MU: 102
% number fibers per MU: [ 50  45  43  51  43  42  50  53  61  71  72  69  78  97  87  87 102  74 101  93]
\begin{figure}
  \centering% \,
  \begin{subfigure}[t]{0.45\textwidth}%
    \centering%
    \includegraphics[width=\textwidth]{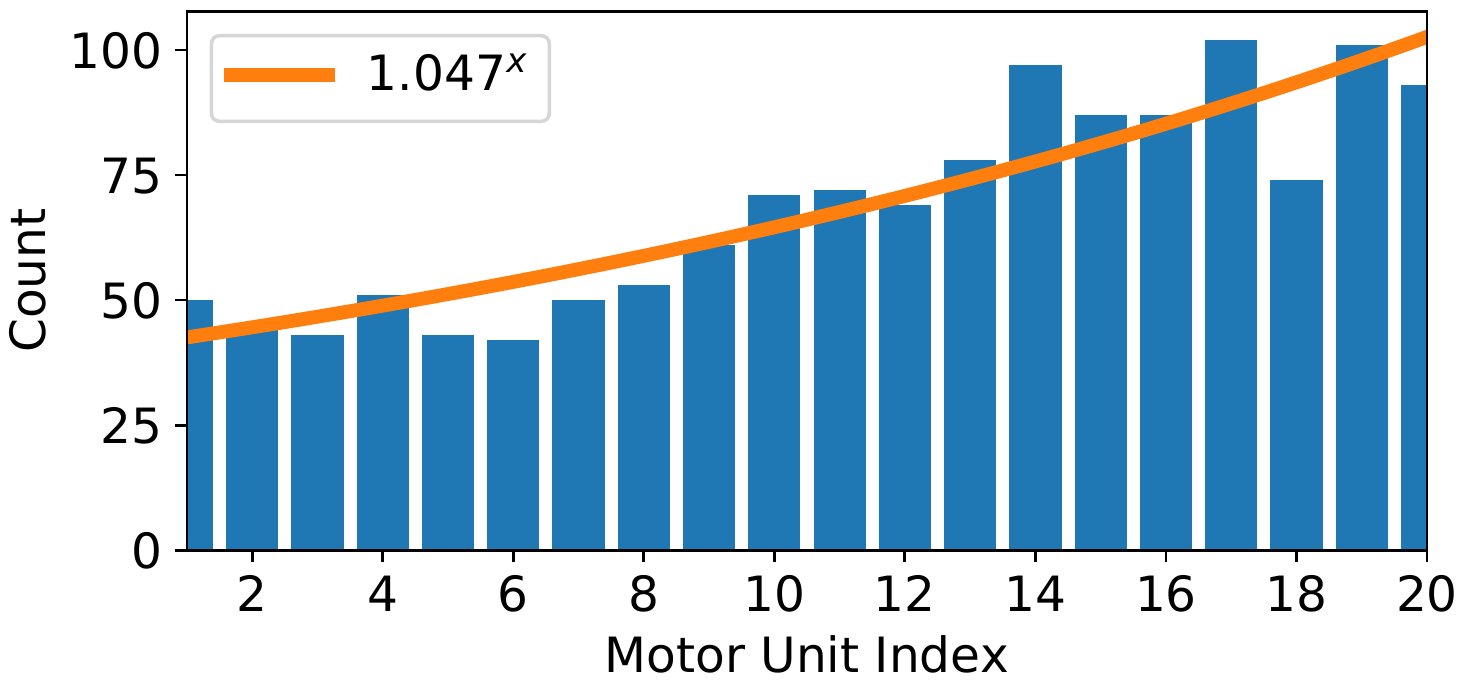}%
    \caption{Progression of MU sizes. The number of fibers per MU is more balanced in this scenario than in \cref{fig:oldmus_progression}.}%
    \label{fig:newmus_progression}%
  \end{subfigure}\hfill
  \begin{subfigure}[t]{0.45\textwidth}%
    \centering%
    \includegraphics[width=\textwidth]{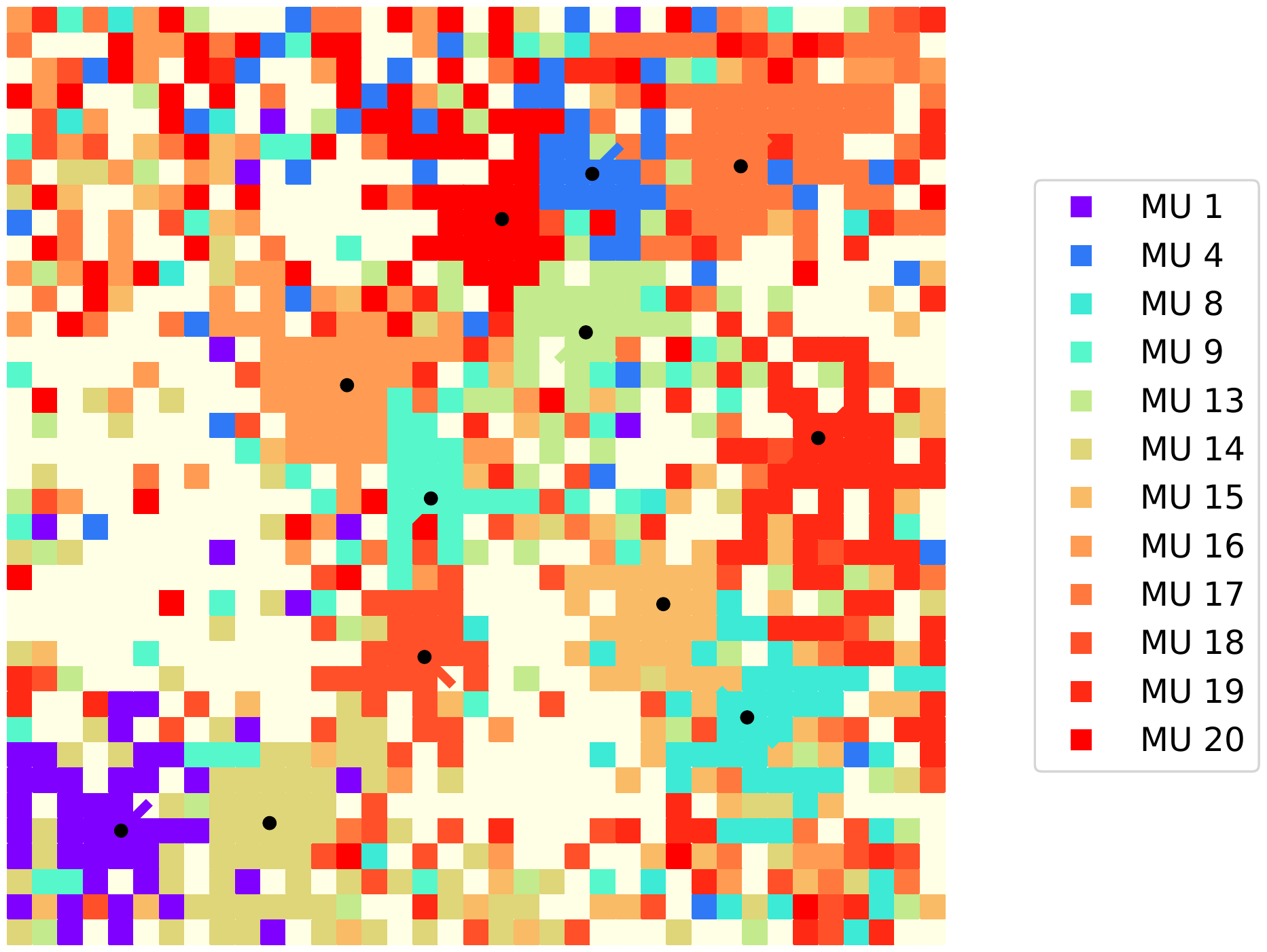}%
    \caption{Spatial fiber arrangement of the MUs that were detected by the EMG decomposition algorithm, with their center points given by black dots.}%3
    \label{fig:newmus_2d}%
  \end{subfigure} 
  \caption{Validation experiment for EMG decomposition based on fiber-based simulations of the biceps brachii muscle: Association of the fibers with motor units for the second scenario with 20 MUs, given in \cref{fig:emg_20mus-40s_new}.}%
  \label{fig:newmus}%
\end{figure}

In this scenario, the association of fibers with MUs followed an exponential MU size progression with a basis of approximately $1.2$, as shown in \cref{fig:oldmus_progression}. The smallest MU contained two fibers and the largest MU had 256 fibers. The method 1 described in \cref{sec:method1_assignment} was used to generate the association between fibers and MUs.

\Cref{fig:oldmus_2d} depicts the location of the four MUs that were detected by DEMUSE. The detected MUs have the indices 16, 18, 19 and 20 and correspond to four of the five largest MUs. It can be seen that MUs 18 to 20 are located mainly in the upper half of the muscle cross-section, in proximity to the electrode array at the top of the diagram.
The MU with the most fibers, MU 20, was detected by the decomposition algorithm even though it is located at the lower left of diagram at a large distance to the skin surface.

Two further scenarios were simulated with the same parameters as the first scenario in \cref{fig:emg_20mus-50s-old2}, but instead with 50 and 100 MUs. In these datasets, DEMUSE was able to detect 8 and 12 MUs, which corresponds to \SI{16}{\percent} and \SI{12}{\percent}. 

% new 20MUs scenario
\begin{figure}
  \centering%
  \includegraphics[width=\textwidth]{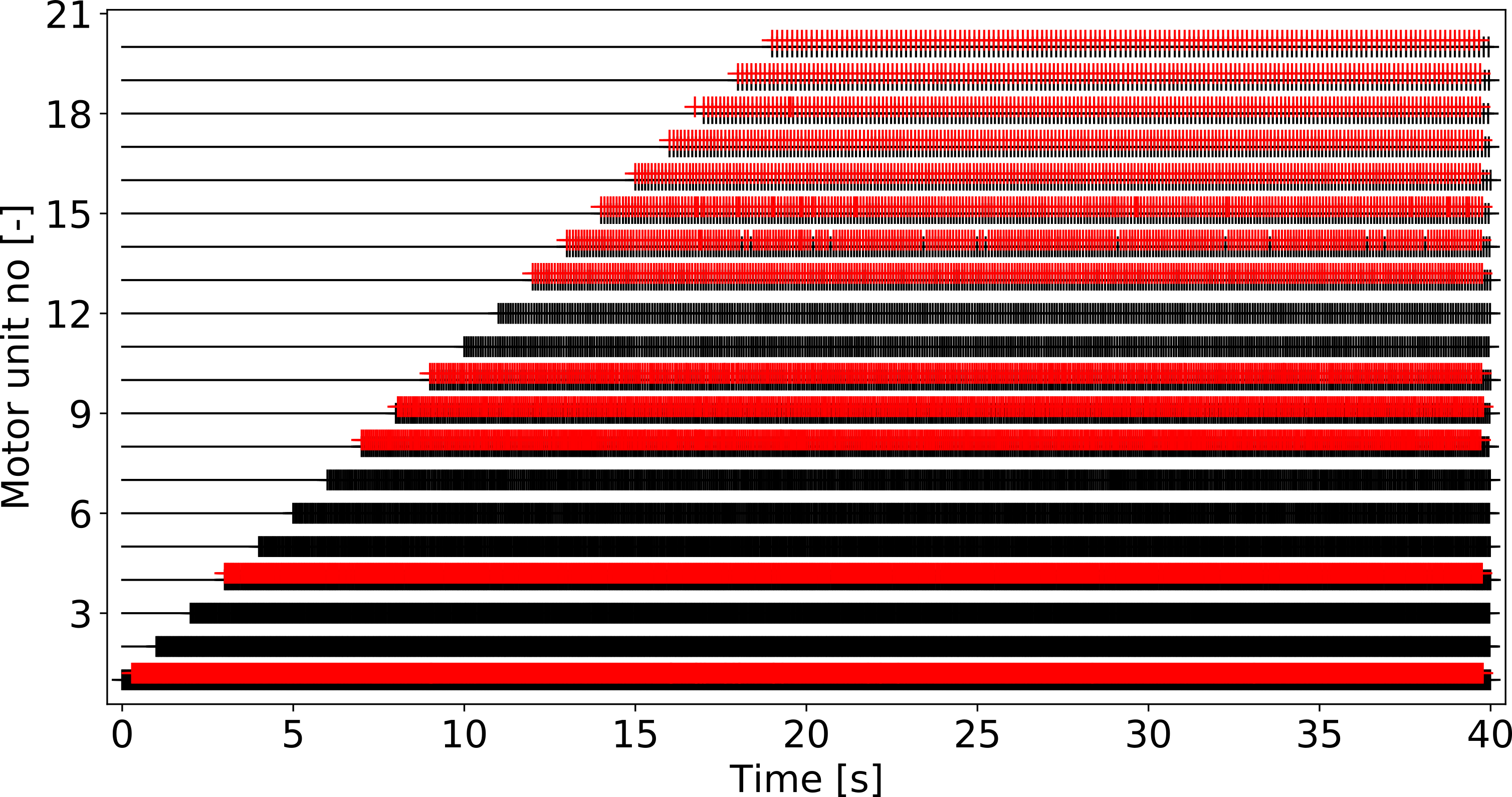}% .pdf
  \caption{Validation experiment for EMG decomposition based on fiber-based simulations of the biceps  muscle: Activation pattern for the second scenario with 20 MUs. The activation times used in the simulation are shown as black markers, the recovered activation pulses of the EMG decomposition algorithm are shown as red markers.}%
  \label{fig:emg_20mus-40s_new}%
\end{figure}

Moreover, another scenario with 20 MUs was computed, but the fat layer was varied to have a thickness of only \SI{2}{\mm} instead of \SI{1}{\cm}. In addition, the association scheme between MUs and fibers was changed to the one shown in \cref{fig:newmus}. The exponential distribution of MU sizes only varied between 42 and 102 fibers per MU, corresponding to a basis in the exponential function of approximately $1.05$ instead of $1.2$.

\Cref{fig:emg_20mus-40s_new} shows the results of the EMG decomposition with the gCKC algorithm for this second scenario with 20 MUs. DEMUSE successfully decomposed the signal into 13 MUs, corresponding to \SI{65}{\percent} of the 20 simulated MUs. DEMUSE also determined two additional MUs, which we do not consider part of the set of successfully recovered MUs. The first dataset only consists of ten innervation pulses, and the second pulse train contains high frequency oscillations.
In this scenario, the software marked only one EMG recording channel as invalid, which means that more data were considered by the decomposition algorithm than in the first scenario with 20 MUs.

Similar to the previously presented scenario, the larger MUs were detected with a higher probability than the smaller MUs. In this scenario, the eight largest MUs were successfully found. 
\Cref{fig:newmus_2d} shows the spatial arrangement of the detected MUs. The area of the muscle cross-section that is occupied by undetected MUs is again located more distantly to the skin surface at the upper boundary. However, the recovered MUs 1 and 14 are nevertheless located at the lower boundary, i.e., in the most distant area from the EMG electrodes. 

%-----

% time shifts
\begin{figure}
  \centering%
  \begin{subfigure}[t]{0.47\textwidth}%
    \centering%
    \includegraphics[width=\textwidth]{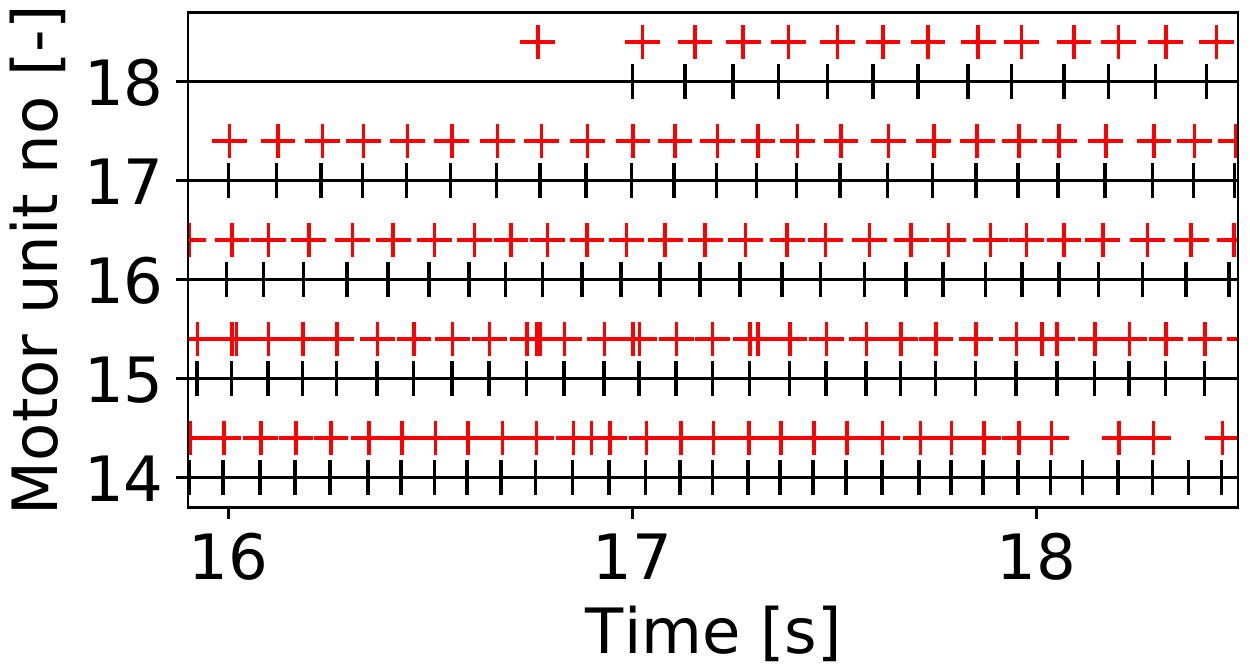}%
    \caption{Original data, where a constant time shift between the simulated (black) and recovered (red) marks can be seen, especially for MUs 16 and 18.}%
    \label{fig:newmus_nocorrection}%
  \end{subfigure}
  \hfill
  \begin{subfigure}[t]{0.47\textwidth}%
    \centering%
    \includegraphics[width=\textwidth]{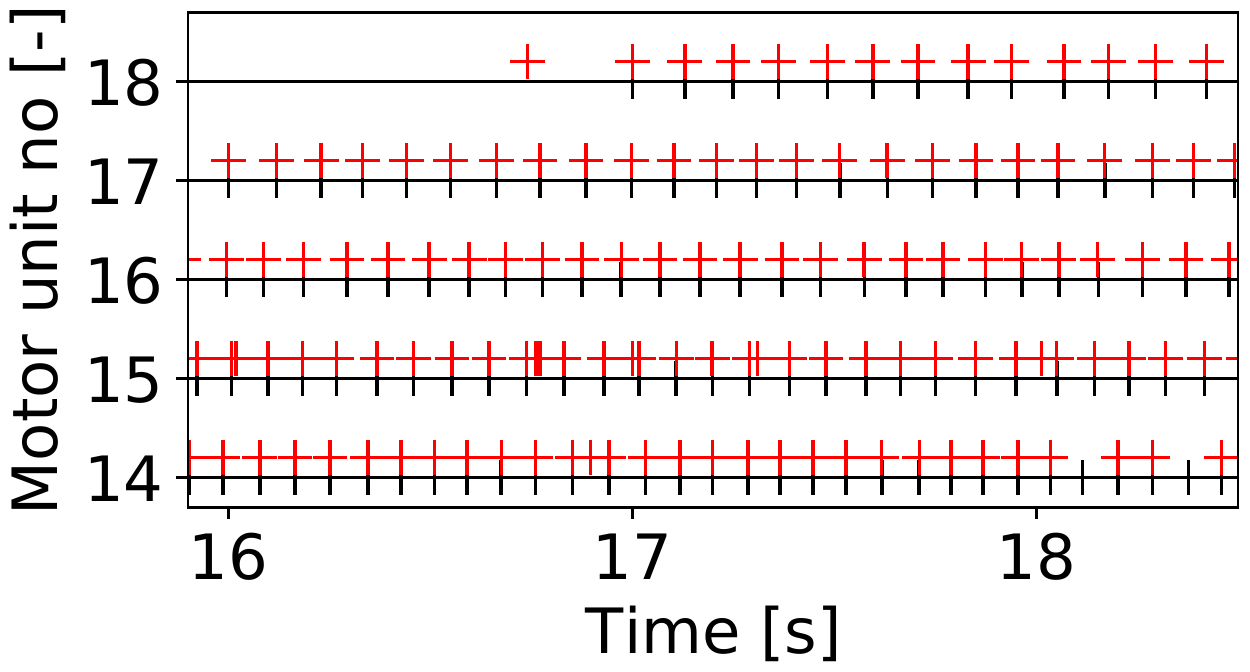}%
    \caption{Same data as in (a) with applied time offset correction. The applied time shifts for MUs 16 and 18 are \SI{-12.9}{\ms} and \SI{-24.9}{\ms}, respectively.}%
    \label{fig:newmus_withcorrection}%
  \end{subfigure}
  \caption{Validation experiment for EMG decomposition based on fiber-based simulations of the biceps  muscle: Excerpts of the detected firing times of MUs 14 to 18 in the second scenario with 20MUs. The stimulation times of the simulation are given by black markers, the recovered times are visualized by red crosses.}%
  \label{fig:newmus_shifting}%
\end{figure}

Next, we evaluate the quality of the innervation pulse trains that were recovered by the gCKC algorithm in our scenarios. We compare the stimulation times calculated by DEMUSE with the stimulation times of the simulation. \Cref{fig:newmus_nocorrection} shows an excerpt of the detected pulse trains of the second scenario with 20 MUs in \cref{fig:emg_20mus-40s_new}, where the gCKC algorithm recovered 13 MUs.  For same MUs, we observe that the recovered stimulation times are consistently shifted in time. This effect is especially visible for MUs 16 and 18. 

The reference times given by the black markers in \cref{fig:newmus_nocorrection} correspond to the times when the fibers were stimulated in the simulation in OpenDiHu. The detected MU activations in DEMUSE, however, correspond to the times when the MU action potential shapes in the EMG recording reached their maximum.
Moreover, the exact times when particular MUs reach particular EMG electrodes  depend on the distance of the electrodes to the innervation points of the MUs. The further the electrodes are away from the neuromuscular junctions along the muscle, the higher is the delay of the recorded spikes to the corresponding innervation pulses. Thus, the constant time shifts in the pulses detected by the gCKC algorithm are valid and have to be accounted for in the evaluation of the decomposition performance.

We correct for these time shifts by adding constant time offsets $\Delta t_k$ to the recovered innervation pulse trains. For every MU $k$, the algorithm finds the matching pairs of simulated and recovered pulses and optimizes the value of $\Delta t_k$ such that the time differences in these pairs after shift correction get minimal.

\Cref{fig:newmus_withcorrection} shows the same extract of MU activity as in \cref{fig:newmus_nocorrection} with applied time offsets. The time offsets for MUs 14 to 18 in this example are given as
\begin{align*}
    \Delta t_{14}&=\SI{-2.4}{\ms}, & \Delta t_{15}&=\SI{-1.1}{\ms}, & \Delta t_{16}&=\SI{-12.9}{\ms},\\
    \Delta t_{17}&=\SI{-2.9}{\ms}, \quad \text{and} & \Delta t_{18}&=\SI{-24.9}{\ms}.
\end{align*}
\Cref{fig:newmus_withcorrection} shows that the recovered pulses now match the simulated data very well. The non-matching pulses are clearly false positive detections.

To compare the recovered MU times between the scenarios, we evaluate metrics such as the rate of agreement. The MU firing times in the simulation serve as the ground truth, to which we compare the recovered MU times. We identify true positive (TP), false positive (FP) and false negative (FN) recovered pulses, depending on whether a matching time to a recovered pulse can or cannot be found in the simulation data within a tolerance of $\eps=\SI{5}{\ms}$. The rate of agreement (RoA) between the gCKC algorithm output and the ground truth data is then computed by%
\begin{align*}
  \textrm{RoA}  = \dfrac{\textrm{TP}}{\textrm{TP} + \textrm{FP} + \textrm{FN}}.
\end{align*}

In the first scenario with 20 MUs in \cref{fig:emg_20mus-50s-old2}, the RoA for MUs 16,18 and 19 is above $\SI{99.7}{\percent}$ and slightly lower at $\SI{82.2}{\percent}$ for MU 20. Here, only 296 of the 334 detected pulses were true positives, corresponding to a precision of $\SI{88.6}{\percent}$.
% MU 16 from matlab: FP: 0, TP: 484, FN: 0 -> RoA: 1.0
% MU 18 from matlab: FP: 1, TP: 409, FN: 0 -> RoA: 0.9975609756097561
% MU 19 from matlab: FP: 1, TP: 370, FN: 0 -> RoA: 0.9973045822102425
% MU 20 from matlab: FP: 38, TP: 296, FN: 26 -> RoA: 0.822222222222222
 
In the second scenario with 20 MUS presented in \cref{fig:emg_20mus-40s_new}, all valid MUs except one have RoA values of above $\SI{94.5}{\percent}$. Five MUs are even detected perfectly with $\SI{100}{\percent}$ rate of agreement. MU 9 is the only detected MU with a degraded RoA of approximately $\SI{57.9}{\percent}$. However, the RoA improves to $\SI{98.1}{\percent}$, if the tolerance $\eps$ for matching pulses is relaxed to $\SI{10}{\ms}$. This shows that the RoA metric also depends on a proper value for the tolerance $\eps$, and that innervation pulse trains detected by DEMUSE can have varying accuracy in a range of less than $\SI{10}{\ms}$.
 
 %MU 9 from matlab: FP: 191, TP: 321, FN: 42 -> RoA: 0.5794223826714802
 
While the gCKC algorithm can be used for EMG decomposition of previously recorded signals in a controlled environment, it is less suited for real-time applications. The separation vectors that decompose the electrode signals and infer the MU innervation pulse trains can be computed in a training phase. 
However, their application on new data requires a certain history of previously captured signals to calculate the decomposed MU pulses. As a consequence, the predictions are delayed, which is usually undesirable in real-time applications. Furthermore, the system is sensitive to noisy data. 

A fundamentally different approach to EMG decomposition is the use of sequence-to-sequence learning methods provided by recurrent neural networks. The authors of \cite{Clarke2021} used a gated recurrent unit (GRU) network for this task. The network was trained using the output of the gCKC algorithm and was subsequently able to decompose surface EMG signals into innervation pulse trains. The approach was shown to be robust and to outperform gCKC for low signal-to-noise ratios.

To assess, whether our simulations of surface EMG can be used for the supervised learning of GRU networks for EMG decomposition, we tried in a first step to reproduce the studies of \cite{Clarke2021}, where the GRU is trained with the output of the gCKC algorithm. Additionally, we trained a GRU network directly on the simulated EMG data. These tasks were carried out in the masters project of Srijay Kolvekar and were supervised by Lena Lehmann and me. For details on the methods and results, we refer to the literature \cite{Clarke2021} and the project report \cite{Srijay}.

In this project, the EMG decomposition of a GRU network trained with raw innervation pulse trains obtained from the gCKC algorithm, similar to the literature, showed many false positive and false negative predictions.
However, a different setup using MU labels instead of raw pulse trains showed promising results. 
Every discrete point in time (according to the EMG sampling frequency) was either associated with the class of the currently active MU or with the background class, when no MU was activated at the time. This classification problem had a large class imbalance, as the background class was active for \SI{86}{\percent} of the timesteps. The issue was mitigated by using class weights. The GRU network was trained with simulation data and yielded per-class rates of agreement of up to \SI{72}{\percent} for the two scenarios with 20 MUs shown in \cref{fig:emg_20mus-50s-old2,fig:emg_20mus-40s_new}, i.e., with the test data set also generated by our simulation.

\begin{figure}
  \centering%
  \includegraphics[width=\textwidth]{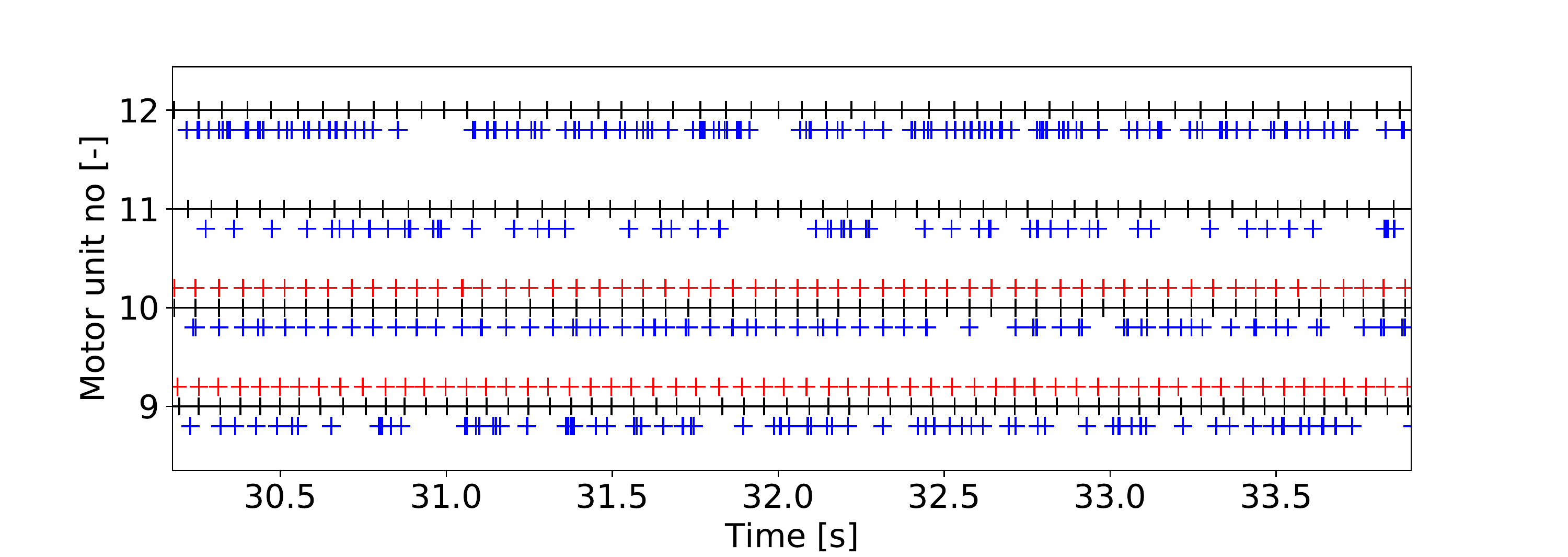}%
  \caption{Comparison of innervation pulse train predictions of the gCKC algorithm (red), a GRU network (blue) and the ground truth data (black).}%
  \label{fig:gru_result}%
\end{figure}

\Cref{fig:gru_result} presents an excerpt of the resulting predictions of a GRU network that was trained with simulation results.
We used the simulation of the second scenario with 20 MUs, which is shown in \cref{fig:emg_20mus-40s_new}. 
The black markers in \cref{fig:gru_result} indicate the stimulation times used in the simulation. 
The red markers correspond to the recovered times by the gCKC algorithm. 
Out of the shown MUs, only MUS 9 and 10 were recovered by the gCKC algorithm.
The blue markers denote the GRU predictions. Correction of time offsets was performed for both the gCKC and GRU outputs.

\Cref{fig:gru_result} shows the best agreement between the two prediction methods for MU 10 with a RoA of \SI{99.6}{\percent} for the gCKC algorithm and \SI{72.2}{\percent} for the GRU network. In contrast to the gCKC algorithm, the GRU network predicts firings for all MUs. However, the quality is only acceptable for MUs that could also be detected by the gCKC algorithm. For MUs 11 and 12, the RoA for the GRU network is around \SI{30}{\percent}.

In future work, the decomposition performance of the GRU networks could be improved by using different training data. For example, the ramp activation in the training data could be replaced by constant tetanic stimulations. Moreover, different network architectures, such as convolutional recurrent neural networks could be investigated.

 %MU 1 from gru: FP: 411, TP: 536, FN: 92 -> RoA: 0.5158806544754572
 %MU 2 from gru: FP: 458, TP: 513, FN: 27 -> RoA: 0.5140280561122245
 %MU 3 from gru: FP: 698, TP: 877, FN: 50 -> RoA: 0.5396923076923077
 %MU 4 from gru: FP: 252, TP: 503, FN: 42 -> RoA: 0.6311166875784191
 %MU 5 from gru: FP: 708, TP: 469, FN: 15 -> RoA: 0.3934563758389262
 %MU 6 from gru: FP: 705, TP: 691, FN: 52 -> RoA: 0.47720994475138123
 %MU 7 from gru: FP: 308, TP: 566, FN: 24 -> RoA: 0.6302895322939867
 %MU 8 from gru: FP: 469, TP: 290, FN: 84 -> RoA: 0.34400948991696323
 %MU 9 from gru: FP: 342, TP: 182, FN: 52 -> RoA: 0.3159722222222222
 %MU 10 from gru: FP: 94, TP: 293, FN: 19 -> RoA: 0.7216748768472906
 %MU 11 from gru: FP: 215, TP: 142, FN: 70 -> RoA: 0.3325526932084309
 %MU 12 from gru: FP: 671, TP: 295, FN: 61 -> RoA: 0.2872444011684518
 %MU 13 from gru: FP: 576, TP: 208, FN: 27 -> RoA: 0.2564734895191122
 %MU 14 from gru: FP: 418, TP: 155, FN: 1 -> RoA: 0.2700348432055749
 %MU 15 from gru: FP: 299, TP: 112, FN: 55 -> RoA: 0.24034334763948498
 %MU 16 from gru: FP: 234, TP: 122, FN: 62 -> RoA: 0.291866028708134
 %MU 17 from gru: FP: 85, TP: 131, FN: 21 -> RoA: 0.5527426160337553
 %MU 18 from gru: FP: 302, TP: 60, FN: 65 -> RoA: 0.1405152224824356
 %MU 19 from gru: FP: 65, TP: 53, FN: 51 -> RoA: 0.3136094674556213
 %MU 20 from gru: FP: 71, TP: 18, FN: 56 -> RoA: 0.12413793103448276

In conclusion, the gCKC algorithm is able to decompose artificially generated surface EMG signals. This means that our simulation can be used to evaluate the performance of EMG decomposition algorithms.

The number of detected MUs depends on the relation between MU sizes and on the distance of the MU territories to the electrodes. If the variance of the sizes of the activated MUs is small, such as in \cref{fig:newmus_progression}, also MUs that are far away from the electrodes are detected. If, in the opposite case, the sizes of active MUs are distributed over a large range such as in \cref{fig:oldmus_progression}, only the largest MUs are detectable.

In addition, the amount of adipose tissue between the electrodes and the muscle influences the number of MUs that can be recovered. In our studies, the performance of EMG decomposition was lower for all scenarios with thicker fat layer than for the scenario with a thin fat layer.

The rate of agreement of the determined pulse trains of the DEMUSE software was above $\SI{95}{\percent}$ in most of the cases. Correspondingly, the rate of false positives was low.
A time shift between the recovered times and the ground truth data was observed for some pulse trains, which can be explained with the delay from first activation to the onset of the EMG signal. As a result, the time shift was corrected for the rate of agreement measurement.

A proof-of-concept implementation of GRU networks showed promising performance for predicting MU firing times from artificial EMG recordings. 
The GRU network predicted firing times also for MUs that were not detected by the gCKC algorithm, however the rate of agreement was low for these MUs.
In future work, the GRU decomposition method has to be refined to be comparative to the gCKC algorithm.

\subsection{Simulation of Electrophysiology with a Phenomenological Fiber Model}\label{sec:sim_rosenfalck}

Instead of the previously presented numerical model of action potential propagation on the muscle fibers, the respective physiological process can also be described with a phenomenological approach.

The model formulated by Rosenfalck \cite{Rosenfalck1969} describes the action potential shape on a 1D domain by the following function:
\begin{align*}
  G(z) = \begin{cases}
    96\,z^3\,\exp(-z) - 9 & \text{for } z \geq 0,\\[4mm]
    -90  & \text{for } z < 0.
  \end{cases}  
\end{align*}

\begin{figure}
  \centering%
  \includegraphics[width=0.6\textwidth]{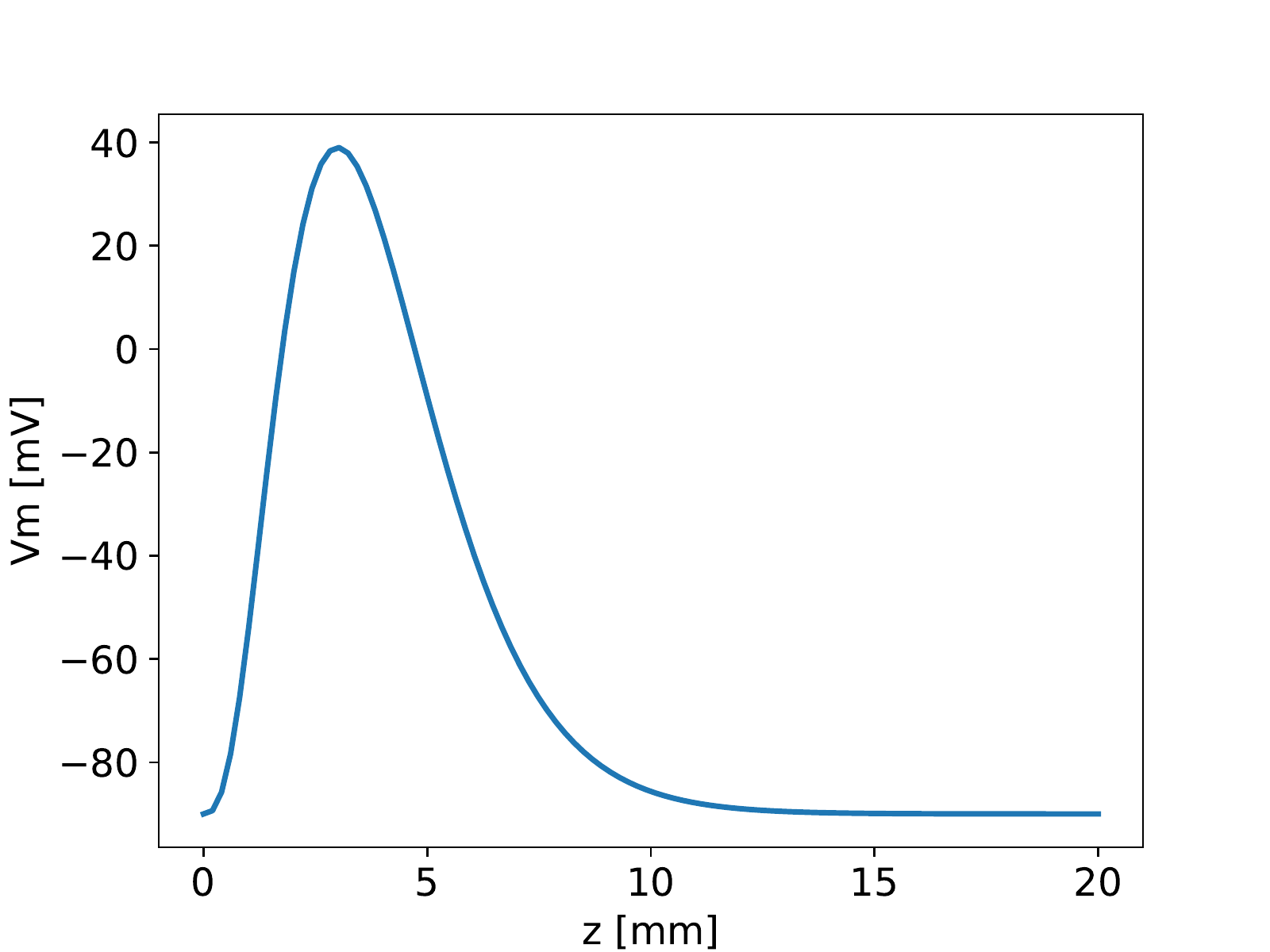}%
  \caption{Phenomenological muscle fiber model: analytic model by Rosenfalck of an action potential shape on a 1D fiber.}%
  \label{fig:rosenfalck_function}%
\end{figure}
\Cref{fig:rosenfalck_function} shows the graph of this function. The resulting value of $G$ specifies the membrane voltage in millivolts. The coordinate $z=x+v\,t$ depends on the distance $x$ to the neuromuscular junction on the fiber, the conduction velocity $v$ and the time $t$ after the last stimulation. In our simulation, we use the proposed propagation velocity of \SI{4}{\meter\per\second}.

The advantage of using an analytic model is its fast calculation compared with the runtime of the numerical model. On the downside, such a model cannot accurately describe fatigue effects in tetanic stimulations or the action potential shape changing properties of more advanced subcellular models.

One use case of an analytic model is to study the effect of muscle fiber arrangements in a muscle volume on the surface EMG signal. In this case, the exact, possibly time-varying shapes of the motor unit action potentials are less important than the location and orientation of the fibers.
We provide an exemplary scenario in OpenDiHu, which uses the Rosenfalck model on multiple 1D muscle fibers that are embedded in a 3D domain. As in the previous sections, the 3D bidomain model given by \cref{eq:bidomain1} is coupled to the fibers and used to simulate EMG recordings on the surface. 

\begin{figure}
  \centering%
  \includegraphics[width=\textwidth]{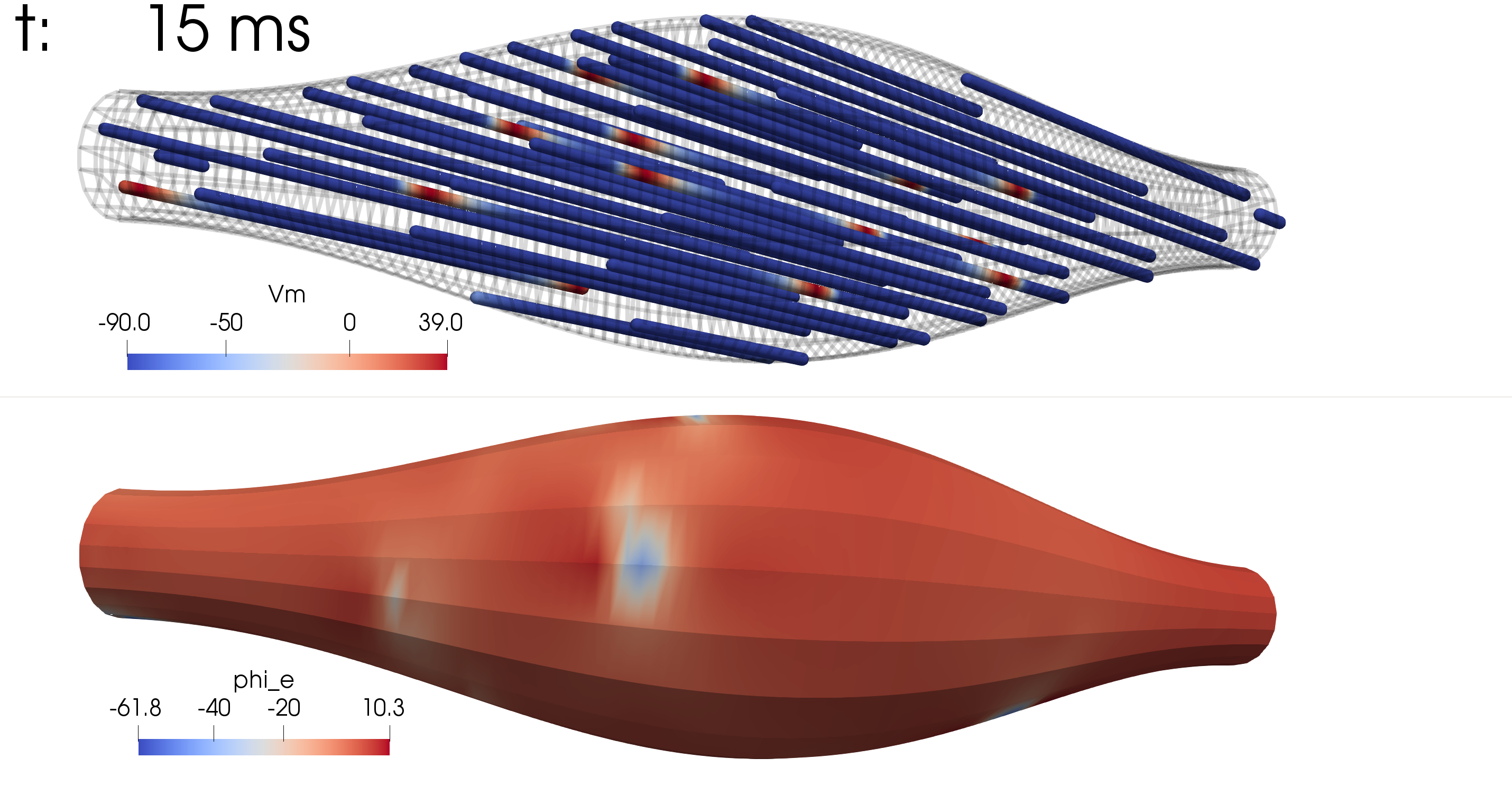}%
  \caption{Simulation of EMG signals on the upper arm; artificial muscle geometry with a phenomenological model of action potential propagation. The upper image shows the muscle fibers, colored according to the transmembrane potential $V_m$. The lower image shows the extracellular potential $\phi_e$ on the surface.
  This scenario can be computed very fast and can, e.g., be useful to investigate the effects of different fiber orientations.}%
  \label{fig:custom_geometry}%
\end{figure}

In addition to the simplified model of action potential propagation, we use a simplified description of the muscle geometry. \Cref{fig:custom_geometry} shows the artificial geometry, which is constructed by rotating a transformed sine curve around the $z$ axis. 
Our program embeds the fibers automatically inside the volume and orients them according to specified spherical coordinates. The orientation angles, the number of fibers and the spacings between the fibers can be adjusted in the Python settings script of the simulation.

\Cref{fig:custom_geometry} shows the location of the fibers inside the artificial muscle belly in the upper image and a simulation result of muscular activity  at $t=\SI{15}{\ms}$ in the lower image. The resulting EMG signal can be seen on the surface.
This simulation scenario provides means to quickly study electrophysiology and generation of EMG for a generic muscle. Solving the model only consists of evaluations of the Rosenfalck function and repeated computation of the 3D model, but no further costly computations of numerical electrophysiology models. Moreover, no mesh file has to be generated and loaded, which simplifies the handling of the scenario.

\begin{reproduce_no_break}
  The simulation in \cref{fig:custom_geometry} can be started as follows.
  \begin{lstlisting}[columns=fullflexible,breaklines=true,postbreak=\mbox{\textcolor{gray}{$\hookrightarrow$}\space}]
    cd $\$$OPENDIHU_HOME/examples/electrophysiology/fibers/analytical_fibers_emg/build_release
    ./analytical_fibers_emg ../settings_analytical_fibers_emg_custom_geometry.py geometry_round.py
  \end{lstlisting}
  The options can be set in the \code{variables/geometry_round.py} settings file. Other artificial geometries are available by using other scripts under \code{variables}.
\end{reproduce_no_break}

\subsection{Conclusion}

In the present section, surface EMG signals were computed using the fiber based electrophysiology model. We showed, how the qualitative nature of the signals depends on the spatial distribution of the MUs, and that the thickness of the body fat layer influences the smoothness of the recorded EMG signals. Furthermore, we investigated the effect of the mesh resolution and the number of fibers in the simulation. The results showed that only a high mesh resolution can resolve the small features in the EMG signal  on the skin surface, which result from fiber action potentials. The number of these features in the result increased with the mesh width and, thus, the two finest discretizations with \num{77e3} \num{274e3} fibers yielded the most accurate results. As a consequence, High Performance Computing simulations are required, if an accurate 2D surface EMG signal should be computed.

Further studies with EMG decomposition algorithms showed another use case of our fiber based surface EMG simulations: They can be used as a validation tool for existing decomposition algorithms, and they can serve as a data generator to develop novel, data based decomposition methods. The decomposition software DEMUSE was tested, and we quantified the rate of agreement of its predictions with our simulation. Similarly, we evaluated a neural network based approach and compared its performance on the simulated data with the previous method.

% ==================
%
% =-------------------
\section{Simulation of the Multidomain Model}\label{sec:solver_multidomain_model}

The multidomain model is an alternative to the fiber based electrophysiology model discussed in the last section.
As introduced in \cref{sec:multidomain_model}, it does not explicitly resolve muscle fibers, but considers activity in the muscle domain in a homogenized view.
On every point in the 3D muscle mesh, separate values $V_m^k$ of the transmembrane potential exist for every MU $k \in \{1,\dots,N_\text{MU}\}$, in addition to the value $\phi_e$ for the extracellular electric potential. 
The computational domain considers the muscle volume $\Omega_M$ and the body domain $\Omega_B$, which represents  adipose tissue on top of the muscle.
To solve the multidomain model, a large linear system has to be solved in every timestep, as described in \cref{sec:discretization_body_domain}.

In the following, \cref{sec:multidomain_components} discusses the model setup for a scenario with four MUs. \Cref{sec:multidomain_simulation_emg} demonstrates a larger simulation scenario with 25 MUs, which can be used to simulate surface EMG signals. We discuss differences between the multidomain approach and the fiber based electrophysiology model in \cref{sec:multidomain_differences}.

\subsection{Components of the Computational Model}\label{sec:multidomain_components}

In the following, we consider a multidomain simulation with muscle and body fat domains and four MUs. We use a muscle mesh with $16 \times 16 \times 74 = \num{18944}$ elements and linear finite element ansatz functions and a fat mesh with $32 \times 4 \times 74 = \num{9472}$ elements, which are partitioned to 128 subdomains for 128 processes. 

The scenario uses the following electric conduction tensors $\bfsigma_i$ and $\bfsigma_e$ for the intra-cellular domain and the extracellular domain, respectively:
\begin{align}\label{eq:multidomain_isotropic_e}
  \bfsigma_i &= \mat{
  8.93 & 0 & 0\\
  0& 0& 0\\
  0& 0& 0}\SI{}{\milli\siemens\per\centi\meter}, &  \bfsigma_e &= \mat{6.7 & 0 & 0\\
        0& 6.7& 0\\
        0& 0& 6.7}\SI{}{\milli\siemens\per\centi\meter}.
\end{align}
In this scenario, we use the subcellular model of Hodgkin and Huxley \cite{Hodgkin1952} and solve it using Heun's method. We discretize the multidomain equations using the Crank-Nicolson scheme with $\theta=\frac12$.
We solve the resulting linear system of equations by a GMRES solver with the parallel incomplete LU factorization preconditioner \emph{Euclid} \cite{euclid} from the HYPRE package \cite{falgout2002hypre}. A tight residual norm tolerance of \num{1e-15} is used in the abortion criterion of the GMRES solver. Such a low tolerance is required, as, for higher tolerances, spurious artificial stimulations can be observed. Timestep widths of $\dt_\text{0D}=\dt_\text{multidomain}=\dt_\text{splitting}=\SI{1e-3}{\ms}$ are used. 
The computation for a simulation end time of $t_\text{end}=\SI{20}{\ms}$ in this scenario takes approximately \SI{27}{\min}.

In the multidomain model, we have to specify which point in the 3D mesh belongs to which MU to which extent. This is achieved by the relative occupancy factors $f_r^{k}$ for MU $k$.
In our implementation, the occupancy factors are computed by Python code in the settings file of the simulation.
The location of a MU in the 3D domain is specified by choosing a 1D muscle fiber, which is considered to be the center of the MU territory. This is possible, as the nodes in the structured 3D mesh can also be interpreted as a set of adjacent 1D fibers.

For every MU $k$, the occupancy factors $f_r^k(x,y,z)$ in every muscle cross-section with fixed $z$ coordinate are defined by a radial function $f(|(x,y)^\top|/d(z))$, which reaches a configurable maximum value at the location of the specified fiber. The argument of the radial function is scaled by the diameter $d(z)$ of the muscle at the considered cross-section. The factor $f_r^k(x,y,z)$ is constant in longitudinal direction of the muscle ($z$ axis). Before the simulation, all factors $f_r^k$ are scaled, such that the maximum of their sum is equal to one:%
\begin{align*}
  \max\limits_{(x,y,z)\in\Omega_M} \sum_{k=1}^{N_\text{MU}} f_r^k(x,y,z) = 1.
\end{align*}

% multidomain fr factors
\begin{figure}
  \centering%
  \begin{subfigure}[t]{0.23\textwidth}%
    \centering%
    \includegraphics[width=\textwidth]{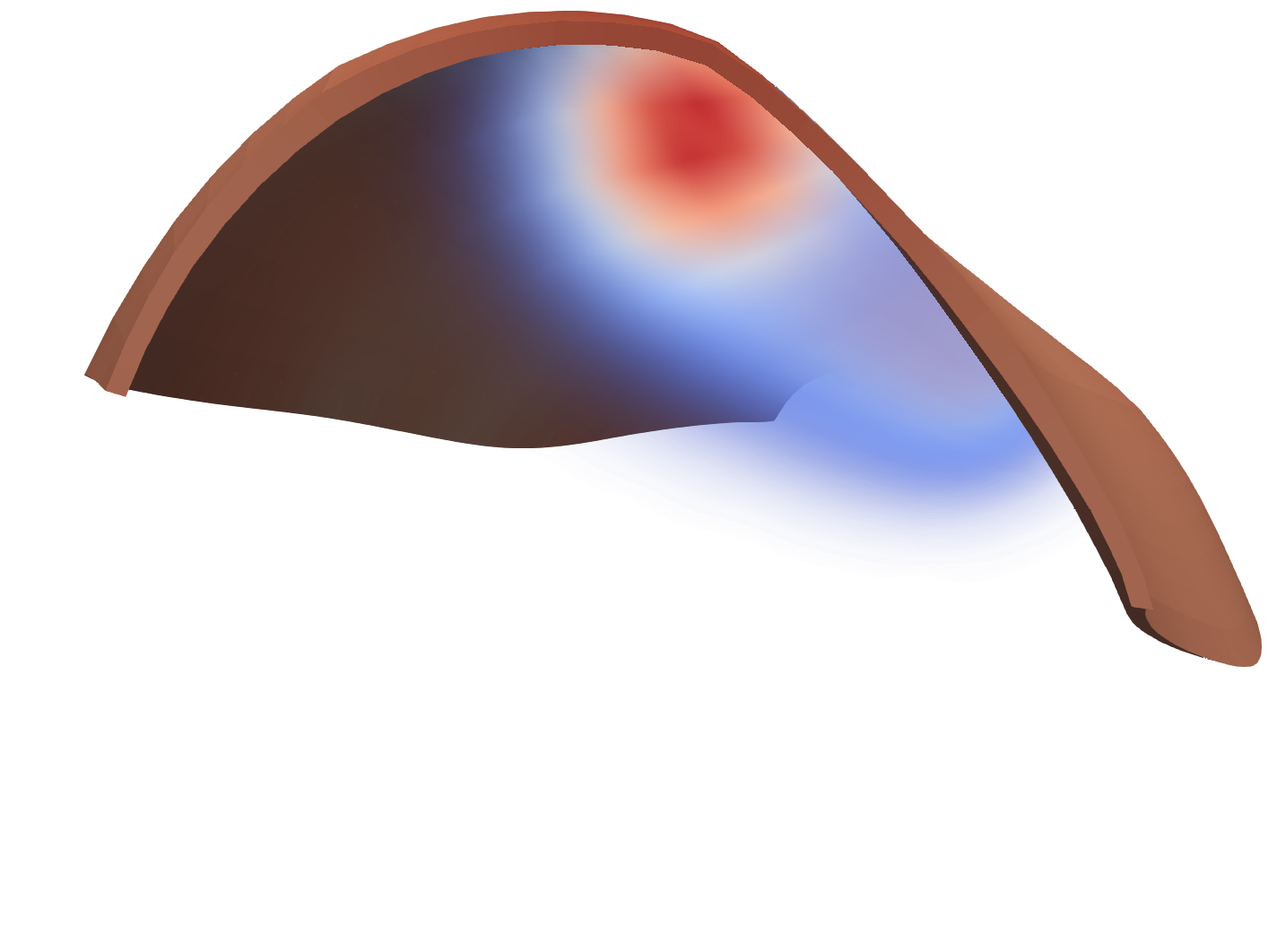}%
    \caption{$f_r^1$}%
    \label{fig:fr0}%
  \end{subfigure}
  \,
  \begin{subfigure}[t]{0.23\textwidth}%
    \centering%
    \includegraphics[width=\textwidth]{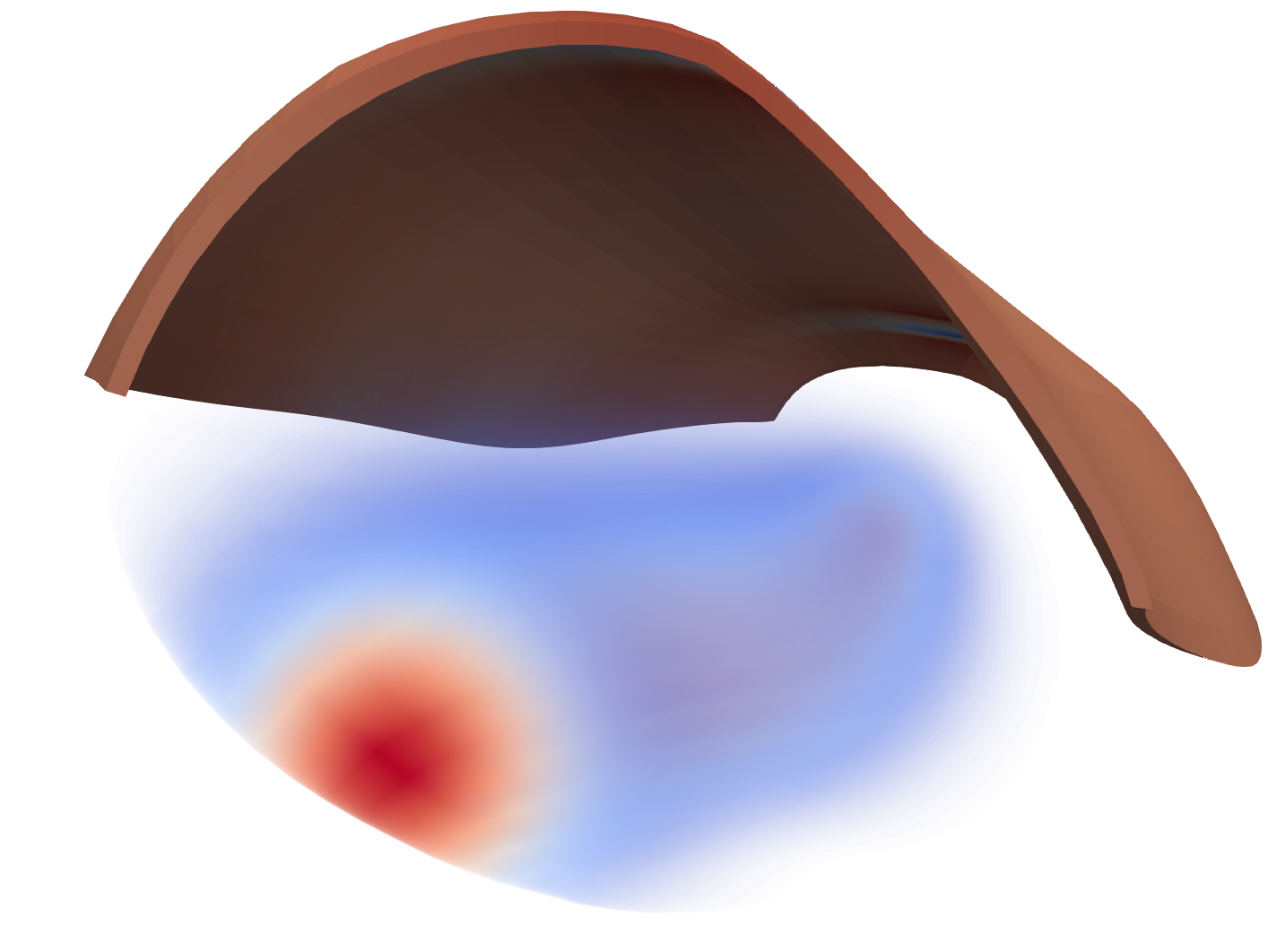}%
    \caption{$f_r^2$}%
    \label{fig:fr1}%
  \end{subfigure}
  \,
  \begin{subfigure}[t]{0.23\textwidth}%
    \centering%
    \includegraphics[width=\textwidth]{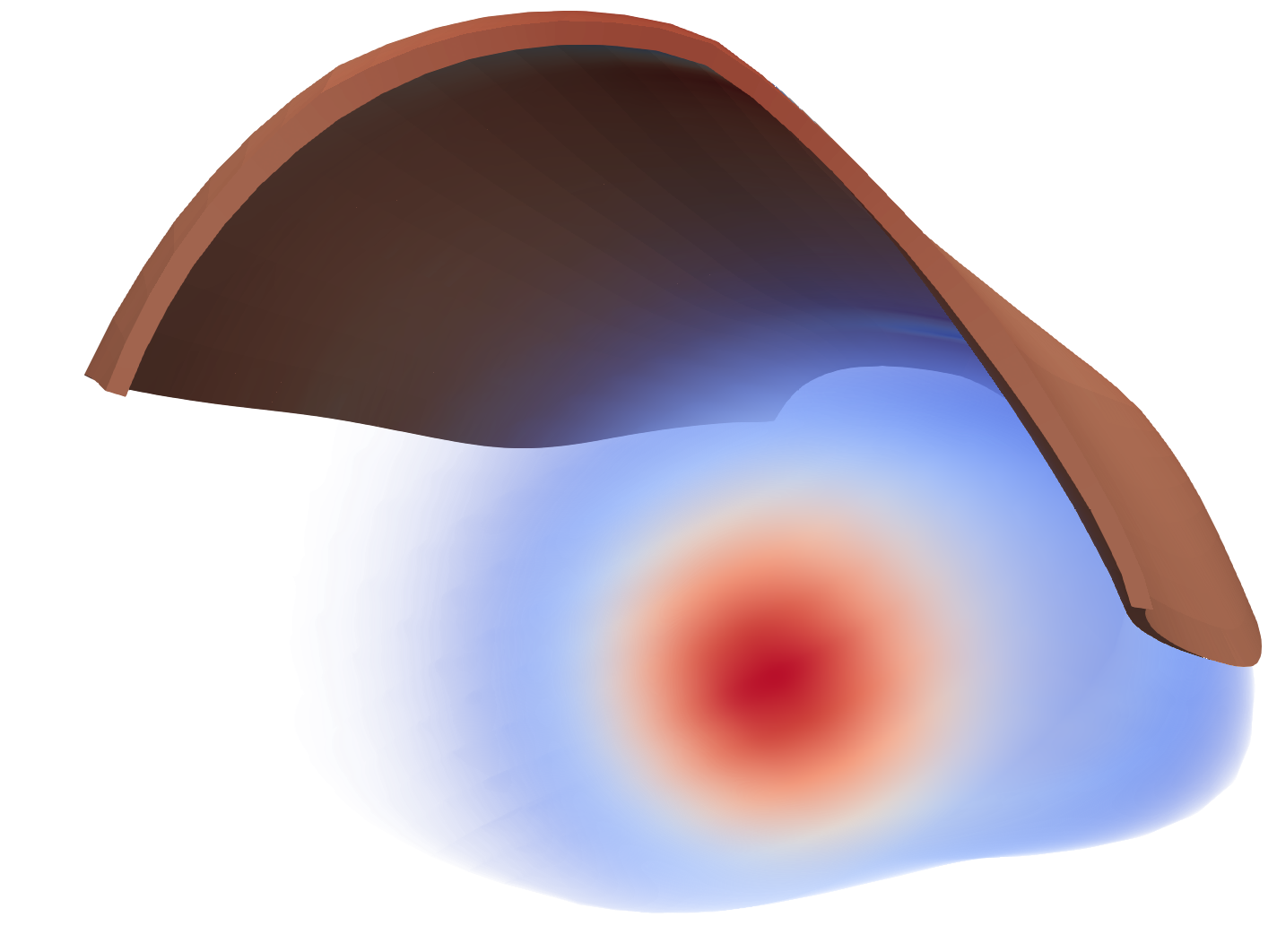}%
    \caption{$f_r^3$}%
    \label{fig:fr2}%
  \end{subfigure}
  \,
  \begin{subfigure}[t]{0.23\textwidth}%
    \centering%
    \includegraphics[width=\textwidth]{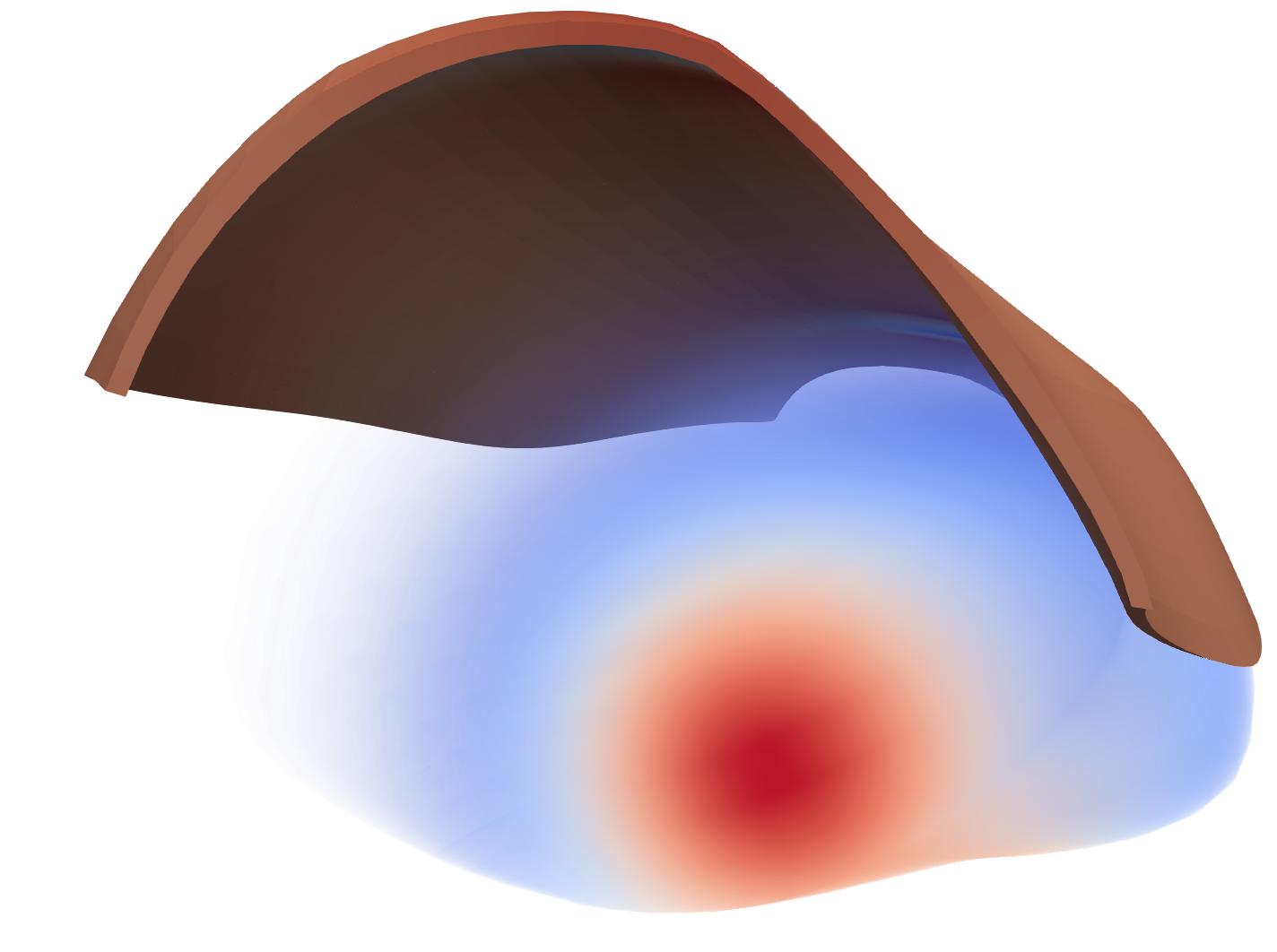}%
    \caption{$f_r^4$}%
    \label{fig:fr3}%
  \end{subfigure}
  \caption{Simulation of electrophysiology with the multidomain model: Value of the occupancy factors $f_r^k$ for MUs 1 to 4. The color coding encodes the maximum value by red color and the decreasing value along the radius by increased transparency and the color transition to blue color. The locations where the factor vanishes, $f_r^k=0$, correspond to full transparency.}%
  \label{fig:multidomain_fr}%
\end{figure}%

\Cref{fig:multidomain_fr} shows the MU occupancy factors for the four MUs in the considered example scenario. It can be seen that the individual MU territories only occupy a small fraction of the muscle domain and are centered around fibers in longitudinal direction of the muscle.

\begin{figure}
  \centering%
  \includegraphics[width=\textwidth]{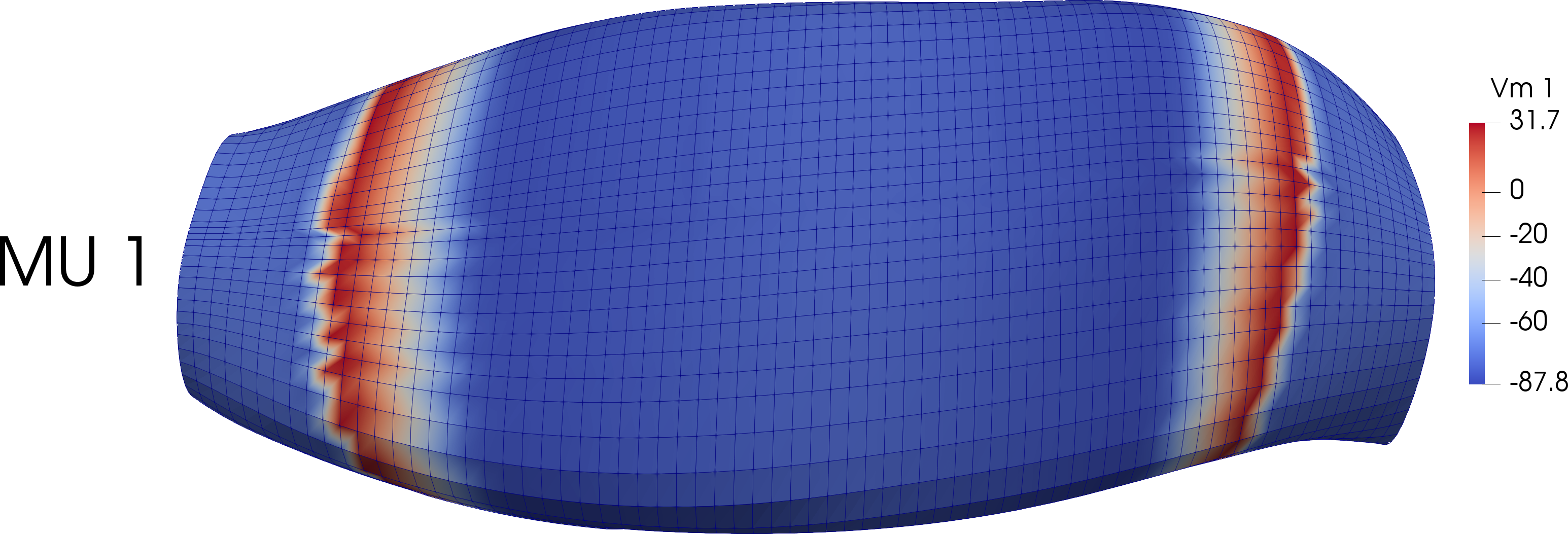}\\[4mm]
  \includegraphics[width=\textwidth]{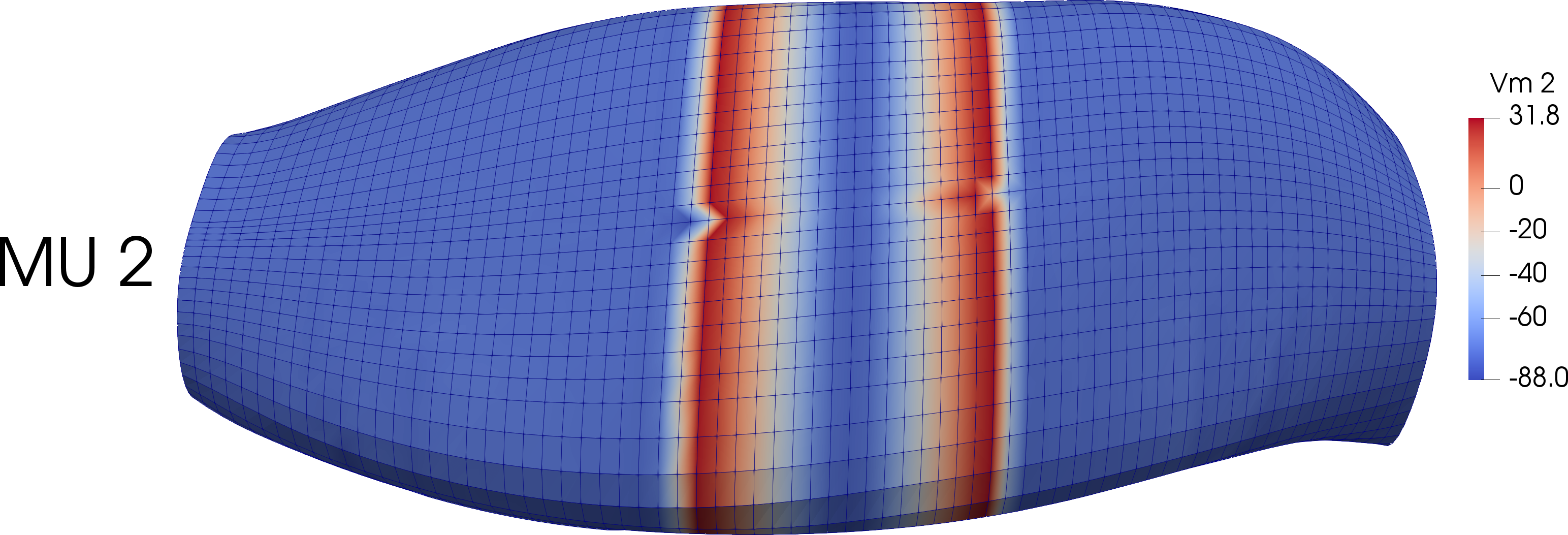}\\[4mm]
  \includegraphics[width=\textwidth]{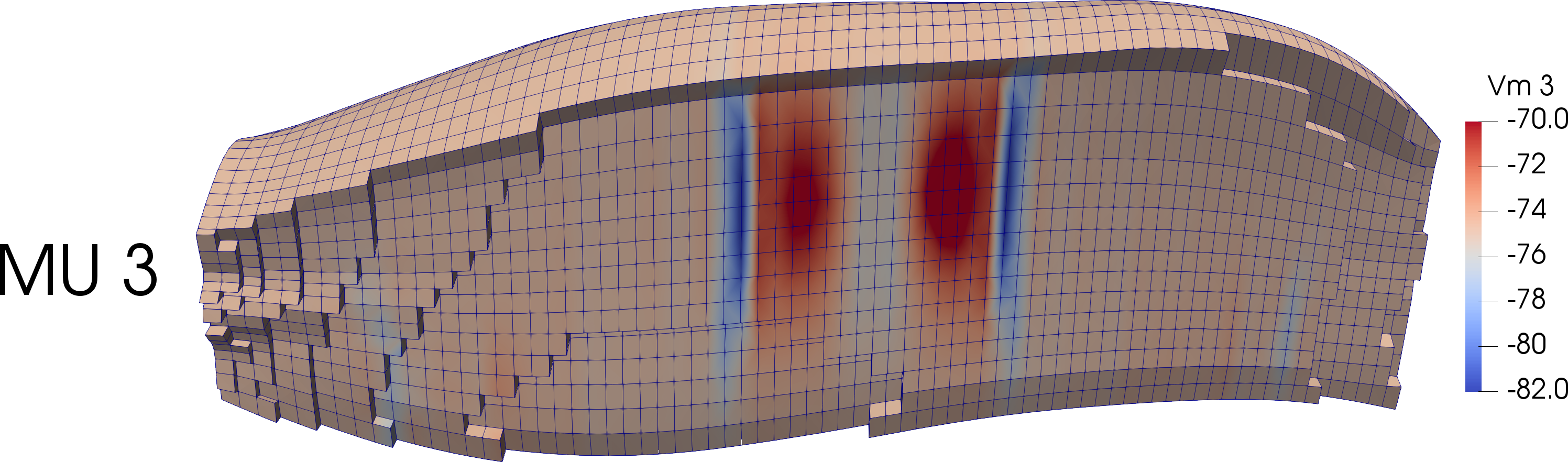}\\[4mm]
  \includegraphics[width=\textwidth]{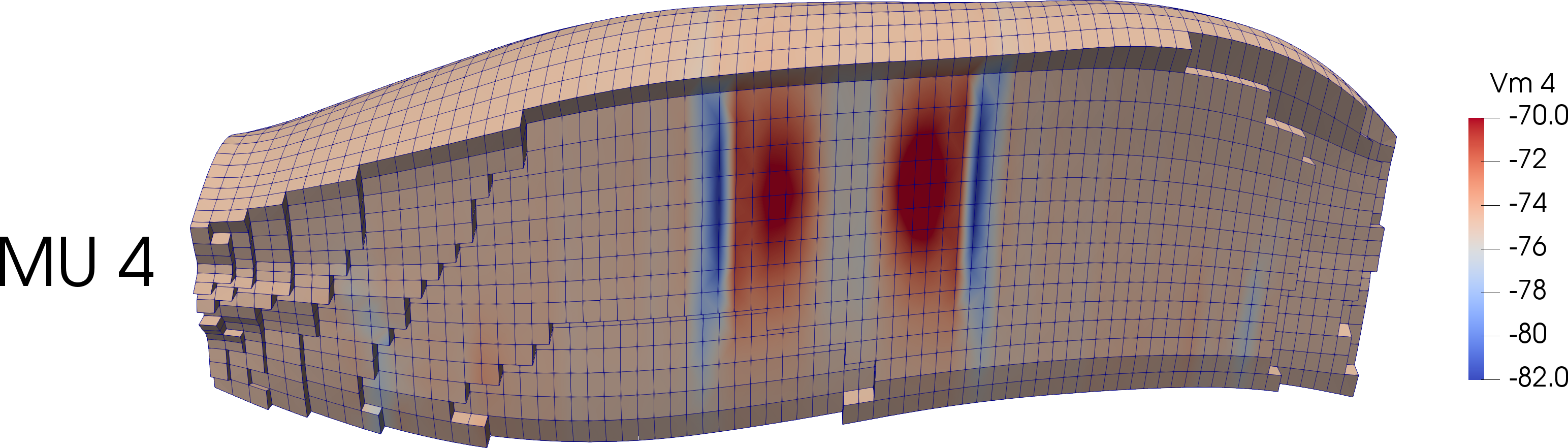}%
  \caption{Simulation of electrophysiology with the multidomain model: Transmembrane potential $V_m^k$ of MUs 1 to 4 at $t=\SI{14}{\ms}$.}%
  \label{fig:multidomain_4mus}%
\end{figure}

Results of the simulation at $t=\SI{14}{\ms}$ are given in \cref{fig:multidomain_4mus}. In the considered scenario, the first and second MU are stimulated at $t=\SI{0}{\ms}$ and $t=\SI{10}{\ms}$, respectively. At the time of the displayed images, MUs 3 and 4 have not yet been stimulated. 

Upon stimulation, we prescribe the membrane voltage $V_m$ for one timestep as $\SI{20}{\milli\volt}$ at the stimulated nodes in the mesh of the respective MU. In this scenario, the stimulated nodes are located in the middle of the muscle in longitudinal direction in three adjacent cross-sectional layers of mesh elements.

The upper two images in \cref{fig:multidomain_4mus} show the locations of the propagated action potential fronts at $t=\SI{14}{\ms}$ for MU 1 and MU 2, given by the values of $V_m^k$. While the action potentials span the entire cross-section of the muscle domain, they contribute to the EMG value scaled by their locally varying occupancy factor $f_r^k$.

% multidomain Vm MUs

As the $V_m^k$ values of all MUs $k\in \{1,\dots,N_\text{MU}\}$ are strongly coupled, the active MUs, MU 1 and MU 2, influence the $V_m^k$ scalar fields of the inactive MUs, MU 3 and MU 4. The lower two images in \cref{fig:multidomain_4mus} show the computational domain of the muscle with several layers of 3D elements removed. 
It can be seen that, at some regions in the interior of the domain, the values of $V_m^3$ and $V_m^4$ correspond to the negated value of $V_m^2$ with a smaller absolute value. Note the different color scales for $V_m^1$, $V_m^3$ and $V_m^4$ in these images.

\begin{figure}
  \centering%
  \begin{subfigure}[t]{\textwidth}%
    \centering%
    \includegraphics[width=\textwidth]{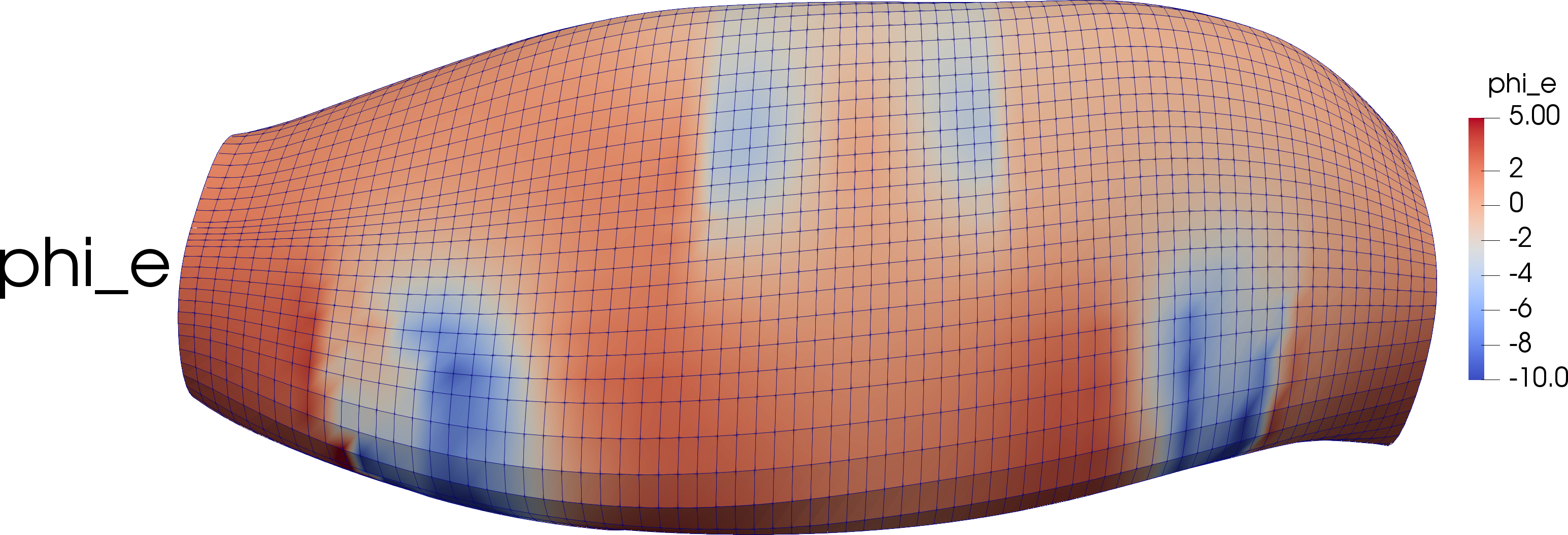}%
    \caption{Extracellular potential $\phi_e$ on the surface of the muscle domain.}%
    \label{fig:multidomain_4mus_body}%
  \end{subfigure} \\
  \begin{subfigure}[t]{\textwidth}%
    \centering%
    \includegraphics[width=\textwidth]{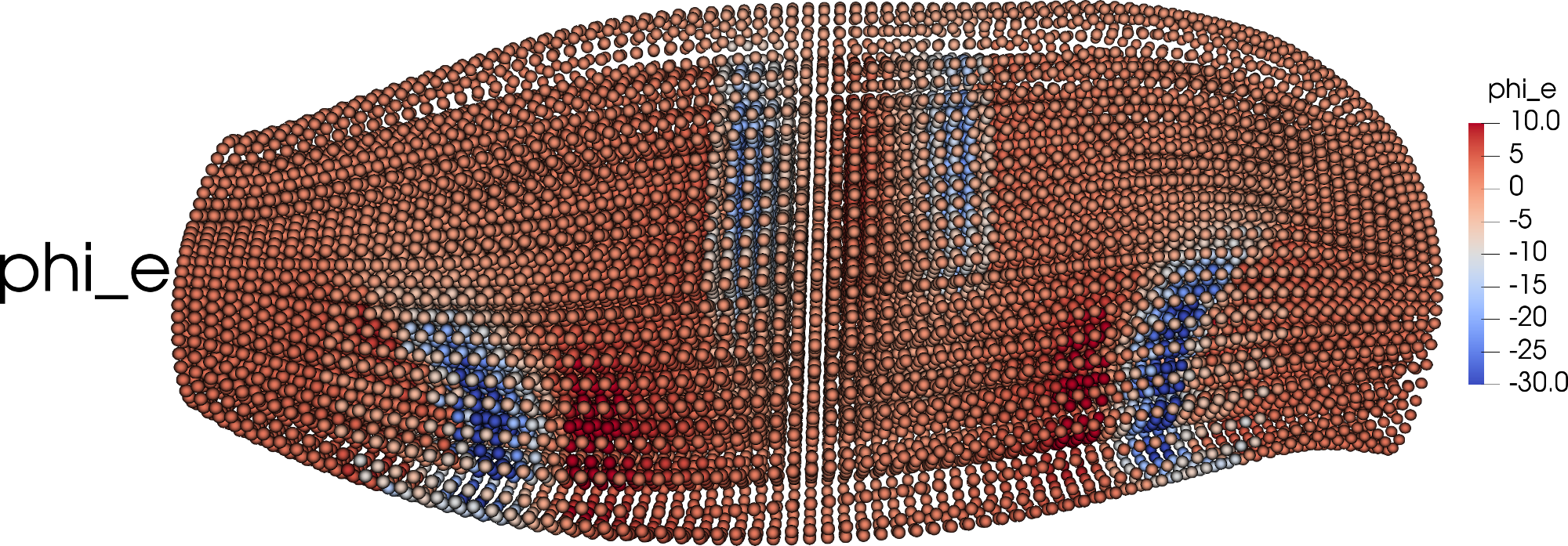}%
    \caption{Extracellular potential $\phi_e$ at points of the 3D muscle mesh.}%
    \label{fig:multidomain_4mus_phi_e_points}%
  \end{subfigure} \\
  \begin{subfigure}[t]{\textwidth}%
    \centering%
    \includegraphics[width=\textwidth]{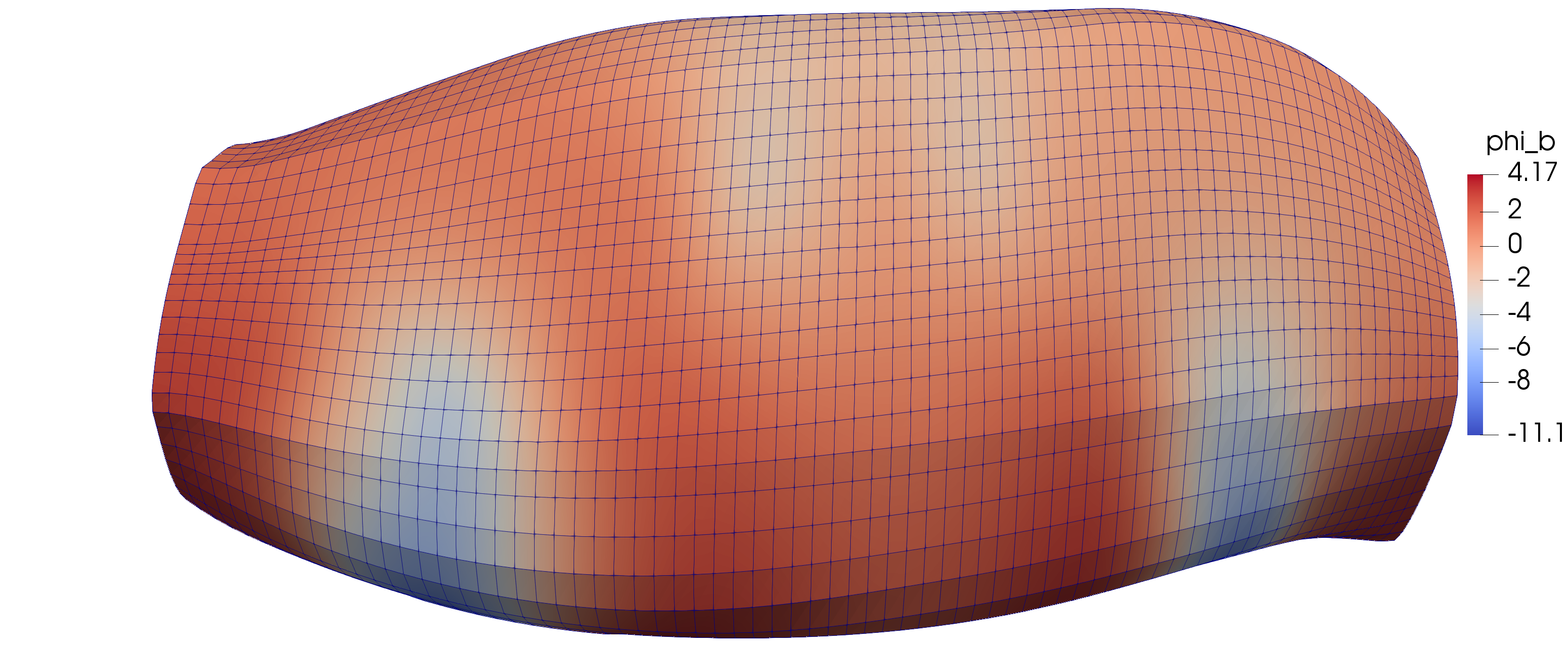}%
    \caption{EMG signal $\phi_b$ on the surface of the body fat domain. The comparison with (a) shows the effect of the fat layer.}%
    \label{fig:multidomain_4mus_emg}%
  \end{subfigure}
  \caption{Simulation of electrophysiology with the multidomain model: Simulation results at $t=\SI{14}{\ms}$ for a scenario with 4 MUs.}%
  \label{fig:multidomain_4mus_2}%
\end{figure}

\Cref{fig:multidomain_4mus_body} shows the values of the extracellular potential $\phi_e$ on the muscle domain. The contributions from the two active MUs can be seen. 
\Cref{fig:multidomain_4mus_phi_e_points} gives an impression of the used mesh and shows the value of $\phi_e$ for all nodes. It can be seen that the $\phi_e$ values span a larger value range in the interior of the domain than on the boundary, as previously shown in \cref{fig:multidomain_4mus_body}. Correspondingly, the color coding in \cref{fig:multidomain_4mus_phi_e_points} uses a larger range than in \cref{fig:multidomain_4mus_body}.

\Cref{fig:multidomain_4mus_emg} shows the EMG values $\phi_b$ on the surface of the body domain mesh. The effect of the body domain is revealed by comparing the electric potential on the boundary of the muscle mesh in \cref{fig:multidomain_4mus_body} with the values in \cref{fig:multidomain_4mus_emg}. The signals get locally smoothed by the fat layer.

% multidomain phi_e

\subsection{Simulation of EMG Signals}\label{sec:multidomain_simulation_emg}

In the second scenario, we simulate a higher number of 25 MUs. The MUs are activated at random times within the first \SI{20}{\ms}. We use a mesh with $12 \times 12 \times 74 = \num{10656}$ elements in the muscle domain and $24 \times 4 \times 74 = 7104$ elements in the body domain. Compared to the mesh in the previous scenario, the spatial resolution in radial direction is chosen slightly coarser, to speed up the computation.
The other model parameters, discretization schemes and solvers are chosen as before.

\begin{figure}
  \centering%
  \begin{subfigure}[t]{\textwidth}%
    \centering%
    \includegraphics[width=11cm]{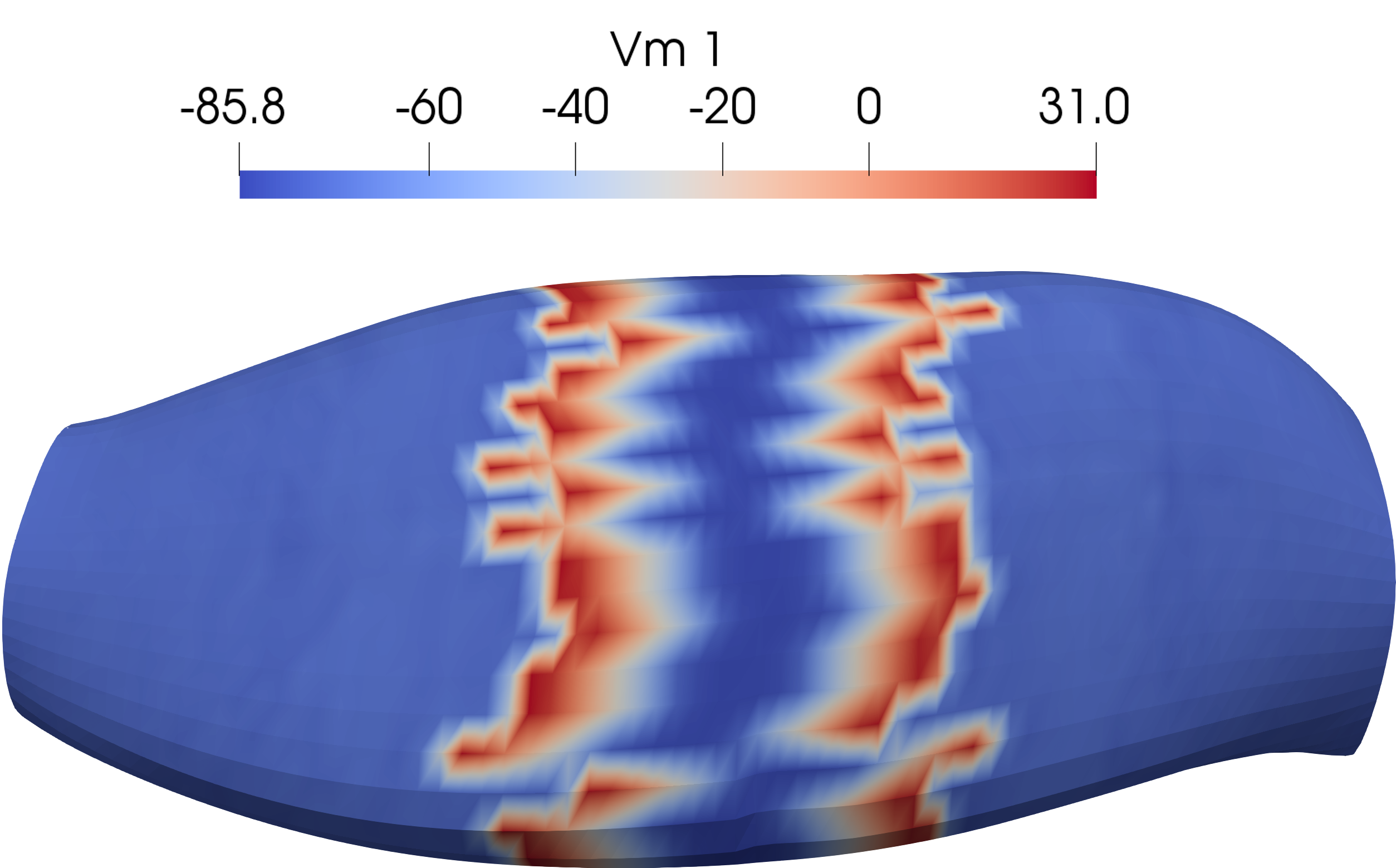}%
    \caption{Transmembrane voltage $V_m^1$ of the first MU.}%
    \label{fig:multidomain_25mus_snapshot}%
  \end{subfigure} \\[4mm]
  \begin{subfigure}[t]{\textwidth}%
    \centering%
    \includegraphics[width=12cm]{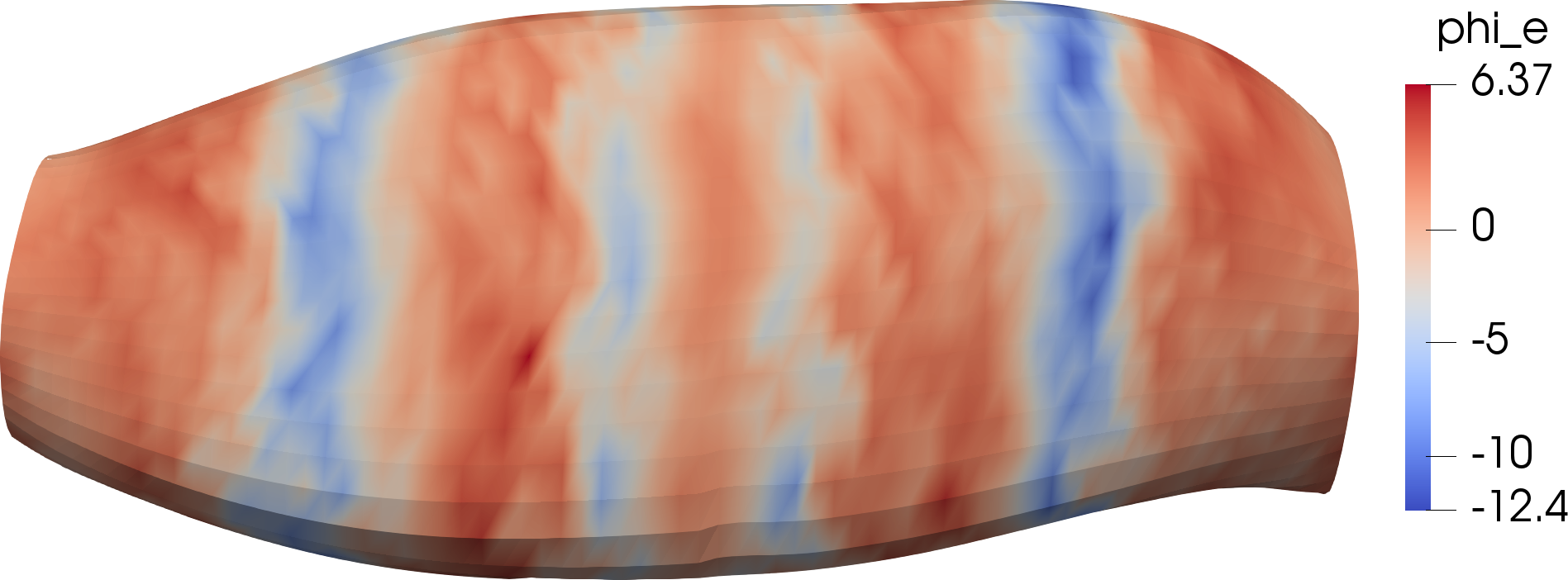}%
    \caption{Extracellular potential $\phi_e$ at the surface of the 3D muscle mesh.}%
    \label{fig:multidomain_25mus2_emg}%
  \end{subfigure} \\[4mm]
  \begin{subfigure}[t]{\textwidth}%
    \centering%
    \includegraphics[width=12cm]{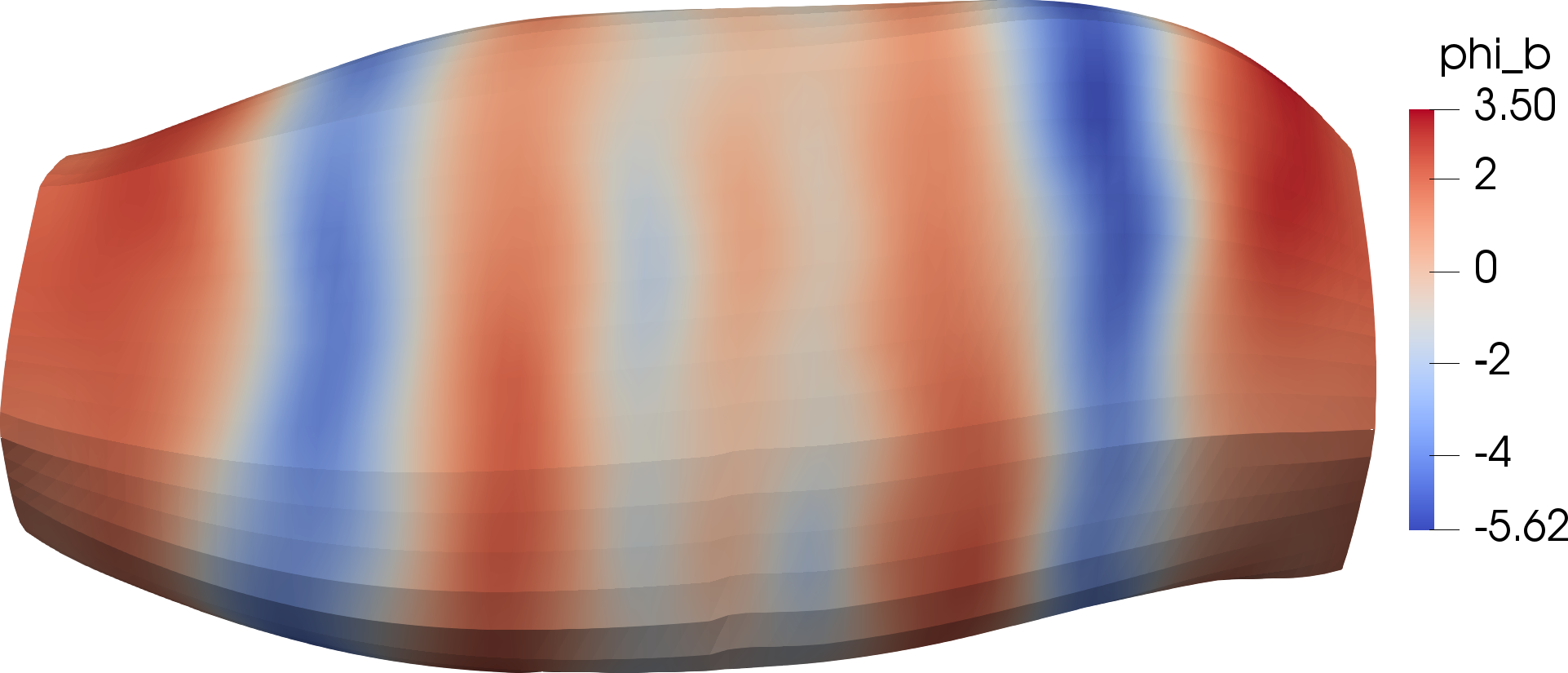}%
    \caption{EMG signal $\phi_b$ at the surface of the body fat domain.}%
    \label{fig:multidomain_25mus2_body}%
  \end{subfigure}
  \caption{Simulation of electrophysiology with the multidomain model: Simulation results at $t=\SI{20}{\ms}$ for a scenario with 25 MUs.}%
  \label{fig:multidomain_25mus2}%
\end{figure}

In this scenario, the stimulated nodes are no longer located in the middle of the muscle, but randomly varied by up to \SI{10}{\percent} of the muscle length using a uniform distribution. This approach to model the neuromuscular junctions is analog to the approach in \cref{sec:simfiber_mu} for the fiber based electrophysiology. 
\Cref{fig:multidomain_25mus_snapshot} shows the membrane voltage $V_m^1$ of MU 1 shortly after the MU has been activated. Because of the different locations of the stimulated points, no uniform action potential \say{front} as in \cref{fig:multidomain_4mus} is seen. Instead, a characteristic 2D MU action potential forms. 

The spike of the depolarized membrane voltage at every stimulated point propagates along the fiber direction and additionally diffuses in  transverse direction. This yields the cone-like structures of lower $V_m$ values as  seen in \cref{fig:multidomain_25mus_snapshot}. The origin of the transverse propagation is the electric conduction in the extracellular space, which is governed by the isotropic conduction tensor $\bfsigma_e$ in \cref{eq:multidomain_isotropic_e}. This isotropic conduction is strongly coupled to the directed action potential propagation within every MU compartment.

The resulting EMG values $\phi_e$ on the muscle boundary and $\phi_b$ on the skin surface are given in \cref{fig:multidomain_25mus2_emg} and \cref{fig:multidomain_25mus2_body}, respectively. The fat layer again smooths out the signal, as observed in the last section and in \cref{sec:simfiber_fat} for the fiber based electrophysiology model.

The two blue vertical stripes in \cref{fig:multidomain_25mus2_emg} with lower $\phi_e$ values correspond to the action potentials of multiple MUs at the respective location. It can be seen that the resulting EMG signal varies more in longitudinal direction than in transverse direction. This can be explained by the wide MU territories in this scenario.

% multidomain Vm MU
% multidomain phi_e

\subsection{Comparison of the Fiber Based Electrophysiology Model and the Multidomain Model}\label{sec:multidomain_differences}

EMG signals on the upper arm can be simulated by both the fiber based electrophysiology model, as demonstrated in \cref{sec:results_fiber_based_electrophysiology} and by the multidomain model, as shown in the previous sections. The two approaches have several similarities and differences.

Both model approaches have in common that they are based on biophysical principles. They both involve a detailed subcellular model, which describes the biochemical processes on the muscle fiber membranes. The subcellular model is solved at discrete points in the 3D domain and the model instances are coupled to a description of electric volume conduction in the muscle domain and the body fat layer.  Both the fiber based approach and the multidomain model are, thus, multi-scale descriptions. 

Both models also resolve the physiological structure of muscle activity given by multiple MUs. Action potential propagation is computed separately for different MUs. In a comprehensive simulation of the neuromuscular system, motor neuron models can be coupled to drive the activation of the MUs.

The two domains of the electrically active muscle tissue and the passive layer of skin and adipose tissue are also considered in both modeling approaches. In the fiber based approach, electric volume conduction is described by the 3D bidomain equation \cref{eq:bidomain1} for the muscle domain and a strongly coupled 3D Laplace equation \cref{eq:body} for the body domain.
The multidomain approach for electric volume conduction generalizes the bidomain model and yields the bidomain equation as a special case, if only one MU is considered. The Laplace equation for the body domain is coupled in the same way as in the fiber based model. In summary, both approaches are very similar regarding the 3D electric conduction part.

The major difference between the models is, that the multidomain approach considers only 3D domains, whereas the fiber based approach resolves individual 1D muscle fibers. Another difference lies in the coupling between the model components. For the fiber based approach, action potential propagation on the 1D fibers is unidirectionally coupled to the 3D volume conduction part of the model. In the multidomain model, all components are bidirectionally coupled. This allows, e.g., to simulate externally applied stimulations by active electrodes on the skin surface. The effects of the external currents on the electric potentials in the 3D muscle volume and down to the 0D subcellular behavior can only be described accurately by the multidomain approach.

The difference in coupling between the action potential propagation model part and the electric volume conduction in the extracellular space can also be seen in the computed EMG signals. A comparison of EMG simulations using the fiber based model, e.g., \cref{fig:emg273529b}, and the multidomain model, e.g., \cref{fig:multidomain_25mus2_emg}, shows that the multidomain approach yields less sharp artifacts in the 2D EMG signal on the muscle surface than the fiber based method. In the fiber based model, the action potentials of individual fibers are visible in the signal. In the multidomain simulations, the regions of similar activity are more clustered in the resulting EMG signals.

Other differences between the two model approaches exist in terms of the computational performance properties of their solvers. In the fiber based approach, action potential propagation can be computed independently for all muscle fibers, which enables large speedups by parallelization and makes large problem sizes with realistic numbers of muscle fibers feasible. For example, \cref{sec:effects_of_the_mesh_width_emg} presents the simulation of \num{270000} muscle fibers with \num{27000} compute cores.
In the multidomain approach, on the other hand, a large linear system of equations has to be solved in every timesteps. This can also be parallelized, but requires communication between the involved processes, which limits the parallel scalability for large problem sizes.

In the multidomain model, the computational effort increases, in good approximation, linearly with both the number of MUs and the number of nodes in the mesh. In the fiber based approach, the amount of computational work mainly corresponds to the number of fibers, not to the number of MUs. 
The 3D problem and, thus, the mesh width of the 3D mesh, typically plays a minor role in the total runtime for the fiber based approach, as the 3D problem is only solved according to the desired EMG sampling frequency. In the multidomain approach, no separate timestep widths can be chosen for the computations of the extracellular and body domain electric potentials, $\phi_e$ and $\phi_b$, as they are computed as a solution of the same linear system of equations.

For example, the computation of \SI{24}{\ms} of the multidomain scenario in \cref{sec:multidomain_simulation_emg} with 25 MUs and 126 processes has a runtime of approximately $\SI{106}{\min}$. The fiber based approach with the same 3D mesh and the same parallel partitioning with 126 processes has a total runtime of \SI{6}{\s} for 169 fibers or \SI{20}{\s} for 1369 fibers. A scenario with 169 fibers leads to a fiber spacing that corresponds to the 3D mesh width in the compared multidomain scenario. The speedup between the models in this case is approximately \num{1000}.
Note that only the computation of the fiber based approach is highly optimized in this work, and a better performance of the multidomain solver could be achieved in future work. However, the structural properties of the models facilitate highly parallel simulations only for the fiber based approach.

As a result, if the simplifications of a unidirectional coupling of the extracellular potential $\phi_e$ from the muscle fibers to the 3D volume can be tolerated, the fiber based approach should be used, as it exhibits significantly lower runtimes. The fiber based approach is (considering the current implementation) the only possible choice for scenarios with at least two of the three requirements (i) long simulation time spans in the range of seconds, (ii) large number of MUs in the range of multiple dozens, and (iii) finely resolved 3D meshes in the range of several $\num{1e6}$ degrees of freedom.
 
The multidomain approach, on the other hand, can describe phenomena that are not accurately captured by the fiber based model, as described earlier.
Moreover, the multidomain model is potentially easier to handle for more irregular geometries, where only a structured 3D mesh and no physiologically oriented fibers are given. Another advantage of the multidomain approach is its ability to fine tune the MU territories. 
The multidomain model can also possibly simulate a given MU distribution with the same accuracy with less 3D points than the fiber based approach. However, investigations in this direction are subject of future research.
If large runtimes are not an issue, the multidomain approach can be used to yield more physically accurate results than the fiber based approach, ultimately advancing the means to describe the neuromuscular system as detailed and accurately as possible.

% common
%   biophysically based, MUs, subcellular models
%   3D bidomain in both

% differences
%   coupling uni->bi
%   computation

% fibers and multidomain
%\begin{figure}[H]
%  \centering%
%  \begin{subfigure}[t]{0.48\textwidth}%
%    \centering%
%    \includegraphics[width=\textwidth]{images/results/application/2_multidomain.png}%
%    \caption{Multidomain}%
%    \label{fig:2_multidomain}%
%  \end{subfigure}
%  \quad
%  \begin{subfigure}[t]{0.48\textwidth}%
%    \centering%
%    \includegraphics[width=\textwidth]{images/results/application/2_fibers.png}%
%    \caption{Fibers}%
%    \label{fig:2_fibers}%
%  \end{subfigure}   
%  \caption{Fibers and Multidomain}%
%  \label{fig:multidomain_fibers}%
%\end{figure}%

\begin{reproduce_no_break}
  The two simulations in this section with 4 and 25 MUs, respectively, which are visualized in \cref{fig:multidomain_4mus,fig:multidomain_4mus_2,fig:multidomain_25mus2}, can be executed by the following commands:
  \begin{lstlisting}[columns=fullflexible,breaklines=true,postbreak=\mbox{\textcolor{gray}{$\hookrightarrow$}\space}]
    cd $\$$OPENDIHU_HOME/examples/electrophysiology/multidomain/multidomain_with_fat/build_release
    mpirun -n 128 multidomain_with_fat ../settings_multidomain_with_fat.py 4mus.py --n_subdomains 8 1 16
    mpirun -n 126 ./multidomain_with_fat_emg ../settings_multidomain_with_fat.py all_active.py --n_subdomains 6 1 21
  \end{lstlisting}
  For other available numbers of processes, the subdomains at the end of the commands have to be adjusted.
\end{reproduce_no_break}

% ----------------
%
% =================

%-----
% ==================
%
% =-------------------
%-----

\section{Simulation of Coupled Electrophysiology and Solid Mechanics}\label{sec:coupled_electrophysiology_and_solid_mechanics}

Simulating muscle contraction with a detailed model, which accurately describes motor recruitment,
yields the basis for new insights into the neuromuscular orchestration of processes that lead to muscle force generation.

We couple the two model approaches for electrophysiology, the fiber based model presented in \cref{sec:results_fiber_based_electrophysiology} and the multidomain model presented in \cref{sec:solver_multidomain_model}, with a solid mechanics model. 
In \cref{sec:solver_solid_mechanics}, we demonstrated the solver for nonlinear hyperelasticity models in simulations of the passive behavior of muscle tissue. The current section aims at simulating active muscle contraction.

\Cref{sec:fiber_based_contraction} couples the fiber based electrophysiology model with a model of muscle contraction. \Cref{sec:prestress_contraction} discusses an algorithm to add prestress to the description. \Cref{sec:multidomain_contraction} demonstrates the coupling of the multidomain model with the model of muscle contraction. \Cref{sec:volume_coupling_contraction} and \cref{sec:surface_coupling_contraction} describe simulations using the numerical coupling library preCICE.

\subsection{Fiber Based Electrophysiology and Muscle Contraction}\label{sec:fiber_based_contraction}
%-----

We begin with coupling the fiber based electrophysiology solver with the solid mechanics model to simulate muscle contraction as a result of the activation of muscle fibers.
We use the subcellular model of Shorten et al. \cite{Shorten2007}. It computes the microscopic activation parameter $\gamma \in [0,1]$, which is related to the concentration of attached cross-bridges in the sarcomeres. The parameter $\gamma$ is mapped and homogenized from the 0D subcellular points to $\bar{\gamma}$ on the 3D mesh. In the macroscopic 3D mechanics description, the factor is multiplied with a maximum active stress parameter $S_\text{max,active}$ and a force-velocity characteristic $f_l(\lambda_f)$, as described in \cref{sec:material_nonlinear_model}.
The 3D mechanics model updates the geometry of the 3D domain and transfers the fiber stretch value $\lambda_f$ and the contraction velocity $\dot{\lambda}_f$ back to the subcellular model.

In this scenario, we aim to simulate a rapid and strong contraction of the biceps muscle.
The scenario contains 169 fibers, which are associated with 15 MUs. This association is generated by method 1 in \cref{sec:method1_assignment}. All MUs are subsequently activated in a ramp in the first $\SI{1.4}{\s}$. 

The muscle geometry is fixed at its lower end and no external forces are considered in this scenario. The dynamic formulation with the transversely isotropic Mooney-Rivlin material is used, as described in \cref{sec:material_nonlinear_model}.
The 3D muscle mesh contains $2 \times 3 \times 18 = 108$ elements with quadratic finite element ansatz functions and 1295 nodes in total and is  partitioned into subdomains for four processes. Time step widths of $\dt_\text{0D} = \dt_\text{1D} = \dt_\text{splitting} = \SI{1e-4}{\ms}$ and $\dt_\text{3D}=\SI{1}{\ms}$ are used. The used numerical solvers and other settings of the electrophysiology and contraction models are equal to the described scenarios in \cref{sec:simfiber_mu,sec:effects_of_the_mesh_width_emg} and \cref{sec:comparison_linear_nonlinear}.

\begin{figure}
  \centering%
  \begin{subfigure}[t]{0.3\textwidth}%
    \centering%
    \includegraphics[height=9cm]{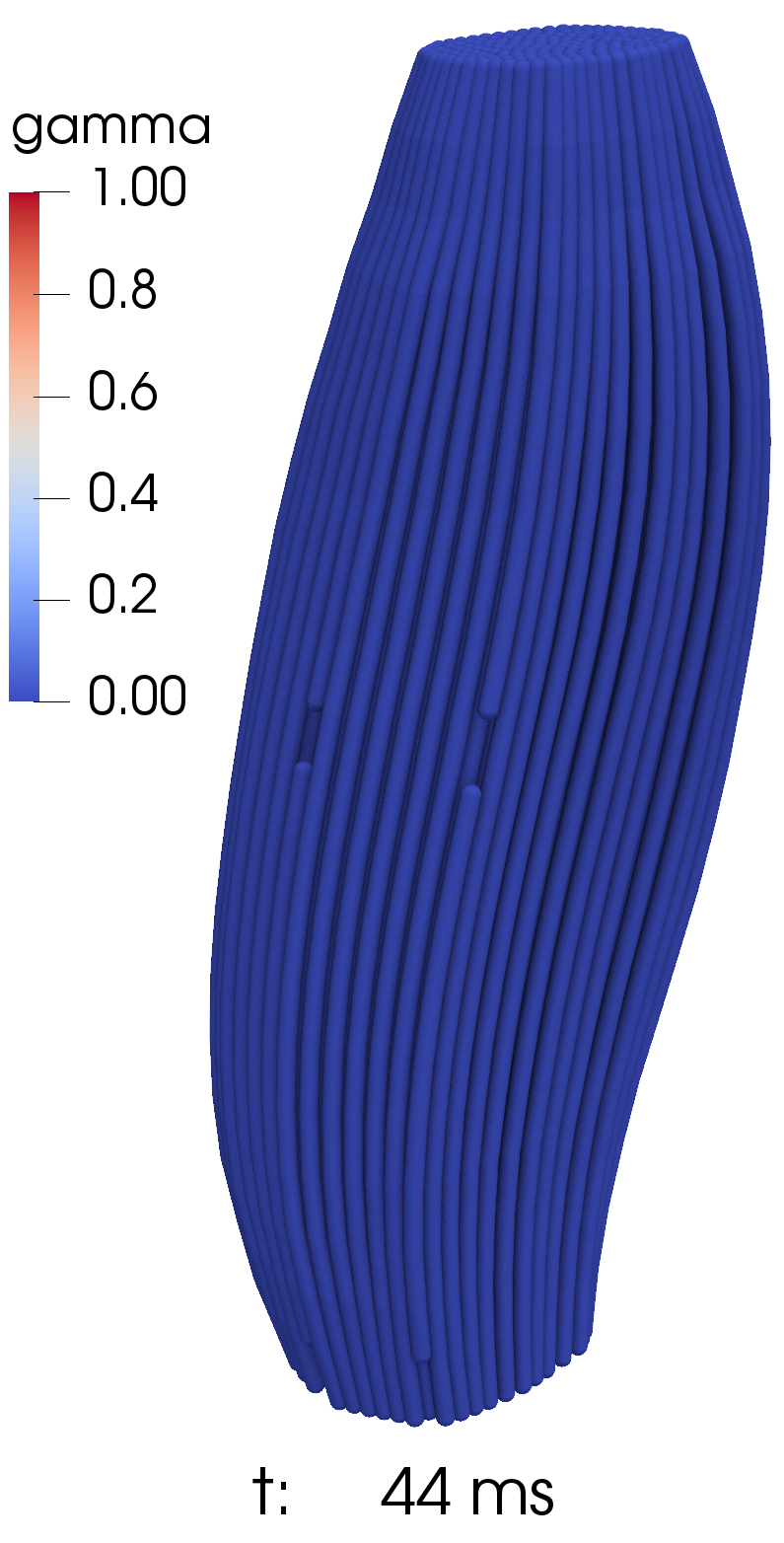}%
    \caption{}%
    \label{fig:contraction_fibers_044}%
  \end{subfigure} \,
  \begin{subfigure}[t]{0.18\textwidth}%
    \centering%
    \includegraphics[height=9cm]{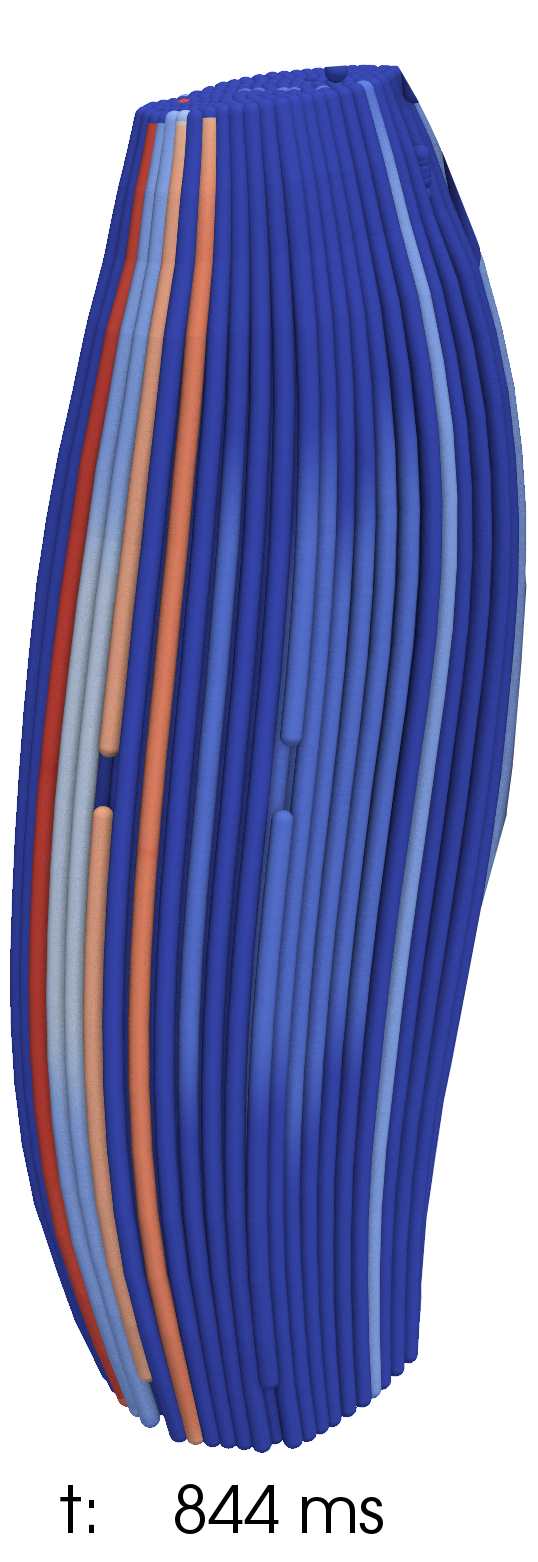}%
    \caption{}%
    \label{fig:contraction_fibers_844b}%
  \end{subfigure}\,
  \begin{subfigure}[t]{0.25\textwidth}%
    \centering%
    \includegraphics[height=9cm]{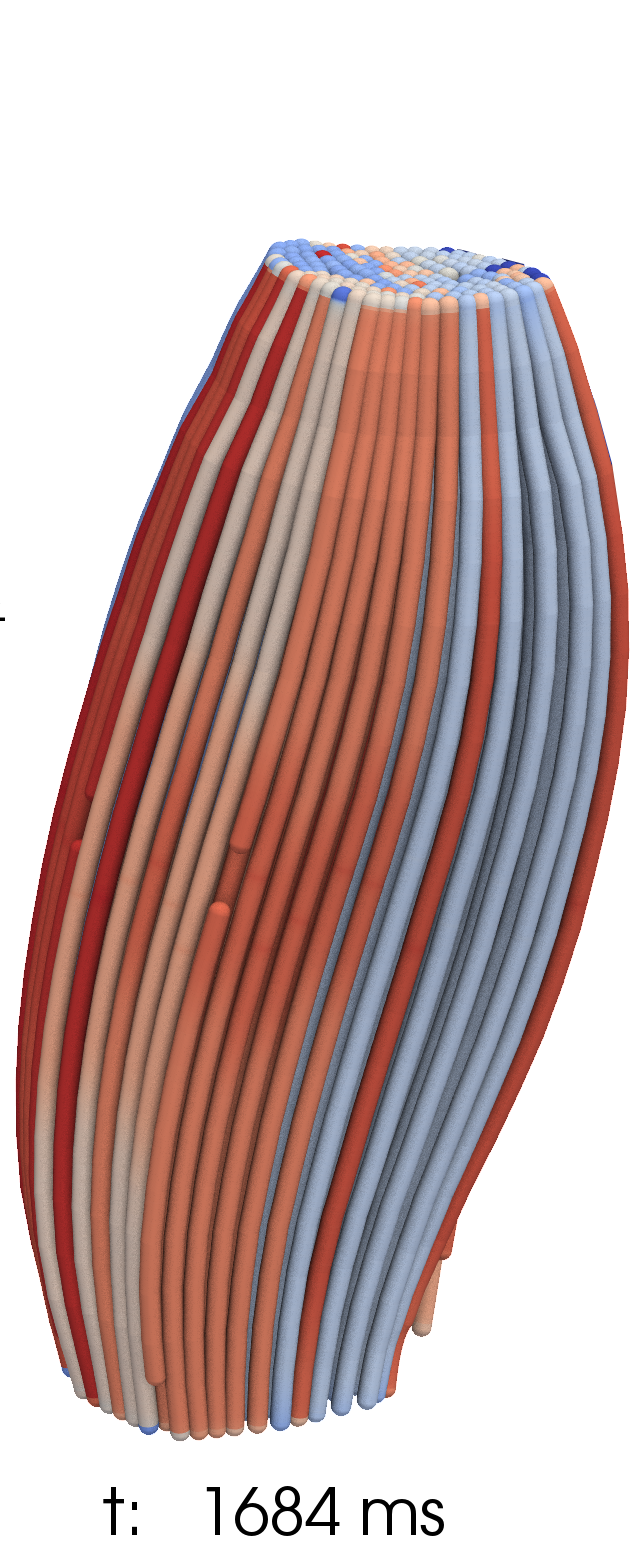}%
    \caption{}%
    \label{fig:contraction_fibers_1684b}%
  \end{subfigure}\,
  \begin{subfigure}[t]{0.2\textwidth}%
    \centering%
    \includegraphics[height=9cm]{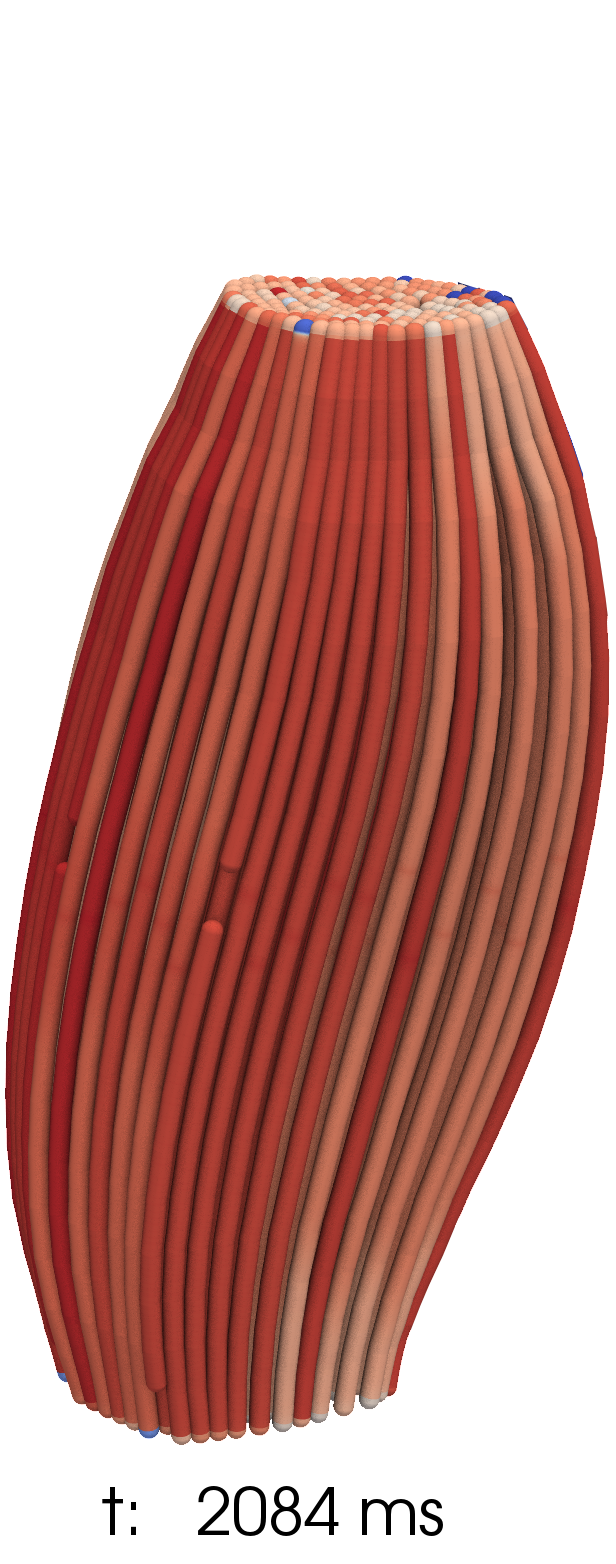}%
    \caption{}%
    \label{fig:contraction_fibers_2084b}%
  \end{subfigure}
  \caption{Simulation of fiber based electrophysiology and muscle contraction: Activation of the muscle fibers and overall deformation at various timesteps.}%
  \label{fig:contraction_fibers_1}%
\end{figure}%

\Cref{fig:contraction_fibers_1} shows the fibers of the contracting muscle at four different timesteps between $t=\SI{44}{\ms}$ and $t=\SI{2.084}{\s}$. The fibers are colored according to the resulting activation parameter $\gamma$, which is a measure for the generated force on the sarcomere level. Between $t=\SI{44}{\ms}$ and $t=\SI{844}{\ms}$, shown in \cref{fig:contraction_fibers_044,fig:contraction_fibers_844b}, the smallest MUs are activated, which, in this example, are mainly located on the left-hand side. As a consequence, the muscle domain initially bends slightly to the left. As more MUs become active at $t=\SI{1684}{\ms}$, depicted in \cref{fig:contraction_fibers_1684b}, the deformation increases and the bending direction is reversed. However, the fibers on the left-hand side still exhibit the highest $\gamma$ value, as they have been stimulated most often at that time.
At $t=\SI{1684}{\ms}$, visualized in \cref{fig:contraction_fibers_2084b}, almost all fibers have a $\gamma$ value close to one, corresponding to full activation.

% contracted state
\begin{figure}
  \centering%
  \begin{subfigure}[t]{0.31\textwidth}%
    \centering%
    \includegraphics[height=77mm]{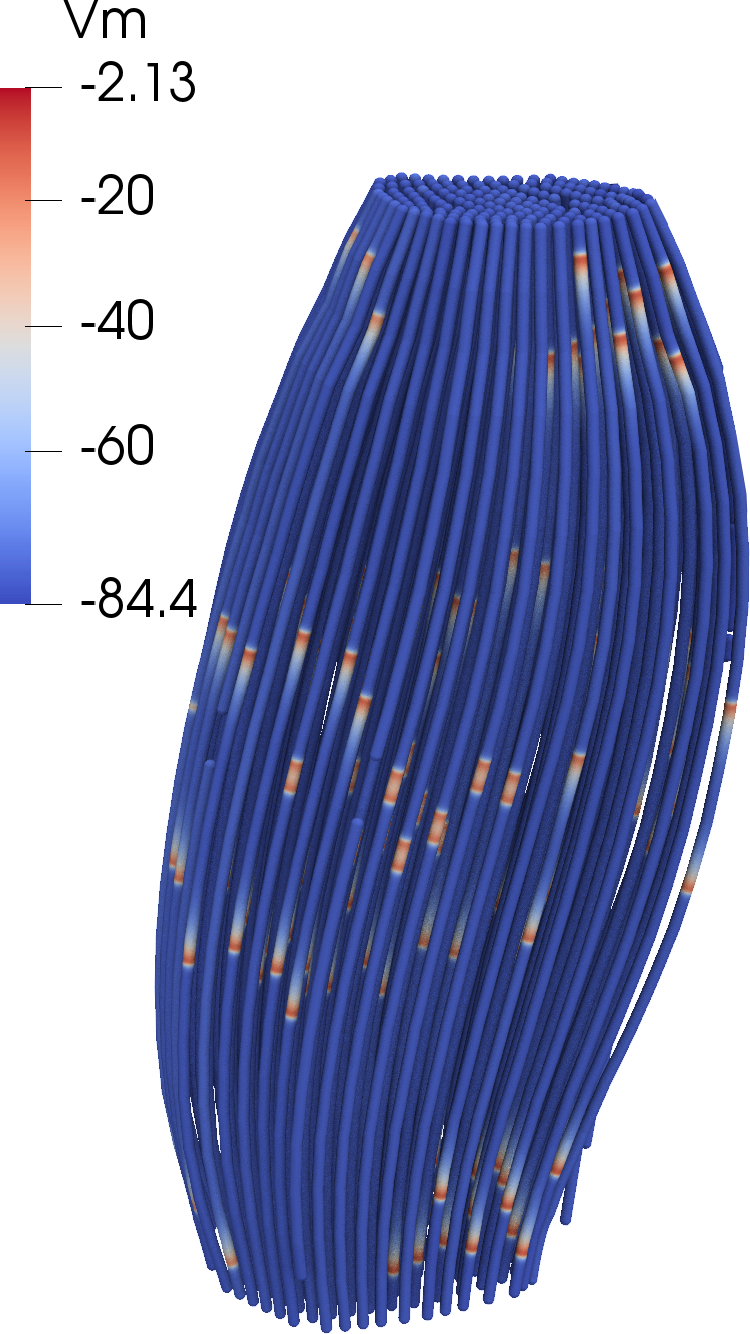}%
    \caption{Action potentials given by the membrane voltage $V_m$ (in millivolts) on the muscle fibers.}%
    \label{fig:contraction_fibers}%
  \end{subfigure}\,
  \begin{subfigure}[t]{0.31\textwidth}%
    \centering%
    \includegraphics[height=8cm]{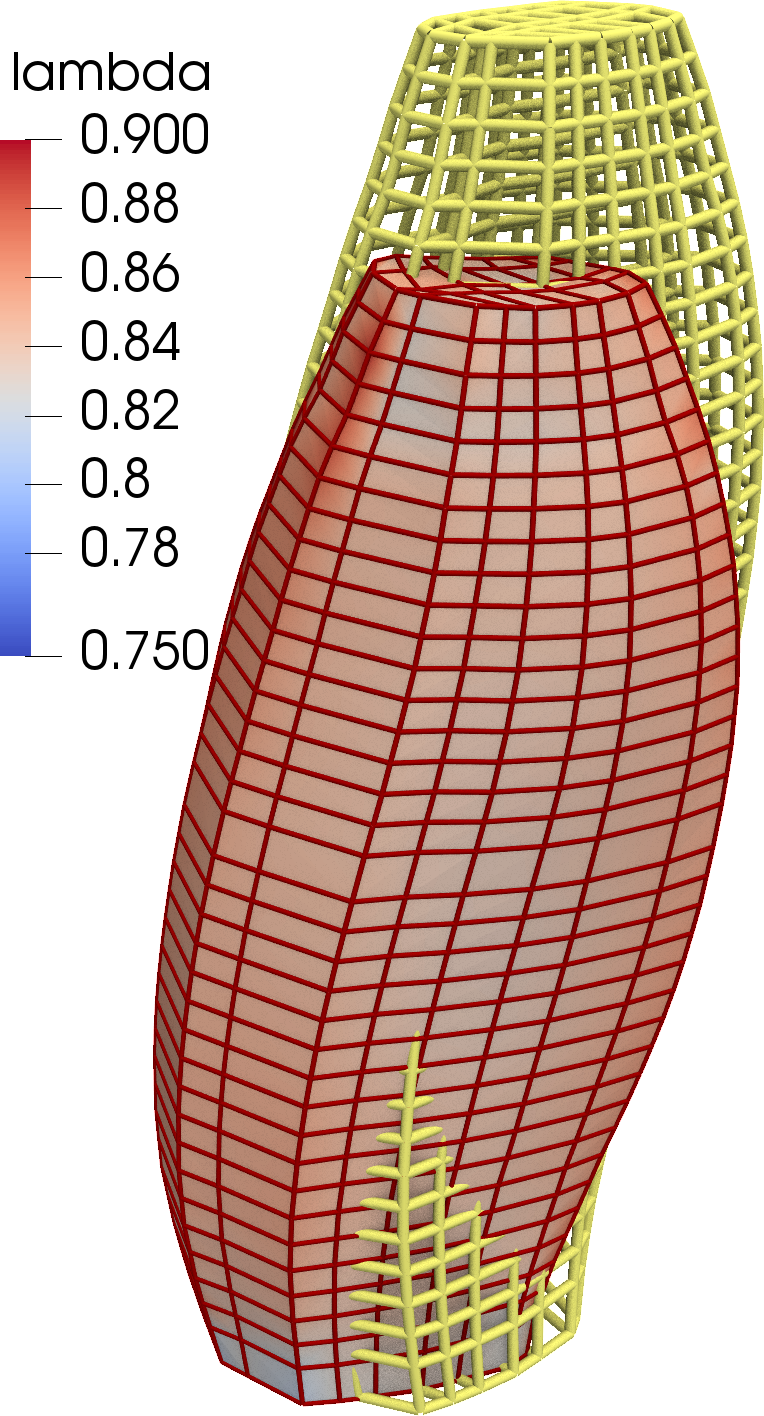}%
    \caption{Stretch parameter $\lambda$ of the deformed 3D muscle domain (red mesh) in comparison to the reference configuration (yellow mesh).}%
    \label{fig:contraction_lambda}%
  \end{subfigure}
  \begin{subfigure}[t]{0.31\textwidth}%
    \centering%
    \includegraphics[height=8cm]{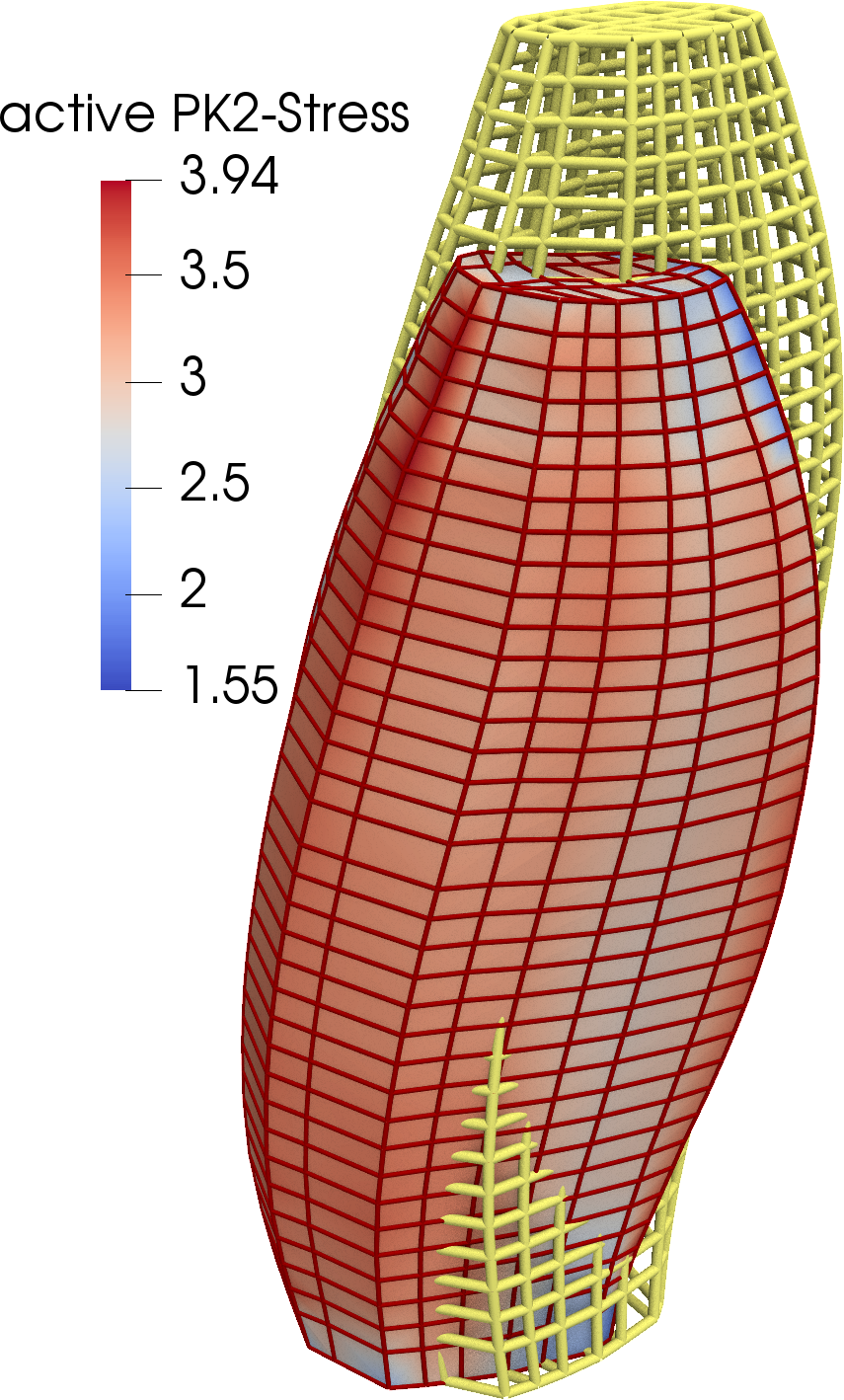}%
    \caption{Value of the active second Piola-Kirchhoff stress.}%
    \label{fig:contraction_active_stress}%
  \end{subfigure}
  \caption{Simulation of fiber based electrophysiology and muscle contraction: Simulation results at $t=\SI{2084}{\ms}$.}%
  \label{fig:contraction_at_end}%
\end{figure}%

\Cref{fig:contraction_at_end} shows several variables at the simulation end time of $t=\SI{2084}{\ms}$. \Cref{fig:contraction_fibers} visualizes the transmembrane voltage $V_m$ on the muscle fibers. Action potentials can be seen on almost all fibers, as the whole muscle is activated at this time. \Cref{fig:contraction_lambda} shows a comparison between the reference configuration given by the yellow mesh and the current configuration given by the red mesh. The muscle domain is colored according to the stretch $\lambda$, which has a nearly constant value of $\lambda \approx \SI{85}{\percent}$ at the end time of this scenario. A similar visualization is given in \cref{fig:contraction_active_stress} for the active stress in the muscle.

Because of the high level of activation and the corresponding active stress distribution in the muscle, our mechanics solver only converges up to the shown simulation time of \SI{2084}{\ms} in this scenario. The aim of the scenario is to simulate the contraction of a fully activated muscle. Other scenarios, where the activation is applied more slowly, allow for a convergence of the mechanics solver during longer simulation time spans.

The presented scenario showed that a fully activated biceps muscle contracted to about \SI{85}{\percent} of its original length. However, in reality, larger contractions are possible. In the shown scenario, the muscle was initially in a stress-free configuration. More realistic scenarios can incorporate pretension forces, where the undeformed reference configuration is subject to a constant stress level in the muscle's direction of the line of action. This is considered in the next scenario.

% muscle contraction
%\begin{figure}
%  \centering%
%  \includegraphics[width=0.5\textwidth]{images/results/application/neuromuscular_muscle_contraction_traction.png}%
 % \caption{contraction traction}%
%  \label{fig:neuromuscular_muscle_contraction_traction}%
%\end{figure}%

\begin{reproduce}
  The simulation in this section can be run as follows:
  \begin{lstlisting}[columns=fullflexible,breaklines=true,postbreak=\mbox{\textcolor{gray}{$\hookrightarrow$}\space}]
    cd $\$$OPENDIHU_HOME/examples/electrophysiology/fibers/fibers_contraction/no_precice/build_release
    mpirun -n 4 ./biceps_contraction ../settings_biceps_contraction.py ramp.py
  \end{lstlisting}
\end{reproduce}

\subsection{Simulation of Prestress}\label{sec:prestress_contraction}
% compressing muscle by external force does not work -> need activation 

To obtain more realistic ranges of muscle contraction, a nonzero, constant prestress can be considered in the undeformed configuration of the muscle. 
In our solid mechanics formulation, the reference configuration always has zero stress. Thus, we need to construct a separate, first reference configuration of a shorter muscle geometry. We stretch it to the original muscle length by applying external forces. The resulting, second configuration resembles the original muscle geometry and has the desired prestress characteristics.

\begin{figure}
  \centering%
  \hfill
  \begin{subfigure}[t]{0.48\textwidth}%
    \centering%
    \includegraphics[height=12cm]{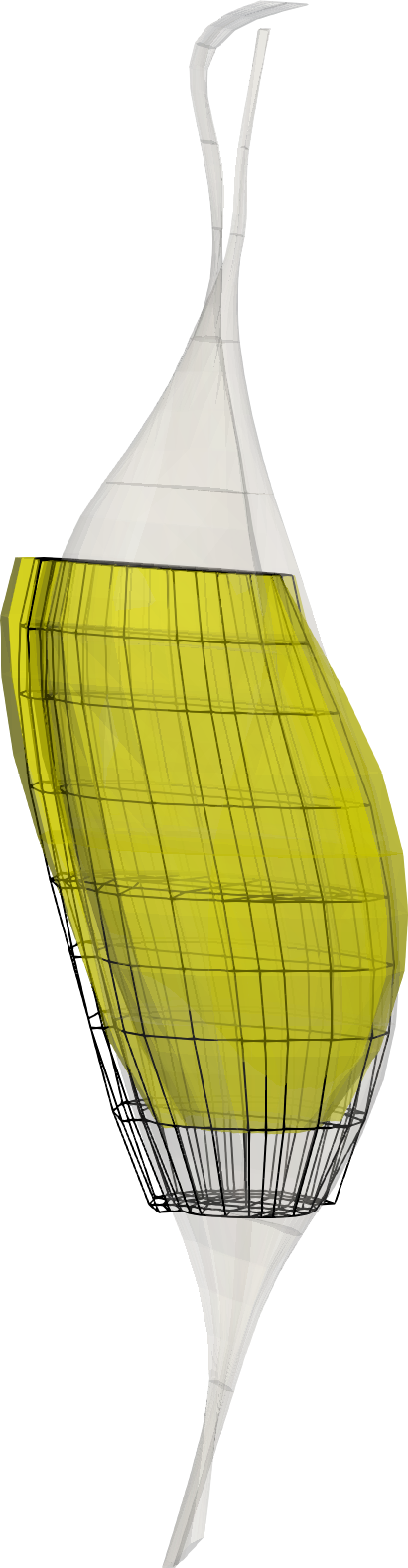}%
    \caption{In the first step, the original mesh (black wireframe) is contracted by an artificial active stress $\alpha_\text{pre}\,S_\text{max,active}$ to yield a shortened muscle (yellow mesh).}%
    \label{fig:neuromuscular_prestretch_1}%
  \end{subfigure}\hfill
  \begin{subfigure}[t]{0.48\textwidth}%
    \centering%
    \includegraphics[height=12cm]{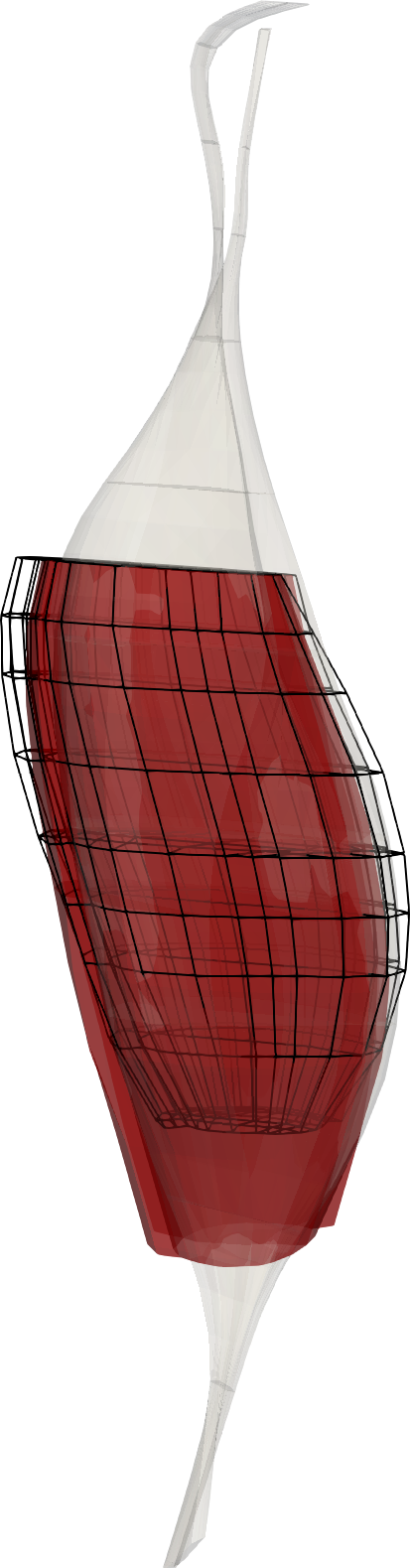}%
    \caption{In the second step, the mesh is extended again by an external surface load. The black wireframe corresponds to the yellow volume in (a), the red volume is the resulting geometry.}%
    \label{fig:neuromuscular_prestretch_2}%
  \end{subfigure}
  \hfill
  \caption{Simulation of biceps muscle geometry with prestress: The two steps of the algorithm to generate a reference geometry with prestress, shown with the geometry of the tendons for reference.}%
  \label{fig:neuromuscular_prestretch}%
\end{figure}

The detailed steps of this algorithm are visualized in \cref{fig:neuromuscular_prestretch} and are described in the following.
We begin with the given geometry of the muscle with body fat layer, which is shown as black wireframe mesh in \cref{fig:neuromuscular_prestretch_1}. In a first static simulation step, a constant active stress $\alpha_\text{pre}\,S_\text{max,active}$ is prescribed in the entire muscle volume. The resulting muscle deformation is computed, using the usual nonlinear hyperelastic muscle material. $S_\text{max,active}$ refers to the maximum active stress value as used in the mechanics model description in \cref{eq:active_stress_term}. The result of this first step is a shortened muscle with the same volume as the original geometry. \Cref{fig:neuromuscular_prestretch} shows the result by the yellow volume for $\alpha_\text{pre}=0.3$. It can be seen that the length of the muscle has shortened by approximately \SI{13}{\percent}.

In the second step, we reuse the computed deformed geometry of the first step as new stress-free reference configuration and re-extend it by applying a constant surface load $F_\text{pre}$ pointing to the bottom on the lower face in the setting of \cref{fig:neuromuscular_prestretch}. The value of $F_\text{pre}$ corresponding to $\alpha_\text{pre}$ has to be estimated by numerical experiments.

This step is again solved as a static problem. The result is a similar muscle geometry as the original one, and contains prestress according to the applied force. The muscle volume is exactly preserved due to the incompressible material formulation. \Cref{fig:neuromuscular_prestretch_2} shows the starting point for the second step by the black wireframe mesh and the resulting geometry for a total applied force of $F_\text{pre}=\SI{30}{\newton}$ by the red volume. The comparison of the original, black mesh in \cref{fig:neuromuscular_prestretch_1} with the red volume in \cref{fig:neuromuscular_prestretch_2} shows a good match of the geometry.
For the subsequent dynamic simulations of, e.g., muscle contraction, the surface load has to be constantly applied. It corresponds to the tendon forces and the loads of the musculoskeletal system acting on the muscle.

The active stress parameter $\alpha_\text{pre}$ and the corresponding preload force $F_\text{pre}$ can be chosen according to the desired amount of prestress. However, the higher these values are chosen, the more difficult is it for the nonlinear solid mechanics solvers to converge to a solution. Especially for irregular or large mechanics meshes, a lower stress factor of, e.g., $\alpha_\text{pre}=0.1$ has to be chosen. To improve convergence, we apply the load in the second step of the algorithm incrementally by several load steps. In addition, reducing the number of unknowns and increasing the mesh width in the mechanics problem can help, as this improves the conditioning of the problem. 
 
% contracted state

\subsection{Coupling of the Multidomain Model and Solid Mechanics Model with Prestress}\label{sec:multidomain_contraction}

In the following, we present a scenario that uses the prestress algorithm of the last section and couples the multidomain and mechanics models to simulate surface EMG signals on the skin surface over a contracting muscle.

We choose $\alpha_\text{pre}=0.1$ and apply the prestretch force $F_\text{pre}=\SI{10}{\newton}$ in three load steps. The multidomain model considers 5 MUs with stimulation frequencies between \SI{7}{\hertz} and \SI{24}{\hertz} and a 3D mesh of $8 \times 8 \times 28 = 1792$ elements. We execute the simulation with four processes. All other parameters and settings of the multidomain model and the solid mechanics model that are not explicitly mentioned in the following are chosen the same as in \cref{sec:multidomain_components} and \cref{sec:comparison_linear_nonlinear}.

\begin{figure}
  \centering%
  \hfill
  \begin{subfigure}[t]{0.31\textwidth}%
    \centering%
    \includegraphics[height=87mm]{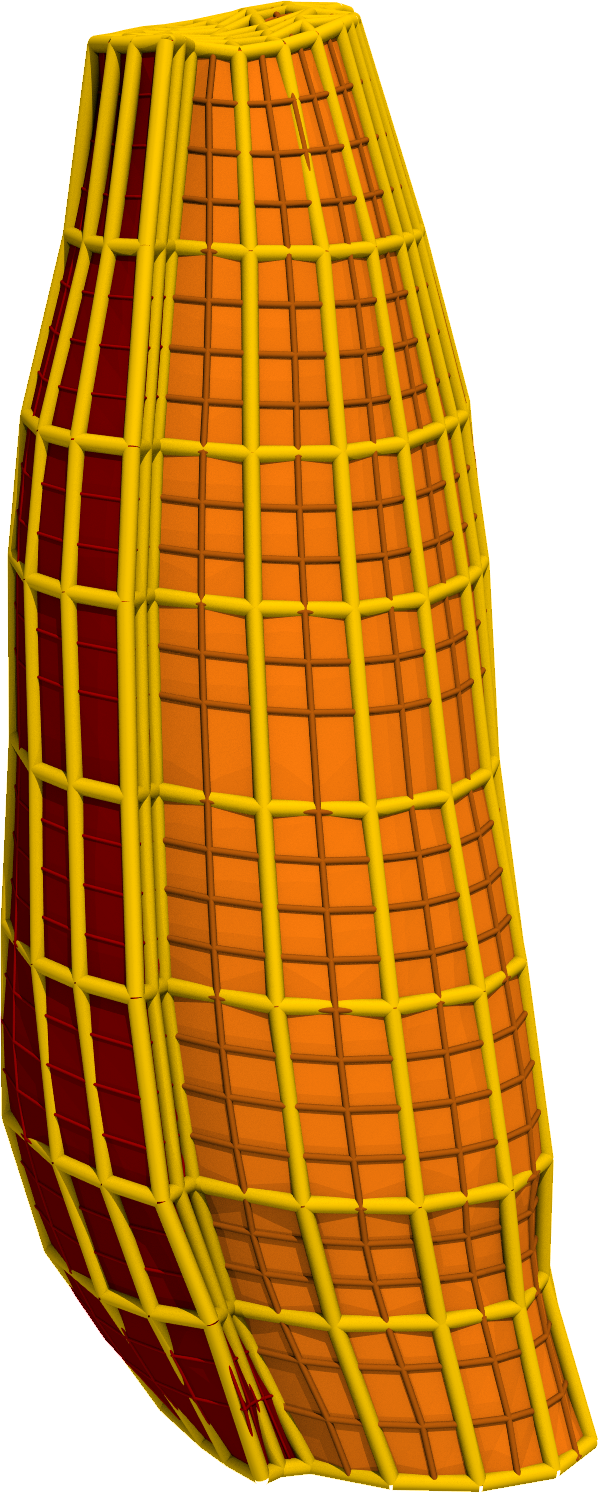}%
    \caption{Multidomain meshes of the muscle domain (orange), body fat domain (red) and the coarser mesh used for the solid mechanics model (yellow).}%
    \label{fig:multidomain_prestretch5}%
  \end{subfigure}
  \begin{subfigure}[t]{0.31\textwidth}%
    \centering%
    \includegraphics[height=95mm]{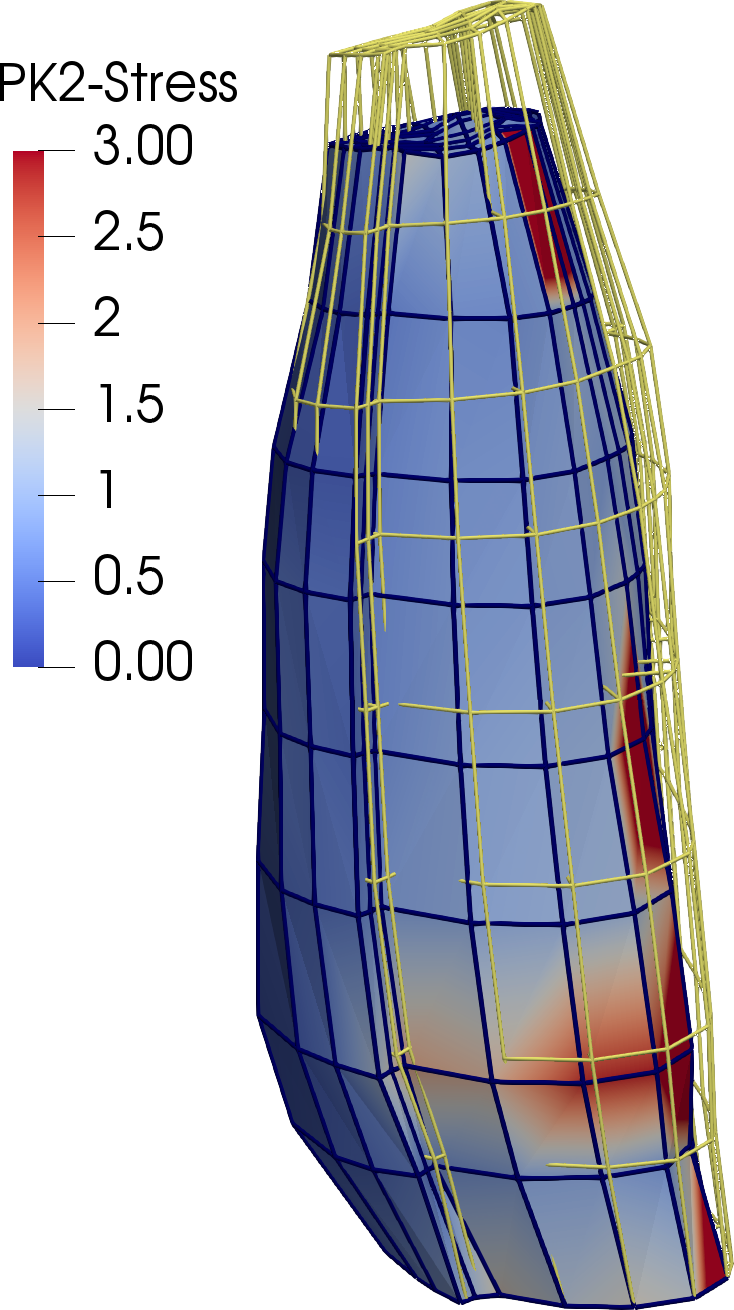}%
    \caption{Reference configuration (yellow mesh) and current configuration of the muscle colored according to the distribution of the second Piola-Kirchhoff stress.}%
    \label{fig:multidomain_prestretch6}%
  \end{subfigure}\qquad
  \begin{subfigure}[t]{0.31\textwidth}%
    \centering%
    \includegraphics[height=87mm]{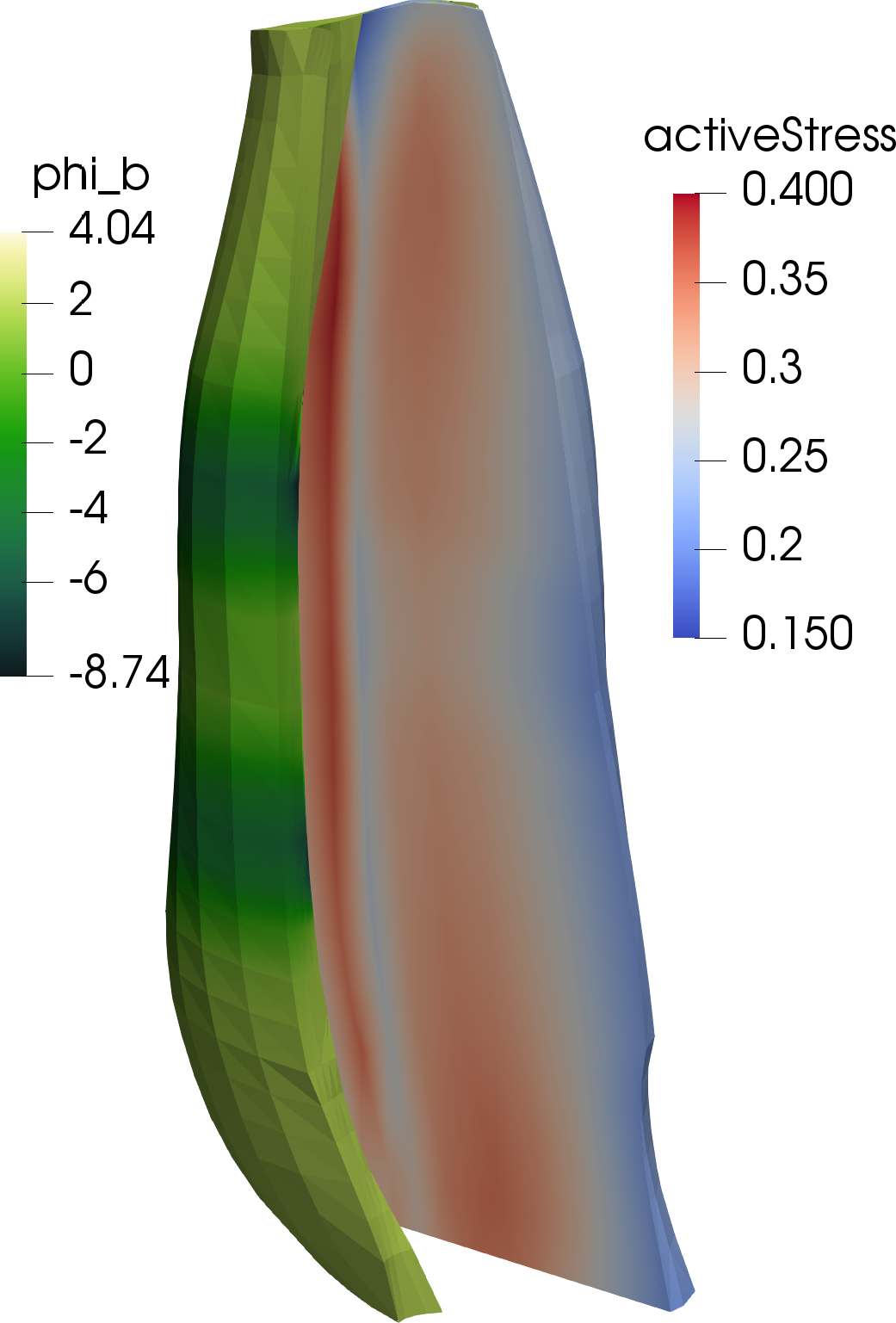}%
    \caption{Electric potential $\phi_b$ in the body domain (green color scale, in millivolts) and active stress in the interior of the muscle (blue-red color scale, in \SI{}{\newton\per\centi\meter\squared}).}%
    \label{fig:multidomain_prestretch2}%
  \end{subfigure}
  \hfill
  \caption{Simulation of muscle contraction based on the multidomain model with prestressed muscle geometry: Used meshes and simulation results at $t=\SI{920}{\ms}$ of a scenario of the multidomain electrophysiology model coupled to the solid mechanics model.}%
  \label{fig:multidomain_prestretch}%
\end{figure}

For the discretization of the mechanics model, we use a coarser mesh than for the multidomain model. Furthermore, we use quadratic elements instead of linear elements. The Python implementation of the settings script of this example contains functionality to create the mechanics mesh by subsampling the multidomain mesh with specified factors. In the current scenario, we set these factors for the $x$, $y$ and $z$ directions to 0.7, 0.7, and 0.3, respectively. As a result, we get meshes with $5 \times 7 \times 9 = 315$ elements for the muscle and $5 \times 1 \times 4 = 20$  elements for the body fat domain. \Cref{fig:multidomain_prestretch5} visualizes all meshes used in this scenario: The orange muscle mesh and the red body mesh are used for the multidomain model, and the yellow mesh is used for the solid mechanics model.

% multidomain prestretch

\Cref{fig:multidomain_prestretch6,fig:multidomain_prestretch2} depict results of the simulation at time $t=\SI{920}{\milli\second}$. 
\Cref{fig:multidomain_prestretch6} shows the reference geometry by the yellow wireframe after applying the prestress. The muscle is colored according to the value of the second Piola-Kirchhoff stress. During this dynamic simulation, the muscle bends elastically slightly to the left and right, as it is only fixed at its bottom in \cref{fig:multidomain_prestretch}. This explains the stress distribution at the snapshot for $t=\SI{920}{\milli\second}$ in \cref{fig:multidomain_prestretch6}, where higher stresses occur on the right-hand side.

\Cref{fig:multidomain_prestretch2} shows the electric potential $\phi_b$ of the body domain by the green color scale on the left of the image. The visible part of the fat layer shows two action potentials, visualized by the two dark green stripes.

Moreover, \cref{fig:multidomain_prestretch2} displays the total active stress  $\bfS_\text{active}$ in the interior of the muscle by the color scale that ranges from blue to red color. In the multidomain model, $\bfS_\text{active}$ is calculated as a weighted sum over the contributions $\bfS_\text{active}^k$ of the MU compartments, scaled by the occupancy factors $f_r^k$ (cf. \cref{sec:multidomain_model}):
\begin{align*}
  \bfS_\text{active} = \sum\limits_{k=1}^{N_\text{MU}} f_r^k\,\bfS_\text{active}^k.
\end{align*}
The muscle domain in \cref{fig:multidomain_prestretch2} is cut open, such that interior distribution at the cut plane can be seen. The image shows two regions of higher active stress, which run vertically through the muscle, given by red color. They are a result on the location of the MUs in this scenario. The legend shows that the active stress inside the muscle is below $\SI{0.4}{\newton\per\centi\meter\squared}$, while the maximum active stress parameter is chosen as $S_\text{max,active}=\SI{7.3}{\newton\per\centi\meter\squared}$. This low activation level is a result of the chosen MU recruitment. As a result, the muscle only slightly contracts, as can be seen in \cref{fig:multidomain_prestretch6}.

In summary, both the fiber based electrophysiology model and the multidomain model can be coupled with the nonlinear solid mechanics model to simulate muscle contraction, as presented in \cref{sec:fiber_based_contraction} and in this section. 
The computational efficiency considerations discussed in the comparison of the fiber based and multidomain approaches in \cref{sec:multidomain_differences} also apply to coupled simulations with muscle contraction. For longer simulation times, the fiber based approach in \cref{sec:fiber_based_contraction} is, therefore, favored.

\begin{reproduce_no_break}
  The simulation can be run with the following commands. Instead of four processes also other numbers are possible. A lot of parameters can be fine-tuned in the \code{../variables/multidomain.py} settings file.
  \begin{lstlisting}[columns=fullflexible,breaklines=true,postbreak=\mbox{\textcolor{gray}{$\hookrightarrow$}\space}]
    cd $\$$OPENDIHU_HOME/examples/electrophysiology/multidomain/multidomain_prestretch/build_release
    mpirun -n 4 ./multidomain_prestretch ../settings_multidomain_prestretch.py multidomain.py
  \end{lstlisting}
\end{reproduce_no_break}

% --------
%
% f==============

% ==================
%
% =-------------------
%-----
\subsection{Coupling of Solid Mechanics Models using the Software preCICE}\label{sec:volume_coupling_contraction}

One problem of multi-scale simulations with solid mechanics models is the limited amount of parallelism, 
if a coarse mechanics mesh with a low number of elements is chosen. The domain can only be partitioned into as many subdomains as there are elements in the 3D mechanics mesh. While this is not an issue for small scale simulations like the ones shown in the previous sections, it prohibits exploitation of High Performance Computing resources, e.g., if numerous muscle fibers are considered as in \cref{sec:effects_of_the_mesh_width_emg}.

The reason for the limited parallelism lies in the partitioning scheme, where every node in the 3D domain corresponds to the subdomain of exactly one process, regardless of the mesh. OpenDiHu does not allow to partition, e.g., the finely resolved 1D muscle fiber meshes differently than the coarse 3D mechanics mesh. However, this restriction can be circumvented by using multiple OpenDiHu programs with different partitioning schemes and by performing the data transfer between the meshes using an external coupling software.

We provide support for the black-box coupling library preCICE \cite{precice}. This open source library allows mapping data between different meshes, can communicate values between subdomains that reside on different processors, and implements implicit numerical coupling schemes with quasi-Newton methods. The implementation is known to scale well on small-scale clusters and supercomputers.
The preCICE library targets a minimally-invasive approach, where the user application implements a preCICE adapter. Multiple, potentially different solver codes can be coupled numerically and compute individual model parts of a joint multi-physics simulation. Moreover, preCICE has an active and growing community where experiences and codes are shared, and open source adapters are available for several popular solvers.

This makes the library suited for our use case. We provide two different types of preCICE adapters in OpenDiHu, one for surface coupling of 2D meshes and one for volume coupling of 3D meshes. These adapters integrate with the structure of nested solvers and can be positioned anywhere in the solver tree (cf. \cref{fig:solver_tree_multidomain_spindles}). The meshes and variables that are exposed to preCICE can be configured in the settings file.

In the current section, we show how to use the volume coupling adapter to resolve the initially stated issue of limited scalability for coupled simulations with electrophysiology and mechanics models. Subsequently, the next section presents a simulation that uses surface coupling. Details can also be found in \cite{hlrs2021}.

We simulate muscle contraction and surface EMG of the biceps muscle using the fiber based electrophysiology model. To fully exploit the  capabilities of an 18-core Intel Core i9-10980XE processor, we compute the electrophysiology model using 16 processes and the mechanics model using 2 processes. The data mapping between the differently partitioned 3D meshes is performed by preCICE.

\Cref{fig:precice_muscle_force} shows the structure of the simulation components with the used meshes and the exchanged variables in this simulation. Two different OpenDiHu programs are executed at the same time, given by the gray boxes. The program corresponding to the left box solves the electric conduction problem, given by the bidomain equation \cref{eq:bidomain1} on the 3D domain and the action potential propagation model, given by the monodomain equation \cref{eq:monodomain} on a large number of 1D muscle fiber meshes. 
The 3D and the 1D mesh in this program are partitioned into 16 subdomains for the 16 processes.

\Cref{fig:precice_muscle_force} visualizes the meshes and their partitioning to the different processes by the colored inset images. It can be seen that the fibers meshes and the 3D mesh used in the OpenDiHu program in the left box have corresponding subdomains.

The second OpenDiHu program visualized by the right box in \cref{fig:precice_muscle_force} only solves the solid mechanics problem using a coarse 3D mesh. The mesh of this problem is partitioned to two processes, as shown by the image. 

The three model parts are numerically coupled and need to exchange several variables. The action potential propagation model, shown at the lower left of \cref{fig:precice_muscle_force}, computes the transmembrane voltage $V_m$ and the activation parameter $\gamma$ and maps them from the 0D points on the fibers to the 3D mesh using the mapping scheme described in \cref{sec:data_mapping_between_meshes}. The activation parameter $\gamma$ is needed in the solid mechanics model. It is transferred between the two OpenDiHu programs using the functionality of preCICE. After the solid mechanics solver has computed a new deformation of the coarse solid mechanics mesh, preCICE maps the node positions to the finer 3D mesh in the left program. The geometries of the 3D and 1D meshes in the left program are updated accordingly.
The preCICE couplings in this example use serial explicit coupling and radial basis functions for the data mapping.

% precice coupling scheme
\begin{figure}
  \centering%
  \def\svgwidth{0.9\textwidth}
  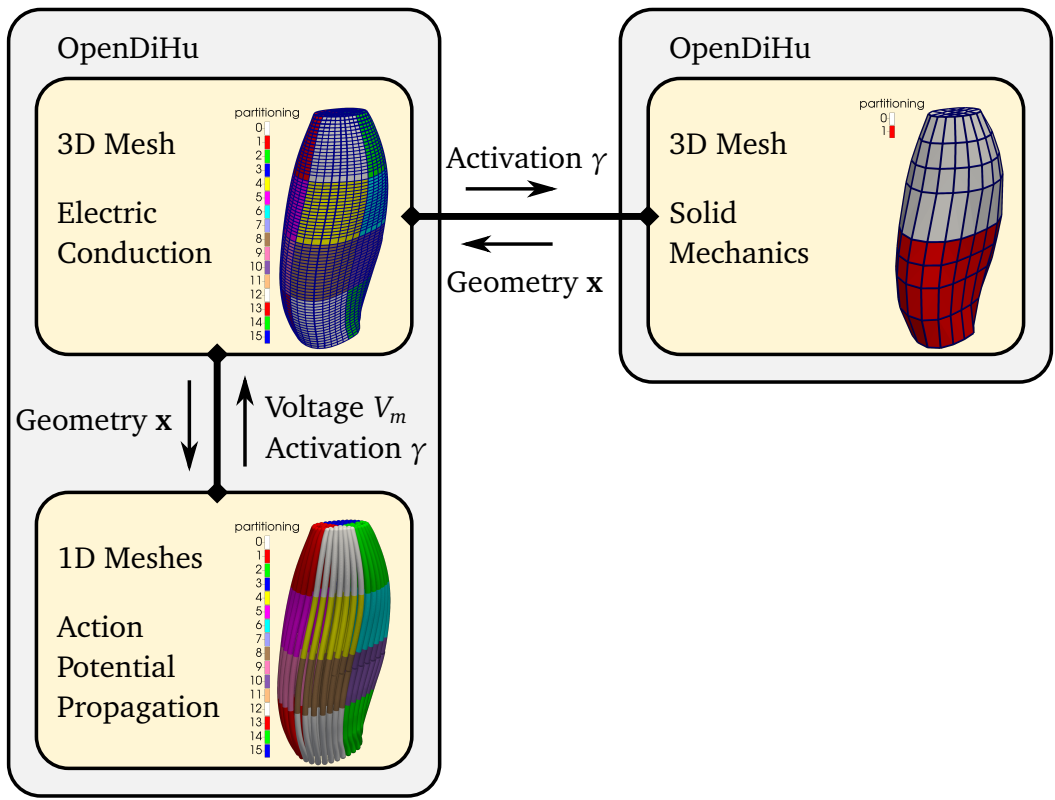%
  \caption{Simulation of muscle contraction: Structure of a coupled simulation with the coupling library preCICE on 18 processes, consisting of the two independent OpenDiHu programs indicated by the gray boxes. The program in the left box solves the electric conduction model using the shown 3D mesh and the action potential propagation model using the shown 1D fiber meshes. Both meshes are partitioned to 16 subdomains as shown by the colors. The program in the right box solves the mechanics problem on a coarse 3D mesh, which is partitioned into 2 subdomains. The arrows between the models indicate the exchanged variables. The coupling within the left gray box is implemented in OpenDiHu, the coupling between the gray boxes is realized using preCICE.}%
  \label{fig:precice_scheme1}%
\end{figure}

The presented scheme in \cref{fig:precice_scheme1} allows us to simulate muscle contraction and surface EMG signals. The volume coupling with preCICE is configured between the two 3D meshes. Even for scenarios where EMG signals are not of interest and only the muscle contraction resulting from the activated muscle fibers should be simulated, the presented approach can be used.

An alternative approach, where preCICE instead couples directly between the fiber meshes and the solid mechanics 3D mesh is also implemented. This approach neither includes the electric conduction model nor the fine 3D mesh for the left program.
However, the mapping between the solid mechanics mesh and the set of 1D fiber meshes is more costly than the mapping between the two 3D meshes, as the fibers contain more data points in total than the 3D mesh of the electric conduction problem. A quantitative analysis of this effect is subject to work in progress.

Apart from ensuring better parallel scalability, the OpenDiHu model setup using preCICE also allows to exchange the solid mechanics solver by a different solver code, e.g., a commercial solver. The black-box approach of preCICE allows to exchange the mechanics solver without any changes to the electrophysiology simulation, contributing to the extensibility goal of combining modular model components.

\begin{reproduce_no_break}
  The two programs with preCICE coupling can be used as follows. Note that the compilation of preCICE has to be enabled in the \code{user-variables.scons.py} configuration file for the \code{scons} build system in the \code{$\$$OPENDIHU_HOME} directory.
  \begin{lstlisting}[columns=fullflexible,breaklines=true,postbreak=\mbox{\textcolor{gray}{$\hookrightarrow$}\space}]
    cd $\$$OPENDIHU_HOME/examples/electrophysiology/fibers/fibers_contraction/with_precice_volume_coupling/build_release
    mpirun -n 2 ./muscle_contraction ../settings_muscle_contraction.py ramp.py
    mpirun -n 16 ./fibers_with_3d ../settings_fibers_with_3d.py ramp.py
  \end{lstlisting}
\end{reproduce_no_break}

%-----
\subsection{Simulation of a Muscle-Tendon Complex using Surface Coupling with preCICE}\label{sec:surface_coupling_contraction}

In all previously presented simulations of muscle contraction, the biceps muscle was considered in isolation. In the following, we present a  physiologically more correct scenario that includes a 3D description of the tendon mechanics.
The simulation consists of four individual solvers in OpenDiHu for the distal tendon, the two proximal tendons, and the muscle belly.
The coupling library preCICE is used to numerically couple the parts.

An advantage of simulations of an entire muscle-tendon complex is their more realistic line of action of the muscle force, compared with a model of the muscle belly without tendons. The goal of the simulation described in this section is to predict the progression of the total muscle force as  result of MU recruitment.

For the simulation of the muscle contraction part, we couple the fiber based electrophysiology model with the nonlinear solid mechanics model as described in \cref{sec:fiber_based_contraction}. The electrophysiology part of the muscle uses the subcellular model of Shorten et al. \cite{Shorten2007}. 
The solid mechanics description of the three tendons uses the hyperelastic Saint-Venant Kirchhoff material, which is the extension of the linear elastic formulation given in \cref{sec:material_linear_model} to the geometrically nonlinear regime.
The proximal tendons are fixed at their insertion points to the skeletal system. We apply corresponding Dirichlet boundary conditions. At the lower end of the distal tendon, a downwards pulling force is applied. We gradually increase the value of this force in the corresponding Neumann boundary conditions from zero up to the maximum value \SI{100}{\newton} during the first \SI{100}{\ms} of the simulation.

The muscle fibers are associated with 10 MUs and activated in a ramp during the first $\SI{1.8}{\s}$. After each MU has been activated for the first time, it fires with a MU specific frequency between \SI{7.66}{\hertz} and \SI{23.92}{\hertz} plus a random jitter value of \SI{10}{\percent}. This setup replicates the progressive recruitment scenario in \cite{Klotz2020}.

The four simulation programs are connected using an implicit Neumann-Dirichlet multi-coupling scheme in preCICE with a constant relaxation factor of 0.5. At the interfaces between the muscle and the tendons, the implicit numerical coupling ensures continuity for the displacements, velocities and stresses.
The tendon solvers send their computed displacement and velocity values to the muscle model, where the corresponding Dirichlet boundary conditions are applied. The muscle model computes traction forces by integrating the stress values over the surface and sends the values to the tendon models, where corresponding Neumann boundary conditions are applied. 

We configure preCICE to use Gaussian radial basis functions for the consistent mapping of the variables between the surface meshes of the muscle and the tendons. 
An error threshold of $\eps=0.1$ for the coupled displacement values is used to terminate the implicit coupling scheme. As a consequence, the scheme requires approximately two iterations per timestep on average to reach the error threshold. The coupling step is repeated with a timestep width of $\dt_\text{coupling} = \SI{1}{\ms}$.

We simulate two scenarios of this model. The first scenario considers a high spatial resolution of 1089 muscle fibers and a simulation time of approximately \SI{1}{\s}, while the second scenario considers only 81 fibers but a longer simulation time span of \SI{10}{\s}. 

In the first scenario, we use a 3D mesh with $9\times 9 \times 21=1701$ nodes, which are partitioned into 160 subdomains.
The meshes of the three tendons each consists of 125 nodes and are each partitioned to four subdomains. We run the computation using 172 processes on four compute nodes of the supercomputer Hawk at the High Performance Computing Center Stuttgart. The hardware is described in more detail in  \cref{sec:effects_of_the_mesh_width_emg}. The simulation time span of $\SI{1}{\s}$ has a runtime of approximately $\SI{7}{\hour}$ $\SI{20}{\minute}$.

\begin{figure}
  \centering%
  \begin{subfigure}[t]{0.34\textwidth}%
    \centering%
    \includegraphics[height=97mm]{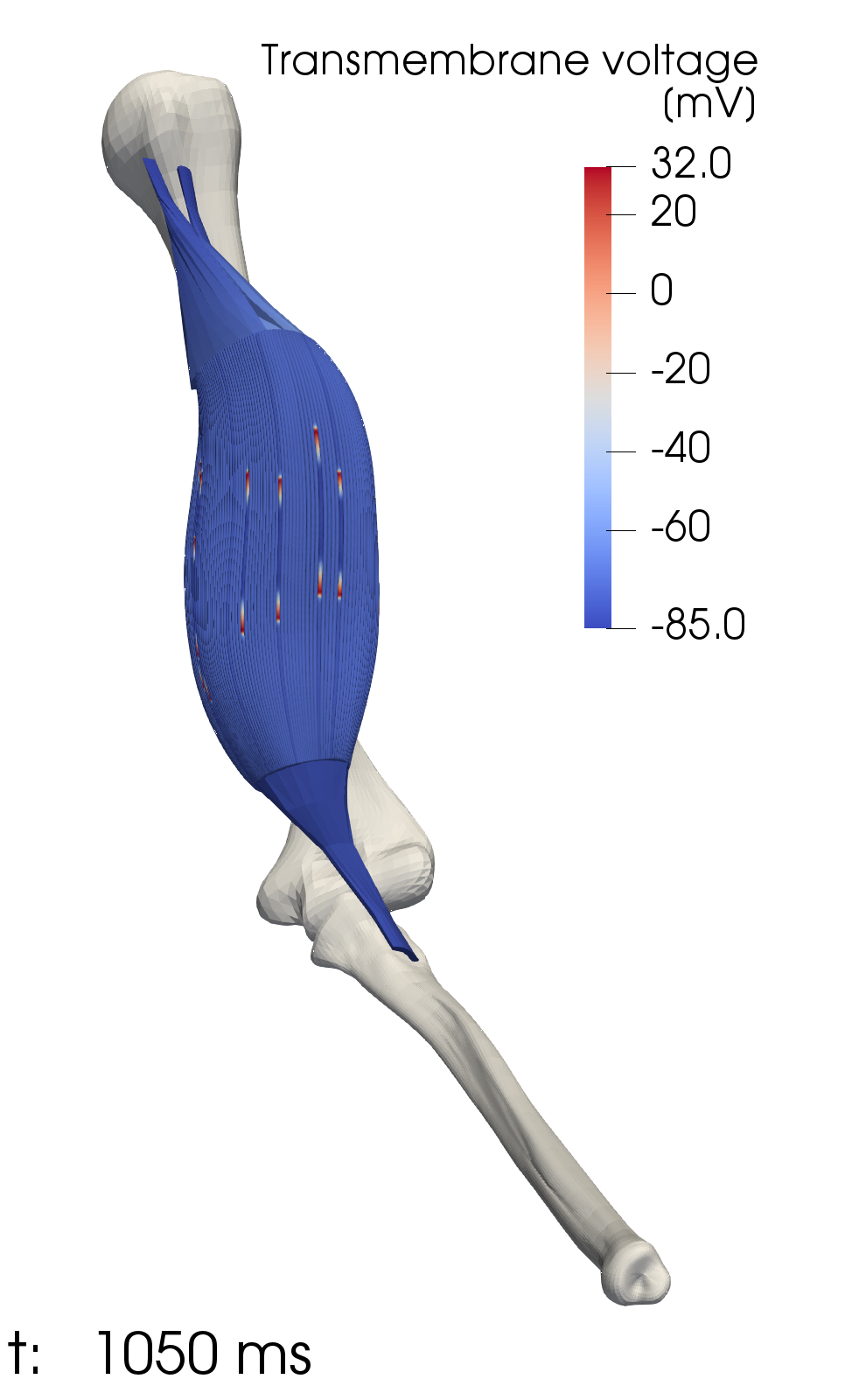}%
    \caption{Muscle fibers colored by the value of the transmembrane potential $V_m$.}%
    \label{fig:precice_activated_muscles_1}%
  \end{subfigure}\,
  \begin{subfigure}[t]{0.28\textwidth}%
    \centering%
    \includegraphics[height=97mm]{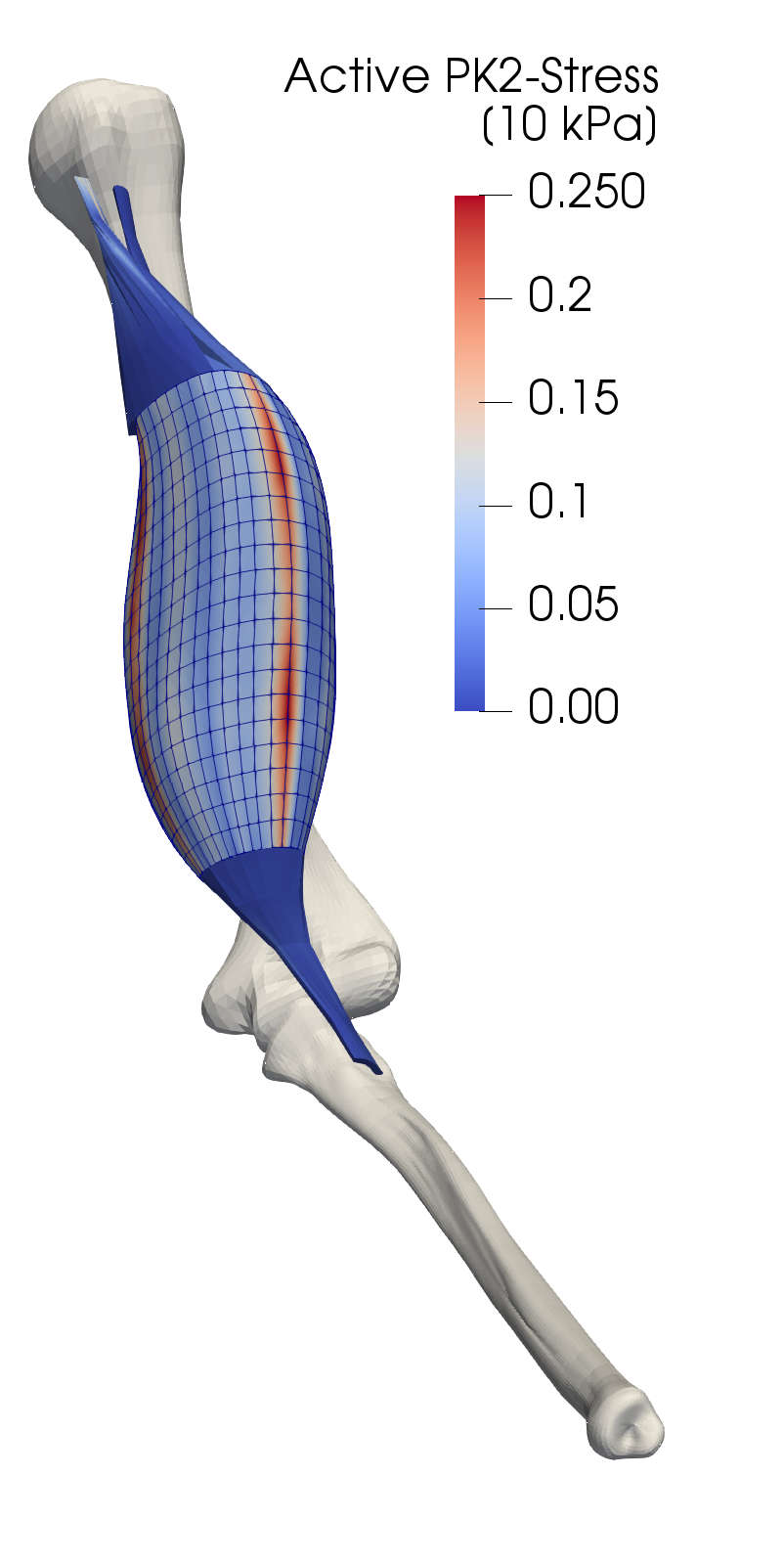}%
    \caption{Active stress in the 3D muscle mesh.}%
    \label{fig:precice_activated_muscles_2}%
  \end{subfigure}
  \begin{subfigure}[t]{0.28\textwidth}%
    \centering%
    \includegraphics[height=97mm]{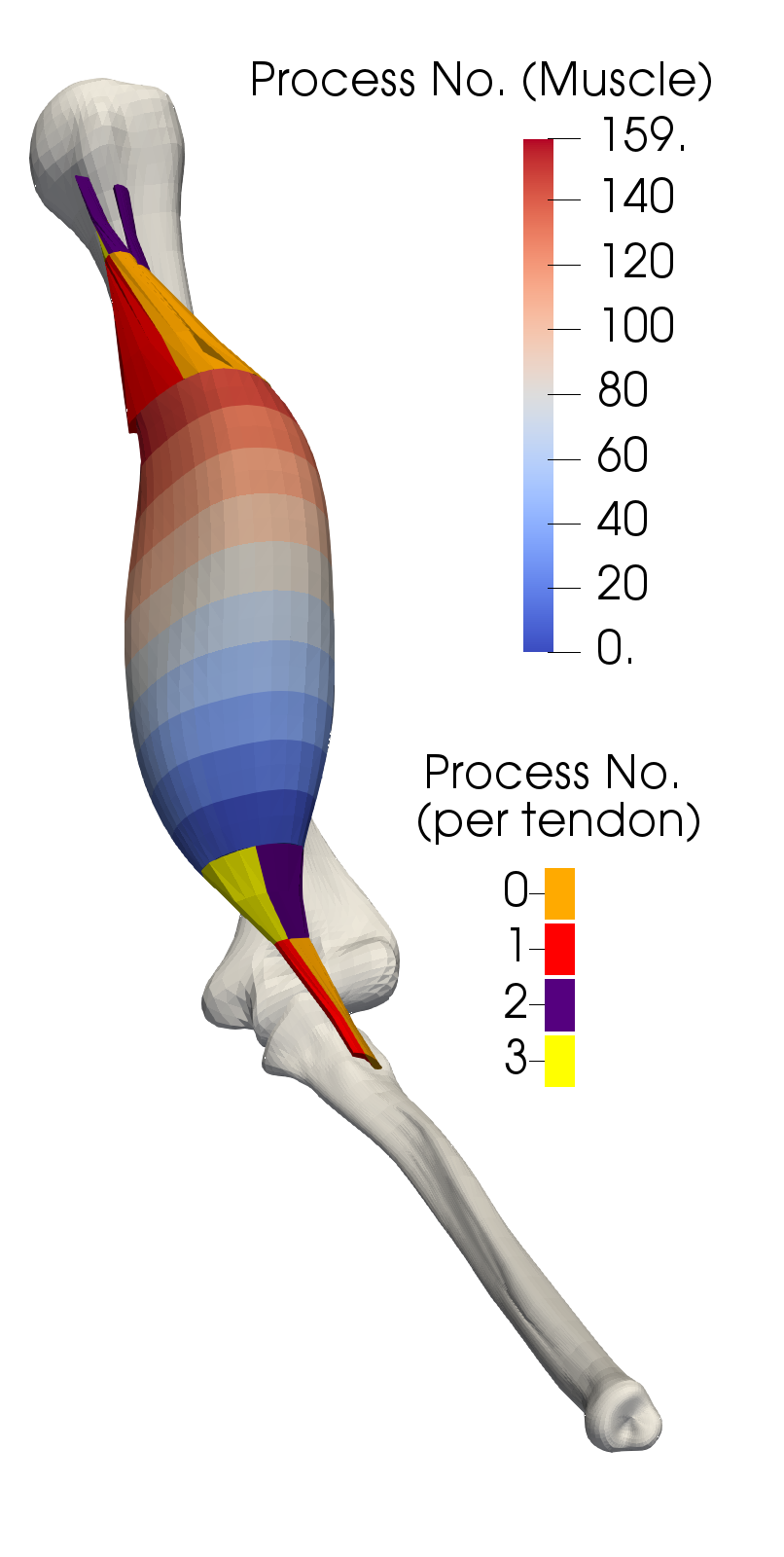}%
    \caption{Parallel partitioning scheme.}%
    \label{fig:precice_activated_muscles_3}%
  \end{subfigure}
  \caption{Simulation of a muscle-tendon complex. Muscle and tendon geometries of the scenario with 1089 muscle fibers, embedded in the skeletal system comprising the ulna bone (lower end) and humerus bone (upper end), result of the simulation at $t=\SI{1.05}{\s}$.}%
  \label{fig:precice_muscle_force}%
\end{figure}

\Cref{fig:precice_muscle_force} presents the simulation results of this scenario at the simulation time $t=\SI{1.05}{\s}$. \Cref{fig:precice_activated_muscles_1} shows the muscle fibers, which are attached to the tendons at both ends. Several action potentials can be seen on the fibers. \Cref{fig:precice_activated_muscles_2} displays the distribution of active stresses in the 3D mesh at the same simulation time. One vertical red line of higher active stress values can be seen at the foreside of the muscle belly, which corresponds to a region of higher muscle activity. This muscle activity results from a MU that is activated early on in the simulation scenario. A corresponding active fiber at that location can also be identified in \cref{fig:precice_activated_muscles_1}.
\Cref{fig:precice_activated_muscles_3} visualizes the parallel partitioning of the 3D domains of muscle and tendons into 160 subdomains for the muscle mesh and 4 subdomains for each of the three tendon meshes.

% contracted state

\begin{figure}
  \centering%
  \includegraphics[width=0.7\textwidth]{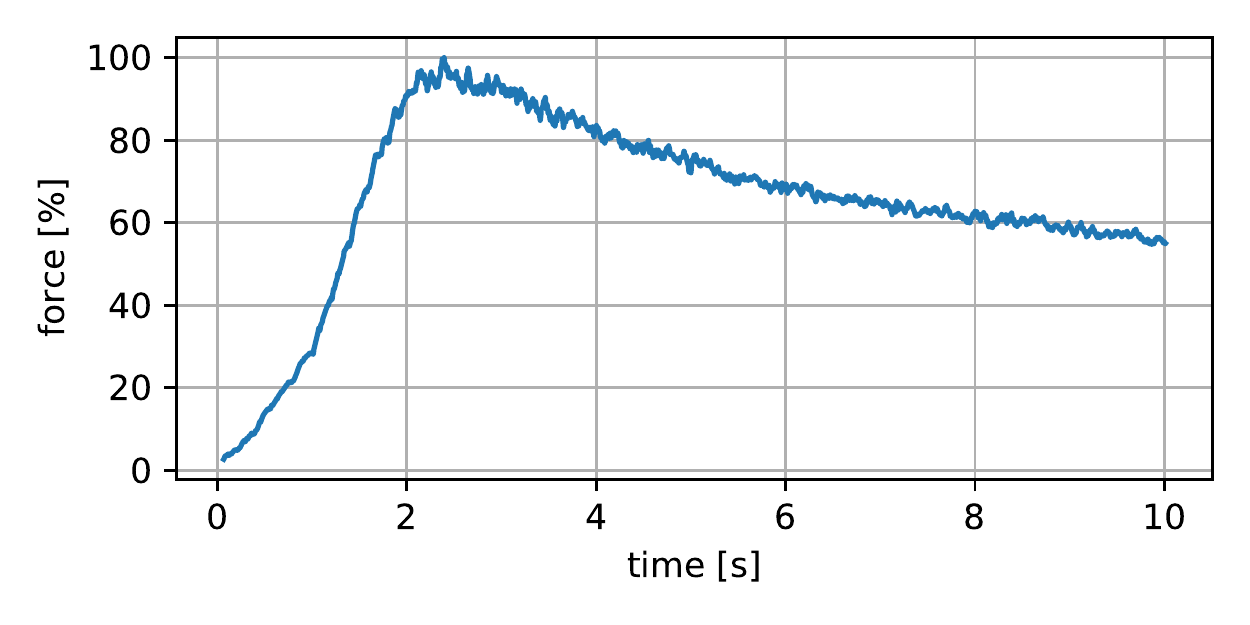}%
  \caption{Simulation of muscle force in a muscle-tendon complex. Resulting relative muscle force of the biceps muscle with attached tendons using the second scenario with 81 muscle fibers.}%
  \label{fig:precice_muscle_force0}%
\end{figure}

The second scenario uses a coarser 3D mesh with 525 nodes and a parallel partitioning into eight subdomains. We simulate the resulting muscle force, measured at the top insertion point of the proximal tendons, over a longer time period of \SI{10}{\s}. \Cref{fig:precice_muscle_force0} shows the resulting relative force progression over time. The plot shows the total force as a moving average function over $\SI{0.1}{\s}$. It can be seen that the force initially increases, as more and more MUs get activated. A short delay between the onset of the last MU at $\SI{1.8}{\s}$ and the maximum force at \SI{2.39}{\s} can be seen. The muscle force exhibits large oscillations during the period of high muscle activation. They result from the lower firing frequencies of the later activated, large MUs and their higher contribution to the overall activity, compared to the smaller MUs. 

% muscle force

\Cref{fig:precice_muscle_force0} also shows that the generated muscle force rapidly decreases after the maximum is reached. This is a result from muscle fatigue, which is described by the Shorten subcellular model. The observed decrease to below \SI{60}{\percent} after \SI{10}{\s} can also be found in experimental studies of healthy subjects, e.g., in \cite{Enoka2008}. 

In conclusion, several biophysical simulation scenarios of MU activity induced muscle contraction have been presented in the previous sections. OpenDiHu allowed us to couple the computationally efficient fiber based electrophysiology description with the solid mechanics model of muscle deformation, as well as the biophysically more accurate multidomain model. An algorithm to include prestresses was presented and the coupling software preCICE was used to numerically couple individual parts of the multi-scale model. 

The last presented scenario simulated the generated force of a muscle-tendon complex for a simulation time span of \SI{10}{\s}. 
It can be used in the future to test hypotheses on the influence of various processes along the pathway from MU recruitment over muscle activation to force generation and macroscopic deformation. The simulated force progression related to the maximum voluntary contraction force is a macroscopic quantity, which can be easily measured in in vivo studies. Thus, a connection between the simulation domain and the experimental domain is given, and the microscopic subcellular processes in the muscle fibers are linked to a quantifiable outcome that can be compared with experiments.

\begin{reproduce_no_break}
  The simulations in this section were carried out on the supercomputer Hawk in Stuttgart. The job scripts can be found in the repository at \\ \href{https://github.com/dihu-stuttgart/performance}{github.com/dihu-stuttgart/performance} in the directory \code{opendihu/15_precice_biceps/with_electrophysiology}. To run similar simulations on other computers, run commands that are similar to the following (adjust the numbers of processes):
  \begin{lstlisting}[columns=fullflexible,breaklines=true,postbreak=\mbox{\textcolor{gray}{$\hookrightarrow$}\space}]
    cd $\$$OPENDIHU_HOME/examples/electrophysiology/fibers/fibers_contraction/with_tendons_precice/multiple_tendons_with_electrophysiology
    mpirun -n 1 muscle_electrophysiology_precice settings_muscle.py ramp.py
    mpirun -n 1 tendon_linear_precice_dynamic settings_tendon_bottom.py
    mpirun -n 1 tendon_linear_precice_dynamic settings_tendon_top_a.py
    mpirun -n 1 tendon_linear_precice_dynamic settings_tendon_top_b.py
  \end{lstlisting}
\end{reproduce_no_break}

% --------------------------------
% studies, performance
%-----
%\section{Strang Splitting}
%-----

\chapter{Performance Analysis}\label{sec:performance_analysis}

In this chapter, we measure the performance of all implemented solvers and evaluate the different actions that were carried out to improve their runtimes and memory characteristics.
We consider the performance of instruction-level parallelism, evaluate parallelization strategies for shared and distributed memory parallelism, offloading to a GPU, and hybrid CPU-GPU  approaches. We measure weak and strong parallel scaling from using small distributed-memory systems up to the large supercomputers Hazel Hen and Hawk at the High Performance Computing Center Stuttgart. Furthermore, we consider the numerical scaling behavior of several solvers.

\Cref{sec:performance_opencmiss_iron} presents numerical studies and improvements in the baseline software OpenCMISS. The numerical properties found in this section are later also used in simulations with OpenDiHu. \Cref{sec:performance_studies_of_the_e} evaluates the runtime performance and various optimization options for the electrophysiology solver in OpenDiHu. The best found optimizations are then compared to the OpenCMISS baseline solver in \cref{sec:parallel_strong_scaling_opencmiss}.
\Cref{sec:performance_gpu} addresses the computation on the GPU and compares the performance with the CPU computations.
\Cref{sec:hpc_emg} conducts parallel weak scaling studies on supercomputers.
\Cref{sec:performance_solid_mechanics} evaluates options and corresponding speedups in the solver of the mechanics model.
\Cref{sec:numerical_studies} conducts numerical studies for the fiber based electrophysiology model and evaluates different solvers for the multidomain model.

\section{Performance Studies with OpenCMISS Iron}\label{sec:performance_opencmiss_iron}

%In this section, we study the performance of the software by evaluating runtimes and parallel scalability for different solvers.
We begin with performance studies on OpenCMISS Iron as the baseline solver, which also implements parts of the multi-scale model considered in this work. The work of \cite{Heidlauf2013} describes the implementation of the fiber based electrophysiology model coupled to a quasi-static hyperelastic material model with OpenCMISS. The implementation is parallelized for a hard-coded number of four processes and serves as the baseline code for the following studies.

We improved the performance of this solver for the multi-scale model by two actions: First, we evaluated and optimized the employed numerical schemes. Second, we implemented parallel partitioning for an arbitrary number of processes and evaluated different parallelization strategies.
These changes were directly implemented in the OpenCMISS code. The improvements were also presented in a publication \cite{Bradley:2018:EDB}. In the following sections, \cref{sec:opencmiss_numeric_improvements,sec:opencmiss_parallel_partitioning}, we describe the numerical improvements and the parallel partitioning strategies. In \cref{sec:opencmiss_memory}, we discuss parallel weak scaling and memory consumption properties.

\subsection{Numerical Improvements}\label{sec:opencmiss_numeric_improvements}

The first numerical improvement is to replace the GMRES solver, which is used to solve the 1D electric conduction problem on the muscle fibers,
by a faster direct solver. 

As observed in \cref{sec:improved_parallel_solver_for_fiber_based}, the 1D electric conduction problem of the monodomain equation yields a tridiagonal system that can be solved with linear time complexity. The baseline solver code employs the restarted GMRES solver of PETSc, which is the default linear system solver in OpenCMISS Iron, as it is a robust choice for arbitrary system matrices. 
However, more efficient solvers for symmetric positive definite systems exist such as the conjugate gradient solver. 
Furthermore, the MUMPS package \cite{mumps2001}, which can be interfaced in PETSc, provides a parallel implementation of a direct, multi-frontal linear solver, which is able to exploit the banded structure of the system matrix.

We study the runtime of these three solvers for different problem sizes of the 1D problem. The monodomain equation is solved on a single muscle fiber and the number of 1D elements is varied from 15 to 2807. The used timestep widths are $\dt_\text{0D}=\SI{1e-4}{\ms}$ and $\dt_\text{1D}=\SI{5e-3}{\ms}$. The end time of the simulation is $\SI{3}{\ms}$, yielding a total of 600 calls to the linear solver in the simulated time. The study is executed on an Intel Xeon E7540 processor with 24 cores, clock frequency of \SI{1064}{\mega\hertz} and \SI{506}{\gibi\byte} RAM.

\begin{figure}
  \centering%
  \includegraphics[width=0.9\textwidth]{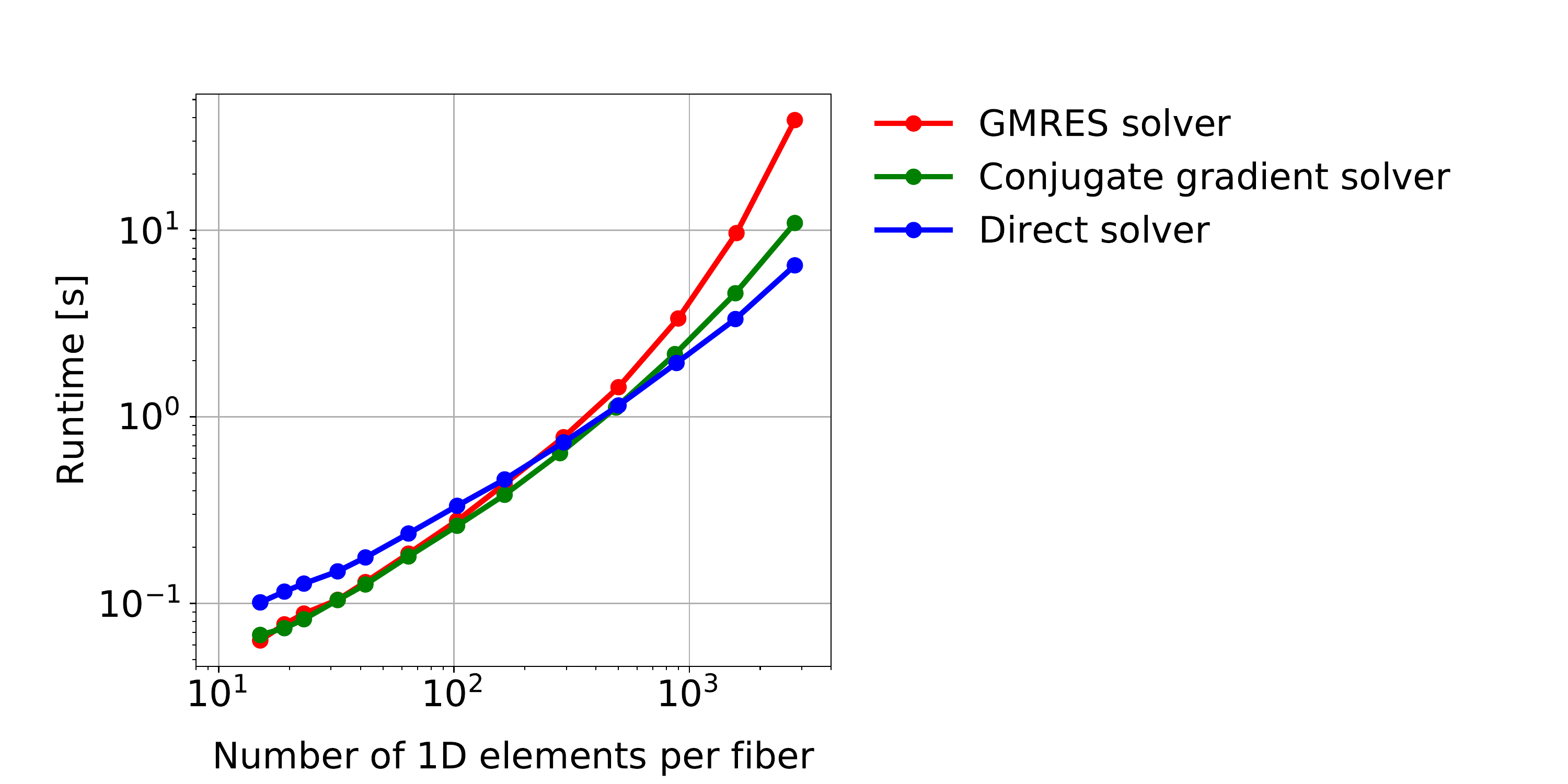}%
  \caption{Numerical improvements in OpenCMISS: Runtime evaluation of different linear system solvers for a single muscle fiber with varying spatial resolution.}%
  \label{fig:opencmiss_linear_solvers}%
\end{figure}

\Cref{fig:opencmiss_linear_solvers} shows the runtimes of GMRES, the conjugate gradient solver and the direct solver for this problem in a double-logarithmic plot.
It can be seen, that, for coarse discretizations with a low number of 1D elements per fiber, GMRES and the conjugate gradient solver are faster than the direct solver. For finer discretizations, the conjugate gradient solver and the direct solver outperform the GMRES solver. For fibers with more than approximately 500 elements, the direct solver has the lowest runtime. Moreover, the direct solver exhibits an almost linear runtime complexity in terms of the problem size. This indicates that the solver is able to exploit the tridiagonal structure of the system matrix.

% linear solvers plot

The second numerical improvement is the exchange of first-order accurate timestepping schemes by second-order schemes. For this exchange, we implemented the Strang operator splitting scheme and use it with the existing Crank-Nicolson implementation in OpenCMISS Iron and a new implementation of the Heun method by Aaron Krämer.

Numerical studies by Aaron Krämer presented in \cite{Bradley:2018:EDB} show that the relation $K=\dt_\text{1D}/\dt_\text{0D}$ between the timestep width $\dt_\text{1D}$ of the 1D electric conduction problem and the timestep width $\dt_\text{0D}$ of the 0D subcellular model has to be set to $K=2$ and $K=5$ for the Godunov and Strang splitting schemes, respectively, such that the errors of the 0D and 1D subproblems are balanced. To achieve a total error for the membrane potential $V_m$ of approximately \num{8e-2}, we can increase the required splitting timestep width $\dt_\text{splitting}$ from $\SI{5e-4}{\ms}$ for the Godunov splitting to $\SI{4e-3}{\ms}$ for the Strang splitting scheme. This results in a runtime speedup  of approximately 7.5.

To evaluate the total speedup of the described numerical improvements, we compare the runtimes without and with the improvements for a complete simulation of the fiber based electrophysiology model coupled with the elasticity model. A cuboid 3D domain is discretized by $2\times 2\times 2=8$ finite elements for the elasticity model, and we embed $6\times 6=36$ 1D fiber meshes. The number of 1D elements per fiber is varied between 576 and \num{239400} to study the scaling behavior of the solvers related to the problem size. The problem is solved in serial to avoid runtime effects introduced by the parallelization.

The baseline implementation uses the Godunov splitting with forward and backward Euler schemes for the 0D subcellular model and the electric conduction model, respectively. The linear system in the 1D problem is solved by a GMRES solver with relative residuum tolerance of \num{1e-5} and restart after 30 iterations. Timestep widths of $\dt_\text{0D}=\SI{1e-4}{\ms}$ and $\dt_\text{splitting}=\dt_\text{1D}=\SI{5e-4}{\ms}$ are used. The improved scheme uses the Strang operator splitting with Heun and Crank-Nicolson schemes and timestep widths of $\dt_\text{0D}=\SI{2e-3}{\ms}$ and $\dt_\text{splitting}=\dt_\text{1D} = \SI{4e-3}{\ms}$. The direct solver is used for the linear system in the 1D problem.
The solver for the 3D elasticity problem is the same for both implementations: A Newton scheme with residual tolerance of \num{1e-8} is used
 and coupled to the 0D and the 1D solver with a coupling timestep width of $\dt_\text{3D}=\SI{1}{\ms}$.

The present study and the studies in the next section are executed on the supercomputer \emph{Hazel Hen} at the High Performance Computing Center Stuttgart. This Cray XC40 system contains compute nodes with two Intel Haswell E5-2680v3 processors with a base frequency of \SI{2.5}{\giga\hertz}, 12 cores per CPU, 24 cores per compute node and \SI{128}{\giga\byte} RAM per node.

% improvements plot
\begin{figure}
  \centering%
  \includegraphics[width=\textwidth]{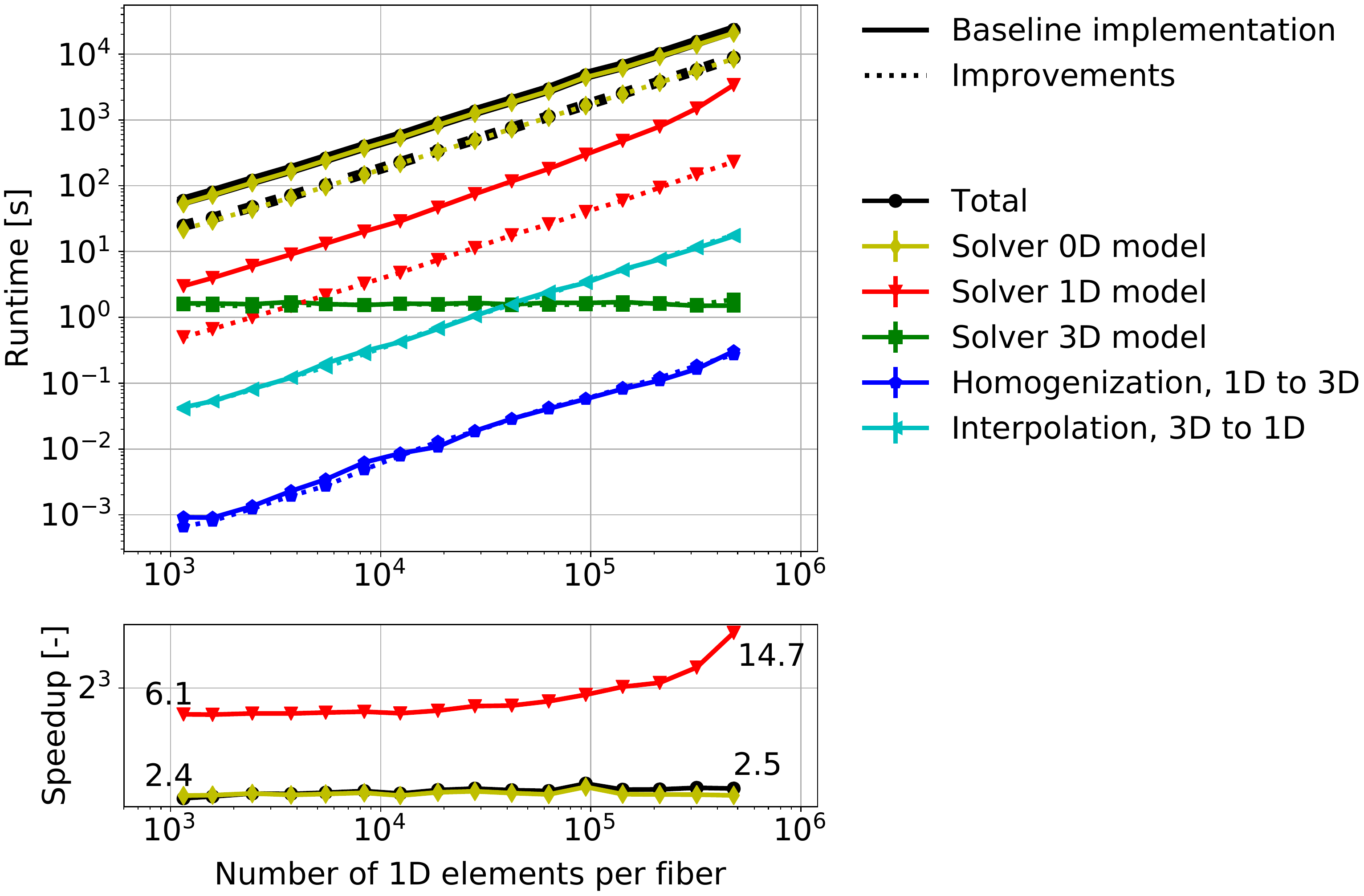}%
  \caption{Numerical improvements in OpenCMISS: Study to evaluate the speedup of the improved implementation of the fiber based electrophysiology and mechanics model in OpenCMISS.}%
  \label{fig:opencmiss_improvements}%
\end{figure}%
% 576.0, 792.0, 1224.0, 1872.0, 2736.0, 4176.0, 6192.0, 9360.0, 14040.0, 21024.0, 31536.0, 47304.0, 70920.0, 106416.0, 159624.0, 239400.0

\Cref{fig:opencmiss_improvements} shows the results of this study. In the upper part, the runtimes for different components of the simulation are indicated by different colors in a plot with double logarithmic scale. The runtimes for the baseline implementation are shown by solid lines and the runtimes of the implementation where the improvements have been incorporated are shown by dashed lines. In the lower plot, the speedups from the baseline to the improved implementation are given.

The total runtime of the simulation is given by the black lines in the upper plot. It can be seen that the total runtime results almost completely from the 0D model solver, which is shown by the yellow lines. The 1D solver, given by the red lines, has the second highest contribution. The effects of the data mapping operations between the 3D mesh and the 1D fibers on the runtime are negligible. These data mapping operations consists of the homogenization step from the 1D fibers to the 3D mesh and the interpolation step from the 3D mesh to the 1D fibers.

The runtimes for almost all problem parts increase linearly for increasing mesh resolution of the 1D fibers. Only the runtime of the 3D problem stays constant, as the 3D mesh is unchanged for the different runs.

Significant runtime improvements of the new implementation compared to the baseline implementation can be seen in the lower plot of   \cref{fig:opencmiss_improvements} for the 0D solver and the 1D solver. The speedup for the 0D solver is constant at approximately 2.5. The speedup resulting from the improved linear system solver in the 1D problem is approximately 6.1 for coarse meshes and increases to 14.7 for the finest mesh. This increase for high mesh resolutions results from the higher runtime of the GMRES solver for large problem sizes in the baseline implementation. The overall speedup is similar to the speedup of the 0D problem, as the 0D solver exhibits the dominant runtime contribution to the overall computation.

This study shows how numerical investigations can help to reduce the total runtime, in this case by a factor of 2.5. Moreover, the solver of the 0D model has the highest potential for improvements that further speed up the computation.

\subsection{Parallel Partitioning Strategies}\label{sec:opencmiss_parallel_partitioning}

To exploit parallelism and, thus, further reduce the computation times, we implemented a generic domain decomposition for the studied problem in OpenCMISS Iron.
Like in OpenDiHu, the 3D mesh can be partitioned to an arbitrary number of $n_x \times n_y \times n_z$ subdomains. The embedded 1D fibers are aligned with the $z$ axis and are partitioned by the same cut planes as the 3D mesh.

% pillars-cubes visualization
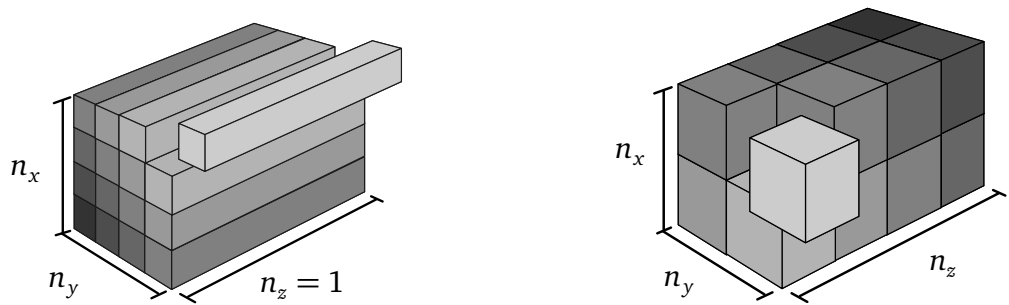
\begin{figure}[H]
  \centering%
  \begin{subfigure}[t]{0.48\textwidth}%
    \centering%
    \def\svgwidth{0.7\textwidth}
    %% Creator: Inkscape inkscape 0.92.3, www.inkscape.org
%% PDF/EPS/PS + LaTeX output extension by Johan Engelen, 2010
%% Accompanies image file 'opencmiss_ddpillar.pdf' (pdf, eps, ps)
%%
%% To include the image in your LaTeX document, write
%%   \input{<filename>.pdf_tex}
%%  instead of
%%   \includegraphics{<filename>.pdf}
%% To scale the image, write
%%   \def\svgwidth{<desired width>}
%%   \input{<filename>.pdf_tex}
%%  instead of
%%   \includegraphics[width=<desired width>]{<filename>.pdf}
%%
%% Images with a different path to the parent latex file can
%% be accessed with the `import' package (which may need to be
%% installed) using
%%   \usepackage{import}
%% in the preamble, and then including the image with
%%   \import{<path to file>}{<filename>.pdf_tex}
%% Alternatively, one can specify
%%   \graphicspath{{<path to file>/}}
%% 
%% For more information, please see info/svg-inkscape on CTAN:
%%   http://tug.ctan.org/tex-archive/info/svg-inkscape
%%
\begingroup%
  \makeatletter%
  \providecommand\color[2][]{%
    \errmessage{(Inkscape) Color is used for the text in Inkscape, but the package 'color.sty' is not loaded}%
    \renewcommand\color[2][]{}%
  }%
  \providecommand\transparent[1]{%
    \errmessage{(Inkscape) Transparency is used (non-zero) for the text in Inkscape, but the package 'transparent.sty' is not loaded}%
    \renewcommand\transparent[1]{}%
  }%
  \providecommand\rotatebox[2]{#2}%
  \newcommand*\fsize{\dimexpr\f@size pt\relax}%
  \newcommand*\lineheight[1]{\fontsize{\fsize}{#1\fsize}\selectfont}%
  \ifx\svgwidth\undefined%
    \setlength{\unitlength}{250.7178297bp}%
    \ifx\svgscale\undefined%
      \relax%
    \else%
      \setlength{\unitlength}{\unitlength * \real{\svgscale}}%
    \fi%
  \else%
    \setlength{\unitlength}{\svgwidth}%
  \fi%
  \global\let\svgwidth\undefined%
  \global\let\svgscale\undefined%
  \makeatother%
  \begin{picture}(1,0.73686381)%
    \lineheight{1}%
    \setlength\tabcolsep{0pt}%
    \put(0,0){\includegraphics[width=\unitlength,page=1]{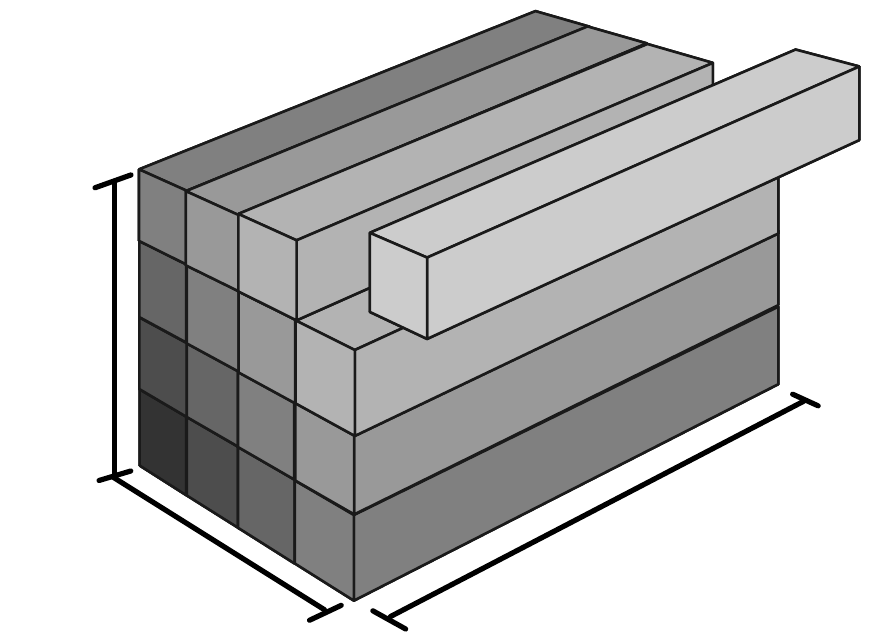}}%
    \put(0.62941279,0.03912118){\color[rgb]{0,0,0}\makebox(0,0)[lt]{\lineheight{0}\smash{\begin{tabular}[t]{l}$n_z=1$\end{tabular}}}}%
    \put(0.0942604,0.0538548){\color[rgb]{0,0,0}\makebox(0,0)[lt]{\lineheight{0}\smash{\begin{tabular}[t]{l}$n_y$\end{tabular}}}}%
    \put(-0.00155194,0.33892603){\color[rgb]{0,0,0}\makebox(0,0)[lt]{\lineheight{0}\smash{\begin{tabular}[t]{l}$n_x$\end{tabular}}}}%
  \end{picture}%
\endgroup%
    \caption{\say{Pillar-like} domain decomposition with $n_z=1$.}%
    \label{fig:opencmiss_ddpillar}%
  \end{subfigure}
  \quad
  \begin{subfigure}[t]{0.48\textwidth}%
    \centering%
    \def\svgwidth{0.7\textwidth}
    %% Creator: Inkscape inkscape 0.92.3, www.inkscape.org
%% PDF/EPS/PS + LaTeX output extension by Johan Engelen, 2010
%% Accompanies image file 'opencmiss_ddcube.pdf' (pdf, eps, ps)
%%
%% To include the image in your LaTeX document, write
%%   \input{<filename>.pdf_tex}
%%  instead of
%%   \includegraphics{<filename>.pdf}
%% To scale the image, write
%%   \def\svgwidth{<desired width>}
%%   \input{<filename>.pdf_tex}
%%  instead of
%%   \includegraphics[width=<desired width>]{<filename>.pdf}
%%
%% Images with a different path to the parent latex file can
%% be accessed with the `import' package (which may need to be
%% installed) using
%%   \usepackage{import}
%% in the preamble, and then including the image with
%%   \import{<path to file>}{<filename>.pdf_tex}
%% Alternatively, one can specify
%%   \graphicspath{{<path to file>/}}
%% 
%% For more information, please see info/svg-inkscape on CTAN:
%%   http://tug.ctan.org/tex-archive/info/svg-inkscape
%%
\begingroup%
  \makeatletter%
  \providecommand\color[2][]{%
    \errmessage{(Inkscape) Color is used for the text in Inkscape, but the package 'color.sty' is not loaded}%
    \renewcommand\color[2][]{}%
  }%
  \providecommand\transparent[1]{%
    \errmessage{(Inkscape) Transparency is used (non-zero) for the text in Inkscape, but the package 'transparent.sty' is not loaded}%
    \renewcommand\transparent[1]{}%
  }%
  \providecommand\rotatebox[2]{#2}%
  \newcommand*\fsize{\dimexpr\f@size pt\relax}%
  \newcommand*\lineheight[1]{\fontsize{\fsize}{#1\fsize}\selectfont}%
  \ifx\svgwidth\undefined%
    \setlength{\unitlength}{237.48424816bp}%
    \ifx\svgscale\undefined%
      \relax%
    \else%
      \setlength{\unitlength}{\unitlength * \real{\svgscale}}%
    \fi%
  \else%
    \setlength{\unitlength}{\svgwidth}%
  \fi%
  \global\let\svgwidth\undefined%
  \global\let\svgscale\undefined%
  \makeatother%
  \begin{picture}(1,0.77581388)%
    \lineheight{1}%
    \setlength\tabcolsep{0pt}%
    \put(0,0){\includegraphics[width=\unitlength,page=1]{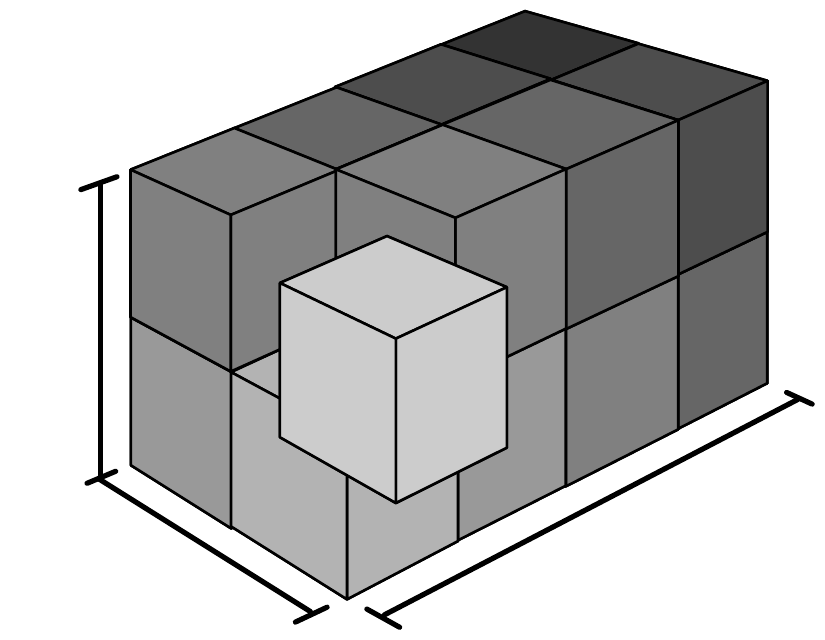}}%
    \put(0.79235619,0.09620438){\color[rgb]{0,0,0}\makebox(0,0)[lt]{\lineheight{0}\smash{\begin{tabular}[t]{l}$n_z$\end{tabular}}}}%
    \put(0.12070548,0.05486903){\color[rgb]{0,0,0}\makebox(0,0)[lt]{\lineheight{0}\smash{\begin{tabular}[t]{l}$n_y$\end{tabular}}}}%
    \put(-0.00163841,0.38231625){\color[rgb]{0,0,0}\makebox(0,0)[lt]{\lineheight{0}\smash{\begin{tabular}[t]{l}$n_x$\end{tabular}}}}%
  \end{picture}%
\endgroup%
    \caption{\say{Cube-like} domain decomposition.}%
    \label{fig:opencmiss_ddcube}%
  \end{subfigure}   
  \caption{Fiber-based electrophysiology and mechanics model in OpenCMISS: Different partitioning strategies for parallelization that have been implemented in OpenCMISS. This figure shows two approaches to partition the domain into 16 subdomains.}%
  \label{fig:opencmiss_dd_annotated}%
\end{figure}%

\Cref{fig:opencmiss_dd_annotated} shows two exemplary partitioning approaches. If the domain is only partitioned in $x$ and $y$ direction, the individual fibers are not split into multiple subdomains. As a result, we get \say{pillar} subdomains as shown in \cref{fig:opencmiss_ddpillar}. An alternative approach is to subdivide the domain in all three coordinate directions, such that the subdomains are approximately cuboid, as shown in \cref{fig:opencmiss_ddcube}.

OpenCMISS Iron already provides the functionality to create parallel partitioned, unstructured meshes. However, every mesh has to be partitioned into non-empty subdomains for all processes. Thus, it is not possible to use individual meshes for the 1D fibers.
In the baseline implementation of the model by \cite{Heidlauf2013}, all 1D fiber meshes are however realized as a single mesh, whose node positions are set according to the positions of the individual fibers. This facilitates the implementation of the 0D subcellular model solvers and 1D model solvers, as the implementation has to deal with only a single mesh. 

To allow for an arbitrary partitioning as in \cref{fig:opencmiss_dd_annotated}, we assigned the 1D elements of the single fiber mesh to the same processes as the corresponding subdomains of the 3D mesh. Furthermore, we reimplemented the data mapping between the 1D mesh and the 3D mesh, which was hard-coded for four processes.

In the following, we investigate the effect of different partitioning strategies on the overall runtime of the solver. The idea is that, for pillar-like partitionings as in \cref{fig:opencmiss_ddpillar}, the 1D problems could potentially be solved faster, as the fibers, which are aligned in $z$-direction, are not subdivided to multiple processes. On the other hand, the partitioning to cubes in \cref{fig:opencmiss_ddcube} requires less communication in the solution of the 3D problem as the cubes minimize the surface of each subdomain and, in consequence, the amount of data to be exchanged. We evaluate how these effects influence the runtimes for the pillar-like partitioning, the cube partitioning and all other possible partitionings specified by numbers of subdomains $n_x \times n_y \times n_z$.

Our test case uses a 3D mesh with $12 \times 12 \times 144$ elements. To reduce the runtime contribution of the 0D/1D electrophysiology problem and the memory consumption of the solver, only two 1D elements per 3D element are included. The numerical parameters are the same as for the improved scenario presented in \cref{fig:opencmiss_improvements}. The simulations are executed on 12 compute nodes of the supercomputer Hazel Hen with 12 processes per node.

\begin{figure}
  \centering%
  \includegraphics[width=\textwidth]{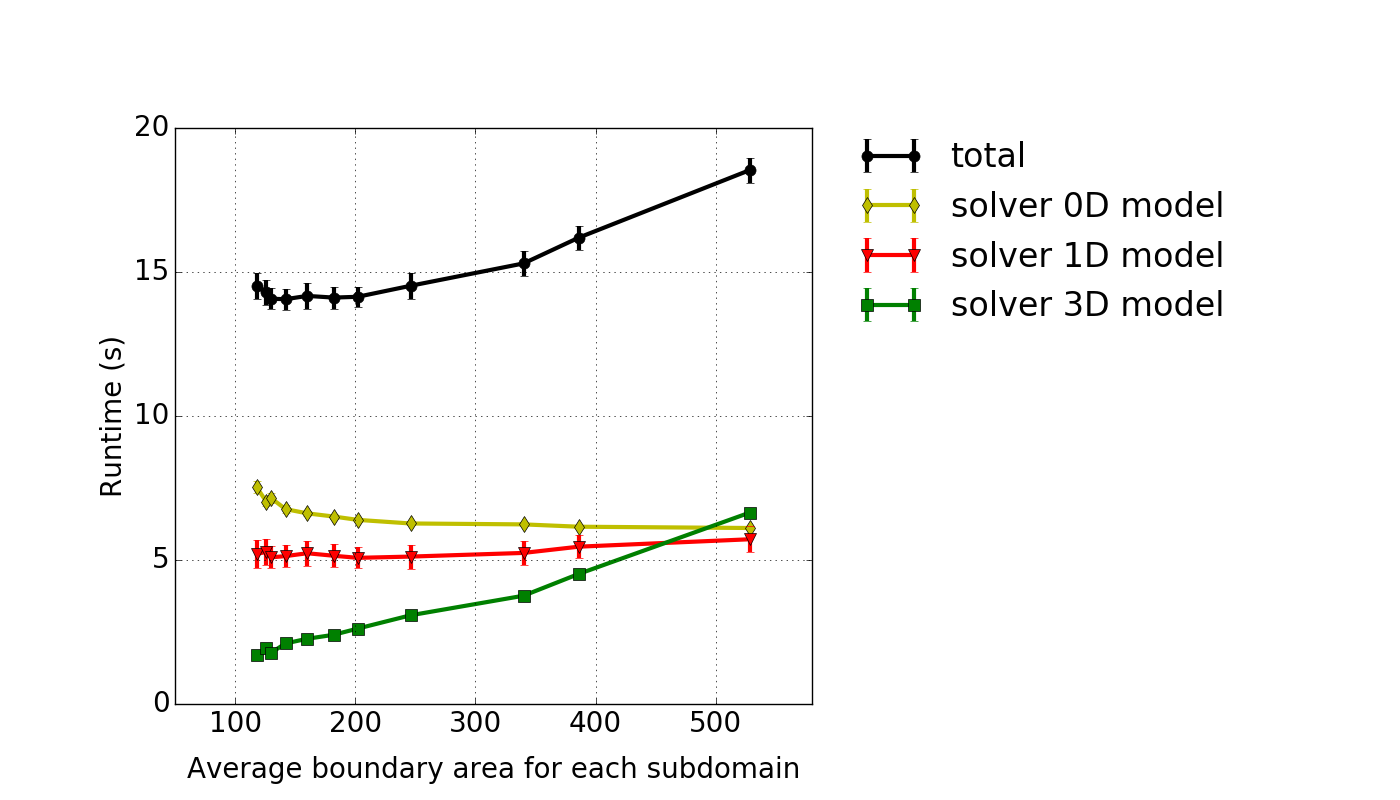}%
  \caption{Fiber-based electrophysiology and mechanics model in OpenCMISS: Runtime of the solvers for different partition shapes, from cube partitions on the left to pillar partitions on the right.\protect\footnotemark}%
  \label{fig:opencmiss_partition_shape}%
\end{figure}

We partition the 3D domain to 144 processes using different combinations of $n_x,n_y$ and $n_z$ such that $n_x\,n_y\,n_z=144$. 
For every partitioning, we compute the average surface area of the boundary of every subdomain. 
\Cref{fig:opencmiss_partition_shape} shows the resulting runtime in relation to this average boundary area.
The pillar-like partitioning uses $12 \times 12 \times 1$ subdomains and exhibits the largest boundary surface area, corresponding to the last point in \cref{fig:opencmiss_partition_shape}. The cube partitioning consists of $6 \times 6 \times 4$ subdomains and corresponds to the first data point with the smallest boundary area.

% plot of partition shapes
\footnotetext{This figure and the following figures have also been published in \cite{Bradley:2018:EDB} under a creative commons license.}

The plot shows that the runtime of the 3D solver increases approximately linearly with the amount of communication, which is expected.
The partitioning with the largest average surface area has a runtime that is approximately four times larger than the runtime for the smallest surface area.

Moreover, the plot shows that the partitioning scheme has no significant influence on the runtime of the 1D solver. The reason is that the implementation does not fully reflect the decoupled nature of the individual problems of the fibers. As noted before, one big linear system has to be solved that contains the degrees of freedom of all fibers. The degrees of freedom are ordered by PETSc, such that the nodes within every subdomain are consecutive. If a subdomain contains (parts of) multiple fibers, the degrees of freedom of a single fiber are not necessarily consecutive in the solution vector and communication is required in the linear solver.

\subsection{Weak Scaling Study and Memory Consumption}\label{sec:opencmiss_memory}

Next, we evaluate the parallel weak scaling properties of the overall solver. We increase the number of elements in the 3D mesh from 1232 to 8640 and the total number of 1D elements in all fibers from \num{14784} to \num{103680}. Correspondingly, the number of processes increases from 24 to 192, such that the amount of work per process stays approximately constant. Each scenario is computed with two different partitioning schemes, once with pillar-like partitioning and once with cuboid partitioning. For the exact problem sizes, numbers of cores and numbers of elements in the partitions, we refer to the paper \cite{Bradley:2018:EDB}. 

% weak scaling runtime
\begin{figure}
  \centering%
  \includegraphics[width=\textwidth]{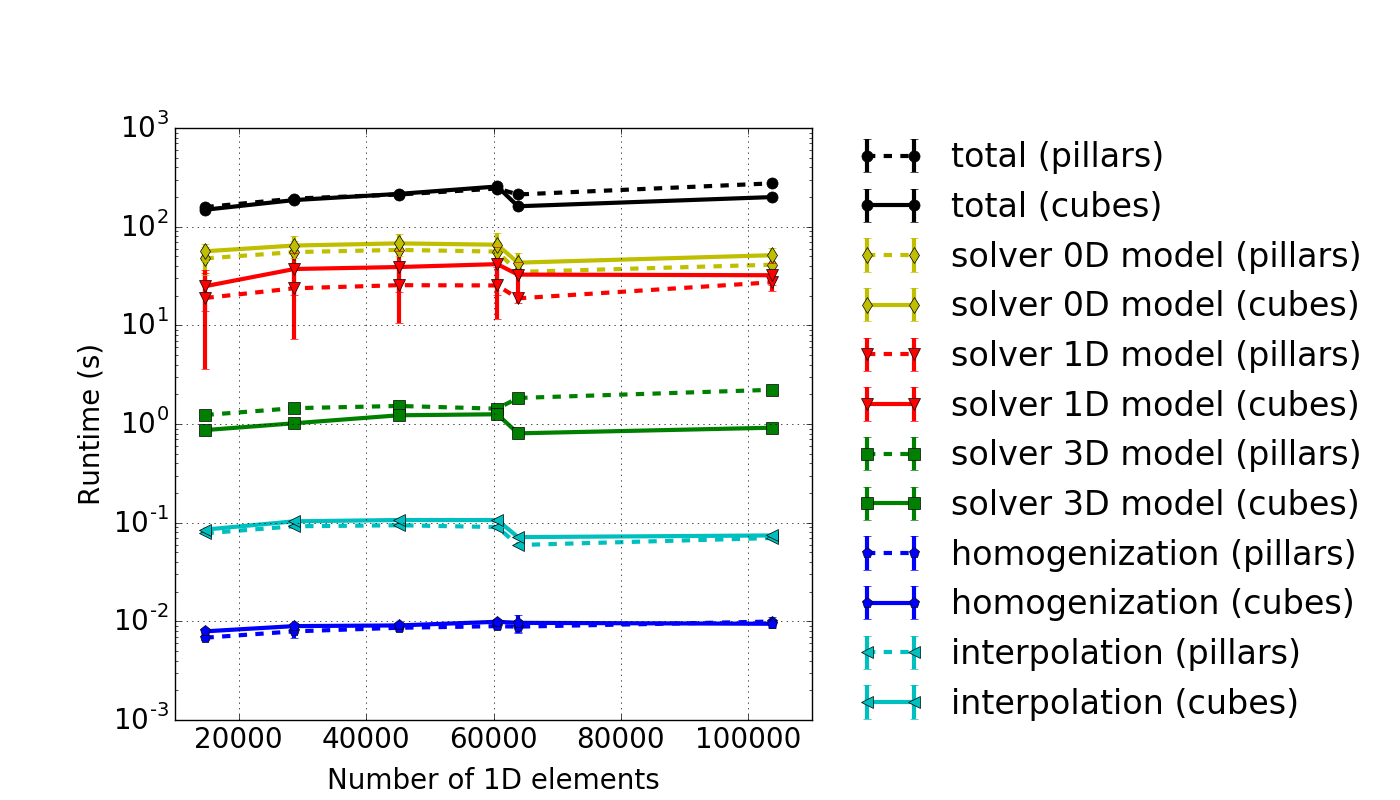}%
  \caption{Fiber-based electrophysiology and mechanics model in OpenCMISS: Parallel weak scaling study of a scenario with the pillar and cube partitionings.}%
  \label{fig:opencmiss_weak_scaling}%
\end{figure}

\Cref{fig:opencmiss_weak_scaling} shows the resulting runtimes of the different components of the simulation. It can be seen that the runtime stays approximately the same for all problem sizes. The observable differences in runtime within the same solver, especially for the last two data points, can be explained by slightly different ratios of element counts to process counts, which result from the goal to use the pillars and cube partitioning schemes while not exceeding the available main memory.

The runtimes of the pillar and cube partitioning schemes are depicted by dashed and solid lines, respectively. The pillar partitioning exhibits shorter runtimes for the 1D solver and longer runtimes for the 3D solver compared to the cube partitioning. In total, the runtime is not significantly different for the different partitioning strategies.

% memory consumption
\begin{figure}
  \centering%
  \includegraphics[width=\textwidth]{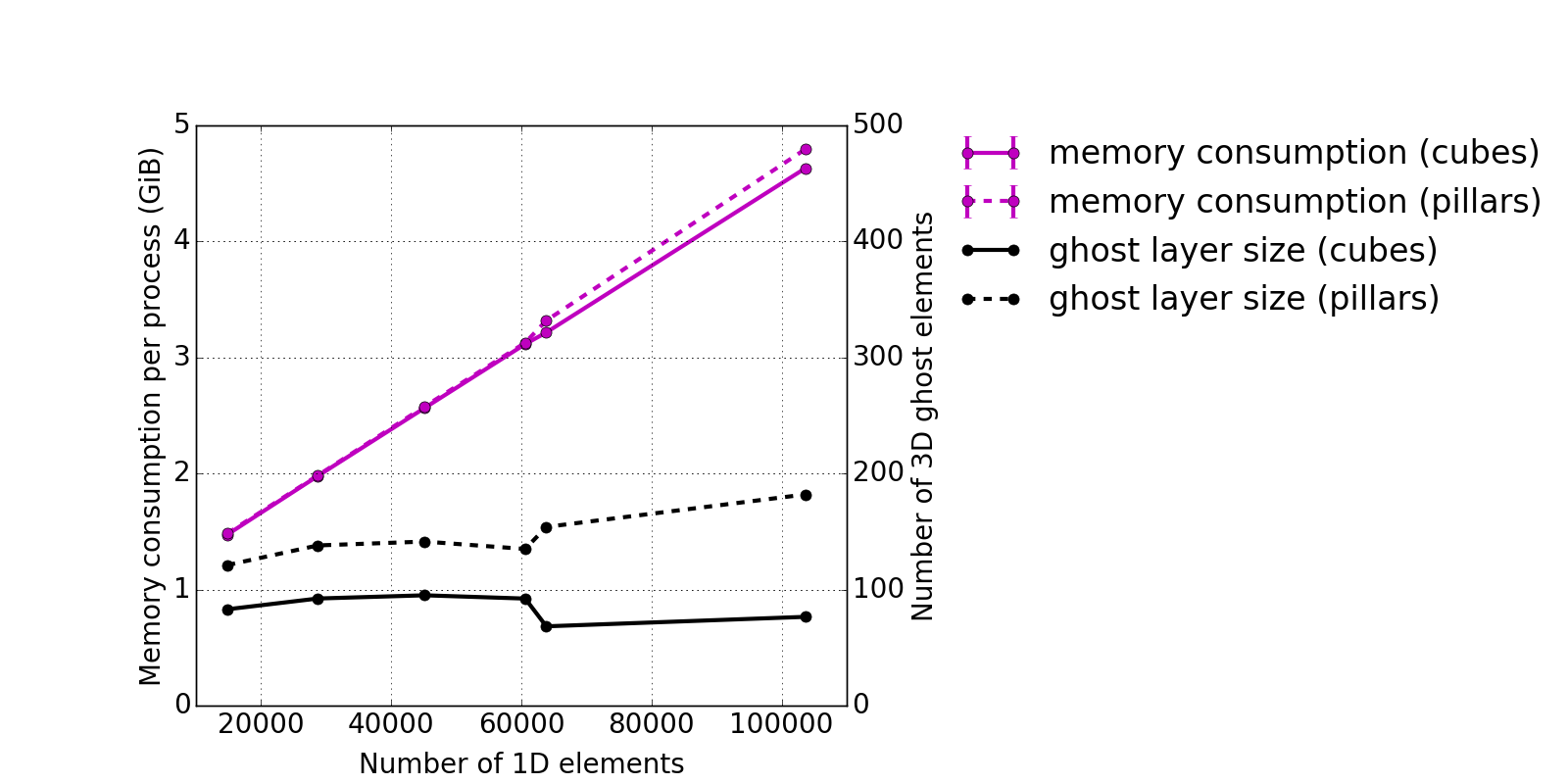}%
  \caption{Fiber-based electrophysiology and mechanics model in OpenCMISS: Memory consumption per process at the end of the simulation corresponding to the weak scaling study of \cref{fig:opencmiss_weak_scaling}} %% previously published in paper
  \label{fig:opencmiss_memory}%
\end{figure}%

A limiting factor for the construction of weak scaling studies with this implementation is the high memory consumption. \Cref{fig:opencmiss_memory} shows the total memory consumption per process at the end of the runtime of the simulations in \cref{fig:opencmiss_weak_scaling}. The used memory is visualized by purple lines. The dashed line again corresponds to the pillar partitioning and the solid line corresponds to the cube partitioning. 

A difference between the pillar partitions and the cube partitions is the size of the subdomain surfaces and the corresponding size of the ghost layer. \cref{fig:opencmiss_memory} shows the number of 3D ghost elements for the scenarios with cubes and pillars by the black lines. In OpenCMISS, a ghost element on a process is an element that contains ghost nodes, which are owned by a different process. The ghost elements serve as data buffers for communication during the assembly of the finite element matrices, similar to OpenDiHu.

The plot in \cref{fig:opencmiss_memory} shows that the number of ghost elements is higher for the pillar partitioning scheme than for the cubes scheme, as expected. As a consequence, the memory consumption per process is also slightly higher for the pillar partitioning.
However, this effect is negligible compared to the high absolute value of the required memory and does not explain this effect.

As can be seen, the memory consumption per process monotonically increases with the total number of 1D elements. 
At the same time, however, the number of elements per process stays approximately constant in this weak scaling setting. The last data point is close to the memory limit of $\SI{128}{\giga\byte} / 24 \approx \SI{4.967}{\gibi\byte}$, which is reached when 24 processes are executed on a compute node of the supercomputer Hazel Hen.

The observed large increase in memory consumption results from the organization of parallel partitioned data in OpenCMISS Iron. On every process, global mesh topology information such as mappings between global indexing and local indexing is stored for the element numbers, node numbers and degree of freedom numbers. While this overhead in storage is negligible for moderately parallel scenarios, it counteracts the domain decomposition approach for higher degrees of parallelism. 

Numerous functions and algorithms in the OpenCMISS Iron code rely on this type of global information. Thus, eliminating the parallelism constraint by reorganizing the data structures is a highly involved task. Especially the initialization of the parallel partitioning heavily uses this global information. This initialization includes, e.g., the distribution of elements and nodes to the subdomains on the processes, the determination of the ghost layers and dofs to send to and receive from neighbor processes, and the setup of local numbers for elements, nodes and degrees of freedom.

We addressed the elimination of this use of global topology information in the initialization steps and developed and implemented appropriate local algorithms in OpenCMISS Iron. This resulted in major code changes that are difficult to oversee, also because of the lacking object orientation in the code base and the difficulty to comprehensively test the functionality. Creating the required set of unit tests for nearly all functionality of OpenCMISS would be a large task that remains to be done. Thus, these code changes could not be merged into the main trunk of OpenCMISS.

Even with these code changes, the memory problem is not yet solved. Another problem prior to the initialization step is that the mesh has to be specified from the user code in a global data structure. It is currently not possible to specify a mesh in a distributed way. Thus, OpenCMISS Iron can only use meshes that initially fit into the main memory on every single core.

Moreover, another issue is concerned with the data structures for matrices. Each process stores its local row indices and additionally a map from global to local row indices for all dofs of the global problem. This global-to-local map also contributes to the bad weak memory scaling and has to be eliminated as well. One possible approach is to use hash maps and only store the relevant portion of the mapping on every process. Work towards resolving this issue has been started by Lorenzo Zanon at the former SimTech Research Group on Continuum Biomechanics and Mechanobiology at the University of Stuttgart. 

One reason for the generic mapping of matrix rows, which uses global information, is that OpenCMISS Iron does not restrict discretization schemes to the finite element method, where the system matrix can be assembled from local element matrices within the subdomains. An example for a different supported scheme is the boundary element method.

In addition, there exist more parts in the code that use a similar global-to-local mapping and would also have to be changed to allow for a constant memory consumption per process, e.g, the boundary condition handling and the data mapping between the 3D mesh and the fibers.

In summary, fixing the issue of non-scaling memory consumption in OpenCMISS Iron, which was revealed in \cref{fig:opencmiss_memory}, corresponds to redeveloping a significant portion of the code. 
To preserve the generic functionality of OpenCMISS, some changes would require new algorithmic considerations and complex workarounds.
This development effort would have to be quick enough to keep up with the independent development of the normal OpenCMISS branch. After completion, the merge back into the main software trunk would only be possible if the branches had not diverged too far and after significant efforts have been put into testing and preserving the feature set of OpenCMISS.

On the other hand, developing the missing functionality from scratch and making sensible restrictions on the generality of the solved problems and used methods requires possibly less effort and allows considering design goals such as performance, usability and extensibility from the beginning.
In this sense, the OpenDiHu software project can be seen as a complement to OpenCMISS Iron with better performance characteristics.  %companion
The mentioned restrictions for OpenDiHu are, e.g., the exclusive use of the finite element method and Cartesian coordinates and the use of parallel partitioned structured meshes instead of the more complex parallelization of unstructured meshes.

%-----

\section{Performance Studies of the Electrophysiology Solver in OpenDiHu}\label{sec:performance_studies_of_the_e}
After the previous studies with OpenCMISS, we now consider the performance of the OpenDiHu software.
In the following sections, we investigate the runtime performance of the solvers for the electrophysiology part of the multi-scale model in OpenDiHu.

\subsection{Evaluation of Compiler Optimizations}

One difference in the data organization in OpenDiHu compared to OpenCMISS Iron lies in the transposed memory layout for the storage of multiple instances of the 0D subcellular model. If the \code{simd} optimization type in the \code{CellmlAdapter} class is used, the components of the state vector $\bfy$ of all 0D model instances are stored consecutively. This storage order is the SoA memory layout, which was described in \cref{sec:optimizations_in_the_generated}. It enables the compiler to automatically employ SIMD instructions and, thus, exploit instruction-level parallelism.

We study the auto-vectorization performance of the GNU, Intel and Cray compilers to determine the effect of these SIMD instructions on the total runtimes of the solver. The simulated scenario consists of one muscle fiber mesh with 2400 nodes, on which the monodomain equation \cref{eq:monodomain} is solved. The subcellular model of Shorten et al. \cite{Shorten2007} is used. The used timestep widths are $\dt_\text{0D} = \SI{1e-3}{\ms}, \dt_\text{1D} = \dt_\text{splitting} = \SI{3e-3}{\ms}$, and the model is computed up to a simulation end time of $t_\text{end} = \SI{20}{\ms}$.

We run the study on one compute node of the supercomputer Hazel Hen at the High Performance Computing Center in Stuttgart. This Cray XC40 system contains two 12-core Intel Haswell E-2680v3 CPUs with clock frequency of $\SI{2.5}{\giga\hertz}$ per dual-socket node, yielding 24 processors per compute node and contains \SI{128}{\giga\byte} memory per compute node.

% compilers performance
\begin{figure}
  \centering%
  \includegraphics[width=0.7\textwidth]{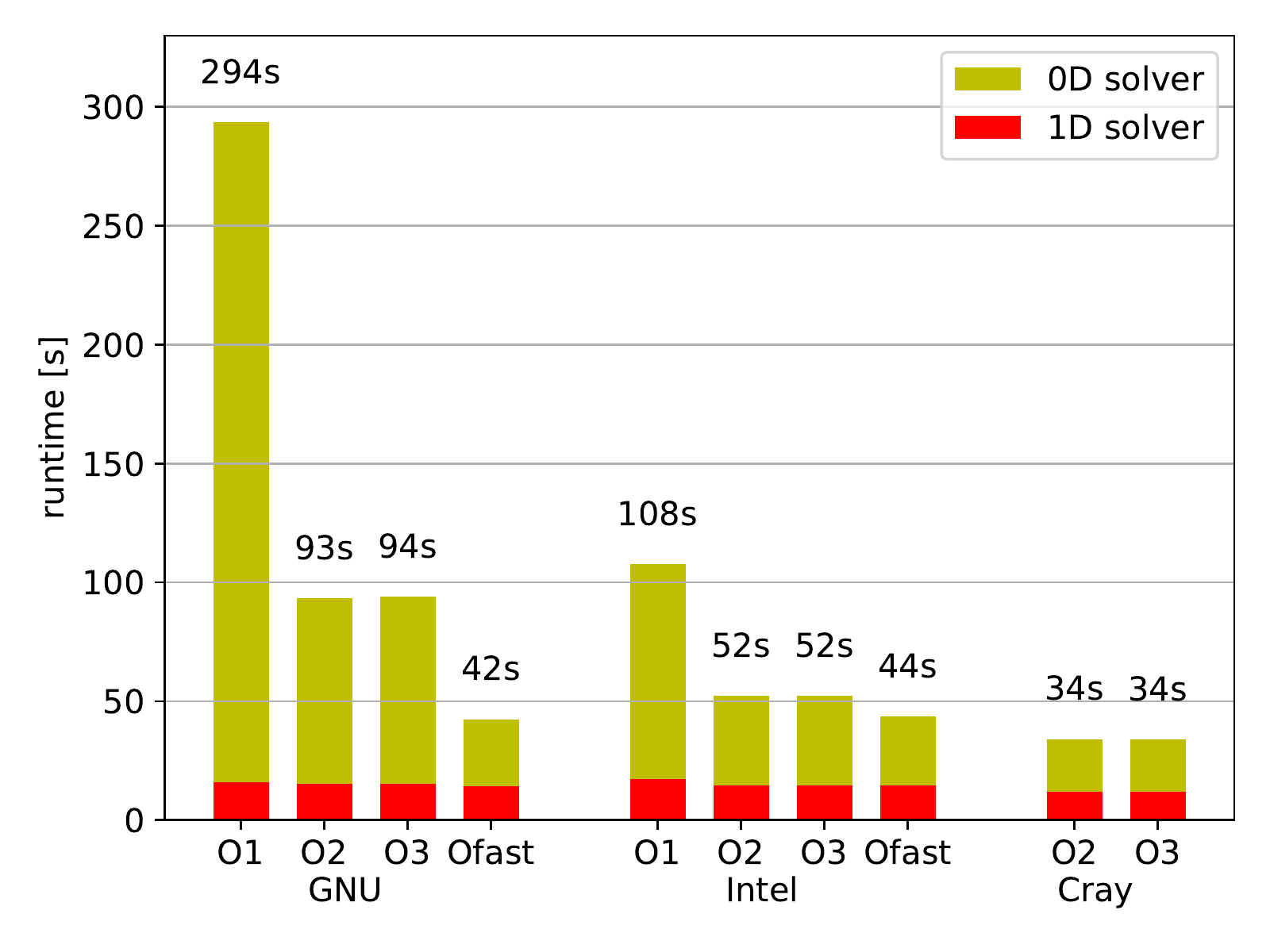}%
  \caption{Electrophysiology Solver in OpenDiHu: Comparison of auto-vectorization in different compilers. Runtime of the 0D and 1D solvers in the fiber based electrophysiology model with \code{simd} optimization type for different compilers and optimization flags.}%
  \label{fig:compilers}%
\end{figure}%

\Cref{fig:compilers} shows the runtime of the 0D and 1D model solvers for the three different compilers with varying optimization flags.
As expected, the runtime of the 1D solver is not affected by the choice of the compiler. The runtime of the 0D solver, however, varies greatly, as the compilers with different optimization flags are able to vectorize the code to a different extent.

For all compilers, the runtime decreases when a higher optimization level is chosen. A significant drop to less than half of the runtime is observed when switching from the \code{O1} to the \code{O2} optimization level for the GNU and for the Intel compiler. This is mainly the result of the SIMD instructions, which are enabled starting from the \code{O2} levels.
The change to the aggressive optimization levels \code{O3}, which enables all available optimizations such as inlining and code transformations does not improve the runtime any further, for all three evaluated compilers. Thus, vectorization is the main driver for good subcellular solver performance.

Another significant decrease in runtime can be observed for the \code{Ofast} optimization flag. For the GNU compiler, the runtime decreases again to less than half of the previous value. For the Intel compiler, the decrease is less prominent with approximately \SI{15}{\percent}. 

The \code{Ofast} level performs optimizations that potentially change the behavior of the code. 
Especially floating-point arithmetic does no longer comply to the standardization rules of IEEE and ISO. Only finite numbers can be represented and the compiler is allowed to perform transformations in formulas that are mathematically correct, but not in terms of propagating rounding errors. The calculated values are correct as long as no invalid operations such as divisions by zero occur. The precision may decrease or even increase compared to \code{O3}. This is usually not an issue for the given simulations, however, divergence of the numerical solvers is not automatically detectable with \code{Ofast} in our code as no infinity values can be represented.

The comparison between the compilers shows that the Intel compiler creates faster assembly code than the GNU compiler, and the Cray compiler creates faster assembly code than the Intel compiler for the same optimization levels. The performance of the \code{Ofast} flag is comparable between the GNU and the Intel compiler. In total, the Cray compiler yields the best performance on the Cray hardware used in this evaluation. 

The Cray compiler has a \say{whole-program mode}, which collects static information about all compilation units and allows, e.g., application-wide inlining during the linking step. The faster runtime is traded for longer compilation times. In our example, the compilation duration increases from approximately $\SI{10}{\min}$ for the GNU and Intel compilers to over $\SI{2}{\hour}$ for the Cray compiler.

For all further simulations, we use the GNU compiler with the \code{Ofast} optimization flag, as it is freely available on all systems, has fast compilation times and showed good performance.

\subsection{Evaluation of Code Generator Optimizations}\label{sec:evaluation_of_code_gen}

Apart from the automatic optimizations by the compiler, the code can also be manually optimized by using efficient data structures and algorithms. \Cref{sec:optimizations_in_the_generated} presents various optimization options in our code generator, which potentially have an influence on the runtime of the subcellular model solver. 
We compare all optimization options for a scenario of a comprehensive surface EMG simulation. 

The considered scenario solves the monodomain equation \cref{eq:monodomain} on every 1D muscle fiber domain and is coupled to a 3D mesh where the bidomain equation \cref{eq:bidomain1} is solved. No body fat domain is considered in this scenario.
We simulate 625 muscle fibers with 1481 nodes per fiber mesh and the subcellular model of Hodgkin and Huxley \cite{Hodgkin1952}. This leads to a total number of \num{3702500} degrees of freedom to be solved for the 0D and 1D models.
We run the code in parallel with 18 processes and a parallel partitioning of the 3D domain into $3 \times 2 \times 3$ subdomains. Thus, every muscle fiber domain is distributed to three different processes.
The 3D mesh contains 5239 nodes. Timestep widths of $\dt_\text{1D} = \SI{1e-3}{\ms}, \dt_\text{3D} = \dt_\text{splitting} = \SI{3e-3}{\ms}$ and an end time of $t_\text{end} = \SI{10}{\ms}$ are used, and file output is disabled for this study.

We use an Intel Core i9-10980XE processor with 18 cores, base frequency of $\SI{3}{\giga\hertz}$, maximum boost frequency of $\SI{4.8}{\giga\hertz}$, cache sizes of 
$\SI{24.8}{\mebi\byte}$, $\SI{18}{\mebi\byte}$ and $\SI{576}{\kibi\byte}$ and \SI{31}{\gibi\byte} main memory. This processor is listed in the upper price segment of consumer hardware and can be considered a typical hardware for individual workstations in scientific research.

% fibers_emg performance hodgkin huxley
\begin{figure}
  \centering%
  \includegraphics[width=\textwidth]{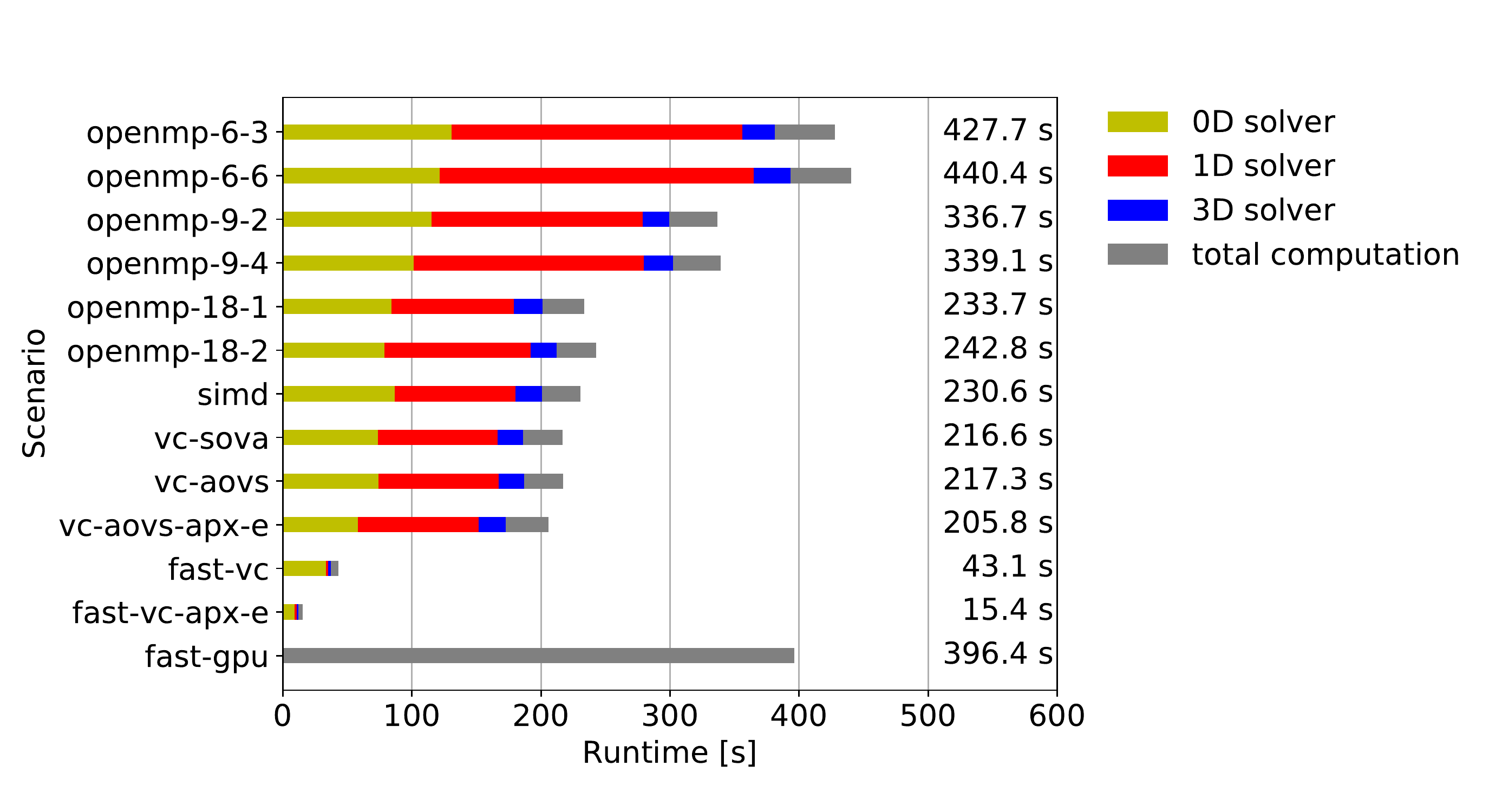}%
  \caption{Electrophysiology Solver in OpenDiHu: Evaluation of various code optimizations for the subcellular model solver. Comparison of runtimes for the 0D, 1D and 3D model solvers with different optimization types in the code generator.}%
  \label{fig:fibers_emg_study}%
\end{figure}%

\Cref{fig:fibers_emg_study} presents the results of the study for all available optimization types in our code generator. For every scenario, the bar chart shows the runtimes of the 0D subcellular solver in yellow color, the runtime of the 1D electric conduction solver in red color, the runtime for the 3D bidomain solver in blue color and the remaining runtime of the coupled solver scheme, which involves, e.g., data transfer between data structures and inter-process communication, in gray color.
The presented runtimes are averaged over several runs and over all processes per run.

The first six bars correspond to the \code{openmp} optimization type, which places OpenMP pragmas in the code and employs thread-based, shared memory parallelism. The scenario \code{openmp-$i$-$j$} refers to $i$ MPI processes in total with $j$ threads on every process. The problem is partitioned into $i$ subdomains and the $j$ OpenMP threads per subdomain simultaneously operate on the shared data structures of the subdomain.
As a result, in the scenarios \code{openmp-6-3}, \code{openmp-9-2} and \code{openmp-18-1}, 18 threads are executed in total on the processor with 18 physical cores. The other scenarios, \code{openmp-6-6}, \code{openmp-9-4} and \code{openmp-18-2}, employ 36 threads.

It can be seen that each set of two scenarios with the same number $i$ of processes and varying number $j$ of threads, i.e., \code{openmp-6-3} and \code{openmp-6-6}, \code{openmp-9-2} and \code{openmp-9-4}, and \code{openmp-18-1} and \code{openmp-18-2} has similar total runtimes. This shows that the runtime is reduced mainly as a result of MPI parallelization. The distribution of the runtime to the solvers allows further insights. Between the two scenarios with the same number of processes, the runtime of the 0D solver decreases. This is a result of the higher number of OpenMP threads that is used to perform the same amount of work. At the same time, the runtimes of the 1D solvers increase, which is due to the multi-threaded solution of the 1D problem in the solver library PETSc, which we consider as a black box.
%For the last two scenarios, \code{openmp-18-0} and \code{openmp-18-1}, the runtime of the 0D solver shows no further decrease, as the 18-core processor is fully occupied as soon as 18 threads are used.

The effect of OpenMP parallelism on the 1D solver is even higher than on the 0D solver in this example. As the code generator using OpenMP parallelism is only responsible for the 0D problem, the performance of the 1D problem depends only on the partition size and workload defined by the parallel partitioning with $i$ MPI processes. A reduction of the MPI parallelism has more impact on the runtime than the resulting increased parallelism of the 0D solver. Thus, the scenarios with high degrees of OpenMP parallelism, e.g., scenario \code{openmp-6-6}, show a worse performance than the scenarios with higher MPI parallelism, e.g., scenario \code{openmp-18-1}.

The next bar in \cref{fig:fibers_emg_study} presents the runtime of the \code{simd} optimization type. The code uses the \code{SoA} memory layout and the program is run with 18 MPI processes. As in all scenarios of this study, the GNU compiler with the \code{Ofast} flag is used and automatically vectorizes the subcellular model equations. The \code{simd} scenario is very similar to the \code{openmp-18-1} scenario, except that the OpenMP pragmas are omitted in the generated code. As a result, the runtimes are also similar to this scenario. A slight reduction in runtime is seen that results from the missing OpenMP initializations before every loop.

While the \code{simd} scenario relies on the auto-vectorization capabilities of the compiler, the \code{vc} scenarios, which are considered next, explicitly employ vector instructions, abstracted by the \emph{Vc} and \emph{std-simd} libraries. 

The \code{vc-sova} scenario uses the Struct-of-Vectorized-Array (SoVA) memory layout and the bar chart shows a slightly lower runtime of the 0D solver compared to the Array-of-Vectorized-Struct (AoVS) memory layout in the \code{vc-aovs} scenario.

The next considered scenario is \code{vc-aovs-apx-e}. It is the same as \code{vc-aovs} except that the exponential function is approximated by $\textrm{exp}^\ast(x)=(1+x/n)^n$ for $n=1024$, as given in \cref{eq:apx-e-function}. The results show that this reduces the runtime of the 0D solver from $\SI{74.24}{\s}$ to \SI{58.02}{\s}, which is a reduction by approximately $\SI{22}{\percent}$.

Instead of generating code only for the 0D subcellular model and solving the 1D subcellular model using a direct solver of PETSc, as in all considered scenarios so far, we can also directly generate combined solver code for the 0D and 1D models and use the Thomas algorithm for the computation of the 1D model. This is done in the \code{fast-vc} scenario and reduces the runtime by a factor of nearly 5. In this approach, the exponential function can also be exchanged by the approximation in \cref{eq:apx-e-function}. This is done in the \code{fast-vc-apx-e} scenario and further decreases the total runtime to now only $\SI{15.4}{\s}$.

The two \code{fast-vc} scenarios demonstrate the performance of the AVX-512 vector instruction set that is available on the used Intel processor. The study shows that its potential is only fully exploited, if the explicit vector instructions are generated in the code, as done in the \code{vc} scenarios.

The solution times for the last two mentioned scenarios can be further reduced if only those subcellular model instances are computed that are not in equilibrium. If enabled, this reduction depends on the activation pattern of the fibers. For the sake of the present study, which aims to compare runtimes of the code generator, this option is not evaluated and, thus, disabled.

The last considered optimization type in the code generator is presented in the scenario \code{fast-gpu}. In this scenario, the program is only run with one MPI process. The total computation of the 0D and 1D models is offloaded to a GPU using OpenMP 4.5 pragmas in the generated code. We use the same simulation scenario and CPU hardware for this run as for the other scenarios. The used computer is equipped with an NVIDIA GeForce RTX 3080 GPU with 8704 CUDA cores, \SI{10}{\giga\byte} of memory and a Thermal Design Power (TDP) of \SI{320}{\watt}. The processing power is \SI{29.77}{\tera\flops} for single precision and \SI{465.1}{\giga\flops} for double precision operations. We use only double precision operations for the computation of the models.

In this scenario, only the total runtime is measured. The bar chart shows a total solver runtime of \SI{396}{\s}, which is slower than the optimized CPU computations. Possible reasons are that the used GPU is targeted at single precision performance, and that the employed GPU code by the OpenMP functionality of the GNU compiler is not optimal.

In the previously considered example, which uses the Hodgkin and Huxley subcellular model with a state vector $\bfy \in \R^4$, the amount of computational work in the 0D and in the 1D solver was in the same range. Other 0D subcellular models exist that have higher workloads. In the next study, we repeat the same measurements as before with the subcellular model of Shorten et al. \cite{Shorten2007}, which has a state vector $\bfy \in \R^{57}$.  Whereas the solver for the model of Hodgkin and Huxley needs to compute 4 ODEs and 9 algebraic equations in every timestep, the solver for the Shorten model computes 57 ODEs and 71 algebraic equations in every timestep.

As the computational effort to solve one instance of the subcellular model increases, we adjust the simulation scenario for the next study. 
We use 49 fibers with 1481 nodes each and a 3D finite element mesh with linear ansatz functions and a total of \num{23696} degrees of freedom. The total number of degrees of freedom in all meshes is \num{4087560}, which is similar to the number \num{3707739} in the previous study. The simulation end time is \SI{3}{\ms}. For this subcellular model, smaller timestep widths of $\dt_\text{splitting}=\dt_\text{1D}=\dt_\text{0D}=\SI{2.5e-05}{\ms}$ and $\dt_\text{3D}=\SI{1e-01}{\ms}$ are used as required to ensure convergence of the solver for this subcellular model.

% fibers_emg performance shorten
\begin{figure}
  \centering%
  \includegraphics[width=\textwidth]{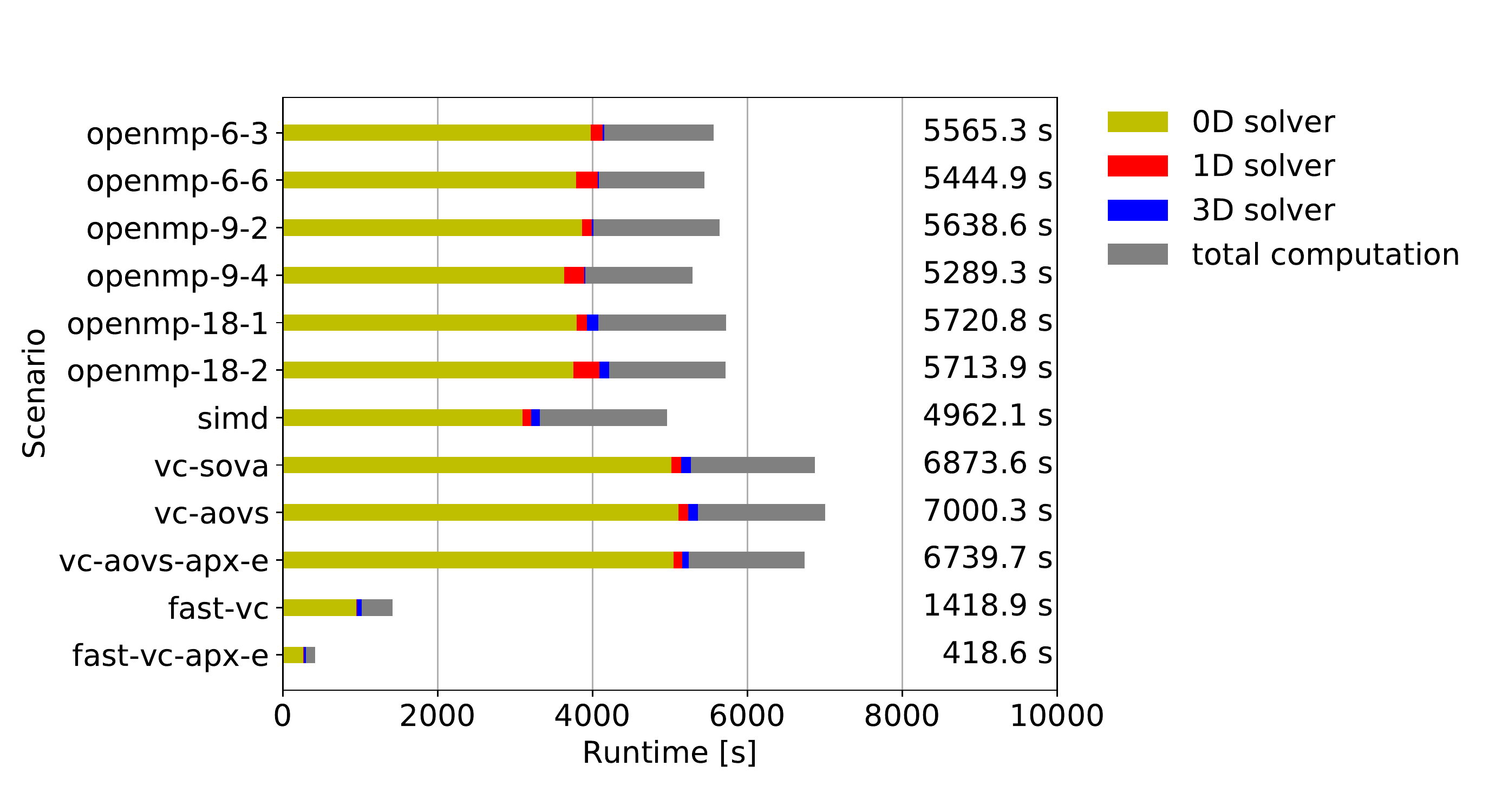}%
  \caption{Electrophysiology Solver in OpenDiHu: Comparison of runtimes for different optimizations in the code generator, for the compute-intense Shorten subcellular model.}%
  \label{fig:fibers_emg_study_shorten}%
\end{figure}%

\Cref{fig:fibers_emg_study_shorten} shows the resulting runtimes for different scenarios in a bar chart analog to \cref{fig:fibers_emg_study}. It can be seen that the solver time for the 0D model now dominates the total runtime in all scenarios. In the \code{openmp-$i$-$j$} scenarios, the runtime for the 0D solver decreases as before, if more threads are used in total. Contrary to the previous study, the total runtime profits from this runtime reduction, as the 0D part is significant enough for the total runtime. Another difference to the results of the previous study is that the durations for the 0D model are nearly the same for every combination of number of MPI processes $i$ and number of OpenMP threads $j$. This shows that the overhead of starting the OpenMP threads, which in the previous study was responsible for larger compute times of the 0D models, is now amortized by the larger overall workload.

The performance in the \code{simd} scenario is, again, comparable to the performance of  the \code{openmp-18-1} scenario and shows a slightly smaller runtime due to the missing OpenMP thread initializations.

A difference to the previous study can be seen for the \code{vc} scenarios. In the present study with the subcellular model of Shorten et al., the runtimes for the \code{vc-sova}, \code{vc-aovs}, and \code{vc-aovs-apx-e} are all higher than for the auto-vectorized scenarios. In contrast, the \code{vc} scenarios showed a large reduction in runtime in the study with the Hodgkin and Huxley subcellular model. 

This effect originates from the operations required to evaluate the subcellular equations. The Shorten model contains many $\log(x)$ function evaluations. These are especially compute intense and, in addition, not supported in the abstraction layer of the AVX-512 instructions provided by the \emph{std-simd} library. Instead, the library employs their non-vectorized counterparts. The auto-vectorization of the compilers, however, is able to employ the respective vectorized functions, which explains the better performance in the \code{openmp} and \code{simd} scenarios. 

We expect that, in the future, the respective functionality will become available in the \emph{std-simd} library, which would automatically increase the performance for these optimization types. For processors without AVX-512 support, but with the AVX2 instruction set, the library \emph{Vc} is used, which supports the respective functions and, thus, yields the expected performance in the \code{vc} scenarios. Whereas AVX-512 has a SIMD lane width of eight double values, AVX2 only supports SIMD lanes with 4 double values.

\begin{figure}
  \centering%
  \includegraphics[width=0.5\textwidth]{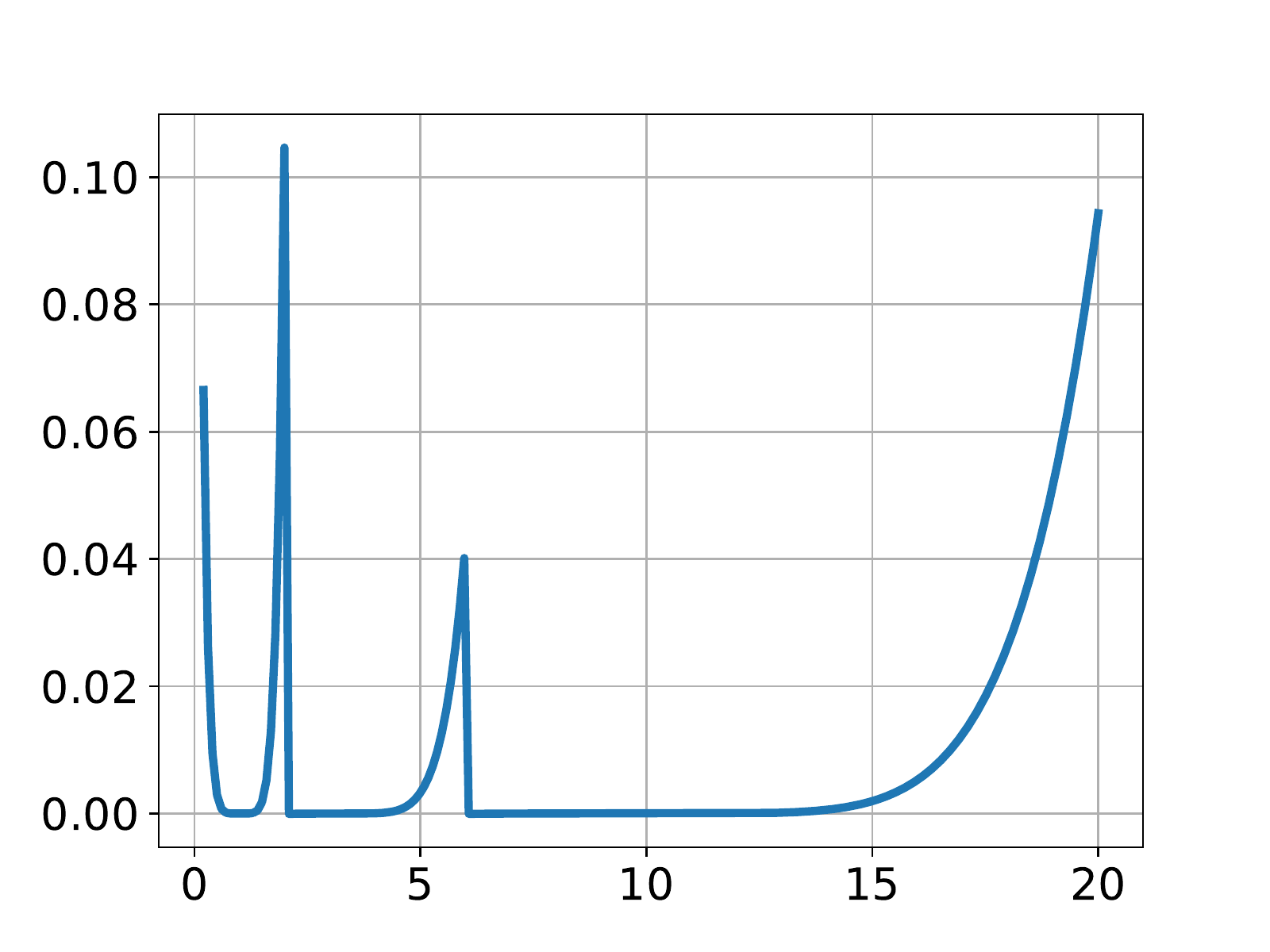}%
  \caption{Electrophysiology Solver in OpenDiHu: Relative error of the piecewise Taylor approximation of the log function as used in the vectorized simulation code.}%
  \label{fig:apxlog}%
\end{figure}%
To mitigate the effect of the missing $\log(x)$ vectorization, we replace the log function by a numerical approximation, in addition to the approximated exp function. We define the approximated logarithm function $\log^\ast(x)$ by its piecewise Taylor polynomials of sixth order around the points $x=1$, 3 and 9 with discontinuities at the points $x=2$ and $x=6$. \Cref{fig:apxlog} shows the absolute relative error for the range between $0.2$ and 20 which, in this range, is bounded by $0.105$. However, better convergence of the 0D-1D problem is achieved, if the approximated log function $\log^\ast$ is the inverse of the approximated exponential function $\exp^\ast$. Therefore, we apply one Newton iteration of the problem %
\begin{align*}
  F(y) = \exp^\ast(y)-x \overset{!}{=} 0  
\end{align*}
to the log value $y$ computed by the Taylor approximation. The Newton iteration consists of subtracting ${(1 - x/\exp^\ast(x))}$ from the computed result $y$. Thus, it only involves one evaluation of the approximated exponential function.

The scenario \code{fast-vc} in \cref{fig:fibers_emg_study_shorten} generates unified solver code for both 0D and 1D models, but does not include this approximation. The approximated exponential and logarithm functions are included in the scenario \code{fast-vc-apx-e}. As a result, it can be seen that the total runtime is largely reduced compared to the auto-vectorized scenarios.

\begin{reproduce_no_break}
  The simulations in this section use the example \code{examples/electrophysiology/fibers/fibers_emg}
   with the variables files \code{optimization_type_study.py} and \code{shorten.py}.
  The commands for the individual runs are executed by the following scripts:
  \begin{lstlisting}[columns=fullflexible,breaklines=true,postbreak=\mbox{\textcolor{gray}{$\hookrightarrow$}\space}]
    cd $\$$OPENDIHU_HOME/examples/electrophysiology/fibers/fibers_emg/build_release
    ../old_scripts/run_optimization_type_study.sh
    ../old_scripts/run_optimization_type_study_shorten.sh
  \end{lstlisting}
  The utility to create the plots from the generated \code{logs/log.csv} files can be found in the repository at \href{https://github.com/dihu-stuttgart/performance}{github.com/dihu-stuttgart/performance}
  in the directory \code{opendihu/18_fibers_emg}:
  \begin{lstlisting}[columns=fullflexible,breaklines=true,postbreak=\mbox{\textcolor{gray}{$\hookrightarrow$}\space}]
    ./plot_optimization_type_study_shorten.py
    ./plot_optimization_type_study.py
  \end{lstlisting}
\end{reproduce_no_break}

% ------------
%
% f===========

% --------------------------------
% studies, performance

% ==============
%
% --------------
\section{Parallel Strong Scaling and Comparison with OpenCMISS Iron}\label{sec:parallel_strong_scaling_opencmiss}

After the performance of different optimization types has been evaluated for a scenario with a single number of MPI processes in the last section, we now conduct a strong scaling study with the optimization type \code{fast-vc-apx-e}, which was found to be the most performant, and compare the runtimes to the reference software OpenCMISS Iron.

\subsection{Evaluation of Runtimes}\label{sec:strong_scaling_runtimes_opencmiss_opendihu}
For a fair comparison, we take care to exactly compute the same scenario with both software packages. The simulated scenario uses the same model as in the previous section: fiber based electrophysiology with the monodomain model given by \cref{eq:monodomain} on every muscle fiber, including the 1D electric conduction along the fibers and the 0D subcellular model of Shorten et al. \cite{Shorten2007}. The fibers are coupled with the bidomain equation in \cref{eq:bidomain1}, which is solved on the 3D domain to yield the EMG signals. No fat layer is considered in this study, as this feature is not available in our OpenCMISS implementation.

The simulated scenario contains 81 fibers with \num{1480} elements each, a coarse 3D mesh with \num{775} nodes and \num{6718591} degrees of freedom in total. 
We use our improved OpenCMISS setup, which is discussed in \cref{sec:opencmiss_numeric_improvements} and employs the second order numerical timestepping schemes and the improved linear solvers:
The monodomain model is solved using a Strang operator splitting with Crank-Nicolson and Heun's methods. A conjugate-gradient solver is used for the linear system of the bidomain equation.

We use timestep widths of $\dt_\text{0D}=\SI{1e-4}{\milli\second}$, $\dt_\text{1D}=\dt_\text{splitting}=\SI{5e-4}{\milli\second}$, $\dt_\text{3D}=\SI{1e-1}{\milli\second}$ and a simulation end time of $t_\text{end}=\SI{2}{\milli\second}$. 
During this time, the resulting values are written to output files after every $\SI{0.1}{\milli\second}$. 
The fibers are assigned to 10 MUs that are activated in a ramp every $\SI{0.2}{\milli\second}$ from $t=\SI{0}{\milli\second}$ to $t=\SI{1.8}{\milli\second}$.

Within the strong scaling study, the same scenario is computed with different numbers of processes, ranging from one to 18 in this case. We use the cubes partitioning strategies presented in \cref{sec:opencmiss_parallel_partitioning} in both OpenCMISS and OpenDiHu.
The study is executed on the same Intel Core i9-10980XE processor as the studies in the previous section.
We measure the total user time of the simulation program, which includes the runtimes for initialization, computation of system matrices and the duration of file output. However, the majority of the runtime in this scenario is spent in the numerical solvers.

In OpenDiHu, we use the setup corresponding to the \code{fast-vc-apx-e} scenario in the last section with enabled approximation of log and exp functions. We measure the runtime of two variants. The first variant computes all subcellular models and performs the same work as the OpenCMISS Iron implementation. In the second variant, the adaptive computation described in \cref{sec:adaptive_computation_for_fiber_based} is enabled, which only computes fibers that have been activated and the subcellular models that are not in equilibrium. For the chosen ramp activation pattern of the MUs, the second variant computes approximately only half of the subcellular model instances.

% weak scaling runtime
\begin{figure}
  \centering%
  \begin{subfigure}[t]{0.49\textwidth}%
    \centering%
    \includegraphics[width=\textwidth]{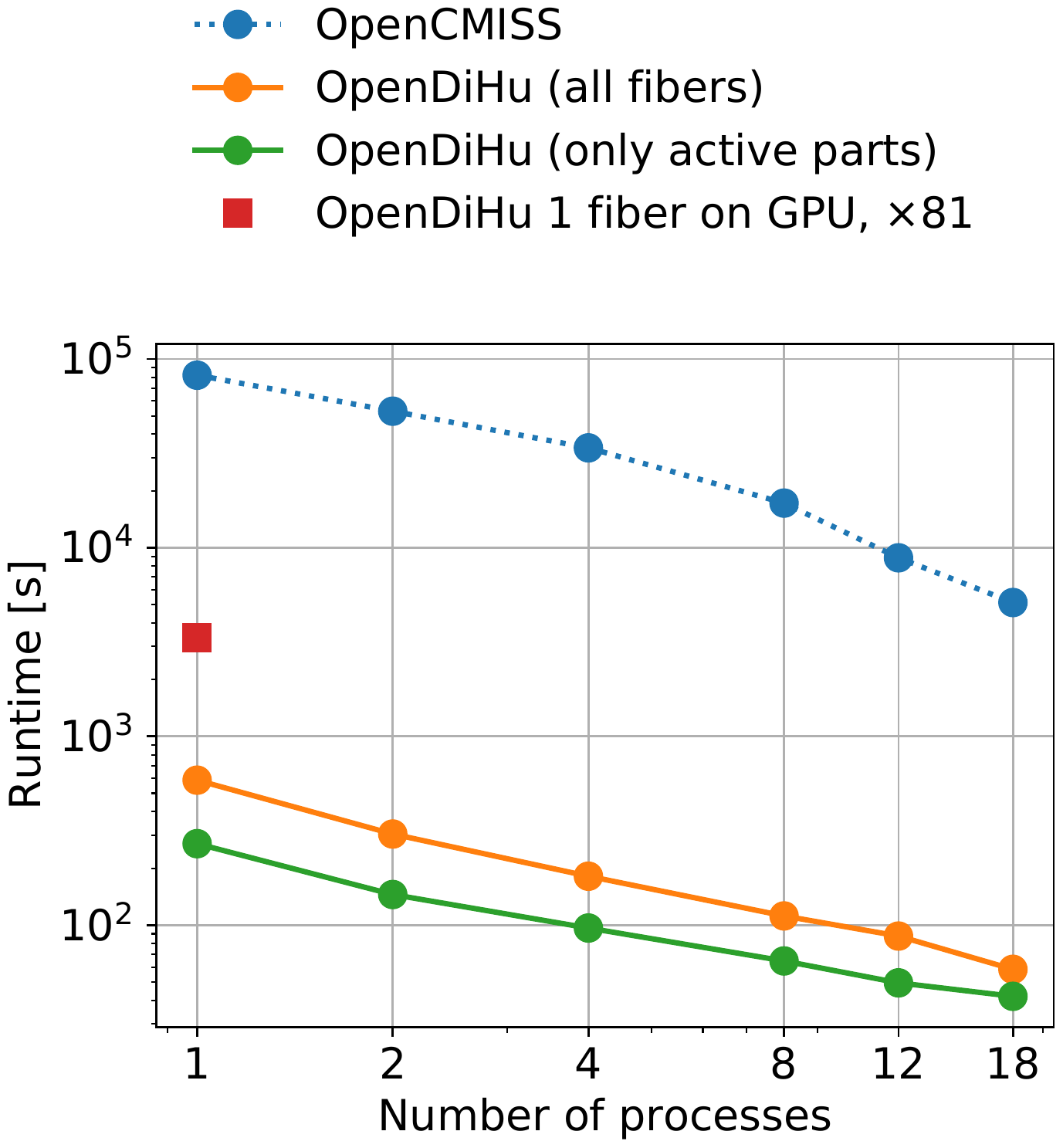}%
    \caption{Runtime of the simulation programs for OpenCMISS (blue) and two variants of OpenDiHu (orange and green), see the description in the text for details. 
    %For OpenDiHu, two variants are shown: the normal computation of all fibers (orange) and the optimization that computes only subcellular model instances that are not in equilibrium (green). In addition, the problem is solved on the GPU (red).
    }%
    \label{fig:0_strong_scaling_runtime}%
  \end{subfigure}
  \quad
  \begin{subfigure}[t]{0.47\textwidth}%
    \centering%
    \includegraphics[width=\textwidth]{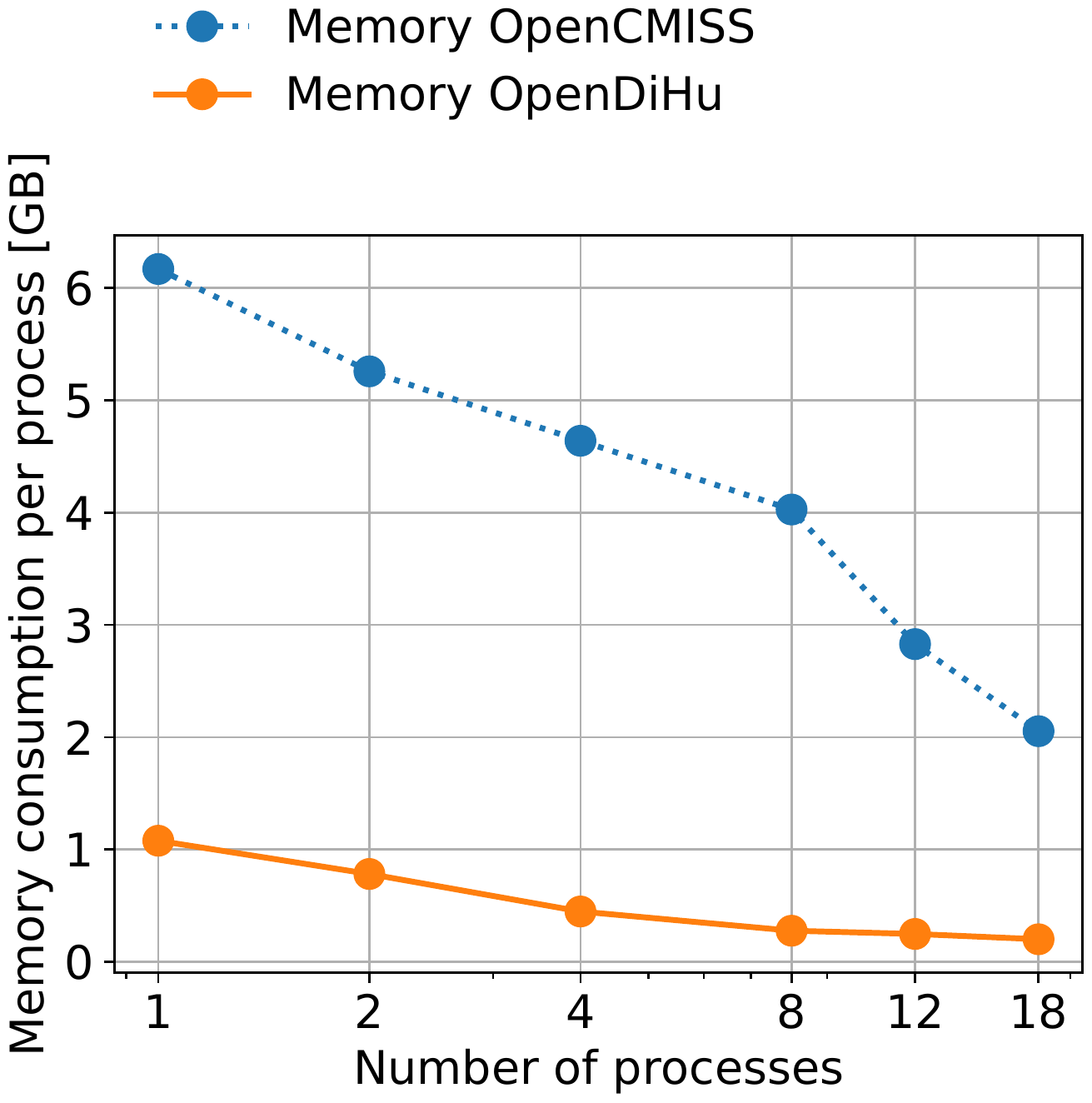}%
    \caption{Memory consumption per process at the end of the program.}%
    \label{fig:0_strong_scaling_memory}%
  \end{subfigure}   
  \caption{Electrophysiology Solver in OpenDiHu: Strong scaling study of fiber based electrophysiology and comparison between the implementations of OpenDiHu and OpenCMISS Iron. The same scenario is solved with both software packages and for increasing numbers of processes from one to 18.}%
  \label{fig:0_strong_scaling}%
\end{figure}%

\Cref{fig:0_strong_scaling_runtime} shows the resulting runtimes in this study. It can be seen that the runtime decreases monotonically for higher numbers of processes for all three tested simulations. The OpenDiHu implementation exhibits lower runtimes for all numbers of processes. The reduction in runtime between OpenCMISS Iron and the first OpenDiHu variant is given by a factor of approximately 100 with a maximum factor of 186 for 4 processes. In addition, the second OpenDiHu variant approximately halves the runtimes as expected, because only half of the subcellular models are computed.

In summary, the improvements to the EMG simulation software, which are described in this work, include the numerical improvements in \cref{sec:opencmiss_numeric_improvements} with a speedup of 2.5, the software improvements with a speedup of over 100 and a measured maximum of 186, and the algorithmic improvement of adaptive 0D model computations, whose speedup factor is scenario dependent. In the present study, the two latter factors, i.e., the speedup between the improved OpenCMISS Iron software and the OpenDiHu scenario with adaptive computation, give a combined maximum speedup of 363 for the measurement with two processes.

Moreover, the computation of this study was also carried out with the \code{gpu} optimization type in OpenDiHu, using the same GPU as in the last section. One process was started on the CPU, which offloaded the computational work of the 0D and 1D problems to the GPU. However, the GPU memory was not sufficient for the computation of all 81 fibers. Therefore, we only compute one fiber, but keep the rest of the simulation scenario equal to the other measurements. The req square in \cref{fig:0_strong_scaling_runtime} shows the measured runtime multiplied by the factor 81 for compensation. As in the studies of the previous section, the computation on the GPU has higher runtimes than the computation on the CPU.

As the memory consumption was a limiting factor for parallelism in OpenCMISS Iron as shown in \cref{sec:opencmiss_memory}, we also measure the memory consumption per process at the end of the simulation in both software frameworks. \Cref{fig:0_strong_scaling_memory} shows the result for OpenCMISS and OpenDiHu. The two variants of OpenDiHu have the same memory consumption characteristics, as the only difference between the variants is that the computation of certain subcellular models is switched on or off. 

It can be seen that the increased parallelism leads to a reduction of the used memory per process in both pieces of software. OpenDiHu approaches a saturation value of $\SI{200}{\mebi\byte}$ for eight and more processes. For OpenCMISS, the memory consumption is higher, but reduces more quickly also for higher numbers of processes. However, the relation between the two curves increases from a value of $\SI{6.168}{\gibi\byte} : \SI{1.078}{\gibi\byte} = 5.7$ for one process to $\SI{2.054}{\gibi\byte} : \SI{0.202}{\gibi\byte} = 10.2$ for 18 processes.

As a result, this study shows a large memory efficiency improvement in OpenDiHu compared with the OpenCMISS Iron software. For OpenCMISS, the memory scaling in this parallel strong scaling scenario is not as bad as in the parallel weak scaling considered in \cref{fig:opencmiss_memory} in \cref{sec:opencmiss_memory}. However, the total memory for all processes still increases to approximately $18 \cdot \SI{2.054}{\gibi\byte} \approx \SI{37}{\gibi\byte}$, which is higher than the main memory capacity of the used processor.

% roofline model
\begin{figure}[H]
  \centering%
  \includegraphics[width=0.8\textwidth]{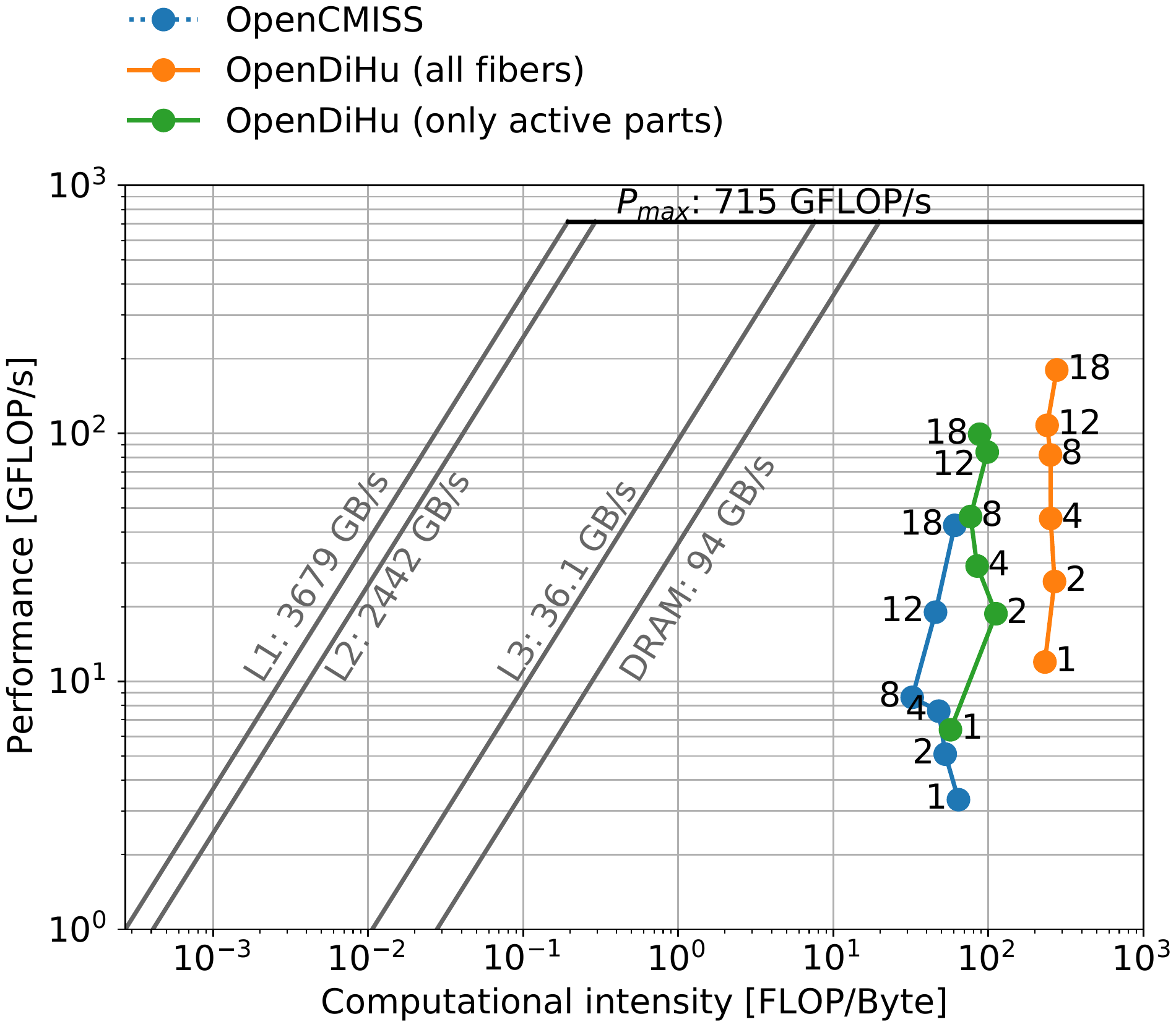}%
  \caption{Electrophysiology model in OpenDiHu: Roofline model of the strong scaling study. The blue, green and orange data points correspond to the runs of the OpenCMISS and OpenDiHu variants, as given in \cref{fig:0_strong_scaling_runtime}.}%
  \label{fig:0_roofline}%
\end{figure}%

\subsection{Roofline Model}\label{sec:roofline_model}
To further investigate the computational behavior, we also present the performance measurements of the solvers in a roofline model. 
\Cref{fig:0_roofline} shows the resulting diagram with the data points of all CPU runs in the strong scaling study of \cref{fig:0_strong_scaling_runtime}. The $x$ axis shows the computational intensity of the simulation, which is measured in double-precision floating-point operations (FLOP) per byte of data that are transferred between the CPU and the main memory and caches. The $y$ axis measures the performance in GFLOP per second. The highest possible performance is given by the peak performance of the processor, which is $P_\text{max} = \SI{715}{\giga\flop\per\second}$ in this case. Furthermore, the performance is limited by the amount of payload data that can be transferred to the CPU over the memory bus. The memory bandwidths of the L1, L2 and L3 caches and the main memory (DRAM) correspond to the shown diagonals in \cref{fig:0_roofline} and form the \say{roofline} of the model.

We measured the memory bandwidths of the Caches and the peak performance using the Empirical Roofline Tool \cite{ert}. The main memory bandwidth was retrieved from the processors' documentation.

To locate the simulation runs of the study in the roofline model, we used hardware counters to count floating point instructions and memory access operations.
For the runs with OpenCMISS Iron and OpenDiHu, we started the hardware counters \SI{90}{\second}, respectively \SI{15}{\second} after the beginning of the simulation, such that the initialization phase was not included in the measured data. The counters were kept active for 15, 30 and 60 seconds, depending on the expected runtimes of the different runs. The counted numbers of events were then divided by the acquisition time to yield the required rates of memory bandwidth and floating-point performance.

\Cref{fig:0_roofline} shows the measured points in the roofline model corresponding to the curves in \cref{fig:0_strong_scaling_runtime}. All data points are located at the right-hand side of the memory bandwidth limits, which indicates that the simulation is compute bound. 
The highest computational intensity and performance are both achieved by the OpenDiHu variant given in orange color, which computes all fibers and subcellular models regardless of their activation state. The values for the adaptive variant given in green color are lower, as fewer computations are performed and a higher portion of the runtime and compute power is spent on determining which subcellular model has to be computed. The two metrics are lowest for the OpenCMISS runs given in blue color.

The roofline diagram shows the data points for all parallel runs and the number of processes is noted in the plot. The run with 18 processes is the most meaningful, as this means that the whole processor is employed. The largest performance for the OpenDiHu run in orange color has a value of \SI{180.157}{\giga\flop\per\second} which corresponds to \SI{25.2}{\percent} of the peak performance and is a very good value. The rated \SI{100}{\percent} of peak performance for processors are practically unreachable. For example, the peak performance assumes only fused multiply add operations and requires a power management that maximizes the employment of the boost clock frequency in the processor. These conditions are not fulfilled in our computations of realistic models and scenarios.
The performance values of the runs with 18 processes for the green and blue data points are \SI{13.9}{\percent} and \SI{6.0}{\percent}, respectively. 

Furthermore, the measurements with lower process counts can also be assessed with a scaled down peak performance according to the fraction of used cores. However, this assessment is slightly off, as, e.g., the CPU can use a higher clock frequency, if only the heat dissipation of one active core has to be compensated. The performance for the OpenDiHu run with one process given by the orange point is \SI{11.966}{\giga\flop\per\second}, which corresponds to \SI{30.1}{\percent} of the fractional peak performance of $\SI{715}{\giga\flop\per\second}/18 = \SI{39.7}{\giga\flop\per\second}$.

\begin{reproduce_no_break}
  The scripts to run the studies in this scenario and to create the plots are available in the repository at \href{https://github.com/dihu-stuttgart/performance}{github.com/dihu-stuttgart/performance}
  in the directory \code{opendihu/20_fibers_emg_avx_opencmiss}:
  \begin{lstlisting}[columns=fullflexible,breaklines=true,postbreak=\mbox{\textcolor{gray}{$\hookrightarrow$}\space}]
    ./0_run.sh
  \end{lstlisting}
  The directory also contains a script that performs all steps to install OpenCMISS, if needed. 
  
  Note, the studies in the previous and the current sections were carried out on the computer with hostname \code{pcsgs05} in the institute network at the time of writing.
\end{reproduce_no_break}

%-----
\section{Performance Measurements on the GPU}\label{sec:performance_gpu}

In the previous two sections, measurements were made on a GPU, which produced worse results than the CPU code. The used GPU was an NVIDIA GeForce RTX 3080, a high-end consumer graphics card, which mainly targets graphics rendering performance using single-precision operations. The ratio between double-precision and single-precision performance is 1:64. However, single-precision calculations were found to not yield a stable subcellular model solver, as the precision is too low. 
% quadro: https://www.techpowerup.com/gpu-specs/quadro-gp100.c2994
% RTX: https://www.techpowerup.com/gpu-specs/geforce-rtx-3080.c3621

In this section, we conduct two studies, where the first study uses the same GPU hardware as before. The second study is executed on an NVIDIA Quadro GP100 GPU, which has a double-precision to single-precision performance ratio of 1:2.
The rated double-precision performance of the Quadro card is \SI{5.168}{\tera\flops}, which is ten times higher than the value of \SI{465.1}{\giga\flops} for the GeForce card. 

The CPU hardware connected with the Quadro card contains a dual-socket CPU with two 12-core Intel Xeon Silver 4116 processors with \SI{2.1}{\giga\hertz} base frequency and \SI{3}{\giga\hertz} maximum turbo frequency, yielding a total core count of 24, and being equipped with main memory of \SI{188}{\gibi\byte}. 

Computational hardware can be compared by its average thermal design power dissipation (TDP). For the studies in the previous two sections, the TDP values for the used CPU and GPU were \SI{165}{\watt} and \SI{320}{\watt}. For the second study in the current section, the values for CPU and GPU are $2\cdot \SI{85}{\watt} = \SI{170}{\watt}$ and \SI{235}{\watt}. This indicates that the employed hardware is in a comparable electrical power range. However, the GPU is more specialized for our double-precision needs.

\subsection{Strong Scaling with the GPU for the Hodgkin-Huxley Model}\label{sec:strong_scaling_gpu_hodgkin_huxley}

While the studies in the last two sections only used one process on the CPU, which managed the offloaded computations on the GPU,
we now additionally consider parallelism on the CPU.

The first study compares the strong scaling with and without GPU acceleration. The runs with GPU acceleration partition the computational domain as usual to multiple subdomains, which are handled by dedicated processes on the CPU. The solution of the 0D and 1D models is performed on the GPU, and every process independently transfers its part of the computational work to the same GPU. The 3D model is fully solved using MPI parallelism on the CPU. We compare this setup with a pure CPU based strong scaling study.

The scenario solves the fiber based electrophysiology model without fat layer with 169 fiber meshes of 1480 elements each and a 3D mesh with 1984 elements. The subcellular model of Hodgkin and Huxley \cite{Hodgkin1952} is used. The computation uses the same numerical parameters as in the first study in \cref{sec:evaluation_of_code_gen}, a 3D solver timestep of $\dt_\text{3D}=\SI{4e-01}{\ms}$ and a simulation end time of $\SI{10}{\ms}$. Moreover, the setup equals the settings of the \code{fast-vc} and \code{fast-gpu} scenarios in \cref{sec:evaluation_of_code_gen}.

\begin{figure}%
  \centering%
  \begin{subfigure}[t]{0.45\textwidth}%
    \centering%
    \includegraphics[height=52mm]{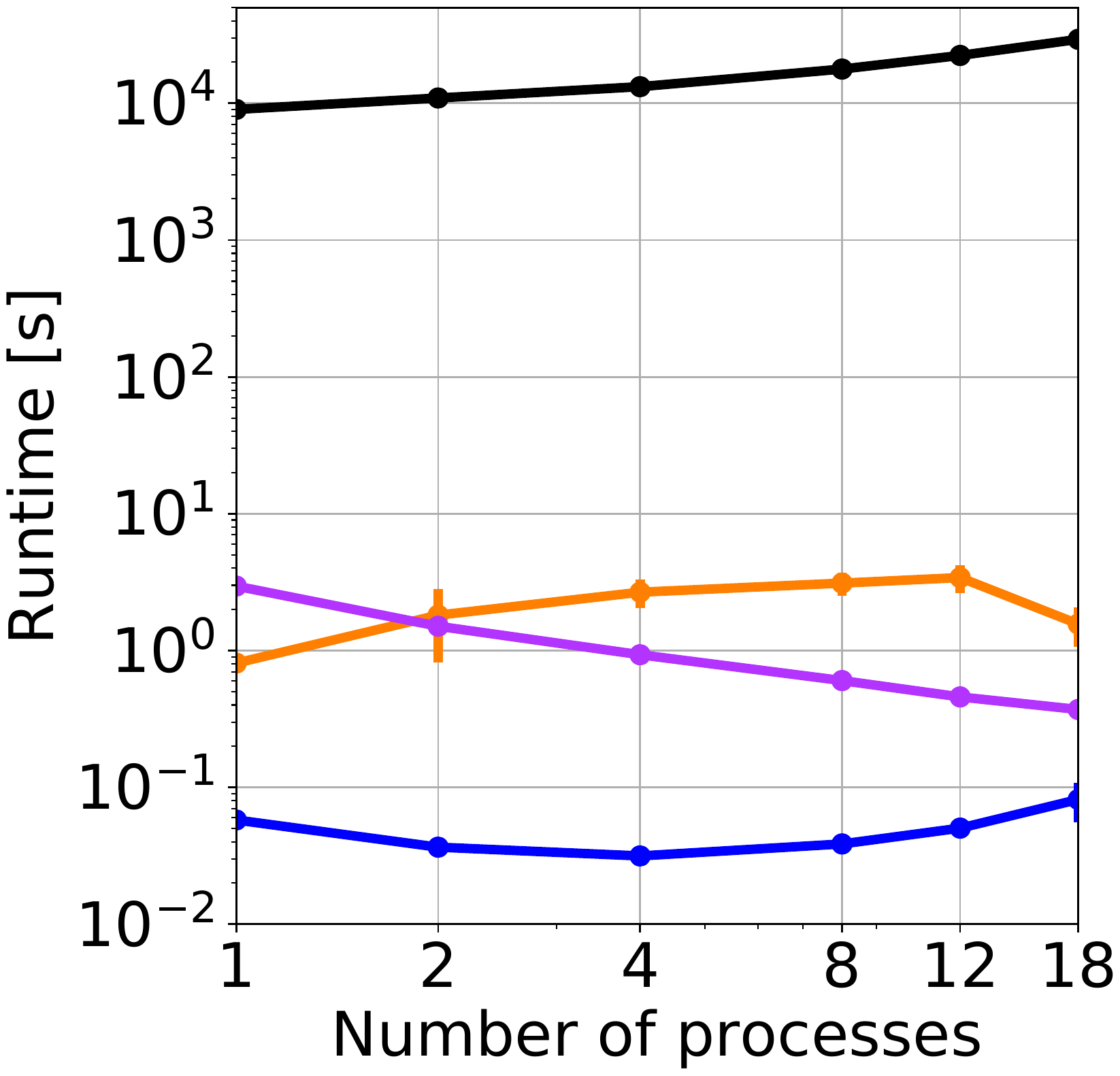}%
    \caption{Study, where every process on the CPU offloads the 0D and 1D model computations to the GPU.}%
    \label{fig:16_hodgkin_huxley_gpu}%
  \end{subfigure}
  \,
  \begin{subfigure}[t]{0.53\textwidth}%
    \centering%
    \includegraphics[height=52mm]{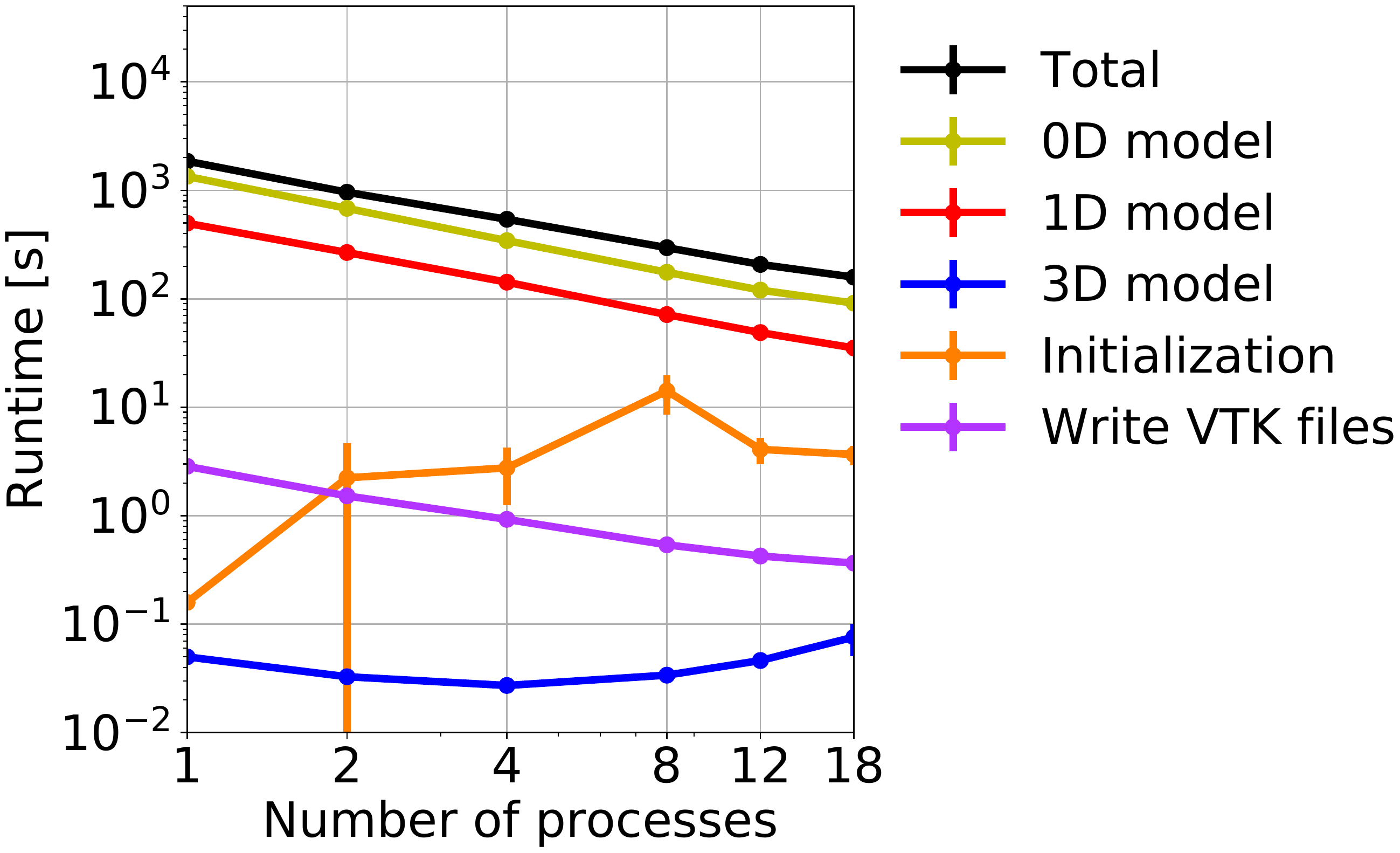}%
    \caption{CPU-only study.}%
    \label{fig:16_hodgkin_huxley_cpu}%
  \end{subfigure}   
  \caption{Electrophysiology model in OpenDiHu: Strong scaling study with and without GPU usage. A scenario with 169 fibers and the subcellular model of Hodgkin and Huxley is simulated. The vertical bars indicate the standard deviation of the runtimes in the set of measurements, which consists of multiple runs and the values of all processors in every run.}%
  \label{fig:16_hodgkin_huxley_cpu_gpu}%
\end{figure}%

\Cref{fig:16_hodgkin_huxley_cpu_gpu} presents the results for the two studies with and without GPU usage. \Cref{fig:16_hodgkin_huxley_cpu} shows the runtimes of different parts in the simulation of the CPU-only strong scaling study. It can be seen that the 0D computations account for most of the runtime, followed by the 1D computations. The 0D computations involve the solution of the Hodgkin-Huxley subcellular model. The 1D computations consists of solving 1D problems in serial using the Thomas algorithm. The 3D solver time is negligible as the 3D problem is only solved every \num{13333} timesteps. The 3D solution is only required right before the VTK file output step for the EMG values. This step occurs every $\SI{0.4}{\ms}$, which corresponds to an EMG sampling frequency of \SI{2.5}{\kilo\hertz}.
The runtimes for initialization and the file output itself are also very low compared with the runtimes of the computations.

\Cref{fig:16_hodgkin_huxley_cpu} shows that the total runtime decreases with higher process counts in this strong scaling study. The parallel efficiency $E_p = T_1/(T_p\,p)$ reaches $E_p=\SI{65.2}{\percent}$ for $p=18$ processes. We observe that the 0D and the 1D solver and the VTK file output have good strong scaling properties, whereas the initialization and the solution of the 3D model contain serial code portions that prohibit optimal scaling.

\Cref{fig:16_hodgkin_huxley_gpu} shows the analog study, where the 0D and 1D computations are offloaded to the GPU. The runtimes of these individual model parts are not explicitly measured, only the total runtime is known.

The plot shows an increasing total runtime for higher CPU parallelism. 
The runtimes for initialization, file output and the 3D solver are equal to the CPU-only study.
The increasing total runtime shows that the GPU is better at solving the complete 0D and 1D problems 
given by one CPU process than at the same computation, but split to several parts and provided by different MPI processes. 
The benefit of using multiple CPU processes to interface the GPU in this study is, thus, only that the VTK output functionality gets parallelized. However, this effect is negligible.

An absolute comparison between the runtimes in \cref{fig:16_hodgkin_huxley_gpu} and \cref{fig:16_hodgkin_huxley_cpu} also reveals that the scenarios for one to 18 processes using the GPU have 4.8 to 181 times longer total runtimes. In this study, the memory transfer between the CPU and the GPU has a low influence on the total runtime, as this transfer only happens before and after the 3D model is solved. The measured runtimes, therefore, correspond to the computation on the GPU.

\subsection{Evaluation of Hybrid CPU and GPU Computation for the Shorten Model}\label{sec:evaluation_hybrid}

Whereas previously, the subcellular model of Hodgkin and Huxley was solved on the GPU, we now switch to the more compute intense model of Shorten et al. \cite{Shorten2007}. This model has higher memory demands, such that it is not possible to solve it with OpenMP 4.5 for a muscle fiber mesh with 1481 nodes on the GeForce RTX 3080 GPU. As noted before, we use the NVIDIA Quadro GP100 GPU for the next study.
This GPU is also not capable of solving the whole set of 1481 models instances per fiber for 169 fibers at the same time. For one instance of the subcellular model, 57 state and rate variables each, and 71 intermediate variables have to be stored, along with other data, such as element lengths for every element.

Thus, we follow a hybrid approach. We parallelize the scenario to 27 processes on the CPU. Only one process offloads its subdomain to the GPU. In this way, both the CPU and the GPU take part in the computation and the available hardware capabilities are fully exploited.

\begin{figure}
  \centering%
  \includegraphics[width=0.9\textwidth]{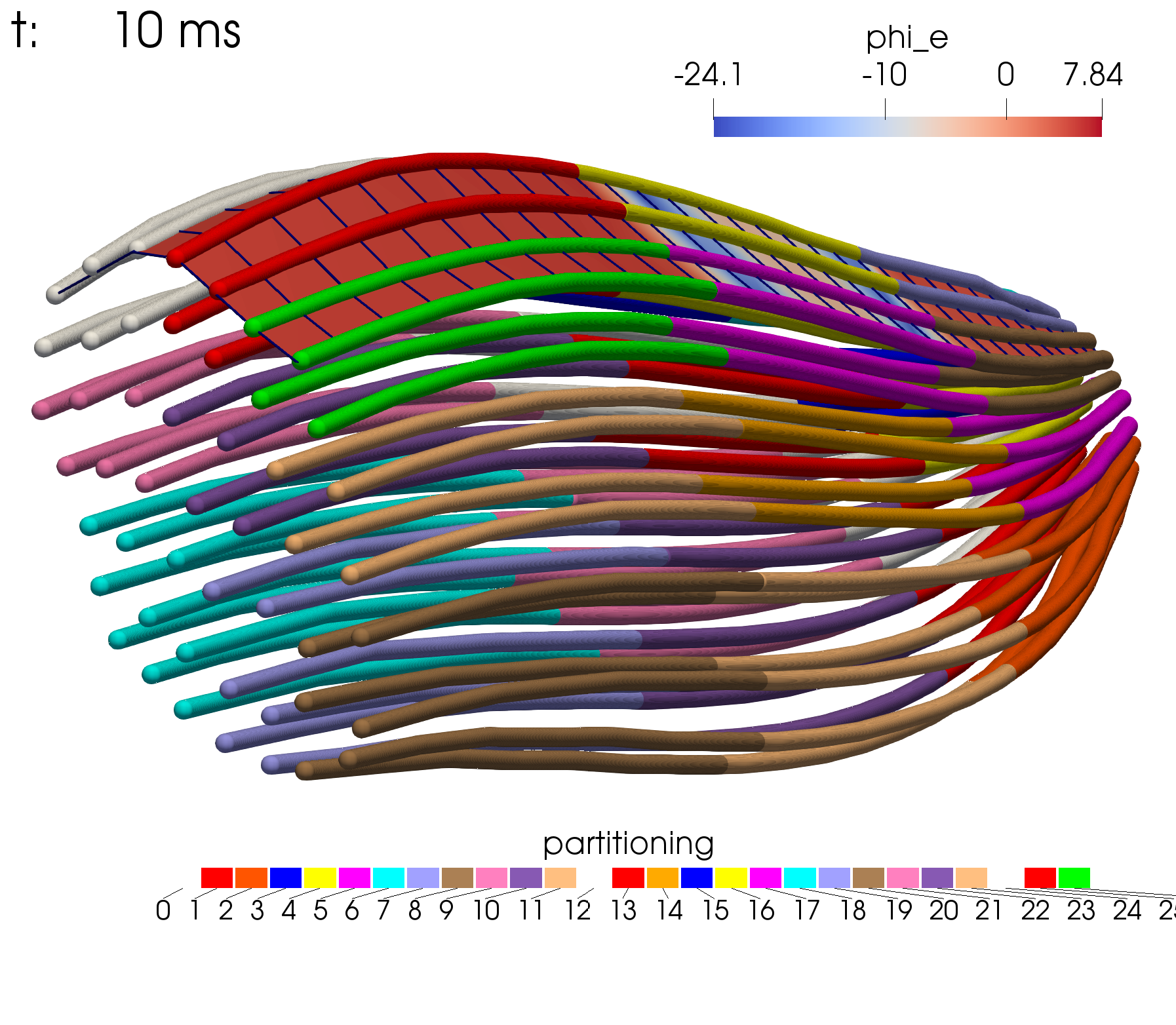}%
  \caption{Electrophysiology solver in OpenDiHu: Partitioning of the 169 fibers to 27 processes used in the runtime study with hybrid CPU-GPU usage. The image also shows the resulting EMG signals $\phi_e$ on the muscle surface.}%
  \label{fig:gpu_parallelization}%
\end{figure}

The scenario and the numerical parameters are the same as described for the study with the Shorten model in \cref{sec:evaluation_of_code_gen}. 49 muscle fibers are used and parallelized to $3\times 3\times 3=27$ subdomains. \Cref{fig:gpu_parallelization} visualizes the partitioning of the fibers by different colors and the EMG values $\phi_e$ on the muscle surface at the simulation end time of $t_\text{end}=\SI{10}{\ms}$.

\begin{figure}
  \centering%
  \includegraphics[width=0.6\textwidth]{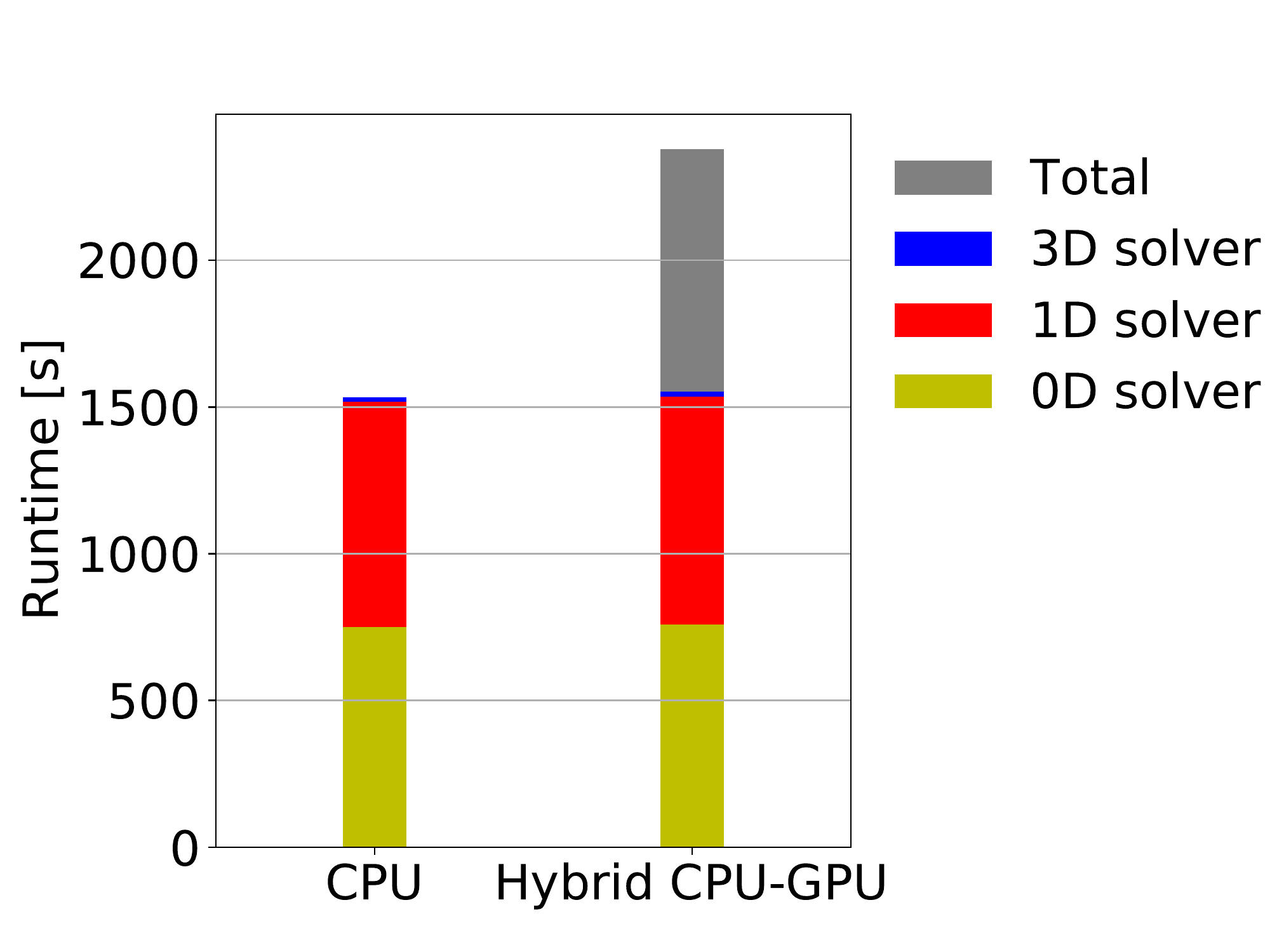}%
  \caption{Electrophysiology model in OpenDiHu: Runtime study for a CPU-only computation and a hybrid parallelization that employs both the CPU and the GPU. For the hybrid approach, one of the 27 partitions was computed on the GPU. The compute-intense subcellular model of Shorten et al. is simulated.}%
  \label{fig:17_shorten_gpu}%
\end{figure}

\Cref{fig:17_shorten_gpu} visualizes the runtimes of two runs. The first bar only employs the CPU and provides the reference for the measured runtimes. The second bar corresponds to the hybrid run, where one process employs the GPU. The 0D and 1D solver runtimes in the second bar are averaged over the CPU computations. The total runtime involves both the CPU and the GPU computations.
It can be seen that the total runtime given by the total bar heights is higher for the hybrid run. In the hybrid runs, the CPU processes have to wait at the synchronization point in the solution of the 3D problem until the GPU process has completed the 0D/1D computations.

\subsection{Conclusion}

Several scenarios with computations of the 0D subcellular and 1D electric conduction models on the GPU have been evaluated. 
\Cref{sec:evaluation_of_code_gen} compared the runtime for the Hodgkin-Huxley subcellular model on 625 fibers on the GPU with implementations on the CPU.
In \cref{sec:strong_scaling_runtimes_opencmiss_opendihu}, the computation on the GPU with the Shorten subcellular model was measured for one fiber.
\Cref{sec:strong_scaling_gpu_hodgkin_huxley} conducted a strong scaling study with the Hodgkin-Huxley model on 169 fibers and \cref{sec:evaluation_hybrid} evaluated a hybrid approach, where the Shorten model on different fibers was computed on the GPU and the CPU at the same time. This last study used a GPU with higher double-precision performance than the previous studies.

In all of these studies, the GPU computations could not compete with their CPU counterparts. The GPU implementation of the models relied on target-specific CUDA code, which was automatically generated by the OpenMP 4.5 pragmas produced by the code generator in OpenDiHu. The CPU computations used highly optimized CPU code with explicit vector instructions, an approach that is close to optimal as shown by the roofline model in \cref{sec:roofline_model}. Thus, the comparison considers different levels of optimization. We cannot conclude in general that the GPU is less suited to solve the fiber based electrophysiology models than the CPU. However, the GPU support in OpenMP is not competitive with our optimized CPU implementation.

The required GPU support of OpenMP in the GNU compiler is functional to the extent needed in our studies only since GCC version 11, which, at the time of writing, is still experimental and not yet released. Further performance gains can be expected in the future as compiler development advances. One problem is also the high memory requirement for the subcellular models, which only allows a certain number of subcellular model instances to be computed on the given hardware. It is also not clear, whether the high memory consumption is also an artifact of the compiler and will reduce with later compiler versions.

Despite the lower performance, it was shown that OpenDiHu can be used to solve the monodomain equation \cref{eq:monodomain} with different subcellular models on the GPU. Switching between the CPU and GPU variants can be accomplished by only changing the \code{optimizationType} parameter between \code{`vc`} and \code{`gpu`}. Hybrid strategies, where some processes use \code{`vc`} and others \code{`gpu`}, have been demonstrated.

In future work, the performance issue of the GPU computations can be addressed by using different technologies to access the compute power of GPUs.
Examples are to directly use the CUDA programming language or the C++ based abstraction layer for acceleration hardware SYCL \cite{sycl}. OpenDiHu already provides a reference implementation for such improvements by the code generator, which outputs model specific code with OpenMP pragmas for GPU offloading. This code implements proper, economical data transfer between the devices and contains hints how to distribute the workload on the GPU by the respective pragma placements. It could be used as a starting point to integrate further \say{optimization types} in the code generator.

%-----

% GPU slower in all studies
% comparison with highly optimized CPU code, near optimal
% GPU OpenMP not yet optimal, requires experimental GCC 11
% problem is also memory requiremnet
% GPU can be easily switched, hybrid is possible in opendihu
% future work: SYCL, CUDA extended code generator

\begin{reproduce_no_break}
  The scripts to run the studies and to create the plots for \cref{fig:16_hodgkin_huxley_cpu_gpu,fig:17_shorten_gpu} are available in the repository at \href{https://github.com/dihu-stuttgart/performance}{github.com/dihu-stuttgart/performance}
  in the directories \code{opendihu/16_hodgkin_huxley_gpu} and  \code{opendihu/17_shorten_gpu}.
\end{reproduce_no_break}

% ------------
%
% f===========

% ==============
%
% --------------
%-----
\section{Parallel Scaling of the EMG Model Using High Performance Computing}\label{sec:hpc_emg}
% in opendihu 2021 paper

After the parallel scaling of the multi-scale model has been investigated in moderately parallel scenarios with up to 27 processes in the previous sections, we now study the parallel scalability in High Performance Computing scenarios with larger degrees of parallelism.
We simulate the fiber based electrophysiology model consisting of a 0D subcellular model, the 1D electric conduction problem and the 3D bidomain equation, as described in \cref{sec:model_equations}.
We conduct these studies on the supercomputer Hawk at the High Performance Computing Center Stuttgart (HLRS). The system contains a total of 5632 compute nodes. Each compute node consists of two AMD EPYC 7742 processors with 64 cores each, a clock frequency of \SI{2.25}{\giga\hertz} and \SI{256}{\giga\byte} memory per node.

In the following, \cref{sec:weak_scaling_hawk} presents a weak scaling study, which scales the problem size up to a realistic number of \num{270000} muscle fibers in a biceps muscle.
Then, \cref{sec:mpi_rank_placement} shows measurements of the scaling behavior for the 1D model solver and gives details on MPI rank placement policies.

\subsection{Weak Scaling of the Fiber Based Electrophysiology Model}\label{sec:weak_scaling_hawk}

We simulate the fiber based electrophysiology model with EMG values on the muscle surface and the subcellular model of Hodgkin and Huxley \cite{Hodgkin1952}.
Corresponding simulation results of this scenario, also for the highly parallel runs, are presented in \cref{sec:effects_of_the_mesh_width_emg}.

In this weak scaling study, the number of fibers and number of processes is varied while their relation is kept approximately constant. The scenarios are constructed such that there are approximately 10 fibers per process, while maintaining a cube partitioning scheme in the 3D domain. The 0D subcellular and the 1D electric conduction problems on the fibers are solved with the \code{FastMonodomainSolver} class and the \code{`vc`} optimization type, which is the fastest available option in OpenDiHu. The Strang operator splitting scheme with Heun's method and the implicit Euler scheme are used. The 3D problem is solved using the conjugate gradient solver of PETSc and a relative tolerance on the residual norm of \num{1e-5}. 

The used timestep widths are $\dt_\text{0D}=\SI{1e-3}{\milli\second}$, $\dt_\text{1D}=\dt_\text{splitting}=\SI{2e-3}{\milli\second}$ and $\dt_\text{3D}=\SI{1}{\milli\second}$. Because only the scaling behavior is of interest in this study, the simulation end time is set to $t_\text{end}=\SI{2}{\milli\second}$.

% Hawk weak scaling
\begin{figure}[H]
  \centering%
  \includegraphics[width=\textwidth]{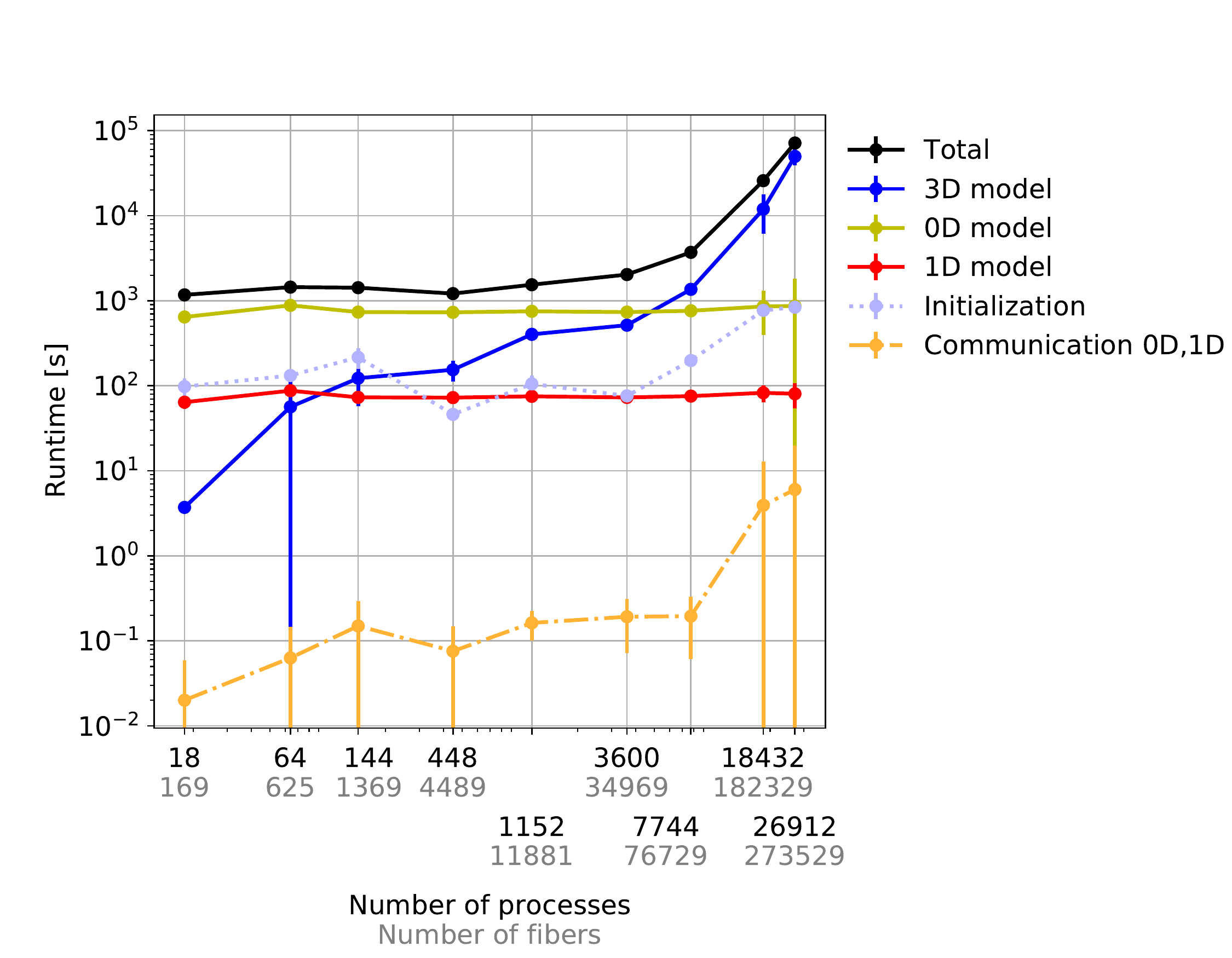}%
  \caption{Weak scaling of the fiber based electrophysiology model on the supercomputer Hawk simulating up to more than \num{270000} muscle fibers.}%
  \label{fig:hawk_weak_scaling}%
\end{figure}

\Cref{fig:hawk_weak_scaling} presents the resulting runtimes for the different parts of the simulation program: the solvers of the 0D, 1D and 3D models, the runtime for initialization and the runtime for the communication in the \code{FastMonodomainSolver}, as explained in \cref{sec:improved_parallel_solver_for_fiber_based}.
To relate the initialization runtime to the runtime of the solvers in a realistic scenario with longer simulation times, all runtimes except for the initialization are scaled to a simulation end time of $\SI{1}{\second}$.

The results show perfect weak scaling properties of the 0D and the 1D solver, given by the yellow and red lines. This is expected due to the construction of the algorithm and the parallel partitioning. The 0D problems are \say{embarrassingly parallel} and are solved independently of each other. In the 1D problem solver, the values are  transferred to a dedicated process, where the serial Thomas algorithm is employed for each fiber as a whole. Thus, the solution of all 1D problems is also performed independently of each other, but an additional communication step is required, before and after running the solver in each implicit time step of the 1D problem.
The plot shows a very small runtime for this communication even for higher parallelism, which is given by the orange dashed line.

The initialization of the computation involves parsing the Python script, which for the last data point requires \SI{35.1}{\s}, parallel file access and read operations of the mesh file, assembly of the 3D stiffness matrix and solution of the potential flow problem to obtain the fiber directions in the 3D mesh, which contains approximately \num{1e8} dofs for the last data point, code generation, compilation, linking and loading of the shared library for the subcellular problem, and initialization of all internal data structures.

Loading the mesh input file from the file system is the part of the initialization, which requires the most runtime.
The dotted light blue line in \cref{fig:hawk_weak_scaling} shows that the initialization time increases to a maximum value of \SI{839}{\s} for the largest problem size.

The runtime of the 3D model is shown by the blue line in \cref{fig:hawk_weak_scaling}. This part of the model is responsible for the highest portion of the total runtime starting from the scenario with \num{7744} fibers and \num{76729} processes. This increase is two-fold: first, the communication cost increases for a larger number of processes. Second, the number of iterations in the conjugate gradient solver increases for a larger number of unknowns.

\begin{table}
  \centering%
  \begin{tabular}{|r|r|r|r|r|r|r|r|r|}
    \hline
    \# processes  & 18 & 64  & 144 & 448  & 1152 & 3600 & 7744 & \num{26912}\\\hline
    \# iterations & 72 & 115 & 176 & 339  & 561  & 1056 & 1636 & 2807\\
    \hline
  \end{tabular}
  \caption{Scaling study of the fiber based muscle model: Number of iterations of the conjugate gradient solver for the 3D bidomain model in the weak scaling study presented in \cref{fig:hawk_weak_scaling}.}%
  \label{tab:cg_solver_iterations}%
\end{table}

In this weak scaling study, the number of conjugate gradient solver iterations increases from 72 for the first data point to \num{2807} for the last data point, as listed in \cref{tab:cg_solver_iterations}. The 3D problem of the last data point has \num{1e8} dofs. The exact numbers of dofs are also listed in \cref{tab:emg_study_parameters} in \cref{sec:effects_of_the_mesh_width_emg}.
Currently, the solution of the 3D problem uses no preconditioner. In future work, a multigrid solver could be employed for preconditioning, which could improve the weak scaling for large problem sizes.

In summary, the solution of the multi-scale model for fiber based electrophysiology without fat layer exhibits a very good weak scalability for up to \num{35000} fibers. For larger problem sizes, the solution of the 3D problem dominates and the weak scaling behavior deteriorates. However, the solution times are still feasible, as such large problems have been successfully solved in \cref{sec:effects_of_the_mesh_width_emg} of this work.

\subsection{Weak Scaling of the 1D Solver and MPI Rank Placement}\label{sec:mpi_rank_placement}

\begin{figure}
  \centering%
  \includegraphics[width=\textwidth]{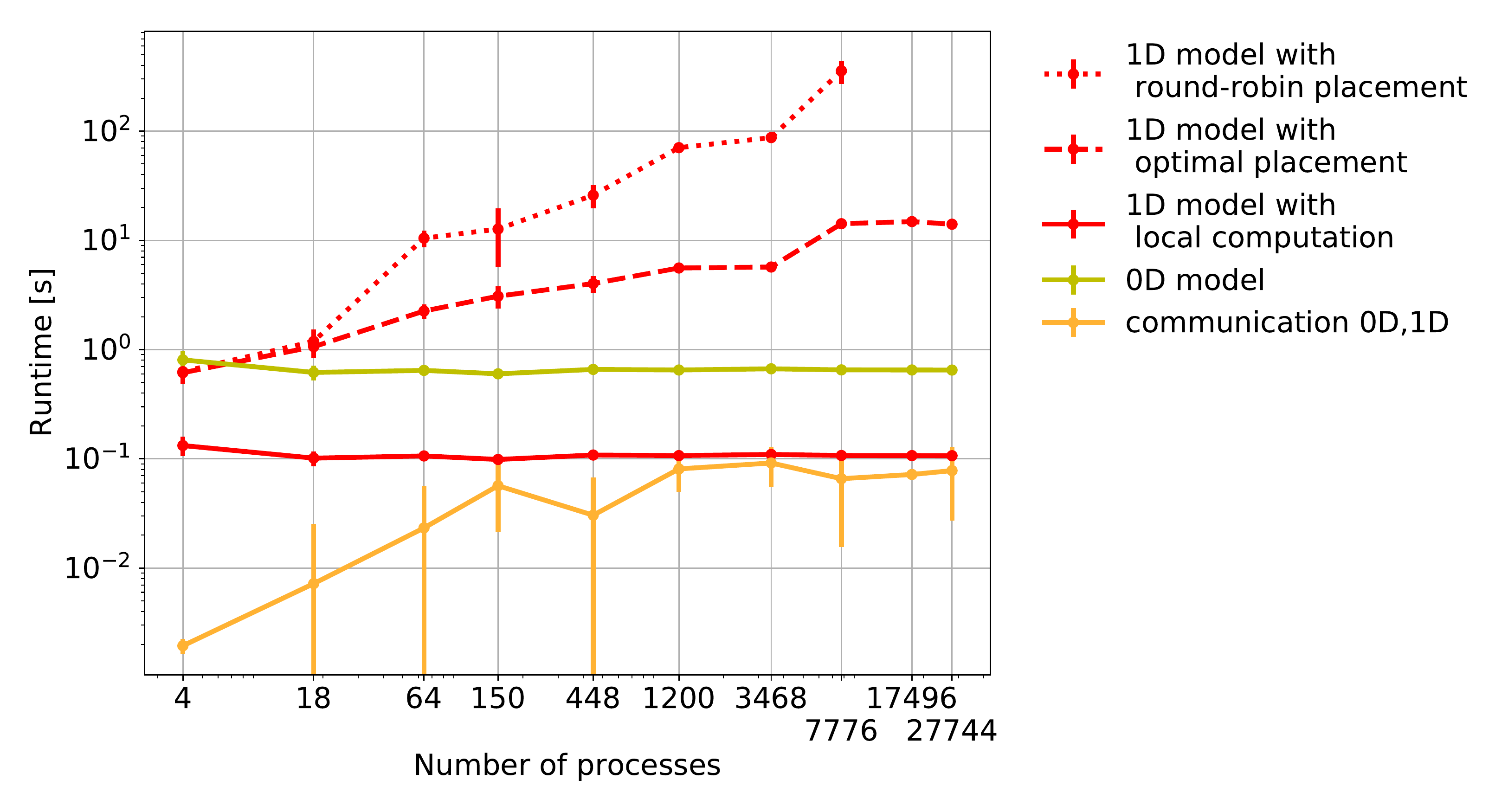}%
  \caption{Scaling study of the fiber based muscle model: Weak scaling behavior of solvers for the 1D problem. Runtimes for the 1D solver with different rank placement strategy (dotted and dashed red lines) and the optimized runtimes of the \code{FastMonodomainSolver}, which combines the solution of the 0D and 1D problem (yellow and solid red lines, respectively).}% 
   \label{fig:hazel_hen_rank_placement}%
\end{figure}

% performance/opendihu/08_0D1D_better_implementation
Next, we compare the different approaches to solve the 1D electric conduction part of the monodomain equation on the fibers in a High Performance Computing setting.
\Cref{fig:hazel_hen_rank_placement} shows a similar weak scaling study to the one in the last section with slightly different process counts. The study was carried out on the supercomputer Hazel Hen, a Cray XC30 system, which was installed until 2020 as the predecessor to Hawk at the High Performance Computing Center Stuttgart. The same problem as in the last section is solved, and, again, the relation between fibers and processes is 10:1. The partitionings were chosen in accordance with the compute nodes of Hazel Hen, which had 24 cores. The detailed problem sizes and partitionings can be found in \cite{Maier2019}.

\Cref{fig:hazel_hen_rank_placement} shows the runtimes of the same 0D and 1D solvers as in the last section by the solid yellow line for the 0D problem, the solid red line below for the 1D problem and the solid orange line for the communication. The perfect weak scaling for these parts is in line with the previous observations.

In addition, we compare the weak scaling of the 1D solution, if a parallel conjugate gradient solver of PETSc is used for the problem on every fiber, instead of the serial Thomas algorithm. This setup corresponds to the \code{vc-aovs} scenario presented in \cref{sec:evaluation_of_code_gen}. As already noted earlier, the performance of this approach is worse. The dashed and dotted lines in \cref{fig:hazel_hen_rank_placement} present the runtimes of this approach for two different MPI rank placement strategies, but for the identical program. It can be seen that the runtimes increase with higher numbers of processes in both curves. This effect is the result of the 1D fiber problems being distributed to more processes, as the total number of processes increases. 

For example, in the scenarios with 1200 and 3468 processes, all fibers are distributed to 12 different processes. For the last three data points of the dashed curve, all fibers are distributed to 24 processes. As a result, the runtime to solve the 1D problems in the measurements with 1200 and 3468 processes is approximately equal, but lower than the runtime for the last three data points, where twice the amount of processors takes part in the solution of a single 1D problem.

The difference between the dotted and the dashed red curves is a different strategy to place the processes on the compute nodes. The dashed curve with the lower runtime corresponds to a placement of all fiber sharing processes on the same compute node. As the subdomain indices in the $n_x \times n_y \times n_z$ partitioning increase fastest in $x$-direction, then in $y$-direction and then in $z$-direction, and the fibers are aligned with the $z$ direction, the set of processes on a compute node contains MPI ranks that are offset with a constant stride of $n_x\,n_y$. This has to be ensured in the job scripts on the supercomputers by explicit MPI rank pinning.

If no such measures are taken, the default placement of MPI ranks on the compute nodes proceeds consecutively by filling the compute nodes in the order of the MPI ranks. This corresponds to a round-robin placement of the fibers on the compute nodes, which is the worst possible way of distributing MPI ranks to compute nodes. All processes that compute a fiber are potentially located on different nodes and have to communicate over compute node boundaries.
\Cref{fig:hazel_hen_rank_placement} shows the resulting runtimes by the upper, dotted red curve. The difference in runtime increases to one order of magnitude for the highest number of cores.

Thus, it is important to properly handle MPI rank placement on compute clusters with multiple compute nodes. As a consequence, we also configured rank placement on the supercomputer Hawk accordingly in all our studies on this system.

% ------------
%
% f===========

\section{Performance Studies of the Solid Mechanics Solver}\label{sec:performance_solid_mechanics}

%(BM TODO read)

Next, we address the performance of the solid mechanics solver. 
Its runtime in OpenDiHu is given by the call to the nonlinear solver of PETSc. PETSc, in turn, calls two functions in OpenDiHu, which evaluate the nonlinear function to be solved and the Jacobian of the nonlinear function for a given vector of unknowns.
In the following, \cref{sec:analytic_numeric_jacobian} compares an analytic and a numerical computation scheme for the Jacobian, and \cref{sec:vectorization_analytic_jacobian} studies the impact of vectorization in this computation.

\subsection{Analytic and Numerical Computations of the Jacobian}\label{sec:analytic_numeric_jacobian}

The computation of the Jacobian can be done in two ways. The first possibility is done by PETSc, which uses finite differences to numerically estimate the value of the Jacobian. The second possibility is to evaluate the respective analytic formulation within OpenDiHu. This analytic formulation is derived in \cref{sec:discretization_mechanics} and uses the SEMT library \cite{semt} to differentiate the mechanics model given by a strain energy function at compile time.

\begin{figure}
  \centering%
  \def\svgwidth{0.6\textwidth}
  %% Creator: Inkscape inkscape 0.92.3, www.inkscape.org
%% PDF/EPS/PS + LaTeX output extension by Johan Engelen, 2010
%% Accompanies image file 'mechanic_scenario.pdf' (pdf, eps, ps)
%%
%% To include the image in your LaTeX document, write
%%   \input{<filename>.pdf_tex}
%%  instead of
%%   \includegraphics{<filename>.pdf}
%% To scale the image, write
%%   \def\svgwidth{<desired width>}
%%   \input{<filename>.pdf_tex}
%%  instead of
%%   \includegraphics[width=<desired width>]{<filename>.pdf}
%%
%% Images with a different path to the parent latex file can
%% be accessed with the `import' package (which may need to be
%% installed) using
%%   \usepackage{import}
%% in the preamble, and then including the image with
%%   \import{<path to file>}{<filename>.pdf_tex}
%% Alternatively, one can specify
%%   \graphicspath{{<path to file>/}}
%% 
%% For more information, please see info/svg-inkscape on CTAN:
%%   http://tug.ctan.org/tex-archive/info/svg-inkscape
%%
\begingroup%
  \makeatletter%
  \providecommand\color[2][]{%
    \errmessage{(Inkscape) Color is used for the text in Inkscape, but the package 'color.sty' is not loaded}%
    \renewcommand\color[2][]{}%
  }%
  \providecommand\transparent[1]{%
    \errmessage{(Inkscape) Transparency is used (non-zero) for the text in Inkscape, but the package 'transparent.sty' is not loaded}%
    \renewcommand\transparent[1]{}%
  }%
  \providecommand\rotatebox[2]{#2}%
  \newcommand*\fsize{\dimexpr\f@size pt\relax}%
  \newcommand*\lineheight[1]{\fontsize{\fsize}{#1\fsize}\selectfont}%
  \ifx\svgwidth\undefined%
    \setlength{\unitlength}{1421.6692915bp}%
    \ifx\svgscale\undefined%
      \relax%
    \else%
      \setlength{\unitlength}{\unitlength * \real{\svgscale}}%
    \fi%
  \else%
    \setlength{\unitlength}{\svgwidth}%
  \fi%
  \global\let\svgwidth\undefined%
  \global\let\svgscale\undefined%
  \makeatother%
  \begin{picture}(1,0.75240378)%
    \lineheight{1}%
    \setlength\tabcolsep{0pt}%
    \put(0,0){\includegraphics[width=\unitlength,page=1]{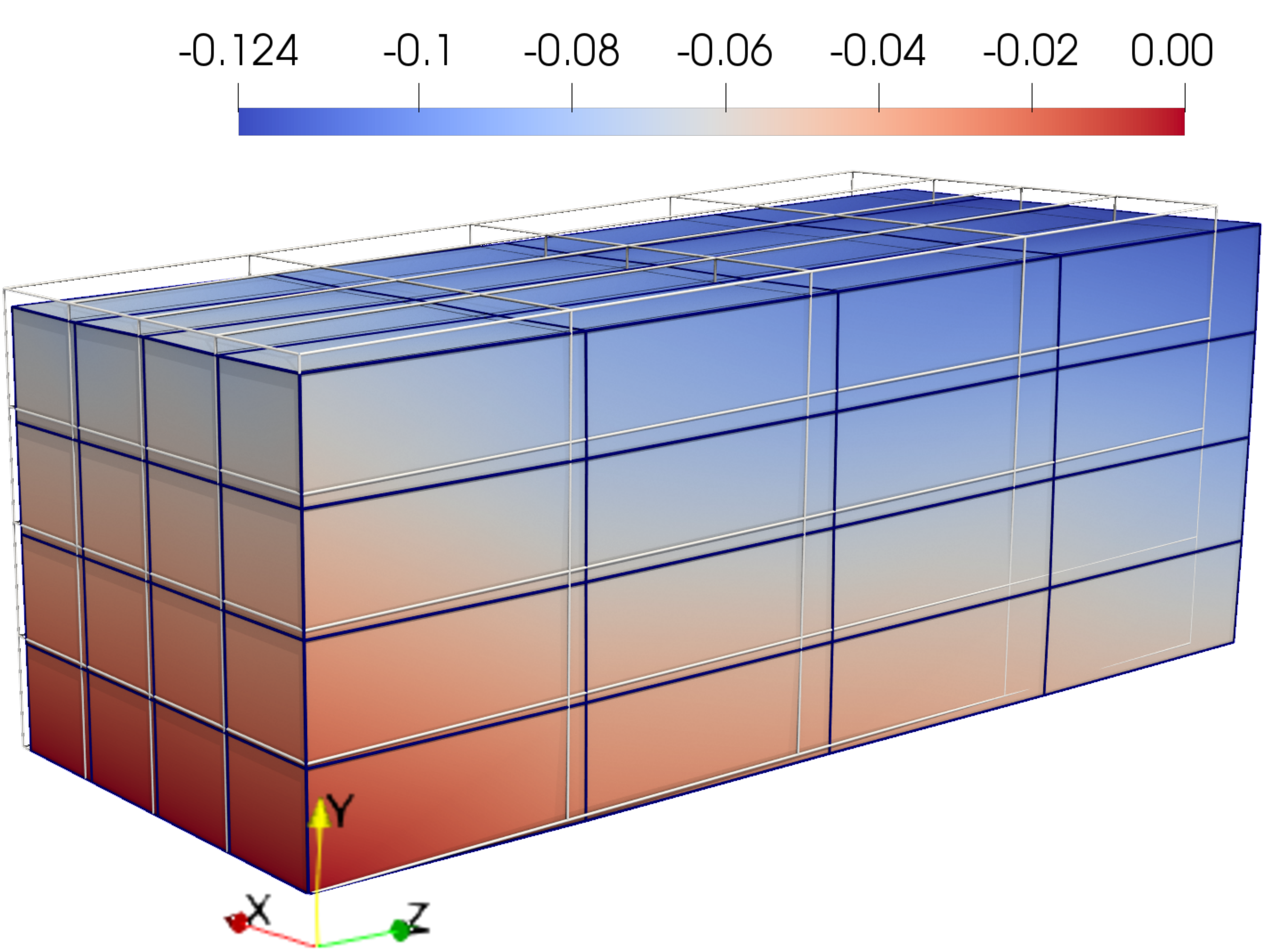}}%
    \put(0.04925542,0.66104813){\color[rgb]{0,0,0}\makebox(0,0)[lt]{\lineheight{1.25}\smash{\begin{tabular}[t]{l}$u_y$\end{tabular}}}}%
  \end{picture}%
\endgroup%
  \caption{Scaling study for solid mechanics: Visualization of the solid mechanics problem used in the weak scaling studies. The box is extended by a surface load at the right end. The wireframe shows the reference configuration, the colored mesh shows the current configuration where the color corresponds to the displacements $u_y$ in vertical direction.}%
  \label{fig:mechanic_scenario}%
\end{figure}

To compare both approaches, we simulate a tension test, where a cuboid domain is extended by an applied force. 
The box is shown in \cref{fig:mechanic_scenario} and has physical dimensions $\SI{2}{\cm} \times \SI{2}{\cm} \times \SI{5}{\cm}$. The left face of the box is fixed at $z=0$ and the two edges of the left face at $x=0$ and $y=0$ are also fixed to prevent rotation of the body. On the right face, a constant surface load with a total force of $10$ is applied.
The material model is the incompressible transversely isotropic Mooney-Rivlin description given in \cref{eq:transiso_mooney_rivlin}, which is also used for the muscle tissue. However, no active stress is considered. The material parameters are chosen as $c_1=2, c_2=3, b = 4$ and $d = 5$. The fiber direction lies in the $y-z$ plane and has an angle of $\SI{40}{\degree}$ to the $z$ axis.
As a result, the box slightly bends in negative $y$ direction, as can be seen in \cref{fig:mechanic_scenario}. In this figure, the volume of the deformed object is colored according to the displacement in $y$ direction.

We solve the problem using the Newton solver of PETSc with a secant line search over the $L_2$ norm of the function. The absolute and relative residual tolerances are set to \SI{1e-5}, which leads to approximately 5 Newton iterations. The parallel direct solver of the MUMPS package \cite{Mumps2017} is used to solve the linear system in every Newton iteration.

We compare the runtimes of numerically and analytically computing the Jacobian in this problem. We vary the number of processes and the problem size in a weak scaling setting, such that there are exactly 8 3D elements per process. A dual-socket AMD EPYC 7742 64-core processor with clock speed of \SI{2.25}{\giga\hertz}, a total core count of 128 and \SI{1.96}{\tera\byte} RAM is used. We vary the number of processes between one and 256 processes. For more than 128 processes, hyperthreading is used.

\begin{figure}
  \centering%
  \includegraphics[width=0.7\textwidth]{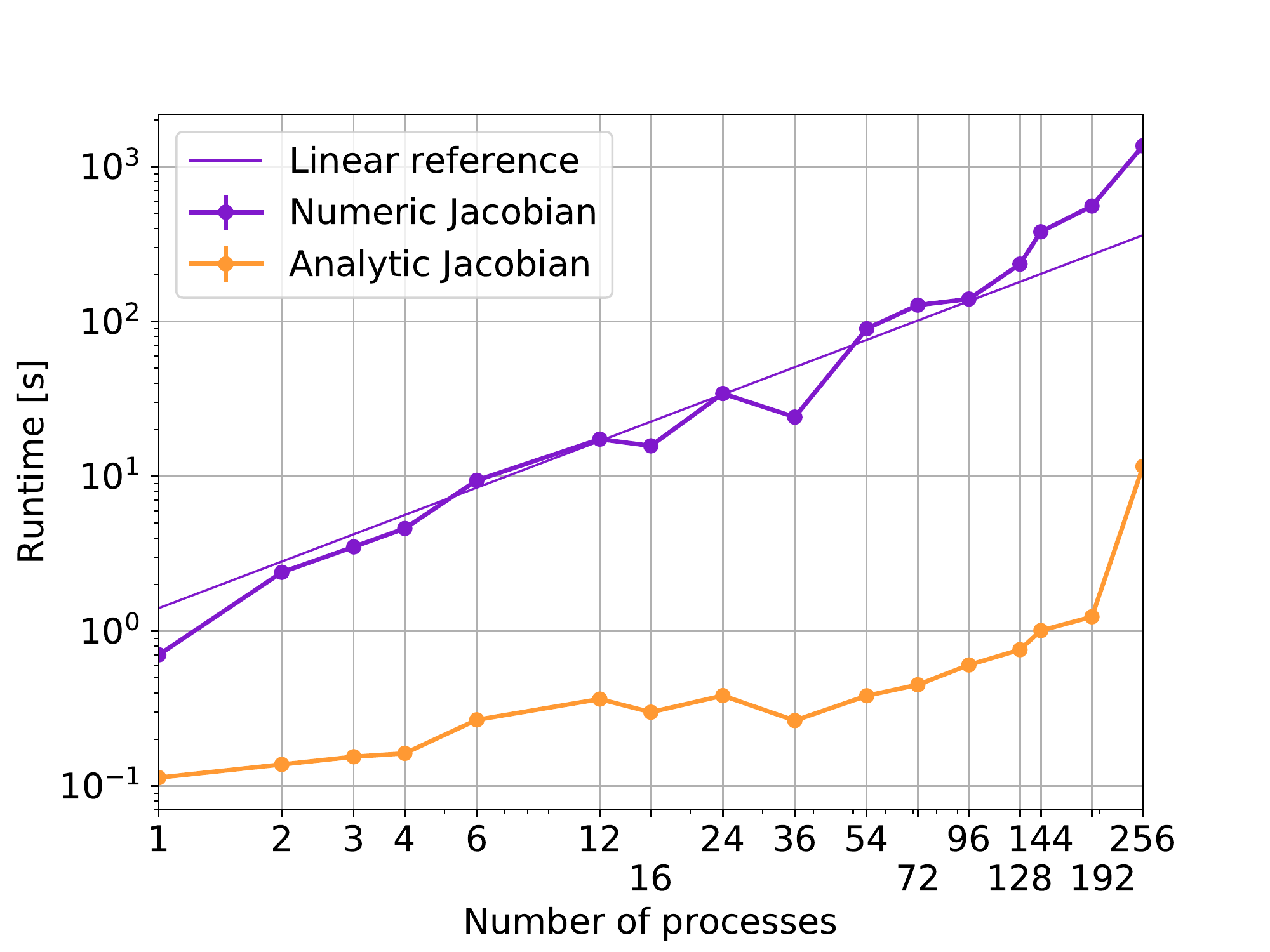}%
  \caption{Weak scaling of the nonlinear mechanics solver. The runtime to solve the nonlinear system equation is given for the numerical and for the analytic computation approaches for the Jacobian matrix.}%
  \label{fig:numeric_analytic}%
\end{figure}

\Cref{fig:numeric_analytic} shows the runtimes to solve the nonlinear system by the orange and purple lines in a double logarithmic plot. An increasing runtime for higher process counts is observed for both computation approaches. The curves have a higher slope for more than 128 processes, when the computation uses more processes than physical cores in the processor.

The computation of the Jacobian requires $\O(n)$ non-zero entries to be calculated as the matrix has a banded sparsity structure, where $n$ is the number of degrees of freedom or elements in the 3D mesh. The number of non-zero Jacobian entries is, thus, constant in the weak scaling setup.
However, \cref{fig:numeric_analytic} shows a linearly increasing runtime for the numerical computation of the Jacobian, which can be seen by the comparison with the plotted linear function. 
The finite difference scheme of PETSc has no information about the sparsity pattern and estimates all $\O(n^2)$ matrix entries.

The runtimes for the analytic computation are only slightly increasing in the weak scaling, which is as expected. This approach only computes the non-zero entries of the Jacobian. Moreover, the values of the material elasticity tensor $\C$ can be computed once and reused for every entry of the Jacobian.

The comparison between the approaches shows lower computation times for the analytic approach by factors between \num{6.2} for one process and \num{450} for 192 processes. 
In conclusion, the analytic formula for the Jacobian, which is implemented in OpenDiHu, allows us to speed up the mechanics computations and to compute larger problem sizes in feasible runtimes.
%[2, 3, 4, 5],  # c1, c2, b1, d1

%performance/opendihu/22_solid_mechanics_vectorization

\subsection{Vectorization of the Analytic Jacobian Computations}\label{sec:vectorization_analytic_jacobian}

In the assembly of finite element system matrices of any kind, contributions are computed on an element level and then combined to form a global system matrix. In OpenDiHu, this algorithm can be vectorized by performing analog but independent computations for multiple elements concurrently in multiple SIMD lanes. This vectorization uses the \emph{Vc} library, similar to the optimizations of the subcellular model solver presented in \cref{sec:performance_studies_of_the_e}. As this vectorization significantly increases compilation times, the feature is turned off by default and has to be enabled before compilation.

In this section, we evaluate the effect of vectorized system matrix computations for the Jacobian matrix of the nonlinear solid mechanics solver. We conduct a similar weak scaling study as before using the same scenario, but an eight times larger number of elements for each measurement. We compare the runtime to solve the nonlinear problem using analytic Jacobian computations with disabled and enabled vectorization.

\begin{figure}
  \centering%
  \includegraphics[width=0.8\textwidth]{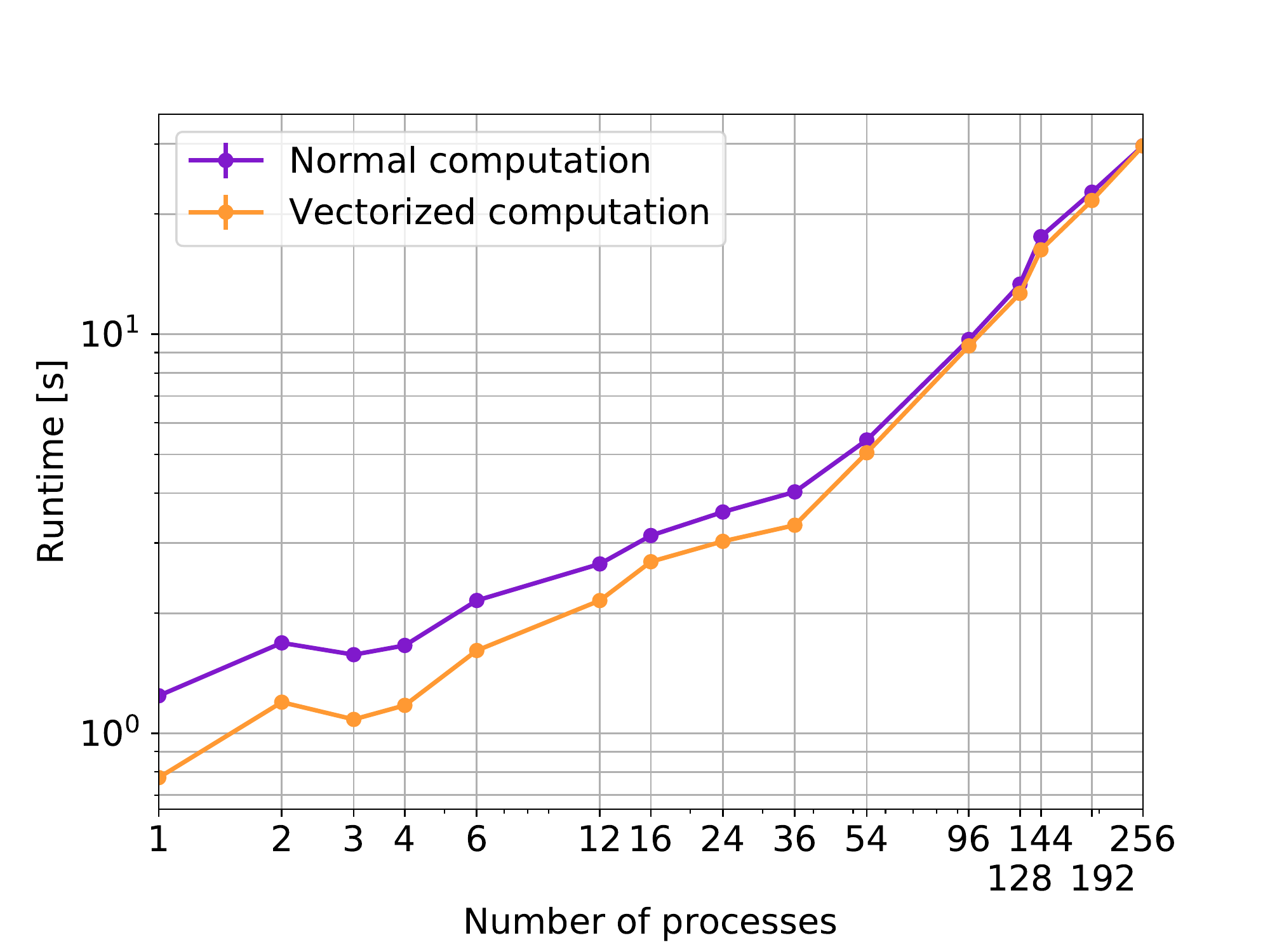}%
  \caption{Scaling Study for the solid mechanics solver: Weak scaling study to evaluate vectorization in the computation of the Jacobian.}%
  \label{fig:vectorized_jacobian}%
\end{figure}

\Cref{fig:vectorized_jacobian} presents the resulting runtimes for the entire nonlinear solver. 
The vectorized computation shows speedups from \num{1.6} for one process to \num{1.06} for 128 processes. The theoretically possible speedup is four, as the processor supports the AVX2 instruction set with a vector register length of four double-precision values. The measured speedups are lower, because the computation of the Jacobian is only one portion of the computations in the nonlinear solver. 
%For some operations during the computation of the Jacobian, the individual values have to be unpacked from the vector registers, e.g., for the check if the jacobian

In summary, the vectorized computation of Jacobian matrices can reduce runtimes for the nonlinear solver. A runtime reduction of \SI{38}{\percent} (corresponding to the speedup of \num{1.6}) was observed for the serial scenario. The performance gain is largest for small problem sizes and small numbers of processes, and gets less prominent for larger degrees of parallelism.

\begin{reproduce_no_break}
  The scripts to run the studies in this scenario and to create the plots of \cref{fig:numeric_analytic,fig:vectorized_jacobian} are available in the repository at \href{https://github.com/dihu-stuttgart/performance}{github.com/dihu-stuttgart/performance}
  in the directory \code{opendihu/22_solid_mechanics_vectorization}.
\end{reproduce_no_break}
% --------------------t
%
% ======================

% numerical investigations
\section{Numerical Studies}\label{sec:numerical_studies}

Next, we perform numerical studies to evaluate mesh resolutions and linear solvers. 
\Cref{sec:action_potential_velocity} addresses the mesh width of the 1D problem. \Cref{sec:multidomain_solvers} evaluates different numerical solver choices for the multidomain model.
\subsection{Effect of the Mesh Width on the Action Potential Propagation Velocity}\label{sec:action_potential_velocity}

The numerical parameters of a simulation scenario such as mesh widths and timestep widths should be chosen, such that the resulting numerical errors are balanced between all model components. In the simulation of activated muscle fibers, the propagation velocity of the action potentials is an important quantity, which also influences the macroscopic outcome of the EMG recordings. Thus, we investigate the effect of numerical parameters on the propagation velocity.

%performance/opendihu/02_propagation_velocity
\begin{figure}%
  \centering%
  \begin{subfigure}[t]{0.45\textwidth}%
    \centering%
    \includegraphics[width=\textwidth]{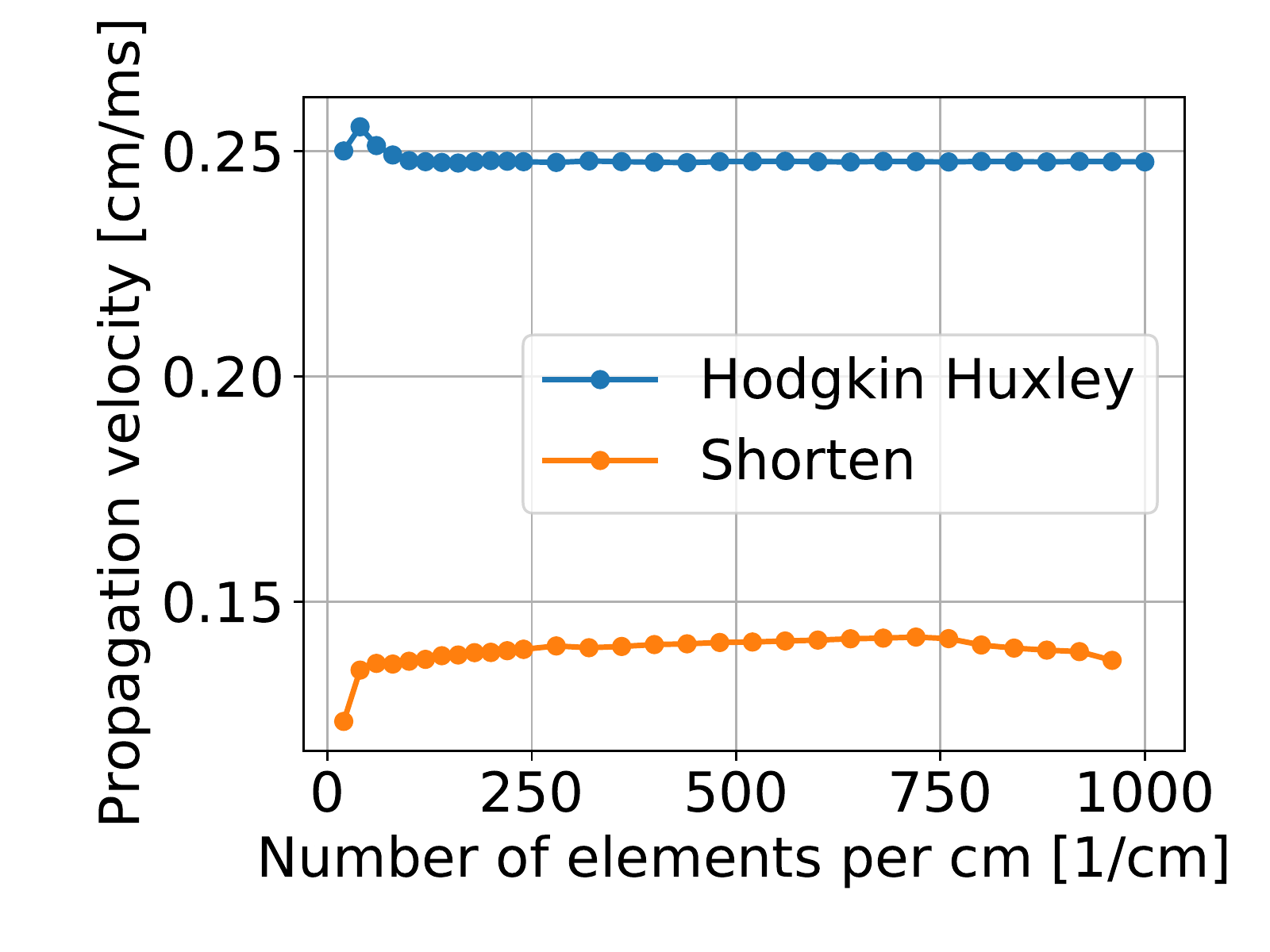}%
    \caption{Propagation velocities over spatial resolution of the 1D mesh.}%
    \label{fig:propagation_velocity_comparison}%
  \end{subfigure}
  \,
  \begin{subfigure}[t]{0.45\textwidth}%
    \centering%
    \includegraphics[width=\textwidth]{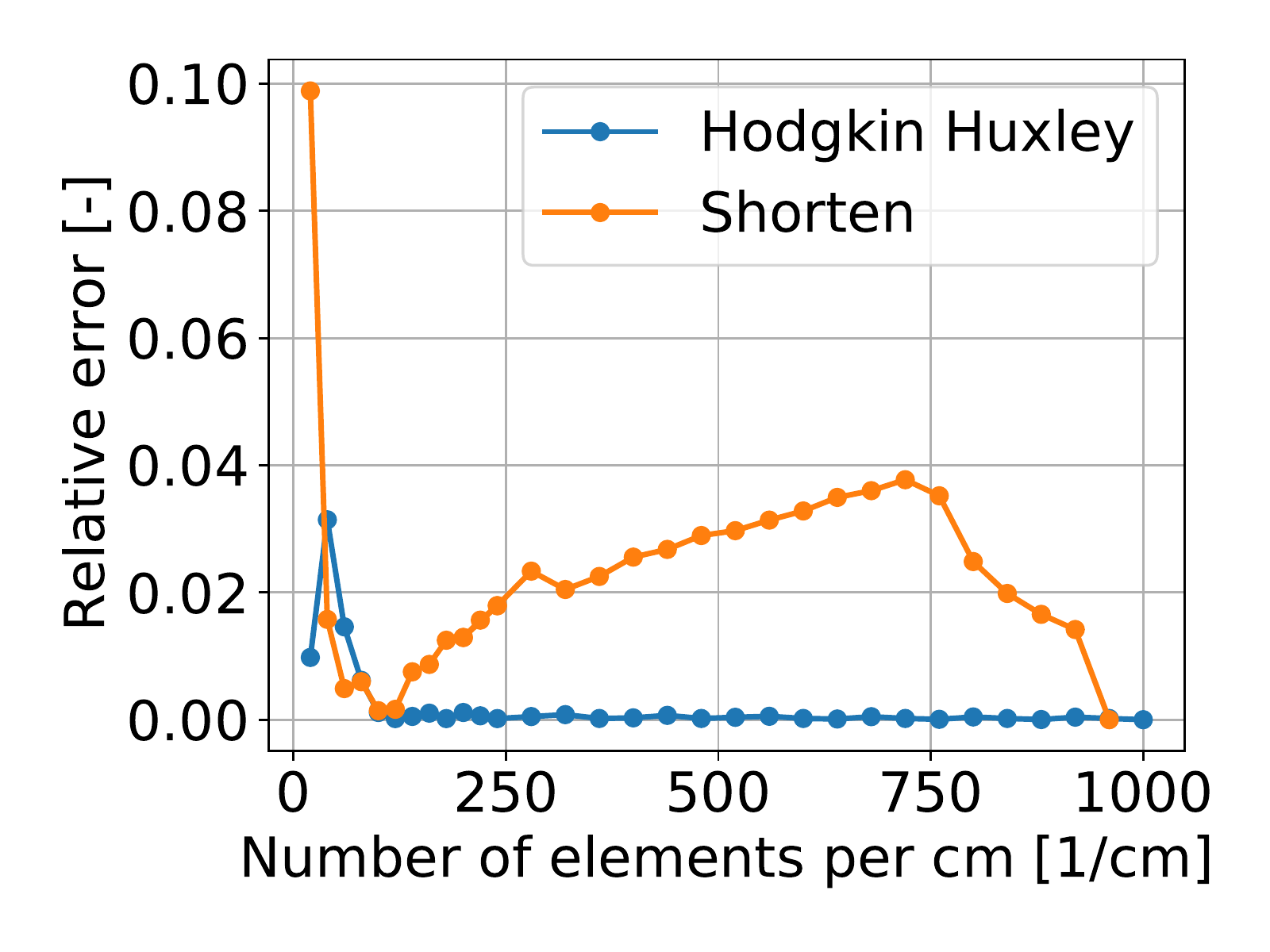}%
    \caption{Relative error of the propagation velocities over spatial resolution of the 1D mesh.}%
    \label{fig:propagation_velocity_rel_error}%
  \end{subfigure}   
  \caption{Influence of the mesh width on the propagation velocity of the action potential for the subcellular models of Shorten and Hodgkin-Huxley in the fiber-based electrophysiology simulation. This study is used to determine the 1D mesh width.}%
  \label{fig:propagation_velocity}%
\end{figure}%

In the first study, we consider a single fiber given by a 1D mesh, where the monodomain equation \cref{eq:monodomain} is used.
We measure the error of the propagation velocity depending on the mesh width. A fiber with a physical length of \SI{4}{\cm} is used and discretized by different numbers of 1D elements. We stimulate the fiber at its center at the beginning of the simulation and run the simulation until an end time of \SI{28}{\ms}. An action potential propagates along the fiber. We determine the location of the propagating peak at the end and compute the propagation velocity.

\Cref{fig:propagation_velocity_comparison} shows the resulting values of the propagation velocity for the Hodgkin-Huxley and Shorten subcellular models for varying mesh resolutions. It can be seen that the velocities level out at a constant value for finer mesh discretizations. The absolute value of the propagation velocity is different for the two subcellular models and depends on various model parameters.

For a quantitative evaluation, we examine the error of the propagation velocity, which is estimated by comparing each run with the value from the finest simulation. 
One issue with comparing propagation velocities on discretized meshes is that the measured velocity can only be calculated as number of elements traversed per time interval. Thus, the measurements for different mesh resolutions have different accuracies, not only due to the usual discretization error in the finite element approach. This issue can be reduced if a long enough time is simulated, such that the action potential propagates a large enough distance in terms of multiples of the element width.

\Cref{fig:propagation_velocity_rel_error} shows the resulting relative error of the propagation velocity. It can be seen that, for the Hodgkin Huxley model, the error gets close to zero for 100 elements per centimeter of fiber length. For the Shorten model, this mesh resolution also exhibits a low relative error. However, the error increases again for higher mesh resolutions before decreasing again starting at approximately 750 elements per centimeter. 

As a result, we use a 1D mesh resolution of 100 elements per centimeter for all muscle fibers in all our simulations. For a biceps muscle of \SI{14.8}{\cm} length, this leads to muscle fiber discretizations with 1480 elements and 1481 nodes.

%performance/opendihu/03_dx_dt_dependence
%\begin{figure}
%  \centering%
%  \includegraphics[width=0.8\textwidth]{images/results/studies/hh_cn_error_propagation_velocity_3d.pdf}%
%  \caption{Error of the action potential propagation velocity for varying mesh width and timestep width.}% 
%   \label{fig:hh_cn_error_propagation_velocity_3d}%
%\end{figure}
%Remove \cref{fig:hh_cn_error_propagation_velocity_3d} as not so clear.

\begin{reproduce_no_break}
  The scripts for this study are available in the repository at \href{https://github.com/dihu-stuttgart/performance}{github.com/dihu-stuttgart/} \href{https://github.com/dihu-stuttgart/performance}{performance} in the directory \code{opendihu/02_propagation_velocity}:
  \begin{lstlisting}[columns=fullflexible,breaklines=true,postbreak=\mbox{\textcolor{gray}{$\hookrightarrow$}\space}]
    ./run.sh
    ./plot_propagation_velocity.py
  \end{lstlisting}
\end{reproduce_no_break}

%-----
\subsection{Linear Solvers for the Multidomain Problem}\label{sec:multidomain_solvers}

% in opendihu 2021 paper
% selected multidomain solvers
\begin{figure}[H]
  \centering%
  \includegraphics[width=\textwidth]{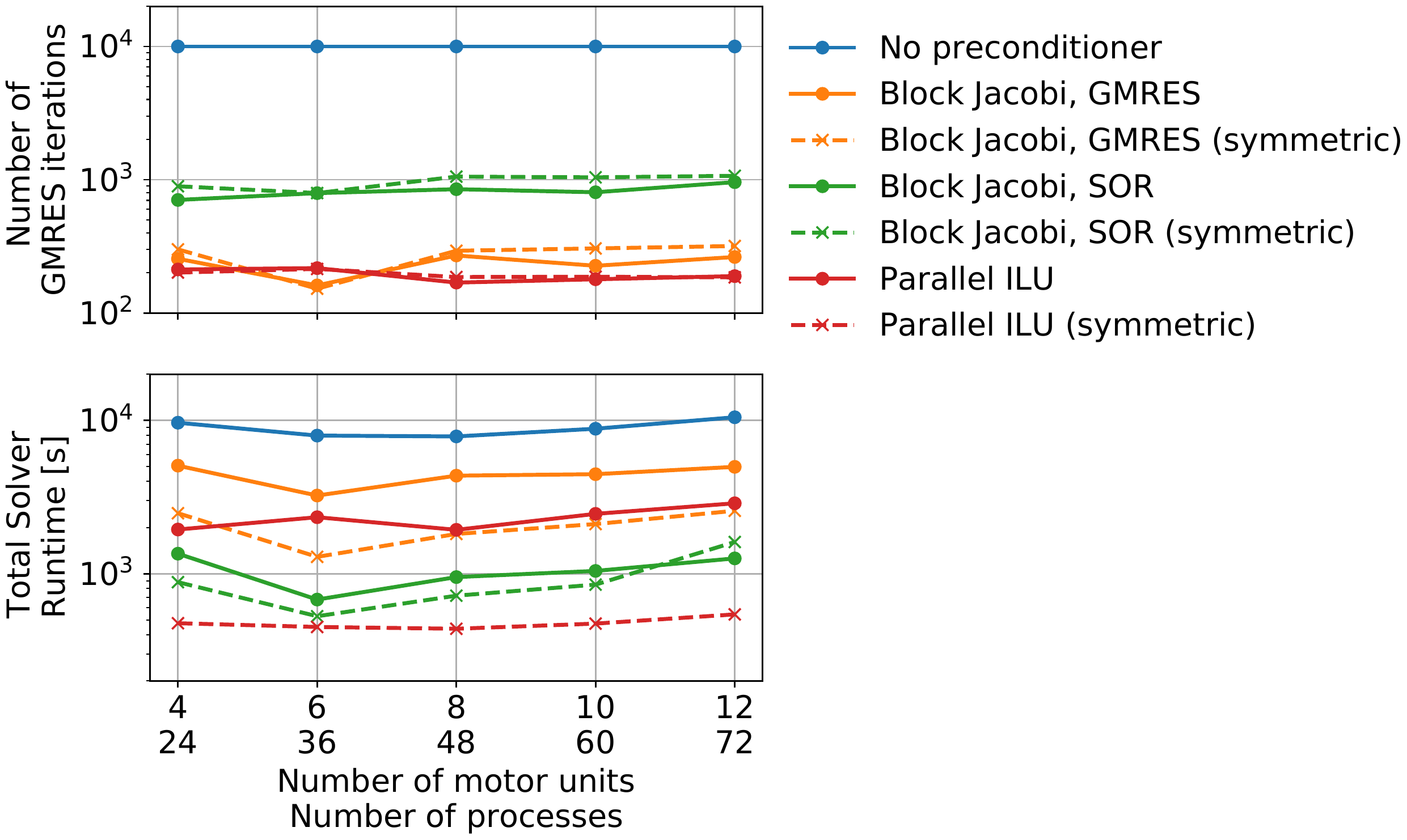}%
  \caption{Evaluation of preconditioners for the multidomain model. The system is solved with a GMRES solver and different preconditioners. The upper plots shows the remaining number of iterations in the GMRES solver, the lower plot measures the total runtime of preconditioning and solution. Three different preconditioners given by orange, green and red lines are compared to the blue reference curve of using no preconditioner.}%
  \label{fig:multidomain_solver}%
\end{figure}

In the next study, we investigate the parallel weak scaling behavior for the multidomain model
with a particular focus on the total runtimes for different choices of the preconditioner that is used in the solution of the linear system of equations. Our goal is to select the fastest solver-preconditioner combination to speed up the computation of the multidomain model.

In contrast to classical weak scaling, we increase the number of MUs with the number of processes, i.e., the size of the blocks in the resulting block structured matrix remains constant, whereas the number of blocks increases.

We simulate the multidomain model with a fat domain as given in \cref{sec:discretization_body_domain} and with the subcellular model of Shorten et al. \cite{Shorten2007}. 
The model is discretized by a Strang operator splitting with Heun's method for the subcellular model and an implicit Euler scheme for the multidomain equations. The timestep widths of all schemes are $\dt_\text{0D}=\dt_\text{multidomain}=\dt_\text{splitting}=\SI{5e-4}{\milli\second}$, and an end time of \SI{1e-1}{\milli\second} is used, corresponding to 200 invocations of the linear solver.

We partition the 3D computational domain, consisting of the muscle mesh with \num{50024} nodes and the fat layer mesh with \num{37000} nodes, to 24, 36, 48, 60 and 72 subdomains. In a weak scaling setup, we correspondingly simulate 4,6,8,10 and 12 MUs.
As the square-shaped system matrix contains one row of blocks for every MU plus one row of blocks for the fat mesh, the total number of rows does not scale exactly linearly with the number of MUs. In consequence, the system matrices in the five scenarios contain \num{279720}, \num{379768}, \num{479816}, \num{579864} and \num{679912} rows and columns. Thus, the problem size per process is only approximately constant in this weak scaling study.

% n dofs
% 24: 279720
% 36: 379768
% 48: 479816
% 60: 579864 
% 72: 679912

We solve the linear system of the multidomain equations by a GMRES solver, because the system matrix is non-symmetric.
The stopping threshold on the residual norm is set to \num{1e-15} and the specified maximum number of iterations is \num{1e4}. 
Different preconditioners are applied and the resulting number of GMRES iterations and the total runtime for preconditioner and solver are measured.
\Cref{fig:multidomain_solver} shows the number of GMRES iterations in the upper plot and the total runtimes in the lower plot.

For every preconditioner, the preconditioning is either performed based on the non-symmetric system matrix (solid lines) or based on a symmetric matrix that is obtained by taking all diagonal blocks of the system matrix (dashed lines), as described in \cref{sec:multidomain_diagonal_matrix}.

The reference measurement is given by the GMRES solver without any preconditioner, visualized by the blue lines in both plots.
The upper plot of \cref{fig:multidomain_solver} indicates the maximum number of \num{1e4} iterations for all measurements of the GMRES solver without preconditioner. This means that the specified tolerance of \num{1e-15} is not reached in the given number of iterations. Thus, a preconditioner is required to obtain an accurate solution.

The first examined preconditioner is the block Jacobi scheme.
A block Jacobi preconditioner divides the system matrix into blocks on the diagonal, yielding smaller problems that can each be solved individually.
The scheme is an iterative solver, which starts with an initial solution $\bfx^{(0)}$ and successively computes  approximations $\bfx^{(i+1)} = \Phi(\bfx^{(i)})$ of the solution until the residual norm reaches the specified threshold.
%The system matrix is approximated by its diagonal.

For a model problem $A\,\bfx = \bfb$, where the system matrix $A=D+L+U$ is decomposed into a matrix $D$ with blocks on the diagonal and lower and upper triangular block matrices $L$ and $U$, respectively, the preconditioning is based on the following iterative computation scheme $\Phi: \bfx^{(i)} \mapsto \bfx^{(i+1)}$:
\begin{align}\label{eq:jacobi_iteration}
  D\,\bfx^{(i+1)} = \bfb - (L+U)\bfx^{(i)}.
\end{align}
Because of the structure of the block diagonal matrix $D$, the system is decoupled and every process can solve its own linear system of equations using the respective diagonal block as system matrix. If the symmetric option for the system matrix used in the preconditioner is chosen, the matrices $L$ and $U$ vanish and the solution of \cref{eq:jacobi_iteration} is trivial.

The preconditioner is constructed using the reordered matrix layout described in \cref{sec:structure_multidomain_system_matrix} and the symmetric matrix is obtained as discussed in \cref{sec:multidomain_diagonal_matrix}. The remaining blocks on the matrix diagonal belong to the subdomains of the parallel partitioning. Each block corresponds to a part of the mesh for all MU compartments.

Two versions of block Jacobi preconditioners provided by the PETSc library are evaluated.
The first variant, shown by the orange lines in \cref{fig:multidomain_solver}, employs a GMRES solver for the resulting smaller linear systems of equations in \cref{eq:jacobi_iteration}.
The second variant, shown by the green lines in \cref{fig:multidomain_solver}, uses a SOR (successive over-relaxation) solver with over-relaxation parameter 
$\omega=1$, i.e., a Gauß-Seidel scheme.

The upper plot in \cref{fig:multidomain_solver} shows that the number of GMRES solver iterations is reduced more for the GMRES solver than for the Gauß-Seidel solver, as the constant number of GMRES iterations in the preconditioner yields a better approximation to the solution than the same number of Gauß-Seidel iterations. The lower plot shows a smaller total runtime for the block Jacobi scheme with Gauß-Seidel solver than for the block Jacobi scheme with GMRES solver. This means that the lower runtime of the Gauß-Seidel solver in the preconditioner outweighs the larger number of GMRES iterations in the solver, compared to the GMRES preconditioning scheme. For both preconditioners, the total runtime for the variant with the symmetric matrix is lower than for the variant with the full system matrix.

Another solver is \emph{Euclid} \cite{euclid} from the HYPRE package, shown by the red lines in \cref{fig:multidomain_solver}. It is a parallel implementation of incomplete LU factorization using graph partitioning and a two-level ordering strategy.

The plots in \cref{fig:multidomain_solver} show a low number of remaining GMRES iterations after the preconditioner has been applied, similar to the block Jacobi scheme with GMRES solver. However, the total runtime of Euclid is significantly lower than the runtime of the GMRES-block Jacobi scheme. If the symmetric system matrix is used for preconditioning, the total runtime is the lowest of all measured preconditioner combinations.

Regarding the parallel weak scaling, the lower plot in \cref{fig:multidomain_solver} shows overall good scaling properties for all considered preconditioners. The runtimes slightly decrease from the first to the second data points, as the system matrix size per process also slightly decreases. Then, a trend of slightly increasing runtimes in the weak scaling setup can be seen, which indicates that the preconditioners and the GMRES solver perform slightly more computations and communication for larger problem sizes. The accuracy of the preconditioning step is not affected by the overall problem size, as can be seen by the constant numbers of GMRES iterations in the upper plot of \cref{fig:multidomain_solver}.

As a result, we use the combination of Euclid preconditioner and GMRES solver in all solutions of the multidomain model in this work, because this is the fastest of the tested combinations. Apart from the three presented preconditioners, more available choices in the software packages PETSc and HYPRE were tested, but yielded worse performance. These include the \emph{Parallel Incomplete Factorization preconditioner} (PILUT) from the HYPRE package and the combinations of the block Jacobi scheme with an algebraic multigrid method or the Euclid preconditioner for the subproblems. Tests with the parallel algebraic multigrid method \emph{BoomerAMG} from the HYPRE package also showed promising results with even lower total runtimes than the Euclid preconditioner, but suffered from occasional long runtimes and divergence in a non-deterministic fashion. However, the full set of possible parameters such as different coarsening and interpolation options and settings for the smoother have not yet been evaluated and, after fine-tuning, corresponding performance improvements could be possible in future work.

\begin{reproduce_no_break}
  The script for this study is available in the repository at \href{https://github.com/dihu-stuttgart/performance}{github.com/dihu-stuttgart/} \href{https://github.com/dihu-stuttgart/performance}{performance} in the directory \code{opendihu/07_multidomain_solver}:
  \begin{lstlisting}[columns=fullflexible,breaklines=true,postbreak=\mbox{\textcolor{gray}{$\hookrightarrow$}\space}]
    ./run_mu.sh
  \end{lstlisting}
\end{reproduce_no_break}

\chapter{Conclusion and Future Work}\label{sec:conclusion_and_future_work}
To conclude this work, we summarize the presented models, algorithms, implementations, studies and the main findings. Moreover, we give an outlook on future work and additional research questions that can be approached building on our work.

\section{Summary of this Work}

The overarching goal of this work was to enable simulations of the neuromuscular system using detailed, biophysical multi-scale models with high resolutions. The simulations should compute numerically accurate results, run efficiently on various hardware and allow parallel scaling to large problem sizes, which should be solved on supercomputers.

As a result, this work established a computational framework for multi-scale modeling of skeletal muscles, their neural activation, muscle contraction and generation of EMG signals on the skin surface. Our approach combined existing models for various parts of the neuromuscular system into a comprehensive multi-scale model framework. Scalability and parallel efficiency of our software were ensured by efficient algorithms, suitable, parallelized numerical schemes and by our accompanying performance analyses.

% ieser Arbeit war ... biophysikalisch, hoch aufgelöst, schnell, skalierbar, genau ...

We described the following topics in this work: After the introduction in \cref{chap:introduction}, we compared two modeling approaches to describe the movement of the upper arm in \cref{chap:comparative_study}. Based on data of experimental trials we conducted during a graduate school workshop, we developed a first, data-driven model using Gaussian process regression and a second model based on a biophysical simulation with two muscle models. The parameters for the biophysical simulation were fitted to experimental training data using numerical optimization. The comparison of the two approaches revealed a slightly better fit for the biophysical simulation model. This approach had the additional benefit of giving biophysical insights into the functioning of the system and provided estimates for subject-specific muscle parameters. While this study used Hill-type muscle models, which describe muscle forces on a 1D line of action, we considered more accurate multi-scale models in the remainder of this work.

\Cref{sec:generation_of_meshes_for_multiscale} dealt with the generation of structured 3D meshes and embedded 1D meshes for muscle fibers. The approach of only using structured meshes, which allowed for a simple domain decomposition proved to be beneficial for the parallel performance of our simulations.  
We described a workflow, how to obtain these meshes from biomedical imaging data. We developed a serial algorithm and a parallel algorithm to construct the required meshes and to ensure a good mesh quality, even for meshes with high resolutions. The algorithms were based on our novel approach of using harmonic maps to transform reference meshes to cross-sectional slices of the muscle mesh.

In \cref{sec:muscle_fibers_and_motor_units}, we described ways to associate muscle fibers with motor units (MUs) in a physiological manner. We developed efficient algorithms for this task for different premises, and employed the algorithms to associate up to \num{270000} muscle fibers to \num{100} MUs for the subsequent use in our simulations.

In \cref{chap:models_and_discretization}, we first described all equations of the state-of-the models that we used, and how they can be combined into a multi-scale description. Then, we described their discretization using the finite element method for the spatial derivative terms and various timestepping and operator splitting schemes for the temporal derivatives. One original contribution is the derivation of the finite element formulation for the multidomain equation. Further, we gave a detailed description of the nonlinear solid mechanics discretization, which we used in our implementation.

Next, we presented details on our simulation software OpenDiHu, which we used to solve various combinations of the described multi-scale model framework to simulate the neuromuscular system. \Cref{chap:usage} gave an introduction to the design and usage of the software and demonstrated its application using various example problems.

\Cref{sec:implementation} described the implementation of OpenDiHu in more detail, motivated various design decisions, introduced the data handling and several algorithms, e.g., to construct a parallel domain decomposition or to map data between meshes, and described the implementation of various solvers for particular parts of the multi-scale model.

\Cref{sec:results} presented numerical results, which were obtained using our simulation software. We simulated the passive mechanical behavior of muscle tissue, subcellular models given in CellML description, electrophysiology on muscle fibers, electric conduction in the muscle and the adipose tissue to obtain surface EMG signals, electrophysiology using the 3D homogenized multidomain description, and coupled scenarios of electrophysiology and muscle contraction. We discussed effects of model and structural parameters and interpreted the obtained simulation results.

In \cref{sec:performance_analysis}, we analyzed the computational performance of our software in general and various solvers in particular. We conducted numerical studies of universal convergence properties with the software OpenCMISS, which also helped to parameterize the numerical solvers in OpenDiHu. Further studies on mesh widths and used linear solvers were carried out directly using OpenDiHu. We evaluated various optimization options in OpenDiHu and compared the most optimized settings in OpenDiHu with the baseline solver OpenCMISS, yielding a high speedup of more than two orders of magnitude. Moreover, we investigated the computational performance of our models on the GPU, and conducted parallel strong scaling and parallel weak scaling tests on small clusters and the supercomputers at the High Performance Computing Center Stuttgart.

\section{Summary of Main Findings}

The present work simulated numerous scenarios with various model combinations, which provided different insights. In the following, we summarize the observed findings. We address the biophysical observations in \cref{sec:observations_biophysics} and results of the performance measurements in \cref{sec:observations_performance}.

\subsection{Observations from the Fields of Biophysics and Biomechanics}\label{sec:observations_biophysics}
The comparison of the linear and nonlinear mechanics models in \cref{sec:solver_solid_mechanics} showed qualitatively different results and demonstrated that the accurate behavior of deforming muscle tissue can only be described by a proper nonlinear anisotropic solid mechanics model. 

Initially, an open question was also how to relate the accuracy of the simulated EMG signals to the number of fibers and the mesh resolution. Our numerical studies in \cref{sec:action_potential_velocity}, which compared the resulting action propagation velocity for different mesh widths of the 1D muscle fiber meshes showed that a mesh width of \SI{100}{\micro\meter} or 100 elements per \SI{}{\centi\meter} gives reasonably accurate results. 

To evaluate the 3D mesh width and the spacing between the muscle fibers, we conducted simulations with different 3D mesh resolutions and numbers of fibers in \cref{sec:effects_of_the_mesh_width_emg}. The number of fibers was scaled up to the realistic number of \num{270000} fibers in a biceps brachii muscle. We concluded that the most accurate solution is obtained for a mesh width as fine as possible, as the EMG results were qualitatively different for every refinement step. This emphasizes the need for highly resolved simulation scenarios (representing the real number of fibers in a muscle accurately) for realistic EMG computations and, as a result, the need for High Performance Computing techniques.

However, if the EMG is to be sampled by electrodes, i.e., if the EMG recording process should also be part of the simulation, lower mesh widths might be possible, as the EMG is only captured at the locations of the electrodes.

One possible approach to reduce the computational effort for EMG simulations would be to only consider the muscle tissue down to a certain depth below the surface with the EMG electrodes. We observed in \cref{sec:simfiber_mu}, that the EMG signal is highly influenced by MUs, whose territories are located close to the electrodes. However, our numerical experiments with EMG decomposition algorithms in \cref{sec:simfiber_decomposition} showed that large MUs located opposite to the EMG electrodes at the deepest muscle tissue layers are detectable in the surface EMG signals. Thus, neglecting the deeper parts of the muscle would remove relevant information from the system and is, therefore, not a valid approach to reduce the computational load.

Furthermore, the layer of adipose tissue on top of the muscle showed a smoothing effect on EMG recordings in our simulations, both with the fiber based approach in \cref{sec:simfiber_fat} and with the multidomain approach in \cref{sec:multidomain_components,sec:multidomain_simulation_emg}. One advantage of our simulations compared to experimental studies is that the thickness of the fat layer is known exactly and can also be adjusted.

Simulations of muscle contraction with coupled electrophysiology and solid mechanics models showed a spatially inhomogeneous contraction  for the biceps muscle while the muscle activation is ramped up. The simulation in \cref{sec:fiber_based_contraction} of an isolated, contracting muscle belly without tendons showed transverse bending, alternating between the left and right-hand sides, as a result of the subsequently activated MUs at the different sides of the muscle. We also simulated the biceps brachii muscle together with its tendons and observed a ripple in the generated muscle force, which is caused by the same inhomogeneous MU activity.

The simulations of muscle contraction also showed that, if the muscle is initially in a stress-free state, the model can only achieve a maximum contraction of approximately \SI{85}{\percent}. However, the muscles of the musculoskeletal system are known to exhibit prestresses in their relaxed states. Accordingly, we added prestress to our simulations. The amount of prestress is adjustable in the simulation settings, and the required amount can be determined by a comparison with experimental studies.

% linear-nonlin
% accuracy -> high mesh resolution
% fibers: effects of fat mesh, distance between fibers to surface, size of MUs (EMG decomposition)
% multidomain: more smoothed
% coupled solid mechanics: inhomogeneous contraction, prestretch required (without only 85% contraction)

\section{Summary of Performance Results}\label{sec:observations_performance}

A major part of the work was also concerned with improving the performance of the simulation software, and, thus, enabling larger simulation scenarios in shorter runtimes.

Previously, literature on biophysical, multi-scale models of skeletal muscles was mainly focused on modelling and interpretation of the results, rather than targeting efficient computations. The work of Röhrle et al. \cite{Roehrle2012} introduced the multi-scale model, which we based our work on, and simulated the tibialis anterior muscle using a 3D mechanics mesh with 12 elements. The work of Heidlauf et al. \cite{Heidlauf2013} considered the same geometry and simulated 400 muscle fibers. The authors parallelized their OpenCMISS based implementation for a fixed number of four processes. We built upon this work with the goal to push the limits of feasible problem sizes, and, in \cref{sec:effects_of_the_mesh_width_emg}, executed our optimized simulation with \num{26912} processes, \num{273529} muscle fibers and a 3D mesh for the electrophysiology model with approximately \num{1e8} degrees of freedom.

The performance analyzes in \cref{sec:performance_analysis} showed that the subcellular model contributes a large portion to the total runtime and, thus, is the most crucial part to optimize. By using proper memory layouts, vectorization is possible. Our approach of using explicit vector instructions outperformed the auto-vectorization capabilities of the compiler. The approximation of the exponential function and an improved parallelization scheme for the 1D electric conduction problem additionally contributed to a high speedup. The comparison to the baseline solver OpenCMISS Iron in a strong scaling study in \cref{sec:strong_scaling_runtimes_opencmiss_opendihu} revealed a maximum speedup of 363 for the purely implementation-specific improvements and an additional speedup of 2.5, shown in \cref{sec:opencmiss_numeric_improvements}, by using more efficient numerical methods.

In addition, the memory characteristics of the solvers were investigated in \cref{sec:strong_scaling_runtimes_opencmiss_opendihu}. 
The linear increase in memory consumption of the baseline solver in a weak scaling setting was improved to a nearly constant scaling. Our analysis using a roofline performance model showed that our solvers are compute bound and achieve a computational performance of approximately \SI{25}{\percent} peak performance, which is a very good value.

Moreover, hybrid shared/distributed memory parallelism and computations on the GPU were investigated, but both approaches were found to be not competitive with our highly optimized distributed memory parallelization. For the GPU, potentially more efficient approaches than our approach using OpenMP exist, such that a performance improvement in the future could be possible.

The modularity of the CellML infrastructure, where computational models can be shared among researchers and are interchangeable in multi-scale simulations was preserved during all optimization endeavors. Our approach was to implement a source-to-source code generator, which transformed the given CellML code into optimized code for the CPU or the GPU.

For the solution of the multidomain model, we evaluated various preconditioners and selected the most performant preconditioner-solver combination for our computations.
One previously unforeseen result is the large discrepancy of required runtime between the fiber based and the multidomain based electrophysiology models, presented in \cref{sec:solver_multidomain_model}. We measured by a factor of 1000 longer computation times for the multidomain model, which result from the structure of the model. Despite the high computational effort, the multidomain model is useful in practice as it can simulate effects that are not captured by the fiber based model. We gave a detailed comparison between both approaches in \cref{sec:multidomain_differences}.

In summary, we provided a computationally efficient and scalable tool for applied biophysics researchers to solve problems in the domains of EMG generation and muscle contraction. For example, the effect of different muscle fiber organizations and MU recruitment strategies can be tested with our software. We demonstrated its use with state-of-the-art EMG decomposition algorithms, which provide the bridge to the experimental domain.
Thus, we hope to contribute one step on the pathway of complementing in vivo with in silico experiments to increase the understanding of the neuromuscular system.

% provide a tool for applied biophysics researchers
% --------------------
% complement in-vivo and in-silico experiments
% test different muscle fiber organizations -> possible
% decomposition of EMG -> tested

% vc better than auto-vec, AVX-512, memory layout
% comparison to OpenCMISS
% preserve CellML modularity -> code generator
% GPU, OpenMP
% proper choice of solvers
% multidomain performance vs fibers

% performance
% --------------
% Röhrle2012: TA 12 elements, MU association
% Heidlauf2013: 400 fibers, TA, OpenCMISS, mechanics
% made new insights possible  

%wir hatten ja mal vorgenommen dass
%tatsächlich ist gelungen:
%überraschenderweise, dass: 

%bewerten mit den Ergebnissen
%was bewahrheitet, 

%unterschiede zwischen grob und hochaufgelöster Sim
%anzahl realitäts anzahl Muskelfaser anzahl

\section{Outlook and Future Work}\label{sec:future_work}
 
The presented work could be extended in multiple directions, spanning performance improvements and model extensions.

First, some ideas for further performance improvements could be implemented and evaluated. 
The monodomain equation could be solved with implicit-explicit (IMEX) schemes, which could potentially achieve higher precision. 
The numerically stiff subcellular model is currently solved explicitly. Implicit schemes could be developed, and the implicit iteration equations could be solved symbolically in a preprocessing step using the parsed CellML code. 

To improve the performance of the multidomain model, the following algorithmic improvements are promising options. 
The 3D problems of action potential propagation in the muscle volume for every compartment could be restricted to the subset of nodes, where the occupancy factors are above a certain threshold, effectively reducing the problem sizes, and reducing the effect of higher MU counts on the runtime. However, this would bring difficulties to ensure a balanced parallel domain decomposition.
Instead of the current parallel partitioning of the domain, the multidomain model could also be parallelized by distributing the MUs to different processes or by a combination of both approaches.

On the numerical side, an extended error analysis could be carried out for all model parts, and the timestep widths, which are currently chosen conservatively, could potentially be increased, while keeping the numerical error below a given threshold. Error estimators could be developed, which would allow an adaptive adjustment of the timestep widths.
The 3D model solvers for the 3D electrophysiology and multidomain problems could be enhanced with geometric or algebraic multigrid preconditioners.

Since all subcellular points in a muscle are usually in similar states at any time, a hybrid approach using analytic descriptions of action potential propagation, as in \cref{sec:sim_rosenfalck}, and a fully numerical treatment could be chosen, and surrogate models could be adaptively added to the computational description.

On the technical side, computations on the GPU could be re-evaluated in the future using the existing OpenMP approach with more mature compiler versions or different accelerator targeting programming technologies.

Second, the range of simulated models could be extended. The simulations could be applied to further muscle geometries such as the triceps brachii or the tibialis anterior muscles. Muscles with more complex geometries and fiber arrangements could be investigated. 
A mechanically coupled problem of agonist-antagonist pair could be considered such as a system of biceps and triceps brachii. Apart from the mechanical coupling, a coupling of the neural recruitment involving sensory organs in the muscles could be implemented and used to approach further biomechanical research questions. Such a neuromuscular feedback loop could also be investigated first for a single muscle, e.g., by extending the preliminary implementation in OpenDiHu for the biceps muscle.

Pathological conditions could be simulated to understand muscular diseases and neuromuscular electrical stimulation of the muscle for stroke rehabilitation could be considered.

By using the preCICE adapters in OpenDiHu, more advanced mechanics solvers could be coupled to an electrophysiology simulation in OpenDiHu, allowing to, e.g., study mechanical effects of surrounding tissue.

On a larger scale, the interplay of more organs could be taken into account. Blood perfusion and muscle metabolism could be added, and coupled by models of the lung and general metabolism in the organism. Thus, a digital human model can be envisioned, which allows to study the effects of anomalies and to develop new therapies, effectively utilizing simulation technology for human wellbeing.

%reduced to 1D problems using appropriate transformations.

% performance improvements
% ----------------------------
% more timestepping methods: CVODE (https://computing.llnl.gov/projects/sundials/cvode), imex
% different parallelisation where not all ranks have to be involved (for multidomain) -> this feature already exists for the fibers with multipleInstances
% more numeric tests on exp function? no
% multidomain: compute 3D problem as 1D problem, or adaptive computation of the parts of the fr factors
% GPU

% extensions
% -------------
% other muscles, more muscles,
% couple more models:
% neuromuscular feedback loop with sensors, 
% 

% bibliography
\printbibliography[%
  % change title
  title=Bibliography,
  % use custom heading
  heading=myheading,
  % print custom note
  postnote=mypostnote,
]

\cleardoublepage

\end{document}